\pdfoutput=1
\documentclass[11pt,a4paper,oneside]{book}

\usepackage[T1]{fontenc}
\usepackage[utf8]{inputenc}
\usepackage{tgpagella} 
\usepackage[a4paper,left=2.50cm,right=2.50cm,top=2.90cm,bottom=2.70cm]{geometry}
\usepackage{setspace}
\usepackage{graphicx}
\usepackage{mathrsfs}
\usepackage{mathalfa}
\usepackage{euscript}
\usepackage{amsmath}
\usepackage[T1]{fontenc}
\usepackage{calligra}
\usepackage{graphicx}

\usepackage{caption}
\usepackage{calligra}
\usepackage{afterpage}
\usepackage{subcaption}
\usepackage{xcolor}
\usepackage{tikz}
\usepackage{array}
\usepackage{tabularx}
\usepackage{booktabs}
\usepackage{titlesec}
\usepackage{tocloft}
\usepackage{enumitem}
\usepackage{inconsolata}
\usepackage{mathtools}
\usepackage{amsmath,amssymb,amsfonts}
\usepackage{mathrsfs}
\usepackage{dsfont}
\usepackage{bbm}
\usepackage{pifont}
\usepackage{multicol}
\usepackage[version=4]{mhchem}
\usepackage{multirow}
\usepackage[customcolors,shade]{hf-tikz}
\usepackage{feynman}
\usepackage[most]{tcolorbox}
\usepackage{longtable}
\usepackage{stmaryrd}
\usepackage{slashed}
\usepackage{bigints}
\usepackage{cancel}
\usepackage{blindtext}
\usepackage{changepage}
\usepackage{lscape}
\usepackage{stackengine}
\usepackage{scalerel}
\usepackage[numbers,sort&compress]{natbib}
\usepackage{etoolbox}
\usepackage{listings}
\usepackage{microtype}
\lstdefinelanguage{Mathematica}{
  morekeywords={
    Table,Flatten,DeleteDuplicates,Cases,Infinity,Reverse,SortBy,Abs,Print
  },
  sensitive=true,
  morecomment=[s]{(*}{*)},
  morestring=[b]"
}

\definecolor{brandburgundy}{RGB}{99,0,60}   
\definecolor{tocred}{RGB}{155,0,20}         
\definecolor{logored}{RGB}{180,31,33}       
\definecolor{paperbg}{RGB}{226,214,182}     
\definecolor{lofblue}{RGB}{155,0,20}        
\colorlet{blue}{black}

\usepackage[
  plainpages=false,
  pdfpagelabels=true,
  hypertexnames=false,
  colorlinks=true,
  linktoc=all,
  allcolors=tocred,
  linkcolor=tocred,
  citecolor=tocred,
  urlcolor=tocred
]{hyperref}
\usepackage{cleveref}

\newcommand{\doilink}[1]{\href{https://doi.org/#1}{\texttt{#1}}}
\newcommand{\arxivlink}[2]{\href{https://arxiv.org/abs/#1}{\texttt{#1 [#2]}}}
\ExplSyntaxOn
\cs_new_protected:Npn \thesis_print_doi:n #1
  {
    \tl_set:Nn \l_tmpa_tl {#1}
    \regex_replace_once:nnN { \A \s* (https?://(dx\.)?doi\.org/)? } {} \l_tmpa_tl
    \regex_replace_all:nnN { \s+ } {} \l_tmpa_tl
    \tl_set_eq:NN \l_tmpb_tl \l_tmpa_tl
    \tl_replace_all:Nnn \l_tmpb_tl {(}{\c_percent_str 28}
    \tl_replace_all:Nnn \l_tmpb_tl {)}{\c_percent_str 29}
    \tl_replace_all:Nnn \l_tmpb_tl {_}{\c_percent_str 5F}
    \tl_set:Nx \l_tmpc_tl { https://doi.org/\tl_to_str:N \l_tmpb_tl }
    \href{\l_tmpc_tl}{\nolinkurl{\tl_to_str:N \l_tmpa_tl}}
  }
\NewDocumentCommand{\thesisdoilink}{m}{\thesis_print_doi:n{#1}}
\ExplSyntaxOff
\providecommand{\doi}[1]{}
\renewcommand{\doi}[1]{doi:\,\thesisdoilink{#1}}
\graphicspath{
  {./}
}
\makeatletter
\def\input@path{
  {./}
}
\makeatother

\makeatletter
\newcommand{\thesisstatictableofcontents}{%
  \begingroup
  \def\@starttoc##1{\makeatletter\contentsline {chapter}{Abstract}{iv}{chapter*.1}%
\contentsline {chapter}{Author's Declaration}{vi}{chapter*.2}%
\contentsline {chapter}{Acknowledgments}{viii}{chapter*.15}%
\contentsline {chapter}{\numberline {1}Introduction}{1}{chapter.16}%
\contentsline {section}{\numberline {1.1}Context and Motivation}{1}{section.20}%
\contentsline {section}{\numberline {1.2}Quantum Field Theory and its Classical Limit}{3}{section.23}%
\contentsline {paragraph}{Magnus expansion and the eikonal generator.}{8}{paragraph*.45}%
\contentsline {section}{\numberline {1.3}Celestial Amplitudes and Eikonalization}{12}{section.64}%
\contentsline {section}{\numberline {1.4}Computational Techniques and Methodology}{15}{section.73}%
\contentsline {section}{\numberline {1.5}Thesis in a nutshell}{22}{section.117}%
\contentsline {chapter}{\numberline {2}WEFT of Black-Hole Binaries with Dark Photon and Axion}{24}{chapter.118}%
\contentsline {section}{\numberline {2.1}Introduction}{24}{section.119}%
\contentsline {section}{\numberline {2.2}Brief review of machinery of EFTs}{28}{section.124}%
\contentsline {section}{\numberline {2.3}Review of worldline effective field theory for a point particle in GR}{30}{section.134}%
\contentsline {section}{\numberline {2.4}Conservative dynamics of binary black holes in a theory of ALP and ULV }{33}{section.147}%
\contentsline {subsection}{\numberline {2.4.1}The Gravitational bound sector}{37}{subsection.162}%
\contentsline {subsection}{\numberline {2.4.2}The Electromagnetic (EM) bound sector}{38}{subsection.169}%
\contentsline {subsection}{\numberline {2.4.3}The Proca bound sector}{39}{subsection.176}%
\contentsline {subsection}{\numberline {2.4.4}The EM-Axion bound sector}{40}{subsection.184}%
\contentsline {subsection}{\numberline {2.4.5}The pure scalar sector}{43}{subsection.194}%
\contentsline {subsection}{\numberline {2.4.6}Proca-electromagnetic interaction sector}{48}{subsection.211}%
\contentsline {section}{\numberline {2.5}The radiative dynamics}{51}{section.221}%
\contentsline {subsection}{\numberline {2.5.1}Radiative sector PN counting}{52}{subsection.229}%
\contentsline {subsection}{\numberline {2.5.2}Scalar interaction and scalar radiation}{53}{subsection.231}%
\contentsline {subsection}{\numberline {2.5.3}Electromagnetic interaction and the multipole decomposition}{60}{subsection.271}%
\contentsline {subsection}{\numberline {2.5.4}Proca interaction and multipole decomposition}{63}{subsection.283}%
\contentsline {subsection}{\numberline {2.5.5}Gravitational radiation and Multipole decomposition}{65}{subsection.293}%
\contentsline {section}{\numberline {2.6}Discussions and outlooks}{76}{section.332}%
\contentsline {section}{\numberline {2.A}Details of computation used in Section~(\ref {Sec4.2})}{77}{section.333}%
\contentsline {section}{\numberline {2.B} Details of the Feynman integral computation used in Section~(\ref {Sec4.3})}{79}{section.342}%
\contentsline {section}{\numberline {2.C} Details of computation of Feynman integral used in (\ref {Sec4}) and (\ref {Sec5})}{82}{section.359}%
\contentsline {section}{\numberline {2.D}Comments on middle vertex contribution to the radiation}{83}{section.366}%
\contentsline {paragraph}{Two-loop Fourier family.}{87}{paragraph*.387}%
\contentsline {paragraph}{\texttt {hard-hard} region.}{89}{paragraph*.405}%
\contentsline {paragraph}{\texttt {hard-soft} and \texttt {soft-hard} regions.}{89}{paragraph*.408}%
\contentsline {paragraph}{\texttt {soft-soft} region.}{90}{paragraph*.412}%
\contentsline {section}{\numberline {2.E}Useful Feynman integrals}{91}{section.425}%
\contentsline {chapter}{\numberline {3}Classical Black-Hole Scattering in Scalar-Tensor Gravity from WQFT}{93}{chapter.432}%
\contentsline {section}{\numberline {3.1}Introduction}{93}{section.433}%
\contentsline {section}{\numberline {3.2}A quick tour to worldline quantum field theory}{96}{section.436}%
\contentsline {section}{\numberline {3.3}Derivation of the Feynman rules}{98}{section.447}%
\contentsline {section}{\numberline {3.4}Computation of impulse}{101}{section.470}%
\contentsline {subsection}{\numberline {3.4.1}1PM contribution to impulse}{101}{subsection.472}%
\contentsline {subsection}{\numberline {3.4.2}2PM contribution to the impulse}{103}{subsection.488}%
\contentsline {section}{\numberline {3.5}Computation of waveform}{120}{section.568}%
\contentsline {subsection}{\numberline {3.5.1}Scalar waveform}{121}{subsection.575}%
\contentsline {subsubsection}{Scalar waveform at 1PM}{122}{subsubsection*.576}%
\contentsline {subsubsection}{Scalar waveform at 2PM}{122}{subsubsection*.582}%
\contentsline {subsection}{\numberline {3.5.2}Gravitational waveform at 2PM: contribution due to extra scalar DOF}{134}{subsection.648}%
\contentsline {section}{\numberline {3.6}Towards massive waveform}{135}{section.656}%
\contentsline {subsection}{\numberline {3.6.1}Stationary Phase (SP) approximation: the need and general procedure}{136}{subsection.663}%
\contentsline {subsection}{\numberline {3.6.2}Dealing with the massive integrals via SP approximation}{137}{subsection.675}%
\contentsline {section}{\numberline {3.7}Discussions and outlook}{143}{section.714}%
\contentsline {section}{\numberline {3.A}Sketching the derivation of the worldline action}{145}{section.715}%
\contentsline {section}{\numberline {3.B}Impulse via Eikonal: a connection to scattering amplitude}{146}{section.721}%
\contentsline {section}{\numberline {3.C}Two master integrals used for the computation of waveform}{151}{section.749}%
\contentsline {section}{\numberline {3.D}Analysis through method of regions for $\lambda _4\varphi ^4$ vertex contribution in the waveform}{155}{section.773}%
\contentsline {section}{\numberline {3.E}Massive integral coming from derivative interaction}{156}{section.778}%
\contentsline {section}{\numberline {3.F}Computation of worldline radiation diagram using large velocity approximation}{159}{section.798}%
\contentsline {section}{\numberline {3.G}Proof of the identity in \eqref {3.54kk}}{161}{section.808}%
\contentsline {section}{\numberline {3.H}Comment on the integral convergence}{162}{section.815}%
\contentsline {section}{\numberline {3.I}Comments on the algebraic properties of the Bessel curve}{166}{section.843}%
\contentsline {paragraph}{One-loop integrals, the Bessel curve, and topological recursion.}{166}{paragraph*.844}%
\contentsline {chapter}{\numberline {4}Spinning Two-Body Problem in dCS Gravity Using WQFT}{169}{chapter.865}%
\contentsline {section}{\numberline {4.1}Introduction}{169}{section.866}%
\contentsline {section}{\numberline {4.2}Chern-Simons gravity and worldline QFT}{171}{section.868}%
\contentsline {section}{\numberline {4.3}Bootstrapping Post-Minkowskian (PM) physics from Post-Newtonian (PN) data: the idea, technology and example}{176}{section.896}%
\contentsline {section}{\numberline {4.4}Computation of eikonal phase}{184}{section.937}%
\contentsline {section}{\numberline {4.5}Conclusion and outlook}{215}{section.1069}%
\contentsline {section}{\numberline {4.A}Useful Feynman integrals}{217}{section.1077}%
\contentsline {section}{\numberline {4.B}Boundary integrals in potential mode relevant for solving the differential equation}{219}{section.1088}%
\contentsline {section}{\numberline {4.C}$A(x,\epsilon )$ in $\epsilon $ factorized form}{221}{section.1099}%
\contentsline {section}{\numberline {4.D}Iterative UT integrals}{222}{section.1101}%
\contentsline {section}{\numberline {4.E}$\boldsymbol {\tilde {h}}$ coefficients}{223}{section.1103}%
\contentsline {chapter}{\numberline {5}Heterotic Footprints in Classical Gravity}{232}{chapter.1104}%
\contentsline {section}{\numberline {5.1}Introduction}{232}{section.1105}%
\contentsline {section}{\numberline {5.2}Setup and methodology}{234}{section.1106}%
\contentsline {section}{\numberline {5.3}Computation of scattering amplitude: Soft-loop expansion and classical limit}{236}{section.1120}%
\contentsline {section}{\numberline {5.4}Post-Minkowskian potential: The Lippmann-Schwinger equation and the Infrared subtraction}{249}{section.1176}%
\contentsline {paragraph}{A digression on Green's functions.}{250}{paragraph*.1183}%
\contentsline {section}{\numberline {5.5}Eikonal exponentiation and the scattering angle}{255}{section.1212}%
\contentsline {section}{\numberline {5.6}Conclusions and Discussion}{258}{section.1235}%
\contentsline {chapter}{\numberline {6}Celestial Amplitude, Shadow, and OPE in Quadratic Gravity}{260}{chapter.1236}%
\contentsline {section}{\numberline {6.1}Introduction}{260}{section.1237}%
\contentsline {section}{\numberline {6.2} Celestial eikonal amplitude for quadratic gravity}{262}{section.1238}%
\contentsline {subsection}{\numberline {6.2.1}Eikonal amplitude in GR}{266}{subsection.1257}%
\contentsline {subsection}{\numberline {6.2.2}Eikonal amplitude in quadratic gravity}{267}{subsection.1261}%
\contentsline {subsection}{\numberline {6.2.3}Analyticity and dispersion relation for celestial eikonal amplitude}{271}{subsection.1276}%
\contentsline {subsection}{\numberline {6.2.4}Dispersion Relations}{274}{subsection.1294}%
\contentsline {section}{\numberline {6.3}Shadowed correlator and operator product expansion (OPE)}{276}{section.1304}%
\contentsline {subsection}{\numberline {6.3.1}Shadowed amplitude corresponding to eikonal amplitude }{277}{subsection.1307}%
\contentsline {subsection}{\numberline {6.3.2}Shadowed amplitude corresponding to celestial Born amplitude}{278}{subsection.1312}%
\contentsline {subsection}{\numberline {6.3.3}Conformal block expansion and partial wave coefficients}{281}{subsection.1330}%
\contentsline {section}{\numberline {6.4}Connection to Carrollian amplitude}{288}{section.1359}%
\contentsline {section}{\numberline {6.5}Conclusion and Discussion}{289}{section.1364}%
\contentsline {section}{\numberline {6.A}Few definitions and conventions}{290}{section.1368}%
\contentsline {chapter}{\numberline {7}Conclusions}{293}{chapter.1378}%
\contentsline {chapter}{References}{296}{chapter.1378}%
\makeatother}%
  \tableofcontents
  \endgroup
}
\newcommand{\thesisstaticlistoffigures}{%
  \begingroup
  \def\@starttoc##1{\makeatletter\addvspace {10\p@ }
\contentsline {figure}{\numberline {1.1}{\ignorespaces Schematic bubble notation for nested Poisson brackets. Popping a bubble produces oriented causal cuts according to the Leibniz rule.}}{9}{figure.caption.47}%
\contentsline {figure}{\numberline {1.2}{\ignorespaces Schematic graphical content of Murua's recursion. The displayed rooted tree has two allowed partitions: \(p_1\), containing only the root-adjacent edges, and \(p_2\), containing all three edges. In general, rooted partitions \(p\in {\rm Pr}(\tau )\) must contain all root-adjacent edges; for non-rooted trees, \(p\in {\rm P}_{s}(\tau )\) must contain all edges adjacent to the chosen semi-root \(s\).}}{11}{figure.caption.57}%
\contentsline {figure}{\numberline {1.3}{\ignorespaces Causality-prescription diagrams for the IR-sensitive master integral. The Magnus/Murua weights assigned to the two oriented orderings cancel the leading \(1/\epsilon ^2\) pole.}}{11}{figure.caption.61}%
\contentsline {figure}{\numberline {1.4}{\ignorespaces Celestial description of a four-dimensional scattering process. A momentum-space \(n\to m\) scattering amplitude in asymptotically flat spacetime is mapped, through a Mellin transform over the external energies, to a conformal correlator on the celestial sphere.}}{14}{figure.caption.70}%
\contentsline {figure}{\numberline {1.5}{\ignorespaces Geometry and contour of nested integrals}}{18}{figure.caption.106}%
\addvspace {10\p@ }
\contentsline {figure}{\numberline {2.1}{\ignorespaces Tree level diagram (left) vs. loop level (one loop) (right) diagram in WEFT}}{29}{figure.caption.129}%
\contentsline {figure}{\numberline {2.2}{\ignorespaces Different scales in the binary inspiral problem. We reproduce the figure from \cite {Porto:2016pyg}.}}{30}{figure.caption.133}%
\contentsline {figure}{\numberline {2.3}{\ignorespaces Diagrams contributing to the gravitational bound sector at 1PN. }}{38}{figure.caption.164}%
\contentsline {figure}{\numberline {2.4}{\ignorespaces Diagrams contributing to the electromagnetic bound sector up to 1PN.}}{39}{figure.caption.171}%
\contentsline {figure}{\numberline {2.5}{\ignorespaces Diagrams contributing to the Proca bound sector up to 1PN.}}{40}{figure.caption.178}%
\contentsline {figure}{\numberline {2.6}{\ignorespaces Scalar-Electromagnetic interaction diagrams which contribute to the bound sector.}}{40}{figure.caption.183}%
\contentsline {figure}{\numberline {2.7}{\ignorespaces 1PN diagrams for the scalar sector with 3 point vertex coming from scalar-graviton interaction.}}{43}{figure.caption.196}%
\contentsline {figure}{\numberline {2.8}{\ignorespaces 1PN scalar diagrams that contribute to the bound sector.}}{44}{figure.caption.199}%
\contentsline {figure}{\numberline {2.9}{\ignorespaces Scalar diagrams coming from the 3-point vertices that contribute at 2 PN order in scalar effective action.}}{45}{figure.caption.203}%
\contentsline {figure}{\numberline {2.10}{\ignorespaces Diagrams contributing to scalar bound sector at 2 PN order coming from worldline vertices. }}{48}{figure.caption.208}%
\contentsline {figure}{\numberline {2.11}{\ignorespaces Diagrams contributing to the bound sector at 1PN order due to the Proca-electromagnetic interaction term.}}{49}{figure.caption.213}%
\contentsline {figure}{\numberline {2.12}{\ignorespaces Self-energy diagram contributing to the effective action for the source term $J$.}}{54}{figure.caption.238}%
\contentsline {figure}{\numberline {2.13}{\ignorespaces Radiative diagrams for the scalar field that contribute up to $N^{(4)}LO$.}}{56}{figure.caption.249}%
\contentsline {figure}{\numberline {2.14}{\ignorespaces Radiative diagrams for electromagnetic field up to $N^{(3)}LO$. }}{61}{figure.caption.277}%
\contentsline {figure}{\numberline {2.15}{\ignorespaces LO radiation coming from the pure gravitational sector.}}{66}{figure.caption.302}%
\contentsline {figure}{\numberline {2.16}{\ignorespaces Radiative diagrams for gravitational fields.}}{68}{figure.caption.310}%
\contentsline {figure}{\numberline {2.17}{\ignorespaces $\alpha (m;r)$ vs $mr$ plot}}{83}{figure.caption.362}%
\contentsline {figure}{\numberline {2.18}{\ignorespaces $\beta (m;r)$ vs $mr$ plot}}{83}{figure.caption.364}%
\contentsline {figure}{\numberline {2.19}{\ignorespaces Correspondence between PN diagrams (left) and one-loop scalar Feynman diagram (right).}}{84}{figure.caption.365}%
\addvspace {10\p@ }
\contentsline {figure}{\numberline {3.1}{\ignorespaces Sketch of the celestial sphere under consideration.}}{160}{figure.caption.803}%
\contentsline {figure}{\numberline {3.2}{\ignorespaces Figure showing dependency of the integral $I(b)$ as a function of $|b|$ for $m=1\,.$}}{165}{figure.caption.842}%
\addvspace {10\p@ }
\addvspace {10\p@ }
\contentsline {figure}{\numberline {5.1}{\ignorespaces Diagrammatic representation of Lippmann-Schwinger equation}}{250}{figure.caption.1181}%
\addvspace {10\p@ }
\contentsline {figure}{\numberline {6.1}{\ignorespaces Figure describing the $s,t,u$-channel diagram respectively, relevant for our computation }}{263}{figure.caption.1243}%
\contentsline {figure}{\numberline {6.2}{\ignorespaces Figure depicting the chosen contour for dispersion relation in the complex-$\omega $ plane. The cross signs depict the location of the poles.}}{275}{figure.caption.1297}%
\contentsline {figure}{\numberline {6.3}{\ignorespaces Plot depicting the matching for extraction of OPE coefficient using BC expansion (\textbf {red}) and OPE inversion for spin-0 (\textbf {blue}). We have set the conformal dimension for the external primaries to be: $\Delta _{\mathcal {O}}=1+i$.}}{287}{figure.caption.1358}%
\addvspace {10\p@ }
\makeatother}%
  \listoffigures
  \endgroup
}
\makeatother

\definecolor{darkbrown}{rgb}{0.787,0.26,0.187}
\definecolor{alizarin}{rgb}{0.82,0.1,0.26}
\definecolor{mygray}{gray}{0.5}
\newcommand{\Rho}{\mathrm{P}}
\DeclareSymbolFont{usualmathcal}{OMS}{cmsy}{m}{n}
\DeclareSymbolFontAlphabet{\mathcal}{usualmathcal}
\DeclareSymbolFont{rmlargesymbols}{OMX}{mdbch}{m}{n}
\DeclareMathSymbol{\thesisintop}{\mathop}{rmlargesymbols}{82}
\DeclareMathSymbol{\thesisointop}{\mathop}{rmlargesymbols}{72}
\let\intop\thesisintop
\let\ointop\thesisointop
\renewcommand{\int}{\intop\nolimits}
\renewcommand{\oint}{\ointop\nolimits}
\newcommand{\rmint}{\int}

\DeclareSymbolFont{rsfs}{U}{rsfs}{m}{n}
\DeclareSymbolFontAlphabet{\mathscrsfs}{rsfs}

\DeclareMathAlphabet\mathbfcal{OMS}{cmsy}{b}{n}

\allowdisplaybreaks      
\newcommand{\ep}{\epsilon}
\newcommand{\sig}{\sigma}
\newcommand{\mpow}[1]{m_p^{#1}}
\newcommand{\SQ}{\sqrt{\sig^2-1}}

\newcommand{\penguinsymbol}[1]{%
  \begin{tikzpicture}[x=1ex,y=1ex,scale=#1,baseline=-0.2ex]
    \fill[black]
      (-0.7,-1.0) .. controls (-0.9,-0.2) and (-0.6,0.9) .. (0,1.1)
      .. controls (0.6,0.9) and (0.9,-0.2) .. (0.7,-1.0) -- cycle;
    \fill[white] (0,-0.2) ellipse (0.45 and 0.7);
    \fill[black] (-0.9,-0.2) .. controls (-1.4,0.2) and (-1.1,-0.6) .. (-0.7,-0.9) -- cycle;
    \fill[black] ( 0.9,-0.2) .. controls ( 1.4,0.2) and ( 1.1,-0.6) .. ( 0.7,-0.9) -- cycle;
    \fill[white] (-0.18,0.75) circle (0.12);
    \fill[white] ( 0.18,0.75) circle (0.12);
    \fill[black] (-0.18,0.75) circle (0.05);
    \fill[black] ( 0.18,0.75) circle (0.05);
    \fill[orange] (0,0.55) -- (-0.12,0.40) -- (0.12,0.40) -- cycle;
    \fill[orange] (-0.30,-1.05) -- (-0.60,-1.25) -- (-0.10,-1.25) -- cycle;
    \fill[orange] ( 0.30,-1.05) -- ( 0.60,-1.25) -- ( 0.10,-1.25) -- cycle;
  \end{tikzpicture}%
}

\DeclareRobustCommand{\penguin}{%
  \mathchoice{\penguinsymbol{1.10}}
             {\penguinsymbol{1.00}}
             {\penguinsymbol{0.80}}
             {\penguinsymbol{0.65}}
}
\newcommand{\filledsquare}[1]{\tikz[baseline=-0.5ex]\draw[fill=#1, draw=black] (0,0) rectangle (0.8ex,0.8ex);}

\newcommand{\ThesisTitle}{Classical Black Hole Scattering to Celestial Amplitudes in Effective Theories of Gravity}
\newcommand{\ThesisSubtitle}{}
\newcommand{\AuthorName}{Saptaswa Ghosh}
\newcommand{\StudentDesignation}{PhD Scholar}
\newcommand{\DegreeName}{Doctor of Philosophy}
\newcommand{\DepartmentName}{Department of Physics}
\newcommand{\InstituteName}{Indian Institute of Technology Gandhinagar}
\newcommand{\InstituteAddress}{Palaj, Gandhinagar, Gujarat 382055, India}
\newcommand{\SubmissionDate}{April, 2026}

\newcommand{\SupervisorName}{Prof.~Arpan Bhattacharyya}

\newcommand{\RollNo}{21310054}
\newcommand{\CertificateSupervisorA}{\SupervisorName}
\newcommand{\CertificateSupervisorB}{}

\newcommand{\thesisbodystretch}{1.13}
\newcommand{\restorethesisbodyformat}{%
  \setstretch{\thesisbodystretch}%
  \selectfont
\setlength{\parindent}{1.3em}%
  \setlength{\parskip}{0pt}%
}
\AtBeginDocument{\restorethesisbodyformat}

\setlist[itemize]{topsep=2pt,itemsep=2pt}
\allowdisplaybreaks[4]
\newcommand{\thesisdisplaymathsize}{\fontsize{10.4}{13.6}\selectfont}
\AtBeginEnvironment{equation}{\thesisdisplaymathsize}
\AtBeginEnvironment{equation*}{\thesisdisplaymathsize}
\AtBeginEnvironment{align}{\thesisdisplaymathsize}
\AtBeginEnvironment{align*}{\thesisdisplaymathsize}
\AtBeginEnvironment{gather}{\thesisdisplaymathsize}
\AtBeginEnvironment{gather*}{\thesisdisplaymathsize}
\AtBeginEnvironment{multline}{\thesisdisplaymathsize}
\AtBeginEnvironment{multline*}{\thesisdisplaymathsize}
\AtBeginEnvironment{flalign}{\footnotesize}
\AtBeginEnvironment{flalign*}{\footnotesize}

\newcommand{\thesischapterbanner}[2]{%
  \begin{center}
    \begin{minipage}{0.88\linewidth}
      \centering
      \color{brandburgundy}
      \makebox[\linewidth]{%
        \rule{0.23\linewidth}{0.85pt}\hfill
        {\scshape\fontsize{12.2}{14.2}\selectfont #1}%
        \hfill\rule{0.23\linewidth}{0.85pt}%
      }\par
      \ifstrempty{#2}{}{%
        \vspace{0.88em}
        {\setlength{\fboxsep}{8.5pt}%
         \fcolorbox{brandburgundy}{paperbg!55!white}{%
           \color{brandburgundy}\bfseries\fontsize{27.5}{30.5}\selectfont #2%
         }}\par
        \vspace{0.72em}
        \rule{0.58\linewidth}{0.85pt}\par
      }%
    \end{minipage}
  \end{center}%
}
\definecolor{paperbg}{RGB}{231,223,205}
\newcommand{\thesischaptertitlefont}{\centering\color{brandburgundy}\bfseries\itshape\fontsize{21.2}{25.4}\selectfont}
\newcommand{\thesissectionfont}{\color{brandburgundy}\bfseries\fontsize{13.2}{16}\selectfont}
\newcommand{\thesissubsectionfont}{\color{brandburgundy!90!black}\bfseries\normalsize}
\newcommand{\thesisleadtopic}[1]{%
  \par\medskip
  \noindent{\color{brandburgundy}\bfseries #1\par}
  \smallskip
}
\tcbset{
  thesisresultbox/.style={
    enhanced,
    breakable,
    width=\linewidth,
    colback=paperbg!34!white,
    colframe=brandburgundy!88!black,
    colbacktitle=brandburgundy,
    coltitle=white,
    fonttitle=\bfseries\itshape,
    boxrule=0.85pt,
    arc=1pt,
    left=1.4mm,
    right=1.4mm,
    top=1.1mm,
    bottom=1.1mm
  },
  thesisstatementbox/.style={
    enhanced,
    breakable,
    colback=paperbg!18!white,
    colframe=brandburgundy!72!black,
    colbacktitle=paperbg!42!white,
    coltitle=brandburgundy!92!black,
    fonttitle=\bfseries,
    boxrule=0.75pt,
    arc=1pt,
    left=1.2mm,
    right=1.2mm,
    top=1mm,
    bottom=1mm
  },
  thesisaccentbox/.style={thesisresultbox},
  thesissoftbox/.style={thesisstatementbox}
}
\newcommand{\thesisresultmathbox}[2][]{%
  \tcboxmath[
    on line,
    nobeforeafter,
    tcbox raise base,
    colback=paperbg!26!white,
    colframe=brandburgundy!82!black,
    boxrule=0.8pt,
    arc=1pt,
    left=1mm,
    right=1mm,
    top=0.6mm,
    bottom=0.6mm,
    #1
  ]{#2}%
}

\newcommand{\thesismathbox}[2][]{%
  \thesisresultmathbox[#1]{#2}%
}
\newenvironment{thesiscompactdisplay}[1][\small]{%
  \begingroup
  #1%
  \setlength{\jot}{2.5pt}%
  \setlength{\abovedisplayskip}{5pt plus 2pt minus 3pt}%
  \setlength{\belowdisplayskip}{5pt plus 2pt minus 3pt}%
  \setlength{\abovedisplayshortskip}{4pt plus 2pt minus 2pt}%
  \setlength{\belowdisplayshortskip}{4pt plus 2pt minus 2pt}%
}{%
  \endgroup
}
\newlength{\thesisInlineDiagramHeight}
\newlength{\thesisAmplitudeDiagramHeight}
\newlength{\thesisPanelDiagramHeight}
\newlength{\thesisAmplitudeLhsWidth}
\NewDocumentCommand{\thesisinlinefeynman}{ O{0.085\linewidth} O{\thesisInlineDiagramHeight} m }{%
  \mathord{%
    \vcenter{%
      \hbox{%
        \makebox[#1][c]{%
          \includegraphics[width=#1,height=#2,keepaspectratio]{#3}%
        }%
      }%
    }%
  }%
}
\NewDocumentCommand{\thesisampdiagram}{ O{0.46\linewidth} O{\thesisAmplitudeDiagramHeight} m }{%
  \mathord{%
    \vcenter{%
      \hbox{%
        \makebox[#1][c]{%
          \includegraphics[width=#1,height=#2,keepaspectratio]{#3}%
        }%
      }%
    }%
  }%
}
\newcommand{\thesisamplhs}[1]{%
  \mathmakebox[\thesisAmplitudeLhsWidth][l]{#1}%
}
\NewDocumentCommand{\thesisdiagrampanel}{ O{\thesisPanelDiagramHeight} m }{%
  \begin{minipage}[c][#1][c]{\linewidth}
    \centering
    #2
  \end{minipage}%
}
\newcommand{\thesisfrontchapterbanner}{%
  \begin{center}
    \begin{minipage}{0.88\linewidth}
      \centering
      \color{brandburgundy}
      \rule{0.62\linewidth}{0.9pt}\par
    \end{minipage}
  \end{center}%
}
\newcommand{\thesischapterpaperbox}[3]{%
  \ifstrempty{#1}{}{%
    \begin{center}
      \begin{tcolorbox}[
        enhanced,
        width=0.9\linewidth,
        boxrule=0.85pt,
        colframe=brandburgundy,
        colback=paperbg!45!white,
        arc=1pt,
        left=12pt,
        right=12pt,
        top=9pt,
        bottom=9pt
      ]
        \centering
        {\color{brandburgundy}\scshape\fontsize{10.5}{12.5}\selectfont Based on the paper}\par
        \vspace{0.35em}
        {\itshape #1\par}
        \vspace{0.45em}
        {\fontsize{10.5}{13}\selectfont #2\par}
        \ifstrempty{#3}{}{%
          \vspace{0.35em}
          {\color{brandburgundy}\fontsize{9.9}{12.2}\selectfont #3\par}
        }%
      \end{tcolorbox}
    \end{center}
    \vspace{1.4em}
  }%
}

\titleformat{\chapter}[display]
  {\normalfont}
  {\thesischapterbanner{\chaptername}{\thechapter}}
  {1.25em}
  {\thesischaptertitlefont}
\titlespacing*{\chapter}{0pt}{0pt}{1.45em}

\titleformat{name=\chapter,numberless}[display]
  {\normalfont}
  {\thesisfrontchapterbanner}
  {1.0em}
  {\thesischaptertitlefont}
  [\vspace{0.85em}\thesisfrontchapterbanner]
\titlespacing*{name=\chapter,numberless}{0pt}{0pt}{1.2em}

\titleformat{\section}
  {\thesissectionfont}
  {\thesection}
  {0.5em}
  {}
\titlespacing*{\section}{0pt}{1.25em}{0.55em}

\titleformat{\subsection}
  {\thesissubsectionfont}
  {\thesubsection}
  {0.45em}
  {}
\titlespacing*{\subsection}{0pt}{1.0em}{0.45em}

\titleformat{name=\section,numberless}[block]
  {\thesissectionfont}
  {}
  {0pt}
  {}
\titlespacing*{name=\section,numberless}{0pt}{1.15em}{0.45em}

\titleformat{name=\subsection,numberless}[block]
  {\thesissubsectionfont}
  {}
  {0pt}
  {}
\titlespacing*{name=\subsection,numberless}{0pt}{0.95em}{0.35em}

\newcommand{\frontchapter}[1]{%
  \chapter*{#1}%
  \addcontentsline{toc}{chapter}{#1}%
}

\makeatletter
\newcommand{\restoremainchapterstyle}{%
  \gdef\@chapapp{\chaptername}%
  \gdef\thechapter{\@arabic\c@chapter}%
  \setcounter{section}{0}%
  \setcounter{subsection}{0}%
  \setcounter{subsubsection}{0}%
  \renewcommand{\thesection}{\thechapter.\arabic{section}}%
  \renewcommand{\thesubsection}{\thesection.\arabic{subsection}}%
  \renewcommand{\theHsection}{chapter.\arabic{chapter}.section.\arabic{section}}%
  \renewcommand{\theHsubsection}{\theHsection.\arabic{subsection}}%
}
\makeatother

\newcommand{\MakeReferenceStyleTitlePage}{%
  \begin{titlepage}
    \sloppy
    \newgeometry{left=2.30cm,right=2.30cm,top=2.00cm,bottom=1.95cm}
    \thispagestyle{empty}
    \begin{tikzpicture}[remember picture,overlay]
      \fill[paperbg] (current page.north west) rectangle (current page.south east);
      \fill[brandburgundy] (current page.north west) rectangle ([yshift=-0.48cm]current page.north east);
      \fill[brandburgundy] ([yshift=0.48cm]current page.south west) rectangle (current page.south east);
      \draw[brandburgundy!70,line width=0.9pt]
        ([xshift=1.05cm,yshift=-0.98cm]current page.north west)
        rectangle
        ([xshift=-1.05cm,yshift=0.98cm]current page.south east);
    \end{tikzpicture}

    \begin{center}
      \vspace*{1.05cm}
      {\color{brandburgundy}\bfseries\fontsize{18}{22}\selectfont \ThesisTitle\par}
      \ifstrempty{\ThesisSubtitle}{}{\vspace{0.20cm}{\fontsize{11}{14}\selectfont\ThesisSubtitle\par}}
      \vspace{0.38cm}
      {\color{brandburgundy!80}\rule{0.76\textwidth}{0.85pt}\par}

      \vspace{1.15cm}
      {\bfseries\fontsize{16}{20}\selectfont \AuthorName
      \par}
      \vspace{0.22cm}
      {\fontsize{13.2}{16.2}\selectfont \StudentDesignation, \DepartmentName\par}
      {\fontsize{13.2}{16.2}\selectfont \InstituteName\par}
\vspace{0.34cm}
        {\fontsize{13.2}{16.2}\selectfont \InstituteAddress\par}
      \vspace{0.95cm}
      \begin{tabularx}{0.84\textwidth}{
        >{\raggedleft\arraybackslash\bfseries}p{0.30\textwidth}
        >{\raggedright\arraybackslash}X
      }
        Advisor: & \SupervisorName
      \end{tabularx}
      \vspace{1.00cm}
      \includegraphics[height=4.65cm]{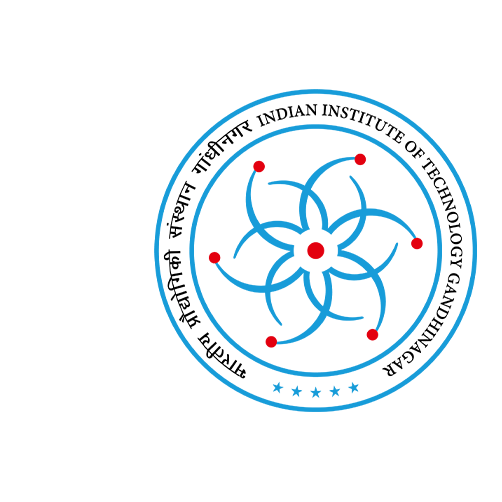}\par

      \vspace{0.90cm}
      \begin{minipage}{0.82\textwidth}
        \centering
        {\fontsize{14.2}{17.2}\selectfont A thesis submitted in partial fulfillment of the requirements for the degree of\par}
        \vspace{0.28cm}
        {\bfseries\fontsize{17}{21}\selectfont \DegreeName\par}
        \vspace{0.44cm}
        {\fontsize{13}{16}\selectfont \SubmissionDate\par}
      \end{minipage}
      \vfill
      {\color{brandburgundy!80}\rule{0.76\textwidth}{0.85pt}\par}
    \end{center}

    \restoregeometry
  \end{titlepage}
}

\newcommand{\MakeCertificatePage}{%
  \begin{titlepage}
    \sloppy
    \newgeometry{left=2.20cm,right=2.20cm,top=2.30cm,bottom=2.10cm}
    \thispagestyle{empty}
    \begin{tikzpicture}[remember picture,overlay]
      \fill[paperbg] (current page.north west) rectangle (current page.south east);
    \end{tikzpicture}

    \begin{center}
      \vspace*{0.9cm}
      {\bfseries\fontsize{20}{24}\selectfont Certificate\par}
      \vspace{0.9cm}
      \rule{0.88\textwidth}{0.45pt}\par
    \end{center}

    \vspace{1.1cm}
    \begin{center}
      \begin{minipage}{0.84\textwidth}
        \centering
        {\fontsize{14}{20}\selectfont
        It is certified that the work contained in the thesis titled:
        {\bfseries \ThesisTitle}, by Mr.~\AuthorName\ (Roll No.~\RollNo),
        has been carried out under my supervision and that this work has not
        been submitted elsewhere for a degree.\par}
      \end{minipage}
    \end{center}

    \vfill
    \begin{center}
      \rule{0.42\textwidth}{0.45pt}\par
      \vspace{0.20cm}
      {\fontsize{13}{16}\selectfont \CertificateSupervisorA, Advisor\par}
      \ifstrempty{\CertificateSupervisorB}{}{\vspace{0.6cm}\rule{0.42\textwidth}{0.45pt}\par
      \vspace{0.20cm}
     
      }
    \end{center}

    \restoregeometry
  \end{titlepage}
}

\newwrite\thesiscrossrefsmarker
\IfFileExists{\jobname.crossrefs-ready}{}{%
  \makeatletter
  \newlabel{1.2}{{1.3}{4}{Quantum Field Theory and its Classical Limit}{equation.24}{}}
\newlabel{fig:eikonal-bubble-poisson}{{1.1}{9}{Schematic bubble notation for nested Poisson brackets. Popping a bubble produces oriented causal cuts according to the Leibniz rule}{figure.caption.47}{}}
\newlabel{fig:eikonal-murua-trees}{{1.2}{11}{Schematic graphical content of Murua's recursion. The displayed rooted tree has two allowed partitions: \(p_1\), containing only the root-adjacent edges, and \(p_2\), containing all three edges. In general, rooted partitions \(p\in {\rm Pr}(\tau )\) must contain all root-adjacent edges; for non-rooted trees, \(p\in {\rm P}_{s}(\tau )\) must contain all edges adjacent to the chosen semi-root \(s\)}{figure.caption.57}{}}
\newlabel{fig:eikonal-ir-causality}{{1.3}{11}{Causality-prescription diagrams for the IR-sensitive master integral. The Magnus/Murua weights assigned to the two oriented orderings cancel the leading \(1/\epsilon ^2\) pole}{figure.caption.61}{}}
\newlabel{fig23}{{1.4}{14}{Celestial description of a four-dimensional scattering process. A momentum-space \(n\to m\) scattering amplitude in asymptotically flat spacetime is mapped, through a Mellin transform over the external energies, to a conformal correlator on the celestial sphere}{figure.caption.70}{}}
\newlabel{eq:exp-ibp-splitting}{{1.47}{16}{}{equation.80}{}}
\newlabel{eq:exp-ibp-identities}{{1.47}{16}{}{equation.81}{}}
\newlabel{lst:finiteflow-ibp}{{1.1}{16}{Solving simple IBPs using FiniteFlow}{lstlisting.82}{}}
\newlabel{7}{{1.50}{17}{}{equation.101}{}}
\newlabel{9}{{1.52}{17}{}{equation.103}{}}
\newlabel{sp1}{{1.5}{18}{Geometry and contour of nested integrals}{figure.caption.106}{}}
\newlabel{Sec2}{{2.2}{28}{Brief review of machinery of EFTs}{section.124}{}}
\newlabel{2.3m}{{2.2}{28}{Brief review of machinery of EFTs}{equation.125}{}}
\newlabel{Stotal}{{2.2}{28}{Brief review of machinery of EFTs}{equation.126}{}}
\newlabel{stotal1}{{2.3}{28}{Brief review of machinery of EFTs}{equation.127}{}}
\newlabel{2.4m}{{2.4}{28}{Brief review of machinery of EFTs}{equation.128}{}}
\newlabel{mfig}{{2.1}{29}{Tree level diagram (left) vs. loop level (one loop) (right) diagram in WEFT}{figure.caption.129}{}}
\newlabel{newfig}{{2.2}{30}{Different scales in the binary inspiral problem. We reproduce the figure from \cite {Porto:2016pyg}}{figure.caption.133}{}}
\newlabel{Sec3}{{2.3}{30}{Review of worldline effective field theory for a point particle in GR}{section.134}{}}
\newlabel{2.1}{{2.8}{30}{Review of worldline effective field theory for a point particle in GR}{equation.135}{}}
\newlabel{eqnew}{{2.10}{31}{Review of worldline effective field theory for a point particle in GR}{equation.137}{}}
\newlabel{2.4}{{2.3}{31}{Review of worldline effective field theory for a point particle in GR}{equation.138}{}}
\newlabel{3.6n}{{2.13}{31}{}{equation.140}{}}
\newlabel{2.8}{{2.15}{32}{}{equation.142}{}}
\newlabel{2.10}{{2.17}{32}{}{equation.144}{}}
\newlabel{KK}{{2.18}{32}{}{equation.145}{}}
\newlabel{Sec4}{{2.4}{33}{Conservative dynamics of binary black holes in a theory of ALP and ULV }{section.147}{}}
\newlabel{2.11}{{2.20}{33}{Conservative dynamics of binary black holes in a theory of ALP and ULV }{equation.148}{}}
\newlabel{ch1:2.13}{{2.21}{34}{Conservative dynamics of binary black holes in a theory of ALP and ULV }{equation.149}{}}
\newlabel{4.3mm}{{2.22}{34}{Conservative dynamics of binary black holes in a theory of ALP and ULV }{equation.150}{}}
\newlabel{24m}{{2.24}{35}{Conservative dynamics of binary black holes in a theory of ALP and ULV }{equation.152}{}}
\newlabel{25m}{{2.26}{35}{Conservative dynamics of binary black holes in a theory of ALP and ULV }{equation.154}{}}
\newlabel{2.28jjj}{{2.28}{35}{Conservative dynamics of binary black holes in a theory of ALP and ULV }{equation.156}{}}
\newlabel{23m}{{2.29}{36}{Conservative dynamics of binary black holes in a theory of ALP and ULV }{equation.157}{}}
\newlabel{4.10}{{2.31}{36}{Conservative dynamics of binary black holes in a theory of ALP and ULV }{equation.159}{}}
\newlabel{table1m}{{2.1}{37}{Table showing the different worldline couplings and the order at which they contribute}{table.caption.161}{}}
\newlabel{fig1a}{{2.3a}{38}{\relax }{figure.caption.164}{}}
\newlabel{sub@fig1a}{{a}{38}{\relax }{figure.caption.164}{}}
\newlabel{fig1b}{{2.3b}{38}{\relax }{figure.caption.164}{}}
\newlabel{sub@fig1b}{{b}{38}{\relax }{figure.caption.164}{}}
\newlabel{fig1c}{{2.3c}{38}{\relax }{figure.caption.164}{}}
\newlabel{sub@fig1c}{{c}{38}{\relax }{figure.caption.164}{}}
\newlabel{fig1d}{{2.3d}{38}{\relax }{figure.caption.164}{}}
\newlabel{sub@fig1d}{{d}{38}{\relax }{figure.caption.164}{}}
\newlabel{fig:my_labela}{{2.3}{38}{Diagrams contributing to the gravitational bound sector at 1PN}{figure.caption.164}{}}
\newlabel{4.13a}{{2.32}{38}{The Gravitational bound sector}{equation.165}{}}
\newlabel{EM section}{{2.4.2}{38}{The Electromagnetic (EM) bound sector}{subsection.169}{}}
\newlabel{mfig4a}{{2.4a}{39}{\relax }{figure.caption.171}{}}
\newlabel{sub@mfig4a}{{a}{39}{\relax }{figure.caption.171}{}}
\newlabel{mfig4b}{{2.4b}{39}{\relax }{figure.caption.171}{}}
\newlabel{sub@mfig4b}{{b}{39}{\relax }{figure.caption.171}{}}
\newlabel{mfig4c}{{2.4c}{39}{\relax }{figure.caption.171}{}}
\newlabel{sub@mfig4c}{{c}{39}{\relax }{figure.caption.171}{}}
\newlabel{mfig4d}{{2.4d}{39}{\relax }{figure.caption.171}{}}
\newlabel{sub@mfig4d}{{d}{39}{\relax }{figure.caption.171}{}}
\newlabel{ch1:fig:em-bound}{{2.4}{39}{Diagrams contributing to the electromagnetic bound sector up to 1PN}{figure.caption.171}{}}
\newlabel{mfig5a}{{2.5a}{40}{\relax }{figure.caption.178}{}}
\newlabel{sub@mfig5a}{{a}{40}{\relax }{figure.caption.178}{}}
\newlabel{mfig5b}{{2.5b}{40}{\relax }{figure.caption.178}{}}
\newlabel{sub@mfig5b}{{b}{40}{\relax }{figure.caption.178}{}}
\newlabel{mfig5c}{{2.5c}{40}{\relax }{figure.caption.178}{}}
\newlabel{sub@mfig5c}{{c}{40}{\relax }{figure.caption.178}{}}
\newlabel{mfig5d}{{2.5d}{40}{\relax }{figure.caption.178}{}}
\newlabel{sub@mfig5d}{{d}{40}{\relax }{figure.caption.178}{}}
\newlabel{ch1:fig:proca-bound}{{2.5}{40}{Diagrams contributing to the Proca bound sector up to 1PN}{figure.caption.178}{}}
\newlabel{fig4an}{{2.6a}{40}{\relax }{figure.caption.183}{}}
\newlabel{sub@fig4an}{{a}{40}{\relax }{figure.caption.183}{}}
\newlabel{fig4cn}{{2.6b}{40}{\relax }{figure.caption.183}{}}
\newlabel{sub@fig4cn}{{b}{40}{\relax }{figure.caption.183}{}}
\newlabel{fig4bn}{{2.6c}{40}{\relax }{figure.caption.183}{}}
\newlabel{sub@fig4bn}{{c}{40}{\relax }{figure.caption.183}{}}
\newlabel{figtheta1}{{2.6}{40}{Scalar-Electromagnetic interaction diagrams which contribute to the bound sector}{figure.caption.183}{}}
\newlabel{Sec4.2}{{2.4.4}{40}{The EM-Axion bound sector}{subsection.184}{}}
\newlabel{2.15}{{2.48}{41}{The EM-Axion bound sector}{equation.188}{}}
\newlabel{4.18n}{{2.49}{41}{The EM-Axion bound sector}{equation.189}{}}
\newlabel{4.31}{{2.50}{42}{The EM-Axion bound sector}{equation.190}{}}
\newlabel{4.33mm}{{2.52}{42}{The EM-Axion bound sector}{equation.192}{}}
\newlabel{fig5aa}{{2.7a}{43}{\relax }{figure.caption.196}{}}
\newlabel{sub@fig5aa}{{a}{43}{\relax }{figure.caption.196}{}}
\newlabel{fig5bb}{{2.7b}{43}{\relax }{figure.caption.196}{}}
\newlabel{sub@fig5bb}{{b}{43}{\relax }{figure.caption.196}{}}
\newlabel{ch1:fig:scalar-graviton-1pn}{{2.7}{43}{1PN diagrams for the scalar sector with 3 point vertex coming from scalar-graviton interaction}{figure.caption.196}{}}
\newlabel{Sec4.3}{{2.4.5}{43}{The pure scalar sector}{subsection.194}{}}
\newlabel{4.24}{{2.54}{43}{The pure scalar sector}{equation.195}{}}
\newlabel{4.25}{{2.55}{43}{The pure scalar sector}{equation.197}{}}
\newlabel{fig6ma}{{2.8a}{44}{\relax }{figure.caption.199}{}}
\newlabel{sub@fig6ma}{{a}{44}{\relax }{figure.caption.199}{}}
\newlabel{fig6mb}{{2.8b}{44}{\relax }{figure.caption.199}{}}
\newlabel{sub@fig6mb}{{b}{44}{\relax }{figure.caption.199}{}}
\newlabel{fig6mc}{{2.8c}{44}{\relax }{figure.caption.199}{}}
\newlabel{sub@fig6mc}{{c}{44}{\relax }{figure.caption.199}{}}
\newlabel{fig6m}{{2.8}{44}{1PN scalar diagrams that contribute to the bound sector}{figure.caption.199}{}}
\newlabel{4.26}{{2.56}{44}{The pure scalar sector}{equation.198}{}}
\newlabel{fig3a}{{2.9a}{45}{\relax }{figure.caption.203}{}}
\newlabel{sub@fig3a}{{a}{45}{\relax }{figure.caption.203}{}}
\newlabel{fig3b}{{2.9b}{45}{\relax }{figure.caption.203}{}}
\newlabel{sub@fig3b}{{b}{45}{\relax }{figure.caption.203}{}}
\newlabel{fig3c}{{2.9c}{45}{\relax }{figure.caption.203}{}}
\newlabel{sub@fig3c}{{c}{45}{\relax }{figure.caption.203}{}}
\newlabel{fig3d}{{2.9d}{45}{\relax }{figure.caption.203}{}}
\newlabel{sub@fig3d}{{d}{45}{\relax }{figure.caption.203}{}}
\newlabel{ch1:fig:scalar-2pn}{{2.9}{45}{Scalar diagrams coming from the 3-point vertices that contribute at 2 PN order in scalar effective action}{figure.caption.203}{}}
\newlabel{4.38}{{2.60}{46}{The pure scalar sector}{equation.204}{}}
\newlabel{fig4a}{{2.10a}{48}{\relax }{figure.caption.208}{}}
\newlabel{sub@fig4a}{{a}{48}{\relax }{figure.caption.208}{}}
\newlabel{fig4b}{{2.10b}{48}{\relax }{figure.caption.208}{}}
\newlabel{sub@fig4b}{{b}{48}{\relax }{figure.caption.208}{}}
\newlabel{fig4}{{2.10}{48}{Diagrams contributing to scalar bound sector at 2 PN order coming from worldline vertices}{figure.caption.208}{}}
\newlabel{4.35}{{2.66}{48}{Proca-electromagnetic interaction sector}{equation.212}{}}
\newlabel{fig9am}{{2.11a}{49}{\relax }{figure.caption.213}{}}
\newlabel{sub@fig9am}{{a}{49}{\relax }{figure.caption.213}{}}
\newlabel{fig9bm}{{2.11b}{49}{\relax }{figure.caption.213}{}}
\newlabel{sub@fig9bm}{{b}{49}{\relax }{figure.caption.213}{}}
\newlabel{fig9cm}{{2.11c}{49}{\relax }{figure.caption.213}{}}
\newlabel{sub@fig9cm}{{c}{49}{\relax }{figure.caption.213}{}}
\newlabel{fig9dm}{{2.11d}{49}{\relax }{figure.caption.213}{}}
\newlabel{sub@fig9dm}{{d}{49}{\relax }{figure.caption.213}{}}
\newlabel{fig9em}{{2.11e}{49}{\relax }{figure.caption.213}{}}
\newlabel{sub@fig9em}{{e}{49}{\relax }{figure.caption.213}{}}
\newlabel{fig:my_label2}{{2.11}{49}{Diagrams contributing to the bound sector at 1PN order due to the Proca-electromagnetic interaction term}{figure.caption.213}{}}
\newlabel{4.52}{{2.71}{50}{Proca-electromagnetic interaction sector}{equation.218}{}}
\newlabel{4.53m}{{2.72}{51}{Proca-electromagnetic interaction sector}{equation.220}{}}
\newlabel{Sec5}{{2.5}{51}{The radiative dynamics}{section.221}{}}
\newlabel{5.1m}{{2.74}{51}{The radiative dynamics}{equation.223}{}}
\newlabel{table2mm}{{2.2}{51}{Table showing the number of diagrams contributing to a specific PN order for bound sector. Also, the orders at which terms to due the axion-photon coupling ($g_{a\gamma \gamma }$) and photon-dark photon kinetic mixing term ($\gamma $)}{table.caption.219}{}}
\newlabel{4.3}{{2.75}{52}{The radiative dynamics}{equation.224}{}}
\newlabel{4.4}{{2.76}{52}{The radiative dynamics}{equation.225}{}}
\newlabel{4.6}{{2.78}{52}{The radiative dynamics}{equation.227}{}}
\newlabel{5.6m}{{2.79}{52}{The radiative dynamics}{equation.228}{}}
\newlabel{EFTcounting}{{2.5.1}{52}{Radiative sector PN counting}{subsection.229}{}}
\newlabel{table3mm}{{2.3}{53}{EFT PN counting for radiative sector}{table.caption.230}{}}
\newlabel{subsec:radiated_scalars}{{2.5.2}{53}{Scalar interaction and scalar radiation}{subsection.231}{}}
\newlabel{mfig12}{{2.12}{54}{Self-energy diagram contributing to the effective action for the source term $J$}{figure.caption.238}{}}
\newlabel{5.10}{{2.82}{54}{Scalar interaction and scalar radiation}{equation.234}{}}
\newlabel{eq:multipole_expansion_scalar}{{2.83}{54}{Scalar interaction and scalar radiation}{equation.235}{}}
\newlabel{5.12mn}{{2.84}{54}{Scalar interaction and scalar radiation}{equation.236}{}}
\newlabel{5.14n}{{2.86}{54}{Scalar interaction and scalar radiation}{equation.239}{}}
\newlabel{4.13}{{2.87}{54}{Scalar interaction and scalar radiation}{equation.241}{}}
\newlabel{4.14}{{2.88}{55}{Scalar interaction and scalar radiation}{equation.242}{}}
\newlabel{4.15}{{2.89}{55}{Scalar interaction and scalar radiation}{equation.243}{}}
\newlabel{5.18n}{{2.90}{55}{Scalar interaction and scalar radiation}{equation.244}{}}
\newlabel{5.21}{{2.93}{55}{Scalar interaction and scalar radiation}{equation.247}{}}
\newlabel{10a}{{2.13a}{56}{\relax }{figure.caption.249}{}}
\newlabel{sub@10a}{{a}{56}{\relax }{figure.caption.249}{}}
\newlabel{10b}{{2.13b}{56}{\relax }{figure.caption.249}{}}
\newlabel{sub@10b}{{b}{56}{\relax }{figure.caption.249}{}}
\newlabel{10c}{{2.13c}{56}{\relax }{figure.caption.249}{}}
\newlabel{sub@10c}{{c}{56}{\relax }{figure.caption.249}{}}
\newlabel{10h}{{2.13d}{56}{\relax }{figure.caption.249}{}}
\newlabel{sub@10h}{{d}{56}{\relax }{figure.caption.249}{}}
\newlabel{10d}{{2.13e}{56}{\relax }{figure.caption.249}{}}
\newlabel{sub@10d}{{e}{56}{\relax }{figure.caption.249}{}}
\newlabel{10e}{{2.13f}{56}{\relax }{figure.caption.249}{}}
\newlabel{sub@10e}{{f}{56}{\relax }{figure.caption.249}{}}
\newlabel{10f}{{2.13g}{56}{\relax }{figure.caption.249}{}}
\newlabel{sub@10f}{{g}{56}{\relax }{figure.caption.249}{}}
\newlabel{10g}{{2.13h}{56}{\relax }{figure.caption.249}{}}
\newlabel{sub@10g}{{h}{56}{\relax }{figure.caption.249}{}}
\newlabel{Figure10}{{2.13}{56}{Radiative diagrams for the scalar field that contribute up to $N^{(4)}LO$}{figure.caption.249}{}}
\newlabel{5.26}{{2.99}{57}{}{equation.254}{}}
\newlabel{5.29}{{2.102}{57}{}{equation.257}{}}
\newlabel{5.46}{{2.104}{58}{}{equation.259}{}}
\newlabel{5.32m}{{2.107}{58}{}{equation.262}{}}
\newlabel{5.36n}{{2.108}{58}{}{equation.263}{}}
\newlabel{emRad}{{2.5.3}{60}{Electromagnetic interaction and the multipole decomposition}{subsection.271}{}}
\newlabel{fig12a}{{2.14a}{61}{\relax }{figure.caption.277}{}}
\newlabel{sub@fig12a}{{a}{61}{\relax }{figure.caption.277}{}}
\newlabel{fig12b}{{2.14b}{61}{\relax }{figure.caption.277}{}}
\newlabel{sub@fig12b}{{b}{61}{\relax }{figure.caption.277}{}}
\newlabel{fig12d}{{2.14c}{61}{\relax }{figure.caption.277}{}}
\newlabel{sub@fig12d}{{c}{61}{\relax }{figure.caption.277}{}}
\newlabel{fig12c}{{2.14d}{61}{\relax }{figure.caption.277}{}}
\newlabel{sub@fig12c}{{d}{61}{\relax }{figure.caption.277}{}}
\newlabel{Fig:12}{{2.14}{61}{Radiative diagrams for electromagnetic field up to $N^{(3)}LO$}{figure.caption.277}{}}
\newlabel{5.42}{{2.118}{61}{Electromagnetic interaction and the multipole decomposition}{equation.275}{}}
\newlabel{5.1}{{2.125}{63}{Proca interaction and multipole decomposition}{equation.284}{}}
\newlabel{5.2}{{2.126}{63}{Proca interaction and multipole decomposition}{equation.285}{}}
\newlabel{5.50}{{2.127}{63}{Proca interaction and multipole decomposition}{equation.286}{}}
\newlabel{sec5.5}{{2.5.5}{65}{Gravitational radiation and Multipole decomposition}{subsection.293}{}}
\newlabel{5.56mm}{{2.135}{65}{Gravitational radiation and Multipole decomposition}{equation.296}{}}
\newlabel{5.57mm}{{2.136}{65}{Gravitational radiation and Multipole decomposition}{equation.297}{}}
\newlabel{5.65}{{2.139}{66}{Gravitational radiation and Multipole decomposition}{equation.300}{}}
\newlabel{5.59m}{{2.142}{66}{Gravitational radiation and Multipole decomposition}{equation.304}{}}
\newlabel{figpq}{{2.15}{66}{LO radiation coming from the pure gravitational sector}{figure.caption.302}{}}
\newlabel{fig13e}{{2.16a}{68}{\relax }{figure.caption.310}{}}
\newlabel{sub@fig13e}{{a}{68}{\relax }{figure.caption.310}{}}
\newlabel{fig13f}{{2.16b}{68}{\relax }{figure.caption.310}{}}
\newlabel{sub@fig13f}{{b}{68}{\relax }{figure.caption.310}{}}
\newlabel{fig13l}{{2.16c}{68}{\relax }{figure.caption.310}{}}
\newlabel{sub@fig13l}{{c}{68}{\relax }{figure.caption.310}{}}
\newlabel{fig13g}{{2.16d}{68}{\relax }{figure.caption.310}{}}
\newlabel{sub@fig13g}{{d}{68}{\relax }{figure.caption.310}{}}
\newlabel{fig13n}{{2.16e}{68}{\relax }{figure.caption.310}{}}
\newlabel{sub@fig13n}{{e}{68}{\relax }{figure.caption.310}{}}
\newlabel{fig13a}{{2.16f}{68}{\relax }{figure.caption.310}{}}
\newlabel{sub@fig13a}{{f}{68}{\relax }{figure.caption.310}{}}
\newlabel{fig13b}{{2.16g}{68}{\relax }{figure.caption.310}{}}
\newlabel{sub@fig13b}{{g}{68}{\relax }{figure.caption.310}{}}
\newlabel{fig13c}{{2.16h}{68}{\relax }{figure.caption.310}{}}
\newlabel{sub@fig13c}{{h}{68}{\relax }{figure.caption.310}{}}
\newlabel{fig13d}{{2.16i}{68}{\relax }{figure.caption.310}{}}
\newlabel{sub@fig13d}{{i}{68}{\relax }{figure.caption.310}{}}
\newlabel{fig13h}{{2.16j}{68}{\relax }{figure.caption.310}{}}
\newlabel{sub@fig13h}{{j}{68}{\relax }{figure.caption.310}{}}
\newlabel{fig13i}{{2.16k}{68}{\relax }{figure.caption.310}{}}
\newlabel{sub@fig13i}{{k}{68}{\relax }{figure.caption.310}{}}
\newlabel{fig13j}{{2.16l}{68}{\relax }{figure.caption.310}{}}
\newlabel{sub@fig13j}{{l}{68}{\relax }{figure.caption.310}{}}
\newlabel{fig13k}{{2.16m}{68}{\relax }{figure.caption.310}{}}
\newlabel{sub@fig13k}{{m}{68}{\relax }{figure.caption.310}{}}
\newlabel{fig13m}{{2.16n}{68}{\relax }{figure.caption.310}{}}
\newlabel{sub@fig13m}{{n}{68}{\relax }{figure.caption.310}{}}
\newlabel{fig13o}{{2.16o}{68}{\relax }{figure.caption.310}{}}
\newlabel{sub@fig13o}{{o}{68}{\relax }{figure.caption.310}{}}
\newlabel{fig13p}{{2.16p}{68}{\relax }{figure.caption.310}{}}
\newlabel{sub@fig13p}{{p}{68}{\relax }{figure.caption.310}{}}
\newlabel{Fig15}{{2.16}{68}{Radiative diagrams for gravitational fields}{figure.caption.310}{}}
\newlabel{5.85n}{{2.162}{73}{Gravitational radiation and Multipole decomposition}{equation.327}{}}
\newlabel{table2m}{{2.4}{74}{Table showing the number of diagrams contributing to each $N^{(n)}LO$ for different radiative fields. We exclude the pure gravitational sector, which is already extensively studied (see \cite {Blanchet:2013haa}), and list only contributions from interactions of other fields with gravity}{table.caption.329}{}}
\newlabel{5.85}{{2.163}{74}{Gravitational radiation and Multipole decomposition}{equation.328}{}}
\newlabel{5.81n}{{2.164}{75}{Gravitational radiation and Multipole decomposition}{equation.330}{}}
\newlabel{5.81m}{{2.165}{75}{Gravitational radiation and Multipole decomposition}{equation.331}{}}
\newlabel{disc}{{2.6}{76}{Discussions and outlooks}{section.332}{}}
\newlabel{ch1:app:A}{{2.A}{77}{Details of computation used in Section~(\ref {Sec4.2})}{section.333}{}}
\newlabel{50}{{2.170}{78}{Details of computation used in Section~(\ref {Sec4.2})}{equation.338}{}}
\newlabel{ch1:app:B}{{2.B}{79}{ Details of the Feynman integral computation used in Section~(\ref {Sec4.3})}{section.342}{}}
\newlabel{A.1}{{2.174}{79}{ Details of the Feynman integral computation used in Section~(\ref {Sec4.3})}{equation.343}{}}
\newlabel{A.3}{{2.176}{79}{ Details of the Feynman integral computation used in Section~(\ref {Sec4.3})}{equation.345}{}}
\newlabel{B.5m}{{2.178}{80}{ Details of the Feynman integral computation used in Section~(\ref {Sec4.3})}{equation.347}{}}
\newlabel{B.6m}{{2.181}{80}{ Details of the Feynman integral computation used in Section~(\ref {Sec4.3})}{equation.350}{}}
\newlabel{B.7m}{{2.182}{81}{ Details of the Feynman integral computation used in Section~(\ref {Sec4.3})}{equation.351}{}}
\newlabel{B.8m}{{2.183}{81}{ Details of the Feynman integral computation used in Section~(\ref {Sec4.3})}{equation.352}{}}
\newlabel{B.9m}{{2.185}{81}{ Details of the Feynman integral computation used in Section~(\ref {Sec4.3})}{equation.354}{}}
\newlabel{A.5}{{2.187}{81}{ Details of the Feynman integral computation used in Section~(\ref {Sec4.3})}{equation.356}{}}
\newlabel{ch1:A.6}{{2.188}{82}{ Details of the Feynman integral computation used in Section~(\ref {Sec4.3})}{equation.357}{}}
\newlabel{ch1:app:C}{{2.C}{82}{ Details of computation of Feynman integral used in (\ref {Sec4}) and (\ref {Sec5})}{section.359}{}}
\newlabel{C.1}{{2.190}{82}{ Details of computation of Feynman integral used in (\ref {Sec4}) and (\ref {Sec5})}{equation.360}{}}
\newlabel{fig12}{{2.17}{83}{$\alpha (m;r)$ vs $mr$ plot}{figure.caption.362}{}}
\newlabel{C3m}{{2.192}{83}{ Details of computation of Feynman integral used in (\ref {Sec4}) and (\ref {Sec5})}{equation.363}{}}
\newlabel{ch1:app:D}{{2.D}{83}{Comments on middle vertex contribution to the radiation}{section.366}{}}
\newlabel{lastfig}{{2.18}{83}{$\beta (m;r)$ vs $mr$ plot}{figure.caption.364}{}}
\newlabel{fig16}{{2.19}{84}{Correspondence between PN diagrams (left) and one-loop scalar Feynman diagram (right)}{figure.caption.365}{}}
\newlabel{E.1}{{2.193}{84}{Comments on middle vertex contribution to the radiation}{equation.367}{}}
\newlabel{E.2}{{2.194}{84}{Comments on middle vertex contribution to the radiation}{equation.368}{}}
\newlabel{2.201j}{{2.204}{86}{Comments on middle vertex contribution to the radiation}{equation.378}{}}
\newlabel{2.205a}{{2.208}{86}{Comments on middle vertex contribution to the radiation}{equation.382}{}}
\newlabel{2.206a}{{2.209}{86}{Comments on middle vertex contribution to the radiation}{equation.383}{}}
\newlabel{ch1:app:E}{{2.E}{91}{Useful Feynman integrals}{section.425}{}}
\newlabel{ch1:tensorfourierindentity}{{2.247}{91}{Useful Feynman integrals}{equation.426}{}}
\newlabel{ch1:eq:1loop}{{2.251}{92}{Useful Feynman integrals}{equation.427}{}}
\newlabel{intro}{{3.1}{93}{Introduction}{section.433}{}}
\newlabel{ch2:sec2}{{3.2}{96}{A quick tour to worldline quantum field theory}{section.436}{}}
\newlabel{ch2:sec3}{{3.3}{98}{Derivation of the Feynman rules}{section.447}{}}
\newlabel{3.1m}{{3.10}{98}{Derivation of the Feynman rules}{equation.448}{}}
\newlabel{3.2}{{3.12}{98}{Derivation of the Feynman rules}{equation.451}{}}
\newlabel{3.3}{{3.13}{98}{Derivation of the Feynman rules}{equation.452}{}}
\newlabel{3.4m}{{3.18}{99}{Derivation of the Feynman rules}{equation.457}{}}
\newlabel{3.8m}{{3.22}{99}{Derivation of the Feynman rules}{equation.461}{}}
\newlabel{tensor}{{3.25}{100}{Derivation of the Feynman rules}{equation.464}{}}
\newlabel{3.12m}{{3.27}{100}{Derivation of the Feynman rules}{equation.466}{}}
\newlabel{3.15m}{{3.30}{100}{Derivation of the Feynman rules}{equation.469}{}}
\newlabel{sec1}{{3.4}{101}{Computation of impulse}{section.470}{}}
\newlabel{4.1w}{{3.31}{101}{Computation of impulse}{equation.471}{}}
\newlabel{e1:barwq2}{{3.34}{102}{1PM contribution to impulse}{equation.476}{}}
\newlabel{e:barwq2}{{3.39}{103}{1PM contribution to impulse}{equation.482}{}}
\newlabel{3.42b}{{3.42}{103}{1PM contribution to impulse}{equation.485}{}}
\newlabel{4.23m}{{3.46}{104}{2PM contribution to the impulse}{equation.490}{}}
\newlabel{e:barwq244}{{3.47}{104}{2PM contribution to the impulse}{equation.491}{}}
\newlabel{e:barwq248}{{3.49}{104}{2PM contribution to the impulse}{equation.493}{}}
\newlabel{4.26m}{{3.51}{105}{2PM contribution to the impulse}{equation.495}{}}
\newlabel{4.19f}{{3.52}{105}{2PM contribution to the impulse}{equation.496}{}}
\newlabel{3.53gg}{{3.53}{105}{2PM contribution to the impulse}{equation.497}{}}
\newlabel{3.54kk}{{3.54}{106}{2PM contribution to the impulse}{equation.498}{}}
\newlabel{4.19c}{{3.56}{106}{2PM contribution to the impulse}{equation.500}{}}
\newlabel{4.21cc}{{3.58}{106}{2PM contribution to the impulse}{equation.502}{}}
\newlabel{4.25j}{{3.59}{107}{2PM contribution to the impulse}{equation.503}{}}
\newlabel{e:barwq246}{{3.62}{107}{2PM contribution to the impulse}{equation.507}{}}
\newlabel{4.37}{{3.67}{108}{2PM contribution to the impulse}{equation.513}{}}
\newlabel{new11}{{3.69}{109}{2PM contribution to the impulse}{equation.515}{}}
\newlabel{eq:4.35}{{3.70}{109}{2PM contribution to the impulse}{equation.516}{}}
\newlabel{eq:4.37}{{3.72}{109}{2PM contribution to the impulse}{equation.518}{}}
\newlabel{eq:4.38}{{3.73}{109}{2PM contribution to the impulse}{equation.519}{}}
\newlabel{e:barwqpp}{{3.74}{109}{2PM contribution to the impulse}{equation.520}{}}
\newlabel{e:barwq24}{{3.76}{110}{2PM contribution to the impulse}{equation.522}{}}
\newlabel{vert1}{{3.77}{110}{2PM contribution to the impulse}{equation.523}{}}
\newlabel{sa}{{3.79}{110}{2PM contribution to the impulse}{equation.525}{}}
\newlabel{4.33a}{{3.81}{111}{2PM contribution to the impulse}{equation.527}{}}
\newlabel{e:barwqp}{{3.84}{111}{2PM contribution to the impulse}{equation.530}{}}
\newlabel{4.51b}{{3.86}{112}{2PM contribution to the impulse}{equation.532}{}}
\newlabel{4.54r}{{3.89}{112}{2PM contribution to the impulse}{equation.535}{}}
\newlabel{4.55m}{{3.90}{113}{2PM contribution to the impulse}{equation.536}{}}
\newlabel{4.56e}{{3.91}{113}{2PM contribution to the impulse}{equation.537}{}}
\newlabel{4.57mm}{{3.93}{113}{2PM contribution to the impulse}{equation.539}{}}
\newlabel{4.58}{{3.94}{114}{2PM contribution to the impulse}{equation.540}{}}
\newlabel{4.58o}{{3.95}{114}{2PM contribution to the impulse}{equation.541}{}}
\newlabel{4.63pp}{{3.97}{114}{2PM contribution to the impulse}{equation.543}{}}
\newlabel{4.55e}{{3.98}{115}{2PM contribution to the impulse}{equation.544}{}}
\newlabel{3.99kk}{{3.99}{115}{2PM contribution to the impulse}{equation.545}{}}
\newlabel{4.57a}{{3.101}{115}{2PM contribution to the impulse}{equation.547}{}}
\newlabel{4.58a}{{3.102}{115}{2PM contribution to the impulse}{equation.548}{}}
\newlabel{4.66t}{{3.104}{116}{2PM contribution to the impulse}{equation.550}{}}
\newlabel{4.6 mm}{{3.105}{116}{2PM contribution to the impulse}{equation.551}{}}
\newlabel{4.37a}{{3.106}{116}{2PM contribution to the impulse}{equation.552}{}}
\newlabel{4.38a}{{3.107}{117}{2PM contribution to the impulse}{equation.553}{}}
\newlabel{3.13}{{3.110}{117}{2PM contribution to the impulse}{equation.556}{}}
\newlabel{3.14}{{3.111}{117}{2PM contribution to the impulse}{equation.557}{}}
\newlabel{eee22}{{3.116}{118}{2PM contribution to the impulse}{equation.562}{}}
\newlabel{4.86}{{3.121}{120}{2PM contribution to the impulse}{equation.567}{}}
\newlabel{ch2:sec5}{{3.5}{120}{Computation of waveform}{section.568}{}}
\newlabel{e:bar}{{3.122}{120}{Computation of waveform}{equation.569}{}}
\newlabel{5.5m}{{3.127}{121}{Computation of waveform}{equation.574}{}}
\newlabel{wave0}{{3.128}{122}{Scalar waveform at 1PM}{equation.578}{}}
\newlabel{5.47 m}{{3.130}{122}{Scalar waveform at 1PM}{equation.580}{}}
\newlabel{wave1}{{3.131}{122}{Scalar waveform at 1PM}{equation.581}{}}
\newlabel{5.11 mn}{{3.132}{123}{Scalar waveform at 2PM}{equation.583}{}}
\newlabel{4PHI}{{3.134}{123}{Scalar waveform at 2PM}{equation.585}{}}
\newlabel{5.16k}{{3.137}{124}{Scalar waveform at 2PM}{equation.588}{}}
\newlabel{5.18k}{{3.139}{124}{Scalar waveform at 2PM}{equation.590}{}}
\newlabel{5.19k}{{3.140}{124}{Scalar waveform at 2PM}{equation.591}{}}
\newlabel{5.7mm}{{3.142}{125}{Scalar waveform at 2PM}{equation.593}{}}
\newlabel{e:ba}{{3.145}{125}{Scalar waveform at 2PM}{equation.596}{}}
\newlabel{5.10 m}{{3.146}{125}{Scalar waveform at 2PM}{equation.597}{}}
\newlabel{7.6}{{3.149}{126}{Scalar waveform at 2PM}{equation.600}{}}
\newlabel{5.24 mmm}{{3.155}{127}{Scalar waveform at 2PM}{equation.606}{}}
\newlabel{5.22 m}{{3.156}{127}{Scalar waveform at 2PM}{equation.607}{}}
\newlabel{ch2:5.23a}{{3.157}{127}{Scalar waveform at 2PM}{equation.608}{}}
\newlabel{5.25mmm}{{3.159}{127}{Scalar waveform at 2PM}{equation.610}{}}
\newlabel{e:::}{{3.161}{127}{Scalar waveform at 2PM}{equation.612}{}}
\newlabel{5.38nm}{{3.162}{128}{Scalar waveform at 2PM}{equation.613}{}}
\newlabel{5.29mmm}{{3.163}{128}{Scalar waveform at 2PM}{equation.614}{}}
\newlabel{Ansatz2}{{3.165}{128}{Scalar waveform at 2PM}{equation.616}{}}
\newlabel{NV2}{{3.166}{128}{Scalar waveform at 2PM}{equation.617}{}}
\newlabel{NIN}{{3.167}{128}{Scalar waveform at 2PM}{equation.618}{}}
\newlabel{m:}{{3.170}{129}{Scalar waveform at 2PM}{equation.621}{}}
\newlabel{CAINT}{{3.171}{129}{Scalar waveform at 2PM}{equation.622}{}}
\newlabel{IMUNU}{{3.173}{129}{Scalar waveform at 2PM}{equation.624}{}}
\newlabel{e:}{{3.174}{129}{Scalar waveform at 2PM}{equation.625}{}}
\newlabel{5.43mm}{{3.175}{129}{Scalar waveform at 2PM}{equation.626}{}}
\newlabel{e::}{{3.176}{129}{Scalar waveform at 2PM}{equation.627}{}}
\newlabel{ex6}{{3.177}{130}{Scalar waveform at 2PM}{equation.628}{}}
\newlabel{5.46 m}{{3.178}{130}{Scalar waveform at 2PM}{equation.629}{}}
\newlabel{ex3}{{3.179}{130}{Scalar waveform at 2PM}{equation.630}{}}
\newlabel{5.57 m}{{3.180}{130}{Scalar waveform at 2PM}{equation.631}{}}
\newlabel{ex}{{3.181}{130}{Scalar waveform at 2PM}{equation.632}{}}
\newlabel{ex1}{{3.182}{130}{Scalar waveform at 2PM}{equation.633}{}}
\newlabel{5.50 m}{{3.183}{131}{Scalar waveform at 2PM}{equation.634}{}}
\newlabel{ex4}{{3.185}{131}{Scalar waveform at 2PM}{equation.636}{}}
\newlabel{5.62x}{{3.186}{131}{Scalar waveform at 2PM}{equation.637}{}}
\newlabel{6.30d}{{3.188}{132}{Scalar waveform at 2PM}{equation.639}{}}
\newlabel{6.33d}{{3.191}{132}{Scalar waveform at 2PM}{equation.642}{}}
\newlabel{5.70l}{{3.194}{133}{Scalar waveform at 2PM}{equation.645}{}}
\newlabel{6.35}{{3.195}{133}{Scalar waveform at 2PM}{equation.646}{}}
\newlabel{5.72l}{{3.196}{134}{Scalar waveform at 2PM}{equation.647}{}}
\newlabel{5.63a}{{3.198}{134}{Gravitational waveform at 2PM: contribution due to extra scalar DOF}{equation.650}{}}
\newlabel{5.55mm}{{3.200}{134}{Gravitational waveform at 2PM: contribution due to extra scalar DOF}{equation.652}{}}
\newlabel{5.56 mm}{{3.201}{134}{Gravitational waveform at 2PM: contribution due to extra scalar DOF}{equation.653}{}}
\newlabel{ch2:e:grav-waveform}{{3.203}{135}{Gravitational waveform at 2PM: contribution due to extra scalar DOF}{equation.655}{}}
\newlabel{sec6}{{3.6}{135}{Towards massive waveform}{section.656}{}}
\newlabel{6.1 m}{{3.204}{135}{Towards massive waveform}{equation.657}{}}
\newlabel{6.11 m}{{3.205}{135}{}{equation.658}{}}
\newlabel{6.11 mm}{{3.206}{135}{}{equation.659}{}}
\newlabel{6.4 mm}{{3.207}{135}{}{equation.661}{}}
\newlabel{6.10 m}{{3.208}{136}{}{equation.662}{}}
\newlabel{6.13 m}{{3.209}{136}{Stationary Phase (SP) approximation: the need and general procedure}{equation.664}{}}
\newlabel{6.14 m}{{3.210}{136}{Stationary Phase (SP) approximation: the need and general procedure}{equation.665}{}}
\newlabel{ch2:e:massive-waveform}{{3.222}{138}{Dealing with the massive integrals via SP approximation}{equation.678}{}}
\newlabel{6.35 m}{{3.226}{138}{Dealing with the massive integrals via SP approximation}{equation.682}{}}
\newlabel{6.36 m}{{3.227}{139}{Dealing with the massive integrals via SP approximation}{equation.683}{}}
\newlabel{6.37 m}{{3.228}{139}{Dealing with the massive integrals via SP approximation}{equation.684}{}}
\newlabel{6.19ab}{{3.230}{139}{Dealing with the massive integrals via SP approximation}{equation.686}{}}
\newlabel{6.20 m}{{3.233}{139}{Dealing with the massive integrals via SP approximation}{equation.689}{}}
\newlabel{ch2:5.23b}{{3.235}{140}{Dealing with the massive integrals via SP approximation}{equation.691}{}}
\newlabel{6.26f}{{3.237}{140}{Dealing with the massive integrals via SP approximation}{equation.693}{}}
\newlabel{6.36f}{{3.247}{142}{Dealing with the massive integrals via SP approximation}{equation.703}{}}
\newlabel{6.38a}{{3.251}{142}{}{equation.707}{}}
\newlabel{sec8}{{3.7}{143}{Discussions and outlook}{section.714}{}}
\newlabel{ch2:app:A}{{3.A}{145}{Sketching the derivation of the worldline action}{section.715}{}}
\newlabel{D.5}{{3.262}{145}{Sketching the derivation of the worldline action}{equation.720}{}}
\newlabel{ch2:app:B}{{3.B}{146}{Impulse via Eikonal: a connection to scattering amplitude}{section.721}{}}
\newlabel{B.8a}{{3.268}{147}{Impulse via Eikonal: a connection to scattering amplitude}{equation.727}{}}
\newlabel{B.9c}{{3.269}{147}{Impulse via Eikonal: a connection to scattering amplitude}{equation.728}{}}
\newlabel{B.12}{{3.273}{148}{Impulse via Eikonal: a connection to scattering amplitude}{equation.732}{}}
\newlabel{B.16ab}{{3.276}{149}{Impulse via Eikonal: a connection to scattering amplitude}{equation.735}{}}
\newlabel{B.16a}{{3.278}{149}{Impulse via Eikonal: a connection to scattering amplitude}{equation.737}{}}
\newlabel{B.15}{{3.282}{149}{Impulse via Eikonal: a connection to scattering amplitude}{equation.741}{}}
\newlabel{B.18b}{{3.284}{150}{Impulse via Eikonal: a connection to scattering amplitude}{equation.743}{}}
\newlabel{B.20a}{{3.286}{150}{Impulse via Eikonal: a connection to scattering amplitude}{equation.745}{}}
\newlabel{B.22aa}{{3.288}{151}{Impulse via Eikonal: a connection to scattering amplitude}{equation.748}{}}
\newlabel{ch2:app:C}{{3.C}{151}{Two master integrals used for the computation of waveform}{section.749}{}}
\newlabel{A.2}{{3.290}{151}{Two master integrals used for the computation of waveform}{equation.751}{}}
\newlabel{C.4 mm}{{3.292}{152}{Two master integrals used for the computation of waveform}{equation.753}{}}
\newlabel{A.4}{{3.293}{152}{Two master integrals used for the computation of waveform}{equation.754}{}}
\newlabel{ch2:A.6}{{3.295}{152}{Two master integrals used for the computation of waveform}{equation.756}{}}
\newlabel{A.8}{{3.297}{153}{Two master integrals used for the computation of waveform}{equation.758}{}}
\newlabel{A.9 m}{{3.298}{153}{Two master integrals used for the computation of waveform}{equation.759}{}}
\newlabel{C.11a}{{3.299}{153}{Two master integrals used for the computation of waveform}{equation.760}{}}
\newlabel{B.16}{{3.304}{154}{Two master integrals used for the computation of waveform}{equation.765}{}}
\newlabel{B.18}{{3.306}{154}{Two master integrals used for the computation of waveform}{equation.767}{}}
\newlabel{A.19 m}{{3.308}{154}{Two master integrals used for the computation of waveform}{equation.769}{}}
\newlabel{ch2:app:D}{{3.D}{155}{\texorpdfstring {Analysis through method of regions for $\lambda _4\varphi ^4$ vertex contribution in the waveform}{Analysis through method of regions for lambda4 phi4 vertex contribution in the waveform}}{section.773}{}}
\newlabel{FstRe1}{{3.312}{155}{\texorpdfstring {Analysis through method of regions for $\lambda _4\varphi ^4$ vertex contribution in the waveform}{Analysis through method of regions for lambda4 phi4 vertex contribution in the waveform}}{equation.774}{}}
\newlabel{FstRe}{{3.313}{155}{\texorpdfstring {Analysis through method of regions for $\lambda _4\varphi ^4$ vertex contribution in the waveform}{Analysis through method of regions for lambda4 phi4 vertex contribution in the waveform}}{equation.775}{}}
\newlabel{2NdRe}{{3.314}{155}{\texorpdfstring {Analysis through method of regions for $\lambda _4\varphi ^4$ vertex contribution in the waveform}{Analysis through method of regions for lambda4 phi4 vertex contribution in the waveform}}{equation.776}{}}
\newlabel{nocont}{{3.315}{156}{\texorpdfstring {Analysis through method of regions for $\lambda _4\varphi ^4$ vertex contribution in the waveform}{Analysis through method of regions for lambda4 phi4 vertex contribution in the waveform}}{equation.777}{}}
\newlabel{ch2:app:E}{{3.E}{156}{Massive integral coming from derivative interaction}{section.778}{}}
\newlabel{A.22 m}{{3.316}{156}{Massive integral coming from derivative interaction}{equation.779}{}}
\newlabel{A.23 m}{{3.317}{156}{Massive integral coming from derivative interaction}{equation.780}{}}
\newlabel{A.24 m}{{3.318}{156}{Massive integral coming from derivative interaction}{equation.781}{}}
\newlabel{A.30 m}{{3.324}{157}{Massive integral coming from derivative interaction}{equation.787}{}}
\newlabel{A.38}{{3.332}{159}{Massive integral coming from derivative interaction}{equation.795}{}}
\newlabel{ch2:app:F}{{3.F}{159}{Computation of worldline radiation diagram using large velocity approximation}{section.798}{}}
\newlabel{7.7}{{3.335}{159}{Computation of worldline radiation diagram using large velocity approximation}{equation.799}{}}
\newlabel{celestial}{{3.1}{160}{Sketch of the celestial sphere under consideration}{figure.caption.803}{}}
\newlabel{ch2:e:collinear-waveform}{{3.342}{161}{Computation of worldline radiation diagram using large velocity approximation}{equation.807}{}}
\newlabel{3.54kkk}{{3.G}{161}{Proof of the identity in \texorpdfstring {\eqref {3.54kk}}{Eq. 3.54kk}}{section.808}{}}
\newlabel{App:3.H}{{3.H}{162}{Comment on the integral convergence}{section.815}{}}
\newlabel{fig:3.2}{{3.2}{165}{Figure showing dependency of the integral $I(b)$ as a function of $|b|$ for $m=1\,.$}{figure.caption.842}{}}
\newlabel{ch3:sec2}{{4.2}{171}{Chern-Simons gravity and worldline QFT}{section.868}{}}
\newlabel{2.4a}{{4.4}{171}{}{equation.872}{}}
\newlabel{2.13k}{{4.13}{173}{}{equation.881}{}}
\newlabel{2.13d}{{4.15}{173}{}{equation.883}{}}
\newlabel{SSC17}{{4.17}{174}{}{equation.886}{}}
\newlabel{2.18a}{{4.20}{175}{}{equation.890}{}}
\newlabel{ch3:2.13}{{4.23}{175}{}{equation.893}{}}
\newlabel{2.19}{{4.24}{175}{}{equation.894}{}}
\newlabel{ch3:sec3}{{4.3}{176}{Bootstrapping Post-Minkowskian (PM) physics from Post-Newtonian (PN) data: the idea, technology and example}{section.896}{}}
\newlabel{3.3a}{{4.28}{177}{Bootstrapping Post-Minkowskian (PM) physics from Post-Newtonian (PN) data: the idea, technology and example}{equation.899}{}}
\newlabel{3.4a}{{4.30}{177}{Bootstrapping Post-Minkowskian (PM) physics from Post-Newtonian (PN) data: the idea, technology and example}{equation.901}{}}
\newlabel{3.7}{{4.33}{178}{Bootstrapping Post-Minkowskian (PM) physics from Post-Newtonian (PN) data: the idea, technology and example}{equation.904}{}}
\newlabel{2.14a}{{4.35}{178}{Bootstrapping Post-Minkowskian (PM) physics from Post-Newtonian (PN) data: the idea, technology and example}{equation.906}{}}
\newlabel{3.10a}{{4.36}{178}{Bootstrapping Post-Minkowskian (PM) physics from Post-Newtonian (PN) data: the idea, technology and example}{equation.907}{}}
\newlabel{3.12}{{4.38}{179}{Bootstrapping Post-Minkowskian (PM) physics from Post-Newtonian (PN) data: the idea, technology and example}{equation.910}{}}
\newlabel{2.17}{{4.40}{179}{Bootstrapping Post-Minkowskian (PM) physics from Post-Newtonian (PN) data: the idea, technology and example}{equation.912}{}}
\newlabel{3.15}{{4.41}{179}{Bootstrapping Post-Minkowskian (PM) physics from Post-Newtonian (PN) data: the idea, technology and example}{equation.913}{}}
\newlabel{3.16}{{4.42}{179}{Bootstrapping Post-Minkowskian (PM) physics from Post-Newtonian (PN) data: the idea, technology and example}{equation.914}{}}
\newlabel{2.23a}{{4.43}{180}{Bootstrapping Post-Minkowskian (PM) physics from Post-Newtonian (PN) data: the idea, technology and example}{equation.915}{}}
\newlabel{MasterM}{{4.3}{181}{Bootstrapping Post-Minkowskian (PM) physics from Post-Newtonian (PN) data: the idea, technology and example}{equation.921}{}}
\newlabel{3.43k}{{4.50}{181}{Bootstrapping Post-Minkowskian (PM) physics from Post-Newtonian (PN) data: the idea, technology and example}{equation.923}{}}
\newlabel{3.29}{{4.52}{182}{Bootstrapping Post-Minkowskian (PM) physics from Post-Newtonian (PN) data: the idea, technology and example}{equation.926}{}}
\newlabel{MasterC}{{4.54}{182}{Bootstrapping Post-Minkowskian (PM) physics from Post-Newtonian (PN) data: the idea, technology and example}{equation.928}{}}
\newlabel{3.37h}{{4.55}{182}{Bootstrapping Post-Minkowskian (PM) physics from Post-Newtonian (PN) data: the idea, technology and example}{equation.929}{}}
\newlabel{MasterL}{{4.58}{183}{Bootstrapping Post-Minkowskian (PM) physics from Post-Newtonian (PN) data: the idea, technology and example}{equation.932}{}}
\newlabel{3.41a}{{4.59}{183}{Bootstrapping Post-Minkowskian (PM) physics from Post-Newtonian (PN) data: the idea, technology and example}{equation.933}{}}
\newlabel{3.4aa}{{4.60}{183}{Bootstrapping Post-Minkowskian (PM) physics from Post-Newtonian (PN) data: the idea, technology and example}{equation.934}{}}
\newlabel{ch3:sec4}{{4.4}{184}{Computation of eikonal phase}{section.937}{}}
\newlabel{4.9}{{4.74}{186}{Computation of eikonal phase}{equation.950}{}}
\newlabel{ch3:4.10}{{4.75}{187}{Computation of eikonal phase}{equation.951}{}}
\newlabel{eq1}{{4.83}{189}{}{equation.961}{}}
\newlabel{eq2}{{4.87}{190}{}{equation.965}{}}
\newlabel{eqt1}{{4.88}{190}{}{equation.966}{}}
\newlabel{eq3}{{4.89}{190}{}{equation.967}{}}
\newlabel{eq4}{{4.90}{190}{}{equation.968}{}}
\newlabel{eq5}{{4.91}{190}{}{equation.969}{}}
\newlabel{eq6}{{4.92}{191}{}{equation.970}{}}
\newlabel{eqt2}{{4.93}{191}{}{equation.971}{}}
\newlabel{eqt3}{{4.95}{191}{}{equation.973}{}}
\newlabel{eq8}{{4.98}{192}{}{equation.976}{}}
\newlabel{eq9}{{4.99}{192}{}{equation.977}{}}
\newlabel{eq10}{{4.100}{192}{}{equation.978}{}}
\newlabel{eq11}{{4.101}{192}{}{equation.979}{}}
\newlabel{eqt4}{{4.102}{193}{}{equation.980}{}}
\newlabel{eqt5}{{4.104}{193}{}{equation.982}{}}
\newlabel{eqt6}{{4.113}{195}{}{equation.991}{}}
\newlabel{eqt7}{{4.121}{197}{}{equation.999}{}}
\newlabel{eqt7b}{{4.130}{201}{}{equation.1008}{}}
\newlabel{eqt8}{{4.132}{201}{}{equation.1010}{}}
\newlabel{eqt9}{{4.136}{202}{}{equation.1014}{}}
\newlabel{eqt10}{{4.137}{202}{}{equation.1015}{}}
\newlabel{eqt11}{{4.140}{203}{}{equation.1018}{}}
\newlabel{4.68j}{{4.143}{203}{}{equation.1021}{}}
\newlabel{eqt12}{{4.148}{204}{}{equation.1026}{}}
\newlabel{eqt13}{{4.150}{205}{}{equation.1028}{}}
\newlabel{eqt14}{{4.152}{205}{}{equation.1030}{}}
\newlabel{eqt15}{{4.154}{206}{}{equation.1032}{}}
\newlabel{eqt16}{{4.156}{207}{}{equation.1034}{}}
\newlabel{eqt17}{{4.159}{207}{}{equation.1037}{}}
\newlabel{eqt18}{{4.164}{209}{}{equation.1042}{}}
\newlabel{eqt19}{{4.166}{209}{}{equation.1044}{}}
\newlabel{eqt20}{{4.169}{210}{}{equation.1047}{}}
\newlabel{eqt21}{{4.171}{211}{}{equation.1049}{}}
\newlabel{eqt22}{{4.173}{211}{}{equation.1051}{}}
\newlabel{eqt23}{{4.177}{212}{}{equation.1055}{}}
\newlabel{3.79}{{4.182}{212}{}{equation.1060}{}}
\newlabel{3.81}{{4.184}{213}{}{equation.1062}{}}
\newlabel{eqt24}{{4.185}{213}{}{equation.1063}{}}
\newlabel{dCSne3waaa}{{4.188}{214}{}{equation.1066}{}}
\newlabel{dC7Snew}{{4.189}{214}{}{equation.1067}{}}
\newlabel{4.133a}{{4.190}{215}{}{equation.1068}{}}
\newlabel{con}{{4.5}{215}{Conclusion and outlook}{section.1069}{}}
\newlabel{6.1k}{{4.191}{216}{Conclusion and outlook}{equation.1070}{}}
\newlabel{6.2k}{{4.192}{216}{Conclusion and outlook}{equation.1071}{}}
\newlabel{5.5a}{{4.197}{217}{Conclusion and outlook}{equation.1076}{}}
\newlabel{ch3:app:A}{{4.A}{217}{Useful Feynman integrals}{section.1077}{}}
\newlabel{ch3:tensorfourierindentity}{{4.198}{217}{Useful Feynman integrals}{equation.1078}{}}
\newlabel{int1}{{4.198}{217}{Useful Feynman integrals}{equation.1078}{}}
\newlabel{int}{{4.199}{217}{Useful Feynman integrals}{equation.1078}{}}
\newlabel{ch3:eq:1loop}{{4.203}{218}{Useful Feynman integrals}{equation.1079}{}}
\newlabel{intextra}{{4.208}{219}{Useful Feynman integrals}{equation.1084}{}}
\newlabel{ch3:app:B}{{4.B}{219}{Boundary integrals in potential mode relevant for solving the differential equation}{section.1088}{}}
\newlabel{2.30}{{4.212}{219}{Boundary integrals in potential mode relevant for solving the differential equation}{equation.1089}{}}
\newlabel{ch3:app:C}{{4.C}{221}{\texorpdfstring {$A(x,\epsilon )$ in $\epsilon $ factorized form}{A(x, epsilon) in epsilon-factorized form}}{section.1099}{}}
\newlabel{ch3:app:D}{{4.D}{222}{Iterative UT integrals}{section.1101}{}}
\newlabel{ch3:app:E}{{4.E}{223}{\texorpdfstring {$\boldsymbol {\tilde {h}}$ coefficients}{tilde-h coefficients}}{section.1103}{}}
\newlabel{ch4:sec2}{{5.2}{234}{Setup and methodology}{section.1106}{}}
\newlabel{1.3h}{{5.4}{234}{}{equation.1110}{}}
\newlabel{ch4:sec3}{{5.3}{236}{Computation of scattering amplitude: Soft-loop expansion and classical limit}{section.1120}{}}
\newlabel{3.28}{{5.42}{241}{}{equation.1150}{}}
\newlabel{3.54j}{{5.65}{248}{}{equation.1173}{}}
\newlabel{ch4:sec4}{{5.4}{249}{Post-Minkowskian potential: The Lippmann-Schwinger equation and the Infrared subtraction}{section.1176}{}}
\newlabel{4.2f}{{5.67}{249}{Post-Minkowskian potential: The Lippmann-Schwinger equation and the Infrared subtraction}{equation.1178}{}}
\newlabel{fig1}{{5.1}{250}{Diagrammatic representation of Lippmann-Schwinger equation}{figure.caption.1181}{}}
\newlabel{4.4g}{{5.69}{250}{Post-Minkowskian potential: The Lippmann-Schwinger equation and the Infrared subtraction}{equation.1180}{}}
\newlabel{4.5h}{{5.70}{250}{Post-Minkowskian potential: The Lippmann-Schwinger equation and the Infrared subtraction}{equation.1182}{}}
\newlabel{eq:Gexp-large-s}{{5.73}{251}{A digression on Green's functions}{equation.1186}{}}
\newlabel{eq:G-ours-leading}{{5.74}{251}{A digression on Green's functions}{equation.1187}{}}
\newlabel{4.20jj}{{5.82}{252}{A digression on Green's functions}{equation.1195}{}}
\newlabel{4.10o}{{5.83}{252}{A digression on Green's functions}{equation.1196}{}}
\newlabel{4.21g}{{5.93}{254}{A digression on Green's functions}{equation.1206}{}}
\newlabel{eq:In_def_refreply}{{5.94}{254}{A digression on Green's functions}{equation.1208}{}}
\newlabel{eq:In_scaling_refreply}{{5.95}{254}{A digression on Green's functions}{equation.1209}{}}
\newlabel{eq:Vn_falloff_refreply}{{5.96}{254}{A digression on Green's functions}{equation.1210}{}}
\newlabel{ch4:sec5}{{5.5}{255}{Eikonal exponentiation and the scattering angle}{section.1212}{}}
\newlabel{del0p}{{5.103}{256}{Eikonal exponentiation and the scattering angle}{equation.1218}{}}
\newlabel{del1p}{{5.105}{256}{Eikonal exponentiation and the scattering angle}{equation.1220}{}}
\newlabel{5.14}{{5.110}{257}{Eikonal exponentiation and the scattering angle}{equation.1225}{}}
\newlabel{5.15}{{5.111}{257}{Eikonal exponentiation and the scattering angle}{equation.1226}{}}
\newlabel{5.17}{{5.113}{257}{Eikonal exponentiation and the scattering angle}{equation.1228}{}}
\newlabel{sec:conclusions}{{5.6}{258}{Conclusions and Discussion}{section.1235}{}}
\newlabel{ch5:sec3}{{6.2}{262}{ Celestial eikonal amplitude for quadratic gravity}{section.1238}{}}
\newlabel{3.1i}{{6.1}{262}{ Celestial eikonal amplitude for quadratic gravity}{equation.1239}{}}
\newlabel{3.2h}{{6.2}{262}{ Celestial eikonal amplitude for quadratic gravity}{equation.1240}{}}
\newlabel{ch5:fig3}{{6.1}{263}{Figure describing the $s,t,u$-channel diagram respectively, relevant for our computation}{figure.caption.1243}{}}
\newlabel{3.9p}{{6.10}{265}{ Celestial eikonal amplitude for quadratic gravity}{equation.1251}{}}
\newlabel{3.8p}{{6.11}{265}{ Celestial eikonal amplitude for quadratic gravity}{equation.1252}{}}
\newlabel{3.10e}{{6.12}{266}{ Celestial eikonal amplitude for quadratic gravity}{equation.1253}{}}
\newlabel{4.18m}{{6.14}{266}{ Celestial eikonal amplitude for quadratic gravity}{equation.1255}{}}
\newlabel{3.19p}{{6.20}{268}{Eikonal amplitude in quadratic gravity}{equation.1264}{}}
\newlabel{3.18j}{{6.23}{269}{Eikonal amplitude in quadratic gravity}{equation.1267}{}}
\newlabel{3.23n}{{6.24}{269}{Eikonal amplitude in quadratic gravity}{equation.1268}{}}
\newlabel{3.19k}{{6.25}{269}{Eikonal amplitude in quadratic gravity}{equation.1269}{}}
\newlabel{3.18p}{{6.26}{269}{Eikonal amplitude in quadratic gravity}{equation.1270}{}}
\newlabel{3.20k}{{6.27}{270}{Eikonal amplitude in quadratic gravity}{equation.1272}{}}
\newlabel{3.23j}{{6.28}{270}{Eikonal amplitude in quadratic gravity}{equation.1273}{}}
\newlabel{3.27y}{{6.30}{271}{Eikonal amplitude in quadratic gravity}{equation.1275}{}}
\newlabel{3.32a}{{6.31}{271}{Analyticity and dispersion relation for celestial eikonal amplitude}{equation.1278}{}}
\newlabel{3.27k}{{6.32}{271}{Analyticity and dispersion relation for celestial eikonal amplitude}{equation.1279}{}}
\newlabel{3.35b1}{{6.34}{272}{Analyticity and dispersion relation for celestial eikonal amplitude}{equation.1282}{}}
\newlabel{3.34p}{{6.38}{273}{Analyticity and dispersion relation for celestial eikonal amplitude}{equation.1288}{}}
\newlabel{3.36m}{{6.42}{274}{Analyticity and dispersion relation for celestial eikonal amplitude}{equation.1292}{}}
\newlabel{eqnnn}{{6.43}{274}{Analyticity and dispersion relation for celestial eikonal amplitude}{equation.1293}{}}
\newlabel{3.26uu}{{6.44}{274}{Dispersion Relations}{equation.1296}{}}
\newlabel{fig:o}{{6.2}{275}{Figure depicting the chosen contour for dispersion relation in the complex-$\omega $ plane. The cross signs depict the location of the poles}{figure.caption.1297}{}}
\newlabel{3.53e}{{6.50}{276}{Dispersion Relations}{equation.1303}{}}
\newlabel{ch5:sec4}{{6.3}{276}{Shadowed correlator and operator product expansion (OPE)}{section.1304}{}}
\newlabel{4.6j}{{6.55}{277}{Shadowed amplitude corresponding to eikonal amplitude }{equation.1310}{}}
\newlabel{4.6l}{{6.56}{277}{Shadowed amplitude corresponding to eikonal amplitude }{equation.1311}{}}
\newlabel{4.28i}{{6.60}{278}{Shadowed amplitude corresponding to celestial Born amplitude}{equation.1316}{}}
\newlabel{s-channel}{{6.64}{279}{Shadowed amplitude corresponding to celestial Born amplitude}{equation.1322}{}}
\newlabel{4.16o}{{6.69}{280}{Shadowed amplitude corresponding to celestial Born amplitude}{equation.1328}{}}
\newlabel{4.40o}{{6.70}{281}{Shadowed amplitude corresponding to celestial Born amplitude}{equation.1329}{}}
\newlabel{4.29}{{6.71}{281}{Conformal block expansion and partial wave coefficients}{equation.1331}{}}
\newlabel{4.36k}{{6.73}{282}{Conformal block expansion and partial wave coefficients}{equation.1333}{}}
\newlabel{5.33u}{{6.75}{282}{Conformal block expansion and partial wave coefficients}{equation.1336}{}}
\newlabel{eq:F1exp}{{6.74}{282}{}{equation.1335}{}}
\newlabel{4.46o}{{6.78}{283}{Conformal block expansion and partial wave coefficients}{equation.1339}{}}
\newlabel{5.35}{{6.79}{283}{}{equation.1341}{}}
\newlabel{4.46p}{{6.82}{284}{}{equation.1345}{}}
\newlabel{4.47y}{{6.83}{284}{}{equation.1346}{}}
\newlabel{4.55u}{{6.84}{284}{}{equation.1348}{}}
\newlabel{approx1}{{6.85}{285}{}{equation.1350}{}}
\newlabel{4.36l}{{6.87}{285}{}{equation.1352}{}}
\newlabel{4.62i}{{6.88}{286}{}{equation.1354}{}}
\newlabel{4.63e1}{{6.89}{286}{}{equation.1355}{}}
\newlabel{4.63e}{{6.90}{287}{}{equation.1357}{}}
\newlabel{fig5f}{{6.3}{287}{Plot depicting the matching for extraction of OPE coefficient using BC expansion (\textbf {red}) and OPE inversion for spin-0 (\textbf {blue}). We have set the conformal dimension for the external primaries to be: $\Delta _{\mathcal {O}}=1+i$}{figure.caption.1358}{}}
\newlabel{ch5:sec5}{{6.4}{288}{Connection to Carrollian amplitude}{section.1359}{}}
\newlabel{4.39}{{6.91}{288}{Connection to Carrollian amplitude}{equation.1360}{}}
\newlabel{5.2w}{{6.92}{288}{Connection to Carrollian amplitude}{equation.1361}{}}
\newlabel{ch5:sec7}{{6.5}{289}{Conclusion and Discussion}{section.1364}{}}
\newlabel{ch5:app:A}{{6.A}{290}{Few definitions and conventions}{section.1368}{}}
\newlabel{eq:F1toOne}{{6.96}{291}{Few definitions and conventions}{equation.1370}{}}
\newlabel{eq:F1toInf}{{6.97}{291}{Few definitions and conventions}{equation.1371}{}}
\newlabel{5.17u}{{6.100}{291}{Few definitions and conventions}{equation.1374}{}}
\bibcite{Porto:2016pyg}{{1}{}{{}}{{}}}
\bibcite{LIGOScientific:2016aoc}{{2}{}{{}}{{}}}
\bibcite{LIGOScientific:2016sjg}{{3}{}{{}}{{}}}
\bibcite{LIGOScientific:2016vlm}{{4}{}{{}}{{}}}
\bibcite{LIGOScientific:2017bnn}{{5}{}{{}}{{}}}
\bibcite{Kokeyama:2020dkg}{{6}{}{{}}{{}}}
\bibcite{LIGOScientific:2019hgc}{{7}{}{{}}{{}}}
\bibcite{Gourgoulhon:2012}{{8}{}{{}}{{}}}
\bibcite{Lehner:2014asa}{{9}{}{{}}{{}}}
\bibcite{LIGOScientific:2014oec}{{10}{}{{}}{{}}}
\bibcite{weyl}{{11}{}{{}}{{}}}
\bibcite{1962JMP.....3..608U}{{12}{}{{}}{{}}}
\bibcite{PhysRevD.16.953}{{13}{}{{}}{{}}}
\bibcite{PhysRev.79.145}{{14}{}{{}}{{}}}
\bibcite{FRADKIN1982469}{{15}{}{{}}{{}}}
\bibcite{Hassan:1991mq}{{16}{}{{}}{{}}}
\bibcite{Holstein:2004dn}{{17}{}{{}}{{}}}
\bibcite{Bern:2019crd}{{18}{}{{}}{{}}}
\bibcite{Bern:2020buy}{{19}{}{{}}{{}}}
\bibcite{Bern:2019nnu}{{20}{}{{}}{{}}}
\bibcite{Bern:2020gjj}{{21}{}{{}}{{}}}
\bibcite{Bern:2021dqo}{{22}{}{{}}{{}}}
\bibcite{Bern:2021xze}{{23}{}{{}}{{}}}
\bibcite{Bern:2022kto}{{24}{}{{}}{{}}}
\bibcite{Bern:2025zno}{{25}{}{{}}{{}}}
\bibcite{Bjerrum-Bohr:2002aqa}{{26}{}{{}}{{}}}
\bibcite{Bjerrum-Bohr:2021wwt}{{27}{}{{}}{{}}}
\bibcite{youtube_484KUavMo0}{{28}{}{{}}{{}}}
\bibcite{Kosower:2018adc}{{29}{}{{}}{{}}}
\bibcite{Kalin:2020mvi}{{30}{}{{}}{{}}}
\bibcite{Kalin:2022hph}{{31}{}{{}}{{}}}
\bibcite{Mogull:2020sak}{{32}{}{{}}{{}}}
\bibcite{Jakobsen:2022psy}{{33}{}{{}}{{}}}
\bibcite{Kim:2024svw}{{34}{}{{}}{{}}}
\bibcite{Magnus:1954}{{35}{}{{}}{{}}}
\bibcite{Murua:2006}{{36}{}{{}}{{}}}
\bibcite{Maldacena:1997re}{{37}{}{{}}{{}}}
\bibcite{Strominger:2013jfa}{{38}{}{{}}{{}}}
\bibcite{Strominger:2017zoo}{{39}{}{{}}{{}}}
\bibcite{Pasterski:2017ylz}{{40}{}{{}}{{}}}
\bibcite{Pasterski:2021raf}{{41}{}{{}}{{}}}
\bibcite{Donnay:2020guq}{{42}{}{{}}{{}}}
\bibcite{He:2014laa}{{43}{}{{}}{{}}}
\bibcite{Adamo:2019ipt}{{44}{}{{}}{{}}}
\bibcite{Gonzalez:2020tpi}{{45}{}{{}}{{}}}
\bibcite{Pasterski:2021fjn}{{46}{}{{}}{{}}}
\bibcite{Gonzo:2022tjm}{{47}{}{{}}{{}}}
\bibcite{Strominger:2021mtt}{{48}{}{{}}{{}}}
\bibcite{Arkani-Hamed:2020gyp}{{49}{}{{}}{{}}}
\bibcite{Pasterski:2016qvg}{{50}{}{{}}{{}}}
\bibcite{Lee:2012cn}{{51}{}{{}}{{}}}
\bibcite{Maierhofer:2017gsa}{{52}{}{{}}{{}}}
\bibcite{Smirnov:2019qkx}{{53}{}{{}}{{}}}
\bibcite{Brunello:2024ibk}{{54}{}{{}}{{}}}
\bibcite{Henn:2013pwa}{{55}{}{{}}{{}}}
\bibcite{Lee:2014ioa}{{56}{}{{}}{{}}}
\bibcite{Lee:2020zfb}{{57}{}{{}}{{}}}
\bibcite{Meyer:2017joq}{{58}{}{{}}{{}}}
\bibcite{Gituliar:2017vzm}{{59}{}{{}}{{}}}
\bibcite{Dlapa:2020cwj}{{60}{}{{}}{{}}}
\bibcite{Bhattacharyya:2023kbh}{{61}{}{{}}{{}}}
\bibcite{Blanchet:2013haa}{{62}{}{{}}{{}}}
\bibcite{Tagoshi:2000zg}{{63}{}{{}}{{}}}
\bibcite{Faye:2006gx}{{64}{}{{}}{{}}}
\bibcite{Blanchet:2006gy}{{65}{}{{}}{{}}}
\bibcite{Blanchet:2004ek}{{66}{}{{}}{{}}}
\bibcite{Damour:2000ni}{{67}{}{{}}{{}}}
\bibcite{Itoh:2003fy}{{68}{}{{}}{{}}}
\bibcite{Boetzel:2019nfw}{{69}{}{{}}{{}}}
\bibcite{Faye:2012we}{{70}{}{{}}{{}}}
\bibcite{Mishra:2013rna}{{71}{}{{}}{{}}}
\bibcite{Fujita:2010xj}{{72}{}{{}}{{}}}
\bibcite{Faye:2014fra}{{73}{}{{}}{{}}}
\bibcite{Blanchet:2023sbv}{{74}{}{{}}{{}}}
\bibcite{Blanchet:2023bwj}{{75}{}{{}}{{}}}
\bibcite{Larrouturou:2021gqo}{{76}{}{{}}{{}}}
\bibcite{Blanchet:2004bb}{{77}{}{{}}{{}}}
\bibcite{Blanchet:2023soy}{{78}{}{{}}{{}}}
\bibcite{Arun:2004hn}{{79}{}{{}}{{}}}
\bibcite{PhysRevD.12.329}{{80}{}{{}}{{}}}
\bibcite{PhysRevD.2.1428}{{81}{}{{}}{{}}}
\bibcite{Kidder:1992fr}{{82}{}{{}}{{}}}
\bibcite{Cho:2022syn}{{83}{}{{}}{{}}}
\bibcite{Steinhoff:2007mb}{{84}{}{{}}{{}}}
\bibcite{Steinhoff:2008ji}{{85}{}{{}}{{}}}
\bibcite{Hergt:2008jn}{{86}{}{{}}{{}}}
\bibcite{Hergt:2010pa}{{87}{}{{}}{{}}}
\bibcite{Porto:2012as}{{88}{}{{}}{{}}}
\bibcite{Enoki:2006kj}{{89}{}{{}}{{}}}
\bibcite{Favata:2011qi}{{90}{}{{}}{{}}}
\bibcite{Munna:2019fjz}{{91}{}{{}}{{}}}
\bibcite{Zhang:2017srh}{{92}{}{{}}{{}}}
\bibcite{AbhishekChowdhuri:2022ora}{{93}{}{{}}{{}}}
\bibcite{Zhang:2018prg}{{94}{}{{}}{{}}}
\bibcite{Saffer:2018jmx}{{95}{}{{}}{{}}}
\bibcite{Lin:2018ken}{{96}{}{{}}{{}}}
\bibcite{Li:2022grj}{{97}{}{{}}{{}}}
\bibcite{Shiralilou:2021mfl}{{98}{}{{}}{{}}}
\bibcite{Julie:2019sab}{{99}{}{{}}{{}}}
\bibcite{Goldberger:2004jt}{{100}{}{{}}{{}}}
\bibcite{Porto:2017dgs}{{101}{}{{}}{{}}}
\bibcite{Porto:2007pw}{{102}{}{{}}{{}}}
\bibcite{Goldberger:2012kf}{{103}{}{{}}{{}}}
\bibcite{Galley:2015kus}{{104}{}{{}}{{}}}
\bibcite{Porto:2010zg}{{105}{}{{}}{{}}}
\bibcite{Levi:2011eq}{{106}{}{{}}{{}}}
\bibcite{Levi:2015uxa}{{107}{}{{}}{{}}}
\bibcite{Levi:2015ixa}{{108}{}{{}}{{}}}
\bibcite{Levi:2014sba}{{109}{}{{}}{{}}}
\bibcite{Maia:2017gxn}{{110}{}{{}}{{}}}
\bibcite{Maia:2017yok}{{111}{}{{}}{{}}}
\bibcite{Foffa:2019yfl}{{112}{}{{}}{{}}}
\bibcite{Porto:2008jj}{{113}{}{{}}{{}}}
\bibcite{Porto:2008tb}{{114}{}{{}}{{}}}
\bibcite{Porto:2007px}{{115}{}{{}}{{}}}
\bibcite{Mandal:2022nty}{{116}{}{{}}{{}}}
\bibcite{Mandal:2022ufb}{{117}{}{{}}{{}}}
\bibcite{Goldberger:2009qd}{{118}{}{{}}{{}}}
\bibcite{Ross:2012fc}{{119}{}{{}}{{}}}
\bibcite{Goldberger:2020fot}{{120}{}{{}}{{}}}
\bibcite{Goldberger:2017ogt}{{121}{}{{}}{{}}}
\bibcite{Goldberger:2006bd}{{122}{}{{}}{{}}}
\bibcite{Goldberger:2005cd}{{123}{}{{}}{{}}}
\bibcite{Goldberger:2022rqf}{{124}{}{{}}{{}}}
\bibcite{Goldberger:2022ebt}{{125}{}{{}}{{}}}
\bibcite{Kuntz:2019zef}{{126}{}{{}}{{}}}
\bibcite{Huang:2018pbu}{{127}{}{{}}{{}}}
\bibcite{Patil:2020dme}{{128}{}{{}}{{}}}
\bibcite{Gupta:2022spq}{{129}{}{{}}{{}}}
\bibcite{GrillidiCortona:2015jxo}{{130}{}{{}}{{}}}
\bibcite{Sanchis-Gual:2022ooi}{{131}{}{{}}{{}}}
\bibcite{Goldstein:2022pxu}{{132}{}{{}}{{}}}
\bibcite{Schutz:2020jox}{{133}{}{{}}{{}}}
\bibcite{Kamionkowski:2014zda}{{134}{}{{}}{{}}}
\bibcite{Adams:2022pbo}{{135}{}{{}}{{}}}
\bibcite{Fukuda:2021drn}{{136}{}{{}}{{}}}
\bibcite{CAPP:2020utb}{{137}{}{{}}{{}}}
\bibcite{Sakhelashvili:2021eid}{{138}{}{{}}{{}}}
\bibcite{Peccei:2006as}{{139}{}{{}}{{}}}
\bibcite{RevModPhys.82.557}{{140}{}{{}}{{}}}
\bibcite{PhysRevLett.40.223}{{141}{}{{}}{{}}}
\bibcite{PhysRevLett.40.279}{{142}{}{{}}{{}}}
\bibcite{Preskill:1982cy}{{143}{}{{}}{{}}}
\bibcite{Gorghetto:2018ocs}{{144}{}{{}}{{}}}
\bibcite{DiLuzio:2021pxd}{{145}{}{{}}{{}}}
\bibcite{Berezhiani:2000gh}{{146}{}{{}}{{}}}
\bibcite{Fukuda:2015ana}{{147}{}{{}}{{}}}
\bibcite{Dimopoulos:2016lvn}{{148}{}{{}}{{}}}
\bibcite{Gherghetta:2016fhp}{{149}{}{{}}{{}}}
\bibcite{Kim:1998va}{{150}{}{{}}{{}}}
\bibcite{DiLuzio:2020wdo}{{151}{}{{}}{{}}}
\bibcite{Conlon:2006tq}{{152}{}{{}}{{}}}
\bibcite{Chakraborty:2021fkp}{{153}{}{{}}{{}}}
\bibcite{Harigaya:2019qnl}{{154}{}{{}}{{}}}
\bibcite{PhysRevLett.38.1440}{{155}{}{{}}{{}}}
\bibcite{PhysRevD.16.1791}{{156}{}{{}}{{}}}
\bibcite{Clowe:2006eq}{{157}{}{{}}{{}}}
\bibcite{Hsu:2004mf}{{158}{}{{}}{{}}}
\bibcite{Agrawal:2017ksf}{{159}{}{{}}{{}}}
\bibcite{Gupta:2020vxb}{{160}{}{{}}{{}}}
\bibcite{PhysRevLett.43.103}{{161}{}{{}}{{}}}
\bibcite{Baker:2006ts}{{162}{}{{}}{{}}}
\bibcite{Zhang:2021mks}{{163}{}{{}}{{}}}
\bibcite{PhysRevLett.120.151301}{{164}{}{{}}{{}}}
\bibcite{Balkin:2022qer}{{165}{}{{}}{{}}}
\bibcite{Cicoli:2013ana}{{166}{}{{}}{{}}}
\bibcite{Svrcek:2006hf}{{167}{}{{}}{{}}}
\bibcite{Hiramatsu:2010yu}{{168}{}{{}}{{}}}
\bibcite{Arias:2012az}{{169}{}{{}}{{}}}
\bibcite{QuilezLasanta:2021yzt}{{170}{}{{}}{{}}}
\bibcite{Ishii:2022lwc}{{171}{}{{}}{{}}}
\bibcite{Cardoso:2018tly}{{172}{}{{}}{{}}}
\bibcite{Brito:2015oca}{{173}{}{{}}{{}}}
\bibcite{Ghosh:2023tyz}{{174}{}{{}}{{}}}
\bibcite{Hill:2015vma}{{175}{}{{}}{{}}}
\bibcite{Flambaum:2019cih}{{176}{}{{}}{{}}}
\bibcite{Zhang:2019eid}{{177}{}{{}}{{}}}
\bibcite{Dar:2018dra}{{178}{}{{}}{{}}}
\bibcite{Yoshida:2017cjl}{{179}{}{{}}{{}}}
\bibcite{Machado:2018nqk}{{180}{}{{}}{{}}}
\bibcite{Machado:2019xuc}{{181}{}{{}}{{}}}
\bibcite{10.21468/SciPostPhys.12.5.171}{{182}{}{{}}{{}}}
\bibcite{Ejlli:2022zah}{{183}{}{{}}{{}}}
\bibcite{Nagano:2021kwx}{{184}{}{{}}{{}}}
\bibcite{Burgess:2020tbq}{{185}{}{{}}{{}}}
\bibcite{Kol:2007bc}{{186}{}{{}}{{}}}
\bibcite{Levi:2018nxp}{{187}{}{{}}{{}}}
\bibcite{Dyadina:2018ryl}{{188}{}{{}}{{}}}
\bibcite{nastase_2019}{{189}{}{{}}{{}}}
\bibcite{poissonwill}{{190}{}{{}}{{}}}
\bibcite{2021EPJC...81.1048L}{{191}{}{{}}{{}}}
\bibcite{Chen:2022qvg}{{192}{}{{}}{{}}}
\bibcite{Liu:2020vsy}{{193}{}{{}}{{}}}
\bibcite{Cardoso:2020iji}{{194}{}{{}}{{}}}
\bibcite{2009GReGr..41.1667H}{{195}{}{{}}{{}}}
\bibcite{Anastasiou:1999ui}{{196}{}{{}}{{}}}
\bibcite{2022arXiv220103593W}{{197}{}{{}}{{}}}
\bibcite{Coogan:2021uqv}{{198}{}{{}}{{}}}
\bibcite{Singh:2022wvw}{{199}{}{{}}{{}}}
\bibcite{Becker:2021ivq}{{200}{}{{}}{{}}}
\bibcite{Yue:2019ozq}{{201}{}{{}}{{}}}
\bibcite{Bhattacharya:2023stq}{{202}{}{{}}{{}}}
\bibcite{Dlapa:2021npj}{{203}{}{{}}{{}}}
\bibcite{Dlapa:2022lmu}{{204}{}{{}}{{}}}
\bibcite{Dlapa:2021vgp}{{205}{}{{}}{{}}}
\bibcite{Dlapa:2023hsl}{{206}{}{{}}{{}}}
\bibcite{Jakobsen:2022fcj}{{207}{}{{}}{{}}}
\bibcite{Jakobsen:2021zvh}{{208}{}{{}}{{}}}
\bibcite{Jakobsen:2021lvp}{{209}{}{{}}{{}}}
\bibcite{Jakobsen:2021smu}{{210}{}{{}}{{}}}
\bibcite{Passarino:1978jh}{{211}{}{{}}{{}}}
\bibcite{Lee:2013mka}{{212}{}{{}}{{}}}
\bibcite{Beneke:1997zp}{{213}{}{{}}{{}}}
\bibcite{smirnov}{{214}{}{{}}{{}}}
\bibcite{Bhattacharyya:2024aeq}{{215}{}{{}}{{}}}
\bibcite{Purrer:2019jcp}{{216}{}{{}}{{}}}
\bibcite{stelle1}{{217}{}{{}}{{}}}
\bibcite{Alexeev:1996vs}{{218}{}{{}}{{}}}
\bibcite{Lehebel:2018zga}{{219}{}{{}}{{}}}
\bibcite{Volkov:2016ehx}{{220}{}{{}}{{}}}
\bibcite{Kunz:2006ca}{{221}{}{{}}{{}}}
\bibcite{Damour:1992we}{{222}{}{{}}{{}}}
\bibcite{Horbatsch:2015bua}{{223}{}{{}}{{}}}
\bibcite{Schon:2021pcv}{{224}{}{{}}{{}}}
\bibcite{Rainer:1996gw}{{225}{}{{}}{{}}}
\bibcite{DeFelice:2011bh}{{226}{}{{}}{{}}}
\bibcite{Gsponer:2021obj}{{227}{}{{}}{{}}}
\bibcite{hyp}{{228}{}{{}}{{}}}
\bibcite{hyp1}{{229}{}{{}}{{}}}
\bibcite{hyp2}{{230}{}{{}}{{}}}
\bibcite{hyp3}{{231}{}{{}}{{}}}
\bibcite{hyp4}{{232}{}{{}}{{}}}
\bibcite{hyp5}{{233}{}{{}}{{}}}
\bibcite{hyp6}{{234}{}{{}}{{}}}
\bibcite{hyp7}{{235}{}{{}}{{}}}
\bibcite{hyp8}{{236}{}{{}}{{}}}
\bibcite{hyp9}{{237}{}{{}}{{}}}
\bibcite{hyp10}{{238}{}{{}}{{}}}
\bibcite{hyp11}{{239}{}{{}}{{}}}
\bibcite{hyp12}{{240}{}{{}}{{}}}
\bibcite{hyp13}{{241}{}{{}}{{}}}
\bibcite{hyp16}{{242}{}{{}}{{}}}
\bibcite{hyp14}{{243}{}{{}}{{}}}
\bibcite{Damour:2016gwp}{{244}{}{{}}{{}}}
\bibcite{Bini:2017wfr}{{245}{}{{}}{{}}}
\bibcite{Bini:2017xzy}{{246}{}{{}}{{}}}
\bibcite{Damour:2017zjx}{{247}{}{{}}{{}}}
\bibcite{Damour:2019lcq}{{248}{}{{}}{{}}}
\bibcite{Bini:2020flp}{{249}{}{{}}{{}}}
\bibcite{Bini:2020uiq}{{250}{}{{}}{{}}}
\bibcite{Damour:2020tta}{{251}{}{{}}{{}}}
\bibcite{Bini:2020rzn}{{252}{}{{}}{{}}}
\bibcite{Bini:2021gat}{{253}{}{{}}{{}}}
\bibcite{Bini:2022enm}{{254}{}{{}}{{}}}
\bibcite{Damour:2022ybd}{{255}{}{{}}{{}}}
\bibcite{Bini:2022wrq}{{256}{}{{}}{{}}}
\bibcite{Rettegno:2023ghr}{{257}{}{{}}{{}}}
\bibcite{Bini:2023fiz}{{258}{}{{}}{{}}}
\bibcite{Ceresole:2023wxg}{{259}{}{{}}{{}}}
\bibcite{Cheung:2020sdj}{{260}{}{{}}{{}}}
\bibcite{Kalin:2020lmz}{{261}{}{{}}{{}}}
\bibcite{Haddad:2020que}{{262}{}{{}}{{}}}
\bibcite{Kalin:2020fhe}{{263}{}{{}}{{}}}
\bibcite{Jinno:2022sbr}{{264}{}{{}}{{}}}
\bibcite{Riva:2021vnj}{{265}{}{{}}{{}}}
\bibcite{PhysRevD.7.2317}{{266}{}{{}}{{}}}
\bibcite{Neill:2013wsa}{{267}{}{{}}{{}}}
\bibcite{Bjerrum-Bohr:2013bxa}{{268}{}{{}}{{}}}
\bibcite{Luna:2017dtq}{{269}{}{{}}{{}}}
\bibcite{Bjerrum-Bohr:2018xdl}{{270}{}{{}}{{}}}
\bibcite{Cristofoli:2021vyo}{{271}{}{{}}{{}}}
\bibcite{DeAngelis:2023lvf}{{272}{}{{}}{{}}}
\bibcite{Brandhuber:2023hhl}{{273}{}{{}}{{}}}
\bibcite{Aoude:2023dui}{{274}{}{{}}{{}}}
\bibcite{Brandhuber:2023hhy}{{275}{}{{}}{{}}}
\bibcite{Georgoudis:2023eke}{{276}{}{{}}{{}}}
\bibcite{Herderschee:2023fxh}{{277}{}{{}}{{}}}
\bibcite{Buonanno:2022pgc}{{278}{}{{}}{{}}}
\bibcite{Cheung:2018wkq}{{279}{}{{}}{{}}}
\bibcite{Cristofoli:2019neg}{{280}{}{{}}{{}}}
\bibcite{Cheung:2020gyp}{{281}{}{{}}{{}}}
\bibcite{ashoke}{{282}{}{{}}{{}}}
\bibcite{ashoke1}{{283}{}{{}}{{}}}
\bibcite{ashoke2}{{284}{}{{}}{{}}}
\bibcite{ashoke3}{{285}{}{{}}{{}}}
\bibcite{ashoke4}{{286}{}{{}}{{}}}
\bibcite{ashoke5}{{287}{}{{}}{{}}}
\bibcite{ashoke6}{{288}{}{{}}{{}}}
\bibcite{Jakobsen:2023ndj}{{289}{}{{}}{{}}}
\bibcite{Jakobsen:2023hig}{{290}{}{{}}{{}}}
\bibcite{Jakobsen:2023pvx}{{291}{}{{}}{{}}}
\bibcite{Bastianelli:2021nbs}{{292}{}{{}}{{}}}
\bibcite{Shi:2021qsb}{{293}{}{{}}{{}}}
\bibcite{Diaz-Jaramillo:2021wtl}{{294}{}{{}}{{}}}
\bibcite{Comberiati:2022cpm}{{295}{}{{}}{{}}}
\bibcite{Buonanno:1998gg}{{296}{}{{}}{{}}}
\bibcite{Strassler:1992zr}{{297}{}{{}}{{}}}
\bibcite{Feal:2022iyn}{{298}{}{{}}{{}}}
\bibcite{Ahmadiniaz:2022yam}{{299}{}{{}}{{}}}
\bibcite{eardley}{{300}{}{{}}{{}}}
\bibcite{Amati:1990xe}{{301}{}{{}}{{}}}
\bibcite{Zhou:2018tva}{{302}{}{{}}{{}}}
\bibcite{Amati:1987wq}{{303}{}{{}}{{}}}
\bibcite{Maybee:2019jus}{{304}{}{{}}{{}}}
\bibcite{smirnov2}{{305}{}{{}}{{}}}
\bibcite{Butzer1997Mellin}{{306}{}{{}}{{}}}
\bibcite{DoNorbury2018}{{307}{}{{}}{{}}}
\bibcite{Bhattacharyya:2024kxj}{{308}{}{{}}{{}}}
\bibcite{NANOGrav:2023gor}{{309}{}{{}}{{}}}
\bibcite{EPTA:2023fyk}{{310}{}{{}}{{}}}
\bibcite{NANOGrav:2023wsz}{{311}{}{{}}{{}}}
\bibcite{Verbiest:2024nid}{{312}{}{{}}{{}}}
\bibcite{PhysRev.138.B988}{{313}{}{{}}{{}}}
\bibcite{Zwiebach:1985uq}{{314}{}{{}}{{}}}
\bibcite{Gross:1986mw}{{315}{}{{}}{{}}}
\bibcite{Jackiw:2003pm}{{316}{}{{}}{{}}}
\bibcite{Alexander:2004us}{{317}{}{{}}{{}}}
\bibcite{Alexander:2004xd}{{318}{}{{}}{{}}}
\bibcite{Schafer:2018kuf}{{319}{}{{}}{{}}}
\bibcite{Warburton:2024xnr}{{320}{}{{}}{{}}}
\bibcite{Wardell:2021fyy}{{321}{}{{}}{{}}}
\bibcite{Brandhuber:2022qbk}{{322}{}{{}}{{}}}
\bibcite{Alessio:2024wmz}{{323}{}{{}}{{}}}
\bibcite{Bern:2024adl}{{324}{}{{}}{{}}}
\bibcite{Bern:2023ity}{{325}{}{{}}{{}}}
\bibcite{Bern:2021yeh}{{326}{}{{}}{{}}}
\bibcite{Brandhuber:2019qpg}{{327}{}{{}}{{}}}
\bibcite{AccettulliHuber:2019jqo}{{328}{}{{}}{{}}}
\bibcite{AccettulliHuber:2020oou}{{329}{}{{}}{{}}}
\bibcite{Barack:2023oqp}{{330}{}{{}}{{}}}
\bibcite{Bjerrum-Bohr:2023iey}{{331}{}{{}}{{}}}
\bibcite{Bjerrum-Bohr:2023jau}{{332}{}{{}}{{}}}
\bibcite{Bjerrum-Bohr:2022ows}{{333}{}{{}}{{}}}
\bibcite{Bjerrum-Bohr:2022blt}{{334}{}{{}}{{}}}
\bibcite{Bjerrum-Bohr:2021vuf}{{335}{}{{}}{{}}}
\bibcite{Bjerrum-Bohr:2020syg}{{336}{}{{}}{{}}}
\bibcite{Bjerrum-Bohr:2019kec}{{337}{}{{}}{{}}}
\bibcite{Bjerrum-Bohr:2016hpa}{{338}{}{{}}{{}}}
\bibcite{Chen:2024mmm}{{339}{}{{}}{{}}}
\bibcite{Alessio:2022kwv}{{340}{}{{}}{{}}}
\bibcite{Alessio:2023kgf}{{341}{}{{}}{{}}}
\bibcite{Gonzo:2024zxo}{{342}{}{{}}{{}}}
\bibcite{Aoude:2023vdk}{{343}{}{{}}{{}}}
\bibcite{Adamo:2024oxy}{{344}{}{{}}{{}}}
\bibcite{Adamo:2023cfp}{{345}{}{{}}{{}}}
\bibcite{Adamo:2022ooq}{{346}{}{{}}{{}}}
\bibcite{Adamo:2021rfq}{{347}{}{{}}{{}}}
\bibcite{Georgoudis:2024pdz}{{348}{}{{}}{{}}}
\bibcite{Bini:2024rsy}{{349}{}{{}}{{}}}
\bibcite{DiVecchia:2023frv}{{350}{}{{}}{{}}}
\bibcite{Jakobsen:2023oow}{{351}{}{{}}{{}}}
\bibcite{Wang:2022ntx}{{352}{}{{}}{{}}}
\bibcite{Klemm:2024wtd}{{353}{}{{}}{{}}}
\bibcite{Driesse:2024xad}{{354}{}{{}}{{}}}
\bibcite{1978ApJ...224...62K}{{355}{}{{}}{{}}}
\bibcite{Bini:2024ijq}{{356}{}{{}}{{}}}
\bibcite{DeVittori:2014psa}{{357}{}{{}}{{}}}
\bibcite{Cheung:2024jpo}{{358}{}{{}}{{}}}
\bibcite{Cheung:2023lnj}{{359}{}{{}}{{}}}
\bibcite{Ivanov:2024sds}{{360}{}{{}}{{}}}
\bibcite{Diedrichs:2023foj}{{361}{}{{}}{{}}}
\bibcite{Loebbert:2020aos}{{362}{}{{}}{{}}}
\bibcite{Mougiakakos:2021ckm}{{363}{}{{}}{{}}}
\bibcite{Mougiakakos:2022sic}{{364}{}{{}}{{}}}
\bibcite{Riva:2022fru}{{365}{}{{}}{{}}}
\bibcite{Bernard:2023eul}{{366}{}{{}}{{}}}
\bibcite{Dlapa:2024cje}{{367}{}{{}}{{}}}
\bibcite{Yunes:2009hc}{{368}{}{{}}{{}}}
\bibcite{Mastrolia:2018uzb}{{369}{}{{}}{{}}}
\bibcite{Brunello:2023rpq}{{370}{}{{}}{{}}}
\bibcite{Brunello:2023fef}{{371}{}{{}}{{}}}
\bibcite{Frellesvig:2023bbf}{{372}{}{{}}{{}}}
\bibcite{Dorigoni:2022npe}{{373}{}{{}}{{}}}
\bibcite{Weinzierl:2007cx}{{374}{}{{}}{{}}}
\bibcite{Loutrel:2018ydv}{{375}{}{{}}{{}}}
\bibcite{Wilson-Gerow:2025xhr}{{376}{}{{}}{{}}}
\bibcite{1967JMP.....8.1591D}{{377}{}{{}}{{}}}
\bibcite{Corinaldesi:1951pb}{{378}{}{{}}{{}}}
\bibcite{1964NCim...34..317D}{{379}{}{{}}{{}}}
\bibcite{Henn:2014qga}{{380}{}{{}}{{}}}
\bibcite{Kotikov:1990kg}{{381}{}{{}}{{}}}
\bibcite{Gehrmann:1999as}{{382}{}{{}}{{}}}
\bibcite{Weinzierl:2022eaz}{{383}{}{{}}{{}}}
\bibcite{Chetyrkin:1981qh}{{384}{}{{}}{{}}}
\bibcite{vonManteuffel:2012np}{{385}{}{{}}{{}}}
\bibcite{Remiddi:1997ny}{{386}{}{{}}{{}}}
\bibcite{Kotikov:1991pm}{{387}{}{{}}{{}}}
\bibcite{Prausa:2017ltv}{{388}{}{{}}{{}}}
\bibcite{Laporta:2000dsw}{{389}{}{{}}{{}}}
\bibcite{Remiddi:1999ew}{{390}{}{{}}{{}}}
\bibcite{goncha}{{391}{}{{}}{{}}}
\bibcite{Mertig:1990an}{{392}{}{{}}{{}}}
\bibcite{Shtabovenko:2020gxv}{{393}{}{{}}{{}}}
\bibcite{Shtabovenko:2023idz}{{394}{}{{}}{{}}}
\bibcite{Kim:2024grz}{{395}{}{{}}{{}}}
\bibcite{Saotome:2012vy}{{396}{}{{}}{{}}}
\bibcite{Parra-Martinez:2020dzs}{{397}{}{{}}{{}}}
\bibcite{Bhattacharyya:2025heterotic}{{398}{}{{}}{{}}}
\bibcite{GIBBONS1988741}{{399}{}{{}}{{}}}
\bibcite{Garfinkle:1990qj}{{400}{}{{}}{{}}}
\bibcite{Sen:1992ua}{{401}{}{{}}{{}}}
\bibcite{Metsaev:1987zx}{{402}{}{{}}{{}}}
\bibcite{Correia:2024jgr}{{403}{}{{}}{{}}}
\bibcite{Faller:2007sy}{{404}{}{{}}{{}}}
\bibcite{Correia:2024yfx}{{405}{}{{}}{{}}}
\bibcite{arpan2}{{406}{}{{}}{{}}}
\bibcite{Caron-Huot:2023vxl}{{407}{}{{}}{{}}}
\bibcite{arpan1}{{408}{}{{}}{{}}}
\bibcite{Bhattacharyya:2025sky}{{409}{}{{}}{{}}}
\bibcite{Osborn:2012vt}{{410}{}{{}}{{}}}
\bibcite{Poland:2018epd}{{411}{}{{}}{{}}}
\bibcite{Lam:2017ofc}{{412}{}{{}}{{}}}
\bibcite{Nandan:2019jas}{{413}{}{{}}{{}}}
\bibcite{Law:2020xcf}{{414}{}{{}}{{}}}
\bibcite{Fan:2021isc}{{415}{}{{}}{{}}}
\bibcite{Fan:2022vbz}{{416}{}{{}}{{}}}
\bibcite{Fan:2022kpp}{{417}{}{{}}{{}}}
\bibcite{Atanasov:2021cje}{{418}{}{{}}{{}}}
\bibcite{De:2022gjn}{{419}{}{{}}{{}}}
\bibcite{Garcia-Sepulveda:2022lga}{{420}{}{{}}{{}}}
\bibcite{Banerjee:2023jne}{{421}{}{{}}{{}}}
\bibcite{FERRARA1972281}{{422}{}{{}}{{}}}
\bibcite{Ferrara:1972kab}{{423}{}{{}}{{}}}
\bibcite{Simmons-Duffin:2012juh}{{424}{}{{}}{{}}}
\bibcite{Caron-Huot:2017vep}{{425}{}{{}}{{}}}
\bibcite{Simmons-Duffin:2017nub}{{426}{}{{}}{{}}}
\bibcite{Mazac:2018qmi}{{427}{}{{}}{{}}}
\bibcite{Stieberger:2018edy}{{428}{}{{}}{{}}}
\bibcite{Chang:2021wvv}{{429}{}{{}}{{}}}
\bibcite{Donnay:2023kvm}{{430}{}{{}}{{}}}
\bibcite{Casali:2022fro}{{431}{}{{}}{{}}}
\bibcite{deGioia:2022fcn}{{432}{}{{}}{{}}}
\bibcite{Banerjee:2023rni}{{433}{}{{}}{{}}}
\bibcite{Ball:2023ukj}{{434}{}{{}}{{}}}
\bibcite{Crawley:2023brz}{{435}{}{{}}{{}}}
\bibcite{Banerjee:2019prz}{{436}{}{{}}{{}}}
\bibcite{Adamo:2024mqn}{{437}{}{{}}{{}}}
\bibcite{Amati:1987uf}{{438}{}{{}}{{}}}
\bibcite{Kabat:1992tb}{{439}{}{{}}{{}}}
\bibcite{Hamber:2009zz}{{440}{}{{}}{{}}}
\bibcite{Abe:2017abx}{{441}{}{{}}{{}}}
\bibcite{Abe:2018rwb}{{442}{}{{}}{{}}}
\bibcite{Abe:2020ikj}{{443}{}{{}}{{}}}
\bibcite{Abe:2022spe}{{444}{}{{}}{{}}}
\bibcite{LLEWELLYNSMITH1973233}{{445}{}{{}}{{}}}
\bibcite{PhysRev.186.1656}{{446}{}{{}}{{}}}
\bibcite{tHooft:1987vrq}{{447}{}{{}}{{}}}
\bibcite{maroun2013generalized}{{448}{}{{}}{{}}}
\bibcite{johnson2000feynman}{{449}{}{{}}{{}}}
\bibcite{Borji:2024pvg}{{450}{}{{}}{{}}}
\bibcite{Dolan:2003hv}{{451}{}{{}}{{}}}
\bibcite{Mukhametzhanov:2018zja}{{452}{}{{}}{{}}}
\bibcite{Fan:2023lky}{{453}{}{{}}{{}}}
\bibcite{Burchnall1940EXPANSIONSOA}{{454}{}{{}}{{}}}
\bibcite{Costa:2012cb}{{455}{}{{}}{{}}}
\bibcite{Liu:2020tpf}{{456}{}{{}}{{}}}
\bibcite{Ancarani:2009zz}{{457}{}{{}}{{}}}
\bibcite{Donnay:2022aba}{{458}{}{{}}{{}}}
\bibcite{Donnay:2022wvx}{{459}{}{{}}{{}}}
\bibcite{Bagchi:2022emh}{{460}{}{{}}{{}}}
\bibcite{Pate:2017fgt}{{461}{}{{}}{{}}}
\bibcite{Banerjee:2022hgc}{{462}{}{{}}{{}}}
\bibcite{Narayanan:2024qgb}{{463}{}{{}}{{}}}
\bibcite{Banerjee:2024hvb}{{464}{}{{}}{{}}}

  \makeatother
  \immediate\openout\thesiscrossrefsmarker=\jobname.crossrefs-ready
  \immediate\write\thesiscrossrefsmarker{ready}
  \immediate\closeout\thesiscrossrefsmarker
}

\begin{document}

\hypersetup{pageanchor=false}
\MakeReferenceStyleTitlePage

\MakeCertificatePage

\hypersetup{pageanchor=true}
\frontmatter
\thispagestyle{plain}
\vspace*{0.42\textheight}
\begin{center}
\itshape ``He who has a why to live can bear almost any how.''\\
-- Friedrich Nietzsche
\end{center}
\newpage
\clearpage
\newpage
\pagecolor{white}
\thispagestyle{empty}

\vspace*{\fill}
\vspace{2 cm}
\hspace{5 cm} {\textcolor{brown}{\Huge\calligra To Maa \& Baba\ldots\ldots}}
\vspace*{\fill}

\noindent
\hfill

\clearpage
\pagecolor{white}
\newpage
\pagecolor{white}
\thispagestyle{empty}

\vspace*{\fill}
\begin{center} \includegraphics[width=1\textwidth]{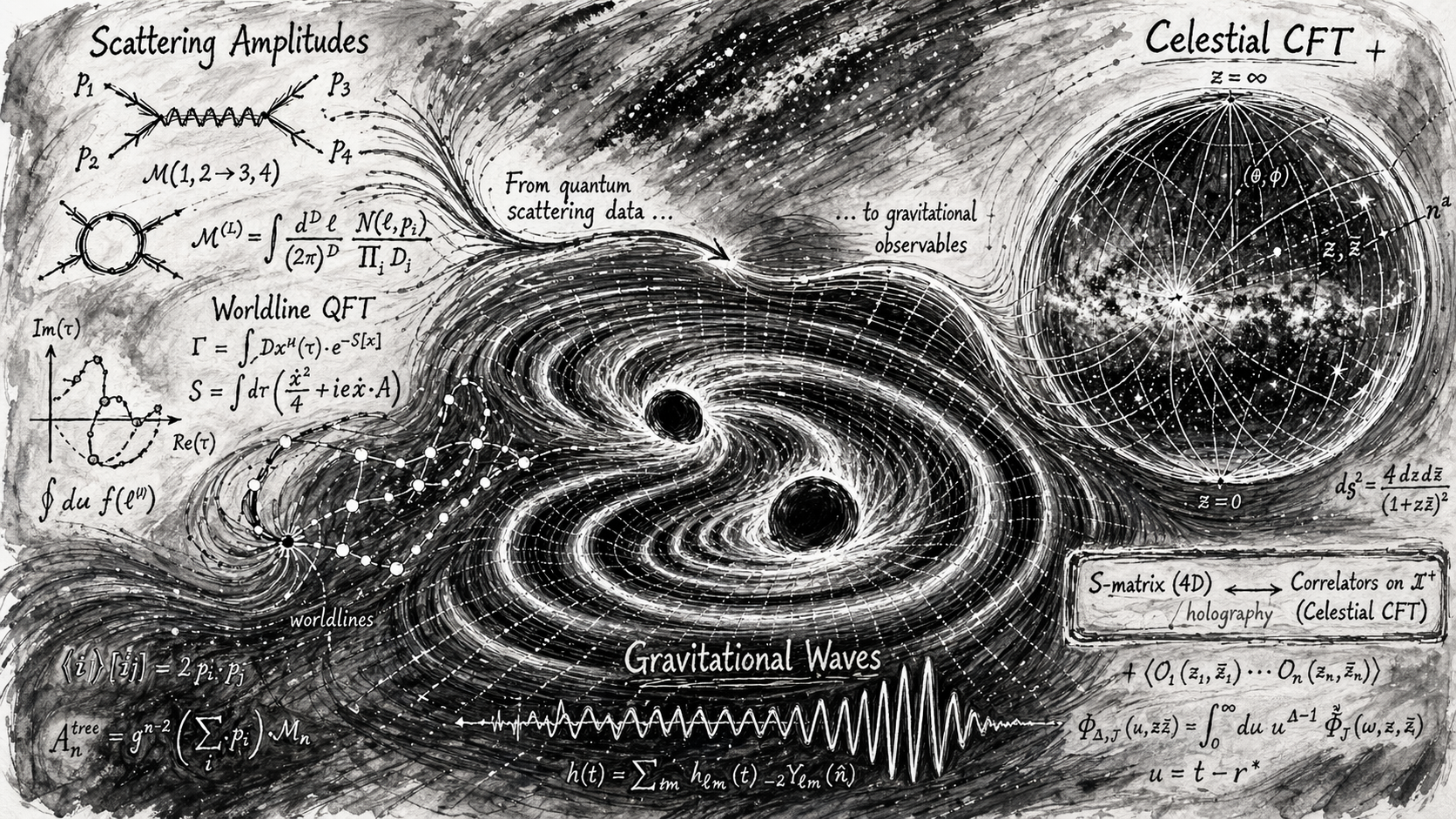}
    \vspace{1 cm}
    {\color{black}\textit{what survives the crossing....?}}
\end{center}
\vspace*{\fill}

\noindent
\hfill
{\color{black}\footnotesize Source: Image idea solely by the author and generated using \href{https://chatgpt.com/}{ChatGPT.} }

\clearpage
\pagecolor{white}
\frontchapter{Abstract}

Modern high-energy physics and gravitational-wave astronomy, despite being very different in their experimental settings, share a common need: precise theoretical predictions. In particle physics, differential cross sections serve this purpose, providing the basis for testing the Standard Model and searching for possible hints of new physics. In gravitational-wave physics, the analogous role is played by the waveforms emitted by coalescing compact binaries, whose accurate modelling is essential both for the detection of signals and for the extraction of the astrophysical information they carry. From a computational point of view, these two subjects also share a significant amount of common structure.

The primary focus of this thesis is the study of the classical gravitational dynamics of two compact objects in different astrophysical situations, in particular the \textit{inspiral} and \textit{scattering} regimes, within effective theories of gravity beyond General Relativity. The central aim is to compute the classically relevant observables in these settings using modern techniques from quantum field theory. The starting point is the recognition that, despite its remarkable success in the weak-field regime, General Relativity is not expected to be the final or UV-complete description of gravity. As a classical theory, it should eventually arise from a consistent quantum framework, while as a low-energy theory, it is naturally understood as the leading term in a more general effective expansion. This viewpoint motivates the study of effective theories of gravity containing higher-curvature corrections and additional degrees of freedom. Such extensions typically introduce new propagating modes beyond the standard massless spin-2 graviton, and therefore lead to departures from the dynamics of pure General Relativity.

On the other hand, Quantum Field Theory (QFT) was originally developed to describe the interactions of fundamental particles. Over the last decade, however, it has become increasingly clear that many techniques borrowed from QFT are also extremely powerful for computing classical observables, provided the external states are prepared appropriately. In the standard particle-physics setting, perturbation theory is often described as an expansion both in the couplings and in $\hbar$. While this viewpoint is useful for scattering of microscopic quantum particles, it is not strictly correct when the external states correspond to macroscopic classical objects such as black holes or neutron stars. In such cases, classical information is not obtained simply by keeping tree-level terms and discarding loops; rather, it emerges from suitable kinematical limits of amplitudes and correlators, for instance through soft-momentum expansions in the large-impact-parameter regime.

There exist several complementary approaches for addressing these problems in different physical regimes. For the inspiral problem, where the orbital motion is non-relativistic, one employs the framework of Post-Newtonian Effective Field Theory (PN-EFT), also known as Non-Relativistic General Relativity (NRGR). The key idea is to exploit the hierarchy of length scales in the binary system, namely the size of the compact object, the orbital separation, and the wavelength of the emitted radiation. This separation of scales allows one to construct a systematic effective field theory description in which the conservative and radiative sectors can be treated in an organized manner.

In contrast, for scattering processes involving compact objects moving at relativistic velocities, the post-Newtonian expansion is no longer adequate. In this regime, it is more natural to use the post-Minkowskian (PM) expansion, which is organized as an expansion in Newton's constant $G$ while retaining the full velocity dependence. Several QFT-based methods have been developed for this purpose. These include direct amplitude-based approaches, such as the PM expansion of amplitudes and the KMOC formalism, as well as worldline-based frameworks such as PM-EFT and Worldline Quantum Field Theory (WQFT), which is closely related to the eikonal scattering amplitude. Together, these methods provide a modern and efficient toolkit for extracting conservative and radiative observables from classical gravitational dynamics beyond General Relativity.

As a further direction, this thesis also explores the celestial representation of scattering amplitudes, in which four-dimensional amplitudes in asymptotically flat spacetime are recast as conformal correlators on the celestial sphere at infinity. A well-known difficulty in this subject is that, in perturbative quantum field theory, especially for UV-incomplete theories, celestial amplitudes are often divergent order by order in perturbation theory. One possible way to improve this behavior is to resum the perturbative series before performing the Mellin transform, a strategy that becomes practically feasible in the eikonal regime. In the context of General Relativity, such an eikonal resummation is known to significantly improve the analytic properties of celestial amplitudes. When higher-derivative corrections to General Relativity are included, the ultraviolet behavior of the theory is modified, and this in turn can lead to an improved analytic structure for the corresponding celestial amplitudes. At the same time, however, constructing eikonal amplitudes beyond General Relativity becomes increasingly challenging. In this thesis, we also address these difficulties by employing techniques borrowed directly from analytic functional analysis.

\frontchapter{Author's Declaration}
I, Saptaswa Ghosh, hereby declare that this thesis is solely authored by me and
where other’s ideas or words incorporated, I have adequately cited and referenced
the original sources. The work presented
herein was carried out during my PhD studies under the supervision of
Prof.~Arpan Bhattacharyya.   Also, I declare that I have adhered to all principles of academic
honesty and integrity and have not falsified or fabricated any idea/fact/source in
this submission. I understand that any violation of the above can cause disciplinary
action by the Institute and can also summon penal action from the sources which
have thus not been properly cited or from whom proper permission has not been
taken. The following manuscripts were co-authored by me
during my PhD and are included in this thesis by reference:

\vspace{0.45em}
\textbf{Published works (included in this thesis)}
\begin{enumerate}[label={[\arabic*]},leftmargin=2.3em,itemsep=0.52em]
  \item A.~Bhattacharyya, S.~Ghosh, and S.~Pal.
  ``Worldline effective field theory of inspiralling black hole binaries in
  presence of dark photon and axionic dark matter.'' In:
  \textit{JHEP} \textbf{08} (2023), 207.
  doi: \doilink{10.1007/JHEP08(2023)207}.
  arXiv: \arxivlink{2305.15473}{hep-th}.

  \item A.~Bhattacharyya, D.~Ghosh, S.~Ghosh, and S.~Pal.
  ``Observables from classical black hole scattering in Scalar-Tensor theory
  of gravity from worldline quantum field theory.'' In:
  \textit{JHEP} \textbf{04} (2024), 015.
  doi: \doilink{10.1007/JHEP04(2024)015}.
  arXiv: \arxivlink{2401.05492}{hep-th}.

  \item A.~Bhattacharyya, D.~Ghosh, S.~Ghosh, and S.~Pal.
  ``Bootstrapping the spinning two body problem in dynamical Chern-Simons
  gravity using worldline QFT.'' In:
  \textit{JHEP} \textbf{04} (2025), 175.
  doi: \doilink{10.1007/JHEP04(2025)175}.
  arXiv: \arxivlink{2407.07195}{hep-th}.
  \item A.~Bhattacharyya, S.~Ghosh, A.~Mishra, and S.~Pal.
  ``Heterotic Footprints in Classical Gravity: PM dynamics from On-Shell soft
  amplitudes at one loop.'' In \textit{JHEP} \textbf{06} (2026), 008.
  doi: \doilink{10.1007/JHEP06(2026)008}. arXiv: \arxivlink{2510.07390}{hep-th}.

  \item A.~Bhattacharyya, S.~Ghosh, and S.~Pal.
  ``The sky remembers everything: Celestial amplitude, shadow and OPE in
  quadratic EFT of gravity.'' In:
  \textit{SciPost Phys.} \textbf{19}, no.2 (2025), 041.
  doi: \doilink{10.21468/SciPostPhys.19.2.041}.
  arXiv: \arxivlink{2505.02899}{hep-th}.
\end{enumerate}

The following manuscripts were co-authored by me during my PhD but are not
included in this thesis:

\vspace{0.45em}
\textbf{Other publications}
\begin{enumerate}[label={[\arabic*]},leftmargin=2.3em,itemsep=0.52em,resume]
  \item S.~Ghosh and A.~Bhattacharyya.
  ``Analytical study of gravitational lensing in Kerr-Newman black-bounce
  spacetime.'' In: \textit{JCAP} \textbf{11} (2022), 006.
  doi: \doilink{10.1088/1475-7516/2022/11/006}.
  arXiv: \arxivlink{2206.09954}{gr-qc}.

  \item A.~Chowdhuri, S.~Ghosh, and A.~Bhattacharyya.
  ``A review on analytical studies in Gravitational Lensing.'' In:
  \textit{Front. Phys.} \textbf{11} (2023), 1113909.
  doi: \doilink{10.3389/fphy.2023.1113909}.
  arXiv: \arxivlink{2303.02069}{gr-qc}.

  \item A.~Bhattacharyya, S.~Ghosh, and S.~Pal.
  ``Aspects of T\={T}+J\={T} deformed Schwarzian: From gravity partition
  function to late-time spectral form factor.'' In:
  \textit{Phys. Rev. D} \textbf{110}, no.12 (2024), 126015.
  doi: \doilink{10.1103/PhysRevD.110.126015}.
  arXiv: \arxivlink{2309.16658}{hep-th}.

  \item A.~Bhattacharyya, S.~Ghosh, P.~Nandi, and S.~Pal.
  ``3D $\mathcal{N}=1$ supergravity from Virasoro TQFT: gravitational partition
  function and Out-of-time-order correlator.'' In:
  \textit{JHEP} \textbf{02} (2025), 027.
  doi: \doilink{10.1007/JHEP02(2025)027}.
  arXiv: \arxivlink{2408.01538}{hep-th}.

  \item A.~Bhattacharyya, S.~Ghosh, S.~Pal, and A.~Vinod.
  ``Finite cutoff JT gravity: Baby universes, Matrix dual, and (Krylov) Complexity.'' arXiv: \arxivlink{2502.13208}{hep-th}. \textcolor{brown}{Accepted in PTEP}.

  \item A.~Bhattacharyya, S.~Ghosh, N.~Kumar, S.~Kumar, and S.~Pal.
  ``Love beyond Einstein: metric reconstruction and Love number in quadratic
  gravity using WEFT.'' In: \textit{JHEP} \textbf{11} (2025), 155.
  doi: \doilink{10.1007/JHEP11(2025)155}.
  arXiv: \arxivlink{2508.02785}{hep-th}.

  \item A.~Bhattacharyya, S.~Ghosh, S.~Pal, and J.~Santara.
  ``On the resolution of categorical symmetries in (Non-) Unitary Rational
  CFTs.'' arXiv: \arxivlink{2511.16363}{hep-th}.
\end{enumerate}

\vfill
\begin{flushright}
Saptaswa Ghosh\\
\textit{Gandhinagar, April 2026.}
\end{flushright}
\frontchapter{Acknowledgments}
 First and foremost, it is my immense pleasure to express my sincere gratitude to my doctoral advisor, Prof. Arpan Bhattacharyya, for constantly suggesting interesting and challenging problems that helped me develop a deeper understanding of the subject. I am particularly thankful for the extraordinary intellectual freedom he gave me to wander across different areas of high energy physics. In the last four and a half years, I can hardly remember hearing a ``no'' from him when I proposed an idea, raised a question, or wished to explore a new direction. His inspiring mentorship, patient guidance, and  encouragement have been invaluable throughout this thesis work.
I have also benefited enormously from our many discussions, not only on physics and mathematics, but also on philosophy and politics, all of which have been deeply enriching. Finally, I owe him thanks for a form of support that was perhaps no less important in practice: an apparently divergent series of restaurant bills and post-dinner orange juices, without which this thesis might have converged much more slowly.\\\\
I would also like to sincerely thank Prof. Somdeb Chakraborty (City College, Kolkata), whose teaching during my Bachelor's days quietly, but quite effectively, nudged me toward high energy theory. I am deeply grateful to Prof. Bobby Ezhuthachan (RKMVERI, Belur) for his beautiful courses on Mathematical Physics, Electrodynamics, and Conformal Field Theory, which made my journey into theoretical physics much smoother. I am fairly certain that without those courses (specially the mathematical physics), the road would have been far more uneven. I am also thankful to Prof. Sanjoy Biswas (RKMVERI, Belur) for his excellent course on Quantum Field Theory, which not only introduced me to the subject but also helped me appreciate its technical depth and elegance. I would also like to thank Suchetan da (IACS, Kolkata) for patiently helping me through many confusions during my MSc days, especially in GR and conformal field theory. Those conversations helped me enormously at the time and continued to guide me in many ways along the journey.\\\\
I am also indebted to my Doctoral Study Committee members Prof. Rusa Mondal, Prof. Tanya Kaushal Srivastava for their critical evaluation and valuable comments on my works during many talks that I presented at IIT Gandhinagar. I am obligated for their unconditional support and guidance throughout my PhD.\\\\
I am deeply grateful to Prof. Jan Plefka at Humboldt University, Germany; Prof. Niels Emil Bjerrum Bohr at the Niels Bohr Institute, Denmark; and Prof. Rafael Porto at DESY, Germany, for hosting me during my academic visits and for the opportunity to present my work. I also sincerely thank Gustav Jacobsen and Maor Ben Sahar at Humboldt University, and Nabha Shah at the Niels Bohr Institute, for many insightful discussions on different topics. I am particularly indebted to the Institut des Hautes Études Scientifiques (IHES), France, for selecting me as an invited researcher in November 2025. I would especially like to thank Julio Parra-Martinez at IHES for many enlightening discussions, for helping to resolve several of my doubts, and for offering a fresh perspective on higher-derivative theories in the context of gravitational-wave physics. I would like to thank Giacomo Brunello at the University of Padua, Italy, for inviting me to give talk at Amplitude Lounge.\\\\
I would especially like to thank Sounak (JT Bhai), who has been not only a close friend but also my primary collaborator throughout the years of this thesis, and happily continues to be so. 
Many of my ideas were sharpened, many confusions were cleared, and many calculations were made less intimidating through our endless discussions. I will particularly miss those late-night conversations in my room, where the wardrobe would quietly assume the role of a blackboard and bear witness to a steady stream of equations, arguments, disagreements, and occasional moments of clarity. Some of the most memorable parts of this journey were shaped in those unplanned hours of thinking aloud together, where physics felt less like work and more like a shared adventure.\\\\
I also thank my other collaborators and friends Jagannath da, Debodirna da, Poulami di, Shailesh da, Abhishek da and  Naman. I have learned a lot of physics from them. I also had the privilege of working with two excellent Master's students, Ankit and Anandu. I am grateful to them for repeatedly confronting me with hard questions, many of which first put me into confusion and only later into clarity. I would also like to thank my other group members and friends, Nilachal, Rishabh, Neha, and Anisha, for sharing many unforgettable moments and spirited arguments along the way. I am also grateful my elderly colleagues Rajes da, Rik da, Abhishek da, Kaushik da for their support in different phases of my PhD. 
\\\\
I would also like to thank my lifelong friends Koustubh, Neelabha, Bibhas, Samriddha, Poulami di, Sayantan, and Subhodeep with whom I shared some of the most vibrant and unforgettable moments of my life at IIT Gandhinagar. Their companionship filled these years with joy, warmth, and a sense of home. I will especially miss those late-night addas and the long conversations on various topics, which remain among my fondest memories of this place. I would especially like to thank Bibhas for helping me install various packages throughout my PhD journey. I would like to thank my friend and football mate Kaushik for teaching me the subtle art of diving on the football field, a skill I might never have learned from textbooks alone. 
\\\\
I am deeply grateful to my old college friends, Jayashish (Joga), Neel, Soham (Singara), Ratul (-master), and Rounak da. Without them, the journey from the very beginning would have been much more difficult. The tours, addas, fun, and the closeness we shared made those days truly special.\\\\
I would like to thank ChatGPT for many useful discussion and Codex for enormous help in coding and formatting.
\\\\
 I would like to thank my family, especially my parents, for their endless love, support and encouragement through the ups and downs of my life. I owe everything to their struggle and sacrifice. Last but not the least, I express my deepest gratitude to Megh, whose love, support, encouragement, and constant faith in me have been my strength and have made my life more meaningful.
\\\\
\textit{\textcolor{brown}{In the end, I also acknowledge the lessons that came from disappointment, reminding me that integrity matters most when it is tested!}}
\newpage
\thesisstatictableofcontents
\clearpage
\thesisstaticlistoffigures
\clearpage

\mainmatter

%
  \restorethesisbodyformat
  \restoremainchapterstyle
  \chapter{Introduction}
  \thesischapterpaperbox{}{}{}
  {%
    \restorethesisbodyformat
    \renewcommand{\appendix}{%
      \setcounter{section}{0}%
      \setcounter{subsection}{0}%
      \setcounter{subsubsection}{0}%
      \renewcommand{\thesection}{\thechapter.\Alph{section}}%
      \renewcommand{\thesubsection}{\thesection.\arabic{subsection}}%
      \renewcommand{\theHsection}{chapter.\arabic{chapter}.appendix.\Alph{section}}%
      \renewcommand{\theHsubsection}{\theHsection.\arabic{subsection}}%
    }%
    \ifstrempty{introduction.tex}{}{%
      \section*{Goal of the Thesis}
General relativity has passed every current precision test in the weak-field regime, while the strong-gravity and high-curvature regimes remain active frontiers. Black-hole binary dynamics, scattering observables, and gravitational radiation are central probes of this regime. Modern collider-inspired methods, especially scattering amplitudes and effective field theory (EFT), now provide a systematic and highly efficient route to compute classical observables that are difficult to obtain using traditional approaches alone.

The broad objective of this thesis is to explore these methodologies in the context of beyond General Relativity:
\begin{enumerate}
\item Classical two-body dynamics in gravity,
\item Amplitude and worldline-based computational frameworks, and
\item Eikonal amplitudes in modified gravity and implication to celestial holography.
\end{enumerate}
The focus is on effective theories of gravity and departures from Einstein gravity, where new interactions can leave signatures in conservative, and radiative dynamics as well as more formal aspects like flat space holography.
\section{Context and Motivation}
The detection of gravitational waves (GWs) \cite{LIGOScientific:2016aoc, LIGOScientific:2016sjg, LIGOScientific:2016vlm, LIGOScientific:2017bnn, Kokeyama:2020dkg, LIGOScientific:2019hgc} has opened a powerful new pathway to test gravity. Alongside its remarkable success in describing the dynamics of spacetime and underpinning our current picture of the Universe’s large-scale structure, Einstein’s theory can now be confronted directly with strong-field, time-dependent data from compact-object sources. Gravitational waves are produced most prominently by compact binaries involving black holes and neutron stars, and the ability to detect these signals motivates increasingly high-precision computations of the source dynamics. The evolution of such a binary is commonly separated into three stages: the \textit{inspiral}, \textit{merger}, and \textit{ringdown}. Probing the merger regime---where gravity is genuinely strong and highly non-linear---typically requires non-perturbative techniques based on numerical relativity \cite{Gourgoulhon:2012,Lehner:2014asa, LIGOScientific:2014oec}.
\par
Beyond quasi-circular inspirals, compact objects can also undergo astrophysical scattering encounters on hyperbolic trajectories; the associated burst-like radiation provides complementary access to source properties and can be used to place constraints on theory parameters. In this broader landscape, it is notable that Quantum Field Theory (QFT), originally developed to describe interactions among subatomic particles, has emerged as an efficient and systematically improvable toolkit for analytic computations in the weak-to-intermediate field regimes relevant to both inspiral and scattering dynamics.
\noindent
\section*{Why Go Beyond General Relativity?}
Despite its remarkable empirical success, General Relativity (GR) is widely regarded as an incomplete framework. By construction, it is a classical field theory governing the gravitational interactions of macroscopic objects. However, like all other fundamental forces in nature, gravity must ultimately be reconciled with quantum theory. Because GR is not ultraviolet (UV) complete, it breaks down at extremely high energies or vanishingly short distances, indicating an absolute need for new physics. As Steven Weinberg pointed out, Einstein's gravity is best viewed not as the final, fundamental theory of nature, but as the leading term in a low-energy effective description. This perspective strongly motivates the study of \emph{effective field theories (EFTs) of gravity}. In an EFT framework, standard GR is augmented by higher-derivative operators that capture the low-energy imprints of high-energy dynamics. The effective action takes the form of an infinite derivative expansion:
\begin{equation}
    S = \int d^4x \sqrt{-g} \left( c_1 R + c_2 \mathcal{R}^2 + c_3 \mathcal{R}^3 + \cdots \right)
\end{equation}
where the leading-order term $R$ is the standard Ricci scalar. From the second order onwards, $\mathcal{R}^n$ schematically represents all possible combinations of curvature invariants (e.g., $R_{\mu\nu}R^{\mu\nu}$ or $R_{\mu\nu\rho\sigma}R^{\mu\nu\rho\sigma}$) at a given derivative order. Each successive term in this expansion probes increasingly smaller length scales, predicting deviations from GR that can be used to build phenomenological models. 

Truncating this expansion at the next-to-leading order yields the simplest non-trivial extension of GR: quadratic gravity,
\begin{equation}
     S_{g}=\int d^4 x \sqrt{-g}\left(\kappa \,{R}+\alpha {R}_{\mu\nu}{R}^{\mu\nu}-\frac{1}{3}(\alpha+\beta){R}^2\right).
\end{equation}
The inclusion of curvature-squared contributions in classical gravity was first proposed by Weyl \cite{weyl}, and their ability to render gravity power-counting renormalizable was subsequently demonstrated \cite{1962JMP.....3..608U, PhysRevD.16.953}. However, this favourable UV behaviour comes at a cost. At the level of the tree-level propagator \eqref{3.2h}, the UV behaviour improves from a $1/q^2$ to a $1/q^4$ fall-off, but this modification unavoidably introduces a massive spin-2 ghost with mass $m = \sqrt{\kappa/\alpha}$. To quantise the theory such that all excitations have positive-definite energy, one must introduce negative-norm states into the Hilbert space, which typically signals a breakdown of perturbative unitarity \cite{PhysRev.79.145}. Consequently, the physical consistency of higher-derivative gravity remains highly contested. Yet, as emphasised by  Fradkin and  Tseytlin \cite{FRADKIN1982469}, unitarity is fundamentally a dynamical property. It cannot be definitively assessed at the tree level or within a strictly perturbative regime; a complete picture requires accounting for loop corrections and non-perturbative radiative effects.

Moreover, GR leaves major puzzles in fundamental physics and cosmology unresolved, such as the origin of dark energy and the exact nature of spacetime singularities, further reinforcing the perspective that Einstein gravity is merely the leading term in a broader effective description. Introducing higher-curvature invariants into the action is theoretically sound, provided that diffeomorphism invariance is rigorously preserved. As suggested by our previous discussion, the impact of these higher-derivative operators becomes highly pronounced in strong-gravity regimes, such as the immediate vicinity of merging black holes. At the classical level, the dimensionful couplings associated with these terms introduce characteristic length scales that dictate exactly when and where these deviations from standard GR become physically significant. Furthermore, upon linearization, higher-derivative terms typically induce new dynamical degrees of freedom beyond the standard massless spin-2 graviton, signaling a fundamental departure from pure GR dynamics. 

Beyond bottom-up phenomenological models, effective field theories can also be constructed via top-down approaches when a complete UV theory is specified. String theory, a premier candidate for a consistent theory of quantum gravity, perfectly exemplifies this paradigm. When deriving the low-energy effective action of string theory, higher-derivative curvature operators naturally and inevitably emerge. Alongside these operators, this top-down derivation universally predicts the presence of additional dynamical degrees of freedom, such as the scalar dilaton and various gauge fields and higher form fields (e.g the Einstein-Dilaton-Higher form theory \cite{Hassan:1991mq}), which further enrich the gravitational sector.
\par
This thesis is primarily devoted to the calculation of \emph{observable} quantities within such effective theories of gravity (in both astrophysical scenarios: inspiral and scattering), with particular emphasis on higher-derivative extensions and related modifications that become relevant in scattering or inspiral processes. Using modern amplitude-based and worldline-QFT(and EFT) techniques, we compute conservative and radiative observables (such as impulse, scattering angle, and waveform) in regimes where precision is essential and where deviations from Einstein gravity can leave measurable imprints.

\section{Quantum Field Theory and its Classical Limit}
In standard undergraduate and graduate presentations of quantum field theory, one often learns that the perturbative expansion of scattering amplitudes may be viewed both as an expansion in the coupling constants and as an expansion in the Planck constant $\hbar$. Within this perspective, tree-level amplitudes, of order $\hbar^{0}$, are identified with the classical result, while higher-loop contributions are interpreted as quantum corrections. While this intuition is often useful, it is not entirely general and can be misleading in certain situations.
This conventional picture is primarily based on particle-physics applications, where the external states are genuinely quantum particles with momenta scaling as
\(k^\mu=\hbar\,\bar{k}^\mu\),
where $\bar{k}^\mu$ is the associated wave vector. With this scaling, the $\hbar$-counting of amplitudes provides a natural way to distinguish classical from quantum effects. However, when quantum field theory is employed to study the scattering of macroscopic compact objects such as black holes or neutron stars, the situation is qualitatively different. These external states correspond to classical bodies, and their momenta do not scale with $\hbar$ in the same manner. One is then led to an important conceptual question: how does one isolate the classical content of a quantum field theoretic amplitude when the external particles are themselves classical objects? Over the past couple of decades, substantial conceptual and computational advances have emerged to address this question. Let us discuss the seemingly different approaches in this context.

\vspace{0.5 cm}
\thesisleadtopic{PM Expansion of Amplitudes \cite{Holstein:2004dn,Bern:2019crd,Bern:2020buy,Bern:2019nnu,Bern:2020gjj,Bern:2021dqo,Bern:2021xze,Bern:2022kto,Bern:2025zno,Bjerrum-Bohr:2002aqa,Bjerrum-Bohr:2021wwt}:}
In scattering processes involving black holes with velocities comparable to the speed of light, the usual post-Newtonian (PN) expansion, which is fundamentally a low-velocity expansion and is well suited to the inspiral regime, ceases to be a good approximation. In such situations, it is more appropriate to employ the \emph{post-Minkowskian} (PM) expansion, which is organised as an expansion in Newton's constant $G$ while retaining the full velocity dependence. Since relativistic scattering amplitudes are naturally arranged as a perturbative series in $G$, keeping all orders in velocity, one defines the PM potential as
\begin{equation}
V(\boldsymbol{p},\boldsymbol{r})=\sum_{n=1}^\infty \left(\frac{G}{|\boldsymbol{r}|}\right)^n c_n(\boldsymbol{p}^2)\, ,
    \label{1.2}
\end{equation}

where the coefficients $c_n(\boldsymbol{p}^2)$ are functions of $\boldsymbol{p}^2 \sim \boldsymbol{v}^2$ and therefore encode arbitrarily high powers of the velocity.

To extract genuinely new information in PM dynamics that can be incorporated into gravitational-wave templates, however, one must carefully understand how scattering data encode the conservative two-body dynamics. The central idea underlying the connection between scattering amplitudes and the bound orbit motion of compact binaries is that both are governed by the same underlying theory, or more precisely, by the same classical effective potential. Thus, by construction, the effective two-body potential in \eqref{1.2} reproduces the same long-range classical physics as the full gravitational theory for kinematics specified by the masses and velocities of the scattering bodies. It then becomes possible to use scattering amplitudes as a tool to determine the effective potential. It was demonstrated long ago that modern field-theoretic methods are particularly powerful for computing scattering amplitudes relevant to such classical problems \cite{Holstein:2004dn}. As emphasised earlier, the main objective of this program is to systematically isolate the classically relevant part of quantum scattering amplitudes.

Let us begin by specifying the theory. A pair of gravitationally interacting spinless compact objects with masses $m_1$ and $m_2$ may be modelled by real scalar fields and described by the action below, neglecting finite-size corrections and spin:
\begin{equation}
        S=\int d^4x \sqrt{-g}\,R
        +\frac{1}{2}\sum_{i=1}^2 \int d^4x \sqrt{-g}\,
        \left(D_\mu \phi_i D^\mu \phi_i - m_i^2 \phi_i^2\right)\, .
\end{equation}
Here, one neglects direct interactions between the scalar fields $\phi_i$, consistent with the classical assumption that the two objects remain widely separated and interact only through gravity.

For conservative scattering, one considers the process in which $\phi_1$ and $\phi_2$ scatter elastically. In the center-of-mass frame, the incoming momenta are $(E_1,\boldsymbol{p})$ and $(E_2,-\boldsymbol{p})$, while the outgoing momenta are $(E_1,\boldsymbol{p}')$ and $(E_2,-\boldsymbol{p}')$. The momentum transfer is then
\begin{equation}
    q=(0,\boldsymbol{q})=(0,\boldsymbol{p}-\boldsymbol{p}')\, .
\end{equation}
The classical regime is characterized by an impact parameter much larger than the de Broglie wavelength, $ |\boldsymbol{b}| \gg \lambda = \frac{1}{|\boldsymbol{p}|}\,,$ or equivalently, by an orbital angular momentum parametrically large compared to $\hbar$, in accordance with the Bohr correspondence principle:
\begin{equation}
    J \gg \hbar \qquad (\text{equivalently, setting }\hbar=1:\; J\gg 1)\, .
\end{equation}
Physically, the condition $J\gg 1$ means that the impact parameter is large compared to the de Broglie wavelength, so that the scattering process admits a semiclassical description.

In this regime, the momentum transfer $q$ is small compared to the hard external scales set by the Mandelstam invariants and the particle masses. The classical contributions then arise from the non-analytic, long-distance dependence of the amplitude on
\begin{equation}
t=q^2\, .
\end{equation}
More precisely, we have a scale hierarchy
\begin{equation}
    s,\; |u|,\; m_i^2 \sim J^2 |t| \gg |t| = |q|^2\, ,
\end{equation}
so that an expansion at large $J$ is equivalent to an expansion in small $|q|$. This is what we refer to as the \emph{soft expansion}: an expansion about small momentum transfer while keeping the hard external kinematics fixed. Within this expansion, the classical part of the amplitude can be systematically isolated, for instance as the leading terms in $|q|$ at fixed $s$ and fixed masses. It is useful to distinguish three classical limits, depending on the degree of relativistic motion:
\begin{equation}
\begin{split}
    &\textrm{Generic classical limit:} \qquad\qquad\qquad\;\; J \gg 1, \textrm{expansion in } \mathcal{O}(|\boldsymbol{q}|)\\
    &\textrm{Near-static (PN) classical limit:} \qquad\,\,\, J \gg 1,\;\; \sigma \sim 1,\\
    &\textrm{High-energy classical limit:} \qquad\qquad\;\; J \gg 1,\;\; \sigma \gg 1,
\end{split}
\end{equation}
where the boost parameter is defined as $ \sigma \equiv \frac{k_1\!\cdot\! k_2}{m_1 m_2}\, .$
This parameter measures the relative velocity of the two bodies: $\sigma \simeq 1$ corresponds to the near-static, post-Newtonian regime, whereas $\sigma \gg 1$ describes ultra-relativistic scattering. In the latter case, the hierarchy of scales becomes even stronger,
\begin{equation}
    s,\; |u| \gg m_i^2 \sim J^2 |t| \gg |t| = |q|^2\, .
\end{equation}
Classical considerations also impose important restrictions on the class of Feynman diagrams that can contribute to the scattering amplitude. These constraints may be summarised as follows (see \cite{youtube_484KUavMo0}):
\begin{itemize}[leftmargin=2em,itemsep=1.2em,topsep=0.6em]

\item
\begin{minipage}[t]{0.64\linewidth}
Any relevant Feynman loop diagram must contain at least one matter line in the loop; in particular, purely messenger loops are excluded from the classical analysis.
\end{minipage}\hfill
\begin{minipage}[t]{0.4\linewidth}
\vspace{-10.3pt}
\hspace{2 cm}
\centering
\includegraphics[width=\linewidth]{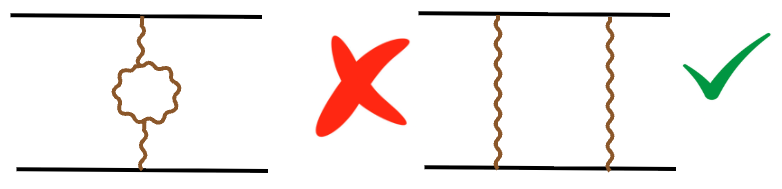}
\end{minipage}

\item
\begin{minipage}[t]{0.64\linewidth}
Diagrams with intersecting matter lines are also excluded. Classically, this is understood as a consequence of the large-separation regime, in which the two bodies do not undergo a head-on collision.
\end{minipage}\hfill
\begin{minipage}[t]{0.4\linewidth}
\vspace{-10.3pt}
\centering
\includegraphics[width=\linewidth]{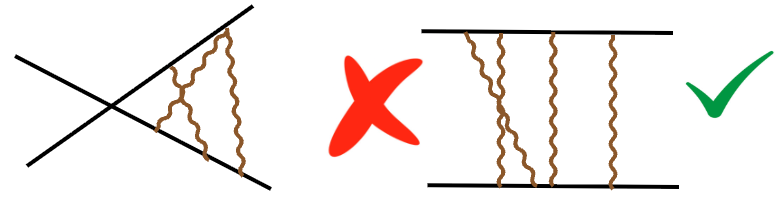}
\end{minipage}
\item
\begin{minipage}[t]{0.64\linewidth}
For lower loop orders, in particular for $L\leq 2$, messenger lines that begin and end on the same matter line do not contribute to the conservative potential. More precisely, diagrams of this type do not enter the conservative dynamics at one and two loops. At two loops, however, they may contribute to the radiative or dissipative sector when the appropriate propagator momenta are taken to be soft and on shell. At higher loop orders ($L\ge 3$), the situation becomes more subtle: multiple radiation lines may simultaneously go on shell, and such configurations can then contribute even to the conservative dynamics.
\end{minipage}\hfill
\begin{minipage}[t]{0.40\linewidth}
\vspace{0pt}
\centering
\includegraphics[width=\linewidth]{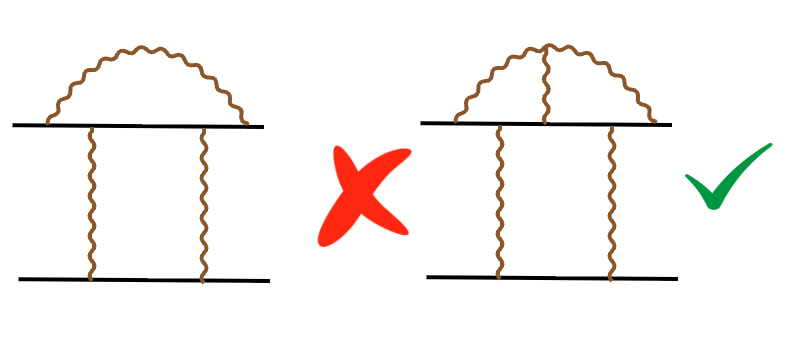}
\end{minipage}
\end{itemize}
\vspace{0.2 cm}
\thesisleadtopic{Kosower-Maybee-O'Connell (KMOC) Formalism \cite{Kosower:2018adc}.} The approach described above is physically quite intuitive: one first extracts an effective two-body potential from the scattering amplitude and then uses this potential to derive the relevant classical observables. However, the subsequent step of obtaining directly measurable quantities, such as those needed for gravitational-wave templates, is a rather different and considerably more involved problem. A major conceptual development in this direction came in 2018, when Kosower, Maybee, and O’Connell proposed a framework in which classical observables—such as the impulse, radiated momentum, and waveform—can be computed directly from scattering amplitudes, without passing through an intermediate effective potential. Although this approach will not be used in the present thesis, it is nonetheless worthwhile to review its basic ideas, since it provides a powerful and complementary perspective on the relation between quantum scattering amplitudes and classical gravitational dynamics. The KMOC formalism provides a direct route from on-shell amplitudes to classical observables, such as the impulse and radiated momentum, without first constructing an effective potential. Its main conceptual point is that the classical limit is not simply the tree-level truncation of the amplitude. Restoring $\hbar$, an $n$-point, $L$-loop amplitude scales as
\begin{equation}
    \mathcal A_{n,L}\sim \hbar^{\,1-\frac{n}{2}-L},
\end{equation}
so loop order alone does not determine whether a contribution is classical. Instead, one takes $\hbar\to0$ while keeping the massive external momenta fixed and scaling the exchanged or emitted massless momenta as $   q^\mu=\hbar\,\bar q^\mu,
    \qquad
    k^\mu=\hbar\,\bar k^\mu.$ Hence, the classical information is isolated from the soft-momentum sector of the amplitude. The incoming state is a semiclassical two-body wave packet with impact parameter $b^\mu$,
\begin{equation}
\begin{split}
    |{\rm in}\rangle_b
    &=
    \int d\Phi(p_1)\,d\Phi(p_2)\,
    \phi_1(p_1)\phi_2(p_2)\,
    e^{\,i b\cdot p_1/\hbar}\,
    |p_1,p_2\rangle, \\
    d\Phi(p)
    &=
    \hat d^4p\,\hat\delta^{(+)}(p^2-m^2),
    \qquad
    \hat d^4p\equiv \frac{d^4p}{(2\pi)^4}.
\end{split}
\end{equation}
Observables are then defined in the out-state, $ \langle \hat{\mathcal O}\rangle
    =
    {}_b\langle{\rm in}|\,S^\dagger \hat{\mathcal O} S\,|{\rm in}\rangle_b.$ For the impulse on particle $1$,
\begin{equation}
\begin{split}
    \Delta p_1^\mu
    &=
    {}_b\langle{\rm in}|\,S^\dagger \hat P_1^\mu S\,|{\rm in}\rangle_b
    -
    {}_b\langle{\rm in}|\,\hat P_1^\mu\,|{\rm in}\rangle_b
    \notag\\
    &=
    I_{(1)}^\mu+I_{(2)}^\mu.
\end{split}
\end{equation}
Here $I_{(1)}^\mu$ is linear in the amplitude, whereas $I_{(2)}^\mu$ is a quadratic unitarity (cut) contribution. Schematically,
\begin{equation}
\begin{split}
    I_{(1)}^\mu
    &\propto
    \int d\Phi(p_1)\,d\Phi(p_2)\,\hat d^4q\,
    e^{-i b\cdot q/\hbar}\,
    i q^\mu\,
    \mathcal A(p_1,p_2\!\to\! p_1+q,p_2-q), \\
    I_{(2)}^\mu
    &\propto
    \sum_X \int \cdots\,
    w_1^\mu\,
    \mathcal A\,\mathcal A^* .
    \end{split}
\end{equation}
Thus, beyond leading order, the classical impulse is an inclusive observable built from both linear and cut terms evaluated in the soft region. Radiation is treated analogously by defining
\begin{equation}
\begin{split}
    R^\mu
    =
    {}_b\langle{\rm in}|\,S^\dagger \hat K^\mu S\,|{\rm in}\rangle_b.
\end{split}
\end{equation}
In the classical limit it takes the form,
\begin{equation}
    R^\mu_{\rm cl}
    =
    \hbar^{-3}
    \sum_X \int d\Phi(k)\,K_X^\mu\,
    \big|\mathcal R(k,r_X)\big|^2 ,
\end{equation}
with the radiation kernel
\begin{equation}
\begin{split}
    \mathcal R(k,r_X)
    &=
    \hbar^{3/2}
    \prod_{i=1}^2
    \int \hat d^4w_i\,
    \hat\delta(2p_i\!\cdot w_i+w_i^2)
    \\
    &\quad\times
    \hat\delta^{(4)}(w_1+w_2-k-r_X)\,
    e^{\,i b\cdot w_1/\hbar}\,
    \mathcal A(p_1+w_1,p_2+w_2\!\to\! p_1,p_2,k,r_X).
\end{split}
\end{equation}
The KMOC formalism, therefore, extracts conservative and radiative observables directly from the $S$-matrix: the classical limit is selected by soft $\hbar$-scaling, not by loop order alone.

\thesisleadtopic{Worldline Formalisms: PM-EFT \cite{Kalin:2020mvi,Kalin:2022hph} and WQFT \cite{Mogull:2020sak,Jakobsen:2022psy}} In addition to direct methods based on scattering amplitudes, there exist complementary worldline approaches for studying the classical two-body problem. One such framework is Post-Minkowskian Effective Field Theory (\textbf{\texttt{PM-EFT}}), which is closely related to the \textbf{\texttt{PN-EFT}}/\textbf{\texttt{NRGR}} methods familiar from the post-Newtonian expansion. In this sense, PM-EFT provides a natural counterpart to the PM expansion of amplitudes discussed earlier. Another important development is Worldline Quantum Field Theory (\textbf{\texttt{WQFT}}), which in many ways complements the KMOC formalism. From the perspective of explicit computations, both PM-EFT and WQFT are especially attractive, since the relevant classical information can be extracted entirely from tree-level worldline diagrams. Put differently, one avoids the need to evaluate diagrams containing pure messenger loops.

Unlike amplitude-based approaches, in PM-EFT the black holes are modeled directly as point particles rather than scalar fields. It is convenient to write the point-particle action in Polyakov form,
\begin{equation}
       S_{\rm pm}
       =
       -\sum_{a=1,2}\frac{m_a}{2}\int d\sigma_a\, e_a
       \left(
       \frac{1}{e_a^2}\,g_{\mu\nu}(x_a)v_a^\mu v_a^\nu+1
       \right).
  \end{equation}
A useful feature of this representation is that, unlike in the usual post-Newtonian treatment, the post-Minkowskian expansion is naturally formulated using proper time to parametrise the worldlines. The conservative two-body effective action is obtained by integrating out the metric fluctuations,
\begin{equation}
    e^{iS_{\rm eff}[x_a]}
    =
    \int \mathcal{D}h_{\mu\nu}\,
    e^{\,iS_{\rm EH}[h]+iS_{\rm GF}[h]+iS_{\rm pp}[x_a,h]},
\end{equation}
and it admits an expansion in powers of Newton's constant,
\begin{equation}
    S_{\rm eff}=\sum_n \int d\tau\, L_n,
    \qquad
    L_n:\mathcal O(G^n).
\end{equation}
Once the effective action is known, classical observables such as the impulse or the deflection angle can be systematically extracted from it. In particular, the impulse on particle \(i\) is given by
\begin{equation}
    \Delta p_i^\mu
    =
    -\eta^{\mu\nu}\sum_n \int_{-\infty}^{\infty} d\tau_i\,
    \frac{\partial L_n}{\partial x_i^\nu}.
\end{equation}
Thus, PM-EFT complements the amplitude-based approach discussed previously: rather than working directly with scattering amplitudes, one reduces the problem to the computation of an effective action, or equivalently an effective potential, from which the conservative dynamics can be derived.

On the other hand, WQFT was introduced as a worldline framework closely connected in spirit to the KMOC program. As in PM-EFT, the compact objects are represented by point particles. However, unlike in standard quantum field theory, one treats not only the messenger fields but also the worldline degrees of freedom as dynamical variables. The WQFT partition function is defined as
\begin{equation}
    Z_{\textrm{WQFT}}^{(n)}
    =
    \mathcal{N}
    \int \mathcal{D}h_{\mu\nu}\,
    e^{iS_{EH}+iS_{gf}}
    \Big(\prod_{i=1}^{n}\mathcal{D}x_i\,e^{iS_{pm}^i}\Big).
\end{equation}
It was further observed that, in the appropriate regime, the WQFT partition function is related to the scattering amplitude through the eikonal phase,
\begin{equation}
    Z_{\textrm{WQFT}}^{(n)}=e^{i\chi},
\end{equation}
where \(\chi\) denotes the eikonal phase. In computing this phase one must keep track of the \(i\varepsilon\)-prescriptions of the worldline propagators. {\color{black}A naive time-symmetric/Feynman prescription may leave spurious infrared poles. Recently in \cite{Kim:2024svw}, those issues has been dealt. We will take a moment to discuss them in the next section.
\paragraph{Magnus expansion and the eikonal generator.}
The relation above should be understood as a shorthand appropriate to the elastic/eikonal sector. A more precise formulation starts from the logarithm of the unitary scattering operator. Following the recent analysis of the classical eikonal from the Magnus expansion~\cite{Kim:2024svw}, one writes
\begin{align}
    \widehat S
    =
    1+\frac{i}{\hbar}\widehat T
    =
    \exp\!\left(\frac{i}{\hbar}\widehat N\right)
    \equiv
    \exp\!\left(\frac{i}{\hbar}\widehat\chi\right),
    \qquad
    \widehat N^\dagger=\widehat N .
\end{align}
Here \(\widehat T\) is the usual transition operator, while \(\widehat N\) is the Hermitian operator appearing in the exponent. The classical eikonal is the classical limit of this logarithm, not the naive amplitude itself. This distinction separates the genuine eikonal generator from the Born iterations already generated by exponentiation. 

\begin{figure}[htb!]
\centering
\color{black}
\begin{tikzpicture}[
    font=\small,
    op/.style={circle,draw=black!80!black,fill=white,inner sep=1.2pt,minimum size=0.42cm,font=\scriptsize},
    root/.style={circle,draw=black!80!black,fill=black!15,inner sep=1pt,minimum size=0.28cm},
    cut/.style={densely dotted,draw=black!80!black,thick,->,>=stealth},
    arr/.style={->,draw=black!80!black,thick}
]
    \node[root] (r0) at (0,0) {};
    \node[op] (a0) at (0.75,0) {1};
    \node[op] (b0) at (1.55,0) {2};
    \draw[cut] (r0) -- (a0);
    \draw[cut] (a0) -- (b0);
    \draw[black!80!black,thick] (0.37,0) ellipse (0.68 and 0.34);

    \draw[arr] (2.15,0) -- (2.75,0);

    \node[root] (r1) at (3.15,0) {};
    \node[op] (a1) at (3.9,0) {1};
    \node[op] (b1) at (4.7,0) {2};
    \draw[cut] (r1) -- (a1);
    \draw[cut] (a1) -- (b1);
    \node at (5.15,0) {\(+\)};
    \node[root] (r2) at (5.55,0) {};
    \node[op] (a2) at (6.22,0.28) {1};
    \node[op] (b2) at (6.22,-0.28) {2};
    \draw[cut] (r2) -- (a2);
    \draw[cut] (r2) -- (b2);

    \node[root] (s0) at (0,-1.5) {};
    \node[op] (s1) at (0.75,-1.5) {1};
    \node[op] (s2) at (1.55,-1.5) {2};
    \node[op] (s3) at (2.35,-1.5) {3};
    \draw[cut] (s0) -- (s1);
    \draw[cut] (s1) -- (s2);
    \draw[cut] (s2) -- (s3);
    \draw[black!80!black,thick] (0.78,-1.5) ellipse (1.12 and 0.38);
    \draw[black!80!black,thick] (0.78,-1.5) ellipse (0.62 and 0.28);

    \draw[arr] (2.95,-1.5) -- (3.45,-1.5);

    \node[root] (t0) at (3.8,-1.5) {};
    \node[op] (t1) at (4.42,-1.5) {1};
    \node[op] (t2) at (5.04,-1.5) {2};
    \node[op] (t3) at (5.66,-1.5) {3};
    \draw[cut] (t0) -- (t1);
    \draw[cut] (t1) -- (t2);
    \draw[cut] (t2) -- (t3);

    \node at (6.05,-1.5) {\(+\)};
    \node[root] (u0) at (6.42,-1.5) {};
    \node[op] (u1) at (7.04,-1.5) {1};
    \node[op] (u2) at (7.66,-1.20) {2};
    \node[op] (u3) at (7.66,-1.80) {3};
    \draw[cut] (u0) -- (u1);
    \draw[cut] (u1) -- (u2);
    \draw[cut] (u1) -- (u3);

    \node at (8.05,-1.5) {\(+\)};
    \node[root] (v0) at (8.42,-1.5) {};
    \node[op] (v1) at (9.04,-1.15) {1};
    \node[op] (v2) at (9.04,-1.85) {2};
    \node[op] (v3) at (9.72,-1.5) {3};
    \draw[cut] (v0) -- (v1);
    \draw[cut] (v0) -- (v2);
    \draw[cut] (v0) -- (v3);
\node at (10.22,-1.5) {\(+\cdots\)};
\end{tikzpicture}
\caption{Schematic bubble notation for nested Poisson brackets. Popping a bubble produces oriented causal cuts according to the Leibniz rule.}
\label{fig:eikonal-bubble-poisson}
\end{figure}
The operator \(\widehat N\) acts as the generator of the scattering process. For any asymptotic observable,
\begin{align}
    \widehat{\mathcal O}_{\rm out}
    =
    \widehat S^\dagger
    \widehat{\mathcal O}_{\rm in}
    \widehat S
    =
    e^{-i\widehat N/\hbar}
    \widehat{\mathcal O}_{\rm in}
    e^{i\widehat N/\hbar}.
\end{align}
Taking the classical limit turns commutators into Poisson brackets,
\begin{align}
    \frac{1}{i\hbar}
    [\widehat A,\widehat B]
    \longrightarrow
    \{A,B\}_{\rm PB},
\end{align}
so that
\begin{align}
    \mathcal O_{\rm out}
    =
    e^{\{N_{\rm cl},\,\cdot\,\}_{\rm PB}}
    \mathcal O_{\rm in}
    =
    \mathcal O_{\rm in}
    +
    \{N_{\rm cl},\mathcal O_{\rm in}\}_{\rm PB}
    +
    \frac{1}{2}
    \{N_{\rm cl},\{N_{\rm cl},\mathcal O_{\rm in}\}_{\rm PB}\}_{\rm PB}
    +\cdots .
\end{align}
Thus the eikonal is best viewed as the canonical generator that maps the incoming free phase space to the outgoing free phase space. This is the conceptual reason why the impulse, spin kick, and other scattering observables should be extracted from the eikonal through a bracket expansion rather than by differentiating a naive phase without accounting for iterations.

A practically feasible way to compute this logarithm is the Magnus expansion~\cite{Magnus:1954,Kim:2024svw}. If
\begin{align}
    \widehat S
    =
    \mathcal T
    \exp\!\left[
      -\frac{i}{\hbar}
      \int_{-\infty}^{\infty}
      dt\,\widehat V_I(t)
    \right]
    =
    \exp\Omega,
\end{align}
then \(\Omega=i\widehat N/\hbar\). The first terms are
\begin{align}
\begin{split}
   & -\widehat N^{(1)}
    =
    \int_{-\infty}^{\infty}
    dt_1\,\widehat V_1,\,\,\qquad
    -\widehat N^{(2)}
    =
    \frac{1}{2i\hbar}
    \int_{t_1>t_2}
    dt_1dt_2\,
    [\widehat V_1,\widehat V_2],\cdots
\end{split}
\end{align}
where \(\widehat V_i\equiv \widehat V_I(t_i)\). After taking the classical limit, these nested commutators become nested Poisson brackets:
\begin{align}
\begin{split}
    -N_{\rm cl}^{(1)}
    &=
    \int dt_1\,V_1,\qquad
    -N_{\rm cl}^{(2)}
    =
    \frac{1}{2}
    \int_{t_1>t_2}
    dt_1dt_2\,
    \{V_1,V_2\}_{\rm PB},\cdots
\end{split}
\end{align}
This is why the Magnus expansion has a smooth classical limit: the factors of \(1/(i\hbar)^{n-1}\) precisely convert the \((n-1)\)-fold nested commutators into Poisson brackets. It also avoids the superclassical intermediate terms that appear in a direct Dyson expansion. The graphical bookkeeping for the nested brackets is summarized in Fig.~(\ref{fig:eikonal-bubble-poisson}).

The same construction admits a compact tree expansion, whose rooted (and non-rooted generalization) is illustrated in Fig.~\eqref{fig:eikonal-murua-trees}. Denoting by \(\tau\) an oriented tree with \(|\tau|=n\) vertices, by \(\sigma(\tau)\) its automorphism/symmetry factor, and by \(I(\tau)\) the associated worldline integral with the appropriate causal propagators, one may write
\begin{align}
    -N_{\rm cl}^{(n)}
    =
    \sum_{|\tau|=n}
    \frac{\omega(\tau)}{\sigma(\tau)}
    I(\tau).
\end{align}
The coefficient \(\omega(\tau)\) is fixed by the Magnus algebra. For rooted trees, Kim et al. use Murua's recursive formula~\cite{Murua:2006,Kim:2024svw},
\begin{align}
    \omega(\tau)
    =
    \sum_{p\in {\rm Pr}(\tau)}
    B_{|p|-1}\,
    e(p')\,
    \omega(\tau\backslash p).
\end{align}
Here \(B_m\) are Bernoulli numbers, \({\rm Pr}(\tau)\) denotes the set of edge partitions that contain all edges attached to the root, \(p'\) is obtained from \(p\) after amputating the root edges, and \(e(\tau)\) is the inverse tree-factorial function. The partition \(p\) represents the new causal cuts generated by the nested Poisson brackets, while \(\tau\backslash p\) represents the lower-order subtrees already present inside the Magnus recursion.

For the non-rooted trees relevant to relativistic WQFT diagrams, one must allow more than one possible semi-root. If \(s\) is a chosen semi-root, the extended Murua formula reads \cite{Kim:2024svw}
\begin{align}
    \omega(\tau)
    =
    \sum_{p\in {\rm P}_{s}(\tau)}
    (-1)^{\ell(p)}
    B_{|p|-1}\,
    e(p')\,
    \omega(\tau\backslash p).
\end{align}
The set \({\rm P}_{s}(\tau)\) contains partitions that include all edges adjacent to the chosen semi-root \(s\). The integer \(\ell(p)\) counts the orientation flips needed to view the partition as rooted at \(s\). The final \(\omega(\tau)\) is independent of this auxiliary semi-root choice. Thus the usual WQFT integrands may be used, while the causal \(i0\)-prescriptions of the master integrals are fixed only after the Magnus weights have been assigned.

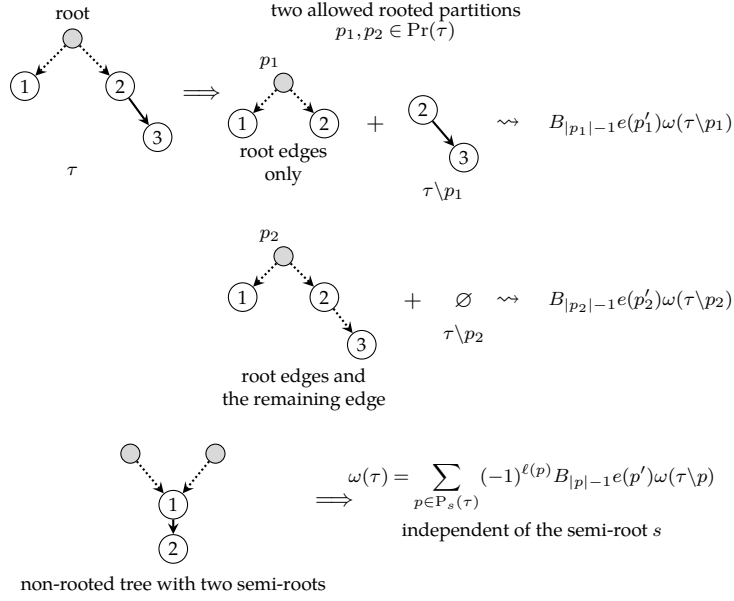
\begin{figure}[htb!]
\centering
\color{black}
\begin{tikzpicture}[
    scale=0.9,
    every node/.style={transform shape},
    font=\small,
    op/.style={circle,draw=black!80!black,fill=white,inner sep=1.2pt,minimum size=0.42cm,font=\scriptsize},
    root/.style={circle,draw=black!80!black,fill=black!15,inner sep=1pt,minimum size=0.28cm},
    cut/.style={densely dotted,draw=black!80!black,thick,->,>=stealth},
    old/.style={draw=black!55!black,thick,->,>=stealth},
    arr/.style={->,draw=black!80!black,thick}
]
    \node[root,label={[font=\scriptsize]above:root}] (r) at (0,0.1) {};
    \node[op] (a) at (-0.7,-0.6) {1};
    \node[op] (b) at (0.7,-0.6) {2};
    \node[op] (c) at (1.25,-1.35) {3};
    \draw[cut] (r) -- (a);
    \draw[cut] (r) -- (b);
    \draw[old] (b) -- (c);
    \node[font=\scriptsize] at (0,-1.85) {\(\tau\)};

    \node at (1.9,-0.72) {\(\Longrightarrow\)};
    \node[align=center,font=\scriptsize] at (4.75,0.35)
        {two allowed rooted partitions\\[-1pt]
        \(p_1,p_2\in{\rm Pr}(\tau)\)};

    \node[font=\scriptsize] at (2.9,-0.25) {\(p_1\)};
    \node[root] (p1r) at (3.1,-0.55) {};
    \node[op] (p1a) at (2.5,-1.15) {1};
    \node[op] (p1b) at (3.7,-1.15) {2};
    \draw[cut] (p1r) -- (p1a);
    \draw[cut] (p1r) -- (p1b);
    \node[align=center,font=\scriptsize] at (3.1,-1.75)
        {root edges\\only};

    \node at (4.45,-1.15) {\(+\)};
    \node[op] (p1rb) at (5.15,-0.95) {2};
    \node[op] (p1rc) at (5.75,-1.65) {3};
    \draw[old] (p1rb) -- (p1rc);
    \node[font=\scriptsize] at (5.45,-2.15) {\(\tau\backslash p_1\)};

    \node at (6.4,-1.15) {\(\leadsto\)};
    \node[align=center,font=\scriptsize] at (8.35,-1.15)
        {\(\displaystyle B_{|p_1|-1}e(p_1')\omega(\tau\backslash p_1)\)};

    \node[font=\scriptsize] at (2.9,-2.8) {\(p_2\)};
    \node[root] (p2r) at (3.1,-3.1) {};
    \node[op] (p2a) at (2.5,-3.7) {1};
    \node[op] (p2b) at (3.7,-3.7) {2};
    \node[op] (p2c) at (4.25,-4.4) {3};
    \draw[cut] (p2r) -- (p2a);
    \draw[cut] (p2r) -- (p2b);
    \draw[cut] (p2b) -- (p2c);
    \node[align=center,font=\scriptsize] at (3.4,-5.05)
        {root edges and\\the remaining edge};

    \node at (5.0,-3.75) {\(+\)};
    \node at (5.75,-3.75) {\(\varnothing\)};
    \node[font=\scriptsize] at (5.75,-4.25) {\(\tau\backslash p_2\)};

    \node at (6.4,-3.75) {\(\leadsto\)};
    \node[align=center,font=\scriptsize] at (8.35,-3.75)
        {\(\displaystyle B_{|p_2|-1}e(p_2')\omega(\tau\backslash p_2)\)};

    \node[root] (s1) at (0.85,-6.0) {};
    \node[root] (s2) at (2.1,-6.0) {};
    \node[op] (m) at (1.48,-6.75) {1};
    \node[op] (d) at (1.48,-7.4) {2};
    \draw[cut] (s1) -- (m);
    \draw[cut] (s2) -- (m);
    \draw[old] (m) -- (d);
    \node[font=\scriptsize] at (1.48,-7.9)
        {non-rooted tree with two semi-roots};

    \node at (3.85,-6.7) {\(\Longrightarrow\)};
    \node[align=center,font=\scriptsize] at (6.75,-6.7)
        {\(\displaystyle
        \omega(\tau)=
        \sum_{p\in{\rm P}_{s}(\tau)}
        (-1)^{\ell(p)}
        B_{|p|-1}e(p')\omega(\tau\backslash p)\)\\[3pt]
        independent of the semi-root \(s\)};

\end{tikzpicture}
\caption{Schematic graphical content of Murua's recursion. The displayed rooted tree has two allowed partitions: \(p_1\), containing only the root-adjacent edges, and \(p_2\), containing all three edges. In general, rooted partitions \(p\in{\rm Pr}(\tau)\) must contain all root-adjacent edges; for non-rooted trees, \(p\in{\rm P}_{s}(\tau)\) must contain all edges adjacent to the chosen semi-root \(s\).}
\label{fig:eikonal-murua-trees}
\end{figure}
This weighting is also what removes the spurious infrared divergence in the 3PM conservative eikonal. In the WQFT calculation of~\cite{Kim:2024svw}, the only relevant master integral (the asymmetric worldline diagrams) whose value depends on the worldline causality prescription enters as
\begin{align}
    I^{(4)}
    =
    w_{++}I^{(4)}_{++}
    +
    w_{+-}I^{(4)}_{+-},
    \qquad
    w_{++}=\frac{2}{3},
    \qquad
    w_{+-}=\frac{1}{3}.
\end{align}
The leading dimensional-regularization behavior is
\begin{align}
    I^{(4)}_{++}
    \propto
    \frac{1}{2\epsilon^2}
    +\mathcal O(\epsilon^{-1}),
    \qquad
    I^{(4)}_{+-}
    \propto
    -\frac{1}{\epsilon^2}
    +\mathcal O(\epsilon^{-1}),
\end{align}
and hence
\begin{align}
    \frac{2}{3}I^{(4)}_{++}
    +
    \frac{1}{3}I^{(4)}_{+-}
    =
    \mathcal O(\epsilon^{-1}).
\end{align}
The dangerous \(1/\epsilon^2\) pole therefore cancels in the Magnus-weighted combination, as shown in Fig.~\eqref{fig:eikonal-ir-causality}. By contrast, the naive time-symmetric assignment \(w_{++}=w_{+-}=1/2\) would leave a \(1/\epsilon^2\) pole in the 3PM eikonal. Thus the Magnus/Murua prescription fixes the causal retarded/advanced weighting of the worldline propagators.

\begin{figure}[htb!]
\centering
\color{black}
\begin{tikzpicture}[
    font=\small,
    op/.style={circle,draw=black!80!black,fill=white,inner sep=1.1pt,minimum size=0.38cm,font=\scriptsize},
    line/.style={draw=black!55!black,thick},
    ret/.style={densely dotted,draw=black!80!black,thick,->,>=stealth},
    adv/.style={densely dotted,draw=black!80!black,thick,->,>=stealth},
    arr/.style={->,draw=black!80!black,thick}
]
    \draw[line] (-0.25,0.52) -- (2.25,0.52);
    \draw[line] (-0.25,-0.52) -- (2.25,-0.52);
    \node[op] (a1) at (0.15,0.52) {1};
    \node[op] (a2) at (1.0,0.52) {2};
    \node[op] (a3) at (1.85,0.52) {3};
    \node[op] (b1) at (0.15,-0.52) {1};
    \node[op] (b2) at (1.0,-0.52) {2};
    \node[op] (b3) at (1.85,-0.52) {3};
    \draw[ret] (a1) -- (b1);
    \draw[ret] (a2) -- (b2);
    \draw[ret] (a3) -- (b3);
    \node[align=center,text=black!80!black] at (1.0,1.12)
        {\(\displaystyle \frac{2}{3}I^{(4)}_{++}\)};
    \node[align=center,font=\scriptsize,text=black!80!black] at (1.0,-1.17)
        {\(\displaystyle I^{(4)}_{++}\propto \frac{1}{2\epsilon^2}\)};

    \node at (3.0,0) {\(+\)};

    \draw[line] (3.75,0.52) -- (6.25,0.52);
    \draw[line] (3.75,-0.52) -- (6.25,-0.52);
    \node[op] (c1) at (4.15,0.52) {1};
    \node[op] (c2) at (5.0,0.52) {2};
    \node[op] (c3) at (5.85,0.52) {3};
    \node[op] (d1) at (4.15,-0.52) {1};
    \node[op] (d2) at (5.0,-0.52) {2};
    \node[op] (d3) at (5.85,-0.52) {3};
    \draw[ret] (c1) -- (d1);
    \draw[adv] (d2) -- (c2);
    \draw[ret] (c3) -- (d3);
    \node[align=center,text=black!80!black] at (5.0,1.12)
        {\(\displaystyle \frac{1}{3}I^{(4)}_{+-}\)};
    \node[align=center,font=\scriptsize,text=black!80!black] at (5.0,-1.17)
        {\(\displaystyle I^{(4)}_{+-}\propto -\frac{1}{\epsilon^2}\)};

    \node[text=black!80!black] at (6.52,0.23) {\(\Longrightarrow\)};
    \node[align=center,text=black!80!black] at (8.85,0.23)
        {\(\displaystyle
        \frac{2}{3}I^{(4)}_{++}
        +
        \frac{1}{3}I^{(4)}_{+-}
        =
        \mathcal O(\epsilon^{-1})\)};
    \node[align=center,font=\scriptsize,text=black!80!black] at (8.85,-0.55)
        {leading \(1/\epsilon^2\) pole cancels};

    \node[align=center,font=\scriptsize,text=black!80!black] at (3.0,-1.95)
        {};
\end{tikzpicture}
\caption{Causality-prescription diagrams for the IR-sensitive master integral. The Magnus/Murua weights assigned to the two oriented orderings cancel the leading \(1/\epsilon^2\) pole.}
\label{fig:eikonal-ir-causality}
\end{figure}
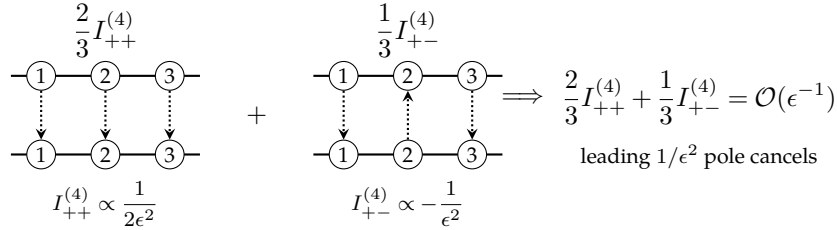
Nonetheless, in our computation of the eikonal phase in dCS theory, we find that the IR divergences cannot be removed by considering only the conservative sector. One might expect that including dissipative corrections would cancel these divergences. However, this expectation may be too naive: IR cancellation is a delicate procedure, and its realization may depend on theory-specific input. If the cancellation prescription is universal for a given class of theories, then this may not pose a serious obstruction. If, however, the prescription is theory dependent, it must be investigated carefully. In this thesis we do not attempt such an analysis, and instead leave the systematic treatment of this issue as an important direction for future work.}\\
Now, within this framework, with suitable care on $i\varepsilon$-prescription, observables are computed as one-point functions,
\begin{equation}
      \mathcal{O}(b_i,v_i)
      :=
      \langle\hat{\mathcal O}(\hat x,\hat h_{\mu\nu})\rangle_{\textrm{WQFT}}
      =
      \frac{1}{Z^{(n)}_{\textrm{WQFT}}}
      \int \mathcal{D}h_{\mu\nu}\,
      e^{iS_{EH}+iS_{gf}}
      \Big(\prod_{i=1}^{n}\mathcal{D}x_i\,e^{iS_{pm}^i}\Big)
      \mathcal O(x_i,h).
\end{equation}
For instance, the impulse and the waveform are obtained from
\begin{equation}
\begin{split}
    \Delta p_i^\mu
    &=
    m_i \int_{-\infty}^{\infty} d\tau_i\,
    \Big\langle \frac{d^2 z_i^\mu}{d\tau_i^2}\Big\rangle_{\text{WQFT}}
    =
    -m_i\omega^2\langle z_i^\mu(\omega)\rangle_{\text{WQFT}}\Big|_{\omega=0},\\
    f_h(k)
    &:=
    \frac{1}{4\pi m_p}\,
    \epsilon^\mu\epsilon^\nu\,
    k^2\langle h_{\mu\nu}(k)\rangle_{\textrm{WQFT}}\Big|_{k^2\to 0}.
\end{split}
\end{equation}
In practice, these one-point functions are evaluated perturbatively using tree-level WQFT diagrams. As in PM-EFT, one must exclude closed loops built solely from messenger and worldline propagators. In this way, WQFT provides an efficient route to classical observables, such as the impulse and waveform, within the framework of perturbative quantum field theory. In this thesis, we use these techniques to compute gravitational observables in several effective theories beyond General Relativity: scalar-tensor gravity, the phenomenological bottom-up EFT known as dynamical Chern-Simons theory, and Einstein-Maxwell-Dilaton theory, which arises as a low-energy effective theory of heterotic string theory.
\section{Celestial Amplitudes and Eikonalization}
{\color{black}
Beyond the study of observables in binary encounters, this thesis also explores the celestial dual of scattering amplitudes in higher-derivative theories of gravity, with particular emphasis on quadratic effective field theories. The motivation for this direction comes from the broader question of flat-space holography. The success of the AdS/CFT correspondence~\cite{Maldacena:1997re} shows that a theory of quantum gravity in an asymptotically anti-de Sitter spacetime can be equivalently described by a conformal field theory living on its boundary. It is therefore natural to ask whether an analogous holographic description exists for asymptotically flat spacetimes. In flat space, however, the most basic observables are not boundary correlation functions but scattering amplitudes defined at null infinity. This suggests that, if a holographic reformulation of flat-space quantum gravity exists, it should organize the usual momentum-space scattering amplitudes into observables of a lower-dimensional theory.

This idea is realized in the framework of Celestial Conformal Field Theory (CCFT)~\cite{Strominger:2013jfa,Strominger:2017zoo,Pasterski:2017ylz,Pasterski:2021raf,Donnay:2020guq}. In this approach, four-dimensional scattering amplitudes in asymptotically flat spacetime are mapped to conformal correlators on the celestial sphere at null infinity. The celestial sphere is the two-sphere of asymptotic null infinity, and each external massless particle is associated with a point on this sphere. The Lorentz group \(SL(2,\mathbb{C})\) acts on the celestial sphere as the global conformal group, while extensions of the asymptotic symmetry group are closely related to supertranslations and superrotations. Consequently, celestial amplitudes provide a bridge between three closely connected structures: flat-space scattering theory, asymptotic symmetries, and holography. This perspective has led to significant developments in the study of soft theorems, memory effects, and the conformal structure of scattering amplitudes~\cite{He:2014laa,Adamo:2019ipt,Gonzalez:2020tpi,Pasterski:2021fjn,Gonzo:2022tjm,Strominger:2021mtt}.

For massless scattering, an external particle is described by a positive energy \(\omega\) and a point \((z,\bar z)\) on the celestial sphere. The complex coordinate \(z\) is a stereographic coordinate on the sphere, and the corresponding null momentum can be parametrized as~\cite{Arkani-Hamed:2020gyp,Pasterski:2016qvg}
\begin{equation}
 p_\mu
 =
 \omega q_\mu(z,\bar z)
 =
 \frac{\omega}{\sqrt{2}}
 \left(
 1+|z|^2,\;
 z+\bar z,\;
 -i(z-\bar z),\;
 1-|z|^2
 \right).
\end{equation}
Here
\begin{equation}
 q_\mu(z,\bar z)
 =
 \frac{1}{\sqrt{2}}
 \left(
 1+|z|^2,\;
 z+\bar z,\;
 -i(z-\bar z),\;
 1-|z|^2
 \right)
\end{equation}
is a null vector specifying the direction of propagation of the particle. Thus the energy \(\omega\) fixes the overall scale of the momentum, while the celestial coordinates \((z,\bar z)\) fix its direction. In this parametrization the usual momentum-space data of each external massless particle are separated into an energy variable and an angular variable on the celestial sphere.

The celestial amplitude is obtained by Mellin transforming the momentum-space amplitude with respect to the external energies. For an \(n\)-point amplitude, one defines
\begin{equation}
    \mathcal{A}_n(\Delta_i,z_i,\bar z_i)
    =
    \int_0^\infty
    \prod_{i=1}^n
    d\omega_i\,\omega_i^{\Delta_i-1}\,
    \mathbb{M}_n(p_1,\ldots,p_n)\,
    \delta^{(4)}
    \left(
    \sum_{i=1}^n
    \epsilon_i\omega_i q_i
    \right),
\end{equation}
where \(q_i\equiv q(z_i,\bar z_i)\) is the null direction vector associated with the \(i\)-th point on the celestial sphere, and \(\omega_i>0\) is the corresponding energy. The sign \(\epsilon_i\) keeps track of whether the particle is outgoing or incoming:
\begin{equation}
    \epsilon_i =
    \begin{cases}
        +1, & \text{for an outgoing particle},\\
        -1, & \text{for an incoming particle}.
    \end{cases}
\end{equation}
The argument of the delta function is therefore the total four-momentum,
\begin{equation}
    \sum_{i=1}^n \epsilon_i p_i^\mu
    =
    \sum_{i=1}^n \epsilon_i \omega_i q_i^\mu,
\end{equation}
and the delta function imposes overall momentum conservation. The Mellin parameter \(\Delta_i\) is interpreted as the conformal dimension of the celestial operator associated with the \(i\)-th external particle. In this way, each external scattering state in four-dimensional momentum space is mapped to a conformal primary insertion on the celestial sphere.

For four-point scattering, the kinematics are especially simple because conformal symmetry fixes much of the coordinate dependence of the celestial correlator. The remaining nontrivial dependence can be expressed in terms of the conformal cross-ratio \(z\), together with its complex conjugate \(\bar z\). In physical scattering regions, momentum conservation imposes constraints on these variables, and different scattering channels correspond to different ranges of the cross-ratio. Thus the usual Mandelstam-channel structure of the momentum-space amplitude is translated into the analytic structure of the celestial correlator as a function of \(z\).
\begin{figure}
    \centering
    \includegraphics[width=0.6\linewidth]{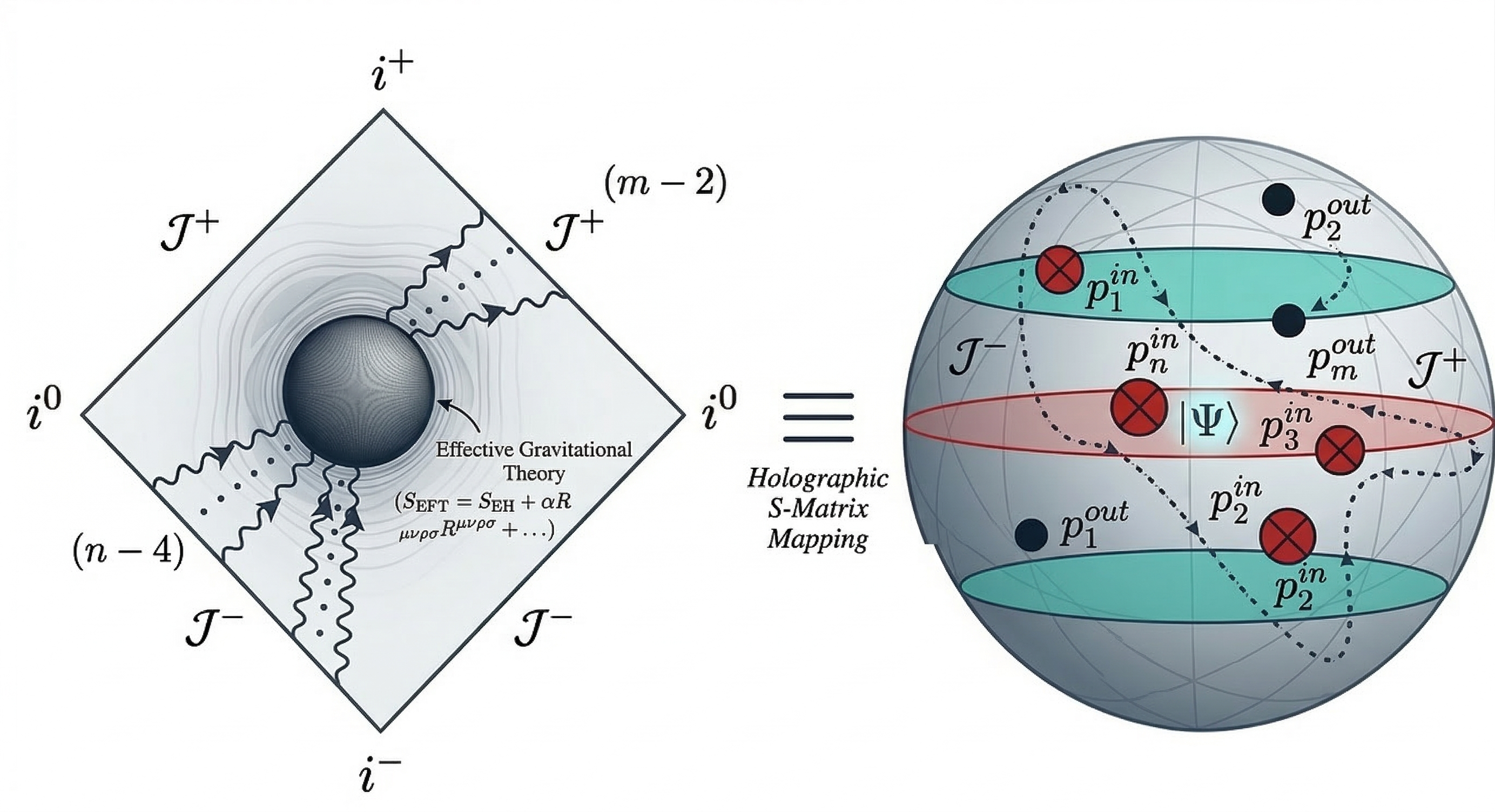}
    \caption{Celestial description of a four-dimensional scattering process. A momentum-space \(n\to m\) scattering amplitude in asymptotically flat spacetime is mapped, through a Mellin transform over the external energies, to a conformal correlator on the celestial sphere.}
    \label{fig23}
\end{figure}

After factoring out the universal kinematical dependence fixed by conformal covariance, the theory-dependent information is contained in a reduced Mellin amplitude. For a four-point process \(ij\to kl\), this reduced amplitude can be written schematically as
\begin{equation}
    \mathbfcal{A}^{ij\to kl}(\gamma,z)
    =
    B^{ij\to kl}(z)
    \int_0^\infty d\omega\,
    \omega^{\gamma-1}\,
    \mathbb{M}
    \big(
    s=\Phi(z)\omega^2,\;
    t=-\phi(z)\omega^2
    \big),
\end{equation}
where
\begin{equation}
    \gamma=\sum_{i=1}^4(\Delta_i-1).
\end{equation}
The functions \(\Phi(z)\) and \(\phi(z)\) are determined by the kinematic channel under consideration, while the prefactor \(B^{ij\to kl}(z)\) contains the universal dependence required by conformal symmetry and by the external quantum numbers. The reduced Mellin amplitude therefore isolates the dynamical content of the underlying four-dimensional theory. In particular, its analytic structure encodes information about the high-energy and fixed-angle behavior of the original momentum-space amplitude.

A central difficulty in celestial holography is that perturbative celestial amplitudes are often divergent when computed order by order in quantum field theory. These divergences arise because the Mellin transform probes both the low-energy and high-energy behavior of the momentum-space amplitude. In theories that are only effective descriptions, the ultraviolet behavior of the amplitude can lead to singularities or ill-defined Mellin integrals. One way to improve this behavior is to resum the perturbative expansion before performing the Mellin transform. This is particularly natural in the eikonal regime, where high-energy small-angle scattering amplitudes exponentiate. In General Relativity, eikonal exponentiation reorganizes the perturbative series and can lead to celestial amplitudes with improved analytic properties.

The inclusion of higher-derivative corrections makes this problem more subtle and more interesting. Quadratic curvature terms modify the momentum dependence of the gravitational amplitude and therefore alter the ultraviolet behavior that enters the celestial Mellin transform. From the celestial perspective, such corrections change the analytic structure of the Mellin amplitude and may soften some of the divergences that appear in pure Einstein gravity. At the same time, constructing the corresponding eikonal amplitudes becomes technically more involved, because higher-derivative interactions introduce additional momentum dependence and new scales into the problem. One of the goals of this thesis is to analyze these issues systematically for quadratic gravitational EFTs and to study how methods from analytic functional analysis can be used to characterize the resulting celestial amplitudes.}
\section{Computational Techniques and Methodology}
Even though QFT enters the classical two-body problem in many seemingly different ways, the technical core is the same. One organises the computation perturbatively and expresses the desired classical observables in terms of multidimensional loop-momentum integrals—Feynman integrals. As one pushes to higher orders, these integrals proliferate and develop increasingly rich analytic structure, which quickly becomes the main bottleneck to precision. This is why advances in amplitude methods and Feynman-integral technology tend to propagate across subfields: the same machinery that was refined for particle scattering now provides a common language connecting collider physics to classical gravitational dynamics.\par

The computation of Feynman Integrals is simplified by powerful algebraic structures. A general Feynman integral in momentum representation takes the form,
\begin{equation}
I \;=\; e^{\,l\epsilon\gamma_E}\,(\mu^2)^{\nu-\frac{lD}{2}}
\int \prod_{r=1}^{l}\frac{d^{D}\ell_r}{i\pi^{D/2}}
\prod_{j=1}^{n_{\text{int}}}\frac{1}{\bigl(q_j^{2}-m_j^{2}\bigr)^{\nu_j}}\,,
\tag{2.138}
\end{equation}
\noindent where
\begin{equation}
\epsilon=\frac{D_{\text{int}}-D}{2},
\qquad
\nu=\sum_{j=1}^{n_{\text{int}}}\nu_j,
\qquad
q_j=\sum_{r=1}^{l}\lambda_{jr}\ell_r+\sum_{r=1}^{n_{\text{ext}}-1}\sigma_{jr}p_r.
\tag{2.139}
\end{equation}
In standard introductory QFT courses, multiloop integrals are often introduced through the Feynman--Schwinger parameterisation. While this method is extremely useful, the integrals that arise in modern precision problems are frequently far more intricate: they depend nontrivially on kinematic invariants and the dimensional regulator $\epsilon$, and they appear in large families whose members proliferate rapidly with perturbative order. In such situations, it is neither practical nor conceptually efficient to evaluate each integral in isolation.  
\subsection*{Integration-by-Parts Identities}
A fundamental simplification in modern amplitude computations stems from the rich algebraic structure of Feynman integrals. The celebrated \textbf{Integration-by-Parts (IBP)} identities provide a systematic mechanism to relate thousands of seemingly distinct integrals within the same kinematic family. By exploiting these relations, an enormous system of integrals can be algebraically projected down to a finite, minimal basis of independent objects known as \textbf{Master Integrals} (MIs). This reduction is the algorithmic backbone of virtually all contemporary multi-loop calculations. The underlying principle relies on a core property of dimensional regularisation:
\begin{tcolorbox}[
  thesisstatementbox,
  title=Theorem,
  before skip=6pt,
  after skip=6pt,
  boxsep=2pt
]
\textit{Within dimensional regularization, the integral of any total derivative with respect to a loop momentum strictly vanishes.}
\begin{equation}
    \int \prod_{r} d^d\ell_r\;
    \frac{\partial}{\partial \ell_i^\mu}
    \left(
      v^\mu
      \prod_{j}\frac{1}{D_j^{\nu_j}}
    \right)=0
\end{equation}
where the vector $v^\mu\in\{p_1,p_2,\ldots,p_{N_{\mathrm{ext}}},\ell_1,\ldots,\ell_r\}$, and the denominator factors are defined as $D_j=q_j^2-m_j^2$.
\end{tcolorbox}

While a variety of highly optimised public codes (such as \textbf{\texttt{LiteRed}}\cite{Lee:2012cn}, \textbf{\texttt{Kira}}\cite{Maierhofer:2017gsa}, or \textbf{\texttt{FIRE}}\cite{Smirnov:2019qkx}) exist to generate and solve these IBP systems automatically, \textcolor{black}{ an important subtlety arises in the specific class of integrals relevant to computing scattering observables like impulse, waveform etc. These observables frequently feature \textbf{exponential factors} in the integrand. Sometimes, it is also a common understanding that treating the Fourier kernel as a part of the integral family may streamline the computation especially when we have massive loops  .}These exponentials modify the standard IBP relations in a way that is entirely invisible to conventional, rational-function-based automated generators. To illustrate this critical subtlety, consider a simple but nice integral family (The usefulness of incorporating the exponential factor into the IBP system was first pointed out in \cite{Brunello:2024ibk} in the context of computing the gravitational waveform in General Relativity.):
\begin{equation}
    I(a_1,a_2)
    =\int d^d\ell\,
    \frac{e^{D_1}}{D_1^{a_1}D_2^{a_2}},
    \qquad
    D_1=i\,\ell\!\cdot\! b,\quad
    D_2=\ell^2+m^2 .
\end{equation}
Applying the fundamental IBP theorem to this integrand yields:
\begin{equation}
\begin{split}
0
&=\int d^d\ell\,
\frac{\partial}{\partial \ell^\mu}
\left(
  \frac{v^\mu\,e^{D_1}}{D_1^{a_1}D_2^{a_2}}
\right)
\nonumber\\
&=\int d^d\ell\,
e^{D_1}\,
\frac{\partial}{\partial \ell^\mu}
\left(
  \frac{v^\mu}{D_1^{a_1}D_2^{a_2}}
\right)
\;+\;
\textcolor{brown}{
\int d^d\ell\,
\bigl(i\,v\!\cdot\! b\bigr)\,
\frac{e^{D_1}}{D_1^{a_1}D_2^{a_2}}
}\,.
\end{split}
\label{eq:exp-ibp-splitting}
\end{equation}
The \textcolor{brown}{second term} is the crucial new mathematical ingredient. It originates directly from the chain rule acting on the exponential $e^{D_1}$ and possesses no analogue in standard, purely rational IBP systems. By choosing specific vectors, such as $v^\mu=\ell^\mu$ and $v^\mu=b^\mu$, this derivative generates a coupled system of difference equations:
\begin{equation}
\begin{split}
&\left(-a_1-2a_2+d\right) I(a_1,a_2)
+2a_2 m^2\, I(a_1,a_2+1)
\textcolor{brown}{+I(a_1-1,a_2)}=0,
\\
&\textcolor{brown}{i b^2\, I(a_1,a_2)}
-i a_1 b^2\, I(a_1+1,a_2)
+2 i a_2\, I(a_1-1,a_2+1)=0.
\end{split}
\label{eq:exp-ibp-identities}
\end{equation}
Because widely used automated packages assume the integrand is a rational function of the loop momenta, they naturally reproduce only the uncoloured rational part of the splitting. The \textcolor{brown}{exponential-induced shift} must be manually injected into the system.
Once the complete IBP system is assembled, it can be solved iteratively using the Laporta algorithm. For modern, highly complex topologies, these custom systems are best tackled using finite-field arithmetic and sparse rational reconstruction techniques, such as those implemented in \textbf{\texttt{FiniteFlow}} (or via simple modifications to existing algebraic packages like \textbf{\texttt{LiteRed}}).
Below is an illustrative snippet demonstrating how to define and solve this modified, exponentially-shifted IBP system using \texttt{FiniteFlow}:

\begin{lstlisting}[language=Mathematica, caption={Solving simple IBPs using FiniteFlow}, label={lst:finiteflow-ibp}]
<< FiniteFlow`
Eq1[a1_, a2_] := (d - 2*a2 - a1)*I[a1, a2] + I[a1-1, a2] + 2*a2*m^2*I[a1, a2+1] == 0
Eq2[a1_, a2_] := bSq*I[a1, a2] + 2*a2*I[a1-1, a2+1] - a2*bSq*I[a1+1, a2] == 0

SeedEqs = Flatten[Table[{Eq1[a1, a2], Eq2[a1, a2]},
  {a1, -2, 1}, {a2, 0, 6}]];
allIntegrals    = DeleteDuplicates[Cases[SeedEqs, _I, Infinity]];
orderedIntegrals = Reverse[SortBy[allIntegrals,
  {Abs[#[[1]]] + Abs[#[[2]]] &, Abs[#[[2]]] &}]];
targets = {I[0, 4]};
solution = FFSparseSolve[SeedEqs, orderedIntegrals,
  "Parameters" -> {d, m, bSq}, "NeededVars" -> targets,
  "MaxPrimes" -> 150, "MaxDegree" -> 500] /. bSq -> b^2;

Print["=== Solution ==="];
Print[solution];
\end{lstlisting}
By solving this system, the infinite tower of integrals collapses. For this specific family, one is left with exactly three Master Integrals: $I(1,0)$, $I(1,1)$, and $I(2,0)$. In general, for any given $L$-loop topology with a defined set of external momenta, \textbf{any} Feynman integral $\mathcal{I}$ within that family can be uniquely expanded as a linear combination of its basis Master Integrals $\mathcal{M}^{(k)}_{\mathrm{MI}}$:
\begin{equation}
   \mathcal{I}=\sum_k c_k(d,s_{ij}) \mathcal{M}^{(k)}_{\mathrm{MI}}
\end{equation}
where the IBP reduction coefficients $c_k(d,s_{ij})$ are strictly rational functions of the spacetime dimension $d$ and the kinematic invariants $s_{ij}=p_i\cdot p_j$.

With the IBP reduction complete, the formidable task of evaluating thousands of individual loop integrals is entirely bypassed. The challenge is now distilled down to its purest form: evaluating just the basis set of Master Integrals. One efficient way to compute the master integral is the method of differential equations. This technique is primarily more useful when the Feynman integrals do not have non-perturbative factorisation in the dimensional regularisation parameter $\epsilon$ and kinematic factors $s,t,u,m^2$. Let's also discuss these techniques (with a simple toy example). Let $\vec{\mathbfcal{J}}=(J_1,J_2,\ldots,J_n)$ denote the vector of Master Integrals (MIs), which in general depend on the set of kinematic invariants
$\boldsymbol{X}$ as well as on the spacetime dimension $d$. Differentiating with respect to $\boldsymbol{X}$ and subsequently applying IBP reduction
ensures that these derivatives can be expressed again as linear combinations of the same master integrals, and they satisfy the following homogeneous system of matrix differential equations.
\begin{equation}
    \partial_{x}\vec{\mathbfcal{J}}=\Omega_{x}\vec{\mathbfcal{J}}\,.
\end{equation}
 Collecting all differentials, one can write the
differential 1-form,
\begin{equation}
    d\vec{\mathbfcal{J}}=\boldsymbol{\Omega}\cdot\vec{\mathbfcal{J}},\qquad \boldsymbol{\Omega}=\sum_{x\in \boldsymbol{X}}\Omega_xdx\,.\label{7}
\end{equation}
By choosing an ordering in which master integrals associated with simpler sub-topologies (with a lesser number of denominators ) are placed first, the differential-equation system can often be arranged into a (block) lower-triangular form. Consistency of the differential system is then encoded in the integrability condition $d^2\vec{\mathbfcal{J}}=0$, which translates into the flatness (Maurer--Cartan) equation for the connection one-form,
\begin{equation}
d\boldsymbol{\Omega}=\boldsymbol{\Omega}\wedge\boldsymbol{\Omega}\, .
\end{equation}
The solution of the differential equation can be formally expressed as a path-ordered integral,
\begin{equation}
    \begin{split}
        \vec{\mathbfcal{J}}(\boldsymbol{x},\epsilon)=\mathcal{P}\exp\left(\int d\boldsymbol\Omega(\boldsymbol{x},\epsilon)\right) \vec{\mathbfcal{J}}(\boldsymbol{x}_0,\epsilon)\label{9}
    \end{split}
\end{equation}
where $\vec{\mathbfcal{J}}(x_0,\epsilon)$ is the vector of the mater integrals determined at a particular kinematic point, sometimes known as the boundary integrals. The boundary integrals can usually be computed using the elementary techniques of evaluating Feynman integrals, such as Feynman and Schwinger parametrisation. However, directly computing the path-ordered integral in \eqref{9} is quite cumbersome, as it is very difficult to determine the order of $\epsilon$. It is possible to make a change of basis from $\vec{\mathbfcal{J}}$ to $\vec{\mathbfcal{H}}$ : $\vec{\mathbfcal{H}}=\mathbf{T}\cdot \vec{\mathbfcal{J}}$, where $\mathbf{T}$ is a $n\times  n$ basis transformation matrix. Therefore, the transformed basis of MIs satisfies a transformed set of differential equations,
\begin{equation}
d\vec{\mathbfcal{H}}=\widetilde{\boldsymbol{\Omega}}\cdot\vec{\mathbfcal{H}},\qquad  \widetilde{\boldsymbol{\Omega}}=\mathbf{T}\cdot \boldsymbol{\Omega}\cdot \mathbf{T}^{-1}-\mathbf{T}d\mathbf{T}^{-1}\,.
\end{equation}
It was conjectured in \cite{Henn:2013pwa} that there exist a particular choice of basis transformation such the the transformed differential equation has $\epsilon$-factorized form: $d\vec{\mathbfcal{H}}=\epsilon\,d{\hat{\boldsymbol{\Omega}}}\cdot\vec{\mathbfcal{H}}$, where, $d{\hat{\boldsymbol{\Omega}}}$ has only logarithmic singularities and can be written in $d\log$-form: $d{\hat{\boldsymbol{\Omega}}}=\sum_{\eta_i\in A}M_i\,d\log\eta_i$, where, $\eta_i$ are called letters, and they only encodes the kinematic dependence of the differential equations, their collection is called Alphabet $A$, and $M_i$ are constant rational matrices. This particular choice of basis is called the canonical basis. The systematic construction of such a basis can be achieved via the Lee algorithm \cite{Lee:2014ioa}. In practice, this complex algebraic transformation is now heavily (semi-)automated and readily available through several specialized public computer algebra packages, including \textbf{\texttt{Libra}} \cite{Lee:2020zfb}, \textbf{\texttt{Canonica}} \cite{Meyer:2017joq}, \textbf{\texttt{Fuchsia}} \cite{Gituliar:2017vzm}, and \textbf{\texttt{INITIAL}} \cite{Dlapa:2020cwj}. In canonical basis, the path order integral can be expanded in a Laurent expansion of $\epsilon$, formally,
\begin{equation}
    \vec{\mathbfcal{H}}=\left(\mathbb{I}+\epsilon\int_{\gamma} d\hat{\boldsymbol{\Omega}}+\epsilon^2\int_{\gamma}d\hat{\boldsymbol{\Omega}}\cdot d\hat{\boldsymbol{\Omega}}+\cdots\right)\vec{\mathbfcal{H}}_0
\end{equation}
\begin{figure}
    \centering
\includegraphics[width=0.25\linewidth]{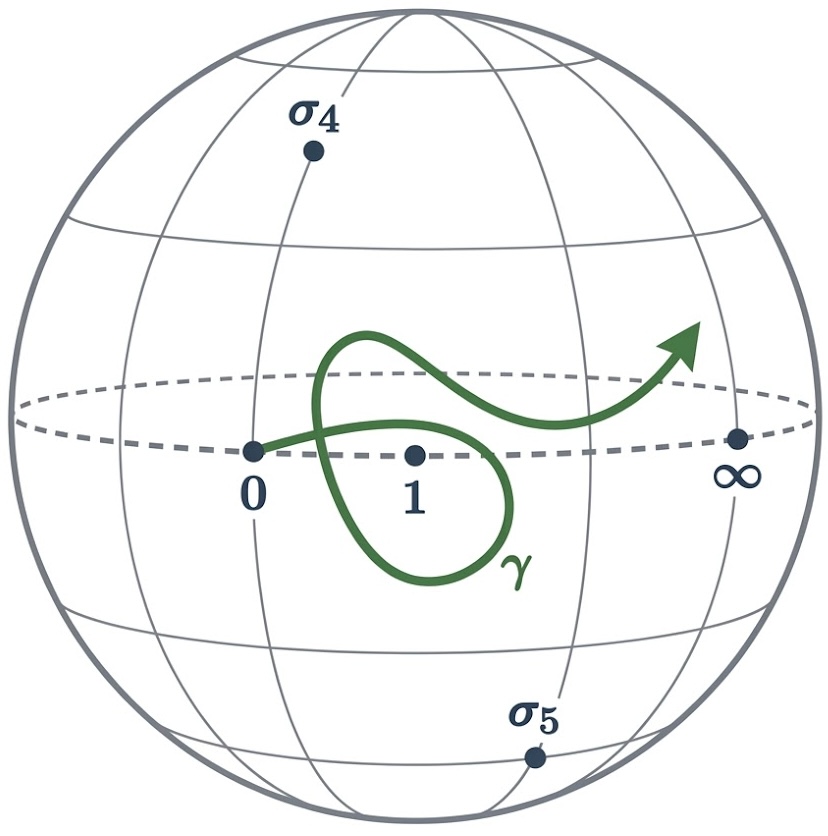}
    \caption{Geometry and contour of nested integrals}
    \label{sp1}
\end{figure}
Each order of an iterated integral corresponds to the successive integration of logarithmic differential 1-forms. Given an alphabet of such forms, typically defined as $\omega_a = d\log(z - a)$, the iterated integral of a sequence of these letters along a specified continuous contour $\gamma$ is expressed as:
\begin{equation}
    \int_{\gamma} \omega_{a_1} \circ \cdots \circ \omega_{a_n}=\int_{\gamma_1}\frac{dz_1}{z_1-a_1}\int_{\gamma_2}\frac{dz_2}{z_2-a_2}\int_{\gamma_3}\frac{dz_3}{z_3-a_3}\cdots,\quad \gamma_i(0)=0,\,\gamma_i(1)=t_{i-1},\, a_i=\{0,1,\sigma_j\}
\end{equation}
Evaluating these iterated integrals with simple logarithmic kernels gives rise to a rich space of functions known as Generalised Polylogarithms (G-MPLs), or Multiple Polylogarithms. Geometrically, these functions are most naturally defined on the \textbf{punctured Riemann sphere}, $\mathbb{C} \cup \{\infty\}$ (see Fig.~(\ref{sp1})). The ``punctures'' on this sphere correspond precisely to the simple poles of the rational logarithmic kernels---the specific points $a_i$ where the differential forms become singular. 
By framing this space on the compact Riemann sphere rather than the open complex plane, the point at infinity is treated on equal footing with the finite singularities. Because the differential 1-forms are closed, the specific value of the resulting GPL is determined entirely by the homotopy class of the integration path $\gamma$. The sphere provides a complete topological canvas for tracking how this contour winds around the singular punctures, elegantly capturing the branch cuts and monodromy properties of the integrals.
To truly appreciate the power of the differential equation method, we must recognise how it shifts the paradigm of computing Feynman integrals. Instead of brute-forcing complicated momentum integrations over massive propagators, this technique transforms the problem into a highly systematic exercise in linear algebra and the analytic continuation of topological paths.
Let us see this in action through an elegant example.

\textbf{The Setup: From Integrals to Algebra.}
Consider the family of two-loop Feynman integrals associated with a Sunrise-type topology with one massive internal line. The integral family is defined as:
\begin{equation}
  J_{a_1,...,a_5}=\begin{minipage}
          [h]{0.40\linewidth}
	\vspace{1.2 pt}
	\scalebox{0.6}{\includegraphics[width=\linewidth]{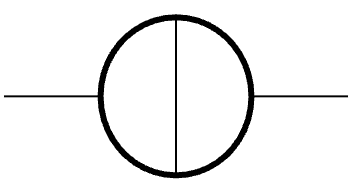}}
      \end{minipage} \hspace{-2 cm}= \int d^d\ell_1 d^d\ell_2 \frac{1}{D_1^{a_1}D_2^{a_2}D_3^{a_3}D_4^{a_4}D_5^{a_5}}
\end{equation}
where the inverse propagators are $D_1=\ell_1^2-m^2$, $D_2=\ell_2^2$, $D_3=(p-\ell_1-\ell_2)^2$, $D_4=(p-\ell_1)^2$, and $D_5=(p-\ell_2)^2$, with the kinematic invariant $p^2=s$. 

Directly integrating this for general powers $a_i$ is exceptionally difficult. However, by leveraging Integration-By-Parts (IBP) identities, the infinite-dimensional space of these integrals is algebraically reduced to a finite set of just seven Master Integrals (MIs):
\begin{equation}
\begin{split}
    \vec{\mathbfcal{J}} = &\{J_{0,0,1,1,1}, J_{1,0,1,0,1}, J_{1,1,0,0,1}, J_{1,1,1,0,0}, J_{2,1,1,0,0}, J_{1,0,1,1,1}, J_{1,1,0,1,1}\}
    \end{split}
\end{equation}
\textbf{The Differential Equation.}
Instead of integrating, we differentiate. Taking the derivative of these MIs with respect to the dimensionless kinematic invariant $x = \frac{s}{m^2}$ yields a closed, coupled system of first-order differential equations:
\begin{equation}
    \partial_{x}\vec{\mathbfcal{J}} = \Omega(x,\epsilon)\cdot \vec{\mathbfcal{J}}
\end{equation}
where the matrix $\Omega(x,\epsilon)$ governs the system's dynamics:
\begin{equation}
\Omega(x,\epsilon) = \left(
\begin{array}{ccccccc}
 \frac{1-2 \epsilon }{x} & 0 & 0 & 0 & 0 & 0 & 0 \\
 0 & 0 & 0 & 0 & 0 & 0 & 0 \\
 0 & 0 & -\frac{\epsilon }{x} & 0 & 0 & 0 & 0 \\
 0 & 0 & 0 & \frac{1-2 \epsilon }{x} & -\frac{1}{x} & 0 & 0 \\
 0 & 0 & 0 & \frac{-6 \epsilon ^2+7 \epsilon -2}{(x-1) x} & \frac{2-(x+3) \epsilon }{(x-1) x} & 0 & 0 \\
 \frac{2-3 \epsilon }{(x-1) x} & \frac{\epsilon -1}{(x-1) x} & 0 & 0 & 0 & \frac{1-(x+1) \epsilon }{(x-1) x} & 0 \\
 0 & 0 & \frac{\epsilon -1}{(x-1) x} & 0 & 0 & 0 & \frac{1-2 x \epsilon }{(x-1) x} \\
\end{array}
\right)
\end{equation}
Notice that in this raw form, the kinematic dependence (the variable $x$) and the dimensional regulator (the variable $\epsilon$) are deeply entangled. Solving this directly remains non-trivial.

\textbf{The Breakthrough: The Canonical Basis.}
This is where the true power of the method emerges. By finding a highly specific algebraic transformation $\mathbf{T}$, we can rotate our basis of MIs from $\vec{\mathbfcal{J}}$ to a ``canonical basis'' $\vec{\mathcal{H}}$, defined by $\vec{\mathcal{H}} = \mathbf{T}\cdot \vec{\mathbfcal{J}}$. For this system, the transformation matrix is:
\begingroup
\setlength{\arraycolsep}{3pt}
\renewcommand{\arraystretch}{1.15}
\begin{equation}
\mathbf{T} =
\resizebox{\textwidth}{!}{$
\begin{pmatrix}
\dfrac{28x(2\epsilon^2-\epsilon)}{5(9\epsilon^2-9\epsilon+2)} & 0 & 0 & 0 & 0 & 0 & 0 \\
0 & -\dfrac{3(2\epsilon-1)}{5(\epsilon-1)} & 0 & 0 & 0 & 0 & 0 \\
0 & 0 & \dfrac{3(2\epsilon-1)}{5(\epsilon-1)} & 0 & 0 & 0 & 0 \\
0 & 0 & 0 & \dfrac{8\epsilon}{9\epsilon^2-9\epsilon+2}\!\left(x-\dfrac{1}{x}\right) &
\dfrac{4(5x\epsilon+7\epsilon-2)}{9\epsilon^2-9\epsilon+2}-\dfrac{16\epsilon}{x(9\epsilon^2-9\epsilon+2)} & 0 & 0 \\
0 & 0 & 0 & \dfrac{8\epsilon}{3\epsilon-1}\!\left(\dfrac{1}{x}-1\right) &
\dfrac{16\epsilon}{x(3\epsilon-1)}-\dfrac{4(8\epsilon-1)}{3\epsilon-1} & 0 & 0 \\
-\dfrac{8}{5x}-\dfrac{4(\epsilon+2)}{5(3\epsilon-1)} & -\dfrac{3}{5x} & 0 & 0 & 0 & \dfrac{1}{x}-1 & 0 \\
0 & 0 & \dfrac{3}{5x} & 0 & 0 & 0 & 1-\dfrac{1}{x}
\end{pmatrix}
$}
\end{equation}
\endgroup
Applying this transformation achieves \textbf{$\epsilon$-factorization}. The new, transformed differential equation takes a remarkably beautiful and simple form, expressed in terms of logarithmic differential 1-forms:
\begin{equation}
    d\vec{\mathcal{H}} = \epsilon \, \hat{\Omega}_{x} \vec{\mathcal{H}}
\end{equation}
where the connection $\hat{\Omega}_{x}$ separates cleanly into constant matrices weighted by pure $d\log$ forms:
\begin{align}
\hat{\Omega}_{x} = \left(
\begin{array}{ccccccc}
 -2 & 0 & 0 & 0 & 0 & 0 & 0 \\
 0 & 0 & 0 & 0 & 0 & 0 & 0 \\
 0 & 0 & -1 & 0 & 0 & 0 & 0 \\
 0 & 0 & 0 & -3 & -6 & 0 & 0 \\
 0 & 0 & 0 & 2 & 4 & 0 & 0 \\
 -\frac{24}{5} & -\frac{3}{5} & 0 & 0 & 0 & 1 & 0 \\
 0 & 0 & -\frac{3}{5} & 0 & 0 & 0 & 0 \\
\end{array}
\right) d\log(x) + \left(
\begin{array}{ccccccc}
 0 & 0 & 0 & 0 & 0 & 0 & 0 \\
 0 & 0 & 0 & 0 & 0 & 0 & 0 \\
 0 & 0 & 0 & 0 & 0 & 0 & 0 \\
 0 & 0 & 0 & -4 & -9 & 0 & 0 \\
 0 & 0 & 0 & 0 & 0 & 0 & 0 \\
 \frac{44}{5} & \frac{3}{5} & 0 & 0 & 0 & -2 & 0 \\
 0 & 0 & \frac{3}{5} & 0 & 0 & 0 & -2 \\
\end{array}
\right) d\log(x-1)
\end{align}
\textbf{The Power of Transport:}
Because $\epsilon$ is factored out, the solution to this matrix differential equation is immediately given as a Dyson series of iterated integrals. The canonical Master Integrals simply possess a Laurent expansion in $\epsilon$:
\begin{equation}
\begin{split}
    \vec{\mathcal{H}} &= \left( \mathbb{I} + \epsilon \int_{\gamma} \hat{\Omega}_x + \epsilon^2 \int_{\gamma} \hat{\Omega}_x \circ \hat{\Omega}_x + \mathcal{O}(\epsilon^3) \right) \vec{\mathcal{H}}_\infty \\ 
    &= \left( \mathbb{I} + \epsilon \sum_{a_i \in \{0,1\}} M_{a_i} \int_{\gamma} d\log(x-a_i) + \epsilon^2 \sum_{a_i, a_j \in \{0,1\}} M_{a_i} M_{a_j} \int_{\gamma} d\log(x-a_i) \circ d\log(x-a_j) + \cdots \right) \vec{\mathcal{H}}_{\infty}
\end{split}
\end{equation}
\textit{This is the ultimate triumph of the method.} The difficult integration over physical space-time momenta has been entirely replaced by a set of nested integrals. Moreover, the boundary vector $\vec{\mathcal{H}}_\infty$ can be evaluated at a kinematically simple point. By taking the Regge limit ($s \to \infty$) or the massless limit ($m \to 0$), the originally formidable massive integrals become trivial to compute (However, one needs to be precise about the regions while expanding  for boundary data; we will discuss these in great detail in the next chapter). The iterated integrals of the $d\log$ forms then seamlessly analytically continue, or transport, this trivial massless result back to the massive kinematic regime. We have effectively computed a complex, massive, multi-loop integral simply by evaluating its simpler massless counterpart and reading the geometric structure of its singularities. In general, different classes of special functions may appear in the computation of Feynman integrals, especially for higher loop computations. To systematically identify the precise class of functions spanned by the final solution, it is highly effective to stratify the family of master integrals into its constituent topological sectors. Because the associated system of differential equations inherently possesses a block-triangular structure, the homogeneous component (usually some non-homogeneous part may be present, which comes from lower sub-topologies) of each diagonal block uniquely dictates the functional alphabet required to evaluate that specific sector. By isolating a specific $k \times k$  block of the homogeneous differential equation system, one can equivalently recast the dynamics of a chosen Master Integral, $J_i$, into a single $k$-th order scalar differential equation:
\begin{equation}
    \mathcal{L}_k J_i(x) = 0, \qquad \mathcal{L}_{k} = \frac{d^k}{dx^k} + \sum_{j=0}^{k-1} c_j(x)\frac{d^j}{dx^j}
\end{equation}
Here, the coefficients $c_j(x)$ are purely rational functions of the kinematic variable $x$. The differential operator $\mathcal{L}_k$ defines the Picard-Fuchs (PF) equation, and its factorisation properties elegantly encode the underlying algebraic geometry of the Feynman integral. 
The structure of this factorisation directly dictates the functional alphabet required to express the solution:
\begingroup
\setlength{\tabcolsep}{0pt}
\renewcommand{\arraystretch}{1.5} 
\begin{center}
\begin{tabularx}{0.94\linewidth}{@{}>{\raggedright\arraybackslash}m{0.78\linewidth}@{\hspace{1.2em}}>{\centering\arraybackslash}m{0.16\linewidth}@{}}

\textbf{Genus-0 rational curve $\implies$ Multiple Polylogarithms.} If the PF operator factorizes entirely into first-order components, such that $\mathcal{L}_{k} = \mathcal{L}_1^{(1)} \circ \mathcal{L}_1^{(2)} \circ \cdots \circ \mathcal{L}_1^{(k)}$, with each $\mathcal{L}_1^{(m)} = \left(\frac{d}{dx} - r_m(x)\right)$, the solution space is strictly spanned by Generalized Multiple Polylogarithms (G-MPLs) and standard iterated integrals. &
\includegraphics[width=\linewidth]{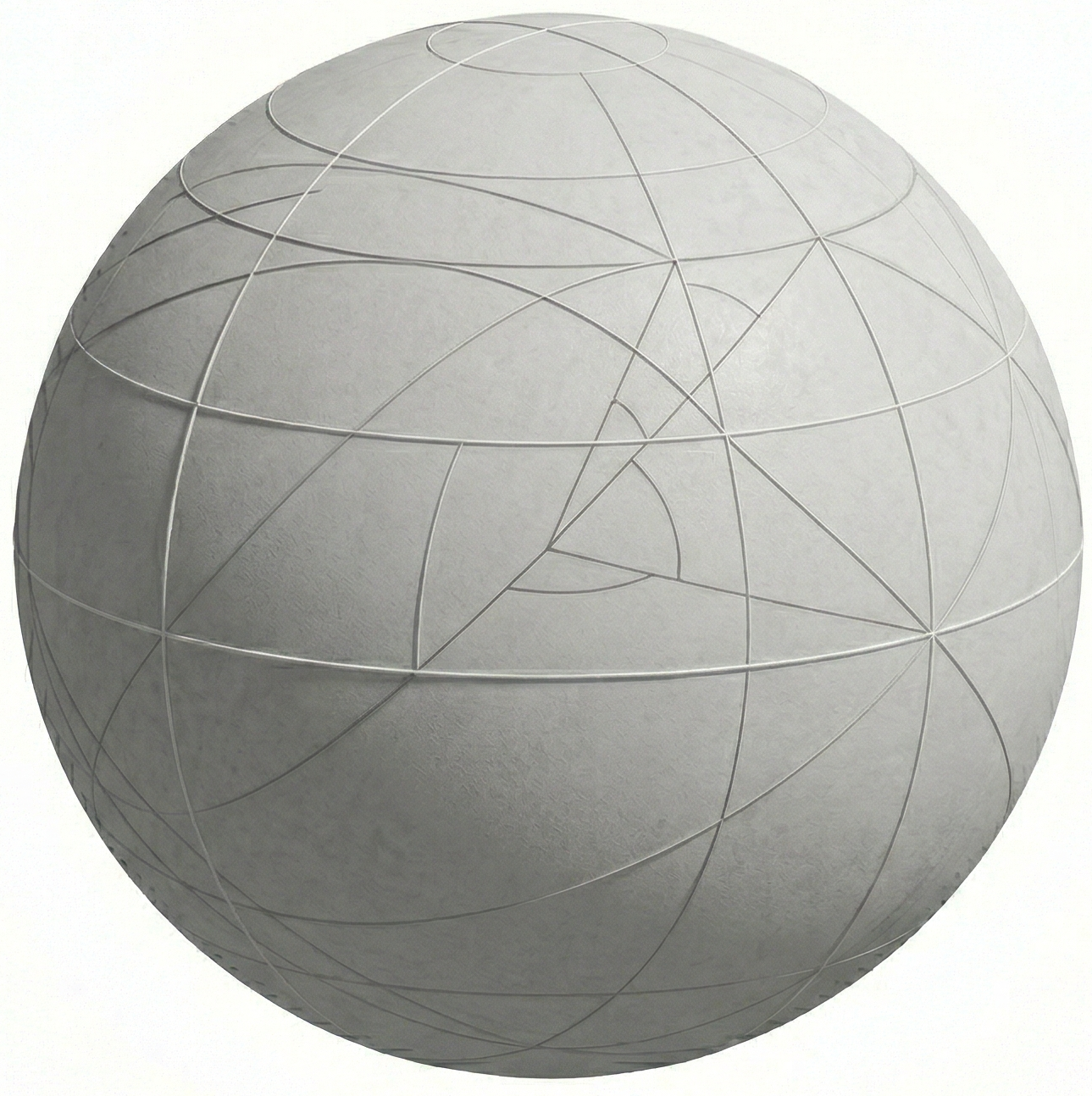}
\\
\textbf{Genus-1 (or torus) irrational curve (Elliptic curve) $\implies$ Elliptic polylogarithms.} If the operator cannot be fully reduced and leaves behind irreducible second-order factors, the geometry transcends simple poles. The corresponding solutions are governed by elliptic curves, requiring integrals over elliptic kernels. &
\includegraphics[width=\linewidth]{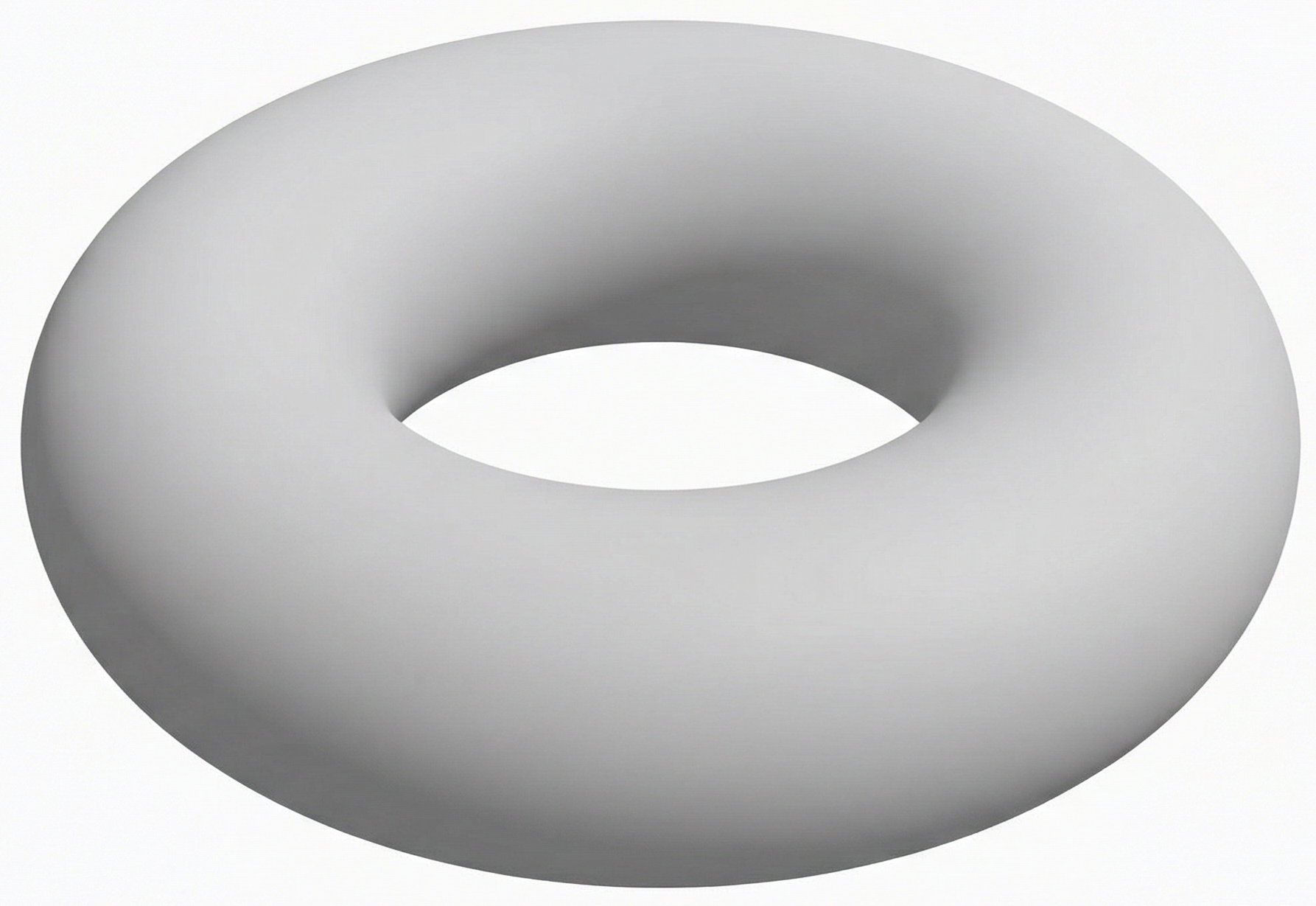}
\\ 
\textbf{Higher genus irrational curves $\implies$ Calabi-Yau periods .} Higher-order irreducible components signal the presence of profoundly more complex geometric structures within the scattering amplitude, such as CY3 surfaces or more general Calabi-Yau manifolds. &
\includegraphics[width=\linewidth]{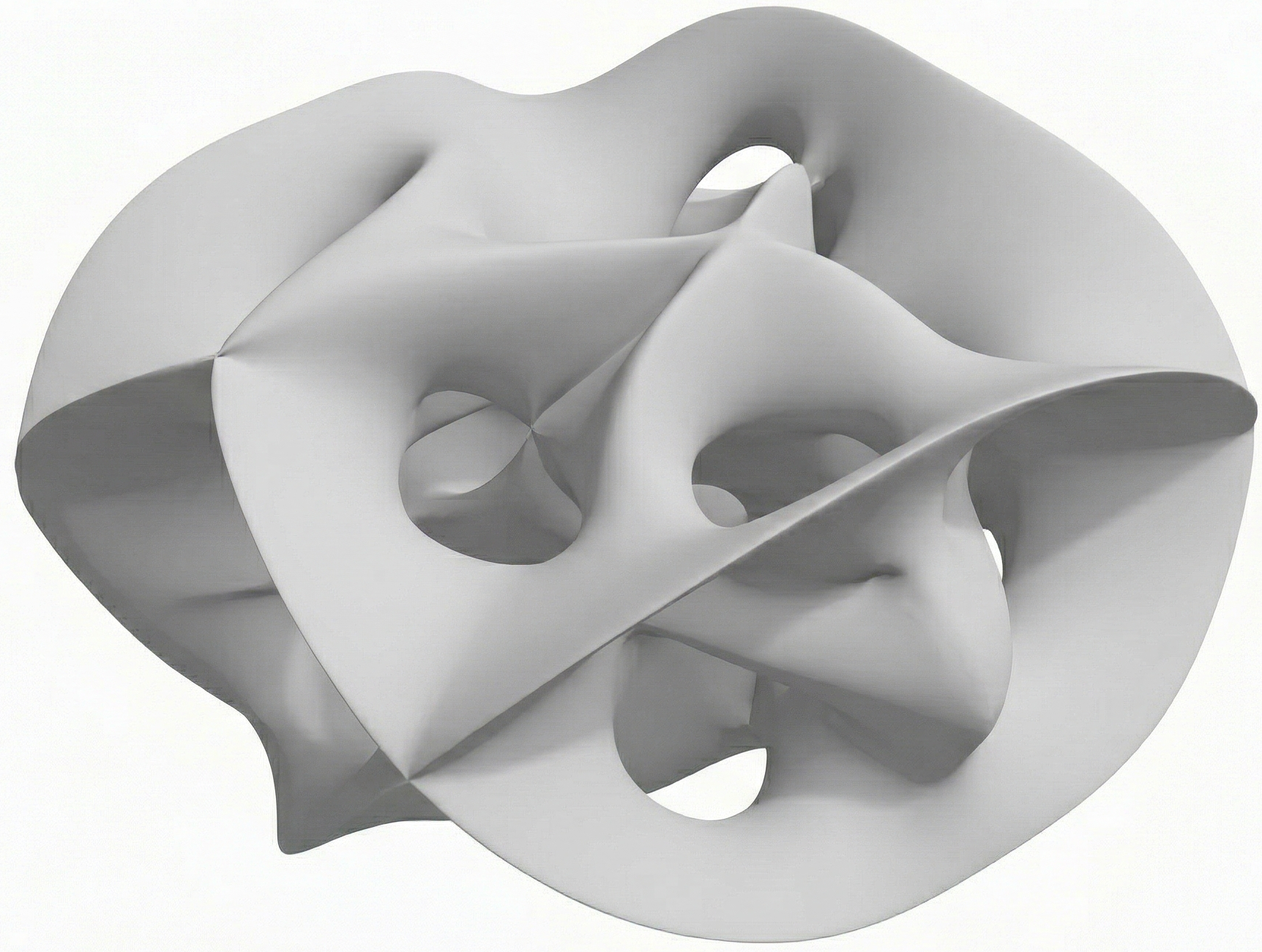}
\end{tabularx}
\end{center}
\endgroup
\textit{Throughout the remainder of this thesis, we will focus exclusively on Feynman integrals whose Picard-Fuchs operators completely factorise at first order, thereby restricting our analytic evaluations entirely to the simplest class: Generalised Multiple Polylogarithms.}
\\
This functional hierarchy provides the organising lens for the technical chapters that follow.
\section{Thesis in a nutshell}
\noindent A chapter-wise summary of the thesis is as follows.
\begin{itemize}[leftmargin=1.75em,topsep=0pt,itemsep=0.18em,parsep=0pt,partopsep=0pt]

  \item \textbf{\texttt{Chapter 2. WEFT of Black-Hole Binaries with Dark Photon and Axion.}} In this chapter, we investigate the correction to the potential that gives rise to the bound orbits and radiation from non-spinning inspiralling binary black holes in a dark matter environment consisting of \textit{axion-like particles} and \textit{dark photons} using the techniques of \textit{Worldline Effective Field Theory}. We compute the conservative dynamics up to $1$PN order for gravitational, electromagnetic, and Proca fields and up to $2$PN order for the scalar field. The effect of axion-electromagnetic coupling ($g_{a\gamma\gamma}$) arises in the conservative dynamics at $2.5$PN order and the kinetic mixing constant ($\gamma$) at $1$PN order. Furthermore, we calculate the radiation due to the various fields present in our theory. We find that the contribution of $g_{a\gamma\gamma}$ to the gravitational radiation appears at $N^{(7)}LO$ and to the scalar radiation at $N^{(5)}LO$. We also find that these radiative corrections due to the coupling $g_{a\gamma\gamma}$ vanish for any orbit confined to a plane because of the existence of a \textit{binormal}-like term in the effective radiative action, but give rise to non-zero contributions for any orbit that lies in \textit{three dimensions}. Last but not least, $\gamma$ contributes to the gravitational radiation at $N^{(2)}LO$ and $N^{(4)}LO$.

  \item \textbf{\texttt{Chapter 3. Classical Black-Hole Scattering in Scalar-Tensor Gravity from WQFT.}} In this chapter, we move to the problem of computing observables in classical black-hole scattering in the context of the simplest modification of General Relativity, namely scalar-tensor gravity. We compute the two observables, impulse and waveform, in a non-spinning black-hole scattering event for the scalar-tensor theory of gravity with a generic scalar potential using the techniques of Worldline Quantum Field Theory. We mainly investigate corrections to the above-mentioned observables arising from the extra scalar degree of freedom. For the computation of the impulse, we consider the most general scenario by making the scalar field massive and then show that each computed diagram has a smooth massless limit. We compute the waveforms for a scalar and a graviton up to 2PM, treating the scalar as massless. Furthermore, we discuss the case in which the scalar has mass and how the radiation integrals become more involved than in the massless case. We also obtain analytical results using the stationary-phase approximation.

  \item \textbf{\texttt{Chapter 4. Spinning Two-Body Problem in dCS Gravity Using WQFT.}} In this chapter, we extend the analysis to the case of spinning black holes in dynamical Chern-Simons gravity. Our main objective is to compute the WQFT partition function, or equivalently, the eikonal phase, for black-hole scattering in this theory using the methods of spinning worldline quantum field theory. We consider the scattering of spinning compact objects and explain in detail the ingredients required to construct the partition function. A central part of the computation involves the evaluation of the relevant two-loop integrals. For this purpose, we derive the $\epsilon$-expansion of the necessary master integrals using Integration-by-Parts (IBP) reduction together with differential equation techniques. These results are then used to determine the eikonal phase at linear order in spin up to third post-Minkowskian (3PM) order. We also analyse how the phase depends on the relative orientations of the black-hole spins, which is crucial for the structure of the resulting observables, since the eikonal phase serves as the basic generating quantity for the classical dynamics. In addition, we discuss the role of asymptotic supersymmetry in the worldline degrees of freedom, which provides a natural way to define the spin of black holes even in the presence of additional propagating fields, such as the dilaton, in addition to the usual massless spin-two graviton. We also comment on the infrared issues that arise in this context and discuss their possible origins and resolutions.

  \item \textbf{\texttt{Chapter 5. Heterotic Footprints in Classical Gravity.}} In this chapter, in a complementary direction, we use amplitude-based approaches to study classical scattering of charged black holes in Einstein-Maxwell-Dilaton (EMD) theory. Working in the classical (Post-Minkowskian) regime, we extract the conservative two-body potential by expanding the one-loop amplitudes in the soft regime. We explicitly show that, as in GR, the relevant soft amplitudes are infrared (IR) finite when long-range interactions are consistently treated via the Lippmann-Schwinger equation and the associated IR subtraction. The scattering angle is then obtained from the eikonal exponentiation of the soft amplitude. Our results track the separate roles of electromagnetic and dilatonic charges in both the conservative dynamics and the eikonal phase, and they smoothly reduce to the GR limit when the charges and the dilaton coupling are switched off. Where applicable, we compare our results with existing literature and find agreement. These findings provide amplitude-based benchmarks for compact-object dynamics in EMD and furnish building blocks for waveform modelling in beyond-GR scenarios.

  \item \textbf{\texttt{Chapter 6. Celestial Amplitude, Shadow, and OPE in Quadratic Gravity.}} In this chapter, as a further investigation into higher-derivative effective field theories, we compute the celestial amplitude arising from higher-curvature corrections to Einstein gravity, incorporating phase dressing. The inclusion of such corrections leads to effective modifications of the theory's ultraviolet (UV) behaviour. In the eikonal limit, we find that, in contrast to Einstein gravity, where the $u$- and $s$-channel contributions cancel, these contributions remain non-vanishing in the presence of higher-curvature terms. We examine the analytic structure of the resulting amplitude and derive a dispersion relation for the phase-dressed eikonal amplitude in quadratic gravity. Furthermore, we investigate the celestial conformal block expansion of the Mellin-transformed conformal shadow amplitude within the framework of celestial conformal field theory (CCFT). As a consequence, we compute the corresponding operator product expansion (OPE) coefficients using the Burchnall-Chaundy expansion. In addition, we evaluate the OPE via the Euclidean OPE inversion formula across various kinematic channels and comment on its applicability and implications. Finally, we briefly explore the Carrollian amplitude associated with the corresponding quadratic EFT.
\end{itemize}
    }%
  }%

  \restorethesisbodyformat
  \restoremainchapterstyle
  \chapter{WEFT of Black-Hole Binaries with Dark Photon and Axion}
  \thesischapterpaperbox{Worldline effective field theory of inspiraling black hole binaries in presence of dark photon and axionic dark matter}{A.~Bhattacharyya, S.~Ghosh, and S.~Pal}{\href{https://doi.org/10.1007/JHEP08(2023)207}{\textit{JHEP} \textbf{08} (2023), 207};  arXiv: \arxivlink{2305.15473}{hep-th}}
  {%
    \restorethesisbodyformat
    \renewcommand{\appendix}{%
      \setcounter{section}{0}%
      \setcounter{subsection}{0}%
      \setcounter{subsubsection}{0}%
      \renewcommand{\thesection}{\thechapter.\Alph{section}}%
      \renewcommand{\thesubsection}{\thesection.\arabic{subsection}}%
      \renewcommand{\theHsection}{chapter.\arabic{chapter}.appendix.\Alph{section}}%
      \renewcommand{\theHsubsection}{\theHsection.\arabic{subsection}}%
    }%
    \ifstrempty{chap1_body.tex}{}{%
\section{Introduction} 
The analysis presented in this chapter is based on Ref.~\cite{Bhattacharyya:2023kbh}.
Detection of gravitational waves (GW) \cite{LIGOScientific:2016aoc, LIGOScientific:2016sjg, LIGOScientific:2016vlm, LIGOScientific:2017bnn, Kokeyama:2020dkg, LIGOScientific:2019hgc} opens a new pathway for testing different theories of gravity. Gravitational waves are generated primarily by binary systems consisting of black holes and neutron stars. These achievements motivate increasingly detailed probes of the structure of the source. The evolution of such binaries has three phases: \textit{inspiral}, \textit{merger}, and \textit{ringdown}. Probing the merger phase, where gravity is strong, requires non-perturbative methods based on numerical simulations \cite{Lehner:2014asa, LIGOScientific:2014oec}. However, the inspiral phase can be studied using perturbative analytic techniques because the orbital velocity of the binaries is small compared to the speed of light, i.e. $\frac{v}{c}\ll 1$. \textcolor{black}{Signals detected by ground-based detectors derive much of their constraining power from the long duration of the inspiral phase}, so this binary-inspiral problem can be studied using the Post-Newtonian approximation \cite{Blanchet:2013haa}. 
\par
The gravitational-wave radiation from non-spinning binaries has been computed up to 3PN order in \cite{Tagoshi:2000zg, Faye:2006gx, Blanchet:2006gy, Blanchet:2004ek, Damour:2000ni, Itoh:2003fy,Boetzel:2019nfw,Faye:2012we,Mishra:2013rna} using traditional methods based on Green's functions. Recently, this computation has been extended up to 4.5PN order in \cite{Fujita:2010xj,Faye:2014fra, Blanchet:2023sbv,Blanchet:2023bwj} \footnote{Interestingly, both UV and IR divergences occur during the computation. The UV divergences appear because the binaries are modelled as point-like particles without internal structure. On the other hand, IR divergences occur due to the insertion of a typically divergent (at long distance) PN expansion in the near zone while defining the source moments. Interested readers are referred to \cite{Larrouturou:2021gqo, Blanchet:2004bb,Blanchet:2023soy} for a detailed discussion of the origin of these divergences and how to regularize them.}. Using these waveform results, parameter estimation based on the Fisher information matrix has been carried out in \cite{Arun:2004hn}. Later, these studies were extended by taking into account the effect of the binary's spin at leading PN order in \cite{PhysRevD.12.329, PhysRevD.2.1428, Kidder:1992fr}. Several subsequent works pushed this analysis to higher PN orders \cite{Cho:2022syn}. The Hamiltonians generating the binary dynamics, including spin effects up to 3PN order, were also derived using the ADM formalism in \cite{Steinhoff:2007mb, Steinhoff:2008ji, Hergt:2008jn, Hergt:2010pa,Porto:2012as}. Gravitational-wave radiation for eccentric orbits has likewise been studied in \cite{Enoki:2006kj, Favata:2011qi, Munna:2019fjz}. Beyond general relativity (GR), these computations are also being extended to alternative theories of gravity \cite{Zhang:2017srh,AbhishekChowdhuri:2022ora, Zhang:2018prg,Saffer:2018jmx, Lin:2018ken,Li:2022grj,Shiralilou:2021mfl,Julie:2019sab}.
\par
In parallel with the traditional method, a more efficient and systematic methodology based on \textit{\textcolor{black}{Worldline} Effective Field Theory} (WEFT) emerged for computing PN corrections. This framework is sometimes called ``Non-Relativistic General Relativity'' (NRGR) \cite{Goldberger:2004jt}. It has been used to compute gravitational-wave radiation at leading order in \cite{Goldberger:2004jt}. \textcolor{black}{In this approach, one can decouple and treat the conservative and radiative degrees of freedom separately. Integrating out the potential modes (heavy modes) yields an effective action for the radiative modes (light modes). The real part of this effective action determines the effective potential and hence the binary dynamics, while the imaginary part gives the total radiated power through an optical theorem.}\footnote{Like the traditional method, various divergences occur while computing various multipole moments of the source using the EFT method. Interested readers are referred to \cite{Porto:2017dgs,Porto:2007pw} for more details about how to handle these divergences (in particular, the IR ones). Also, the renormalizability of EFTs of binary systems and the necessity of a time-dependent mass counterterm have been discussed in \cite{Goldberger:2012kf,Galley:2015kus}.} Subsequently, the method was developed further to include effects of the binary's spin (up to quadratic order) in the radiative multipole moments up to \cite{Porto:2010zg} 
\textcolor{black}{3PN order}, and then it has been extended to higher PN  orders in \cite{Levi:2011eq, Levi:2015uxa, Levi:2015ixa, Levi:2014sba}. Apart from these, this EFT based approach has been used to study various phenomena like self-force effect on the curved spacetime, gravitational tail effects and also inspecting conservative and radiative dynamics of the binary inspiral problem (via computing effective Hamiltonian/Lagrangian) \cite{Maia:2017gxn,Maia:2017yok,Foffa:2019yfl,Kalin:2022hph,Porto:2008jj,Porto:2008tb,Porto:2007px,Mandal:2022nty,Mandal:2022ufb,Goldberger:2009qd,Ross:2012fc,Goldberger:2020fot,Goldberger:2017ogt,Goldberger:2006bd,Goldberger:2005cd}
\footnote{The list is by no means exhaustive. Interested readers are referred to the following reviews \cite{Goldberger:2022rqf,Goldberger:2022ebt} and references therein for more details.}
 Finally, the EFT approach has been extended to compute the conservative dynamics as well as power radiation due to a massive (as well as massless) scalar field minimally coupled to GR up to 1PN order \cite{Kuntz:2019zef,Huang:2018pbu}, and to study the conservative dynamics of the electromagnetic field minimally coupled to GR in \cite{Patil:2020dme,Gupta:2022spq}.
\par
Ultra-light bosonic fields are promising candidates for dark-matter \cite{GrillidiCortona:2015jxo, Sanchis-Gual:2022ooi, Goldstein:2022pxu, Schutz:2020jox} and  dynamical dark-energy \cite{Kamionkowski:2014zda}
\footnote{The list is by no means exhaustive.  Interested readers are referred to this Snowmass review \cite{Adams:2022pbo} and references therein for more details.}. One famous example of such ultra-light bosonic fields is the QCD axion. It arises from the Peccei--Quinn mechanism \cite{Fukuda:2021drn,CAPP:2020utb,Sakhelashvili:2021eid, Peccei:2006as, RevModPhys.82.557,PhysRevLett.40.223,PhysRevLett.40.279,Preskill:1982cy,Gorghetto:2018ocs,DiLuzio:2021pxd,Berezhiani:2000gh,Fukuda:2015ana,Dimopoulos:2016lvn,Gherghetta:2016fhp,Kim:1998va,DiLuzio:2020wdo,Conlon:2006tq,Chakraborty:2021fkp,Harigaya:2019qnl}, which was proposed as a solution to the strong CP problem \cite{PhysRevLett.38.1440,PhysRevD.16.1791,PhysRevLett.40.279,Clowe:2006eq,Hsu:2004mf,Agrawal:2017ksf,Gupta:2020vxb,PhysRevLett.43.103}. Experimental bounds on the neutron electric dipole moment imply that the strong CP angle must be much smaller than $10^{-10}$, while the CP angle in the CKM matrix is known to be $\mathcal{O}(1)$ \cite{Baker:2006ts}. This tension can be resolved by introducing the axionic coupling \cite{Huang:2018pbu, Zhang:2021mks}.$$\frac{a}{f_a}\frac{g_{a\gamma\gamma}^2}{32\pi^2}F^{\mu\nu}\tilde{F}^{\mu\nu},$$
where $g_{a\gamma\gamma}$ is the strong coupling constant, $\tilde{F}^{\mu\nu}$ is the dual field strength tensor, and $f_a$ is the axionic decay constant. QCD Axions have several constraints.  For example, the ADMX experiment showed that the first constraint on the QCD axion parameter space in the $\mu$eV mass range \cite{PhysRevLett.120.151301}.  Also, it was suggested that one could possibly obtain constraints on the parameter space for QCD axions by studying its effects on the stellar configuration of a white dwarf \cite{Balkin:2022qer}.  There are other ultra-light bosons with similar properties to the QCD axion, with only the difference that their mass is not related to the decay constant.  They are often termed as \textit{Axion-like particles (ALP)}. They can arise from different considerations, e.g., from string compactifications  \cite{Cicoli:2013ana,Svrcek:2006hf,Hiramatsu:2010yu}.  There exist phenomenological constraints on the mass of these ultra-light bosons stemming from dark-matter phenomenology, black hole superradiance  \cite{Arias:2012az, DiLuzio:2021pxd, QuilezLasanta:2021yzt, Ishii:2022lwc, 
 Cardoso:2018tly, Brito:2015oca,Ghosh:2023tyz} as well as on its anomalous coupling with electromagnetic field ($g_{a\gamma\gamma}$) by using the induced oscillating electric dipole moment of the electron as advocated in \cite{Hill:2015vma}.
Besides axions (ultra-light boson fields), \textit{ultra-light vector fields} (ULV) are also possible candidates for dark matter \cite{Cardoso:2018tly}.  One example of ultra-light vector fields is the so-called \textit{dark photon} \cite{Flambaum:2019cih}.  This can often be thought of as a \textit{hidden U(1)} massive vector field which can interact with the photon \cite{Cardoso:2018tly}.
\par
Both the ultra-light boson and vector fields can couple with gravity, e.g., axions can couple with gravity minimally and non-minimally.  The effect of axions on
the detected waveform is negligible until the binaries are separated by roughly a Compton wavelength of the axion \cite{Zhang:2019eid}. As the orbital length scale decays to the Compton wavelength, scalar radiation may become an important source
of orbital energy loss \cite{Huang:2018pbu}, especially for large Compton wavelengths.  Scalar radiation is detectable for both NS-NS and NS-BH binaries \cite{Dar:2018dra}.  Furthermore, one can write down Chern-Simons type terms both in the electromagnetic and gravitational sectors by which the massive scalar (axion) can couple with both electromagnetic and gravitational fields.  This is natural in the context when we consider the axion to play the role of dark matter \cite{Yoshida:2017cjl}.  If the axions interact with gravity via the Chern-Simons term, GW waves induce axion decays into gravitons \cite{Yoshida:2017cjl}.  Apart from this, in \cite{Machado:2018nqk, Machado:2019xuc}, it was shown that if axions couple with the dark photon with an unbroken $U(1)$ symmetry, there is a possibility of the generation of a stochastic gravitational wave when it experiences a tachyonic instability.  The resulting signal can be detected by ground-based GW detectors  \cite{Machado:2018nqk, Machado:2019xuc,10.21468/SciPostPhys.12.5.171}.  Last but not least, as advocated in \cite{ Ejlli:2022zah, Nagano:2021kwx}, polarimetry experiments on GW signals may provide a novel way to probe the axion-like particles.
\par 
Previously, several works have taken steps toward constraining \textcolor{black}{ALP couplings} using GW observations \cite{Huang:2018pbu, Zhang:2021mks}. This is done by considering a massive scalar field minimally coupled to GR and then computing the additional contribution to the total radiated power at leading PN order, together with the correction to the phase of the emitted GW due to the presence of the scalar field. In this paper, we consider a more generic scenario. Motivated by the dark-matter model discussed in \cite{Cardoso:2018tly}, we study a setup containing both the ALP, in the form of a massive scalar field with an anomalous coupling to the electromagnetic field via a Chern--Simons-type term, and the ULV, in the form of a massive vector field (Proca field) that also couples to the electromagnetic field through a kinetic-mixing term. For simplicity, we neglect the coupling of the ALP through the gravitational Chern--Simons term. Our broader goal is to investigate whether GW observations can constrain, in addition to the masses of the scalar and Proca fields, the axion-photon coupling parameter $g_{a\gamma\gamma}$ as well as the Proca-electromagnetic coupling constant $\gamma$ arising from kinetic mixing. To that end, we take a first step by computing the power radiation from a binary system in this dark-matter environment, which is essential for determining the phase of the gravitational waveform within the WEFT approach, and we eventually comment on the possibility of detecting these couplings.
\par
This chapter is organised as follows: In Sec.~(\ref{Sec2}), we briefly review the general machinery of the EFT approach. We also discuss various length scales associated with the EFT of inspiralling binaries, which is crucial for separating various degrees of freedom. In Sec.~(\ref{Sec3}) we discuss how the binaries can be modelled by a point particle action. We review the procedure of integrating out the potential modes to get effective action for Einstein's gravity. We also give the EFT power counting for the bound sector. In Sec.~(\ref{Sec4}), we introduce our model, which consists of a scalar field (axion), electromagnetic (photon), and Proca field (dark photon) minimally coupled with Einstein gravity and an interaction term between photon and dark photon as well between axion and photon. Then we focus on the bound sector, which gives rise to the corrections to the effective potential. \textit{Interestingly, we observe that the correction due to the axion-photon coupling ($g_{a\gamma\gamma}$) enters at the 2.5PN order.} In Sec.~(\ref{Sec5}), we discuss the dissipative sector. We used the familiar optical theorem to calculate the power radiation. We again discuss the EFT power counting for the radiative sector. We describe the corrections to the gravitational radiation coming from different field vertices. We reproduce the well-known expression for the power radiation coming from the pure gravitational and scalar sectors at the leading order. \par 
We extend the computation to the $N^{(4)}LO$ and also showed that the $g_{a\gamma\gamma}$ appears at $N^{(5)}LO$ for the  scalar sector. Due to the presence of other fields, we obtain new contributions to the power radiation from the gravitational sector at different orders of perturbations. We observe that the contribution due to the axion-photon coupling ($g_{a\gamma\gamma}$) enters at the $N^{(7)}LO$ in the expression for the power radiation from the gravitational sector. Furthermore, we also compute the contribution of electromagnetic and Proca fields to the total power radiation at the leading order.  We also comment on the possibility of detecting axion coupling ($g_{a\gamma\gamma}$) through the gravitational flux. Lastly, the kinetic mixing constant $\gamma$ starts contributing from $1PN$ in conservative sector and appears in radiative sector at $N^{(2)}LO$ and $N^{(4)}LO$ through gravitational radiation.\par
Finally, in Sec.~(\ref{disc}), we summarise our main findings and conclude with some future directions. Some details regarding the computation of a few integrals are given in Appendix ~(\ref{ch1:app:A}), (\ref{ch1:app:B}), (\ref{ch1:app:C}) and (\ref{ch1:app:E}), while comments on the middle vertex contribution to radiation are collected in Appendix~(\ref{ch1:app:D}).

For the reader, the chapter is easiest to follow in three passes. First, the introductory sections identify the model, the relevant scales, and the EFT power counting. Second, the conservative and radiative sections explain how each physical sector modifies the orbit or the emitted power. Third, the appendices collect the more repetitive integral reductions and tensor algebra. In the main text, we therefore emphasise the physical role of each class of diagrams, work out one representative computation whenever a new contraction pattern appears, and quote the remaining contributions in a compact form.
\subsection*{\textbf{ \textit{Notations and conventions:}}}

\begin{multicols}{2}
\begin{itemize}
    \item Metric signature: (-,+,+,+).
    \item $\rmint \mathcal{D}\hat\xi \,e^{iS_{\text{quad}}}\rightarrow \rmint \Bar{\mathcal{D}}\hat\xi$.
    \item $\rmint \frac{d^3\boldsymbol{k}}{(2\pi)^3}\rightarrow \rmint_{\boldsymbol{k}}.$
    \item ${\boldsymbol{x}}_0=\frac{\boldsymbol{x}_1-\boldsymbol{x}_2}{2}.$
    \item Reduced mass: $\mu=\frac{m_1m_2}{M}$\,\,\text{with},\,\,$M=m_1+m_2$.
    \item Mass ratio: $\nu=\frac{\mu}{m_1+m_2}$.
    \item Planck mass: $m_p:=\frac{1}{\sqrt{8\,\pi\, G}},\textstyle{with\, (\hbar,\,c)=1.}$
\end{itemize}
\end{multicols}
\begin{itemize}
\item Levi-Civita symbol convention: $\epsilon^{\mu\nu\rho\sigma}\equiv\frac{\Hat{\epsilon}^{\,\mu\nu\rho\sigma}}{\sqrt{-g}}$, with $\Hat{\epsilon}^{\,0123}=1$.
\item $nPN$ in conservative sector symbolizes $\sim \mathcal{O}(Lv^{2n})$, where $L$ denotes angular momentum.
 $LO$ (Leading Order) symbolizes $\sim\mathcal{O}$($L^{1/2}v^{1/2}$) in radiation sector. Then the subsequent orders are denoted by $N^{(n)}LO\sim \mathcal{O}(L^{1/2}v^{n+1/2}).$

\end{itemize}

\section{Brief review of machinery of EFTs}\label{Sec2}
Consider a QFT with a general tensor field $\xi_{\alpha\beta..}$ in 3+1 dimensions described by the action $S[\xi]$. In order to compute the Effective action, one could first write down the generating functional and then take the logarithm. Usually, in QFT, one writes down the generating functions corresponding to the connected diagrams and, by Legendre transformation, can systematically evaluate the effective action as,
\begin{eqnarray}
        \mathcal{Z}[J]&=&\rmint \mathcal{D}\xi_{\alpha\beta..} \, e^{iS[\xi]+\rmint J\cdot \xi},\nonumber\\
      e^{i W[J]}&=& \mathcal{Z}[J],\\
      S_{\text{eff}}[\xi_{\text{cl}}]&=& W[J]-\rmint d^4x J\,\xi_{\text{cl}}.\nonumber \label{2.3m}
\end{eqnarray}
Here $W[J]$ have contributions only from connected diagrams and $\xi_{\text{cl}}$ is defined as $\xi_{\text{cl}}=\frac{\delta W[J]}{\delta J}$. 
 Effective action in (\ref{2.3m}) is complex in general. \textcolor{black}{The real part gives the conservative dynamics of the system, and the imaginary part gives the radiative dynamics of the system. Hence, for a binary inspiral problem, the real part gives the equation of motion, i.e., the orbit equation of the binary system, and the imaginary part gives the power radiation from the system.} The perturbative computation in $\hbar$ also leads to effective action at the classical and quantum levels.
 \begin{align}
     \begin{split}
S_{\text{eff}}^{\text{tot}}=\sum \underbrace{\text{Tree-level diagrams}}_{\mathcal{O}(\hbar^0)(\text{classical contribution})}+ \sum \underbrace{\text{loop level diagram}}_{\mathcal{O}(\hbar^n)(\text{quantum corrections})}. \label{Stotal}
     \end{split}
 \end{align}
 \par
 At this point, we assume that the theory has two different energy scales, a high energy scale ($\Lambda$) and a low energy scale ($\epsilon$). As we are interested in the low energy dynamics of the quantum field theory, it is convenient to decompose the field in those two modes: $\xi_{\alpha\beta...}=\hat{\xi}_{\alpha\beta..}+\Bar{\xi}_{\alpha\beta..}$ such that,
\begin{itemize}
    \item $\Bar{\xi}_{\alpha\beta..}$ are the light modes with energy scales $\epsilon$.
    \item  $\hat{\xi}_{\alpha\beta..}$ are the heavy modes with energy scale $\Lambda>>\epsilon$.
\end{itemize}
The effective theory of the light modes $\Bar{\xi}_{\alpha\beta..}$ can then be determined by integrating out the heavy modes $\hat{\xi}_{\alpha\beta..}$. This can be done using a path integral in the following way,
\begin{align}
    \begin{split}
    \mathcal{Z}[\Bar{\xi}_{\alpha\beta..}]:=    e^{i\mathcal{S}_{\text{eff}}[\Bar{\xi}_{\alpha\beta..}]}=\rmint \mathcal{D}\hat{\xi}_{\alpha\beta..}\,e^{i\,\mathcal{S}^{\mathcal{O}(\hbar^0)}_{\text{eff}}[\Bar{\xi}_{\alpha\beta..},\hat{\xi}_{\alpha\beta..}]}\,,
        \end{split} \label{stotal1}
\end{align}
where $S^{\mathcal{O}(\hbar^0)}_{eff}$ is the classical part of the action mentioned in (\ref{Stotal}).
The path integral eventually gives the effective action for the light modes as follows,
\begin{align}
    \begin{split}
       \mathcal{S}_{\text{eff}}[\Bar{\xi}_{\alpha\beta..}]= \rmint d^4x\Big[\frac{1}{2}\partial_{\mu}\Bar{\xi}_{\alpha\beta..}\partial^{\mu}\Bar{\xi}^{\alpha\beta..}+
        \sum_{n}C_{n}\mathcal{O}_{n}(\Bar{\xi}_{\alpha\beta..})
\Big]\,.
    \end{split}\label{2.4m}
\end{align}
The coefficients $C_{n}$ are sometimes called Wilson coefficients, and they contain information about the UV sector of the theory. This approach is called the \textit{top-down} approach. One can also proceed in the opposite direction. In that case, one first writes down an effective action with the $C_{n}'s$ left unfixed, and then determines them by comparing with experimental data or with the computation from the \textit{top-down} approach. This is called the \textit{bottom-up} approach. We must therefore compute the effective action perturbatively in the small parameter $\frac{\epsilon}{\Lambda}$ and determine how many $\mathcal{O}_n$ contribute at a given order. This is fixed by the EFT power-counting rules, which tell us how the relevant quantities scale with $\frac{\epsilon}{\Lambda}$. This is discussed systematically in Sec.~(\ref{Sec4}) for our case. \par 
\par
 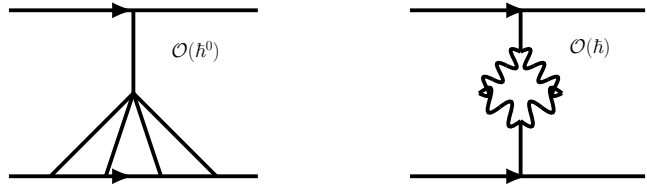
\begin{figure}
     \centering
    \thesisdiagrampanel{\scalebox{0.36}{\begin{feynman}
    \fermion[lineWidth=4, showArrow=false]{5.40, 4.00}{5.80, 5.20}
    \fermion[showArrow=false, lineWidth=4]{5.80, 5.20}{7.00, 4.00}
    \fermion[lineWidth=4, showArrow=false]{11.40, 5.80}{11.40, 6.40}
    \electroweak[flip=true, lineWidth=4]{10.80, 5.20}{11.40, 5.80}
    \fermion[lineWidth=4]{9.80, 6.40}{13.20, 6.40}
    \fermion[showArrow=false, lineWidth=4, label=$\mathcal{O}(\hbar^0)$]{5.80, 5.20}{5.80, 6.40}
    \fermion[lineWidth=4]{4.00, 4.00}{7.60, 4.00}
    \fermion[showArrow=false, lineWidth=4]{4.60, 4.00}{5.80, 5.20}
    \fermion[lineWidth=4, showArrow=false]{6.20, 4.00}{5.80, 5.20}
    \fermion[lineWidth=4]{9.80, 4.00}{13.20, 4.00}
    \electroweak[lineWidth=4]{11.40, 4.80}{10.80, 5.20}
    \fermion[lineWidth=4]{4.00, 6.40}{7.60, 6.40}
    \electroweak[label=$\mathcal{O}(\hbar)$, lineWidth=4]{11.40, 5.80}{12.00, 5.20}
    \electroweak[flip=true, lineWidth=4]{11.40, 4.80}{12.00, 5.20}
    \fermion[showArrow=false, lineWidth=4]{11.40, 4.00}{11.40, 4.80}
\end{feynman}
}}
     \caption{Tree level diagram (left) vs. loop level (one loop) (right) diagram in WEFT}
     \label{mfig}
 \end{figure}
At this point, we are not interested in the quantum effective action, since the quantum corrections are highly subleading. Hence, we neglect the $\mathcal{O}(\hbar^n)$ terms on the right-hand side of (\ref{stotal1}) and focus only on the tree-level diagrams, i.e. the $\mathcal{O}(\hbar^0)$ diagrams shown in Fig.~(\ref{mfig}). Furthermore, we expand the classical effective action in powers of $(v/c)$. This expansion is known as the Post-Newtonian (PN) expansion, and the $n$-PN order scales as $\mathcal{O}(\frac{v}{c})^{2n}$. It is worth noting that, in order to compute the tree-level, i.e. $\mathcal{O}(\hbar^0)$, PN diagrams, one still needs to evaluate several Feynman loop integrals. These tree-level PN diagrams are sometimes called ``Classical Loops.'' For more details, interested readers are referred to \cite{Burgess:2020tbq,Porto:2016pyg}. As explained before, EFT approaches are especially useful when there is a clear separation of scales, which is precisely the case for the binary inspiral problem. The relevant length scales are:
\begin{itemize}
    \item The size of the astrophysical object $R\sim G\,M$.
    \item The orbital radius r.
    \item The wavelength of radiation: $\lambda$.
\end{itemize}
These three scales are not independent; rather, they are related through the relative velocity of the binaries, with $\frac{R}{r}\sim v^2$. Keeping in mind the sensitivity of gravitational-wave detectors, one may roughly assume $\frac{1}{\lambda}\sim \omega_{GW}\sim \frac{v}{r}=\Omega_{orbit}$. These relations show that the different orders in the PN expansion probe different physical length scales, with the hierarchy $R\ll r \ll\lambda$. Hence, in EFT language, the heavy and light field modes are decoupled, and their dynamics can be treated separately.  
\par
\textcolor{black}{In the Worldline Effective Field Theory (WEFT) approach used to study the binary inspiral problem in this paper, apart from field degrees of freedom we do have worldline degrees of freedom $\{x^{\mu}_{a}\}$. We treat the worldline degrees of freedom as non-dynamical DOFs. Then integrating out the heavy modes, we get the total effective action: $\mathcal{S}_{\text{eff}}^{\text{tot}}[\boldsymbol{x}_a,\Bar{\xi}]$ which has the following form,
\begin{align}
    \begin{split}
        \mathcal{S}_{\text{eff}}^{\text{tot}}[\boldsymbol{x}_a,\Bar{\xi}]=-i\,\log\,\mathcal{Z}[\boldsymbol{x}_a,\Bar{\xi}]=\underbrace{\mathcal{S}_{\text{eff}}^{\text{cons}}[\boldsymbol{x}_a]}_{\text{real}}+\underbrace{\mathcal{S}_{\text{eff}}^{\text{rad}}[\boldsymbol{x}_a,\bar\xi]}_{\text{complex}}.
    \end{split}
\end{align}
Extremizing $\mathcal{S}_{\text{eff}}^{\text{cons}}[\boldsymbol{x}_a]$ we get the conservative dynamics of the binary system:
\begin{align}
    \begin{split}
     &    \frac{\delta}{\delta \boldsymbol{x}_a(t)}\,\mathcal{S}_{\text{eff}}^{\text{cons}}[\boldsymbol{x}_a]\Big |_{\mathcal{O}(\hbar^0)}=0 \implies \text{Orbit equation of the binary system.}
    \end{split}
\end{align}
and, $ \mathcal{S}_{\text{eff}}^{\text{rad}}[\boldsymbol{x}_a,\bar\xi]$ is in general complex and related to the radiated power from the system via optical theorem as,
\begin{align}
    \begin{split}
        \,\text{Im}[\gamma_{\text{eff}}[\boldsymbol{x}_a]]\equiv \text{Im}[-i\log\rmint \mathcal{D}\bar\xi \,e^{i\mathcal{S}_{\text{eff}}^{\text{rad}}[\boldsymbol{x}_a,\bar\xi]}]=\frac{\mathcal{T}}{2}\rmint dE\,d\Omega \frac{d^2\Gamma}{dEd\Omega},\,\text{with radiated power},\, dP=E\,d\Gamma.
    \end{split}
\end{align}}

The hierarchy of EFT for the binary inspiral problem is shown in Fig.~(\ref{newfig}).  
\begin{figure}
    \centering
    \includegraphics[scale=0.20]{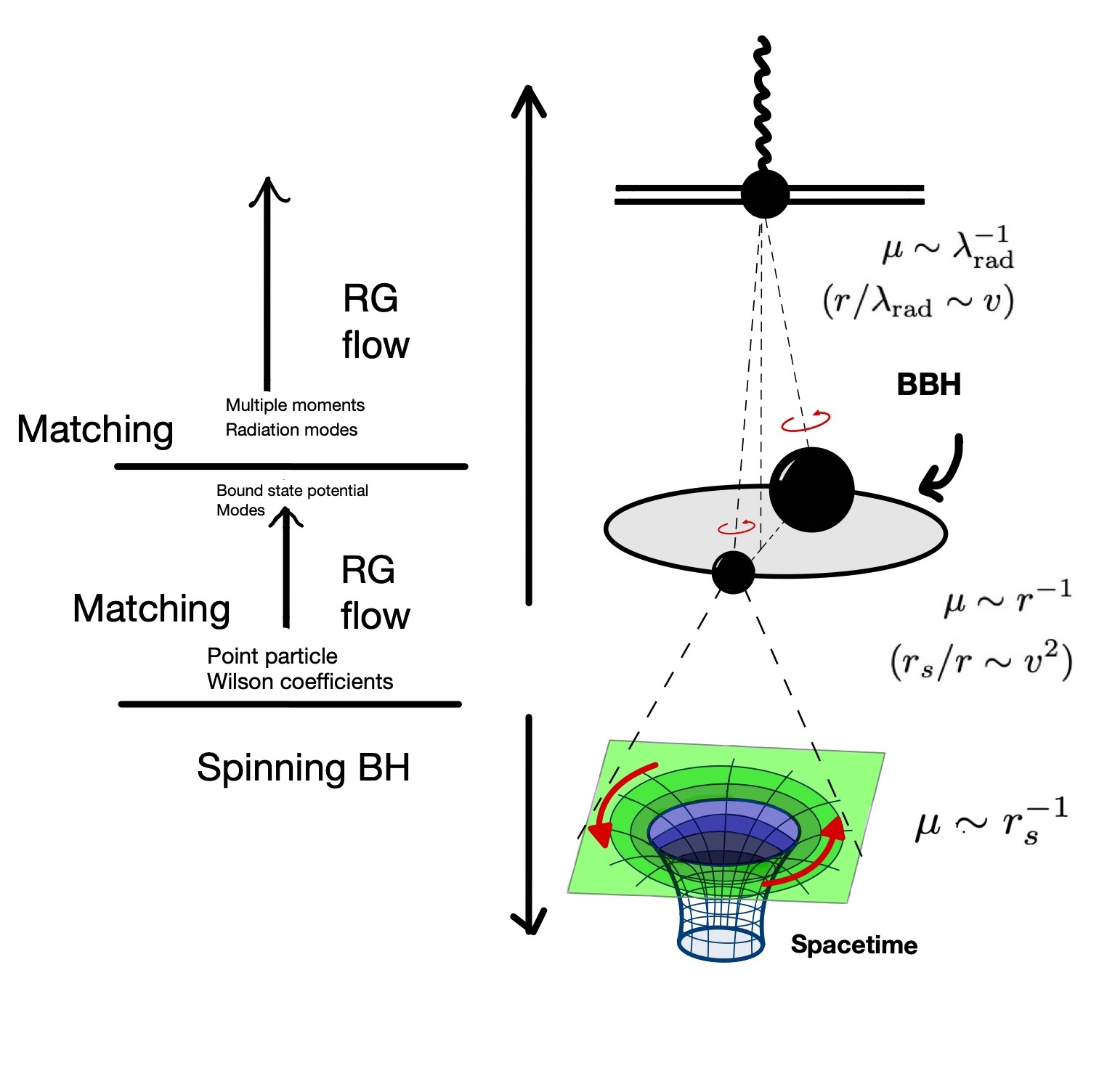}
    \caption{Different scales in the binary inspiral problem. We reproduce the figure from \cite{Porto:2016pyg}.}
    \label{newfig}
\end{figure}
\section{Review of worldline effective field theory for a point particle in GR}\label{Sec3}
In this section, we briefly discuss the action of an inspiralling non-spinning black hole binary in Einstein theory. The background metric is chosen to be in the Kaluza-Klein form, consisting of a non-relativistic gravitational field. 

We start with the following action, 
\begin{equation}
    S=S_{\text{EH}}+S_{\text{pp}} \label{2.1}
\end{equation}
where the Einstein-Hilbert (EH) action is given by,
\begin{eqnarray}
    S_{EH}=\frac{1}{16\pi G}\rmint d^4x \sqrt{-g}\,R
\end{eqnarray}
and describes the dynamics of the graviton. $S_{\text{pp}}$ models the binary (determines the dynamics of the two-body system (black holes or any other compact objects)) and takes the following form, 
\begin{eqnarray} \label{eqnew}
    S_{pp}=\sum_{a=1}^2 m_a \rmint d\tau_{a}+\sum_{a=1}^2 C^{\text{R}}_a \rmint d\tau_{a} R({x_a})+\sum_{a=1}^2 C_{a}^{\text{V}}\rmint d\tau_a R_{\mu\nu}(x_a)\,\dot{x}_a^\mu \,\dot{x}_\nu^a+.....\,.
\end{eqnarray}

Here, $\tau_a$ denotes the proper time along the particle's world line and  $C^R$, $C^V$ are the Wilson coefficients. In this paper, we will only consider non-spinning binaries, hence we can ignore the finite size effects. Henceforth, we will not consider the proportional to $C^R$, $C^V$ from (\ref{eqnew}).  \par
A gravitational wave detector such as LIGO measures the gravitational waveform. The first step towards computing such waveforms is to calculate the radiated energy flux from the binary system. Given the action in (\ref{2.1}) we can calculate this quantity by considering the perturbation around a flat metric, i.e., $g_{\mu\nu}=\eta_{\mu\nu}+h_{\mu\nu}$. Integrating out the graviton field, one can obtain the effective action for the point particles,
\begin{eqnarray}
    e^{i\,\mathcal{S}_{\text{eff}}}=\rmint \mathcal{D}h_{\mu\nu} \exp{[iS_{EH}+iS_{pp}]}\,.
\end{eqnarray}\label{2.4}
The real part of the effective action describes the dynamics of the two-body system, and the imaginary part measures the graviton emission from the binary system over a large time. In principle, one can directly evaluate the full relativistic path integral using the Feynman rules for the Einstein-Hilbert action, which involves tree-level as well as loop-level diagrams. However, after calculating the amplitudes, it is not easy to take the non-relativistic limit. In order to overcome the problem, following \cite{Kol:2007bc}, one can decompose the metric in terms of the non-relativistic gravitational fields and then perform the calculations. We will only focus on the tree-level diagrams relevant to classical results. The methodology is discussed in detail in the subsequent sections. \par

\subsubsection*{Non relativistic general relativity}
In this section, we will briefly discuss how one can arrive at the effective action in a non-relativistic regime \cite{Goldberger:2004jt,Levi:2018nxp,Porto:2016pyg}. We expand the point particle action in powers of the particle's three-velocities.
\begin{eqnarray}
    S_{pp}=\sum_{a=1}^2\Big[\frac{1}{2}v_a^2-\frac{1}{2}h_{00}-h_{0i}\,v^a_{i}-\frac{1}{2}h_{00}v_{a}^2-\frac{1}{2}h_{ij}v^{i}_a\,v^{j}_a+\frac{1}{8}v_a^4+.......\Big]\,.
\end{eqnarray}
Here we should remember that all components of the metric perturbations $h_{\mu\nu}$ are evaluated at the particular particle world line.\par 
However, at this point the propagator for $h_{\mu\nu}$ is fully relativistic. Hence we can not distinguish between the potential gravitons (heavy modes) and the long-wavelength radiation gravitons (light modes). We are not interested in potential gravitons as the gravitational wave detector can not detect them. So due to our interest in finding out the effective action in the non-relativistic regime, we integrate over the potential gravitons. Now, to connect with the discussion in Sec.~(\ref{Sec2}) about integrating out the heavy modes we decompose the metric perturbation as follows,
\begin{eqnarray}
    h_{\mu\nu}=\Bar{h}_{\mu\nu}+
\mathcal{H}_{\mu\nu}\,,\label{3.6n}
\end{eqnarray}
where $\mathcal{H}_{\mu\nu}$ is the potential gravitons with the properties $\partial_{i}\mathcal{H}_{\mu\nu}\sim\frac{\mathcal{H}_{\mu\nu}}{r},\partial_{0}\mathcal{H}_{\mu\nu}\sim{\mathcal{H}_{\mu\nu}}\frac{v}{r}$. And $\Bar{h}_{\mu\nu}$ is the radiation gravitons with $\partial_{\beta}\bar{h}_{\mu\nu}\sim{\bar{h}_{\mu\nu}}\frac{v}{r}$. It is more convenient to express $\mathcal{H}_{\mu\nu}$ as,
\begin{eqnarray}
    \mathcal{H}_{\mu\nu}(x^\mu)=\rmint\frac{d^3\boldsymbol{k}}{(2\pi)^3}\,e^{i\boldsymbol{k}.\boldsymbol{x}}\,\mathcal{H}^{k}_{\mu\nu}(x^0)\,.
\end{eqnarray}
Now, the conservative (or radiative) effective action for the binaries $S_{\text{NRGR}}$ can be computed by integrating over the non-relativistic potential gravitons,
\begin{eqnarray}
    e^{i\mathcal{S}_{\text{eff}}[x_a,\Bar{h}_{\mu\nu}]}=\rmint \mathcal{D}\mathcal{H}_{\mu\nu}\,\exp{[iS[\Bar{h}+\mathcal{H},x_a]+iS_{GF}]}\label{2.8}
\end{eqnarray}
where, $S_{GF}$ is the gauge fixing term. While evaluating the path integral, ghost terms arise in the effective action, but at the classical level, they do not contribute. Now, to preserve the gauge invariance, we choose the gauge fixing term as \cite{Goldberger:2004jt},
\begin{eqnarray}
    S_{GF}=\rmint d^4x\sqrt{\Bar{g}}\,\Gamma_\mu \Gamma^\mu
\end{eqnarray}
which is invariant under the general coordinate transformation of the background metric $\Bar{g}_{\mu\nu}=\eta_{\mu\nu}+\Bar{h}_{\mu\nu}$ and $\Gamma_\mu=\nabla_{\alpha}\mathcal{H}^{\alpha}_{\mu}-\frac{1}{2}\nabla_{\mu}\mathcal{H}_{\alpha}^{\alpha}.$ From the EH Lagrangian (up to $\mathcal{O}(\mathcal{H}^2)$) in momentum space, one can write down the \textcolor{black}{orbital} graviton propagator as \cite{Goldberger:2004jt},
\begin{eqnarray}
    \Big\langle\mathcal{H}^{k}_{\mu\nu}(x^0)\mathcal{H}^{q}_{\alpha\beta}(0)\Big\rangle=-(2\pi)^3 \delta^{(3)}(\boldsymbol{k}+\boldsymbol{q})\delta(x^0)\frac{i}{\boldsymbol{k}^2}P_{\mu\nu,\alpha\beta}\label{2.10}
\end{eqnarray}
with, $P_{\mu\nu,\alpha\beta}=\frac{1}{2}[\eta_{\mu\alpha}\eta_{\nu\beta}+\eta_{\mu\beta}\eta_{\nu\alpha}-\frac{2}{d-2}\eta_{\mu\nu}\eta_{\alpha\beta}]$. So in order to find $S_{NRGR}$ one have to compute the functional integral in (\ref{2.8}) using the (\ref{2.10}). To do this, one has to sum over the relevant Feynman diagram, which has the following topological properties \cite{Goldberger:2022ebt, Goldberger:2004jt},
\begin{itemize}
    \item All diagrams must be connected with stripped-off particle worldlines.
    \item Diagram can contain internal graviton potential modes ($\mathcal{H}_{\mu\nu}$) but can not have external potential mode.
    \item Radiative graviton modes ($\Bar{h}_{\mu\nu}$) are only appear in the external lines.
\end{itemize}
As there is a gauge redundancy, it is better to choose a particular gauge from the beginning and use the proper decomposition of the non-relativistic graviton field \cite{Kol:2007bc}. This helps us to identify the correct diagrams at each PN order and avoid any confusion arising due to the mixing of various components of $h_{\mu\nu}$. Hence, following \cite{Kol:2007bc}, we choose the standard Kaluza-Klein (KK) decomposition,
\begin{align}
    \begin{split}
        ds^2=g_{\mu\nu}dx^{\mu}dx^{\nu}=-e^{2\psi}(dt-\tilde{\mathcal{A}_i}dx^{i})^2+e^{-2{\psi}}\gamma_{ij}dx^{i}dx^{j}\,.
    \end{split} \label{KK}
\end{align}
In the component form:
$$g_{\mu\nu}=
\begin{pmatrix}
-e^{2\psi} & e^{2\psi} \tilde{\mathcal{A}}_{j}\\
e^{2\psi} \tilde{\mathcal{A}}_{i} & \,\,e^{-2\psi}\gamma_{ij}-e^{2\psi}\tilde{\mathcal{A}}_{i}\tilde{\mathcal{A}}_{j}\\
\end{pmatrix}$$
where $\psi,\tilde{\mathcal{A}_i},\gamma_{ij}=\delta_{ij}+\sigma_{ij}$ are the set of \textit{non-relativistic graviton field} (NRG), sometimes called as Kol-Smolkin variables, are Newtonian potential, gravitomagnetic potential, and metric 3-tensor, respectively \cite{Kol:2007bc}.
For our future computations to make all NRG fields have mass-dimension one, one can rescale the fields as follows,
\begin{align}
    \begin{split}
        \psi\rightarrow\frac{\psi}{m_p},\,\,\tilde{\mathcal{A}}_{i}\rightarrow\frac{\tilde{\mathcal{A}}_{i}}{m_p},\,\,\sigma_{ij}\rightarrow \frac{\sigma_{ij}}{m_p}.
    \end{split}
\end{align}
\section{Conservative dynamics of binary black holes in a theory of ALP and ULV }\label{Sec4}
In this section, we describe the action for inspiralling binary black holes in a theory of \textit{Axion-like particles and Ultra-light vectors}. Apart from the usual EH term, it contains a Maxwell field, a Proca field, and a massive scalar field.  We have the following action for this theory \cite{Cardoso:2018tly}:
\begin{align}
\begin{split}
S_{\text{field}}=\rmint d^4x\sqrt{-g}&\Big(\frac{m_{p}^2}{2}R-\frac{1}{4}F_{\mu\nu}F^{\mu\nu}-\frac{1}{2}\partial_\mu\phi\partial^\mu\phi-\frac{1}{2}m^2\phi^2+\frac{g_{a\gamma\gamma}}{4\,m_{p}}\phi F_{\mu\nu}^*F^{\mu\nu}-\frac{1}{4}B_{\mu\nu}B^{\mu\nu}\\ &
+\frac{\gamma}{2}F_{\mu\nu}B^{\mu\nu}
-\frac{1}{2}\mu_{\gamma}^2B_{\mu}B^{\mu}\Big)\,.
\label{2.11}
\end{split}
\end{align}
The action in (\ref{2.11}) describes the Axion-like-particles (ALPs) (scalar), massless photon and massive (often known as \textit{dark photon}) gauge field (Proca) minimally coupled with gravity. In (\ref{2.11}) $R,\, \phi $ are the Ricci scalar and ALP. Also, 
$$F_{\mu\nu}=\partial_{\mu} {A}_{\nu}-\partial_{\nu} {A}_{\mu},\, B_{\mu\nu}=\partial_{\mu} {B}_{\nu}-\partial_{\nu} {B}_{\mu},$$ where ${A}_{\mu}$ and ${B}_{\mu}$ are Electromagnetic and Proca field respectively. $\gamma$ is the coupling constant between photon and dark photon. Furthermore, $g_{a\gamma\gamma}$ is the coupling constant for axion-photon, and it has a negative mass dimension, implying the theory is not renormalizable, and the term $\phi F^{*} F$ is sometimes called the \textit{Theta term}. However, that is not a problem for our case as we finally intend to write down a classical effective action for the light modes in powers of $(v/c)$ as mentioned in Sec.~(\ref{Sec2}). A good EFT does not need to be renormalized. We are only concerned about the tree-level diagrams as we want classical results. Eventually, we will encounter loop integrals while computing the effective action, and we call them classical loops as they are proportional to $\hbar ^0$. \par

Next we compute the effective action at $\mathcal{O}{(\hbar^0)}$ and organize it as an expansion in $\mathcal{O}(\frac{v}{c})^n$. We therefore work in the non-relativistic regime, choose the background metric in KK form as mentioned in (\ref{KK}), and keep the non-trivial worldline couplings of the non-relativistic gravitational (NRG) fields. Our conservative analysis extends up to 1PN order in the gravitational, electromagnetic, and Proca sectors, and up to 2PN order in the scalar sector. We will also show that the leading correction from the CP-violating $g_{a\gamma\gamma}$ term in (\ref{2.11}) enters the conservative dynamics only at \textit{2.5PN}. \par

The calculation follows a uniform pattern throughout the chapter. We first identify the quadratic action and the relevant worldline operators, which determine the propagators and the vertices. We then use EFT power counting to decide which connected diagrams contribute at a given PN order. Whenever a new contraction pattern appears, we display one explicit derivation; for the remaining diagrams of the same type we quote the reduced amplitudes directly so that the physical structure of the result stays visible in the main text. The EFT of radiation fields, in KK parametrization, is obtained by decomposing fields into potential and radiation sectors as $\psi=\Tilde{\psi}+\Bar{\psi},\Tilde{\mathcal{A}}_i=\Hat{\mathcal{A}}_i+\Bar{\mathcal{A}}_i,\sigma_{ij}=\zeta_{ij}+\Bar{\sigma}_{ij}\,,\phi=\varphi+\Bar{\phi}, A_{\mu}=\mathcal{A}_{\mu}+\Bar{a}_{\mu}$ and $B_{\mu}=\mathcal{B}_{\mu}+\Bar{b}_{\mu}$ and then integrating over the potential modes. The partition function takes the form
\begin{align}
\begin{split}
  \mathcal{Z}[\boldsymbol{x}_a,\Bar{\psi},\Bar{{\mathcal{A}}}_i,\Bar{\sigma}_{ij},\Bar{a}_{\mu},\Bar{b}_{\mu}]  &:=e^{i\mathcal{S}_{\text{eff}}[\boldsymbol{x}_a,\Bar{\psi},\Bar{{\mathcal{A}}}_i,\Bar{\sigma}_{ij},\Bar{a}_{\mu},\Bar{b}_{\mu}]}\,,\\ &
  =\rmint \underbrace{\mathcal{D}\tilde{\psi} \,\mathcal{D}\hat{\mathcal{A}}_{i}\,\mathcal{D}\varphi \,\mathcal{D}\mathcal{A}_{\mu}\,\mathcal{D}\mathcal{B}_{\nu}\,
  \mathcal{D}\boldsymbol{\zeta}_{ij}\,}_{\text{integrating over potential modes}\,(\mathcal{D}\hat{\xi})}\,e^{iS_{\text{field}}+iS_{\text{pp}}}\,.\label{ch1:2.13}
\end{split}
\end{align}
We can compute the effective action in (\ref{ch1:2.13}) by evaluating the  amplitudes of the suitable Feynman diagrams by taking the logarithm as,
\begin{align}
    \begin{split}
        i\mathcal{S}_{\text{eff}}[\boldsymbol{x}_a,\Bar{\xi}]&=\log\Big[{\rmint\mathcal{D}\hat\xi\,e^{iS_{\text{field}}+iS_{\text{pp}}}}\Big]\,.\label{4.3mm}
    \end{split}
\end{align}
One can compute (\ref{4.3mm}) perturbatively by evaluating the connected Feynman diagrams. We need two kinds of vertices, worldline vertices and field vertices. To get the worldline vertices, we need to consider the point particle action, including all the fields.
Now we can write down the field action by, $S_{\text{field}}=S_{\text{quad}}+S_{\text{int}}$ and use,
\begin{align}
    \begin{split}
        \Big\langle\mathcal{O}_{1}(\boldsymbol{x}_1(t_1))...\mathcal{O}_{n}(\boldsymbol{x}_n(t_n))\Big\rangle=\rmint\mathcal{D}\hat\xi \,e^{iS_{\text{quad}}}\,\underbrace{\mathcal{O}_{1}(\boldsymbol{x}_1(t_1),\hat\xi)...\mathcal{O}_{n}(\boldsymbol{x}_n(t_n),\hat\xi)}_{\text{got expanding \,$e^{iS_{\text{pp}}+iS_{\text{int}}}$}}
    \end{split}
\end{align}
and the wick contraction to reduce the higher point correlation function into the two-point functions to compute the effective action.
\par
In order to evaluate the connected diagrams, one next needs the Feynman propagators (two-point functions). We therefore record the quadratic actions for the relevant field configurations and the resulting non-relativistic propagators. The purpose of the next formulas is mainly to collect the ingredients that repeatedly enter the later diagram evaluation, rather than to dwell on algebra that is identical from sector to sector.

\textbf{Propagators of different fields:} We list the two-point functions, i.e. the propagators of different fields in our theory. We also do the conventional non-relativistic expansion of the propagator, assuming $|k_0|\ll |\boldsymbol{k}|$. In this limit, we expand the relativistic Feynman propagator, which contains an infinite sum of the delta function and its time derivatives (even). We only consider the non-relativistic contribution and the first-order relativistic corrections.

\newpage
\underline{\textit{Scalar propagator:}}

 In KK parametrization the scalar action has the following form:
 \begin{align}
    \begin{split}
        S_{\phi}&=\rmint d^4x \sqrt{-g}\Big[-\frac{1}{2}\partial_{\mu}\phi\partial^{\mu}\phi-\frac{1}{2}m^2 \phi^2\Big]\,,\\ &
        =\rmint d^3x \,dt \sqrt{\gamma}\Big[\frac{e^{-4\psi}}{2}\partial_{0}\phi\partial_{0}\phi-2\Hat{\mathcal{A}}^{i}\partial_{i}\phi\partial_{0}\phi-\frac{1}{2}\gamma^{ij}\partial_{i}\phi\partial_{j}\phi-\frac{1}{2}e^{-2\psi}m^2\phi^2 \nonumber
        \Big] \,,\end{split}
\end{align}
\begin{align}
    \begin{split}
\,\,\,\,\,\,\,\,\,\,\,\,\,\,\,\,\,\,\,\,\,\,\,\,\,\,\,=\rmint d^3x dt \,\Big(1+\frac{\text{Tr}(\sigma)}{2}\Big)&\Big[\frac{1}{2}\partial_{0}\phi\partial_{0}\phi-2\psi \partial_{0}\phi\partial_{0}\phi-2\Hat{\mathcal{A}}^{i}\partial_{i}\phi\partial_{0}\phi-\frac{1}{2}\delta^{ij}\partial_{i}\phi\partial_{j}\phi \\&+ \frac{1}{2}\sigma^{ij}\partial_{i}\phi\partial_{j}\phi-\frac{1}{2}(1-2\psi)m^2\phi^2\Big]\,.\label{24m}
    \end{split}
\end{align}
From the action (\ref{24m}) one can identify the scalar propagator as follows,
 \begin{align}
    \begin{split}
\Big\langle\varphi(x_1)\varphi(x_2)\Big\rangle &=\rmint \frac{d^4 k}{(2\pi)^4} \frac{e^{ik\cdot(x_1-x_2)}}{-k^2-m^2}\\&=\rmint \frac{dk_0}{2\pi}\,e^{-ik_0 (t_1-t_2)}\rmint \frac{d^3k}{(2\pi)^3}\frac{e^{i\boldsymbol{k}\cdot(\boldsymbol{x}_1-\boldsymbol{x}_2)}}{\boldsymbol{k}^2+m^2}\Big[1+\frac{k_0^2}{\boldsymbol{k}^2+m^2}+...\Big]
        \\ &
        =\delta(t_1-t_2)\rmint \frac{d^3k}{(2\pi)^3}\frac{e^{i\boldsymbol{k}\cdot\boldsymbol{r}}}{\boldsymbol{k}^2+m^2}+\partial_{t_1}\partial_{t_2}\delta(t_1-t_2)\rmint \frac{d^3k}{(2\pi)^3}\frac{e^{i\boldsymbol{k}\cdot\boldsymbol{r}}}{(\boldsymbol{k}^2+m^2)^2}+...\,. 
    \end{split}
\end{align}
\underline{\textit{ Electromagnetic propagator:}}

The action for free Maxwell theory is given by,
\begin{align}
    \begin{split}
        S_{\text{EM}}=-\frac{1}{4}\rmint d^3x dt \sqrt{\gamma}&\Big[-F_{0i}F_{0i}+e^{2\psi}\gamma^{ij}\gamma^{kl}F_{ik}F_{jl}+e^{2\psi}\Hat{\mathcal{A}}^{i}\Hat{\mathcal{A}}^{j}F_{0j}F_{i0}+e^{2\psi}\Hat{\mathcal{A}}^{i}\gamma^{jk}F_{0j}F_{ik}\Big]\\ &
        +(\text{higher order in fields})\,.\label{25m}
    \end{split}
\end{align}
From the quadratic part of the action (\ref{25m}), one can identify the propagator of the electromagnetic field as follows,
\begin{align}
    \begin{split}
&\Big\langle\mathcal{A}_{0}(x_1)\mathcal{A}_0(x_2)\Big\rangle=\textcolor{black}{-}\Big[\delta(t_1-t_2)\rmint \frac{d^3k}{(2\pi)^3}\frac{e^{i\boldsymbol{k}\cdot\boldsymbol{r}}}{\boldsymbol{k}^2}+\partial_{t_1}\partial_{t_2}\delta(t_1-t_2)\rmint \frac{d^3k}{(2\pi)^3}\frac{e^{i\boldsymbol{k}\cdot\boldsymbol{r}}}{\boldsymbol{k}^4}...\Big]\,,\\ &
        \Big\langle\mathcal{A}_{i}(x_1)\mathcal{A}_k(x_2)\Big\rangle=\Big[\delta(t_1-t_2)\rmint \frac{d^3k}{(2\pi)^3}\frac{e^{i\boldsymbol{k}\cdot\boldsymbol{r}}}{\boldsymbol{k}^2}+\partial_{t_1}\partial_{t_2}\delta(t_1-t_2)\rmint \frac{d^3k}{(2\pi)^3}\frac{e^{i\boldsymbol{k}\cdot\boldsymbol{r}}}{\boldsymbol{k}^4}...\Big]\delta_{ik}\,.\\ &
    \end{split}
\end{align}
\underline{\textit{ Proca propagator:}}

The action for the Proca field is the same as (\ref{25m}) with a mass term, hence the propagator has the following form:
\begin{align}
    \begin{split}
    &\Big\langle\mathcal{B}_{0}(x_1)\mathcal{B}_0(x_2)\Big\rangle=\textcolor{black}{-}\Big[\delta(t_1-t_2)\rmint \frac{d^3k}{(2\pi)^3}\frac{e^{i\boldsymbol{k}\cdot\boldsymbol{r}}}{\boldsymbol{k}^2+\mu_{\gamma}^2}+\partial_{t_1}\partial_{t_2}\delta(t_1-t_2)\rmint \frac{d^3k}{(2\pi)^3}\frac{e^{i\boldsymbol{k}\cdot\boldsymbol{r}}}{(\boldsymbol{k}^2+\mu_{\gamma}^2)^2}+.....\Big]\,.\\ &
\Big\langle\mathcal{B}_{i}(x_1)\mathcal{B}_k(x_2)\Big\rangle=\Big[\delta(t_1-t_2)\rmint \frac{d^3k}{(2\pi)^3}\frac{e^{i\boldsymbol{k}\cdot\boldsymbol{r}}}{\boldsymbol{k}^2+\mu_{\gamma}^2}+\partial_{t_1}\partial_{t_2}\delta(t_1-t_2)\rmint \frac{d^3k}{(2\pi)^3}\frac{e^{i\boldsymbol{k}\cdot\boldsymbol{r}}}{(\boldsymbol{k}^2+\mu_{\gamma}^2)^2}...\Big]\delta_{ik}\,.\label{2.28jjj}
    \end{split}
\end{align}
The propagator given in \eqref{2.28jjj} is not the full expression; in general, it also contains a correction term of the form \(\frac{k^\mu k^\nu}{\mu_{\gamma}^2}\). However, this additional piece contributes only at subleading order to both the potential and the waveform. Since, for the massive modes, our interest is restricted to the leading contribution, we neglect these extra terms.\\\\
\underline{\textit{Graviton propagators:}}

Lastly, the gravitational action takes the following form in KK parametrization:
\begin{align}
    \begin{split}
        S_{\text{G}}=\frac{m_{p}^2}{2}\rmint d^3x dt \,\sqrt{\gamma}&\Big[R[\gamma]-2\gamma^{ij}\partial_{i}\psi\partial_{j}\psi+\frac{e^{4\psi}}{4}\Hat{F}_{ij}\Hat{F}_{kl}\gamma^{ik}\gamma^{jl}+\frac{e^{-4\psi}}{4}[\Dot{\gamma}_{ij}\Dot{\gamma}_{kl}\gamma^{ik}\gamma^{jl}-(\gamma^{ij}\Dot{\gamma}_{ij})^2]\\ &
-4\gamma^{ij}\Dot{\Hat{\mathcal{A}}}\,\partial_{j}\psi
        +e^{-4\psi}(2\Dot{\psi}\gamma^{ij}\Dot{\gamma}_{ij}-6\Dot{\psi}^2)\Big]\,.\label{23m}
    \end{split}
\end{align}
For the action in (\ref{23m}) one can identify the static graviton propagator and its relativistic time corrections as:
\begin{align}
  \begin{split}
&\Big\langle\tilde{\psi}(x_1)\tilde{\psi}(x_2)\Big\rangle=\frac{1}{2}\Big[\delta(t_1-t_2)\rmint \frac{d^3k}{(2\pi)^3}\frac{e^{i\boldsymbol{k}\cdot\boldsymbol{r}}}{\boldsymbol{k}^2}+\partial_{t_1}\partial_{t_2}\delta(t_1-t_2)\rmint \frac{d^3k}{(2\pi)^3}\frac{e^{i\boldsymbol{k}\cdot\boldsymbol{r}}}{\boldsymbol{k}^4}...\Big]\,,\\ &
        \Big\langle\Hat{\mathcal{A}}_{i}(x_1)\Hat{\mathcal{A}}_k(x_2)\Big\rangle=-2\Big[\delta(t_1-t_2)\rmint \frac{d^3k}{(2\pi)^3}\frac{e^{i\boldsymbol{k}\cdot\boldsymbol{r}}}{\boldsymbol{k}^2}+\partial_{t_1}\partial_{t_2}\delta(t_1-t_2)\rmint \frac{d^3k}{(2\pi)^3}\frac{e^{i\boldsymbol{k}\cdot\boldsymbol{r}}}{\boldsymbol{k}^4}...\Big]\delta_{ik}\,,\\ &
        \Big\langle\boldsymbol{\zeta}_{ij}(x_1)\boldsymbol{\zeta}_{kl}(x_2)\Big\rangle=4\Big[\delta(t_1-t_2)\rmint \frac{d^3k}{(2\pi)^3}\frac{e^{i\boldsymbol{k}\cdot\boldsymbol{r}}}{\boldsymbol{k}^2}+\partial_{t_1}\partial_{t_2}\delta(t_1-t_2)\rmint \frac{d^3k}{(2\pi)^3}\frac{e^{i\boldsymbol{k}\cdot\boldsymbol{r}}}{\boldsymbol{k}^4}...\Big]P_{ij,kl} \,.
    \end{split}
\end{align}
We denote the relativistic time correction to the non-relativistic propagator as $\Big\langle....\Big\rangle_{\varoslash}$.

\textbf{Point Particle action:} In our analysis, we model the binaries as charged point particles, ignoring any internal structure of the astrophysical objects under consideration. Our first task is to identify the worldline operators. For that, we first write down the point particle worldline action. As we have a massive scalar field in the theory, we expect the objects' masses to depend on the scalar field \cite{Dyadina:2018ryl}. We also assume that the binary possesses electromagnetic and Proca charges (dark charges). So we include the coupling between the charges (electromagnetic (EM) and Proca) with the relevant fields. Here, we ignore the finite-size effects terms. 
The action is therefore given (at leading order in gauge fields) by,
\begin{align}
    \begin{split}
        S_{pp}=&-\sum_{a}\rmint m_a(\phi)d\tau_{a}+\sum_a\rmint\,Q_a \,d\tau_a \textcolor{black}{A^\mu}\,v_\mu+\sum_a\rmint\,Q_a'\, d\tau_a B^\mu\,v_\mu \,,\\ &
        =\sum_{a}\, m_a\rmint dt\Big[-1+\frac{1}{2}v_a^2-\frac{\psi}{m_{p}}-\frac{\hat{\mathcal{A}}_{i}}{m_{p}}\,v^i_{a}-\frac{\psi^2}{2m_{p}^2}-\frac{3}{2}\frac{\psi}{m_{p}} v_{a}^2\textcolor{black}{+}\frac{1}{2}\frac{\sigma_{ij}}{m_{p}}v^{i}_a\,v^{j}_a+\frac{1}{8}v_a^4+\mathcal{O}(v^6)\Big]\Tilde{\mathcal{L}} \\ &
        +\sum_a \rmint dt \Big[Q_a\,A_{0}(1-\frac{\psi}{m_{p}}-\frac{\psi^2}{2m_{p}^2}) +Q_a\,v^{i}_a A_{i}(1-\frac{\psi}{m_{p}}-\frac{\psi^2}{2m_p^2})+\cdots\Big]\\ &
        \sum_a \rmint dt \Big[Q_a'\,B_{0}(1-\frac{\psi}{m_{p}}-\frac{\psi^2}{2m_{p}^2}) +Q_a'\,v^{i}_a B_{i}(1-\frac{\psi}{m_{p}}-\frac{\psi^2}{2m_p^2})+\cdots\Big]\,, \label{4.10}
    \end{split}
 \end{align}
 where, $\Tilde{\mathcal{L}}=\Big[1+s_a\,\frac{\phi}{m_{p}}+g_a\,\frac{\phi^2}{m_{p}^2}\Big].$ Also, $a=1,2$ denotes the two objects in the binary system. We have expanded $m_a(\phi)$ in a Taylor series up to second order, with $s_a=\frac{1}{m_a}\frac{\partial m_a(\phi)}{\partial \phi}|_{\phi=0}$ and $g_a=\frac{1}{m_a}\frac{\partial^2 m_a(\phi)}{\partial \phi^2}|_{\phi=0} $. \textit{ $Q_1'$ and $Q_2'$ denote two dark charges. They are not associated with an electric or magnetic field. They are sometimes also referred to as `hidden' charges because they do not interact with ordinary matter in the same way that electric and magnetic charges do.}
\footnote{Terms with time derivatives in fields will contribute as corrections to the non-relativistic propagator.  One can show that taking the 2-point self-interacting temporal vertices produces the same corrections to the non-relativistic propagator. 
}

\begin{table}[htb!]
\centering
\scalebox{0.70}{
\setlength{\arrayrulewidth}{0.3mm}
\setlength{\tabcolsep}{20pt}
\renewcommand{\arraystretch}{2.2}
\begin{tabular}{|p{3cm}|p{3cm}|p{3cm}|p{3cm}|p{3cm}|}
\hline
\multicolumn{4}{|c|}{EFT PN COUNTING} \\
\hline
Fields & Worldline-Coupling & Diagram &order \\
\hline
$\hat{\mathcal{A}}_i$& -$\frac{m_a}{m_{p}} \rmint dt\,\Hat{\mathcal{A}}^i\,v_i^a$ & \scalebox{0.3}{\begin{feynman}
    \gluon[lineWidth=4,color=fcb900,label=$\Hat{\mathcal{A}}^i$]{4.00, 5.20}{5.80, 5.20}
    \fermion[lineWidth=4, showArrow=true, flip=false]{4.00, 4.00}{4.00, 6.40}
   
\end{feynman}}& $\sim v$\\
\hline
$ \mathcal{A}_i$& $Q\rmint dt \mathcal{A}_i v^i$& \scalebox{0.3}{\begin{feynman}
    \fermion[lineWidth=4, showArrow=false]{4.00, 4.00}{4.00, 6.20}
    \electroweak[color=9900ef, lineWidth=4, label=$\mathcal{A}_i$]{4.00, 5.00}{5.40, 5.00}
\end{feynman}}& $\sim v$\\
\hline
$ \mathcal{B}_i$& $Q'\rmint dt\, \mathcal{B}_i \,v^i$& \scalebox{0.2}{\begin{feynman}
    \gluon[lineWidth=4, flip=true, endcaps=false, color=eb144c, label=$\mathcal{B}_i$]{4.00, 5.20}{5.60, 5.20}
    \fermion[lineWidth=4, showArrow=true, flip=false]{4.00, 4.00}{4.00, 6.40}
    
\end{feynman}}& $\sim v$\\
\hline
$ \mathcal{A}_0$& $Q\rmint dt\,  \mathcal{A}_0$& \scalebox{0.3}{\begin{feynman}
    \dashed[lineWidth=4, showArrow=false, label=$\mathcal{A}_0$,flip=true]{4.00, 5.00}{6.00, 5.00}
    \fermion[lineWidth=4]{4.00, 4.00}{4.00, 6.00}
    
\end{feynman}}& $\sim v^0$\\
\hline
$\tilde{\psi}$& $\frac{-m_a}{m_{p}}\rmint dt\, \,\tilde{\psi}$& \scalebox{0.3}{\begin{feynman}
    \electroweak[lineWidth=4, label=$\boldsymbol{\tilde{\psi}}$]{4.00, 5.20}{6.00, 5.20}
    \fermion[lineWidth=4]{4.00, 4.00}{4.00, 6.20}
\end{feynman}
}& $\sim v^0$\\
\hline
$ \mathcal{B}_0$& $Q'\,\rmint dt\,  \mathcal{B}_0$ & \scalebox{0.3}{\begin{feynman}
    \dashed[lineWidth=4, showArrow=false, color=9900ef, label=$\mathcal{B}_0$]{4.00, 5.20}{5.40, 5.20}
    \fermion[lineWidth=4, showArrow=true, flip=false]{4.00, 4.00}{4.00, 6.40}
   
\end{feynman}}& $\sim v^0$\\
\hline
$\varphi$& -$s_a\, \frac{m_a}{m_{p}} \rmint dt\, \varphi $ & \scalebox{0.3}{\begin{feynman}
    \electroweak[lineWidth=4, color=0693e3, label=$\boldsymbol{\varphi}$, flip=true]{4.00, 5.00}{5.40, 5.00}
    \fermion[lineWidth=4]{4.00, 4.00}{4.00, 6.20}
\end{feynman}}& $\sim v^0$\\
\hline
\end{tabular}
}
\caption{Table showing the different worldline couplings and the order at which they contribute.}
\label{table1m}
\end{table}

\textbf{EFT counting for different fields:}
As discussed in Sec.~(\ref{Sec2}), we write the effective action for the light modes up to a given PN order. This leads directly to EFT power counting. Hence, we need to know how the different field couplings to the worldline scale with $v$. We use Table~(\ref{table1m}) to determine how the different terms in the conservative effective action scale with the velocities of the black holes. 
The contribution of effective action corresponding to every diagram has the following form $Lv^{2n}$, which we call the $n PN$ diagram.
\subsection{The Gravitational bound sector}
With the propagators and power counting rules in hand, the remaining task is to identify the connected diagrams that contribute to the real part of the effective action. That real part determines the corrected orbit. In what follows, we therefore organize the calculation sector by sector. For readability, we work out one representative Wick contraction whenever a new structure appears and then collect the remaining diagrams in their reduced form. We begin with the gravitational bound sector. Its 1PN result is standard \cite{Kuntz:2019zef}, but it also provides the template for the electromagnetic and Proca sectors discussed next. Figure~(\ref{fig:my_labela}) shows the four graphs that contribute at this order.\footnote{To do the integrals, we take the help of the master integrals given in Appendix~(\ref{ch1:app:E}).}
\begin{figure}
    \centering
    \begin{subfigure}[t]{0.22\textwidth}
        \centering
\thesisdiagrampanel{\scalebox{0.28}{\begin{feynman}
    \electroweak[lineWidth=6, label=$\tilde{\psi}$]{5.60, 4.00}{5.60, 5.00}
    \fermion[lineWidth=6]{4.00, 6.40}{7.20, 6.40}
    \electroweak[label=$\tilde{\psi}$, lineWidth=6]{5.60, 5.40}{5.60, 6.40}
    \fermion[lineWidth=6]{4.00, 4.00}{7.20, 4.00}
    \parton{5.60,5.20}{0.20}
\end{feynman}
}}
        \caption{}
        \label{fig1a}
    \end{subfigure}
    \begin{subfigure}[t]{0.22\textwidth}
        \centering
\thesisdiagrampanel{\scalebox{0.36}{\begin{feynman}
    \electroweak[label=$\Tilde{\psi}$, lineWidth=4]{4.40, 4.00}{5.20, 5.60}
    \fermion[lineWidth=4]{4.00, 5.60}{6.40, 5.60}
    \electroweak[flip=true, label=$\Tilde{\psi}$, lineWidth=4]{6.00, 4.00}{5.20, 5.60}
    \fermion[lineWidth=4]{4.00, 4.00}{6.40, 4.00}
\end{feynman}}}
        \caption{}
        \label{fig1b}
    \end{subfigure}
    \begin{subfigure}[t]{0.22\textwidth}
        \centering
\thesisdiagrampanel{\scalebox{0.36}{\begin{feynman}
    \gluon[label=$\mathcal{A}_i$, color=fcb900, lineWidth=4]{5.20, 4.00}{5.20, 5.60}
    \fermion[lineWidth=4]{4.00, 5.60}{6.40, 5.60}
    \fermion[lineWidth=4]{4.00, 4.00}{6.40, 4.00}
\end{feynman}}}
        \caption{}
        \label{fig1c}
    \end{subfigure}
    \begin{subfigure}[t]{0.22\textwidth}
        \centering
\thesisdiagrampanel{\scalebox{0.36}{\begin{feynman}
    \fermion[lineWidth=4]{4.00, 5.60}{6.60, 5.60}
    \electroweak[lineWidth=4, label=$\Tilde{\psi}$]{5.20, 4.00}{5.20, 5.40}
    \fermion[lineWidth=4]{4.00, 4.00}{6.60, 4.00}
    \parton{5.20,5.60}{0.20}
\end{feynman}
}}
        \caption{}
        \label{fig1d}
    \end{subfigure}
    \caption{Diagrams contributing to the gravitational bound sector at 1PN. }
    \label{fig:my_labela}
\end{figure}
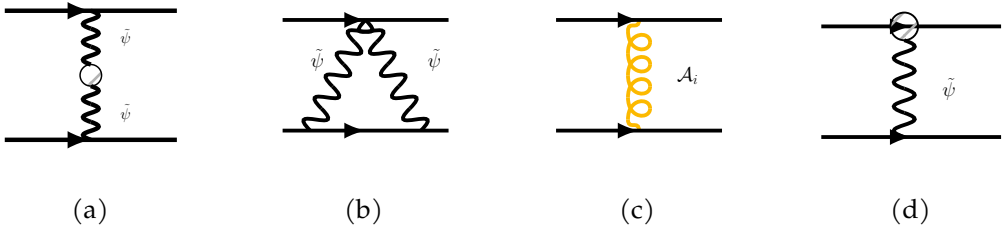
\noindent
The four graphs in Fig.~(\ref{fig:my_labela}) capture four distinct 1PN effects: the relativistic correction to the $\tilde{\psi}$ propagator, the quadratic worldline coupling $\tilde{\psi}^2$, exchange of the gravito-magnetic field $\hat{\mathcal{A}}_i$, and the velocity correction to the $\tilde{\psi}$ worldline vertex. The first graph is representative of the basic EFT step:
\begin{align}
    \begin{split}
   \mathcal{S}_{\text{eff}}\Big|_{\text{fig}(\ref{fig1a})}& =\frac{m_1 m_2}{m_p^2}\rmint \Bar{\mathcal{D}}\hat\xi\rmint dt_1\,\tilde{\psi}(\boldsymbol{x}_1
    (t_1))\rmint dt_2\,\tilde{\psi}(\boldsymbol{x}_2(t_2))
  \\ & =\frac{m_1m_2}{m_{p}^2}\rmint dt_1dt_2\Big{\langle}\tilde{\psi}(\boldsymbol{x}_1
    (t_1))\tilde{\psi}(\boldsymbol{x}_2(t_2))\Big{\rangle}_{\varoslash}\,.\\ &
    =\frac{ m_1m_2}{16 \pi m_p^2}\rmint dt\, \frac{(\boldsymbol{v}_1\cdot \boldsymbol{v}_2)-(\boldsymbol{v}_1\cdot \boldsymbol{r})(\boldsymbol{v}_2\cdot \boldsymbol{r})}{r}\sim \mathcal{O}(Lv^2)\,. \label{4.13a}
    \end{split} 
\end{align}
Here the passage from the first line to the second is simply the Wick contraction of the two potential gravitons. The remaining three diagrams use the same logic and reduce to
\begin{align}
    \mathcal{S}_{\text{eff}}\Big|_{\text{fig}(\ref{fig1b})}
    &=-\frac{m_1m_2^2}{128 \pi^4 m_p^4}\rmint \frac{ dt}{r^2}+(1\leftrightarrow 2)\,\sim \mathcal{O}(Lv^2)\,,\\
    \mathcal{S}_{\text{eff}}\Big|_{\text{fig}(\ref{fig1c})}
    &=-\frac{m_1m_2}{2\pi m_p^2}\rmint dt \,\frac{(\boldsymbol{v}_1\cdot \boldsymbol{v}_2)}{r}+(1\leftrightarrow 2)\,\sim \mathcal{O}(Lv^2)\,,\\
    \mathcal{S}_{\text{eff}}\Big|_{\text{fig}(\ref{fig1d})}
    &=-\frac{3  m_1m_2}{16\pi m_p^2}\rmint dt\, \frac{(v_1^2+v_2^2)}{r}\,\sim \mathcal{O}(Lv^2)\,.
\end{align}
Together, these four pieces reproduce the familiar 1PN gravitational correction to the conservative action.
\subsection{The Electromagnetic (EM) bound sector} \label{EM section}
Now, the EM and Proca sectors are left to be investigated. The EM bound sector consists of the following terms :
\begin{align}
    \begin{split}
  F_{\mu\nu}F^{\mu\nu}&=(\nabla_{\mu} A_{\nu}-\nabla_{\nu} A_{\mu})(\nabla^{\mu} A^{\nu}-\nabla^{\nu} A^{\mu})\\ &
=(\partial_0 A_i\partial_0 A_i-\partial_0 A_i\partial_iA_0+\partial_lA_i\partial_l A_i-\partial_l A_i\partial_iA_l-\partial_k A_0\partial_k A_0+\partial_kA_0\partial_0 A_k)\,.
      \end{split}
\end{align}

\begin{figure}
    \centering
    \begin{subfigure}[t]{0.22\textwidth}
        \centering
\thesisdiagrampanel{\scalebox{0.36}{\begin{feynman}
    \dashed[lineWidth=4, showArrow=false, label=$\mathcal{A}_{0}$]{5.20, 4.00}{5.20, 5.60}
    \fermion[lineWidth=4]{4.00, 4.00}{6.60, 4.00}
    \fermion[lineWidth=4]{4.00, 5.60}{6.60, 5.60}
\end{feynman}}}
        \caption{}
        \label{mfig4a}
    \end{subfigure}
      \begin{subfigure}[t]{0.22\textwidth}
        \centering
\thesisdiagrampanel{\scalebox{0.36}{\begin{feynman}
    \electroweak[lineWidth=4, label=$\mathcal{A}_{i}$, color=9900ef]{5.20, 4.00}{5.20, 5.60}
    \fermion[lineWidth=4]{4.00, 4.00}{6.60, 4.00}
    \fermion[lineWidth=4]{4.00, 5.60}{6.60, 5.60}
\end{feynman}}}
        \caption{}
        \label{mfig4b}
    \end{subfigure}
      \begin{subfigure}[t]{0.22\textwidth}
        \centering
\thesisdiagrampanel{\scalebox{0.34}{\begin{feynman}
    \dashed[showArrow=false, lineWidth=4, label=$\mathcal{A}_{0}$]{4.40, 4.00}{5.20, 5.60}
    \electroweak[lineWidth=4, label=$\Tilde{\psi}$]{6.20, 4.00}{5.20, 5.40}
    \fermion[lineWidth=4]{4.00, 4.00}{6.60, 4.00}
    \fermion[lineWidth=4]{4.00, 5.60}{6.60, 5.60}
\end{feynman}}}
        \caption{}
        \label{mfig4c}
    \end{subfigure}
    \begin{subfigure}[t]{0.22\textwidth}
        \centering
\thesisdiagrampanel{\scalebox{0.34}{\begin{feynman}
    \dashed[showArrow=false, lineWidth=4, label=$\mathcal{A}_{0}$]{5.20, 4.00}{5.20, 5.60}
    \fermion[lineWidth=4]{4.00, 4.00}{6.60, 4.00}
    \fermion[lineWidth=4]{4.00, 5.60}{6.60, 5.60}
    \parton{5.20,4.80}{0.20}
\end{feynman}}}
        \caption{}
        \label{mfig4d}
    \end{subfigure}
    \caption{Diagrams contributing to the electromagnetic bound sector up to 1PN.}
    \label{ch1:fig:em-bound}
\end{figure}
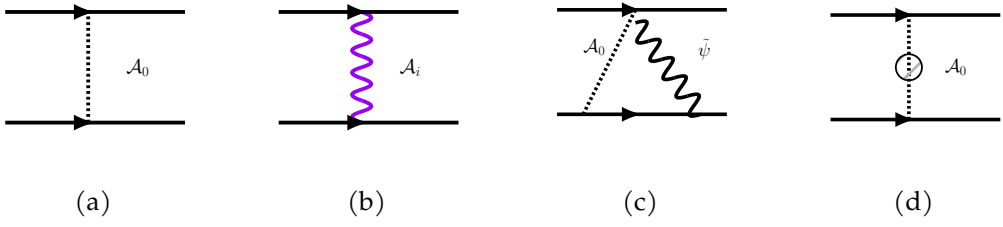
Remember that terms quadratic in a single field species belong to the propagator and do not appear again as independent interaction vertices. The gauge-fixing term $(\partial_\mu A^\mu)^2$ likewise does not contribute at 1PN order. The electromagnetic sector, therefore, contains one Coulomb exchange and three 1PN corrections. Since the Wick contractions are the same as in the gravitational example, it is enough to display the leading exchange explicitly,
\begin{align}
    \begin{split}
     \mathcal{S}_{\text{eff}}\Big|_{\text{fig}(\ref{mfig4a})}
     &=Q_1Q_2\rmint dt_1 dt_2\,\Big\langle\mathcal{A}_{0}(\boldsymbol{x}_1(t_1))\mathcal{A}_{0}(\boldsymbol{x}_2(t_2))\Big\rangle\\ &
     =-\frac{Q_1Q_2}{4\pi}\rmint \frac{dt}{r}+(1\leftrightarrow 2)\sim \mathcal{O}(Lv^0)\,.
    \end{split}
\end{align}
and summarize the remaining 1PN contributions as
\begin{align}
    \mathcal{S}_{\text{eff}}\Big|_{\text{fig}(\ref{mfig4b})}
    &=\frac{Q_1 Q_2}{4\pi}\rmint dt\, \frac{\boldsymbol{v}_1\cdot \boldsymbol{v}_2}{r}+(1\leftrightarrow 2)\sim \mathcal{O}(Lv^2)\,,\\
    \mathcal{S}_{\text{eff}}\Big|_{\text{fig}(\ref{mfig4c})}
    &=-\frac{m_2Q_1Q_2}{32\pi^2 m_p^2}\rmint \frac{dt}{r^2}+(1\leftrightarrow 2)\sim \mathcal{O}(Lv^2)\,,\\
    \mathcal{S}_{\text{eff}}\Big|_{\text{fig}(\ref{mfig4d})}
    &=\frac{Q_1Q_2}{8\pi}\rmint dt\,\Big[\frac{\boldsymbol{v}_1\cdot \boldsymbol{v}_2}{r}-\frac{(\boldsymbol{v}_1\cdot \Hat{n})(\boldsymbol{v}_2\cdot \Hat{n})}{r}\Big]\sim \mathcal{O}(Lv^2)\,.
\end{align}
The three 1PN pieces can be read respectively as the spatial-photon exchange, the graviton dressing of the Coulomb interaction, and the relativistic correction to the $\mathcal{A}_0$ propagator.
\subsection{The Proca  bound sector}
The Proca bound sector consists of the following term:
\begin{align}
    \begin{split}
  B_{\mu\nu}B^{\mu\nu}&=\frac{1}{2}(\nabla_\mu B_\nu-\nabla_\nu B_\mu)(\nabla^\mu B^\nu-\nabla^\nu B^\mu)\\ &
=(\partial_0B_i\partial_0B_i-\partial_0B_i\partial_iB_0+\partial_lB_i\partial_lB_i-\partial_lB_i\partial_iB_l-\partial_kB_0\partial_kB_0+\partial_kB_0\partial_0B_k)\,.
      \end{split}
\end{align}

We show the effective action for the Proca field (coupled to gravity) up to 1PN order below. There are a total of four contributing diagrams. The computation is similar to that of Sec.~(\ref{EM section}).
\begin{figure}
    \centering
    \begin{subfigure}[t]{0.22\textwidth}
        \centering
\thesisdiagrampanel{\scalebox{0.35}{\begin{feynman}
    \dashed[showArrow=false, lineWidth=4, label=$\mathcal{B}_{0}$, color=9900ef]{5.20, 4.00}{5.20, 5.60}
    \fermion[lineWidth=4]{4.00, 4.00}{6.60, 4.00}
    \fermion[lineWidth=4]{4.00, 5.60}{6.60, 5.60}
\end{feynman}}}
        \caption{}
        \label{mfig5a}
    \end{subfigure}
         \begin{subfigure}[t]{0.22\textwidth}
        \centering
\thesisdiagrampanel{\scalebox{0.35}{\begin{feynman}
    \fermion[lineWidth=4]{4.00, 4.00}{6.60, 4.00}
    \gluon[lineWidth=4, color=eb144c, label=$\mathcal{B}_i$]{5.20, 4.00}{5.20, 5.60}
    \fermion[lineWidth=4]{4.00, 5.60}{6.60, 5.60}
\end{feynman}}}
        \caption{}
        \label{mfig5b}
    \end{subfigure}
      \begin{subfigure}[t]{0.20\textwidth}
        \centering
\thesisdiagrampanel{\scalebox{0.35}{\begin{feynman}
    \electroweak[lineWidth=4, label=$\Tilde{\psi}$]{6.20, 4.00}{5.20, 5.60}
    \fermion[lineWidth=4]{4.00, 4.00}{6.60, 4.00}
    \dashed[lineWidth=4, showArrow=false, color=9900ef, label=$\mathcal{B}_{0}$]{4.40, 4.00}{5.20, 5.60}
    \fermion[lineWidth=4]{4.00, 5.60}{6.60, 5.60}
\end{feynman}}}
        \caption{}
        \label{mfig5c}
    \end{subfigure}
      \begin{subfigure}[t]{0.20\textwidth}
        \centering
\thesisdiagrampanel{\scalebox{0.35}{\begin{feynman}
    \fermion[lineWidth=4]{4.00, 4.00}{6.60, 4.00}
    \dashed[lineWidth=4, showArrow=false, color=9900ef, label=$\mathcal{B}_{0}$]{5.20, 4.00}{5.20, 5.60}
    \fermion[lineWidth=4]{4.00, 5.60}{6.60, 5.60}
    \parton{5.20,4.80}{0.20}
\end{feynman}}}
        \caption{}
        \label{mfig5d}
    \end{subfigure}
    \caption{Diagrams contributing to the Proca bound sector up to 1PN.}
    \label{ch1:fig:proca-bound}
\end{figure}
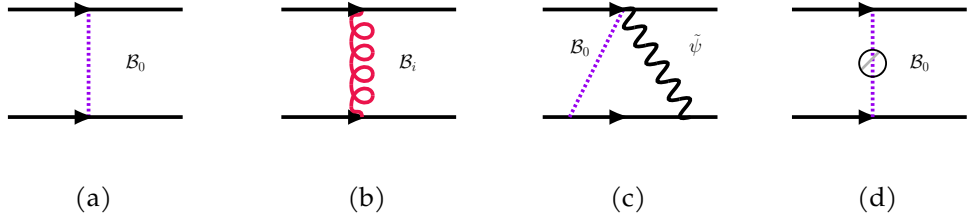
The Proca sector is structurally identical to the electromagnetic one, with the massless propagator simply replaced by the Yukawa propagator. It is therefore convenient to keep only the leading Yukawa exchange explicitly,
\begin{align}
    \begin{split}
     \mathcal{S}_{\text{eff}}\Big|_{\text{fig}(\ref{mfig5a})}
     &=Q_1'Q_2'\rmint dt_1 dt_2\,\Big\langle\mathcal{B}_{0}(\boldsymbol{x}_1(t_1))\mathcal{B}_{0}(\boldsymbol{x}_2(t_2))\Big\rangle\\ &
     =-\frac{Q_1'Q_2'}{4\pi}\rmint dt\,\frac{e^{-\mu_{\gamma}r}}{r}+(1\leftrightarrow 2)\sim \mathcal{O}(Lv^0)\,,
    \end{split}
\end{align}
while the remaining 1PN terms are
\begin{align}
    \mathcal{S}_{\text{eff}}\Big|_{\text{fig}(\ref{mfig5b})}
    &=\frac{Q_1' Q_2'}{4\pi}\rmint dt\, \frac{(\boldsymbol{v}_1\cdot \boldsymbol{v}_2)e^{-\mu_{\gamma}r}}{ r}+(1\leftrightarrow 2)\sim \mathcal{O}(Lv^2)\,,\\
    \mathcal{S}_{\text{eff}}\Big|_{\text{fig}(\ref{mfig5c})}
    &=-\frac{m_2Q_1'Q_2'}{32\pi^2 m_p^2}\rmint dt \,\frac{e^{-\mu_{\gamma}r}}{r^2}+(1\leftrightarrow 2)\sim \mathcal{O}(Lv^2)\,,\\
    \mathcal{S}_{\text{eff}}\Big|_{\text{fig}(\ref{mfig5d})}
    &=-\frac{Q_1'Q_2'}{8\pi}\rmint dt \,e^{-\mu_{\gamma}r}\Big[\frac{\boldsymbol{v}_1\cdot\boldsymbol{v}_2}{r}-\frac{(\boldsymbol{v}_1\cdot\Hat{n})(\boldsymbol{v}_2\cdot\Hat{n})}{r}(1+\mu_{\gamma}r)\Big]+(1\leftrightarrow 2)\sim\mathcal{O}(Lv^2)\,.
\end{align}
This makes the close analogy between the electromagnetic and Proca sectors manifest: the new ingredient is the Yukawa suppression and the extra $(1+\mu_{\gamma}r)$ structure that accompanies the longitudinal part of the massive propagator.

\begin{figure}[htb!]
    \centering 
    \begin{subfigure}[t]{0.2\textwidth}
   \thesisdiagrampanel{\scalebox{0.2}{ \begin{feynman}
    \electroweak[flip=true, lineWidth=8, label=$\mathcal{A}_{i}$, color=9900ef]{5.20, 4.00}{6.40, 5.80}
    \electroweak[lineWidth=8, label=$\mathcal{A}_{i}$, color=9900ef]{6.40, 5.80}{7.60, 4.00}
    \fermion[lineWidth=8]{4.00, 7.60}{9.40, 7.60}
    \electroweak[lineWidth=8, label=$\varphi$, color=0693e3]{6.40, 5.80}{6.40, 7.60}
    \fermion[lineWidth=8]{4.00, 4.00}{9.40, 4.00}
\end{feynman}}}
\caption{}
\label{fig4an}
\end{subfigure}
\begin{subfigure}[t]{0.2\textwidth}
    \centering
     \thesisdiagrampanel{\scalebox{0.2}{\begin{feynman}
    \electroweak[flip=true, lineWidth=8, label=$\mathcal{A}_{i}$, color=9900ef]{5.20, 4.00}{6.40, 5.80}
    \fermion[lineWidth=8]{4.00, 7.60}{9.40, 7.60}
    \electroweak[lineWidth=8, label=$\varphi$, color=0693e3]{6.40, 5.80}{6.40, 7.60}
    \fermion[lineWidth=8]{4.00, 4.00}{9.40, 4.00}
    \dashed[showArrow=false, lineWidth=8, label=$\mathcal{A}_0$]{6.40, 5.80}{7.60, 4.00}
\end{feynman}}}
    \caption{}
    \label{fig4cn}
\end{subfigure}
\begin{subfigure}[t]{0.2\textwidth}
    \centering 
\thesisdiagrampanel{\scalebox{0.2}{\begin{feynman}
    \electroweak[flip=true, lineWidth=8, label=$\mathcal{A}_{i}$, color=9900ef]{6.40, 4.00}{6.40, 5.80}
    \electroweak[lineWidth=8, label=$\mathcal{A}_i$, color=9900ef]{6.40, 5.80}{7.60, 7.60}
    \fermion[lineWidth=8]{4.00, 7.60}{9.40, 7.60}
    \electroweak[lineWidth=8, label=$\varphi$, color=0693e3]{6.40, 5.80}{5.20, 7.60}
    \fermion[lineWidth=8]{4.00, 4.00}{9.40, 4.00}
\end{feynman}
}}
\caption{}
\label{fig4bn}
\end{subfigure}
    \caption{Scalar-Electromagnetic interaction diagrams which contribute to the bound sector.}
    \label{figtheta1}
\end{figure}
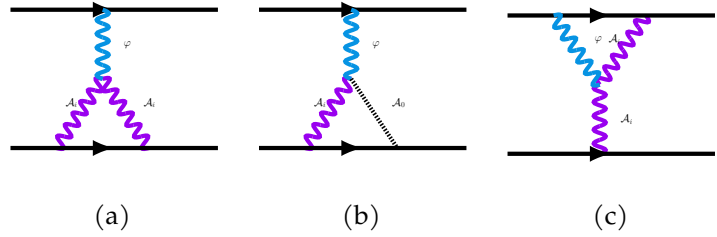

\subsection{The EM-Axion bound sector}\label{Sec4.2}
In this section, we mainly focus on the \textit{Theta term} as we want to investigate the effect of $g_{a\gamma\gamma}$ on the bound potential as well as the power radiation as discussed in Sec.~(\ref{sec5.5}). \textcolor{black}{The bulk interaction vertex which is relevant for this is the following,
\begin{equation}
    S_{int}^{\text{Theta}}=
\frac{g_{a\gamma\gamma}}{4m_p}\rmint d^4x \sqrt{-g}\,\phi\,F_{\mu\nu}\,\Tilde{F}^{\mu\nu}. 
\end{equation}
}
The contribution comes from this term at 2.5 PN order and has the following form\footnote{From the subsequent sections we replace $\hat\epsilon\rightarrow \epsilon$.},
\begin{equation}
    S_{int}^{\text{Theta}}\Big |_{2.5PN}= \frac{g_{a\gamma\gamma}}{m_p}\rmint d^4x [\hat\epsilon^{\,0ikm}\phi\,\partial_{0}\mathcal{A}_{i}\partial_{k}\mathcal{A}_{m}+ \hat\epsilon^{\,i0km}\phi\,\partial_{i}\mathcal{A}_{0}\partial_{k}\mathcal{A}_{m}]\,.
\end{equation}
 Three possible diagrams will contribute to this specific interaction. The contributing diagrams have the two-photon line and one dynamical scalar line as shown in Fig.~(\ref{figtheta1}).
\begin{itemize}
\item The first diagram that contributes has two electromagnetic propagators from the same worldline as shown in Fig.~(\ref{fig4an}) and Fig.~(\ref{fig4cn}) \textcolor{black}{with vertex factors $\phi\,\partial_{0}\mathcal{A}_{i}\partial_{k}\mathcal{A}_{m}$ and $\phi\,\partial_{i}\mathcal{A}_{0}\partial_{k}\mathcal{A}_{m}$ respectively}. The amplitude is given by, $  \mathcal{S}_{\text{eff}}=\frac{g_{a\gamma\gamma}}{m_p}Q_2^2\,m_1\,s_1\epsilon^{i0km}(  \mathcal{S}^{\text{eff}}_{ikm}+\Bar{  \mathcal{S}}^{\text{eff}}_{ikm}).$ The first and second terms are defined in (\ref{2.15}) and (\ref{4.18n})\,.
\begin{align}
   \begin{split}
      \mathcal{S}_{\text{eff}}^{ikm}\Big|_{\text{fig}(\ref{fig4an})}&=\rmint \Bar{\mathcal{D}}\hat{\xi}\rmint dt_2  \mathcal{A}_{j}(\boldsymbol{x}_2(t_2))v_{2}^{j}(t_2)\rmint dt_{3}  \mathcal{A}_{l}(\boldsymbol{x}_2(t_3))v_2^l(t_3)\rmint dt_1  \frac{\varphi(\boldsymbol{x}_1(t_1))}{m_p}\\ &
      \hspace{1.2 cm}\rmint d^4x\,  \varphi\,\partial_0  \mathcal{A}_{i}\partial_{k}\mathcal{A}_{m}(x)\,, \\ &
    =\frac{1}{m_p}\rmint \prod_{i=1}^{3}dt_i\,  dt\, v_{2}^{j}(t_2) v_2^l(t_3) \rmint d^3x\, \partial_{0}\Big\langle \mathcal{A}_{j}(\boldsymbol{x}_2(t_2))\mathcal{A}_{i}(x)\Big\rangle\,\partial_{k}\boldsymbol{\Big\langle}\mathcal{A}_{l}(\boldsymbol{x}_2(t_3))\mathcal{A}_{m}(x)\Big\rangle\,\\&\hspace{6.6cm}\Big\langle\varphi(\boldsymbol{x}_1(t_1))\varphi({x})\Big\rangle,\\ &
     =-\frac{i}{m_p}\rmint dt \,v_2^{m}(t)\Big[a_{2}^{i}(t)\rmint_{k_1,k_2}\frac{k_2^{k}}{\boldsymbol{k}_1^2\boldsymbol{k}_2^2[(\boldsymbol{k}_1+\boldsymbol{k}_2)^2+m^2]}e^{i(\boldsymbol{k}_1+\boldsymbol{k}_2)\cdot \boldsymbol{r}}\\ &
   \,\,\,\,\,\,\,\,\,\,\,\,\,\,  +v_2^{i}(t)v_2^{a}(t)\rmint_{k_1,k_2}\frac{i\,k_2^k\,k_1^a\,e^{i(\boldsymbol{k}_1+\boldsymbol{k}_2)\cdot \boldsymbol{r}}}{\boldsymbol{k}_1^2\boldsymbol{k}_2^2[(\boldsymbol{k}_1+\boldsymbol{k}_2)^2+m^2]}\Big]
   +(1\leftrightarrow 2)
   \label{2.15}
    \end{split}
\end{align}
and
\begin{align}
    \begin{split}
        \mathcal{ \Bar{S}_{\text{eff}}}^{ikm}\Big |_{\text{fig}(\ref{fig4cn})}&=-\rmint\Bar{\mathcal{D}}\hat{\xi}\rmint dt_2 \mathcal{A}_{0}(\boldsymbol{x}_2(t_2))\rmint dt_3 \mathcal{A}_{j}(\boldsymbol{x}_2(t_3))v^{j}(t_3)\rmint dt_1\,\varphi(\boldsymbol{x}_1(t_1))\\ &
        \hspace{4.3 cm}\rmint d^4x \,\varphi(x)\partial_{i}\mathcal{A}_{0}\partial_{k}\mathcal{A}_{m}\,,\\ &
        =-\rmint dt\, dt_1 dt_2 dt_3 \,v^{j}(t_3)\rmint d^3 x\,\partial_{i}\Big\langle \mathcal{A}_{0}(\boldsymbol{x}_2(t_2))\mathcal{A}_{0}(x)\Big\rangle \partial_{k}\Big\langle \mathcal{A}_{j}(\boldsymbol{x}_2(t_3))\mathcal{A}_{m}(x) \Big\rangle\\ &
       \hspace{5.8 cm} \Big\langle 
\varphi(\boldsymbol{x}_1(t_1))\varphi(x)\Big\rangle\,,\\&
        =\rmint dt \,v^{m}(t)\rmint_{\boldsymbol{k}_1,\boldsymbol{k}_2}\frac{k_1^i\,k_2^k\,e^{i(\boldsymbol{k}_1+\boldsymbol{k}_2)\cdot\boldsymbol{r}(t)}}{\boldsymbol{k}_1^2\boldsymbol{k}_2^2[(\boldsymbol{k}_1+\boldsymbol{k}_2)^2+m^2]}\,,\\ &
        =-\frac{1}{64}\rmint dt\, v^m(t)\Big[f(r)n^in^k+g(r)\delta^{ik}\Big]\,.\label{4.18n}
    \end{split}
\end{align}
Now due to the symmetric nature of $n^in^k$ and $\delta^{ik},$  $\mathcal{ \Bar{S}_{\text{eff}}}^{ikm}\Big |_{\text{eff}(\ref{fig4cn})}$ does not contribute to the amplitude after getting contracted with the $\epsilon_{i0km}.$
Therefore only $ \mathcal{S}_{\text{eff}}^{ikm}\Big|_{\text{fig}(\ref{fig4an})}$ contributes to the amplitude. The details of the computations are given in Appendix~(\ref{ch1:app:A}).
\begin{align}
    \begin{split}
          \mathcal{S}_{\text{eff}}\Big|_{\text{fig}(\ref{fig4an})}&=\frac{g_{a\gamma\gamma}}{m_{p}^2}\,Q_2^2 \,m_1\,s_1\,\epsilon^{i0km}\,  \mathcal{S}^{\text{eff}}_{ikm}\,,\\ &
= \frac{\pi^{3/2}\, g_{a\gamma\gamma}Q_2^2m_1s_1 }{8m_p^2}\Bigg[\rmint dt\, \Vec{a_2}\cdot(\hat{n}\times\boldsymbol{v}_2)\Bigg\{\frac{1}{2} \, m \,G_{1,3}^{2,1}\left(\frac{m^2 r^2}{4}\Big|
\begin{array}{c}
 -\frac{1}{2} \\
 -\frac{1}{2},-\frac{1}{2},0 \\
\end{array}
\right)\\&\hspace{5.3cm}-\frac{\, G_{1,3}^{2,1}\left(\frac{m^2 r^2}{4}\Big|
\begin{array}{c}
 \frac{1}{2} \\
 \frac{1}{2},\frac{1}{2},0 \\
\end{array}
\right)}{m\, r^2}\Bigg\}\Bigg] +1\leftrightarrow 2
\sim 
 \mathcal{O}(Lv^5).
\
\label{4.31}
\end{split}
\end{align}
Also we can easily check that, when one takes $m\rightarrow 0$ limit, (\ref{4.31}) smoothly goes to the following, 

\begin{align}
    \begin{split}
          \mathcal{S}_{\text{eff}}\Big|_{\text{fig}(\ref{fig4an})}&
        = \frac{ g_{a\gamma\gamma}\,Q_2^2\,m_1\,s_1\,}{32\pi^2\,m_p^2}\rmint dt\,\frac{\Vec{a_2}.(\hat{n}\times\boldsymbol{v}_2)}{r}+(1\leftrightarrow2)\sim \mathcal{O}(Lv^5)\,.
    \end{split}
\end{align}
\par

\item  There is another diagram that contributes  consists of two electromagnetic fields coupled with two different worldline vertices  $\rmint dt Q_a v^{i}_{a}\mathcal{A}_{i}.$ The corresponding \textcolor{black}{amplitude} is given by,
\begin{align}
    \begin{split}
\mathcal{S}_{\text{eff}}^{ikm}\Big |_{\text{fig}(\ref{fig4bn})}&=\frac{g_{a\gamma\gamma Q_1Q_2m_1s_1}}{m_p^2}{\rmint\Bar{\mathcal{D}}\Hat{\xi}\rmint dt_1\,v_1^{j}(t_1) \mathcal{A}_{j}(\boldsymbol{x}_1(t_1))\rmint dt_2\,\varphi(\boldsymbol{x}_1(t_2))\rmint dt_3 v_2^{l}(t_1)\mathcal{A}_{l}(\boldsymbol{x}_2(t_3))}\\ & \hspace{7 cm}
\rmint d^4x \,\varphi \partial_{0}  \mathcal{A}_{i}\partial_{k}\mathcal{A}_{m}(x)\,,\\ &
        =\frac{g_{a\gamma\gamma}}{m_p^2}\rmint dt\prod_{i=1}^{3}dt_{i}\,v_1^{j}(t_1)v_2^{l}(t_3)\rmint d^3x \partial_{t}\Big\langle \mathcal{A}_{j}(\boldsymbol{x}_1(t_1))\mathcal{A}_{i}(x)\Big\rangle\partial_{k}\Big\langle \mathcal{A}_{l}(\boldsymbol{x}_2(t_3))\mathcal{A}_{m}(x)\Big\rangle \\& \hspace{5cm} \Big\langle\varphi(\boldsymbol{x}_1(t_2))\varphi(x,t)\Big\rangle\,,\\ &
        =\frac{2ig_{a\gamma\gamma}Q_1Q_2m_1s_1}{m_p^2}\rmint dt v_2^{m}(t)a_1^{i}(t)\rmint_{k}\frac{k^k\,e^{i\boldsymbol{k}\cdot\boldsymbol{r}}}{\boldsymbol{k}^2}\textcolor{black}{\rmint_{k_3}\frac{1}{(\boldsymbol{k}-\boldsymbol{k}_3)^2(\boldsymbol{k}_3^2+m^2)}}\,,\\ &
  =\frac{ g_{a\gamma\gamma}Q_1Q_2m_1s_1}{2\pi m_p^2}\rmint dt \, v_2^m (t)a_1^i(t)\,\partial_r\alpha(m;r)n^k\,,\label{4.33mm}
    \end{split}
\end{align}
Therefore, the contracted amplitude contributing to the effective action is given by,
\begin{align}
    \begin{split}
        \mathcal{S}_{\text{eff}}\Big|_{\text{fig}(\ref{fig4bn})}=\frac{ g_{a\gamma\gamma}Q_1Q_2m_1s_1}{2\pi m_p^2}\rmint dt \,(\Vec{a}_1 \times \Hat{n})\cdot \boldsymbol{v}_2\, \partial_{r}\alpha(m;r)\sim \mathcal{O}(Lv^5).
    \end{split}
\end{align}
\end{itemize}
\textcolor{black}{where the function $\alpha(m;r)$ is defined in (\ref{C3m}) of Appendix(\ref{ch1:app:C}).}
\subsection{The pure scalar sector}\label{Sec4.3}
\textcolor{black}{The contributing terms in this case is simply given by expanding the 3rd and 4th term of the Lagrangian mentioned in (\ref{2.11}).}
The non-relativistic decomposition of the action is given by,
\begin{align}
    \begin{split}
         S[{\varphi}]_{\text{int}}&
        =\frac{1}{m_p}\rmint d^3x dt \,[-2\psi \partial_{0}\varphi\partial_{0}\varphi-2\Hat{\mathcal{A}}^{i}\partial_{i}\varphi\partial_{0}\varphi-\frac{1}{4}\sigma^{lm}\delta_{lm}\partial_{i}\varphi\partial_{i}\phi+\frac{1}{2}\sigma^{ij}\partial_{i}\varphi\partial_{j}\varphi+\psi m^2 \varphi^2]\,.\label{4.24}
    \end{split}
\end{align}
Note that all the interaction terms except the last one will contribute to the effective action at the order of $L\,v^4$, i.e., of 2PN order. The last term in (\ref{4.24}) corresponds to scalar-graviton interaction at 1PN. There are two diagrams contributing to the 1PN effective action that comes from the 3-point vertex as shown in Fig.~(\ref{ch1:fig:scalar-graviton-1pn}). Next, we give the details of the amplitudes corresponding to these diagrams.
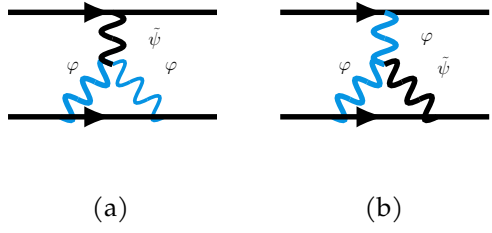
\begin{figure}[t!]
    \centering
    \begin{subfigure}[t]{0.22\textwidth}
\thesisdiagrampanel{\scalebox{0.34}{\begin{feynman}
    \electroweak[flip=true, label=$\varphi$, lineWidth=6, color=0693e3]{4.80, 4.00}{5.60, 4.80}
    \fermion[lineWidth=6]{4.00, 5.60}{7.20, 5.60}
    \electroweak[label=$\varphi$, lineWidth=4, color=0693e3]{5.60, 4.80}{6.40, 4.00}
    \electroweak[lineWidth=6, label=$\Tilde{\psi}$]{5.60, 4.80}{5.60, 5.60}
    \fermion[lineWidth=6]{4.00, 4.00}{7.20, 4.00}
\end{feynman}}}
        \caption{}
        \label{fig5aa}
    \end{subfigure}
    \begin{subfigure}[t]{0.22\textwidth}
        \centering
\thesisdiagrampanel{\scalebox{0.34}{\begin{feynman}
    \electroweak[flip=true, label=$\varphi$, lineWidth=6, color=0695e3]{4.80, 4.00}{5.60, 4.80}
    \fermion[lineWidth=6]{4.00, 5.60}{7.20, 5.60}
    \electroweak[lineWidth=6, label=$\varphi$, color=0693e3]{5.60, 4.80}{5.60, 5.60}
    \electroweak[lineWidth=6, label=$\Tilde{\psi}$]{5.60, 4.80}{6.40, 4.00}
    \fermion[lineWidth=6]{4.00, 4.00}{7.20, 4.00}
\end{feynman}
}}
        \caption{}
        \label{fig5bb}
    \end{subfigure}
    \caption{1PN diagrams for the scalar sector with 3 point vertex coming from scalar-graviton interaction.}
    \label{ch1:fig:scalar-graviton-1pn}
\end{figure}
\begin{itemize}
    \item The amplitude corresponds to the diagram in Fig.~(\ref{fig5aa}) with the bulk interaction vertex $\rmint d^4 x \,\Tilde{\psi}\varphi\varphi(x)$ where two scalar fields are contracted with the same worldline:
    \begin{align}
        \begin{split}
\mathcal{S}_{\text{eff}}\Big |_{\text{fig}(\ref{fig5aa})}&=\frac{m^2m_1 m_2^2 s_2^2}{m_{p}^4}\rmint\Bar{\mathcal{D}}\Hat{\xi}\rmint dt_1 \,\Tilde{\psi}(\boldsymbol{x}_1(t_1))\rmint dt_2 \,\varphi(\boldsymbol{x}_2(t_2))\rmint dt_3 \varphi(\boldsymbol{x}_2(t_3))\rmint d^4 x \,\Tilde{\psi}\varphi\varphi(x)\,,\\ &
            =\frac{m^2 m_1 m_2^2 s_2^2}{m_p^4} \rmint dt\prod_{i=1}^3 dt_i \Big\langle\Tilde{\psi}(\boldsymbol{x}_1(t_1))\Tilde{\psi}(x)\Big\rangle \Big\langle\varphi(\boldsymbol{x}_2(t_2))\varphi(x) \Big\rangle\Big\langle\varphi(\boldsymbol{x}_2(t_3))\varphi(x) \Big\rangle\,,\\ &
            =\frac{m^2 m_1 m_2^2 s_2^2}{2m_p^4} \rmint dt \rmint_{\boldsymbol{k},\boldsymbol{k}_1}\,\frac{e^{i\boldsymbol{k}\cdot \boldsymbol{r}}}{\boldsymbol{k}^2(\boldsymbol{k}_1^2+m^2)[(\boldsymbol{k}-\boldsymbol{k}_1)^2+m^2]}\,,\\ &
            =\frac{m^2 m_1 m_2^2 s_2^2}{8\pi m_p^4}\rmint dt \rmint \frac{e^{i\boldsymbol{k}\cdot \boldsymbol{r}}}{\boldsymbol{k}^3}\arctan(|\boldsymbol{k}|/2m)\,,\\&
            =\frac{m_1 m_2^2 s_2^2}{64\pi^2 m_p^4}\rmint dt \,\frac{m-2m^2r\,\text{Ei}(-2mr)-m\,e^{-2mr}}{r}+(1\leftrightarrow 2)\sim\mathcal{O}(Lv^2)\,.
            \label{4.25}
        \end{split}
    \end{align}
    In (\ref{4.25}), $\text{Ei}(z)=-\rmint_{-z}^{\infty}dt\,\frac{e^{-t}}{t}$ is the exponential integral function.
 \item The amplitude corresponds to the diagram in Fig.~(\ref{fig5bb}) with the bulk interaction vertex $\rmint d^4 x \,\Tilde{\psi}\varphi\varphi(x)$ where two scalar fields are contracted with two different worldlines:
    \begin{align}
        \begin{split}
            \mathcal{S}_{\text{eff}}\Big |_{\text{fig}(\ref{fig5bb})}&=\frac{m^2m_1 m_2^2 s_2^2}{m_{p}^4}\rmint\Bar{\mathcal{D}}\Hat{\xi}\rmint dt_1 \,\varphi(\boldsymbol{x}_1(t_1))\rmint dt_2 \,\Tilde{\psi}(\boldsymbol{x}_2(t_2))\rmint dt_3 \varphi(\boldsymbol{x}_2(t_3))\rmint d^4 x \,\Tilde{\psi}\varphi\varphi(x)\,,\\ &=\frac{m^2m_1s_1 m_2^2 s_2}{m_p^4} \rmint d^3x\,dt\prod_{i=1}^3 dt_i \Big\langle\Tilde{\psi}(\boldsymbol{x}_2(t_2))\Tilde{\psi}(x)\Big\rangle \Big\langle\varphi(\boldsymbol{x}_1(t_1))\varphi(x) \Big\rangle\Big\langle\varphi(\boldsymbol{x}_2(t_3))\varphi(x) \Big\rangle\,,\\ &
            =\frac{m^2m_1s_1 m_2^2 s_2}{m_p^4}\rmint_{k}\frac{e^{i\boldsymbol{k}\cdot \boldsymbol{r}}}{\boldsymbol{k}^2+m^2}\rmint_{k_1}\frac{1}{\boldsymbol{k}_1^2[(\boldsymbol{k}_1-\boldsymbol{k})^2+m^2]}\,,\\ &
           = \frac{m^2 m_1s_1 m_2^2 s_2}{4\pi m_p^4}\rmint dt \,\beta(m;r)+(1\leftrightarrow 2)\,\sim \mathcal{O}(Lv^2).\label{4.26}
        \end{split}
    \end{align}
    In (\ref{4.26}), $\beta(m;r)$ has no closed-form expression. Its behaviour is shown in Fig.~(\ref{lastfig}) in Appendix~(\ref{ch1:app:B}).
\end{itemize}


   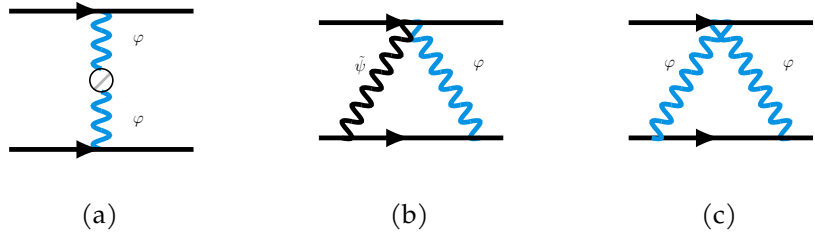
\begin{figure}[t!]
    \centering
    \begin{subfigure}[t]{0.25\textwidth}
       \centering
\thesisdiagrampanel{\scalebox{0.3}{\begin{feynman}
    \electroweak[label=$\varphi$, lineWidth=6, color=0693e3]{5.60, 4.00}{5.60, 5.00}
    \electroweak[label=$\varphi$, lineWidth=6, color=0693e3]{5.60, 6.40}{5.60, 5.40}
    \fermion[lineWidth=6]{4.00, 6.40}{7.20, 6.40}
    \fermion[lineWidth=6]{4.00, 4.00}{7.20, 4.00}
    \parton{5.60,5.20}{0.20}
\end{feynman}}}
        \caption{}
        \label{fig6ma}
    \end{subfigure}
    \begin{subfigure}[t]{0.25\textwidth}
        \centering
\thesisdiagrampanel{\scalebox{0.3}{\begin{feynman}
    \electroweak[lineWidth=6, color=0693e3, label=$\varphi$]{5.60, 6.00}{6.80, 4.00}
    \fermion[lineWidth=6]{4.00, 4.00}{7.20, 4.00}
    \electroweak[flip=true, lineWidth=6, label=$\Tilde{\psi}$]{4.40, 4.00}{5.60, 6.00}
    \fermion[lineWidth=6]{4.00, 6.00}{7.20, 6.00}
\end{feynman}}}
        \caption{}
        \label{fig6mb}
    \end{subfigure}
    \begin{subfigure}[t]{0.25\textwidth}
        \centering
\thesisdiagrampanel{\scalebox{0.3}{\begin{feynman}
    \electroweak[lineWidth=6, color=0693e3, label=$\varphi$]{5.60, 6.00}{6.80, 4.00}
    \fermion[lineWidth=6]{4.00, 4.00}{7.20, 4.00}
    \electroweak[flip=true, lineWidth=6, label=$\varphi$, color=0693e3]{4.40, 4.00}{5.60, 6.00}
    \fermion[lineWidth=6]{4.00, 6.00}{7.20, 6.00}
\end{feynman}}}
        \caption{}
        \label{fig6mc}
    \end{subfigure}
    \caption{1PN scalar diagrams that contribute to the bound sector.}
    \label{fig6m}
\end{figure}
There are three more diagrams contributing to 1PN coming purely from the worldline coupling operators, one is from the correction to the scalar propagator, and another is from the $\Tilde{\psi}\varphi$ coupling at any of the two worldlines. 
\begin{itemize}
\item The amplitude corresponds to the diagram in Fig.~(\ref{fig6ma}) with worldline vertices $\rmint dt\,\varphi$ and with the corrected scalar propagator:
\begin{align}
\begin{split}
    \mathcal{S}_{\text{eff}}\Big|_{\text{fig}(\ref{fig6ma})}&=\frac{m_1s_1m_2s_2}{m_{p}^2}\rmint\Bar{\mathcal{D}}\Hat{\xi}\rmint dt_1 \,\varphi(\boldsymbol{x}_1(t_1)) \rmint dt_2 \,\varphi(\boldsymbol{x}_2(t_2)),\\ &
    =\frac{m_1s_1m_2s_2}{m_p^2}\rmint dt_1 dt_2\,\Big\langle\varphi(\boldsymbol{x}_1(t_1))\varphi(\boldsymbol{x}_2(t_2))\Big\rangle_{\varoslash}\,,\\ &
    =\frac{m_1s_1m_2s_2}{m_p^2}\rmint dt_1 dt_2\, \partial_{t_1}\partial_{t_2}\delta(t_1-t_2)\rmint_{k}\frac{e^{i\boldsymbol{k}\cdot(\boldsymbol{x}_1-\boldsymbol{x}_{2})}}{(\boldsymbol{k}^2+m^2)^2}\,,\\ &
    =\frac{m_1s_1m_2s_2}{8\pi m_p^2}\rmint dt \,e^{-mr}\Big[\frac{\boldsymbol{v}_1\cdot\boldsymbol{v}_2}{r}-\frac{(\boldsymbol{v}_1\cdot\Hat{n})(\boldsymbol{v}_2\cdot\Hat{n})}{r}(1+mr)\Big]\,\sim \mathcal{O}(Lv^2).
    \end{split}
\end{align}
\item The amplitude corresponds to the diagram as shown in Fig.~(\ref{fig6mb}) with the worldline vertices $\rmint dt \,\Tilde{\psi}\varphi,$  $\rmint dt \,\Tilde{\psi}$ and  $\rmint dt \,\varphi$ : 
\begin{align}
    \begin{split}
        \mathcal{S}_{\text{eff}}\Big |_{\text{fig}(\ref{fig6mb})}&=\frac{m_1s_1m_2^2s_2}{m_{p}^4}\rmint\Bar{\mathcal{D}}\Hat{\xi}\rmint dt_1 \,\varphi(\boldsymbol{x}_1(t_1))\Tilde{\psi}(\boldsymbol{x}_1(t_1)) \rmint dt_2 \,\varphi(\boldsymbol{x}_2(t_2))\rmint dt_3 \Tilde{\psi}(\boldsymbol{x}_2(t_3)) \,,\\ &=\frac{m_1 s_1 m_2^2s_2}{m_p^4}\rmint dt_1 dt_2 dt_3 \Big\langle\varphi(\boldsymbol{x}_1(t_1))\varphi(\boldsymbol{x}_2(t_2))\Big\rangle \Big\langle\Tilde{\psi}(\boldsymbol{x}_1(t_1))\Tilde{\psi}(\boldsymbol{x}_2(t_3))\Big\rangle\,,\\ &
        =\frac{m_1 s_1 m_2^2s_2}{2 m_p^4}\rmint dt \rmint_{k_1}\frac{e^{i\boldsymbol{k}_1\cdot\boldsymbol{r}}}{\boldsymbol{k}_1^2+m^2}\rmint_{k_2}\frac{e^{i\boldsymbol{k}_2\cdot\boldsymbol{r}}}{\boldsymbol{k}_2^2}\,,\\ &
        =\frac{m_1 s_1 m_2^2s_2}{32 \pi^2 m_p^4}\rmint dt \frac{e^{-mr}}{r^2}+(1\leftrightarrow 2)\,\,\sim \mathcal{O}(Lv^2).
    \end{split}
\end{align}
\item The amplitude corresponds to the diagram in Fig.~(\ref{fig6mc}) with the worldline vertices $\rmint dt\, \varphi\varphi$ and $\rmint dt \,\varphi$ : 
\begin{align}
    \begin{split}
\mathcal{S}_{\text{eff}}\Big |_{\text{fig}(\ref{fig6mc})}&=\frac{m_1g_1m_2^2s_2^2}{m_{p}^4}\rmint\Bar{\mathcal{D}}\Hat{\xi}\rmint dt_1 \,\varphi(\boldsymbol{x}_1(t_1))\varphi(\boldsymbol{x}_1(t_1)) \rmint dt_2 \,\varphi(\boldsymbol{x}_2(t_2))\rmint dt_3 \varphi(\boldsymbol{x}_2(t_3)) \,,\\ & =\frac{m_1 g_1 m_2^2 s_2^2}{m_p^4}\rmint dt_1 dt_2 dt_3 \Big\langle\varphi(\boldsymbol{x}_1(t_1))\varphi(\boldsymbol{x}_2(t_2))\Big\rangle \Big\langle\varphi(\boldsymbol{x}_1(t_1))\varphi(\boldsymbol{x}_2(t_3))\Big\rangle\,,\\ &
         =\frac{m_1 g_1 m_2^2 s_2^2}{16\pi^2 m_p^4}\rmint dt \frac{e^{-2mr}}{r^2}+(1\leftrightarrow 2)\,\,\sim \mathcal{O}(Lv^2).
    \end{split}
\end{align}
\end{itemize}
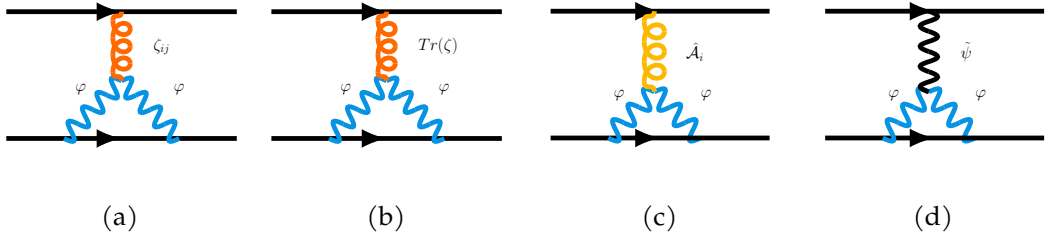
\begin{figure}
\centering
\begin{subfigure}[t]{.22\textwidth}
  \centering
  \thesisdiagrampanel{\scalebox{0.3}{\begin{feynman}
    \fermion[lineWidth=6]{4.00, 6.20}{8.00, 6.20}
    \fermion[lineWidth=6]{4.00, 4.00}{8.00, 4.00}
    \electroweak[lineWidth=6, color=0693e3, label=$\varphi$]{6.00, 5.00}{7.00, 4.00}
    \electroweak[flip=true, lineWidth=6, color=0693e3, label=$\varphi$]{5.00, 4.00}{6.00, 5.00}
    \gluon[lineWidth=6, color=ff6900, label=$\zeta_{ij}$]{6.00, 5.00}{6.00, 6.20}
\end{feynman}}}
  \caption{}
  \label{fig3a}
\end{subfigure}%
\begin{subfigure}[t]{.22\textwidth}
  \centering
   \thesisdiagrampanel{\scalebox{0.3}{\begin{feynman}
    \fermion[lineWidth=6]{4.00, 6.20}{8.00, 6.20}
    \fermion[lineWidth=6]{4.00, 4.00}{8.00, 4.00}
    \electroweak[lineWidth=6, color=0693e3, label=$\varphi$]{6.00, 5.00}{7.00, 4.00}
    \electroweak[flip=true, lineWidth=6, color=0693e3, label=$\varphi$]{5.00, 4.00}{6.00, 5.00}
    \gluon[lineWidth=6, color=ff6900, label=$Tr(\zeta)$]{6.00, 5.00}{6.00, 6.20}
\end{feynman}
}}
  \caption{}
  \label{fig3b}
\end{subfigure}
\begin{subfigure}[t]{.22\textwidth}
\centering
\thesisdiagrampanel{\scalebox{0.3}{\begin{feynman}
    \electroweak[lineWidth=6, color=0693e3, label=$\varphi$]{5.80, 4.80}{6.60, 4.00}
    \fermion[lineWidth=6]{4.00, 6.20}{7.80, 6.20}
    \fermion[lineWidth=6]{4.00, 4.00}{7.80, 4.00}
    \electroweak[flip=true, lineWidth=6, color=0693e3, label=$\varphi$]{5.00, 4.00}{5.80, 4.80}
    \gluon[lineWidth=6, label=$\Hat{\mathcal{A}}_{i}$, color=fcb900]{5.80, 4.80}{5.80, 6.20}
\end{feynman}
}}
    \caption{}
  \label{fig3c}
\end{subfigure}
\begin{subfigure}[t]{0.22\textwidth}
  \centering
  \thesisdiagrampanel{\scalebox{0.3}{\begin{feynman}
    \electroweak[lineWidth=6, color=0693e3, label=$\varphi$]{5.80, 4.80}{6.60, 4.00}
    \fermion[lineWidth=6]{4.00, 6.20}{7.80, 6.20}
    \fermion[lineWidth=6]{4.00, 4.00}{7.80, 4.00}
    \electroweak[flip=true, lineWidth=6, color=0693e3, label=$\varphi$]{5.00, 4.00}{5.80, 4.80}
    \electroweak[lineWidth=6, label=$\Tilde{\psi}$]{5.80, 4.80}{5.80, 6.20}
\end{feynman}}}
\caption{}
\label{fig3d}
\end{subfigure}
\caption{Scalar diagrams coming from the 3-point vertices that contribute at 2 PN order in scalar effective action.}
\label{ch1:fig:scalar-2pn}
\end{figure}
\textcolor{black}{For the scalar sector, we find the effective action up to 2PN order. At 2PN order, corrections appear due to the scalar-graviton interaction vertices shown in Fig.~(\ref{ch1:fig:scalar-2pn}).} Below we list their contributions.
\begin{itemize}
\item The amplitude corresponds to the diagram in Fig.~(\ref{fig3a}) with the following bulk interaction vertex $\rmint d^4 x\frac{1}{2}\boldsymbol{\zeta}^{lm}\partial_{l}\varphi\partial_{m}\varphi(x)$:
\begin{align}
    \begin{split}
        \mathcal{S}_{\text{eff}}\Big|_{\text{fig}(\ref{fig3a})}&=-\frac{m_1}{2m_p^4}\rmint\bar {\mathcal{D}}{\hat\xi}\,\rmint dt_1 \boldsymbol{\zeta}_{ij}(\boldsymbol{x}_1(t_1))v_1^i v_1^j\rmint dt_2 m_2 s_2 \varphi(\boldsymbol{x}_2(t_2))\rmint dt_2 m_2 s_2 \varphi(\boldsymbol{x}_2(t_3))\\&\hspace{5cm}\rmint d^3x \,dt\frac{1}{2}\boldsymbol{\zeta}^{lm}\partial_{l}\varphi\partial_{m}\varphi(x)\,,\\ &
        =-\frac{ m_1 m_2^2 s_2^2}{4m_p^4}\rmint \prod_{i=1}^3 dt_i dt \,v_1^{i}(t_1)v_1^{j}(t_1)\rmint d^3x \,\partial_{l}\Big\langle\varphi(\boldsymbol{x}_2(t_2))\varphi(x)\Big\rangle \partial_{m}\Big\langle\varphi(\boldsymbol{x}_2(t_3))\varphi(x)\Big\rangle \\& \hspace{8.5cm}
        \Big\langle\boldsymbol{\zeta}_{ij}(\boldsymbol{x}_1(t_1))\boldsymbol{\zeta}^{lm}(x)\Big\rangle\,,\\ &
        =-\frac{2 m_1m_2^2 s_2^2}{m_p^4}\rmint dt\, v_1^{i}(t)v_1^{j}(t)\rmint_{\boldsymbol{k},\boldsymbol{k}_1}\frac{e^{i\boldsymbol{k}\cdot \boldsymbol{r}}k_{1l}(k_{1m}-k_m)}{(\boldsymbol{k}_1^2+m^2)[(\boldsymbol{k}-\boldsymbol{k}_1)^2+m^2]\boldsymbol{k}^2}\\&\,\,\,\,\,\,\,\,\,\,\,\,\,\,\,\,\,\,\,\,\,\,\,\,\,\,\,\,\,\,\,\,\,\,\,\,\,\,\,\,\,\,\,\,\,\,\,\,\,\,\,\,\,\,\,\,\,\,\,\,\,\,\,\,\, \,\,\,\,\,\,\,\,\,\,\,\,\,\,\,\,\,\,\,\,\,\,\,\,\,\,\,\,\,\,\,\,\,\,\,\,\,\,\,\,\,\,\,\,\,\,\,\,\,\,\,\,\,\,\,\,\,\,\,\,\,\,\,\,\,\,\,\,\,\,\,\,\,\,\,\,(\delta_{i}^{l}\delta_j^m+\delta_{i}^m\delta_j^l-2\delta_{ij}\delta^{lm})\,,\\ &
        =-\frac{4m_1 m_2^2 s_2^2}{m_p^4}\rmint dt \Big[(-2\lambda_1+2\chi_1-\lambda_2+\chi_2)v_1^2+(\lambda_2-\chi_2)(\boldsymbol{v}_1\cdot\Hat{n})^2\Big]\\&\,\,\,\,\,\,\,\,\,\,\,\,\,\,\,\,\,\,\,\,\,\,\,\,\,\,\,\,\,\,\,\,\,\,\,\,\,\ +(1\leftrightarrow2)\sim \mathcal{O}(Lv^4)\,.
        \label{4.38}
    \end{split}
\end{align}
\vspace{-0.4cm}
The calculation and the functions $\lambda_{1,2}(r)$, $\chi_{1,2}(r)$ are explicitly given in Appendix~ (\ref{ch1:app:B}).
\vspace{0.2cm}
\item The amplitude corresponds to the diagram in Fig.~(\ref{fig3b}) with the following bulk interaction vertex $\rmint d^4x\,\frac{1}{4}\boldsymbol{\zeta}^{lm}\delta_{lm}\partial_{k}\varphi\partial_{k}\varphi(x)$:
\begin{align}
    \begin{split}
          \mathcal{S}_{\text{eff}}\Big |_{\text{fig}(\ref{fig3b})}&=-\frac{m_1}{2m_p^4}\rmint\Bar{\mathcal{D}}\Hat{\xi}\rmint dt_1 \boldsymbol{\zeta}_{ij}(\boldsymbol{x}_1(t_1))v_1^i v_1^j\rmint dt_2 m_2 s_2 \varphi(\boldsymbol{x}_2(t_2))\rmint dt_2 m_2 s_2 \varphi(\boldsymbol{x}_2(t_3))\\&\hspace{5cm}\rmint d^3x \,dt\frac{1}{4}\boldsymbol{\zeta}^{lm}\delta_{lm}\partial_{k}\varphi\partial_{k}\varphi(x)\,,\\ &
        =-\frac{m_1 m_2^2 s_2^2}{8 m_p^4}\rmint dt_1 dt_2 dt_3 dt v_{1}^i(t_1)v_1^j(t_1) \rmint d^3x \,\partial_{k}\Big\langle\varphi(\boldsymbol{x}_2(t_2))\varphi(x)\Big\rangle \partial_{k}\Big\langle\varphi(\boldsymbol{x}_2(t_3))\varphi(x)\Big\rangle
       \\&\hspace{8.5cm} \Big\langle\boldsymbol{\zeta}_{ij}(\boldsymbol{x}_1(t_1))\boldsymbol{\zeta}^{lm}(x)\Big\rangle \delta_{lm}\,,\\ &
        =-\frac{m_1 m_2^2 s_2^2}{ m_p^4}\rmint dt v_1^{i}(t)v_1^j(t)\rmint_{k,k_1}\frac{e^{i\boldsymbol{k}\cdot \boldsymbol{r}}(\boldsymbol{k}_1^2-\boldsymbol{k}\cdot \boldsymbol{k}_1)}{(\boldsymbol{k}_1^2+m^2)[(\boldsymbol{k}_1-\boldsymbol{k})^2+m^2]\boldsymbol{k}^2}\delta_{ij}\,,\\ &
        =-\frac{m_1 m_2^2 s_2^2}{m_p^4}\rmint dt\, v_1^2\Big(3(\lambda_1-\chi_1)+(\lambda_2-\chi_2)\Big)+(1\leftrightarrow 2)\sim \mathcal{O}(Lv^4)\,.
    \end{split}
\end{align}
Again, all the details of the integrals are given in Appendix~(\ref{ch1:app:B}). The functions $\lambda_{1,2},\chi_{1,2}$ are defined in  (\ref{B.5m}) of Appendix~(\ref{ch1:app:B}).
\item The amplitude corresponds to the diagram in Fig.~(\ref{fig3c}) with the bulk interaction vertex $\rmint d^4x\,\Hat{\mathcal{A}}^{k}\partial_{k}\varphi\partial_{0}\varphi(x)$:
\begin{align}
    \begin{split}
          \mathcal{S}_{\text{eff}}\Big|_{\text{fig}(\ref{fig3c})}&=2m_1\rmint\Bar{\mathcal{D}}\Hat{\xi}\rmint dt_1\, \frac{\Hat{\mathcal{A}}_{i}(\boldsymbol{x}_1(t_1))}{m_{p}}v_1^{i}(t_1)\rmint dt_2\,\frac{m_2 s_2}{m_p}\varphi(\boldsymbol{x}_2(t_2))\rmint dt_3\,\frac{m_2 s_2}{m_p}\varphi(\boldsymbol{x}_2(t_3))\\&\hspace{5cm}\rmint d^3x dt \,\frac{\Hat{\mathcal{A}}^{k}}{m_p}\partial_{k}\varphi\partial_{0}\varphi(x)\,,\\ &
        =\frac{2m_1 m_2^2 s_2^2}{m_p^4}\rmint dt dt_1 dt_2 dt_3 v_1^i(t_1)\rmint d^3x \Big\langle\Hat{\mathcal{A}}_{i}(\boldsymbol{x}_1(t_1))\mathcal{\Hat{A}}^{k}(x)\Big\rangle \partial_{k}\Big\langle\varphi(\boldsymbol{x}_2(t_2))\varphi(x)\Big\rangle \\&\hspace{9.5cm}\partial_{0}\Big\langle\varphi(\boldsymbol{x}_2(t_3))\varphi(x  )\Big\rangle\,,\\ &
        =-\frac{4m_2 m_1^2 s_1^2}{m_p^4}\rmint dt v_1^i(t)v_2^{l}(t)\rmint_{k_2,k_3}\frac{k_2^i\,k_3^l\,e^{-i(\boldsymbol{k}_2+\boldsymbol{k}_3)\cdot \boldsymbol{r}}}{(\boldsymbol{k}_2+\boldsymbol{k}_3)^2(\boldsymbol{k}_2^2+m^2)(\boldsymbol{k}_3^2+m^2)}\,,\\ &
        =-\frac{4m_2 m_1^2s_1^2}{m_p^4}\rmint dt[(\lambda_1+\chi_1)\boldsymbol{v}_1\cdot\boldsymbol{v}_2+(\lambda_2+\chi_2)(\boldsymbol{v}_1\cdot \Hat{n})(\boldsymbol{v}_2\cdot \Hat{n})]+(1\leftrightarrow 2)\sim \mathcal{O}(L v^4)\,.
    \end{split}
\end{align}
Again all the  details of the computation and the functions $\lambda_{1,2},\chi_{1,2}$ are given in Appendix~(\ref{ch1:app:B}).
\item The amplitude corresponds to the diagram in Fig.~(\ref{fig3d}) with the bulk interaction vertex $\rmint d^4x\, \frac{\Tilde{\psi}}{m_p}\partial_{0}\varphi\partial_{0}\varphi$:
\begin{align}
    \begin{split}
          \mathcal{S}_{\text{eff}}\Big |_{\text{fig}(\ref{fig3d})}&=2m_1\rmint\Bar{\mathcal{D}}\Hat{\xi}\rmint dt_1 \frac{\Tilde{\psi}(\boldsymbol{x}_1(t_1))}{m_p}\rmint dt_2 \frac{m_2 s_2}{m_p}\varphi(\boldsymbol{x}_2(t_2))\rmint dt_3 \frac{m_2 s_2}{m_p}\varphi(\boldsymbol{x}_2(t_3))\\& \,\,\,\,\,\,\,\,\,\,\,\,\,\,\,\,\,\,\,\,\,\,\,\,\,\,\,\,\,\,\,\,\,\,\,\,\,\,\,\,\,\,\,\,\,\,\,\,\,\,\,\,\,\,\,\,\,\,\,\,\,\,\,\,\,\,\,\,\,\,\,\,\,\,\,\,\,\,\,\,\,\,\,\,\,\,\,\,\,\,\,\,\,\,\,\,\,\,\,\,\,\,\,\,\,\,\,\,\,\,\,\,\,\,\,\,\,\,\,\,\,\,\,\, \rmint d^3x\,dt\, \frac{\Tilde{\psi}}{m_p}\partial_{0}\varphi\partial_{0}\varphi\,.\\ &
        =\textstyle{\frac{2m_1 m_2^2 s_2^2}{m_p^4}\rmint dt_1 dt_2 dt_3 dt\rmint d^3x \Big\langle\Tilde{\psi}(\boldsymbol{x}_1(t_1))\Tilde{\psi}(x)\Big\rangle\partial_{0}\Big\langle\varphi(\boldsymbol{x}_2(t_2))\varphi(x)\Big\rangle \partial_{0}\Big\langle\varphi(\boldsymbol{x}_2(t_3))\varphi(x  )\Big\rangle}\,,\\ &
        =\frac{m_1 m_2^2 s_2^2}{m_p^4}\rmint dt \, v_2^a(t)v_2^b(t)\rmint_{k,k_1}\frac{k_{1a}(k_{1b}-k_b)e^{i\boldsymbol{k}\cdot \boldsymbol{r}}}{\boldsymbol{k}^2(\boldsymbol{k}_1^2+m^2)[(\boldsymbol{k}-\boldsymbol{k}_1)^2+m^2]}\,,\\ &
        =\frac{m_1 m_2^2 s_2^2}{m_p^4}\rmint dt \, [v_2^2(\lambda_1-\chi_1)+(\boldsymbol{v}_2\cdot \Hat{n})^2(\lambda_2-\chi_2)]+(1\leftrightarrow 2)\sim \mathcal{O}(Lv^4)\,.
    \end{split}
\end{align}
\end{itemize}
At 2PN there are several more terms that are of the order $\mathcal{O}(Gv^4)$. Below we list out the contribution of those diagrams to the effective action. 
\begin{figure}[t!]
    \centering
    \begin{subfigure}[t]{0.25\textwidth}
        \centering
        \thesisdiagrampanel{\scalebox{0.3}{\begin{feynman}
    \electroweak[lineWidth=4, label=$\varphi$, color=0693e3]{5.80, 4.00}{5.80, 6.20}
    \fermion[lineWidth=4, label=$v^4$]{4.00, 6.20}{7.60, 6.20}
    \fermion[lineWidth=4, label=$v^0$]{4.00, 4.00}{7.60, 4.00}
\end{feynman}}}
\caption{}
\label{fig4a}
    \end{subfigure}
    \begin{subfigure}[t]{0.25\textwidth}
        \centering
        \thesisdiagrampanel{\scalebox{0.3}{\begin{feynman}
    \electroweak[lineWidth=4, label=$\varphi$, color=0693e3]{5.80, 4.00}{5.80, 6.20}
    \fermion[lineWidth=4, label=$v^2$]{4.00, 6.20}{7.60, 6.20}
    \fermion[lineWidth=4, label=$v^2$]{4.00, 4.00}{7.60, 4.00}
\end{feynman}}}
\caption{}
\label{fig4b}
    \end{subfigure}
    \caption{Diagrams contributing to scalar bound sector at 2 PN order coming from worldline vertices. }
    \label{fig4}
\end{figure}
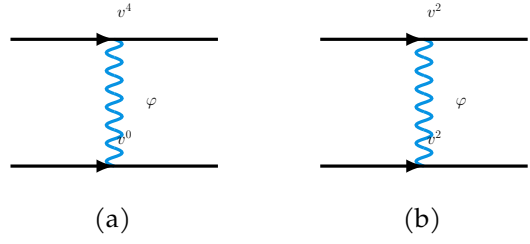
\begin{itemize}
    \item One term that contributes consists of the worldline vertex $\frac{1}{8}m_1 s_1\rmint dt \,v^4\frac{\varphi}{m_p}$ as shown in the Fig.~(\ref{fig4a}). The corresponding amplitude is given by,
    \begin{align}
        \begin{split}
              \mathcal{S}_{\text{eff}}\Big|_{\text{fig}(\ref{fig4a})}&=-\frac{m_1 s_1 m_2 s_2 }{8m_p^2}\rmint\Bar{\mathcal{D}}\Hat{\xi}\rmint dt_1\,v_1^4(t_1) \varphi(\boldsymbol{x}_1(t_1))\rmint dt_2 \varphi(\boldsymbol{x}_2(t_2))\,,\\ &
            =-\frac{m_1 s_1 m_2 s_2 }{8m_p^2}\rmint dt_1 \,dt_2\,v_1^4(t_1)\Big\langle\varphi(\boldsymbol{x}_1(t_1))\varphi(\boldsymbol{x}_2(t_2))\Big\rangle\,,\\ &
            =-\frac{m_1 s_1 m_2 s_2}{32\pi m_p^2}\rmint dt \,v_1^4(t)\,\frac{e^{-mr}}{r}+(1\leftrightarrow2)\sim \mathcal{O}(Lv^4)\,.
        \end{split}
    \end{align}
    \item Another term that contributes at this order consists of the worldline vertex vertex:$\rmint dt \frac{1}{2m_p}v^2\,m_a s_a \varphi$ as shown in Fig.~(\ref{fig4b}). The corresponding amplitude is given by,
    \begin{align}
        \begin{split}
              \mathcal{S}_{\text{eff}}\Big|_{\text{fig}(\ref{fig4b})}&=\frac{m_1 s_1 m_2 s_2}{4m_p^2}\rmint\Bar{\mathcal{D}}\Hat{\xi}\rmint dt_1 v_1^2(t_1)\varphi(\boldsymbol{x}_1(t_1))\rmint dt_2 v_2^2(t_2)\varphi(\boldsymbol{x}_2(t_2))\,,\\ &
            =\frac{m_1 s_1 m_2 s_2}{4m_p^2}\rmint dt_1 \, dt_2 v_1^2(t_1)v_2^2(t_2)\Big\langle\varphi(\boldsymbol{x}_1(t_1))\varphi(\boldsymbol{x}_2(t_2))\Big\rangle\,,\\ &
            =\frac{m_1 s_1 m_2 s_2}{16 \pi m_p^2}\rmint dt\,v_1^2(t)v_2^2(t)\frac{e^{-mr}}{r}+(1\leftrightarrow 2)\,\sim \mathcal{O}(Lv^4)\,.
        \end{split}
    \end{align}
\end{itemize} 
\subsection{Proca-electromagnetic interaction sector}
Now we focus on the Proca-electromagnetic interaction sector. This sector contains an interaction term of the form $F_{\mu\nu}B^{\mu\nu}$. This term can be expanded \textcolor{black}{up to 1PN order }in the following fashion.
\begin{align}
    \begin{split}
  \frac{\gamma}{2} F_{\mu\nu}B^{\mu\nu}&=\frac{\gamma}{2}(\nabla_\mu A_\nu-\nabla_\nu A_\mu)(\nabla^\mu B^\nu-\nabla^\nu B^\mu)\\ &
=\gamma(\partial_0A_i\partial_0B_i-\partial_0A_i\partial_iB_0+\partial_lA_i\partial_lB_i-\partial_lA_i\partial_iB_l-\partial_kA_0\partial_kB_0+\partial_kA_0\partial_0B_k)\,.\label{4.35}
      \end{split}
\end{align}
In the action mentioned in (\ref{4.35}), the two fields are different (although they are quadratic), and they provide us with two-point interaction vertices. 
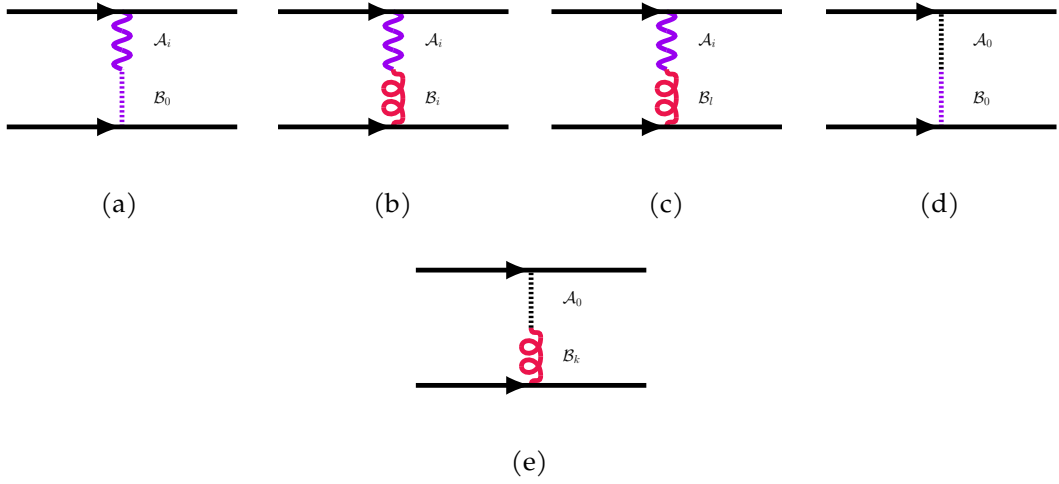
\begin{figure}[t!]
    \centering
    \begin{subfigure}[t]{0.22\textwidth}
    \centering
    \thesisdiagrampanel{\scalebox{0.3}{ \begin{feynman}
    \electroweak[lineWidth=6, color=9900ef, label=$\mathcal{A}_i$]{6.00, 5.00}{6.00, 6.00}
    \dashed[showArrow=false, color=9900ef, lineWidth=6, label=$\mathcal{B}_0$]{6.00, 4.00}{6.00, 5.00}
    \fermion[lineWidth=6]{4.00, 4.00}{8.00, 4.00}
    \fermion[lineWidth=6]{4.00, 6.00}{8.00, 6.00}
\end{feynman}}}
        \caption{}
        \label{fig9am}
    \end{subfigure}
    \begin{subfigure}[t]{0.22\textwidth}
        \centering
        \thesisdiagrampanel{\scalebox{0.3}{\begin{feynman}
    \electroweak[lineWidth=6, color=9900ef, label=$\mathcal{A}_i$]{6.00, 5.00}{6.00, 6.00}
    \gluon[flip=true, lineWidth=6, color=eb144c, label=$\mathcal{B}_i$]{6.00, 4.00}{6.00, 5.00}
    \fermion[lineWidth=6]{4.00, 4.00}{8.00, 4.00}
    \fermion[lineWidth=6]{4.00, 6.00}{8.00, 6.00}
\end{feynman}
}}
\caption{}
\label{fig9bm}
    \end{subfigure}
    \begin{subfigure}[t]{0.22\textwidth}
        \centering
        \thesisdiagrampanel{\scalebox{0.3}{\begin{feynman}
    \electroweak[lineWidth=6, color=9900ef, label=$\mathcal{A}_i$]{6.00, 5.00}{6.00, 6.00}
    \gluon[flip=true, lineWidth=6, color=eb144c, label=$\mathcal{B}_l$]{6.00, 4.00}{6.00, 5.00}
    \fermion[lineWidth=6]{4.00, 4.00}{8.00, 4.00}
    \fermion[lineWidth=6]{4.00, 6.00}{8.00, 6.00}
\end{feynman}
}}
\caption{}
\label{fig9cm}
    \end{subfigure}
    \begin{subfigure}[t]{0.22\textwidth}
        \centering
        \thesisdiagrampanel{\scalebox{0.3}{\begin{feynman}
    \dashed[showArrow=false, lineWidth=6, color=9900ef, label=$\mathcal{B}_0$]{6.00, 4.00}{6.00, 5.00}
    \fermion[lineWidth=6]{4.00, 4.00}{8.00, 4.00}
    \fermion[lineWidth=6]{4.00, 6.00}{8.00, 6.00}
    \dashed[showArrow=false, lineWidth=6, label=$\mathcal{A}_0$]{6.00, 5.00}{6.00, 6.00}
\end{feynman}}}
\caption{}
\label{fig9dm}
    \end{subfigure}
    \begin{subfigure}[t]{0.22\textwidth}
        \centering
        \thesisdiagrampanel{\scalebox{0.3}{\begin{feynman}
    \gluon[flip=true, color=eb144c, lineWidth=6, label=$\mathcal{B}_k$]{6.00, 4.00}{6.00, 5.00}
    \fermion[lineWidth=6]{4.00, 4.00}{8.00, 4.00}
    \fermion[lineWidth=6]{4.00, 6.00}{8.00, 6.00}
    \dashed[showArrow=false, lineWidth=6, label=$\mathcal{A}_0$]{6.00, 5.00}{6.00, 6.00}
\end{feynman}}}
\caption{}
\label{fig9em}
    \end{subfigure}
    \caption{Diagrams contributing to the bound sector at 1PN order due to the Proca-electromagnetic interaction term.}
    \label{fig:my_label2}
\end{figure}
Now we need to calculate the amplitude of all the diagrams shown in Fig.~(\ref{fig:my_label2}) that contribute to 1PN order in the effective action. Below we list all such contributions. 
\begin{itemize}
\item The amplitude corresponding to the diagram in Fig.~(\ref{fig9am}) with bulk interaction vertex $\rmint d^4 x\,\partial_0\mathcal{A}_i\partial_i\mathcal{B}_0$:
\begin{align}
    \begin{split}
   \mathcal{S}_{\text{eff}}\Big |_{{\text{fig}(\ref{fig9am})}}&=\gamma \,Q_1 Q_2'\rmint\Bar{\mathcal{D}}\Hat{\xi}\rmint dt_1v_1^j(t_1)\mathcal{A}_j(\boldsymbol{x}_1(t_1))\rmint dt_2\, \mathcal{B}_0(\boldsymbol{x}_2(t_2))\rmint d^4x\,\partial_0\mathcal{A}_i(x)\partial_i\mathcal{B}_0(x),\\ &=\gamma \,Q_1 Q_2'\rmint dt_1dt_2v_1^j(t_1)\rmint d^4x\,\partial_0\Big{\langle}\mathcal{A}_j(\boldsymbol{x}_1(t_1))\mathcal{A}_i(x)\Big{\rangle}\partial_i\Big{\langle}\mathcal{B}_0(\boldsymbol{x}_2(t_2))\mathcal{B}_0(x)\Big{\rangle},\\ &
=i\gamma\,Q_1 Q_2'\rmint dt \Bigg[a^{i}_{1}(t)\rmint_{\boldsymbol{k}}\frac{k^i\,e^{-i\,\boldsymbol{k}\cdot\boldsymbol{r}}}{\boldsymbol{k}^2(\boldsymbol{k}^2+\mu_{\gamma}^2)}+v_1^{i}(t)v_{1}^{a}(t)\frac{i\,k^a\,k^i\,e^{-i\boldsymbol{k}\cdot\boldsymbol{r}}}{\boldsymbol{k}^2(\boldsymbol{k}^2+\mu_{\gamma}^2)}\Bigg]\,,\\ &
=-\frac{\gamma\,Q_1 Q_2'}{2\pi}\rmint dt \Bigg[(\Vec{a}_1\cdot \Hat{n})\Big\{\frac{  e^{-\mu_{\gamma} r}}{2 \mu_{\gamma} \,r}-\frac{1 -  e^{-\mu_{\gamma} r}}{2\, \mu_{\gamma}^2\, r^2}\Big\}+\boldsymbol{v}_1^2 \,\frac{e^{-r \mu _{\gamma }} \left(-r \mu _{\gamma }+e^{r \mu _{\gamma }}-1\right)}{  r^3 \mu _{\gamma }^2}\\ &
\,\,\,\,\,\,\,\,\,\,\,\,\,\,\,\,\,\,\,\,\,\,\,\,\,\,\,\,\,
+(\boldsymbol{v}_1\cdot \Hat{n})^2\,\frac{e^{-r \mu _{\gamma }} \Big(r \mu _{\gamma } \left(r \mu _{\gamma }+3\right)-3 e^{r \mu _{\gamma }}+3\Big)}{  r^3 \mu _{\gamma }^2}
\Bigg]+(1\leftrightarrow 2)\,\sim \mathcal{O}(L v^2)\,.
\end{split}
\end{align}
\vspace{-0.5cm}
\item The amplitude corresponding to the diagram in Fig.~(\ref{fig9bm}) with bulk interaction vertex $\rmint d^4x\,\partial_l\mathcal{A}_i\partial_l\mathcal{B}_i$:
\begin{align}
    \begin{split}
   \mathcal{S}_{\text{eff}}\Big|_{\text{fig}(\ref{fig9bm})}&=\gamma \,Q_1 Q_2'\rmint\Bar{\mathcal{D}}\Hat{\xi}\rmint dt_1v_1^j(t_1)\mathcal{A}_j(\boldsymbol{x}_1(t_1))\rmint dt_2\, \mathcal{B}_m(\boldsymbol{x}_2(t_2))v_2^m(\boldsymbol{x}_2(t_2))\\ & \hspace{6.5 cm}\rmint d^4x\,\partial_l\mathcal{A}_i(x)\partial_l\mathcal{B}_i(x),\\&=\gamma Q_1 Q_2'\rmint dt_1dt_2v_1^j(t_1)v_2^m(t_2)\rmint d^4x\partial_l\Big{\langle}\mathcal{A}_j(\boldsymbol{x}_1(t_1))\mathcal{A}_i(x)\Big{\rangle}\partial_l\Big{\langle}\mathcal{B}_m(\boldsymbol{x}_2(t_2))\mathcal{B}_i(x)\Big{\rangle}\,,\\ &
=\frac{\gamma Q_1 Q_2'}{4\pi}\rmint dt\, (\boldsymbol{v}_1\cdot \boldsymbol{v}_2) \frac{e^{-\mu_{\gamma}r}}{r}+(1\leftrightarrow 2)\,\sim \mathcal{O}(L v^2)\,.
\end{split}
\end{align}
\item The amplitude corresponding to the diagram in Fig.~(\ref{fig9cm}) with bulk interaction vertex $\rmint d^4x\,\partial_l\mathcal{A}_i\partial_i\mathcal{B}_l$:
\begin{align}
    \begin{split}
   \mathcal{S}_{\text{eff}}\Big|_{\text{fig}(\ref{fig9cm})}&=\gamma \,Q_1 Q_2'\rmint\Bar{\mathcal{D}}\Hat{\xi}\rmint dt_1v_1^j(t_1)\mathcal{A}_j(\boldsymbol{x}_1(t_1))\rmint dt_2\, \mathcal{B}_m(\boldsymbol{x}_2(t_2))v_2^m(\boldsymbol{x}_2(t_2))\\ &\hspace{6.5 cm}\rmint d^4x\,\partial_l\mathcal{A}_i(x)\partial_i\mathcal{B}_l(x),\\&=\gamma Q_1 Q_2'\rmint dt_1dt_2v_1^j(t_1)v_2^m(t_2)\rmint d^4x\partial_l\Big{\langle}\mathcal{A}_j(\boldsymbol{x}_1(t_1))\mathcal{A}_i(x)\Big{\rangle}\partial_i\Big{\langle}\mathcal{B}_m(\boldsymbol{x}_2(t_2))\mathcal{B}_l(x)\Big{\rangle}\,,\\ &
=\gamma Q_1Q_2'\rmint dt\, \,\Big[v_1^i(t) v_2^l(t)\
\rmint \frac{k_i k_l\,e^{-i\boldsymbol{k}\cdot \boldsymbol{r}}}{k^2(\boldsymbol{k}^2+\mu^2)}\Bigg]\,,                 \\ &
=\gamma Q_1 Q_2'\rmint dt \, \Big[(\boldsymbol{v}_1\cdot \boldsymbol{v}_2)\frac{e^{-r \mu _{\gamma }} \left(-r \mu _{\gamma }+e^{r \mu _{\gamma }}-1\right)}{2 \pi  r^3 \mu _{\gamma }^2}\\ &
\,\,\,\,\,\,\,\,\,\,\,\,\,\,\,\,\,\,\,\,\,\,\,\,\,\,\,\,\,\,\,\,\,+(\boldsymbol{v}_1\cdot \Hat{n})(\boldsymbol{v}_2\cdot \Hat{n})\frac{e^{-r \mu _{\gamma }} \Big(r \mu _{\gamma } \left(r \mu _{\gamma }+3\right)-3 e^{r \mu _{\gamma }}+3\Big)}{2 \pi  r^3 \mu _{\gamma }^2}\Big]+(1\leftrightarrow 2) \sim \mathcal{O}(L v^2)\,.
\end{split}
\end{align}
\item The amplitude corresponding to the diagram in Fig.~(\ref{fig9dm}) with bulk interaction vertex $\rmint d^4x\,\partial_k \mathcal{A}_0\partial_k \mathcal{B}_0$:
\begin{align}
    \begin{split}
       \mathcal{S}_{\text{eff}}\Big|_{\text{fig}(\ref{fig9dm})}&=\gamma \,Q_1 Q_2'\rmint\Bar{\mathcal{D}}\Hat{\xi}\rmint dt_1 \mathcal{A}_0(\boldsymbol{x}_1(t_1))\rmint dt_2\, \mathcal{B}_0(\boldsymbol{x}_2(t_2))\rmint d^4x\,\partial_k\mathcal{A}_0(x)\partial_k\mathcal{B}_0(x),\\&
       =\gamma Q_1 Q_2'\rmint dt_1dt_2\rmint d^4x\partial_k\Big{\langle}\mathcal{A}_0(\boldsymbol{x}_1(t_1))\mathcal{A}_0(x)\Big{\rangle}\partial_k\Big{\langle}\mathcal{B}_0(\boldsymbol{x}_2(t_2))\mathcal{B}_0(x)\Big{\rangle}\,,\\ &
=\gamma Q_1Q_2'\rmint dt\, \,\Big[\rmint \frac{e^{-i\boldsymbol{k}\cdot \boldsymbol{r}}}{(\boldsymbol{k}^2+\mu^2)}\Bigg]    \,,             \\ &
=\gamma Q_1Q_2'\rmint dt\,\frac{e^{-\mu_{\gamma}r}}{4\pi r}+(1\leftrightarrow 2)\,\sim \mathcal{O}(L v^2)\,.
    \end{split}
\end{align}
\item The amplitude corresponding to the diagram in Fig.~(\ref{fig9em}) with the bulk interaction vertex $\rmint d^4x \,\partial_k\mathcal{A}_0\partial_0 \mathcal{B}_k$:
\begin{align}
    \begin{split}
   \mathcal{S}_{\text{eff}}\Big|_{\text{fig}(\ref{fig9em})}&=\gamma \,Q_1 Q_2'\rmint\Bar{\mathcal{D}}\Hat{\xi}\rmint dt_1\mathcal{A}_0(\boldsymbol{x}_1(t_1))\rmint dt_2\, \mathcal{B}_j(\boldsymbol{x}_2(t_2))v_2^j(\boldsymbol{x}_2(t_2))\rmint d^4x\,\partial_k\mathcal{A}_0(x)\partial_0\mathcal{B}_k(x),\\&=\gamma Q_1 Q_2'\rmint dt_1dt_2v_2^j(t_2)\rmint d^4x\partial_k\Big{\langle}\mathcal{A}_0(\boldsymbol{x}_1(t_1))\mathcal{A}_0(x)\Big{\rangle}\partial_0\Big{\langle}\mathcal{B}_j(\boldsymbol{x}_2(t_2))\mathcal{B}_k(x) \Big{\rangle}\,,\\ &
=i\gamma Q_1 Q_2'\rmint dt \Bigg[a^{k}_{2}(t)\rmint_{\boldsymbol{k}}\frac{k^k\,e^{-i\,\boldsymbol{k}\cdot\boldsymbol{r}}}{\boldsymbol{k}^2(\boldsymbol{k}^2+\mu^2)}-v_2^{k}(t)v_{2}^{a}(t)\frac{i\,k^a\,k^k\,e^{-i\boldsymbol{k}\cdot\boldsymbol{r}}}{\boldsymbol{k}^2(\boldsymbol{k}^2+\mu^2)}\Bigg]\,,\\ &
=-\frac{\gamma Q_1 Q_2'}{2\pi}\rmint dt \Bigg[(\Vec{a}_2\cdot \Hat{n})\Big\{\frac{  e^{-\mu_{\gamma} r}}{2 \mu_{\gamma} \,r}-\frac{1 -  e^{-\mu_{\gamma} r}}{2\, \mu_{\gamma}^2\, r^2}\Big\}+\boldsymbol{v}_2^2 \,\frac{e^{-r \mu _{\gamma }} \left(-r \mu _{\gamma }+e^{r \mu _{\gamma }}-1\right)}{  r^3 \mu _{\gamma }^2}\\ &
\,\,\,\,\,\,\,\,\,\,\,\,\,\,\,\,\,\,\,\,\,\,\,\,\,\,\,\,\,
+(\boldsymbol{v}_2\cdot \Hat{n})^2\,\frac{e^{-r \mu _{\gamma }} \Big(r \mu _{\gamma } \left(r \mu _{\gamma }+3\right)-3 e^{r \mu _{\gamma }}+3\Big)}{  r^3 \mu _{\gamma }^2}
\Bigg]+(1\leftrightarrow 2)\,\sim \mathcal{O}(L v^2)\, \label{4.52}
\end{split}
\end{align}
\end{itemize}

Before we conclude this section, we summarize our results in Table~(\ref{table2mm}). We show the number of diagrams responsible for the corrections to the conservative dynamics due to the various fields and their interactions, apart from the pure gravity contributions. 

\begin{table}[b!]
 \centering
 \scalebox{0.91}{\begin{tabular}{|c|c|c|c|c|}
 \hline
 & $0 PN$ & $1PN$ &  $2PN$ &  $2.5PN$ \\
  \hline
 \textit {scalar} & - & 5 & 6 &-\\
  \hline
 \textit{electromagnetic} & 1 & 3 & - &-\\
 \hline 
 \textit{proca} & 1 & 3 & - &- \\
 \hline 
 \textit{gravitational} & - & 4 & - &-\\
 \hline
 \textit{scalar-em interaction}
 & - & - & - &2($\sim g_{a\gamma\gamma}$)\\
 \hline
 \textit{proca-em interaction}&- & 4 $(\sim \gamma)$ & - &-\\
 \hline
 \end{tabular}}
 \caption{Table showing the number of diagrams contributing to a specific PN order for bound sector. Also, the orders at which terms to due the axion-photon coupling ($g_{a\gamma\gamma}$) and photon-dark photon kinetic mixing term ($\gamma$).}
 \label{table2mm}
 \end{table}
\vspace{0.5cm}
\textcolor{black}{{\bf Corrections to the bound orbit}: \textcolor{black}{Now we can add all the contributions mentioned in 
 (\ref{4.13a})-(\ref{4.52}) and that gives us the real part of the total effective action. To find out the orbit equation, one needs to extremize this.}
\begin{eqnarray}
  &  \frac{\delta}{\delta \boldsymbol{x}_a(t)}\,\text{Re}\,\mathcal{S}_{\text{eff}}^{\text{tot}}[\boldsymbol{x}_a]\Big |_{\mathcal{O}(\hbar^0)}=0\,.\label{4.53m}
\end{eqnarray}}
The leading order contribution to effective action is the Newtonian potential. That gives rise to the Keplerian orbit as the solution of (\ref{4.53m}). This we use to study the radiative dynamics in Sec.~(\ref{Sec5}). Note that we will get corrections to the usual Keplerian orbit solution due to other terms in the effective action at higher-order PN. Although one should consider these corrections and find out the corrected orbit equation and then use it to study the radiative dynamics, in this paper, we will neglect such corrections and work with the leading order solution, i.e. Keplerian orbit solution (in fact, the circular orbit).

\section{The radiative dynamics}\label{Sec5}
Till now, we have studied the conservative dynamics, where no external radiative line is attached to the diagrams. In the radiative sector, we keep one long-wavelength field external and integrate out only the potential modes. For a generic field $\xi=\Hat{\xi}+\Bar{\xi}$, this produces an effective action for the radiation field of the form
\begin{equation}
\mathcal{S}_{\text{eff}}^{\Bar{\xi}}[\boldsymbol{x}_a,\Bar{\xi}]=S_{\text{free}}^{\Bar{\xi}}+\rmint d^4 x\,J(x)\Bar{\xi}(x)\,,
\end{equation}
\noindent where $J(x)$ is the source that encodes the binary dynamics relevant for emission. By comparing with (\ref{2.4m}) one can identify this source with the corresponding Wilson coefficient. Expanding around $\bar\xi=0$,
\begin{align}
    \begin{split}
       & \mathcal{S}_{\text{eff}}^{\Bar{\xi}}[\boldsymbol{x}_a,\Bar{\xi}]= \mathcal{S}_{0}[\boldsymbol{x}_a]+\mathcal{S}_{1}[\boldsymbol{x}_a,\Bar{\xi}]+\mathcal{S}_{2}[\boldsymbol{x}_a,\Bar{\xi}]+\mathcal{O}(\bar\xi^3).\label{5.1m}
    \end{split}
\end{align}
The first term in (\ref{5.1m}) is the conservative action, the second determines the radiation source, and the higher powers describe self-interactions of the radiative field. To compute the emitted power we integrate out the radiation field itself and obtain the effective action for the source,
\begin{align}
\begin{split}
   \mathcal{Z}[J]:= e^{i{\gamma}_{\text{eff}}[J]}&=\rmint \mathcal{D}\Bar{\xi}\rmint \mathcal{D}\Hat{\xi}\,e^{iS[{\xi}]}\,,\\ &
   =\rmint \mathcal{D}\Bar{\xi}\, e^{i\mathcal{S}_{\text{eff}}^{\Bar{\xi}}[\Bar{\xi},J]}\,.
   \label{4.3}
   \end{split}
\end{align}
Again, in the in-out formalism $\mathcal{Z}[J]$ is the overlap between in and out states,
\begin{eqnarray}
    \mathcal{Z}[J]=\Big\langle0_+|0_-\Big\rangle.\label{4.4}
\end{eqnarray}
using the two relations (\ref{4.3}) and (\ref{4.4}) we get \cite{nastase_2019},
\begin{eqnarray}
   | \langle0_+|0_-\rangle|^2=e^{-2\text{Im}[\gamma_{\text{eff}}]}\,.
\end{eqnarray}
Expanding in small $\text{Im}[\gamma_{\text{eff}}]$ we will get,
\begin{align}
    \begin{split}
        2\,\text{Im}[\gamma_{\text{eff}}]=\mathcal{T}\rmint dE\,d\Omega \frac{d^2\Gamma}{dEd\Omega}\label{4.6}
    \end{split}
\end{align}
and the radiated power is given by,
\begin{align}
    \begin{split}
        P=\rmint dE\,d\Omega\,E\,\frac{d^2\Gamma}{dEd\Omega}\label{5.6m}
    \end{split}
\end{align}
where, $\mathcal{T}$ is the interaction time, $d\Gamma$ is the rate of particle emission and $P$ is the radiated power. The practical workflow in every radiative subsection is therefore the following:
\begin{itemize}
    \item identify the diagrams with one external radiation line,
    \item read off the source term or the corresponding multipole moments,
    \item use $\text{Im}\,\gamma_{\text{eff}}$ together with (\ref{5.6m}) to obtain the radiated power.
\end{itemize}
\subsection{Radiative sector PN counting} \label{EFTcounting}
 Now, to compute the radiated power up to a given \textcolor{black}{$N^{(n)}LO$}, we need to systematically assign EFT power-counting rules to the radiative fields. To make things more compact, we display the scaling of the radiation fields $\bar{\xi}\equiv\{\bar\psi,\bar{{\mathcal{A}}}_i,\bar\sigma_{ij},\bar{\phi},\Bar{b}_{\mu}, \Bar{a}_{\mu}\}$ in Table~(\ref{table3mm}). One should note that, for the potential fields, one must perform a partial Fourier transform. Therefore,
$$\hat\xi(x^0,\boldsymbol{x})=\rmint_k e^{i\boldsymbol{k}\cdot \boldsymbol{x}}\,\hat\xi_k(x^0)$$
where $\xi_k\equiv \{\psi_k,\Hat{\mathcal{A}}_{{\boldsymbol{k}}i},\zeta_{{\boldsymbol{k}}ij}, \mathcal{A}_{\boldsymbol{k}i},\mathcal{B}_{\boldsymbol{k}i},\varphi_{\boldsymbol{k}}\}$ are the potential fields. Here in this sector, we define the diagrams as $N^{(n)}LO$, i.e., leading order, next to leading order, and so on. The specification of scaling for the radiative sector is denoted in terms of these notations, as scaling rules are a bit subtle and different in the radiative sector. 
\begin{table}[ht]
\begin{center}
\begin{tabular}{|c|c|}
    \hline
   \multicolumn{2}{|c|}{EFT scaling} \\
\hline
Fields & scaling\\
\hline
t &  $\sim\frac{r}{v}$\\
\hline
  $\delta x$ &$\sim r$\\
  \hline
  $\frac{m}{M_{p}}$ & $\sim L^\frac{1}{2}v^\frac{1}{2}$\\
  \hline
  $Q^2$ & $ \sim L\,v$\\
  \hline
  $\xi_k$ & $\sim r^2v^\frac{1}{2}$\\
  \hline
$\bar{\Xi}$  & $\sim\frac{v}{r}$\\
\hline
$\partial_0\xi_k$ & $\sim\frac{v}{r}\xi_k$\\
\hline
  $\partial_{\mu}\bar{\Xi}$ & $\sim\frac{v}{r}\bar{\Xi}$\\
  \hline 
\end{tabular}
\end{center}
\caption{EFT PN counting for radiative sector}
\label{table3mm}
\end{table} 
\par  \textcolor{black}{Now expanding the metric at the linear level the  relation between NRG fields and the linearized metric components in the radiative sector is given by:$$
\bar g_{\mu\nu}=\eta_{\mu\nu}+\frac{1}{m_p}
\begin{pmatrix}
-2\bar\psi & \bar{\mathcal{A}_j}\\
\bar{\mathcal{A}_i} & \,\,\bar{\sigma_{ij}}-2\delta_{ij}\bar\psi\\
\end{pmatrix}\equiv\eta_{\mu\nu}+\frac{1}{m_p}
\begin{pmatrix}
\bar h_{00} & \bar h_{0j}\\
\bar h_{i0} & \bar h_{ij}\\
\end{pmatrix}
$$
The gravitational radiation formula we used in the later section in (\ref{5.65}) is basically considering the $\bar\psi$ radiation. One subtle point to note here is that when we calculate radiation usually we consider the TT gauge along with the harmonic gauge condition. But from the beginning, we can't set $\bar h_{00}=0$ in the effective Lagrangian for radiation demanded by \textit{Transverse Traceless (TT)} gauge condition.}

\subsection{Scalar interaction and scalar radiation}\label{subsec:radiated_scalars}

We begin with the scalar sector because it already contains all the conceptual ingredients that reappear later: a non-trivial source, a multipole expansion, and the extraction of power through the optical theorem. The strategy is to first derive a compact emission formula in terms of scalar multipoles and then compute the source $J_\phi$ order by order in the PN expansion. In the main text, we therefore keep the multipole logic explicit while compressing the more repetitive tensor algebra. To begin with, we integrate the potential scalar modes and get the effective action for the radiating scalars.  
\begin{equation}
\mathcal{S}_{\text{eff}}^{\Bar{\phi}}=  \rmint d^4x  \left(\underbrace{ -\frac{1}{2} \eta^{\mu \nu}  \partial_\mu \bar{\phi} \partial_\nu \bar{\phi} 
-\frac{1}{2}m^2\bar{\phi}^2 }_{\mathcal{S}_{\rm free}^{\bar\phi}} + \underbrace{\frac{1}{m_p} J   \bar{\phi}}_{\mathcal{S}_{\text{int}}^{\bar\phi}} \right)\;, 
\end{equation}
which leads to the following equation of motion 
\begin{equation}
\textcolor{black}{(\square-m^2 )\bar{\phi} = \frac{J}{m_p} \;, \qquad \square \equiv \eta^{\mu \nu}  \partial_\mu  \partial_\nu \;.}
\end{equation}
We will now calculate the imaginary part of the effective action $\gamma_{\text{eff}}^{\bar\phi}$ for the source obtained by integrating out the radiation scalars as discussed in Eq.~(\ref{4.3}) 
and then using the optical theorem, we obtain the power radiation for the scalar sector. We will work out the power radiation essentially for circular orbits. Now,
\begin{equation}
\begin{split}
\mathcal{S}_{\rm int}^{\bar\phi} = \rmint d^4 x \frac{J \bar \phi }{m_p}= \rmint dt \, \rmint d^3 x \frac{J (t, \boldsymbol{x})}{m_p}  \bigg( \bar{\phi}(t, \boldsymbol{0}) + x^i \partial_i \bar{\phi}(t,\boldsymbol{0}) + \frac{1}{2} x^i x^j \partial_i \partial_j \bar{\phi}(t, \boldsymbol{0})  \\
+ \frac{1}{3!} x^i x^j x^k \partial_i \partial_j \partial_k \bar{\phi}(t, \boldsymbol{0}) + \ldots \bigg) \;,
\label{5.10}
\end{split}
\end{equation}
From which we can write,
\begin{align}
\begin{split}
S_{\rm int}^{\bar\phi} = \frac{1}{m_p} \rmint dt \left( I_\phi \bar{\phi} + I_\phi^i \partial_i \bar{\phi} + \frac{1}{2} I_\phi^{ij} \partial_i \partial_j \bar{\phi}  + \ldots \right) \;,
\end{split}
\label{eq:multipole_expansion_scalar}
\end{align}
where 
\begin{equation}
\label{5.12mn}
\textcolor{black}{I_\phi  \equiv  \rmint d^3 x \left( J + \frac{1}{6}  (\partial_t^2+m^2) J  x^2 \right) \;, \quad I_\phi^i  \equiv \rmint d^3x \, x^i \left(J + \frac{1}{10} (\partial_t^2+m^2) J x^2 \right) \;, \quad  I_\phi^{ij} \equiv \rmint d^3x J  Q^{ij} \;}
\end{equation}
are respectively the scalar monopole, dipole and quadrupole.
\textcolor{black}{ To get the irreducible forms of the multipole moments, instead of writing $x^ix^j$ one should use,
\begin{align}
    \begin{split}
        {Q}^{ij}=x^ix^j-\frac{1}{3}x^2\delta^{ij}\,.
    \end{split}
\end{align}}
Then we integrate over the scalar radiation field and get the effective action for the source, which is equivalent to evaluating the self-energy diagram shown in Fig.~(\ref{mfig12}).
\begin{figure}
    \centering
\thesisdiagrampanel{\scalebox{0.36}{\begin{feynman}
    \electroweak[flip=true, lineWidth=4]{4.60, 4.10}{5.60, 4.80}
    \fermion[]{4.00, 4.00}{7.20, 4.00}
    \electroweak[lineWidth=4, label=$\bar\phi$]{5.60, 4.80}{6.60, 4.10}
    \fermion[]{4.00, 4.10}{7.20, 4.10}
\end{feynman}
}}
    \caption{Self-energy diagram contributing to the effective action for the source term $J$.}
    \label{mfig12}
\end{figure}
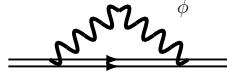
To quadratic order in the source moments, the path integral reduces to
\begin{align}
\begin{split}
i\gamma_{\text{eff}}^{\bar\phi}[J]\sim -\frac{1}{2m_p^2}\rmint dt_1 dt_2 \Big[&
I_{\phi}(t_1)I_{\phi}(t_2)\Big\langle\Bar{\phi}(t_1,\boldsymbol{0})\Bar{\phi}(t_2,\boldsymbol{0})\Big\rangle\\ &
+I_{\phi}^{i}(t_1)I_{\phi}^{j}(t_2)\Big\langle\partial_{i}\Bar{\phi}(t_1,\boldsymbol{0})\partial_{j}\Bar{\phi}(t_2,\boldsymbol{0})\Big\rangle\\ &
+I_{\phi}^{ij}(t_1)I_{\phi}^{kl}(t_2)\Big\langle\partial_{i}\partial_{j}\Bar{\phi}(t_1,\boldsymbol{0})\partial_{k}\partial_{l}\Bar{\phi}(t_2,\boldsymbol{0})\Big\rangle+\ldots\Big]\,.
\label{5.14n}
\end{split}
\end{align}
Odd-point correlators vanish because we expand around a background with $\langle 0 |\bar\phi|0\rangle=0$.\footnote{It would be interesting to revisit this computation when the radiative scalar has a non-zero vacuum expectation value, i.e. $\langle 0 |\bar\phi|0\rangle\ne 0$.}
Using the Feynman propagator
\begin{align}
    \begin{split}
        \Big\langle T \Bar{\phi}(t_1,\boldsymbol{x}_1)\Bar{\phi}(t_2,\boldsymbol{x}_2)\Big\rangle=\rmint \frac{d^4k}{(2\pi)^4}\frac{i}{-k^2-m^2+i\epsilon}e^{-ik\cdot(x_1-x_2)}\,,\label{4.13}
    \end{split}
\end{align}
and rotational invariance, the derivative correlators reduce to
\begin{align}
    \begin{split}
        \Big\langle T \partial_i \Bar{\phi}(t_1,\boldsymbol{0})\partial_j \Bar{\phi}(t_2,\boldsymbol{0})\Big\rangle&=\rmint \frac{d^4k}{(2\pi)^4}\frac{i}{-k^2-m^2+i\epsilon}e^{ik_0(t_1-t_2)}[\partial_{i}^{(x_1)}e^{-i\boldsymbol{k}.\boldsymbol{x}_1}][\partial_{j}^{(x_2)}e^{i\boldsymbol{k}\cdot\boldsymbol{x}_2}]\big|_{\boldsymbol{x}_1,\boldsymbol{x}_2=0}\\ &
        =\rmint \frac{d^4k}{(2\pi)^4}\frac{i}{-k^2-m^2+i\epsilon}e^{ik_0(t_1-t_2)}\langle\langle k_{i}\,k_{j}\rangle\rangle\,,\\ &
        =\frac{1}{3}\rmint \frac{d^4k}{(2\pi)^4}\frac{i}{-k^2-m^2+i\epsilon}e^{ik_0(t_1-t_2)}|\boldsymbol{k}|^2\delta_{ij}\,,\label{4.14}
    \end{split}
\end{align}
\begin{align}
    \begin{split}
          \Big\langle T \,\partial_i\partial_j \Bar{\phi}(t_1,\boldsymbol{0})\,\partial_k\partial_l \Bar{\phi}(t_2,\boldsymbol{0})\Big\rangle&=\rmint \frac{d^4k}{(2\pi)^4}\frac{i}{-k^2-m^2+i\epsilon}e^{ik_0(t_1-t_2)}[\partial_{i}\partial_j^{(x_1)}e^{-i\boldsymbol{k}.\boldsymbol{x}_1}][\partial_{k}\partial_l^{(x_2)}e^{i\boldsymbol{k}\cdot\boldsymbol{x}_2}]\big|_{\boldsymbol{x}_1,\boldsymbol{x}_2=0}\,,\\ &
          =\rmint \frac{d^4k}{(2\pi)^4}\frac{i}{-k^2-m^2+i\epsilon}e^{ik_0(t_1-t_2)}\langle\langle k_ik_jk_kk_l\rangle\rangle\,,\\ &
          =\rmint \frac{d^4k}{(2\pi)^4}\frac{i}{-k^2-m^2+i\epsilon}e^{ik_0(t_1-t_2)}|\boldsymbol{k}|^4 T_{ij,kl}\,,\label{4.15}
    \end{split}
\end{align}
where $T_{ij,kl}$ is constructed from $\langle\langle n_in_jn_kn_l\rangle\rangle$ and $\langle\langle n_in_j \rangle\rangle$ which is symmetric in $ij$ and $kl$ indices \cite{poissonwill}. Then finally we get, 
\begin{equation}
  \gamma_{\text{eff}}^{(\bar\phi)} = \frac{1}{2 m_p^2} \rmint \frac{d^4 k}{(2 \pi)^4} \frac{1}{-k^2-m^2 +i \epsilon} \bigg( | \mathcal{I}_\phi (k_0)|^2 + \frac{1}{3} |\boldsymbol{k}|^2   | \mathcal{I}_\phi^i (k_0)|^2 + \frac{1}{30} |\boldsymbol{k}|^4   | \mathcal{I}_\phi^{ij} (k_0)|^2  \bigg) \;, \label{5.18n}
\end{equation}
where, 
\begin{equation}
\mathcal{I}_\phi(k_0) = \rmint dt \,\mathcal{I}_\phi (t) e^{i k_0 t} \;, \mathcal{I}_{\phi}^{i}(k_0) = \rmint dt \,\mathcal{I}_\phi^i (t) e^{i k_0 t} \;, \mathcal{I}_{\phi}^{ij}(k_0) = \rmint dt \,\mathcal{I}_\phi^{ij} (t) e^{i k_0 t} \;.
\end{equation}
To extract the imaginary part, we need to find  the principal value of the propagator i.e.,
\begin{align}
    \begin{split}
        \frac{1}{-k^2-m^2+i\epsilon}=PV\Big(\frac{1}{-k^2-m^2}\Big)-i\pi\,\delta(-k^2-m^2)\,.
    \end{split}
\end{align}

Therefore, besides the quadrupole,  the monopole and the dipole also contribute to the scalar radiation. 
But for the massive case, we will have some changes in the quadrupole formula compared to \cite{Kuntz:2019zef}.
\begin{align}
    \begin{split}
        P_{\phi}^m=\frac{1}{4\pi^2 m_p^2\mathcal{T}}\sum_{l}\frac{1}{l!(2l+1)!!}\rmint_0^\infty d\omega\,\omega\,(\omega^2-m^2)^{l+1/2}|\mathcal{I}^{m}_{(L)}(\omega)|^2 .
        \label{5.21}
    \end{split}
\end{align}

We will truncate our computation up to $l=1$, i.e., we will compute only the dipole radiation. Here in (\ref{5.21}) $`L$' denotes the \textit{symmetric trace free (STF)} indices $i,j,k,\cdots.$

\subsubsection*{Computing the source term $J_{\phi}$:}
We now construct the scalar source only up to the orders needed for the dipole-radiation formula. The diagrams in Fig.~(\ref{Figure10}) can be grouped by PN order, and in the main text we quote the reduced amplitudes after the time and tensor algebra has been simplified. This makes it easier to see which physical effects first enter the source at each order.
\begin{itemize}
\item The $LO$ contribution comes from the worldline vertex $\rmint dt \,m_a s_a \Bar{\phi}.$ The amplitude corresponding to the diagram in Fig.~(\ref{10a}):
\begin{align}
    \begin{split}
\mathcal{S}_{\text{eff}}\Big|_{\text{fig}(\ref{10a})}=-\sum_{a=1}^{2}\rmint dt\,\frac{m_a\,s_a}{m_p}\,\Bar{\phi}(\boldsymbol{x}_a(t))\,\sim\mathcal{O}(L^{1/2}v^{1/2}).
    \end{split}
\end{align}
\begin{figure}
    \centering
  \begin{subfigure}[t]{0.22\textwidth}
\thesisdiagrampanel{\scalebox{0.3}{\begin{feynman}
    \electroweak[lineWidth=6, color=0693e3, label=$\Bar{\phi}$]{6.00, 6.00}{7.00, 7.00}
    \fermion[lineWidth=6]{4.00, 6.00}{8.00, 6.00}
    \fermion[lineWidth=6]{4.00, 4.00}{8.00, 4.00}
\end{feynman}
}}
      \caption{}
      \label{10a}
  \end{subfigure}
   \begin{subfigure}[t]{0.22\textwidth}
 \thesisdiagrampanel{\scalebox{0.3}{\begin{feynman}
    \electroweak[lineWidth=6, color=0693e3, label=$\varphi$]{5.60, 4.00}{5.60, 6.00}
    \electroweak[lineWidth=6, label=$\Bar{\phi}$, color=0693e3]{5.60, 6.00}{6.80, 6.80}
    \fermion[lineWidth=6]{4.00, 6.00}{7.60, 6.00}
    \fermion[lineWidth=6]{4.00, 4.00}{7.60, 4.00}
\end{feynman}}}
      \caption{}
      \label{10b}
  \end{subfigure}
   \begin{subfigure}[t]{0.20\textwidth}
 \thesisdiagrampanel{\scalebox{0.25}{\begin{feynman}
    \fermion[lineWidth=6]{4.00, 4.00}{7.80, 4.00}
    \fermion[lineWidth=6]{4.00, 6.40}{7.80, 6.40}
    \electroweak[lineWidth=6, label=$\tilde{\psi}$]{5.80, 6.40}{5.80, 4.00}
    \electroweak[lineWidth=6, label=$\Bar{\phi}$, color=0693e3]{5.80, 6.40}{7.60, 7.60}
\end{feynman}
}}
      \caption{}
      \label{10c}
  \end{subfigure}
  \begin{subfigure}[t]{0.22\textwidth} \centering
        \thesisdiagrampanel{\scalebox{0.3}{
\begin{feynman}
    \electroweak[label=$\Bar{\phi}$, lineWidth=6, color=0693e3]{5.40, 5.00}{7.60, 5.00}
    \fermion[lineWidth=6]{4.00, 4.00}{6.80, 4.00}
    \electroweak[label=$\varphi$, lineWidth=6, color=0693e3]{5.40, 6.00}{5.40, 5.00}
    \electroweak[label=$\Tilde{\psi}$, lineWidth=6]{5.40, 5.00}{5.40, 4.00}
    \fermion[lineWidth=6]{4.00, 6.00}{6.80, 6.00}
\end{feynman}}}
\caption{}
\label{10h}
\end{subfigure}
  \begin{subfigure}[t]{0.20\textwidth}
 \thesisdiagrampanel{\scalebox{0.25}{\begin{feynman}
    \electroweak[label=$\varphi$, lineWidth=6, color=0693e3]{5.60, 4.00}{5.60, 5.00}
    \electroweak[label=$\bar{\phi}$, color=0693e3, lineWidth=6]{5.60, 6.40}{7.40, 7.80}
    \electroweak[label=$\varphi$, lineWidth=6, color=0693e3]{5.60, 6.40}{5.60, 5.40}
    \fermion[lineWidth=6]{4.00, 6.40}{7.20, 6.40}
    \fermion[lineWidth=6]{4.00, 4.00}{7.20, 4.00}
    \parton{5.60,5.20}{0.20}
\end{feynman}
}}
      \caption{}
      \label{10d}
  \end{subfigure}
     \begin{subfigure}[t]{0.22\textwidth}
        \centering
        \thesisdiagrampanel{\scalebox{0.25}{
\begin{feynman}
    \electroweak[color=0693e3, lineWidth=6, label=$\phi$]{5.60, 5.20}{5.60, 4.00}
    \fermion[lineWidth=6, showArrow=false]{4.00, 6.40}{7.20, 6.40}
    \electroweak[color=0693e3, lineWidth=6, label=$\bar{\phi}$]{5.60, 5.20}{8.20, 5.20}
    \electroweak[lineWidth=6, label=$\psi$]{5.60, 6.40}{5.60, 5.20}
    \fermion[lineWidth=6, showArrow=false]{4.00, 4.00}{7.20, 4.00}
\end{feynman}
}}
\caption{}
\label{10e}
\end{subfigure}
    \begin{subfigure}[t]{0.22\textwidth}
        \centering
        \thesisdiagrampanel{\scalebox{0.25}{\begin{feynman}
    \electroweak[color=0693e3, lineWidth=6, label=$\varphi$]{5.60, 5.20}{5.60, 4.00}
    \fermion[lineWidth=6, showArrow=false]{4.00, 6.40}{7.20, 6.40}
    \gluon[lineWidth=6, label=$\hat{\mathcal{A}_i}$, color=fcb900]{5.60, 6.40}{5.60, 5.20}
    \electroweak[color=0693e3, lineWidth=6, label=$\bar{\phi}$]{5.60, 5.20}{8.20, 5.20}
    \fermion[lineWidth=6, showArrow=false]{4.00, 4.00}{7.20, 4.00}
\end{feynman}
}}
\caption{}
\label{10f}
\end{subfigure}
\begin{subfigure}[t]{0.22\textwidth} \centering
        \thesisdiagrampanel{\scalebox{0.3}{
 \begin{feynman}
    \electroweak[label=$A_m$, color=9900ef, lineWidth=6]{5.40, 5.00}{5.40, 4.00}
    \electroweak[label=$A_i$, color=9900ef, lineWidth=6]{5.40, 6.00}{5.40, 5.00}
    \fermion[lineWidth=6]{4.00, 6.00}{7.00, 6.00}
    \fermion[lineWidth=6]{4.00, 4.00}{7.00, 4.00}
    \electroweak[label=$\bar{\phi}$, lineWidth=6, color=0693e3]{5.40, 5.00}{7.60, 5.00}
\end{feynman}}}
\caption{}
\label{10g}
\end{subfigure}
\caption{Radiative diagrams for the scalar field that contribute up to $N^{(4)}LO$.}
\label{Figure10}
\end{figure}
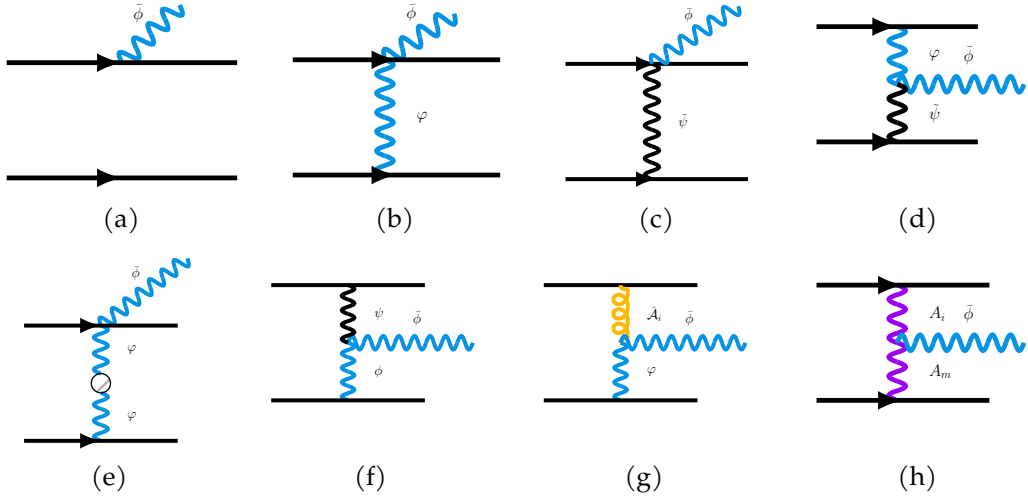
\item The $N^{(1)}LO$ contribution comes from the worldline vertex $\rmint dt\, \frac{1}{2}\,v_a^2\, m_as_a \frac{\phi}{m_p}$, and its contribution to the effective action has the following form,
\begin{align}
    \begin{split}
        \mathcal{S}_{\text{eff}}=\frac{1}{2}\sum_{a}m_a s_a v_a^2\,\Bar{\phi}(\boldsymbol{x}_a(t_a))\sim \mathcal{O}(L^{1/2}v^{3/2}).
    \end{split}
\end{align}

\item The $N^{(2)}LO$ contribution comes from the vertex: $-\rmint dt\, g_a \varphi\, \Bar{\phi}$ and $-\rmint dt\,s_a \varphi$. The amplitude corresponds to the diagram in Fig.~(\ref{10b}):
\begin{align}
    \begin{split}
          \mathcal{S}_{\text{eff}}\Big|_{\text{fig}(\ref{10b})}&=\frac{m_1 m_2}{m_p^2}\rmint\Bar{\mathcal{D}}\Hat{\xi}\rmint dt_1\,g_1 \varphi(\boldsymbol{x}_1(t_1))\frac{\Bar{\phi}(\boldsymbol{x}_1(t_1))}{m_p}\rmint dt_2\,s_2\,\varphi(\boldsymbol{x}_2(t_2))\,,\\ &
        =\frac{4 \pi  g_1 s_2 m_1 m_2}{m_p^2} \rmint dt\rmint_{\boldsymbol{k}}\frac{e^{i\boldsymbol{k}\cdot \boldsymbol{r}}}{\Vec{k^2}+m^2}\frac{\Bar{\phi}(\boldsymbol{x}_1(t_1))}{m_p}\,,\\ &
        =\frac{g_1 s_2 m_1 m_2}{m_p^2}\rmint dt \frac{e^{-mr}}{r}\frac{\Bar{\phi}(\boldsymbol{x}_1(t))}{m_p}+(1\leftrightarrow 2)\,\sim\mathcal{O}(L^{1/2}v^{5/2}).
    \end{split}
\end{align}

\item \textcolor{black}{Another $N^{(2)}LO$ contribution for the source comes from the worldline vertex:
-$\rmint dts_a m_a\bar{\phi}\tilde{\psi}\,.$}
The amplitude corresponds to the diagram in Fig.~(\ref{10c}):

\begin{align}
    \begin{split}
        \mathcal{S}_{\text{eff}}^{0}\Big|_{\text{fig}(\ref{10c})}&=\frac{s_1m_1m_2}{m_p^3}\rmint\Bar{\mathcal{D}}\Hat{\xi}\rmint dt_1 \,\frac{\Bar{\phi}(\boldsymbol{x}_1(t_1))}{m_p}\tilde{\psi}(\boldsymbol{x}_1(t_1))\rmint dt_2 \tilde{\psi}(\boldsymbol{x}_2(t_2))\,,\\ &
        =\frac{s_1m_1m_2}{m_p^3}\rmint dt_1\,dt_2\,\Bar{\phi}(\boldsymbol{x}_1(t_1))\Big\langle\tilde{\psi}(\boldsymbol{x}_1(t_1))\tilde{\psi}(\boldsymbol{x}_2(t_2))\Big\rangle\,, \\ &
        =\frac{s_1m_1m_2}{8\pi m_p^2}\rmint dt\, \frac{1}{r} \frac{\Bar{\phi}(\boldsymbol{x}_1(t))}{m_p}+(1\leftrightarrow 2)\,\,\sim\mathcal{O}({L^{1/2}v^{5/2}}).
    \end{split}
\end{align}
\item  The $N^{(3)}LO$ contribution coming from vertex $m^2 \rmint d^4x \Tilde{\psi}\varphi\Bar{\phi}(x)$ corresponding to the diagram in Fig.~(\ref{10h}), contributes to the effective action in the following way:
\begin{align}
    \begin{split}
\mathcal{S}_{\text{eff}}\Big |_{\text{fig}(\ref{10h})}&=\frac{m^2 s_1m_1m_2}{m_p^3}\rmint dt_1dt_2\rmint d^4x \Big\langle \psi(\boldsymbol{x}_2(t_2))\psi(x)\Big\rangle \Big\langle\varphi(\boldsymbol{x}_1(t_1))\varphi(x) \Big\rangle \bar{\phi}(x_0)\,,\\&
=\frac{s_1m_1m_2}{8\pi m_p^2}\rmint dt \frac{(1-e^{-mr})}{r}\frac{\bar{\phi}(\boldsymbol{x}_0(t))}{m_p}+1\leftrightarrow 2\,\sim\mathcal{O}(L^{1/2}v^{7/2}).
    \end{split}
\end{align}
\item \textcolor{black}{One $N^{(4)}LO$ correction comes from the relativistic time correction shown in Fig.~(\ref{10d}). After the time integrations and an integration by parts, the reduced contribution is}
\begin{align}
    \begin{split}
\mathcal{S}_{\text{eff}}\Big|_{\text{fig}(\ref{10d})}
=\frac{m_1 m_2 g_1 s_2}{16 \pi m_p^2}\rmint dt \,
\mathcal{K}_{10d}(t)\,
\frac{\Bar{\phi}(\boldsymbol{x}_1(t))}{m_p}+ T(\Dot{\Bar{\phi}})+ (1 \leftrightarrow 2)\,\sim\mathcal{O}(L^{1/2}v^{9/2})\,.
\label{5.26}
    \end{split}
\end{align}
with
\begin{align}
    \begin{split}
        \mathcal{K}_{10d}(t)=e^{-mr}\Bigg[
        \frac{\boldsymbol{v}_1\cdot\boldsymbol{v}_2}{r}
        -\frac{(\boldsymbol{v}_1\cdot \hat{n})(\boldsymbol{v}_2\cdot \hat{n})}{r}
        +\frac{m}{2}\,(\boldsymbol{v}_2\cdot \Hat{n})(\boldsymbol{v}_1\cdot \hat{n})
        \Bigg].
    \end{split}
\end{align}
Here, 
\begin{align}
    \begin{split}
        T(\Dot{\Bar{\phi}})&=-\frac{m_1 m_2 g_1 s_2}{16\pi m_p^2}\rmint dt\, e^{-mr}(\boldsymbol{v}_2\cdot \Hat{n})\frac{ \Dot{\Bar{\phi}}(\boldsymbol{x}_1(t))}{m_p},\\ &
        =\frac{m_1 m_2 g_1 s_2}{16\pi m_p^2}\rmint dt \frac{e^{-mr}}{r}(\boldsymbol{v}_2\cdot \Dot{\boldsymbol{r}}+\Vec{a}_2\cdot \boldsymbol{r})\,\frac{\Bar{\phi}(\boldsymbol{x}_1(t))}{m_p}\,.
    \end{split}
\end{align}
We have ignored the total derivative term by assuming the time variation of $\Bar{\phi}$ is very small and $|\Dot{\boldsymbol{r}}|=0$ for circular orbits.
\item Another $N^{(4)}LO$ term comes from the bulk interaction $\rmint d^4x \psi\partial_0\phi\partial_0\bar{\phi}$ shown in Fig.~(\ref{10e}). Writing $\Delta\boldsymbol{v}:=\boldsymbol{v}_2-\boldsymbol{v}_1$ and $\Delta\Vec{a}:=\Vec{a}_2-\Vec{a}_1$, its reduced contribution is
\begin{align}
    \begin{split}
\mathcal{S}_{\text{eff}}\Big |_{\text{fig}(\ref{10e})}
=\frac{-s_2m_1m_2}{8\pi m^2 m_p^2}\rmint dt \,
\mathcal{K}_{10e}(t)\,
\frac{\bar\phi(\boldsymbol{x}_0(t))}{m_p}+(1\leftrightarrow 2)\sim\mathcal{O}(L^{1/2}v^{9/2})\,,
\label{5.29}
    \end{split}
\end{align}
where
\begin{align}
    \begin{split}
        \mathcal{K}_{10e}(t)&=\partial_{0}f(r)\,\Delta\boldsymbol{v}\cdot \Hat{n}+f(r)\Bigg\{\Delta\Vec{a}\cdot \Hat{n}
        +\Delta\boldsymbol{v}\cdot
        \Bigg(
        \Hat{n}\frac{(\boldsymbol{v}_1-\boldsymbol{v}_2)\cdot \Hat{n}}{r}
        +\frac{\boldsymbol{v}_2-\boldsymbol{v}_1}{r}
        \Bigg)\Bigg\}.
    \end{split}
\end{align}
where in (\ref{5.29}) the function $f(r)$ is given by,
\begin{align}
    \begin{split}
        f(r)=\frac{e^{-mr}(mr-e^{mr}+1)}{r^2}.
        \label{5.46}
    \end{split}
\end{align}
\vspace{-0.925cm}
\item A third diagram contributes at  $N^{(4)}LO$ through the bulk interaction vertex $\rmint d^4x\,\hat{\mathcal{A}}_i\,\,\partial_i\varphi\,\partial_0\bar{\phi}(x)$ shown in Fig.~(\ref{10f}). In reduced form,
\begin{align}
\begin{split}
S_{\textrm{eff}}\Big |_{\text{fig}(\ref{10f})}
=\frac{-1}{2\pi }\frac{m_1m_2s_2}{m_p^2m^2}\rmint dt\,
\mathcal{K}_{10f}(t)\,
\frac{\Bar{\phi}(\boldsymbol{x}_0(t))}{m_p}  +(1\leftrightarrow 2)\sim\mathcal{O}(L^{1/2}v^{9/2})\,.
\end{split}
\end{align}
with
\begin{align}
\begin{split}
\mathcal{K}_{10f}(t)&=(\boldsymbol{v}_1\cdot \Hat{n}) \partial_0f(r) \\ &
\quad +\Bigg\{
(\Vec{a}_1\cdot \Hat{n})
+\boldsymbol{v}_1\cdot \Bigg[
\frac{(\boldsymbol{v}_1-\boldsymbol{v}_2)\cdot \Hat{n}}{r}\,\Hat{n}
+\frac{\boldsymbol{v}_2-\boldsymbol{v}_1}{r}
\Bigg]
\Bigg\}f(r).
\end{split}
\end{align}
Here $f(r)$ is defined in Eq.~(\ref{5.46}).

\item \textcolor{black}{A bulk vertex term causing scalar radiation at $N^{(5)}LO$ is $g_{a\gamma\gamma}\,\epsilon^{0ikm}\rmint d^4x \partial_{0}\mathcal{A}_{i}\partial_{k}\mathcal{A}_{m}\Bar{\phi}$, coming from the \textit{Theta} term. The corresponding reduced amplitude for Fig.~(\ref{10g}) is}
\begin{align}
    \begin{split}
          \mathcal{S}_{\text{eff}}\Big|_{\text{fig}(\ref{10g})}
          =\frac{Q_1 Q_2 g_{a\gamma\gamma}}{8\pi}\rmint dt \,[\Vec{a}_1\cdot (\boldsymbol{v}_2\times \Hat{n})+1\leftrightarrow 2]\,\frac{\Bar{\phi}(\boldsymbol{x}_0(t))}{m_p}\,\sim\mathcal{O}(L^{1/2}v^{11/2})\,.
\label{5.32m}
    \end{split}
\end{align}
\end{itemize}
Although we get a non-trivial contribution to the radiative effective action from the \textit{theta} term in (\ref{5.32m}), it vanishes for any orbit confined to a \textit{plane}. 
\textcolor{black}{Apart from the above-mentioned terms, we can have some more 3-point bulk interaction vertices with a radiating field of the following form,
\begin{equation}
\rmint d^4x \, \boldsymbol{\zeta}_{qr}\partial_{i}\varphi\partial_{j}\Bar{\phi}. \label{5.36n}
\end{equation}
But they would not contribute to the source term for the following reason. To identify the source term, the effective action should have the following form $\rmint d^4x J(x)\Bar{\phi}(\boldsymbol{x}_0,t)$. 
 Now, to write down the interaction action in the specified form, we need to  integrate by parts:
 \begin{equation}
    \rmint dt \rmint d^3x \, \boldsymbol{\zeta}_{qr}\partial_{i}\varphi\partial_{j}\Bar{\phi}=-\rmint dt\Bar{\phi}(\boldsymbol{x}_0(t))\rmint d^3x \partial_{j}\{\boldsymbol{\zeta}_{qr}\partial_{i}\varphi\}\,.
 \end{equation} 
 After that, demanding that the potential fields vanish at spatial infinity, we get the contribution to zero $\implies \text{the source term}\,J=0$ for that specific interaction.}
\\
Now we will calculate the power radiation in the centre of mass frame defined by, $$\boldsymbol{x}_1=\frac{m_2}{m_1+m_2}\boldsymbol{r} \textrm{ and } \boldsymbol{x}_2=\frac{-m_1}{m_1+m_2}\boldsymbol{r}\,.$$
It is evident that the scalar monopole moment does not contribute. We are considering a circular orbit parameterized by, $$x=r \cos(\Omega t),\quad y=r \sin(\Omega t), \quad z=0\,,$$ with the orbital frequency $\Omega$.
The source term can be calculated from the effective action by omitting the radiative fields as defined in (\ref{5.10}). The source term contributing to scalar radiation has the form
\begin{align}
    \begin{split}
        J_{\phi}=& -\sum_{a=1}^{2}m_a\,s_a\delta(\boldsymbol{x}-\boldsymbol{x}_a)+\frac{m_1m_2}{8\pi m_p^2 r}\sum_{a=1}^{2}s_a \delta(\boldsymbol{x}-\boldsymbol{x}_a)+\frac{1}{2}\sum_{a}m_as_av_a^2\delta(\boldsymbol{x}-\boldsymbol{x}_a)\\&
       +\underbrace{ \frac{m_1^2 m_2^2}{16 \pi \,m_p^2(m_1+m_2)^2}}_{\frac{\mu^2}{16\pi}}e^{-mr}r^2\Omega^2\sum_{a\ne b}^{2}g_a s_{b}\,\delta(\boldsymbol{x}-\boldsymbol{x}_a)\\ &+\frac{m_1m_2}{m_p^2} \frac{e^{-mr}}{r}\Big[1+\frac{1}{16}\Big(\frac{m_1}{M}-\nu\Big)\Omega^2\,r^2\Big]\sum_{a\ne b}g_a s_b \delta(\boldsymbol{x}-\boldsymbol{x}_a)\\&+\underbrace{\Big\{[\tilde{c}_1+\tilde{c}_2]r^2\Omega^2+\Big[\frac{\gamma_2m_1-\gamma_1m_2}{M}\Big]r^2\Omega^2+[\tilde{G}_1+\tilde{G}_2]+(\tilde{d}_1-\tilde{d}_2)\Omega^2+\Big[\frac{\Gamma_2m_1-\Gamma_1m_2}{M}\Big]\Omega^2\Big\}}_{\eta(r,\Omega)}\\& \,\,\,\,\,\,\,\,\delta(x-\boldsymbol{x}_0)\,,
    \end{split}
\end{align}
where the functions are defined below,
\begin{align}
\begin{split}
&\tilde{c}_{1,2}=\frac{-s_{2,1}\,\mu\, M f(r)}{8\pi m^2m_p^2r}\,\,\,,
    \tilde{G}_{1,2}=\frac{s_{1,2}\,\mu\, M}{8\pi m_p^2}\frac{(1-e^{-mr})}{r}\,\, ,\,\gamma_{1,2}(r)=-\frac{1}{2\pi}\frac{\mu \,M\, s_{2,1}\,f(r)}{m_p^2 m^2r},\,\\& \tilde{d}_{1,2}=\tilde{c}_{1,2}\, r,\,\,\,\,\,
    \Gamma_{1,2}=\gamma_{1,2}\, r.
    \end{split}
\end{align}
and  $f(r)$ is defined in equation (\ref{5.29}). \par
 Now the dipole moment is essentially given by,
\begin{align}
    \begin{split}
        I_{\phi}^{i}(t)&=\rmint d^3 x\,x^{i} J_{\phi}+\textcolor{black}{\frac{m^2}{10} \rmint d^3 x\,  x^i x^2\,J_{\phi} }\,,\\ &
        =\sum_{a\ne b}\underbrace{[-m_a s_a(1-\frac{v_a^2}{2})+\frac{ m_1 m_2}{8\pi m_p^2r}s_a +\frac{\mu^2 (g_a s_b)}{16\pi m_p^2}e^{-mr}r^2 \Omega^2+\frac{m_1m_2}{m_p^2}\frac{e^{-mr}}{r}\Big[1+\frac{1}{16}\Big(\frac{m_1}{M}-\nu\Big)\Omega^2\,r^2\Big]g_a s_b]}_{\mathcal{C}_{a}(r)}x_a^{i}\\ &
        +\textcolor{black}{\frac{m^2}{10}\sum_{a\ne b}\mathcal{C}_a x_a^2 x_a^{i}}+\frac{m^2}{10}\eta(r,\Omega){x}_0^i\,\boldsymbol{x}_0^2+\eta(r,\Omega){x}_0^i\,.
    \end{split}
\end{align}
In the frequency domain, the moments are given by,
\begin{align}
    \begin{split}
      &  I_{\phi}^{i}(\omega)=\rmint dt e^{i\omega t}I_{\phi}^{i}(t)\,,\\ &
        \implies I_{\phi}^{x}(\omega)=\Big[ \frac{\mathcal{C}_1 m_2-\mathcal{C}_2 m_1}{m_1+m_2}+\textcolor{black}{\frac{m^2r^2}{10}\frac{\mathcal{C}_1m_2^3-\mathcal{C}_2 m_1^3}{(m_1+m_2)^3} }+\frac{\eta(r,\Omega)}{2}+\frac{m^2}{80}\eta(r,\Omega)r^2\Big]r\,\sqrt{\frac{\pi}{2}}\delta(\omega-\Omega)\,,\\&\quad \text{and}\quad \,I_{\phi}^{y}(\omega)=i I_{\phi}^{x}(\omega),\,.
    \end{split}
\end{align}
Then the radiated power is given by \footnote{The formula for scalar power radiation matches up to some numerical factor with the result of \cite{Huang:2018pbu} due to the different choices of overall normalisation of the gravitational action. Also, the mass-dependent second term, i.e., the $\frac{m^2r^2}{10}(...)$ is not there. This is because of the mass dependence of the multipole moment, coming from the equation of motion, in (\ref{5.12mn}). },\par
\begin{tcolorbox}[thesisresultbox,, title=Scalar Radiation Power]
\restorethesisbodyformat
\begin{equation}
\begin{aligned}
        P_{\phi}&=\frac{1}{12\pi m_p^2 }\Big(\frac{\mathcal{C}_1 m_2-\mathcal{C}_2 m_1}{m_1+m_2}+\textcolor{black}{\frac{m^2 r^2}{10}\frac{\mathcal{C}_1m_2^3-\mathcal{C}_2 m_1^3}{(m_1+m_2)^3} }+\frac{\eta(r,\Omega)}{2}+\frac{m^2}{80}\eta(r,\Omega)r^2\Big)^2\\&\,\,\,\,\,\,\,\,\,\,\,\,\,\,\,\,\,\,\,\,\,\,\,\,\,\,\,\,\,\,\,\,\,\,\,\,\,\,\,\,\,\,\,\,\,\,\,\,\,\,\,\,\,\,\,\,\,\,\,\,\,\,\,\,\,\,\,\,\,\,\,\,\,\,\,\,\,\,\,\,\,\,\,\,\,\,\,\,\,\,\,\,\,\,\,\,\,\,\,\,\,\,\,\,\,\,\,\,\,\,\,\,\,\,\,\,\,\,\,\,\,\,\,\,\,\,\,\,\,\,\,\,\rmint d\omega \,\omega (\omega^2-m^2)^{3/2} r^2\delta(\omega-\Omega)\,,\\ &
        =\frac{1}{12\pi m_p^2 }\Big(\frac{\mathcal{C}_1 m_2-\mathcal{C}_2 m_1}{m_1+m_2}+\textcolor{black}{\frac{m^2r^2}{10}\frac{\mathcal{C}_1m_2^3-\mathcal{C}_2 m_1^3}{(m_1+m_2)^3} }+\frac{\eta(r,\Omega)}{2}+\frac{m^2}{80}\eta(r,\Omega)r^2\Big)^2\,r^2 \, \Omega^4 (1-\frac{m^2}{\Omega^2})^{3/2}\,.
\end{aligned}
\end{equation}
\end{tcolorbox}
Before we end this section, some comments are in order. 
\begin{itemize}
   \item  \textit{To the best of our knowledge, this is the first calculation of scalar radiation up to $N^{(4)}LO$. Leading-order computations were carried out in \cite{Huang:2018pbu, Kuntz:2019zef}}.
    \item The contribution coming from the $g_{a\gamma\gamma}$ term drops for the  orbits which are confined to a plane (e.g. for the circular orbit that we have considered in this paper), but it would have contributed otherwise for orbits embedded in 3-dimensions (e.g. some helical orbits embedded in 3d \cite{2021EPJC...81.1048L,Chen:2022qvg,Liu:2020vsy}).
\end{itemize}
\subsection{Electromagnetic interaction and the multipole decomposition}\label{emRad}
Now we consider the radiative photons with the source.  
The effective action can be written as
\begin{align}
    \begin{split}
        \mathcal{S}_{\text{eff}}=\rmint d^4 x \Big(-\frac{1}{4}F_{\mu\nu}F^{\mu\nu}+J^{\mu}\Bar{a}_{\mu}\Big)\,.
    \end{split}
\end{align}
We impose the Lorenz gauge: $$\partial_{\mu}\Bar{a}^{\mu}=0.$$ The equation of motion takes the following form,
\begin{eqnarray}
    \Box\Bar{a}_{\mu}=-J_{\mu}\,.
\end{eqnarray}

To compensate for the extra term we need to add an extra term $\frac{1}{6}\rmint d^3 x\, x^2 J^{\mu}\nabla^2\Bar{a}_{\mu}$ with the monopole term. Now use $\Box \Bar{a}_{\mu}=0$ (as we are measuring outside the source) we can see the term can be written as
\begin{align}
    \begin{split}
        S_{int}'&=\frac{1}{6}\rmint dt \rmint d^3 x\, x^2\,J^\mu (\Box+\partial_{t}^2)\Bar{a}_{\mu}\\ &
        =\frac{1}{6}\rmint dt\rmint d^3x\,\Bar{a}_{\mu}\, \partial_{t}^2 J^{\mu} x^2\,.
    \end{split}
\end{align}
We compute the effective action for the source  following the same approach (in Sec.~(\ref{subsec:radiated_scalars})) as scalar field case. By doing a similar computation like the scalar field case the imaginary part of the effective action for the source in terms of multipole moments can be found as  \cite{Kuntz:2019zef}, 
\begin{align}
    \begin{split}
       \text{Im} [\gamma_\mathrm{eff}^{(\bar a_\mu)}]&=\frac{1}{8\pi^2}\rmint dk^0\rmint d|\boldsymbol{k}|\,|\boldsymbol{k}|^2 \, \delta(k_0^2-\boldsymbol{k}^2)\Big[|\mathcal{I}^{\mu}(k_0)|^2+\frac{1}{3}|\mathcal{I}^{(\mu)ij}(k_0)|^2|\boldsymbol{k}|^2+\frac{1}{30}|\mathcal{I}^{(\mu)ij}(k_0)|^2|\boldsymbol{k}|^4\Big]\,,\\ &
    =\frac{1}{8\pi^2}\rmint dk^0\rmint d|\boldsymbol{k}|\,|\boldsymbol{k}|^2 \frac{1}{2|k^0|}[\delta(k^0+|\boldsymbol{k}|)+\delta(k^0-|\boldsymbol{k}|)]\Big[|\mathcal{I}^{\mu}(k_0)|^2+\frac{1}{3}|\mathcal{I}^{(\mu)ij}(k_0)|^2|\boldsymbol{k}|^2\\ &\hspace{3cm}
    \,\,\,\,\,\,\,\,\,\,+\frac{1}{30}|\mathcal{I}^{(\mu)ij}(k_0)|^2|\boldsymbol{k}|^4\Big]\,,\\ &
    =\frac{1}{8\pi^2}\rmint_0^{\infty} d\omega\,\Big[|\mathcal{I}^{\mu}(\omega)|^2\omega+\frac{1}{3}|\mathcal{I}^{(\mu)i}(\omega)|^2\omega^3
    +\frac{1}{30}|\mathcal{I}^{(\mu)ij}(\omega)|^2\omega^5\Big]\,.
    \label{5.42}
    \end{split}
\end{align}
 We next compute this effective action up to LO. The diagrams that contribute at this order are shown in Fig.~(\ref{Fig:12}).
\begin{itemize}
\item The $LO$ contribution comes from a diagram with the worldline vertex $\rmint dt Q_a \Bar{a}_{0}$ as shown in Fig.~(\ref{fig12a}) has the following form,
\begin{align}
    \begin{split}
        \mathcal{S}_{\text{eff}}\Big |_{\text{fig}(\ref{fig12a})}=\sum_{a}\rmint dt \,Q_a \,\Bar{a}_{0}\sim \mathcal{O}(L^{1/2}v^{1/2}).
    \end{split}
\end{align}
\begin{figure}
    \centering
    \begin{subfigure}[t]{0.22\textwidth}
      \thesisdiagrampanel{\scalebox{0.3}{\begin{feynman}
    \fermion[lineWidth=6]{4.00, 5.80}{7.40, 5.80}
    \fermion[lineWidth=6]{4.00, 4.00}{7.40, 4.00}
    \dashed[showArrow=false, lineWidth=6, label=$\Bar{a_0}$]{5.60, 5.80}{7.40, 7.20}
\end{feynman}
}}
    \caption{}
    \label{fig12a}
    \end{subfigure}
\begin{subfigure}[t]{0.22\textwidth}
    \centering
\thesisdiagrampanel{\scalebox{0.3}{\begin{feynman}
    \fermion[lineWidth=6]{4.00, 5.80}{7.40, 5.80}
    \fermion[lineWidth=6]{4.00, 4.00}{7.40, 4.00}
    \electroweak[color=9900ef, lineWidth=6, label=$\Bar{a_i}$]{5.60, 5.80}{7.40, 7.20}
\end{feynman}
}}
    \caption{}
    \label{fig12b}
\end{subfigure}
\begin{subfigure}[t]{0.22\textwidth}
    \centering
 \thesisdiagrampanel{\scalebox{0.25}{
\begin{feynman}
    \fermion[lineWidth=6]{4.00, 6.20}{7.40, 6.20}
    \electroweak[lineWidth=6, label=$\Tilde{\psi}$]{5.60, 6.20}{5.60, 4.00}
    \dashed[showArrow=false, lineWidth=6, label=$\Bar{a_0}$]{5.60, 6.20}{7.60, 7.80}
    \fermion[lineWidth=6]{4.00, 4.00}{7.40, 4.00}
\end{feynman}}}
 \caption{}
    \label{fig12d}
\end{subfigure}
\begin{subfigure}[t]{0.22\textwidth}
    \centering
 \thesisdiagrampanel{\scalebox{0.3}{\begin{feynman}
    \electroweak[label=$\tilde{\psi}$, lineWidth=6]{5.60, 5.80}{5.60, 4.00}
    \fermion[lineWidth=6]{4.00, 5.80}{7.40, 5.80}
    \fermion[lineWidth=6]{4.00, 4.00}{7.40, 4.00}
    \electroweak[color=9900ef, label=$\Bar{a}_i$, lineWidth=6]{5.60, 5.80}{7.40, 7.20}
\end{feynman}
}}
    \caption{}
    \label{fig12c}
\end{subfigure}
\caption{Radiative diagrams for electromagnetic field up to $N^{(3)}LO$. }
\label{Fig:12}
\end{figure}
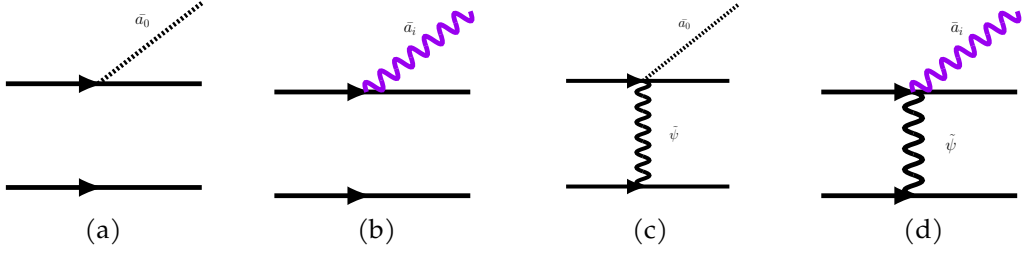

\item The diagram shown in the Fig.~(\ref{fig12b}) contributes at $N^{(1)}LO.$ It consists of a worldline vertex $\rmint dt\,Q_av_a^{i}\Bar{a}_{i}$ and the corresponding amplitude is given by,
\begin{align}
    \begin{split}
          \mathcal{S}_{\text{eff}}\Big |_{\text{fig}(\ref{fig12b})}=\sum_{a}\rmint dt \, Q_a\,v_a^{i}\, \Bar{a}_{i}(\boldsymbol{x}_a(t))\,\sim\mathcal{O}(L^{1/2}v^{3/2}).
    \end{split}
\end{align}
\item \textcolor{black}{The $N^{(2)}LO$ contribution for the source of $J^{0}$ comes from the diagram shown in Fig.~(\ref{fig12d}) with the worldline vertex:
$-\rmint dt Q_a\,\Bar{a}_{0}\tilde{\psi}$\,.} The amplitude corresponds to the diagram is: 
\begin{align}
    \begin{split} \mathcal{S}_{\text{eff}}\Big |_{\text{fig}(\ref{fig12d})}&=\frac{Q_1m_2}{m_p^2}\rmint\Bar{\mathcal{D}}\Hat{\xi}\rmint dt_1 \Bar{a}_{0}(x_1,t_1)\tilde{\psi}(x_1,t_1)\rmint dt_2 \tilde{\psi}(x_2,t_2)\,,\\ &
        =\frac{Q_1m_2}{m_p^2}\rmint dt_1\,dt_2\,\Bar{a}_{0}(x_1,t_1)\Big\langle\tilde{\psi}(x_1,t_1)\tilde{\psi}(x_2,t_2)\Big\rangle\,, \\ &
        =\frac{1}{8\pi m_p^2}\,\sum_{a\ne b}Q_a m_b\rmint dt\, \frac{1}{r}\,\Bar{a}_{0}(x_a,t)\,\sim\mathcal{O}(L^{1/2}v^{5/2}).
    \end{split}
\end{align}
\item Other radiation diagram consisting of the following world line vertices, $-\rmint dt Q_a v_a^{i}\tilde{\psi}\Bar{a}^{i}$ and $\rmint dt \,\tilde{\psi}$ as shown in Fig.~(\ref{fig12c}) contributes at $N^{(3)}LO\,.$ The corresponding amplitude is given by,
\begin{align}
    \begin{split}
          \mathcal{S}_{\text{eff}}\Big |_{\text{fig}(\ref{fig12c})}&=-Q_1m_2\rmint\Bar{\mathcal{D}}\Hat{\xi}\rmint dt_1\, v_1^{i}(t_1)\tilde{\psi}(\boldsymbol{x}_1(t_1))\Bar{a}_{i}(\boldsymbol{x}_1(t_1))\rmint dt_2 \, \tilde{\psi}(\boldsymbol{x}_2(t_2))\,,\\ &
    =-\frac{Q_1m_2}{2m_p^2}\rmint dt v_1^{i}\rmint_{\boldsymbol{k}}\frac{e^{i\boldsymbol{k}\cdot\boldsymbol{r}}}{\boldsymbol{k}^2}\,\Bar{a}_{i}(\boldsymbol{x}_1(t))\,,\\ &
    =-\frac{Q_1m_2}{2m_p^2}\rmint dt v_1^{i}\, \frac{1}{4\pi r}\,\Bar{a}_{i}(\boldsymbol{x}_1(t))+(1\leftrightarrow 2)\,\sim\mathcal{O}(L^{1/2}v^{7/2}).
    \end{split}
\end{align}
\end{itemize}
By adding all these contributions, we get the total effective action for the radiation sector. Then we can calculate the source term, and it is given by,
\begin{align}
    \begin{split}
      &  J_{a}^{0}=\frac{1}{8\pi m_p^2\,r}\sum_{a\ne b}Q_{a}m_b\delta(\boldsymbol{x}-\boldsymbol{x}_a)+\sum_{a}Q_a\delta(\boldsymbol{x}-\boldsymbol{x}_a)\,,\\ &
J_{a}^{i}=\sum_{a}Q_{a}\,v_a^{i}\delta(\boldsymbol{x}-\boldsymbol{x}_a)-\frac{1}{8\pi m_p^2\,r}\sum_{a\ne b}Q_a\, m_b \,v_a^{i}\delta(\boldsymbol{x}-\boldsymbol{x}_a)\,.
    \end{split}
\end{align}
Finally, we compute the multipole moment using this source term, and then, using (\ref{5.42}) (as well as using the optical theorem), we get the following power radiation formula. Here we have confined ourselves to the dipole term (i.e., $l=1$).\\
\begin{tcolorbox}[thesisresultbox, title=Electromagnetic Dipole Radiation]
\restorethesisbodyformat
\begin{equation}
\begin{aligned}
        P_{em}=&\frac{1}{12\pi}\Bigg\{\Bigg[\frac{(Q_1m_2-Q_2 m_1)}{M}r+\frac{(\mathcal{C}_1m_2^2-\mathcal{C}_2m_1^2)}{M}\Bigg]^2\Omega^4\\ &
        +\Bigg[-\frac{(\mathcal{D}_1m_2^2+\mathcal{D}_2m_1^2)}{M^2}r
+\frac{(\tilde{d_1}m_2^4+\tilde{d_2}m_1^4)}{M^4}r^3\Omega^2
+\frac{\Delta_1m_2^3+\Delta_2m_1^3}{M^2}\\ &
+\frac{(G\tilde{d_1}m_2^5+G\tilde{d_2}m_1^5)}{M^4}r^2\Omega^2
 \Bigg]^2 16r^2\Omega^6\Bigg\}\,.
\end{aligned}
\end{equation}
\end{tcolorbox}
 
 Here, 
 $$ \mathcal{C}_a=\frac{Q_a}{8\pi m_p^2},\mathcal{D}_a=\frac{Q_a}{2},\tilde{d_a}=\frac{Q_a}{20},\Delta_a=\frac{\mathcal{D}_a}{8\pi m_p^2}, $$ with $a=1,2.$\par 
\textit{Again, to the best of our knowledge, the electromagnetic dipole radiation has been computed previously only up to $LO$ \cite{Cardoso:2020iji}. Here we have computed the same up to  $N^{(3)}LO$.}

\subsection{Proca interaction and multipole decomposition}
Now we will focus on the Proca sector. In order to derive the power radiation formula we need to calculate the imaginary part of the effective action of the source. Following the procedure outlined in Sec.~(\ref{subsec:radiated_scalars}) we get, 
\begin{align}
    \begin{split}
        i \gamma_\mathrm{eff}^{(\bar b_\mu)}&\sim \rmint dt_1\rmint d t_2 \Big[\mathcal{I}^{(\mu)}_{b}\mathcal{I}^{(\nu)}_{b}\Big\langle T \Bar{b}_{\mu}(t_1,\boldsymbol{0})\Bar{b}_{\nu}(t_2,\boldsymbol{0})\Big\rangle+\mathcal{I}^{(\mu)i}_{b}\mathcal{I}^{(\nu)j}_{b}\Big\langle T \partial_i \Bar{b}_{\mu}(t_1,\boldsymbol{0})\partial_j \Bar{b}_{\nu}(t_2,\boldsymbol{0})\Big\rangle+\\ &
        \,\,\,\,\,\,\,\,\,\,\,\,\,\,\,\,\,\,\,\,\,\,\,\,\,\,\,\,\,\,\,\,\,\,\,\,\,\,\,\,\frac{1}{4}\mathcal{I}^{(\mu)ij}_{a}\mathcal{I}^{(\nu)kl}_{b}\Big\langle T \partial_i\partial_j\Bar{b}_{\mu}(t_1,\boldsymbol{0})\partial_k\partial_l\Bar{b}_{\nu}(t_2,\boldsymbol{0})\Big\rangle\Big]\,.\label{5.1}
    \end{split}
\end{align}
The main difference between Proca and an electromagnetic propagator is that, in the case of a Proca field, the propagator is a massive propagator. The Feynman propagator for this case is given below,
\begin{align}
    \begin{split}
        \Big\langle T \Bar{b}_{\mu}(t_1,\boldsymbol{x}_1)\Bar{b}_{\nu}(t_2,\boldsymbol{x}_2)\Big\rangle=\rmint \frac{d^4k}{(2\pi)^4}\frac{-i\,\left(g_{\mu\nu}-\frac{k_\mu k_\nu}{\mu_\gamma ^2}\right)}{-k^2-\mu_{\gamma}^2+i\epsilon}e^{-ik\cdot(x_1-x_2)}\,.\label{5.2}
    \end{split}
\end{align}
Therefore, following the same method discussed in Sec.~(\ref{subsec:radiated_scalars}), we get the imaginary part of the effective action.
\begin{align}
    \begin{split}
        \text{Im}[\gamma^{\bar{b}_{\mu}}_{\text{eff}}]\Big|_{(1)}&=\frac{1}{8\pi^2}\rmint dk_{0}\rmint d|\boldsymbol{k}|\,|\boldsymbol{k}|^2\delta(k_0^2-\boldsymbol{k}^2-\mu_{\gamma}^2)\Big[|\mathcal{I}^{\mu}(k_0)|^2+\frac{1}{3}|\mathcal{I}^{(\mu)i}(k_0)|^2|\boldsymbol{k}|^2\\ &
   \hspace{7cm}+\frac{1}{30}|\mathcal{I}^{(\mu)ij}(k_0)|^2|\boldsymbol{k}|^4\Big]
   \,,\\ &
    =\frac{1}{8\pi^2}\rmint dk_0\rmint d|\boldsymbol{k}|\,|\boldsymbol{k}|^2\frac{1}{2\sqrt{k_0^2-\mu_{\gamma}^2}}\Big[\delta(-|\boldsymbol{k}|+\sqrt{k_0^2-\mu_{\gamma}^2})+\delta(|\boldsymbol{k}|+\sqrt{k_0^2-\mu_{\gamma}^2})\Big]\Big[|\mathcal{I}^{\mu}(k_0)|^2\\ &
   \hspace{7cm}+\frac{1}{3}|\mathcal{I}^{(\mu)i}(k_0)|^2|\boldsymbol{k}|^2+\frac{1}{30}|\mathcal{I}^{(\mu)ij}(k_0)|^2|\boldsymbol{k}|^4\Big]\,,\\ &
    =\frac{1}{8\pi^2}\rmint d\omega \, \sum_{L}\frac{1}{l!(2l+1)!!}\,(\omega^2-\mu_{\gamma}^2)^{1/2+l}\,|\mathcal{I}^{(\mu)}_{L}(\omega)|^2\,.
    \label{5.50}
    \end{split}
\end{align}
The other part comes from the second part of the propagator,
\begin{align}
    \begin{split}
\text{Im}[\gamma^{\bar{b}_{\mu}}_{\text{eff}}]\Big|_{(2)}&=-\frac{1}{8\pi^2}\int dt_1 dt_2 I^{\mu}(t_1)I^{\mu}(t_2)\int dk_0\frac{e^{ik_0\,(t_1-t_2)}}{2\sqrt{k_0^2-\mu_{\gamma}^2}}\int d^3k \,\delta\left(-|\boldsymbol{k}|+\sqrt{k_0^2-\mu_{\gamma}^2}\right)\frac{k_\mu k_\nu}{\mu_{\gamma}^2}\\ &
=-\frac{1}{8\pi^2}\int \frac{d\omega}{2\sqrt{\omega^2-\mu_\gamma^2}} I^{\mu}(\omega)I^{*\nu}(\omega)\int d^3k \,\delta\left(-|\boldsymbol{k}|+\sqrt{k_0^2-\mu_{\gamma}^2}\right)\frac{k_\mu k_\nu}{\mu_{\gamma}^2}\\ &
=-\frac{1}{16\pi^2\mu_{\gamma^2}}\int d\omega \,\sqrt{\omega^2-\mu_\gamma^2}\left(\omega^2|I^{0}(\omega)|^2-\frac{1}{3}(\omega^2-\mu_\gamma^2)|I^{i}(\omega)|^2\right)
\end{split}
\end{align}
However, in our following computations, we ignore the second part of the radiative effective action as we are only looking for the leading contribution for this massive mode. Although the correction can be straightforwardly done.
Therefore, the power radiation is given by,
\begin{align}
    \begin{split}
        P_{pr}=\frac{1}{4\pi^2\mathcal{T}}\sum_{l}\frac{1}{l!(2l+1)!!}\rmint_0^\infty d\omega\,\omega\,(\omega^2-\mu_{\gamma}^2)^{l+1/2}|\mathcal{I}^{\mu}_{(L)}(\omega)|^2\,.
    \end{split}
\end{align}
 We need to calculate the two source terms, $J_{b}^{0}$ and $J_{b}^{i}$, and the computation is exactly analogous to the electromagnetic case discussed in Sec.~(\ref{emRad}). Again, we restrict ourselves to \textcolor{black}{$N^{(3)}LO$}. Below, we quote the results for the source terms. 
\begin{align}
    \begin{split}
      &  J_{b}^{0}=\frac{1}{8\pi m_p^2 r}\sum_{a\neq b}Q_{a}'m_b\delta(\boldsymbol{x}-\boldsymbol{x}_a)+\sum_{a}Q_a'\delta(\boldsymbol{x}-\boldsymbol{x}_a)\,\\ &
      J_{b}^{i}=\sum_{a}Q_{a}'\,v_a^{i}\delta(\boldsymbol{x}-\boldsymbol{x}_a)-\frac{1}{8\pi m_p^2 r}\sum_{a\ne b}Q_a'\, m_b \,v_a^{i}\delta(\boldsymbol{x}-\boldsymbol{x}_a)\,.
    \end{split}
\end{align}

Now that we have the source terms, we can use (\ref{5.50}) to calculate the multipole moments up to $N^{(3)}LO$ as follows,
 \footnote{ We use $\delta(x-a)^2=\delta(x-a)\delta(0)$.}
\begin{align}
\begin{split}
 |\mathcal{I}^{\mu}_{(L)}(\omega)|^2= &
\Bigg[\frac{({Q}_1'm_2-{Q}_2'm_1)r}{M}+\frac{\mu_{\gamma}^2}{10}\frac{({Q}_1'm_2^3-{Q}_2'm_1^3)}{M^3}r^2+\frac{(\mathcal{C}_1'm_2^2-\mathcal{C}_2'm_1^2)}{M}+\frac{\mu_{\gamma}^2}{10}\frac{(\mathcal{C}_1'm_2^3-\mathcal{C}_2'm_1^3)}{M^3}r^2\Bigg]^2\\&{\frac{\pi\delta(\omega-\Omega)\delta(0)}{2}}+\Bigg[- \frac{(\mathcal{D}_1'm_2^2+\mathcal{D}_2'm_1^2)}{M^2}r
+\frac{(\tilde{d_1}'m_2^4+\tilde{d_2}'m_1^4)}{M^4}r^3\Omega^2\textcolor{black}{(1-\frac{\mu_{\gamma}^2}{\Omega^2}})
\\&+\frac{\Delta_1'm_2^3+\Delta_2'm_1^3}{M^2} +\frac{(\tilde{d_1}'m_2^5+\tilde{d_2}'m_1^5)}{8\pi m_p^2 M^4}r^2\Omega^2\textcolor{black}{(1-\frac{\mu_{\gamma}^2}{\Omega^2}})
 \Bigg]^2\frac{\pi\, r^2\Omega^2}{2}\delta(\omega-2\Omega)\delta(0).
 \end{split}
 \end{align}

\textcolor{black}{Finally, we get the expression for the power radiation  for the Proca sector from the dipole part ($l=1$)}
 \begin{tcolorbox}[thesisresultbox, title=Proca Dipole Radiation]
 \restorethesisbodyformat
 \begin{equation}
 \begin{aligned}
      P_{\text{pr}}=&\frac{1}{12\pi} \Bigg\{\Bigg[\frac{({Q}_1'm_2-{Q}_2'm_1)r}{M}+\frac{\mu_{\gamma}^2}{10}\frac{({Q}_1'm_2^3-{Q}_2'm_1^3)}{M^3}r^2+\frac{(\mathcal{C}_1'm_2^2-\mathcal{C}_2'm_1^2)}{M}+\frac{\mu_{\gamma}^2}{10}\frac{(\mathcal{C}_1'm_2^4-\mathcal{C}_2'm_1^4)}{M^3}r^2\Bigg]^2\\ &
       \Omega^4(1-\frac{\mu_{\gamma}^2}{\Omega^2})^{\frac{3}{2}} +\Bigg[-\frac{(\mathcal{D}_1' m_2^2+\mathcal{D}_2' m_1^2)}{M^2}r
+\frac{(\tilde{d_1}'m_2^4+\tilde{d_2}'m_1^4)}{M^4}r^3\Omega^2\textcolor{black}{(1-\frac{\mu_{\gamma}^2}{\Omega^2}})
+\frac{\Delta'_1m_2^3+\Delta'_2m_1^3}{M^2}\\ &
+\frac{(\tilde{d_1}'m_2^5+\tilde{d_2}' m_1^5)}{8\pi m_p^2 M^4}r^2\Omega^2\textcolor{black}{(1-\frac{\mu_{\gamma}^2}{\Omega^2}})
 \Bigg]^2 16r^2\Omega^6(1-\frac{\mu_{\gamma}^2}{4\Omega^2})^{\frac{3}{2}}\Bigg\}\,,
\end{aligned}
\end{equation}
\end{tcolorbox}
 
 where $$ \mathcal{C}_a'=\frac{Q_a'}{8\pi m_p^2},\mathcal{D}_a'=\frac{Q_a'}{2},\tilde{d_a}'=\frac{Q_a'}{20},\Delta_a'=\frac{\mathcal{D}_a}{8\pi m_p^2}. $$\par 
\textit{Again, to the best of our knowledge, we have computed for the first time the Proca dipole radiation up to  $N^{(3)}LO$.}
\subsection{Gravitational radiation and Multipole decomposition}\label{sec5.5}
Finally, we focus on the gravitational sector. This is already studied in the \cite{Kuntz:2019zef, Levi:2018nxp}. We provide the computation for the gravitational power radiation for completeness, and also compute the corrections coming from different field vertices due to their coupling with the gravitational field at different orders of perturbation, which are explicitly mentioned. 
First, we will discuss the EFT  power counting similar to the one discussed in Sec.~(\ref{EFTcounting}) to understand which diagrams contribute to a \textcolor{black}{specific order}.\par
To see how the multipole moments scale, we need to know the scaling of the pseudo stress-energy tensor $T^{\mu\nu}.$

\begin{equation}\rmint d^3xT^{\mu\nu}\sim \begin{cases}
\begin{array}{ll}
M & (\mu=0,\nu=0),\\
Mv & (\mu=0,\nu=i),\\
Mv^2 & (\mu=i,\nu=j),\\
\end{array}
\end{cases}
\end{equation}
where $M$ denotes the mass-scale. From this we get,

\begin{equation}\rmint d^3x T^{\mu\nu}x^L\sim \begin{cases}
\begin{array}{ll}
L^{\frac{1}{2}}r^lv^{\frac{1}{2}} & (\mu=0,\nu=0),\\
L^{\frac{1}{2}}r^lv^{\frac{3}{2}} & (\mu=0,\nu=i),\\
L^{\frac{1}{2}}r^lv^{\frac{5}{2}} & (\mu=i,\nu=j),\\
\end{array}
\end{cases}
\end{equation}

where $L$ denotes the length-scale.

The formula for power radiation can be obtained in the same way as discussed in (\ref{5.18n}). First, we write down the imaginary part of the effective action.
\begin{align}
    \begin{split}
        i\gamma_{\text{eff}}^{\text{g}}=-\frac{1}{32 m_p^2}\rmint dt_1 dt_2\, I^{ij}_{g}(t_1)I^{kl}_{g}(t_2)\,\Big\langle T\ddot{\Bar{h}}_{ij}^{\text{TT}}(t_1,\boldsymbol{0})\ddot{\Bar{h}}_{kl}^{\text{TT}}(t_2,\boldsymbol{0})\Big\rangle+\cdots,\label{5.56mm}
    \end{split}
\end{align}
where ${\Bar{h}}_{ij}^{\text{TT}}$ is the radiative gravitational field in \textit{transverse-traceless (TT)} gauge and ($\cdots$) denote the higher order moments. Now use the following propagator for ${\Bar{h}}_{ij}^{\text{TT}}$ \cite{Goldberger:2004jt} to evaluate the effective action.
\begin{align}
    \begin{split}
        \Big\langle T\ddot{\Bar{h}}_{ij}^{\text{TT}}(t_1,\boldsymbol{0})\ddot{\Bar{h}}_{kl}^{\text{TT}}(t_2,\boldsymbol{0})\Big\rangle=\frac{8}{5}\Big[\frac{1}{2}(\delta_{ik}\delta_{jl}+\delta_{ik}\delta_{jl}-\frac{1}{3}\delta_{ij}\delta_{kl})\Big]\rmint \frac{d^4k}{(2\pi)^4}\frac{k_0^4}{-k^2-i\epsilon}e^{ik_0(t_1-t_2)},\label{5.57mm}
    \end{split}
\end{align}
Then using (\ref{5.57mm}) into (\ref{5.56mm}) we can write the effective action as \cite{Goldberger:2004jt,Kuntz:2019zef},
\begin{align}
    \begin{split}
        \gamma_{\text{eff}}^{\text{g}}=\frac{1}{20m_p^2}\rmint \frac{d^4 k}{(2\pi)^4}\frac{k_0^4}{-k^2-i\epsilon}|I^{ij}(k_0)|^2+\cdots.
    \end{split}
\end{align}
From this, we can write down the  power radiation in terms of \textit{quadrupole} moment,
\begin{align}
    \begin{split}
        P_{g}=\frac{2G}{5\mathcal{T}}\rmint \frac{d\omega}{2\pi}\omega^6\,|I^{ij}_{g}(\omega)|^2.
    \end{split}
\end{align}
In the rest of the section, we mainly use this expression to compute the power radiation. This can be extended systematically by including contributions from higher multipole moments. Interested readers are referred to \cite{Goldberger:2004jt} for more details. For completeness, we also write down the expression for power radiation up to the \textit{octupole} moment in the following way \cite{Goldberger:2004jt},
\begin{align}
    \begin{split}
        \mathcal{P}_{g}=\frac{2G}{5\mathcal{T}}\rmint \frac{d\omega}{2\pi}\, \Bigg\{\omega^6\, |{I}_{h}^{ij}(\omega)|^2+\frac{16\,\omega^6}{45}|J^{ij}(\omega)|^2+\frac{\omega^8}{189}|I^{ijk}(\omega)|^2+.........\Bigg\}\,.
        \label{5.65}
    \end{split}
\end{align}
 \textcolor{black}{ The explicit forms of ${I}^{ij}_{h}, \,J^{ij}$ and $I^{ijk}$ are given by:
\begin{align}
    \begin{split}
      &   {I}^{ij}_{h}(t)=\rmint d^3x \,T^{00}Q^{ij}.\\ &
      J^{ij}=-\frac{1}{2}\rmint d^3x (\epsilon^{ikl}T^{0k}x^ix^j+\epsilon^{jkl}T^{0k}x^ix^l)+.........\\ &
      I^{ijk}=\rmint d^3x (T^{00}+T^{ll})[x^ix^jx^k]^{STF}\,.
    \end{split}
\end{align}
}
\begin{figure}[b!]
\centering
\thesisdiagrampanel[2.3cm]{\scalebox{0.3}{
\begin{feynman}
    \electroweak[lineWidth=4, label=$\Bar{\psi}$]{5.60, 6.00}{7.60, 7.20}
    \fermion[lineWidth=4]{4.00, 4.00}{7.40, 4.00}
    \fermion[lineWidth=4]{4.00, 6.00}{7.40, 6.00}
\end{feynman}}}
\caption{LO radiation coming from the pure gravitational sector.}
\label{figpq}
\end{figure}
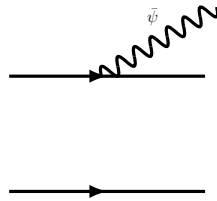

\underline{\textit{{$LO$ result for pure gravity}}}:

First, we will briefly discuss the $LO$ power radiation from the pure gravitational sector (i.e., no other field couplings are present).  Only one diagram consists of the vertices $-\sum_a m_a\rmint dt\,\psi$ as shown in Fig.~(\ref{figpq}) contributes to this case. The corresponding amplitude is given by: 
\begin{align}
    \begin{split}
S_{\text{eff}}\Big|_{{\text{fig}(\ref{figpq})}}=-\sum_{a}m_a\rmint dt\, \frac{\Bar{\psi}(\boldsymbol{x}_a(t))}{m_p}\sim\mathcal{O}(L^{1/2}v^{1/2})\,.
    \end{split}
\end{align}

Now, to compute power radiation, we first define the centre of mass velocities (at leading order) in the following way: $$\boldsymbol{v}_1=\frac{m_2}{m_1+m_2} \boldsymbol{v}\quad \& \quad \boldsymbol{v}_2=-\frac{m_1}{m_1+m_2}  \boldsymbol{v}.$$\\

Then we compute the quadrupole moment,
\begin{align}
\begin{split}
   \mathcal{I}^{ij}_h&=-\rmint d^3 x\rmint dt\underbrace{\sum_a m_a\delta(\boldsymbol{x}-\boldsymbol{x}_a)}_{T^{00}}[x^ix^j]_{STF}\,,\\&
    =-\rmint dt{\sum_a m_a}[x_a^ix_a^j]_{STF}\,.\label{5.59m}
\end{split}
\end{align}
\vspace{-0.1cm}

Using (\ref{5.59m}) one can obtain the formula for $LO$ power radiation as ,
\begin{align}
\begin{split}
{P}_g&=\frac{2G}{5}\rmint \frac{d\omega}{2\pi}\omega^6\underbrace{(\frac{\pi}{2}\mu^2r^4\delta(\omega-2\Omega))}_{|\mathcal{I}^{ij}_h|^2}\,,\\&
=\frac{32G}{5}\mu^2r^4\Omega^6.
\end{split}
\end{align}
which agrees with our known GR radiation formula in leading order \cite{2009GReGr..41.1667H}. As there are couplings of the gravitational field with other fields for our case, we will also have contributions from them to the gravitational power radiation. Next, we calculate those non-trivial corrections of the gravitational power radiation to their respective leading orders.

\vspace{0.2 cm}
\underline{\textit{{Radiation from Pure  gravity at higher order ($N^{(2)}LO$)}}}:
\begin{itemize}
\item \textcolor{black}{The diagram that contributes at $N^{(2)}LO$ is shown Fig.~($\ref{figpq}$). The corresponding amplitude is given by,
\begin{align}
\begin{split}
\mathcal{S}_\text{eff}\Big |_{\text{fig}(\ref{figpq})}=-\sum_a m_a \rmint dt \,\frac{3}{2}v_a^2\,\frac{\Bar{\psi}(\boldsymbol{x}_a(t))}{m_p}\sim\mathcal{O}(L^{1/2}v^{5/2})\,.
\end{split}
\end{align}}
\item The amplitude corresponding to the diagram in Fig.~(\ref{fig13e}) contributing at $N^{(2)}LO$ with worldline coupling $\frac{m_a}{2}\rmint dt\,\tilde{\psi}\Bar{\psi} :$
\begin{align}
    \begin{split}
         \mathcal{S}_{\text{eff}}\Big |_{\text{fig}(\ref{fig13e})}&=\frac{m_1m_2}{2m_p^2}\rmint \Bar{\mathcal{D}}\hat{\xi}\rmint dt_1 \Tilde{\psi}(\boldsymbol{x}_1(t_1))\rmint dt_2\Tilde{\psi}(\boldsymbol{x}_2(t_2))\frac{\Bar{\psi}(\boldsymbol{x}_1(t_1))}{m_p}\,\\&
         =\frac{m_1m_2}{2m_p^2}\rmint dt_1 dt_2 \Big{\langle}\tilde{\psi}(\boldsymbol{x}_1(t_1))\Tilde{\psi}(\boldsymbol{x}_2(t_2))\Big{\rangle}\frac{\Bar{\psi}(\boldsymbol{x}_1(t_1))}{m_p}\,,\\&
      =\frac{m_1m_2}{16m_p^2\pi}\rmint dt\frac{1}{r}\frac{\Bar{\psi}(\boldsymbol{x}_1(t))}{m_p}+(1\leftrightarrow 2)\,,\sim\mathcal{O}(L^{1/2}v^{5/2}).
    \end{split}
\end{align}
\end{itemize}
\underline{\textit{Radiation due to other field couplings with gravity}}\footnote{There are many diagrams contributing to the pure gravitational radiation at different $N^{(n)}LO$. But we don't calculate all of them explicitly as the results are already there in literature \cite{Blanchet:2013haa}.}:
\begin{itemize}
 \item The amplitude corresponding to the diagram in Fig.~(\ref{fig13f}) with worldline interaction vertex $-Q_a'\rmint dt\,\mathcal{B}_0\Bar{\psi}$ at $N^{(2)}LO :$
\begin{align}
    \begin{split}
         \mathcal{S}_{\text{eff}}\Big |_{\text{fig}(\ref{fig13f})}&=-Q_1'Q_2'\rmint\Bar{\mathcal{D}}\Hat{\xi}\rmint dt_1\mathcal{B}_0(\boldsymbol{x}_1(t_1))\rmint dt_2\,\mathcal{B}_0(\boldsymbol{x}_2(t_2))\frac{\Bar{\psi}(\boldsymbol{x}_1(t_1))}{m_p}\\&
       =-Q_1'Q_2'\rmint dt_1 dt_2\Big{\langle}\,\mathcal{B}_0(\boldsymbol{x}_1(t_1))\mathcal{B}_0(\boldsymbol{x}_2(t_2))\Big{\rangle}\frac{\Bar{\psi}(\boldsymbol{x}_1(t_1))}{m_p}\,, \\&
       =Q_1'Q_2'\rmint dt\frac{e^{-\mu_{\gamma} r}}{4\pi r}\frac{\Bar{\psi}(\boldsymbol{x}_1(t))}{m_p}+(1\leftrightarrow 2)\,\,\sim\mathcal{O}(L^{1/2}v^{5/2})\,.
    \end{split}
\end{align}
\end{itemize}
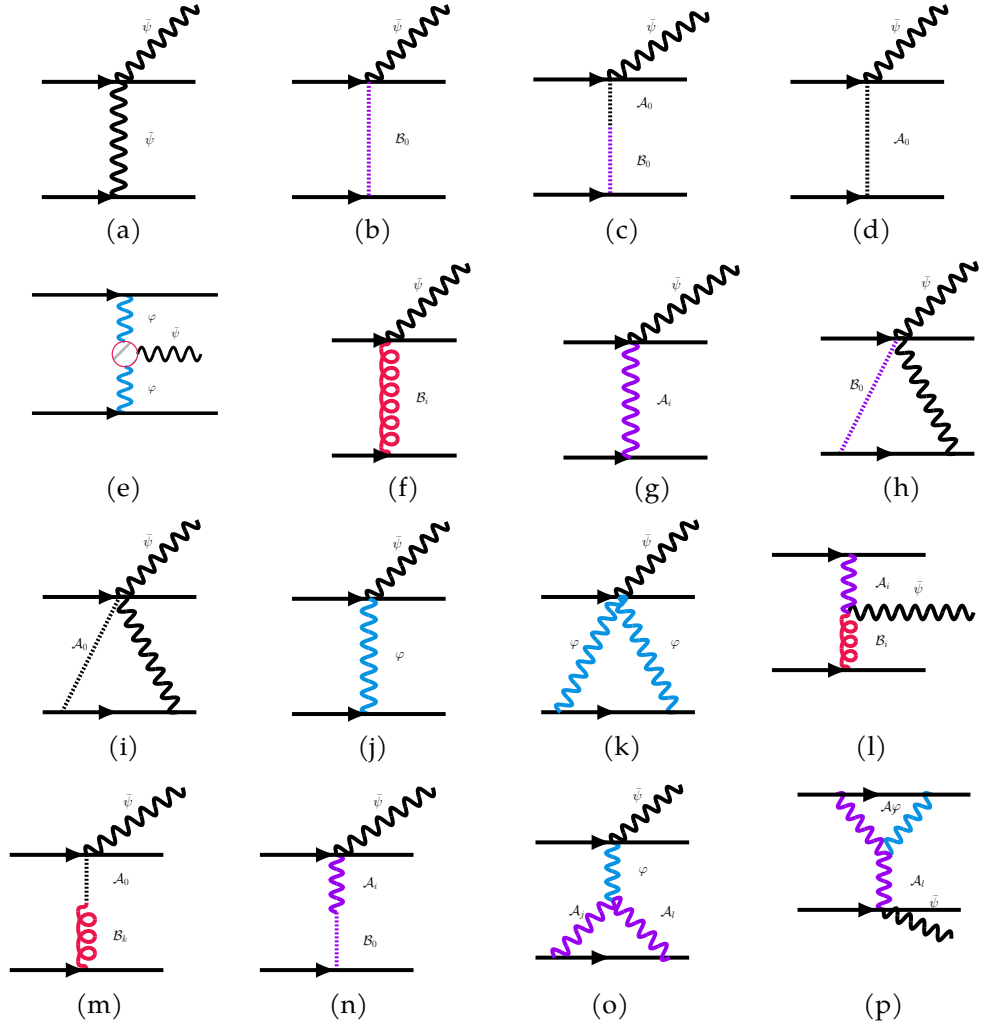
\begin{figure}
    \centering
    \begin{subfigure}[t]{0.2\textwidth}
        \centering
        \thesisdiagrampanel{\scalebox{0.25}{\begin{feynman}
    \fermion[lineWidth=6]{4.00, 6.40}{7.20, 6.40}
    \fermion[lineWidth=6]{4.00, 4.00}{7.20, 4.00}
    \electroweak[lineWidth=6, label=$\tilde{\psi}$]{5.60, 6.40}{5.60, 4.00}
\electroweak[label=$\Bar{\psi}$, lineWidth=6]{5.60, 6.40}{7.20, 8.00}
\end{feynman}
}}
\caption{}
\label{fig13e}
    \end{subfigure}
    \begin{subfigure}[t]{0.2\textwidth}
        \centering
        \thesisdiagrampanel{\scalebox{0.25}{
    \begin{feynman}
    \fermion[lineWidth=6]{4.00, 6.40}{7.20, 6.40}
    \fermion[lineWidth=6]{4.00, 4.00}{7.20, 4.00}
    \dashed[showArrow=false, lineWidth=6, color=9900ef, label=$\mathcal{B}_0$]{5.60, 6.40}{5.60, 4.00}
\electroweak[label=$\Bar{\psi}$, lineWidth=6]{5.60, 6.40}{7.20, 8.00}
\end{feynman}
}}\caption{}
\label{fig13f}
    \end{subfigure}
      \begin{subfigure}[t]{0.2\textwidth}
        \centering
        \thesisdiagrampanel{\scalebox{0.25}{
\begin{feynman}
    \dashed[showArrow=false, lineWidth=6, color=9900ef, label=$\mathcal{B}_0$]{5.60, 5.40}{5.60, 4.00}
    \fermion[lineWidth=6]{4.00, 6.40}{7.20, 6.40}
    \electroweak[lineWidth=6, label=$\Bar{\psi}$]{5.60, 6.40}{7.60, 7.80}
    \fermion[lineWidth=6]{4.00, 4.00}{7.20, 4.00}
    \dashed[lineWidth=6, showArrow=false, label=$\mathcal{A}_0$]{5.60, 6.40}{5.60, 5.40}
\end{feynman}
}}
\caption{}
\label{fig13l}
\end{subfigure}
\begin{subfigure}[t]{0.2\textwidth}
        \centering
        \thesisdiagrampanel{\scalebox{0.25}{
\begin{feynman}
    \fermion[lineWidth=6]{4.00, 6.40}{7.20, 6.40}
    \dashed[showArrow=false, lineWidth=6, label=$\mathcal{A}_0$]{5.60, 6.40}{5.60, 4.00}
    \fermion[lineWidth=6]{4.00, 4.00}{7.20, 4.00}
\electroweak[label=$\Bar{\psi}$, lineWidth=6]{5.60, 6.40}{7.20, 8.00}
\end{feynman}
}}
\caption{}
\label{fig13g}
    \end{subfigure}
    \begin{subfigure}[t]{0.25\textwidth}
    \centering
\thesisdiagrampanel{\scalebox{0.22}{\begin{feynman}
\electroweak[label=$\varphi$, lineWidth=6, color=0693e3]{6.20, 5.10}{6.20, 4.00}
    \electroweak[lineWidth=5, label=$\Bar{\psi}$]{6.50, 5.40}{8.00, 5.40}
    \fermion[lineWidth=6]{4.00, 4.00}{8.40, 4.00}
    \electroweak[lineWidth=6, color=0693e3, label=$\varphi$]{6.20, 5.70}{6.20, 6.80}
    \fermion[lineWidth=6]{4.00, 6.80}{8.40, 6.80}
    \parton[color=eb144c]{6.20,5.40}{0.30}
\end{feynman}}}
    \caption{}
\label{fig13n}
\end{subfigure}
    \begin{subfigure}[t]{0.2\textwidth}
        \centering
        \thesisdiagrampanel{\scalebox{0.25}{
        \begin{feynman}
    \gluon[color=eb144c, label=$\mathcal{B}_i$, lineWidth=6, flip=true]{5.20, 6.40}{5.20, 4.00}
    \fermion[lineWidth=6]{4.00, 4.00}{6.60, 4.00}
    \fermion[lineWidth=6]{4.00, 6.40}{6.60, 6.40}
    \electroweak[lineWidth=6, label=$\bar{\psi}$]{5.20, 6.40}{6.80, 8.00}
\end{feynman}
}}
\caption{}
\label{fig13a}
    \end{subfigure}
    \begin{subfigure}[t]{0.2\textwidth}
        \centering
        \thesisdiagrampanel{\scalebox{0.25}{
        \begin{feynman}
    \fermion[lineWidth=6]{4.00, 4.00}{7.00, 4.00}
    \electroweak[lineWidth=6, color=9900ef, label=$\mathcal{A}_i$]{5.40, 6.40}{5.40, 4.00}
    \electroweak[lineWidth=6, label=$\Bar{\psi}$]{5.40, 6.40}{7.60, 8.00}
    \fermion[lineWidth=6]{4.00, 6.40}{7.00, 6.40}
\end{feynman}
}}\caption{}
\label{fig13b}
    \end{subfigure}
\begin{subfigure}[t]{0.2\textwidth}
        \centering
        \thesisdiagrampanel{\scalebox{0.25}{
\begin{feynman}
    \fermion[lineWidth=6]{4.00, 6.40}{7.20, 6.40}
    \dashed[showArrow=false, label=$\mathcal{B}_0$, color=9900ef, lineWidth=6]{4.40, 4.00}{5.60, 6.40}
    \fermion[lineWidth=6]{4.00, 4.00}{7.20, 4.00}
    \electroweak[lineWidth=6]{5.60, 6.40}{6.80, 4.00}
\electroweak[label=$\Bar{\psi}$, lineWidth=6]{5.60, 6.40}{7.20, 8.00}
\end{feynman}
}}
\caption{}
\label{fig13c}
    \end{subfigure}
\begin{subfigure}[t]{0.2\textwidth}
        \centering
        \thesisdiagrampanel{\scalebox{0.25}{
    \begin{feynman}
    \fermion[lineWidth=6]{4.00, 6.40}{7.20, 6.40}
    \dashed[showArrow=false, lineWidth=6, label=$\mathcal{A}_0$]{4.40, 4.00}{5.60, 6.40}
    \fermion[lineWidth=6]{4.00, 4.00}{7.20, 4.00}
    \electroweak[lineWidth=6]{5.60, 6.40}{6.80, 4.00}
\electroweak[label=$\Bar{\psi}$, lineWidth=6]{5.60, 6.40}{7.20, 8.00}
\end{feynman}
}}\caption{}
\label{fig13d}
    \end{subfigure}
\begin{subfigure}[t]{0.2\textwidth}
        \centering
        \thesisdiagrampanel{\scalebox{0.25}{
    \begin{feynman}
    \fermion[lineWidth=6]{4.00, 6.40}{7.20, 6.40}
    \electroweak[lineWidth=6, color=0693e3, label=$\varphi$]{5.60, 6.40}{5.60, 4.00}
    \fermion[lineWidth=6]{4.00, 4.00}{7.20, 4.00}
\electroweak[label=$\Bar{\psi}$, lineWidth=6]{5.60, 6.40}{7.20, 8.00}
\end{feynman}
}}\caption{}
\label{fig13h}
    \end{subfigure}
\begin{subfigure}[t]{0.2\textwidth}
        \centering
        \thesisdiagrampanel{\scalebox{0.25}{
\begin{feynman}
    \fermion[lineWidth=6]{4.00, 6.40}{7.20, 6.40}
    \electroweak[lineWidth=6, color=0693e3, label=$\varphi$]{5.60, 6.40}{6.80, 4.00}
    \fermion[lineWidth=6]{4.00, 4.00}{7.20, 4.00}
    \electroweak[lineWidth=6, label=$\varphi$, color=0693e3]{4.40, 4.00}{5.60, 6.40}
\electroweak[label=$\Bar{\psi}$, lineWidth=6]{5.60, 6.40}{7.20, 8.00}
\end{feynman}
}}
\caption{}
\label{fig13i}
    \end{subfigure}
\begin{subfigure}[t]{0.2\textwidth}
        \centering
        \thesisdiagrampanel{\scalebox{0.25}{
\begin{feynman}
    \gluon[flip=true, lineWidth=6, label=$\mathcal{B}_i$, color=eb144c]{5.60, 5.20}{5.60, 4.00}
    \fermion[lineWidth=6]{4.00, 6.40}{7.20, 6.40}
    \electroweak[lineWidth=6, label=$\Bar{\psi}$]{5.60, 5.20}{8.20, 5.20}
    \fermion[lineWidth=6]{4.00, 4.00}{7.20, 4.00}
    \electroweak[lineWidth=6, color=9900ef, label=$\mathcal{A}_i$]{5.60, 6.40}{5.60, 5.20}
\end{feynman}
}} 
\caption{}
\label{fig13j}
\end{subfigure}
\begin{subfigure}[t]{0.2\textwidth}
        \centering
        \thesisdiagrampanel{\scalebox{0.25}{\begin{feynman}
    \gluon[lineWidth=6, label=$\mathcal{B}_k$, color=eb144c, flip=true]{5.60, 5.40}{5.60, 4.00}
    \fermion[lineWidth=6]{4.00, 6.40}{7.20, 6.40}
    \dashed[lineWidth=6, label=$\mathcal{A}_0$, showArrow=false]{5.60, 6.40}{5.60, 5.40}
    \electroweak[lineWidth=6, label=$\Bar{\psi}$]{5.60, 6.40}{7.60, 7.80}
    \fermion[lineWidth=6]{4.00, 4.00}{7.20, 4.00}
\end{feynman}
}}
\caption{}
\label{fig13k}
\end{subfigure}
\begin{subfigure}[t]{0.2\textwidth}
        \centering
        \thesisdiagrampanel{\scalebox{0.25}{
\begin{feynman}
    \dashed[showArrow=false, lineWidth=6, color=9900ef, label=$\mathcal{B}_0$]{5.60, 5.20}{5.60, 4.00}
    \electroweak[lineWidth=6, label=$\mathcal{A}_i$, color=9900ef]{5.60, 5.20}{5.60, 6.40}
    \fermion[lineWidth=6]{4.00, 6.40}{7.20, 6.40}
    \electroweak[lineWidth=6, label=$\Bar{\psi}$]{5.60, 6.40}{7.60, 7.80}
    \fermion[lineWidth=6]{4.00, 4.00}{7.20, 4.00}
\end{feynman}
}}
\caption{}
\label{fig13m}
\end{subfigure}
\begin{subfigure}[t]{0.22\textwidth}
    \centering
\thesisdiagrampanel{\scalebox{0.25}{
\begin{feynman}
    \electroweak[lineWidth=6, label=$\bar{\psi}$]{5.60, 6.40}{7.00, 7.60}
    \fermion[lineWidth=6]{4.00, 4.00}{7.20, 4.00}
    \electroweak[color=9900ef, lineWidth=6, label=$\mathcal{A}_l$]{5.60, 5.20}{6.80, 4.00}
    \electroweak[color=0693e3, lineWidth=6, label=$\varphi$]{5.60, 6.40}{5.60, 5.20}
    \electroweak[color=9900ef, lineWidth=6, label=$\mathcal{A}_j$]{4.40, 4.00}{5.60, 5.20}
    \fermion[lineWidth=6]{4.00, 6.40}{7.20, 6.40}
\end{feynman}}}
 \caption{}
\label{fig13o}
\end{subfigure}
\begin{subfigure}[t]{0.22\textwidth}
        \centering
        \thesisdiagrampanel{\scalebox{0.25}{\begin{feynman}
    \electroweak[flip=true, lineWidth=6, label=$\varphi$, color=0693e3]{5.80, 5.80}{6.80, 7.00}
    \electroweak[lineWidth=6, color=9900ef, label=$\mathcal{A}_l$]{5.80, 5.80}{5.80, 4.60}
    \electroweak[color=9900ef, lineWidth=6, label=$\mathcal{A}_j$]{4.80, 7.00}{5.80, 5.80}
    \fermion[lineWidth=6]{4.00, 4.60}{7.40, 4.60}
    \electroweak[lineWidth=6, label=$\bar{\psi}$, flip=true]{5.80, 4.60}{7.20, 4.00}
    \fermion[lineWidth=6]{4.00, 7.00}{7.60, 7.00}
\end{feynman}
}}
 \caption{}
\label{fig13p}
\end{subfigure}
\caption{Radiative diagrams for gravitational fields.}
\label{Fig15}
\end{figure}

\begin{itemize}

\item  The amplitude corresponding to the diagram in Fig.~(\ref{fig13l}) contributing at $N^{(2)}LO$ with the following bulk interaction vertex $\rmint d^4x\,\partial_k \mathcal{A}_0\partial_k \mathcal{B}_0 :$ 
\vspace{-0.5cm}
 \begin{align}
    \begin{split}
\mathcal{S}_{\text{eff}}\Big|_{\text{fig}(\ref{fig13l})}&=-\gamma Q_1 Q_2'\rmint \Bar{\mathcal{D}}\hat{\xi}\rmint dt_1\mathcal{A}_0(\boldsymbol{x}_1(t_1))\rmint dt_2\mathcal{B}_0(\boldsymbol{x}_2(t_2))\rmint d^4x\,\partial_k\mathcal{A}_0(x)\partial_k\mathcal{B}_0(x)\frac{\Bar{\psi}}{m_p}\,,\\&=-\gamma Q_1 Q_2'\rmint dt_1dt_2\rmint d^4x\partial_k\Big{\langle}\mathcal{A}_0(x)\mathcal{A}_0(\boldsymbol{x}_1(t_1))\Big{\rangle}\partial_k\Big{\langle}\mathcal{B}_0(x)\mathcal{B}_0(\boldsymbol{x}_2(t_2))\Big{\rangle}\frac{\Bar{\psi}}{m_p}\,,\\ &
=-\gamma Q_1Q_2'\rmint dt\,\frac{e^{-\mu_{\gamma}r}}{4\pi r}\Big(\frac{\Bar{\psi}(\boldsymbol{x}_1(t))}{m_p}+\frac{\Bar{\psi}(\boldsymbol{x}_2(t))}{m_p}\Big)+(1\leftrightarrow 2)\,\,\sim \mathcal{O}(L^{1/2}v^{5/2}).
    \end{split}
\end{align}
\item The amplitude corresponding to the diagram in Fig.~(\ref{fig13g})   contributing at $N^{(2)}LO$ with the worldline interaction vertex $-Q_a\rmint dt\,\mathcal{A}_0\Bar{\psi} :$
\begin{align}
    \begin{split}
         \mathcal{S}_{\text{eff}}\Big|_{\text{fig}(\ref{fig13g})}&=-Q_1Q_2\rmint \Bar{\mathcal{D}}\hat{\xi}\rmint dt_1dt_2\mathcal{A}_0(\boldsymbol{x}_1(t_1))\mathcal{A}_0(\boldsymbol{x}_2(t_2))\frac{\Bar{\psi}(\boldsymbol{x}_1(t_1))}{m_p}\,,\\&
          =-Q_1Q_2\rmint dt_1 dt_2\Big{
       \langle}\mathcal{A}_0(\boldsymbol{x}_1(t_1))\mathcal{A}_0(\boldsymbol{x}_2(t_2))\Big{\rangle}\frac{\Bar{\psi}(\boldsymbol{x}_1(t_1))}{m_p} \,,\\&
       =Q_1Q_2\rmint dt\frac{1}{4\pi r}\frac{\Bar{\psi}(\boldsymbol{x}_1(t))}{m_p}\,+(1\leftrightarrow 2)\,\sim\mathcal{O}(L^{1/2}v^{5/2})\,.
    \end{split}
\end{align}
\item Diagram as shown in Fig.~(\ref{fig13n}) consists of a scalar propagator with relativistic correction, contributing to the gravitational radiation at $N^{(4)}LO.$ The amplitude is given by, 
\begin{align}
    \begin{split}
  \mathcal{S}_{\text{eff}}^{ikm}\Big |_{{\text{fig}(\ref{fig13n})}} &=\frac{m_1m_2 s_1 s_2}{m_p^3}\rmint \Bar{\mathcal{D}}\hat\xi\rmint dt_1 \varphi(\boldsymbol{x}_1(t_1))\bar{\psi}(x_0) \rmint d^4 x\partial_0\varphi(x)\partial_0\varphi(x)\rmint dt_2 \varphi(\boldsymbol{x}_2(t_2))\,, \\&
=\frac{m_1 m_2 s_1 s_2}{16 \pi m_p^2}\rmint dt \,e^{-mr}\Bigg\{\Big(\frac{\boldsymbol{v}_1\cdot\boldsymbol{v}_2}{r}-\frac{(\boldsymbol{v}_1\cdot \hat{n})(\boldsymbol{v}_2\cdot \hat{n})}{r}\Big)+\frac{m}{2}\,(\boldsymbol{v}_2\cdot \Hat{n})(\boldsymbol{v}_1\cdot \hat{n})\Bigg\}\,\frac{\Bar{\psi}(\boldsymbol{x}_0(t)}{m_p}\\&+T(\Dot{\Bar{\psi}})+ (1 \leftrightarrow 2)
\sim \mathcal{O}({L^{1/2}v^{9/2}}),
\end{split}
\end{align}
where $T(\dot{\bar{\psi}})$ is given in  equation (\ref{5.26}) and we just need to replace $\phi$ by $\psi$. 
\begin{align}
    \begin{split}
        T(\Dot{\Bar{\psi}})&=-\frac{m_1 m_2 s_1 s_2}{16\pi m_p^2}\rmint dt\, e^{-mr}(\boldsymbol{v}_2\cdot \Hat{n})\frac{ \Dot{\Bar{\psi}}(\boldsymbol{x}_0(t))}{m_p},\\ &
        =\frac{m_1 m_2 s_1 s_2}{16\pi m_p^2}\rmint dt \frac{e^{-mr}}{r}(\boldsymbol{v}_2\cdot \Dot{\boldsymbol{r}}+\Vec{a}_2\cdot \boldsymbol{r})\,\frac{\Bar{\psi}(\boldsymbol{x}_0(t))}{m_p}\,.
    \end{split}
\end{align}
\item \textcolor{black}{The amplitude corresponding to the diagram with worldline interaction vertex $Q_a'\rmint dt\,v_a\Bar{\psi}\mathcal{B}_i$ in Fig.~(\ref{fig13a}) which contributes at $N^{(4)}LO :$ }
\begin{align}
    \begin{split}
         \mathcal{S}_{\text{eff}}\Big |_{\text{fig}(\ref{fig13a})}&=-Q_1' Q_2'\rmint\Bar{\mathcal{D}}\Hat{\xi}\rmint dt_1\,v_1^i(t_1)\mathcal{B}_i(\boldsymbol{x}_1(t_1))\rmint dt_2 \,v_2^j(t_2)\mathcal{B}_j(\boldsymbol{x}_2(t_2))\frac{\Bar{\psi}(\boldsymbol{x}_1(t_1))}{m_p}\,,\\ &
         =-Q_1'Q_2'\rmint dt_1 dt_2 v_1^i(t_1)v_2^{j}(t_2)\Big\langle\mathcal{B}_i(\boldsymbol{x}_1(t_1))\mathcal{B}_j(\boldsymbol{x}_2(t_2))\Big\rangle\,\frac{\bar\psi(\boldsymbol{x}_1(t_1))}{m_p},\\ &
       =\frac{-Q_1'Q_2'}{4\pi }\rmint dt \,(\boldsymbol{v}_1 \cdot \boldsymbol{v}_2)\frac{e^{-\mu r}}{r}\frac{\Bar{\psi}(\boldsymbol{x}_1(t))}{m_p}+(1\leftrightarrow 2)\,\sim \mathcal{O}(L^{1/2}v^{9/2}).
    \end{split}
\end{align}
\item The amplitude corresponding to the diagram with the worldline vertex $Q_a\rmint dt\,v_a\Bar{\psi}\mathcal{A}_i$ in Fig.~(\ref{fig13b}) which contributes at $N^{(4)}LO :$ 
\begin{align}
    \begin{split}
         \mathcal{S}_{\text{eff}}\Big |_{\text{fig}(\ref{fig13b})}&=-Q_1 Q_2\rmint\Bar{\mathcal{D}}\Hat{\xi}\rmint dt_1\,v_1^i(t_1)\mathcal{A}_i(\boldsymbol{x}_1(t_1))\rmint dt_2 \,v_2^j(t_2)\mathcal{A}_j(\boldsymbol{x}_2(t_2))\frac{\Bar{\psi}(\boldsymbol{x}_1(t_1))}{m_p}\,,\\ &
         =-Q_1Q_2\rmint dt_1 dt_2 v_1^i(t_1)v_2^{j}(t_2)\Big\langle\mathcal{A}_i(\boldsymbol{x}_1(t_1))\mathcal{A}_j(\boldsymbol{x}_2(t_2))\Big\rangle\,\frac{\bar\psi(\boldsymbol{x}_1(t_1))}{m_p},\\ &
       =\frac{-Q_1Q_2}{4\pi }\rmint dt \,\frac{\boldsymbol{v}_1 \cdot \boldsymbol{v}_2}{r}\frac{\Bar{\psi}(\boldsymbol{x}_1(t))}{m_p}+(1\leftrightarrow 2)\,\sim \mathcal{O}(L^{1/2}v^{9/2}).
    \end{split}
\end{align}
\item The amplitude corresponding to the diagram in Fig.~(\ref{fig13c}) with worldline interaction vertex $Q_a'\rmint dt\,\mathcal{B}_0\Tilde{\psi}\Bar{\psi}$ which contributes at $N^{(4)}LO :$ 
\begin{align}
    \begin{split}
         \mathcal{S}_{\text{eff}}\Big |_{\text{fig}(\ref{fig13c})}&=-\frac{m_2Q_1'Q_2'}{m_p^2}\rmint\Bar{\mathcal{D}}\Hat{\xi}\rmint dt_1\mathcal{B}_0(\boldsymbol{x}_1(t_1))\Tilde{\psi}(\boldsymbol{x}_1(t_1))\rmint dt_2\Tilde{\psi}(\boldsymbol{x}_2(t_2)\rmint dt_3\mathcal{B}_0(\boldsymbol{x}_2(t_3))\frac{\Bar{\psi}(\boldsymbol{x}_1(t_1))}{m_p}\\&
         =-\frac{m_2Q_1'Q_2'}{m_p^2}'\rmint dt_1dt_2dt_3\Big{\langle}\Tilde{\psi}(\boldsymbol{x}_2(t_2)\Tilde{\psi}(\boldsymbol{x}_1(t_1))\Big{\rangle}\Big{\langle}\mathcal{B}_0(\boldsymbol{x}_2(t_3))\mathcal{B}_0(\boldsymbol{x}_1(t_1))\Big{\rangle}\frac{\Bar{\psi}(\boldsymbol{x}_1(t_1))}{m_p}\,,\\&
       =-\frac{m_2Q_1'Q_2'}{32m_p^2\pi^2}\rmint dt \frac{e^{-\mu r}}{r^2}\frac{\Bar{\psi}(\boldsymbol{x}_1(t))}{m_p}\,+(1\leftrightarrow 2)\,\sim\mathcal{O}(L^{1/2}v^{9/2}).
    \end{split}
\end{align}
\item The amplitude corresponding to the diagram in Fig.~(\ref{fig13d}) with worldline coupling $Q_a\rmint dt\,\mathcal{A}_0\tilde\psi\Bar{\psi}$ which contributes at $N^{(4)}LO :$ 
\begin{align}
    \begin{split}
         \mathcal{S}_{\text{eff}}\Big |_{\text{fig}(\ref{fig13d})}&=-\frac{m_2Q_1Q_2}{m_p^2}\rmint\Bar{\mathcal{D}}\Hat{\xi}\rmint dt_1\mathcal{A}_0(\boldsymbol{x}_1(t_1))\Tilde{\psi}(\boldsymbol{x}_1(t_1))\rmint dt_2\Tilde{\psi}(\boldsymbol{x}_2(t_2)\rmint dt_3\mathcal{A}_0(\boldsymbol{x}_2(t_3))\frac{\Bar{\psi}(\boldsymbol{x}_1(t_1))}{m_p}\\&=-\frac{m_2Q_1Q_2}{m_p^2}\rmint dt_1dt_2dt_3\Big{\langle}\tilde{\psi}(\boldsymbol{x}_2(t_2)\tilde{\psi}(\boldsymbol{x}_1(t_1))\Big{\rangle}\Big{\langle}\mathcal{A}_0(\boldsymbol{x}_2(t_3)\mathcal{A}_0(\boldsymbol{x}_1(t_1))\Big{\rangle}\frac{\Bar{\psi}(\boldsymbol{x}_1(t_1))}{m_p}\,,\\&
       =-\frac{m_2Q_1Q_2}{32m_p^2\pi^2}\rmint dt \frac{1}{r^2}\frac{\Bar{\psi}(\boldsymbol{x}_1(t))}{m_p}+(1\leftrightarrow 2)\,\,\sim\mathcal{O}(L^{1/2}v^{9/2})\,.
    \end{split}
\end{align}
 \item The amplitude corresponding to the diagram in Fig.~(\ref{fig13h}) is given by: 
\begin{align}
    \begin{split}
         \mathcal{S}_{\text{eff}}\Big |_{\text{fig}(\ref{fig13h})}&=\frac{m_1m_2s_1s_2}{m_p^2}\rmint \Bar{\mathcal{D}}\hat{\xi}\rmint dt_1  \Big(\frac{1}{m_p}+\frac{3v_1^2}{2m_p}\Big)\varphi(\boldsymbol{x}_1(t_1)){\Bar{\psi}(\boldsymbol{x}_1(t_1))}\rmint dt_2\varphi(\boldsymbol{x}_1(t_2))\,,\\&
       =\frac{m_1m_2s_1s_2}{m_p^2}\rmint dt_1 dt_2 \Big(\frac{1}{m_p}+\frac{3v_1^2}{2m_p}\Big) \Big{\langle}\varphi(\boldsymbol{x}_1(t_1)) \varphi(\boldsymbol{x}_2(t_2))\Big{\rangle}{\Bar{\psi}(\boldsymbol{x}_1(t_1))}\,,\\&
       =\frac{m_1m_2s_1s_2}{m_p^2} \rmint dt_1 dt_2 \Big(\frac{1}{m_p}+\frac{3v_1^2}{2m_p}\Big)\delta(t_1-t_2)\rmint \frac{d^3k}{(2\pi)^3}\frac{e^{i\boldsymbol{k}\cdot\boldsymbol{r}}}{\boldsymbol{k}^2+m^2}{\Bar{\psi}(\boldsymbol{x}_1(t_1))}\,,\\ &
       =\frac{m_1m_2s_1s_2}{m_p^2} \rmint dt \Big(1+\frac{3v_1^2}{2}\Big) \frac{1}{4\pi r}e^{-mr}\frac{\Bar{\psi}(\boldsymbol{x}_1(t))}{m_p}+(1\leftrightarrow 2)\,,\\&\sim \mathcal{O}(L^{1/2}v^{5/2})+\mathcal{O}(L^{1/2}v^{9/2})\,.
    \end{split}
\end{align}
Note that, a part of it contributes at $N^{(2)}LO$ and the other part at $N^{(4)}LO.$ 
 \item Amplitude corresponding to the diagram in Fig.~(\ref{fig13i})  with the radiation coming from a triangle vertex $\phi\phi\Bar{\psi}$  is given by: 
\begin{align}
    \begin{split}
         \mathcal{S}_{\text{eff}}\Big |_{\text{fig}(\ref{fig13i})}&=\frac{m_1m_2^2s_2^2g_1}{m_p^4}\rmint \Bar{\mathcal{D}}\hat{\xi}\rmint dt_1\Big(1+\frac{3}{2}v_1^2\Big)\varphi(\boldsymbol{x}_1(t_1))\varphi(\boldsymbol{x}_1(t_1))\rmint dt_2 \varphi(\boldsymbol{x}_2(t_2)) \,\, \\& \,\,\,\,\,\,\,\,\,\,\,\,\,\,\,\,\,\,\,\,\,\,\,\,\,\,\,\,\,\,\,\,\,\,\,\,\,\,\,\,\,\,\,\,\,\,\,\,\,\,\,\,\,\,\,\,\,\,\,\,\,\,\,\,\,\,\,\,\,\,\,\,\,\,\,\,\,\,\,\,\,\,\,\,\,\,\,\,\,\,\,\,\,\,\,\,\,\,\,\,\,\,\,\,\,\,\,\,\,\,\,\,\,\,\,\,\,\,\,\,\,\,\,\,\,\,\rmint dt_3\varphi(\boldsymbol{x}_2(t_3)) \frac{\Bar{\psi}(\boldsymbol{x}_1(t_1))}{m_p}\,,\\&
         =\frac{m_1m_2^2s_2^2g_1}{m_p^4}\rmint dt_1 dt_2 dt_3\Big(1+\frac{3}{2}v_1^2\Big)\Big{\langle}\varphi(\boldsymbol{x}_1(t_1)) \varphi(\boldsymbol{x}_2(t_2))\Big{\rangle}\Big{\langle}\varphi(\boldsymbol{x}_1(t_1)) \varphi(\boldsymbol{x}_2(t_3))\Big{\rangle}\\& \,\,\,\,\,\,\,\,\,\,\,\,\,\,\,\,\,\,\,\,\,\,\,\,\,\,\,\,\,\,\,\,\,\,\,\,\,\,\,\,\,\,\,\,\,\,\,\,\,\,\,\,\,\,\,\,\,\,\,\,\,\,\,\,\,\,\,\,\,\,\,\,\,\,\,\,\,\,\,\,\,\,\,\,\,\,\,\,\,\,\,\,\,\,\,\,\,\,\,\,\,\,\,\,\,\,\,\,\,\,\,\,\,\,\,\,\,\,\,\,\,\,\,\,\,\, \frac{\Bar{\psi}(\boldsymbol{x}_1(t_1))}{m_p}\,,\\&
       =\frac{m_1m_2^2s_2^2g_1}{16\pi^2m_p^4}\rmint dt \Big(1+\frac{3}{2}v_1^2\Big)\frac{e^{-2mr}}{r^2}\frac{\Bar{\psi}(\boldsymbol{x}_1(t))}{m_p}+(1\leftrightarrow 2)\,,\\&\sim \mathcal{O}(L^{1/2}v^{9/2})+\mathcal{O}(L^{1/2}v^{13/2}).
    \end{split}
\end{align}
Again note that a part of it is contributing at $N^{(4)}LO$ and the other part is contributing at $N^{(6)}LO.$
\item The amplitude corresponds to the diagram in Fig.~(\ref{fig13j}) which contributes at $N^{(4)}LO$ with the bulk vertex $\rmint d^4x\,\psi \partial_l\mathcal{A}_i\partial_l\mathcal{B}_i :$ 
\vspace{-0.3cm}
\begin{align}
    \begin{split}
   \mathcal{S}_{\text{eff}}\Big |_{\text{fig}(\ref{fig13j})}&=-4\gamma Q_1Q_2'\rmint \Bar{\mathcal{D}}\hat{\xi}\rmint dt_1\mathcal{A}_j(\boldsymbol{x}_1(t_1))v_1^j\rmint dt_2\mathcal{B}_m(\boldsymbol{x}_2(t_2))v_2^m(t_2)\,\\&  \,\,\,\,\,\,\,\,\,\,\,\,\,\,\,\,\,\,\,\,\,\,\,\,\,\,\,\,\,\, \,\,\,\,\,\,\,\,\,\,\,\,\,\,\,\,\,\,\,\,\,\,\,\,\,\,\,\,\,\, \,\,\,\,\,\,\,\,\,\,\,\,\,\,\,\,\,\,\,\,\,\,\,\,\,\,\,\,\,\, \,\,\,\,\,\,\,\,\,\,\,\,\,\,\,\,\,\,\,\,\,\,\,\, \rmint d^4x\partial_l\mathcal{A}_i\partial_l\mathcal{B}_i \frac{ \Bar{\psi}(\boldsymbol{x}_0(t))}{m_p}\\&=-4\gamma Q_1Q_2'\rmint \prod_{i=1}^2dt_iv_1^j(t_1)v_2^m(t_2)\rmint d^4x\partial_l\Big{\langle}\mathcal{A}_i(x)\mathcal{A}_j(\boldsymbol{x}_1(t_1))\Big{\rangle}\\&\hspace{6cm}\partial_l\Big{\langle}\mathcal{B}_i(x)\mathcal{B}_m(\boldsymbol{x}_2(t_2))\Big{\rangle}\frac{\Bar{\psi}(\boldsymbol{x}_0(t))}{m_p}\,,\\ &
=-4\gamma Q_1Q_2'\rmint dt\, \,\Bigg[\boldsymbol{v}_1(t)\cdot \boldsymbol{v}_2(t)\rmint_k \frac{e^{-i\boldsymbol{k}\cdot \boldsymbol{r}}}{(\boldsymbol{k}^2+\mu^2)}\Bigg]                \frac{\Bar{\psi}(\boldsymbol{x}_0(t))}{m_p}\,,\\ &
=-\frac{\gamma Q_1 Q_2'}{\pi}\rmint dt\, \boldsymbol{v}_1\cdot \boldsymbol{v}_2\, \frac{e^{-\mu_{\gamma}r}}{r}\Big(\frac{\Bar{\psi}(\boldsymbol{x}_{0}(t))}{m_p}\Big)+(1\leftrightarrow 2)\,\sim \mathcal{O}(L^{1/2}v^{9/2}).
\end{split}
\end{align}
\item \textcolor{black}{ The amplitude that governs the effective action corresponding to the bulk interaction vertex $\rmint d^4x\,\psi \partial_l\mathcal{A}_i\partial_i\mathcal{B}_l$ at $N^{(4)}LO$,} 
\begin{align}
    \begin{split}
   \mathcal{S}_{\text{eff}}\Big |_{\text{fig}(\ref{fig13j})}&=-4\gamma Q_1Q_2'\rmint \Bar{\mathcal{D}}\hat{\xi}\rmint dt_1\mathcal{A}_j(\boldsymbol{x}_1(t_1))v_1^j\rmint dt_2\mathcal{B}_m(\boldsymbol{x}_2(t_2))v_2^m(t_2)\\& \,\,\,\,\,\,\,\,\,\,\,\,\,\,\,\,\,\,\,\,\,\,\,\,\,\,\,\,\,\, \,\,\,\,\,\,\,\,\,\,\,\,\,\,\,\,\,\,\,\,\,\,\,\,\,\,\,\,\,\, \,\,\,\,\,\,\,\,\,\,\,\,\,\,\,\,\,\,\,\,\,\,\,\,\,\,\,\,\,\,\,\,\,\,\,\,\,\,\,\,\,\,\,\,\,\,\,\,\,\,\,\,\,\,\,\,\,\,\,\,  \rmint d^4x\,\partial_l\mathcal{A}_i\partial_i\mathcal{B}_l \frac{ \Bar{\psi}(\boldsymbol{x}_0(t))}{m_p}\\&
   =-4\gamma\textstyle{Q_1 Q_2'\rmint dt_1dt_2v_1^j(t_1)v_2^m(t_2)\rmint d^4x\partial_l\Big{\langle}\mathcal{A}_i(x)\mathcal{A}_j(\boldsymbol{x}_1(t_1))\Big{\rangle}}\\&\hspace{6cm}\textstyle{\partial_i\Big{\langle}\mathcal{B}_l(x)\mathcal{B}_m(\boldsymbol{x}_2(t_2))\Big{\rangle}\frac{\Bar{\psi}(\boldsymbol{x}_0(t))}{m_p}}\,,\\ &
=-4\gamma Q_1 Q_2'\rmint dt \, \Big[\boldsymbol{v}_1\cdot \boldsymbol{v}_2\frac{e^{-r \mu_{\gamma }} \left(-r \mu _{\gamma }+e^{r \mu _{\gamma }}-1\right)}{2 \pi  r^3 \mu_{\gamma }^2}\\ & \hspace{1cm}
+(\boldsymbol{v}_1\cdot \Hat{n})(\boldsymbol{v}_2\cdot \Hat{n})\frac{e^{-r \mu _{\gamma }} \Big(r \mu _{\gamma } \left(r \mu _{\gamma }+3\right)-3 e^{r \mu _{\gamma }}+3\Big)}{2 \pi  r^3 \mu _{\gamma }^2}\Big]\Big(\frac{\Bar{\psi}(\boldsymbol{x}_0(t))}{m_p}\Big)+(1\leftrightarrow 2)\,\\&\sim\mathcal{O}(L^{1/2}v^{9/2}).
\end{split} 
\end{align}
\vspace{-0.6cm}
\item The amplitude corresponds to the diagram in Fig.~(\ref{fig13k}) that contributes at $N^{(4)}LO$ with the bulk interaction vertex $\rmint d^4x\,\partial_k\mathcal{A}_0\partial_0 \mathcal{B}_k :$
\begin{align}
   \mathcal{S}_{\text{eff}}\Big|_{\text{fig}(\ref{fig13k})}
   &=-\gamma Q_1Q_2'\rmint \Bar{\mathcal{D}}\hat{\xi}\rmint dt_1\mathcal{A}_0(\boldsymbol{x}_1(t_1))\rmint dt_2\mathcal{B}_j(\boldsymbol{x}_2(t_2))v_2^j(t_2)\rmint d^4x \partial_k\mathcal{A}_0\partial_0\mathcal{B}_k(x) \frac{ \Bar{\psi}}{m_p}\,,\notag\\
   &=-\gamma Q_1Q_2'\rmint dt_1dt_2v_2^j(t_2)\rmint d^4x\partial_k\Big{\langle}\mathcal{A}_0(x)\mathcal{A}_0(\boldsymbol{x}_1(t_1))\Big{\rangle}\partial_0\Big{\langle}\mathcal{B}_k(x) \mathcal{B}_j(\boldsymbol{x}_2(t_2))\Big{\rangle}\frac{\Bar{\psi}}{m_p}\,,\notag\\
   &=\frac{\gamma Q_1Q_2'}{2\pi}\rmint dt \Bigg[(\Vec{a}_2\cdot \Hat{n})\Big\{\frac{  e^{-\mu_{\gamma} r}}{2 \mu_{\gamma} \,r}-\frac{1 -  e^{-\mu_{\gamma} r}}{2\, \mu_{\gamma}^2\, r^2}\Big\}+v_2^2 \,\frac{e^{-r \mu _{\gamma }} \left(-r \mu _{\gamma }+e^{r \mu _{\gamma }}-1\right)}{  r^3 \mu _{\gamma }^2}\notag\\
   &
\,\,\,\,\,\,\,\,\,\,\,\,\,\,\,\,\,\,\,\,\,\,\,\,\,\,\,\,\,
(\boldsymbol{v}_2\cdot \Hat{n})^2\,\frac{e^{-r \mu _{\gamma }} \Big(r \mu _{\gamma } \left(r \mu _{\gamma }+3\right)-3 e^{r \mu _{\gamma }}+3\Big)}{  r^3 \mu _{\gamma }^2}
\Bigg]\Big(\frac{\Bar{\psi}(\boldsymbol{x}_1(t))}{m_p}+\frac{\Bar{\psi}(\boldsymbol{x}_2(t))}{m_p}\Big)\notag\\
   &\quad +(1\leftrightarrow 2)\,\sim\mathcal{O}(L^{1/2}v^{9/2}).
\end{align}
\vspace{-0.7cm}
\item  The amplitude corresponds to the diagram in Fig.~(\ref{fig13m}) that contributes at $N^{(4)}LO$ with bulk interaction vertex $\rmint d^4x\,\partial_0\mathcal{A}_i\partial_i\mathcal{B}_0$ :\footnote{It is important to remember that we have diagrams with asymmetric vertices, i.e., vertices connected with two different fields. Hence, along with shuffling the worldlines, we also have to consider the radiation from the other field vertex (e.g., if we only shuffle the worldlines indices, it will generate a diagram producing radiation from the same field but from different worldlines). But for asymmetric vertices, it demands considering the same diagram with radiation from the other field coupled to one of the worldlines before shuffling the indices).}
\begin{align}
   \mathcal{S}_{\text{eff}}\Big |_{{\text{fig}(\ref{fig13m})}}
   &=-\gamma Q_1Q_2'\rmint \Bar{\mathcal{D}}\hat{\xi}\rmint dt_1v_1^j(t_1)\mathcal{A}_j(\boldsymbol{x}_1(t_1))\rmint dt_2\mathcal{B}_0(\boldsymbol{x}_2(t_2))\rmint d^4x\,\partial_0\mathcal{A}_i(x)\partial_i\mathcal{B}_0(x)\frac{\Bar{\psi}}{m_p}\,,\notag\\
   &=-\gamma Q_1Q_2'\rmint dt_1dt_2v_1^j(t_1)\rmint d^4x\,\partial_0\Big{\langle}\mathcal{A}_i(x)\mathcal{A}_j(\boldsymbol{x}_1(t_1))\Big{\rangle}\partial_i\Big{\langle}\mathcal{B}_0(x)\mathcal{B}_0(\boldsymbol{x}_2(t_2))\Big{\rangle}\frac{\Bar{\psi}}{m_p}\,,\notag\\
   &=\frac{\gamma Q_1Q_2'}{2\pi}\rmint dt \Bigg[(\Vec{a}_1\cdot \Hat{n})\Big\{\frac{  e^{-\mu_{\gamma} r}}{2 \mu_{\gamma} \,r}-\frac{1 -  e^{-\mu_{\gamma} r}}{2\, \mu_{\gamma}^2\, r^2}\Big\}+v_1^2 \,\frac{e^{-r \mu _{\gamma }} \left(-r \mu _{\gamma }+e^{r \mu _{\gamma }}-1\right)}{  r^3 \mu _{\gamma }^2}\notag\\
   &
\,\,\,\,\,\,\,\,\,\,\,\,\,\,\,\,\,\,\,\,\,\,\,\,\,\,\,\,\,
(\boldsymbol{v}_1\cdot \Hat{n})^2\,\frac{e^{-r \mu _{\gamma }} \Big(r \mu _{\gamma } \left(r \mu _{\gamma }+3\right)-3 e^{r \mu _{\gamma }}+3\Big)}{  r^3 \mu _{\gamma }^2}
\Bigg]\Big(\frac{\Bar{\psi}(\boldsymbol{x}_1(t))}{m_p}+\frac{\Bar{\psi}(\boldsymbol{x}_2(t))}{m_p}\Big)\notag\\
   &\quad +(1\leftrightarrow 2)\,,\sim\mathcal{O}(L^{1/2}v^{9/2})\,.\end{align}
\item The amplitude corresponds to the diagram in Fig.~(\ref{fig13o}) that contributes at $N^{(7)}LO$ with the  3-point bulk interaction vertex $\rmint d^4x\,\varphi\,\partial_{0}\mathcal{A}_{i}\partial_{k}\mathcal{A}_{m}(x) :$ 
\begin{align}
          \mathcal{S}_{\text{eff}}^{ikm}\Big |_{{\text{fig}(\ref{fig13o})}}
          &=-\frac{Q_2^2m_1s_1}{m_p^2}\rmint\Bar{\mathcal{D}}\Hat{\xi}\rmint dt_2\, v_2^{j}(t_2)\mathcal{A}_{j}(\boldsymbol{x}_2(t_2))\rmint dt_3 \,v_2^{l}(t_3)\mathcal{A}_{l}(\boldsymbol{x}_2(t_3))\notag\\
          &\quad \rmint dt_1\,\varphi(\boldsymbol{x}_1(t_1))\frac{\Bar{\psi}}{m_p}(\boldsymbol{x}_1(t_1))\rmint d^4x \,\varphi\,\partial_0 \mathcal{A}_{i}\partial_{k} \mathcal{A}_{m}(x)\,,\notag\\
        &=-\frac{Q_2^2m_1s_1}{m_p^2}\rmint dt_1dt_2dt_3dt\,v_2^{j}(t_2)v_2^{l}(t_3)\frac{\Bar{\psi}(\boldsymbol{x}_1(t_1))}{m_p}\rmint d^3 x\, \partial_{k}\Big\langle\mathcal{A}_{j}(\boldsymbol{x}_2(t_2))\mathcal{A}_{m}(x)\Big\rangle \notag\\
        &\qquad \partial_{0}\Big\langle \mathcal{A}_{l}(\boldsymbol{x}_2(t_3))\mathcal{A}_{i}(x)\Big\rangle \Big\langle \varphi(\boldsymbol{x}_1(t_1))\varphi(\boldsymbol{x}_2(t_2))\Big\rangle\,,\notag\\
    &=-\frac{Q_2^2m_1s_1}{m_p^2}\rmint dt\, v_2^{m}(t)a_2^{i}(t)\Bigg[\frac{1}{16} \pi^{3/2}\, m \,G_{1,3}^{2,1}\left(\frac{m^2 r^2}{4}\Big|
\begin{array}{c}
 -\frac{1}{2} \\
 -\frac{1}{2},-\frac{1}{2},0 \\
\end{array}
\right)\notag\\
&\qquad -\frac{\,\pi^{3/2}\, G_{1,3}^{2,1}\left(\frac{m^2 r^2}{4}\Big|
\begin{array}{c}
 \frac{1}{2} \\
 \frac{1}{2},\frac{1}{2},0 \\
\end{array}
\right)}{8 \,m\, r^2}\Bigg]n^k\,
\frac{\Bar{\psi}(\boldsymbol{x}_1(t))}{m_p}\notag\\
&\quad +\text{term symmetric in $i,m$}+1\leftrightarrow 2\,\sim \mathcal{O}(L^{1/2} v^{15/2}).
\end{align}
 \item \textcolor{black}{A similar type of interaction arises from two different electromagnetic worldline couplings: $\rmint dt Q_1v_1^{i}\mathcal{A}_{i}$ and $-\rmint dt Q_2v_2^{i}\mathcal{A}_{i}\Bar{\psi}$. This is shown in Fig.~(\ref{fig13p}) and it contributes at $N^{(7)}LO.$
Corresponding amplitude is given by (to calculate this amplitude, we need to invoke the one-loop massive scalar integral \cite{Anastasiou:1999ui,2022arXiv220103593W})}: 
\begin{align}
\begin{split}
    \mathcal{S}_{\text{eff}}^{ikm}\Big |_{{\text{fig}(\ref{fig13p})}}&=-\frac{1}{m_p^2}\rmint \Bar{\mathcal{D}}\hat{\xi}\rmint dt_1 Q_1v_1^{j}(t_1)\mathcal{A}_{j}(\boldsymbol{x}_1(t_1))\rmint dt_2 m_1s_1 \varphi(\boldsymbol{x}_1(t_2))\rmint dt_3 Q_2 v_2^{l}(t_3)\mathcal{A}_{l}(\boldsymbol{x}_2(t_3))\\&\,\,\,\,\,\,\,\,\,\,\,\,\,\,\,\,\,\,\,\,\,\,\,\,\,\,\frac{\Bar{\psi}(\boldsymbol{x}_2(t_2))}{m_p}\rmint d^4x \varphi\partial_0 \mathcal{A}_{i}\partial_{k} \mathcal{A}_{m}(x)\,,\\ &
        =-\frac{Q_1Q_2m_1s_1}{m_p^2}\rmint dt_1dt_2dt_3dt\,v_1^{j}(t_1)v_2^{l}(t_3)\frac{\Bar{\psi}(\boldsymbol{x}_2(t_2))}{m_p}\rmint d^3 x \,\partial_{t}\Big\langle\mathcal{A}_{j}(\boldsymbol{x}_1(t_1))\mathcal{A}_{i}(x)\Big\rangle\\ &
\hspace{6cm}\partial_{k}\Big\langle\mathcal{A}_{l}(\boldsymbol{x}_2(t_3))  \mathcal{A}_{m}(x)  \Big\rangle\Big\langle \varphi(x_1(t_2))\varphi(x)\Big\rangle\,,\\ &
    =-\frac{Q_1Q_2m_1s_1}{m_p^2}\rmint dt\, dt_1 v_1^{i}(t_1)v_2^{m}(t)\frac{\Bar{\psi}(\boldsymbol{x}_2(t_2))}{m_p}\partial_{t}\delta(t-t_1)\,\\&\hspace{2cm} \rmint_{\boldsymbol{x}} \rmint_{\boldsymbol{k}_1,\boldsymbol{k}_2,\boldsymbol{k}_3}\frac{-ik_2^{k}}{\boldsymbol{k}_1^2\boldsymbol{k}_2^2(\boldsymbol{k}_3^2+m^2)}e^{i\boldsymbol{k}_1\cdot[\boldsymbol{x}_1(t_1)-\boldsymbol{x}(t)]}e^{i\boldsymbol{k}_2\cdot[\boldsymbol{x}_2(t)-\boldsymbol{x}(t)]} 
       e^{i\boldsymbol{k}_3\cdot[\boldsymbol{x}_1(t)-\boldsymbol{x}(t)]}\,,\\ &
  =\frac{Q_1 Q_2 m_1 s_1}{4\pi\,m_p^2}\rmint dt \, v_2^m (t)a_1^i(t)\,\partial_r\alpha(m;r)n^k \frac{\bar{\psi}}{m_p}(\boldsymbol{x}_2(t))\,\sim \mathcal{O}(L^{1/2}v^{15/2})+(1\leftrightarrow 2)\,.\label{5.85n}
  \end{split}
\end{align}
The details of computations are given in Appendix~(\ref{ch1:app:C}). Also the function  $\alpha(m;r)$ is defined (\ref{C3m}). One can obtain the contribution to the effective action by contracting with $\epsilon^{0ikm}$.
\end{itemize}
\color{black}
Then from this, as discussed in the previous subsections we can read out the source term i.e. $T^{00}$ and from that, we can compute the total power radiation. Before writing that we make the following comment about the contribution coming from axion-photon coupling ($g_
{a\gamma\gamma}$) to the gravitational radiation.  
\vspace{0.2cm}

\underline{{\textit{ The $g_{a\gamma\gamma}$ contribution}}}:\par
The source term coming from the diagrams (\ref{fig13o}) and (\ref{fig13p}), captures the effect of the \textit{theta} term.  
\begin{align}
\begin{split}
T^{00}&\sim
\frac{2Q_1Q_2m_1s_1g_{a\gamma \gamma}}{4\pi\,m_p^2}\, [(\hat n\times \boldsymbol{v}_2)\cdot \Vec{a}_1]
\partial_r\alpha(m;r) \delta(\boldsymbol{x}-\boldsymbol{x}_2)\\&
   + \frac{\pi^{3/2}}{4\,m_p^2}g_{a\gamma\gamma}Q_2^2\,m_1\,s_1\,[\Vec{a_2}\cdot(\hat{n}\times\boldsymbol{v}_2)]\Bigg[\frac{1}{2}\, m \,G_{1,3}^{2,1}\left(\frac{m^2 r^2}{4}\Big|
\begin{array}{c}
 -\frac{1}{2} \\
 -\frac{1}{2},-\frac{1}{2},0 \\
\end{array}
\right)-\frac{ G_{1,3}^{2,1}\left(\frac{m^2 r^2}{4}\Big|
\begin{array}{c}
 \frac{1}{2} \\
 \frac{1}{2},\frac{1}{2},0 \\
\end{array}
\right)}{ \,m\, r^2}\Bigg]\\&\delta(\boldsymbol{x}-\boldsymbol{x}_1)+(1\leftrightarrow 2)\label{5.85}
\end{split}
\end{align}
But the source term in (\ref{5.85}) is zero for any orbit that lies in two dimensions. Hence, the term does not contribute to the gravitational power radiation. This term will only contribute to the power radiation expression for orbits that lie in three dimensions instead.\par

\begin{table}[t!]
 \centering
\scalebox{0.87}{\begin{tabular}{|c|c|c|c|c|c|c|c|c|}
 \hline
  Fields & $LO$ & $N^{(1)}LO$ &  $N^{(2)}LO$ &  $N^{(3)}LO$ &   $N^{(4)}LO$&   $N^{(5)}LO$&   $N^{(6)}LO$ &   $N^{(7)}LO$\\
  \hline
 \textit {scalar} & 1 & 1 & 2 & 1 & 3 & 1 $\sim g_{a\gamma\gamma}$ & - & -\\
  \hline
 \textit{electromagnetic} & 1 & 1 & 1 & 1 & - &-& -& -\\
 \hline 
 \textit{proca} & 1 & 1 & 1 & 1 & - &-& -& -\\
 \hline 
 \textit{gravitational} & 1 & - & 1($\sim \gamma$)+5=6 & - & 4($\sim \gamma$)+7=11 & - & 1 & 2 $\sim g_{a\gamma\gamma}$\\
 \hline
 \end{tabular}}
\caption{Table showing the number of diagrams contributing to each $N^{(n)}LO$ for different radiative fields. We exclude the pure gravitational sector, which is already extensively studied (see \cite{Blanchet:2013haa}), and list only contributions from interactions of other fields with gravity.}
 \label{table2m}
 \end{table}
 
\textit{Before we end this section, we summarize our results in Table~(\ref{table2m}). We show the number of diagrams that contribute to gravitational radiation and arise due to the (minimal) coupling of different fields with the gravitational field. We also highlight the order in which the two couplings constant $g_{a\gamma\gamma}$ and $\gamma$ contribute to the gravitational radiation. Again we emphasize that the pure gravitational sector is already well-studied \cite{Blanchet:2013haa} as mentioned before. Here we mainly focus on the leading order contributions to the gravitational radiation due to the coupling of scalar, em and Proca fields with the gravitational field. To the best of our knowledge, these are some of the new results.} 
Finally, we write down the total gravitational power for \textit{circular orbit} radiation considering the term which arises at $LO$  for pure gravity and terms which arise at their respective leading orders due to the interaction between other fields (scalar, em and Proca) with the gravitational field.
\begin{tcolorbox}[thesisresultbox, title=Gravitational Radiation Power]
\restorethesisbodyformat
\begin{equation}
\begin{aligned}
     P_g=&\frac{32G}{5}\Omega^6\Bigg[\mu r^2+\frac{3}{2}\Bigg\{\frac{m_2^3+m_1^3}{M^2}\Bigg\}\nu r^4\Omega^2+\\&\mathcal{D}\Bigg\{\frac{m_1^2+m_2^2}{M^2}\Bigg\}\nu r^4\Omega^2+\Bigg\{\frac{\mathcal{E}_2m_1^2+\mathcal{E}_1m_2^2}{M^2}\Bigg\}r^2+\mathcal{D}(r)r^2\frac{(m_2^2+m_1^2)}{M^2}+\\&\nu r^2\Bigg\{G_1(r)m_2^3(1+\frac{3m_2^2r^2\Omega^2}{2M^2})+G_2(r)m_1^3(1+\frac{3m_1^2r^2\Omega^2}{2M^2})\Bigg\}+\\&\nu r^4\Omega^2\Bigg\{A_1(r)(\frac{m_2+m_1}{2M})^2+A_2(r)(\frac{m_2+m_1}{2M})^2\Bigg\}+\Bigg\{(\frac{H_1(r)m_2^2}{ M^2}+\frac{H_2(r)m_1^2}{ M^2})\frac{(m_1^2+m_2^2)}{M^2}(r^4\Omega^2)\Bigg\}+\\&\Bigg\{\Big(\frac{H_1(r)m_1^2}{ M^2}+\frac{H_2(r)m_2^2}{ M^2}\Big)\frac{(m_1^2+m_2^2)}{M^2}(r^4\Omega^2)\Bigg\}+\Bigg\{\frac{\mathcal{P}_2m_1^2+\mathcal{P}_1m_2^2}{M^2}r^2\Bigg\}+\\&
       \Bigg\{\frac{\mathcal{P}_1m_1^2+\mathcal{P}_2m_2^2}{M^2}r^2\Bigg\}+\Bigg\{\frac{V_2m_1^3}{M^3}-\frac{V_1m_2^3}{M^3}\Bigg\}r^3\Omega^2-
       \Bigg\{\frac{V_1m_1-V_2m_2}{M}\Bigg\}(\nu r^3 \Omega^2)\\&+\Bigg\{\frac{{V}_1m_1^3}{M^3}-\frac{{V}_2m_2^3}{M^3}\Bigg\}r^3\Omega^2+
       \Bigg\{\frac{{V}_1m_2-{V}_2m_1}{M}\Bigg\}(\nu r^3 \Omega^2)+R(r)\Bigg\{\frac{m_1-m_2}{M}\Bigg\}r^4\Omega^2\Bigg]^2
\end{aligned}
\label{5.81n}
\end{equation}
\end{tcolorbox}

where, 
\begin{align}
\begin{split}
  & \mathcal{D}(r)= \Bigg\{\frac{Q_1Q_2}{4\pi } \frac{1}{r}+\frac{Q_1'Q_2'}{4\pi} \frac{e^{-\mu_{\gamma} r}}{r}\Bigg\}\,,\quad
   G_1(r)=\frac{s_2^2g_1}{16\pi^2m_p^4}\frac{e^{-2mr}}{r^2}\,,\\& 
\mathcal{E}_1(r)=\Bigg[\frac{m_1m_2s_1s_2}{m_p}  \Big(\frac{1}{m_p}+\frac{3\frac{m_2^2}{M^2}r^2\Omega^2}{2m_p}\Big) \frac{e^{-mr}}{4\pi r}
+\frac{m_1m_2}{16\pi m_p^2 r}-\frac{m_2 Q_1Q_2}{32\pi^2m_p^2} \frac{1}{r^2}-\frac{m_2 Q_1'Q_2'}{32\pi^2m_p^2} \frac{e^{-\mu_{\gamma} r}}{r^2}\Bigg]\,,\\&
A_1(r)=\Bigg\{\frac{-\gamma Q_1Q_2'}{2\pi}\Big[  \left(\frac{  e^{-\mu_{\gamma} r}}{ r}+\frac{4e^{-r\mu_{\gamma}}\left(-1-r\mu_{\gamma}+e^{\mu_{\gamma} r}\right)}{\mu_{\gamma}^2 r^3}\right) \Big]+\frac{m_1 m_2 s_1 s_2}{16 \pi m_p^2}\,\frac{e^{-m\,r}}{r}\Bigg\}\,,\\&
\mathcal{P}_1(r)=-\frac{\gamma Q_1Q_2'}{4\pi r}e^{-\mu_{\gamma} r}\,,\quad 
H_1(r)=\frac{\gamma Q_1Q_2'}{2\pi}\Bigg[\,\frac{e^{-r \mu _{\gamma }} \left(-r \mu _{\gamma }+e^{r \mu _{\gamma }}-1\right)}{ r^3 \mu _{\gamma }^2}
\Bigg]\,,\\&
V_1(r)=\gamma Q_1Q_2'\frac{1}{2\pi} \Bigg[\Big\{\frac{  e^{-\mu_{\gamma} r}}{2 \mu_{\gamma} \,r}-\frac{1 -  e^{-\mu_{\gamma} r}}{2\, \mu_{\gamma}^2\, r^2}\Big\}\Bigg]\,,\quad 
R(r)=\frac{m_1m_2s_1s_2e^{-mr}}{64\pi m_p^2r}\,.\label{5.81m}
\end{split}
\end{align}
We have used an index subscript for the functions which are not symmetric in two indices (denoting the two worldlines), i.e., one and two, and no indices for the symmetric ones. Also, $\mathcal{E}_2,G_2,A_2,\mathcal{P}_2,H_2,V_2$ can be defined by just replacing the index one by two. There are a few comments that are in order. 

\begin{itemize}
    \item As expected, the total power radiation is a function of the binary separation.
\item In (\ref{5.81m}). Also, we have a term like $\frac{m_1m_2}{16\pi m_p^2 r}$, which doesn't go away even when we set the em charges, dark charges, and screening terms to zero. So it is a pure gravitational radiation correction at $N^{(2)}LO$. We can easily check that it matches the $X^2$ term in  \cite{Huang:2018pbu}. This also serves as a consistency check of our computation. 
\end{itemize}

\section{Discussions and outlooks}\label{disc}
Motivated by recent prospects for searching for dark matter using GW observations \cite{Zhang:2021mks, Coogan:2021uqv, Singh:2022wvw, Becker:2021ivq, Yue:2019ozq, Cardoso:2020iji,Bhattacharya:2023stq}, in this paper we consider a black-hole binary in a dark-matter environment modelled by an \textit{ultra-light particle} (in the form of a massive scalar field with an anomalous coupling to the EM field via a Chern--Simons-type term) and an \textit{ultra-light vector field} (in the form of a massive vector field, or Proca field, that also couples to the EM field through kinetic mixing). Ultimately, we would like to understand whether GW observations can constrain, in addition to the masses of the scalar and Proca fields, the axion-photon coupling parameter $g_{a\gamma\gamma}$ and the Proca-electromagnetic coupling constant $\gamma$, and how those constraints compare with those obtained from other phenomenological considerations mentioned earlier. To that end, we took a first step in this paper by computing the power radiation from a binary system in this dark-matter environment, which is a key ingredient in determining the phase of the gravitational waveform. We use the WEFT formalism to carry out the computation. The study presented in this paper mainly focuses on the regime in which gravity is perturbative. This is approximately true during the adiabatic binary inspiral phase, which is especially important for gravitational-wave detection. A set of length scales is relevant in this regime, ranging from the gravitational radius to the size of the compact objects, their typical orbital separation, and finally the wavelength of the emitted radiation. Keeping these scales in mind and integrating over the different potential fields, we derive the orbital and dissipative dynamics truncated at a given PN order. We then compute the power radiation from a binary moving in a \textit{circular orbit}, which is the first step toward determining the phase of the GW waveform. Below we summarize our results. 
\begin{itemize}
\item We calculated the electromagnetic and Proca bound sectors up to $1PN \sim \mathcal{O}(Lv^2)$. 

\item We also investigated the \textit{em-axion} bound sector, which contributes to the conservative dynamics at $2.5PN\sim \mathcal{O}(Lv^5)$, and the pure scalar bound sector up to $2PN \sim \mathcal{O}(Lv^4)$. \textit{We have provided the details of the 2PN scalar bound sector and of the radiative sector up to $N^{(4)}LO$, which is, to the best of our knowledge, a new result.}

\item \textit{Electromagnetic and Proca radiation has been investigated up to $N^{(3)}LO$. We restricted ourselves to the dipole radiation only. Again this is a new result to the best of our knowledge.} 

\item We also calculated the gravitational radiation due to different field couplings with the gravitational field (results for the pure gravitational sector are well known, and we do not reproduce them here beyond $LO$). \textit{We show that the effect of kinetic term mixing (between photon and dark photon) coupling,  $\gamma$ appears at $N^{(2)}LO (L^{1/2}v^{5/2})$ and $N^{(4)}LO (L^{1/2}v^{9/2}) $ order and
axion-photon, $g_{a\gamma\gamma}$ appears at $\sim N^{(7)}LO(L^{1/2}v^{15/2})$}.  
        
\item \textit{We obtained a novel result showing that the parity-violating axionic term ($g_{a\gamma\gamma}$) leads to detectable power radiation only for genuinely three-dimensional orbits in the case of non-spinning binaries; it vanishes for orbits confined to a two-dimensional plane.} However, it is easy to see that if the two binaries are spinning and their spins are not parallel, this may induce orbital oscillations and hence open up the novel possibility of detecting axions in radiation at infinity.

\end{itemize}

We conclude by mentioning some future directions. First and foremost, we would like to compute the phase of the GW waveform using our results. This will make it possible to estimate various parameters of the theory along the lines of \cite{Zhang:2021mks}, thereby constraining the couplings $g_{a\gamma\gamma}$ and $\gamma$. In turn, this should help constrain this dark-matter model and the parameter space of ultra-light scalar and vector fields. We hope to report on this soon. Furthermore, it will be interesting to generalize our computations to spinning binaries, which would help us build a more complete waveform model. \par 

Another extension of the computation presented in this paper will be to generalize the results for non-Abelian gauge fields and supersymmetric dark matter models. By computing the conservative and radiative dynamics, one will be able to comment on the dependence of our main results on the \textit{color} indices of the gauge fields. One can also introduce finite size effects (tidal deformation etc.)  and investigate the effects of those in our setup. \par 
Last but not least, it will be interesting to investigate the binary dynamics in the Post-Minkowskian (PM) regime. One can use Schwinger-Keldysh `in-in' formalism to study the conservative as well as the dissipative (radiation-reaction) dynamics in a unified framework \cite{Dlapa:2021npj,Kalin:2022hph,Dlapa:2022lmu,Dlapa:2021vgp,Dlapa:2023hsl}. This will enable us to do a full relativistic computation. In recent times, a method known as \textit{Worldline Quantum Field Theory} (WQFT) has been developed to perform such full relativistic computation \cite{Mogull:2020sak,Jakobsen:2022fcj,Jakobsen:2021zvh,Jakobsen:2021lvp,Jakobsen:2021smu,Jakobsen:2022psy} where not only the fields are quantized but also the worldline deflections are quantized. 
Using the WQFT method, it would be interesting to compute the full relativistic waveform, i.e., $\langle h_{\mu\nu}\rangle, \langle A_{\mu}\rangle$ etc. and then taking the non-relativistic result one can compare it to our result. It would be a consistency check of the two different formalisms. Another interesting formalism based on scattering amplitude, namely KMOC formalism \cite{Kosower:2018adc}, has been proposed to calculate the gravitational radiation from a binary system. Again in \cite{Mogull:2020sak}, it is shown that the gravitational waveform calculated from KMOC formalism is related to the one-point function in WQFT. Hence it is interesting to find the correspondence between three completely different `new' formalisms using different examples. We hope to report on some of these issues in the near future. 


\appendix 
\section{Details of computation used in Section~(\ref{Sec4.2})}\label{ch1:app:A}
In order to evaluate the amplitude in (\ref{Sec4.2}) we need to do two  Feynman loop integrals (\ref{2.15}). Although the integrals are not in the form of Feynman master integrals, one can reduce the integrals by Fourier transforming them into the known master integral form. We have the following two integrals,
\begin{align}
    \begin{split}
       & \mathcal{I}(r)=\rmint_{\boldsymbol{k}_1\Vec{{k_2}}} \frac{k_2^k}{\boldsymbol{k}_1^2\boldsymbol{k}_2^2[(\boldsymbol{k}_1+\boldsymbol{k}_2)^2+\textcolor{black}{m^2}]}e^{i(\boldsymbol{k}_1+\boldsymbol{k}_2).\boldsymbol{r}(t)},\,\Tilde{\mathcal{I}}(r)=\rmint_{\boldsymbol{k}_1\boldsymbol{k}_2}\frac{\,k_2^k\, k_1^a\,e^{i(\boldsymbol{k}_1+\boldsymbol{k}_2).\boldsymbol{r}(t)}}{\boldsymbol{k}_1^2\boldsymbol{k}_2^2[(\boldsymbol{k}_1+\boldsymbol{k}_2)^2+\textcolor{black}{m^2}]}\,.
    \end{split}
\end{align}
We start with the first one,
\begin{align}
    \begin{split}
      &  \mathcal{I}(r)=\rmint_{\boldsymbol{k}_1\Vec{{k_2}}} \frac{k_2^k}{\boldsymbol{k}_1^2\boldsymbol{k}_2^2[(\boldsymbol{k}_1+\boldsymbol{k}_2)^2+\textcolor{black}{m^2}]}e^{i(\boldsymbol{k}_1+\boldsymbol{k}_2).\boldsymbol{r}(t)}\,.
    \end{split}
\end{align}
After doing the Fourier transform we get,
\begin{align}
    \begin{split}
        \hat{I}(\boldsymbol{k})&=\rmint d^3r e^{-i\boldsymbol{k}.\boldsymbol{r}}\,\mathcal{I}(r)\,,\\ &
        =\rmint_{\boldsymbol{k}_1}\frac{k^k-k_1^k}{\boldsymbol{k}_1^2\,(\boldsymbol{k}^2+\textcolor{black}{m^2})\,(\boldsymbol{k}_1-\boldsymbol{k})^2}\,.
    \end{split}
\end{align}
Now do the inverse Fourier transform,
\begin{align}
    \begin{split}
        \mathcal{I}(r)&=\iint_{\boldsymbol{k},\boldsymbol{k}_1}e^{i\boldsymbol{k}.\boldsymbol{r}}\,\Big(\frac{k^k-k_1^k}{\boldsymbol{k}_1^2\,(\boldsymbol{k}^2+\textcolor{black}{m^2})\,(\boldsymbol{k}_1-\boldsymbol{k})^2}\Big)\,,\\ &
    =\iint_{\boldsymbol{k},\boldsymbol{k}_1} e^{i\boldsymbol{k}.\boldsymbol{r}}\frac{k^k}{\boldsymbol{k}_1^2\,(\boldsymbol{k}^2+\textcolor{black}{m^2})\,(\boldsymbol{k}_1-\boldsymbol{k})^2}-\iint_{\boldsymbol{k},\boldsymbol{k}_1} e^{i\boldsymbol{k}.\boldsymbol{r}}\frac{k_1^k}{\boldsymbol{k}_1^2\,(\boldsymbol{k}^2+\textcolor{black}{m^2})\,(\boldsymbol{k}_1-\boldsymbol{k})^2}\,,\\ &
=\frac{1}{8}\rmint_{\boldsymbol{k}_1}\frac{k_1^k \,e^{i\boldsymbol{k}_1.\boldsymbol{r}}}{|\boldsymbol{k}_1|(\boldsymbol{k}_1^2+\textcolor{black}{m^2})}-\frac{1}{16}\rmint_{\boldsymbol{k}_1}\frac{k_1^k \,e^{i\boldsymbol{k}_1.\boldsymbol{r}}}{|\boldsymbol{k}_1|(\boldsymbol{k}_1^2+\textcolor{black}{m^2})}\,,\\ &
=\textstyle{\Bigg[\frac{i}{16} \pi^{3/2}\, m \,G_{1,3}^{2,1}\left(\frac{m^2 r^2}{4}\Big|
\begin{array}{c}
 -\frac{1}{2} \\
 -\frac{1}{2},-\frac{1}{2},0 \\
\end{array}
\right)-\frac{i\,\pi^{3/2}\, G_{1,3}^{2,1}\left(\frac{m^2 r^2}{4}\Big|
\begin{array}{c}
 \frac{1}{2} \\
 \frac{1}{2},\frac{1}{2},0 \\
\end{array}
\right)}{8 \,m\, r^2}\Bigg]n^k}\,.
    \end{split}
\end{align}
If we expand the integral in a small mass limit, we will get,
\begin{align}
    \begin{split}
        -i\mathcal{I}(r)\approx\frac{1}{r}+\frac{1}{3} m^2 r [\log \left(mr\right)+ \gamma_E -\frac{4}{3}]\,.\label{50}
    \end{split}
\end{align}\\
We can see there is a smooth massless limit of the expression: $-i\mathcal{I}(r)\approx\frac{1}{r}$ as expected. Another interesting point to note is that the integral in (\ref{50}) in the small mass limit has the same form as a scalar one-loop integral without the divergent term. That implies that although we are doing some complicated loop integral, the results are convergent. That ensures they are not quantum loop rather we can call them as ``Classical Loops''.\par
Now do to the second integral,
\begin{equation}
    \Tilde{\mathcal{I}}(r)=\rmint_{\boldsymbol{k}_1\boldsymbol{k}_2}\frac{\,k_2^k\, k_1^a\,e^{i(\boldsymbol{k}_1+\boldsymbol{k}_2).\boldsymbol{r}(t)}}{\boldsymbol{k}_1^2\boldsymbol{k}_2^2[(\boldsymbol{k}_1+\boldsymbol{k}_2)^2+\textcolor{black}{m^2}]}\,.
\end{equation}
Proceeding as before, we first Fourier transform it.
\begin{align}
    \begin{split}
        \hat{\Tilde{\mathcal{I}}}(\boldsymbol{k})&=\rmint d^3 r\, e^{-i\boldsymbol{k}.\boldsymbol{r}}\, \Tilde{\mathcal{I}}(r)\,,\\ &
        =\rmint_{k_1}\frac{k_1^a(k^k-k_1^k)}{\boldsymbol{k}_1^2\,(\boldsymbol{k}^2+\textcolor{black}{m^2})\,(\boldsymbol{k}_1-\boldsymbol{k})^2}\,.\\ &
    \end{split}
\end{align}
Now do the inverse Fourier transform and get,
\begin{align}
\resizebox{\linewidth}{!}{$
    \begin{aligned}
    \Tilde{\mathcal{I}}(r)&=\iint_{\boldsymbol{k},\boldsymbol{k}_1}\frac{k_1^a\,k^ke^{i\boldsymbol{k}.\boldsymbol{r}}}{\boldsymbol{k}_1^2\,(\boldsymbol{k}^2+\textcolor{black}{m^2})\,(\boldsymbol{k}_1-\boldsymbol{k})^2}-\iint_{\boldsymbol{k},\boldsymbol{k}_1}\frac{k_1^a\,k_1^ke^{i\boldsymbol{k}.\boldsymbol{r}}}{\boldsymbol{k}_1^2\,(\boldsymbol{k}^2+\textcolor{black}{m^2})\,(\boldsymbol{k}_1-\boldsymbol{k})^2}\,,\\
         &=\rmint_{\boldsymbol{k}_1}\frac{k_1^k \,e^{i\boldsymbol{k}_1.\boldsymbol{r}}}{\boldsymbol{k}_1^2+\textcolor{black}{m^2}}\rmint_{\boldsymbol{k}}\frac{k^a}{\boldsymbol{k}^2(\boldsymbol{k}-\boldsymbol{k}_1)^2}-\rmint_{\boldsymbol{k}_1}\frac{e^{i\boldsymbol{k}_1.\boldsymbol{r}}}{\boldsymbol{k}_1^2+\textcolor{black}{m^2}}\rmint_{\boldsymbol{k}}\frac{k^a\,k^k}{\boldsymbol{k}^2(\boldsymbol{k}-\boldsymbol{k}_1)^2}\,,\\
         &=\frac{1}{16}\rmint_{\boldsymbol{k}_1}\frac{k_1^k\,k_1^a\,e^{i\boldsymbol{k}_1.\boldsymbol{r}}}{|\boldsymbol{k}_1|(\boldsymbol{k}_1^2+\textcolor{black}{m^2})}+\frac{1}{16}\rmint_{\boldsymbol{k}_1}\frac{e^{i\boldsymbol{k}_1.\boldsymbol{r}}}{|\boldsymbol{k}_1|(\boldsymbol{k}_1^2+\textcolor{black}{m^2})}\Bigg[\frac{1}{4}\boldsymbol{k}_1^2\delta^{ak}-\frac{3}{4}k_1^a\,k_1^k\Bigg]\,,\\
         &=\frac{1}{64}\rmint_{\boldsymbol{k}_1}\frac{k_1^k\,k_1^a\,e^{i\boldsymbol{k}_1.\boldsymbol{r}}}{|\boldsymbol{k}_1|(\boldsymbol{k}_1^2+\textcolor{black}{m^2})}+\mathcal{F}\,(|\boldsymbol{r}|)\delta^{ak}\,,\\
         &=-\frac{1}{64}\frac{\partial^2}{\partial x^{k}\partial x^{a}}\rmint_{\boldsymbol{k}_1}\frac{e^{i\boldsymbol{k}_1.\boldsymbol{r}}}{|\boldsymbol{k}_1|(\boldsymbol{k}_1^2+\textcolor{black}{m^2})}+\mathcal{F}\,(|\boldsymbol{r}|)\delta^{ak}\,,\\
         &=-\frac{1}{64}\Bigg[ \frac{1}{r}(\delta^{ak}-n^a\,n^k)\Bigg\{\pi ^{3/2} m G_{1,3}^{2,1}\left(\frac{m^2 r^2}{4}|
\begin{array}{c}
 -\frac{1}{2} \\
 -\frac{1}{2},-\frac{1}{2},0 \\
\end{array}
\right)-\frac{2 \pi ^{3/2} G_{1,3}^{2,1}\left(\frac{m^2 r^2}{4}|
\begin{array}{c}
 \frac{1}{2} \\
 \frac{1}{2},\frac{1}{2},0 \\
\end{array}
\right)}{m r^2}\Bigg\}+\\
    &\quad n^a n^k\Bigg\{\frac{\pi ^{3/2} \left(-2 m^2 r^2 G_{1,3}^{2,1}\left(\frac{m^2 r^2}{4}|
\begin{array}{c}
 -\frac{1}{2} \\
 -\frac{1}{2},-\frac{1}{2},0 \\
\end{array}
\right)+8 G_{1,3}^{2,1}\left(\frac{m^2 r^2}{4}|
\begin{array}{c}
 \frac{1}{2} \\
 \frac{1}{2},\frac{1}{2},0 \\
\end{array}
\right)+ m^4 r^4 G_{1,3}^{2,1}\left(\frac{m^2 r^2}{4}|
\begin{array}{c}
 -\frac{3}{2} \\
 -\frac{3}{2},-\frac{3}{2},0 \\
\end{array}
\right)\right)}{2 m r^3}\Bigg\}\\
    &\quad +\mathcal{F}(|\boldsymbol{r}|)\delta^{ak}\Bigg]\,.
    \end{aligned}
$}
\end{align}
Finally using these results we can write down the amplitude mentioned in (\ref{4.31}).
\section{ Details of the Feynman integral computation  used in Section~(\ref{Sec4.3})}\label{ch1:app:B}
The integral in (\ref{4.38}) can be evaluated as follows:
\begin{align}
    \begin{split}
         \mathcal{I}_1&= \rmint_{k,k_1}\frac{k_{1i}k_{1j}\,e^{i\boldsymbol{k}\cdot \boldsymbol{r}}}{\boldsymbol{k}^2\,[(\boldsymbol{k}-\boldsymbol{k}_1)^2+m^2](\boldsymbol{k}_{1}^2+m^2)}\,,\\ &
         =\rmint_{k}\frac{e^{i\boldsymbol{k}\cdot \boldsymbol{r}}}{\boldsymbol{k}^2}\underbrace{\rmint_{k_1}\frac{k_{1i}k_{1j}}{(\boldsymbol{k}_1^2+m^2)[(\boldsymbol{k}_1-\boldsymbol{k})^2+m^2]}}_{k_i k_j B_{21}+\delta_{ij}B_{22}}\,.\label{A.1}
    \end{split}
\end{align}
Then the tensor integral in (\ref{A.1}) can be reduced to a scalar integral using Passarino-Veltman reduction \cite{Passarino:1978jh,2022arXiv220103593W} where $B_{21,22}$ is given by,
 \begin{align}
     \begin{split}
          & B_{21}=\frac{1}{2\boldsymbol{k}^2}\Big[\frac{3}{2}\,\boldsymbol{k}^2\,B_{1}+m^2\,B_0-\frac{1}{2}A_0(im)\Big]\,,\\ &
          B_{22}=\frac{1}{4}\Big[-B_1\boldsymbol{k}^2-2m^2B_{0}-A_{0}(im)\Big]\,,
     \end{split}
 \end{align}
 where
 \begin{align}
     \begin{split}
      B_{0}&=\rmint_{k_1}\frac{1}{(\boldsymbol{k}_1^2+m^2)[(\boldsymbol{k}_1-\boldsymbol{k})^2+m^2]}\\ &
      =\frac{1}{4\pi|\boldsymbol{k}|}\arctan(|\boldsymbol{k}|/2m),
    \label{A.3}
     \end{split}
 \end{align}
 and $B_1,A_0(im)$ are given by,
 \begin{align}
     \begin{split}
           &  B_1=\frac{1}{2}B_0=\frac{1}{8\pi|\boldsymbol{k}|}\arctan(|\boldsymbol{k}|/2m)\,,\\ &
       A_0(im)=\frac{m}{4\pi}\,.
     \end{split}
 \end{align}
Therefore the integral in (\ref{A.1}) can be written as,
\begin{align}
    \begin{split}
    \mathcal{I}_{1}&=\rmint_{k}\frac{e^{i\boldsymbol{k}\cdot\boldsymbol{r}}}{\boldsymbol{k}^2}\Big[\frac{3}{4}\frac{1}{8\pi|\boldsymbol{k}|}\arctan(|\boldsymbol{k}|/2m)+\frac{m^2}{8\pi|\boldsymbol{k}|^3}\arctan(|\boldsymbol{k}|/2m)-\frac{m}{16\pi|\boldsymbol{k}|^2}\Big]k_ik_j\\ &
   \,\,\,\, +\rmint_{k}\frac{e^{i\boldsymbol{k}\cdot \boldsymbol{r}}}{\boldsymbol{k}^2}\Big[-\frac{1}{32\pi}|\boldsymbol{k}|\arctan(|\boldsymbol{k}|/2m)-\frac{m^2}{8\pi |\boldsymbol{k}|}\arctan(|\boldsymbol{k}|/2m)-\frac{m}{16\pi}\Big]\delta_{ij}\,,
   \\ & =\frac{3}{32\pi}\Big[\delta_{ij}\frac{1-e^{-2 m r}}{8 \pi  m r^3}+n_in_j\frac{e^{-2m r} \left(2m r-3 e^{2m r}+3\right)}{8 \pi  m r^3}\Big]+\frac{m^2}{8\pi}\Big[\Big\{-\frac{2 m^2 r^2+e^{-2 m r} (2 m r+1)-1}{32 \pi  m^3 r^3}\,n_in_j\Big\}\\ &
  \,\,\,\,\,\, +\Big\{\frac{e^{-2 m r} \left(e^{2 m r} \left(-8 m^3 r^3 \text{Ei}(-2 m r)+6 m^2 r^2-1\right)-4 m^2 r^2+2 m r+1\right)}{96 \pi  m^3 r^3}\delta_{ij}\Big\}\Big]
   \\&-\frac{m}{16\pi}\Big[\frac{1}{8\pi r}\Big(\delta_{ij}-n_i n_j\Big)\Big]
  \,
  +\Big[\frac{e^{-2mr}}{4\pi r^2}-\frac{m}{64 \pi^2 r}\Big(1-e^{-2mr}-2mr \text{Ei}(-2mr)\Big)-\frac{m}{128\pi^2 r}\Big]\delta_{ij}\,,\\ &
   =\lambda_1(r)\delta_{ij}+\lambda_2(r)n_i n_j\,.
    \end{split}\label{B.5m}
\end{align}
The explicit forms of the functions $\lambda_1(r)$ and $\lambda_2(r)$ are given by,
\begin{align}
    \begin{split}
        \lambda_1(r)=&\frac{3-3e^{-2mr}}{256\pi^2m r^3}+\frac{e^{-2 m r} \left(e^{2 m r} \left(-8 m^3 r^3 \text{Ei}(-2 m r)+6 m^2 r^2-1\right)-4 m^2 r^2+2 m r+1\right)}{768\pi^2m r^3}\\ &
        -\frac{m}{128\pi^2 r}+\frac{e^{-2mr}}{4\pi r^2}-\frac{m}{64 \pi^2 r}\Big(1-e^{-2mr}-2mr \text{Ei}(-2mr)\Big)-\frac{m}{128\pi^2 r}\,,
    \end{split}
\end{align}
and,

\begin{align}
    \begin{split}
        \lambda_2(r)=&\frac{e^{-2m r} \left(2m r-3 e^{2m r}+3\right)}{256\pi^2 m r^3}-\frac{2m^2r^2+e^{-2mr}(2mr+1)-1}{256\pi^2m r^3}+\frac{m}{128\pi^2 r}\,.
    \end{split}
\end{align}
The integral in (\ref{B.5m}) can be done as follows,
\begin{align}
    \begin{split}
        \mathcal{I}_{c}(x;r)&=\rmint_{k}\frac{e^{i\boldsymbol{k}\cdot \boldsymbol{r}}}{\boldsymbol{k}^5}\arctan(|\boldsymbol{k}|/x)\,,\\ &
        =\frac{1}{2\pi^2 r x^3}\underbrace{\rmint_{0}^{\infty}dk \, \frac{\sin(\gamma k)}{k^4}\arctan(k)}_{\Hat{\delta}(\gamma)}\,\,,\gamma=x\,r\,.\\ &
    \end{split}\label{B.6m}
\end{align}
The second integral in (\ref{B.5m}) is IR finite. But direct computation is a hard task. So, we first compute the integral in (\ref{B.6m}) and then take darivatives two times w.r.t $x_i.$  But the integral $\Hat{\delta}(\gamma)$ has \textcolor{black}{IR divergence} but one can extract out the finite part by the Feynman trick. For that, we first take double derivative with respect to $\gamma$, and then integrate twice to find out the mother integral with some infinite constant.
\begin{align}
    \begin{split}
         \frac{d^2 \Hat{\delta}(\gamma)}{d\gamma^2}&=-\rmint_{0}^{\infty}dk \, \frac{\sin(\gamma k)}{k^2}\arctan(k)\,,\\ &
         =\frac{1}{2} \pi  \left(\gamma\,  \text{Ei}(-\gamma )+e^{-\gamma }-1\right)\,.
    \end{split}\label{B.7m}
\end{align}
Integrating (\ref{B.7m}) we will have,
\begin{align}
    \begin{split}
\Hat{\delta}(\gamma)&=\rmint^{\gamma}d\gamma'\rmint^{\gamma'}d\gamma''\,\frac{d^2 \Hat{\delta}(\gamma'')}{d\gamma''^2}+C_1 \gamma+C_{2},\,(\text{$C_1, C_2$ are $\gamma$ independent constant})\,,\\ &
        =\frac{\pi}{12}  e^{-\gamma } \Big(\left(1-3 e^{\gamma }\right) \gamma ^2-\gamma +e^{\gamma } \gamma ^3 \text{Ei}(-\gamma )+2\Big)+C_1\,\gamma+C_2.
    \end{split}\label{B.8m}
\end{align}
In order to find the constants, one needs to use the following initial conditions: $\Hat{\delta}(0)=0\,\text{and}\,|\Hat{\delta}'(0)|\rightarrow \infty$. Hence using equation (\ref{B.8m}) one can find that, 
\begin{align}
    \begin{split}
      &  0=\Hat{\delta}(0)=\frac{\pi}{6}+C_2\implies C_2=-\frac{\pi}{6}.\\ &
      \text{and},\,\infty=|\Hat{\delta}'(0)|=|-\frac{\pi}{4}+C_1|\implies |C_1|\rightarrow \infty\,.
    \end{split}
\end{align}
Therefore,
\begin{align}
    \begin{split}
        \mathcal{I}_c=\frac{e^{- x\,r} \Big[(r^2 x^2\, e^{x\,r}\, (x\,r\, \text{Ei}(-x\,r)-3)+r^2 x^2-x\,r+2\Big]}{24 \pi  r x^3}-\frac{1}{12\pi r x^3}+ \text{$`r$' independent infinite constant.}\label{B.9m}
    \end{split}
 \end{align}
Eventually, using the result of (\ref{B.9m}), one can compute the following integral.
\begin{align}
    \begin{split}
     \mathcal{I}_m &= \rmint_{k}\frac{e^{i\boldsymbol{k}\cdot \boldsymbol{r}}}{\boldsymbol{k}^5}\arctan(|\boldsymbol{k}|/x)k_ik_j,\,\,x=2m\\ &
     =-\partial^{x}_{i}\partial^{x}_{j}\,\mathcal{I}_c(2m;r)\,,\\ &
     =\frac{e^{-2 m r} \left(e^{2 m r} \left(-8 m^3 r^3 \text{Ei}(-2 m r)+6 m^2 r^2-1\right)-4 m^2 r^2+2 m r+1\right)}{96 \pi  m^3 r^3}\delta_{ij}\\ &
     \,\,\,\,\,\,\,\,\,\,-\frac{2 m^2 r^2+e^{-2 m r} (2 m r+1)-1}{32 \pi  m^3 r^3}\,n_in_j\,.
    \end{split}
\end{align}
One can obtain the two coefficients by matching the coefficients of $\delta_{ij}$ and $n_i n_j$. Furthermore, we can omit the infinite constant because it does not contribute to the equation of motion obtained from the variation of the effective action.\par 

Another integral relevant for evaluating (\ref{4.38}) is:
\begin{align}
    \begin{split}
         \mathcal{I}_2&= \rmint_{k,k_1}\frac{k_{1i}k_{j}\,e^{i\boldsymbol{k}\cdot \boldsymbol{r}}}{\boldsymbol{k}^2\,[(\boldsymbol{k}-\boldsymbol{k}_1)^2+m^2](\boldsymbol{k}_{1}^2+m^2)}\,,\\ &
         =\rmint_{k}\frac{e^{i\boldsymbol{k}\cdot \boldsymbol{r}}k_j}{\boldsymbol{k}^2}\underbrace{\rmint_{k_1}\frac{k_{1i}}{(\boldsymbol{k}_1^2+m^2)[(\boldsymbol{k}_1-\boldsymbol{k})^2+m^2]}}_{k_{i}B_{1}}\,,\label{A.5}
    \end{split}
\end{align}
where $B_1$ is given in (\ref{A.3}). Therefore the integral in (\ref{A.5}) can be written as,
\begin{align}
    \begin{split} \mathcal{I}_2&=\rmint_{k}\frac{e^{i\boldsymbol{k}\cdot \boldsymbol{r}}k_i k_j}{\boldsymbol{k}^2}\frac{1}{16|\boldsymbol{k}|}\arctan(|\boldsymbol{k}|/2m)\,,\\ &
    =\delta_{ij}\frac{1-e^{-2 m r}}{128 \pi  m r^3}+n_in_j\frac{e^{-2m r} \left(2m r-3 e^{2m r}+3\right)}{128 \pi  m r^3}\,,\\ &
    =\chi_1(r)\delta_{ij}+\chi_2(r)n_i n_j\,.
        \label{ch1:A.6}
    \end{split}
\end{align}
Again the two coefficients $\chi_{1,2}$ can be obtained by matching the coefficients of $\delta_{ij}$ and $n_i n_j$ in (\ref{ch1:A.6}) as,
\begin{align}
    \begin{split}
        &\chi_1(r)=\frac{1-e^{-2mr}}{128\pi m r^3}\,,\\ &
        \text{and,}\\ &
        \chi_2(r)=\frac{e^{-2mr}(2mr-3e^{2mr}+3)}{128\pi m r^3}\,.
    \end{split}
\end{align}
\section{ Details of computation of Feynman integral used in (\ref{Sec4}) and (\ref{Sec5})}\label{ch1:app:C}
In (\ref{4.33mm}) we have the following integral to evaluate:
\begin{align}
    \begin{split}
        I&=\rmint_{k}\frac{k^a e^{i\boldsymbol{k}\cdot \boldsymbol{r}}}{\boldsymbol{k}^2}\underbrace{\rmint_{k_3}\frac{1}{(\boldsymbol{k}-\boldsymbol{k}_3)^2(\boldsymbol{k}_3^2+m^2)}}\\ &
        =\rmint_{k}\frac{k^a e^{i\boldsymbol{k}\cdot \boldsymbol{r}}}{\boldsymbol{k}^2}
        \begin{cases}
        \frac{1}{8|\boldsymbol{k}|}\Big[1-\frac{2}{\pi}\arctan(m/|\boldsymbol{k}|)\Big] & ,k^2> m^2\\
        \frac{1}{4\pi|\boldsymbol{k}|}\arctan(|\boldsymbol{k}|/m) &, k^2<m^2
        \end{cases}\\ &
        =\frac{1}{4\pi}\rmint_{k}\frac{k^a e^{i\boldsymbol{k}\cdot\boldsymbol{r}}}{\boldsymbol{k}^3}\arctan(|\boldsymbol{k}|/m)\,,\\ &
        =-\frac{i}{4\pi}n^{a}\partial_{r}\,\underbrace{\rmint_{k}\frac{e^{i\boldsymbol{k}\cdot\boldsymbol{r}}}{\boldsymbol{k}^3}\arctan(|\boldsymbol{k}|/m)}_{\alpha(m;r)}\,.\label{C.1}
    \end{split}
\end{align}
In (\ref{C.1}) we have used the identity, $\arctan(x)+\arctan(1/x)=\pi/2, \forall x>0$ and $\alpha(m;r)$ has the following form,
\begin{align}
    \begin{split}
        \alpha(m;r)&=\frac{4\pi}{mr}\rmint_{0}^{\infty}dk \frac{\sin(mrk)}{k^2}\arctan(k)\,,\\ &
        =\frac{2 \pi ^2 r \left(-m r \text{Ei}(-m r)-e^{-m r}+1\right)}{m}\,.
    \end{split}
\end{align}
\begin{figure}
    \centering
    \includegraphics[scale=0.5]{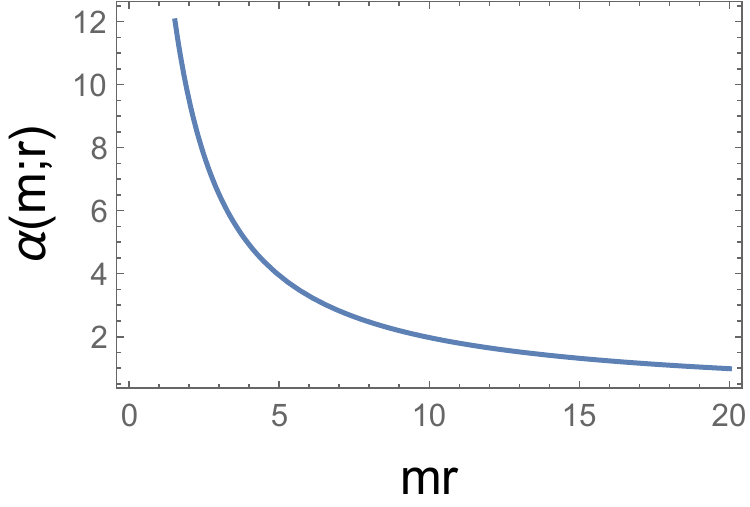}
    \caption{$\alpha(m;r)$ vs $mr$ plot}
    \label{fig12}
\end{figure}
 We numerically integrate and show the nature of the integral in Fig~(\ref{fig12}).\par
Another useful relevant integral for (\ref{4.26}) has the following form,
\begin{align}
    \begin{split}
        \beta(m;r)&=\rmint_{k}\frac{e^{i\boldsymbol{k} \cdot \vec r}}{\boldsymbol{k}^2+m^2}\rmint_{k_1}\frac{1}{\boldsymbol{k}_1^2[(\boldsymbol{k}_1-\boldsymbol{k})^2+m^2]}\\ &
        = \rmint_{k}\frac{e^{i\boldsymbol{k} \cdot \vec r}}{\boldsymbol{k}^2+m^2}
        \begin{cases}
                \frac{1}{8|\boldsymbol{k}|}[1-\frac{2}{\pi}\arctan(m/|\boldsymbol{k}|)]\, & \boldsymbol{k}^2>m^2\\
                \frac{1}{4\pi|\boldsymbol{k}|}\arctan(|\boldsymbol{k}|/m)\, & \boldsymbol{k}^2<m^2
            \end{cases}\,,\\ &
 =\rmint_{k}\frac{e^{i\boldsymbol{k}\cdot\boldsymbol{r}}}{(\boldsymbol{k}^2+m^2)|\boldsymbol{k}|}\arctan(|\boldsymbol{k}|/m)\\ &
 =\frac{1}{2\pi^2 mr}\rmint_{0}^{\infty}dk\,\frac{\sin(mrk)}{k^2+1}\arctan(k)\,.\label{C3m}
 \end{split}
\end{align}
The integral in (\ref{C3m}) does not have any closed form expression. One can numerically integrate it and the nature of the function $\beta(m;r)$ is shown in Fig.~(\ref{lastfig}).
\begin{figure}[b!]
    \centering
    \includegraphics[scale=0.4]{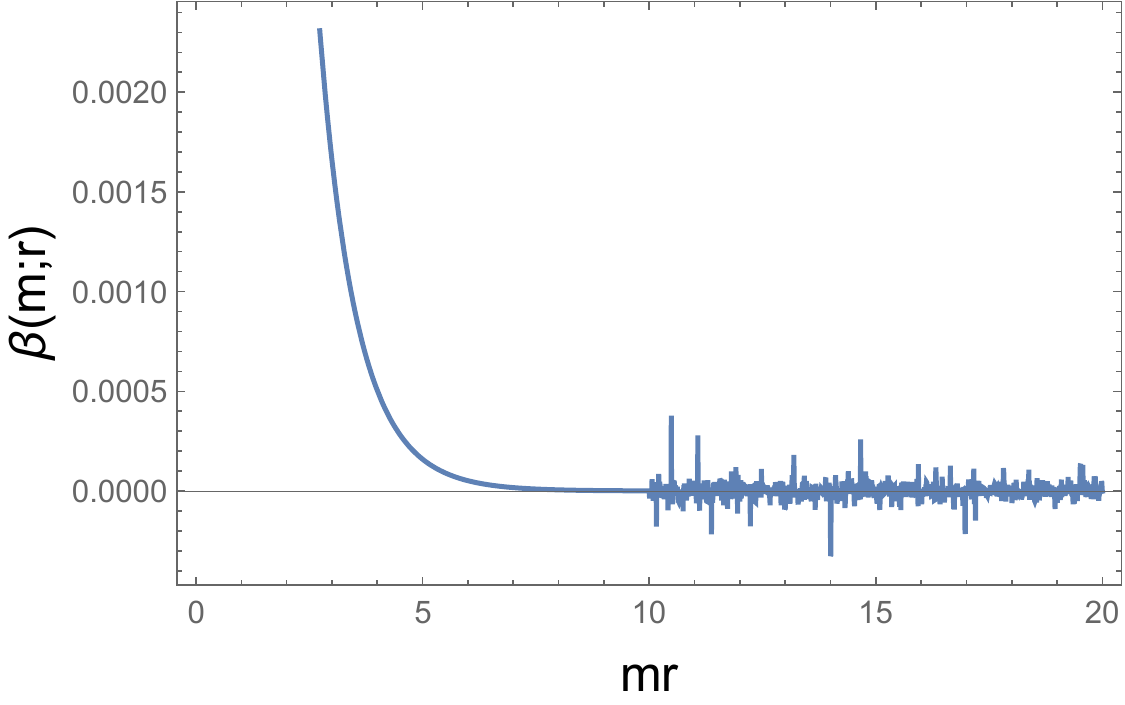}
    \caption{$\beta(m;r)$ vs $mr$ plot}
    \label{lastfig}
\end{figure}

\begin{figure}[t!]
    \centering
    \thesisdiagrampanel{\scalebox{0.36}{\begin{feynman}
    \fermion[label=$$, lineWidth=4]{10.40, 4.80}{11.40, 4.80}
    \fermion[showArrow=false, lineWidth=4]{8.00, 4.40}{7.60, 4.80}
    \fermion[lineWidth=4]{12.60, 4.80}{13.60, 4.80}
    \dashed[showArrow=false, color=eb144c]{4.80, 4.00}{6.00, 4.80}
    \dashed[showArrow=false, color=eb144c]{5.20, 4.00}{6.00, 4.40}
    \dashed[showArrow=false, color=eb144c]{4.80, 4.80}{6.00, 5.60}
    \dashed[showArrow=false, color=eb144c]{4.80, 4.40}{6.00, 5.20}
    \fermion[showArrow=false]{6.00, 4.00}{6.00, 5.60}
    \fermion[showArrow=false, lineWidth=4]{9.20, 5.20}{9.60, 4.80}
    \fermion[lineWidth=4, label=$$]{4.00, 4.00}{6.80, 4.00}
    \fermion[showArrow=false, lineWidth=4]{9.20, 4.40}{9.60, 4.80}
    \fermion[lineWidth=4]{4.00, 5.60}{6.80, 5.60}
    \dashed[showArrow=false, color=eb144c]{4.80, 5.20}{5.60, 5.60}
    \fermion[showArrow=false]{4.80, 4.00}{4.80, 5.60}
    \fermion[showArrow=false, lineWidth=4]{7.60, 4.80}{8.00, 5.20}
    \fermion[lineWidth=5, showArrow=false]{7.60, 4.80}{9.60, 4.80}
    \parton[color=eb144c]{12.00,4.80}{0.60}
\end{feynman}}}
    \caption{Correspondence between PN diagrams (left) and one-loop scalar Feynman diagram (right).}
    \label{fig16}
\end{figure}
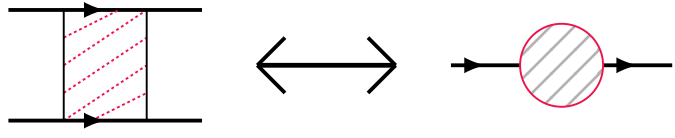

\section{Comments on middle vertex contribution to the radiation}\label{ch1:app:D}
In the radiative diagrams, where one radiative field is coming out from the vertex, we implicitly assume that the radiation is coming out from an average position $\boldsymbol{x}_0$. But in general, the radiative field is integrated over all the spacetime i.e, the radiation could be from anywhere in the spacetime. \\

\textbf{Scalar diagrams consisting of radiative bulk vertices:}

1)\,Fig.~(\ref{10h}):

\begin{align}
    \begin{split}
\mathcal{S}_{\text{eff}}\Big |_{\text{fig}(\ref{10h})}&=\frac{m^2 s_1m_1m_2}{m_p^3}\rmint dt_1dt_2\rmint d^4x \Big\langle \psi(\boldsymbol{x}_2(t_2))\psi(x)\Big\rangle \Big\langle\varphi(\boldsymbol{x}_1(t_1))\varphi(x) \Big\rangle \Big\{\bar{\phi}(\vec 0,t)+x^i\partial_{i}\bar\phi(\vec 0,t)+\cdots\Big\}\,\label{E.1}
    \end{split}
\end{align}
The first and second terms in (\ref{E.1}) give the monopole and dipole terms, respectively. First, compute the dipole term. Here, it is difficult to isolate the source term $J$, but one can directly compute the dipole moment by isolating the coefficient of the $\partial_i \bar \phi(\vec 0,t)$. Hence, the effective action has the following, 
\begin{align}
    \begin{split}
        \mathcal{S}_{\text{eff}}\Big |_{\text{fig}(\ref{10h})}&=\frac{m^2 s_1m_1m_2}{2 m_p^3}\rmint dt\,\partial_i\bar\phi(\vec 0,t)\rmint d^3x\,x^i \rmint_{\boldsymbol{k}_1,\boldsymbol{k}_2}\frac{e^{i\boldsymbol{k}_1\cdot (\boldsymbol{x}_2-\boldsymbol{x})}e^{i\boldsymbol{k}_2\cdot (\boldsymbol{x}_1-\boldsymbol{x})}}{\boldsymbol{k}_1^2(\boldsymbol{k}_2^2+m^2)}\,.\label{E.2}
    \end{split}
\end{align}
Now, from \eqref{E.2}, it is clear that the dipole moment has the following form,
\begin{align}
    \begin{split}
        I_{\phi}^{i}\Big|_{\text{fig.} (\ref{10h})}&=\frac{m^2 s_1m_1m_2}{2 m_p^3}\rmint d^3x\,x^i \rmint_{\boldsymbol{k}_1,\boldsymbol{k}_2}\frac{e^{i\boldsymbol{k}_1\cdot (\boldsymbol{x}_2-\boldsymbol{x})}e^{i\boldsymbol{k}_2\cdot (\boldsymbol{x}_1-\boldsymbol{x})}}{\boldsymbol{k}_1^2(\boldsymbol{k}_2^2+m^2)}\,,\\ &
        =\frac{m^2 s_1m_1m_2}{2 m_p^3}\Big[r^i\frac{(m^2r^2-2)+2e^{-mr}(mr)}{4\pi m^4 r^3}+x_2^i\frac{1-e^{-mr}}{4\pi m^2 r}\big]\,.
    \end{split}
\end{align}
2) Fig~(\ref{10e}): Here, the computation is more involved as we have a time derivative over the radiation field.
\begin{align}
    \begin{split}
        \mathcal{S}_{\text{eff}}\Big |_{\text{fig}(\ref{10e})}&=\frac{m_1m_2s_2}{m_p^2}\rmint\Bar{\mathcal{D}}\Hat{\xi}\rmint dt_1dt_2\tilde{\psi}(\boldsymbol{x}_1(t_1))\phi(\boldsymbol{x}_2(t_2))\rmint d^4x \,\frac{\psi}{m_p}\partial_0\phi\partial_0\bar{\phi}\\ &
        =-\frac{m_1m_2s_2}{m_p^2}\rmint \Bar{\mathcal{D}}\Hat{\xi}\rmint dt_1dt_2\tilde{\psi}(\boldsymbol{x}_1(t_1))\phi(\boldsymbol{x}_2(t_2))\rmint d^4x \,\Big[\partial_0\tilde\psi\partial_0 \varphi+\tilde\psi\partial_0^2 \varphi\Big]\Big\{\bar\phi(\vec 0,t)+x^i \partial_i \bar\phi(\vec 0,t)\Big\}
    \end{split}
\end{align}
Again, focus on the dipole term only,
\begin{align}
    \begin{split}
    \mathcal{S}_{\text{eff}}\Big |_{\text{fig}(\ref{10e})}&=\frac{m_1m_2s_2}{m_p^2}\rmint dt_1 dt_2 \rmint d^4x \Big[\partial_0\Big\langle\tilde\psi(\boldsymbol{x}_1(t_1))\tilde\psi(x)\Big\rangle \partial_0\Big\langle\varphi(\boldsymbol{x}_2(t_2))\varphi(x)\Big\rangle\,,\\ &
    \hspace{3 cm}+\Big\langle\tilde\psi(\boldsymbol{x}_1(t_1))\tilde\psi(x)\Big\rangle \partial_0^2\Big\langle\varphi(\boldsymbol{x}_2(t_2))\varphi(x)\Big\rangle\Big]x^{i}\partial_i\bar\phi(\vec 0,t)\,.
    \end{split}
\end{align}
Considering the first term only,
\begin{align}
    \begin{split}
         \mathcal{S}_{\text{eff}}\Big |_{\text{fig}(\ref{10e})}^{(1)}&=\frac{m_1m_2s_2}{m_p^2}\rmint dt_1 dt_2 \rmint d^4x \partial_0\Big\langle\tilde\psi(\boldsymbol{x}_1(t_1))\tilde\psi(x)\Big\rangle \partial_0\Big\langle\varphi(\boldsymbol{x}_2(t_2))\varphi(x)\Big\rangle x^{i}\partial_i \bar\phi(\vec 0,t)\,,\\ &
         =\frac{m_1m_2s_2}{2m_p^2}\rmint dt \,\partial_{i}\bar\phi(\vec 0,t) \rmint d^3x \rmint_{k_1}\frac{(i\boldsymbol{k}_1\cdot \boldsymbol{v}_1) }{\boldsymbol{k}_1^2}\Big[x_1^{i}+i\partial_{k_1^i}\Big]e^{i\boldsymbol{k}_1\cdot (\boldsymbol{x}_1-\boldsymbol{x})}\rmint_{k_2}(i\boldsymbol{k}_2\cdot \boldsymbol{v}_2)\frac{e^{ik_2\cdot (\boldsymbol{x}_2-\boldsymbol{x})}}{\boldsymbol{k}_2^2+m^2}\,.
    \end{split}
\end{align}
Therefore, the dipole moment can be isolated as,
\begin{align}
    \begin{split}
        I_{\phi}^{i}(t)\Big|_{\text{fig.}(\ref{10e})}&=-v_1^a v_2^b\rmint d^3x \rmint_{k_1}\Big[\frac{k_1^a\,x_1^i}{\boldsymbol{k}_1^2} -i\Big(\frac{\delta_{ai}}{\boldsymbol{k}_1^2}-\frac{2k_1^ik_1^a}{\boldsymbol{k}_1^4}\Big)\Big] e^{i\boldsymbol{k}_1\cdot (\boldsymbol{x}_1-\boldsymbol{x})}\rmint_{k_2}k_2^b\,\frac{e^{i\boldsymbol{k}_2\cdot(\boldsymbol{x}_2-\boldsymbol{x})}}{\boldsymbol{k}_2^2+m^2}\\ &
        =v_1^a v_2^b\rmint_{k_1}\Big[\frac{k_1^a\,x_1^i}{\boldsymbol{k}_1^2} -i\Big(\frac{\delta_{ai}}{\boldsymbol{k}_1^2}-\frac{2k_1^ik_1^a}{\boldsymbol{k}_1^4}\Big)\Big]\frac{k_1^b}{\boldsymbol{k}_1^2+m^2}e^{i\boldsymbol{k}_1\cdot \vec r}\\ &
        =v_1^a v_2^b\Bigg[x_1^i\rmint_{\boldsymbol{k}_1}\frac{k_1^a k_1^b}{\boldsymbol{k}_1^2(\boldsymbol{k}_1^2+m^2)}-i\delta_{ai}\rmint_{k_1}\frac{k_1^b}{\boldsymbol{k}_1^2(\boldsymbol{k}_1^2+m^2)}+2i\rmint_{k_1}\frac{k_1^i k_1^a k_1^b}{\boldsymbol{k}_1^4(\boldsymbol{k}_1^2+m^2)}\Bigg]e^{i\boldsymbol{k}\cdot \vec r}
    \end{split}
\end{align}
The integrals can be done by investigating two separate families of Feynman-Fourier integrals that we discuss in the subsequent subsection. However, we are not showing the explicit results of these integrals, as they can be straightforwardly evaluated using the results we provide in the subsequent section.\\\\
\textbf{IBP with exponential and systematic computation of the Fourier integrals.}\\ 
Let's take a moment to systematically compute the Fourier Feynman Integrals.\\
\textbf{\texttt{One-loop Fourier family}:}\,Let us define a family of integrals,
\begin{align}
    \begin{split}
        {J}_{a_1,a_2}= \int d^dk\, \frac{e^{i\boldsymbol{k}\cdot \boldsymbol r}}{(i \boldsymbol{k}\cdot \boldsymbol r)^{a_1}(\boldsymbol{k}^2+m^2)^{a_2}}.
    \end{split}
\end{align}
After solving the IBP relations (with a simple modified version of \textbf{\texttt{LiteRed}}\cite{Lee:2013mka}) we have three master integrals for the first family: ${J}^{}_{0,1},{J}^{}_{0,2}, {J}^{}_{1,1}$. First do the following rescaling, $\boldsymbol{k}\to \boldsymbol{\ell}=\,\boldsymbol{k}/m$ and $\boldsymbol{r}\to \boldsymbol{s}=\boldsymbol{r}m$. We left with the integral,
\begin{align}
    J_{a_1,a_2}=(m^2)^{d/2-a_2}\int d^d\boldsymbol{\ell} \frac{e^{i\boldsymbol{\ell}\cdot\boldsymbol{s}}}{(i\boldsymbol{\ell}\cdot \boldsymbol{s}-i\varepsilon)^{a_1}(\boldsymbol{\ell}^2+1)^{a_2}}
\end{align}
The differential equation satisfied by the master integral is given by (with $x=s^2$),
\begin{align}
    \partial_{x}\boldsymbol{\mathbfcal{J}}=\boldsymbol{\Omega}_{x}\cdot \boldsymbol{\mathbfcal{J}},\qquad   \boldsymbol{\Omega}_{x}=\left(
\begin{array}{ccc}
 -\frac{d-2}{2 x} & -\frac{1}{x} & 0 \\
 -\frac{1}{4} & 0 & 0 \\
 \frac{1}{2 x} & 0 & -\frac{1}{2 x} \\
\end{array}
\right)
\end{align}
Therefore, we have three coupled differential equations to be solved,
\begin{align}
    \begin{split}&\partial_{x}J_{0,1}=\frac{2-d}{2x}J_{0,1}-\frac{1}{x}J_{0,2},\\&
    \partial_{x}J_{0,2}=-\frac{1}{4}J_{0,1},\\ &
    \partial_{x}J_{1,1}=\frac{1}{2x}J_{0,1}-\frac{1}{2x}J_{1,1}.
    \end{split}
\end{align}
The solution is given by,
\begin{align}
\begin{split}
&    J_{0,1}=x^{\frac{2-d}{4}}\left(C_1 I_{\frac{d-2}{2}}(\sqrt{x})+C_2 K_{\frac{d-2}{2}}(\sqrt{x})\right),\\ &
J_{0,2}
=
\frac{1}{2}\,x^{\frac{4-d}{4}}
\left(
-\,C_1 I_{\frac{d-4}{2}}(\sqrt{x})
+
C_2 K_{\frac{d-4}{2}}(\sqrt{x})
\right),\\ &
J_{1,1}(x)
=
\frac{1}{\sqrt{x}}
\left[
C_3
+
\int^{\sqrt{x}}
t^{-\frac{d-2}{2}}
\left(
C_1 I_{\frac{d-2}{2}}(t)
+
C_2 K_{\frac{d-2}{2}}(t)
\right)\,dt
\right].\label{2.201j}
    \end{split}
\end{align}
The constants $C_1, C_2$ can be fixed using the asymptotic expansion around $x\to 0$ of the integral. In this asymptotic, the integral has two distinct momentum scalings,
\begin{align}
   \texttt{hard:} |\boldsymbol{\ell}|\sim\frac{1}{s},\,\,  \texttt{soft:} |\boldsymbol{\ell}|\sim1
\end{align}
There are therefore two regions: from the method of regions \cite{Beneke:1997zp},\\
\textbf{\texttt{hard region}.} In this region the propagator expanded as massless, $\frac{1}{\boldsymbol{\ell}^2+1}\sim \frac{1}{\boldsymbol{\ell}^2}$ and hence,
\begin{align}
    J_{0,1}^{(h)}(x)=2^{d-2}\pi^{d/2} \Gamma\left(\frac{d}{2}-1\right)x^{1-\frac{d}{2}}
\end{align}
\textbf{\texttt{soft region}.} In this region, the exponent has a soft expansion, and considering the leading order, one has,
\begin{align}
    J_{0,1}^{(s)}(x)=\int d^d\boldsymbol{\ell} \frac{1}{\boldsymbol{\ell}^2+1}=\pi^{d/2}\Gamma\left(1-\frac{d}{2}\right)
\end{align}
Collecting all the leading contributions from the asymptotic region,
\begin{align}
    J_{0,1}\overset{x\to 0}{\sim} 2^{d-2}\pi^{d/2}\Gamma\left(\frac{d}{2}-1\right) x^{d/2-1}+\pi^{d/2}\Gamma\left(1-\frac{d}{2}\right)+\cdots\label{2.205a}
\end{align}
Now, expanding $J_{0,1}$ in \eqref{2.201j} we get,
\begin{align}
   J_{0,1}\overset{x\to 0}{\sim}  \frac{2^{-\frac{d}{2}} \left(2 C_1+C_2 \Gamma \left(\frac{d}{2}\right) \Gamma \left(1-\frac{d}{2}\right)\right)}{\Gamma \left(\frac{d}{2}\right)}+C_2 2^{\frac{d}{2}-2} x^{1-\frac{d}{2}} \Gamma \left(\frac{d-2}{2}\right)\label{2.206a}
\end{align}
Now, matching \eqref{2.205a} and \eqref{2.206a} we get,
\begin{align}
    C_1=0,\,C_2=(2\pi)^{d/2}
\end{align}
In a similar way, $C_3$ can be fixed from the asymptotic expansion of  $J_{1,1}$, in soft region,
\begin{align}
\begin{split}
    J_{1,1}^{(s)}&\overset{x\to0}{\sim} \int d^d\ell \frac{1-i\boldsymbol{\ell}\cdot \boldsymbol{s}}{(i\boldsymbol{\ell\cdot s-i\varepsilon})(\boldsymbol{\ell}^2+1)}+\cdots\\ &
    \sim \int d^d\ell \frac{1}{(i\boldsymbol{\ell\cdot s-i\varepsilon})(\boldsymbol{\ell}^2+1)}+\cdots=\frac{\pi^{\frac{d+1}{2}}\Gamma\left(\frac{3-d}{2}\right)}{\sqrt{x}\,}+\cdots
    \end{split}
\end{align}
Comparing we have, $C_3={\pi^{\frac{d+1}{2}}}\Gamma\left(\frac{3-d}{2}\right)$
Therefore, the integrals become,
\begin{align}
\begin{split}
 &   J_{0,1}=(2\pi)^{d/2} \left(\frac{m}{r}\right)^{d/2-1}K_{\frac{d-2}{2}}(mr),\\ &
    J_{0,2}=\frac{1}{2}(m^2)^{d/2}\left(mr\right)^{\frac{4-d}{4}} K_{\frac{d-4}{2}}(mr)\\ &
    J_{1,1}=\frac{1}{\sqrt{m r}}\Bigg[2^{\frac{d}{2}-3} \pi ^d \csc \left(\frac{\pi  d}{2}\right) \sqrt[4]{m r} \Big(4 \pi ^{3/2} \, _1\tilde{F}_2\left(\frac{1}{2};\frac{d}{2},\frac{3}{2};\frac{\sqrt{m r}}{4}\right)\\ &-\pi  2^d \Gamma \left(\frac{3}{2}-\frac{d}{2}\right) (m r)^{\frac{2-d}{4}} \, _1\tilde{F}_2\left(\frac{3}{2}-\frac{d}{2};2-\frac{d}{2},\frac{5}{2}-\frac{d}{2};\frac{\sqrt{m r}}{4}\right)\Big)+\pi ^{\frac{d+1}{2}} \Gamma \left(\frac{3}{2}-\frac{d}{2}\right)\Bigg]
    \end{split}
\end{align}
\paragraph{Two-loop Fourier family.}
We consider the two-loop family
\begin{align}
G_{a_1,a_2,a_3,a_4,a_5}(\boldsymbol{r})
=
\int d^d k\, d^d k_1\,
\frac{e^{D_1}}{D_1^{a_1} D_2^{a_2} D_3^{a_3} D_4^{a_4} D_5^{a_5}},
\end{align}
where
\begin{align}
D_1 = i\,\boldsymbol{k}\!\cdot\!\boldsymbol{r},
\qquad
D_2 = i\,\boldsymbol{k}_1\!\cdot\!\boldsymbol{r},
\qquad
D_3 = \boldsymbol{k}^2,
\qquad
D_4 = \boldsymbol{k}_1^2 + m^2,
\qquad
D_5 = (\boldsymbol{k}_1-\boldsymbol{k})^2 + m^2.
\end{align}
After solving the IBP relations, the family reduces to $17$ master integrals. For our purposes, it is sufficient to consider the subset
\begin{align}
\mathcal G(x)
=
\bigl(
G_1(x),G_2(x),G_3(x),G_4(x),G_5(x),G_6(x)
\bigr)^T,
\end{align}
with
\begin{align}
G_1 &= G_{0,0,0,1,1},
\qquad
G_2 = G_{0,0,0,2,1},
\qquad
G_3 = G_{0,0,0,1,2},
\\
G_4 &= G_{0,0,-1,1,1},
\qquad
G_5 = G_{0,0,1,0,1},
\qquad
G_6 = G_{0,0,1,1,0}.
\end{align}
and we define the dimensionless variable
\begin{align}
x = (m r)^2,
\qquad r = |\boldsymbol{r}|.
\end{align}
The differential system satisfied by these master integrals is
\begin{align}
\partial_x \mathbfcal G(x) = \Delta_x\, \mathbfcal G(x),\quad \Delta_x =
\begin{pmatrix}
\dfrac{2-d}{x} & -\dfrac{1}{x} & -\dfrac{1}{x} & 0 & 0 & 0 \\[6pt]
\dfrac{(d-2)^2}{4x} & 0 & \dfrac{d-2}{2x} & \dfrac{1}{8} & 0 & 0 \\[6pt]
\dfrac{(d-2)^2}{4x} & \dfrac{d-2}{2x} & 0 & \dfrac{1}{8} & 0 & 0 \\[6pt]
\dfrac{2(d-3)}{x} & \dfrac{4}{x} & \dfrac{4}{x} & \dfrac{1-d}{x} & 0 & 0 \\[6pt]
0 & 0 & 0 & 0 & -\dfrac{d-2}{2x} & 0 \\[6pt]
0 & 0 & 0 & 0 & 0 & -\dfrac{d-2}{2x}
\end{pmatrix}.
\end{align}
By symmetry, one has
\begin{align}
G_2(x)=G_3(x),
\qquad
G_5(x)=G_6(x),
\end{align}
and therefore the system reduces to four coupled equations:
\begin{align}
\begin{split}
G_1'(x) &= \frac{2-d}{x}\,G_1(x) - \frac{2}{x}\,G_2(x), \\
G_2'(x) &= \frac{(d-2)^2}{4x}\,G_1(x) + \frac{d-2}{2x}\,G_2(x) + \frac{1}{8}\,G_4(x), \\
G_4'(x) &= \frac{2(d-3)}{x}\,G_1(x) + \frac{8}{x}\,G_2(x) + \frac{1-d}{x}\,G_4(x), \\
G_5'(x) &= \frac{2-d}{2x}\,G_5(x).
\end{split}
\end{align}
Eliminating $G_2$ and $G_4$ from the first three equations, one finds that $G_1$ satisfies the third-order differential equation
\begin{align}
2x^2 G_1'''(x) + 3d\,x\,G_1''(x) + \bigl(d^2-d-2x\bigr) G_1'(x) + (1-d) G_1(x) = 0.
\end{align}
Its general solution can be written as
\begin{align}
G_1(x) = c_1 u_1(x) + c_2 u_2(x) + c_3 u_3(x),
\end{align}
where
\begin{align}
\begin{split}
u_1(x) &= {}_1F_2\!\left(\frac{d-1}{2};\, d-1,\frac{d}{2};\, x\right), \\
u_2(x) &= x^{1-\frac{d}{2}}\,
{}_1F_2\!\left(\frac{1}{2};\, 2-\frac{d}{2},\frac{d}{2};\, x\right), \\
u_3(x) &= x^{2-d}\,
{}_1F_2\!\left(\frac{3-d}{2};\, 3-d,2-\frac{d}{2};\, x\right).
\end{split}
\end{align}
Consequently,
\begin{align}
\begin{split}
G_2(x) &= (2-d)\,G_1(x) - x\,G_1'(x), \\
G_4(x) &= 8\,G_2'(x) - \frac{2(d-2)^2}{x}\,G_1(x) - \frac{4(d-2)}{x}\,G_2(x), \\
G_5(x) &= c_4\,x^{1-\frac{d}{2}}.
\end{split}
\end{align}
Although $G_1$ can be identified with the square of the one-loop master integral discussed previously, let us assume that this relation is not known a priori and determine the constants $c_1,c_2,c_3$ directly from boundary data. To this end, we introduce the dimensionless variables
\begin{align}
\boldsymbol{s} = m \boldsymbol{r},
\qquad
x = \boldsymbol{s}^2,
\end{align}
and rescale the loop momenta according to
\begin{align}
\boldsymbol{q} = \frac{\boldsymbol{k}_1}{m},
\qquad
\boldsymbol{p} = \frac{\boldsymbol{k}_1-\boldsymbol{k}}{m}.
\end{align}
Then the dimensionless master integral can be written as
\begin{align}
G_1(x)
=
m^{4-2d} G_{0,0,0,1,1}(\boldsymbol{r})
=
\int d^d q\, d^d p\,
\frac{e^{i(\boldsymbol{q}-\boldsymbol{p})\cdot \boldsymbol{s}}}{(\boldsymbol{q}^2+1)(\boldsymbol{p}^2+1)}.
\end{align}
To determine the small-$x$ behavior of $G_1$, we use the method of regions in the limit $x \to 0$, or equivalently $s \to 0$. The relevant momentum scalings are
\begin{align}
\texttt{hard:}\quad |\boldsymbol{q}|,|\boldsymbol{p}| \sim \frac{1}{s},
\qquad
\texttt{soft:}\quad |\boldsymbol{q}|,|\boldsymbol{p}| \sim 1.
\end{align}
There are therefore four asymptotic regions.

\paragraph{\texttt{hard-hard} region.}
In this region, both propagators may be expanded as massless,
\begin{align}
\frac{1}{\boldsymbol{q}^2+1} \sim \frac{1}{\boldsymbol{q}^2},
\qquad
\frac{1}{\boldsymbol{p}^2+1} \sim \frac{1}{\boldsymbol{p}^2},
\end{align}
and hence
\begin{align}
G_1^{(hh)}(x)
&=
\int d^d q\,\frac{e^{i\boldsymbol{q}\cdot \boldsymbol{s}}}{\boldsymbol{q}^2}
\int d^d p\,\frac{e^{-i\boldsymbol{p}\cdot \boldsymbol{s}}}{\boldsymbol{p}^2}
\nonumber\\
&=
2^{2d-4}\pi^d
\Gamma\!\left(\frac{d}{2}-1\right)^2
x^{2-d}.
\end{align}

\paragraph{\texttt{hard-soft} and \texttt{soft-hard} regions.}
Consider, for instance, the region $|\boldsymbol{q}| \sim s^{-1}$ and $|\boldsymbol{p}| \sim 1$. The soft exponential is expanded as
\begin{align}
e^{-i\boldsymbol{p}\cdot \boldsymbol{s}}
=
1 - i\,\boldsymbol{p}\cdot \boldsymbol{s} - \frac{1}{2}(\boldsymbol{p}\cdot \boldsymbol{s})^2 + \cdots.
\end{align}
The odd terms vanish upon angular integration, and at leading order, we obtain
\begin{align}
G_1^{(hs)}(x)
&=
\int d^d q\,\frac{e^{i\boldsymbol{q}\cdot \boldsymbol{s}}}{\boldsymbol{q}^2}
\int d^d p\,\frac{1}{\boldsymbol{p}^2+1}.
\end{align}
Including the symmetric \texttt{soft-hard} contribution, one finds
\begin{align}
G_1^{(hs)}(x) + G_1^{(sh)}(x)
=
2^{d-1}\pi^d
\Gamma\!\left(\frac{d}{2}-1\right)
\Gamma\!\left(1-\frac{d}{2}\right)
x^{1-\frac{d}{2}}.
\end{align}

\paragraph{\texttt{soft-soft} region.}
In this case both loop momenta are of order unity, and the exponential can be expanded as
\begin{align}
e^{i(\boldsymbol{q}-\boldsymbol{p})\cdot \boldsymbol{s}} = 1 + \mathcal O(s).
\end{align}
The leading term is therefore
\begin{align}
G_1^{(ss)}(x)
&=
\left(
\int d^d p\,\frac{1}{\boldsymbol{p}^2+1}
\right)^2
=
\pi^d \Gamma\!\left(1-\frac{d}{2}\right)^2.
\end{align}
Collecting the leading contributions from all regions, we obtain the asymptotic expansion
\begin{align}
G_1(x)
\sim
2^{2d-4}\pi^d
\Gamma\!\left(\frac{d}{2}-1\right)^2
x^{2-d}
+
2^{d-1}\pi^d
\Gamma\!\left(\frac{d}{2}-1\right)
\Gamma\!\left(1-\frac{d}{2}\right)
x^{1-\frac{d}{2}}
+
\pi^d
\Gamma\!\left(1-\frac{d}{2}\right)^2
+\cdots.
\end{align}
Matching this expansion against the Frobenius basis $\{u_1,u_2,u_3\}$ immediately yields
\begin{align}
\begin{split}
c_1 &= \pi^d \Gamma\!\left(1-\frac{d}{2}\right)^2, \\
c_2 &= 2^{d-1}\pi^d
\Gamma\!\left(\frac{d}{2}-1\right)
\Gamma\!\left(1-\frac{d}{2}\right), \\
c_3 &= 2^{2d-4}\pi^d
\Gamma\!\left(\frac{d}{2}-1\right)^2.
\end{split}
\end{align}
We consider another the two-loop family that arises in our prebious computations,
\begin{align}
Q_{a_1,a_2,a_3,a_4,a_5}(\boldsymbol{r})
=
\int d^d k\, d^d k_1\,
\frac{e^{D_1}}{D_1^{a_1} D_2^{a_2} D_3^{a_3} D_4^{a_4} D_5^{a_5}},
\end{align}
where
\begin{align}
D_1 = i\,\boldsymbol{k}\!\cdot\!\boldsymbol{r},
\qquad
D_2 = i\,\boldsymbol{k}_1\!\cdot\!\boldsymbol{r},
\qquad
D_3 = \boldsymbol{k}^2+m^2,
\qquad
D_4 = \boldsymbol{k}_1^2 ,
\qquad
D_5 = (\boldsymbol{k}_1-\boldsymbol{k})^2.
\end{align}
Solving IBP identities, we were left with 10 master integrals. However, for our purpose, we need three of them,
\begin{align}
    \mathbfcal{Q}:Q_{0,0,0,1,1},\,Q_{0,0,1,1,1},\, Q_{0,0,2,1,1}.
\end{align}
The differential equation satisfied by them is given by,
\begin{align}
    \partial_{x} \mathbfcal{Q}(x)=\mathbf\Lambda_x \mathbfcal{Q}(x), \qquad \mathbf\Lambda_x =\left(
\begin{array}{ccc}
 \frac{2-d}{x} & 0 & 0 \\
 0 & \frac{3-d}{x} & -\frac{1}{x} \\
 \frac{1}{4} & \frac{2 (d-7) d-x+24}{4 x} & \frac{d-4}{2 x} \\
\end{array}
\right)
\end{align}
Therefore, the system reduces to three coupled differential equations,
\begin{align}
    \begin{split}
        &Q_1'(x)=\frac{2-d}{x}Q_1,\\ &
        Q_2'(x)=\frac{3-d}{x}Q_1-\frac{1}{x}Q_2,\\ &
        Q_3'(x)=\frac{1}{4}Q_1(x)+\frac{2 (d-7) d-x+24}{4 x}Q_2+\frac{d-4}{2x}Q_3.
    \end{split}
\end{align}
As we see, $Q_1$ is decoupled and can be integrated to,
\begin{align}
    Q_1(x)=c_1 x^{2-d}.
\end{align}
After massaging the equations a bit, we find that $Q_2$ satisfies the following differential equation,
\begin{align}
    4 x^2 Q_2'''(x)+(4-6 d)\, x\, Q_2''(x)+(x-2 (d-2) d) Q_2'(x)+(d-2) Q_2(x)=0
\end{align}
 The general solution takes the form,
 \begin{align}
     Q_2(x)=x^{\frac{2-d}{4}}\left(c_1 I_{\frac{d-2}{2}}\left(\sqrt{x}\right)+c_2 K_{\frac{d-2}{2}}(\sqrt{x})\right)+c_3 x^{3-d}\, _1F_2\left(1;4-d,3-\frac{d}{2};\frac{x}{4}\right)
 \end{align}
 Similarly, $Q_3=(3-d)Q_1-x\, Q_2'(x)$. The constants $c_i$ can be fixed by studying the asymptotic expansion of the integrals, as we showed previously.
\section{Useful Feynman integrals} \label{ch1:app:E}
We have seen that there is a direct connection between PN diagrams and the one-loop scalar Feynman diagrams, Fig.~(\ref{fig16}). Below, we list master integrals that will be useful in our context. We take the integrals from \cite{Levi:2011eq}.
\begingroup
\small
\begin{eqnarray}
\label{ch1:tensorfourierindentity}
I^i\equiv\rmint\frac{d^d\bf{k}}{(2\pi)^d}\frac{k^ie^{i\bf{k}\cdot\bf{r}}}{({\bf{k}}^2)^\alpha}&=&\frac{i}{(4\pi)^{d/2}}\frac{\Gamma(d/2-\alpha+1)}{\Gamma(\alpha)}\left(\frac{{\bf{r}}^2}{4}\right)^{\alpha-d/2-1/2}n^i,\\
I^{ij}\equiv\rmint\frac{d^d\bf{k}}{(2\pi)^d}\frac{k^ik^je^{i\bf{k}\cdot\bf{r}}}{({\bf{k}}^2)^\alpha}&=&\frac{1}{(4\pi)^{d/2}}\frac{\Gamma(d/2-\alpha+1)}{\Gamma(\alpha)}\left(\frac{{\bf{r}}^2}{4}\right)^{\alpha-d/2-1}\left(\frac{1}{2}\delta^{ij}+(\alpha-1-d/2)n^in^j\right),\\
I^{ijl}\equiv\rmint\frac{d^d\bf{k}}{(2\pi)^d}\frac{k^ik^jk^le^{i\bf{k}\cdot\bf{r}}}{({\bf{k}}^2)^\alpha}&=&\frac{i}{(4\pi)^{d/2}}\frac{\Gamma(d/2-\alpha+2)}{\Gamma(\alpha)}\left(\frac{{\bf{r}}^2}{4}\right)^{\alpha-d/2-3/2}\nonumber\\
&&\quad
\times\left(\frac{1}{2}\left(\delta^{ij}n^l+\delta^{il}n^j+\delta^{jl}n^i\right)+(\alpha-d/2-2)n^in^jn^l\right),\\
I^{ijlm}\textstyle{\equiv\rmint\frac{d^d\bf{k}}{(2\pi)^d}\frac{k^ik^jk^lk^me^{i\bf{k}\cdot\bf{r}}}{({\bf{k}}^2)^\alpha}}&=&\textstyle{\frac{1}{(4\pi)^{d/2}}\frac{\Gamma(d/2-\alpha+2)}{\Gamma(\alpha)}\left(\frac{{\bf{r}}^2}{4}\right)^{\alpha-d/2-2}\left(\frac{1}{4}\left(\delta^{ij}\delta^{lm}+\delta^{il}\delta^{jm}+\delta^{im}\delta^{jl}\right)\right.} \\
&&
\textstyle{+\frac{\alpha-d/2-2}{2}\left(\delta^{ij}n^ln^m+\delta^{il}n^jn^m+\delta^{im}n^jn^l+\delta^{jl}n^in^m+\delta^{jm}n^in^l+\delta^{lm}n^in^j\right) }\nonumber\\
&&\left.\quad
+(\alpha-d/2-2)(\alpha-d/2-3)n^in^jn^ln^m\right).\nonumber
\end{eqnarray}
\endgroup
We use the d-dimensional master formula for one-loop scalar integrals given by ,
\begin{align}
\begin{split}
   & J\equiv\rmint \frac{d^d\bf{k}}{(2\pi)^d}\frac{1}{\left[{\bf{k}}^2\right]^\alpha\left[({\bf{k}-\bf{q}})^2\right]^\beta}=  \frac{1}{(4\pi)^{d/2}}\frac{\Gamma(\alpha+\beta-d/2)}{\Gamma(\alpha)\Gamma(\beta)}\frac{\Gamma(d/2-\alpha)\Gamma(d/2-\beta)}{\Gamma(d-\alpha-\beta)}\left(q^2\right)^{d/2-\alpha-\beta}.\label{ch1:eq:1loop}
    \end{split}
\end{align}
The d-dimensional master formula for one-loop tensor integrals is taken from \cite{Levi:2011eq}. 
Similarly, one can also derive the following d-dimensional formulas for the one-loop tensor integrals: 
\begin{align}
\begin{split}
J^i\equiv\rmint \frac{d^d\bf{k}}{(2\pi)^d}\frac{k^i}{\left[{\bf{k}}^2\right]^\alpha\left[({\bf{k}-\bf{q}})^2\right]^\beta}=\frac{1}{(4\pi)^{d/2}}\frac{\Gamma(\alpha+\beta-d/2)}{\Gamma(\alpha)\Gamma(\beta)}\frac{\Gamma(d/2-\alpha+1)\Gamma(d/2-\beta)}{\Gamma(d-\alpha-\beta+1)}\\ \left(q^2\right)^{d/2-\alpha-\beta}q^i\end{split}
\end{align}
\begin{align}
\begin{split}
J^{ij}\equiv\rmint \frac{d^d\bf{k}}{(2\pi)^d}\frac{k^ik^j}{\left[{\bf{k}}^2\right]^\alpha\left[({\bf{k}-\bf{q}})^2\right]^\beta}=\frac{1}{(4\pi)^{d/2}}\frac{\Gamma(\alpha+\beta-d/2-1)}{\Gamma(\alpha)\Gamma(\beta)}\frac{\Gamma(d/2-\alpha+1)\Gamma(d/2-\beta)}{\Gamma(d-\alpha-\beta+2)}\left(q^2\right)^{d/2-\alpha-\beta}\\
\times\left(\frac{d/2-\beta}{2}q^2\delta^{ij}+(\alpha+\beta-d/2-1)(d/2-\alpha+1)q^iq^j\right),\end{split}
\end{align}
\begin{align}
\begin{split}
J^{ijl}\equiv\rmint \frac{d^d\bf{k}}{(2\pi)^d}\frac{k^ik^jk^l}{\left[{\bf{k}}^2\right]^\alpha\left[({\bf{k}-\bf{q}})^2\right]^\beta}=\frac{1}{(4\pi)^{d/2}}\frac{\Gamma(\alpha+\beta-d/2-1)}{\Gamma(\alpha)\Gamma(\beta)}\frac{\Gamma(d/2-\alpha+2)\Gamma(d/2-\beta)}{\Gamma(d-\alpha-\beta+3)}\left(q^2\right)^{d/2-\alpha-\beta} \\ 
\times\left(\frac{d/2-\beta}{2}q^2\left(\delta^{ij}q^l+\delta^{il}q^j+\delta^{jl}q^i\right)\right.  \left. 
+(\alpha+\beta-d/2-1)(d/2-\alpha+2)q^iq^jq^l\right),
\end{split}
\end{align}

\begin{align}
\begin{split}
&J^{ijlm}\equiv\rmint \frac{d^d\bf{k}}{(2\pi)^d}\frac{k^ik^jk^lk^m}{\left[{\bf{k}}^2\right]^\alpha\left[({\bf{k}-\bf{q}})^2\right]^\beta}=\frac{1}{(4\pi)^{d/2}}\frac{\Gamma(\alpha+\beta-d/2-2)}{\Gamma(\alpha)\Gamma(\beta)}\\& \frac{\Gamma(d/2-\alpha+2)\Gamma(d/2-\beta)}{\Gamma(d-\alpha-\beta+4)}\left(q^2\right)^{d/2-\alpha-\beta}\,\,\,\,
\times\left(\frac{(d/2-\beta)(d/2-\beta+1)}{4}q^4\left(\delta^{ij}\delta^{lm}+\delta^{il}\delta^{jm}+\delta^{jl}\delta^{im}\right)\right.\\
& \textstyle{
+(\alpha+\beta-d/2-2)(d/2-\alpha+2)\frac{d/2-\beta}{2}q^2 
\times\left(\delta^{ij}q^lq^m+\delta^{il}q^jq^m+\delta^{im}q^jq^l+\delta^{jl}q^iq^m+\delta^{jm}q^iq^l+\delta^{lm}q^iq^j\right)}\\
&\left.\,\,\,\,\,\,\,\,\,
+(\alpha+\beta-d/2-2)(\alpha+\beta-d/2-1)(d/2-\alpha+2)(d/2-\alpha+3)\right.
\left.
\times q^iq^jq^lq^m\right).
\end{split}
\end{align}
For more details on the Feynman integral, especially the massive integrals, can be found in \cite{smirnov}.
    }%
  }%

  \restorethesisbodyformat
  \restoremainchapterstyle
  \chapter{Classical Black-Hole Scattering in Scalar-Tensor Gravity from WQFT}
  \thesischapterpaperbox{Observables from classical black hole scattering in Scalar-Tensor theory of gravity from worldline quantum field theory}{A.~Bhattacharyya, D.~Ghosh, S.~Ghosh, and S.~Pal}{\href{https://doi.org/10.1007/JHEP04(2024)015}{\textit{JHEP} \textbf{04} (2024), 015}; arXiv: \arxivlink{2401.05492}{hep-th}}
  {%
    \restorethesisbodyformat
    \renewcommand{\appendix}{%
      \setcounter{section}{0}%
      \setcounter{subsection}{0}%
      \setcounter{subsubsection}{0}%
      \renewcommand{\thesection}{\thechapter.\Alph{section}}%
      \renewcommand{\thesubsection}{\thesection.\arabic{subsection}}%
      \renewcommand{\theHsection}{chapter.\arabic{chapter}.appendix.\Alph{section}}%
      \renewcommand{\theHsubsection}{\theHsection.\arabic{subsection}}%
    }%
    \ifstrempty{chap2_body.tex}{}{%
\section{Introduction} \label{intro}
The analysis presented in this chapter is based on Ref.~\cite{Bhattacharyya:2024aeq}.
The previous chapter focused on gravitational observables for an inspiralling compact binary, where it is typically well-justified to work in a non-relativistic (post-Newtonian) expansion. In this chapter, we instead study an astrophysical \emph{scattering} event: two compact objects (black holes, or black hole--neutron star systems) approach from asymptotic infinity, interact gravitationally, and re-separate on hyperbolic trajectories. Such burst-like encounters are particularly relevant for the emerging low-frequency gravitational-wave program. Future space-based detectors promise access to the milli-Hz band, while pulsar timing arrays probe nano-Hz frequencies, potentially constraining burst signals from hyperbolic scattering. Beyond their astrophysical content, these observations provide a new arena to test Einstein’s general relativity (GR) in a strong-field, dynamical regime.

From the theory side, extracting information from scattering (or near-scattering) events demands high-precision analytical control over both the conservative dynamics (classical potential) and the emitted radiation \cite{Purrer:2019jcp}. Conceptually, this task mirrors precision computations of scattering cross-sections in particle physics: one is ultimately computing observables derived from amplitudes, organized perturbatively, and matched onto classical physics. In this spirit, perturbative quantum-gravity methods—particularly on-shell and effective-field-theory tools—have proven remarkably efficient for isolating the \emph{classical} gravitational interaction of compact objects.

There is also strong motivation to extend these high-precision predictions beyond GR. Since the first direct detections of gravitational waves, there has been renewed interest in testing GR across all accessible length scales. While GR has been extraordinarily successful, it is widely expected to be incomplete: a consistent UV-complete quantum theory of gravity should exist, and GR does not by itself explain the observed late-time acceleration of the Universe (dark energy). A conservative viewpoint is therefore that GR should emerge as an effective theory within some broader framework. Two broad routes to modifying GR are (i) adding higher-curvature operators, and (ii) introducing extra gravitational degrees of freedom (DOF) \cite{stelle1,Alexeev:1996vs,Lehebel:2018zga,Volkov:2016ehx,Kunz:2006ca}. The simplest extension of the second type is scalar--tensor gravity, where a scalar field accompanies the massless spin-2 graviton \cite{Damour:1992we,Horbatsch:2015bua,Schon:2021pcv,Rainer:1996gw,DeFelice:2011bh}. Such a scalar is phenomenologically motivated: it can appear as an effective description of dark-sector physics, and scalar dynamics can contribute to cosmic acceleration \cite{DeFelice:2011bh,Gsponer:2021obj}. Most importantly for gravitational-wave phenomenology, an additional scalar DOF generically introduces extra radiative channels and polarizations, affecting both energy loss and waveform structure. In this chapter, our primary target is precisely this: to compute the scalar polarization content of radiation emitted during a scattering process.

Theoretical interest in scalar DOF also has a long history. Early unification attempts by Weyl and Kaluza \cite{weyl} initiated a line of ideas in which extra fields arise naturally from higher-dimensional or generalized geometric frameworks. In particular, Kaluza’s five-dimensional proposal (later developed as Kaluza--Klein theory) yields additional lower-dimensional fields upon compactification, and it helped motivate scalar--tensor models such as Jordan’s theory, where the scalar couples non-minimally to the tensor sector. These motivations, together with the observational prospects, make scattering in scalar--tensor gravity a natural testing ground for modern amplitude-based methods.

Classical gravitational scattering has been extensively studied in both post-Newtonian and post-Minkowskian regimes
\cite{hyp,hyp1,hyp2,hyp3,hyp4,hyp5,hyp6,hyp7,hyp8,hyp9,hyp10,hyp11,hyp12,hyp13,hyp16,hyp14,Damour:2016gwp,Bini:2017wfr,Bini:2017xzy,Damour:2017zjx,Damour:2019lcq,Bini:2020flp,Bini:2020uiq,Damour:2020tta,Bini:2020rzn,Bini:2021gat,Bini:2022enm,Damour:2022ybd,Bini:2022wrq,Rettegno:2023ghr,Bini:2023fiz,Ceresole:2023wxg}.
Motivated by the QFT perspective and catalyzed by Damour’s proposal \cite{Damour:2017zjx}, the amplitude community has developed a powerful toolkit for the two-body problem. On the EFT side, a worldline effective-field-theory approach based on the post-Minkowskian (PM) expansion—organizing results in powers of Newton’s constant while resumming PN effects at fixed PM order—was developed for conservative dynamics in \cite{Kalin:2020mvi}, and subsequently extended to include finite-size, spin, and tidal effects \cite{Bini:2020flp,Cheung:2020sdj,Kalin:2020lmz,Haddad:2020que}. Higher-PM developments appear in \cite{Kalin:2020fhe,Dlapa:2021npj,Dlapa:2021vgp,Kalin:2022hph,Jinno:2022sbr,Dlapa:2022lmu,Dlapa:2023hsl,Riva:2021vnj}. In parallel, more direct on-shell amplitude techniques have been used to compute classical gravitational observables
\cite{PhysRevD.7.2317,Holstein:2004dn,Neill:2013wsa,Bjerrum-Bohr:2013bxa,Luna:2017dtq,Bjerrum-Bohr:2018xdl,Kosower:2018adc,Cristofoli:2021vyo,DeAngelis:2023lvf,Brandhuber:2023hhl,Aoude:2023dui,Brandhuber:2023hhy,Georgoudis:2023eke,Herderschee:2023fxh}\footnote{These references are by no means exhaustive; see the citations collected in \cite{Buonanno:2022pgc} for a broader entry point.},
including the 2PM and 3PM conservative potential \cite{Bjerrum-Bohr:2018xdl,Cheung:2018wkq,Cristofoli:2019neg,Cheung:2020gyp,Bern:2019nnu,Bern:2019crd}\footnote{In the context of soft theorems, two-body gravitational waveforms up to 3PM order have been discussed in \cite{ashoke,ashoke1,ashoke2,ashoke3,ashoke4,ashoke5,ashoke6}.}.

A natural question then arises: why do the worldline-EFT and on-shell-amplitude approaches agree so precisely? This gap was bridged by Plefka \emph{et al.} in \cite{Mogull:2020sak}. They showed that the Feynman--Schwinger representation of a graviton-dressed scalar propagator provides the key link between (i) scattering amplitudes of massive scalars and (ii) operator expectation values in \emph{Worldline Quantum Field Theory} (WQFT). WQFT is closely related to worldline EFT (WEFT), but differs in that the worldline degrees of freedom—most notably the worldline fluctuations—are quantized. This framework has enabled detailed cross-checks and new computations: first for the non-spinning case \cite{Jakobsen:2021smu}, then for spinning systems \cite{Jakobsen:2021lvp,Jakobsen:2021zvh,Jakobsen:2023ndj,Jakobsen:2023hig}, and further into higher-PM orders including tidal effects \cite{Jakobsen:2023pvx}. WQFT has also been applied to gravitational lensing \cite{Bastianelli:2021nbs}, and to aspects of the classical double copy \cite{Shi:2021qsb, Diaz-Jaramillo:2021wtl,Comberiati:2022cpm}.

\textit{Most existing scattering analyses focus on GR. In this chapter, we take a step beyond GR by computing scattering observables in a modified theory: a massive scalar--tensor model with a non-trivial cubic and quartic scalar potential. Our broader goal is to assess whether gravitational-wave observations can constrain not only the scalar mass, but also the cubic ($\lambda_3$) and quartic ($\lambda_4$) self-couplings.} As a first step, we compute two central observables in a scattering event: the \emph{impulse} and the \emph{waveform}. The impulse directly determines the scattering angle and interfaces naturally with bound-state information through the EOB map \cite{Buonanno:1998gg}. The waveform controls the radiative sector and ultimately the gravitational-wave phase. Technically, the main challenge in the WQFT computation of radiation is the evaluation of nontrivial worldline integrals. Here we make explicit progress: we compute a set of massive waveform integrals and develop analytic approximation strategies. \textit{To the best of our knowledge, these integral results and the associated methods are new.}

The chapter is organized as follows. In Section~(\ref{ch2:sec2}) we review the WQFT formalism and summarize the key results of \cite{Mogull:2020sak}. In Section~(\ref{ch2:sec3}) we describe how these results are modified by the presence of an additional massive scalar, and we derive the complete set of worldline Feynman rules for scalar and graviton interactions. In Section~(\ref{sec1}) we compute the scalar-induced corrections to the impulse, including diagrams involving scalar self-interactions and scalar--graviton couplings. In Section~(\ref{ch2:sec5}) we construct the radiation integrands and compute the time-domain waveform in the massless-scalar limit. In Section~(\ref{sec6}) we treat the massive case: we explain how the radiation integrals are complicated by more involved phase factors, list the massive radiation integrands, and evaluate the relevant integrals using the stationary-phase method (accurate in the $|x|\to\infty$ regime), discussing subtleties along the way. Appendix~(\ref{ch2:app:A}) derives the worldline effective action with an extra scalar DOF. Appendix~(\ref{ch2:app:B}) explains the relation between impulse and the scattering amplitude via the eikonal phase. Appendices~(\ref{ch2:app:C}), (\ref{ch2:app:D}), (\ref{ch2:app:E}), (\ref{3.54kk}) and (\ref{App:3.H}) provide additional details about some integrals, while Appendix~(\ref{ch2:app:F}) presents an alternative approximation based on a large-velocity expansion. 
\subsection*{Notations and Conventions}
\begin{itemize}[leftmargin=1.55em,itemsep=0.42em,topsep=0.45em,label=\textcolor{brandburgundy}{\small\textbullet}]
\item Metric sign convention: $(+,-,-,-)$.
\item All computations are done in units where $(c,\hbar)=1$.
\item The impact parameter $b$ is purely spacelike, $b^2=b^\mu b_\mu=-|b|^2$ and $b^{\mu}=(0,0,|b|,0)$.
\item Black-hole velocity parametrization: $v_1=(\gamma,\gamma\beta,0,0)$ and $v_2=(1,0,0,0)$.
\item Planck mass: $m_p=\frac{1}{\sqrt{8\pi G_N}}$, where $G_N$ is Newton's constant.
\item Scaled delta function: $\hat\delta^{(D)}(\cdots)\equiv (2\pi)^{D}\delta^{(D)}(\cdots)$.
\item Incomplete beta function: $B_{z}(a,b):=\rmint_0^z dt\,t^{a-1}(1-t)^{b-1}$.
\item $J_n$ and $K_n$ denote the Bessel function of the first kind and the modified Bessel function of the second kind, respectively.
\item The \textbf{Meijer G} function is defined as
\begin{align}
    G_{p q}^{m n} \left(z,r\Big|
\begin{array}{c}
 a_1\text{...}.a_p \\
 b_1\text{...} b_q \\
\end{array}\right)
=\frac{r}{2 \pi  i}\rmint_{-i\infty}^{i\infty} \frac{   \Gamma  \left(b_1+r s\right) \Gamma\left(-a_n-r s+1\right) \Gamma  \left(-a_1-r s+1\right) \Gamma\left(b_m+r s\right)}{  \Gamma  \left(a_p+r s\right) \left(-b_q-r s+1\right) \Gamma  \left(a_{n+1}+r s\right) \Gamma  \left(-b_{m+1}-r s+1\right)}z^{-s}\,ds\,.\nonumber
\end{align}
\item The Bickley–Naylor function is defined as $ \textrm{Ki}_{n}(x)=\int_0^{\pi/2}  d\theta\,e^{-\frac{x}{\cos\theta}}\cos^{n-1}\theta$.
\end{itemize}
\section{A quick tour to worldline quantum field theory}\label{ch2:sec2}
In this section, we briefly summarize the results of the Worldline Quantum Field Theory (WQFT) approach to the scattering problem. In \cite{Mogull:2020sak}, precise correspondence has been derived between the scalar-graviton S-matrix element and the one-point functions of the worldline operators. In this string-inspired approach, one maps black hole scattering in worldline theory to field scattering in the field theory approach. Using the gravitationally dressed Green's functions, one can write S-matrix elements as the expectation value of operators present in the worldline theory.

Now, for spinless black holes in GR, the action can be written in the EFT framework as\footnote{To get the linearized action one should expand $\sqrt{-g}$ as,
\begin{align}
    \begin{split}
        \sqrt{-g}=1+\frac{1}{2m_p} h-\frac{1}{4m_p^2} h^{\mu\nu}h_{\mu\nu}+\frac{1}{8m_p^2}h^2,\,\, h=\text{Tr}( h_{\mu\nu}).\nonumber
    \end{split}
\end{align}},
$$S=S_{EH}+S_{gf}+\sum_i S_{pm}^i\,,$$
where $S_{EH}$ is the usual Einstein-Hilbert action, and $S_{gf}$ is the gauge-fixing action, which in the weak field approximation is given by,

$$S_{gf}=\rmint d^D x\Big(\partial_\nu h^{\mu\nu}-\frac{1}{2}\partial^\mu h^{\nu}_{\nu}\Big)^2\,.$$

This imposes the de-Donder gauge condition. Now, for an extended object, the worldline action consists of the Wilson coefficients $c_v$ and $c_R$ as follows,
\begin{align}S_{pm}=-m\rmint d\tau +c_R\rmint d\tau R(x)+c_v\rmint d\tau R_{\mu\nu}(x)\dot{x}^\mu \dot{x}^\nu+\cdots\,.\end{align}
In our paper, we will drop the second and third terms above to remove the complicacy due to the finite-size effects. In principle, there will be an infinite tower of terms consisting of higher-order derivatives of the metric represented as dots.

Now, assuming a fixed background, we can write Green's function of the field theory as a two-point correlator : 
\begin{align}
G_i(x,x')=\mathcal{Z}_i^{-1}\rmint \mathcal{D}[\phi_i]\phi_i(x)\phi_i^{\dagger}(x')e^{i S_i},
\end{align}
where, $S_i=\rmint d^D x \sqrt{-g}(g^{\mu\nu}\partial_\mu\phi_i^{\dagger}\partial_\nu\phi_i-m^2\phi_i^{\dagger}\phi_i-\zeta R\phi_i^{\dagger}\phi_i)$.\par
 To this end, we have to integrate out certain field degrees of freedom with respect to a relevant scale of the problem. This will give rise to the one-loop effective action, which can be represented by a worldline path integral \cite{Strassler:1992zr,Feal:2022iyn,Ahmadiniaz:2022yam}. Then, from this path integral, we can identify the point particle action in the background of $h_{\mu\nu}$.
Now, for our interest, we evaluate the S-matrix element of two scalars with or without a final state graviton in certain \textit{classical limit} as,
\begin{align}
&\langle \Omega|T\{h_{\mu\nu}\phi_1(x_1)\phi_1^{\dagger}(x_1')\phi_2(x_2)\phi_2^{\dagger}(x_2')\}|\Omega\rangle\\&
=\mathcal{Z}_i^{-1}\rmint \mathcal{D}[h_{\mu\nu},\phi_i]h_{\mu\nu}(x)\prod_{i=1}^2\phi_i(x_i)\phi_i^{\dagger}(x_i')e^{i S'}\,.
\end{align}
Fourier transforming this relation and using LSZ reduction in the $\hbar \rightarrow 0 $ limit we get \cite{Mogull:2020sak},
\begin{align}&\langle \phi_1\phi_2(+h)|S|\phi_1\phi_2\rangle\\&
=\mathcal{Z}^{-1}\rmint d^D[x_i,x_i',x]e^{ip\cdot x-ip_i'\cdot x_i'-ik\cdot x}\times \rmint [\mathcal{D}h_{\alpha\beta}]\,\epsilon^{\mu\nu}(k)h_{\mu\nu}(x)\prod_{i=1}^2 G_i(x_i,x_i')e^{i S_{EH}+S_{gf}}\bigg{|}_{\textrm{Amputated,connected}}\,.
\end{align}
The study of the S-matrices consists of applying the LSZ reduction. Cutting the propagators on external legs, we convert correlators into S-matrices and send the external legs to infinity, where they interact weakly. We put the scalar legs on-shell, and the integration over the finite time domain extends $\tau \in (-\infty, \infty)$.
The key relation connecting the QFT form factor to the WQFT correlator is given by \cite{Mogull:2020sak},
\begin{align}
\begin{split}
&\frac{\Xi(b,v;\{\epsilon^{l},k_{l}\})}{\Xi_0}=\delta\Bigg(\sum_{l=1}^N k_l\cdot v\Bigg) \exp\Big({\sum_{l=1}^N k_l\cdot b} \Big)\mathcal{F}(p,p'|\{\epsilon^{l},k_{l}\})
\end{split}
\end{align}
\newcommand{\chthreeinlinefeynwidth}{0.13\linewidth}
\newcommand{\chthreeleadfeynwidth}{0.32\linewidth}

where,
\begin{align}
  &  \Xi(b,v;\{\epsilon^{l},k_{l}\})=\rmint \mathcal{D}[x]\rmint \mathcal{D}[a,b,c]\exp[-i\rmint d\sigma \Big[\frac{1}{4}g_{\mu\nu}(\dot{x}^\mu\dot{x}^\nu+a^\mu a^\nu+b^\mu c^\nu)\Big].
\end{align}
Having the gravitationally dressed propagator in momentum space, we put the external scalar legs on-shell to perform the LSZ reduction. Effectively, it can be represented as,

\begin{center}
\includegraphics[width=\chthreeleadfeynwidth]{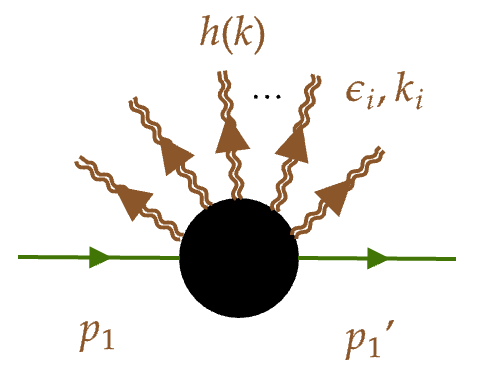}
\end{center}
\begin{align}
\mathcal{F}(p,p'|\{\epsilon^{l},k_{l}\})=\langle p'|\prod_{i=1}^N\epsilon_i\cdot h(k_i)|p\rangle\,.
\end{align}
For details, we refer the reader to \cite{Mogull:2020sak}.
In the following section, we will discuss how to derive the Feynman rules for the scattering events of two black holes (including extra scalar modes), which are considered to be point-like objects and, hence, spinless.

\section{Derivation of the Feynman rules}\label{ch2:sec3}
One of the key ingredients to compute the observables in a scattering problem is the partition function. 
We compute the partition function for a binary system in the Scalar-Tensor theory of gravity with a generic scalar potential. The gravitational action is given by,
\begin{align}
    \begin{split}
        S_{g}=\rmint d^4x \sqrt{- g}\Big(-\frac{m_p^2}{2} R+\frac{1}{2} {g}^{\alpha\beta}\partial_{\alpha}{\varphi}\partial_{\beta}{\varphi}-\frac{1}{2}m^2{\varphi}^2-\frac{\lambda_3}{\textcolor{black}{3!}} m_p\, \varphi^3-\frac{\lambda_4}{\textcolor{black}{4!}}{\varphi}^4\Big)\in S_{EH}+S_{scalar}.\label{3.1m}
    \end{split}
\end{align}
In this theory, the mass of the point particle which is the avatar of the scalar field $\phi_i$, is not constant but rather depends on the extra scalar degree of freedom $\varphi$ \cite{eardley}. Hence the matter action has the following form,
\begin{align}
    \begin{split}
        S_m=\sum_{i=1}^2\rmint d^4 x \sqrt{-g}(g^{\mu\nu}\partial_\mu\phi_i^{\dagger}\partial_\nu\phi_i-m_i(\varphi)^2\phi_i^{\dagger}\phi_i-\zeta R\phi_i^{\dagger}\phi_i)\,.
    \end{split}
\end{align}
Therefore, the WQFT partition function takes the form \footnote{A short derivation of the worldline effective action in the presence of extra scalar degrees of freedom appearing in the partition function is given in Appendix~(\ref{ch2:app:E}).} ,
\begin{align}
    \begin{split}
        {Z}_{\textrm{WQFT}}=\mathcal{N}\times \rmint \mathcal{D}[h_{\mu\nu},\,{\varphi}]\rmint \mathcal{D}[x_i,a_i,b_i,c_i]e^{iS_{g}}\exp\Big\{-i\sum_{i=1}^{2}\rmint d\tau_{i}\frac{m_i(\varphi)}{2}g_{\mu\nu}(\dot x_{i}^{\mu}\dot x_{i}^{\nu}+a_{i}^{\mu}a_{i}^{\nu}+b_{i}^{\mu}c_{i}^{\nu})\Big\}\label{3.2}
    \end{split}
\end{align}
where $a_i,b_i$ s are Lee-Yang Ghost and can be ignored in the classical limit. Therefore, the $n$-body partition function (non-spinning) in WQFT can be written as,
\begin{align}
    \begin{split}
        Z_{\textrm{WQFT}}^{(n)}=\mathcal{N}\rmint \mathcal{D}h_{\mu\nu}\mathcal{D}\varphi\,e^{iS_{EH}+iS_{scalar}}\Big(\prod_{i=1}^{n}\mathcal{D}x_{i}\,e^{iS_{pm}^i}\Big)\,.\label{3.3}
    \end{split}
\end{align}
It has been noticed that the partition function is related to the Eikonal phase  $\chi$ by the following exponentiation \cite{Amati:1990xe, Mogull:2020sak}.
\begin{align}
    \begin{split}
Z_{\textrm{WQFT}}:=e^{i\chi}\,.
    \end{split}
\end{align}
Now $\chi$ can be calculated by computing the connected Feynman diagrams. In order to compute the path integral in \eqref{3.3} one could start by demanding that a gravitational field is a fluctuation over Minkowski space, i.e., one can decompose the metric $g_{\mu\nu}$ as,
\begin{align}
    \begin{split}
g_{\mu\nu}=\eta_{\mu\nu}+\frac{h_{\mu\nu}}{m_p}.
    \end{split}
\end{align}
and, eventually, the worldline degree of freedom can be expanded as,
\begin{align}
    \begin{split}
        x^{\mu}(\tau)=b^{\mu}+v^{\mu}\tau+z^{\mu}(\tau)
    \end{split}
\end{align}
where, $z^{\mu}(\tau)$ is the worldline fluctuation about the straightline geodesic. Once we have the well-defined partition function, one could compute the classical observables as a correlation function in WQFT.
\begin{align}
    \begin{split}
        \mathcal{O}(b_i,v_i):=\langle\hat{\mathcal{O}}(\hat{x},\hat{h}_{\mu\nu},\hat\varphi)\rangle=\frac{1}{Z^{(n)}_{\textrm{WQFT}}}\rmint \mathcal{D}h_{\mu\nu}\mathcal{D}\varphi\,e^{iS_{EH}+iS_{scalar}}\Big(\prod_{i=1}^{n}\mathcal{D}x_{i}e^{iS_{pm}^i}\Big)\mathcal{O}(x_i,h,\varphi)
    \end{split}
\end{align}
where the worldline action can be written in Polyakov form.
The point particle action has the following form,
\begin{align}
    \begin{split}
        S_{pm}=-\sum_{i=1}^{n}\rmint_{-\infty}^{\infty}d\tau_i\frac{m_i( \varphi)}{2}(g_{\mu\nu}\dot x^{\mu}_i\dot{x}^{\nu}_i+1).\label{3.4m}
    \end{split}
\end{align}
We intend to compute the two main observables, impulse and waveform, schematically defined as,
\begin{align}
    \begin{split}
      \langle \hat h_{\mu\nu}(k),\hat\varphi(k),\hat z^{\mu}(\omega)\rangle=Z_{\textrm{WQFT}}^{-1}\rmint\mathcal{D}h_{\mu\nu}\mathcal{D}\varphi\,e^{iS_{EH}+iS_{scalar}}\Big(\prod_{i=1}^{2}\mathcal{D}x_{i}e^{iS_{pm}^i}\Big)\{h_{\mu\nu}(k),\varphi(k),z^{\mu}(\omega)\}\,.
    \end{split}
\end{align}
Before going to the computations of the partition function, we first derive all the Feynman rules. In Scalar-Tensor theory, the violation of the equivalence principle automatically implies that the mass of the binaries depends on the extra gravitational polarisation ${\varphi}$. As there is a self-interaction term so one can expand $m_{a}(\varphi)$ around one of the chosen vacuums ${\varphi}_{0}$, which for our case is zero, as,
\textcolor{black}{
\begin{align}
    \begin{split}
        m_a({\varphi})=m_a(0)\Big \{1+s_a\frac{\varphi}{m_p}+g_a\frac{\varphi^2}{m_p^2}+\mathcal{O}(\varphi^3)\Big\}.
    \end{split}
\end{align}}
In order to derive the Feynman rules it would be better to decompose the fields  ($\chi(x)\in(h_{\mu\nu},\varphi)$) in their Fourier mode assuming the black hole worldline has following the fluctuation due to the radiation reaction: $x_{(i)}^{\mu}=b_{(i)}^{\mu}+v_{(i)}^{\mu}\tau+z_{(i)}^{\mu}(\tau)$. where $z_{i}^{\mu}(\tau)$ is the worldline fluctuations about the straight line geodesic. Now, we expand the point particle action at different orders in $z$ to get the worldline vertices. In our theory we have gravitational field ($ h_{\mu\nu}$) and scalar field (${\varphi}$). One can decompose the fields in Fourier modes, when it couples with worldline, as,
\begin{align}
    \begin{split}
        &  \chi(x_i)=\rmint_{k}e^{ik\cdot(b_i+v_i\tau+z_i)} \chi(-k)\,.
    \end{split}
\end{align}
Now, expanding the exponential at different orders in $z$ and is given by,
\begin{align}
    \begin{split}
         \chi (x_i)=\rmint_{k}e^{ik\cdot(b_i+v_i \tau)}\prod_{n=0}^{\infty}\frac{i^{n}}{n!}(k\cdot z)^{n}  \chi(-k)\,.\label{3.8m}
    \end{split}
\end{align}
Again, we decompose the worldline fluctuations as,
\begin{align}
    \begin{split}
        z^{\mu}(\tau)=\rmint_{\omega}e^{i\omega \tau}z^{\mu}(-\omega)\,.
    \end{split}\end{align}
From the quadratic part of Einstein-Hilbert action, one can identify the graviton propagator as,
\begin{align}
   \langle  h_{\mu\nu}(x_1) h_{\rho\sigma}(x_2)\rangle\equiv \thesisinlinefeynman[\chthreeinlinefeynwidth]{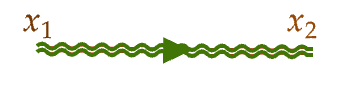}\hspace{0.45cm}=iP_{\mu\nu;\rho\sigma}\rmint_{k}\frac{e^{-ik\cdot(x_1-x_2)}}{k^2+i\epsilon}
\end{align}
where,
\begin{align}
    \begin{split} \label{tensor}
        P_{\mu\nu,\rho\sigma}=\textcolor{black}{\frac{1}{2}}\Big(\eta_{\mu\rho}\eta_{\nu\sigma}+\eta_{\mu\sigma}\eta_{\nu\rho}-\eta_{\mu\nu}\eta_{\rho\sigma}\Big).
    \end{split}
\end{align}
and propagator for the scalar degrees of freedom has the following form,
\begin{align}
    \begin{split}
        \langle \varphi(x_1)\varphi(x_2)\rangle\equiv \thesisinlinefeynman[\chthreeinlinefeynwidth]{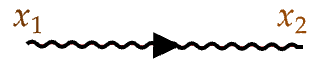}\hspace{0.50cm}= i\rmint_{k}\frac{e^{-ik\cdot(x_1-x_2)}}{k^2-m^2+i\epsilon} \,.
    \end{split}
\end{align}
Apart from the field degrees of freedom, we have another dynamical degree of freedom: worldline fluctuations $z^{\mu}(\omega)$. One can identify the propagator by inspecting the quadratic part of the worldline action (\ref{3.4m}).
\begin{align}
    \begin{split}
        S_{\textrm{pm}}=-m(\varphi_{0})\rmint d\tau \Big[1+\eta_{\mu\nu}v^{\mu}\dot z^{\nu}+\frac{1}{2}\eta_{\mu\nu}\dot z^{\mu}\dot z^{\nu}\Big]\label{3.12m}
    \end{split}
\end{align}
The propagator of $z^{\mu}$ can be identified from the quadratic part of (\ref{3.12m}).
\begin{align}
    \begin{split}
      \Big  \langle z^{\mu}(\tau_1)z^{\nu}(\tau_2)\Big\rangle\equiv \thesisinlinefeynman[\chthreeinlinefeynwidth]{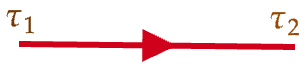}\hspace{0.35cm}=-i\frac{\eta^{\mu\nu}}{m}\rmint_{\omega}\frac{e^{-i \omega (\tau_1-\tau_2)}}{(\omega\pm i\epsilon)^2}\,.
    \end{split}
\end{align}
Now, we are in a position to derive the worldline Feynman rules. The key ingredient is the worldline action $S_{\textrm{pm}}$ in (\ref{3.4m}).
\begin{align}
    \begin{split}
        S_{\textrm{pm}}\Big|_{\textrm{int.}}\in \rmint d\tau \Big[\Big(-\frac{m_a}{2}-\frac{m_a s_a}{2m_p} \varphi-\frac{m_a g_a}{2m_p^2} \varphi^2\Big)(\eta_{\mu\nu}+\frac{1}{m_p}h_{\mu\nu})(v^{\mu}+\dot z^{\mu})(v^{\nu}+\dot z^{\nu})\Big]\,.
    \end{split}
\end{align}
We can rewrite (\ref{3.8m}) by inserting the Fourier mode of the worldline fluctuation in the following way.
\begin{align}
    \begin{split}
         \chi=\sum_{m=0}^{\infty}\frac{i^m}{m!}\rmint_{k,\omega_1,...,\omega_m}e^{ik\cdot b}e^{i(k\cdot v+\sum_{i=1}^{m} \omega_{i})\tau}\prod_{i=1}^{m}[k\cdot z(-\omega_i)] \chi(-k).\label{3.15m}
    \end{split}
\end{align}
The expansion in \eqref{3.15m} helps derive the Feynman rules up to $\mathcal{O}(z^{j})$. We now list the Feynman rules coming from the scalar-worldline interaction:
\vspace{-0.4 cm}
\subsection*{\underline{Vertex for scalar field:}}
{\scriptsize
    \textbullet $\,\,$ $\mathcal{O}{(z^0)}:- i\frac{m_a s_a}{2m_p} \,e^{ik\cdot b_a}\,\hat\delta(k\cdot v)\rightarrow  \thesisinlinefeynman[\chthreeinlinefeynwidth]{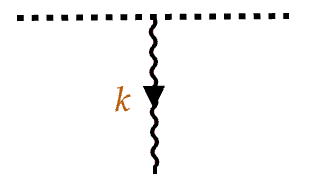}$

 \textbullet $\,\,$ $\mathcal{O}(z):\frac{m_a s_a}{2m_p}\,e^{ik\cdot b}\,\hat{\delta}(k\cdot v+\omega)(2\omega v_{\rho}+k_{\rho})\rightarrow \thesisinlinefeynman[\chthreeinlinefeynwidth]{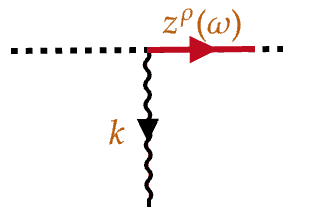}$

 \textbullet $\,\,$ $ \mathcal{O}(z): -\frac{m_a s_a}{2m_p^2}e^{i(k_1+k_2)\cdot b_a}\hat\delta(k_1\cdot v_1+k_2\cdot v_1+\omega)\{(k_{1\rho}+k_{2\rho})v^{\mu}v^{\nu}+2\omega v^{(\mu}\delta_{\rho}^{\nu)}\}\rightarrow
\thesisinlinefeynman[\chthreeinlinefeynwidth]{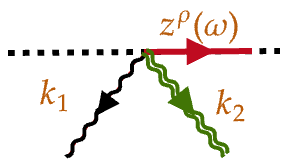}$

 \textbullet $\,\,$ $ \mathcal{O}(z): -\frac{m_a g_a}{2m_p^2}e^{i(k_1+k_2)\cdot b_a}\hat\delta(k_1\cdot v_1+k_2\cdot v_1+\omega)\{(k_{1\rho}+k_{2\rho})v^{\mu}v^{\nu}+2\omega v_\rho\}\rightarrow
\thesisinlinefeynman[\chthreeinlinefeynwidth]{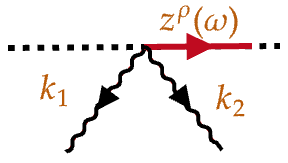}$

   \textbullet $\,\,$ $\mathcal{O}{(z^2)}:i\frac{m_a s_a}{2m_p} \,e^{ik\cdot b_a}\,\hat\delta(k\cdot v+\omega_1+\omega_2)  (\frac{1}{2} k_{\rho_1} k_{\rho_2} +\omega_1 k_{\rho_2}v_{\rho_1}+\omega_2 k_{\rho_1}v_{\rho_2}+\omega_1\omega_2 \eta_{\rho_1\rho_2})\rightarrow  \thesisinlinefeynman[\chthreeinlinefeynwidth]{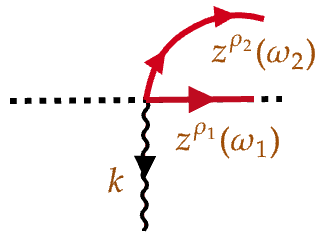}$
\subsection*{\underline{Vertex for gravitational field:}}
     \textbullet $\,\,$ $\mathcal{O}{(z^0)}:- i\frac{m_a }{2 m_p} \,e^{ik\cdot b_a}\,\hat\delta(k\cdot v)v_a^\mu v_a^\nu \rightarrow  \thesisinlinefeynman[\chthreeinlinefeynwidth]{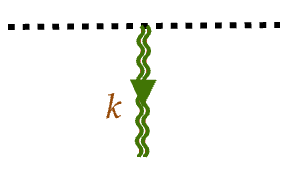}$

 \textbullet $\,\,$ $\mathcal{O}(z^1):\frac{m_a }{2 m_p}\,e^{ik\cdot b}\,\hat{\delta}(k\cdot v+\omega)(2\omega v^{(\mu}\delta^{\nu)}_\rho+v^\mu v^\nu k_{\rho})\rightarrow \thesisinlinefeynman[\chthreeinlinefeynwidth]{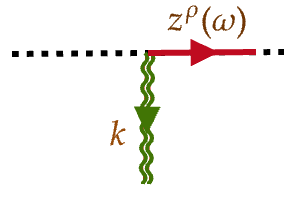}$

 \textbullet $\,\,$ $\mathcal{O}{(z^2)}:i\frac{m_a }{m_p} \,e^{ik\cdot b_a}\,\hat\delta(k\cdot v+\omega_1+\omega_2) (\frac{1}{2} k_{\rho_1} k_{\rho_2} v^\mu v^\nu +\omega_1 k_{\rho_2}v^{(\mu}\delta^{\nu)}_{\rho_1}+\omega_2 k_{\rho_1}v^{(\mu}\delta^{\nu)}_{\rho_2}+\omega_1\omega_2 \delta^{(\mu}_{\rho_1}\delta^{\nu)}_{\rho_2})\rightarrow  \thesisinlinefeynman[\chthreeinlinefeynwidth]{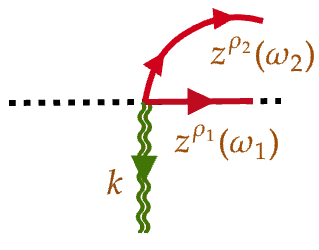}$
}
\section{Computation of impulse}\label{sec1} 
In this section, we will compute the impulse in a purely conservative setting. We mainly focus on the corrections to the impulse coming from the purely scalar degree of freedom or the interaction between scalar and graviton.
The impulse is defined as,
\begin{align}
    \begin{split}
        \Delta p^{\mu}_{i}=m_i \rmint_{-\infty}^{\infty}d\tau_{i}\Big\langle\frac{d^2 z_{i}^{\mu}}{d\tau_{i}^2}\Big\rangle_{\text{WQFT}}=-m_i \omega^2\langle z_{i}^{\mu}(\omega)\rangle|_{\text{WQFT}}\Big |_{\omega=0}.\label{4.1w}
    \end{split}
\end{align}
We now compute (\ref{4.1w}) order by order in Newton's constant $G_N$ up to 2PM order. 
\subsection{1PM contribution to impulse}
\textbullet $\,\,$ The simplest diagram at  1 PM order consists only of a scalar field that has the following form, 
\begin{align}
    \begin{split}
        [\Delta p_1^{\mu}]_{(a)}=\thesisinlinefeynman[\chthreeinlinefeynwidth]{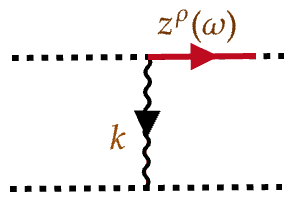}
\hspace{0.35cm}&=-i\frac{m_1s_1m_2s_2}{4m_p^2}\rmint_{k_1,k_2}e^{ik_1\cdot b_1+ik_2\cdot b_2}\frac{\hat\delta(k_2\cdot v_2)\hat\delta(\omega-k_1\cdot v_1)}{k_1^2-m^2}\hat\delta^{(4)}(k_1+k_2)\\&  \hspace{0.8cm}\times(k_1^{\mu}+2\omega v_1^\mu)\Big|_{\omega=0}\,,\\ &
=\frac{\partial}{\partial b_{1\mu}}\underbrace{\Big[-\frac{m_1s_1m_2s_2}{4m_p^2}\rmint_{k}e^{ik\cdot(b_1-b_2)}\frac{\delta(k\cdot v_1)\hat\delta(k\cdot v_2)}{k^2-m^2}\Big]}_{\chi}\,.
    \end{split}
\end{align}

To do this integral, we need to choose the coordinates suitably. We study the dynamics from the frame of the second black hole. Hence in the mentioned parametrization of velocities
satisfying $v_1\cdot v_2=\gamma$.
Hence \footnote{Note that $(2\pi)$ is associated with each one-dimensional delta function and $(2\pi)^4$ factor with a four-dimensional delta function. Now, for each of the momentum integrals, there is a $\frac{1}{(2\pi)^4}$ factor associated with the integration measure. One has to take into account all of these factors. Furthermore, there will be a $(2\pi)$ factor whenever we get a ${\bf K_0}$ or ${\bf J_0}$  and $(2\pi)^2$ factor whenever we get a $\boldsymbol{\arctan}$ after doing an integral. One needs to consider these to get the correct factor of $\pi$ at the end. In all the subsequent integrals, we have taken this into account. },
\begin{align}
    \begin{split}
    \chi&=-\frac{m_1m_2s_1s_2}{4m_p^2}\rmint_{k}\frac{\hat{\delta}{(\gamma k^{(0)}-\gamma\beta k^{(1)})}\hat\delta(k^{0})}{k^2-m^2}e^{ik\cdot b}\,,\\ &
    =\frac{\pi ^2 m_1m_2s_1s_2}{m_p^2\sqrt{\gamma^2-1}}
   \rmint_{\tilde k}\frac{e^{-i\tilde k \cdot b}}{\tilde k^2+m^2}=\frac{m_1m_2s_1s_2}{8\pi m_p^2\sqrt{\gamma^2-1}}K_{0}(m|b|)\,.
    \end{split}
\end{align}
Therefore,
\begin{align}
\begin{split} \label{e1:barwq2}
  [\Delta p_1^{\mu}]_{(a)}&=-\frac{m_1m_2s_1s_2}{8\pi m_p^2\sqrt{\gamma^2-1}} \frac{b^\mu}{|b|}m K_1(m|b|)\,.
\end{split}
\end{align} 
In the massless limit the impulse took the form,
\begin{align}
    \begin{split}
          [\Delta p_1^{\mu}]_{(a)}\Big|_{m\to 0}=-\frac{m_1m_2s_1s_2}{8\pi m_p^2\sqrt{\gamma^2-1}} \frac{b^\mu}{|b|^2}.
    \end{split}
\end{align}
\vspace{0.5cm}
\textbullet $\,\,$ One more diagram comes from the self-interaction vertex, which contributes at 1 PM order.
\begin{align}
    \begin{split}
        \mathcal{O}(z,\lambda_2):-\frac{\lambda_3}{\textcolor{black}{3!}} m_p\rmint d^4x\,\varphi(x)^3.
    \end{split}
\end{align}
Then\footnote{\textcolor{black}{Again note that the combinatorial factor associated with this diagram is 3!. We have multiplied it by that. For the subsequent diagrams, we will also multiply by the suitable combinatorial factors from the beginning.} },
\begin{align}
    \begin{split}
        [\Delta p_1^{\mu}]_{(b)}=\thesisinlinefeynman[\chthreeinlinefeynwidth]{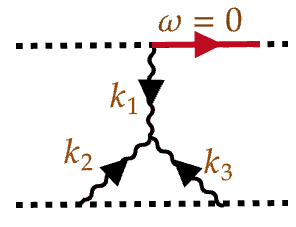}&=-i\lambda_3\frac{m_1s_1}{2}\times \Big(\frac{m_2s_2}{2m_p}\Big)^2\rmint_{k_i}\hat\delta^{(4)}\Big(\sum_ik_i\Big)\frac{k_1^\mu\hat\delta(k_1\cdot v_1)\hat\delta(k_2\cdot v_2)\hat\delta(k_3\cdot v_2)}{\prod_{i=1}^3(k_i^2-m^2)}e^{ik_1\cdot b_1}e^{i(k_2+k_3)\cdot b_2}\,,\\ &
=-i\lambda_3\frac{m_1s_1}{2}\times \Big(\frac{m_2s_2}{2m_p}\Big)^2\rmint_{k_{1,2}}\frac{k_1^\mu\hat\delta(k_1\cdot v_1)\hat\delta(k_2\cdot v_2)\hat\delta(k_1\cdot v_2)}{(k_1^2-m^2)(k_2^2-m^2)[(k_1+k_2)^2-m^2]}e^{ik_1\cdot (b_1-b_2)}\,.
    \end{split}
\end{align}
Now using the proper velocity parametrization and the integral results derived in \cite{Bhattacharyya:2023kbh}, we get,
\begin{align}
    \begin{split}
         [\Delta p_1^{\mu}]_{(b)}&=-i\lambda_3\frac{m_1s_1}{2(2\pi)^3}\times \Big(\frac{m_2s_2}{2m_p}\Big)^2\frac{1}{\gamma\beta}\rmint d^2 l_1\frac{l_1^\mu e^{il_1\cdot b}}{(\vec l_1^2+m^2)|\vec l_1|}\arctan\Big(\frac{|\vec l_1|}{2m}\Big)\,,\\ &
        =\lambda_3\frac{m_1s_1}{2(2\pi)^2}\times \Big(\frac{m_2s_2}{2m_p}\Big)^2\frac{1}{\gamma\beta}\frac{\partial}{\partial b_{1\mu}}\underbrace{\rmint_{0}^{\infty}dl_1\,\frac{J_{0}(|b|l_1)}{ l_1^2+m^2}\arctan\Big(\frac{l_1}{2m}\Big)}_{I_3(m,b)}\,,\\ &
        =\lambda_3\frac{m_1s_1}{2(2\pi)^2}\times \Big(\frac{m_2s_2}{2m_p}\Big)^2\frac{1}{\sqrt{\gamma^2-1}}\frac{b^\mu}{|b|}\frac{\partial I_1(m,|b|)}{\partial |b|}\,.
    \end{split}
\end{align}  
Finally, we get, 

\begin{equation}\label{e:barwq2}
  [\Delta p_1^{\mu}]_{(b)}=\lambda_3\frac{m_1s_1}{8\pi^2}\times \Big(\frac{m_2s_2}{2m_p}\Big)^2\frac{1}{\sqrt{\gamma^2-1}}\frac{b^\mu}{|b|}\frac{\partial I_1(m,|b|)}{\partial |b|}\,.
\end{equation}\\
\textcolor{black}{To the best of our knowledge this integral does not have any closed-form expression. One can perform a numerical analysis to solve the integral. Moreover, the integral has a smooth massless limit.} In the massless limit, the contribution gives,
\begin{align}
    \begin{split}
         [\Delta p_1^{\mu}]_{(b)}\Big|_{m\to 0}=-\lambda_3 \frac{\pi m_1s_1}{8\pi}\times \Big(\frac{m_2s_2}{4m_p}\Big)^2\frac{1}{\sqrt{\gamma^2-1}}\frac{b^\mu}{|b|}\,.
    \end{split}
\end{align}
Then, the total impulse (due to the scalar field) at 1PM order is the sum of (\ref{e1:barwq2}) and (\ref{e:barwq2})  as well as the terms that come from interchanging the worldline one and two\,.
\begin{equation}
     \Delta p_1^{\mu}\Big|^{\textrm{1PM}, \textrm{Total}}_{\textrm{scalar}}=   [\Delta p_1^{\mu}]_{(a)}+ [\Delta p_1^{\mu}]_{(b)}+1\leftrightarrow 2\,.
\end{equation}
Finally, collecting all individual expressions we get, 
\textcolor{black}{
\begin{align}
\begin{split}
\Delta p_1^{\mu}\Big|^{\textrm{1PM}, \textrm{Total}}_{\textrm{scalar}}= &-\frac{m_1m_2s_1s_2}{8\pi m_p^2\sqrt{\gamma^2-1}} \frac{b^\mu}{|b|}m K_1(m|b|)+\lambda_3\frac{m_1s_1}{8\pi^2}\times \Big(\frac{m_2s_2}{2m_p}\Big)^2\frac{1}{\sqrt{\gamma^2-1}}\frac{b^\mu}{|b|}\frac{\partial I_1(m,|b|)}{\partial |b|}\,+1\leftrightarrow 2,\label{3.42b}
\end{split}
\end{align}
where, $$I_1(m,|b|)=\rmint_{0}^{\infty}dx\,\frac{J_{0}(|b|x)}{ x^2+m^2}\arctan\Big(\frac{x}{2m}\Big)\,.$$}
The expression in \eqref{3.42b} is not a convenient one; further massaging, we get the following,
\begin{align}
   \partial_{b} I_1=- \int _0^\infty\frac{x\,J_1(|b|x)}{x^2+m^2}\arctan\left(\frac{x}{2m}\right)=-2\int_0^1 ds \frac{K_1(m|b|)-\frac{2}{s}K_1\left(\frac{2m|b|}{s}\right)}{4-s^2}
\end{align}
which can be achieved by realising,
\begin{align}
    \arctan\left(\frac{x}{2m}\right)=2mx\int_0^1 \frac{ds}{x^2s^2+4m^2}.
\end{align}
In the next subsection, we extend the impulse computation to 2PM order.

\subsection{2PM contribution to the impulse}
In this subsection, we compute the 2PM contribution to impulse, mainly focusing on the scalar field contribution.

\textbullet $\,\,$  At $\mathcal{O}(z)$ simplest diagram comes from the graviton-scalar interaction vertex:
\begin{align}
    \begin{split}
        \mathcal{O}(z): -\frac{1}{2}m^2\rmint d^4x \frac{h}{2m_p}\varphi^2\,.
    \end{split}
\end{align}
The corresponding contribution to the impulse is given by,
\begin{align}
    \begin{split}
        [\Delta p_1^{\mu}]_{(c)}& = \thesisinlinefeynman[\chthreeinlinefeynwidth]{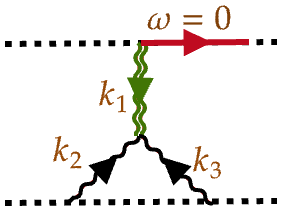}\hspace{0.7 cm}\\ &
=-im^2\frac{m_1 m_2^2s_2^2}{8m_p^4}\rmint_{k_i}\hat\delta^{(4)}\Big(\sum_{i=1}^{3}k_i\Big)\frac{\hat\delta(k_1\cdot v_1)\hat\delta(k_2\cdot v_2)\hat\delta(k_3\cdot v_2)}{k_1^2(k_2^2-m^2)(k_3^2-m^2)}e^{ik_1\cdot b_1}e^{ik_2\cdot b_2}e^{ik_3\cdot b_2}k_1^\mu \underbrace{P_{\rho\rho,\sigma\delta}\,v_1^\sigma v_1^\delta}_{-1}\,,\\ &
=im^2\frac{m_1 m_2^2s_2^2}{8m_p^4}\rmint_{k_{1,2}}\frac{k_1^\mu\hat\delta(k_1\cdot v_1)\hat\delta(k_2\cdot v_2)\hat\delta(k_1\cdot v_2)}{k_1^2(k_2^2-m^2)[(k_1+k_2)^2-m^2]}e^{ik_1\cdot (b_1-b_2)}\,,\\ &
=m^2\frac{m_1m_2^2s_2^2}{32\pi^2 m_p^4}\frac{1}{\sqrt{\gamma^2-1}}\frac{b^\mu}{|b|}\partial_{|b|}\underbrace{\rmint _{0}^{\infty}dl\,\frac{J_{0}(bl)}{l^2}\arctan\Big(\frac{l}{2m}\Big)}_{I_2(m,|b|)}.\label{4.23m}
    \end{split}
\end{align}
Therefore, the contribution to the impulse from this diagram has the following form.

\begin{equation}\label{e:barwq244}
  [\Delta p_1^{\mu}]_{(c)}=m^2\frac{m_1m_2^2s_2^2}{32 \pi^2 m_p^4}\frac{1}{\sqrt{\gamma^2-1}}\frac{b^\mu}{|b|}\partial_{|b|}{I_2(m,|b|)}\,.
\end{equation}\\
Like (\ref{e:barwq2}), it also does not have any closed-form expression. Furthermore, it is evident from \eqref{4.23m} that it becomes identically zero in the massless limit as it is proportional to $m^2$.

\textbullet $\,\,$ Another contributing diagram may appear from the scalar-graviton interaction vertex where one scalar and one graviton line connect with one worldline, and the other scalar line connects with another worldline.
\begin{align}
    \begin{split}
         [\Delta p_1^{\mu}]_{(d)}& =
        \thesisinlinefeynman[\chthreeinlinefeynwidth]{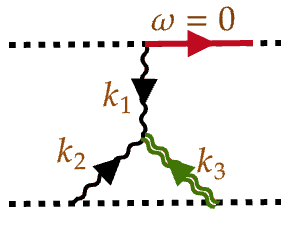}\\ &
        =im^2 \frac{m_1 s_1 m_2^2 s_2}{8 m_p^4}\rmint_{k_{1,2}}\frac{k_1^\mu\hat\delta(k_1\cdot v_1)\hat\delta(k_1\cdot v_2)\hat\delta(k_2\cdot v_2)}{(k_1^2-m^2)(k_2^2-m^2)(k_1+k_2)^2}e^{ik_1\cdot b}\\ &
        =m^2 \frac{m_1 s_1 m_2^2 s_2}{32 \pi^2 m_p^4\sqrt{\gamma^2-1}}\frac{b^\mu}{|b|}\partial_{|b|} \underbrace{\rmint_{k_1}dk \frac{ J_{0}(|b|k_1)}{k_1^2+m^2}\arctan\Big(\frac{k_1}{m}\Big)}_{\Bar{I}_2(m,|b|)}
    \end{split}
\end{align}
Hence, the contribution to the impulse from this diagram has the following form:

\begin{equation}\label{e:barwq248}
 [\Delta p_1^{\mu}]_{(d)}=m^2 \frac{m_1 s_1 m_2^2 s_2}{32 \pi^2 m_p^4\sqrt{\gamma^2-1}}\frac{b^\mu}{|b|}\partial_{|b|} \bar{I}_2(m,|b|)
\,.
\end{equation}

\textbullet $\,\,$  Now, we concentrate on the scalar interaction diagrams. We compute the contribution to the impulse at 2PM order coming from the scalar self-interaction vertex:
\begin{align}
    \begin{split}
        \mathcal{O}(z):-\frac{\lambda_4}{\textcolor{black}{4!}}\rmint d^4x\,\varphi^4(x)\,.
    \end{split}
\end{align}
The corresponding contribution to the impulse is shown below,
\begin{align}
    \begin{split}
          [\Delta p_1^{\mu}]_{(e)}&=\thesisinlinefeynman[\chthreeinlinefeynwidth]{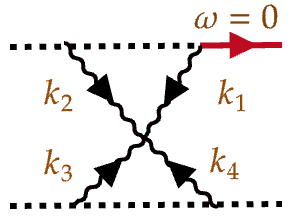}\\ & =-i\lambda_4\Big(\frac{m_1s_1m_2s_2}{4m_p^2}\Big)^2\rmint_{\{k_i\}}\hat\delta^{(4)}\Big(\sum_{i=1}^{4}k_{i}\Big)\hat\delta(k_1\cdot v_1)\hat\delta(k_2\cdot v_1)\hat\delta(k_3\cdot v_2)\hat\delta(k_4\cdot v_2)\prod_{j=1}^{4}\frac{1}{k_j^2-m^2}\\ & 
\hspace{0.8cm}\times k_1^{\mu} \,e^{i(k_1+k_2)\cdot b_2}e^{i(k_3+k_4)\cdot b_2}\,,\\ &
=-i\lambda_4\Big(\frac{m_1s_1m_2s_2}{4m_p^2}\Big)^2\rmint_{k_i\ne k_4}\hat\delta(k_1\cdot v_1)\hat\delta(k_2\cdot v_1)\hat\delta(k_3\cdot v_2)\hat\delta(k_1\cdot v_2+k_2\cdot v_2)\\ &\hspace{0.8cm}\times\frac{k_1^\mu e^{i(k_1+k_2)\cdot (b_1-b_2)}}{\prod_{j=1}^{3}(k_j^2-m^2)[(k_1+k_2+k_3)^2-m^2]}\,,\\ &
\xrightarrow[]{k_1+k_2\rightarrow q}-i\lambda_4\Big(\frac{m_1s_1m_2s_2}{4m_p^2}\Big)^2\rmint_{q,k_1,k_3}\frac{k_1^\mu\hat\delta(k_1\cdot v_1)\hat\delta(q\cdot v_1)\hat\delta(k_3\cdot v_2)\hat\delta(q\cdot v_2)}{(k_1^2-m^2)(k_3^2-m^2)[(q-k_1)^2-m^2][(q+k_3)^2-m^2]}e^{iq\cdot b}\,.\\ &
   \label{4.26m} \end{split}
\end{align}
In \eqref{4.26m} one can see that the delta function constraints set $q$ and $k_3$ two and three-dimensional vectors, respectively and $k_1$ a three-dimensional vector with scaled variable: $
        \bar k_1^{(1)}=\frac{k_1^{(1)}}{\gamma}.
   $ 
Therefore, the integral in  (\ref{4.26m}) can be re-written as,
\begin{align}
\begin{split}
   &   [\Delta p_1^{\mu}]_{(e)}\\ &=-i\lambda_4\Big(\frac{m_1s_1m_2s_2}{4m_p^2}\Big)^2 \int e^{iq\cdot b}\,\frac{q^\mu}{q^2}\,\hat \delta(q\cdot v_1)\hat\delta(q\cdot v_2)\int_{k_3,k_1}\frac{\hat\delta(k_1\cdot v_1)\hat\delta(k_3\cdot v_2)\left((k_1^2-m^2)+q^2-((k_1-q)^2-m^2)\right)}{(k_1^2-m^2)(k_3^2-m^2)[(q-k_1)^2-m^2][(q+k_3)^2-m^2]}\,, \\ &
   =-i\lambda_4\Big(\frac{m_1s_1m_2s_2}{4m_p^2}\Big)^2\int e^{iq\cdot b} \frac{q^\mu}{q^2}\,\hat \delta(q\cdot v_1)\hat\delta(q\cdot v_2)\Bigg(\int_{k_1}\frac{\hat\delta(k_1\cdot v_1)}{(k_1-q)^2-m^2}\int_{k_3}\frac{\hat\delta(k_3\cdot v_2)}{(k_3^2-m^2)((k_3+q)^2-m^2)}\\ & \hspace{1 cm}+q^2\int_ {k_1}\frac{\hat\delta(k_1\cdot v_1)}{(k_1^2-m^2)((k_1-q)^2-m^2)}\int _{k_3}\frac{\hat\delta(k_3\cdot v_2)}{(k_3^2-m^2)((k_3+q)^2-m^2)}-\int_{k_1}\frac{\hat\delta(k_1\cdot v_1)}{k_1^2-m^2}\int_{k_3}\frac{\hat\delta(k_3\cdot v_2)}{(k_3^2-m^2)((k_3+q)^2-m^2)}\Bigg)\,,\\ &
   =-i\lambda_4\Big(\frac{m_1s_1m_2s_2}{16\pi m_p^2}\Big)^2\int e^{iq\cdot b} \frac{q^\mu}{-q^2}\,\hat \delta(q\cdot v_1)\hat\delta(q\cdot v_2)\arctan^2\left(\frac{\sqrt{-q^2}}{2m}\right)\,,\\ &
  =- \lambda_4\Big(\frac{m_1s_1m_2s_2}{16\pi m_p^2}\Big)^2\frac{b^\mu}{\sqrt{\gamma^2-1}|b|} \int_0^\infty dq\, J_{1}(q\,b)\arctan^2\left(\frac{q^2}{2m}\right)\,,
   \label{4.19f}
    \end{split}
\end{align}
The integral in \eqref{4.19f} can be rearranged in a nicer form as,
\begin{align}
    [\Delta p_1^{\mu}]_{(e)}=-2 \lambda_4\Big(\frac{m_1s_1m_2s_2}{16\pi m_p^2}\Big)^2\frac{b^\mu}{\sqrt{\gamma^2-1}|b|^2}\int_0^1 ds\frac{K_0(2m|b|)-K_0\left(\frac{2m|b|}{s}\right)}{1-s^2}\,.\label{3.53gg}
\end{align}
Although not instructive, the integral in \eqref{3.53gg} can be written as a closed form expression in terms of Bickley–Naylor functions as,
\begin{align}
  \int_0^1 ds\frac{K_0(2m|b|)-K_0\left(\frac{2m|b|}{s}\right)}{1-s^2}  =\frac{1}{2} (\log (4m |b|)+\gamma_{E} ) K_0(2 m|b|)-\partial_\alpha \textrm{Ki}_{\alpha}(2m|b|)|_{\alpha\to0};\label{3.54kk}
\end{align}
The derivation is somewhat instructive. We put it in the Appendix~(\ref{3.54kkk}).
The massless counterpart takes the following form,
\begin{align}
    \begin{split}
        [\Delta p_1^{\mu}]_{(e)}=- \lambda_4\Big(\frac{m_1s_1m_2s_2}{32m_p^2}\Big)^2\frac{b^\mu}{\sqrt{\gamma^2-1}|b|^2}
    \end{split}
\end{align}

\textbullet $\,\,$ Another diagram involving $\lambda_4 \varphi^4$ vertex will contribute to the impulse, where we will have three scalar lines connected with one worldline, and the other scalar line connects with another worldline.
\begin{align}
    \begin{split}
         [\Delta p_1^{\mu}]_{(f)}&=\thesisinlinefeynman[\chthreeinlinefeynwidth]{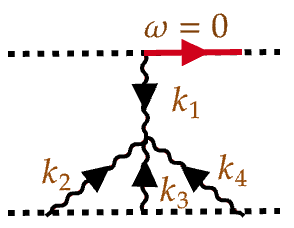}\\ &=-i\lambda_4 \Big(\frac{m_2 s_2}{2m_p}\Big)^3\Big(\frac{m_1 s_1}{2m_p}\Big)\rmint_{\{k_i\}}\hat\delta^{(4)}\Big(\sum_{i=1}^{4}k_{i}\Big)\hat\delta(k_1\cdot v_1)\hat\delta(k_2\cdot v_2)\hat\delta(k_3\cdot v_2)\hat\delta(k_4\cdot v_2)\prod_{j=1}^{4}\frac{1}{k_j^2-m^2}\\ & 
\hspace{0.8cm}\times k_1^{\mu} \,e^{ik_1\cdot b_1}e^{i(k_2+k_3+k_4)\cdot b_2}\,,\\ &
=-i\lambda_4 \Big(\frac{m_2 s_2}{2m_p}\Big)^3\Big(\frac{m_1 s_1}{2m_p}\Big)\rmint e^{ik_1\cdot b}k_1^\mu \,\frac{\hat\delta(k_1\cdot v_1)\hat\delta(k_1\cdot v_2)}{k_1^2-m^2}\rmint \frac{\hat\delta(q\cdot v_2)\hat\delta(k_3\cdot v_2)}{[(q-k_1)^2-m^2](k_3^2-m^2)[(q+k_3)^2-m^2]}\,,\\ &
=\frac{\lambda_4}{(2\pi)^2} \Big(\frac{m_2 s_2}{2m_p}\Big)^3\Big(\frac{m_1 s_1}{2 m_p}\Big)\frac{1}{\sqrt{\gamma^2-1}}\frac{b^\mu}{|b|}\partial_{|b|}\rmint  d^2 k_1\,\frac{e^{-i\vec k_1\cdot \vec b}}{\vec k_1^2+m^2}\,\hat{I}(|k_1|,m)\,.
\label{4.19c}
    \end{split}
\end{align}
We can further simplify \eqref{4.19c} using Schwinger parametrization.
\begin{align}
    \begin{split}
       \hat{I}(|k_1|,m)=\frac{1}{(2\pi)^4}\rmint d^3 q\frac{1}{|\vec q\,|} \rmint_{0}^{\infty}d\alpha\, e^{-\alpha (\vec q-\vec k_1)^2}e^{-\alpha m^2}\arctan\Big(\frac{|\vec q\,|}{2m}\Big)\,.
    \end{split}
\end{align}
Now, in \eqref{4.19c}, we first perform the angular part of the $q$ integral,
\begin{align}
    \begin{split}
         [\Delta p_1^{\mu}]_{(f)} &\sim \frac{1}{(2\pi)^3}\rmint d\alpha\rmint d^3q \frac{e^{-\alpha (q^2+m^2)}}{|q|}\arctan\Big(\frac{|\vec q\,|}{2m}\Big)\rmint_{0}^\infty dk_1\frac{k_1}{ k_1^2+m^2}J_{0}(k_1|b|)e^{-\alpha k_1^2+2\alpha \vec q\cdot \vec k_1}\,,\\ &
        =\frac{1}{(2\pi)^2}\rmint d\alpha \,\frac{e^{-\alpha m^2}}{\alpha}\rmint dk_1\,\frac{1}{ k_1^2+m^2}J_{0}(k_1|b|)e^{-\alpha k_1^2}\rmint dq\,\sinh(2q k_1 \alpha)\arctan\Big(\frac{q}{2m}\Big)e^{-\alpha q^2}\,,\\ &
        =\frac{1}{(2\pi)^2}\rmint_0^\infty dq \,dk_1\,\frac{J_{0}(k_1 b)}{k_1^2+m^2}\arctan\Big(\frac{q}{2m}\Big)\textrm{arctanh}\Big(\frac{2qk_1}{m^2+q^2+k_1^2}\Big)\,.\label{4.21cc}
    \end{split}
\end{align}
The $q$ integral can be done by taking the \textbf{logarithmic} representation of the \textbf{ArcTan} and \textbf{ArcTanh} in the following way,
\begin{align}
    \begin{split}
       & \rmint dq \arctan\Big(\frac{q}{2m}\Big)\,\textrm{arctanh}\Big(\frac{2q k_1}{q^2+k_1^2+m^2}\Big)\,,\\ &
       =\frac{i}{4}\rmint dq\, \log\Bigg[\frac{1-\frac{iq}{2m}}{1+\frac{iq}{2m}}\Bigg]\log\Big(1+\frac{4q k_1}{m^2+(q-k_1)^2}\Big)\,,\\ &
       =\frac{\pi}{2}\Big[2k_1-2m\arctan\Big(\frac{k_1}{m}\Big)-6m\arctan\Big(\frac{k_1}{3m}\Big)-k_1\log\Big(1+\frac{k_1^2}{m^2}\Big)-k_1\log\Big(1+\frac{k_1^2}{9m^2}\Big)-2k_1 \log(3m^2)\Big]\,.\label{4.25j}
    \end{split}
\end{align}
Therefore, the impulse becomes \footnote{As one can see, the impulse has a logarithmic behaviour w.r.t $|b|$. Hence, one should divide it by a UV cutoff $b_0$ to make it dimensionless. However, it does not contribute to the finite part of the impulse. The same thing can be said whenever some logarithmic terms appear. },
\begin{align}
    \begin{split}
         [\Delta p_1^{\mu}]_{(f)}&\sim \frac{1}{32\pi^3}\rmint dk_1\frac{J_{0}(k_1b)}{k_1^2+m^2}\Big[2k_1-2m\arctan\Big(\frac{k_1}{m}\Big)-6m\arctan\Big(\frac{k_1}{3m}\Big)\\ &
        \hspace{0.8cm}-k_1\log\Big(1+\frac{k_1^2}{m^2}\Big)-k_1\log\Big(1+\frac{k_1^2}{9m^2}\Big)-2k_1 \log(3m^2)\Big]\,,\\ &
        =\frac{1}{32\pi^3}\Big[2(1-\log3m^2)K_{0}(|b|m)-\rmint dk_1\frac{J_{0}(k_1 b)}{k_1^2+m^2}\Theta(k_1,m)\Big]\,,
    \end{split}
\end{align}
where,
\begin{align}
    \begin{split}
        \Theta(k_1,m)=-2m\arctan\Big(\frac{k_1}{m}\Big)-6m\arctan\Big(\frac{k_1}{3m}\Big)-k_1 \log\Big[\Big(1+\frac{k_1^2}{m^2}\Big)\Big(1+\frac{k_1^2}{9m^2}\Big)\Big]\,.
    \end{split}
\end{align}
After inserting the prefactors, we get

%

\begin{equation}\label{e:barwq246}
 [\Delta p_1^{\mu}]_{(f)}=\frac{\lambda_4}{32\pi^3} \Big(\frac{m_2 s_2}{2m_p}\Big)^3\Big(\frac{m_1 s_1}{2m_p }\Big)\frac{1}{\sqrt{\gamma^2-1}}\frac{b^\mu}{|b|}\partial_{|b|}\Big(2(1-\log(3m^2/\mu^2))K_{0}(|b|m)-\rmint_0^{\infty} dk_1\frac{J_{0}(k_1 |b|)}{k_1^2+m^2}\Theta(k_1,m)\Big)\,.\,
\end{equation}\\
where $\mu$ is the UV cutoff.\textcolor{black}{ As can be seen, the integral in \eqref{4.21cc} does not have any trivial closed-form, but one can derive the massless limit}. Although the integral can be rearranged as,
\begin{align}
   \begin{split}
   & \Big(2(1-\log(3m^2/\mu^2))K_{0}(|b|m)-\rmint_0^{\infty} dk_1\frac{J_{0}(k_1 |b|)}{k_1^2+m^2}\Theta(k_1,m)\Big)\\ &
   \hspace{3 cm}=2\left(1-\log\frac{9m^2}{16\mu^2}\right)K_0(m|b|)+2\int_{0}^1 \frac{ds}{s(s+1)}K_0\left(\frac{m|b|}{s}\right)+18 \int_0^1 ds \frac{1-s}{s(9-s^2)}K_0\left(\frac{3m|b|}{s}\right)\,.
    \end{split}
\end{align}

The massless limit is given by,
\begin{align}
    \begin{split}
      [\Delta p_1^{\mu}]_{(f)}\Big|_{m\rightarrow 0}   &=\frac{\lambda_4 }{32\pi^3}\Big(\frac{m_2 s_2}{2m_p}\Big)^3\Big( \frac{m_1 s_1}{2 m_p}\Big)\frac{1}{\sqrt{\gamma^2-1}}\frac{b^\mu}{|b|}\Big[-\frac{2}{|b|}+4\partial_{|b|}\rmint_0^\infty dk_1\frac{ \,J_{0}(k_1 |b|)}{k_1}\log(k_1)\Big]\,,\\ &
     =\frac{\lambda_4}{8\pi^3} \Big(\frac{m_2 s_2}{2m_p}\Big)^3\Big( \frac{m_1 s_1}{2 m_p}\Big)\frac{1}{\sqrt{\gamma^2-1}}\frac{b^\mu}{|b|^2}\Big[\log (|b|/b_0)+\gamma_E-\frac{1}{2}-\log 2\Big]\,.
    \end{split}
\end{align}
\textbullet $\,\,$ Now we will deal with the derivative interactions. The simplest scalar-gravitation derivative interaction, which contributes to the impulse at 2PM order:
\begin{align}
    \begin{split}
        \mathcal{O}(z):\rmint d^4x \frac{h^{\alpha\beta}}{m_p}\,\partial_{\alpha}\varphi\partial_{\beta}\varphi\,.
    \end{split}
\end{align}
\vspace{-0.53 cm}
The vertex factor has the following form,
\begin{align}
    \begin{split}
        \mathcal{V}_{\alpha\beta}(k_1,k_2,k_3)=  -\rmint_{\{k_{i}\}}  \hat\delta\Big(\sum_{i=1}^{3}k_{i}\Big)   (k_2)_{\alpha} (k_3)_{\beta}\,.
    \end{split}
\end{align}
The contribution to the impulse is given by\footnote{Time-symmetric Feynman propagators imply an elastic scattering of the black holes. In principle, one can also begin with the retarded graviton propagators to take care of radiative effects, but we will use the first one.},
\begin{align}
    \begin{split}
         [\Delta p_1^{\mu}]_{(g)} &=\thesisinlinefeynman[\chthreeinlinefeynwidth]{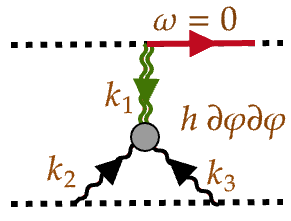}\hspace{0.7 cm}\\ &
=i\Big(\frac{m_2s_2}{2m_p}\Big)^2\times \frac{m_1}{ m_p^2}\rmint_{k_i}\hat\delta^{(4)}\Big(\sum_{i}k_i\Big)k_2^\alpha k_3^\beta\,\frac{k_1^\mu P_{\sigma\delta;\alpha\beta }v_1^\sigma v_1^\delta\,\hat\delta(k_1\cdot v_1)\hat\delta(k_2\cdot v_2)\hat\delta(k_3\cdot v_2)}{k_1^2\prod_{i=2,3}(k_i^2-m^2)}e^{ik_1\cdot b_1}e^{i(k_2+k_3)\cdot b_2}\,,\\ &
=\Big(\frac{m_2s_2}{2m_p}\Big)^2\times \frac{m_1}{2m_p^2}\\ &
\hspace{2 cm}\times\frac{\partial}{\partial b_{1\mu}} \underbrace{\rmint_{k_{1,2}}\Big[2(k_2\cdot v_1)(k_1+k_2)\cdot v_1-\,k_2\cdot (k_1+k_2)\Big]\frac{\hat\delta(k_1\cdot v_1)\hat\delta(k_2\cdot v_2)\hat\delta(k_1\cdot v_2)}{k_1^2(k_2^2-m^2)[(k_1+k_2)^2-m^2]}e^{ik_1\cdot b}}_{I_{4}(m,|b|)}\,.\label{4.37} 
    \end{split}
\end{align}
The computation in \eqref{4.37} is an involved one and should be done carefully. We show below the details of the integration for individual terms.
\begin{align}
    \begin{split}
        I_4(m,|b|)\Big|_{(1)}&\equiv 2\rmint_{k_{1,2}}(k_2\cdot v_1)^2\frac{\hat\delta(k_1\cdot v_1) \hat\delta(k_2\cdot v_2)\hat\delta(k_1\cdot v_2)}{k_1^2(k_2^2-m^2)[(k_1+k_2)^2-m^2]}e^{i k_1\cdot b}\,,\\ &
         =(v_1)_{\mu}(v_1)_{\nu}\rmint_{k_{1,2}}k_2^{\mu}\,k_2^{\nu}\frac{\hat\delta(\gamma k_1^{(0)}-\gamma\beta k_1^{(1)})\hat\delta( k_2^0)\hat\delta(k_1^0)}{k_1^2(k_2^2-m^2)[(k_1+k_2)^2-m^2]}e^{i k_1 \cdot b}\,,\\ &
    =\textcolor{black}{\frac{1}{(2\pi)^5}}\frac{v_{1i}v_{1j}    }{\gamma\beta}\rmint d^2 k_1\frac{e^{i k_1 \cdot b}}{\vec k_1^2}
    \underbrace{\rmint d^3 k \frac{k_i k_j}{(\vec k^2+m^2)[(\vec k+\vec k_1)^2+m^2]}}_{\mathcal{C}_{ij}}\,.
    \end{split}
\end{align}
Using the Passarino-Veltman reduction one can deduce the form of $\mathcal{C}_{ij}$,
\begin{align}
    \begin{split}
\mathcal{C}_{ij}=&\underbrace{\Big[\textcolor{black}{(2\pi)^2}\frac{3}{8 |\vec k_1|}\arctan\Big(\frac{|\vec k_1|}{2m}\Big)+\textcolor{black}{(2\pi)^2}\frac{m^2}{2 |\vec k_1|^3}\arctan\Big(\frac{|\vec k_1|}{2m}\Big)-\textcolor{black}{\frac{\pi^2}{2}}\frac{m}{|\vec k_1|^2}\Big]}_{\Xi_1(|k_1|,m)}k_1^i k_1^j\\ &
       \underbrace{- \Big[|\vec k_1|\textcolor{black}{\frac{(2\pi)^2}{8}}\arctan\Big(\frac{|\vec k_1|}{2m}\Big)+\textcolor{black}{\frac{(2\pi)^2}{2}}\frac{m^2}{ |\vec k|}\arctan\Big(\frac{|\vec k_1|}{2m}\Big)+m\textcolor{black}{\frac{\pi^2}{2}}\Big]}_{\Xi_2 (|k_1|,m)}\delta^{ij}\,,\\ &
={\Xi}_1(|k_1|,m)k_1^i k_1^j+{\Xi}_{2}(|k_1|,m)\delta^{ij}\,.  \label{new11}
    \end{split}
\end{align}
Therefore,
\begin{align}
    \begin{split} \label{eq:4.35}
        I_{4}(m,|b|)\Big|_{(1)}=\frac{2(2\pi)}{\sqrt{\gamma^2-1}}\Bigg[-v_{1i}v_{1j}\partial_{b_i}\partial_{b_j}\rmint_{0}^{\infty}dl\,\frac{J_{0}(bl)}{l}\,{\Xi}_1\Big(l,m\Big)+(\gamma^2-1)\rmint_{0}^\infty dl\,\frac{J_{0}(bl)}{l}{\Xi}_2\Big(l,m\Big)\Bigg]
    \end{split}
\end{align}
where ${\Xi}_1(l,m)$ and ${\Xi}_2(l,m)$ are defined in (\ref{new11}). 
Other parts of the integrals $I_4(m,b)$ can be computed as follows,
\begin{align}
    \begin{split}
        I_{4}(m,|b|)\Big|_{(2)}\equiv 2\rmint_{k_{1,2}}(k_2\cdot v_1)(k_1\cdot v_1)\hat\delta(k_1\cdot v_1)\cdots \rightarrow 0.
    \end{split}
\end{align}
and,
\begin{align}
    \begin{split}
        I_{4}(m,|b|)\Big|_{(3)}&\equiv- \rmint_{k_{1,2}}k_1\cdot k_2\,\frac{\hat\delta(k_1\cdot v_1) \hat\delta(k_2\cdot v_2)\hat\delta(k_1\cdot v_2)}{k_1^2(k_2^2-m^2)[(k_1+k_2)^2-m^2]}e^{i k_1\cdot b}\,,\\ &
        =\textcolor{black}{\frac{1}{(2\pi)^5}}\frac{1}{\sqrt{\gamma^2-1}}\rmint d^2 k_1 \,\frac{k_1^i e^{ik_1\cdot b}}{k_1^2}\rmint d^3 k_2 \frac{k_2^i}{(\vec k_2^2+m^2)[(\vec k_1+\vec k_2)^2+m^2]}\,,\\ &
        =\textcolor{black}{\frac{1}{(2\pi)^2}}\frac{1}{2\sqrt{\gamma^2-1}}\rmint_{0}^{\infty} dl \,J_{0}(|b|l)\arctan\Big(\frac{l}{2m}\Big) \label{eq:4.37}
    \end{split}
\end{align}
 and,
 \begin{align}
     \begin{split}
         I_{4}(m,|b|)\Big|_{(4)}&\equiv -\rmint_{k_{1,2}}k_2^2\,\frac{\hat\delta(k_1\cdot v_1) \hat\delta(k_2\cdot v_2)\hat\delta(k_1\cdot v_2)}{k_1^2(k_2^2-m^2)[(k_1+k_2)^2-m^2]}e^{i k_1\cdot b}\,,\\ &
         =-\textcolor{black}{\frac{1}{(2\pi)^5}}\frac{1}{\sqrt{\gamma^2-1}}\rmint d^2 k_1\frac{e^{-i\vec k_1\cdot \vec b }}{-\vec k_1^2}\rmint d^3 k_2 \frac{-\vec k_2^2}{(\vec k_2^2+m^2)[(\vec k_1+\vec k_2)^2+m^2]}\,,\\ &
         =-\textcolor{black}{\frac{1}{(2\pi)^4}}\frac{1}{\sqrt{\gamma^2-1}}\rmint_0^{\infty} dl\, \frac{J_{0}(bl)}{l}\Big(l^2\,{\Xi}_1(l,m)+3\,{\Xi}_2(l,m)\Big)\,, \label{eq:4.38}
     \end{split}
 \end{align}
 where ${\Xi}_1(l,m)$ and ${\Xi}_2(l,m)$ are defined in (\ref{new11}). 
\textcolor{black}{After (\ref{eq:4.35}), (\ref{eq:4.37}) and (\ref{eq:4.38}) we get the full answer for $I_{4}(m,|b|)$ mentioned in (\ref{4.37}). Then we get, the }

\begin{equation}\label{e:barwqpp}
 [\Delta p_1^{\mu}]_{(g)}=\Big(\frac{m_2s_2}{2m_p}\Big)^2\times \frac{m_1}{2m_p^2}\frac{b^\mu}{|b|}\,\partial_{|b|} I_{4}(m,|b|)\,.
\end{equation}
Similiarly, the massless limit can be taken by using the fact that $\arctan(\infty)=\frac{\pi}{2}$ and taking the massless limit of $\Xi(l,m)$.

\textbullet $\,\,$ Another interesting diagram will contribute to the impulse from the previously mentioned derivative interaction where one scalar line connects with one worldline and the other scalar line and the graviton line connects with the other worldline.
\begin{align}
    \begin{split}
         [\Delta p_1^{\mu}]_{(h)}&=\thesisinlinefeynman[\chthreeinlinefeynwidth]{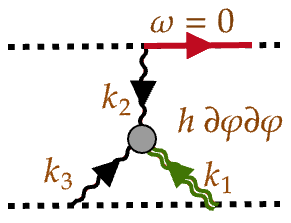}\\ &=i\frac{m_1 s_1 m_2^2 s_2}{4m_p^4}\rmint_{k_i}\hat\delta^{(4)}(\sum_i k_i)k_2^\alpha k_3^\beta \frac{k_2^\mu P_{\alpha\beta;\sigma\delta}v_2^\sigma v_2^\delta\hat\delta(k_2\cdot v_1)\hat\delta(k_1\cdot v_2)\hat\delta(k_3\cdot v_2)}{k_1^2(k_2^2-m^2)(k_3^2-m^2)}e^{ik_2\cdot b_1}e^{i(k_1+k_3)\cdot b_2}\,,\\  &
        =-i\frac{m_1 s_1 m_2^2 s_2}{8m_p^4}\rmint_{k_3,k_2}e^{ik_2\cdot b}k_2^\mu\,\hat\delta(k_2\cdot v_1)\hat\delta(k_2\cdot v_2)\frac{k_2\cdot k_3\,\hat\delta(k_3\cdot v_2)}{(k_2^2-m^2)(k_2+k_3)^2(k_3^2-m^2)}\,,\\ &
        =\frac{m_1 s_1 m_2^2 s_2}{128 \pi^2 m_p^4}\frac{1}{\sqrt{\gamma^2-1}}\frac{b^\mu}{|b|}\partial_{|b|}\underbrace{\rmint_0^\infty dk \frac{k^2}{k^2+m^2}J_{0}(k|b|)\arctan\Big(\frac{k}{m}\Big)}_{\bar{I}_4(m,|b|)}\,.
    \end{split}
\end{align}
Hence, the contribution from the diagram reads,

\begin{equation}\label{e:barwq24}
  [\Delta p_1^{\mu}]_{(h)}=\frac{m_1 s_1 m_2^2 s_2}{128 \pi^2 m_p^4}\frac{1}{\sqrt{\gamma^2-1}}\frac{b^\mu}{|b|}\partial_{|b|}\bar{I}_4(m,|b|)
\,.
\end{equation}
\textbullet $\,\,$ Another interesting 3-point worldline vertex, which comes from the derivative interaction,  contributing to the impulse at 2PM order, 
\begin{align}
    \begin{split} \label{vert1}
        \mathcal{O}(z): -\frac{m_1s_1}{2m_p^2}\rmint_{k_1,k_2,\omega}e^{i(k_1+k_2)\cdot b_1}\hat\delta(k_1\cdot v_1+k_2\cdot v_1+\omega)\{(k_{1\rho}+k_{2\rho})v^{\mu}v^{\nu}+2\omega v^{(\mu}\delta_{\rho}^{\nu)}\}\varphi(-k_1) h_{\mu\nu}(-k_2)z^{\rho}(-\omega)\,.
    \end{split}
\end{align}
The corresponding contribution to the impulse is given by, 
\begin{align}
    \begin{split}
         [\Delta p_1^{\mu}]_{(i)}=\thesisinlinefeynman[\chthreeinlinefeynwidth]{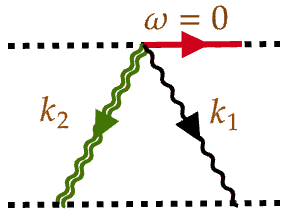}\hspace{0.7 cm}&=i\frac{m_1s_1}{2m_p^2}\times \frac{m_2}{2m_p}\times \frac{m_2s_2}{2m_p}\\ &
\hspace{0.35cm}\rmint_{k_{1,2}}(k_1^\mu+k_2^\mu)e^{i(k_1+k_2)\cdot b}\frac{\hat\delta(k_1\cdot v_1+k_2\cdot v_1)\hat\delta(k_1\cdot v_2)\hat\delta(k_2\cdot v_2)}{k_2^2(k_1^2-m^2)}P_{\alpha\beta;\rho\sigma}v_1^\alpha v_1^\beta v_2^\rho v_2^\sigma\,,\\ &=\frac{i m_1s_1m_2^2s_2}{8m_p^4}P_{\alpha\beta;\rho\sigma}v_1^\alpha v_1^\beta v_2^\rho v_2^\sigma\rmint_{q,k_2}\frac{q^\mu\hat\delta(q\cdot v_1)\hat\delta(q\cdot v_2)\hat\delta(k_2\cdot v_2)}{k_2^2[(q-k_2)^2-m^2]}e^{iq\cdot b}\,,\\ &
=\frac{m_1 s_1 m_2^2 s_2}{64\pi^2 m_p^4\sqrt{\gamma^2-1}}(2\gamma^2-1)\frac{b^\mu}{|b|}\partial_{|b|}\rmint_0^{\infty} dl\,J_{0}(|b|l)\arctan\Big(\frac{l}{m}\Big)\,.
    \end{split}
\end{align}
\begin{equation}\label{sa}
 [\Delta p_1^{\mu}]_{(i)}=\frac{m_1 s_1 m_2^2 s_2}{64\pi^2 m_p^4\sqrt{\gamma^2-1}}(2\gamma^2-1)\frac{b^\mu}{|b|}\partial_{|b|}\rmint_0^{\infty} dl\,J_{0}(|b|l)\arctan\Big(\frac{l}{m}\Big)\,.\vspace{0.5cm}
\end{equation}\\\
\textcolor{black}{The integral in \eqref{sa} has a closed-form expression, which can be seen by taking \textbf{logarithmic }representation of  \textbf{ArcTan} which gives,
\begin{align}
    \begin{split}
        [\Delta p_1^{\mu}]_{(i)}&=\frac{m_1 s_1 m_2^2 s_2}{64\pi m_p^4\sqrt{\gamma^2-1}}(2\gamma^2-1)\frac{b^\mu}{8\pi|b|^2}\\ &\Bigg[2 G_{3,1}^{1,3}\left(\frac{i}{{m\,b}},\frac{1}{2}\Big|
\begin{array}{c}
 1,1,\frac{3}{2} \\
 \frac{3}{2} \\
\end{array}
\right)+2 G_{3,1}^{1,3}\left(-\frac{i}{{m\,b}},\frac{1}{2}\Big|
\begin{array}{c}
 1,1,\frac{3}{2} \\
 \frac{3}{2} \\
\end{array}
\right)+G_{4,2}^{1,4}\left(\frac{i}{{m\,b}},\frac{1}{2}\Big|
\begin{array}{c}
 1,1,\frac{3}{2},\frac{3}{2} \\
 \frac{3}{2},\frac{1}{2} \\
\end{array}
\right)\\ &+G_{4,2}^{1,4}\left(-\frac{i}{{m\,b}},\frac{1}{2}\Big|
\begin{array}{c}
 1,1,\frac{3}{2},\frac{3}{2} \\
 \frac{3}{2},\frac{1}{2} \\
\end{array}
\right)\Bigg]\,.
    \end{split}
\end{align}}
Here have again used that fact that $$P_{\alpha\beta;\rho\sigma}v_1^\alpha v_1^\beta v_2^\rho v_2^\sigma=\frac{2\gamma^2-1}{2}\,.$$
Again, we have a finite massless counterpart of the diagram which gives,
\begin{align}
    \begin{split}
         [\Delta p_1^{\mu}]_{(i)}\Big|_{m\to 0}=-\frac{m_1 s_1 m_2^2 s_2}{64\pi^2 m_p^4\sqrt{\gamma^2-1}}(2\gamma^2-1)\frac{b^\mu}{|b|^3}\,.\label{4.33a}
    \end{split}
\end{align}
\textbullet $\,\,$  Lastly, there will be another diagram where $h_{\mu\nu}$ in (\ref{vert1}), is replaced by $\varphi$ in the worldline vertex. The worldline vertex under consideration is,
\begin{align}
    \begin{split}
          \mathcal{O}(z):& -\frac{m_1g_1}{2m_p^2}\rmint_{k_1,k_2,\omega}e^{i(k_1+k_2)\cdot b_1}\hat\delta(k_1\cdot v_1+k_2\cdot v_1+\omega)\{(k_{1\rho}+k_{2\rho})v^{\mu}v^{\nu}+2\omega v^{(\mu}\delta_{\rho}^{\nu)}\}\\&\hspace{0.8cm}\times\varphi(-k_1)\varphi(-k_2) \eta_{\mu\nu}z^{\rho}(-\omega)\,.
    \end{split}
\end{align}
Therefore, the impulse is given by,
\begin{align}
    \begin{split}
         [\Delta p_1^{\mu}]_{(j)}=\thesisinlinefeynman[\chthreeinlinefeynwidth]{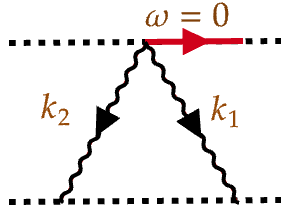}\hspace{0.7 cm}&=i\frac{m_1g_1}{m_p^2}\times \frac{m_2^2 s_2^2}{2m_p^2}\rmint_{q,k_2}\frac{q^\mu\hat\delta(q\cdot v_1)\hat\delta(q\cdot v_2)\hat\delta(k_2\cdot v_2)}{(k_2^2-m^2)[(q-k_2)^2-m^2]}e^{iq\cdot b}\,,\\ &
=\frac{m_1g_1 m_2^2 s_2^2}{8\pi^2 m_p^4\sqrt{\gamma^2-1}}\frac{b^\mu}{|b|}\partial_{|b|}\rmint_0^\infty dl\,J_{0}(|b|l)\arctan\Big(\frac{l}{2m}\Big)\,.
    \end{split}
\end{align}
Therefore, the contribution to the impulse from the above diagram is given by,\\\
 
\begin{equation}\label{e:barwqp}
  [\Delta p_1^{\mu}]_{(j)}=\frac{m_1g_1 m_2^2 s_2^2}{8\pi^2 m_p^4\sqrt{\gamma^2-1}}\frac{b^\mu}{|b|}\partial_{|b|}\rmint_0^\infty dl\,J_{0}(|b| l)\arctan\Big(\frac{l}{2m}\Big)\,.\vspace{0.5cm}
\end{equation}
\textcolor{black}{Note that, similar to (\ref{sa}), this integral can also be recast in terms of {\bf MeijerG}. } 
The massless limit of  \eqref{4.33a} can be taken and it gives the following,
\begin{align}
    \begin{split}
         [\Delta p_1^{\mu}]_{(j)}\Big|_{m\to 0}=\frac{m_1g_1 m_2^2 s_2^2}{16 \pi m_p^4\sqrt{\gamma^2-1}}\frac{b^\mu}{|b|^3}\,.
    \end{split}
\end{align}

Finally, there will be 2PM diagrams coming from the bulk $\varphi^3$  interaction vertex. Below we will give the details of the contribution to the impulse coming from these vertices.

\textbullet $\,\,$ First, we consider the following Feynman diagram.
\begin{align}
    \begin{split}
       [\Delta p_1^{\mu}]_{(k)}&=\thesisinlinefeynman[\chthreeinlinefeynwidth]{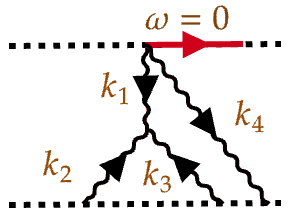}\\&=-i\lambda_3 m_p\Big(\frac{m_1 g_1}{2m_p^2}\Big)\Big(\frac{m_2^3s_2^3}{8m_p^3}\Big)\rmint_{k_i}\hat\delta\Big(\sum_{i=1}^3 k_i\Big)\frac{(k_1+k_4)^\mu\hat\delta(k_1\cdot v_1+k_4\cdot v_1)\hat\delta(k_2\cdot v_2)\hat\delta(k_3\cdot v_2)\hat\delta(k_4\cdot v_2)}{\prod_{i}^4 (k_i^2-m^2)}\\ &\hspace{0.8cm}\times e^{i(k_1+k_4)\cdot b_1}e^{i(k_2+k_3)\cdot b_2}e^{-ik_4\cdot b_2}\,.\label{4.51b}
    \end{split}
\end{align}
\vspace{-0.3 cm}
Focusing only on the integral, we get,
\begin{align}
    \begin{split}
        [\Delta p_1^{\mu}]_{(k)}^\mu&\sim\rmint_{k_i}\frac{(k_1+k_4)^\mu \hat\delta(k_1\cdot v_1+k_4\cdot v_1)\hat\delta(k_2\cdot v_2)\hat\delta(k_1\cdot v_2)\hat\delta(k_4\cdot v_2)}{(k_1^2-m^2)(k_2^2-m^2)[(k_1+k_2)^2-m^2](k_4^2-m^2)} e^{i(k_1+k_4)\cdot b}\,,\\ &
        \xrightarrow[]{k_1+k_4\to q}\rmint_{k_i,q}\frac{q^\mu\hat\delta(q\cdot v_1)\hat\delta(k_2\cdot v_2)\hat\delta(k_1\cdot v_2)\hat\delta(q\cdot v_2)}{(k_1^2-m^2)(k_2^2-m^2)[(k_1+k_2)^2-m^2][(q-k_1)^2-m^2]}e^{i q\cdot b}\,,\\ &
        =\rmint _q q^\mu \hat{\delta}(q\cdot v_1)\hat\delta(q\cdot v_2)e^{i q \cdot b}\rmint_{k_1,k_2}\frac{\hat\delta(k_1\cdot v_2)\hat\delta(k_2\cdot v_2)}{(k_1^2-m^2)(k_2^2-m^2)[(k_1+k_2)^2-m^2][(q-k_1)^2-m^2]}\,,\\ &
        =\frac{1}{\sqrt{\gamma^2-1}}\frac{b^\mu}{|b|}\partial_{|b|}\rmint \hat d^2 q\, e^{-i \vec q\cdot \vec b}\, \hat L(\vec q,m)
    \end{split}
\end{align}
where,
\begin{align}
    \begin{split}
        \hat L(\vec q,m)&=\rmint_{\vec k_1,\vec k_2}\frac{1}{(\vec k_1^2+m^2)[(\vec k_1-\vec q)^2+m^2]|\vec k_1|}\arctan\Big(\frac{|\vec k_1|}{2m}\Big)\,,\\ &
        =\rmint_{0}^\infty d\alpha \,e^{-\alpha m^2}\rmint_{\vec k_1}\frac{1}{|\vec k_1|(\vec k_1^2+m^2)} e^{-\alpha(\vec k_1-\vec q)^2}\arctan\Big(\frac{|\vec k_1|}{2m}\Big)\,,\\ &
        =\frac{2\pi}{q} \rmint dk_1 \frac{1}{k_1^2+m^2}\tanh ^{-1}\left(\frac{2 k_1 q}{k_1^2+m^2+q^2}\right)\arctan\Big(\frac{|\vec k_1|}{2m}\Big)\,.
    \end{split}
\end{align}
Therefore, the impulse is (after restoring all the prefractors),

\begin{equation}
 [\Delta p_1^{\mu}]_{(k)}=\frac{\lambda_3}{(2\pi)^4} \Big(\frac{m_1 g_1}{2m_p}\Big)\Big(\frac{m_2^3s_2^3}{8m_p^3}\Big)\frac{1}{\sqrt{\gamma^2-1}}\frac{b^\mu}{|b|}\partial_{|b|}\rmint_0^{\infty} dq \,dk_1 \,\frac{J_{0}(q|b|)}{k_1^2+m^2}\,\textrm{arctanh}\Big(\frac{2 k_1 q}{k_1^2+q^2+m^2}\Big)\arctan\Big(\frac{k_1}{2m}\Big)\label{4.54r}. \vspace{0.5cm}
\end{equation}
The integral in \eqref{4.54r} can be further simplified using the logarithmic representation of ArcTan as shown in \eqref{4.25j}. The massless limit also can be taken as \eqref{4.25j}.

\textbullet $\,\,$Another Feynman diagram that we will contribute can be obtained by replacing one scalar propagator in (\ref{4.51b}) with a graviton propagator. The impulse is given by : 
\begin{align}
    \begin{split}
         [\Delta p_1^{\mu}]_{(l)}&=\thesisinlinefeynman[\chthreeinlinefeynwidth]{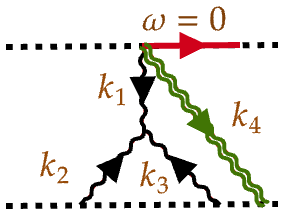}\\ &= \frac{2\gamma^2-1}{2}\frac{\lambda_3}{(2\pi)^4}\Big(\frac{m_1 s_1}{2m_p}\Big)\Big(\frac{m_2^3 s_2^2}{8m_p^3}\Big)\frac{1}{\sqrt{\gamma^2-1}}\frac{b^\mu}{|b|}\partial_{|b|}\rmint_0^\infty dq  dk_1 \frac{J_0(q|b|)}{k_1^2+m^2}\arctan\Big(\frac{k_1}{2m}\Big)\, \textrm{arctanh}\Big(\frac{2k_1 q}{k_1^2+q^2}\Big)\,.\label{4.55m}
\end{split}
\end{align}
Again this integral in \eqref{4.55m} can be further simplified using the logarithmic representation of ArcTan as \eqref{4.25j}.

\textbullet $\,\,$ Another Feynman topology contributing to the  impulse at 2PM order is given by: 
\begin{align}
    \begin{split}
         [\Delta p_1^{\mu}]_{(m)}&=\thesisinlinefeynman[\chthreeinlinefeynwidth]{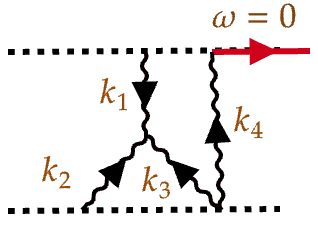}\\ &= -i\lambda_3\Big(\frac{m_1^2 s_1^2}{4m_p}\Big)\Big(\frac{m_2 s_2 m_2 g_2 }{4m_p^3}\Big)\rmint_{k_i}\frac{k_1^\mu \hat\delta(k_1\cdot v_1)\hat\delta(k_2\cdot v_2)\hat\delta(k_3\cdot v_2+k_4\cdot v_2)\hat\delta(k_4\cdot v_1)}{\prod_i (k_i^2-m^2)}\hat\delta^{(4)}(k_1+k_2+k_3)\\ &
\hspace{0.8cm}\times e^{i(k_1-k_4)\cdot b_1}e^{i(k_2+k_3+k_4)\cdot b_2}\,,\\ &
        =-i\lambda_3\Big(\frac{m_1^2 s_1^2}{4m_p}\Big)\Big(\frac{m_2 s_2 m_2 g_2 }{4m_p^3}\Big)\rmint_{k_i}\frac{k_1^\mu \hat\delta(k_1\cdot v_1)\hat\delta(k_2\cdot v_2)\hat\delta(-k_1\cdot v_2-k_2\cdot v_2+k_4\cdot v_2)\hat\delta(k_4\cdot v_1)}{(k_1^2-m^2)(k_2^2-m^2)[(k_1+k_2)^2-m^2](k_4^2-m^2)}e^{i(k_1-k_4)\cdot b}\,,\\ &
        \xrightarrow[]{k_1-k_4\to q}-i\lambda_3\Big(\frac{m_1^2 s_1^2}{4m_p}\Big)\Big(\frac{m_2 s_2 m_2 g_2 }{4m_p^3}\Big)\rmint_{k_i,q}\frac{k_1^\mu\hat\delta(k_1\cdot v_1)\hat\delta(k_2\cdot v_2)\hat\delta(q\cdot v_2)\hat\delta(q\cdot v_1)}{(k_1^2-m^2)(k_2^2-m^2)[(k_1+k_2)^2-m^2][(q-k_1)^2-m^2]}e^{iq\cdot b}\,.
    \end{split}\label{4.56e}
\end{align}
Now, doing the $k_2$ integral, we will have (omitting the prefactors),
\begin{align}
    \begin{split}
         [\Delta p_1^{\mu}]_{(m)}\propto \rmint_{k_i,q}\frac{k_1^\mu\hat\delta(k_1\cdot v_1)\hat\delta(q\cdot v_2)\hat\delta(q\cdot v_1)}{(k_1^2-m^2)[(q-k_1)^2-m^2]|\vec k_1|}\arctan\Big(\frac{|\vec k_1|}{2m}\Big)e^{iq\cdot b}\,.
    \end{split}
\end{align}
\textcolor{black}{We see that due to the asymmetric structure of this diagram, it is difficult to obtain a closed-form expression, like the case of $\lambda_4 \varphi^4$ vertex. We first do the $q$ integral. It is clear from the delta function constraints that $q$ and $k_1$ are two-dimensional and three-dimensional vectors, respectively.} Therefore, after performing the $q$ integral, the result will only depend on the second and third components of $\vec k_1$. Therefore, we are left with the following, 
\begin{align}
    \begin{split}
          [\Delta p_1^{\mu}]_{(m)}&\propto 2 \pi \frac{1}{\sqrt{\gamma^2-1}}\rmint_{k_1} e^{-i ({b^{(2)}} {k_1^{(2)}}+{b^{(3)}}{k_1^{(3)}})} K_0\left(\sqrt{{b^{(2)}}^2+{b^{(3)}}^2} \sqrt{\vec k_1^2-{k_1^{(2)}}^2-{k_1^{(3)}}^2+m^2}\right)\\ &\hspace{0.8cm}\times\frac{k_1^\mu}{(k_1^2-m^2)|\vec k_1|}\arctan\Big(\frac{|\vec k_1|}{2m}\Big)\,.\label{4.57mm}
    \end{split}
\end{align}
Now, note that in our parametrisation, the impact parameter takes the form: $b^{\mu}=(0,0,1,0)$. Therefore, restoring the prefactors integral in \eqref{4.57mm} can be recasted as,
\begin{align}
    \begin{split}
         [\Delta p_1^{\mu}]_{(m)}&=\frac{\lambda_3}{(2\pi)^4}\Big(\frac{m_1^2 s_1^2}{4m_p}\Big)\Big(\frac{m_2 s_2 m_2 g_2 }{4m_p^3}\Big)\frac{1}{\sqrt{\gamma^2-1}}\rmint d^4 k_1 e^{ik_1\cdot b}\,K_{0}\Big(|b||\sqrt{k_{1(1)}^2+m^2}|\Big)\\ &
        \hspace{0.8cm}\times\frac{k_1^\mu\hat\delta(k_1\cdot v_1)}{(k_1^2-m^2)|\vec k_1|}\arctan\Big(\frac{|\vec k_1|}{2m}\Big)\,,\\ &
        =\frac{\lambda_3}{(2\pi)^4}\Big(\frac{m_1^2 s_1^2}{4m_p}\Big)\Big(\frac{m_2 s_2 m_2 g_2 }{4m_p^3}\Big)\frac{1}{\gamma\sqrt{\gamma^2-1}}\frac{b^\mu}{|b|}\partial_{|b|}\rmint d^3 k_1 e^{-i \vec k_1\cdot \vec b}K_{0}\Big(|b||\sqrt{k_{1(1)}^2+m^2}|\Big)\\ &
        \hspace{0.8cm}\times\frac{1}{(\bar k_1^2+m^2)|\vec k_1|}\arctan\Big(\frac{|\vec k_1|}{2m}\Big)\,.\label{4.58}
    \end{split}
\end{align}
Hence, the contribution has the following form,

\begin{align}
\begin{split}
  [\Delta p_1^{\mu}]_{(m)}&=\frac{\lambda_3}{(2\pi)^4}\Big(\frac{m_1^2 s_1^2}{4m_p}\Big)\Big(\frac{m_2 s_2 m_2 g_2 }{4m_p^3}\Big)\frac{1}{\gamma\sqrt{\gamma^2-1}}\frac{b^\mu}{|b|}\partial_{|b|}\rmint_{-\infty}^{\infty} d^3 k_1 e^{-i \vec k_1\cdot \vec b}K_{0}\Big(|b||\sqrt{k_{1(1)}^2+m^2}|\Big)\\ &\hspace{0.8cm}\times \frac{1}{(\bar k_1^2+m^2)|\vec k_1|}\arctan\Big(\frac{|\vec k_1|}{2m}\Big)\label{4.58o} \vspace{0.5cm}
\end{split}
\end{align}
where, $\bar k_1=(k_1^{(1)}/\gamma,k_1^{(2)},k_1^{(3)})$. \textcolor{black}{Again, we can see that the integral in \eqref{4.58} has an asymmetry in the different components of the momentum integration and therefore, to the best of our knowledge, does not have a closed-form. Further, in Appendix~\eqref{App:3.H} we progress on computation of the integral more systematically. We also comment on the convergence of the integral.} Although, for the massless case, we also do not have any closed-form due to the asymmetry, but one can take a smooth massless limit.

\textbullet $\,\,$ Similar to \eqref{4.56e} we will have another contributing diagram where the $k_4$ scalar propagator will be replaced by a graviton propagator. Corresponding contribution to the impulse takes the following form,
\begin{align}
    \begin{split}
         [\Delta p_1^{\mu}]_{(n)} &= \thesisinlinefeynman[\chthreeinlinefeynwidth]{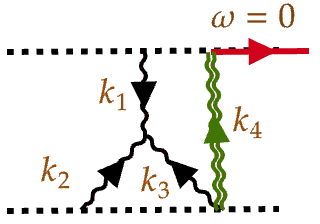} \\&=\frac{\lambda_3}{(2\pi)^4}\Big(\frac{m_1^2 s_1^2}{4m_p}\Big)\Big(\frac{m_2^2 s_2^2  }{4m_p^3}\Big)\frac{2\gamma^2-1}{2\gamma\sqrt{\gamma^2-1}} \frac{b^\mu}{|b|}\partial_{|b|}\rmint d^3 k_1 e^{-i \vec k_1\cdot \vec b}K_{0}\Big(|b||{k_{1(1)}}|\Big)\frac{1}{(\bar k_1^2+m^2)|\vec k_1|}\arctan\Big(\frac{|\vec k_1|}{2m}\Big)\,.
    \end{split}
\end{align}

\begin{equation}
  [\Delta p_1^{\mu}]_{(n)}=\frac{\lambda_3}{(2\pi)^4}\Big(\frac{m_1^2 s_1^2}{4m_p}\Big)\Big(\frac{m_2^2 s_2^2  }{4m_p^3}\Big)\frac{2\gamma^2-1}{2\gamma\sqrt{\gamma^2-1}} \frac{b^\mu}{|b|}\partial_{|b|}\rmint_{-\infty}^{\infty} d^3 k_1 e^{-i \vec k_1\cdot \vec b}K_{0}\Big(|b||{k_{1(1)}}|\Big)\frac{1}{(\bar k_1^2+m^2)|\vec k_1|}\arctan\Big(\frac{|\vec k_1|}{2m}\Big)\label{4.63pp}. \vspace{0.5cm}
\end{equation}
Again, due to the reason mentioned above, this integral does not possess any closed-form expression. One has to do it numerically componentwise. We will comment on the derivation in the Appendix.

\textbullet $\,\,$ We have another type of vertex $\lambda_3 h \varphi^3$, which also contributes to the 2PM diagrams. Calculations are similar to the computations of $\lambda_4 \varphi^4$ vertex \eqref{4.26m}.
\begin{align}
    \begin{split}
         [\Delta p_1^{\mu}]_{(o)}&=\thesisinlinefeynman[\chthreeinlinefeynwidth]{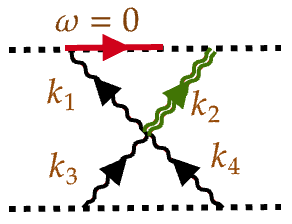}\\ &=-i\lambda_3\Big(\frac{m_1^2 s_1 (m_2 s_2)^2}{16m_p^4}\Big)P^{\alpha}_{\,\,\alpha,\rho\sigma}v_1^\rho v_1^\sigma\rmint_{\{k_i\}}\hat\delta^{(4)}\Big(\sum_{i=1}^{4}k_{i}\Big)\hat\delta(k_1\cdot v_1)\hat\delta(k_2\cdot v_1)\hat\delta(k_3\cdot v_2)\hat\delta(k_4\cdot v_2)\\ &\hspace{0.8cm}\times\prod_{j=1,3,4}\frac{1}{(k_j^2-m^2)k_2^2} 
 k_1^{\mu} \,e^{i(k_1+k_2)\cdot b_2}e^{i(k_3+k_4)\cdot b_2}\,.\label{4.55e}
    \end{split}
\end{align}
Now, \eqref{4.55e} has the same form as \eqref{4.26m} except that one denominator is massless. The result is given by,

\begin{align}
\begin{split}
  [\Delta p_1^{\mu}]_{(o)}&=-\frac{\lambda_3}{(2\pi)^3}\Big(\frac{m_1^2 s_1 (m_2 s_2)^2}{16m_p^4}\Big)\frac{1}{\sqrt{\gamma^2-1}}\frac{b^\mu}{|b|}\rmint_0^{\infty} dx {J_{1}(x|b|)}\arctan\Big(\frac{x}{2m}\Big)\arctan\Big(\frac{x}{m}\Big)\,,\\ &
  =-\frac{\lambda_3}{(2\pi)^3}\Big(\frac{m_1^2 s_1 (m_2 s_2)^2}{8 m_p^4}\Big)\frac{1}{\sqrt{\gamma^2-1}}\frac{b^\mu}{|b|^2}\int_0^1 ds\,\left(\frac{K_0(2m|b|)-K_0\left(\frac{m|b|}{s}\right)}{1-4s^2}+\frac{K_0(m|b|)-K_0\left(\frac{2m|b|}{s}\right)}{4-s^2}\right)\,.\label{3.99kk}
  \end{split}
\end{align}
\textcolor{black}{Like before, although this integral does not possess any closed-form expression, it has a smooth massless limit. In the massless limit, the above diagram can be calculated exactly as},
\begin{align}
    \begin{split}
         [\Delta p_1^{\mu}]_{(o)}\Big|_{m\rightarrow 0}=-{\lambda_3}\Big(\frac{m_1^2 s_1 (m_2 s_2)^2}{1024 \pi m_p^4}\Big)\frac{1}{\sqrt{\gamma^2-1}}\frac{b^\mu}{|b|^3}\,.
        \end{split}
    \end{align} 
\textbullet $\,\, $ There could be another possibility like \eqref{4.19c} where we have one line (graviton) connected with the first worldline and the other three (scalars) are connected with the second worldline. Now, the corresponding contribution to the impulse reads,
\begin{align}
    \begin{split}
         [\Delta p_1^{\mu}]_{(p)}&=\thesisinlinefeynman[\chthreeinlinefeynwidth]{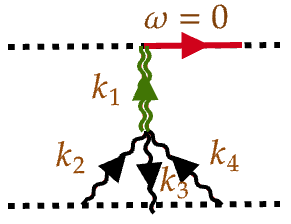}\\ &= -i\lambda_3\Big(\frac{m_1  (m_2 s_2)^3}{16m_p^4}\Big)P^{\alpha}_{\,\,\alpha,\rho\sigma}v_1^\rho v_1^\sigma \rmint_{\{k_i\}}\hat\delta^{(4)}\Big(\sum_{i=1}^{4}k_{i}\Big)\hat\delta(k_1\cdot v_1)\hat\delta(k_2\cdot v_2)\hat\delta(k_3\cdot v_2)\hat\delta(k_4\cdot v_2)\\ &\hspace{0.8cm}\prod_{j=2}^{4} \frac{1}{(k_j^2-m^2)k_1^2}
\times k_1^{\mu} \,e^{ik_1\cdot b_1}e^{i(k_2+k_3+k_4)\cdot b_2}\,.\label{4.57a}
    \end{split}
\end{align}
Following the derivation of \eqref{4.19c} we will get,

\begin{equation}
 [\Delta p_1^{\mu}]_{(p)}=\frac{\lambda_3}{(2\pi)^3}\Big(\frac{m_1  (m_2 s_2)^3}{16m_p^4}\Big)\frac{1}{\sqrt{\gamma^2-1}}\frac{b^\mu}{|b|}\partial_{|b|}\rmint_0^\infty dq \,dk_1\,\frac{J_{0}(k_1 |b|)}{k_1^2}\arctan\Big(\frac{q}{2m}\Big)\textrm{arctanh}\Big(\frac{2qk_1}{m^2+q^2+k_1^2}\Big)\,.\label{4.58a} \vspace{0.5cm}
\end{equation}
The integral in \eqref{4.58a} can be further simplified using the logarithmic representation of ArcTan. The massless limit also can be taken as \eqref{4.25j}.

\textbullet $\,\,$ There will be another diagram topologically equivalent to \eqref{4.57a} contributing to the impulse where one line (scalar) connects with one worldline and the other three (one graviton and two scalars) connect with the other worldline. The corresponding contribution is,
\begin{align}
    \begin{split}
      [\Delta p_1^{\mu}]_{(q)} &=\thesisinlinefeynman[\chthreeinlinefeynwidth]{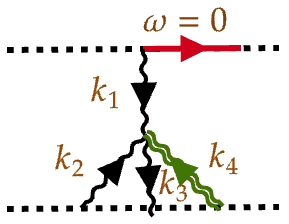}\\ &= -i\lambda_3\Big(\frac{m_1s_1 m_2 (m_2 s_2)^2}{16m_p^4}\Big)P^{\alpha}_{\,\,\alpha,\rho\sigma}v_2^\rho v_2^\sigma \rmint_{\{k_i\}}\hat\delta^{(4)}\Big(\sum_{i=1}^{4}k_{i}\Big)\hat\delta(k_1\cdot v_1)\hat\delta(k_2\cdot v_2)\hat\delta(k_3\cdot v_2)\hat\delta(k_4\cdot v_2)\\ & 
\hspace{0.8cm}\times\prod_{j=2}^{4}\frac{1}{(k_j^2-m^2)k_1^2} k_1^{\mu} \,e^{ik_1\cdot b_1}e^{i(k_2+k_3+k_4)\cdot b_2}\,,
    \end{split}
\end{align}
Similarly, following the derivation of \eqref{4.19c} we have,

\begin{equation}
[\Delta p_1^{\mu}]_{(q)} =\frac{\lambda_3}{(2\pi)^3}\Big(\frac{m_1s_1 m_2 (m_2 s_2)^2}{16m_p^4}\Big)\frac{1}{\sqrt{\gamma^2-1}}\frac{b^\mu}{|b|}\partial_{|b|}\rmint_0^\infty dq \,dk_1\,\frac{J_{0}(k_1 |b|)}{k_1^2+m^2}\arctan\Big(\frac{q}{m}\Big)\textrm{arctanh}\Big(\frac{2qk_1}{m^2+q^2+k_1^2}\Big).\label{4.66t} \vspace{0.5cm}
\end{equation}
The integral in \eqref{4.66t} can be further simplified using the logarithmic representation of ArcTan. The massless limit also can be taken as \eqref{4.25j}.

\textbullet $\,\,$ So far, the diagrams that contributed to the impulse at 2PM order require expanding the point-particle action up to $\mathcal{O}({z})\,.$ However, there is one additional diagram that must be included to complete the 2PM computation, and it requires expanding the point-particle action up to $\mathcal{O}({z}^2)$. 
\begin{align}
    \begin{split}
       [\Delta p_1^{\mu}]_{(r)} =\thesisinlinefeynman[\chthreeinlinefeynwidth]{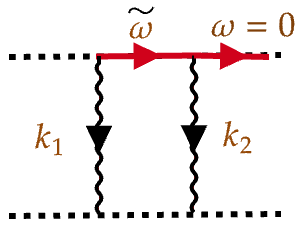}\hspace{0.7 cm}&= -im_1\Big(\frac{s_1m_2 s_2}{2\sqrt{2}m_p^2}\Big)^2\rmint_{k_i,\tilde\omega}\hat\delta(k_1\cdot v_2)\hat\delta(k_2\cdot v_2)\hat\delta(-k_1\cdot v_1+\tilde\omega)\hat\delta(-k_2\cdot v_1-\tilde\omega+\omega)\\ &\hspace{0 cm}\times (2\tilde\omega v_1^{{\rho}}-k_1^{\rho}) (\frac{1}{2}k_{2\rho}k_{2}^\mu+\tilde\omega k_{2}^\mu v_{1\rho}-\omega k_{2\rho}v_{1}^\mu-\omega\,\tilde\omega\delta^\mu_{\rho})\frac{e^{i(k_1+k_2)\cdot b}}{(k_1^2-m^2)(k_2^2-m^2)\tilde\omega^2}\Big |_{\omega=0},\\ &
        =im_1\Big(\frac{s_1m_2 s_2}{4m_p^2}\Big)^2\textcolor{black}{\rmint_{k_1,k}\frac{(k-k_1)^{\mu}\hat\delta(k_1\cdot v_2)\hat\delta(k\cdot v_2)\hat\delta(k\cdot v_1)e^{ik\cdot b}}{(k_1^2-m^2)[(k-k_1)^2-m^2](k_1\cdot v_1)^2}k_1\cdot(k-k_1)}\,.\label{4.6 mm}
    \end{split}
\end{align}
One can write down the integral as ,
\begin{align}
    \begin{split}
       [\Delta p_1^{\mu}]_{(r)} &=im_1\Big(\frac{s_1m_2 s_2}{4m_p^2}\Big)^2\rmint \hat d^4k\,\hat\delta(k\cdot v_1)\hat\delta(k\cdot v_2)e^{ik\cdot b}\rmint \hat d^4 k_1\,\hat\delta(k_1\cdot v_2)\frac{k_1\cdot(k-k_1)(k-k_1)^{\mu}}{(k_1^2-m^2)[(k-k_1)^2-m^2](k_1\cdot v_1)^2}\,,\\ &
        =im_1\Big(\frac{ s_1m_2 s_2}{4m_p^2}\Big)^2\rmint \hat d^4k\,\hat\delta(k\cdot v_1)\hat\delta(k\cdot v_2)e^{ik\cdot b}\mathcal{K}^{\mu}\label{4.37a}
    \end{split}
\end{align}
where $\mathcal{K}^{\mu}(k,v_i)$ can be reduced by Passarino-Veltman reduction as,
\begin{align}
    \begin{split}
        \mathcal{K}^{\mu}=a_1 k^{\mu}+a_2 v_1^{\mu}+a_3 v_2^{\mu}\,.\label{4.38a}
    \end{split}
\end{align}
It is evident that with the support $k\cdot v_1 =0\,\text{and } k\cdot v_2=0$, one can write $ v_{2\mu}\mathcal{K}^{\mu}=0$, implying, $a_2\gamma+a_3=0$. Hence, we left with the following,
\begin{align}
    \begin{split}
\mathcal{K}^{\mu}=a_2(v_1^{\mu}-\gamma v_2^{\mu})+a_1k^\mu\,.
    \end{split}
\end{align}
From \eqref{B.22aa} it is clear that the $a_1$ coefficient is non-zero for 
and is given by,
\begin{align}
    \begin{split}
     a_1=   \frac{1}{4}(-\vec k^2-2m^2)\hat\chi_k(|\vec k|,m)
    \end{split}
\end{align}
where, $\hat\chi_{k}(k,m)$ is defined in \eqref{B.16ab}. Similarly,
$a_2$ can be fixed by contracting both side by $v_1^{\mu}$,
\begin{align}
    \begin{split}
        a_2&=\frac{1}{\gamma^2-1}\rmint \hat d^{4}k_1 \frac{\hat\delta(k_1\cdot v_2)\,k_1\cdot(k-k_1)}{(k_1^2-m^2)[(k-k_1)^2-m^2](k_1\cdot v_1+i\epsilon)}\,,\\ &
        =\frac{1}{2(\gamma^2-1)}\rmint \hat d^4 k_1\hat\delta(k_1\cdot v_2)\Big[-\frac{1}{(k_1^2-m^2)(k_1\cdot v_1+i\epsilon)}-\frac{1}{(k-k_1)^2-m^2](k_1\cdot v_1+i\epsilon)}\\\ &
        \hspace{0.8cm} \frac{k^2}{(k_1^2-m^2)[(k-k_1)^2-m^2](k_1\cdot v_1+i\epsilon)}
      -\frac{2m^2}{(k_1^2-m^2)[(k-k_1)^2-m^2](k_1\cdot v_1+i\epsilon)}  \Big]\,.\label{3.13}
    \end{split}
\end{align}
The first two terms will not contribute. Now doing a change of variable, $k_1-k\rightarrow -q$, $a_2$ can be written as,
\begin{align}
    \begin{split}
        a_2=-\frac{1}{2(\gamma^2-1)}\rmint \hat d^4q \frac{(k^2-2m^2)\hat\delta(q\cdot v_2)}{(q^2-m^2)[(q-k)^2-m^2](q\cdot v_1-i\epsilon)}\,.\label{3.14}
    \end{split}
\end{align}
Now adding (\ref{3.13}) and (\ref{3.14}) and using the fact,
\begin{align}
    \begin{split}
       - i\hat \delta(x)=\frac{1}{x+i\epsilon}-\frac{1}{x-i\epsilon}
    \end{split}
\end{align}
one gets,
\begin{align}
    \begin{split}
        a_2=\frac{-i}{4(\gamma^2-1)}\rmint \hat d^4 q \frac{(k^2-2m^2)\hat\delta(q\cdot v_1)\hat\delta(q\cdot v_2)}{(q^2-m^2)[(q-k)^2-m^2]}\,.
    \end{split}
\end{align}
Thereafter, $\Delta p_1^{\mu}$ can be written as,
\begin{align}
    \begin{split}
         \hspace{0cm}[\Delta p_1^{\mu}]_{(r)} &= -m_1\Big(\frac{s_1m_2 s_2}{8\sqrt{2}\sqrt{\gamma^2-1}m_p^2}\Big)^2\rmint \hat d^4 k\,\hat d^4 q\,\hat\delta(k\cdot v_1)\hat\delta(k\cdot v_2)\hat\delta(q\cdot v_1)\hat\delta(q\cdot v_2)\,e^{i k\cdot b}\,\frac{k^2-2m^2}{(q^2-m^2)[(q-k)^2-m^2]}\\ &
\hspace{0.8cm}\times (v_1^{\mu}-\gamma v_2^{\mu})+im_1\Big(\frac{s_1m_2 s_2}{4m_p^2}\Big)^2\rmint_{k}\hat\delta(k\cdot v_1)\hat\delta(k\cdot v_2)e^{ik\cdot b}k^{\mu}a_1\,,
         \\&
        =-\frac{m_1}{(2\pi)^4}\Big(\frac{s_1m_2 s_2}{8\sqrt{2}m_p^2}\Big)^2\frac{1}{(\gamma^2-1)^3}\rmint  d^2 k\,e^{-ik\cdot b}(-k^2-2m^2)\rmint  d^2 q\,\frac{1}{(q^2+m^2)[(q-k)^2+m^2]}(v_1^{\mu}-\gamma v_2^{\mu})\\ &
      \hspace{0.8cm}  -m_1\Big(\frac{s_1m_2 s_2}{8m_p^2}\Big)^2\frac{1}{\pi^{3/2}(\gamma^2-1)^{3/2}} \frac{b^\mu}{|b|}\Big(m K_1(2 |b| m)\Big)
        \,.
    \end{split}
\end{align}
The $`q$' integral can be done using dimensional regularisation with $d=2-\epsilon$. We need to be cautious about the UV (or IR) divergences that may appear while doing the integral. The integral can be written as,
\begin{align}
    \begin{split}
      I&=  \rmint {d}^{2-\epsilon}q \frac{1}{(q^2+m^2)[(q-k)^2+m^2]}\,,
      \\ &
       \xrightarrow[]{\epsilon \rightarrow 0} (2\pi)\frac{2 \log \left(\sqrt{\frac{k^2}{4 m^2}+1}+\frac{k}{2 m}\right)}{m\,k  \sqrt{\frac{k^2}{4 m^2}+1}}\,=(2\pi)\frac{4[\log(k+\sqrt{k^2+4m^2})-\log(2m)]}{k\sqrt{k^2+4m^2}}\,.
    \end{split}
\end{align}
Now, the integral $\Delta p_1^{\mu}$ can be expressed as,
\begin{align}
\begin{split} \label{eee22}
[\Delta p_1^{\mu}]_{(r)} =&-\frac{m_1}{(2\pi)^2}\Big(\frac{s_1m_2 s_2}{8\sqrt{2}\,m_p^2}\Big)^2\frac{1}{(\gamma^2-1)^3}(v_1^\mu-\gamma v_2^\mu)\rmint_{0}^{\infty}{d}k\,k\, J_{0}(k|b|)(k^2+2m^2)
      \\&\hspace{0.8cm} \Bigg( \frac{4[\log(k+\sqrt{k^2+4m^2})-\log(2m)]}{k\sqrt{k^2+4m^2}}\Bigg)\\ &
      \hspace{0 cm}-m_1\Big(\frac{s_1m_2 s_2}{8m_p^2}\Big)^2\frac{1}{\pi^{3/2}\gamma^3(\gamma^2-1)^{3/2}} \frac{b^\mu}{|b|}\Big(m K_1(2 |b| m)\Big)\,.
\end{split}
\end{align}
In the massless limit we have,
\begin{align}
    \begin{split}
     [\Delta p_1^{\mu}]_{(r)} \Big|_{m\to 0}&= -\frac{m_1}{(2\pi)^2}\Big(\frac{s_1m_2 s_2}{4\sqrt{2}\,m_p^2}\Big)^2\frac{1}{(\gamma^2-1)^3}(v_1^\mu-\gamma v_2^\mu)\Bigg(\rmint_{0}^{\infty}dk \, k\,J_{0}(k|b|) \log(2 k)-4\frac
     {\log(2\epsilon)J_{1}(|b|\Lambda)}{|b|}\Bigg)\\ &
       \hspace{0.8cm}-m_1\Big(\frac{s_1m_2 s_2}{8m_p^2}\Big)^2\frac{1}{\pi^{3/2}(\gamma^2-1)^{3/2}} \frac{b^\mu}{|b|^2}
    \end{split}
\end{align}
\vspace{-0 cm}
where $\epsilon$ and $\Lambda$ are the IR and UV cut-off, respectively. The second term is a divergent one. In fact, UV/IR  divergence is mixed, i.e., it is divergent in both the limit  ${\epsilon}\rightarrow 0 \,\& \,\Lambda\rightarrow \infty\,.$ So we ignore this in the classical limit. The finite part takes the following form,
\begin{align}
    \begin{split}
    [\Delta p_1^{\mu}]_{(r)} \Big|_{m\to 0}\sim N_1\frac{(v_1^{\mu}-\gamma v_2^{\mu})}{4(\gamma^2-1)^3} \frac{1}{|b|^2}\textcolor{black}{+N_2 \frac{b^\mu}{(\gamma^2-1)^{3/2}|b|^2}}\,.
    \end{split}
\end{align}
The total impulse (due to the scalar field) at 2PM order is the sum of  (\ref{e:barwq244}), (\ref{e:barwq248}), (\ref{4.19f}), (\ref{e:barwq246}), (\ref{e:barwqpp}), (\ref{e:barwq24}), (\ref{sa}), (\ref{e:barwqp}), (\ref{4.54r}), (\ref{4.55m}), (\ref{4.58o}), (\ref{4.63pp}), (\ref{3.99kk}), (\ref{4.58a}), (\ref{4.66t}) and (\ref{eee22}) as well as the terms that come from interchanging the worldline one and two. 
\begin{align}
\begin{split}
     \Delta p_1^{\mu}\Big|^{\textrm{2PM}, \textrm{Total}}_{\textrm{scalar}}= & \sum_{\upsilon=c}^{r}[\Delta p_1^{\mu}]_{(\upsilon)}\,.
     \end{split}
\end{align}
\newpage
Finally, collecting all individual expressions we get,
\begin{align}
\Delta p_1^{\mu}\Big|^{\textrm{2PM},\textrm{total}}_{\textrm{scalar}}
&=\frac{m^2\, m_1m_2^2 s_2}{32\pi^2\,m_p^4\,\sqrt{\gamma^2-1}}\frac{b^\mu}{|b|}\partial_{|b|}\Big(s_2 I_2(m,|b|)+s_1 \bar{I}_2(m,|b|)\Big)\notag\\
&\quad+\frac{m_1 m_2^2 s_2}{8\,m_p^4}\frac{b^{\mu}}{|b|}\partial_{|b|}\Big(s_2 I_{4}(m,|b|)+\frac{s_1}{16 \pi^2 \sqrt{\gamma^2-1}}\bar{I}_4(m,|b|)\Big)\notag\\
&\quad+\frac{m_1 m_2^2 s_2}{8\pi^2m_p^4\sqrt{\gamma^2-1}}\frac{b^\mu}{|b|}\partial_{|b|}\Bigg[\frac{s_1\,(2\gamma^2-1)}{8}\rmint dl\,J_{0}(|b|l)\arctan\Big(\frac{l}{m}\Big)\notag\\
&\qquad\qquad+g_1 s_2\rmint_0^\infty dl\,J_{0}(|b|l)\arctan\Big(\frac{l}{2m}\Big)\Bigg]\notag\\
&\quad-\frac{m_1}{(2\pi)^2}\Big(\frac{s_1m_2 s_2}{4\sqrt{2}\,m_p^2}\Big)^2\frac{1}{(\gamma^2-1)^3}(v_1^\mu-\gamma v_2^\mu)\rmint_{0}^{\infty}{d}k\,k\,J_{0}(k|b|)(k^2+2m^2)\notag\\
&\qquad\qquad\times\Bigg(\frac{\log(k+\sqrt{k^2+4m^2})-\log(2m)}{k\sqrt{k^2+4m^2}}\Bigg)\notag\\
&\quad-m_1\Big(\frac{s_1m_2 s_2}{8m_p^2}\Big)^2\frac{1}{\pi^{3/2}(\gamma^2-1)^{3/2}} \frac{b^\mu}{|b|}\Big(m K_1(2 |b| m)\Big)\notag\\
&\quad+\frac{\lambda_4\,m_1s_1 (m_2s_2)^2}{128\pi^3m_p^4\sqrt{\gamma^2-1}}\frac{b^\mu}{|b|}\partial_{|b|}\Bigg[m_1s_1\,I_3(m,|b|)\notag\\
&\qquad\qquad+\frac{m_2 s_2}{4\gamma}\Big(2(1-\log(3m^2))K_{0}(|b|m)-\rmint dk_1\frac{J_{0}(k_1 |b|)}{k_1^2+m^2}\Theta(k_1,m)\Big)\Bigg]\notag\\
&\quad+\frac{\lambda_3}{(2\pi)^4}\Big(\frac{m_2^3s_2^3}{16m_p^4}\Big)\frac{1}{\sqrt{\gamma^2-1}}\frac{b^\mu}{|b|}\partial_{|b|}\Bigg[m_1 g_1 I_{5}(m,|b|)+\frac{(2\gamma^2-1)\,m_1 s_1}{2}\bar{I}_{5}(m,|b|)\Bigg]\notag\\
&\quad+\frac{\lambda_3}{(2\pi)^3}\Big(\frac{m_1 (m_2 s_2)^2}{16m_p^4}\Big)\frac{1}{\sqrt{\gamma^2-1}}\frac{b^\mu}{|b|}\partial_{|b|}\Bigg[-m_1s_1 I_{6}(m,|b|)+\frac{m_2 s_2}{\gamma}\bar{I}_{6}(m,|b|)+\frac{m_2 s_1}{\gamma}\bar{I}_{6}(m,|b|)\Bigg]\notag\\
&\quad+\frac{\lambda_3}{(2\pi)^4}\Big(\frac{m_1^2 s_1^2}{4m_p}\Big)\Big(\frac{m_2 s_2}{4m_p^3}\Big)\frac{1}{\gamma\,\sqrt{\gamma^2-1}}\frac{b^\mu}{|b|}\partial_{|b|}\Bigg[m_2g_2\,I_{7}(m,|b|)+\Big(\frac{m_2 s_2\,(2\gamma^2-1)}{2}\Big)\bar{I}_{7}(m,|b|)\Bigg]\notag\\
&\quad+(1\leftrightarrow 2)\,.
\end{align}
where, \begin{align}
    \begin{split}
    &I_2(m,|b|)=\rmint _{0}^{\infty}dx\,\frac{J_{0}(|b|x)}{x^2}\arctan\Big(\frac{x}{2m}\Big)\,,\quad 
    \bar{I}_{2}(m,|b|)=\rmint_{0}^\infty dx\frac{J_{0}(b|x|)}{x^2+m^2}\arctan\Big(\frac{x}{m}\Big)\,,\\ &
    I_{3}(m,|b|)=\rmint_{0}^\infty dx\, \frac{J_{0}(x|b|)}{x}\arctan^2\Big(\frac{x}{2m}\Big),\quad
    \bar{I}_{4}(m,|b|)=\rmint_0^\infty dx \frac{x^2}{x^2+m^2}J_{0}(x|b|)\arctan\Big(\frac{x}{m}\Big)\,,\\&
    I_{5}(m,|b|)=\rmint_0^{\infty} dx \,dz \,\frac{J_{0}(x|b|)}{z^2+m^2}\,\textrm{arctanh}\Big(\frac{2 x\,z}{x^2+z^2+m^2}\Big)\arctan\Big(\frac{z}{2m}\Big)\,,\\&
    \bar{I}_{5}(m,|b|)=\rmint_0^\infty dx  dz \frac{J_0(x|b|)}{z^2+m^2}\arctan\Big(\frac{z}{2m}\Big)\, \textrm{arctanh}\Big(\frac{2x\,z}{x^2+z^2}\Big)\,,\\&
    I_{6}(m,|b|)=\rmint_0^{\infty} dx \frac{J_{0}(x|b|)}{x}\arctan\Big(\frac{x}{2m}\Big)\arctan\Big(\frac{x}{m}\Big)\,,\\&
    \bar{I}_{6}(m,|b|)=\rmint_0^\infty dx \,dz\,\frac{J_{0}(z |b|)}{z^2}\arctan\Big(\frac{x}{2m}\Big)\textrm{arctanh}\Big(\frac{2x\,z}{m^2+x^2+z^2}\Big)\,,\\&
    \bar{\bar{I}}_6(m,|b|)=\rmint_0^\infty dx\,dz\,\frac{J_{0}(z |b|)}{z^2+m^2}\arctan\Big(\frac{x}{m}\Big)\textrm{arctanh}\Big(\frac{2x\,z}{m^2+x^2+z^2}\Big)\,,\\&
   I_{7}(m,|b|)= \rmint_{-\infty}^{\infty} d^3 z e^{-i \vec z\cdot \vec b}K_{0}\Big(|b||\sqrt{z_{(1)}^2+m^2}|\Big)\frac{1}{(\bar z^2+m^2)|\vec z|}\arctan\Big(\frac{|\vec z|}{2m}\Big)\,,\\&
   \bar{I}_{7}(m,|b|)=\rmint_{-\infty}^{\infty} d^3 z e^{-i \vec z\cdot \vec b}K_{0}\Big(|b||{z_{(1)}}|\Big)\frac{1}{(\vec{z}^2+m^2)|\vec z|}\arctan\Big(\frac{|\vec z|}{2m}\Big) \label{4.86}
    \end{split}
\end{align}
and $I_{4}(m,|b|)$ is defined in (\ref{4.37}). \textcolor{black}{As discussed previously, to the best of our knowledge these integrals do not possess any closed-form expression (except $I_5(m,|b|),\bar{I}_5(m,|b|), \bar{I}_6(m,|b|)$ and $\bar{\bar{I}}_6(m,|b|)$ as they can be done to some extent using method discussed around (\ref{4.25j}). They can be evaluated numerically and possess smooth behaviour w.r.t. $|b|\,.$ They also admit smooth massless limits as discussed earlier case by case basis.} \par
Note that the expression for the total impulse also receives a contribution from the pure Einstein-Hilbert part. However, that result is already well known in the literature \cite{Mogull:2020sak}. Hence, instead of reproducing it here, we focus only on the new contributions from the various vertices involving the scalar field up to 2PM order. We also show the corresponding results for the massless case. The fall-off with respect to $|b|$ is analogous to the gravitational case.

Before ending this section, we note that although we have computed the impulse up to 2PM, we have not yet discussed its connection with the scattering amplitude. \textcolor{black}{We elaborate on this connection in Appendix~(\ref{ch2:app:B}).}
\section{Computation of waveform}\label{ch2:sec5}
In this section, we will compute the frequency domain gravitational waveform due to different field configurations. In the wave zone, $f_{\varphi,h}(k)$ is defined by the one point function as follows,
\begin{equation}\label{e:bar}\begin{split}
&  f_{h}(k):=\frac{1}{4\pi m_p}\epsilon^{\mu}\epsilon^{\nu}k^2\langle h_{\mu\nu}(k)\rangle\Big|_{k^2\rightarrow 0}\,, \\ &
        f_{\varphi}(k) := (k^2-m^2)\frac{\langle \varphi(k)\rangle}{m_p}\Big|_{k^2\rightarrow m^2}\\
        \end{split}
\end{equation}

where $k^\mu=\Omega\, n^\mu$ describes the on-shell momentum of the graviton/scalar. As, on-shell graviton is massless then, $n_\mu n^\mu=0$  and for the scalar $n_\mu n^\mu=\frac{m^2}{\Omega^2}$. $n^\mu$ can be parameterized as, 
\begin{align}
    \begin{split}
n^\mu\Big|_{g,h}=\Big(1,\sqrt{1-\frac{m_s^2}{\Omega^2}}\boldsymbol{\hat{x}}\Big)
    \end{split}
\end{align}
where,
\begin{align}
    \begin{split}
\boldsymbol{\hat{x}}^{\mu}=e_{1}^{\mu}\cos\theta+\sin(\theta)(e_{2}^{\mu}\cos\phi+e_3^{\mu}\sin\phi),\,e_{i}^{\mu}=(0,\hat\xi_{i})\,.
    \end{split}
\end{align}
with $\hat \xi\in (\hat i,\hat j,\hat k)$. In the parametrization mentioned above, the $n$ can be written as follows,
\begin{align}
    \begin{split}
        n=(1,\cos(\theta),\sin(\theta)\cos(\phi),\sin(\theta)\sin(\phi))\,.
    \end{split}
\end{align}
First, we will compute the corrections to the waveform from the scalar degrees of freedom. For the sake of simplicity, we first take the scalar field to be massless, and in the next section, we will discuss the massive integrals and how the integrals get complicated in the massive case. It has been shown in \cite{Mogull:2020sak} that in a two-body scattering problem, the scattering amplitude with on-shell external graviton (or scalar) is related to the WQFT correlator. In general, the scattering amplitude has the following schematic form.
\begin{center}
\includegraphics[width=0.46\linewidth]{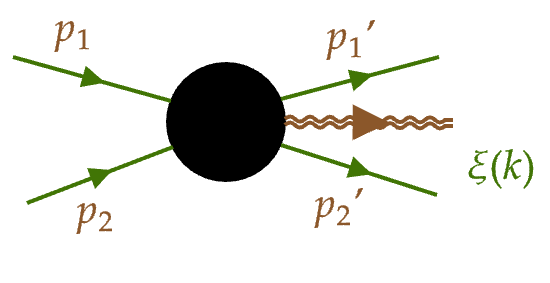}
\end{center}
\begin{align}
    \mathcal{M}[g,\varphi]\equiv \text{S-matrix topology shown above.}
\end{align}
Using the result established in \cite{Mogull:2020sak}, one can show that the connection between the S-matrix element $\langle\phi_1,\phi_2\Big |\,\mathcal{S}\,\Big |\phi_1,\phi_2,\chi\rangle$ with the WQFT correlator,
\begin{align}
    \begin{split}
     (k^2-m_{\chi}^2)   \langle \chi(k)\rangle \propto \rmint_{k_1,k_2} d\mu_{1,2}(k_i,k)\lim_{\hbar\rightarrow 0}\mathcal{M}[g,\varphi](p_i,p_i',k),\,\chi\in (h_{\mu\nu},\varphi).\label{5.5m}
    \end{split}
\end{align}
where, in \eqref{5.5m} $k_i=p_i-p_i'$ and $d\mu_{1,2}$ is the measure of integration depends on the interaction.

\subsection{Scalar waveform}
We now initiate the computation of the scalar waveform, which we evaluate up to 2PM order. Next, we list all the diagrams contributing at this order and evaluate the corresponding expressions.

\subsubsection{Scalar waveform at 1PM}
In this subsection, we list the diagrams that contribute to the 1PM scalar waveform.

\textbullet $\,\,$ We start by considering the self-interaction vertices and their contribution to the waveform. First, we consider the $\frac{\lambda_3}{\textcolor{black}{3!}} m_p\,\varphi^3$ vertex. Its contribution to the one-point function has the following form and appears at 1PM order \footnote{\textcolor{black}{Again note that the combinatorial factor associated with this diagram is 3!. We have multiplied the result by this factor. For the subsequent diagrams, we will likewise include the appropriate combinatorial factors from the outset.}}. 
\begin{align}
    \begin{split} \label{wave0}
        k^2\Big\langle\varphi(k)\Big\rangle\equiv \thesisinlinefeynman[\chthreeinlinefeynwidth]{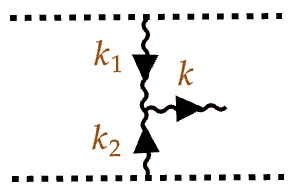}\hspace{1.2 cm}=\lambda_3\Big(\frac{m_1s_1m_2s_2}{4m_p}\Big)\rmint \frac{d\mu_{1,2}(k)}{k_1^2k_2^2}\,.
    \end{split}
\end{align}
where the integral measure is given by,
\begin{align}
    \begin{split}
        \rmint d\mu_{1,2}(k)=\rmint_{k_1,k_2}e^{ik_1\cdot b_1}e^{ik_2\cdot b_2}\hat\delta(k_1\cdot v_1)\hat\delta(k_2\cdot v_2)\hat\delta^{(4)}(k_1+k_2-k)\,.
    \end{split}
\end{align}
Now, our focus is to compute the time domain waveform, which is the Fourier transform of the frequency domain waveform. Hence, we have to do a Fourier transformation of (\ref{wave0}). This gives the following, 
\begin{align}
    \begin{split}
        f_{\varphi}(x)&= \lambda_3\Big(\frac{m_1s_1m_2s_2}{4m_p^2}\Big)\rmint_{\Omega}e^{-ik\cdot x}\rmint_{k_i}e^{ik_1\cdot b_1}e^{ik_2\cdot b_2}\frac{\hat\delta(k_1\cdot v_1)\hat\delta(k_2\cdot v_2)}{k_1^2k_2^2}\hat\delta^{(4)}(k_1+k_2-k)\,,\\ &
     =   \lambda_3\Big(\frac{m_1s_1m_2s_2}{4m_p^2}\Big) \rmint_{\Omega}e^{-ik\cdot (x-b_1)}\underbrace{\rmint_{k_2}e^{-ik_2\cdot b}\frac{\hat\delta(k\cdot v_1-k_2\cdot v_1)\hat\delta(k_2\cdot v_2)}{k_2^2(k-k_2)^2}}_{J_{(0)}(k)}\,.
     \label{5.47 m}
    \end{split}
\end{align}
The integral in \eqref{5.47 m} can be evaluated using the results derived in \eqref{A.2} and \eqref{A.8}.

\begin{equation}\label{wave1}\begin{split}
  f_{\varphi}(x)&= \lambda_3\Big(\frac{m_1s_1m_2s_2}{4m_p^2}\Big) \rmint_{\Omega}e^{-ik\cdot (x-b_1)}\,J_{(0)}(k),\\ &
        =-\frac{\lambda_3}{(2\pi)^3}\Big(\frac{m_1s_1m_2s_2}{4m_p^2}\Big) \frac{|b|}{4\gamma}\rmint_{0}^{1}dy\,\frac{1}{\Bar\Delta(y)}\sqrt{\frac{l^2}{|b|^2\bar\Delta^2}+1},\,l:=n\cdot(x-b_1+y\,b)\\
        \end{split}
\end{equation}
\vspace{0.5 cm}

where, $\Bar\Delta$ is defined in \eqref{A.4} and \eqref{ch2:A.6}. We have restored the factors of $\pi$ that come from the integration measures and the delta functions. So at 1PM order, (\ref{wave1}) gives the entire contribution to the scalar waveform. Next, we will extend our study to a 2PM order.

\subsubsection{Scalar waveform at 2PM}
In this subsection, we intend to compute the 2PM scalar waveform coming from the purely scalar sector and the scalar-graviton interaction sector.

\textbullet $\,\,$ \textit{We start with the self-interacting vertex, namely,  $\frac{\lambda_4}{\textcolor{black}{4!}}\varphi^4$. We will show that it doesn't contribute to the waveform.} To show that, we first write down the one-point function for this case. 
\begin{align}
    \begin{split}\label{5.11 mn}
        k^2\Big\langle\varphi(k)\Big\rangle\equiv    \thesisinlinefeynman[\chthreeinlinefeynwidth]{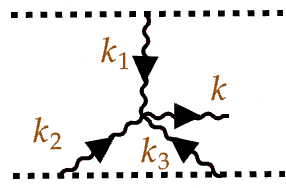}\hspace{1.4 cm}=\rmint \frac{d\mu_{1,2}(k)}{k_1^2 k_2^2 k_3^2}\,.
    \end{split}
\end{align}
where , in this case, the integral measure has the following form
\begin{align}
    \begin{split}
       \rmint d\mu_{1,2}(k)=\rmint_{k_1,k_2,k_3}e^{i k_1\cdot b_1}e^{ik_2\cdot b_2}e^{i k_3 \cdot b_2}\hat\delta(k_1\cdot v_1)\hat\delta(k_2\cdot v_2)\hat\delta(k_3\cdot v_2)\hat\delta^{(4)}(k_1+k_2+k_3-k)\,.
    \end{split}
\end{align}
Hence, the corresponding contribution to the waveform has the following form,
\begin{align}
    \begin{split}
  \label{4PHI}      f_{\varphi}(x)&=\rmint_{\Omega}e^{-ik\cdot x}\rmint_{k_i}e^{ik_1\cdot b_1}e^{i k_2\cdot b_2}e^{i k_3\cdot b_2}\frac{\hat\delta(k_1\cdot v_1)\hat\delta(k_2\cdot v_2)\hat\delta(k_3\cdot v_2)}{k_1^2 k_2^2 k_3^2}\hat\delta^{(4)}(k_1+k_2+k_3-k)\,,\\ &
=\rmint_{\Omega}e^{-ik\cdot x}\rmint_{k_2,k_3}\frac{\hat\delta((k-k_2-k_3)\cdot v_1)\hat\delta(k_2\cdot v_2)\hat\delta{(k_3\cdot v_2)}}{k_2^2 k_3^2 (k-k_2-k_3)^2}\,e^{i (k-k_2-k_3)\cdot b_1}e^{i(k_2+k_3)\cdot b_2} \,,\\ &
\xrightarrow[]{q=k_2+k_3}\rmint_{\Omega}e^{-i k\cdot x}\rmint_{k_3,q} \frac{\hat\delta(k\cdot v_1-q\cdot v_1)\hat\delta(k_3\cdot v_2)\hat\delta(k_3\cdot v_2-q\cdot v_2)}{k_3^2 (q-k)^2(q-k_3)^2}e^{i (k-q)\cdot b_1}e^{i(q-k_3)\cdot b_2}e^{ik_3\cdot b_2}\,.
    \end{split}
\end{align}
We have the following integral to be solved which can be systematically done by introducing $\alpha$ parametrization,
\begin{align}
    \begin{split}
        f_{\varphi}(x)&=\rmint_{\Omega}e^{-i k\cdot x+i k\cdot b_{1}}\rmint_{k_3,q} \frac{\hat\delta(k\cdot v_1-q\cdot v_1)\hat\delta(k_3\cdot v_2)\hat\delta(q\cdot v_2)}{k_3^2 \, (q-k_3)^2(q-k)^2}e^{-i q\cdot b}\,,\\ &
        =\int_{\Omega}e^{-i k\cdot(x-b_1)}\int_{0}^\infty d\alpha\, d\beta\, d\gamma\, \exp\Big[-i\alpha (q-k)^2-i\beta(q-k_3)^2-i\gamma k_3^2 \Big]\\ &
        \hspace{0.8cm}\times  \hat\delta(k\cdot v_1-q\cdot v_1)\hat\delta(k_3\cdot v_2)\hat\delta(q\cdot v_2) e^{-iq\cdot b}\,,\\ &
=\int_{0}^{\infty}d\hat\alpha\,d\hat\beta\,d\hat\gamma \int_{\Omega}e^{-i\Omega \,n\cdot (x-b_1)}\int_{\vec k_3,\vec q}\hat\delta(\Omega\,n\cdot v_1+\vec q\cdot \vec v_1)\,\exp\Big[i\hat\alpha(\vec q^2-2\Omega \vec q\cdot \vec n)\\ &
\hspace{0.8cm}+i\hat\beta (\vec q^2-2\vec q\cdot \vec k_3+\vec k_3^2)+{i\hat\gamma \vec k_3^2}\Big]e^{i\vec q\cdot \vec b}
    \end{split}
\end{align}
Now doing the $\Omega $ integral we have,
\begin{align}
    \begin{split}
        f_{\varphi}(x)&=\frac{1}{n\cdot v_1}\int_{0}^{\infty}d\hat\alpha\,d\hat\beta\,d\hat\gamma \int_{\vec k_3,\vec q} \exp\Big[i\frac{\vec q\cdot \vec v_1}{n\cdot v_1}n\cdot \tilde x\Big]\exp\Big[i(\hat\alpha+\hat\beta)\,\vec q^2+2i\hat\alpha \Big(\frac{\vec q\cdot \vec v_1}{n\cdot v_1}\Big)(\vec q\cdot \vec n)+i\vec q\cdot \vec b\Big]\\ &
        \hspace{0.8cm}\times \exp\Big[i(\hat\beta+\hat\gamma)\vec k_3^2-2i\hat\beta \vec k_3\cdot \vec q\Big]\,,\Tilde{x}\equiv x-b_1
    \end{split}
\end{align}
Now we do the $\vec k_3$ integral by dimensional regularisation,
\begin{align}
    \begin{split}
        f_{\varphi}(x)&=\frac{1}{n\cdot v_1}\int_{0}^{\infty}d\hat\alpha\,d\hat\beta\,d\hat\gamma \int_{\vec q}\exp\Big[i\frac{\vec q\cdot \vec v_1}{n\cdot v_1}n\cdot \tilde x\Big]\exp\Big[i(\hat\alpha+\hat\beta)\,\vec q^2+2i\hat\alpha \Big(\frac{\vec q\cdot \vec v_1}{n\cdot v_1}\Big)(\vec q\cdot \vec n)+i\vec q\cdot \vec b\Big]\\ &
    \hspace{0.8cm} \times e^{-i\frac{\pi}{4}+i\frac{\epsilon}{2}}\pi^{\frac{3}{2}-\epsilon}(\hat\beta+\hat\gamma)^{-\frac{3}{2}+\epsilon}\exp\Big(-i \frac{\hat\beta^2 \vec q^2}{\hat\beta +\hat\gamma}\Big)\\ &
    =\frac{1}{n\cdot v_1}e^{-i\frac{\pi}{4}+i\frac{\epsilon}{2}}\pi^{\frac{3}{2}-\epsilon}\int_{0}^{\infty}d\hat\alpha\,d\hat\beta\,d\hat\gamma \,(\hat\beta+\hat\gamma)^{-\frac{3}{2}+\epsilon}\int d^3q \exp\Big(i\lambda_1 \vec q^2+2i \lambda_2 (\vec q\cdot \vec v_1)(\vec q\cdot \vec n)+i\vec q\cdot \vec \lambda_3\Big)\label{5.16k}
    \end{split}
\end{align}
where,
\begin{align}
    \begin{split}
&\lambda_1(\hat\alpha,\hat\beta,\hat\gamma)=\hat\alpha+\hat\beta -\frac{\hat\beta^2}{\hat\beta +\hat\gamma}\\ &
\lambda_2(\hat\alpha,\hat\beta,\hat\gamma)=\frac{\hat\alpha }{n\cdot v_1}\\ &
\vec \lambda_3=\frac{n\cdot \tilde x}{n\cdot v_1}\vec v_1+\vec b\equiv \Upsilon(x)\vec v_1+\vec b
    \end{split}
\end{align}
The non-triviality comes in the $q$ integral due to the presence of $\vec q\cdot \vec n\,\vec q\cdot \vec v_1$ term. For the sake of simplicity, we choose $ n=(1,1,0,0)$ and the $\vec q$ integral can be done component wise,
\begin{align}
    \begin{split}
        f_{\varphi}(x)&=\frac{1}{n\cdot v_1}e^{-i\frac{\pi}{4}+i\frac{\epsilon}{2}}\pi^{\frac{3}{2}-\epsilon}\int_{0}^{\infty}d\hat\alpha\,d\hat\beta\,d\hat\gamma \,(\hat\beta+\hat\gamma)^{-\frac{3}{2}+\epsilon}\int dq_{(1)}\exp\Big(i(\lambda_1+2\lambda_2\gamma\beta\cos\epsilon)q_{(1)}^2+i q_{(1)}\Upsilon(x)\,\gamma\beta\Big)\\ &
        \hspace{0.8cm}\times \int dq_{(2)}\exp\Big(i\lambda_1 q_{(2)}^2+iq_{(2)}b\Big)\int dq_{(3)} \exp(i\lambda_1 q_{(3)}^2)\\ & =  \frac{1}{n\cdot v_1}e^{-i\frac{\pi}{4}+i\frac{\epsilon}{2}}\pi^{\frac{3}{2}-\epsilon}\int_{0}^{\infty}d\hat\alpha\,d\hat\beta\,d\hat\gamma \,(\hat\beta+\hat\gamma)^{-\frac{3}{2}+\epsilon}\Bigg(\sqrt{\frac{\pi}{\lambda_1 +2\lambda_2\gamma\beta}}\\ &
        \hspace{0.8cm}\times\exp\Big(i\frac{\pi}{4}-i\frac{\Upsilon(x)^2\gamma^2\beta^2}{4\lambda_1 +8\lambda_2\gamma\beta}\Big)\Bigg)\times \Bigg(\sqrt{\frac{\pi}{\lambda_1}}\exp\Big(i\frac{\pi}{4}-i\frac{b^2}{4\lambda_1}\Big)\Bigg)\times \sqrt{\frac{\pi}{\lambda_1}}\exp\Big(i\frac{\pi}{4}\Big)\\ &
        =\frac{\pi^3}{n\cdot v_1}\int_0^\infty d\hat\alpha \, d\hat\beta\,d\hat\gamma\,(\hat\beta+\hat\gamma)^{-3/2}\frac{1}{\lambda_1\sqrt{\lambda_1+2\lambda_2\gamma\beta}}\exp\Big(-i\frac{b^2}{4\lambda_1}-i\frac{\Upsilon(x)^2\gamma^2\beta^2}{4\lambda_1 +8\lambda_2\gamma\beta}\Big)\label{5.18k}
    \end{split}
\end{align}
The integral in \eqref{5.18k} does not admit any closed form. One has to perform the remaining integrals numerically and see whether there are any finite contribution.\par  However, we can extract some information by doing an asymptotic analysis,
\begin{align}
    \begin{split}
        \textrm{Reg.}|f_{\varphi}(x)|\le \frac{\pi^3}{n\cdot v_1}\textrm{Reg.}\int_0^\infty d\hat\alpha \, d\hat\beta\,d\hat\gamma\,(\hat\beta+\hat\gamma)^{-3/2}\frac{1}{\lambda_1\sqrt{\lambda_1+2\lambda_2\gamma\beta}}\label{5.19k}
    \end{split}
\end{align}
Now, expanding the integrand \eqref{5.19k} around two limit ($0,\infty$) we get,
\begin{align}
     \textrm{Reg.}|f_{\varphi}(x)|\le 0\implies  \textrm{Reg.}|f_{\varphi}(x)|\to 0
\end{align}
To make this statement more concrete, we make an analysis by using the method of region in Appendix~(\ref{ch2:app:D}). But a more precise numerical analysis is required to see whether there is any finite contribution from this integral, which we leave for future investigations.  In a similar fashion, one can, in principle, compute the waveform corresponding to $\lambda_3h\varphi^3$ vertex.

\textbullet $\,\,$ The simplest 2PM contribution comes from quadratic scalar field coupling in the worldline.
\begin{align}
    \begin{split}
        k^2\Big\langle\varphi(k)\Big\rangle\equiv \thesisinlinefeynman[\chthreeinlinefeynwidth]{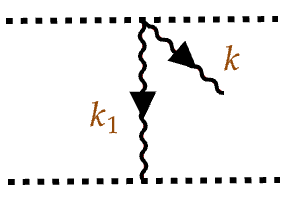}\hspace{0.35cm}=\Big(\frac{m_1g_1}{2m_p^2}\Big)\times \Big(\frac{m_2s_2}{2m_p}\Big)\rmint d\mu_{1,2}(k)\frac{1}{k_1^2}\label{5.7mm}
    \end{split}
\end{align}
where the integral measure has the following form,
\begin{align}
    \begin{split}
        d\mu_{1,2}(k)=\rmint_{k_1}e^{-ik_1\cdot b}e^{ik\cdot b_1}\hat\delta(k_1\cdot v_2)\hat\delta(k\cdot v_1-k_1\cdot v_1)\,.
    \end{split}
\end{align}
Therefore, the  time domain waveform has the following form,
\begin{align}
    \begin{split}
        f_{\varphi}(x)&=\Big(\frac{m_1g_1}{2m_p^2}\Big)\times \Big(\frac{m_2s_2}{2m_p}\Big)\frac{1}{n\cdot v_1}\rmint_{\Omega} e^{-ik\cdot (x-b_1)}\rmint_{k_1} e^{-i k_1\cdot b}\frac{\hat\delta(k_1\cdot v_2)\hat\delta\Big(\Omega-\frac{k_1\cdot v_1}{n\cdot v_1}\Big)}{k_1^2}\,,\\ &
         =  \Big(\frac{m_1g_1}{2m_p^2}\Big)\times \Big(\frac{m_2s_2}{2m_p}\Big)\frac{1}{n\cdot v_1}\rmint_{k_1}e^{-ik_1\cdot w_1}\frac{\hat\delta(k_1\cdot v_2)}{k_1^2},\,\textrm{with}, \, w_1\equiv \frac{n\cdot (x-b_1)}{n\cdot v_1}v_1+b,\nonumber
         \end{split}
\end{align}
\begin{align}
\begin{split}
  \hspace{0cm}  & 
      = -\Big(\frac{m_1g_1}{2m_p^2}\Big)\times \Big(\frac{m_2s_2}{2m_p}\Big)\frac{1}{n\cdot v_1}\frac{1}{4\pi|\vec w_1|}\,.
    \end{split}
\end{align}
Therefore, the final result of the integrals looks,

\begin{equation}\label{e:ba}\begin{split}
 f_{\varphi}(x)&=-\Big(\frac{m_1g_1}{2m_p^2}\Big)\times \Big(\frac{m_2s_2}{2m_p}\Big)\frac{1}{n\cdot v_1}\frac{1}{4\pi|\vec w_1|}\,.
\\
        \end{split}
\end{equation}\\
\textbullet $\,\,$ Another 2PM contribution comes from the following:
\begin{align}
    \begin{split}
        k^2\Big\langle\varphi(k)\Big\rangle\Big|_{k^2\rightarrow 0}\equiv \thesisinlinefeynman[\chthreeinlinefeynwidth]{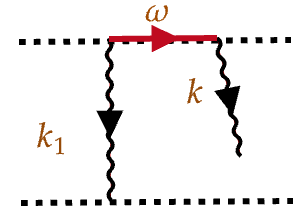}\hspace{0.35cm}=m_1\Big(\frac{s_1}{2m_p}\Big)^2 \Big(\frac{m_2s_2}{2m_p}\Big)\rmint d\mu_{1,2}(k)\frac{\{2\omega (v_1)_{\rho}-(k_1)_{\rho}\}\{-2\omega (v_1)_{\rho}+ k_{\rho}\}}{\omega^2 k_1^2}\label{5.10 m}
    \end{split}
\end{align}
where, the measure $d\mu_{1,2}(k)$ has the following form,
\begin{align}
    \begin{split}
        \rmint d\mu_{1,2}(k)=\rmint_{k_1,\omega}e^{i(k-k_1)\cdot b_1}e^{i k_1\cdot b_2}\hat\delta(k_1\cdot v_1-\omega)\hat\delta(k \cdot v_1-\omega)\hat\delta(k_1\cdot v_2)\,.
    \end{split}
\end{align}
Now its contribution to the time-domain waveform is,
\begin{align}
    \begin{split}
        f_{\varphi}(x)& = -m_1\Big(\frac{s_1}{2m_p}\Big)^2 \Big(\frac{m_2s_2}{2m_p}\Big) \rmint_{\Omega} e^{-ik\cdot x} \rmint d\mu_{(a)}(k) \frac{k\cdot k_1}{{\omega}^2\,k_1^2}\,,\\ &
        =-m_1\Big(\frac{s_1}{2m_p}\Big)^2 \Big(\frac{m_2s_2}{2m_p}\Big)\rmint_{\Omega} e^{-ik\cdot x}\rmint_{k_1} e^{-ik_1\cdot b}e^{i k\cdot b_1}\frac{k\cdot k_1\hat\delta(k_1\cdot v_2)}{k_1^2\,(k_1\cdot v_1+i\epsilon)^2}\hat\delta[(k-k_1)\cdot v_1]\,.
    \end{split}
\end{align}
Therefore, the whole integral can be written as,
\begin{align}
\begin{split}
      f_{\varphi}(x)&= -m_1\Big(\frac{s_1}{2m_p}\Big)^2 \Big(\frac{m_2s_2}{2m_p}\Big)\rmint_{\Omega}e^{-ik\cdot (x-b_1)}\rmint_{\omega,k_1}e^{-i k_1\cdot b}\hat{\delta}(k_1\cdot v_2)\hat{\delta}(\Omega\, n\cdot v_1-\omega)\hat\delta(k_1\cdot v_1-\omega)\frac{\Omega(n\cdot k_1)}{\omega^2 k_1^2}\,,\\ &
       = -m_1\Big(\frac{s_1}{2m_p}\Big)^2 \Big(\frac{m_2s_2}{2m_p}\Big)\frac{1}{(n\cdot v_1)^2}\rmint_{\Omega}e^{-ik\cdot (x-b_1)}\rmint_{k_1}e^{-ik_1\cdot b}\hat\delta(k_1\cdot v_2)\hat{\delta}(\Omega-\frac{k_1\cdot v_1}{n\cdot v_1})\frac{n\cdot k_1}{(k_1\cdot v_1+i\epsilon)k_1^2}\,,\\ &
     = -m_1\Big(\frac{s_1}{2m_p}\Big)^2 \Big(\frac{m_2s_2}{2m_p}\Big)\frac{n^{\mu}}{(n\cdot v_1)^2}\rmint_{k_1}e^{-i k_1\cdot w_1}\hat{\delta}(k_1\cdot v_2)\frac{k_{1\mu}}{(k_1\cdot v_1+i\epsilon)k_1^2},\,\, w_1\equiv\frac{n\cdot (x-b_1)}{n\cdot v_1}v_1+b \,.\label{7.6}
      \end{split}
\end{align}
The integral in (\ref{7.6}) can be easily done from the frame of the second particle as follows,
\begin{align}
    \begin{split}
        \mathcal{J}^{\mu}:=\rmint_{k_1}e^{-i k_1\cdot w_1}\frac{k_1^{\mu}\,\hat\delta(k_1\cdot v_2)}{(k_1\cdot v_1+i\epsilon)k_1^2}&=-i\rmint d\tau \,\theta(\tau)\,\rmint_{k_1}\hat{\delta}(k_1\cdot v_2) e^{-ik_1\cdot (w_1-\tau\, v_1)}\frac{k_1^{\mu}}{k_1^2}\,,\\ &
        =-i\rmint d\tau\,\theta(\tau)\rmint_{k_1}\hat\delta{(k_1^{0})} \exp[{-ik_1\cdot \underbrace{(w_1-\tau\, v_1)}_{\tilde{w}_1}}]\frac{k_1^{\mu}}{k_1^2}\,,\\ &
        \xrightarrow[]{k_1^0\rightarrow 0}-\rmint d\tau\, \theta(\tau)\rmint_{\vec k_1}\frac{k_1^i}{-\vec k_1^2}\,e^{i \vec{k}_1\cdot \vec {\tilde w}_1}\,,\\ &
        =-\rmint_{-\infty}^{\infty}d\tau \, \theta(\tau)\frac{(\vec w_1-\tau\,\vec v_1)^{i}}{|\vec w_1-\tau\,\vec v_1|^{3}}\,.
    \end{split}
\end{align}
Now from (\ref{7.6}), it is clear that we need to compute the following quantity,
\begin{align}
    \begin{split}
        n_{\mu}\mathcal{J}^{\mu}\rightarrow n_{i}\mathcal{J}^{i}=\rmint_{-\infty}^{\infty}d\tau\, \theta(\tau)\frac{\vec n\cdot (\vec w_1-\tau\, \vec v_1)}{|\vec w_1-\tau \vec v_1|^3}\,.
    \end{split}
\end{align}
To proceed further, we have to properly parameterize the impact parameter $b_1,\,b_2$. As we are in the frame of the second particle, it is convenient to choose $b_1=(0,0,b,0)$ and $b_2=0$.
\begin{align}
    \begin{split}
       \vec n\cdot (\vec w_1-\tau \vec v_1)&\rightarrow \Big[\frac{n\cdot (x-b_1)}{n\cdot v_1}-\tau\Big]\vec n\cdot \vec v_1+\vec n\cdot \vec b\\ &
       =(u_1-\tau)\gamma\tilde\beta+\chi,\,\,\text{with},\chi:=b\sin\theta\cos\varphi,\,\tilde\beta:=\beta \cos\theta
    \end{split}
\end{align}
and,
\begin{align}
    \begin{split}
        |\vec w_1-\tau \vec v_1|^3\rightarrow (\gamma^2-1)^{3/2}\Big[\tau^2-u_1^2-2\tau u_1+\frac{|b|^2}{\gamma^2-1}\Big]^{3/2}.
    \end{split}
\end{align}
Hence, 
\begin{align}
    \begin{split}
    n_{\mu}\mathcal{J}^{\mu}=\frac{\gamma  \tilde{\beta } \left(2 u_1^2-\frac{|b|^2}{\left(\gamma ^2-1\right)^2}\right)+\chi  \left(\sqrt{\frac{|b|^2}{\left(\gamma ^2-1\right)^2}-u_1^2}+u_1\right)}{(\gamma^2-1)^{3/2}\left(\frac{|b|^2}{\left(\gamma ^2-1\right)^2}-2 u_1^2\right) \sqrt{\frac{|b|^2}{\left(\gamma ^2-1\right)^2}-u_1^2}}
    \end{split}
\end{align}
where, we define, $u_i=\frac{n\cdot (x-b_1)}{n\cdot v_i}$.Therefore, restoring the factors of $\pi$ from delta functions and integration measures, the contribution to the waveform from the particular diagram in  (\ref{5.10 m}) has the following,
\begin{align}\begin{split}\label{5.24 mmm}
 f_{\varphi}(x)= -\frac{m_1}{(2\pi)^2}\Big(\frac{s_1}{2m_p}\Big)^2 \Big(\frac{m_2s_2}{2m_p}\Big)\frac{1}{\gamma^2(1-\tilde\beta)^2(\gamma^2-1)^{3/2}}\frac{\gamma  \tilde{\beta } \left(2 u_1^2-\frac{|b|^2}{\left(\gamma ^2-1\right)^2}\right)+\chi  \left(\sqrt{\frac{|b|^2}{\left(\gamma ^2-1\right)^2}-u_1^2}+u_1\right)}{\left(\frac{|b|^2}{\left(\gamma ^2-1\right)^2}-2 u_1^2\right) \sqrt{\frac{|b|^2}{\left(\gamma ^2-1\right)^2}-u_1^2}}\,.
\\
 \end{split}
\end{align}
\textbullet $\,\,$ Finally, another 3-point scalar-graviton interaction vertex involving a derivative contributes to 2PM radiation. It is of the following form: $h^{\mu\nu}\partial_{\mu}\varphi\partial_{\nu} \varphi$  As we are computing the one-point function of the scalar field, it would be useful to partially integrate over the interaction Lagrangian and separate one of the scalar fields as,
\begin{align}
    \begin{split}
        S_{\textrm{int.}}=\frac{1}{m_p}\rmint d^4x \,h^{\mu\nu} \partial_{\mu}\varphi\partial_{\nu}\varphi=-\frac{1}{m_p}\rmint d^4 x \Big[\underbrace{h^{\mu\nu}\partial_{\mu}\partial_{\nu}\varphi}_{\textrm{Term I}}+\underbrace{\partial_{\mu}h^{\mu\nu}\partial_{\nu}\varphi}_{\textrm{Terrm II}}\Big]\,\varphi.\label{5.22 m}
    \end{split}
\end{align}
Now, the contribution to the scalar one-point function from the Term I is given by,
\begin{align}
    \begin{split}
        k^2\Big\langle \varphi(k)\Big\rangle\equiv \thesisinlinefeynman[\chthreeinlinefeynwidth]{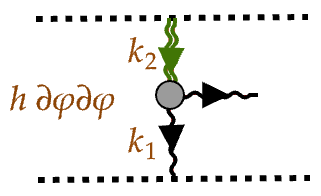}\hspace{0.35cm}=\Big(\frac{m_1m_2s_2}{2m_p^3}\Big)\rmint d\mu_{1,2}(k)\frac{v_1^\alpha v_1^\beta\,P_{\mu\nu;\alpha\beta}k_2^\mu k_2^\nu}{k_1^2k_2^2}\label{ch2:5.23a}
    \end{split}
\end{align}
where the integral measure has the following form,
\begin{align}
    \begin{split}
     \rmint   d\mu_{1,2}(k)=\rmint_{k_1,k_2}e^{ik_1\cdot b_1}e^{ik_2 \cdot b_2}\hat\delta(k_1\cdot v_1)\hat\delta(k_2\cdot v_2)\hat\delta^{(4)}(k_1+k_2-k)\,.
    \end{split}
\end{align}
Hence, the time domain waveform has the following form,
\begin{align}
    \begin{split}
        f_{\varphi}(x)\Big|_1=\Big(\frac{m_1m_2s_2}{2m_p^3}\Big)\rmint_{\Omega}e^{-ik\cdot x}\rmint_{k_1,k_2}e^{ik_1\cdot b_1}e^{ik_2\cdot b_2}\hat\delta(k_1\cdot v_1)v_1^{\alpha}v_1^{\beta}\hat\delta(k_2\cdot v_2)\frac{k_2^{\mu}k_2^{\nu}P_{\mu\nu,\alpha\beta}}{k_1^2k_2^2}\delta^{(4)}(k_1+k_2-k)\,.\label{5.25mmm}
    \end{split}
\end{align}
The integral in \eqref{5.25mmm} can be done using the partial fraction approach, where we can separate the denominator with undefined momentum as,
\begin{align}
    \begin{split}
        \frac{1}{k_1^2 k_2^2}=-\frac{1}{2}\frac{1}{k_1^2(k_1\cdot k)}-\frac{1}{2}\frac{1}{k_2^2(k_2\cdot k)}\,.
    \end{split}
\end{align}
Therefore, the waveform can be reduced into two independent momentum integrals,
\begin{align}
    \begin{split} \label{e:::}
        f_{\varphi}(x)\Big|_{1}=\Big(\frac{m_1m_2s_2}{4 m_p^3}\Big)\Big(f_{\varphi}(x)\Big|_{1}^{(1)}+f_{\varphi}(x)\Big|_{1}^{(2)}\Big)\,.
    \end{split}
\end{align}
where,
\begin{align}
    \begin{split} \label{5.38nm}
        f_{\varphi}(x)\Big|_{1}^{(1)}&=v_1^{\alpha}v_1^{\beta}P_{\mu\nu;\alpha\beta}\rmint_{\Omega}e^{-ik\cdot x}\rmint_{k_1,k_2}e^{ik_1\cdot b_1}e^{ik_2\cdot b_2}\frac{k_2^\mu k_2^\nu\hat\delta(k_1\cdot v_1)\hat\delta(k_2\cdot v_2)}{k_1^2(k_1\cdot k)}\delta^{(4)}(k_1+k_2-k)\,,\\ &
=v_1^{\alpha}v_1^{\beta}P_{\mu\nu;\alpha\beta}\rmint_{\Omega}e^{-ik\cdot (x-b_2)}\rmint_{k_1}e^{ik_1\cdot b}\frac{\hat\delta(k_1\cdot v_1)\hat\delta(\Omega-k_1\cdot v_2)}{k_1^2(k_1\cdot n)\Omega}(k-k_1)^{\mu}(k-k_1)^{\nu}\,,\\ &
=v_1^{\alpha}v_1^{\beta}P_{\mu\nu;\alpha\beta}\rmint_{k_1}e^{-i k_1\cdot \tilde w_1}\frac{\hat\delta(k_1\cdot v_1)}{k_1^2 (k_1\cdot n)(k_1\cdot v_2)}\Big[n^\mu n^\nu (k_1\cdot v_2)^2-2 (k_1\cdot v_2)k_1^{(\mu}n^{\nu)}+k_1^{\mu}k_1^{\nu}\Big]\,.
    \end{split}
\end{align}
The integral of concern is the following,
\begin{align}
    \begin{split}
        I^{\mu\nu}&=\rmint_{k_1}\frac{e^{-ik_1\cdot \tilde w_1}\hat\delta(k_1\cdot v_1)}{k_1^2(k_1\cdot n)(k_1\cdot v_2)}k_1^{\mu}k_1^{\nu},\, \tilde w_1=\frac{n\cdot (x-b_2)}{n\cdot v_2}v_2-b\,.\label{5.29mmm}
    \end{split}
    \end{align}
{\color{black}One can see that because of the delta function constraint, one can replace  $\tilde w_1$ with $\hat w_2:=-b+u_2 (v_2-\gamma v_1)$. Also, $I^{\mu\nu}$ is orthogonal to $\hat w_2$ and $v_1$. Therefore, it can be written in terms of a basis plane orthogonal to  $\hat w_2$ and $v_1$.} $I^{\mu\nu}$ is orthogonal to $\hat w_2$ and $v_1$.  Therefore,
    \begin{align}
        \begin{split}
I^{\mu\nu}&=\Pi^{\mu}_{\alpha}\,\Pi^{\nu}_{\beta}\Big[c_{vv} v_2^{\alpha}v_2^{\beta}+c_{nn} n^{\alpha}n^{\beta}+c_{nv} v_2^{(\alpha}n^{\beta)}\Big]\,.
        \end{split}
    \end{align}
One can notice that $I^{\mu\nu}$ is symmetric under the exchange of $v_2 \leftrightarrow n$ implies $c_{nn}=c_{vv}$. Now, the integral can be done using the integral reduction technique by Passarino and Veltman, redefining the basis and writing the following ansatz, 
\begin{align}
        \begin{split}
\label{Ansatz2}I^{\mu\nu}&=c_{a}\Pi_{1}^{\mu\nu}+ 2c_{b}\Big(\Pi_{1}.v_{2}\Big)^{(\mu}\Big(\Pi_{1}.n\Big)^{\nu)}
\end{split}
    \end{align}
where $c_{a}$ and $c_{b}$ are two new constants and $\Pi_{1}^{\mu\nu}= |w_{1}|^{2}\,P_{1}^{\mu\nu}\,+ w_{1}^{\mu}\,w_{1}^{\nu}$ is the projection operator orthogonal to $w_{1}^{\mu}$ and $v_{1}^{\mu}$. One can evaluate the constants from the following two equations:
\begin{align}
\label{NV2}  &  n\cdot I\cdot v_2=\rmint_{k_1}e^{-ik_1\cdot \hat w_1}\frac{\hat\delta(k_1\cdot v_1)}{k_1^2}= c_{a}\Big(n\cdot \Pi\cdot v_{2}\Big)+ c_{b}\,\Big[\Big(n\cdot \Pi\cdot v_{2}\Big)\,\Big(v_{2}\cdot \Pi\cdot n\Big)+ \Big(v_{2}\cdot \Pi\cdot v_{2}\Big)\,\Big(n\cdot \Pi\cdot n\Big)\Big]\,,\\ &
\label{NIN}  n\cdot I \cdot n= n_\mu \rmint_{k_1}e^{-ik_1\cdot \hat w_2}\frac{k_1^\mu\hat\delta(k_1\cdot v_1)}{k_1^2 (k_1\cdot v_2)}= c_{a}\Big(n\cdot \Pi\cdot n\Big)+ 2c_{b}\Big(n\cdot \Pi\cdot v_{2}\Big)\,\Big(n\cdot \Pi\cdot n\Big)\,.
\end{align}
The first equality of (\ref{NIN}) gives,
\begin{align}
\begin{split}
   n_\mu \rmint_{k_1}e^{-ik_1\cdot \hat w_2}\frac{k_1^\mu\hat\delta(k_1\cdot v_1)}{k_1^2 (k_1\cdot v_2)}&\to -i n_\mu\, \rmint d\tau\,\theta(\tau)\rmint_{ k_1} e^{-i k_1\cdot (\hat{w}_2-\tau v_2)}\frac{{k}_1^\mu}{k_1^2}\hat\delta(k_1\cdot v_1)\,,\\ &
   =-i\, \grave{n}_i\rmint d\tau\,\theta(\tau)
   \rmint_{\bar k_1}e^{i \bar k_1\cdot \grave{\underbar{$w$}}_2}\frac{\bar k_1^i}{-\bar k_1^2}\,,\\ &
   =-\grave n_i\rmint d\tau \,\theta(\tau)\frac{\grave{\underbar{$w$}}(\tau)^i}{|\grave{\underbar{$w$}}(\tau)|^3},\,\grave{\underbar{$w$}}\equiv \hat{w}_2-\tau v_2.
   \end{split}
\end{align}
where,
\begin{align}
    \begin{split}
        \grave n_i\equiv \Big\{\gamma(n^{(1)}-\beta n^{(0)}),n^{(2)},n^{(3)}\Big\}\,.
    \end{split}
\end{align}
Comparing with the second equality of (\ref{NIN}), we get,

\begin{align}\begin{split}\label{m:}
  &c_a\to -\frac{-(n\cdot \Pi\cdot v_2) (n\cdot I\cdot n) (v_2\cdot \Pi\cdot n)+2(n\cdot \Pi\cdot v_2)(n\cdot \Pi\cdot n) (n\cdot I\cdot v_2)-(v_2\cdot \Pi\cdot v_2) (n\cdot \Pi\cdot n)(n\cdot I \cdot n)}{-2 (n\cdot \Pi\cdot v_2)^2+(n\cdot \Pi\cdot v_2) (v_2\cdot \Pi\cdot n)+(v_2\cdot \Pi\cdot v_2) (n\cdot \Pi\cdot n)},\\&
 c_b\to \frac{(n\cdot \Pi\cdot n) (n\cdot I\cdot v_2)-(n\cdot \Pi\cdot v_2) (n\cdot I\cdot n)}{(n \cdot \Pi\cdot n) \left(-2 (n\cdot \Pi\cdot v_2)^2+(n\cdot \Pi\cdot v_2) (v_2\cdot \Pi\cdot n)+(v_2\cdot \Pi\cdot v_2) (n\cdot \Pi\cdot n)\right)}.\\
\end{split}
\end{align}
Next, we compute the integral of the R.H.S of the first equality of (\ref{NV2}) to get,
\begin{eqnarray}
\label{CAINT} \rmint_{k_1}e^{-ik_1\cdot \hat w_1}\frac{\hat\delta(k_1\cdot v_1)}{k_1^2} =-\frac{1}{4\pi}\frac{1}{|\grave{w}_2|}
\end{eqnarray}
where $\grave{w}_2$ is a three vector defined as,
\begin{align}
    \grave{w}_2=\Big\{\gamma(\hat{w}_2^{(1)}-\beta \hat{w}_2^{(0)}),\hat{w}_2^{(2)},\hat{w}_2^{(3)}\Big\}\,.
\end{align}
Thus, the integral $I^{\mu\nu}$ can be finally written in a concise form from the ansatz (\ref{Ansatz2}) as,
\begin{align}
\begin{split} 
\hspace{0cm}\label{IMUNU}I^{\mu\nu} =-\Big[\frac{-(n\cdot \Pi\cdot v_2) (n\cdot I\cdot n) (v_2\cdot \Pi\cdot n)+2(n\cdot \Pi\cdot v_2)(n\cdot \Pi\cdot n) (n\cdot I\cdot v_2)-(v_2\cdot \Pi\cdot v_2) (n\cdot \Pi\cdot n)(n\cdot I \cdot n)}{-2 (n\cdot \Pi\cdot v_2)^2+(n\cdot \Pi\cdot v_2) (v_2\cdot \Pi\cdot n)+(v_2\cdot \Pi\cdot v_2) (n\cdot \Pi\cdot n)}\Big]\,\Pi^{\mu\nu} \\
+ 2\,\Big[\frac{(n\cdot \Pi\cdot n) (n\cdot I\cdot v_2)-(n\cdot \Pi\cdot v_2) (n\cdot I\cdot n)}{(n \cdot \Pi\cdot n) \left(-2 (n\cdot \Pi\cdot v_2)^2+(n\cdot \Pi\cdot v_2) (v_2\cdot \Pi\cdot n)+(v_2\cdot \Pi\cdot v_2) (n\cdot \Pi\cdot n)\right)}\Big]\Big(\Pi \cdot v_{2}\Big)^{(\mu}\Big(\Pi\cdot n\Big)^{\nu)}\,.
\end{split}
\end{align}
Once we have the closed-form expression for $I^{\mu\nu}\,,$ one can write down the exact expression for the first part of the waveform after restoring the factor of $\pi$.
\begin{align}\begin{split}\label{e:}
f_{\varphi}(x)\Big|_{1}^{(1)}=-\,\frac{v_1^\alpha v_1^\beta}{4\pi^2}P_{\mu\nu;\alpha\beta}\Big[n^{\mu}n^\nu I^{\rho\sigma}v_{2\rho}v_{2\sigma}-2v_{2\rho}I^{\rho(\mu}n^{\nu)}+I^{\mu\nu}\Big].\\
\end{split}
\end{align}
Now we focus on the other part.
\begin{align}
    \begin{split}
        f_{\varphi}(x)\Big|_{1}^{(2)}&=-\,v_1^\alpha v_1^\beta P_{\mu\nu;\alpha\beta}\rmint_{\Omega}e^{-ik\cdot x}\rmint_{k_1,k_2}e^{ik_1\cdot b_1}e^{ik_2\cdot b_2}\frac{k_2^\mu k_2^\nu \hat\delta(k_1\cdot v_1)\hat\delta(k_2\cdot v_2)}{k_2^2 (k_2\cdot k)}\hat\delta^{(4)}(k_1+k_2-k)\,,\\ &
        =-\,v_1^\alpha v_1^\beta P_{\mu\nu;\alpha\beta}\rmint_{\Omega}e^{-i \Omega \,n \cdot x}\rmint_{k_2}e^{i (k-k_2)\cdot b_1} e^{ik_2\cdot b_2} \frac{k_2^\mu k_2^\nu \,\hat\delta(k_2\cdot v_2)\hat\delta(\Omega\,n\cdot v_1-k_2\cdot v_1)}{k_2^2 (k_2\cdot n)\Omega}\,,\\ &
        =-\,v_1^\alpha v_1^\beta P_{\mu\nu;\alpha\beta} \underbrace{\rmint_{k_2} e^{-i k_2 \cdot w_1}k_2^\mu k_2^\nu\, \frac{\hat\delta(k_2\cdot v_2)}{k_2^2(k_2\cdot n)(k_2\cdot v_1)}}_{\bar{I}^{\mu\nu}} \textrm{with},\,w_1\equiv \frac{n\cdot (x-b_1)}{n\cdot v_1}v_1+b\,.\label{5.43mm}
    \end{split}
\end{align}
The integral in \eqref{5.43mm} can be similar to \eqref{5.29mmm} and is given by,
\begin{align}
    \begin{split} \label{e::}
f_{\varphi}(x)\Big|_{1}^{(2)}=-\,v_1^{\alpha}v_1^\beta P_{\mu\nu;\alpha\beta} \bar I^{\mu\nu}.\\
    \end{split}
\end{align}
where $\bar I^{\mu\nu}$ is defined (\ref{5.43mm}) and can be calculated as \eqref{5.29mmm}. So adding (\ref{e:}) and  (\ref{e::}) we get the full expression for $f_{\phi}(x)\Big|_1$ mentioned in (\ref{e:::}).

Now, we deal with the Term II of the interaction Lagrangian of \eqref{5.22 m}. The corresponding contribution to the waveform is given by,
\begin{align}
    \begin{split} \label{ex6}
        f_{\varphi}(x)\Big|_{2}=\Big(\frac{m_1 m_2 s_2}{2m_p^3}\Big)v_1^\alpha v_1^\beta P_{\mu\nu;\alpha\beta}\rmint_{\Omega} e^{-ik\cdot x}\rmint_{k_1,k_2}e^{ik_1\cdot b_1}e^{ik_2\cdot b_2}\frac{k_1^\mu k_2^\nu\hat\delta(k_1\cdot v_1)\hat\delta(k_2\cdot v_2)}{k_1^2 k_2^2}\hat\delta^{(4)}(k_1+k_2-k)\,.
    \end{split}
\end{align}
Now, doing the integration over $k_1$ using the delta function, we will get,
\begin{align}
    \begin{split}
         f_{\varphi}(x)\Big|_{2}=\Big(\frac{m_1 m_2 s_2}{2m_p^3}\Big)v_1^\alpha v_1^\beta P_{\mu\nu;\alpha\beta}\rmint_{\Omega}e^{-ik\cdot (x-b_1)}\rmint_{k_2}e^{-i k_2 \cdot b}\frac{(k-k_2)^\mu \,k_2^\nu}{k_2^2\,(k-k_2)^2}\hat\delta(k\cdot v_1-k_2\cdot v_1)\hat\delta(k_2\cdot v_2)\,.\label{5.46 m}
    \end{split}
\end{align}
Again, this can be split into two parts and can be dealt with separately. 
\begin{align}
    \begin{split} \label{ex3}
    f_{\varphi}(x)\Big|_{2}=\Big(\frac{m_1 m_2 s_2}{2m_p^3}\Big)\Big(f_{\varphi}(x)\Big|^{(1)}_{2}+f_{\varphi}(x)\Big|^{(2)}_{2}\Big)\,.
     \end{split}
\end{align}
The first part gives,
\begin{align}
    \begin{split}
        f_{\varphi}(x)\Big|_{2}^{(1)}=\underbrace{(v_1\cdot P\cdot v_1)_{\mu\nu}\,n^{\mu}}_{T_{\nu}}\rmint_{\Omega}e^{-ik\cdot (x-b_1)}\,\Omega\underbrace{\rmint_{k_2}\frac{k_2^\nu}{k_2^2(k-k_2)^2}\hat\delta(k\cdot v_1-k_2\cdot v_1)\hat\delta(k_2\cdot v_2)}_{J_{(1)}^\nu}\,.\label{5.57 m}
    \end{split}
\end{align}
The integral in \eqref{5.57 m} can be done using the results in \eqref{A.9 m}. We know that $J_{(1)}^{\mu}$ has two parts. We first concentrate on the first part, i.e. $\mathcal{J}^{\nu}$ in \eqref{A.9 m}. Therefore, the first term of $f_{\varphi}(x)\Big|^{(1)}_{2}$ takes the following form:
\begin{align}
    \begin{split} \label{ex}
        \widetilde {f_{\varphi}(x)}\Big|_{2}^{(1)}&=T_{\nu}\rmint_{\Omega}e^{-ik\cdot (x-b_1)}\Omega\,\mathcal{J}^{\nu}(\Omega)\,,\\ &
        =i\,\frac{\partial}{\partial l}\,\Big(f_1+f_2+f_3\Big)
    \end{split}
\end{align}
where, $f_i$'s are defined in \eqref{A.19 m}. Apart from the $\mathcal{J}^{\nu}$ part we have another part in form of $J_{(1)}^\nu$, as shown in (\ref{A.9 m}), which gives,
\begin{align}
    \begin{split} \label{ex1}
      \widetilde{ \widetilde{f_{\varphi}(x)}}\Big|_{2}^{(1)}&= \frac{|b| \,n^\nu}{\gamma}T_{\nu}\rmint_{\Omega}e^{-ik\cdot (x-b_1)}\,\Omega^2 \rmint_{0}^1dy\,y\,\,e^{-iyk\cdot b}\frac{K_1(|b|\Delta(k,y))}{4\pi\Delta(k,y)}\,, \\ &
       =\frac{|b|^2 \,T\cdot n}{4\gamma}\rmint_0^\infty dy\,y\rmint d\Omega\, e^{-i\Omega l}\,\Omega^2\, \rmint_{0}
^\infty \frac{dt}{t^2}\,\exp\Big(-t-\frac{\Omega^2|b|^2\bar\Delta(y)^2}{4t}\Big)\,,\\ &
=\frac{\sqrt{\pi} \,T\cdot n}{|b|^3\gamma}\rmint_{0}^1dy\,\frac{y}{\bar\Delta(y)^5}\rmint_0^\infty dt\,t^{-1/2}(|b|^2\bar\Delta^2-2l^2t)\exp\Big(-\frac{l^2 t}{|b|^2\bar\Delta^2}-t\Big)\,,\\ &
=\frac{{\pi} \,T\cdot n}{|b|^3\gamma}\rmint_0^1 dy\,\frac{y}{\Bar \Delta}\times \Bigg[\frac{b^2\Bar\Delta^2}{\sqrt{\frac{l^2}{|b|^2 \Bar\Delta ^2}+1}}-l^2\frac{1}{ \left(\frac{l^2}{|b|^2 \Bar\Delta ^2}+1\right)^{3/2}}\Bigg],\,l:=n\cdot (x-b_1-yb)\,.
\end{split}
\end{align}
Then adding (\ref{ex}) and (\ref{ex1}) we get the full expression for $f_{\phi}(x)\Big|_{2}^{(1)}\,.$
The second part in \eqref{ex3} is given by,
\begin{align}
    \begin{split}
    f_{\varphi}(x)\Big|_{2}^{(2)}=(v_1\cdot P\cdot v_1)_{\mu\nu}\rmint_{\Omega}e^{-ik\cdot (x-b_1)}\rmint_{k_2}e^{-ik_2\cdot b}\frac{k_2^\mu\,k_2^\nu}{k_2^2(k-k_2^2)}\hat\delta(k\cdot v_1-k_2\cdot v_1)\hat\delta(k_2\cdot v_2)\,.\label{5.50 m}
    \end{split}
\end{align}
Now, we will rewrite \eqref{5.50 m} to match with \eqref{5.25mmm}. To do this, we introduce an auxiliary variable $k_1$ rewrite (\ref{5.50 m}) using a four-dimensional delta function in the following way.

\begin{align}
\begin{split}f_{\varphi}(x)\Big|_{2}^{(2)}=(v_1\cdot P\cdot v_1)_{\mu\nu}\rmint_{\Omega}e^{-ik\cdot x}\rmint_{k_2}e^{ik_1\cdot b_1}e^{ik_2\cdot b_2}\frac{k_2^\mu\,k_2^\nu}{k_2^2 k_1^2}\hat\delta(k_1\cdot v_1)\hat\delta(k_2\cdot v_2)\hat\delta^{(4)}(k_1+k_2-k)\,.\end{split}
\end{align}
Then we can proceed just like the case of (\ref{5.38nm}). Finally we get,

\begin{equation}\label{ex4}\begin{split}
 f_{\varphi}(x)\Big|_{2}^{(2)}&
      =(v_1\cdot P\cdot v_1)_{\mu\nu}\Big[n^{\mu}n^\nu I^{\rho\sigma}v_{2\rho}v_{2\sigma}-2v_{2\rho}I^{\rho(\mu}n^{\nu)}+I^{\mu\nu}+\bar I^{\mu\nu}\Big].
\\
        \end{split}
\end{equation}\\
Here $I^{\mu\nu}$ and $\bar{I}^{\mu\nu}$ are defined in (\ref{IMUNU}) and (\ref{5.43mm}) respectively. Then  adding (\ref{ex4}) with (\ref{ex}) and (\ref{ex1}) we get the entire expression for $f_{\phi}(x)\Big|_2$ defined in (\ref{ex3}). Finally, adding (\ref{ex3}) with (\ref{5.25mmm}) we get the entire contribution to the waveform coming from this 3-point vertex at 2PM order.

\textbullet $\,\,$ Apart from the above-mentioned diagrams, there will be three more diagrams coming from the $\lambda_3\varphi^3$ vertex, which will contribute to the scalar wave form.
\begin{align}
    \begin{split}
        f(x) &\equiv\thesisinlinefeynman[\chthreeinlinefeynwidth]{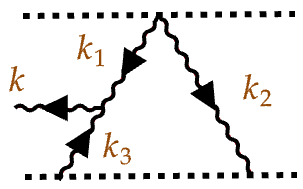}\\ &=\lambda_3\frac{m_1 g_1 m_2^2 s_2^2}{8m_p^3}\rmint e^{-ik\cdot x}\rmint_{k_i}e^{i(k_1+k_2)\cdot b_1}e^{i(k_3-k_2)\cdot b_2}\frac{\hat\delta(k_1\cdot v_1+k_2\cdot v_1)\hat\delta(k_2\cdot v_2)\hat\delta(k_3\cdot v_2)}{k_1^2 k_2^2 k_3^2}\hat\delta^{(4)}(k-k_1-k_3)\,,\\ &
        =\lambda_3\frac{m_1 g_1 m_2^2 s_2^2}{8m_p^3}\rmint_{k}e^{-ik\cdot (x-b_1)}\rmint_{k_2} e^{ik_2\cdot b}\frac{\hat\delta(k_2\cdot v_2)}{k_2^2}\rmint_{k_3}\hat\delta(k_3\cdot v_2)\hat\delta(k_3\cdot v_1-k\cdot v_1-k_2\cdot v_1)\frac{e^{-ik_3\cdot b}}{k_3^2(k_3-k)^2}.\label{5.62x}
    \end{split}
\end{align}
The $k_3$ integral can be further simplified using Feynman parametrization.
\begin{align}
    \begin{split}
        I_{k_3}&=\rmint_{k_3}\hat\delta(k_3\cdot v_2)\hat\delta(k_3\cdot v_1-k\cdot v_1-k_2\cdot v_1)\frac{e^{-ik_3\cdot b}}{k_3^2(k_3-k)^2},\quad \textrm{with}\quad k^2=0\,,\\ &
        =\rmint_0^1 dy\, e^{-iy k\cdot b}\rmint_{0}^\infty dt\,t\,\rmint_{\tilde k_3}\exp\Big[i \tilde k_3\cdot b-t\, \tilde k_3^2-t\Sigma(y,k,k_2\cdot v_1)^2\Big]\,,\\ &
    =\frac{|b|}{\gamma}\rmint_0^1dy\,e^{-iyk\cdot b}\frac{K_1\Big[|b|\Sigma(k,y,k_2\cdot v_1)\Big]}{4\pi \Sigma(k,y,k_2\cdot v_1)}\,.
    \end{split}
\end{align}
 Note that in our velocity parametrization, $\frac{k_2\cdot v_1}{\gamma\beta}=-k_2^{(1)}$, which implies that the function $\Sigma$ depends only on one component of $k_2$. Hence, we can perform the integral over the other two components of $k_2$, which reads,
 \begin{align}
     \begin{split}
         f(x)&\sim\frac{|b|}{4\pi\gamma}\rmint_0^1 dy \rmint_{k}e^{-ik\cdot (x-b_1+yb)}\rmint dz K_0(|z||b|)\frac{K_1\Big[|b|\Sigma(k,y,k_2\cdot v_1)\Big]}{\Sigma(k,y,k_2\cdot v_1)}\\ &
         =\frac{|b|}{4\pi\gamma}\rmint_{0}^1 dy\rmint_{-\infty}^{\infty} d \Omega\rmint_{-\infty}^\infty dz \,e^{-i \Omega \lambda(x,y,b)}K_{0}(|z||b|)\frac{K_1\Big[|b|\Sigma(\Omega,y,z)\Big]}{\Sigma(\Omega,y,z)},\,\lambda\equiv n\cdot(x-b_1+yb)\label{6.30d}
     \end{split}
 \end{align}
where,
\begin{align}
    \begin{split}
        \Sigma=\sqrt{\frac{\Omega^2\Big[y^2(n\cdot v_2)^2+(1-y)^2(n\cdot v_1)^2+2y(1-y)\gamma (n\cdot v_2)(n\cdot v_1)\Big]}{\gamma^2-1}+z^2-2z\,\Omega\,\Big(\frac{y}{\beta}n\cdot v_2+\frac{1-y}{\gamma\beta}n\cdot v_1\Big)}\,.
    \end{split}
\end{align}
The integral in \eqref{6.30d} can be further simplified by using the integral representation of Bessel functions,
\begin{align}
    \begin{split}
        f(x)&\sim  \frac{|b|}{4\pi\gamma}\rmint_{0}^1 dy\rmint_{-\infty}^{\infty} d \Omega\rmint_{-\infty}^\infty dz \,e^{-i \Omega \lambda(x,y,b)}K_{0}(|z||b|)\frac{K_1\Big[|b|\Sigma(\Omega,y,z)\Big]}{\Sigma(\Omega,y,z)}\,,\\ &
      =  \frac{|b|^2}{16\pi\gamma}\rmint_{0}^1 dy\rmint_{-\infty}^{\infty} d \Omega\rmint_{-\infty}^\infty dz \,e^{-i \Omega \lambda(x,y,b)}\rmint dt_1\, \frac{1}{2t_1}\exp\Big(-t_1-\frac{z^2 |b|^2}{4t_1}\Big)\rmint \frac{dt_2}{t_2^2}\exp\Big(-t_2-\frac{|b|^2\Sigma^2}{4t_2}\Big)\,,\\ &
      =\frac{|b|^2}{32\pi \gamma}\rmint_0^1 dy \rmint_{-\infty}^{\infty} d\Omega e^{-i \Omega\lambda(x,y,b)} \rmint_0^\infty \frac{dt_1 dt_2}{t_1 t_2^2}e^{-(t_1+t_2)}\rmint_{-\infty}^{\infty}dz \exp\Big(-\frac{z^2|b|^2}{4t_1}-\frac{(z^2-2z\theta_1+\theta_2)|b|^2}{4t_2}\Big)\,.
    \end{split}
\end{align}
After doing the $z$ integral we left with,
\begin{align}
    \begin{split}
        f(x)&\sim \frac{|b|^2}{32\pi \gamma}\int_0^1 dy \rmint_{-\infty}^{\infty} d\Omega e^{-i \Omega\lambda(x,y,b)} \rmint_0^\infty \frac{dt_1 dt_2}{t_1 t_2^2}e^{-(t_1+t_2)}\frac{2 \sqrt{\pi } \exp\Big({\frac{|b|^2 \theta _1^2 t_1}{4 t_2^2+4 t_1 t_2}}-\frac{\theta_2|b|^2}{4t_2}\Big)}{b \sqrt{\frac{1}{t_2}+\frac{1}{t_1}}}\,,\\ &
        =\frac{|b|\textcolor{black}{\sqrt{\pi}}}{32\pi \gamma}\rmint_0^1 dy\rmint_0^\infty \frac{dt_1 dt_2}{t_1t_2^2}\frac{e^{-(t_1+t_2)}}{\sqrt{\frac{1}{t_1}+\frac{1}{t_2}}}\rmint_{-\infty}^{\infty} d\Omega\,e^{-i\Omega \lambda}\exp\Big[-\Omega^2\Big(\frac{\hat{\theta}_2|b|^2}{4t_2}-\frac{|b|^2\hat\theta_1^2t_1}{4t_2^2+4t_1t_2}\Big)\Big]\,.\\ &
       \label{6.33d}
    \end{split}
\end{align}
In \eqref{6.33d} the $\Omega$ integral is convergent if,
$$\hat{\theta}_2>\hat{\theta}_1^2\frac{t_1}{t_1+t_2}\implies \underbrace{(\hat\theta_2-\hat\theta_1^2)}_{-y^2}t_1+\hat\theta_2 t_2>0\,,$$
which sets a bound on the $t_1,\, t_2$ integral. The region of integral lies in $t_1< \frac{\hat\theta_2}{y^2}\,t_2$. Now within this region one can do the $\Omega$ integral which reads,
\begin{align}
\begin{split}
      f(x)&   =\frac{|b|}{16 \gamma}\rmint_0^1 dy\rmint_0^\infty \frac{dt_2}{t_2^2}\rmint_0^{\frac{\hat\theta_2}{y^2}t_2}\frac{ dt_1}{t_1}\frac{e^{-(t_1+t_2)}}{\sqrt{\frac{1}{t_1}+\frac{1}{t_2}}}\frac{1}{\sqrt{\frac{\hat{\theta}_2|b|^2}{4t_2}-\frac{|b|^2\hat\theta_1^2t_1}{4t_2^2+4t_1t_2}}}\exp\Big[-\frac{\lambda^2}{4}\frac{1}{\frac{\hat{\theta}_2|b|^2}{4t_2}-\frac{|b|^2\hat\theta_1^2t_1}{4t_2^2+4t_1t_2}}\Big]\\ &\hspace{0.8cm}+ \textrm{pure divergence coming from $\Omega$ integral.}\\ &
        \xrightarrow[]{\textrm{finite part}}\frac{1}{8 \gamma}\rmint_0^1 dy\rmint_0^\infty dt_2 \rmint_{0}^{\frac{\hat\theta_2}{y^2}t_2}dt_1\frac{e^{(-t_1-t_2)}}{t_2\sqrt{t_1}\sqrt{(\hat\theta_2-\hat\theta_1^2)t_1+\hat\theta_2 t_2}}\exp\Big[-\frac{\lambda^2}{|b|^2}\frac{t_2(t_1+t_2)}{(\hat\theta_2-\hat\theta_1^2)t_1+\hat\theta_2 t_2}\Big]
    \end{split}
\end{align}
where,
\begin{align}
    \begin{split}
        &\theta_1(y)=\Omega\Big(\frac{y}{\beta}n\cdot v_2+\frac{1-y}{\gamma\beta}n\cdot v_1\Big):=\Omega\, \hat\theta_1\,,\\ &
        \theta_2(y)=\frac{\Omega^2}{{\gamma^2-1}}\Big({y^2(n\cdot v_2)^2+(1-y)^2(n\cdot v_1)^2+2y (1-y)\gamma(n\cdot v_2)(n\cdot v_1)}\Big):=\Omega^2\,\hat\theta_2\,.
    \end{split}
\end{align}
Restoring the prefactors the waveform has the following form,

\begin{align}
\begin{split}
    f(x)&=\frac{\lambda_3}{(2\pi)^5}\frac{\sqrt{\pi}m_1 g_1 m_2^2 s_2^2}{64\gamma m_p^3}\rmint_0^1 dy\rmint_0^\infty dt_2 \rmint_{0}^{\frac{\hat\theta_2}{y^2}t_2}dt_1\frac{e^{-(t_1+t_2)}}{t_2\sqrt{t_1}\sqrt{(\hat\theta_2-\hat\theta_1^2)t_1+\hat\theta_2 t_2}}\\ &\hspace{0.8cm}\times\exp\Big[-\frac{\lambda^2}{|b|^2}\frac{t_2(t_1+t_2)}{(\hat\theta_2-\hat\theta_1^2)t_1+\hat\theta_2 t_2}\Big]\,.\label{5.70l}
    \end{split}
\end{align}
The integral in \eqref{5.70l}, to the best of our knowledge, does not have a closed-form, and one can, in principle, do the integral numerically while doing further investigations.

\textbullet $\,\,$ Another waveform diagram has the following form:
\begin{align}
    \begin{split}
        f(x)&\sim\thesisinlinefeynman[\chthreeinlinefeynwidth]{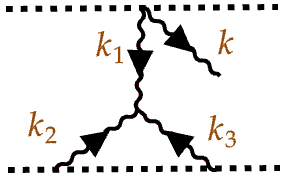}\\ &=\lambda_3\frac{m_1 g_1 m_2^2 s_2^2}{8m_p^3} \rmint_{k}e^{-ik\cdot x}\rmint_{k_i}\hat\delta^{(4)}\Big(\sum_ik_i\Big)\frac{\hat\delta(k_1\cdot v_1+k\cdot v_1)\hat\delta(k_2\cdot v_2)\hat\delta(k_3\cdot v_2)}{k_1^2 k_2^2 k_3^2}e^{i(k+k_1)\cdot b_1}e^{i(k_2+k_3)\cdot b_2}\,,\\ &
        =\lambda_3\frac{m_1 g_1 m_2^2 s_2^2}{8m_p^3}\rmint_{\Omega}e^{-i\Omega\,n\cdot(x-b_1)}\rmint_{k_1,k_2}\frac{\hat\delta(k\cdot v_1+k_1\cdot v_1)\hat\delta(k_2\cdot v_2)\hat\delta(k_1\cdot v_2)}{k_1^2 k_2^2 (k_1+k_2)^2}e^{ik_1\cdot b}\,,\\ &
        =\lambda_3\frac{m_1 g_1 m_2^2 s_2^2}{8m_p^3}\frac{1}{n\cdot v_1}\rmint_{k_1,k_2}e^{-i k_1\cdot \delta_1}\frac{\hat\delta(k_1\cdot v_2)\hat\delta(k_2\cdot v_2)}{k_1^2 k_2^2 (k_1+k_2)^2}\,,\delta_1\equiv \frac{n\cdot(x-b_1)}{n\cdot v_1}v_1-b\,,\\ &
        =\lambda_3\frac{m_1 g_1 m_2^2 s_2^2}{64m_p^3(n\cdot v_1)}\rmint_{\vec k_1}\frac{e^{i \vec k_1\cdot \vec \delta_1}}{|\vec k_1|^3}\,.\label{6.35}
    \end{split}
\end{align}
The integral in \eqref{6.35} can be done using the dimensional regularisation and we get,\vspace{0.5 cm}
\begin{align}
    \begin{split}
           f(x)=\frac{\lambda_3}{\textcolor{black}{(2\pi)^3}}\frac{m_1 g_1 m_2^2 s_2^2}{8m_p^3} \frac{1}{32\pi^2(n\cdot v_1)}\Big[{-\log (\vec\delta_1 ^2/\Lambda^2)-\gamma_E +\log (4)}\Big]\,.\label{5.72l}
    \end{split}
\end{align}

\textit{Last but not the least, the total contribution to the scalar waveform at 2PM order is sum of (\ref{e:ba}), (\ref{5.24 mmm}), (\ref{5.25mmm}),(\ref{ex6}), \eqref{5.70l} and \eqref{5.72l}.} \textcolor{black}{Also, note that we need to replace worldline 1 with worldline 2 in all of our results and add them to get the total contribution, as the final result has to be symmetric under this exchange. }
\subsection{Gravitational waveform at 2PM: contribution due to extra scalar DOF}
Now, we will also spell out the correction to the gravitational waveform due to the presence of the extra scalar field. The leading order correction comes from bulk scalar graviton interaction with interacting action:
\begin{align}
    S_{int}=\frac{1}{m_p}\rmint d^4x \,h^{\mu\nu}\partial_{\mu}\varphi\partial_{\nu}\varphi\,.
\end{align}
The corresponding graviton one-point function has the following form:

\begin{align}
    \begin{split}
        k^2\Big\langle h_{\mu\nu}(k)\Big\rangle\Big|_{k^2\rightarrow 0}= \thesisinlinefeynman[\chthreeinlinefeynwidth]{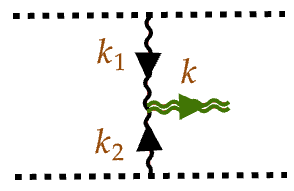}\hspace{0.35cm}=\Big(\frac{m_1 s_1 m_2 s_2}{2m_p}\Big)^2\rmint d\mu_{1,2}(k) \frac{k_1^\mu k_2^\nu} {k_1^2 k_2^2}\,.\label{5.63a}
    \end{split}
\end{align}
where the integral measure takes the form,
\begin{align}
    \begin{split}
        \rmint d\mu_{1,2}(k)= \rmint_{k_1,k_2}e^{ik_1\cdot b_1} e^{i k_2\cdot b_2}\hat\delta(k_1\cdot v_1)\hat\delta(k_2\cdot v_2)\hat\delta^{(4)}(k_1+k_2-k)
    \end{split}
\end{align}
Therefore, the correction to the time domain waveform for this particular interaction is given by,
\begin{align}
    \begin{split}
        f_{h}&=\Big(\frac{m_1 s_1 m_2 s_2}{2m_p}\Big)^2\frac{1}{4\pi m_p}\epsilon^\mu \epsilon^\nu \rmint_{\Omega}e^{-ik\cdot x}\rmint d\mu_{1,2}(k)\,\frac{k_{1\mu}k_{2\nu}}{k_1^2\,k_2^2}\,,\\ &
        =\Big(\frac{m_1 s_1 m_2 s_2}{2m_p}\Big)^2\frac{1}{4\pi m_p}\epsilon^\mu \epsilon^\nu \rmint_{\Omega}e^{-ik\cdot x}\rmint_{k_2}e^{i(k-k_2)\cdot b_1}e^{ik_2\cdot b_2}\frac{(k-k_2)_{\mu}k_{2\nu}}{k_2^2(k-k_2)^2}\,\hat\delta\Big(k\cdot v_1-k_2\cdot v_1\Big)\hat\delta(k_2\cdot v_2)\,.\label{5.55mm}
    \end{split}
\end{align}
The total waveform in \eqref{5.55mm} can be separated into two parts, and the parts will be treated separately.
\begin{align}
    \begin{split}
        f_h(x)\Big|_{1}=\Big(\frac{m_1 s_1 m_2 s_2}{2m_p}\Big)^2\frac{\epsilon^\mu \epsilon^\nu}{4\pi m_p}\rmint_{\Omega}e^{-ik\cdot (x-b_1)}k_{\mu}\rmint_{k_2}e^{-ik_2\cdot b}\frac{k_{2\nu}}{k_2^2(k-k_2)^2}\,\hat\delta\Big(k\cdot v_1-k_2\cdot v_1\Big)\hat\delta(k_2\cdot v_2)\,.\label{5.56 mm}
    \end{split}
\end{align}
It is evident from \eqref{5.56 mm} that $f_h\Big|_{1}\sim \epsilon\cdot k$ and we know that $\epsilon\cdot k=0$. Therefore, the first part will not contribute to the waveform. Now, the second part of the waveform takes the following form,
\begin{align}
    \begin{split}
        f_{h}(x)\Big|_{2}=-\Big(\frac{m_1 s_1 m_2 s_2}{2m_p}\Big)^2\frac{\epsilon^\mu \epsilon^\nu}{4\pi m_p}\rmint_{\Omega}e^{-ik\cdot x}\rmint_{k_1,k_2}\frac{k_{2\mu}k_{2\nu}}{k_1^2k_2^2}\hat\delta(k_1\cdot v_1)\hat\delta(k_2\cdot v_2)\hat\delta^{(4)}(k_1+k_2-k)\,.
    \end{split}
\end{align}
Restoring the factors of $\pi$ from the delta functions and the integration  measures, this integral is the same as \eqref{5.25mmm} and results in,

\begin{equation}\label{ch2:e:grav-waveform}
\begin{split}
 f_{h}\Big|_{2}(x)=\Big(\frac{m_1 s_1 m_2 s_2}{2m_p}\Big)^2\frac{1}{2(2\pi)^2 \pi m_p }\epsilon^{\mu}\epsilon^{\nu}(I_{\mu\nu}+\bar I_{\mu\nu}).
\\
        \end{split}
\end{equation}\\
where we again used the fact that $n\cdot \epsilon=0$ and $I^{\mu}$ and $\bar{I}^{\mu\nu}$ are defined in (\ref{IMUNU}) and (\ref{5.43mm}) respectively.
\section{Towards massive waveform}\label{sec6}
As mentioned in the previous section, we could get exact results only when we set the mass of the scalar field to zero. In this section, we will discuss how the radiation integrals become more complicated when we make the scalar field massive. Then, we provide an approximate way to evaluate those integrals and get an analytic result. \par 
For massive scalar, the radiated momentum has to be parametrized as follows,
\begin{align}
    k^{\mu}=\Omega n^\mu,\, n^\mu=(1,\sqrt{1-\frac{m^2}{\Omega^2}}\boldsymbol{\hat{x}})\,.\label{6.1 m}
\end{align}
\textit{From \eqref{6.1 m} it is clear that the unit vector towards the direction of the observed scalar depends on the observed frequency $\Omega$. We need to integrate over all possible frequencies $\Omega$ for computing the time domain waveform. There lies the difficulty due to the presence of complicated phase factors in the time-domain waveform integrand. }
\subsection*{Radiation integrals for massive scalar waveform}
Before proceeding further, we first list all the relevant massive integrands having a massless counterpart, which we evaluated in the previous section.
\begin{itemize}
  \item The massive integrated corresponding to $ \lambda_3 \varphi^3$ in \eqref{wave0} vertex has the following form:
    \begin{align}
        \begin{split}
          f_{\varphi}\propto \rmint_{\Omega}e^{-ik\cdot x}\rmint_{k_i}e^{ik_1\cdot b_1}e^{ik_2\cdot b_2}\frac{\hat\delta(k_1\cdot v_1)\hat\delta(k_2\cdot v_2)}{(k_1^2-m^2)(k_2^2-m^2)}\hat\delta^{(4)}(k_1+k_2-k)\,. \label{6.11 m}
        \end{split}
    \end{align}
\item The massive integrated corresponding to $\lambda_4 \varphi^4$ in \eqref{5.11 mn} vertex has the following form:
\begin{align} 
    \begin{split} \label{6.11 mm}
        f_{\varphi}\propto \rmint_{\Omega}e^{-ik\cdot x}\rmint_{k_i}e^{ik_1\cdot b_1}e^{i k_2\cdot b_2}e^{i k_3\cdot b_2}\frac{\hat\delta(k_1\cdot v_1)\hat\delta(k_2\cdot v_2)\hat\delta(k_3\cdot v_2)}{(k_1^2-m^2)(k_2^2 -m^2)(k_3^2-m^2)}\hat\delta^{(4)}(k_1+k_2+k_3-k)\,.
    \end{split} 
\end{align}
\item  
Massive integral corresponding to worldline radiation in \eqref{5.24 mmm} \footnote{Another way to approximate this integral is to take a large velocity limit. We have discussed that in detail in Appendix~(\ref{ch2:app:F}). Although this approximation helps get an approximate closed-form result for this diagram, it is unclear whether a large velocity approximation will also be useful for all other massive integrals. Hence, we focused on the stationary-phase method in the main text.}:
\begin{align}\label{6.4 mm}
f_{\varphi}(x)&\propto\rmint_{\Omega}e^{-ik\cdot (x-b_1)}\rmint_{\omega,k_1}e^{-i k_1\cdot b}\hat{\delta}(k_1\cdot v_2)\hat{\delta}[\Omega\, n(m,\Omega)\cdot v_1-\omega]\hat\delta(k_1\cdot v_1-\omega)\frac{\Omega(n\cdot k_1)}{\omega^2 (k_1^2-m^2)}\,.
\end{align}
     \item The massive integrated corresponding to the diagram \eqref{5.22 m} coming from the derivative interaction term $\rmint d^4x h^{\mu\nu}\partial_{\mu}\varphi\partial_\nu \varphi$ has the following form:
    \begin{align}
        \begin{split}
            f_{\varphi}\propto \rmint_{\Omega}e^{-ik\cdot x}\rmint_{k_1,k_2}e^{ik_1\cdot b_1}e^{ik_2\cdot b_2}\hat\delta(k_1\cdot v_1)v_1^{\alpha}v_1^{\beta}\hat\delta(k_2\cdot v_2)\frac{k_2^{\mu}k_2^{\nu}P_{\mu\nu,\alpha\beta}}{k_1^2(k_2^2-m^2)}\delta^{(4)}(k_1+k_2-k) \,.\label{6.10 m}
        \end{split}
    \end{align}
\end{itemize}
{\color{black}
\subsection{Stationary Phase (SP) approximation: the need and general procedure}
We mentioned in the previous subsection how the massive integrals get complicated. Now, we will evaluate the integrals mentioned above using the \textit{stationary phase approximation} for large observation distance (and time) $|x|\rightarrow \infty$. This will help us to get an approximate analytical result. Note that the idea behind the stationary phase approximation stems from the fact that sinusoids with rapidly varying phases interfere destructively. However, they can be added constructively if they have the same phases. Physically, in a scattering scenario, we expect a burst signal when the emitted radiation gets detected around some frequencies. Hence, we believe that using an approximation method, such as the stationary phase method, to evaluate these integrals that will give the waveform is reasonable. 
We have to evaluate the following type of integrals.
\begin{align}
\begin{split}
        f_{\varphi}\propto \rmint_{\Omega}\exp\Big({-i\underbrace{\Omega \,n(\Omega)\cdot (x-b_1)}_{l(\Omega)}}\Big)g(\Omega)\,.\label{6.13 m}
    \end{split}
\end{align}
The stationary point can be obtained by,
\begin{align}
    \begin{split}
        \frac{d}{d\Omega}\Big(\Omega \,n(\Omega)\cdot (x-b_1) \Big)\Big |_{\Omega_0}=0\,.\label{6.14 m}
    \end{split}
\end{align}
Therefore, the integral can be approximated to,
\begin{align}
    \begin{split}
        f_{\varphi}\propto g(\Omega_0) \,e^{-i l(\Omega_0)}e^{\frac{-i\pi}{4}\textrm{sgn}(l''(\Omega_0))}\sqrt{\frac{\pi}{|l''(\Omega_0)|}}+\mathcal{O}\Big(\frac{1}{|x-b_1|}\Big)
    \end{split}
\end{align}
{\color{black}
\textbf{\textit{Validity of the stationary phase approximation:}}
The stationary phase approximation is justified only under a few physical assumptions. First, the detector must be placed in the far radiation zone, namely
\begin{align}
    |x-b_1|\gg b,
\end{align}
where \(b\) denotes the characteristic length scale of the scattering process, such as the impact parameter. Second, the prefactor \(g(\Omega)\) should vary slowly in the neighbourhood of the saddle point. More precisely, if \(\Omega_0\) denotes the stationary point, then one requires
\begin{align}
    g(\Omega)\simeq g(\Omega_0)
\end{align}
within the saddle width
\begin{align}
    \Delta\Omega \sim \frac{1}{\sqrt{|l''(\Omega_0)|}} .
\end{align}
For a massive waveform, the spatial momentum satisfies  $ k(\Omega)=\sqrt{\Omega^2-m^2}.$ Thus different frequency modes propagate with different group velocities,
\begin{align}
    v_g(\Omega)
    =
    \frac{k(\Omega)}{\Omega}
    =
    \frac{\sqrt{\Omega^2-m^2}}{\Omega}.
\end{align}
Consider the phase factor in the form
\begin{align}
    l(\Omega)=\Omega T-k(\Omega)R,
    \qquad
    R=|\vec{x}-\vec{b}_1| .
\end{align}
The saddle point is determined by $   l'(\Omega_0)=0 .$ Now using $   \frac{dk}{d\Omega}
    =
    \frac{\Omega}{\sqrt{\Omega^2-m^2}}
    =
    \frac{1}{v_g(\Omega)},$ the saddle condition gives
\begin{align}
    T
    =
    R\frac{\Omega_0}{k(\Omega_0)}
    =
    \frac{R}{v_g(\Omega_0)} .
\end{align}
Therefore, the saddle chooses the frequency mode whose group velocity carries the radiation from the source to the detector in the observed time \(T\). Since a massive field has subluminal group velocity,  $  0<v_g(\Omega)<1,$, the above condition implies   $ T>R .$ Thus the massive waveform is observed inside the future light cone, rather than exactly on null infinity.
Finally, the stationary phase approximation also requires the phase evaluated at the saddle to be large. Solving the saddle condition gives
\begin{align}
    \Omega_0
    =
    \frac{mT}{\sqrt{T^2-R^2}},
    \qquad
    k(\Omega_0)
    =
    \frac{mR}{\sqrt{T^2-R^2}} .
\end{align}
Hence
\begin{align}
    l(\Omega_0)
    =
    \Omega_0 T-k(\Omega_0)R
    =
    m\sqrt{T^2-R^2}.
\end{align}
Therefore, the large-phase condition becomes   $ m\sqrt{T^2-R^2}\gg 1,$ or equivalently  $  T^2-R^2\gg \frac{1}{m^2}.$ In this regime, the integral is dominated by a small neighbourhood of the saddle point, and the stationary phase approximation is physically well justified.}}
\subsection{Dealing with the massive integrals via SP approximation}
We start with the integral mentioned in (\ref{6.11 m}) we get, 
\begin{align}
    g(\Omega_0)=\rmint_{k_2}e^{-ik_2\cdot b}\frac{\hat\delta(k_0\cdot v_1-k_2\cdot v_1)\hat\delta(k_2\cdot v_2)}{[(k_2-k_0)^2-m^2](k_2^2-m^2)},\,k_0\equiv \Omega_0 \,n({\Omega_0})\,.
\end{align}
Then the final integral over $k_2$ can be done using Feynman parametrization in the following way, 
\begin{align}
    \begin{split}
        g(\Omega_0)&=\rmint_{0}^1 dy\rmint e^{-ik_2\cdot b}\frac{\hat\delta(k_2\cdot v_1-k_0\cdot v_1)\hat\delta(k_2\cdot v_2)}{[(k_2-yk_0)^2-(1-y+y^2)m^2]^2},\,k_0^2=m^2\,,\\ &
        \xrightarrow[]{k_2-yk_0\rightarrow \bar q}\rmint_0^1 dy e^{-iy\,k_0\cdot b}\rmint \frac{e^{-i \bar q\cdot b}}{[\bar q^2-(1-y+y^2)m^2]^2}\hat\delta(\bar q\cdot v_1-(1-y)k_0\cdot v_1)\hat\delta(\bar q\cdot v_2+y k_0\cdot v_2)\,,\\ &
        =\frac{1}{\gamma}\rmint_0^1\,dy\,e^{-iy\,k_0\cdot b}\rmint \frac{dt}{4\pi}\exp\Big(-\frac{|b|^2}{4t}-t\Delta(k_0,y)^2-t(1-y+y^2)m^2\Big)\,,\\ &
        =\frac{b}{4\pi \gamma}\rmint_0^1
dy\,e^{-iy\,k_0\cdot b}\frac{K_1\Big( b\sqrt{\Delta(y)^2+(1-y+y^2)m^2}\Big)}{\sqrt{\Delta(y)^2+(1-y+y^2)m^2}}\,.
\end{split}
\end{align}
Then, the contribution to the waveform takes the following form,
\begin{equation}\label{ch2:e:massive-waveform}\begin{split}
f_{\varphi}\propto \frac{b}{4\pi\gamma}\sqrt{\frac{\pi}{|l''(\Omega_0)|}}\rmint_0^1 dy\,e^{-i\sigma(b,\Omega_0,m,y)}\frac{K_1\Big( b\sqrt{\Delta(y)^2+(1-y+y^2)m^2}\Big)}{\sqrt{\Delta(y)^2+(1-y+y^2)m^2}},\,\sigma=y k_0\cdot b+l(\Omega_0)-\frac{\pi}{4}\,.
\\
        \end{split}
\end{equation}

Next, we consider the integral in \eqref{6.4 mm} and apply  the method of stationary phase to it. It gives the following,
\begin{align}
    \begin{split}
        f(\varphi_0)\propto g(\Omega_0)e^{-il(\Omega_0)}e^{\frac{i\pi}{4}}\sqrt{\frac{\pi}{|l''(\Omega_0)|}}
    \end{split}
\end{align}
where $g(\Omega_0)$ is given by,
\begin{align}
    \begin{split}
        g(\Omega_0)&=\frac{1}{\Omega_0^2 \,\Big(n(\Omega_0)\cdot v_1 \Big)^2}\rmint_{k_1}e^{-ik_1\cdot b}\frac{-k_1\cdot k_0}{(k_1^2-m^2)}\hat\delta(k_1\cdot v_1-k_0\cdot v_1)\hat\delta(k_1\cdot v_2)\,,\\ &
        =-\frac{k_0^\mu}{\Omega_0^2 \,\Big(n(\Omega_0)\cdot v_1 \Big)^2}\underbrace{\rmint_{k_1}e^{-ik_1\cdot b}\frac{k_1^\mu}{k_1^2-m^2}\hat\delta(k_1\cdot v_1-k_0\cdot v_1)\hat\delta(k_1\cdot v_2)}_{g^\mu(\Omega_0)}\,.
    \end{split}
\end{align}
Here, the $g_\mu(\Omega_0)$ can be evaluated using Passarino-Veltman reduction.
\begin{align}
    \begin{split}
        g^{\mu}(\Omega_0)= \lambda_b b^\mu+\lambda_1 v_1^\mu+\lambda_2 v_2^\mu\,.
    \end{split}
\end{align}
To extract the constants, one needs to contract the index structure in the RHS with the LHS, which gives the following:

\textbullet $\,\,$ Contracting with $b^\mu$ gives,
\begin{align}
    \begin{split}
        -\lambda_b \,|b|^2&=\rmint_{k_1} e^{-i k_1\cdot b} \frac{k_1\cdot b}{k_1^2-m^2}\hat\delta(k_1\cdot v_1-k_0\cdot v_1)\hat\delta(k_1\cdot v_2)\,,\\ &
        =i\lim_{\kappa\to 1}\frac{\partial}{\partial \kappa}\rmint_{0}^\infty dl\,l\,\frac{J_{0}(\kappa\,l|b|)}{l^2+\hat m^2},\,\hat m^2\equiv m^2+\Big(\frac{k_0\cdot v_1}{\sqrt{\gamma^2-1}}\Big)^2\,.\label{6.35 m}
    \end{split}
\end{align}
\textbullet $\,\,$ Contracting with $v_1^\mu$ we will get,
\begin{align}
    \begin{split}
        \lambda_1+\gamma\lambda_2&= \rmint_{k_1} e^{-i k_1\cdot b} \frac{k_1\cdot v_1}{k_1^2-m^2}\hat\delta(k_1\cdot v_1-k_0\cdot v_1)\hat\delta(k_2\cdot v_2)\,,\\ &
        =k_0\cdot v_1\rmint_{k_1}e^{-ik_1\cdot b}\frac{\hat\delta(k_1\cdot v_1-k_0\cdot v_1)\hat\delta(k_1\cdot v_2)}{k_1^2-m^2}\,,\\ &
        =k_0\cdot v_1\rmint_0^\infty dl\,l\,\frac{J_{0}(l|b|)}{l^2+\hat m^2
        }\,.\label{6.36 m}
    \end{split}
\end{align}
\textbullet $\,\,$ Contracting with $v_2^\mu$ we will get,
\begin{align}
    \begin{split}
        \gamma \lambda_1+\lambda_2=0,\,\textrm{as}\,k_1\cdot v_2=0\,.\label{6.37 m}
    \end{split}
\end{align}
\textcolor{black}{Solving \eqref{6.36 m} and \eqref{6.37 m} we will get,
\begin{align}
    \begin{split}
      &  \lambda_1=\frac{1}{1-\gamma^2}\,k_0\cdot v_1\rmint_0^\infty dl\,l\,\frac{J_{0}(l|b|)}{l^2+\hat m^2}=\frac{1}{1-\gamma^2}\,(k_0\cdot v_1)K_0\left(|b| \hat{m}\right)\,,\\ &
      \lambda_2=\frac{\gamma}{\gamma^2-1} k_0\cdot v_1\,\rmint_0^\infty dl\,l\,\frac{J_{0}(l|b|)}{l^2+\hat m^2}=\frac{\gamma}{\gamma^2-1} (k_0\cdot v_1)\,K_0\left(|b| \hat{m}\right)
    \end{split}
\end{align}
and from \eqref{6.35 m} we will get,
\begin{align}
    \lambda_b=\frac{i}{|b|}\rmint_0^\infty dl\,l^2\,\frac{J_1(l|b|)}{l^2+\hat m^2}=\frac{i}{|b|}\hat{m}\, K_1\left(|b| \hat{m}\right)\,.\label{6.19ab}
\end{align}
}
Therefore, its contribution to the waveform looks like,

\begin{align}
    \begin{split}
    f(x)\propto -\frac{1}{\Omega_0^2 \,\Big(n(\Omega_0)\cdot v_1 \Big)^2} \,e^{-il(\Omega_0)}e^{\frac{i\pi}{4}}\sqrt{\frac{\pi}{|l''(\Omega_0)|}}\,k_0\cdot[\lambda_b b+\lambda_1 v_{1}+\lambda_2 v_{2}]\,.
    \end{split}
\end{align}
Now, we analyze the integral, which comes from the derivative interaction as mentioned in \eqref{6.10 m}. It takes the following form,
\begin{align}
    \begin{split}
       f_{\varphi}\propto (v_1\cdot P\cdot v_1)_{\mu\nu} \rmint_{\Omega}\exp\Big(-i\Omega\,n(\Omega)\cdot (x-b_1)\Big)J_{(2)}^{\mu\nu}(\Omega)\,,
    \end{split}
\end{align}
where,
\begin{align}
    \begin{split}
        J_{(2)}^{\mu\nu}(\Omega)=\rmint_{q}\hat\delta(q\cdot v_1-k\cdot v_1)\hat\delta(q\cdot v_2)\frac{q^\mu q^\nu}{(q-k)^2(q^2-m^2)}e^{-iq \cdot b}\,.\label{6.20 m}
    \end{split}
\end{align}
The integral over $q$ in \eqref{6.20 m} has been done in \eqref{A.22 m}. Then using the method of the stationary phase, we get,
\begin{align}
    \begin{split}
        f_{\varphi}\propto (v_1\cdot P\cdot v_1)_{\mu\nu}\sqrt{\frac{\pi}{|l''(\Omega_0)|}}e^{-il(\Omega_0)}e^{\frac{i\pi}{4}}\,J^{\mu\nu}_{(2)}(\Omega_0)\,.
    \end{split}
\end{align}
where $l(\Omega)$ and the saddle point $\Omega_0$ is defined in \eqref{6.13 m} and \eqref{6.14 m} respectively.

\textbullet $\,\,$ Apart from the integrals listed in (\ref{6.11 m}), (\ref{6.4 mm}) and (\ref{6.10 m}), one can possibly have one term proportional to scalar mass $m$  contributing to the waveform at 2PM order. 
\textcolor{black}{$$S_{int}=-\frac{1}{2}\frac{m^2}{m_{p}}\rmint d^4x\,h\varphi^2\,.$$}
The corresponding contribution to the scalar one-point function is , 
\begin{align}
   \begin{split}
      k^2\Big\langle \varphi(k)\Big\rangle=m^2\Big(\frac{m_1m_2s_2}{8m_p^3}\Big)\rmint d\mu_{1,2}(k)\frac{v_1^\alpha v_1^\beta\,P_{\mu\nu;\alpha\beta}\,\eta^{\mu\nu}}{k_1^2(k_2^2-m^2)}\,.\label{ch2:5.23b}
    \end{split}
\end{align}
where, the integral measure has the following form,
\begin{align}
    \begin{split}
        d\mu_{1,2}(k)&=\rmint_{k_1,k_2}\hat{\delta}^{(4)}(k_1+k_2-k)e^{ik_1\cdot b_1+ik_2\cdot b_2 }\hat\delta(k_1\cdot v_1)\hat\delta(k_2\cdot v_2)\,,\\ &
        =\rmint_{k_2}e^{ik\cdot b_1}e^{-ik_2\cdot b}\hat\delta(k\cdot v_1-k_2\cdot v_1)\hat\delta(k_2\cdot v_2)\,.
    \end{split}
\end{align}
Therefore the corresponding contrubution to scalar waveform has the following form,
\begin{align}
    \begin{split}
        f_{\varphi}(x)&\propto\thesisinlinefeynman[\chthreeinlinefeynwidth]{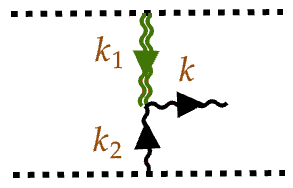}\\ &=-m^2\Big(\frac{m_1m_2s_2}{4m_p^3}\Big)\rmint_{\Omega}e^{-ik\cdot(x-b_1)}\rmint_{k_2}e^{-ik_2\cdot b}\frac{\hat\delta(k_2\cdot v_2)\hat\delta(k\cdot v_1-k_2\cdot v_1)}{(k-k_2)^2(k_2^2-m^2)}\,,\\ &
        =-m^2\Big(\frac{m_1m_2s_2}{4m_p^3}\Big)\frac{|b|}{4\pi\gamma}\sqrt{\frac{\pi}{|l''(\Omega_0)|}}\rmint_0^1 dy\,e^{-i\sigma(|b|,\Omega_0,m,y)}\frac{K_1\Big(|b|\sqrt{\Delta(y)^2+(1-y)^2m^2}\Big)}{\sqrt{\Delta(y)^2+(1-y)^2m^2}}\,.\label{6.26f}
    \end{split}
\end{align}
One can see that the contribution to the waveform from \eqref{6.26f} identically vanishes when one takes the scalar to be massless. Hence, the contribution to the massive scalar waveform takes the form:

\begin{align}
    \begin{split} 
           f(x)\propto -m^2\Big(\frac{m_1m_2s_2}{8m_p^3}\Big)\frac{|b|}{4\pi\gamma}\sqrt{\frac{\pi}{|l''(\Omega_0)|}}\rmint_0^1 dy\,e^{-i\sigma(|b|,\Omega_0,m,y)}\frac{K_1\Big(|b|\sqrt{\Delta(y)^2+(1-y)^2m^2}\Big)}{\sqrt{\Delta(y)^2+(1-y)^2m^2}}\,.
    \end{split}
\end{align}
\textbullet $\,\,$ The massive counterpart of \eqref{5.62x} will take the form.
\begin{align}
    \begin{split}
        f(x)&\propto \rmint_{k}e^{-ik\cdot (x-b_1)}\rmint_{k_2} e^{ik_2\cdot b}\frac{\hat\delta(k_2\cdot v_2)}{k_2^2-m^2}\rmint_{k_3}\hat\delta(k_3\cdot v_2)\hat\delta(k_3\cdot v_1-k\cdot v_1-k_2\cdot v_1)\frac{e^{-ik_3\cdot b}}{(k_3^2-m^2)[(k_3-k)^2-m^2]}\,,\\ &
        =e^{-il(\Omega_0)}e^{i\frac{\pi}{4}}\sqrt{\frac{\pi}{|l''(\Omega_0)|}}g(\Omega_0)
    \end{split}
\end{align}
where,
\begin{align}
    \begin{split}
        g(\Omega_0)=\rmint_{k_2}\frac{e^{i k_2\cdot b}\,\hat\delta(k_2\cdot v_2)}{k_2^2-m^2}I_{k_3}(m,k_2,b,k_0)\,.
    \end{split}
\end{align}
Now, $I_{k_3}(m,k_2,b,k_0)$ takes the form,
\begin{align}
    \begin{split}
        I_{k_3}(m,k_2,b,k_0)&=\frac{|b|}{\gamma}\rmint_0^1dy\,e^{-iyk_0\cdot b}\frac{K_1\Big[|b|\sqrt{\Sigma(k_0,y,k_2\cdot v_1)^2+(1-y+y^2)m^2}\Big]}{4\pi\sqrt{ \Sigma(k_0,y,k_2\cdot v_1)^2+(1-y+y^2)m^2}}\,.\\ &
    \end{split}
\end{align}
Therefore, 
\begin{align}
    \begin{split}
        g(\Omega_0)=\frac{|b|}{\gamma}\rmint_0^1 dy\,e^{-iy k_0\cdot b}\rmint_{-\infty}^{\infty} dz\, K_{0}(\sqrt{z^2+m^2}|b|)\frac{K_1\Big[|b|\sqrt{\Sigma(k_0,y,z)^2+(1-y+y^2)m^2}\Big]}{4\pi\sqrt{ \Sigma(k_0,y,z)^2+(1-y+y^2)m^2}}\,.
    \end{split}
\end{align}
Therefore, the full waveform proportional to,
\begin{align}
    \begin{split}
     f(x)\propto e^{-il(\Omega_0)}e^{i\frac{\pi}{4}}\sqrt{\frac{\pi}{|l''(\Omega_0)|}}\frac{|b|}{\gamma}\rmint dy\,e^{-iy k_0\cdot b}\rmint_{-\infty}^{\infty} dz\, K_{0}(\sqrt{z^2+m^2}|b|)\frac{K_1\Big[|b|\sqrt{\Sigma(k_0,y,z)^2+(1-y+y^2)m^2}\Big]}{4\pi\sqrt{ \Sigma(k_0,y,z)^2+(1-y+y^2)m^2}}.\,
    \end{split}
\end{align}
\textbullet $\,\,$ The massive counter part of \eqref{6.35} takes the form.
\begin{align}
    \begin{split}
        f(x)&\propto e^{-il(\Omega_0)}e^{i\frac{\pi}{4}}\sqrt{\frac{\pi}{|l''(\Omega_0)|}}\rmint_{k_1,k_2}\frac{\hat\delta(k_0\cdot v_1+k_1\cdot v_1)\hat\delta(k_2\cdot v_2)\hat\delta(k_1\cdot v_2)}{(k_1^2-m^2) (k_2^2-m^2) [(k_1+k_2)^2-m^2]}e^{ik_1\cdot b}\,.
    \end{split}
\end{align}
After integrating over the loop momenta, the full waveform takes the following form,

\begin{align}
    \begin{split}
    f(x)\sim e^{-il(\Omega_0)}e^{i\frac{\pi}{4}}\sqrt{\frac{\pi}{|l''(\Omega_0)|}} \rmint_0^\infty dl\,l\,\frac{J_{0}(l|b|)}{(l^2+\hat{m}^2)\sqrt{l^2+\hat{m}^2-m^2}}\arctan\Big(\frac{l^2+\hat{m}^2-m^2}{2m}\Big).\,
    \end{split}
\end{align}
\textbullet $\,\,$ The massive counterpart of \eqref{5.11 mn}, which is written in \eqref{6.11 mm} can be cast again by the stationary phase approximation and is given by,
\begin{align}
    \begin{split}
        f_{\varphi}(x)&\propto e^{-il(\Omega_0)}e^{i\frac{\pi}{4}}\sqrt{\frac{\pi}{|l''(\Omega_0)|}}\rmint_{k_3,q} \frac{\hat\delta(k\cdot v_1-q\cdot v_1)\hat\delta(k_3\cdot v_2)\hat\delta(q\cdot v_2)}{(k_3^2-m^2) \, [(q-k_3)^2-m^2][(q-k)^2-m^2]}e^{-i q\cdot b}\\ &
        = e^{-il(\Omega_0)}e^{i\frac{\pi}{4}}\sqrt{\frac{\pi}{|l''(\Omega_0)|}}\int_0^\infty d\hat\alpha\,d\hat\beta\,d\hat\gamma\int_{\vec k_3,\vec q}\hat\delta(\Omega_0\,n\cdot v_1+\vec q\cdot \vec v_1)\,\exp\Big[i\hat\alpha(\vec q^2-2\Omega_0 \vec q\cdot \vec n)\\ &
\hspace{0.8cm}+i\hat\beta (\vec q^2-2\vec q\cdot \vec k_3+\vec k_3^2)+{i\hat\gamma \vec k_3^2}+i(\hat\beta+\hat\gamma)m^2\Big]e^{i\vec q\cdot \vec b}\,.
    \end{split}
\end{align}
Now, from the delta function constraint, we can do the integral over $q_{(1)}$, and we are left with,
\begin{align}
    \begin{split}
       f_{\varphi}(x)&\sim e^{-il(\Omega_0)}e^{i\frac{\pi}{4}}\sqrt{\frac{\pi}{|l''(\Omega_0)|}}\frac{1}{\sqrt{\gamma^2-1}}\int_0^\infty d\hat\alpha\,d\hat\beta\,d\hat\gamma \,e^{i(\hat\beta+\hat\gamma) m^2}\exp\Big[i(\hat\alpha+\hat\beta)\Big(\frac{\Omega_0 n\cdot v_1}{\gamma\beta}\Big)^2+2i\hat\alpha\frac{\Omega_0^2n\cdot v_1}{\gamma\beta}\Big]\\ &
       \hspace{0.8cm}\times  e^{-i\frac{\pi}{4}}\pi^{\frac{3}{2}}(\hat\beta+\hat\gamma)^{-\frac{3}{2}}\int_{\tilde q}\exp\Big(-i \frac{\hat\beta^2 \vec q^2}{\hat\beta +\hat\gamma}\Big)\exp\Big(i(\hat\alpha +\hat\beta)\tilde q^2+i\tilde q\cdot b\Big)\\ &
       = \pi^{\frac{5}{2}}e^{-il(\Omega_0)}\sqrt{\frac{\pi}{|l''(\Omega_0)|}}\frac{1}{\sqrt{\gamma^2-1}}\int_0^\infty d\hat\alpha\,d\hat\beta\,d\hat\gamma \,e^{i(\hat\beta+\hat\gamma) m^2}\exp\Big[i(\hat\alpha+\hat\beta)\Big(\frac{\Omega_0 n\cdot v_1}{\gamma\beta}\Big)^2+2i\hat\alpha\frac{\Omega_0^2n\cdot v_1}{\gamma\beta}\\ &
       \hspace{0.8cm}-i\frac{\hat\beta^2}{\hat\beta+\hat\gamma}\frac{\Omega_0^2(n\cdot v_1)^2}{\gamma^2(\gamma^2-1)}\Big]  (\hat\beta+\hat\gamma)^{-\frac{3}{2}}\frac{e^{\frac{-ib^2}{4\lambda_1}}}{\lambda_1},\,\textrm{with},\,\lambda_1\equiv \hat\alpha+\hat\beta-\frac{\hat\beta^2}{\hat\beta+\hat\gamma}\,.\label{6.36f}
    \end{split}
\end{align}
The integral in \eqref{6.36f}, to the best of our knowledge, does not have any closed form and should be done numerically. One can analogously find the contribution from $\lambda_3 h\varphi^3$ interaction vertex.

\subsection*{Radiation integrals for gravitational waveform due to massive scalar}
Finally, we will also discuss diagrams which contribute to the gravitational waveform due to the presence of a massive scalar.

\textbullet $\,\,$ The correction to the gravitational waveform comes from bulk scalar graviton interaction with the interacting action:
\begin{align}
    S_{int}=-\frac{m^2}{2m_p}\rmint d^4x \,h\varphi^2\,.
\end{align}
Note that this diagram does not have any massless counterpart, since the vertex is proportional to the scalar mass. Its contribution to the graviton one-point function takes the following form:
\begin{align}
    \begin{split}
        k^2\Big\langle h_{\mu\nu}(k)\Big\rangle\Big|_{k^2\rightarrow 0}= \thesisinlinefeynman[\chthreeinlinefeynwidth]{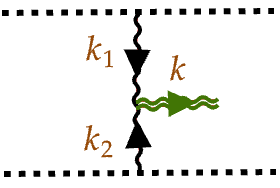}\hspace{1.4 cm}=\rmint d\mu_{1,2}(k) \frac{\eta^{\mu\nu}} {(k_1^2-m^2) (k_2^2-m^2)}\,,
    \end{split}
\end{align}
where the integral measure takes the form,
\begin{align}
    \begin{split}
        \rmint d\mu_{1,2}(k)= \rmint_{k_1,k_2}e^{ik_1\cdot b_1} e^{i k_2\cdot b_2}\hat\delta(k_1\cdot v_1)\hat\delta(k_2\cdot v_2)\hat\delta^{(4)}(k_1+k_2-k)\,.
    \end{split}
\end{align}

Therefore, the correction to the time domain waveform for this particular interaction is given by,
\begin{align}
    \begin{split}
        f_{h}&=\frac{1}{4\pi m_p}\epsilon^\mu \epsilon^\nu \rmint_{\Omega}e^{-ik\cdot x}\rmint d\mu_{1,2}(k)\,\frac{\eta_{\mu\nu}}{(k_1^2-m^2)\,(k_2^2-m^2)}\,.\\ &
       \label{6.38a}
    \end{split}
\end{align}
But since $\epsilon$ is null, i.e. $\epsilon^2=0$, the index structure of the above integral ensures that its contribution to the waveform vanishes.

\textbullet $\,\,$ Next, we focus on the contribution to the gravitational waveform from a massive scalar field through derivative interaction. The massless counterpart of it is shown in \eqref{5.63a}. The time domain waveform has the following form,
\begin{align}
    \begin{split}
          f_{h}(x)&
        \propto\frac{1}{4\pi m_p}\epsilon^\mu \epsilon^\nu \rmint_{\Omega}e^{-ik\cdot x}\rmint_{k_2}e^{i(k-k_2)\cdot b_1}e^{ik_2\cdot b_2}\frac{(k-k_2)_{\mu}k_{2\nu}}{(k_2^2-m^2)[(k-k_2)^2-m^2]}\,\hat\delta\Big(k\cdot v_1-k_2\cdot v_1\Big)\hat\delta(k_2\cdot v_2)\,,\\ &
=-\frac{1}{4\pi m_p}\epsilon^\mu \epsilon^\nu \rmint_{\Omega}e^{-ik\cdot x}\rmint_{k_2}e^{i(k-k_2)\cdot b_1}e^{ik_2\cdot b_2}\frac{k_{2\mu}k_{2\nu}}{(k_2^2-m^2)[(k-k_2)^2-m^2]}\,\hat\delta\Big(k\cdot v_1-k_2\cdot v_1\Big)\hat\delta(k_2\cdot v_2)\,,\\ &
=-\frac{1}{4\pi m_p}\epsilon\cdot \hat J_{(2)}\cdot \epsilon\,.
    \end{split}
\end{align}
where,
\begin{align}
    \begin{split}
        \hat J_{\mu\nu}^{(2)}=J_{\mu\nu}^{(2)}[(1-y)^2\to (1-y+y^2)].
    \end{split}
\end{align}
where $J_{\mu\nu}^{(2)}$ is defined in (\ref{A.22 m}).

\textbullet $\,\,$
The other diagram corresponding to the graviton radiation from worldline-1 with bulk $\lambda \varphi^3$ vertex is given by,

\begin{align}
    \begin{split}
        k^2\Big\langle h_{\mu\nu}(k)\Big\rangle\Big|_{k^2\rightarrow 0}= \thesisinlinefeynman[\chthreeinlinefeynwidth]{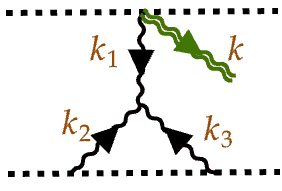}\hspace{1.4 cm}=\rmint d\mu_{1,2}(k)\frac{1}{(k_1^2-m^2)(k_2^2-m^2)(k_3^2-m^2)}\,,
    \end{split}
\end{align}
where the measure looks like,
\begin{align}
    \begin{split}
        d\mu_{1,2}(k)=\rmint_{k_i}\hat\delta^{(4)}\Big(\sum_{i}k_i\Big)\hat\delta(k_1\cdot v_1+k\cdot v_1)\hat\delta(k_2\cdot v_2)\hat\delta(k_3\cdot v_2)e^{i(k+k_1)\cdot b_1}e^{i(k_2+k_3)\cdot b_2}\,.
    \end{split}
\end{align}
Therefore, omitting the prefactors, the corresponding waveform takes the form,
\begin{align}
    \begin{split}
        f(x)&\sim \rmint_{\Omega}e^{-ik\cdot x}\rmint_{k_i}\frac{\hat\delta^{(4)}\Big(\sum_{i}k_i\Big)\hat\delta(k_1\cdot v_1+k\cdot v_1)\hat\delta(k_2\cdot v_2)\hat\delta(k_3\cdot v_2)}{(k_1^2-m^2)(k_2^2-m^2)(k_3^2-m^2)}e^{i(k+k_1)\cdot b_1}e^{i(k_2+k_3)\cdot b_2}\,,\\ &
        =\rmint_{\Omega}e^{-i\Omega n\cdot (x-b_1)}\rmint_{k_1,k_2}\frac{\hat\delta(k_1\cdot v_1+k\cdot v_1)\hat\delta(k_2\cdot v_2)\hat\delta(k_1\cdot v_2)}{(k_1^2-m^2)(k_2^2-m^2)[(k_1+k_2)^2-m^2]}e^{ik_1\cdot b}\,,\\ &
        =-\frac{4\pi(2\pi)^2}{(n\cdot v_1)|\vec \delta_1|}\rmint_{0}^{\infty} dk_1 \frac{\sin{(k_1|\vec \delta_1|)}}{(k_1^2+m^2)}\arctan\Big(\frac{|\vec k_1|}{2m}\Big)\,.
    \end{split}
\end{align}
Now, restoring the prefactors, the waveform can be recast as,
\begin{align}
 &   \begin{split}
    f(x)=-\frac{\lambda_3 }{2(2\pi)^3}\Big(\frac{m_1 s_1m_2^2 s_2^2}{4m_p^3}\Big)\frac{1}{(n\cdot v_1)|\vec \delta_1|}\rmint_{0}^{\infty} dy \frac{\sin{(y|\vec \delta_1|)}}{(y^2+m^2)}\arctan\Big(\frac{y}{2m}\Big)\,.    
    \end{split}
\end{align}

\section{Discussions and outlook}\label{sec8}
Inspired by the recent developments in scattering amplitude techniques in QFT, we consider a scalar-tensor theory of gravity. After briefly discussing the novel WQFT formalism \cite{Mogull:2020sak}, we compute the two main observables in a scattering event for scalar-tensor theory. We compute the impulse and waveform coming from the scalar contribution to the gravitational field by computing the one-point correlators for the fields (scalar and graviton) and worldline degrees of freedom. \textit{To the best of our knowledge, this is the first study regarding applying WQFT to compute impulse and waveform for a non-GR theory, namely the Scalar-Tensor theory. Also, note that this study helps us extend the computation of the gravitational waveform in the post-Minkwoskian regime for a theory beyond GR.} Below, we summarize the main findings of our paper.
\begin{itemize}
    \item First we compute the impulse $\Delta p^\mu$ in the massive scalar tensor theory with scalar potential $V(\varphi)\sim \lambda_3 \varphi^3+\lambda_4 \varphi^4$ up to 2PM. We mainly concentrate on the corrections to the impulse coming from the scalar degree of freedom. Also, we have provided the corresponding expressions when we take the massless limit. \textcolor{black}{Note that to the best of our knowledge some of the massive integrals as listed in (\ref{4.86}) do not possess any closed-form expression. They can be evaluated numerically and can be shown to have a smooth behaviour w.r.t. $|b|.$ Also, they possess smooth massless limits as discussed in the main text. Hence, these massless expressions serve as a consistency check of our results.}  The fall-off in that case w.r.t. $|b|$ is analogous to the gravitational case. Nevertheless, even in the one-loop case, where a fully closed-form expression is not available, the result can still be represented in terms of a $0$--$1$ parameter integral with a Bessel kernel:
\begin{align}
    I_{\text{1-loop}}
    \sim
    \int_0^1 ds\, \rho(s)\, K_0(\cdots).\nonumber
\end{align}
Moreover, we have verified explicitly in a nontrivial two-loop example that the impulse may be expressed as
\begin{align}
    I_{\text{2-loop}}
    \sim
    \int_0^1 ds \int_0^1 du\,
    \rho_1(u,s)\,\rho_2(s,u)\,
    K_0^2(\cdots)+\cdots.\nonumber
\end{align}
While establishing this form in general is technically challenging, these examples suggest that the leading $n$-loop contribution may admit a representation of the schematic form
\begin{align}
    I_{\text{$n$-loop}}
    \sim
    \prod_{i=1}^{n}\int_0^1 ds_i\;
    \rho_i(\{s_j\})\, K_0^{\,n}(\cdots)+\cdots\nonumber
\end{align}
It would be very interesting to clarify whether this pattern persists to all loop orders, to understand the underlying algebraic--geometric structure of these Bessel curve, and ultimately to identify the natural function space to which these observables belong \cite{Zhou:2018tva}.

\item Next, we compute the radiation integrals coming from the field one-point function $\langle\chi(k)\rangle$ and time domain waveform in the massless case for the scalar waveform as well as the correction due to the gravitational waveform due to the presence of bulk scalar interaction vertices. Note that the total waveform will also have contributions from the GR term, which is already known in the literature. As discussed in the introduction (\ref{intro}), the simplest way to add extra degrees of freedom is to introduce a scalar field. Further motivation was provided for considering the Scalar-Tensor theory there. Keeping this in mind, we have focused only on the correction due to the presence of extra scalar degrees of freedom. To the best of our knowledge, this is the first study of PM waveform for a Scalar-Tensor theory. 
\item Eventually, we proceed to compute the waveform where the scalar field has non-zero mass. In this case, we find that the analytical (exact) computation of the waveform becomes significantly difficult due to the presence of a complicated phase structure. \textit{We propose a procedure to handle those integrals using the stationary phase approximation.} As we expect a burst kind of signal from scattering events, this is a reasonable approximation one can make. 
    \item Apart from that, we would like to emphasize that we show different approaches to computing different kinds of Feynman integrals and discuss the underlying subtleties. Some of these integrals do not appear in the corresponding GR computation. We hope that this will help while exploring WQFT methods for other non-GR theories of gravity. 
\end{itemize}
The connection to the scattering amplitude is trivial in this process. One need not compute the effective action by integrating out the different energy modes, similar to the effective field theory techniques \cite{Bhattacharyya:2023kbh}. To this end, one eventually encounters the loop integrals that one will encounter in scattering processes with intermediate loops.
To the best of our knowledge, we tried to compute diagrams of radiation and impulse with massive propagators in several places, which is new to the literature of WQFT for scalar-tensor theories. It will also be quite an interesting follow-up to incorporate the spin of each black hole by considering finite-size effects in the worldline action. In fact, WQFT of $\mathcal{N}=1 $ \textit{supersymmetric spinning particles} exist in literature \cite{Jakobsen:2021lvp}. Eventually, as a follow-up, one can also try to calculate 3PM three-body radiation. As an advantage of this procedure, one can do this without taking the classical limit ($\hbar\rightarrow 0$) explicitly in each diagram computation. Last but not least, as the connection to the scattering amplitude is apparent, it is important to investigate the Double-copy setup to make it clear in this formalism. We hope to report on some of these issues in the near future. 


\appendix 
\section{Sketching the derivation of the worldline action}\label{ch2:app:A}
In our case, the matter Lagrangian has the following form:
\begin{align}
    \begin{split}
        \mathcal{L}=g^{\mu\nu}\partial_\mu\phi_i^{\dagger}\partial_\nu\phi_i-m_i(\varphi)^2\phi_i^{\dagger}\phi_i
    \end{split}
\,.\end{align}
We start by representing the partition function (in Euclidean signature)  using the Schwinger proper time parametrization.
\begin{align}
    \Gamma[g,\varphi]=\log\Big[\rmint \mathcal{D}[\phi,\phi^{\dagger}]\,e^{-S}\Big] &=-\log\Big[\det(\nabla_{\mu}\nabla^\mu+m(\varphi)^2)\Big]\,,\\ &
    =-\textrm{Tr}\log\Big[\nabla_{\mu}\nabla^\mu+m(\varphi)^2\Big]\,,\\ &
    =\rmint_0^\infty \frac{dT}{T}\rmint \frac{d^4 k}{(2\pi)^4}\exp\Big[-\frac{1}{2}eT\Big(g_{\mu\nu}k^\mu k^\nu+m(\varphi)^2\Big)\Big]
\end{align}
where, $e$ is the einbein. Now we convert the result into the path integral over $x(\tau)$.
\begin{align}
    \begin{split}
        \Gamma[g,\phi]=\rmint_0^\infty \frac{dT}{T}\mathcal{N}(T)\rmint \mathcal{D}[x]\exp\Big[-\rmint d\tau (\frac{1}{2e}g_{\mu\nu}\dot x^\mu\dot x^\nu+\frac{e}{2}m(\varphi)^2)\Big]\,.\label{D.5}
    \end{split}
\end{align}
The result in \eqref{D.5} is a one-dimensional field theory of $x^\mu(\tau)$. For the  case of gravitational field, the integral measure becomes metric dependent which causes the existence of Lee-Yang ghost. However in classical limit the ghost fields are irrelevant and can be ignored as shown in \cite{Mogull:2020sak}.
\section{Impulse via Eikonal: a connection to scattering amplitude}\label{ch2:app:B}
In section~(\ref{sec1}) we computed the impulse up to 2PM. We now discuss its connection with the scattering amplitude. It was shown in \cite{Amati:1987wq, Amati:1990xe} that the eikonal phase $\chi$ is related to the four-point scattering amplitude as,
\begin{align}
    \begin{split}
e^{i\chi}&\equiv 1+\frac{i}{4m_1 m_2}\rmint e^{iq\cdot b}\hat\delta(q\cdot v_1)\hat\delta(q\cdot v_2)\lim_{\hbar\rightarrow 0}\mathcal{M}_4(\phi_1,\phi_2\rightarrow \phi_1,\phi_2)\,,\\ &
=1+\frac{i}{4m_1 m_2}\rmint_{\boldsymbol{q}_{\perp}}e^{i \boldsymbol{q}_{\perp}\cdot b}\lim_{\hbar\rightarrow 0}\mathcal{M}_4(\phi_1,\phi_2\rightarrow \phi_1,\phi_2)\,.
    \end{split}
\end{align}
It has been demonstrated in \cite{Bern:2020buy,Bjerrum-Bohr:2018xdl} that, in the centre-of-mass frame, the eikonal phase is related to the impulse through 2PM order and takes the following form,
\begin{align}
    \begin{split}
        \Delta p_{\perp}= \frac{\partial \chi}{\partial b}\,.
    \end{split}
\end{align}
Later, it has been shown that the result can be extended to higher PM order \cite{Maybee:2019jus,Bern:2020buy}. In WQFT, one can identify the classical part of $\chi$ to be the free energy of the WQFT at tree level and hence given by,
\begin{align}
    \begin{split}
    e^{i\chi(\hat b_i,\hat v_i)}= Z_{\textrm{WQFT}}:=\mathcal{N}\rmint \mathcal{D}h_{\mu\nu}\mathcal{D}\varphi\,\rmint\prod_{k=1}^2\mathcal{D}z_{k}\exp\Big(iS_{g}+iS^k_{pm}\Big)
    \end{split}
\end{align}
where, $\hat b$ and $\hat v$ can be related to the averaged incoming momenta $\hat p_i$ which satisfies, $\hat p_1\Delta p_1=0$. Therefore, the formula for impulse is given by,
\begin{align}
    \begin{split}
        \Delta p_{1\mu}=-\frac{\partial \chi}{\partial \hat b_1^\mu}.
    \end{split}
\end{align}
If one can compute the impulse using the Eikonal method, one may lose some extra terms in the impulse, which are proportional to $v_i^\mu$. If one blindly computes the impulse, for example, for the following diagram, 
\begin{align}
    \begin{split}
        \chi&\propto \thesisinlinefeynman[\chthreeinlinefeynwidth]{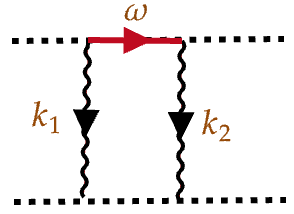} \\ &
= \rmint  \check{\mathcal{D}}[h_{\mu\nu}, \varphi,\{z_{i}\}]\rmint_{\{k_i,\omega_j\}}\hat{\delta}(k_1\cdot v_2)\hat{\delta}(k_2\cdot v_2) \varphi(k_1) \varphi(k_2) e^{i(k_1+k_2)\cdot b_2}\\ &
        \hspace{0.8cm}\hat\delta(k_3\cdot v_1+\omega_1)\hat\delta(k_4\cdot v_1+\omega_2)\varphi(k_3)\varphi(k_4)z^{\rho}(\omega_1)z^{\sigma}(\omega_2)\{2\omega_1 (v_1)_{\rho}+(k_{3})_{\rho}\}\\ &
        \hspace{0.8cm}
        \{2\omega_2(v_1)_{\sigma}+(k_4)_{\sigma}\}\hat\delta^{(4)}(k_2+k_4)\hat\delta^{(4)}(k_1+k_3)\delta(\omega_1+\omega_2)\,,\\&
        =\rmint_{\{k_i,\omega_j\}}\hat{\delta}(k_1\cdot v_2)\hat{\delta}(k_2\cdot v_2)\hat\delta(k_3\cdot v_1+\omega_1)\hat\delta(k_4\cdot v_1+\omega_2)\hat\delta^{(4)}(k_1+k_3)\hat\delta^{(4)}(k_2+k_4)\delta(\omega_1+\omega_2)\\ &
       \hspace{2 cm} \frac{1}{(k_1^2-m^2)(k_2^2-m^2)\omega_1^2}\{4\omega_1\omega_2+2\omega_1 v_1\cdot k_4+2 \omega_2 v_1\cdot k_3+k_3\cdot k_4\}e^{i(k_1+k_2)\cdot b_2}e^{i(k_3+k_4)\cdot b_1}\,,\\ &
=\rmint_{k,k_1,\omega_1}\frac{\hat\delta(k_1\cdot v_2)\hat\delta(k\cdot v_2)\hat\delta(k\cdot v_1)\hat\delta(\omega_1-k_1\cdot v_1)}{(k_1^2-m^2)[(k-k_1)^2-m^2]\omega_1^2}e^{ik \cdot b}\\&\hspace{0.8cm}[-4\omega_1^2+2\omega_1v_1\cdot(k_1-k)+2\omega_1v_1\cdot k_1+k_1\cdot (k-k_1)]\,.
    \end{split}
\end{align}
Integrating over $\delta(\omega_1-k_1\cdot v_1)$ and using the fact that the integral has support at $k\cdot v_1=0$, we left with the following integral,
{
}
The integral can be done 
\begin{align}
    \begin{split}
\chi&\sim\rmint_{k,k_1}\frac{\hat\delta(k_1\cdot v_2)\hat\delta(k\cdot v_2)\hat\delta(k\cdot v_1)}{(k_1^2-m^2)[(k-k_1)^2-m^2](k_1\cdot v_1)^2}[k_1\cdot(k-k_1)]e^{ik\cdot b}\,,\\ &
        =\frac{1}{2}\rmint_{k} e^{ik\cdot b}{\hat\delta(k\cdot v_1)\hat\delta(k\cdot v_2)}(k^2-2m^2)\underbrace{\rmint_{k_1}\frac{\hat\delta (k_1\cdot v_2)}{(k_1^2-m^2)[(k-k_1)^2-m^2](k_1\cdot v_1)^2}}_{\hat\chi_{k}}\,.\label{B.8a}
    \end{split}
\end{align}
One can see that the delta function inside the $k_1$ integral reduces the four dimensional integral into a three dimensional integral which reads,
\begin{align}
    \begin{split}
        \hat\chi_{k}\sim 4\rmint_{\vec k_1} \frac{1}{(\vec k_1^2+m^2)[(\vec k-\vec k_1)^2+m^2](2\,\vec k_1\cdot \vec v_1)^2}\,.\label{B.9c}
    \end{split}
\end{align}
The integral can be done by introducing the alpha parametrization \cite{smirnov2}, 
\begin{align}
    \frac{1}{(-A)^\lambda}=\frac{i^\lambda}{\Gamma(\lambda)}\rmint_{0}^{\infty}d\alpha \, \alpha^{\lambda-1}e^{iA\alpha}
\end{align}
which gives,
\begin{align}
    \begin{split}
      (2\pi)^3  \hat\chi_{k}&=\frac{4}{\Gamma(1)^2\Gamma(2)}\rmint d^{3-2\epsilon}k_1\rmint_0^\infty  \prod_{i=1}^3 d\alpha_i\,\alpha_3 \exp\Big[i( \vec k_1^2+m^2)\alpha_1+i[(\vec k_1-\vec k)^2+m^2]\alpha_2 +i\,2(\vec k_1\cdot \vec v_1)\alpha_3\Big]\,,\\ &
        =\frac{4}{\Gamma(1)^2\Gamma(2)}\rmint_0^\infty \prod_{i=1}^3 d\alpha_i\,\alpha_3\,e^{i(\alpha_1+\alpha_2)m^2}e^{i\alpha_2\vec k^2}\rmint d^{3-2\epsilon}k_1\exp\Big[i\vec k_1^2(\alpha_1+\alpha_2)-2\vec k_1\cdot (\alpha_2 \vec k-\alpha_3 \vec v_1)\Big]\,,\\ &
        =\frac{4}{\Gamma(1)^2\Gamma(2)}e^{i\frac{\pi}{2}(\epsilon-\frac{1}{2})}\pi^{\frac{3}{2}-\epsilon}\rmint_0^\infty  \prod_{i=1}^3 d\alpha_i\,\alpha_3\,e^{i(\alpha_1+\alpha_2)m^2}(\alpha_1+\alpha_2)^{\epsilon-\frac{3}{2}}\exp\Big(i\frac{\alpha_1\alpha_2\vec k^2-\alpha_3^2\vec v_1^2}{\alpha_1+\alpha_2}\Big)\,.
    \end{split}
\end{align}
Now doing the integral over $\alpha_3$ we get,
\begin{align}
    \begin{split}
      (2\pi)^3  \hat\chi_{k}&= \frac{4}{\Gamma(2)\Gamma(1)^2}e^{i\frac{\pi}{2}(\epsilon-\frac{1}{2})}\pi^{\frac{3}{2}-\epsilon}\rmint_0^\infty  \prod_{i=1}^2 d\alpha_i \,(\alpha_1+\alpha_2)^{\epsilon-\frac{3}{2}}\,e^{i(\alpha_1+\alpha_2)m^2}\exp\Big(i\frac{\vec{k}^2\alpha_1\alpha_2-\alpha_3^2 \gamma^2\beta^2}{\alpha_1+\alpha_2}\Big)\,,\\ &
        =-\frac{4i}{\gamma^2\beta^2}e^{i\frac{\pi}{2}(\epsilon-\frac{1}{2})}\pi^{\frac{3}{2}-\epsilon}\rmint_0^\infty  \prod_{i=1}^2 d\alpha_i (\alpha_1+\alpha_2)^{\epsilon-\frac{1}{2}}\underbrace{e^{i(\alpha_1+\alpha_2)m^2}}_{\textrm{extra term coming due to non zero mass}}\exp\Big(i\frac{\vec{k}^2\alpha_1\alpha_2}{\alpha_1+\alpha_2}\Big)\,.
    \end{split}
\end{align}
Now doing the following variable change,
$$\alpha_1+\alpha_2\to \eta\,\,\,\textrm{and}\,\,\, \frac{\alpha_1}{\alpha_1+\alpha_2}\to \xi$$ we get,
\begin{align}
    \begin{split}
     (2\pi)^3   \hat\chi_{k}=&-\frac{4i}{(\gamma^2-1)} e^{i\frac{\pi}{2}(\epsilon-\frac{1}{2})}\pi^{3/2-\epsilon}\rmint_0^1 d\xi \rmint_0^\infty d\eta \,\eta^{\epsilon+\frac{1}{2}}\exp\Big[i\eta\{m^2+\vec{k}^2\xi(1-\xi)\}\Big]\,,\\ &
        =\frac{4}{(\gamma^2-1)}\lim_{\epsilon\to 0}e^{i\epsilon\pi}\pi^{5/2-\epsilon}\frac{    \sec (\pi  \epsilon ) \left(B_{z_1}\left(-\epsilon -\frac{1}{2},-\epsilon -\frac{1}{2}\right)-B_{z_2}\left(-\epsilon -\frac{1}{2},-\epsilon -\frac{1}{2}\right)\right)}{k\,\left(k^2+4 m^2\right)^{\epsilon +1}  \Gamma \left(-\epsilon -\frac{1}{2}\right)}\label{B.12}
    \end{split}
\end{align}
where,
\begin{align}
   z_1= \frac{1}{2}-\frac{|\vec{k}|}{2 \sqrt{\vec{k}^2+4 m^2}} \,\textrm{and,}\,\, z_2=\frac{1}{2}+\frac{|\vec{k}|}{2 \sqrt{\vec{k}^2+4 m^2}}\,.
\end{align}
The beta functions can be written in a simplified manner to obtain,
\begin{align}
\begin{split}
    & B_{z_1}(-\epsilon-\frac{1}{2} ,-\epsilon-\frac{1}{2} )-B_{z_2}(-\epsilon-\frac{1}{2} ,-\epsilon-\frac{1}{2} )\\ &=-\rmint_{z_1}^{z_2}dt \,t^{-\epsilon-\frac{3}{2}}(1-t)^{-\epsilon-\frac{3}{2}}\,,\\ &
      =\frac{1}{\epsilon +\frac{1}{2}}\left(\frac{2m^2}{\vec k^2+4m^2}\right)^{-\epsilon -\frac{1}{2}} \Big[\left(\frac{|\vec k|}{\sqrt{\vec k^2+4 m^2}}+1\right)^{\epsilon +\frac{1}{2}} \,\, _2F_1\left(-\epsilon -\frac{1}{2},\epsilon +\frac{3}{2};\frac{1}{2}-\epsilon ;\frac{1}{2}-\frac{|\vec k|}{2 \sqrt{ \vec k^2+4 m^2}}\right)\\ &-\left(1-\frac{|\vec k|}{\sqrt{\vec k^2+4 m^2}}\right)^{\epsilon +\frac{1}{2}} \, _2F_1\left(-\epsilon -\frac{1}{2},\epsilon +\frac{3}{2};\frac{1}{2}-\epsilon ;\frac{|\vec k|}{2 \sqrt{\vec k^2+4 m^2}}+\frac{1}{2}\right)\Big].
     \end{split}
     \end{align}
     Expanding around $\epsilon=0$ we get,
     \begin{align}
         \begin{split}
             (2\pi)^3\hat\chi_{k}&=\frac{8\,\pi^{5/2}}{(\gamma^2-1)}\frac{1}{m\,(\vec k^2+4m^2)}.\label{B.16ab}
         \end{split}
     \end{align}
     The integral $\hat\chi_k$ defined in \eqref{B.8a} can be done using Integral-by-Parts (IBP) reduction in a more sophisticated way. We have,
     \begin{align}
      \chi_k=   \int_{k_1}\frac{\hat\delta (k_1\cdot v_2)}{(k_1^2-m^2)[(k-k_1)^2-m^2](k_1\cdot v_1)^2}
     \end{align}
By unitarity cut the integral can be re written as,
\begin{align}
    \begin{split}
        \chi_k\sim \int_{k_1} \frac{1}{(k_1^2-m^2)[(k-k_1)^2-m^2](k_1\cdot v_1)^2}\Bigg[\frac{1}{k_1\cdot v_2+i\sigma}-\frac{1}{k_1\cdot v_2-i\sigma}\Bigg]\label{B.16a}
    \end{split}
    \end{align}
  The integral in \eqref{B.16a} belongs to the following family of integral,
  \begin{align}
      \begin{split}
          \mathcal{G}^{(\pm)}_{n_1,n_2,n_3,n_4}=\int_{k_1}\frac{1}{D_1^{n_1}D_2^{(\pm)n_2}D_3^{n_3}D_4^{n_4}}
      \end{split}
  \end{align}
  where,
  \begin{align}
      D_1=k_1\cdot v_1,\,D_2^{(\pm)}=k_1\cdot v_2\pm i\sigma,\,D_3=k_1^2-m^2, \,\textrm{and},\,D_4=(k_1-k)^2-m^2\,.
  \end{align}
  Hence it is evident that  the integral $\chi_k= \mathcal{G}_{2,1,1,1}$ can be evaluated  in terms of master integrals using \textbf{LiteRed} \cite{Lee:2012cn} as follows,
  \begin{align}
  \begin{split}
    \chi_k&\sim  \mathcal{G}^{(+)}_{2,1,1,1}- \mathcal{G}^{(-)}_{2,1,1,1}\,,\\ &
    =-\frac{2}{m^2(\gamma^2-1)(4m^2-k^2)}\Big[\mathcal{G}_{0,1,0,1}^{(+),\textrm{master}}-\mathcal{G}_{0,1,0,1}^{(-),\textrm{master}}\Big]+\mathcal{O}(\sigma)\,,\\ &
    =-\frac{2}{m^2(\gamma^2-1)(4m^2-k^2)}\int_{k_1}\frac{\hat\delta(k_1\cdot v_2)}{(k_1-k)^2-m^2}\,\,(\textrm{via reverse unitarity})\,,\\ &
    \propto \frac{2}{m(\gamma^2-1)(4m^2+\vec k^2)}
    \end{split}
  \end{align}
  which exactly matches with the result given in \eqref{B.16ab}. In the last line use use the fact that $\hat\delta(k\cdot v_i)=0\,,\forall i=1,2$. Finally, the Eikonal phase is given by,
     \begin{align}
         \begin{split}
             \chi&\sim\frac{1}{2} \rmint_0^\infty dk\,k\, J_{0}(k|b|)(k^2+2m^2)\chi_k(m)\,,\\ &
             =\frac{1}{m\sqrt{\pi}(\gamma^2-1)}\rmint_0^\infty dk\, k\,J_{0}(k|b|)\frac{(k^2+2m^2)}{k^2+4m^2}\,,\\ &
             =\frac{1}{m\sqrt{\pi}(\gamma^2-1)}\rmint_0^\infty dk\, k\,J_{0}(k|b|)\Big[1-\frac{2m^2}{k^2+4m^2}\Big]\,.\label{B.15}
         \end{split}
     \end{align}
     The first term in \eqref{B.15} does not have any non-zero finite value (purely UV divergent) and hence can be ignored in the classical limit. Now the second term gives,
     \begin{align}
         \begin{split}
             \chi\sim -\frac{2 m}{\sqrt{\pi}(\gamma^2-1)}K_{0}(2 m|b|)\,.
         \end{split}
     \end{align}
We can see that the massive expression in \eqref{B.15} is non-zero in general.
 We will now show that there will be a non-vanishing. contribution to the $a_1$ 
 start with:
\begin{align}
    \begin{split}
        \mathcal{K}^\mu= a_1k^\mu+a_2 v_1^\mu+a_3 v_2^\mu.\label{B.18b}
    \end{split}
\end{align}
$a_2$ and $a_3$ s are already derived. Now contracting both side of \eqref{B.18b} with $k^\mu$ we get,
\begin{align}
    \begin{split}
        k\cdot \mathcal{K}= a_1 k^2.
    \end{split}
\end{align}
Therefore,
\begin{align}
    \begin{split}
        a_1&=\frac{1}{k^2}\rmint_{k_1}\hat\delta(k_1\cdot v_2)\frac{k_1\cdot (k-k_1)\,k\cdot (k-k_1)}{(k_1^2-m^2)[(k_1-k)^2-m^2](k_1\cdot v_1+i\epsilon)^2}\,,\\ &
        =\frac{1}{2k^2}\rmint_{k_1}\hat\delta(k_1\cdot v_2)k_1\cdot (k-k_1)\Big[\frac{k^2}{(k_1^2-m^2)[(k_1-k)^2-m^2](k_1\cdot v_1+i\epsilon)^2}+\frac{1}{(k_1^2-m^2)(k_1\cdot v_1+i\epsilon)^2}\\ &
        \hspace{0.8cm}-\frac{1}{[(k_1-k)^2-m^2](k_1\cdot v_1+i\epsilon)^2}\Big]\,.\label{B.20a}
    \end{split}
\end{align}
The second and third terms in \eqref{B.20a} will cancel each other by virtue of relabelling the third term by $k_1-k\to -k_1$ and taking into account $\hat\delta(k\cdot v_i)$\footnote{Note that, implicitly, we take $\epsilon\to 0$ beforehand. But if we take this limit at the end, it will result in the same. For  specified $i\epsilon$ prescription the last two terms for \eqref{B.20a} takes the form,

{\footnotesize
\[
\begin{aligned}
\int_{k_1}\hat\delta(k_1\cdot v_2)\frac{k_1\cdot (k-k_1)}{k_1^2-m^2}\hat\delta'(k_1\cdot v_1)
&\to -\int_{\vec k_1}\frac{d}{dk_1^{(1)}}\hat\delta(k_1^{(1)})\\
&\quad +\int_{ k_1^{(2,3)}}(-m^2+\vec k_1\cdot \vec k)\int_{k_1^{(1)}}\frac{1}{k_1^{(1)2}+k_1^{(2)2}+k_1^{(3)2}+m^2}\frac{d}{dk_1^{(1)}}\hat\delta(k_1^{(1)})\\
&\to 0
\end{aligned}
\]
}}. Therefore, we are left with,
\begin{align}
    \begin{split}
        a_1&=\frac{1}{2}\rmint_{k_1}\hat{\delta}(k_1\cdot v_2)\frac{k_1\cdot(k-k_1)}{(k_1^2-m^2)[(k_1-k)^2-m^2](k_1\cdot v_1+ i \epsilon)^2}\,,\\ &
        =\frac{1}{4}(k^2-2m^2)\hat\chi(k,m)\,.
    \end{split}
\end{align}
Hence, the correction to the impulse takes the following form,
\begin{align}
    \begin{split}
        \Delta p_1^\mu\Big|_{\textrm{corr.}}&=im_1\Big(\frac{s_1m_2 s_2}{8m_p^2}\Big)^2\rmint_{k}e^{ik\cdot b}\hat\delta(k\cdot v_1)\hat\delta(k\cdot v_2)k^\mu(k^2-2m^2)\,\hat\chi_{k}(k,b)\,,\\ &
        =-\frac{m_1}{2\pi}\Big(\frac{s_1m_2 s_2}{8m_p^2}\Big)^2\frac{1}{\sqrt{\gamma^2-1}}\frac{b^\mu}{|b|}\partial_{|b|}\rmint_{0}^{\infty} dk\,k\,J_{0}(k|b|)(k^2+2m^2)\,\hat\chi_{k}(k,m)\,,\\ &
        =-m_1\Big(\frac{s_1m_2 s_2}{8m_p^2}\Big)^2\frac{1}{\pi^{3/2}(\gamma^2-1)^{3/2}} \frac{b^\mu}{|b|}\Big(m K_1(2 |b| m)\Big)\,.\label{B.22aa}
    \end{split}
\end{align}
\textcolor{black}{
\eqref{B.22aa} explicitly indicates that if one compute the impulse from the Eikonal phase, one may get only a part of the impulse which is proportional to $b^\mu$ and other parts (proportional to $v_i^\mu$) will be lost. Along with that \eqref{B.22aa} has a smooth massless limit which reads, $$ \Delta p_1^\mu|_{\textrm{corr.}}\sim \frac{b^\mu}{b^2}\,.$$ }
However, the result in the massless limit direct contradicts with the result in \cite{Kalin:2020mvi}, where they showed $a_1=0$ by ignoring the radiation poles. But it seems there is no such radiation region due to the presence of delta function.
\section{Two master integrals used for the computation of  waveform}\label{ch2:app:C}
We analyze the following two integrals,
\begin{align}
    \begin{split}
     &   I^{\mu_1\cdots\mu_n}_n:=\rmint_{q}\hat\delta(q\cdot v_1-k\cdot v_1)\hat\delta(q\cdot v_2)\frac{e^{-iq\cdot b}}{q^2}q^{\mu_1}\cdots q^{\mu_n}\,,\\ &
     J^{\mu_1\cdots\mu_n}_n:=\rmint_{q}\hat\delta(q\cdot v_1-k\cdot v_1)\hat\delta(q\cdot v_2)\frac{e^{-iq\cdot b}}{q^2(k-q)^2}q^{\mu_1}\cdots q^{\mu_n}\,.
    \end{split}
\end{align}
The two particular cases that we used in the main text are: $J^{\mu}_{1}$ and $J_{0}$. We first start with $J_{0}\,.$

\begin{align}
    \begin{split}
        J_{(0)}(k)=\rmint_{q}\hat\delta(q\cdot v_1-k\cdot v_1)\hat\delta(q\cdot v_2)\frac{e^{-iq\cdot b}}{q^2(k-q)^2}\,.\label{A.2}
    \end{split}
\end{align}
We will do this integral by implementing a certain choice of frame,
\begin{align}
    v_2^\mu=\delta^\mu_0,\,v_1^\mu=(\gamma,\gamma\beta \,\hat{e}_v)\,b=(0,|b|\,\hat{e}_b)\,,
\end{align}
such that, the unit vectors $\hat{e}_v$ and $\hat{e}_b$ are mutually orthogonal i.e. $\hat{e}_b \cdot \hat{e}_v=0$.
We further proceed with this integral by using the Feynman parametrization method along with the on-shell condition: $k^2=0$. Therefore, $J_0$ can be written as,
\begin{align}
    \begin{split}
        J_{(0)}(k)&=\rmint_0^1dy\rmint_q \hat\delta(q\cdot v_1-k\cdot v_1)\hat\delta(q\cdot v_2)\frac{e^{-iq\cdot b}}{(q-yk)^4},\,k^2=0\,,\\ &
        \xrightarrow[]{\bar q\rightarrow q-yk}\rmint_0^1 dy e^{-iy k\cdot b} \rmint_{\bar q}\hat\delta(\bar q\cdot v_1-(1-y)k\cdot v_1)\hat\delta(\bar q\cdot v_2+y k\cdot v_2)\frac{e^{-i\bar q\cdot b}}{\bar q^4}\,,\\ &
        =\frac{1}{\gamma}\rmint_{0}^1dy\,e^{-iyk\cdot b}\rmint_0^{\infty}dt\,t\rmint_{\tilde q}\exp\Big[i\tilde q\cdot b-t\tilde q^2-t\Delta(y)^2\Big]\,,\\ &
        =\frac{1}{\gamma}\rmint_{0}^1dy\,e^{-iyk\cdot b}\rmint_0^{\infty}\frac{dt}{4\pi}\, \exp(-\frac{|b|^2}{4t}-t\Delta(k,y)^2)\,,\\ &
       = \frac{b}{\gamma}\rmint_{0}^1dy\,e^{-iyk\cdot b}\frac{ K_1\left(|b| {\Delta(k,y) }\right)}{4 \pi  {\Delta(k,y) }}\label{C.4 mm}
    \end{split}
\end{align}
where, 
\begin{align}
    \begin{split}
       \Delta(k,y)&=\sqrt{-y^2(k\cdot v_2)^2+\frac{y^2}{\beta^2}(k\cdot v_2)^2+\frac{(1-y)^2}{\gamma^2\beta^2}(k\cdot v_1)^2+2\frac{y(1-y)}{\gamma\beta^2}(k\cdot v_2)(k\cdot v_1)}\,,\\ &
        =\frac{1}{\sqrt{\gamma^2-1}}\sqrt{y^2(k\cdot v_2)^2+(1-y)^2(k\cdot v_1)^2+2y(1-y)\gamma (k\cdot v_2)(k\cdot v_1)}\,.\label{A.4}
    \end{split}
\end{align}
In thee waveform calculation, we have to further  perform a integeral over $k$, which in general has the following form,
\begin{align}
    \begin{split}
        f(x):=\rmint_{k}e^{-ik\cdot x}T_{\mu_1\cdots \mu_n}J^{\mu_1\cdots \mu_n}(k)\,.
    \end{split}
\end{align}
In the case of $\varphi^3$ interaction we have,
\begin{align}
    \begin{split}
        f(x)&=\rmint_{k}e^{-ik\cdot x}J_{0}(k), \,k=\Omega n\,,\\ &
        =\frac{b}{4\pi\gamma}\rmint_{\Omega}e^{-i\Omega n\cdot x}\rmint_{0}^1 dy\,e^{-i\,y\,\Omega(n\cdot b)}\frac{K_1[|\Omega|\,|b|\bar\Delta(y)]}{|\Omega|\bar\Delta(y)}\,\,,\bar\Delta(y)\equiv\frac{\Delta(k,y)}{|\Omega|}\,.\label{ch2:A.6}
    \end{split}
\end{align}
Now to do the integral over $\Omega$, use the following representation of modified Bessel function:
\begin{align}
    \begin{split}
        K_{\nu}(x)=\frac{1}{2}(\frac{x}{2})^{\nu}\rmint_{0}^{\infty}dt\,\exp\Big(-t-\frac{x^2}{4t}\Big)\frac{1}{t^{\nu+1}}\,.
    \end{split}
\end{align}
Therefore,
\begin{align}
    \begin{split}
        f(x)&=\frac{|b|}{4\pi\gamma}\rmint_{0}^{1}dy\, \frac{1}{\Bar\Delta(y)}\rmint_{\Omega}e^{-i\Omega n\cdot(x+yb)}\frac{|b|}{4}\Bar{\Delta(y)}\rmint \frac{dt}{t^2}\,\exp\Big(-t-\frac{\Omega^2|b|^2\bar\Delta(y)^2}{4t}\Big)\,,\\ &
        =\frac{|b|^2}{16\pi\gamma}\rmint_{0}^{1}dy\, \rmint_{0}^{\infty} \frac{dt}{t^2}e^{-t}\rmint d\Omega\,e^{-i\,\Omega\, l}\exp\Big(-\frac{\Omega^2|b|^2\bar\Delta(y)^2}{4t}\Big)\,,\\ &
         =\frac{|b|^2}{16\pi\gamma}\rmint_{0}^{1}dy\, \rmint_{0}^{\infty} \frac{dt}{t^2}e^{-t}\frac{2 \sqrt{\pi } \sqrt{t} e^{-\frac{l^2 t}{|b|^2 \bar\Delta(y) ^2}}}{|b| \bar\Delta(y) }\,,\\ &
          =-\frac{|b|}{4\gamma}\rmint_{0}^{1}dy\,\frac{1}{\Bar\Delta(y)}\sqrt{\frac{l^2}{|b|^2\bar\Delta^2}+1},\,l:=n\cdot(x+yb)\,.\label{A.8}
    \end{split}
\end{align}
Now come discuss the other important vector integral:
\begin{align}
    \begin{split}
J_{(1)}^{\mu}&=\rmint_q\hat\delta(q\cdot v_1-k\cdot v_1)\hat\delta(q\cdot v_2)e^{-iq\cdot b}\frac{q^{\mu}}{q^2(k-q)^2}\,,\\ &
=\rmint_{0}^1dy\, e^{-iyk\cdot b}\rmint_{q}\hat\delta(q\cdot v_1-(1-y))\hat\delta(q\cdot v_2+yk\cdot v_2)\frac{e^{-iq\cdot b}}{q^4}(q^\mu+yk^\mu)\,,\\ &
:=\mathcal{J}^{\mu}+\frac{|b|\,k^\mu}{\gamma}\rmint_{0}^1dy\,y\,\,e^{-iyk\cdot b}\frac{K_1(|b|\Delta(k,y))}{4\pi\Delta(k,y)}\,.\label{A.9 m}
    \end{split}
\end{align}
$\mathcal{J}^{\mu}$ can be solved by reducing the vector integral in scalar integral as follows,
\begin{align}
    \begin{split}
\mathcal{J}^{\mu}=\rmint_{0}^1dy\,e^{-iyk\cdot b}\,[\lambda_b b^{\mu}+\lambda_1 v_1^{\mu}+\lambda_2v_2^\mu]\,,\label{C.11a}
    \end{split}
\end{align}
Now we need to solve the following equations:\par
    \textbullet $\,\,$ Contracting with $ b_{\mu}$ in the both side of \eqref{C.11a} we will get,
\begin{align}
    \begin{split}
 \rmint_{0}^1dy\, e^{-iyk\cdot b}\rmint_{q}\hat\delta(q\cdot v_1-(1-y))\hat\delta(q\cdot v_2+yk\cdot v_2)\frac{e^{-iq\cdot b}}{q^4}q\cdot b=\mathcal{J}\cdot b=-\rmint_0^1 dy\,e^{-iyk\cdot b}b^2\lambda_b\,.
    \end{split}
\end{align}
To carry out the $q$ integral on the left-hand side, we introduce a new parameter $\kappa$ as,
\begin{align}
    \begin{split}
        -|b|^2\lambda_b&=\lim_{\kappa \rightarrow 1}\rmint_{q}\hat\delta(q\cdot v_1-(1-y))\hat\delta(q\cdot v_2+yk\cdot v_2)\frac{e^{-i\kappa q\cdot b}}{q^4}q\cdot b\,,\\ &
        =i\lim_{\kappa \rightarrow 1}\frac{\partial }{\partial \kappa}\rmint_{q}\hat\delta(q\cdot v_1-(1-y))\hat\delta(q\cdot v_2+yk\cdot v_2)\frac{e^{-i\kappa q\cdot b}}{q^4}\,,\\ &
=i\lim_{\kappa\rightarrow 1}\frac{\partial}{\partial \kappa}\hat{J}_0(\kappa |b|),\,\hat J_{0}(|b|)\equiv \frac{|b|}{\gamma}\frac{K_{1}(|b|\Delta(k,y))}{4\pi \Delta(k,y)}\,.
    \end{split}
\end{align}
Hence, 
\begin{align}
    \begin{split}
        \lambda_b=\frac{ i\, K_0(|b| \Delta(k,y) )}{4\pi\gamma }\,.
    \end{split}
\end{align}
 \textbullet $\,\,$ Contracting with $ v_{1\mu}$ in the both side of \eqref{C.11a} we will get,
\begin{align}
    \begin{split}
  \rmint_0^1 dy\,e^{-iyk\cdot b}\rmint_{q}\hat\delta(q\cdot v_1-(1-y)k\cdot v_1)\hat\delta(q\cdot v_2+yk\cdot v_2)\frac{e^{-iq\cdot b}}{q^4}q\cdot v_1=\mathcal{J}\cdot v_1=  \rmint_0^1 dy\,e^{-iyk\cdot b}[\lambda_1+\gamma \lambda_2]\,.
    \end{split}
\end{align}
Again, the $q$ integral in LHS can be done again by introducing an auxiliary parameter $\kappa$ as,
\begin{align}
    \begin{split} \label{B.16}
        \lambda_1+\gamma \lambda_2&=\rmint_{q}\hat\delta(q\cdot v_1-(1-y)k\cdot v_1)\hat\delta(q\cdot v_2+yk\cdot v_2)\frac{e^{-iq\cdot b-i \kappa q\cdot v_1}}{q^4}q\cdot v_1\,,\\ &
      =  \rmint_{q}\hat\delta(q\cdot v_1-(1-y)k\cdot v_1)\hat\delta(q\cdot v_2+yk\cdot v_2)\frac{e^{-iq\cdot b}}{q^4}(1-y)k\cdot v_1\,,\\ &
      =(1-y)k\cdot v_1 \hat J_0\,.
    \end{split}
\end{align}
 \textbullet $\,\,$ Contracting with $ v_{2\mu}$ in the both side of \eqref{C.11a} we will get,
\begin{align}
    \begin{split}
    \rmint_0^1 dy\,e^{-iyk\cdot b}\rmint_{q}\hat\delta(q\cdot v_1-(1-y)k\cdot v_1)\hat\delta(q\cdot v_2+yk\cdot v_2)\frac{e^{-iq\cdot b}}{q^4}q\cdot v_2=\mathcal{J}\cdot v_2=  \rmint_0^1 dy\,e^{-iyk\cdot b}[\gamma\lambda_1+\lambda_2]\,.
    \end{split}
\end{align}
Hence,
\begin{align}
    \begin{split} \label{B.18}
        \gamma \lambda_1+\lambda_2&=-\rmint_{q}\hat\delta(q\cdot v_1-(1-y)k\cdot v_1)\hat\delta(q\cdot v_2+yk\cdot v_2)\frac{e^{-iq\cdot b}}{q^4}y(k\cdot v_2)\,,\\ &
        =-y k\cdot v_2\hat J_0 \,.
    \end{split}
\end{align}
Now solving (\ref{B.16}) and (\ref{B.18}) we get,
\begin{align}
    \begin{split}
      &  \lambda_2=\frac{y(k\cdot v_2)+\gamma (1-y)k\cdot v_1}{\gamma^2-1}\hat{J}_0\,,\\ &
      \lambda_1=\frac{-\gamma y k\cdot v_2-(1-y)k\cdot v_1}{\gamma^2-1}\hat{J}_0\,.
    \end{split}
\end{align}
Therefore, the contribution to the waveform takes the following form,
\begin{align}
   \begin{split}
       f(x)&=T_{\mu}\rmint_{\Omega}e^{-ik\cdot x}\mathcal{J}^{\mu}\,,\\ &
=T_{\mu}\rmint_{\Omega}e^{-ik\cdot x}\rmint_0^1dy\, e^{-iyk\cdot b}[\lambda_b b^\mu+\lambda_1 v_1^\mu+\lambda_2 v_2^\mu]\,,\\ &
=f_1+f_2+f_3\,,\label{A.19 m}
   \end{split}
\end{align}
where, 
\begin{align}
    \begin{split}
        f_1(x|l)&=\frac{-iT\cdot b}{4\pi \gamma}\rmint_{\Omega}e^{-i\Omega\,(n\cdot x)}\rmint_{0}^1dy\,K_0(|b||\Omega|\bar\Delta(y))\,,\\ &
        =\frac{-iT\cdot b}{4\pi \gamma}\rmint_{0}^1 dy\rmint_{0}^\infty \frac{dt}{2t}e^{-t}\rmint_{\Omega}d\Omega\,e^{-i\Omega l}\exp(-\frac{|b|^2\Omega^2\bar\Delta^2}{4t})\,,\\ &
       =\frac{-iT\cdot b}{4 \gamma |b|}\rmint_{0}^1 dy\frac{1}{\bar\Delta(y)}\sqrt{\frac{|b|^2\bar\Delta^2}{|b|^2\bar\Delta^2+l^2}}\,,
    \end{split}
\end{align}
\begin{align}
    \begin{split}
        f_2(x|l)&=T\cdot v_1\, \bar\lambda_1\rmint_{\Omega}\Omega\, e^{-i\Omega (n\cdot x)}\rmint_0^1 dy\,e^{-i\Omega y(n\cdot b)}\frac{b}{\gamma}\frac{K_{1}(b|\Omega|\Delta(k,y))}{4\pi |\Omega|\bar\Delta(y)}\,,\\ &
        =\frac{|b|^2\, T\cdot v_1\, \bar\lambda_1}{16\pi\gamma}\rmint_0^1 dy\rmint_{0}^\infty dt\frac{e^{-t}}{t^2}\rmint_{\Omega} e^{-i\Omega l}\Omega \,\exp\Big(-\frac{\Omega^2 |b|^2 \bar\Delta(y)^2}{4t}\Big)\,,\\ &
        =-\frac{i |b|^2\, T\cdot v_1\, \bar\lambda_1}{4\gamma}\rmint_0^1 dy\, \frac{l}{|b|^3 \bar\Delta^3}\frac{1}{\sqrt{\frac{l^2}{|b|^2 \Delta ^2}+1}}\,.
    \end{split}
\end{align}
and,
\begin{align}
    \begin{split}
        f_3(x|l)= -\frac{i \, T\cdot v_2\, \bar\lambda_2}{4\gamma |b|} \rmint_0^1 dy\, \frac{l}{ \bar\Delta^3}\frac{1}{\sqrt{\frac{l^2}{|b|^2 \Delta ^2}+1}}\,,
    \end{split}
\end{align}
where, $\bar\lambda_i=\frac{\lambda_i}{\Omega}$.
\section{\texorpdfstring{Analysis through method of regions for $\lambda_4\varphi^4$ vertex contribution in the waveform}{Analysis through method of regions for lambda4 phi4 vertex contribution in the waveform}}\label{ch2:app:D}
We analyze the integral mentioned in (\ref{5.18k}) using the \textit{method of regions}. The method of regions \cite{Beneke:1997zp} is a universal technique for expanding Feynman integrals in various limits of momenta and masses.
We split the integration into two regions, one where $q\sim b^{-1}\gg k_{3}$ and $k_{3}\gg q\sim b^{-1}$. The integral \eqref{4PHI} for the region $q\gg k_{3}$ reduces to 
\begin{align}
    \begin{split}
    \label{FstRe1}f_{\varphi}(x)&=\rmint_{\Omega}e^{-i k\cdot x+i k\cdot b_{1}}\rmint_{k_3,q} \frac{\hat\delta(k\cdot v_1-q\cdot v_1)\hat\delta(k_3\cdot v_2)\hat\delta(k_3\cdot v_2-q\cdot v_2)}{k_3^2 \, q^2(q-k)^2}e^{i q\cdot b}\,.
    \end{split}
\end{align}
The above integration can be done easily in the rest frame of the second particle. Hence, taking $v^{\mu}_{2}=(1,0,0,0)$, the above integral simplifies to 
\begin{align}
    \begin{split}
  \label{FstRe}   f_{\varphi}(x)&=\rmint_{\Omega}e^{-i k\cdot x+i k\cdot b_{1}}\rmint_{\vec{k}_3,\vec{q}} \frac{\hat\delta(k\cdot v_1-\vec{q}\cdot \vec{v}_1)}{|\vec{k}_3^2 |\, \vec{q}^2(q-k)^2|_{q^{0}=0}}e^{i \vec{q}\cdot \vec{b}}\,.
    \end{split}
\end{align}
The integral \eqref{FstRe} is a pure divergent for the $\vec{k}_3$ integral. Similarly, for the limit $k_3\gg q$, the integral \eqref{4PHI} reduces to 
\begin{align}
    \begin{split}
    \label{2NdRe}f_{\varphi}(x)&=\rmint_{\Omega}e^{-i k\cdot x+i k\cdot b_{1}}\rmint_{k_3,q} \frac{\hat\delta(k\cdot v_1-q\cdot v_1)\hat\delta(k_3\cdot v_2)\hat\delta(k_3\cdot v_2-q\cdot v_2)}{k_3^4 \,(q-k)^2}e^{i q\cdot b}
    \end{split}
\end{align}
and solving it in the rest frame of the second particle, the $\vec{k}_{3}$ integral and the $\vec{q}$ integral factorizes and the $\vec{k}_{3}$ integral equates to zero. Thus, 
\begin{align}
    \begin{split} \label{nocont}
    f_{\varphi}(x) =&
\rmint_{\Omega}e^{-i k\cdot x}\rmint_{k_3\gg q} \frac{\hat\delta(k\cdot v_1-q\cdot v_1)\hat\delta(k_3\cdot v_2)\hat\delta(k_3\cdot v_2-q\cdot v_2)}{k_3^2 (q-k)^2(q-k_3)^2}e^{i (k-q)\cdot b_1}e^{i(k_3-q)\cdot b_2}e^{ik_3\cdot b_2}\, + \\
&\rmint_{\Omega}e^{-i k\cdot x}\rmint_{q\gg 
k_3} \frac{\hat\delta(k\cdot v_1-q\cdot v_1)\hat\delta(k_3\cdot v_2)\hat\delta(k_3\cdot v_2-q\cdot v_2)}{k_3^2 (q-k)^2(q-k_3)^2}e^{i (k-q)\cdot b_1}e^{i(k_3-q)\cdot b_2}e^{ik_3\cdot b_2}
 \end{split}
\end{align}
is purely divergent. 

\section{Massive integral coming from derivative interaction} \label{ch2:app:E}
The integral of interest is the following,
\begin{align}
    \begin{split}
        J_{(2)}^{\mu\nu}&= \rmint_{q}\hat\delta(q\cdot v_1-k\cdot v_1)\hat\delta(q\cdot v_2)\frac{q^\mu q^\nu\,e^{-i q\cdot b}}{(q-k)^2 (q^2-m^2)}\,,\\ &
        =\rmint_0^1 dy\,e^{-iy\,k\cdot b}\rmint_q \hat\delta(q\cdot v_1-(1-y)k\cdot v_1)\hat\delta(q\cdot v_2+yk\cdot v_2)\frac{e^{-i q\cdot b}}{[q^2-(1-2y+y^2)m^2]^2}(q^{\mu}+yk^\mu)(q^{\nu}+yk^\nu)\,.\label{A.22 m}
    \end{split}
\end{align}
In \eqref{A.22 m}, the term proportional to $k^\mu k^\nu$ and $q^{\mu} k^\nu$ can be done using the same procedure as discussed in Appendix~(\ref{ch2:app:C}). We will concentrate on the term which is proportional to $q^\mu q^\nu$, which can be expanded in terms of basis vectors.
\begin{align}
    \begin{split}
        \mathcal{J}_{(2)}^{\mu\nu}(\sim q^\mu q^\nu)=\rmint_0^1 dy\, e^{-iy\,k\cdot b}[\lambda_{\eta}\eta^{\mu\nu}+\lambda_{bb}b^\mu b^\nu+\lambda_{1b}b^{(\mu}v_1^{\nu)}+\lambda_{2b}b^{(\mu}v_2^{\nu)}+\lambda_{11}v_1^\mu v_1^\nu+\lambda_{22}v_2^\mu v_2^\nu+\lambda_{12}v_1^{(\mu}v_2^{\nu)}]\,.\label{A.23 m}
    \end{split}
\end{align}
Now to extract the coefficients in \eqref{A.23 m} one needs to contract the LHS with the tensor structure of the RHS. The equations we need to solve are the following,

    \textbullet $\,\,$ Contracting with $b_\mu b_\nu$ gives:
    \begin{align}
       \rmint_q \hat\delta(q\cdot v_1-(1-y)k\cdot v_1)\hat\delta(q\cdot v_2+yk\cdot v_2)\frac{e^{-i q\cdot b}}{[q^2-(1-2y+y^2)m^2]^2} (q\cdot b)^2=\mathcal{X}_1=-\lambda_{\eta}|b|^2+\lambda_{bb}|b|^4\,.\label{A.24 m}
    \end{align}
    \textbullet $\,\,$ Contracting with $b_{(\mu}v_{1\nu)}$ gives:
    \begin{align}
    \begin{split}
        \rmint_q \hat\delta(q\cdot v_1-(1-y)k\cdot v_1)\hat\delta(q\cdot v_2+yk\cdot v_2)\frac{e^{-i q\cdot b}}{[q^2-(1-2y+y^2)m^2]^2} (q\cdot b)(q\cdot v_1)&=\mathcal{X}_2\\ &\hspace{0cm}=-\lambda_{1b}\frac{|b|^2}{2}-\lambda_{2b}\frac{\gamma |b|^2}{2}\,.
        \end{split}
    \end{align}
    \textbullet $\,\,$ Contracting with $b_{(\mu}v_{2\nu)}$ gives:
    \begin{align}
        \begin{split}
        \rmint_q \hat\delta(q\cdot v_1-(1-y)k\cdot v_1)\hat\delta(q\cdot v_2+yk\cdot v_2)\frac{e^{-i q\cdot b}}{[q^2-(1-2y+y^2)m^2]^2} (q\cdot b)(q\cdot v_2)&=\mathcal{X}_3\\ &=-\frac{|b|^2}{2}(\lambda_{2b}+\gamma \lambda_{1b})\,.
        \end{split}
    \end{align}
     \textbullet $\,\,$ Contracting with $v_{1\mu} v_{1\nu}$ gives:
    \begin{align}
        \begin{split}
             \rmint_q \hat\delta(q\cdot v_1-(1-y)k\cdot v_1)\hat\delta(q\cdot v_2+yk\cdot v_2)\frac{e^{-i q\cdot b}}{[q^2-(1-2y+y^2)m^2]^2}(q\cdot v_1)^2&=\mathcal{X}_4\\ &
           \hspace{0cm}  =\lambda_{\eta}+\lambda_{11}+\lambda_{22}\gamma^2+\lambda_{12}\gamma\,.
        \end{split}
    \end{align}
   \textbullet $\,\,$ Contracting with $v_{2\mu} v_{2\nu}$ gives:
    \begin{align}
        \begin{split}
              \rmint_q \hat\delta(q\cdot v_1-(1-y)k\cdot v_1)\hat\delta(q\cdot v_2+yk\cdot v_2)\frac{e^{-i q\cdot b}}{[q^2-(1-2y+y^2)m^2]^2}(q\cdot v_2)^2&=\mathcal{X}_5\\ &
              \hspace{0cm}=\lambda_\eta+\lambda_{22}+\lambda_{11} \gamma^2+\lambda_{12}\gamma\,.
        \end{split}
    \end{align}
    \textbullet $\,\,$ Contracting with $v_{1(\mu}v_{2\nu)}$ gives:
    \begin{align}
        \begin{split}
             \rmint_q \hat\delta(q\cdot v_1-(1-y)k\cdot v_1)\hat\delta(q\cdot v_2+yk\cdot v_2)\frac{e^{-i q\cdot b}}{[q^2-(1-2y+y^2)m^2]^2}(q\cdot v_1)(q\cdot v_2)&=\mathcal{X}_6\\ &
              \hspace{0cm}=\lambda_{\eta}\gamma+\gamma(\lambda_{11}+\lambda_{22})+\frac{\lambda_{12}}{2}(\gamma^2+1)\,.
        \end{split}
    \end{align}
      \textbullet $\,\,$ Contracting with $\eta_{\mu\nu}$ gives:
      \begin{align}
          \begin{split}
                \rmint_q \hat\delta(q\cdot v_1-(1-y)k\cdot v_1)\hat\delta(q\cdot v_2+yk\cdot v_2)\frac{e^{-i q\cdot b}}{[q^2-(1-2y+y^2)m^2]^2}q^2&=\mathcal{X}_7\\ &
                \hspace{0cm}=4\lambda_\eta-\lambda_{bb}|b|^2+\lambda_{11}+\lambda_{22}+\lambda_{12}\gamma\,.\label{A.30 m}
          \end{split}
      \end{align}
Solving \eqref{A.24 m} to \eqref{A.30 m} we will get,
\begin{align}
\lambda _{\eta }
&\to \frac{\gamma ^2 \mathcal{X}_{7}-2 \gamma  \mathcal{X}_{6}+\mathcal{X}_{4}+\mathcal{X}_{5}-\mathcal{X}_{7}}{\gamma ^2-1}+\frac{\mathcal{X}_{1}}{|b|^2},\notag\\
\lambda _{\text{bb}}
&\to \frac{2 \mathcal{X}_{1}}{|b|^4}+\frac{\gamma ^2 \mathcal{X}_{7}-2 \gamma  \mathcal{X}_{6}+\mathcal{X}_{4}+\mathcal{X}_{5}+\mathcal{X}_{7}}{|b|^2 \left(\gamma ^2-1\right)},\notag\\
\lambda _b
&\to -\frac{2 \left(\gamma  \mathcal{X}_{3}-\mathcal{X}_{2}\right)}{|b|^2 \left(\gamma ^2-1\right)},\qquad
\lambda _{2 b}\to -\frac{2 \left(\gamma  \mathcal{X}_{2}-\mathcal{X}_{3}\right)}{|b|^2 \left(\gamma ^2-1\right)},\notag\\
\lambda _{12}
&\to \frac{2 \left(-|b|^2 \left(\gamma ^3 \left(-\mathcal{X}_{7}\right)+3 \gamma ^2 \mathcal{X}_{6}-2 \gamma  \mathcal{X}_{4}-2 \gamma  \mathcal{X}_{5}+\gamma  \mathcal{X}_{7}+\mathcal{X}_{6}\right)+\gamma  \left(\gamma ^2-1\right) \mathcal{X}_{1}\right)}{|b|^2 \left(\gamma ^2-1\right)^2},\notag\\
\lambda _{11}
&\to \frac{\frac{\left(\gamma ^2-1\right) \mathcal{X}_{1}}{|b|^2}+\gamma ^2 \mathcal{X}_{7}+\left(\gamma ^2+1\right) \mathcal{X}_{5}-4 \gamma  \mathcal{X}_{6}+2 \mathcal{X}_{4}-\mathcal{X}_{7}}{\left(\gamma ^2-1\right)^2},\notag\\
\lambda _{22}
&\to \frac{\frac{\left(\gamma ^2-1\right) \mathcal{X}_{1}}{|b|^2}+\gamma ^2 \mathcal{X}_{7}+\left(\gamma ^2+1\right) \mathcal{X}_{4}-4 \gamma  \mathcal{X}_{6}+2 \mathcal{X}_{5}-\mathcal{X}_{7}}{\left(\gamma ^2-1\right)^2}\,.
\end{align}
Next, we list the values of the integrals $\mathcal{X}_i\,.$
\begin{align}
    \begin{split}
        \mathcal{X}_1&= -\lim_{\kappa\to 1}\frac{\partial^2}{\partial \kappa^2}   \rmint_q \hat\delta(q\cdot v_1-(1-y)k\cdot v_1)\hat\delta(q\cdot v_2+yk\cdot v_2)\frac{e^{-i\kappa q\cdot b}}{[q^2-(1-2y+y^2)m^2]^2}\,,\\ &
        =-\frac{\kappa |b|}{4\pi \gamma}\lim_{\kappa\to 1}\frac{\partial^2}{\partial \kappa^2}\frac{K_1(\kappa |b|\sqrt{\Delta^2+(1-2y+y^2)m^2})}{\sqrt{\Delta^2+(1-2y+y^2)m^2}}\,,\\ &
        =-\frac{|b|}{4\pi \gamma}\Big[\frac{[|b|^2 \left(\Delta ^2+m^2 (y-1)^2\right)+2]K_1\left(|b| \sqrt{m^2 (y-1)^2+\Delta ^2}\right)}{\sqrt{\Delta ^2+m^2 (y-1)^2}}-|b| K_2\left(|b| \sqrt{m^2 (y-1)^2+\Delta ^2}\right)\Big]\,.
    \end{split}
\end{align}
\begin{align}
    \begin{split}
        \mathcal{X}_2&=i(1-y)k\cdot v_1\lim_{\kappa\to 1}\frac{\partial}{\partial \kappa}\rmint_q \hat\delta(q\cdot v_1-(1-y)k\cdot v_1)\hat\delta(q\cdot v_2+yk\cdot v_2)\frac{e^{-i\kappa q\cdot b}}{[q^2-(1-2y+y^2)m^2]^2}\,,\\ &
        =-i(1-y)k\cdot v_1\frac{|b|^2 K_0\left(|b| \sqrt{m^2 (y-1)^2+\Delta ^2}\right)}{4 \pi  \gamma }\,.
    \end{split}
\end{align}
\begin{align}
    \begin{split}
        \mathcal{X}_3&=-iy\,k\cdot v_2\lim_{\kappa\to 1}\frac{\partial}{\partial \kappa}\rmint_q \hat\delta(q\cdot v_1-(1-y)k\cdot v_1)\hat\delta(q\cdot v_2+yk\cdot v_2)\frac{e^{-i \kappa\,q\cdot b}}{[q^2-(1-2y+y^2)m^2]^2}\,,\\ &
        =i y \,k\cdot v_2 \frac{|b|^2 K_0\left(|b| \sqrt{m^2 (y-1)^2+\Delta ^2}\right)}{4 \pi  \gamma }\,.
    \end{split}
\end{align}
\begin{align}
    \begin{split}
        \mathcal{X}_4&=(1-y)^2 (k\cdot v_1)^2\rmint_q \hat\delta(q\cdot v_1-(1-y)k\cdot v_1)\hat\delta(q\cdot v_2+yk\cdot v_2)\frac{e^{-i q\cdot b}}{[q^2-(1-2y+y^2)m^2]^2}\,,\\ &
        =(1-y)^2 (k\cdot v_1)^2\frac{|b|}{4\pi \gamma}\frac{K_1( |b|\sqrt{\Delta^2+(1-2y+y^2)m^2})}{\sqrt{\Delta^2+(1-2y+y^2)m^2}}\,.
    \end{split}
\end{align}
\begin{align}
    \begin{split}
         \mathcal{X}_5&=y^2 (k\cdot v_2)^2\rmint_q \hat\delta(q\cdot v_1-(1-y)k\cdot v_1)\hat\delta(q\cdot v_2+yk\cdot v_2)\frac{e^{-i q\cdot b}}{[q^2-(1-2y+y^2)m^2]^2}\,,\\ &
        =y^2 (k\cdot v_2)^2\frac{|b|}{4\pi \gamma}\frac{K_1( |b|\sqrt{\Delta^2+(1-2y+y^2)m^2})}{\sqrt{\Delta^2+(1-2y+y^2)m^2}}\,.
    \end{split}
\end{align}
\begin{align}
    \begin{split}
        \mathcal{X}_{6}&=-y(1-y)(k\cdot v_1)(k\cdot v_2)\rmint_q \hat\delta(q\cdot v_1-(1-y)k\cdot v_1)\hat\delta(q\cdot v_2+yk\cdot v_2)\frac{e^{-i q\cdot b}}{[q^2-(1-2y+y^2)m^2]^2}\,,\\ &
        =-y(1-y)(k\cdot v_1)(k\cdot v_2)\frac{|b|}{4\pi \gamma}\frac{K_1( |b|\sqrt{\Delta^2+(1-2y+y^2)m^2})}{\sqrt{\Delta^2+(1-2y+y^2)m^2}}\,.
    \end{split}
\end{align}
The integral of $\mathcal{X}_7$ is a little bit involved. One can do this integral in the following way,
\begin{align}
    \begin{split}
        \mathcal{X}_7&=\rmint_q \hat\delta(q\cdot v_1-(1-y)k\cdot v_1)\hat\delta(q\cdot v_2+yk\cdot v_2)\frac{e^{-i q\cdot b}}{[q^2-(1-2y+y^2)m^2]^2}\Big[q^2-(1-y)^2m^2+(1-y)^2 m^2\Big]\,.\label{A.38}
    \end{split}
\end{align}
The integral in \eqref{A.38} has two parts and can be evaluated separately,
\begin{align}
    \begin{split}
        \mathcal{X}_7^{(1)}&\equiv  \rmint_q \hat\delta(q\cdot v_1-(1-y)k\cdot v_1)\hat\delta(q\cdot v_2+yk\cdot v_2)\frac{e^{-i q\cdot b}}{[q^2-(1-2y+y^2)m^2]}\,,\\ &
        =\frac{1}{\gamma}\rmint_0^\infty \frac{dt}{4\pi}\,\frac{1}{t}\exp\Big(-\frac{|b|^2}{4t}-t\Delta^2-t(1-y)^2m^2\Big)\,,\\ &
        =\frac{1}{2\pi \gamma}K_{0}(|b|\sqrt{\Delta^2+(1-y)^2m^2})\,.
    \end{split}
\end{align}
and the second term takes the form,
\begin{align}
    \begin{split}
        \mathcal{X}_7^{(2)}&=(1-y)^2 m^2\,\rmint_q \hat\delta(q\cdot v_1-(1-y)k\cdot v_1)\hat\delta(q\cdot v_2+yk\cdot v_2)\frac{e^{-i q\cdot b}}{[q^2-(1-2y+y^2)m^2]^2}\,,\\ &
        =(1-y)^2 m^2\frac{|b|}{4\pi\gamma} \frac{K_1( |b|\sqrt{\Delta^2+(1-2y+y^2)m^2})}{\sqrt{\Delta^2+(1-2y+y^2)m^2}}\,.
    \end{split}
\end{align}

\section{Computation of worldline radiation diagram using large velocity approximation}\label{ch2:app:F}
We start with \eqref{6.4 mm} (omitting the overall constant, which is irrelevant to explain our point).
\begin{align}
    \begin{split}
        f_{\varphi}(x)&\propto\rmint_{\Omega}e^{-ik\cdot (x+b_1)}\rmint_{\omega,k_1}e^{-i k_1\cdot b}\hat{\delta}(k_1\cdot v_2)\hat{\delta}[\Omega\, n(m,\Omega)\cdot v_1-\omega]\hat\delta(k_1\cdot v_1-\omega)\frac{\Omega(n\cdot k_1)}{\omega^2 (k_1^2-m^2)}\,.\label{7.7}
    \end{split}
\end{align}
To do the integration over $\Omega$ in (\ref{7.7}) we first find the roots of the equation $f(\Omega):=\Omega\, n(m,\Omega)\cdot v_1-\omega=0$, which gives,
\begin{align}
    \begin{split}
       & \hat{\delta}[\Omega\, n(m,\Omega)\cdot v_1-\omega]=\frac{\hat\delta(\Omega-\Omega_1)}{|f'(\Omega_1)|}+\frac{\hat\delta(\Omega-\Omega_2)}{|f'(\Omega_2)|},\,\, \\ &
        \text{where,}\,\, \Omega_{1}(\omega)=\thesismathbox{\displaystyle\frac{\tilde\beta  \sqrt{\left(\tilde\beta ^2-1\right) \gamma ^2 m^2+\omega ^2}+\omega }{\left(1-\tilde\beta ^2\right) \gamma }},\\ &
        \text{if }\gamma  m<\omega \lor \frac{\omega }{\gamma }<m<\frac{\omega }{\sqrt{1-\tilde\beta ^2} \gamma },\\ &
        \text{and}\quad \Omega_2(\omega)=\thesismathbox{\displaystyle\frac{\tilde\beta  \sqrt{\left(\tilde\beta ^2-1\right) \gamma ^2 m^2+\omega ^2}-\omega }{\left(\tilde\beta ^2-1\right) \gamma }},\\ &
        \text{if }\frac{\omega }{\gamma }<m<\frac{\omega }{\sqrt{1-\tilde\beta ^2} \gamma }\,.
    \end{split}
\end{align}
As the observed frequency cannot be negative, only the $\Omega_1$ solution is relevant to us. Therefore, the integral in (\ref{7.7}) can be written as (with $\tilde{x}=x-b_1$),
\begin{align}
    \begin{split}
        f_{\varphi}(x)&\propto  \sum_{a=1}^2\rmint_{k_1}e^{-i \Omega_a\, n(m,\Omega_a)\cdot (x-b_1)}e^{-ik_1\cdot b}\hat\delta(k_1\cdot v_2)\frac{\Omega_a\,n(m,\Omega_a)\cdot k_1}{|f'(\Omega_a)|(k_1\cdot v_1+i\epsilon)^2(k_1^2-m^2)}\,.
    \end{split}
\end{align}
\textcolor{black}{The integral could be significantly simplified by imposing a suitable IR cutoff on the $\omega$ and $\Omega$ integrals in the regime $\frac{m^2}{\Omega^2}\ll 1$. The argument in favour of this statement is the following: $\frac{1}{m}$ is of cosmological scale, while $\Omega$ for a compact binary with an orbital period of roughly an hour is of the order of $10^{-3}$ Hz (for $1/m \sim 1$ Mpc, one has $m \sim 10^{-14}$ Hz)\,. One point to note is that placing a cutoff on the $\omega$ and $\Omega$ integrals automatically implies that there should also be a cutoff on the $k_1$ integral (coming from the condition $k_1\cdot v_1=\omega$). However, one could, in principle, argue that instead of imposing the cutoff directly on the $k_1$ integral, one should impose a suitable condition on $|v_1|$ such that $k_1\cdot v_1\gg m$, even if $|k_1|\rightarrow 0$. The quantity $\sqrt{(k_1\cdot v_1)^2-m^2\gamma^2(1-\tilde\beta^2)}$ coming from the solutions for $\Omega$ takes the following form in terms of the parametrization used here: $\sqrt{\gamma^2\beta^2 k_1^{(x)2}-m^2\gamma^2(1-\tilde\beta^2)}=\gamma\beta \sqrt{k_1^{(x)2}-\frac{m^2}{\gamma^2\beta^2}+m^2\sin^2\theta}$. Now we assume that we can place our detector at a particular direction on the celestial sphere, as shown in Fig.~(\ref{celestial}), such that $\sin^2\theta=\frac{1}{\gamma^2\beta^2}\,.$ For further simplification, we choose $\gamma\beta\rightarrow \infty$ so that $\sin \theta\rightarrow 0$.}
Hence, the integration that we have to perform is of the form as shown below,
\begin{align}
    \begin{split}
        f_{\varphi}(x)\sim \rmint_{\Omega \in| \Omega|_{\textrm{IR}}\gg m}d\Omega\rmint_{\omega \in |\omega|_{\textrm{IR}}\gg m}d\omega\rmint_{k_1,|k_1\cdot v_1|\gg m}\cdots\,.
    \end{split}
\end{align}

\begin{figure}
    \centering
    \includegraphics[scale=0.15]{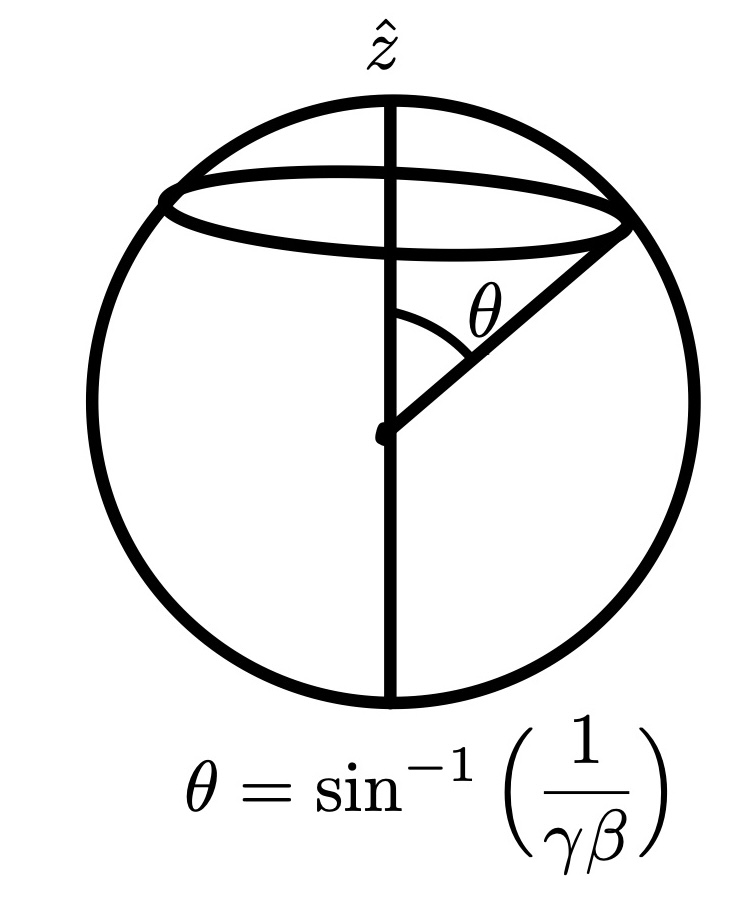}
    \caption{Sketch of the celestial sphere under consideration.}
    \label{celestial}
\end{figure}
Finally we get,
\begin{align}
    \begin{split}
        f_{\varphi}(x)\propto\frac{n^{\mu}}{(n\cdot v_1)^2}\underbrace{\rmint_{k_1}e^{-i k_1\cdot w_1}\hat{\delta}(k_1\cdot v_2)\frac{k_{1\mu}}{(k_1\cdot v_1+i\epsilon)(k_1^2-m^2)}}_{\mathcal{K}_{\mu}^{(2)}},\,\, w_1\equiv\frac{n\cdot (x-b_1)}{n\cdot v_1}v_1+b
    \end{split}
\end{align}  
where,
\begin{align}
    \begin{split}
        \mathcal{K}_{\mu}^{(2)}\to-\rmint_{-\infty}^{\infty}d\tau\,\theta(\tau)\frac{(\vec w_1-\tau \vec v_1)_{i}}{|\vec w_1-\tau \vec v_1|^3}\,[1+m |\vec w_1-\tau \vec v_1|]\,\exp(-m |\vec w_1-\tau \vec v_1|)\,.
    \end{split}
\end{align}
Then we proceed as follows:
\begin{align}
    \begin{split}
        n\cdot \mathcal{K}^{(2)}&\rightarrow -\rmint d\tau\,\theta(\tau) \frac{\vec n\cdot (\vec w_1-\tau \vec v_1)}{|\vec w_1-\tau v_1|^3}[1+m |\vec w_1-\tau v_1|]\,\exp(-m |\vec w_1-\tau \vec v_1|)\,,\\ &
        =\rmint_{-\infty}^{\infty} d\tau\, \theta(\tau)\frac{(u_1-\tau)\gamma\tilde\beta+\chi}{(\gamma^2-1)^{3/2}\Big[\tau^2-u_1^2-2\tau u_1+\frac{b^2}{\gamma^2-1}\Big]^{3/2}}\Bigg[1+m(\gamma^2-1)^{1/2}\sqrt{\tau^2-u_1^2-2\tau u_1+\frac{|b|^2}{\gamma^2-1}}\Bigg]\\ &
        \hspace{2 cm}\exp\Big[-m(\gamma^2-1)^{1/2}\sqrt{\tau^2-u_1^2-2\tau u_1+\frac{|b|^2}{\gamma^2-1}}\Big]\,.
    \end{split}
\end{align}
In the limit of $\sin\theta\rightarrow 0$, the integral has a closed-form and smooth massless limit.
\begin{equation}\label{ch2:e:collinear-waveform}\begin{split}
f_{\varphi}(x)\propto \frac{\gamma}{(n\cdot v_1)^2(\gamma^2-1)} \frac{ \exp\Big({-m\sqrt{\gamma ^2-1}  \sqrt{\frac{|b|^2}{\gamma ^2-1}-u_1^2}}\,\Big)}{\sqrt{\frac{|b|^2}{\gamma ^2-1}-u_1^2}}\,.
        \end{split}
\end{equation}
\section{Proof of the identity in \texorpdfstring{\eqref{3.54kk}}{Eq. 3.54kk}}\label{3.54kkk}
The claim we made in the main text is,
\begin{align}
     \int_0^1 ds\frac{K_0(2m|b|)-K_0\left(\frac{2m|b|}{s}\right)}{1-s^2}  =\frac{1}{2} (\log (4m |b|)+\gamma_{E} ) K_0(2 m|b|)-\partial_\alpha \textrm{Ki}_{\alpha}(2m|b|)|_{\alpha\to0};
\end{align}
\textbf{Proof:} We have,
\begin{align}
    i(a)= \int_0^1 ds\frac{K_0(a)-K_0\left(\frac{a}{s}\right)}{1-s^2}\,.
\end{align}
The idea is to first take the Mellin transformation in $a$, which gives,
\begin{align}
\begin{split}
    \mathcal{M}_{a}(p)&=\int_0^\infty da a^{p-1} \int_0^1 ds\frac{K_0(a)-K_0\left(\frac{a}{s}\right)}{1-s^2}\,,\\ &
    =2^{p-2}\Gamma^2\left(\frac{p}{2}\right)\int_0^1 ds \frac{1-s^p}{1-s^2}\,,\\ &
    =2^{p-3} \Gamma \left(\frac{p}{2}\right)^2 \left(H_{\frac{p-1}{2}}+\log (4)\right)\,.
    \end{split}    
\end{align}
Now consider the following integral,
\begin{align}
    \begin{split}
        i'(a)=\frac{1}{2}(\log(2a)+\gamma_E)K_0(a)+\int_0^\infty dt e^{-a \cosh t}\log(\cosh t)\,.
    \end{split}
\end{align}
The Mellin transformation gives,
\begin{align}
    \begin{split}
        \mathcal{M}'_a(p)&=\int_0^\infty da
        \,a^{p-1} i'(a)\,,\\ &
        =\frac{1}{4} \Gamma \left(\frac{p}{2}\right)^2 \left(\psi ^{(0)}\left(\frac{p}{2}\right)+\gamma_E +\log (2)\right)+\Gamma(p){\int_0^\infty  dt \cosh^{-p}t\, \log(\cosh t)}\,,\\ &
        =\frac{1}{4} \Gamma \left(\frac{p}{2}\right)^2 \left(\psi ^{(0)}\left(\frac{p}{2}\right)+\gamma_E +\log (2)\right)-\Gamma(p)\partial_p\underbrace{\int_0^\infty dt \cosh ^{-p}t}_{\frac{\sqrt{\pi } \Gamma \left(\frac{p}{2}\right)}{2 \Gamma \left(\frac{p+1}{2}\right)}}\,,\\ &
        =2^{p-3} \Gamma \left(\frac{p}{2}\right)^2 \left(\psi ^{(0)}\left(\frac{p}{2}\right)+\gamma +\log (4)\right)-\Gamma(p)\frac{\sqrt{\pi } \Gamma \left(\frac{p}{2}\right) \left(\psi ^{(0)}\left(\frac{p}{2}\right)-\psi ^{(0)}\left(\frac{p+1}{2}\right)\right)}{4 \Gamma \left(\frac{p+1}{2}\right)}\,,\\ &
        =2^{p-3} \Gamma \left(\frac{p}{2}\right)^2 \left(H_{\frac{p-1}{2}}+\log (4)\right)=\mathcal{M}_a(p)\,.
        \end{split}
\end{align}
Now, from the Mellin uniqueness theorem \cite{Butzer1997Mellin}, one can assert that $i(a)=i'(a)$.
Moreover, $\int_0^\infty dt e^{-a \cosh t}\log(\cosh t)$ can be written as derivative of Bickley–Naylor function $\textrm{Ki}_n$ as,
\begin{align}
    \textrm{Ki}_n(x)=\int_0^\infty dt e^{-x \cosh t} \textrm{sech}^nt\implies  -\partial_n\textrm{Ki}_n(x)|_{n=0}=\int_0^\infty dt e^{-a \cosh t}\log(\cosh t),
\end{align}
which proves our claim. One can also numerically cross-check.
\section{Comment on the integral convergence}\label{App:3.H}
The integral we would like to analyze is
\begin{align}
I^{\mu}(|b|)
=
\int_{k_i,q}
\frac{
k_1^\mu\,
\hat\delta(k_1\!\cdot\! v_1)\,
\hat\delta(k_2\!\cdot\! v_2)\,
\hat\delta(q\!\cdot\! v_2)\,
\hat\delta(q\!\cdot\! v_1)
}{
(k_1^2-m^2)(k_2^2-m^2)\bigl[(k_1+k_2)^2-m^2\bigr]\bigl[(q-k_1)^2-m^2\bigr]
}
\,e^{iq\cdot b}.
\end{align}
Our goal is to reduce this expression to a form where the integrations can be performed in a systematic way. The delta functions already impose strong kinematical constraints, and it is therefore useful to exploit them from the outset by choosing variables adapted to the geometry of the problem.
To this end, we first introduce the spacelike unit vector
\begin{align}
u_1^\mu
=
\frac{\gamma v_1^\mu-v_2^\mu}{\zeta},
\qquad
\zeta=\sqrt{\gamma^2-1}\,,
\qquad
u_1^2=-1\,,
\qquad
u_1\cdot v_1=0\,.
\end{align}
In the frame we are using, this vector takes the explicit form
\begin{align}
u_1^\mu=(\zeta,\gamma,0,0)\,.
\end{align}
The importance of $u_1^\mu$ is that, together with the directions transverse to the scattering plane, it provides a natural basis for decomposing momenta constrained by the delta functions.

Indeed, the condition $\hat\delta(k_1\!\cdot\! v_1)$ implies that $k_1^\mu$ lies entirely in the three-dimensional hyperplane orthogonal to $v_1^\mu$. It is therefore natural to write
\begin{align}
k_1^\mu=a\,u_1^\mu+\ell^\mu,
\qquad
\ell\cdot v_1=0,
\qquad
\ell\cdot v_2=0,
\end{align}
where $\ell^\mu$ is purely transverse. Similarly, the pair of delta functions
\begin{align}
\hat\delta(q\!\cdot\! v_1)\hat\delta(q\!\cdot\! v_2)
\end{align}
forces the exchanged momentum $q^\mu$ to be purely transverse as well, so that
\begin{align}
q^\mu=q_\perp^\mu.
\end{align}
Thus, both $\ell^\mu$ and $q^\mu$ live in the transverse two-dimensional subspace, which will later allow us to trade the angular integrations for Bessel functions. With this parametrization, the relevant propagators simplify considerably. Writing
\begin{align}
\ell^\mu=(0,0,\boldsymbol{\ell}_\perp),
\qquad
q^\mu=(0,0,\boldsymbol{q}_\perp),
\end{align}
we find
\begin{align}
k_1^2-m^2
&=-(a^2+\boldsymbol{\ell}_\perp^2+m^2),\\
(q-k_1)^2-m^2
&=-\bigl(\boldsymbol{q}_\perp-\boldsymbol{\ell}_\perp\bigr)^2-a^2-m^2.
\end{align}
In this form, the $q$-dependence is entirely transverse, and the Fourier transform with respect to the impact parameter can be performed in the standard way. One obtains
\begin{align}
\int e^{iq\cdot b}\,
\frac{\hat\delta(q\!\cdot\! v_1)\hat\delta(q\!\cdot\! v_2)}
{(q-k_1)^2-m^2}
\propto
-\frac{1}{\sqrt{\gamma^2-1}}\,
e^{-i\boldsymbol{\ell}_\perp\cdot\boldsymbol{b}}\,
K_0\!\left(|\boldsymbol{b}|\sqrt{a^2+m^2}\right),
\end{align}
where the modified Bessel function $K_0$ arises from the two-dimensional massive Fourier transform. At this stage, the original integral is reduced to a mixed integral over the longitudinal variable $a$ and the transverse momentum $\ell$. The next nontrivial ingredient is the bubble subintegral involving $k_2$,
\begin{align}
\mathcal{B}(a,\ell)
=
\int
\frac{\hat\delta(k_2\!\cdot\! v_2)}
{(k_2^2-m^2)\bigl[(k_1+k_2)^2-m^2\bigr]}\,.
\end{align}
Since $v_2^\mu=(1,0,0,0)$, the delta function sets $k_2^0=0$, so it is convenient to parametrize
\begin{align}
k_2^\mu=(0,c,\boldsymbol{k}).
\end{align}
The bubble integral then becomes
\begin{align}
\begin{split}
\mathcal{B}(a,\ell)
&=
\int
\frac{dc\,d^2\boldsymbol{k}}{(2\pi)^3}
\frac{1}{
(c^2+\boldsymbol{k}^2+m^2)
\left[(a\gamma+c)^2+(\boldsymbol{\ell}_\perp+\boldsymbol{k})^2+m^2-a^2\zeta^2\right]
}.
\end{split}
\end{align}
This is now a standard Euclidean-type bubble integral in three dimensions. Introducing a Feynman parameter $s$, combining denominators, and performing the Gaussian integrations gives
\begin{align}
\mathcal{B}(a,\ell)
=
\frac{1}{8\pi}\int_0^1 ds\,
\frac{1}{
\sqrt{
m^2+s(1-s)\boldsymbol{\ell}_\perp^2+(1-s)(s\gamma^2-\zeta^2)a^2+i0
}
}.
\end{align}
The parameter integral may be further evaluated in closed form, yielding
\begin{align}
\mathcal{B}(a,\ell)
=
\frac{1}{32\pi^2\sqrt{a^2\gamma^2+\boldsymbol{\ell}_\perp^2}}
\arctan\!\left(
\frac{\sqrt{a^2\gamma^2+\boldsymbol{\ell}_\perp^2}}
{m+\sqrt{m^2-a^2\zeta^2}}
\right).
\end{align}
For later convenience, it is useful to rewrite the same result in the equivalent representation
\begin{align}
\mathcal{B}(a,\ell)
=
\frac{1}{32\pi^2}\,
M_a
\int_0^1 \frac{ds}{s^2}\,
\frac{1}{\dfrac{M_a^2}{s^2}+P_\ell(a)^2},
\end{align}
where we have defined
\begin{align}
M_a=m+\sqrt{m^2-a^2\zeta^2},
\qquad
P_\ell(a)=\sqrt{a^2\gamma^2+\boldsymbol{\ell}_\perp^2}.
\end{align}
This form is particularly convenient because it separates the $\ell$-dependence in a way that combines naturally with the remaining propagator.

Substituting these ingredients back into the original expression, we are left with
\begin{align}
I^\mu
\propto
\int_{-\infty}^{\infty} da
\int d^2\boldsymbol{\ell}_\perp\,
e^{-i\boldsymbol{\ell}_\perp\cdot\boldsymbol{b}}
\frac{a\,u_1^\mu+\ell^\mu}{\boldsymbol{\ell}_\perp^2+a^2+m^2}\,
\mathcal{B}(a,\ell)\,
K_0\!\left(|\boldsymbol{b}|\sqrt{a^2+m^2}\right).
\end{align}
Now the decomposition of $k_1^\mu$ becomes especially useful. The term proportional to $a\,u_1^\mu$ is odd under $a\to -a$, while the remaining part of the integrand is even in $a$. Consequently, that term integrates to zero. Only the transverse piece survives, and we obtain
\begin{align}
\begin{split}
I^\mu
&\propto
\frac{1}{\sqrt{\gamma^2-1}}
\int_{-\infty}^{\infty} da
\int d^2\boldsymbol{\ell}_\perp\,
e^{-i\boldsymbol{\ell}_\perp\cdot\boldsymbol{b}}
\frac{\ell^\mu}{\boldsymbol{\ell}_\perp^2+a^2+m^2}\,
\mathcal{B}(a,\ell)\,
K_0\!\left(|\boldsymbol{b}|\sqrt{a^2+m^2}\right).
\end{split}
\end{align}
Rotational invariance in the transverse plane then implies that the vector structure must be proportional to $b^\mu$. Using this fact, together with the representation of $\mathcal{B}(a,\ell)$ above, we can write
\begin{align}
\begin{split}
I^\mu
&=
-\frac{i\,b^\mu}{\sqrt{\gamma^2-1}|\boldsymbol{b}|}\partial_{|\boldsymbol{b}|}
\int_0^1\frac{ds}{s^2}
\int_{-\infty}^{\infty} da\,M_a\,
K_0\!\left(|\boldsymbol{b}|\sqrt{a^2+m^2}\right)
\\
&\hspace{2.2cm}\times
\int d^2\boldsymbol{\ell}_\perp\,
e^{-i\boldsymbol{\ell}_\perp\cdot\boldsymbol{b}}
\frac{1}{
(\boldsymbol{\ell}_\perp^2+a^2+m^2)
\left(\boldsymbol{\ell}_\perp^2+a^2\gamma^2+\dfrac{M_a^2}{s^2}\right)}\propto b^\mu \partial_{|\boldsymbol{b}|
}I(b).
\end{split}
\end{align}
The remaining transverse integral is radial. Passing to polar coordinates in the transverse plane gives
\begin{align}
\begin{split}
I
&\propto
-\frac{1}{\sqrt{\gamma^2-1}}
\int_0^1\frac{ds}{s^2}
\int_{-\infty}^{\infty} da\,M_a\,
K_0\!\left(|\boldsymbol{b}|\sqrt{a^2+m^2}\right)
\int d\ell\,\ell\,
\frac{J_0(\ell|\boldsymbol{b}|)}
{(\ell^2+\Lambda_a^2)(\ell^2+\Lambda_{a,s}^2)},
\end{split}
\end{align}
where we have introduced
\begin{align}
\Lambda_a=\sqrt{a^2+m^2}\,,
\qquad
\Lambda_{a,s}=\sqrt{a^2\gamma^2+\frac{M_a^2}{s^2}}\,.
\end{align}
The appearance of the Bessel function $J_0$ is a standard consequence of the angular integration in two dimensions. The radial integral can then be evaluated using known Fourier--Bessel identities, leading to
\begin{align}
\begin{split}
I
&\propto
-2\,\frac{1}{\sqrt{\gamma^2-1}}
\int_0^1\frac{ds}{s^2}
\int_0^\infty da\,M_a\,
\frac{
K_0^2(\Lambda_a|\boldsymbol{b}|)
-
K_0(\Lambda_{a,s}|\boldsymbol{b}|)\,K_0(\Lambda_a|\boldsymbol{b}|)
}{
\Lambda_{a,s}^2-\Lambda_a^2
}.
\end{split}
\end{align}
This is already a rather compact representation: all transverse momentum integrations have been carried out, and the answer is expressed in terms of modified Bessel functions.

Finally, it is convenient to map the semi-infinite $a$-integration domain to the finite interval $u\in[0,1]$ through the change of variables
\begin{align}
a=m\,\frac{\sqrt{1-u^2}}{u}.
\end{align}
After this substitution, the integral becomes
\begin{align}
\begin{split}
I
\propto
\int_0^1
\frac{du}{u^2\sqrt{1-u^2}}
\int_0^1 ds\,
\frac{1+\rho(u)}{(1+\rho(u))^2-s^2\rho(u)^2}
\Biggl[
K_0^2\!\left(\frac{m|\boldsymbol{b}|}{u}\right)
-
K_0\!\bigl(m|\boldsymbol{b}|\,L(u,s)\bigr)
K_0\!\left(\frac{m|\boldsymbol{b}|}{u}\right)
\Biggr],
\end{split}
\end{align}
where
\begin{align}
\rho(u)=\frac{\sqrt{\gamma^2u^2-\zeta^2+i0}}{u},
\qquad
L(u,s)=
\sqrt{
\frac{\gamma^2(1-u^2)}{u^2}
+
\frac{(1+\rho(u))^2}{s^2}
}.
\end{align}
This final form is particularly useful for further analytical study and numerical evaluation, since both integrations now run over finite ranges and the entire tensor structure has been reduced to the expected overall factor $b^\mu$. Although the integral is convergent, the two-loop structure allows the propagator $(k_1+k_2)^2-m^2$ to go on shell. Consequently, the integral acquires both real and imaginary parts. The conservative part of the impulse is extracted from the real part alone. The imaginary part contributes to the radiative dynamics. The dependency on the impact parameter is shown in Fig~\eqref{fig:3.2}. In a similar fashion other two loop asymmetric integrals can be done.
\begin{figure}
    \centering
    \includegraphics[width=0.5\linewidth]{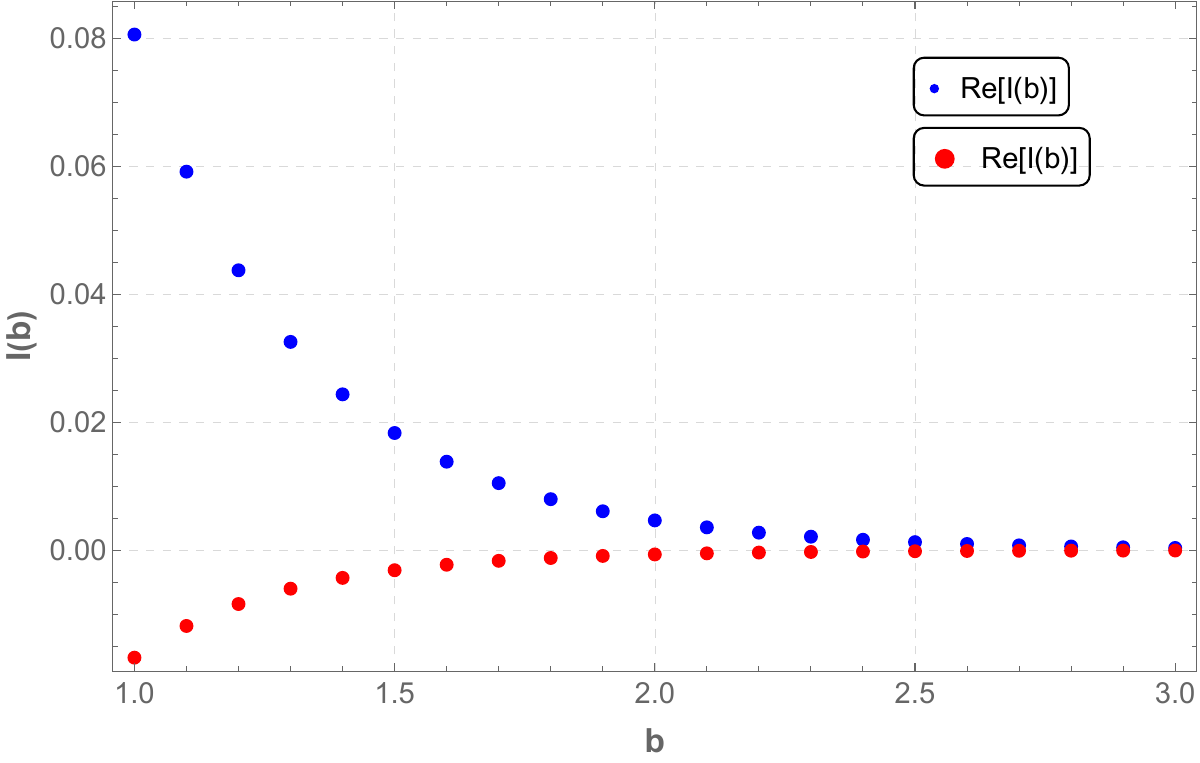}
    \caption{Figure showing dependency of the integral $I(b)$ as a function of $|b|$ for $m=1\,.$}
    \label{fig:3.2}
\end{figure}
\section{Comments on the algebraic properties of the Bessel curve}
\paragraph{One-loop integrals, the Bessel curve, and topological recursion.}
The massive one-loop integrals encountered in our analysis have the schematic form
\begin{align}
    I_{\mathrm{1\mbox{-}loop}}
    \sim
    \sum_r \int_0^1 ds\, \rho_r(s)\, K_0\!\big(\alpha_r(s)\,b\big),
\end{align}
where \(\rho_r(s)\) is a rational function of \(s\). Thus the special-function content is entirely carried by the modified Bessel kernel \(K_0\), which may be viewed as the wavefunction associated with the quantization of the Bessel curve.

We define the classical Bessel curve as the rational spectral curve \cite{DoNorbury2018},
\begin{align}
    x(z)=\frac{z^2}{2},
    \qquad
    y(z)=\frac{1}{z},
    \qquad\Longrightarrow\qquad
    2xy^2-1=0.
\end{align}
We quantize this curve by promoting the classical variables \(x,y\) to non-commuting operators
\begin{align}
    \hat x=x,
    \qquad
    \hat y=\hbar\,\partial_x,
    \qquad
    [\hat x,\hat y]=\hbar.
\end{align}
With the symmetric ordering prescription, the quantum curve associated with
\(
H(x,y)=2xy^2-1=0
\)
is
\begin{align}
    \hat H\,\psi=(2\hat y\,\hat x\,\hat y-1)\psi=0.
\end{align}
In the \(x\)-polarization this becomes
\begin{align}
    \left(2(\hbar\partial_x)\,x\,(\hbar\partial_x)-1\right)\psi(x)=0.
\end{align}
Rewriting the equation in terms of the spectral parameter \(z\), with \(x=z^2/2\), one finds
\begin{align}
    \hbar^2 z^2\psi_0''(z)+\hbar^2 z\psi_0'(z)-z^2\psi_0(z)=0.
\end{align}
This is precisely the modified Bessel equation of order zero, and its decaying solution is
\begin{align}
    \psi_0(z)\propto K_0\!\left(\frac{z}{\hbar}\right).
\end{align}

The same wavefunction can be reconstructed from the Eynard--Orantin topological recursion on the Bessel curve. In the quantum-curve/WKB formalism one writes
\begin{align}
    \psi_0(z,\hbar)
    =
    \exp\!\left(
        \frac{1}{\hbar}S_0(z)+S_1(z)+\hbar S_2(z)+\hbar^2 S_3(z)+\cdots
    \right),
\end{align}
where
\begin{align}
    S_0(z)=\int^z y(z')\,dx(z'),
    \qquad
    S_1(z)=-\frac{1}{2}\log\frac{dx}{dz},
\end{align}
and, for \(k\ge 2\),
\begin{align}
    S_k(z)
    =
    \sum_{2g-1+n=k}
    \frac{(-1)^n}{n!}
    \int_{\infty}^{z}\cdots\int_{\infty}^{z}
    \omega_{g,n}(z_1,\ldots,z_n).
\end{align}
Here \(\omega_{g,n}\) are the symmetric meromorphic multidifferentials on \((\mathbb{CP}^1)^n\) produced recursively by the topological recursion. For the Bessel curve one finds, on the decaying branch,
\begin{align}
    S_0(z)=-z,
    \qquad
    S_1(z)=-\frac{1}{2}\log z,
    \qquad
    S_2(z)=-\frac{1}{8z},
    \qquad
    S_3(z)=\frac{1}{16z^2},
\end{align}
up to additive constants that only affect the overall normalization. Therefore
\begin{align}
    \psi_0(z,\hbar)
    \sim
    \exp\!\left(
        -\frac{z}{\hbar}
        -\frac{1}{2}\log z
        -\hbar\frac{1}{8z}
        +\hbar^2\frac{1}{16z^2}
        +\cdots
    \right).
\end{align}
Equivalently,
\begin{align}
    \psi_0(z,\hbar)
    \sim
    z^{-1/2}e^{-z/\hbar}
    \left(
        1-\frac{\hbar}{8z}
        +\frac{9\hbar^2}{128z^2}
        +O\!\left(\frac{\hbar^3}{z^3}\right)
    \right),
\end{align}
which matches the standard large-argument asymptotic expansion of the modified Bessel function,
\begin{align}
    K_0\!\left(\frac{z}{\hbar}\right)
    \sim
    \sqrt{\frac{\pi\hbar}{2z}}\,
    e^{-z/\hbar}
    \left(
        1-\frac{\hbar}{8z}
        +\frac{9\hbar^2}{128z^2}
        +O\!\left(\frac{\hbar^3}{z^3}\right)
    \right).
\end{align}
Thus the asymptotic expansion of the Bessel kernel may be recovered directly from the topological-recursion data of the Bessel curve.

In our one-loop integrals, the kernel appears in the combination
\begin{align}
    K_0\!\big(\alpha_r(s)\,b\big),
\end{align}
so each term may be interpreted as the Bessel-curve wavefunction evaluated at the parameter-dependent point
\begin{align}
    z=\alpha_r(s)\,b,
    \qquad
    x=\frac{z^2}{2}=\frac{\alpha_r(s)^2b^2}{2}.
\end{align}
Hence the one-loop amplitude is naturally viewed as an integral over parameter space of wavefunctions of the quantized Bessel curve.

At this point it is important to identify the correct semiclassical parameter. The integration variable \(s\) should not itself be identified with \(\hbar\). Rather, the natural semiclassical regime of the full integral is the large-impact-parameter limit \(b\to\infty\) with \(m\) fixed. In this regime one may introduce the effective small parameter
\begin{align}
    \hbar_{\mathrm{eff}}:=\frac{1}{mb},
\end{align}
so that, schematically,
\begin{align}
    K_0\!\big(\alpha_r(s)\,b\big)
    =
    K_0\!\left(\frac{\alpha_r(s)}{m}\,\frac{1}{\hbar_{\mathrm{eff}}}\right).
\end{align}
Thus the one-loop integral is semiclassical in the limit \(\hbar_{\mathrm{eff}}\to0\), namely for \(mb\gg1\). In particular, for a kernel of the form
\begin{align}
    K_0\!\left(\frac{2mb}{s}\right),
\end{align}
the argument is large throughout the large-\(b\) regime and becomes even larger near the endpoint \(s\to0\). Therefore the WKB expansion of the Bessel wavefunction yields a controlled large-\(b\) asymptotic expansion of the integral:
\begin{align}
    K_0\!\left(\frac{2mb}{s}\right)
    \sim
    \sqrt{\frac{\pi s}{4mb}}\,
    \exp\!\left(-\frac{2mb}{s}\right)
    \left(
        1-\frac{s}{16mb}
        +\frac{9s^2}{512m^2b^2}
        +O\!\left(\frac{s^3}{(mb)^3}\right)
    \right).
\end{align}
This makes clear that the role of the Bessel curve is to supply the local semiclassical structure of the kernel, while the large-impact-parameter limit controls the asymptotics of the full one-loop integral.\\
    }%
  }%

  \restorethesisbodyformat
  \restoremainchapterstyle
  \chapter{Spinning Two-Body Problem in dCS Gravity Using WQFT}
  \thesischapterpaperbox{Bootstrapping the spinning two body problem in dynamical Chern-Simons gravity using worldline QFT}{A.~Bhattacharyya, D.~Ghosh, S.~Ghosh, and S.~Pal}{\href{https://doi.org/10.1007/JHEP04(2025)175}{\textit{JHEP} \textbf{04} (2025), 175}; arXiv: \arxivlink{2407.07195}{hep-th}}
  {%
    \restorethesisbodyformat
    \renewcommand{\appendix}{%
      \setcounter{section}{0}%
      \setcounter{subsection}{0}%
      \setcounter{subsubsection}{0}%
      \renewcommand{\thesection}{\thechapter.\Alph{section}}%
      \renewcommand{\thesubsection}{\thesection.\arabic{subsection}}%
      \renewcommand{\theHsection}{chapter.\arabic{chapter}.appendix.\Alph{section}}%
      \renewcommand{\theHsubsection}{\theHsection.\arabic{subsection}}%
    }%
    \ifstrempty{chap3_body.tex}{}{%
      \section{Introduction}
The analysis presented in this chapter is based on Ref.~\cite{Bhattacharyya:2024kxj}.
Building on the previous chapter—where we used the WQFT framework to analyze hyperbolic scattering beyond GR in a scalar--tensor setup and focused on observables such as impulse and waveform—this chapter turns to a different (and intrinsically spin-sensitive) deformation of Einstein gravity: \emph{parity-violating} dynamical Chern--Simons (dCS) gravity. Observationally, compact-object encounters remain a prime target for current and future gravitational-wave experiments. The network of ground-based detectors has established gravitational-wave astronomy in the high-frequency band \cite{LIGOScientific:2014oec,LIGOScientific:2016aoc,LIGOScientific:2016sjg,LIGOScientific:2016vlm,LIGOScientific:2017bnn,LIGOScientific:2019hgc}, and these tests will strengthen further with the next generation of ground- and space-based detectors \cite{hyp10}. Complementarily, pulsar timing arrays probe the nano-Hz window \cite{NANOGrav:2023gor,EPTA:2023fyk,NANOGrav:2023wsz,Verbiest:2024nid}, where low-frequency radiation from wide systems and burst-like signals from hyperbolic encounters can become accessible. In addition to astrophysical information, such measurements sharpen constraints on deviations from GR, particularly in regimes where strong-field dynamics and radiation are simultaneously relevant.

From the effective-theory viewpoint, GR is expected to receive corrections at higher energies/curvatures. In Weinberg’s EFT approach \cite{PhysRev.138.B988}, one systematically augments the Einstein--Hilbert action by higher-derivative operators. One may also extend the gravitational sector by introducing additional degrees of freedom (scalars, vectors, \emph{etc.}) \cite{Damour:1992we,Horbatsch:2015bua,Schon:2021pcv,Rainer:1996gw,DeFelice:2011bh}, or consider both mechanisms at once. Two widely studied higher-curvature extensions are Einstein--dilaton--Gauss--Bonnet (EdGB) and dynamical Chern--Simons (dCS) gravity \cite{Zwiebach:1985uq,Gross:1986mw,Jackiw:2003pm}. The EdGB interaction naturally appears in the low-energy effective action of heterotic string theory via a scalar non-minimally coupled to the Gauss--Bonnet invariant \cite{Zwiebach:1985uq,Gross:1986mw}. Dynamical Chern--Simons gravity instead introduces a parity-violating correction through a scalar coupled to a topological quadratic curvature invariant, yielding an effective CP-violating deformation of four-dimensional GR and featuring prominently in string-motivated discussions \cite{Alexander:2004us,Alexander:2004xd}. The crucial point for the present chapter is that, for \emph{non-spinning} black holes, the parity-violating dCS sector does not correct the conservative scattering dynamics at leading orders; hence \emph{spin is essential} to expose the dCS imprint in classical observables. This chapter therefore focuses on \emph{spinning} black-hole scattering in dCS gravity.

Our target observable is the \emph{spinning eikonal}, computed through \emph{3PM} order. This choice is motivated by two considerations. First, within the post-Minkowskian expansion (an expansion in Newton’s constant valid to all orders in velocity), the eikonal phase provides a compact encoding of conservative scattering data, directly tied to gauge-invariant quantities such as the scattering angle and its spin dependence (and hence relevant for EOB connections \cite{Buonanno:1998gg}). Second, already at 3PM the problem probes genuinely multiloop structures, making it a sharp testing ground for both formalism and technology.

High-precision results in GR—especially for waveform modeling and radiation reaction—have a vast literature \cite{Blanchet:2013haa,Schafer:2018kuf,Buonanno:1998gg,Blanchet:2004ek,Blanchet:2006gy,Blanchet:2023bwj,Blanchet:2023sbv,Blanchet:2023soy,Warburton:2024xnr,Levi:2018nxp,Wardell:2021fyy}, driven in part by template accuracy requirements \cite{Purrer:2019jcp}. In parallel, classical two-body observables have been efficiently extracted using QFT-inspired methods: direct amplitude-based approaches \cite{Brandhuber:2022qbk,Brandhuber:2023hhy,Kosower:2018adc,DeAngelis:2023lvf,Bjerrum-Bohr:2018xdl,Bjerrum-Bohr:2013bxa,Alessio:2024wmz,Brunello:2024ibk,Bern:2019crd,Bern:2019nnu,Bern:2020buy,Bern:2024adl,Bern:2023ity,Bern:2022kto,Bern:2021yeh,Bern:2021dqo,Bern:2020gjj,Brandhuber:2019qpg,AccettulliHuber:2019jqo,AccettulliHuber:2020oou,Barack:2023oqp,Bjerrum-Bohr:2023iey,Bjerrum-Bohr:2023jau,Bjerrum-Bohr:2022ows,Bjerrum-Bohr:2022blt,Bjerrum-Bohr:2021wwt,Bjerrum-Bohr:2021vuf,Bjerrum-Bohr:2020syg,Bjerrum-Bohr:2019kec,Cristofoli:2019neg,Bjerrum-Bohr:2016hpa,Chen:2024mmm,Alessio:2022kwv,Alessio:2023kgf,ashoke1,ashoke5,Gonzo:2024zxo,Aoude:2023vdk,Aoude:2023dui,Brandhuber:2023hhl,Adamo:2024oxy,Adamo:2023cfp,Adamo:2022ooq,Adamo:2021rfq,Georgoudis:2023eke,Georgoudis:2024pdz,Bini:2024rsy}\footnote{For a broad overview and further references, see \cite{DiVecchia:2023frv}.},
worldline/QFT hybrids such as WQFT \cite{Mogull:2020sak,Jakobsen:2021lvp,Jakobsen:2021zvh,Jakobsen:2023ndj,Jakobsen:2023hig,Jakobsen:2022psy,Jakobsen:2023oow,Wang:2022ntx,Klemm:2024wtd,Bhattacharyya:2024aeq,Driesse:2024xad,Adamo:2024oxy,DeAngelis:2023lvf,Cristofoli:2021vyo},
direct field-equation approaches \cite{1978ApJ...224...62K,Damour:1992we,Damour:2016gwp,Damour:2019lcq,Bini:2020rzn,Damour:2022ybd,Bini:2024ijq,DeVittori:2014psa,hyp16},
and EFT-based methods \cite{Porto:2007pw,Porto:2008jj,Porto:2012as,Levi:2018nxp,Cheung:2024jpo,Cheung:2023lnj,Cheung:2020gyp,Ivanov:2024sds,Bhattacharyya:2023kbh,Huang:2018pbu,Diedrichs:2023foj,Loebbert:2020aos,Mougiakakos:2021ckm,Riva:2021vnj,Mougiakakos:2022sic,Riva:2022fru,Bernard:2023eul,Porto:2007px,Porto:2017dgs,Dlapa:2024cje}.
In this chapter we adopt the WQFT viewpoint, leveraging the operator/S-matrix correspondence established by Mogull--Plefka--Steinhoff \cite{Mogull:2020sak,Jakobsen:2021lvp}, which has proven effective for computing classical scattering observables while bypassing certain bookkeeping subtleties that arise in more traditional amplitude extractions. The same philosophy has already enabled explorations beyond GR in scalar--tensor theories \cite{Bhattacharyya:2023kbh}; here we apply it to a higher-curvature, parity-violating deformation.

The genuinely new ingredient is \emph{spin}. In WQFT, spin is incorporated through additional anti-commuting worldline variables $\psi^{a}$, whose bilinears realize the spin tensor, schematically $S^{ab}\sim\bar\psi^{[a}\psi^{b]}$. Jakobsen \emph{et al.} showed that the spinning worldline theory underlying GR enjoys a hidden $\mathcal N=2$ structure \cite{Jakobsen:2021lvp,Jakobsen:2021zvh}, which constrains the couplings and clarifies the relationship to standard spinning worldline actions used in EFT treatments. For scattering, this formulation is particularly natural: classical conservative dynamics is efficiently packaged by the \emph{spinning eikonal}, and in the present dCS context it is precisely the spin sector that activates the parity-violating correction. A qualitative theme we emphasize is that the resulting dCS contribution can depend sensitively on spin orientation in the scattering configuration—an imprint of the CP-violating structure \cite{Yunes:2009hc}.

Technically, pushing PM calculations to higher orders confronts the multiloop integration bottleneck. Recent progress has revealed that high-PM integrals can probe rich analytic structures and connect to modern mathematical tools: intersection theory \cite{Mastrolia:2018uzb,Brunello:2023rpq,Brunello:2023fef}, Calabi--Yau geometries \cite{Frellesvig:2023bbf,Klemm:2024wtd}, modular graph forms \cite{Dorigoni:2022npe}, and function classes extending beyond multiple polylogarithms \cite{Weinzierl:2007cx}. In this chapter, we work at 3PM and develop a practical analytic pipeline based on integration-by-parts (IBP) reduction and differential equations for master integrals, which suffices to obtain closed-form control over the spinning eikonal in the dCS deformation.

The chapter is organised as follows. In Section~(\ref{ch3:sec2}) we formulate spinning WQFT in the presence of the effective parity-violating (dCS) interaction and derive the corresponding Feynman rules, including the vertices involving the extra spin degrees of freedom. Section~(\ref{ch3:sec3}) develops the technical tools used throughout, focusing on IBP reduction and differential-equation methods for the relevant multiloop master integrals. In Section~(\ref{ch3:sec4}) we apply these ingredients to compute the spinning eikonal through 3PM and connect it to the exponentiated WQFT partition function. Useful integral details are collected in the appendices.
\section{Chern-Simons gravity and worldline QFT}
\label{ch3:sec2}
In this section, we will review the dynamical Chern-Simons (dCS) model and derive all relevant Feynman rules. We also comment on the main differences between the higher curvature EFTs and Einstein GR.
\subsection*{Bulk gravitational action:}
The bulk action for dCS theory is given by,
\begin{align}
    \begin{split}
        S_{\textrm{dCS}}=\rmint d^4 x\sqrt{-g}\Big[-\frac{m_p^2}{2}R+\frac{1}{2}g^{\alpha\beta}\partial_{\alpha}\varphi\partial_{\beta}\varphi-\frac{l_{dCS}^2}{m_p^3}\varphi\, {}^{*}R R\Big]\,.
    \end{split}
\end{align}
Now we expand the bulk gravitational action around flat metric as,
\begin{align}
    \begin{split}
        g_{\mu\nu}=\eta_{\mu\nu}+\frac{h_{\mu\nu}}{m_p},\,\, \varphi\to 0+\varphi,
    \end{split}
\end{align}
Hence, the Chern-Simons term can be expanded in order of fluctuation $h$ as,
\begin{align}
    *RR=\epsilon^{\chi\varepsilon\mu\nu}\Big(\partial_\mu\partial_\beta h_{\nu}^{\,\sigma}-\partial_\mu \partial^\sigma h_{\nu\beta}\Big)\Big(\partial_{\chi}\partial_{\sigma}h_{\delta}^{\,\beta}-\partial_{\chi}\partial^{\beta}h_{\delta\sigma}\Big)+\mathcal{O}(h^3).
\end{align}
Before going to the worldline action for the higher curvature EFTs, we first briefly review the case of General relativity.
\subsection*{SUSY worldline action in General relativity:}
As we will be considering spinning binaries in this paper, following \cite{Jakobsen:2021zvh}, we start with the $\mathcal{N}=2$ supersymmetric worldline action. It  has the following form:
\begin{align}
    \begin{split}
        S_{\mathcal{N}=2}=-\sum_{k=1}^2 
        \int d\tau\, \Big[\frac{1}{2e}g_{\mu\nu}\dot x_k^{\mu}\dot x_k^{\nu}+i\bar \psi_{k,a}\frac{\mathtt{D}\psi^a_k}{\mathtt{D}\tau}+\frac{e}{2}R_{abcd}\,\bar \psi^a_k \psi^b_k \bar \psi^c_k \psi^d_k+\frac{e}{2}m_k^2\Big]
    \end{split}\label{2.4a}
\end{align}
where,
\begin{align}
    \frac{\mathtt{D}\psi ^a_k}{\mathtt{D}\tau}= \dot\psi_k^a+\dot x^\mu\, \omega_{\mu}^{\,\,ab}\psi_{k,b}\,.
\end{align}
Moreover, in a convenient fashion, we can make the following gauge choice $e=1/m_k$ and make a rescaling of the fermions \cite{Jakobsen:2021zvh}:
\begin{align}
    \psi_k^a \to \sqrt{m_k}\,\psi_k^a,\,\,\bar\psi_k^a \to \sqrt{m_k}\,\bar\psi_k^a,
\end{align}
The action in \eqref{2.4a} is invariant under the following asymptotic $\mathcal{N}=2$ supersymmetry transformation.
\begin{align}
    \begin{split}
    &    \delta x^\mu=ie^{\mu}_a\,(\bar\epsilon\psi^a+\epsilon\bar\psi^ a),\,\,\,\delta \psi^a=-\epsilon e^{a}_\mu \dot x^\mu-\delta x^\mu\,{\omega_{\mu}}^{\,\,a}_{\,\,\,\,\,\,\,b}\,\psi^b\\ &
    \textrm{and,}\,\textcolor{black}{\delta}\bar\psi^a=-\bar\epsilon e^{a}_\mu \dot x^\mu-\delta x^\mu\,{\omega_{\mu}}^{\,\,a}_{\,\,\,\,\,\,\,b}\,\bar\psi^b.
    \end{split}
\end{align}
It also has global $U(1)$ symmetry.
\begin{align}
    \delta_{\hat\epsilon}\psi^a=i\hat\epsilon \psi^a,\,\,\delta_{\hat\epsilon}\bar\psi^a=i\hat\epsilon \bar\psi^a,\,\,\delta_{\hat\epsilon}x^\mu=0.
\end{align}
The Noether charges derived from these global transformations give different conservation laws along the worldline \cite{Jakobsen:2021zvh}.\par
In a perturbative framework, one can not distinguish between tetrad indices $(a,b,..)$ and curved indices $(\mu,\nu,..)$. In order to describe the scattering process, one can expand worldline DOF around undeflected trajectories as,
\begin{align}
    \begin{split}
        x_k^\mu\to b_k^\mu+v^\mu_k\tau+z_{k}^{\mu}(\tau),\,\,\, \psi^{a}_k(\tau)\to\Psi_k^{a}+{\psi}_k^a(\tau)\,.
    \end{split}
\end{align}
One can also introduce the constant spin tensor as,
\begin{align}
    \begin{split}
        \mathcal{S}_k^{ab}=-2i\bar \Psi_k^{[a}\Psi_k^{b]}\,.
    \end{split}
\end{align}
From the quadratic part of the action, one can find out the field and worldline propagators,
\begin{align}
    \begin{split}
     &    \begin{minipage}[h]{0.075\linewidth}
	\vspace{4pt}
	\scalebox{1.8}{\includegraphics[width=\linewidth]{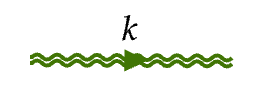}}
\end{minipage}\hspace{0.45cm}=\frac{iP_{\mu\nu;\rho\sigma}}{k^2+i\varepsilon},\,\begin{minipage}[h]{0.075\linewidth}
	\vspace{4pt}
	\scalebox{1.6}{\includegraphics[width=\linewidth]{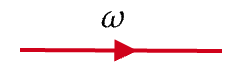}}
\end{minipage}\hspace{0.45cm}=-\frac{i \eta^{\mu\nu}}{2m_k}\left(\frac{1}{(\omega+i\varepsilon)^2}+\frac{1}{(\omega-i\varepsilon)^2}\right),\\ &
 \begin{minipage}[h]{0.075\linewidth}
	\vspace{4pt}
	\scalebox{1.8}{\includegraphics[width=\linewidth]{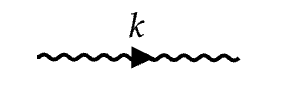}}
\end{minipage}\hspace{0.45cm}=\frac{i}{k^2+i\varepsilon},\,
        \begin{minipage}[h]{0.075\linewidth}
	\vspace{4pt}
	\scalebox{1.6}{\includegraphics[width=\linewidth]{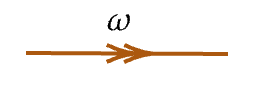}}
\end{minipage}\hspace{0.45cm}=-\frac{i \eta^{\mu\nu}}{2m_k}\left(\frac{1}{\omega+i\varepsilon}+\frac{1}{\omega-i\varepsilon}\right).
    \end{split}
\end{align}
\textcolor{black}{Note that in this paper, we intend to compute the Eikonal phase, and hence we choose a time-symmetric/Feynman propagator. However, if one wants to compute the observables, then one needs to choose the retarded propagators, which gives the causally sensible classical solutions.}

One can eventually expand the tetrad and spin connection as,
{\small
\begin{align}
    \begin{split}
& e^a_{\,\mu}=\eta^{a\nu}\Big(\eta_{\mu\nu}+\frac{1}{2m_p}h_{\mu\nu}-\frac{1}{8m_p^2}h_{\mu\rho}h^{\rho}_{\,\nu}+\mathcal{O}(1/m_p^3)\Big)\,,\\ &
\omega_{\mu}^{\,\,\,ab}=-\frac{1}{m_p}\partial^{[a}h^{b]}_{\,\,\,\mu}-\frac{1}{2m_p^2}h^{\nu[a}\left(\partial^{b]}h_{\mu \nu}-\partial_\nu h^{b]}_{\,\,\,\,\mu}+\frac{1}{2}\partial_\mu h^{b]}_{\,\,\,\,\nu}\right)+\mathcal{O}(1/m_p^3).
    \end{split}
\end{align}
}
\subsection*{What happens for modified theories of gravity?}
In some EFT of gravity involving extra scalar degrees of freedom the scalar dipole moment must couple to the scalar field in the worldline Lagrangian. One way, mentioned in \cite{Loutrel:2018ydv}, is to demand the new canonical momenta $\mathcal{P}^\mu$ is now not only a function of graviton degrees of freedom but also a function of the extra degrees of freedom:
\begingroup
\small
\begin{align}
    \begin{split}
        \mathcal{P}_\mu=p_{\mu}+(\textrm{pure gravity contributions})+\frac{\partial \mathcal{P_{\mu}}}{\partial{\varphi}}\Bigg|_{\varphi=0}\,\varphi+\frac{\partial \mathcal{P_{\mu}}}{\partial{\nabla_{\alpha}\varphi}}\Bigg|_{\nabla_\alpha\varphi=0}\,\nabla_{\alpha}\varphi+\cdots\label{2.13k}
    \end{split}
\end{align}
\endgroup
In dCS theory, we have an extra bulk degree of freedom (in the form of a scalar field) along with graviton, and so we need to promote the masses of the black holes as a function of that scalar field $\varphi$, which  does not preserve the SUSY invariance for the following re-scaled fermions as,
\begingroup
\small
\begin{align}
    \psi_k^a \to \sqrt{m_k(\varphi)}\,\psi^a,\,\,\bar\psi_k^a \to \sqrt{m_k(\varphi)}\,\bar\psi^a,
\end{align}
\endgroup
 Extrapolating \cite{Loutrel:2018ydv}, the worldline action takes the form (up-to linear in spin),
\begin{align}
    \begin{split}
        \mathcal{S}_k\subset -\int d\tau \Big[\frac{m_k(\varphi)}{2}g_{\mu\nu}\dot x_k^\mu \dot x_k^\nu+i m_{k}(\varphi)\bar\psi_a^k\, D_{\tau}\psi^a_k-i\,\mathcal{C}_{dCS} \,\dot{x}^{\mu} \nabla_{\alpha}\varphi\,\epsilon_{\mu\,\,\,\rho\,\sigma}^{\,\,\alpha}\bar{\psi}_k^a\,\psi_k^{b}e^{\rho}_a e^{\sigma}_{b}\Big]\label{2.13d}
    \end{split}
\end{align}
where $\mathcal{C}_{dCS}=\frac{\partial \mathcal{P_{\mu}}}{\partial{\nabla_{\alpha}\varphi}}\Bigg|_{\nabla_\alpha\varphi=0}$ is an undetermined coefficient and refers to as scalar-dipole constant and is dependent on the coupling constant(s) of the theory. However, in our computation, we will ignore the finite size corrections. Hence, we can avoid the dipole term from the worldline action.  \par 
Now expanding the mass around $\varphi=0$, we get \footnote{In \cite{Wilson-Gerow:2025xhr}, the author asserts that the dynamical Chern-Simons (dCS) term contributes to the spinless scattering angle by treating the sensitivity parameters as functions of the theory's coupling. However, some contradictory claims regarding our computation have been made in \cite{Wilson-Gerow:2025xhr}, and we would like to provide a few clarifications to avoid any potential confusion. First and foremost, we do not compute the scattering angle, contrary to the assertion in \cite{Wilson-Gerow:2025xhr}. Moreover, deriving the scattering angle from the Eikonal phase is nontrivial at 3PM order and requires the careful considerations discussed in Section~\eqref{con}. Second, as we understand it, the author's approach primarily relies on the on-shell behaviour of the scalar field in the far region to establish a link between the sensitivity parameters and the couplings of the underlying theory. While it is true that all parameters in the model are implicitly (or explicitly) dependent on the theory's coupling, the precise functional relationship between these parameters and the theory itself remains ambiguous. For this reason, we adopt a specific prescription in our analysis, treating these parameters as independent. 
Our focus is on the diagrams where the explicit coupling of the dCS term is directly incorporated (and it can be easily seen that it contributes only for spinning case due its structure), and we defer the investigation of potential indirect effects emerging from the broader theoretical framework to future work. If one wishes to carry out the computation using an alternative prescription, one need only apply the transformation:  
   $ \{s_i, g_i\} \to \{s_i(l_{\textrm{dCS}}), g_i(l_{\textrm{dCS}})\}$
where the sensitivity parameters and couplings are explicitly treated as functions of the dCS coupling and get the appropriate results from the expressions provided in our paper. Also, in the \textit{spinless} part of the eikonal there are contributions form UT-2 polynomial as presented in \eqref{dCSne3waaa} which seems to be absent from the result presented in \cite{Wilson-Gerow:2025xhr}.  },
\begin{align}
    \begin{split}
        m_{k}(\varphi)=m_{k}(0)\Big(1+s_k \varphi +\cdots)
    \end{split}
\end{align}
Upon applying the SUSY transformation, we get, $\delta m_{k}(\varphi)=s_k \partial_{\mu}\varphi \delta x^\mu+\cdots$ and demanding that the sensitivity parameter $s_1$ is very small, we get $\delta m_k(\varphi)\approx \mathcal{O}(s_1)$ and the SUSY is preserved in an approximate sense. However, the breaking of SUSY is not a problem as the action in \eqref{2.13d} eventually recovers the standard spinning worldline action with dynamical mass \cite{Levi:2018nxp}. 
\par
\par
\subsubsection*{Primary differences between GR and modified EFTs:}
For GR, following \cite{Jakobsen:2021zvh}, one can argue that the worldline action is invariant under the SUSY transformation along the entire trajectory of the black holes as well as in the asymptotic regions, which eventually implies that $p_k^2,\ p_k\cdot \bar\psi_k,\,p_k\cdot \psi_k,\,\textrm{and,}\,\psi_k\cdot \bar\psi_k$ (asymptotic charges) are conserved between the initial and final asymptotic states. Using those constraints together with the extra condition $v_k\cdot \Psi_k=v_k\cdot \bar\Psi_k=0$ then implies that $p_{k,\mu} \cdot \mathcal{S}_k^{\mu\nu}$ is conserved throughout the scattering event, and without loss of generality we can choose it to vanish, which essentially recovers the SSC \cite{1967JMP.....8.1591D,Corinaldesi:1951pb,1964NCim...34..317D}:
\begin{align}
    \textrm{\bf SSC}: \, p_{k,\mu} \cdot \mathcal{S}_k^{\mu\nu}=0,\,\, \textrm{with,}\,\,
     p_k^\mu=m_k v_k^\mu+\mathcal{O}(\mathcal{S}^2).
    \label{SSC17}
\end{align}
\textcolor{black}{In our case, the worldline SUSY invariance is broken (or only approximate if we demand that the bulk scalar varies very slowly with respect to the proper time), and hence the conservation of supercharges is only approximate. As a consequence, the conservation of $p_k\cdot \mathcal{S}_k^{\mu\nu}(\tau)\,$ is also only approximate. Thus, the SSC throughout the dynamical process is only approximate, as shown in \cite{Loutrel:2018ydv}\footnote{If one chooses $m_k(\varphi)$ to be the renormalized mass such that it satisfies $D_{\tau} m_{k}(\varphi)=\mathcal{O}(S^4)$, then the SUSY invariance is recovered. Interested readers are referred to Appendix (B) of \cite{Loutrel:2018ydv}.}. However, the theory still enjoys asymptotic (background) SUSY invariance, since there is no gravity in the asymptotic states. Following \cite{Loutrel:2018ydv}, the SSC defined in \eqref{SSC17} holds for the asymptotic states, i.e. $v_{k,\mu} \cdot \mathcal{S}_k^{\mu\nu}(-\infty)=0$, which we will use in the subsequent computations, and the observables or eikonal depend only on the asymptotic parameters. Therefore, we can still use the WQFT formalism to compute the scattering observables. We will also work to linear order in the spin and in the small-coupling ($l_{dCS}$) limit so that we can replace the canonical momenta $p_k^\mu\to m\,v_k^\mu$. 
}
\par
\subsection*{Main goal and computational procedure:}
In WQFT formalism $h_{\mu\nu}(x),\varphi(x),x^\mu(\tau),\psi^\mu(\tau)$ are promoted to the quantum operators. For this case, the partition function looks like,
\begin{align}
    \begin{split}
        Z_{\textrm{S-WQFT}}\equiv e^{i\chi}=\mathcal{N}&\times\rmint \boldsymbol{\mathcal{D}}[h_{\mu\nu},\varphi]\rmint \prod_{i=1}^n \boldsymbol{\mathcal{D}}[z_i,\psi_i,\bar{\psi}_i,\mathfrak{a}_i,\mathfrak{b}_i,\mathfrak{c}_i]\,,\\ &
        \times \exp\Big[i\Big(S_{EH}+S_{\textrm{dCS}}+\sum_i (S_{\textrm{sp}}^{i}+S^{i}_{\textrm{ghost}})\Big)\Big],
    \end{split}
\end{align}
where, $\chi$ is the eikonal phase and the extra ``ghost'' term comes from the metric dependent worldline measure,
\begin{align}
    \begin{split}
       \boldsymbol{ \mathcal{D}}[x]&=D[x]\prod_{0\le \sigma\le T}\sqrt{-g[x(\sigma)]}\,,\\ &
        =D[x]\underbrace{\rmint \boldsymbol{\mathcal{D}}[\mathfrak{a},\mathfrak{b},\mathfrak{c}]\exp \Big[-i\rmint d\tau\Big(\frac{1}{2}g_{\mu\nu}(\mathfrak{a}^\mu\mathfrak{a}^\mu+\mathfrak{b}^\nu\mathfrak{c}^\nu)\big)\Big]}_{\textrm{ghost term}}\,.
    \end{split}
\end{align}
In the classical computations, the ghost field does not appear and can be ignored. The relevant question, then, is how the WQFT partition function is related to the scattering amplitude. It has been shown that in the classical limit the eikonal phase is given by \cite{Amati:1987wq,Amati:1990xe},
\begin{align}
    e^{i\chi}=\frac{1}{4m_1 m_2}\int  \frac{d^4 q}{(2\pi)^4} \, \hat\delta(q\cdot v_1)\hat\delta(q\cdot v_2) e^{iq\cdot b}\,\langle \,\textrm{final}\,|\mathbb{S}|\,\textrm{initial}\,\rangle\xrightarrow[]{} Z_{\textrm{WQFT}}\label{2.18a}
\end{align}
So, the WQFT partition function is equivalent to computing the eikonal phase, which is nothing but the $2\to2$ scattering amplitude, Fourier transformed into impact parameter space.
Now, we have all the ingredients to compute the Feynman rules. Our primary goal in this paper is to compute the eikonal phase for the spinning binaries, especially to investigate the contribution due to the extra degrees of freedom (scalar field) present in the theory and see how the dCS modifies the phase. 
We will mainly concentrate on the spin part of the worldline action as the spinless part (from the scalar field) has already been done in \cite{Bhattacharyya:2024aeq}. For simplicity, we only focus on the terms where spins are linear. Now, to read off the interaction vertices from the supersymmetric worldline action, write the bulk fields in Fourier space.
\begin{align}
    \begin{split}
        \chi(x_i)=\sum_{n-0}^{\infty}\frac{i^n}{n!}\rmint_{k,\omega_1,\cdots \omega_n}e^{ik\cdot b_i}e^{i(k\cdot v_i+\sum_{j=1}^{n}\omega_j)\tau}\Big(\prod_{j=1}^n k\cdot z_i(-\omega_j)\Big)\chi(-k)
    \end{split}
\end{align}
and the worldline fields can be written in energy space as,
\begin{align}
    \begin{split}
        z_{i}^{\mu}(\tau)=\rmint_{\omega} e^{i\omega \tau}z_i^\mu{(-\omega)},\,\,\,\,\,\,\breve{\psi}_i^\mu(\tau)=\rmint_{\omega}e^{i\omega \tau}\breve{\psi}_i^\mu(-\omega)\,.
    \end{split}
\end{align}
The linear spin term in the worldline action takes the following form,
\begin{align}
    \begin{split}
        S_{sp}\sim& -i\rmint d\tau \Big(m_k(0)+\frac{m_k s_k}{m_p} \varphi+\frac{m_k g_k }{m_p^2}\varphi^2\Big)\Big(\bar\psi_{k,a}\dot \psi_k^a+\dot x_k^\mu \omega_{\mu}^{\,\,ab}\bar \psi_{k,a} \psi_{k,b}\Big)\\ &+(\textrm{dCS contribution to the worldline action})\,.\label{ch3:2.13}
    \end{split}
\end{align}
Note that in \eqref{ch3:2.13} at leading order ($\mathcal{O}(\varphi^0)$) we can add a total derivative term $-1/2\, d_{\tau}(\bar\psi^a\,\psi_a)$ which preserves the SUSY transformation. But when we go to the linear (or higher order in $\varphi$), this addition will be important so that we can have both $\varphi-\psi$ and $\varphi-\bar\psi$ vertices. Again, as mentioned earlier, the SUSY invariance is approximate, and we restrict ourselves to the linear order of $\varphi\,.$ 
Keeping terms that are linear in $\varphi$ we have the following,
\begin{align}
    \begin{split}
        S_{sp}\subset -i\frac{m_i s_i}{2 m_p}\int d\tau\,\varphi\, \,\,\Big(\bar\psi _a \dot \psi^a-\dot{\bar\psi}_a \psi^a\Big)\,. \label{2.19}
    \end{split}
\end{align}
\\We can now compute the Feynman rules in $\mathcal{O}(z^n,{\psi}^n,\chi^n)$. We will mainly focus on the diagrams in which we have an external $\varphi$ field.

\subsection*{Feynman rules:}
 We list the additional vertices that we need in our computations. Other relevant vertices for the graviton-worldline coupling are derived in \cite{Jakobsen:2021zvh}, while the vertices involving the extra scalar and scalar-graviton couplings are derived in \cite{Bhattacharyya:2024aeq}.
\begingroup
\scriptsize
\begin{align}
    \begin{split}
      &  \mathcal{V}_1:   \begin{minipage}[h]{0.08\linewidth}	
\scalebox{1.5}{\includegraphics[width=\linewidth]{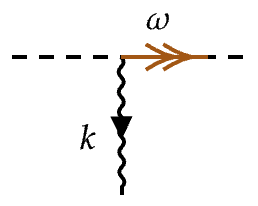}}
\end{minipage}\hspace{0.35cm}
\equiv 
\hspace{0 cm}i\frac{m_a s_a}{m_p} \,\bar\Psi^a_{\eta}\,\omega\,\hat\delta(\omega+k\cdot v_a) e^{ik\cdot b_a}\,,\\ &
\mathcal{V}_2:   \begin{minipage}[h]{0.08\linewidth}	
\scalebox{1.5}{\includegraphics[width=\linewidth]{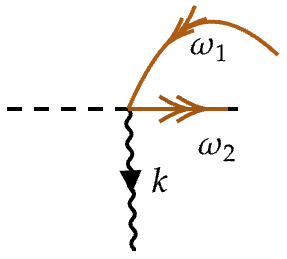}}
\end{minipage}\hspace{0.35cm}
\equiv 
\hspace{0 cm}i\frac{m_a s_a}{2 m_p}\hat\delta(k\cdot v_a-\omega_1+\omega_2 )\,(\omega_2+\omega_1) e^{ik\cdot b_a}\,,\\ &
\mathcal{V}_3:   \begin{minipage}[h]{0.08\linewidth}	
\scalebox{1.3}{\includegraphics[width=\linewidth]{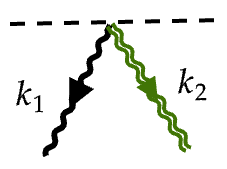}}
\end{minipage}\hspace{0.35cm}
\equiv 
\hspace{0 cm}-i\frac{m_a s_a}{2m_p^2}e^{i(k_1+k_2)\cdot b}\,\hat\delta(k_1\cdot v_a+k_2\cdot v_a)\Big(v_a^\mu v_a^\nu+i (k_2\cdot \mathcal{S}_a)^{(\mu}v_a^{\nu)}\Big)\,,
\\ &
\hspace{0cm}\mathcal{V}_4:   \begin{minipage}[h]{0.08\linewidth}	
\scalebox{1.6}{\includegraphics[width=\linewidth]{h_z_2.png}}
\end{minipage}\hspace{0.35cm}
\equiv \, i\frac{m_a }{m_p} \,e^{ik\cdot b_a}\,\hat\delta(k\cdot v_a+\omega_1+\omega_2) \Bigg(\frac{1}{2} k_{\rho_1} k_{\rho_2} v_a^\mu v_a^\nu +\omega_1 k_{\rho_2}v_a^{(\mu}\delta^{\nu)}_{\rho_1}+\omega_2 k_{\rho_1}v_a^{(\mu}\delta^{\nu)}_{\rho_2}+\omega_1\omega_2 \delta^{(\mu}_{\rho_1}\delta^{\nu)}_{\rho_2}\\&
\hspace{0.8cm}+\left.\frac{i}{2}\Bigg(\left(\omega _1 \,k\cdot S^{(\mu }\delta ^{\nu )}_{\rho _1}k_{\rho _2}+\omega _2\,k \cdot S^{(\mu }\delta ^{\nu )}_{\rho _2}k_{\rho _1}\right)+\frac{1}{2}k\cdot S^{(\eta }\delta ^{\eta )}_{(\nu }v_{a,\mu )}k_{\rho _1}k_{\rho _2}\Bigg)\Bigg)\right]
    \end{split}
\end{align}
\endgroup
\section{Bootstrapping Post-Minkowskian (PM) physics from Post-Newtonian (PN) data: the idea, technology and example}
\label{ch3:sec3}
In this section, we briefly discuss the idea of bootstrapping of Post-Minkowskian (relativistic) observables from Post-Newtonian (non-relativistic) data \cite{Dlapa:2023hsl,Henn:2014qga},  which primarily based on differential equation technique for evaluating multiloop Feynman integrals \cite{Kotikov:1990kg,Gehrmann:1999as} and also illustrate the method with a specific two loop example relevant for our computations. In the upcoming sections, we widely encounter two loop integrals, which can be solved using IBP relations. In this section, we will present the results for the master integrals that we will use for our computations of the eikonal phase in the subsequent section.

\textbf{Identifying the integrands:} We encounter a general family of n-loop integrals that has the following form, which is essentially equivalent to computing $2\to 2$ scattering amplitude,
\begin{align}
           &\langle \,\textrm{final}\,|\mathbb{S}|\,\textrm{initial}\,\rangle\sim\mathbfcal{M}^{a_1\cdots a_n;\pm\cdots \pm}_{\alpha_1\cdots \alpha_n;\beta_1\cdots \beta_m}(|q|,\gamma)=\Bigg(\prod_{i=1}^n\int_{\ell_i}\frac{\hat\delta(\ell_i\cdot v_{a_i})}{(\pm \ell_i\cdot v_{\cancel{a}_i}+i\varepsilon)^{\alpha_i}}\Bigg)\frac{1}{D_1^{\beta _1 }D_2^{\beta_2}\cdots D_m^{\beta_m}}\,,
\end{align}
where $a_i$s are particle index label with $\cancel{1}=2,\,\cancel{2}=1$. $D_i$s are the graviton (scalar) propagators along with irreducible scalar products that form the numerators of the loops integrals and have the following form,
$$D_i=P_i^2,\,\,P_i=\tilde\Upsilon_{ij}\,l_j+\tilde\Upsilon_i\, q,\,\,\textrm{with},\,\,\tilde\Upsilon_{ij},\tilde\Upsilon_i\in(0,\pm 1)\,\,\forall\,\,\, 1\le i,j\le m$$
where $q$ is the external momentum and $v_{1,2}$ are the velocities of the two astrophysical objects. Furthermore, one would notice that each loop integral is supported by a Dirac-$\delta$ function, $\hat\delta(l_i\cdot v_{a_i})$. At this point, to utilize the multiloop collider physics techniques \cite{Weinzierl:2022eaz}, it is convenient to represent the $\delta$-functions by reverse-unitarity (sometimes called velocity-cut),
\begin{align}
    \begin{split}
       \frac{i}{(-1)^{s+1}} \hat\delta^{(s)}(D_i)=\frac{1}{(D_i+i\varepsilon)^{s+1}}-\frac{1}{(D_i-i\varepsilon)^{s+1}}
    \end{split}
\end{align}                 
which allows us to identify the $\delta$-functions as cut propagators and one can freely apply the Integration-by-parts (IBP) technique to evaluate the integrals.

\textbf{IBP reduction:}  After identifying the integrands, the next task is to find the IBP identities, first introduced in\cite{Chetyrkin:1981qh}, which help to identify the master integrals in a given family of Feynman integrals. IBP identities can be obtained by demanding that under dimensional regularisation, loop integrals of any total derivative identically vanish \cite{Weinzierl:2022eaz},
\begin{align}
    \int_{l_i}\frac{\partial}{\partial l_k^\mu}\Bigg(\frac{\eta^{\mu}}{(\pm l_i\cdot v_{\slashed{a}_i}+i\varepsilon)^{\alpha_i}\cdots D_1^{\beta_1}\,D_2^{\beta_2}\cdots \slashed{D}_1^{\gamma_1}\slashed{D}_2^{\gamma_2}\cdots}\Bigg)=0\,,\label{3.3a}
\end{align}
where, $l_k$s are loop momenta, $\eta^\mu$ is the linear combination of loop momenta and external momenta and $\slashed{D}$ are the `cut' propagators coming from the $\delta$-functions. The IBP reduction procedure generates a large number of homogeneous linear systems of equations of form,
\begin{align}
    \sum_k \gamma_k(|q|,d,\gamma)\mathbfcal{M}^{(\pm\pm)i}_{\alpha_1+\kappa_{k,1},\alpha_2+\kappa_{k,2},\cdots}(|q|,\gamma,d)=0\,.
\end{align}
However, all the equations are not independent but rather related by various symmetry relations. Ultimately, one needs to solve the equations in unique sectors. Fortunately, there are several publicly available packages for doing this task \textbf{LiteRed} \cite{Lee:2012cn,Lee:2013mka}, \textbf{Kira} \cite{Maierhofer:2017gsa}, \textbf{Reduze 2} \cite{vonManteuffel:2012np} and \textbf{FIRE6} \cite{Smirnov:2019qkx}. For our computations, we extensively use \textbf{LiteRed} .

\textbf{Master integrals and differential equations:}
One of the main tasks of the IBP procedure is to expand a general Feynman integral as a linear combination of master integrals. For example, let us take a look at the two-loop case. A general two-loop integral has the following form,
\begin{align}
    \begin{split}
        \mathbfcal{M}^{\pm,\pm,\textrm{2-loop}}_{\alpha_1,\alpha_2;\beta_1\cdots \beta_{5}}=\int_{l_{1,2}}\Bigg(\frac{\hat\delta^{(\gamma_1)}(\slashed D_1)\hat\delta^{(\gamma_2)}(\slashed D_2)}{(\pm D_1+i\varepsilon)^{\alpha_1}(\pm D_2+i\varepsilon)^{\alpha_2}}\Bigg)\frac{1}{D_3^{\beta_1}\cdots D_7^{\beta_5}}\label{3.4a}
    \end{split}
\end{align}
where, $\{D_i,\{\slashed D_k\}\}\in\Big(\ell_{1}\cdot v_1,\,\,\ell_{2}\cdot v_2,\,,\,\ell_{1}^2,\,\, \ell_{2}^2,\,\, (\ell_{1}+\ell_{2}-q)^2,\,\, (\ell_{1}-q)^2,\,\, (\ell_{2}-q)^2,\,\, \{\ell_{1}\cdot v_2,\,\, \ell_{2}\cdot v_1\}\Big) $. The set $\{D_i,\slashed D_k\}$ s are called basis function set and the number of independent basis functions (scalar-product) can be determined by noting the number of loop momenta and external momenta: $N_{\textrm{basis}}=\frac{L(L+1)}{2}+LE$, where $L$ and $E$ are the number of loop integrals and external momenta respectively. For the two-loop case, we have $2$ loop momenta and $3$ external momenta, hence the number of basis functions is, indeed, $N_{\textrm{basis}}^{2-\textrm{loop}}=9$. Once we have the basis function, we solve the IBP relation using \textbf{LiteRed}; after solving, one gets the unique master integral for this particular family of integrals. Now using built-in command \textbf{IBPReduce} \cite{Lee:2012cn,Lee:2013mka}, one can write the general two-loop integrals as,
\begin{align}
    \begin{split}
         \mathbfcal{M}^{\textrm{2-loop}}=\sum_{n=1}^{N_{m}} a_{n}(\gamma,d)\,\mathbfcal{M}^{\textrm{master}}_n(\gamma, d)\,.
    \end{split}
\end{align}
We face 16 independent master integrals (excluding the different signs of $i\varepsilon$-prescription, which can be derived further using appropriate field redefinition and invoking symmetry relations) after solving 17 unique sectors using \textbf{LiteRed}.

\textbf{Solving master integrals using differential equation:} In general, the master integrals do not factorize into the kinematic variable $\gamma$ and dimensional regularisation parameter $\epsilon$. In this case, it is difficult to find any closed-form expression. So, one needs to find the solutions order by order in $\epsilon$. For this purpose, the method of differential equations is very useful \cite{Remiddi:1997ny,Kotikov:1991pm, Henn:2013pwa}. We first define a column vector consisting of relevant master integrals and redefine the kinematic variable $\gamma$ to $x$ as $\gamma\to \frac{x^2+1}{2x}$,
\begin{align}
    \begin{split}
        \vec f(x,\epsilon)=\{\mathbfcal{M}_1^{\textrm{master}},\cdots, \mathbfcal{M}_{N_m}^{\textrm{master}} \}
    \end{split}
\end{align}
which satisfies the following differential equation,
\begin{align}
      \partial_x \vec f(x,\epsilon)=A(x,\epsilon)
     \vec f(x,\epsilon)\,.\label{3.7}
\end{align}
For all practical purposes, the differential equation is quite hard to solve. But it is convenient to find a suitable transformation \cite{Henn:2014qga},
\begin{align}
    \begin{split}
        \vec f(x,\epsilon)=\mathbb{T}(x,\epsilon)\vec{g}(x,\epsilon),
    \end{split}
\end{align}
such that the differential system can be brought to the $\epsilon$-form, which is easier to solve. Now, the system of differential equations has the form,
\begin{align}
    \begin{split}
        \partial_x \vec g(x,\epsilon)=\mathbb{T}^{-1}(A \mathbb{T}-\partial_x \mathbb{T})\vec g(x,\epsilon)\equiv \epsilon\, \mathbb{S}(x) \vec g(x,\epsilon)\,.\label{2.14a}
    \end{split}
\end{align}
The goal is to find the basis transformation matrix $\mathbb{T}(x,\epsilon)$ such that $\mathbb{T}(A \mathbb{T}-\partial_x \mathbb{T})=\epsilon \,\mathbb{S}(x)$. Note that $\epsilon$ is now completely factorized from the matrix $\mathbb{S}(x)\,.$, It was first conjectured by Henn \cite{Henn:2013pwa}. The $\mathbf {\epsilon}$-\textbf{factorization} can be systematically done using Lee's algorithm \cite{Lee:2014ioa} and is (semi-) automatized in the following packages \textbf{Fuchsia} \cite{Gituliar:2017vzm}, \textbf{epsilon} \cite{Prausa:2017ltv}, \textbf{Libra} \cite{Lee:2020zfb} and \textbf{CANONICA} \cite{Meyer:2017joq} and the matrix equation in $\bf{\epsilon}$-form has a general solution in terms of path-ordered exponential (Bootstrap equation), which can be solved systematically using \textbf{Libra} \cite{Lee:2020zfb},
\begin{align}
    \begin{split}
        \vec g(x,\epsilon)=\underbrace{\Bigg(\mathbfcal{P}e^{\epsilon\int_{x_0}^{x}S(x')\,dx'}\Bigg)}_{\mathbb{B}(x,x_0)}\vec g(x_0,\epsilon).\label{3.10a}
    \end{split}
\end{align}
Once we have \eqref{3.10a}, we need to use the boundary condition to fix $\vec g(x_0,\epsilon)\,.$ For relativistic two-body problem it is easy to find the solution of integration at `\textit{soft}' (near-static) limit which is at $\gamma\to 1^{+}\,(x\to 1^{-})$, which is nothing but the solution of the master integrals in Post-Newtonian (non-relativistic) regime: \textit{this is nothing but bootstrapping the complete relativistic integrals (PM-integrals) using the non-relativistic integrals (PN-integrals) \cite{Dlapa:2023hsl}.}

\par

\noindent
For illustration, we show the details of the computation in the \textbf{Sectors:\{0,0,0,0,1,1,1\}\&\{1,1,0,0,1,1,1\}}, where we encounter the following {\bf four} master integrals (the original basis is sometimes called the Laporta basis \cite{Laporta:2000dsw}). We take the following vector \footnote{For the sake of convenience, we will henceforth drop the subscript ``master'' from the $\mathbfcal{M}$'s, and the $|q|$ dependence can be recovered from dimensional analysis, since $|q|$ is the dimensionful scale.},
\begin{align}
\begin{split}
    \vec f(\gamma,\epsilon)=&\{\mathbfcal{M}_{0,0,0,0,1,1,1,\slashed 1,\slashed 1},\mathbfcal{M}_{0,0,0,0,2,1,1,\slashed 1,\slashed 1},\mathbfcal{M}_{0,0,0,0,1,2,1,\slashed 1,\slashed 1}, \mathbfcal{M}_{1,1,0,0,1,1,1,\slashed 1,\slashed 1}\}^{T},
    \end{split}
\end{align}
which satisfies, $\partial_x \vec f(x,\epsilon)=A(x,\epsilon)\vec f(x,\epsilon)$ with,
\begin{align}
      A(x,\epsilon)=  \left(
\begin{array}{cccc}
 \frac{3 \left(x^4+6 x^2+1\right) \epsilon -\left(x^2+1\right)^2}{x \left(x^4-1\right)} & -\frac{x^2-1}{2 \left(x^3+x\right)} & \frac{2 x (2 \epsilon +1)}{\epsilon -x^4 \epsilon } & 0 \\
 \frac{6 \left(x^2-1\right) \epsilon ^2}{x^3+x} & \frac{x^4 (\epsilon +1)+x^2 (6 \epsilon +2)+\epsilon +1}{x-x^5} & \frac{4 x (2 \epsilon +1)}{x^4-1} & 0 \\
 -\frac{6 \left(x^2-1\right) \epsilon ^2}{x^3+x} & \frac{\epsilon -x^2 \epsilon }{x^3+x} & \frac{\left(x^2-1\right) (2 \epsilon +1)}{x^3+x} & 0 \\
 0 & -\frac{4}{x^2-1} & 0 & \frac{2 \left(x^2+1\right)}{x-x^3} \\
\end{array}
\right)\,.\label{3.12}
\end{align}
The matrix equation can be brought to $\epsilon$-form (canonical form) using suitable basis transformation using the publicly available packages mentioned earlier. In the canonical basis/Uniform Transcendental (UT) basis, the differential equation takes the form, $\partial_x \vec g(x,\epsilon)=\epsilon \mathbb{S}(x)\vec g(x,\epsilon)$, where $\mathbb{S}(x)$ and the transformation matrix $\mathbb{T}(x,\epsilon)$ are given by,
\begin{align}
\begin{split}
   &\mathbb{T}(x,\epsilon)= \left(
\begin{array}{cccc}
 -\frac{7 x}{x^2-1} & \frac{x}{2-2 x^2} & \frac{3 x}{x^2-1} & 0 \\
 -\frac{6 x \epsilon }{x^2-1} & -\frac{5 x \epsilon }{x^2-1} & \frac{6 x \epsilon }{x^2-1} & 0 \\
 \frac{12 \left(x^2-1\right) \epsilon ^2}{2 x \epsilon +x} & \frac{2 \left(x^2-1\right) \epsilon ^2}{2 x \epsilon +x} & -\frac{12 x \epsilon ^2}{2 \epsilon +1} & 0 \\
 0 & 0 & 0 & \frac{2 x^2}{\left(x^2-1\right)^2} \\
\end{array}
\right),\quad\textrm{and}\\ &
\mathbb{S}(x)=
\left(
\begin{array}{cccc}
 \frac{9 \left(2 x^2+1\right)}{2 \left(x^3-x\right)} & \frac{2 x^2+3}{4 \left(x^3-x\right)} & -\frac{6 x}{x^2-1} & 0 \\
 -\frac{3 \left(2 x^2+5\right)}{x^3-x} & -\frac{6 x^2+5}{2 \left(x^3-x\right)} & \frac{12 x}{x^2-1} & 0 \\
 \frac{6}{x} & -\frac{1}{3 x} & -\frac{2}{x} & 0 \\
 \frac{12}{x} & \frac{10}{x} & -\frac{12}{x} & 0 \\
\end{array}
\right)=\sum_{i=1}^3\frac{\mathbb{S}_i}{x-x_i},\,(x_i=1,-1,0)
\end{split}
\end{align}
where the matrix residues take the form,
\begin{align}
\mathbb{S}_0=\left(\begin{array}{cccc}
         -  \frac{9}{2}  & -\frac{3}{4}& 0 &0  \\
            15 & \frac{5}{2} & 0 &0\\
            6&-\frac{1}{3}&-2&0\\
            12 & 10 & -12 &0 
     \end{array}
     \right)\,, \quad    
    \mathbb{S}_{\pm 1}=\left(\begin{array}{cccc}
           \frac{27}{4}  & \frac{5}{8}& -3 &0  \\
            -\frac{21}{2} & -\frac{11}{4} & 6 &0\\
            0&0&0&0\\
            0 & 0 & 0 &0 
     \end{array}
     \right)\,.  \label{2.17} 
\end{align}
In this fuchsified form, the equation looks like this,
\begin{align}
    \begin{split}
        \partial_x\vec g(x,\epsilon)=\epsilon \Bigg(\frac{S_0}{x}+\frac{S_1}{x+1}+\frac{S_{-1}}{x-1}\Bigg)\vec g(x,\epsilon)\label{3.15}
    \end{split}
\end{align}
which has the solution in terms of a path ordered exponential as mentioned in \eqref{3.10a}. Finally, in  the Laporta basis, the formal solution takes the form,
\begin{align}
    \begin{split}
        \vec f(x,\epsilon)=\mathbb{T}(x,\epsilon)\,\Bigg(\mathbfcal{P}e^{\epsilon\int_{1^{-}}^x \,\mathbb{S}(x')\,dx'}\Bigg)\,\mathbb{T}^{-1}(1^{-},\epsilon) \,\vec f(1^{-},\epsilon)\,.\label{3.16}
    \end{split}
\end{align}
Taking the boundary limit is a bit tricky, and one needs to take the limit very carefully.First we expand the $\mathbb T^{-1}(x_0=1-v_{\infty})$ matrix  and  also expand the boundary vector $\vec f(x_0,\epsilon)$ around $v_\infty=0$ using the method of regions. For generic two-loop integrals, we will have both the contribution from potential and the radiation region \cite{Dlapa:2023hsl}. \textit{ However, at 3PM, the full conservative dynamics of the boundary integrals are captured by the contribution from the potential region only \cite{Dlapa:2023hsl}.}\\

\textbf{\textit{Potential region:}} In momentum space, the potential region is characterized by the following loop momenta scaling: $l_{i}\sim |\boldsymbol{q}|(v_{\infty},1)$ and, the velocity of black holes are parametrized as \cite{Dlapa:2023hsl}, 
\begin{align}
    \begin{split}
        v_2=(1,0,0,0),\,\, v_1=(1,v_{\infty},0,0).\label{2.23a}
    \end{split}
\end{align}
Using the above-mentioned scaling, the integral is reduced to,
\begin{align}
    \begin{split}
     \mathbfcal{M}\Big|_{v_{\infty}\to 0}^{\textrm{pot.}}\sim    v_{\infty}^{-\sum n_i}\Bigg(\prod_{i}\int_{\vec \ell_i}\frac{1}{(\pm\vec \ell_i\cdot \boldsymbol{n}+i\varepsilon)^{n_i}}\Bigg)\frac{1}{\boldsymbol {D_1}^{m_1}\cdots\boldsymbol{D_k}^{m_k}},\,\boldsymbol D\equiv \textrm{spatial part of the squared propagators.}
    \end{split}
\end{align}
For our case, we are interested in the following two loop integrals in potential regions.
\begin{align}
    \begin{split}
\mathbfcal{M}^{\textrm{2-loop},(\mp)}_{n_1,n_2;m_1,..,m_5}\Big|_{v_{\infty}\to 0}^{\textrm{pot.}}&\sim\int_{\ell_{1},\ell_{2}}\frac{\hat\delta(\ell_{1}^0)\hat\delta(\ell_{2}^0-v_{\infty}\ell_{2}^{(x)})}{(\ell_{1}^0-v_{\infty}\ell_{1}^{(x)}+i\varepsilon)^{n_1}(\ell_{2}^0+i\varepsilon)^{n_2}}\frac{1}{\boldsymbol{D_1}^{m_1}\cdots\boldsymbol{D_5}^{m_5} }+\mathcal{O}(v_{\infty}^{-1})\\ &
=v_{\infty}^{-n_1-n_2}\int_{\boldsymbol{\ell_1},\boldsymbol{\ell_2}}\frac{1}{(-\boldsymbol{\ell_1}\cdot \boldsymbol{n}+i\varepsilon)^{n_1}(\boldsymbol{\ell_2}\cdot \boldsymbol{n}+i\varepsilon)^{n_2}}\frac{1}{\boldsymbol{D_1}^{m_1}\cdots\boldsymbol{D_5}^{m_5} }+\mathcal{O}(v_{\infty}^{1-n_1-n_2})\,.
    \end{split}
\end{align}
The details of the computation of potential mode master integrals are given in Appendix~(\ref{ch3:app:B}).\par

\noindent
Now, the solution \eqref{3.16} to the differential equation to the differential equation can be presented in terms of multiple
polylogarithms (MPLs) \cite{Remiddi:1999ew,goncha},
\begin{align}
    G(a_1,a_2,\cdots a_n;z)=\int_{0}^z \frac{dt}{t-a_1}G(a_2,\cdots,a_n;t),\,\, G(\,;z)=1,\,\,\forall a_i,z\in \mathbb{C}\,.
\end{align}
If all $a_i's$ are zero, 
\begin{align}
    G(\vec 0_n;z)=\frac{1}{n!}\,\log^{n}(z)
\end{align}
and, for equal $a_i=a$ one can obtain,
\begin{align}
    G(\vec a_n,z)=\frac{1}{n! }\log^{n}\Big(1-\frac{z}{a}\Big),\,\,\forall a\ne 0\,.
\end{align}
Apart from ordinary logarithms, MPLs also contain classical polylogarithms, which are defined as,
\begin{align}
    \begin{split}
        G(\vec 0_{n-1},1;z)=- \textrm{{\bf Li}}_2(z)\,.
    \end{split}
\end{align}
Now, after collecting the ingredients discussed above and suitably regularizing the logarithmic divergences, we obtain the solutions for the Laporta master integrals needed in our subsequent computations, and we present the epsilon-form matrix $\mathbb{S}(x)$ in Appendix~\eqref{ch3:app:C} \footnote{{We match all the listed master integrals in \eqref{MasterM} with those of \cite{Jakobsen:2022fcj}. At leading order, up to which the integrals are given in \cite{Jakobsen:2022fcj}, the results agree perfectly with ours except for $\mathbfcal{M}_{0,0;1,1,2,1,1}$. However, we have cross-checked our result by explicitly matching it to the boundary value. S.G. would like to thank Gustav Jacobsen for confirming this point. }}.
\begin{adjustwidth}{0cm}{0cm}
\begin{tcolorbox}[thesisresultbox, title=\textit{Master Integrals}, left=2.1mm, right=2.1mm]
\restorethesisbodyformat
\resizebox{0.955\linewidth}{!}{$\begin{aligned}
&
(-q^2)^{2\epsilon}\mathbfcal{M}_{0,0;0,0,1,1,1}=-\frac{x \log (x)}{32 \pi ^2 \left(x^2-1\right) \epsilon }+\frac{3x \left(\log^2(x)+\textbf{Li}_2(1-x^2)\right)}{32 \pi ^2 \left(x^2-1\right)}-\frac{ x \epsilon}{192 \pi ^2 \left(x^2-1\right)}  \Bigg[-7 \pi ^2 \log(x)\\ &\hspace{1.4 cm}+216 \boldsymbol{\mathscr{J}}(\{-1,-1,0\},x)-216 \boldsymbol{\mathscr{J}}(\{-1,0,0\},x)+216 \boldsymbol{\mathscr{J}}(\{-1,1,0\},x)-216 \boldsymbol{\mathscr{J}}(\{0,-1,0\},x)\\ &\hspace{-0.3 cm}+120 \boldsymbol{\mathscr{J}}(\{0,0,0\},x)-216 \boldsymbol{\mathscr{J}}(\{0,1,0\},x)+216 \boldsymbol{\mathscr{J}}(\{1,-1,0\},x)-216 \boldsymbol{\mathscr{J}}(\{1,0,0\},x)+216 \boldsymbol{\mathscr{J}}(\{1,1,0\},x)\Bigg]+\mathcal{O}(\epsilon^2)\,,\\ &
(-q^2)^{2\epsilon+1}\mathbfcal{M}_{0,0;0,0,1,2,1}=-\frac{x^2+1}{64 \pi ^2 x}+\frac{x^2+(1-x^2) \log (x)+1}{32 \pi ^2 x}\epsilon+\frac{\epsilon^2}{384 \pi ^2 x}\Big[120 \left(x^2-1\right) (\textbf{Li}_2(1-x)-\textbf{Li}_2(-x))\\ &\hspace{2.3 cm}-3 \left(8+\pi ^2\right) x^2+12 \log (x) \left(\left(9 x^2-1\right) \log (x)+\left(x^2-1\right) (2-10 \log (x+1))\right)+17 \pi ^2-24\Big]\,,\\ &
(-q^2)^{2\epsilon+1}\mathbfcal{M}_{0,0;0,0,2,1,1}=-\frac{x \log (x)}{16 \pi ^2 \left(x^2-1\right)}-\epsilon  \frac{x   \left(\log^2(x)+\textbf{Li}_2(1-x^2)\right)}{16 \pi ^2 \left(x^2-1\right)}+\mathcal{O}(\epsilon^2)\,,\\ &
     (-q^2)^{2\epsilon+1} \textcolor{black}{\mathbfcal{M}^{(++)}_{1,1;0,0,1,1,1}}=\frac{x^2}{8 \pi ^2 \left(x^2-1\right)^2 \epsilon ^2}-\frac{x^2 \left(\pi ^2-6 \log ^2(x)\right)}{48 \left(\pi ^2 \left(x^2-1\right)^2\right)}+\frac{x^2 \epsilon }{24 \pi ^2 \left(x^2-1\right)^2}\Big[ (-12 (\boldsymbol{\mathscr{J}}(\{0,-1,0\},x)\\ &\hspace{0.8cm}-\boldsymbol{\mathscr{J}}(\{0,0,0\},x)+\boldsymbol{\mathscr{J}}(\{0,1,0\},x))-32 \zeta (3))\Big]+O\left(\epsilon ^2\right)=  (-q^2)^{2\epsilon+1}\frac{1}{2}\mathbfcal{M}^{(+-)}_{1,1;0,0,1,1,1}\,,
 \\     &(-q^2)^{1+2\epsilon}\mathbfcal{M}^{(+\pm)}_{0,0;1,1,0,1,1}=\frac{1}{(4\pi)^{3-2\epsilon}}\frac{\Gamma^4(1/2-\epsilon)\Gamma^2(1/2+\epsilon)}{\Gamma^2(1-2\epsilon)},\\ &
(-q^2)^{2+2\epsilon}\mathbfcal{M}^{(+\pm)}_{0,0;1,1,1,1,1}=\frac{x \log (x)}{16 \pi ^2 (x^2-1) \epsilon }+\frac{\Big(\log^2(x)+\textbf{Li}_2(1-x^2)\Big)}{16\pi^2(x^2-1)}-\frac{\epsilon }{96 \pi ^2 \left(x^2-1\right)} \Bigg(24 x (3 \boldsymbol{\mathscr{J}}(\{-1,-1,0\},x)\\&\hspace{2.4 cm} -3 \boldsymbol{\mathscr{J}}(\{-1,0,0\},x)+3 \boldsymbol{\mathscr{J}}(\{-1,1,0\},x)-3 \boldsymbol{\mathscr{J}}(\{0,-1,0\},x)+\boldsymbol{\mathscr{J}}(\{0,0,0\},x)-3 \boldsymbol{\mathscr{J}}(\{0,1,0\},x)\\ & \hspace{2.4 cm}+3 \boldsymbol{\mathscr{J}}(\{1,-1,0\},x)-3 \boldsymbol{\mathscr{J}}(\{1,0,0\},x)+3 \boldsymbol{\mathscr{J}}(\{1,1,0\},x))-5 \pi^2 x \log (x)\Bigg)+\mathcal{O}(\epsilon^2)\,,\\ &
(-q^2)^{3+2\epsilon}\mathbfcal{M}^{(+\pm)}_{0,0;1,1,2,1,1}=\frac{x \left(x^4-4 x^2 \log (x)-1\right)}{16 \pi ^2 \left(x^2-1\right)^3 \epsilon }+\frac{x}{16 \pi ^2 \left(x^2-1\right)^3}\Bigg[5 \left(x^4-1\right)-2 \left(x^4+4 x^2+1\right) \log (x)\\ & \hspace{0.2 cm}+4x^2 \Big(\log^2 x+\textbf{Li}_2(1-x^2)\Big)\Bigg]+\epsilon\frac{x}{96 \pi ^2 \left(x^2-1\right)^3}\Big[288 x^2\Bigg( \boldsymbol{\mathscr{J}}(\{-1,-1,0\},x)- \boldsymbol{\mathscr{J}}(\{-1,0,0\},x)+ \boldsymbol{\mathscr{J}}(\{-1,1,0\},x)\\ &\hspace{0.2 cm}- \boldsymbol{\mathscr{J}}(\{0,-1,0\},x)+ \frac{1}{3} \boldsymbol{\mathscr{J}}(\{0,0,0\},x)- \boldsymbol{\mathscr{J}}(\{0,1,0\},x)+ \boldsymbol{\mathscr{J}}(\{1,-1,0\},x)- \boldsymbol{\mathscr{J}}(\{1,0,0\},x)+ \boldsymbol{\mathscr{J}}(\{1,1,0\},x)\Bigg)\\ &\hspace{0.2 cm}+24 \left(x^4+12 x^2+1\right) \textbf{Li}_2(1-x)-24 \left(x^4+12 x^2+1\right) \textbf{Li}_2(-x)+54 \left(x^4-1\right)+36 x^4 \log ^2(x)-24 x^4 \log (x) \log (x+1)\\ &\hspace{0.2 cm}-96 x^4 \log (x)+\pi ^2 \left(3 x^2 \left(x^2-8\right)-7\right)+144 x^2 \log ^2(x)-288 x^2 \log (x) \log (x+1)-20 \pi ^2 x^2 \log (x)\\ &\hspace{1.2 cm}+24 x^2 \log (x)-12 \log ^2(x)-24 \log (x) \log (x+1)-96 \log (x)\Big]+\mathcal{O}(\epsilon^2)\,,\\ &
 (-q^2)^{2\epsilon} \textcolor{black}{\mathbfcal{M}^{(+\pm)}_{0,0;0,1,1,0,1}}=0,\,\,
 (-q^2)^{2\epsilon} \textcolor{black}{\mathbfcal{M}^{(+\pm)}_{0,1;0,0,1,1,1}}=-\frac{i\,x}{32\pi \epsilon (x^2-1)}+\mathcal{O}(\epsilon^0)
\end{aligned}$}\label{MasterM}
\end{tcolorbox}
\restorethesisbodyformat
\end{adjustwidth}
\vspace{0.15cm}
The iterative integral in \eqref{MasterM} is defined as,
\begin{align}
    \boldsymbol{\mathscr{J}}(\{i,j,k\},x)\equiv \int_{1^-}^x \frac{dt}{t-i}\int_{1^-}^{t}\frac{dt'}{t'-j}\int_{1^-}^{t'}\frac{dt''}{t''-k} \boldsymbol{\mathscr{J}}(\{\},t'');~~~~~~\boldsymbol{\mathscr{J}}(\{\},t'')=1,\label{3.43k}
\end{align}
and, the relevant 3-point iterative UT integrals are defined in Appendix~\eqref{ch3:app:D}\footnote{Note that the integrals defined in \eqref{3.43k} are not exactly Goncharov MPLs (which is defined for iterative integrals with lower limit `$0$') because here the lower limit is `$1$'. However one can write \eqref{3.43k} as a difference of Goncharov MPLs.}. 
In \eqref{MasterM} we extensively use the following identity,
\begin{align}
    \textbf{Li}_2(-x)-\textbf{Li}_2(1-x)=-\frac{\pi^2}{12}-\log(x)\log(1+x)-\frac{1}{2}\textbf{Li}_2(1-x^2),
\end{align}
which can be derived from the fundamental identities of dilogarithm, namely \textit{Reflection formula and Abel's duplication formula}  \cite{goncha}. From now on, we make the $(++)$ sign of the last master integral of \eqref{MasterM} implicit for all computations. All the derivations are done by implicitly making the $i\varepsilon$ prescription positive for all linear propagators using proper field redefinition.\par
Now, we provide a bit of detail regarding the computation of master integrals. Let's focus on the following $\mathbfcal{M}_{0,1;0,0,1,1,1}$ and $\mathbfcal{M}_{0,0;0,1,0,1,1}$ that we do not include in the differential system mentioned in \eqref{3.12}. Fortunately, the first integral can be found separately using the differential equation method as it closes itself, and the second one identically vanishes because of the tadpole nature of $\ell_{1}$ loop integral,
\begin{align}
    \begin{split}
     &   \frac{d}{d\gamma}\mathbfcal{M}_{0, 1; 0, 0, 1, 1, 1, \slashed 1, \slashed 1}(\gamma,\epsilon)=-\frac{\gamma}{\gamma^2-1}\mathbfcal{M}_{0, 1; 0, 0, 1, 1, 1, \slashed 1, \slashed 1}(\gamma,\epsilon)+\frac{1}{\gamma^2-1}{\mathbfcal{M}_{0, 0; 0, 0, 1, 1, 1, \slashed 2, \slashed 1}}\,.
     \label{3.29}
    \end{split}
\end{align}
It appears that the master $\mathbfcal{M}_{0, 1; 0, 0, 1, 1, 1, \slashed 1, \slashed 1}(\gamma,\epsilon)$ does not close, but we now sketch that the master $\mathbfcal{M}_{0, 0; 0, 0, 1, 1, 1, \slashed 2, \slashed 1}$, that appears in the second term in \eqref{3.29} does not contribute and can be proved by noting that the integral has a flip symmetry $\ell_i\to -\ell_i,\,q\to -q$.
\begin{align}
\begin{split}
    \mathbfcal{M}_{0, 0; 0, 0, 1, 1, 1, \slashed 2, \slashed 1}(|q|,\gamma)&\propto \int_{\ell_{1},\ell_{2}}\frac{\hat\delta(\ell_{2}\cdot v_1)}{\ell_{1}^2 \ell_{2}^2 (\ell_{1}+\ell_{2}-q)^2}\Bigg(\frac{1}{(\ell_{1}\cdot v_2+i\varepsilon)^2}-\frac{1}{(\ell_{1}\cdot v_2-i\varepsilon)^2}\Bigg)\,,\\ &\xrightarrow[l_i\to -\ell_{1}]{q\to -q}\int_{\ell_{1},\ell_{2}}\frac{\hat\delta(\ell_{2}\cdot v_1)}{\ell_{1}^2 \ell_{2}^2 (\ell_{1}+\ell_{2}-q)^2}\Bigg(\frac{1}{(\ell_{1}\cdot v_2-i\varepsilon)^2}-\frac{1}{(\ell_{1}\cdot v_2+i\varepsilon)^2}\Bigg)\,,\\ &
    =-  \mathbfcal{M}_{0, 0; 0, 0, 1, 1, 1, \slashed 2, \slashed 1}\implies   \mathbfcal{M}_{0, 0; 0, 0, 1, 1, 1, \slashed 2, \slashed 1}\to 0\,.
    \end{split}
\end{align}
Therefore, the solution can be written straightforwardly and is given by,
\begin{align}
    \begin{split}
     &   \mathbfcal{M}_{0, 1; 0, 0, 1, 1, 1}(\gamma,\epsilon)=\frac{C_1(\epsilon,\gamma_0)}{\sqrt{\gamma^2-1}},
     \\ &
   \mathbfcal{M}_{0,0;0,1,0,1,1}(\gamma,\epsilon)  =0. \label{MasterC}
    \end{split}
\end{align}
To fix the boundary data, we evaluate the integrals in the following limit,
\begin{align}
    \begin{split}
   \lim_{\gamma\to 1^{++}}     \sqrt{\gamma^2-1} \,\mathbfcal{M}_{0, 1; 0, 0, 1, 1, 1, \slashed 1, \slashed 1}^{(++)}(\gamma,\epsilon)&=\lim_{\gamma\to 1^{+}}\sqrt{\gamma^2-1}\int_{\ell_{1},\ell_{2}}\frac{{\hat\delta(\ell_{1}^{(0)})\hat\delta(\gamma \ell_{2}^{0}-\gamma\beta \ell_{2}^{(x)})}}{(\ell_{2}^{(0)}+i\varepsilon)(\ell_{1}+\ell_{2}-q)^2(\ell_{1}-q)^2(\ell_{2}-q)^2}\,,\\ &
   =-\int_{\boldsymbol{\ell_1},\boldsymbol{\ell_2}}\frac{1}{(\boldsymbol{\ell_2}\cdot \boldsymbol{n}+ i \delta)\,\boldsymbol{\ell_1}^2\,\boldsymbol{\ell_2}^2 \,(\boldsymbol{\ell_1}+\boldsymbol{\ell_2}-\boldsymbol{q})^2}\,,\\ &
   =\frac{i 4^{2 \epsilon -3} \pi ^{2 \epsilon -\frac{5}{2}} \Gamma \left(\frac{1}{2}-2 \epsilon \right) \Gamma (-\epsilon ) \Gamma \left(2 \epsilon +\frac{1}{2}\right)
   \Gamma^2(\frac{1}{2}-\epsilon)}{\Gamma \left(\frac{1}{2}-3 \epsilon \right) \Gamma(1-2\epsilon)}\frac{1}{(-q^2)^{1/2+2\epsilon}}\,.\label{3.37h}
    \end{split}
\end{align}
The appearance of negative sign in \eqref{3.37h} can be justified by noticing that $\mathbfcal{M}_{0,1;0,0,1,1,1}^{(+)}(+i\varepsilon)=-\mathbfcal{M}_{0,1;0,0,1,1,1}^{(-)}(-i\varepsilon)$, which can be achieved by momentum flip $\ell_i\to -\ell_i,\,q\to -q$. 
Now, we can extract the constant $C_1$ mentioned in \eqref{MasterC} and is given by,
\begin{align}
    \begin{split}
        C_1(\gamma_0\to 1^{+},\epsilon)=\frac{i\times 4^{2 \epsilon -3} \pi ^{2 \epsilon -\frac{5}{2}} \Gamma \left(\frac{1}{2}-2 \epsilon \right) \Gamma (-\epsilon ) \Gamma \left(2 \epsilon +\frac{1}{2}\right)
   \Gamma^2(\frac{1}{2}-\epsilon)}{\Gamma \left(\frac{1}{2}-3 \epsilon \right) \Gamma(1-2\epsilon)}\,.
    \end{split}
\end{align}
Therefore, the master integral is given by,
\begin{align}
    \begin{split}
       (-q^2)^{1/2+2\epsilon} \mathbfcal{M}_{0, 1; 0, 0, 1, 1, 1}^{(++)}(|q|,\gamma,\epsilon)&=\frac{i\times 4^{2 \epsilon -3} \pi ^{2 \epsilon -\frac{5}{2}}\Gamma \left(\frac{1}{2}-2 \epsilon \right) \Gamma (-\epsilon ) \Gamma \left(2 \epsilon +\frac{1}{2}\right)
   \Gamma^2(\frac{1}{2}-\epsilon)}{\sqrt{\gamma^2-1}\Gamma \left(\frac{1}{2}-3 \epsilon \right) \Gamma(1-2\epsilon)}\,,\\ &
       =-\frac{i}{64\pi \epsilon\sqrt{\gamma^2-1}}+\mathcal{O}(\epsilon^0)=  (-q^2)^{1/2+2\epsilon} \mathbfcal{M}_{0, 1; 0, 0, 1, 1, 1}^{(+-)}\,.
    \end{split}
\end{align}
\par 
At two loops we encounter another family of integrals defined in \eqref{3.4a} with basis functions,
 $\{D_i,\slashed D_k\}\in\Big(\ell_{1}\cdot v_1,\,\,\ell_{2}\cdot v_1,\,,\,\ell_{1}^2,\,\, \ell_{2}^2,\,\, (\ell_{1}+\ell_{2}-q)^2,\,\, (\ell_{1}-q)^2,\,\, (\ell_{2}-q)^2,\,\, \ell_{1}\cdot v_2,\,\, \ell_{2}\cdot v_2\Big) $, where the last two entries are the cut propagators. They are just `potential region' counterparts of the $\mathbfcal{M}$-type master integrals discussed earlier. After carrying out the  IBP reduction, we found 8 master integrals, and they are completely factorized in $\epsilon$ and $\gamma$, and we do not need differential equation techniques to solve them. Of these, we will need only two for our subsequent computations, and we list the results for those two below:
 \begin{align}
     \begin{split}
& (-q^2)^{2\epsilon}\mathbfcal{L}_{0,0;0,0,1,1,1}\equiv \frac{1}{(4\pi)^{3-2\epsilon}}\frac{\Gamma^3(1/2-\epsilon)\Gamma(2\epsilon)}{\Gamma(3/2-3\epsilon)}\,,\\ &
(-q^2)^{2\epsilon+1}\mathbfcal{L}^{(++)}_{1,1;0,0,1,1,1}=(-q^2)^{2\epsilon+1}2\mathbfcal{L}^{(+-)}_{1,1;0,0,1,1,1}\equiv \frac{1}{(4\pi)^{2-2\epsilon}}\frac{\Gamma^3(-\epsilon)\Gamma(1+2\epsilon)}{3(\gamma^2-1)\Gamma(-3\epsilon)}\,. \label{MasterL}
\end{split}
 \end{align}
At one loop, we encounter the following family of integral (we always look for the case when $\gamma=1$),
\begin{align}
    \begin{split}
        \mathbfcal{G}^{\pm,\textrm{1-loop}}_{\alpha_1;\beta_1 \beta_{2}}=\int_{l_{1,2}}\frac{\hat\delta^{(\gamma_1)}(\slashed D_1)}{(\pm D_1+i\varepsilon)^{\alpha_1}}\frac{1}{D_2^{\beta_1}D_3^{\beta_2}}\label{3.41a}
    \end{split}
\end{align}
with $\{D_i\,\lceil\slashed D_k\rceil\}\in \Big(\ell_{1}\cdot v_1,\,\ell_{1}^2,\,(\ell_{1}-q)^2,\,\lceil \ell_{1}\cdot v_2\rceil\Big)$. After IBP reduction we found two master integrals: ($\mathbfcal{G}_{0;1,1},\,\mathbfcal{G}_{1;1,1}$). We will require only one of them, and it is given by,
\begin{align}
    \begin{split}
        &(-q^2)^{\epsilon+1/2}\mathbfcal{G}_{0;1,1}=\frac{2^{2 \epsilon -3} \pi ^{\epsilon -\frac{3}{2}} \Gamma \left(\frac{1}{2}-\epsilon \right)^2 \Gamma \left(\epsilon +\frac{1}{2}\right)}{\Gamma (1-2 \epsilon )}\,. \label{3.4aa}
    \end{split}
\end{align}
{\color{black}
\textbf{\textit{Comment on the \(i\varepsilon\) prescription for the bulk propagators:}}

In the computation of the eikonal phase, and in its relation to classical observables such as the impulse and the waveform, the choice of \(i\varepsilon\) prescription plays an important role. The \(i\varepsilon\) is not merely a regulator; it specifies how the poles of the propagators are bypassed and therefore determines the boundary conditions of the corresponding classical solution. In the conservative sector, which is naturally associated with the elastic eikonal phase, one uses the standard in-out prescription with Feynman propagators,
\begin{equation}
    \frac{1}{P^2+i\varepsilon}\, .
\end{equation}
The conservative contribution is then obtained from the real part of the in-out effective action or, equivalently, from the real part of the corresponding scattering amplitude. This prescription is appropriate for time-symmetric observables such as the conservative scattering angle and the conservative part of the eikonal phase.
For causal observables, such as the full impulse and the gravitational waveform, the situation is more subtle. These observables depend on the physical radiation emitted during the scattering process and therefore require retarded boundary conditions in their fully causal formulation. In the worldline QFT description this is implemented through the Schwinger--Keldysh, or in-in, formalism. The corresponding graviton propagators are retarded or advanced propagators, whose momentum-space prescription may be written schematically as
\begin{equation}
    P^2
    \;\longrightarrow\;
    (P^0\pm i\varepsilon)^2-\mathbf{P}^2 ,
\end{equation}
where the sign selects the retarded or advanced Green's function, depending on the Fourier-transform convention.

However, for the conservative part of the impulse at two-loop order, it is sufficient to use the Feynman prescription for the bulk propagators. In this case, diagrams in which a single internal propagator goes on shell are associated with radiative, dissipative dynamics, rather than with the conservative impulse. Thus the conservative two-loop impulse may still be extracted from the Feynman-prescribed in-out amplitude, after taking the appropriate real, time-symmetric contribution.

Starting at three loops and beyond, the distinction becomes more delicate. Multiple internal propagators can go on shell simultaneously, and such multi-cut contributions may also enter the conservative dynamics. In that regime, the relation between the Feynman prescription, the retarded prescription, and the separation into conservative and radiative sectors must be handled with greater care. Equivalently, the dissipative part of the impulse can be obtained either from the full in-in result with retarded propagators after subtracting the conservative part, or by explicitly accounting for the difference between Feynman and retarded Green's functions~\cite{Jakobsen:2022psy}.

Therefore, throughout this thesis, whenever the eikonal phase is discussed we use the Feynman prescription appropriate to the conservative elastic sector. For the conservative impulse at two loops, the same bulk Feynman prescription is sufficient. For causal observables, such as the full impulse and waveform, one must impose the retarded prescription required by causality, while at three loops and higher the possible conservative contribution of multiple on-shell internal modes requires a more careful treatment of the \(i\varepsilon\) prescription.
}
\section{Computation of eikonal phase}
\label{ch3:sec4}
In this section, we provide the detailed computation of the eikonal phase up to third Post-Minkowskian (PM) order. We mainly focus on the scalar-field and dCS corrections to the phase, since the contribution of the Einstein-Hilbert term has already been investigated in \cite{Jakobsen:2022fcj, Jakobsen:2021zvh}. We restrict our analysis to linear order in spin, leaving the study beyond that order for future work.

\textbf{Spinning eikonal at 2PM:}
In this subsection, we compute the eikonal phase at 2PM order by considering only the scalar-graviton interaction terms by noting that spin only appears when there are scalar-graviton vertices or scalar-fermion vertices in the worldline. The spinless diagrams involving scalar and scalar-graviton have been done in \cite{Bhattacharyya:2024aeq}.\\

\textbullet $\,\,$ We start with the diagram of the following topology where we have a scalar-superfield vertex,
\begin{align}
    \begin{split}
         \begin{minipage}[h]{0.08\linewidth}
	\vspace{4pt}
	\scalebox{1.8}{\includegraphics[width=\linewidth]{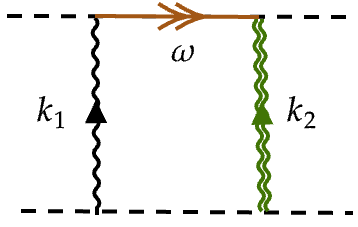}}
 \end{minipage}&
    \end{split}
\end{align}
The corresponding amplitude is given by
\begin{align}
    \begin{split}
      i \chi\Big|_{\mathcal{S}^1}= -i\frac{s_1 s_2 m_1 m_2^2}{8 m _p^4} \int e^{-i(k_1+k_2)\cdot b}\frac{\hat\delta(\omega-k_1\cdot v_1)\hat\delta(\omega+k_2\cdot v_1)\hat\delta(k_1\cdot v_2)\hat\delta(k_2\cdot \textcolor{black}{v_2})}{\omega\,k_1^2\, k_2^2}\mathcal{N}
    \end{split}
\end{align}
where the numerator takes the form:
\begin{align}
    \begin{split}
        \mathcal{N}=\omega \,\bar\Psi_{1\eta}\,\Psi_1^{\sigma}\eta^{\eta\delta}k_{2[\delta}\delta_{\sigma]}^{(\mu}v_1^{\nu)}P_{\mu\nu;\alpha\beta}v_2^\alpha v_2^\beta\,.
    \end{split}
\end{align}
Now, using the spin-supplementary condition (\ref{SSC17}) and integrating over the worldline energy we get,
\begin{align}
    \begin{split}
      i  \chi\Big|_{\mathcal{S}^1}&= \frac{\gamma s_1 s_2 m_1 m_2^2}{32 m _p^4} v_{2\eta}\mathcal{S}_1^{\eta\delta}\int e^{-i(k_1+k_2)\cdot b}\frac{\hat\delta(k_1\cdot v_1+k_2\cdot v_1)\hat\delta(k_1\cdot v_2)\hat\delta(k_2\cdot v_2)\,k_{2\delta}}{k_1^2 k_2^2}\,,\\ &
        =\frac{\gamma s_1 s_2 m_1 m_2^2}{32 m _p^4}v_{2\eta}\mathcal{S}_1^{\eta\delta}\int_{q} e^{-iq\cdot b}\hat\delta(q\cdot v_1)\hat\delta(q\cdot v_2)\int_{k_1}\frac{\hat\delta(k_1\cdot v_2)\,(q-k_1)_\delta}{k_1^2(k_1-q)^2}\,.
    \end{split}
\end{align}
Now, using Passarino-Veltman (PV) reduction and inserting the vertex factors, we get,

\begin{align}
    \begin{split} 
      i  \chi\Big|_{\mathcal{S}^1}&= \frac{\gamma s_1 s_2 m_1 m_2^2}{512 m _p^4}v_{2\eta}\mathcal{S}_1^{\eta i}\int e^{-iq\cdot b}\frac{\hat\delta(q\cdot v_1)\hat\delta(q\cdot v_2)\, q_i}{(-q^2)^{1/2}}\,,\\ &
        =-i\frac{ s_1 s_2 m_1 m_2^2}{1024 \pi m _p^4|\boldsymbol{b}|^2}\Big(\frac{1+x^2}{1-x^2}\Big)\Big(\hat b\cdot \mathcal{S}_1\cdot v_2\Big)\,.
        \end{split}
\end{align}
Here we have replaced $\gamma$ by $\frac{x^2+1}{2 x}.$ All the subsequent expressions for the eikonal phase will be expressed in terms of $x\,.$ For the $q$ integral, we have used \eqref{int} of Appendix~(\ref{ch3:app:A}). In all the subsequent computations, whenever we will encounter this kind of vector integral, we will use the result mentioned in \eqref{int}.

\textbullet $\,\,$ Another possible diagram in this order can be obtained by replacing the graviton line with the scalar line in the previous diagram.
\begin{align}
    \begin{split}
         \begin{minipage}[h]{0.08\linewidth}
	\vspace{4pt}
	\scalebox{1.8}{\includegraphics[width=\linewidth]{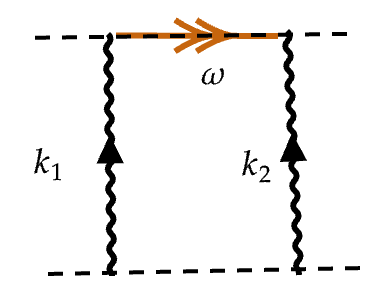}}
 \end{minipage}&
    \end{split}
\end{align}
The corresponding phase takes the following form,
\begin{align}
    \begin{split}
        i\chi\sim\int e^{-i(k_1+k_2)\cdot b}\frac{\hat\delta(\omega-k_1\cdot v_1)\hat\delta(\omega+k_2\cdot v_1)\hat\delta(k_1\cdot v_2)\hat\delta(k_2\cdot v_2)}{\omega\,k_1^2\,k_2^2}\mathcal{N}
    \end{split}
\end{align}
where the numerator takes the form,
\begin{align}
    \begin{split}
        \mathcal{N}=\omega^2\, \bar\Psi_{\eta}\Psi_{\delta}\,\eta^{\eta\delta}\,.
    \end{split}
\end{align}
Now, using the fact that another diagram with opposite spinor flow will eventually add up to this diagram such that we will eventually have terms that are antisymmetric in of $(\eta,\delta)$ and therefore, this diagram will not contribute to the eikonal phase.\textit{This structure eventually ensures that only the scalar line along with worldline propagators does not contribute to the spinning eikonal.}
\par
\textbullet $\,\,$ Now we show the detailed computation of the following 2PM diagram with $\mathbb{U}$-type topology. Note that we only focus on the spinning part of the diagram as the spinless part has been done in \cite{Bhattacharyya:2024aeq}.
\begin{align}
    \begin{split}
         \begin{minipage}[h]{0.08\linewidth}
	\vspace{4pt}
	\scalebox{1.7}{\includegraphics[width=\linewidth]{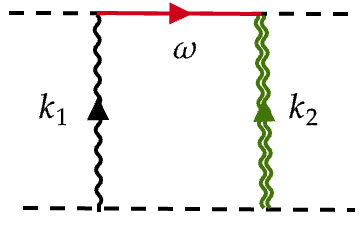}}
 \end{minipage}&
    \end{split}
\end{align}
The contribution to the eikonal phase from this diagram takes the following form,
\begin{align}
    \begin{split}
       i \chi\Big|_{\mathcal{S}^1} =-i\frac{s_1 s_2 m_1 m_2^2}{16m_p^4} \int e^{-i(k_1+k_2)\cdot b}\frac{\hat\delta(k_1\cdot v_1+k_2\cdot v_1)\hat\delta(k_1\cdot v_2)\hat\delta(k_2\cdot v_2)}{\omega^2 \,k_1^2\, k_2^2}\mathcal{N}\,,
    \end{split}
\end{align}
where the numerator $\mathcal{N}$ has the following form,
\begin{align}
    \begin{split}
        \mathcal{N}=&i\gamma (k_2\cdot \mathcal{S}_1\cdot v_2)\Big[k_1\cdot k_2-2 (k_1\cdot v_1)^2-2\, k_1\cdot v_1\,k_2\cdot v_1\Big]-4i\gamma \,(k_2\cdot \mathcal S_2\cdot v_1)(k_1\cdot v_1)^2+i\gamma (k_2\cdot \mathcal{S}_2\cdot k_1)(k_1\cdot v_1)\\ &
        -\frac{i}{2}(k_2\cdot \mathcal{S}_1\cdot k_1)(k_1\cdot v_1)\,.
    \end{split}
\end{align}
Now, redefining $k_1+k_2\to q$ and ignoring the tadpoles, we will get \footnote{We will always throw away the tadpoles before performing the IBP reduction. From now on, we will not mention it explicitly. It should be understood.},
\begin{align}
    \begin{split}
      i  \chi\Big|_{\mathcal{S}^1}\sim \int e^{-iq\cdot b}\hat\delta(q\cdot v_1)\hat\delta(q\cdot v_2)&\int \frac{\hat\delta(k_1\cdot v_2)}{(k_1\cdot v_1)^2\,k_1^2\,(k_1-q)^2}\Bigg[\frac{\gamma}{2}q^2\Big(q\cdot \mathcal{S}_1\cdot v_2-k_1\cdot \mathcal{S}_1\cdot v_2\Big)-4\gamma\Big(q\cdot \mathcal{S}_2\cdot v_1-k_1\cdot \mathcal S_2\cdot v_1\Big)\\ &
    \times (k_1\cdot v_1)^2+\gamma(q\cdot \mathcal S_2\cdot k_1)(k_1\cdot v_1) -\frac{1}{2}(q\cdot \mathcal{S}_1\cdot k_1)(k_1\cdot v_1)   \Bigg]\,.\label{4.9}
    \end{split}
\end{align}
To proceed further, we need to evaluate the following tensor integrals:
\begin{align}
    \begin{split}
        \theta_1^{\eta}=\int_{k_1}\frac{\hat\delta(k_1\cdot v_2)\,k_1^\eta}{(k_1\cdot v_1)^2\,k_1^2\,(k_1-q)^2}&=\frac{v_1^\eta-\gamma v_2^\eta}{1-\gamma^2}\int_{k_1}\frac{\hat\delta(k_1\cdot v_2)}{(k_1\cdot v_1)\,k_1^2\,(k_1-q)^2}+\frac{q^\eta}{2}\int_{k_1}\frac{\hat\delta(k_1\cdot v_2)}{(k_1\cdot v_1)^2\,k_1^2\,(k_1-q)^2}\,,\\ &
        =\frac{v_1^\eta-\gamma v_2^\eta}{1-\gamma^2} \,\mathbfcal{G}_{1;1,1}+\frac{q^\eta}{2}\frac{8 \epsilon } {(\gamma^2 -1) q^2}\mathbfcal{G}_{0;1,1}\,,\\ &
        \xrightarrow[]{\epsilon\to 0}\frac{v_1^\eta-\gamma v_2^\eta}{1-\gamma^2} \,\mathbfcal{G}_{1;1,1},\\ &
        \hspace{0cm}
        \theta_2^\eta =\int_{k_1}\frac{\hat\delta(k_1\cdot v_2)\,k_1^\eta}{(k_1\cdot v_1)\,k_1^2\,(k_1-q)^2}=\frac{v_1^\eta-\gamma v_2^\eta}{1-\gamma^2}\mathbfcal{G}_{0;1,1}+\frac{q^\eta}{2} \mathbfcal{G}_{1;1,1},\,\,\\&  \hspace{0cm}\theta_3^\eta=\int_{k_1}\frac{\hat\delta(k_1\cdot v_2)\,k_1^\eta}{k_1^2\,(k_1-q)^2}=\frac{q^\eta}{2}\mathbfcal{G}_{0;1,1}\,.
    \end{split}\label{ch3:4.10}
\end{align}
 In \eqref{ch3:4.10}, we have used the IBP reduction using a suitable choice of basis functions and write the full integral in terms of two master integrals: ($\mathbfcal{G}_{1;1,1},\,\mathbfcal{G}_{0;1,1}$). These are defined in \eqref{3.41a}. 
Now, using \eqref{ch3:4.10} as well as the anti-symmetric nature of the spin tensor and the spin supplementary condition {\eqref{SSC17}}, we get,

\begin{align}
    \begin{split}
    i    \chi\Big|_{\mathcal{S}^1}=&\frac{s_1 s_2 m_1 m_2^2}{16m_p^4} \int e^{-iq\cdot b}\hat\delta(q\cdot v_1)\hat\delta(q\cdot v_2)\Bigg(\frac{\gamma}{1-\gamma^2}\,(q\cdot \mathcal{S}_2\cdot v_1)\,+\frac{\gamma}{2(1-\gamma^2)}(q\cdot \mathcal{S}_1\cdot v_2)\Bigg)\mathbfcal{G}_{0;1,1}\,,\\&
    =i\frac{ s_1 s_2 m_1 m_2^2\,x^2(1+x^2)}{64\, \pi\, |\boldsymbol{b}|^2 m_p^4\left(1-x^2\right)^3} \Big(\hat b\cdot \mathcal{S}_2\cdot v_1 +\frac{1}{2}\hat{b}\cdot \mathcal{S}_1\cdot v_2\Big)\,.
    \end{split}
\end{align}
As we can see, although the intermediate step is tedious, the result is remarkably simple and depends on only one master integral $\mathbfcal{G}_{0;1,1}$ defined in \eqref{3.4aa} \footnote{Note that we have only shown the regularization independent finite part after doing an expansion in small $\epsilon\,.$ We will follow this same strategy for all the diagrams.}. We will use a similar methodology to evaluate the rest of the diagrams (at 3PM also).

\textbullet $\,\,$ Another scalar-graviton worldline interaction vertex having a $\mathbb{V}$-type topology will contribute to the 2PM eikonal phase. Again, for this case, the spinless part has been done in \cite{Bhattacharyya:2024aeq}\,.
\begin{align}
    \begin{split}
         \begin{minipage}[h]{0.08\linewidth}
	\vspace{4pt}
	\scalebox{1.7}{\includegraphics[width=\linewidth]{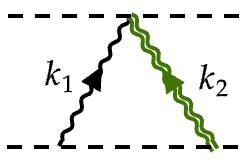}}
 \end{minipage}&
    \end{split}
\end{align}
The eikonal phase contribution is given by,
\begin{align}
    \begin{split}
        i\,\chi\Big|_{\mathcal{S}^1} & =-i\frac{s_1 s_2 m_1 m_2^2}{8m_p^4} \int e^{-i(k_1+k_2)\cdot b}\frac{\hat\delta(k_1\cdot v_1+k_2\cdot v_1)\hat\delta(k_1\cdot v_2)\hat\delta(k_2\cdot v_2)}{k_1^2 k_2^2}\\ &
        \hspace{0.8cm}\times\Big(v_1^\mu v_1^\nu-i(k_2\cdot \mathcal{S}_1)^{(\mu}v_1^{\nu)}\Big)P_{\mu\nu;\alpha\beta}\Big(v_2^\alpha v_2^\beta+i (k_2\cdot \mathcal{S}_2)^{(\alpha}v_2^{\beta)}\Big)\,,\\ &
        =\frac{ \gamma s_1 s_2 m_1 m_2^2}{8m_p^4} \int e^{-iq \cdot b}\hat\delta(q\cdot v_1)\hat\delta(q\cdot v_2)\int \frac{\hat\delta(k_1\cdot v_2)}{k_1^2 (k_1-q)^2}(-k_1\cdot 
    \mathcal{S}_2\cdot v_1+k_1\cdot \mathcal{S}_1\cdot v_2+q\cdot \mathcal{S}_2\cdot v_1-q\cdot \mathcal{S}_1\cdot v_2),\\ &
    =\frac{ \gamma s_1 s_2 m_1 m_2^2}{16m_p^4} \int e^{-iq\cdot b}\hat\delta(q\cdot v_1)\hat\delta(q\cdot v_2)\Big((q\cdot \mathcal S_2\cdot v_1)-(q\cdot \mathcal S_1\cdot v_2)\Big)\,\mathbfcal{G}_{0;1,1}\,,\\&
    =i\frac{ s_1 s_2 m_1 m_2^2\,(1+x^2)}{256\,\pi\,  |b|^2 m_p^4\left(1-x^2\right)} \Big(\hat b\cdot \mathcal{S}_2\cdot v_1 -\hat{b}\cdot \mathcal{S}_1\cdot v_2\Big)\,.
    \end{split}
\end{align}
Hence the final form of the eikonal phase is,

\begin{align}
    \begin{split}
i\chi\Big|_{\mathcal{S}^{1}}=  i\frac{ s_1 s_2 m_1 m_2^2\,(1+x^2)}{256\,\pi\,  |\boldsymbol{b}|^2 m_p^4\left(1-x^2\right)} \Big(\hat b\cdot \mathcal{S}_2\cdot v_1 -\hat{b}\cdot \mathcal{S}_1\cdot v_2\Big)\,.
    \end{split}
\end{align}
At this point, we would like to remind the readers that one also needs to add the contributions from 2PM diagrams obtained by interchanging the worldlines one and two in all the diagrams mentioned above. We will not show them explicitly as the contributions from those diagrams will be similar to those we have already computed and can be obtained easily by exchanging labels one and two.

Finally adding all the $\mathcal{O}(\mathcal{S}^1)$ at 2PM contributions we have,

\begin{tcolorbox}[thesisresultbox, title=\textit{Spinning Eikonal at 2PM}]
\restorethesisbodyformat
\begin{align*}   \chi\Big|_{\mathcal{S}^1}^{\textrm{2PM}}=&\frac{m_1m_2^2}{m_p^4}\Bigg[\Rho_{01}\Big(\hat b\cdot \mathcal{S}_{1}\cdot v_2\Big)+\Rho_{02}\Big(\hat b\cdot \mathcal{S}_2\cdot v_1 \Big)\Bigg]+(1\leftrightarrow 2)\,,
\end{align*}
\end{tcolorbox}
\restorethesisbodyformat
where, $$\Rho_{01}=\frac{s_1\,s_2\,\left(x^2+1\right) \left(5 x^4-18 x^2+5\right)}{1024 \pi  |\boldsymbol{b}|^2  \left(x^2-1\right)^3}\,,\quad \Rho_{02}=-\frac{s_1\,s_2\,\left(x^2+1\right)^3}{256 \pi  |\boldsymbol{b}|^2  \left(x^2-1\right)^3}\,.$$
\subsection*{Spinning eikonal at 3PM}
Now, we will compute the contributions to the eikonal phase at 3PM order. We will focus on the correction to the eikonal phase due to the extra scalar degree of freedom.

\textbullet $\,\,$ First, we will compute the Feynman diagrams coming purely from the scalar part. We start with the first comb-type diagram. 
\begin{center}
\includegraphics[width=0.38\linewidth]{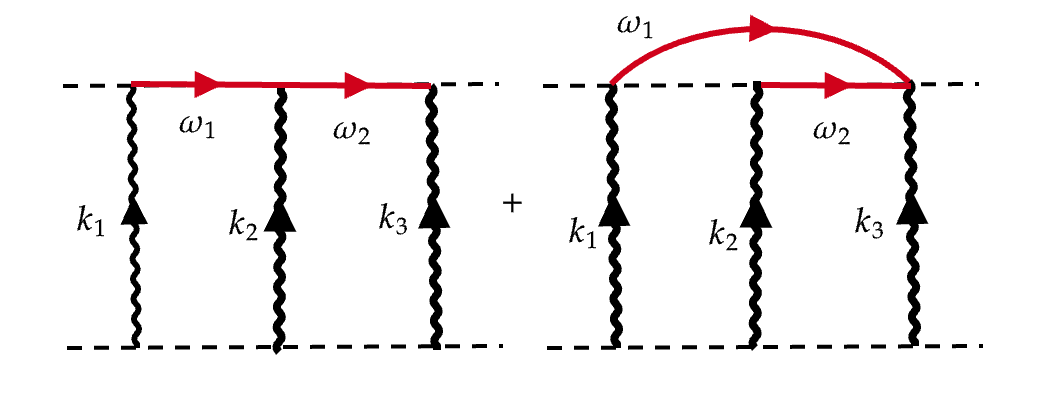}
\end{center}
The contribution to the eikonal phase looks like,
\begin{align}
    \begin{split}
       i\, \chi\Big|_{\mathcal{S}^0}&=-i \frac{s_1^3 s_2^3 m_1 m_2^3}{64 m_p^6} \rmint_{k_i,\omega_i}\hat\delta(k_1\cdot v_2)\hat\delta(k_2\cdot v_2)\hat\delta(k_3\cdot v_2)\hat\delta(-k_1\cdot v_1+\omega_1)\hat\delta(-k_2\cdot v_1+\omega_2-\omega_1)\hat\delta(\omega_2+k_3\cdot v_1)\\ &
       \hspace{2 cm} \times\frac{ \mathcal{N}}{\prod_i k_i^2}\left(\sum_{(\pm)} \frac{1}{(\pm\omega_1+i\varepsilon)^2(\pm\omega_2+i\varepsilon)^2}\right)\,e^{-i(k_1+k_2+k_3)\cdot b}
    \end{split}
\end{align}
Now relabeling the momenta $k_1\to \ell_1,\,k_2\to q-\ell_1-\ell_2\,,k_3\to\ell_2$ we can write the numerator as,\footnote{We have performed all the tensor contractions used in this paper using \textbf{FeynCalc} \cite{Mertig:1990an, Shtabovenko:2020gxv,Shtabovenko:2023idz}\,.},
\begin{align}
    \begin{split}
        \mathcal{N}=&4(\ell_1\cdot v_1)(\ell_2\cdot v_1)\left(\ell_1\cdot v_1+\ell_2\cdot v_1\right)^2
    \end{split}
\end{align}
\noindent
Now, taking care of the $\pm i\varepsilon$ prescription and then doing IBP reduction using \textbf{LiteRed} \cite{Lee:2012cn,Lee:2013mka}, one can find the eikonal phase in terms of two master integrals and is given by, 
\begin{align}
    \begin{split}
      &  i\chi\Big|_{\mathcal{S}^0}= -i \frac{s_1^3 s_2^3 m_1 m_2^3}{64 m_p^6}  \int_{q}e^{-iq\cdot b}\hat\delta(q\cdot v_1)\hat\delta(q\cdot v_2)\Bigg[  \boldsymbol{\tilde h_1}\mathbfcal{L}_{0,0;0,0,1,1,1}(q) -\boldsymbol{\boldsymbol{\tilde h_2}}\,q^2\mathbfcal{L}^{(+-)}_{1,1;0,0,1,1,1}(q)\Bigg]\,.
    \end{split}
\end{align}
where, $\boldsymbol{\tilde h_1}$ and $\boldsymbol{\tilde h_2}$ are defined in Appendix~(\ref{ch3:app:E})\,.
Then using \eqref{MasterL}, we finally get,

\begin{align}
    \begin{split} \label{eq1}
\chi\Big|_{\mathcal{S}^0}=\frac{ m_1 m_2^3}{m_p^6}\Bigg(\frac{s_1^3 s_2^3 x}{256 \pi ^3 |\boldsymbol{b}|^2 \left(x^2-1\right)}\Bigg) ,
    \end{split}
\end{align}
where $\gamma_E$ is the Euler-Mascheroni constant. For the $q$ integral, we have used \eqref{int1} of Appendix~(\ref{ch3:app:A}). In all the subsequent computations, whenever we will encounter this kind of scalar integral, we will use the result mentioned in \eqref{int1}.
The second diagram gives the following contribution,
\begin{align}
    \begin{split}
      i\chi\Big|_{\mathcal{S}^0}=-i \frac{s_1^3 s_2^3 m_1 m_2^3}{256 m_p^6} \int e^{-iq\cdot b}\hat\delta(q\cdot v_1)\hat\delta(q\cdot v_2)\int \sum_{(\pm)}\frac{\hat\delta(\ell_1\cdot v_2)\hat\delta(\ell_2\cdot v_2)\,\mathcal{N}}{(\pm\ell_1\cdot v_1+i\varepsilon)^2\,(\pm\ell_2\cdot v_1+i\varepsilon)^2\,\ell_1^2\,\ell_2^2\,(\ell_1+
        \ell_2-q)^2}
    \end{split}
\end{align}
where the numerator has the following form,
\begin{align}
    \begin{split}
        \mathcal{N}=&\frac{1}{8}\Bigg[(\ell_1-q)^2(\ell_2-q)^2+4q^2 (\ell_1\cdot v_1)(\ell_2\cdot v_1)\Bigg]\,.
    \end{split}
\end{align}
Therefore, the eikonal phase is given by,
\begin{align}
    \begin{split} 
    i\chi\Big|_{\mathcal{S}^0}=-i \frac{s_1^3 s_2^3 m_1 m_2^3}{64 m_p^6}\int e^{-iq\cdot b}\hat\delta(q\cdot v_1)\hat\delta(q\cdot v_2)\Bigg[\boldsymbol{\tilde h}_3\,\mathbfcal{L}_{0,0;0,0,1,1,1}+\frac{1}{4}\boldsymbol{\tilde h}_4\,q^2\,\mathbfcal{L}^{(++)}_{1,1;0,0,1,1,1}\Bigg]
    \end{split}
\end{align}
where, $ \boldsymbol{\tilde h_3},  \boldsymbol{\tilde h_4}$ are defined in Appendix~(\ref{ch3:app:E}).
Using \eqref{MasterL}, we finally get,

\begin{align}
    \begin{split} \label{eq2}
 \chi\Big|_{\mathcal{S}^0}=\frac{m_1 m_2^3}{m_p^6}\Bigg(\frac{ s_1^3 s_2^3 x^3 \left(x^4+1\right) (6 \log (|\boldsymbol{b}|)+5 \gamma_E +\log \left(\frac{\pi }{16}\right))}{512 \pi ^3 |\boldsymbol{b}|^2 \left(x^2-1\right)^5}\Bigg)\,.
    \end{split}
\end{align}
Then, the total contribution to the eikonal phase coming from these two diagrams after summing (\ref{eq1}) and (\ref{eq2}) is the following, 
\begin{align}
    \begin{split} \label{eqt1}
 \chi\Big|_{\mathcal{S}^0}=\frac{m_1 m_2^3}{m_p^6}\Delta_{\mathfrak{01}}^{\textrm{poly}}\,,
    \end{split}
\end{align}
where, $$\Delta_{\mathfrak{01}}^{\textrm{poly}}=\Bigg(\frac{s_1^3 s_2^3 x \left(\left(x^6+x^2\right) \log \left(\frac{\pi  |\boldsymbol{b}|^6}{16}\right)+2 x^8+(5 \gamma_E -8) x^6+12 x^4+(5 \gamma_E -8) x^2+2\right)}{512 \pi ^3 |\boldsymbol{b}|^2 \left(x^2-1\right)^5}\Bigg)\,.$$

\textbullet $\,\,$ Apart from the previous one, we have the following two comb-type diagrams where one of the scalar propagators is replaced by the graviton propagator. 
\begin{center}
\includegraphics[width=0.38\linewidth]{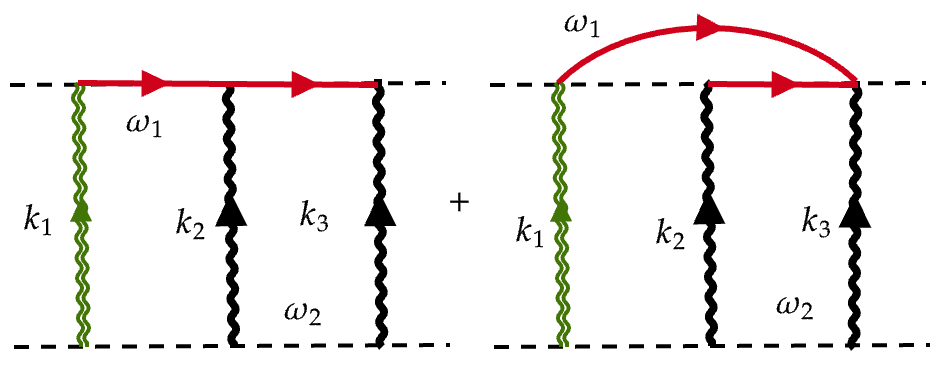}
\end{center}
We proceed as before. The non-spinning part of the eikonal phase is given by, 
\begin{align}
    \begin{split} \label{eq3}
      i\chi\Big|_{\mathcal{S}^0}=-i\frac{m_1m_2^3 s_1^2 s_2^2}{32 m_p^6}\int e^{-iq\cdot b}\hat\delta(q\cdot v_1)\hat\delta(q\cdot v_2)\,\Bigg[ \boldsymbol{\tilde h}_5 \,\mathbfcal{L}_{0,0;0,0,1,1,1}-\frac{1}{2}\boldsymbol{\tilde h}_6 \,q^2\,\mathbfcal{L}^{(+-)}_{1,1;0,0,1,1,1}\Bigg]
    \end{split}
\end{align}
$ \boldsymbol{\tilde h_5},  \boldsymbol{\tilde h_6}$ are defined in Appendix~(\ref{ch3:app:E}).
Similarly, the spinning part is given by, 
\begin{align}
    \begin{split}  \label{eq4}
      i\chi\Big|_{\mathcal{S}^1}=-i\frac{m_1m_2^3 s_1^2 s_2^2}{32 m_p^6}\int e^{-iq\cdot b}\hat\delta(q\cdot v_1)\hat\delta(q\cdot v_2)\,\Bigg[ &i\,\Bigg(\boldsymbol{\tilde h}_7 \,\mathbfcal{L}_{0,0;0,0,1,1,1}-\boldsymbol{\tilde h}_8 \,q^2\,\mathbfcal{L}^{(+-)}_{1,1;0,0,1,1,1}\Bigg)(q\cdot \mathcal{S}_1\cdot v_2)+\\&i\,\Bigg(\boldsymbol{\tilde h}_9\,\mathbfcal{L}_{0,0;0,0,1,1,1}-\frac{1}{2}\boldsymbol{\tilde h}_{10}\,q^2\,\mathbfcal{L}^{(+-)}_{1,1;0,0,1,1,1}\Bigg)(q\cdot \mathcal{S}_2\cdot v_1)\Bigg]\,.
    \end{split}
\end{align}

Now, we compute the contribution to the eikonal phase from the second diagram.
The spinless part of the eikonal phase is given by,
\begin{align}
    \begin{split}\label{eq5}
        i\chi\Big|_{\mathcal{S}^0}=-i\frac{m_1m_2^3 s_1^2 s_2^2}{32 m_p^6}\int e^{-iq\cdot b}\hat\delta(q\cdot v_1)\hat\delta(q\cdot v_2)\,\Bigg[\boldsymbol{\tilde h}_{11}\,\mathbfcal{L}_{0,0;0,0,1,1,1}+\frac{1}{4}\boldsymbol{\tilde h}_{12}q^2\,\mathbfcal{L}^{(++)}_{1,1;0,0,1,1,1}\,\Bigg]\,.
    \end{split}
\end{align}
Similarly, the spinning part is given by, 
\begin{align}
    \begin{split} \label{eq6}
      i\chi\Big|_{\mathcal{S}^1}=-i\frac{m_1m_2^3 s_1^2 s_2^2}{32 m_p^6}\int e^{-iq\cdot b}\hat\delta(q\cdot v_1)\hat\delta(q\cdot v_2)\,\Bigg[ &i\,\Bigg(\boldsymbol{\tilde h}_{13} \,\mathbfcal{L}_{0,0;0,0,1,1,1}+ \frac{1}{4}\boldsymbol{\tilde h}_{14} \,q^2\,\mathbfcal{L}^{(++)}_{1,1;0,0,1,1,1}\Bigg)(q\cdot \mathcal{S}_1\cdot v_2)+\\ &i\,\Bigg(\boldsymbol{\tilde h}_{15}\,\mathbfcal{L}_{0,0;0,0,1,1,1}+ \frac{1}{4}\boldsymbol{\tilde h}_{16}\,q^2\,\mathbfcal{L}^{(++)}_{1,1;0,0,1,1,1}\Bigg)(q\cdot \mathcal{S}_2\cdot v_1)\Bigg]\,.
    \end{split}
\end{align}
Again all the $ \boldsymbol{\tilde h_i}'s\,, i=7,\cdots 16\,.$ are defined in Appendix~(\ref{ch3:app:E}). Note that, at $\mathcal{O}(\mathcal{S})$, we encounter tensor integrals. We first have to perform a Veltman-Passarino Reduction (PV) to write it in terms of a set of scalar integrals. Then, we do the IBP reduction for each of those scalar integrals. We will use this same strategy throughout this paper.  \par
Finally, the total contribution at $\mathcal{O}(\mathcal{S}^0)$ to the eikonal phase coming from these two diagrams after summing (\ref{eq3}) and (\ref{eq5}) is the following, 
\begin{align}
    \begin{split}  \label{eqt2}
   \chi\Big|_{\mathcal{S}^0} =\frac{m_1m_2^3}{m_p^6}\Delta_{\mathfrak{02}}^{\textrm{poly}}\,,
 \end{split}
\end{align}
\enlargethispage{2\baselineskip}
where, 
\begin{thesiscompactdisplay}
\begin{align}
    \begin{split} 
\Delta_{\mathfrak{02}}^{\textrm{poly}}= &\Bigg(\frac{s_1^2s_2^2}{12288 \pi ^3 |\boldsymbol{b}|^2 x \left(x^2-1\right)^7}\Bigg)\Bigg[6 \left(-x^{12}+10 x^{10}+x^8+44 x^6+x^4+10 x^2-1\right) x^2 \log \left(\pi  |\boldsymbol{b}|^6\right)\\&\hspace{0.8cm}-9 x^{16}+x^{14} (-36 \gamma_E +56+\log (4096))+4 x^{12} (82 \gamma_E  -25-46 \log (2))\\&\hspace{0.8cm}-4 x^{10} (7 \gamma_E -6+35 \log (2))+2 x^8 (696 \gamma_E  -35-456 \log (2))\\&\hspace{0.8cm}-4 x^6 (7 \gamma_E  -6+35 \log (2))+4 x^4 (82 \gamma_E  -25-46 \log (2))\\&\hspace{0.8cm}+x^2 (-36 \gamma_E  +56+\log (4096))-9\Bigg]
     \end{split}
\end{align}
\end{thesiscompactdisplay}
and the total contribution at $\mathcal{O}(\mathcal{S}^1)$ to the eikonal phase coming from these two diagrams after summing (\ref{eq4}) and (\ref{eq6}) is the following, 
\begin{align}
    \begin{split}   \label{eqt3}
     \chi\Big|_{\mathcal{S}^1}=\frac{m_1m_2^3}{m_p^6}\Bigg[\Delta_{\mathfrak{03}}^{\textrm{poly}}(\hat{b}\cdot \mathcal{S}_1\cdot v_2)+\Delta_{\mathfrak{04}}^{\textrm{poly}}(\hat{b}\cdot \mathcal{S}_2\cdot v_1)\Bigg]\,,
     \end{split}
\end{align}
\enlargethispage{2\baselineskip}
where, 
\begin{thesiscompactdisplay}
\begin{align}
    \begin{split}
    \Delta_{\mathfrak{03}}^{\textrm{poly}}= \Bigg(\frac{s_1^2s_2^2\,x} {3072 \pi ^3 |\boldsymbol{b}|^3 \left(x^2-1\right)^5 }\Bigg)&\Bigg[-6 \left(x^8+6 x^7-2 x^6-26 x^5-6 x^4-26 x^3-2 x^2+6 x+1\right) \log (|\boldsymbol{b}|)\\&\hspace{0 cm}+5 x^8+x^8 \log (64)-x^8 \log (\pi )+x^7 \log (4096)-6 x^7 \log (\pi )-6 x^6\\ &+2 x^6 \log \left(\frac{\pi }{64}\right)-81 x^5+10 x^5 \log (4)-26 x^5 \log \left(\frac{16}{\pi }\right)+2 x^4\\ &-6 x^4 \log (64)+6 x^4 \log (\pi )-81 x^3+10 x^3 \log (4)-26 x^3 \log \left(\frac{16}{\pi }\right)-6 x^2\\ &+2 x^2 \log \left(\frac{\pi }{64}\right)-4 \gamma_E  \left(x^8+9 x^7-2 x^6-35 x^5-6 x^4-35 x^3-2 x^2+9 x+1\right)\\ &+x \log \left(\frac{4096}{\pi ^6}\right)+5+\log (64)-\log (\pi )\Bigg]\,,
        \end{split}
\end{align}
\begin{align}
    \begin{split}    \Delta_{\mathfrak{04}}^{\textrm{poly}}=\Bigg(\frac{s_1^2s_2^2\,x^2(1+x^2)} {768  \pi ^3 |\boldsymbol{b}|^3 \left(x^2-1\right)^7 }\Bigg)&\Bigg[12 \left(x^8+x^6-8 x^4+x^2+1\right) \log (|\boldsymbol{b}|)+13 x^8+x^8 \log \left(\frac{\pi ^2}{64}\right)-159 x^6\\ &-12 x^6 \log (2)+2 x^6 \log (\pi )+392 x^4+68 x^4 \log (2)-16 x^4 \log (\pi )\\ &-159 x^2-12 x^2 \log (2)+2 x^2 \log (\pi )+\gamma_{E}  \left(11 x^8+8 x^6-78 x^4+8 x^2+11\right)\\ &+13+\log \left(\frac{\pi ^2}{64}\right)\Bigg]\,.
    \end{split}
\end{align}
\end{thesiscompactdisplay}
\par
\textbullet $\,\,$ We have another two comb-type diagrams with two graviton line 
\begin{center}
\includegraphics[width=0.38\linewidth]{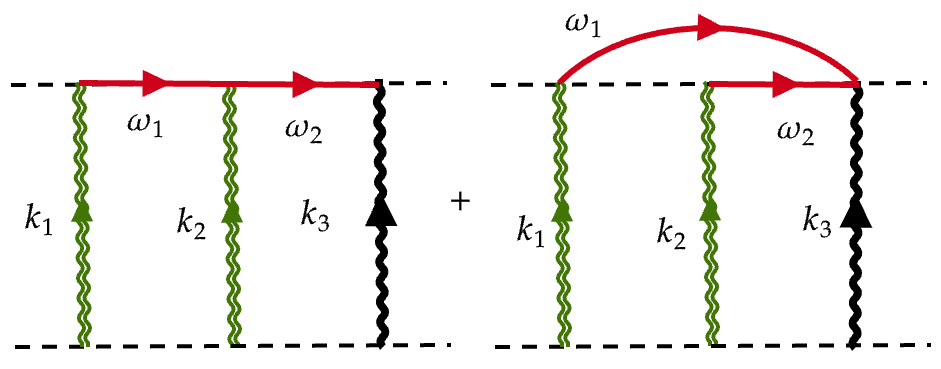}
\end{center}


For the first diagram, we have, 

\begin{align}
    \begin{split} \label{eq8}
        i\chi\Big|_{\mathcal{S}^0} =i\frac{m_1 m_2^3 s_1 s_2}{64 m_p^6}\int e^{-iq\cdot b}\hat\delta(q\cdot v_1)\hat\delta(q\cdot v_2)\Bigg( \boldsymbol{\tilde h}_{17}\mathbfcal{L}_{0,0;0,0,1,1,1}-\frac{1}{2}\boldsymbol{\tilde h}_{18}q^2\,\mathbfcal{L}^{(+-)}_{1,1;0,0,1,1,1}\Bigg)\,,
    \end{split}
\end{align}
and 
\begin{align}
    \begin{split} \label{eq9}
        i\chi\Big|_{\mathcal{S}^1} =i\frac{m_1 m_2^3 s_1 s_2}{64 m_p^6}\int e^{-iq\cdot b}\hat\delta(q\cdot v_1)\hat\delta(q\cdot v_2)&\Bigg[i\Bigg( \boldsymbol{\tilde h}_{19}\mathbfcal{L}_{0,0;0,0,1,1,1}-\frac{1}{2}\boldsymbol{\tilde h}_{20}q^2\,\mathbfcal{L}^{(+-)}_{1,1;0,0,1,1,1}\Bigg)(q\cdot \mathcal{S}_1\cdot v_2)\\&i\,\Bigg( \boldsymbol{\tilde h}_{21}\mathbfcal{L}_{0,0;0,0,1,1,1}-\frac{1}{2}\boldsymbol{\tilde h}_{22}q^2\,\mathbfcal{L}^{(+-)}_{1,1;0,0,1,1,1}\Bigg)(q\cdot \mathcal{S}_2\cdot v_1)\Bigg]\,,
    \end{split}
\end{align}

We now compute the eikonal phase for the second diagram.
The spinless eikonal phase is given by,
\begin{align}
    \begin{split} \label{eq10}
        \chi\Big|_{\mathcal{S}^0} =-i\frac{m_1 m_2^3 s_1 s_2}{64 m_p^6} \int e^{-iq\cdot b}\hat\delta(q\cdot v_1)\,\hat\delta(q\cdot v_2)\,\Bigg[\boldsymbol{\tilde h}_{23} \,\mathbfcal{L}_{0,0;0,0,1,1,1}+\frac{1}{4}\,\boldsymbol{\tilde h}_{24}\,q^2\,\mathbfcal{L}^{(++)}_{1,1;0,0,1,1,1}\Bigg]
    \end{split}
\end{align}
The spinning eikonal phase is given by,
\begin{align}
    \begin{split} \label{eq11}
        i\chi\Big|_{\mathcal{S}^1} =-i\frac{m_1 m_2^3 s_1 s_2}{64 m_p^6}\int e^{-iq\cdot b}\hat\delta(q\cdot v_1)\hat\delta(q\cdot v_2)&\Bigg[i\Bigg( \boldsymbol{\tilde h}_{25}\mathbfcal{L}_{0,0;0,0,1,1,1}+\frac{1}{4}\boldsymbol{\tilde h}_{26}q^2\,\mathbfcal{L}^{(++)}_{1,1;0,0,1,1,1}\Bigg)(q\cdot \mathcal{S}_1\cdot v_2)\\&i\,\Bigg( \boldsymbol{\tilde h}_{27}\mathbfcal{L}_{0,0;0,0,1,1,1}+\frac{1}{4}\boldsymbol{\tilde h}_{28}q^2\,\mathbfcal{L}^{(++)}_{1,1;0,0,1,1,1}\Bigg)(q\cdot \mathcal{S}_2\cdot v_1)\Bigg]\,,
    \end{split}
\end{align}
Again, all the coefficients $\boldsymbol{\tilde h}'s$ are defined in Appendix ~(\ref{ch3:app:E}).
Finally, the total contribution at $\mathcal{O}(\mathcal{S}^0)$ to the eikonal phase coming from these two diagrams after summing (\ref{eq8}) and (\ref{eq10}) is the following, 
\begin{align}
    \begin{split}  \label{eqt4}
     \chi\Big|_{\mathcal{S}^0}=\frac{m_1m_2^3}{m_p^6}\Delta_{\mathfrak{05}}^{\textrm{poly}}
     \end{split}
\end{align}
where,
\begin{align}
    \begin{split}
\Delta_{\mathfrak{05}}^{\textrm{poly}}=\frac{s_1s_2}{24576 \pi ^3 |\boldsymbol{b}|^2 x \left(x^2-1\right)^5}\Bigg[&24 \left(x^8+x^4\right) \log \left(\pi  |\boldsymbol{b}|^6\right)+x^2 \Big(-5 x^6 (37+20 \log (2))-4 x^4 (\log (4)-47)\\ &-5 x^2 (37+20 \log (2))+x^8 \left(x^2 (\log (16)-43)+134+\log (16)\right)\\ &+134+\log (16)\Big)+2 \gamma_{E}  \left(x^{12}+x^{10}+59 x^8-2 x^6+59 x^4+x^2+1\right)\\ &-43+\log (16)\Bigg]\,.
        \end{split}
\end{align}
The total contribution at $\mathcal{O}(\mathcal{S}^1)$ to the eikonal phase coming from these two diagrams after summing (\ref{eq9}) and (\ref{eq11}) is the following,
\begin{align}
    \begin{split}  \label{eqt5}
     \chi\Big|_{\mathcal{S}^1}=\frac{m_1m_2^3}{m_p^6}\Bigg[\Delta_{\mathfrak{06}}^{\textrm{poly}}(\hat{b}\cdot \mathcal{S}_1\cdot v_2)+\Delta_{\mathfrak{07}}^{\textrm{poly}}(\hat{b}\cdot \mathcal{S}_2\cdot v_1)\Bigg]\,,
     \end{split}
\end{align}
where, 
\begin{align}
    \begin{split}   &\Delta_{\mathfrak{06}}^{\textrm{poly}}=\frac{s_1s_2}{147456 \pi ^3 |\boldsymbol{b}|^3  \left(x^2-1\right)^7}\Bigg[3\alpha_0-3\alpha_1(x^2+x^{12})+3\alpha_2x^{14}-3\alpha_3(x^6+x^{8})+3\alpha_4(x^4+x^{10})\\&\hspace{0.8cm}-13824 x^{11}+41472 x^9-55296 x^7+41472 x^5-13824 x^3\Bigg]\,,\\
& \Delta_{\mathfrak{07}}^{\textrm{poly}}=-\frac{s_1 s_2}{24576 \pi ^3 |\boldsymbol{b}|^3 \left(x^2-1\right)^7 }\Bigg[\hat{\alpha}_0+8\hat{\alpha}_1(x^5+x^{9})+\hat{\alpha}_0x^{14}+\alpha_2(x^2+x^{12})+\hat{\alpha}_3(x^4+x^{10})\\&\hspace{0.8cm}+\hat{\alpha}_4(x^6+x^{8})+8\hat{\alpha}_5(x^3+x^{11})-360 x^{13}-360 x\\&\hspace{0.8cm}-16 x^7 (18 \log (|\boldsymbol{b}|)+14 \gamma_E -161-14 \log (2)+3 \log (\pi ))\Bigg]
\end{split}
\end{align}
\enlargethispage{2\baselineskip}
with, 
\begin{thesiscompactdisplay}
\begin{align}
    \begin{split} 
   & \alpha_0= 8 \log \left(\frac{\pi  |\boldsymbol{b}|^6}{16}\right)+72 \log ^2(b)+24 \left(5 \gamma_E -3+\log \left(\frac{\pi }{16}\right)\right) \log (|\boldsymbol{b}|)+50 \gamma_E ^2+\pi ^2-19 \gamma_E +89+\\&\hspace{1cm} 2 \log ^2(\pi )+48 \log (2)+16 \log (2) \log (4)+\log (4)+20 \gamma_E  \log \left(\frac{\pi }{16}\right)-16 \log (2) \log (\pi )-12 \log (\pi )\,,\\&
   \alpha_1=24 \log (|\boldsymbol{b}|) (3 \log (b)+5 \gamma_E +24)+4 \log \left(\frac{\pi }{16}\right) (6 \log (b)+5 \gamma_E )+\pi ^2+\gamma_E  (473+50 \gamma )+573+32 \log ^2(2)\\&\hspace{1cm}-2 (199+8 \log (\pi )) \log (2)+2 \log (\pi ) (48+\log (\pi ))\,,\\
    &\alpha_2=24 \left(\log \left(\frac{\pi  |\boldsymbol{b}|^3}{16}\right)+5 \gamma_E -1\right) \log (|\boldsymbol{b}|)+50 \gamma_E ^2+\pi ^2+89+\log (4) \left(9+8 \log \left(\frac{4}{\pi }\right)\right)+2 (\log (\pi )-2) \log (\pi )\\&\hspace{1cm}+\gamma_E  (-19-80 \log (2)+20 \log (\pi ))\,,\nonumber
     \end{split}
\end{align}
 \begin{align}
    \begin{split} 
& \alpha_3=-504 \log \left(\frac{\pi  |\boldsymbol{b}|^6}{16}\right)+24 \log (|\boldsymbol{b}|) (3 \log (b)+5 \gamma_E -3)+4 \log \left(\frac{\pi }{16}\right) (6 \log (|\boldsymbol{b}|)+5 \gamma_E)+\gamma_E (50 \gamma_E -2583)+\pi ^2\\&\hspace{1cm}+2245+(21+16 \log (2)) \log (4)+2 \log (\pi ) (-6-8 \log (2)+\log (\pi ))\,,\\&
\alpha_4=-20 \log \left(\frac{\pi  |\boldsymbol{b}|^6}{16}\right)+36 \left(6 \log (|\boldsymbol{b}|) (3 \log (|\boldsymbol{b}|)+5 \gamma_E -3)+\log \left(\frac{\pi }{16}\right) (6 \log (|\boldsymbol{b}|)+5 \gamma_E )\right)+450 \gamma_E ^2+9 \pi ^2-651 \gamma_E \\&\hspace{1cm}+2697+(205+144 \log (2)) \log (4)+18 \log (\pi ) (-6-8 \log (2)+\log (\pi ))\,,\\&\hat{\alpha}_0=11 \log \left(\frac{\pi  |\boldsymbol{b}|^6}{16}\right)+65 \gamma_E +54+10 \log (4)\,,\\&\hat{\alpha}_1=72 \log ^2(|\boldsymbol{b}|)+12 (10 \gamma_E-7-8 \log (2)+2 \log (\pi )) \log (|\boldsymbol{b}|)+50 \gamma_E ^2+\pi ^2-271+4 \log (2) (13+\log (256))+\\&\hspace{1cm}2 \left(\log \left(\frac{\pi }{256}\right)-7\right) \log (\pi )-4 \gamma_E  (18+20 \log (2)-5 \log (\pi))\,,\\&\hat{\alpha}_2=32 \log \left(\frac{\pi }{16}\right) (6 \log (|\boldsymbol{b}|)+5 \gamma_E)+6 \log (|\boldsymbol{b}|) (96 \log (|\boldsymbol{b}|)+160 \gamma_E -85)+400 \gamma_E ^2+8 \pi ^2-435 \gamma_E +684+\\&\hspace{1cm}16 \left(\log ^2(16)+\log ^2(\pi )+\log (2) (20-8 \log (\pi ))\right)-85 \log (\pi )\,,\\&\hat{\alpha}_3=-275 \log \left(\frac{\pi  |\boldsymbol{b}|^6}{16}\right)-64 \left(6 \log (|\boldsymbol{b}|) (3 \log (|\boldsymbol{b}|)+5 \gamma_E -3)+\log \left(\frac{\pi }{16}\right) (6 \log (|\boldsymbol{b}|)+5 \gamma_E )\right)-\\&\hspace{1cm}\gamma_E  (429+800 \gamma_E )-4 \left(4 \pi ^2+398+\log (2) (199+128 \log (2))+8 \log (\pi ) (-6-8 \log (2)+\log (\pi ))\right)\,,\\&\hat{\alpha}_4=-131 \log \left(\frac{\pi |\boldsymbol{b}|^6}{16}\right)-160 \left(6 \log (|\boldsymbol{b}|) (3 \log (|\boldsymbol{b}|)+5 \gamma_E -3)+\log \left(\frac{\pi }{16}\right) (6 \log (|\boldsymbol{b}|)+5 \gamma_E )\right)+\\&\hspace{1cm}\gamma_E  (1759-2000 \gamma_E)-40 \pi ^2+1814-4 \log (2) (473+320 \log (2))-80 \log (\pi ) (-6-8 \log (2)+\log (\pi ))\,,\\&
\hat{\alpha}_5=\log \left(\frac{\pi |\boldsymbol{b}|^6}{4}\right)+6 \gamma_E +163\,.
    \end{split}
\end{align}
\end{thesiscompactdisplay}
\textbullet $\,\,$ Next we compute the following two diagrams. 
\begin{center}
\includegraphics[width=0.38\linewidth]{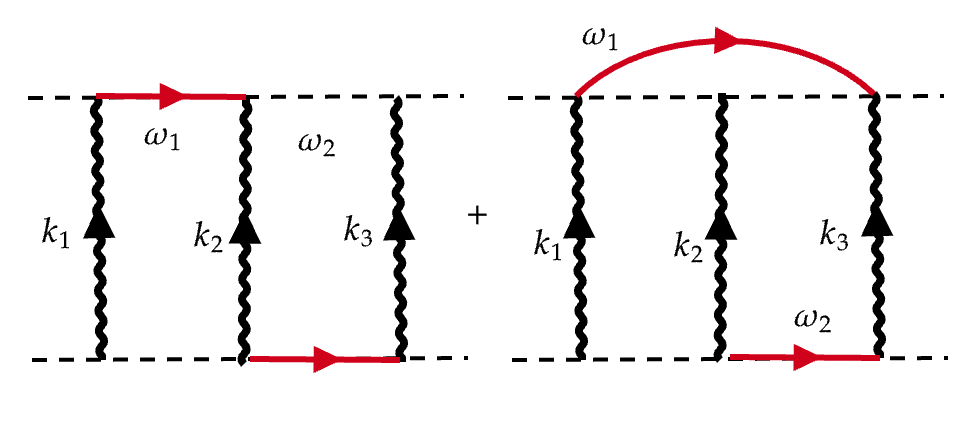}
\end{center}
The amplitude for the first diagram is given by,
\begin{align}
    \begin{split}
        i\chi&=  i\,\frac{s_1^3 s_2^3 m_1^2 m_2^2} {128 m_p^6}\int_q e^{-iq\cdot b}\hat\delta(q\cdot v_1)\hat\delta(q\cdot v_2)\int_{\ell_{1},\ell_{2}}\left(\sum_{(\pm)}\frac{1}{(\pm\omega_1+i\varepsilon)^2(\pm\omega_2+i\varepsilon)^2}\right)\frac{\hat\delta(\ell_{1}\cdot v_2)\hat\delta(\ell_{2}\cdot v_1)}{\ell_{1}^2 \ell_{2}^2(q-\ell_{1}-\ell_{2})^2}\\&\hspace{0.8cm}\times\Big((q-\ell_{1}-\ell_{2})\cdot \ell_{1}\Big)\Big( (q-\ell_{1}-\ell_{2})\cdot \ell_{2}\Big)\Bigg|_{\omega_1\to \ell_1\cdot v_1,\omega_2\to\ell_2\cdot v_2}\,.\\ &
    \end{split}
\end{align}
After IBP reduction, we are left with the following integral,
\begin{align}
    \begin{split}
       i\, \chi\Big|_{\mathcal{S}^0} &=
         i\,\frac{s_1^3 s_2^3 m_1^2 m_2^2} {128 m_p^6}\int_q e^{-iq\cdot b}\hat\delta(q\cdot v_1)\hat\delta(q\cdot v_2)\Big[\boldsymbol{\tilde h}_{29}\mathbfcal{M}_{0,0;0,0,1,1,1}+\boldsymbol{\tilde h}_{30}\,q^2\,\mathbfcal{M}_{0,0;0,0,2,1,1}  +\boldsymbol{\tilde h}_{31}\,q^2\,\mathbfcal{M}_{0,0;0,0,1,2,1}\\ &\hspace{0.8cm}-\frac{1}{2}\boldsymbol{\tilde h}_{32} q^2\,\mathbfcal{M}^{(++)}_{1,1;0,0,1,1,1} \Big]\,.
    \end{split}
\end{align}
The contribution to the phase from the second diagram has the following form,
\begin{align}
\begin{split}
i \chi\Big|_{\mathcal{S}^0}
=&\,i\frac{s_1^3 s_2^3 m_1^2 m_2^2} {128 m_p^6} \int_{k_i}e^{-i(k_1+k_2+k_3)\cdot b}\hat\delta(k_1\cdot v_2)\hat\delta(k_2\cdot v_1)\hat\delta(k_1\cdot v_1+k_3\cdot v_1)\hat\delta(k_2\cdot v_2+k_3\cdot v_2)\\
&\times\frac{\mathcal{N}}{(k_1\cdot v_1+i\varepsilon)^2(k_3\cdot v_2+i\varepsilon)^2\,k_1^2 \,k_2^2\,k_3^2}\,.
\end{split}
\end{align}
Now, doing the following variable transformation, $k_1\to \ell_1-q,\,k_2\to (\ell_2-q)\,,k_3\to (q-\ell_1-\ell_2)$, we will get,
\begin{align}
    \begin{split}
   \hspace{0cm}   i\,  \chi \Big|_{\mathcal{S}^0}=i\frac{s_1^3 s_2^3 m_1^2 m_2^2} {128 m_p^6}  \int e^{i q\cdot b}\hat\delta(q\cdot v_1)\hat\delta(q\cdot v_2)\int_{\ell_1,\ell_2}\sum_{\mp}\frac{\hat\delta(\ell_1\cdot v_2)\hat\delta(\ell_2\cdot v_1)\mathcal{N}}{(\pm\ell_1\cdot v_1+i\varepsilon)^2(\mp\ell_2\cdot v_2+i\varepsilon)^2\,(\ell_1-q)^2(\ell_2-q)^2(\ell_1+\ell_2-q)^2},
    \end{split}
\end{align}
where, the numerator $\mathcal{N}$ takes the form (ignoring the tadpoles),
\begin{align}
    \begin{split}
        \mathcal{N}=\frac{1}{8}\ell_1^2\,\ell_2^2-\frac{1}{16}(\ell_1^4+\ell_2^4)+\frac{q^4}{16}\,.
    \end{split}
\end{align}
Therefore, the contribution to the eikonal phase from the second diagram is given by,
\begin{align}
    \begin{split}
       i\, \chi\Big|_{\mathcal{S}^0}&=i\frac{s_1^3 s_2^3 m_1^2 m_2^2} {128 m_p^6}  \int e^{iq\cdot b}\hat\delta(q\cdot v_1)\hat\delta(q\cdot v_2)\Bigg[\boldsymbol{\tilde h}_{33}\,\mathbfcal{M}_{0,0;0,0,1,1,1}+\boldsymbol{\tilde h}_{34}\,q^2 \mathbfcal{M}_{0,0;0,0,2,1,1}+\boldsymbol{\tilde h}_{35}\,q^2 \mathbfcal{M}_{0,0;0,0,1,2,1}\\ &\hspace{0.8cm}+\frac{1}{4}\boldsymbol{\tilde h}_{36}\,q^2 \mathbfcal{M}^{(+-)}_{1,1;0,0,1,1,1}\Bigg]\,.
    \end{split}
\end{align}
We finally get the total contribution to the eikonal phase from these two diagrams at $\mathcal{O}(\mathcal{S}^0)$, 
\begin{align}
\begin{split} \label{eqt6}
 \chi\Big|_{\mathcal{S}^0} =\frac{m_1^2m_2^2}{m_p^6}\Bigg(\Delta_{\mathfrak{08}}^{\textrm{poly}}+\Delta_{\mathfrak{01}}^{\textrm{log}}\log (x)\Bigg)
\end{split}
\end{align}
where,
\begin{align}
\begin{split}
&\Delta_{\mathfrak{08}}^{\textrm{poly}}= -\frac{ s_1^3 s_2^3 x^4 \left(x^2+1\right) (258 \log (|\boldsymbol{b}|)+215 \gamma_E -12-172 \log (2)+43 \log (\pi ))}{6144 \pi ^3 |\boldsymbol{b}|^2 \left(x^2-1\right)^5 }\,,\\&
\Delta_{\mathfrak{01}}^{\textrm{log}}=-\frac{ s_1^3 s_2^3 x^4}{512 \pi ^3 |\boldsymbol{b}|^2 \left(x^2-1\right)^4 }\,.
\end{split}
\end{align}
We have used the results for the master integrals mentioned in \eqref{MasterM}. 
\par
\textbullet $\,\,$ Now, we move on to calculating  Feynman topology with two scalar lines and one graviton line, which appear at 3PM. In these diagrams, the spin of the black holes appears because of the presence of graviton. 
\begin{center}
\includegraphics[width=0.38\linewidth]{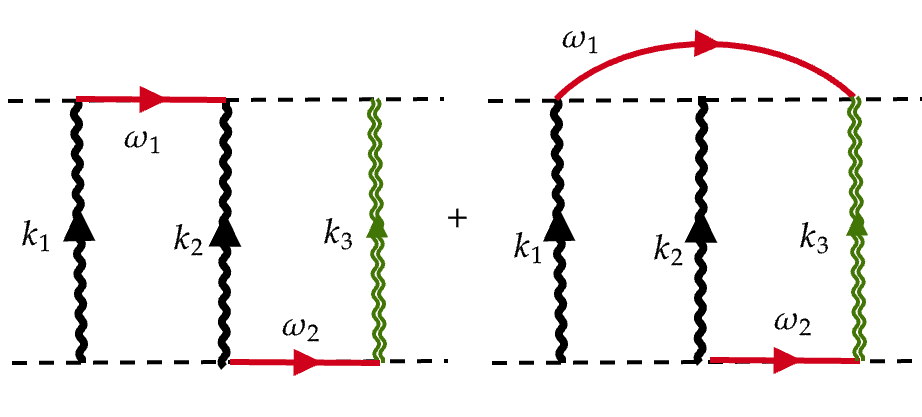}
\end{center}
The corresponding phase is given by,
\begin{align}
    \begin{split}
        i\chi&=i\,\frac{s_1^2 s_2^2 m_1^2 m_2^2} {128 m_p^6} \int_{k_i,\omega_i}e^{-i(k_1+k_2+k_3)\cdot b}\hat\delta(k_1\cdot v_2)\hat\delta(k_3\cdot v_1)\hat\delta(\omega_1-k_1\cdot v_1)\hat\delta(\omega_1+k_2\cdot v_1)\\ &
\hspace{2 cm}\times \hat\delta(\omega_2+k_2\cdot v_2)\hat\delta(\omega_2-k_3\cdot v_2)\,\frac{\mathcal{N}}{k_1^2 \,k_2^2\, k_3^2}\left(\sum_{(\pm)}\frac{1}{(\pm\omega_1+i\varepsilon)^2(\pm\omega_2+i\varepsilon)^2}\right)
    \end{split}
\end{align}
Now redefining $k_1\to \ell_{1}, k_3\to \ell_{2}, q=\ell_{1}+k_2+\ell_{2}$, we can write the numerator up to linear order in spin in the following way ($\mathcal{N}= \mathcal{N}|_{\mathcal{S}^0}+ \mathcal{N}|_{\mathcal{S}^1}$).
\begin{align}
    \begin{split}
    &   \mathcal{N}\Big|_{\mathcal{S}^0}= -\frac{\gamma^2(d-2)-2}{d-2}(\ell_2-q)^2(\ell_2\cdot v_2)^2\,,\\ &
     \mathcal{N}\Big|_{\mathcal{S}^1}=i\gamma \Big(\ell_2\cdot \mathcal{S}_1\cdot v_2\Big)(\ell_2\cdot v_2)^2(\ell_2-q)^2\,.
    \end{split}
\end{align}
The spinless part of the eikonal phase is given by,
\begin{align}
    \begin{split}
        i\chi\Big|_{\mathcal{S}^0}&
        =-i\,\frac{s_1^2 s_2^2 m_1^2 m_2^2} {128 m_p^6} \frac{\gamma^2(d-2)-2}{d-2}\int e^{-iq\cdot b}\hat\delta(q\cdot v_1)\hat\delta(q\cdot v_2)\Bigg(\boldsymbol{\tilde h}_{37}\mathbfcal{M}_{0,0;0,0,1,1,1}+\boldsymbol{\tilde h}_{38}\,q^2\mathbfcal{M}_{0,0;0,0,2,1,1}\\ &\hspace{0.8cm}+\boldsymbol{\tilde h}_{39}\,q^2\mathbfcal{M}_{0,0;0,0,1,2,1}\Bigg)\,.
    \end{split}
\end{align}
At $\mathcal{O}(\mathcal{S})$ we get, 
\begin{align}
    \begin{split}
      i\chi\Big|_{\mathcal{S}^1}  = i\,\frac{s_1^2 s_2^2 m_1^2 m_2^2} {128 m_p^6} \int_{q}e^{-i q\cdot b}\hat\delta(q\cdot v_1)\,\hat\delta(q\cdot v_2)
      &\Bigg[i\gamma\Bigg(\boldsymbol{\tilde h}_{40}\mathbfcal{M}_{0,0;0,0,1,1,1}+\boldsymbol{\tilde h}_{41} q^2\,\mathbfcal{M}_{0,0;0,0,2,1,1}\\&+\boldsymbol{\tilde h}_{42}q^2\,\mathbfcal{M}_{0,0;0,0,    1,2,1}\Bigg)\Bigg]\Big(q\cdot \mathcal{S}_1\cdot v_2\Big)\,,\\&
    \end{split}
\end{align}
The contribution from the second diagram to the eikonal phase is given by,
\begin{align}
    \begin{split}
i\chi\Big|_{\mathcal{S}^0}=-\frac{i m_1^2 m_2^2 s_1^2 s_2^2}{64 m_p^6}\,\int e^{-i q \cdot b}\, \hat\delta(q\cdot v_1)\hat\delta(q\cdot v_2)&\Big(\boldsymbol{\tilde h}_{43}\,\,\mathbfcal{M}_{0,0;0,0,1,1,1}+\boldsymbol{\tilde h}_{44}\,\,q^2\mathbfcal{M}_{0,0;0,0,2,1,1}\\&+\boldsymbol{\tilde h}_{45}\,\,q^2\mathbfcal{M}_{0,0;0,0,1,2,1}+\frac{1}{4}\boldsymbol{\tilde h}_{46}\,\,q^2\mathbfcal{M}^{(+-)}_{1,1;0,0,1,1,1}\Big)\,,
   \end{split}
\end{align}
Similarly,
\begin{align}
    \begin{split}
       i\, \chi\Big|_{\mathcal{S}^1}=-i \frac{s_1^2 s_2^2 m_1^2 m_2^2}{64 m_p^6} \int_{q}e^{i q\cdot b}\hat\delta(q\cdot v_1)\,\hat\delta(q\cdot v_2)&\Bigg[i\,\Bigg(\boldsymbol{\tilde h}_{47}\,\,\mathbfcal{M}_{0,0;0,0,1,1,1}+\boldsymbol{\tilde h}_{48}\,\,q^2\mathbfcal{M}_{0,0;0,0,2,1,1}\\&+\boldsymbol{\tilde h}_{49}\,\,q^2\mathbfcal{M}_{0,0;0,0,1,2,1}+\frac{1}{4}\boldsymbol{\tilde h}_{50}\,\,q^2\mathbfcal{M}^{(+-)}_{1,1;0,0,1,1,1}\Bigg)\big(q\cdot\mathcal{S}_1\cdot v_2\big)+\\&i\,\Bigg(\boldsymbol{\tilde h}_{51}\,\,\mathbfcal{M}_{0,0;0,0,1,1,1}+\boldsymbol{\tilde h}_{52}\,\,q^2\mathbfcal{M}_{0,0;0,0,2,1,1}\\&+\boldsymbol{\tilde h}_{53}\,\,q^2\mathbfcal{M}_{0,0;0,0,1,2,1}+\frac{1}{4}\boldsymbol{\tilde h}_{54}\,\,q^2\mathbfcal{M}^{(+-)}_{1,1;0,0,1,1,1}\Bigg)\big(q\cdot\mathcal{S}_2\cdot v_1\big)\Bigg]
    \end{split}
\end{align}
We finally get the total contribution to the eikonal phase from these two diagrams at $\mathcal{O}(\mathcal{S}^0)$, 
\begin{align}
\begin{split} \label{eqt7}
 \chi\Big|_{\mathcal{S}^0} =\frac{m_1^2m_2^2}{m_p^6}\Bigg(\Delta_{\mathfrak{09}}^{\textrm{poly}}+\Delta_{\mathfrak{02}}^{\textrm{log}}\log (x)\Bigg)
\end{split}
\end{align}
\enlargethispage{2\baselineskip}
where, 
\begin{thesiscompactdisplay}
\begin{align}
\begin{split}
&\Delta_{\mathfrak{09}}^{\textrm{poly}}=-\frac{s_1^2 s_2^2}{4096 \pi ^3 |\boldsymbol{b}|^2 \left(x^2-1\right)^5}\Bigg(\left(x^2+1\right) \big(-12 \left(11 x^4-6 x^2+11\right) x^2 \log (|\boldsymbol{b}|)+5 x^8-22 x^6 \left(5 \gamma_E +2+\log \left(\frac{\pi }{16}\right)\right)\\&\hspace{0.8cm}+6 x^4 (10 \gamma_E +9-8 \log (2)+2 \log (\pi ))-22 x^2 \left(5 \gamma_E +2+\log \left(\frac{\pi }{16}\right)\right)+5\big)\Bigg)\,,\\&
\Delta_{\mathfrak{02}}^{\textrm{log}}=-\frac{s_1^2 s_2^2 \left(x^4+6 x^2+1\right)}{512 \pi ^3 |\boldsymbol{b}|^2 \left(x^2-1\right)^2}\,.
\end{split}
\end{align}
\end{thesiscompactdisplay}
Also the total contribution at $\mathcal{O}(\mathcal{S}^1)$ is, 
\enlargethispage{2\baselineskip}
\begin{thesiscompactdisplay}[\scriptsize]
\begin{align}
\begin{split}
\chi\Big|_{\mathcal{S}^1} =\frac{m_1^2m_2^2}{m_p^6}&\Bigg[\Big(\Delta_{\mathfrak{10}}^{\textrm{poly}}+\Delta_{\mathfrak{03}}^{\textrm{log}}\log (x)+\Delta_{\mathfrak{01}}^{\textrm{Polylog}}\left(\textbf{Li}_2(1-x^2)+\log^2(x)\right)+\boldsymbol{\mathfrak{U}}_{\mathfrak{1}}^{\textrm{LI}_3}\Big\{54\, \textbf{Li}_3\left(x^2\right)-216\, \textbf{Li}_3\left(\frac{x+1}{1-x}\right)
\\ &+432\, \textbf{Li}_3(1-x)+216\, \textbf{Li}_3\left(\frac{x+1}{x-1}\right)+432\, \textbf{Li}_3(x+1)+216 \Big(-\textbf{Li}_2\left(\frac{x+1}{1-x}\right)-\textbf{Li}_2(1-x)\\ &+\textbf{Li}_2\left(\frac{x+1}{x-1}\right)+\textbf{Li}_2(x+1)\Big) \log \left(\frac{2}{x+1}-1\right)-54 \left(\pi ^2-2 \log ^2(x)\right) \log \left(1-x^2\right)-20 \log ^3(x)\\ &+108 i \pi  \log ^2(1-x)+108 i \pi  \log ^2(x+1)-11 \pi ^2 \log (x)-432 \zeta (3)\Big\}\Big)\Big(\hat{b}\cdot \mathcal{S}_1\cdot v_2\Big)+\\&\Big(\Delta_{\mathfrak{11}}^{\textrm{poly}}+\Delta_{\mathfrak{04}}^{\textrm{log}}\log (x)+\Delta_{\mathfrak{02}}^{\textrm{Polylog}}\left(\textbf{Li}_2(1-x^2)+\log^2(x)\right)+\boldsymbol{\mathfrak{U}}_{\mathfrak{2}}^{\textrm{LI}_3}\Big\{54\, \textbf{Li}_3\left(x^2\right)-216\, \textbf{Li}_3\left(\frac{x+1}{1-x}\right)
\\ &+432\, \textbf{Li}_3(1-x)+216\, \textbf{Li}_3\left(\frac{x+1}{x-1}\right)+432\, \textbf{Li}_3(x+1)+216 \Big(-\textbf{Li}_2\left(\frac{x+1}{1-x}\right)-\textbf{Li}_2(1-x)\\ &+\textbf{Li}_2\left(\frac{x+1}{x-1}\right)+\textbf{Li}_2(x+1)\Big) \log \left(\frac{2}{x+1}-1\right)-54 \left(\pi ^2-2 \log ^2(x)\right) \log \left(1-x^2\right)-20 \log ^3(x)\\ &+108 i \pi  \log ^2(1-x)+108 i \pi  \log ^2(x+1)-11 \pi ^2 \log (x)-432 \zeta (3)\Big\}\Big)\Big)\Big(\hat{b}\cdot \mathcal{S}_2\cdot v_1\Big)\Bigg]
\end{split}
\end{align}
\end{thesiscompactdisplay}
\enlargethispage{2\baselineskip}
where,
\begin{thesiscompactdisplay}[\scriptsize]
\begin{align}
\begin{split}
\Delta_{\mathfrak{10}}^{\textrm{poly}}=\frac{xs_1^2 s_2^2}{12288 \pi ^3 |\boldsymbol{b}|^3 \left(x^2-1\right)^5 }\Bigg[&-12 \left(12 x^6+77 x^5+72 x^4+154 x^3+72 x^2+77 x+12\right) x \log(|\boldsymbol{b}|)\\
&-13 x^8-120 \gamma_E  x^7+60 x^7+24 x^7 \log \left(\frac{16}{\pi }\right)-770 \gamma_E  x^6+398 x^6+154 x^6 \log \left(\frac{16}{\pi }\right)\\
&-720 \gamma_E  x^5+372 x^5+144 x^5 \log \left(\frac{16}{\pi }\right)-1540 \gamma_E  x^4+630 x^4+308 x^4 \log \left(\frac{16}{\pi }\right)\\
&-720 \gamma_E  x^3+372 x^3+144 x^3 \log \left(\frac{16}{\pi }\right)-770 \gamma_E  x^2+398 x^2+154 x^2 \log \left(\frac{16}{\pi }\right)\\
&-120 \gamma_E  x+60 x+24 x \log \left(\frac{16}{\pi }\right)-13\Bigg]\,,\\ &
\hspace{-4.5 cm}\Delta_{\mathfrak{3}}^{\textrm{log}}=\frac{1}{18432 \pi ^3 |\boldsymbol{b}|^3 \left(x^2-1\right)^6 \left(x^2+1\right) }\Bigg[-48(1+x^{12})+(-24 x^8-24 x^4)\beta_1+2(x^5+x^7)\beta_2+3(x^3+x^9)\beta_3\\
& +(x+x^{11})\beta_4-24x^6\beta_5+6(x^2+x^{10})\beta_6\Bigg]\,,\nonumber
\end{split}
\end{align}
\begin{align}
\begin{split}
&\Delta_{\mathfrak{01}}^{\textrm{Polylog}}=\frac{x^2s_1^2 s_2^2}{6144 \pi ^3 |\boldsymbol{b}|^3 \left(x^2-1\right)^6 \left(x^2+1\right) }\Bigg[72 \left(x^{10}+24 x^9+3 x^8-10 x^6-48 x^5-10 x^4+3 x^2+24 x+1\right) \log(|\boldsymbol{b}|)\\ &\hspace{0.8cm}+9 x^{10}-12 x^{10} \log \left(\frac{16}{\pi }\right)-1236 x^9-288 x^9 \log \left(\frac{16}{\pi }\right)+9 x^8\\ &\hspace{0.8cm}-36 x^8 \log \left(\frac{16}{\pi }\right)-392 x^7+222 x^6+120 x^6 \log \left(\frac{16}{\pi }\right)+3224 x^5\\ &\hspace{0.8cm}+576 x^5 \log \left(\frac{16}{\pi }\right)+222 x^4+120 x^4 \log \left(\frac{16}{\pi }\right)-392 x^3+9 x^2\\ &\hspace{0.8cm}-36 x^2 \log \left(\frac{16}{\pi }\right)+60 \gamma_E  (x^{10}+24 x^9+3 x^8-10 x^6-48 x^5-10 x^4\\ &\hspace{0.8cm}+3 x^2+24 x+1)-1236 x-288 x \log \left(\frac{16}{\pi }\right)+9-12 \log \left(\frac{16}{\pi }\right)\Bigg]\,,\\
&\boldsymbol{\mathfrak{U}_1}^{\textrm{LI}_3}=\frac{ \pi ^2 s_1^2 s_2^2 x}{9216 \pi ^5 |\boldsymbol{b}|^3 \left(x^2-1\right)^6}\Bigg[x^8+24 x^7+2 x^6-24 x^5-12 x^4-24 x^3+2 x^2+24 x+1\Bigg],
\\ &
\Delta_{\mathfrak{04}}^{\textrm{log}}=\frac{1}{36864 \pi ^3 |\boldsymbol{b}|^3 (x^2-1)^6 \left(x^2+1\right) }\Bigg[96(1+x^{12})+48(x^3+x^{9})\delta_1+3(x^2+x^{10})\delta_2+48(x^5+x^7)\delta_3\\ &\hspace{0.8cm}+4(x^4+x^8)\delta_4+x^6\delta_5\Bigg]\,,\\&
\end{split}
\end{align}
\begin{align}
\Delta_{\mathfrak{11}}^{\textrm{poly}}
&= \frac{x^2 s_1^2 s_2^2}{6144 \pi ^3 |\boldsymbol{b}|^3 \left(x^2-1\right)^5 }\Bigg[
12 (42 x^4-36 x^3+89 x^2-36 x+42) x \log(|\boldsymbol{b}|)-12 x^6+420 \gamma_E  x^5-323 x^5
\notag\\
&\quad -84 x^5 \log \left(\frac{16}{\pi }\right)-360 \gamma_E  x^4+108 x^4+72 x^4 \log \left(\frac{16}{\pi }\right)+890 \gamma_E  x^3-558 x^3
\notag\\
&\quad -178 x^3 \log \left(\frac{16}{\pi }\right)-360 \gamma_E  x^2+108 x^2+72 x^2 \log \left(\frac{16}{\pi }\right)+420 \gamma_E  x-323 x
\notag\\
&\quad -84 x \log \left(\frac{16}{\pi }\right)-12\Bigg]\,, \notag\\[2pt]
\Delta_{\mathfrak{02}}^{\textrm{Polylog}}
&= \frac{x^3 s_1^2 s_2^2}{768 \pi ^3 |\boldsymbol{b}|^3 (x^2-1)^6 \left(x^2+1\right) }\Bigg[
18 \left(12 x^8-7 x^6-12 x^5-38 x^4-12 x^3-7 x^2+12\right) \log (|\boldsymbol{b}|)-150 x^8
\notag\\
&\quad -36 x^8 \log \left(\frac{16}{\pi }\right)-15 x^7+30 x^6+21 x^6 \log \left(\frac{16}{\pi }\right)+39 x^5+36 x^5 \log \left(\frac{16}{\pi }\right)
\notag\\
&\quad +554 x^4+114 x^4 \log \left(\frac{16}{\pi }\right)+39 x^3+36 x^3 \log \left(\frac{16}{\pi }\right)+30 x^2+21 x^2 \log \left(\frac{16}{\pi }\right)
\notag\\
&\quad +15 \gamma_E  \left(12 x^8-7 x^6-12 x^5-38 x^4-12 x^3-7 x^2+12\right)-15 x-150
\notag\\
&\quad -36 \log \left(\frac{16}{\pi }\right)\Bigg]\,, \notag\\[2pt]
\boldsymbol{\mathfrak{U}}_{\mathfrak{2}}^{\textrm{LI}_3}
&= \frac{s_1^2 s_2^2 x^2}{4608 \pi ^3 |\boldsymbol{b}|^3 \left(x^2-1\right)^6}\Bigg[
12 x^6-19 x^4-12 x^3-19 x^2+12\Bigg]
\end{align}
\end{thesiscompactdisplay}
with,
\begin{align}
    \begin{split}
       & \beta_1=90 \log (|\boldsymbol{b}|)+75 \gamma_E -26-15 \log \left(\frac{16}{\pi }\right),\\ &
       \beta_2=-1080 \log ^2(|\boldsymbol{b}|)-1800 \gamma_E  \log (|\boldsymbol{b}|)+666 \log (|\boldsymbol{b}|)+3 \log \left(\frac{\pi }{16}\right) \left(-120 \log (|\boldsymbol{b}|)+37+10 \log \left(\frac{16}{\pi }\right)\right)\\ &\hspace{0.35cm}-750 \gamma_E ^2-25 \pi ^2+555 \gamma_E +216+300 \gamma_E  \log \left(\frac{16}{\pi }\right),\\ &
       \beta_3=216 \log ^2(|\boldsymbol{b}|)+18 (20 \gamma_E +1-16 \log (2)+4 \log (\pi )) \log (|\boldsymbol{b}|)+5 \pi ^2+176+15 \gamma_E  (10 \gamma_E +1-16 \log (2)\\ &\hspace{0.35cm}+4 \log (\pi ))+3 \log \left(\frac{16}{\pi }\right) \left(\log \left(\frac{256}{\pi ^2}\right)-1\right)\,,\\ &
       \beta_4=216 \log ^2(|\boldsymbol{b}|)+18 (20 \gamma_E +3-16 \log (2)+4 \log (\pi )) \log (|\boldsymbol{b}|)+5 \pi ^2+144+15 \gamma_E  (10 \gamma_E +3-16 \log (2)\\ &\hspace{0.35cm}+4 \log (\pi ))+3 \log \left(\frac{16}{\pi }\right) \left(\log \left(\frac{256}{\pi ^2}\right)-3\right)\,,\\ &
       \beta_5=432 \log ^2(|\boldsymbol{b}|)+6 (120 \gamma_E -137-96 \log (2)+24 \log (\pi )) \log (|\boldsymbol{b}|)+10 \pi ^2+151+12 \log ^2\left(\frac{\pi }{16}\right)+548 \log (2)\\ &\hspace{0.35cm}-137 \log (\pi )+5 \gamma_E  (60 \gamma_E -137-96 \log (2)+24 \log (\pi ))\,,\\ &
       \beta_6=2 \Big(432 \log ^2(|\boldsymbol{b}|)+6 (120 \gamma_E -103-96 \log (2)+24 \log (\pi )) \log (|\boldsymbol{b}|)+10 \pi ^2+12 \log ^2\left(\frac{\pi }{16}\right)\\ &\hspace{0.35cm}+103 (1+\log (16)-\log (\pi ))+5 \gamma_E  (60 \gamma_E -103-96 \log (2)+24 \log (\pi ))\Big)
    \end{split}
\end{align}
and,
\begin{align}
    \begin{split}
        &\delta_1=30 \log(|\boldsymbol{b}|)+25 \gamma_E +6-5 \log \left(\frac{16}{\pi }\right),\\ &
        \delta_2=-192 \log(|\boldsymbol{b}|) (18 \log(|\boldsymbol{b}|)+30 \gamma_E -25)+96 \log \left(\frac{16}{\pi }\right) \left(12 \log(|\boldsymbol{b}|)+10 \gamma_E +\log \left(\frac{\pi }{16}\right)\right)\\ &\hspace{0.7 cm}-800 \gamma_E  (3 \gamma_E -5)-80 \pi ^2-1199-800 \log \left(\frac{16}{\pi }\right)\,,\\ &
        \delta_3=216 \log ^2(|\boldsymbol{b}|)+6 (60 \gamma_E -13-48 \log (2)+12 \log (\pi )) \log(|\boldsymbol{b}|)+5 \left(\gamma_E  (30 \gamma_E -13)+\pi ^2-14\right)\\ &
        \hspace{0.7 cm}+\log \left(\frac{\pi }{16}\right) (60 \gamma_E -13-24 \log (2)+6 \log (\pi ))\,,\\ &
        \delta_4=84 \log \left(\frac{\pi }{16}\right) (6 \log(|\boldsymbol{b}|)+5 \gamma_E )+72 \log(|\boldsymbol{b}|) (21 \log(|\boldsymbol{b}|)+35 \gamma_E -10)+5 \left(30 \gamma_E  (7 \gamma_E -4)+7 \pi ^2-54\right)\\ &\hspace{0.7 cm}+6 \log \left(\frac{16}{\pi }\right) (20+28 \log (2)-7 \log (\pi ))\,,\\ &
        \delta_5=96 \left(19 \log \left(\frac{\pi }{16}\right) (6 \log(|\boldsymbol{b}|)+5 \gamma_E )+6 \log(|\boldsymbol{b}|) (57 \log(|\boldsymbol{b}|)+95 \gamma_E -91)\right)+760 \pi ^2+240 \gamma_E  (95 \gamma_E -182)\\ &\hspace{0.7 cm}+9642+14592 \log ^2(2)+7296 \log (8)+3264 \log \left(\frac{16}{\pi }\right)+912 \left(\log \left(\frac{\pi }{256}\right)-6\right) \log (\pi )\,.
    \end{split}
\end{align}
\textbullet $\,\,$ We now compute the 3PM diagrams with a graviton propagator in the middle. The topology has the following form,
\begin{center}
\includegraphics[width=0.38\linewidth]{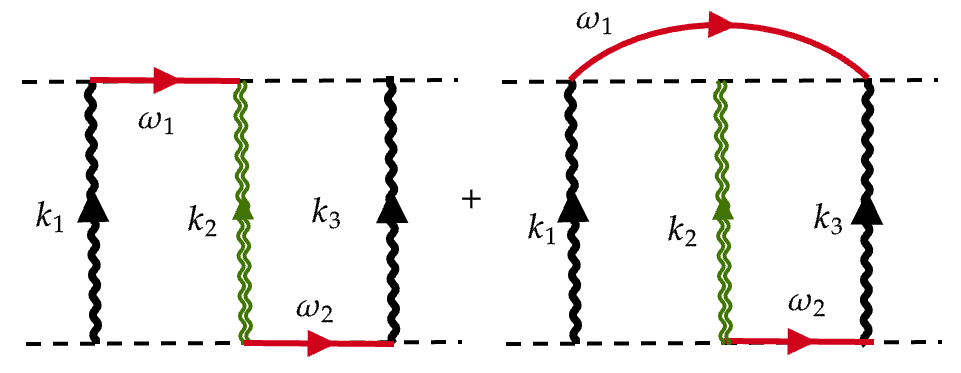}
\end{center}
 At $\mathcal{O}(\mathcal{S}^0)$ we get the total contribution from these two diagrams after doing the IBP reduction, 
\begin{align}
    \begin{split}
\chi\Big|_{\mathcal{S}^0}=-\frac{ m_1^2 m_2^2 s_1^2 s_2^2}{256 m_p^6}\,\int e^{-i q \cdot b}\, \hat\delta(q\cdot v_1)\hat\delta(q\cdot v_2)&\Big(\boldsymbol{\tilde h}_{55}\,\,\mathbfcal{M}_{0,0;0,0,1,1,1}+\boldsymbol{\tilde h}_{56}\,\,q^2\mathbfcal{M}_{0,0;0,0,2,1,1}\\&+\boldsymbol{\tilde h}_{57}\,\,q^2\mathbfcal{M}_{0,0;0,0,1,2,1}+\boldsymbol{\tilde h}_{58}\,\,q^2\mathbfcal{M}^{(++)}_{1,1;0,0,1,1,1}\Big)\,,
   \end{split}
\end{align}
and proceeding as before we get at $\mathcal{O}(\mathcal{S}^1)$, 
\begin{align}
    \begin{split}
\chi\Big|_{\mathcal{S}^1}=-\frac{ m_1^2 m_2^2 s_1^2 s_2^2}{64m_p^6}\,\int e^{-i q \cdot b}\, \hat\delta(q\cdot v_1)\hat\delta(q\cdot v_2)&\Bigg[i\,\Big(\boldsymbol{\tilde h}_{59}\,\,\mathbfcal{M}_{0,0;0,0,1,1,1}+\boldsymbol{\tilde h}_{60}\,\,q^2\mathbfcal{M}_{0,0;0,0,2,1,1}\\&+\boldsymbol{\tilde h}_{61}\,\,q^2\mathbfcal{M}_{0,0;0,0,1,2,1}-\frac{1}{2}\boldsymbol{\tilde h}_{62}\,\,q^2\mathbfcal{M}^{(++)}_{1,1;0,0,1,1,1}\\&-\frac{1}{4}\boldsymbol{\tilde h}_{63}\,\,q^2\mathbfcal{M}^{(+-)}_{1,1;0,0,1,1,1}\Big)\big(q\cdot\mathcal{S}_1\cdot v_2\big)\\&+i\,\Big(\boldsymbol{\tilde h}_{64}\,\,\mathbfcal{M}_{0,0;0,0,1,1,1}+\boldsymbol{\tilde h}_{65}\,\,q^2\mathbfcal{M}_{0,0;0,0,2,1,1}\\&+\boldsymbol{\tilde h}_{66}\,\,q^2\mathbfcal{M}_{0,0;0,0,1,2,1}-\frac{1}{2}\boldsymbol{\tilde h}_{67}\,\,q^2\mathbfcal{M}_{1,1;0,0,1,1,1}\\&-\frac{1}{4}\boldsymbol{\tilde h}_{68}\,\,q^2\mathbfcal{M}^{(+-)}_{1,1;0,0,1,1,1}\Big)\big(q\cdot\mathcal{S}_2\cdot v_1\big)\Bigg]\,.
 \end{split}
\end{align}
Finally, using \eqref{MasterM} we get,

\begin{align}
    \begin{split} \label{eqt7b}
        \chi\Big|_{\mathcal{S}^0}=\frac{m_1^2m_2^2}{m_p^6}&\Bigg[\Big(\Delta_{\mathfrak{12}}^{\textrm{poly}}+\Delta_{\mathfrak{05}}^{\textrm{log}}\log (x)\Bigg] 
        \end{split}
\end{align}
where, 
\begin{align}
    \begin{split}
   & \Delta_{\mathfrak{12}}^{\textrm{poly}}=-\frac{s_1^2s_2^2 (1+x^2)}{8192 \pi ^3 |\boldsymbol{b}|^2 \left(x^2-1\right)^5}\Bigg[48 \left(3 x^4-2 x^2+3\right) x^2 \log (|\boldsymbol{b}|)+x^8+24 x^6 \left(5 \gamma_E +1+\log \left(\frac{\pi }{16}\right)\right)\\&\hspace{0.8cm}-2 x^4 (40 \gamma_E +17-32 \log (2)+8 \log (\pi ))+24 x^2 \left(5 \gamma_E +1+\log \left(\frac{\pi }{16}\right)\right)+1\Bigg]\,,\\&
   \Delta_{\mathfrak{05}}^{\textrm{log}}=-\frac{ s_1^2 s_2^2 \left(3 x^4+22 x^2+3\right)}{6144 \pi ^3 |\boldsymbol{b}|^2 \left(x^2-1\right)^2 }\,.
      \end{split}
\end{align}
The total contribution at $\mathcal{O}(\mathcal{S}^1)$ is, 
\begin{align}
\begin{split} \label{eqt8}
\chi\Big|_{\mathcal{S}^1} =\frac{m_1^2m_2^2}{m_p^6}&\Bigg[\Big(\Delta_{\mathfrak{13}}^{\textrm{poly}}+\Delta_{\mathfrak{06}}^{\textrm{log}}\log (x)\Big)\Big(\hat{b}\cdot \mathcal{S}_1\cdot v_2\Big)+\Big(\Delta_{\mathfrak{14}}^{\textrm{poly}}+\Delta_{\mathfrak{07}}^{\textrm{log}}\log (x)\Big)\Big(\hat{b}\cdot \mathcal{S}_2\cdot v_1\Big)\Bigg]
\end{split}
\end{align}
where,
\begin{align}
\begin{split}
&\Delta_{\mathfrak{13}}^{\textrm{poly}}=\frac{s_1^2s_2^2\,x}{3072 \pi ^3 |\boldsymbol{b}|^3 \left(x^2-1\right)^5}\Bigg[132 \left(x^3+x\right)^2 \log (|\boldsymbol{b}|)+3 x^8-6 x^7+2 x^6 (55 \gamma_E -38-44 \log (2)+11 \log (\pi ))\\&\hspace{0.8cm}-18 x^5+22 x^4 (10 \gamma_E -5-8 \log (2)+2 \log (\pi ))-18 x^3\\&\hspace{0.8cm}+2 x^2 (55 \gamma_E -38-44 \log (2)+11 \log (\pi ))-6 x+3\Bigg]\,,\\&
\Delta_{\mathfrak{14}}^{\textrm{poly}}=-\frac{s_1^2s_1^2 x}{6144 \pi ^3 |\boldsymbol{b}|^3 \left(x^2-1\right)^5 }\Bigg[48 \left(48 x^4+55 x^2+48\right) x^2 \log (|\boldsymbol{b}|)+15 x^8+2 x^6 (960 \gamma_E -359-768 \log (2)\\&\hspace{0.8cm}+192 \log (\pi ))-24 x^5+10 x^4 (220 \gamma_ E -49-176 \log (2)+44 \log (\pi ))-24 x^3\\&\hspace{0.8cm}+2 x^2 (960 \gamma_E -359-768 \log (2)+192 \log (\pi ))+15\Bigg]\,,\\&
\Delta_{\mathfrak{6}}^{\textrm{log}}=\frac{3s_1^2 s_2^2 x^3 \left(x^4-1\right)}{256 \pi ^3 |\boldsymbol{b}|^3 \left(x^2-1\right)^5 }\,,\Delta_{\mathfrak{7}}^{\textrm{log}}=-\frac{s_1^2 s_2^2 x \left(x^8-10 x^6+x^4\right)}{128 \pi ^3 |\boldsymbol{b}|^3 \left(x^2-1\right)^6 \left(x^2+1\right) }\,.
\end{split}
\end{align}
\textbullet $\,\,$ Now we consider the following diagram. It comes from the 3-point interaction vertex between the graviton and the kinetic term of the scalar field. 
\begin{align}
    \begin{split}
         \begin{minipage}[h]{0.08\linewidth}
	\vspace{4pt}
	\scalebox{1.65}{\includegraphics[width=\linewidth]{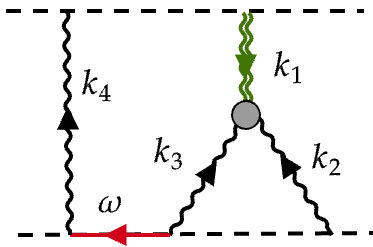}}
 \end{minipage}&
    \end{split}
\end{align}
Proceeding as before, we get the following,
\begin{align}
    \begin{split}
i\chi\Big|_{\mathcal{S}^{0}}&=i\frac{s_1 s_2^3m_1^2 m_2^2}{128 m_p^6}\int e^{-iq\cdot b} \,\hat\delta(q\cdot v_1)\hat\delta(q\cdot v_2)\int_{\ell_{1},\ell_{2}}\left(\frac{1}{(\ell_2\cdot v_2+i\varepsilon)^2}+\frac{1}{(\ell_2\cdot v_2-i\varepsilon)^2}\right)\frac{\hat\delta(\ell_{1}\cdot v_2)\hat\delta(\ell_{2}\cdot v_1)\,}{\ell_{1}^2 \ell_{2}^2 (\ell_{1}+\ell_{2}-q)^2(\ell_{2}-q)^2}\\ &
         \hspace{0.8cm}\Bigg((\ell_{1}\cdot v_1)^2\Big[(q-\ell_{1})^2-\ell_2^2-(\ell_1+\ell_2-q)^2\Big]-\frac{1}{d-2}\ell_1^2(\ell_1-q)^2\Bigg)\,,\\ &
        =i\frac{s_1 s_2^3m_1^2 m_2^2}{64 m_p^6}\int e^{-iq\cdot b} \,\hat\delta(q\cdot v_1)\hat\delta(q\cdot v_2)\Bigg[\boldsymbol{\tilde h}_{69}\,\mathbfcal{M}_{0,0;0,0,1,1,1}+\boldsymbol{\tilde h}_{70}\,\,q^2
      \mathbfcal{M}_{0,0;0,0,2,1,1} +\boldsymbol{\tilde h}_{71}\,\,q^2\mathbfcal{M}_{0,0;0,0,1,2,1}\Bigg]\,.
         \end{split}
\end{align}
Then using \eqref{MasterM}. we get, 
\begin{align}
    \begin{split}  \label{eqt9}      \chi\Big|_{\mathcal{S}^{0}}&=\frac{m_1^2m_2^2}{m_p^6}\Delta_{\mathfrak{15}}^{\textrm{poly}}\,,\,\, \Delta_{\mathfrak{15}}^{\textrm{poly}}=\Bigg(\frac{s_{1}s_{2}^3 \left(x^2+1\right)}{4096 \pi ^3 |\boldsymbol{b}|^2 \left(x^2-1\right)}\Bigg)\,.
         \end{split}
\end{align}
Another term that appears in the spinless part from a term proportional to $\ell_{1}^2$ in the numerator. In that case, the integral will be proportional $\mathbfcal{M}_{0,0;0,1,0,1,1}$ which vanishes identically.\par 
Now, the spinning part of the eikonal phase is given by,
\begin{align}
    \begin{split}
       \chi\Big|_{\mathcal{S}^1}=&-\frac{s_1 s_2^3m_1^2 m_2^2}{128 m_p^6}\, S_{1}^{\eta\delta}\int e^{-iq\cdot b} \,\hat\delta(q\cdot v_1)\hat\delta(q\cdot v_2)\int_{\ell_{1},\ell_{2}}\sum_{(\pm)}\frac{\hat\delta(\ell_{1}\cdot v_2)\hat\delta(\ell_{2}\cdot v_1)\,(\ell_{1}\cdot v_1) \,(\ell_2-q)_{[\eta} \, \ell_{1\delta]}}{(\ell_2\cdot v_2\pm i\varepsilon)^2\ell_{1}^2 \,\ell_{2}^2\,(\ell_{1}+\ell_{2}-q)^2}\,\Big[(\ell_{1}-q)^2\\ &
        \hspace{0.8cm}-(\ell_1+\ell_2-q)^2-\ell_2^2\Big]\,,\\&
        =-\frac{s_1 s_2^3m_1^2 m_2^2}{64 m_p^6}\int e^{-iq\cdot b} \,\hat\delta(q\cdot v_1)\hat\delta(q\cdot v_2)\Bigg[i \Bigg(\boldsymbol{\tilde h}_{72}\,\mathbfcal{M}_{0,0;0,0,1,1,1}+\boldsymbol{\tilde h}_{73}\,\,q^2
      \mathbfcal{M}_{0,0;0,0,2,1,1}\\& \hspace{0.8cm}+\boldsymbol{\tilde h}_{74}\,\,q^2\mathbfcal{M}_{0,0;0,0,1,2,1}+\boldsymbol{\tilde h}_{75}\,\,q^2\mathbfcal{M}^{(++)}_{1,1;0,0,1,1,1}
      \Bigg)\big(q\cdot\mathcal{S}_1\cdot v_2\big)\Bigg]\,,\\&=\frac{m_1^2m_2^2}{m_p^6}\Bigg[\Delta_{\mathfrak{16}}^{\textrm{poly}}+\Delta_{\mathfrak{08}}^{\textrm{log}}\log(x)\Bigg],\label{eqt10}      
    \end{split}
\end{align}
where,
\begin{align}
    \begin{split}
    &\Delta_{\mathfrak{16}}^{\textrm{poly}}=\Bigg(\frac{s_1 s_2^3}{131072 \pi ^3 |\boldsymbol{b}|^3 \left(x^2-1\right)^6 \left(x^2+1\right)}\Bigg)\Bigg[x\bigg\{6 \left(65 x^{12}-387 x^8+387 x^4-65\right) \log (|\boldsymbol{b}|)\\ & \hspace{0.8cm}+\left(x^4-1\right) \big(x^8 (86-260 \log (2)+65 \log (\pi ))\\& \hspace{0.8cm}+1712 x^6+2 x^4 (1466+644 \log (2)-161 \log (\pi ))\\&\hspace{0.8cm}+1712 x^2+5 \gamma_E  \left(65 x^8-322 x^4+65\right)+86-260 \log (2)\\& \hspace{0.8cm}+65 \log (\pi )\bigg\}\Bigg], \\&\,\Delta_{\mathfrak{08}}^{\textrm{log}}=\Bigg(\frac{s_1 s_2^3}{131072 \pi ^3 |\boldsymbol{b}|^3 \left(x^2-1\right)^6 \left(x^2+1\right)}\Bigg)\Bigg[8 x\Bigg\{20 x^{12}+239 x^{10}+700 x^8+1154 x^6+700 x^4+239 x^2+20\Bigg\}\Bigg]\,.
    \end{split}
\end{align}
\par
\textbullet $\,\,$ The next diagrams mentioned below come from the two scalar-graviton interaction vertices (and graviton three vertex) and have $\mathbb{H}$-type topological structure. This diagram will only contribute to the spinless part of the 3PM eikonal phase. For $\mathbb{H}$-type diagram, the worldline propagator will not appear in the amplitude, and we get five independent propagators: $\ell_{1}^2,\,\ell_{2}^2,\,(\ell_{1}+\ell_{2}-q)^2,\,(\ell_{1}-q)^2,\,(q-\ell_{2})^2$. The diagrams are in the following form,
\begin{center}
\includegraphics[width=0.38\linewidth]{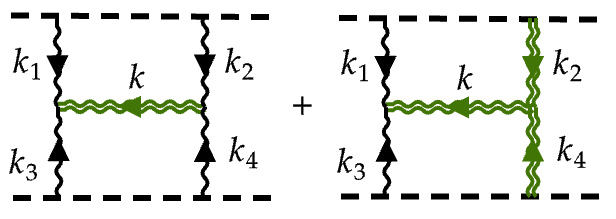}
\end{center}
From the first diagram, the contribution to the eikonal phase is given by,
\begin{align}
    \begin{split}
     i   \chi\Big|_{\mathcal{S}^0}
        =i \frac{s_1^2 s_2^2 m_1^2 m_2^2}{64 m_p^6} \int e^{-iq\cdot b}\hat\delta(q\cdot v_1)\hat\delta(q\cdot v_2)\,\Bigg[&\frac{\epsilon }{4 (\epsilon -1)}\mathbfcal{M}_{0,0;0,0,1,1,1}-\frac{3 q^2}{8 \epsilon }\mathbfcal{M}_{0,0;0,0,2,1,1}\\&+\frac{q^2 (3 \epsilon -4)}{16 (\epsilon -1)}\mathbfcal{M}_{0,0;1,1,0,1,1}+\frac{q^4}{4}\mathbfcal{M}_{0,0;1,1,1,1,1}\Bigg]\,,
    \end{split}
\end{align}
Then using \eqref{MasterM} we get,

\begin{align}
    \begin{split} \label{eqt11}   
     \chi\Big|_{\mathcal{S}^0}=-\Big(\frac{ m_1^2 m_2^2}{64 m_p^6}\Big)\Delta_{\mathfrak{17}}^{\textrm{poly}}\,, \Delta_{\mathfrak{17}}^{\textrm{poly}}=-\frac{5 x^2s_1^2 s_2^2 \log (x)}{32 \pi ^3 |\boldsymbol{b}|^2 \left(x^2-1\right)^2}\,.
        \end{split}
\end{align}
 The second one with $\mathbb{H}$-type topology may appear at 3PM where we have one scalar-scalar-graviton and one graviton 3-point vertex. This diagram contributes to the spinless as well as the spinning part of the eikonal phase because of the.
The graviton 3-point vertex takes the form,
\begin{align}
    S_{EH}\Big|_{h^3}=\int d^4x \,\,\,\mathcal{U}^{\mu\nu\,\,\alpha\beta\rho\,\,\gamma\delta\sigma}\,h_{\mu\nu}\,\partial_{\rho}h_{\alpha\beta}\partial_{\sigma}h_{\gamma\delta}\,.
\end{align}
In the Fourier space, the vertex factor takes the following form,
\begin{align}
    \begin{split}
        \mathcal{V}(k_1,k_2,k_3)=-\hat\delta^{(4)}(k_1+k_2+k_3)\,\mathcal{U}^{\mu\nu\,\,\alpha\beta\rho\,\,\gamma\delta\sigma}\,k_{2\rho}k_{3\sigma}h_{\mu\nu}(-k_1)\,h_{\alpha\beta}(-k_2)\,h_{\gamma\delta}(-k_3)\,.
    \end{split}
\end{align}
Correspondingly, the contribution to the eikonal phase is given by,
\begin{align}
    \begin{split}
        i\chi=i\frac{m_1^2 m_2^2 s_1 s_2}{64 m_p^6}\int e^{-i q\cdot b}\hat\delta(q\cdot v_1)\hat\delta(q\cdot v_2)\int_{\ell_{1},\ell_{2}}\frac{\hat\delta(\ell_{1}\cdot v_2)\hat\delta(\ell_{2}\cdot v_1)}{\ell_{1}^2 \ell_{2}^2 (\ell_{1}+\ell_{2}-q)^2 (\ell_{1}-q)^2 (\ell_{2}-q)^2}\mathcal{N}\label{4.68j}
    \end{split}
\end{align}
where the numerator $\mathcal{N}$ has the following form,
\begin{align}
    \begin{split}
        \mathcal{N}= \Gamma_{(3)}^{\mu\nu\,\alpha\beta\,\gamma\delta}(k_2,k_4,k)\,k_1^a k_3^b,P_{a b;\mu\nu}P_{\alpha\beta,c d}P_{\gamma\delta;e f}(v_1^{c}v_1^d+i(k_2\cdot \mathcal{S}_1))^{(c}v_1^{d)}\,\,(v_2^{e}v_2^f+i(k_4\cdot \mathcal{S}_2)^{(e}v_2^{f)}),
    \end{split}
\end{align}
where the 3-point vertex function $\Gamma_{(3)}$ takes the form,
\begin{align}
    \begin{split}
    \Gamma^{\mu\nu\,\alpha\beta\,\gamma\delta}=&\textrm{sym}\Big[\frac{1}{2}k_{2}^\mu k_{4}^\nu\eta^{\alpha\beta}\eta^{\gamma\delta}-\frac{1}{4}k_2\cdot k_4\,\eta^{\mu\nu}\eta^{\alpha\beta}\eta^{\gamma\delta}+k_2\cdot k_4 \eta^{\alpha\nu}\eta^{\beta\mu}\eta^{\gamma\delta}-k_2^\delta k_4^\gamma \eta^{\alpha\mu}\eta^{\beta\nu}\\ &
        +\frac{1}{4}(k_2\cdot k_4)\eta^{\mu\nu}\eta^{\alpha\gamma}\eta^{\beta\delta}-k_2^\nu\,k_4^\beta \eta^{\gamma\delta}\eta^{\mu\alpha}-k_{2}^\mu k_4^\nu \eta^{\alpha\delta}\eta^{\beta\gamma}+\frac{1}{2}\,k_2^\alpha\,k_4^\beta\,\eta^{\mu\nu}\eta^{\gamma\delta}\\ &
        -k_2\cdot k_4\,\eta^{\alpha\mu}\eta^{\beta\delta}\eta^{\nu\gamma}-\frac{1}{2}k_2^\gamma\,k_4^\beta\,\eta^{\mu\nu}\eta^{\alpha\delta}+k_2^\gamma k_4^\alpha\,\eta^{\mu\beta}\eta^{\nu\delta}-\frac{1}{2}\,k_2^\mu k_4^\nu \eta^{\alpha\delta}\eta^{\beta \gamma}+2 k_{2}^\nu k_4^\alpha\eta^{\beta\gamma}\eta^{\mu\delta}+\textrm{permutations}\Big]\,.
    \end{split}
\end{align}
Then, proceeding as before, we get, 
\begin{align}
    \begin{split}
        \chi\Big|_{\mathcal{S}^0}= \frac{m_1^2 m_2^2 s_1 s_2}{64 m_p^6}\int e^{-i q\cdot b}\hat\delta(q\cdot v_1)\hat\delta(q\cdot v_2)\Bigg(&\boldsymbol{\tilde h}_{76}\mathbfcal{M}_{0,0;0,0,1,1,1}+\boldsymbol{\tilde h}_{77}\,q^2\mathbfcal{M}_{0,0;0,0,2,1,1}+\boldsymbol{\tilde h}_{78}\,q^2\mathbfcal{M}_{0,0;0,0,1,2,1}\\ &+\boldsymbol{\tilde h}_{79}\,q^2\mathbfcal{M}_{0,0;1,1,0,1,1}+\boldsymbol{\tilde h}_{80}\,q^4\mathbfcal{M}_{0,0;1,1,1,1,1}+\boldsymbol{\tilde h}_{81}\,q^6\mathbfcal{M}_{0,0;1,1,2,1,1}\Bigg)
    \end{split}
\end{align}
and,
\begin{align}
    \begin{split}
        \chi\Big|_{\mathcal{S}^1}= \frac{m_1^2 m_2^2 s_1 s_2}{64 m_p^6}\int e^{-i q\cdot b}\hat\delta(q\cdot v_1)\hat\delta(q\cdot v_2)\Bigg[i\,\Bigg(&\boldsymbol{\tilde h}_{82}\mathbfcal{M}_{0,0;0,0,1,1,1}+\boldsymbol{\tilde h}_{83}\,q^2\mathbfcal{M}_{0,0;0,0,2,1,1}\\ &+\boldsymbol{\tilde h}_{84}\,q^2\mathbfcal{M}_{0,0;0,0,1,2,1}+\boldsymbol{\tilde h}_{85}\,q^2\mathbfcal{M}_{0,0;1,1,0,1,1}\\&+\boldsymbol{\tilde h}_{86}\,q^4\mathbfcal{M}_{0,0;1,1,1,1,1}+\boldsymbol{\tilde h}_{87}\,q^6\mathbfcal{M}_{0,0;1,1,2,1,1}\Bigg)\big(q\cdot\mathcal{S}_1\cdot v_2\big)+\\&i\,\Bigg(\boldsymbol{\tilde h}_{88}\mathbfcal{M}_{0,0;0,0,1,1,1}+\boldsymbol{\tilde h}_{89}\,q^2\mathbfcal{M}_{0,0;0,0,2,1,1}\\ &+\boldsymbol{\tilde h}_{90}\,q^2\mathbfcal{M}_{0,0;0,0,1,2,1}+\boldsymbol{\tilde h}_{91}\,q^2\mathbfcal{M}_{0,0;1,1,0,1,1}\\&+\boldsymbol{\tilde h}_{92}\,q^4\mathbfcal{M}_{0,0;1,1,1,1,1}+\boldsymbol{\tilde h}_{93}\,q^6\mathbfcal{M}_{0,0;1,1,2,1,1}\Bigg)\big(q\cdot\mathcal{S}_2\cdot v_1\big)\Bigg]\,.
    \end{split}
\end{align}
Then using \eqref{MasterM} we get,

\begin{align}
    \begin{split} \label{eqt12}   
        \chi\Big|_{\mathcal{S}^0}=\frac{m_1^2m_2^2}{m_p^6}\Bigg[\Delta_{\mathfrak{18}}^{\textrm{poly}}+\Delta_{\mathfrak{09}}^{\textrm{log}}\log(x)+\Delta_{\mathfrak{03}}^{\textrm{Polylog}}\Big(\log(x)^2+ \textbf{Li}_2\left(1-x^2\right)\Big)\Bigg]\,,
          \end{split}
\end{align}
\enlargethispage{2\baselineskip}
where, 
\begin{thesiscompactdisplay}
\begin{align}
    \begin{split} 
  &  \Delta_{\mathfrak{18}}^{\textrm{poly}}= \frac{s_1s_2 (3-3 x^4)}{16384 \pi ^3 |\boldsymbol{b}|^2 \left(x^2-1\right)^2}\,,   \Delta_{\mathfrak{09}}^{\textrm{log}}=\frac{s_1s_2\bigg\{4 \left(-2 x^2 \log \left(\pi  |\boldsymbol{b}|^6\right)+5 x^4+x^2 (-10 \gamma_E -4+\log (256))+5\right)\bigg\}}{16384 \pi ^3 |\boldsymbol{b}|^2 \left(x^2-1\right)^2}\,,   \\& \Delta_{\mathfrak{03}}^{\textrm{Polylog}}=\frac{4\,s_1s_2\,(x-1) x}{16384 \pi ^3 |\boldsymbol{b}|^2 \left(x^2-1\right)^2}
      \end{split}
\end{align}
\end{thesiscompactdisplay}
and 
\begin{align}
    \begin{split} \label{eqt13}  
        \hspace{0cm}\chi\Big|_{\mathcal{S}^1}=&\frac{m_1^2m_2^2}{m_p^6}\Bigg\{\Bigg[\Delta_{\mathfrak{19}}^{\textrm{poly}}+\Delta_{\mathfrak{10}}^{\textrm{log}}\log(x)+\Delta_{\mathfrak{04}}^{\textrm{Polylog}}\Big(\log(x)^2+ \textbf{Li}_2\left(1-x^2\right)\Big)\Bigg]\big(\hat{b}\cdot\mathcal{S}_1\cdot v_2\big)\\&\hspace{1cm}+\Bigg[\Delta_{\mathfrak{20}}^{\textrm{poly}}+\Delta_{\mathfrak{11}}^{\textrm{log}}\log(x)+\Delta_{\mathfrak{05}}^{\textrm{Polylog}}\Big(\log(x)^2+ \textbf{Li}_2\left(1-x^2\right)\Big)\Bigg]\big(\hat{b}\cdot\mathcal{S}_2\cdot v_1\big)\Bigg\}
          \end{split}
\end{align}
\enlargethispage{2\baselineskip}
where,
\begin{thesiscompactdisplay}
\begin{align}
\begin{split}
&\Delta_{\mathfrak{19}}^{\textrm{poly}}=\frac{s_{1} s_{2} \left(x^6-10 x^5+3 x^4-49 x^3+3 x^2-10 x+1\right)}{8192 \pi ^3 |\boldsymbol{b}|^3 \left(x^2-1\right)^3},\\ &
\Delta_{\mathfrak{4}}^{\textrm{Polylog}}=\frac{7 x^2\,s_1s_2\, (x-1) x^2 \left(x^4+4 x^2+1\right)}{8192 \pi ^3 |\boldsymbol{b}|^3 \left(x^2-1\right)^4 \left(x^2+1\right)},\\ &
\Delta_{\mathfrak{10}}^{\textrm{log}}=-\Bigg(\frac{s_1s_2}{{16384 \pi ^3 |\boldsymbol{b}|^3 \left(x^2-1\right)^4 \left(x^2+1\right)}}\Bigg)\Bigg[x \big(28 \left(x^6+4 x^4+x^2\right) \log \left(\pi  |\boldsymbol{b}|^6\right)-37 x^8+8 x^7+4 x^6 (35 \gamma_E -8\\&-28 \log (2))+24 x^5+x^4 (560 \gamma_E -278-448 \log (2))+24 x^3+4 x^2 (35 \gamma_E -8-28 \log (2))+8 x-37\big)\Bigg]\,,\\ &
\Delta_{\mathfrak{11}}^{\textrm{log}}=\Bigg(\frac{s_1s_2}{16384 \pi ^3 |\boldsymbol{b}|^3 \left(x^2-1\right)^6 \left(x^2+1\right)^2}\Bigg)\Bigg[2 \big(48 \left(x^2-1\right)^2 \left(x^6+5 x^4+5 x^2+1\right) x^3 \log ( |\boldsymbol{b}|)-18 x^{16}-13 x^{15}+96 x^{14}\\
&+x^{13} (40 \gamma_E -33-32 \log (2)+8 \log (\pi ))+280 x^{12}+x^{11} (120 \gamma_E -47-96 \log (2)+24 \log (\pi ))\\
&-992 x^{10}+x^9 (-160 \gamma_E +93+128 \log (2)-32 \log (\pi ))-2604 x^8\\
&+x^7 (-160 \gamma_E +93+128 \log (2)-32 \log (\pi ))-992 x^6+x^5 (120 \gamma_E -47-96 \log (2)+24 \log (\pi ))\\
&+280 x^4+x^3 (40 \gamma_E -33-32 \log (2)+8 \log (\pi ))+96 x^2-13 x-18\big)\Bigg]\,,\\ &
\Delta_{\mathfrak{5}}^{\textrm{Polylog}}=-\frac{8 (x-1) x^2 \left(x^6+5 x^4+5 x^2+1\right)\, s_1s_2}{16384 \pi ^3 |\boldsymbol{b}|^3 \left(x^2-1\right)^4 \left(x^2+1\right)^2}\,,\\ &
\Delta_{\mathfrak{20}}^{\textrm{poly}}=\frac{x  \left(19 x^4-218 x^2+19\right)\,s_1s_2}{16384 \pi ^3 |\boldsymbol{b}|^3 \left(x^2-1\right)^3 }\,.
\end{split}
\end{align}
\end{thesiscompactdisplay}

\textbullet $\,\,$ At 3PM we face a $\mathbb{N}$-type topology involving scalar and graviton lines. It gives the following three diagrams. 
\begin{center}
\includegraphics[width=0.46\linewidth]{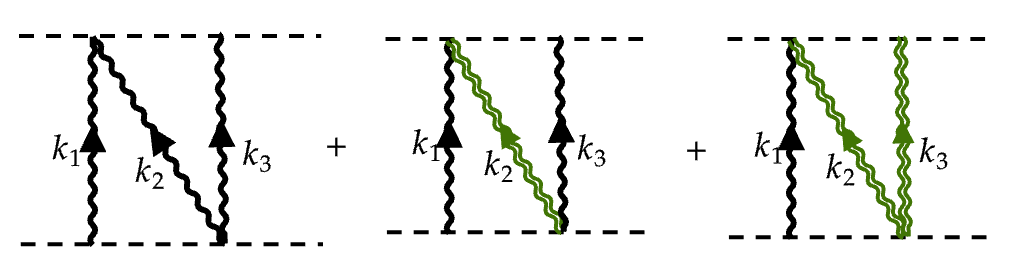}
\end{center}
The spinless contributions come from the first and the second diagrams only. We get, 
\begin{align}
\begin{split} \label{eqt14}  
    \chi\Big|_{\mathcal{S}^0}&=\frac{m_1^2 m_2^2}{16\,m_p^6}\int e^{-i q\cdot b}\hat\delta(q\cdot v_1)\hat\delta(q\cdot v_2)\Bigg(s_1s_2g_1g_2+s_1^2s_2^2\,\boldsymbol{\tilde h}_{94}\Bigg)\mathbfcal{M}_{0,0;0,0,1,1,1}\,,\\&
    =\frac{m_1^2 m_2^2}{m_p^6}\Delta_{\mathfrak{12}}^{\textrm{log}}\,,\,\,\Delta_{\mathfrak{12}}^{\textrm{log}}=\Bigg(\frac{s_{1}s_{2} \log (x) \left(8 g_{1} g_{2} x^2+s_{1} s_{2} \left(x^4+1\right)\right)}{1024 \pi ^3 |\boldsymbol{b}|^2 \left(x^2-1\right)^2}\Bigg)\,.
    \end{split}
\end{align}




The spinning contributions come from the second and third diagrams. We get,
\begin{align}
    \begin{split}
       & \chi\Big|_{\mathcal{S}^{1}}=-\frac{m_1^2 m_2^2 s_1^2  s_2^2}{32 m_p^6}\int e^{-i q\cdot b}\hat\delta(q\cdot v_1)\hat\delta(q\cdot v_2)\Bigg[i\,\Bigg(\boldsymbol{\tilde h}_{95}\mathbfcal{M}_{0,0;0,0,1,1,1}+\boldsymbol{\tilde h}_{96}\,q^2\mathbfcal{M}_{0,0;0,0,1,2,1}\\ &\hspace{0.8cm}+\boldsymbol{\tilde h}_{97}\,q^2\mathbfcal{M}_{0,0;0,0,2,1,1}\Bigg)\big(q\cdot\mathcal{S}_1\cdot v_2\big)+\\&\hspace{0.8cm}i\,\Bigg\{\Big(\frac{1}{2 s_1 s_2}\boldsymbol{\tilde h}_{98}-\boldsymbol{\tilde h}_{95}\Big)\mathbfcal{M}_{0,0;0,0,1,1,1}+\Big(\frac{1}{2 s_1 s_2}\boldsymbol{\tilde h}_{99}-\boldsymbol{\tilde h}_{97}\Big)\,q^2\mathbfcal{M}_{0,0;0,0,2,1,1}\\ &\hspace{0.8cm}+\Big(\frac{1}{2 s_1 s_2}\boldsymbol{\tilde h}_{100}-\boldsymbol{\tilde h}_{96}\Big)\,q^2\mathbfcal{M}_{0,0;0,0,1,2,1}\Bigg)\big(q\cdot\mathcal{S}_2\cdot v_1\big)\Bigg\}\Bigg]\,.
    \end{split}
\end{align}
Now using (\ref{MasterM}) we get, 
\begin{align}
    \begin{split} \label{eqt15}     \chi\Big|_{\mathcal{S}^{1}}=\frac{m_1^2m_2^2}{m_p^6}\Bigg\{&\Bigg[\Delta_{\mathfrak{13}}^{\textrm{log}}\log(x)+\Delta_{\mathfrak{06}}^{\textrm{Polylog}}\Big(\log(x)^2+ \textbf{Li}_2\left(1-x^2\right)\Big)\Bigg]\big(\hat{b}\cdot\mathcal{S}_1\cdot v_2\big)+\\&\Bigg[\Delta_{\mathfrak{21}}^{\textrm{poly}}+\Delta_{\mathfrak{14}}^{\textrm{log}}\log(x)+\Delta_{\mathfrak{07}}^{\textrm{Polylog}}\Big(\log(x)^2+ \textbf{Li}_2\left(1-x^2\right)\Big)\Bigg]\big(\hat{b}\cdot\mathcal{S}_2\cdot v_1\big)\Bigg\}\,,
         \end{split}
\end{align}
where, 
\begin{align}
    \begin{split}
 &   \Delta_{\mathfrak{13}}^{\textrm{log}}=-\Bigg(\frac{s_{1}^2 s_{2}^2}{2048 \pi ^3 |\boldsymbol{b}|^3 \left(x^2-1\right)^4}\Bigg)\Bigg[x \left(x^2+1\right) \big(-12 \left(x^4+1\right) \log (|\boldsymbol{b}|)+x^4-10 \gamma_E  \left(x^4+1\right)+8 x^4 \log (2)\\&\hspace{0.8cm}-2 \left(x^4+1\right) \log (\pi )+18 x^2+1+\log (256)\big)\Bigg]\,,\\&
    \Delta_{\mathfrak{06}}^{\textrm{Polylog}}=-\frac{3 s_{1}^2 s_{2}^2 x \left(x^2+1\right) \left(x^4+1\right)}{1024 \pi ^3 |\boldsymbol{b}|^3 \left(x^2-1\right)^4}\,,
     \Delta_{\mathfrak{21}}^{\textrm{poly}}=-\frac{41\,s_{1} s_{2} x \left(x^2+1\right)^2}{2048 \pi ^3 |\boldsymbol{b}|^3 \left(x^2-1\right)^3}\,,\\&
       \Delta_{\mathfrak{14}}^{\textrm{log}}=\Bigg(\frac{s_{1}s_{2}}{4096 \pi ^3 |\boldsymbol{b}|^3 \left(x^2-1\right)^4}\Bigg)\Bigg[-x \left(x^2+1\right) \big(12 \left(x^4+1\right) \log (|\boldsymbol{b}|) (2\,s_{1}s_{2}+1)+10 \gamma_E  \left(x^4+1\right) (2\,s_{1} s_{2}+1)\\&\hspace{0.8cm}+2 \log (\pi ) \left(2\,s_{1} s_{2} \left(x^4+1\right)+x^4\right)-2\,s_{1}s_{2} \left(x^4+8 \left(x^4+1\right) \log (2)+18 x^2+1\right)\\&\hspace{0.8cm}+x^4 (5-8 \log (2))-18 x^2+5-8 \log (2)+2 \log (\pi )\big)\Bigg]\,,\\&
       \Delta_{\mathfrak{07}}^{\textrm{Polylog}}=\frac{3 \,s_{1} s_{2} x \left(x^2+1\right) \left(x^4+1\right) (2 s_{1}s_{2}+1)}{2048 \pi ^3 |\boldsymbol{b}|^3 \left(x^2-1\right)^4 }\,.
      \end{split}
\end{align}
\textbullet $\,\,$ We face the following topology with single 3-point kind of interaction vertex,
\begin{center}
\includegraphics[width=0.46\linewidth]{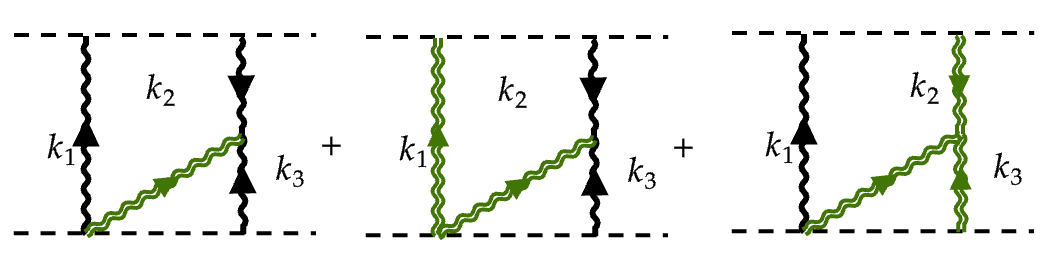}
\end{center}
Before quoting the result for the contribution to the eikonal phase, we note that because of the presence of graviton two-point worldline vertex in the second diagram, this diagram only contributes to the spinning eikonal, and also, there is no term linear in the spin of the first black hole coming from this diagram. Now, we first write down the total spinless contribution from these three diagrams, 
\begin{align}
    \begin{split}  \label{eqt16} 
   \chi\Big|_{\mathcal{S}^0}&=\frac{m_1^2 m_2^2}{m_p^6}\int e^{-i q\cdot b}\hat\delta(q\cdot v_1)\hat\delta(q\cdot v_2)\Bigg[\Bigg(\frac{s_1^2 s_2^2}{128 (1-\epsilon)}-\frac{s_1 s_2}{32} \boldsymbol{\tilde h}_{101}\Bigg)\mathbfcal{M}_{0,0;0,0,1,1,1}-\frac{s_1 s_2}{32}\boldsymbol{\tilde h}_{102}q^2\mathbfcal{M}_{0,0;0,0,1,2,1}\Bigg)\Bigg]\,,\\&
   =\frac{m_1^2 m_2^2}{m_p^6}\Bigg(\Delta_{\mathfrak{22}}^{\textrm{poly}}+\Delta_{\mathfrak{15}}^{\textrm{log}}\log(x)\Bigg)
  \end{split}
\end{align}    
where, 
\begin{align}
    \begin{split}
      \Delta_{\mathfrak{22}}^{\textrm{poly}}= \frac{s_{1} s_{2} \left(-x^8-x^6+x^2+1\right)}{4096 \pi ^3 |\boldsymbol{b}|^2 x^2 \left(x^2-1\right)^2}\,,\Delta_{\mathfrak{15}}^{\textrm{log}}=-\frac{s_{1} s_{2} \left(x^2 (5-s_{1}s_{2})+2 x^4+2\right)}{1024 \pi ^3 |\boldsymbol{b}|^2 \left(x^2-1\right)^2}\,.
    \end{split}
\end{align}
The total contribution to the spinning part is
\begin{align}
\begin{split}
\chi\Big|_{\mathcal{S}^1}
=&\frac{m_1^2 m_2^2}{m_p^6}\Big(-i\frac{s_1s_2}{32}\Big)\int e^{-i q\cdot b}\hat\delta(q\cdot v_1)\hat\delta(q\cdot v_2)\Bigg[\Bigg(\boldsymbol{\tilde h}_{103}\mathbfcal{M}_{0,0;0,0,1,1,1}
+\boldsymbol{\tilde h}_{104}q^2\mathbfcal{M}_{0,0;0,0,2,1,1}\\
&+\boldsymbol{\tilde h}_{105}q^2\mathbfcal{M}_{0,0;0,0,1,2,1}\Bigg)\big(q\cdot\mathcal{S}_1\cdot v_2\big)\\
&+\Bigg((\boldsymbol{\tilde h}_{106}+\boldsymbol{\tilde h}_{109})\mathbfcal{M}_{0,0;0,0,1,1,1}
+(\boldsymbol{\tilde h}_{107}+\boldsymbol{\tilde h}_{110})q^2\mathbfcal{M}_{0,0;0,0,2,1,1}\\
&+(\boldsymbol{\tilde h}_{108}+\boldsymbol{\tilde h}_{111})q^2\mathbfcal{M}_{0,0;0,0,1,2,1}\Bigg)\big(q\cdot\mathcal{S}_2\cdot v_1\big)\Bigg]\,,\\
=&\frac{m_1^2 m_2^2}{m_p^6}\Bigg\{\Bigg(\Delta_{\mathfrak{23}}^{\textrm{poly}}+\Delta_{\mathfrak{16}}^{\textrm{log}}\log(x)\Bigg)(\hat b\cdot \mathcal{S}_1\cdot v_2)\\
&\qquad\qquad\qquad\;
+\Bigg(\Delta_{\mathfrak{24}}^{\textrm{Poly}}+\Delta_{\mathfrak{17}}^{{\textrm{log}}}\log(x)\Bigg)(\hat b\cdot \mathcal{S}_2\cdot v_1)\Bigg\}\,.
\end{split}
\end{align}
where, 
\begin{align}
    \begin{split}
   & \Delta_{\mathfrak{23}}^{\textrm{poly}}=-\frac{s_{1}s_{2} \left(x^{10}+9 x^8+32 x^6-32 x^4-9 x^2-1\right)}{2048 \pi ^3 |\boldsymbol{b}|^3 x \left(x^2-1\right)^4 }\,,\Delta_{\mathfrak{16}}^{\textrm{log}}=-\frac{s_{1} s_{2} x \left(x^2+1\right)^3}{128 \pi ^3 |\boldsymbol{b}|^3 \left(x^2-1\right)^4 }\,,
   \\& \Delta_{\mathfrak{24}}^{\textrm{poly}}=\frac{s_{1} s_{2} \left(113 x^6-32 x^5+555 x^4+555 x^2-32 x+113\right)}{32768 \pi ^3 |\boldsymbol{b}|^3 \left(x^2-1\right)^3 }\,,\\&\Delta_{\mathfrak{17}}^{{\textrm{log}}}=\frac{s_{1} s_{2} \left(17 x^8-8 x^7+153 x^6-8 x^5+328 x^4-8 x^3+153 x^2-8 x+17\right)}{8192 \pi ^3 |\boldsymbol{b}|^3 \left(x^2-1\right)^4}\,.  \label{eqt17} 
     \end{split}
\end{align} 
\textbullet $\,\,$ We have the following topologies both involving scalar-graviton and graviton 3-point vertices.
\begin{center}
\includegraphics[width=0.38\linewidth]{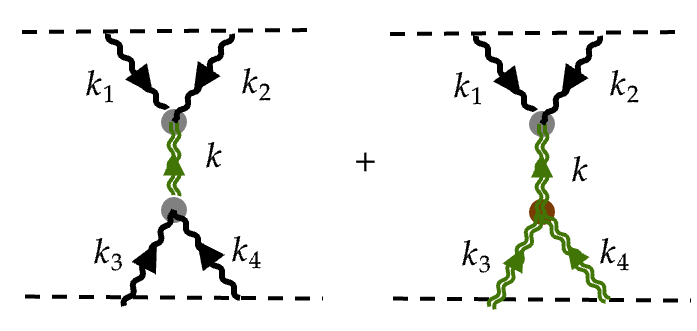}
\end{center}
We can easily check that, after evaluating the corresponding integrals for both the spinless and spin-dependent parts, they vanish identically up to $\mathcal{O}(\epsilon^0)\,.$ Hence, this class of diagrams does not contribute to the eikonal phase in our case. 
\par
\textbullet $\,\,$
We now discuss the scalar-graviton diagrams where we have spin contribution coming from the scalar-fermion vertex. The diagrams look like the following. 
\begin{center}
\includegraphics[width=0.38\linewidth]{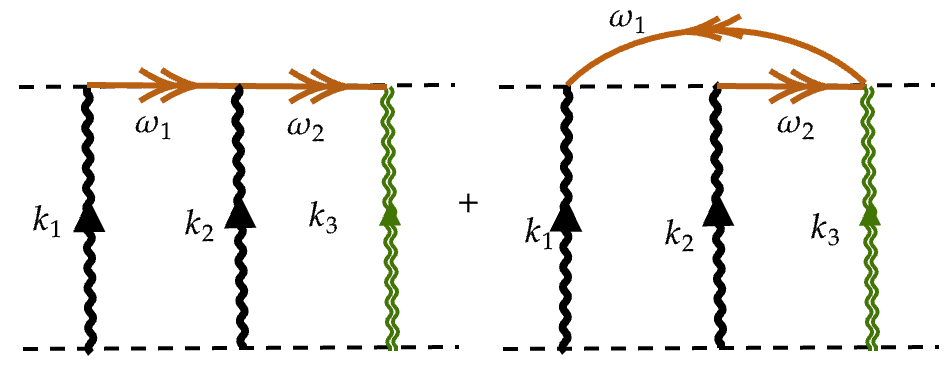}
\end{center}
One may notice that, unlike other diagrams, this one consists of $\bar\Psi^\mu\Psi^\nu$. To recover the spin, one needs to take the spinor flow in the opposite direction. So, for this particular diagram, we need to take the antisymmetric combination of $\bar\Psi^\mu\Psi^\nu$ to recover the classical spin.  Moreover, we can easily check that the contribution coming from the first diagram to the eikonal phase gets cancelled by the second diagram. Hence, together, they don't contribute. 
\par
\textbullet $\,\,$ Another 3PM spinor diagram that will contribute has the topology of the following form,
\begin{center}
\includegraphics[width=0.38\linewidth]{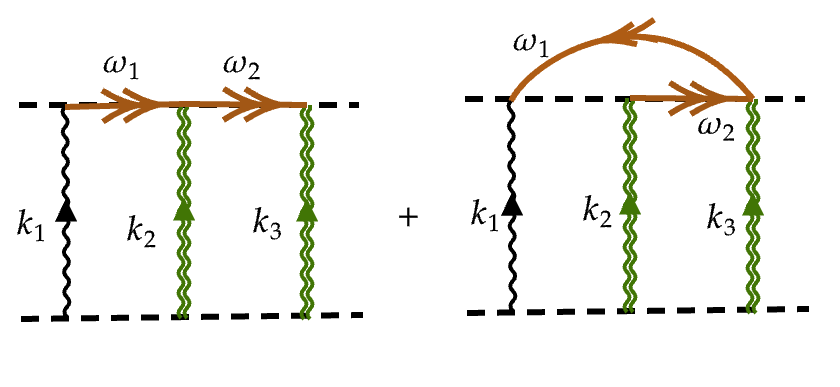}
\end{center}
Let's start our computations from the first diagram.   The 3-point spin-spin-graviton interaction vertex which contributes to this term comes from the following part of the action,
\begin{align}
    \begin{split}
        S\subset& \,i\frac{m_i}{m_p}\int  d\tau\, \dot x^\mu \omega^{ab}_{\mu}\bar\psi_{a} \psi_{b}\,,\\ &
        = \frac{i m_i}{m_p}v^\mu\int d\tau\, \partial^{[a}h^{b]}_\mu \bar\psi_{[a}\psi_{b]} \xrightarrow[]{F.S.}-\frac{m_a}{m_p}\int_{k,\omega_i}e^{ik\cdot b}\hat\delta(k\cdot v-\omega_1+\omega_2) v^\mu\,k^{[a}h^{b]}_\mu (-k)\bar\psi_{[a}(-\omega_1)\,\psi_{b]}(-\omega_2)  \,.
        \end{split}
\end{align}
Therefore, the vertex factor is given by
\begin{align}
    V: e^{ik\cdot b}\hat\delta(k\cdot v-\omega_1+\omega_2)\,v_i^\mu \,k^a+\textrm{other 3 combinations}\,.
\end{align}
Schematically, the amplitude looks like
\begin{align}
    \begin{split}
       \mathcal{A}\sim\int_{k_i,\omega_i}\Big(\prod\boldsymbol{\delta}^{(6)}\Big)\bar\Psi^\eta\Psi^\sigma\omega_1\langle \psi_\eta\,\bar\psi_{[[a}\rangle_{\omega_1}k_{2[a}\langle  h_{b]\mu}h_{\alpha\beta}\rangle_{k_2} \langle \psi_{b]]}\,\bar\psi_\rho\rangle k_{3[\rho}\delta^{(\chi}_{\sigma]}v_1^{\pi)} \langle  h_{\chi\pi}h_{\kappa\theta}\rangle_{k_3}v_2^\alpha v_2^\beta v_2^\kappa v_2^\theta\,.\nonumber
    \end{split}
\end{align}
Here, $\prod \boldsymbol\delta^6$ stands for the product of all relevant delta functions along with $e^{ik\cdot b}$ coming from the worldline vertices. \textcolor{black}{Note that we take only the antisymmetric combination of $\eta,\sigma$ to relate the combination of background superfield to classical spin, which is eventually important to define the Pauli-Lubanski vector.} 
Then proceeding as before we get, 
\begin{align}
    \begin{split}
i\,\chi\Big|_{\mathcal{S}^1}&=\frac{ m_1 m_2^3 s_1 s_2}{16 m_p^6}\int e^{-iq\cdot b} \hat\delta(q\cdot v_1)\hat\delta(q\cdot v_2)\boldsymbol{\tilde h}_{112} \mathbfcal{L}_{0,0;0,0,1,1,1}\big(q\cdot\mathcal{S}_1\cdot v_2\big)\,.
       \end{split}
\end{align}
Correspondingly, the contribution from the second diagram has the following form,
\begin{align}
    \begin{split}
i\,\chi\Big|_{\mathcal{S}^1}&=\frac{ m_1 m_2^3 s_1 s_2}{8 m_p^6}\int e^{-iq\cdot b} \hat\delta(q\cdot v_1)\hat\delta(q\cdot v_2)\boldsymbol{\tilde h}_{113} \mathbfcal{L}_{0,0;0,0,1,1,1}\big(q\cdot\mathcal{S}_1\cdot v_2\big)\,.
       \end{split}
\end{align}
Then, the total contribution to the spinning part is,
\begin{align}
    \begin{split} \label{eqt18} 
 \chi\Big|_{\mathcal{S}^1}&=\frac{m_1m_2^3}{m_p^6}\Delta_{\mathfrak{25}}^{\textrm{poly}}\,,\quad\,\,\Delta_{\mathfrak{25}}^{\textrm{poly}}=\frac{ s_{1} s_{2}\,9 \left(x^2+1\right) \left(7 x^4+2 x^2+7\right)}{32768 \pi ^3 |\boldsymbol{b}|^3 \left(x^2-1\right)^3}\,.
      \end{split}
\end{align}
Here we have used (\ref{MasterL}).

\textbullet $\,\,$ Another 3PM diagram having one worldline and one super-field propagator that will contribute has the following topology 
\begin{center}
\includegraphics[width=0.38\linewidth]{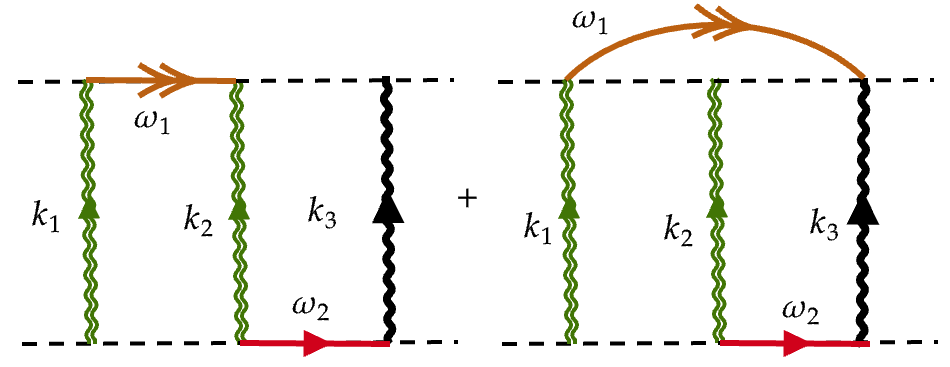}
\end{center}
It is clear from the structure of this diagram that due to the presence of a spinning worldline propagator, we need to take only the spinless terms from the other vertices. Like the previous diagram, this one also has $\bar\Psi^\mu \Psi^\nu$, but for this case, we can add another counterpart, which will have opposite spinor flow. So, in this case, we can also take the antisymmetric combination of $\bar\Psi^\mu \Psi^\nu$ to obtain the classical spin.  Taking all of those things into account, the total contribution to the eikonal phase coming from these two diagrams is given by,
\begin{align}
    \begin{split}
     \chi\Big|_{\mathcal{S}^1}=\frac{m_1^2 m_2^2}{m_p^6}\Big(-i\frac{s_1s_2}{16}\Big)\int e^{-i q\cdot b}\hat\delta(q\cdot v_1)\hat\delta(q\cdot v_2)&\Bigg[\Bigg(\boldsymbol{\tilde h}_{114}\mathbfcal{M}_{0,0;0,0,1,1,1}+\boldsymbol{\tilde h}_{115}q^2\mathbfcal{M}_{0,0;0,0,2,1,1}\\&\hspace{0.25cm}+\boldsymbol{\tilde h}_{116}q^2\mathbfcal{M}_{0,0;0,0,1,2,1}-\frac{1}{2}\boldsymbol{\tilde h}_{117}q^2\mathbfcal{M}^{(++)}_{1,1;0,0,1,1,1}\Bigg)\big(q\cdot\mathcal{S}_1\cdot v_2\big)\,.
     \end{split}
\end{align}
Finally, we get, 
\begin{align}
    \begin{split}  \label{eqt19} 
     \chi\Big|_{\mathcal{S}^1}=\frac{m_1^2 m_2^2}{m_p^6}\Bigg(\Delta_{\mathfrak{26}}^{\textrm{poly}}+\Delta_{\mathfrak{18}}^{\textrm{log}}\log(x)\Bigg)\big(\hat b\cdot\mathcal{S}_1\cdot v_2\big)\,,
      \end{split}
\end{align} 
where, 
\begin{align}
    \begin{split}
  &\Delta_{\mathfrak{26}}^{\textrm{poly}}=\frac{s_1s_2}{32768 \pi ^3 |\boldsymbol{b}|^3 \left(x^2-1\right)^5}\Bigg[-192 \left(x^4-1\right)^2 x \log (|\boldsymbol{b}|)-48 x^{10}+x^9 (-160 \gamma_E +407+128 \log (2)\\&\hspace{0.8cm}-32 \log (\pi ))-144 x^8+1520 x^7+192 x^6+x^5 (320 \gamma_E +2226-256 \log (2)\\&\hspace{0.8cm}+64 \log (\pi ))+192 x^4+1520 x^3-144 x^2+x (-160 \gamma_E +407+128 \log (2)\\&\hspace{0.8cm}-32 \log (\pi ))-48\Bigg]\,,\\&
\Delta_{\mathfrak{18}}^{\textrm{log}}=\frac{s_{1} s_{2} x}{4096 \pi ^3 |\boldsymbol{b}|^3 \left(x^2-1\right)^6 \left(x^2+1\right)}\Bigg(20 x^{12}-32 x^{11}+239 x^{10}-32 x^9+700 x^8+64 x^7+1154 x^6\\&\hspace{0.8cm}+64 x^5+700 x^4-32 x^3+239 x^2-32 x+20\Bigg)\,.
      \end{split}
\end{align}
\textbullet $\,\,$ Two 3PM diagrams coming from graviton-scalar and scalar-scalar worldline interaction vertex have the following topology,
\begin{center}
\includegraphics[width=0.38\linewidth]{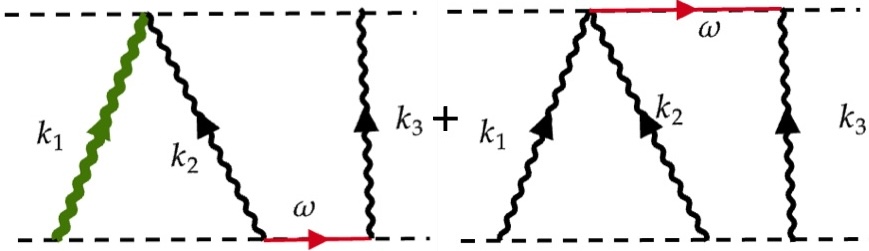}
\end{center}
For the first diagram, we get the following contribution for the spinless part of the eikonal phase,

 \begin{align}
     \begin{split}
         i\chi\Big|_{\mathcal{S}^{0}}=& -i\frac{s_1^2 s_2^2 m_1^2 m_2^2}{32 m_p^6} \frac{\gamma^2(d-2)-1}{2(d-2)}\int e^{-iq\cdot b}\hat\delta(q\cdot v_1)\hat\delta(q\cdot v_2)\Bigg(\frac{\epsilon  (2 \epsilon +1)}{2 \epsilon -1}\mathbfcal{M}_{0,0;0,0,1,1,1}+\frac{q^2 (2 \epsilon +1)}{2-4 \epsilon }\mathbfcal{M}_{0,0;0,0,2,1,1}\\ &\hspace{0.8cm}+\frac{q^2 (2 \epsilon +1)}{2 \left(\gamma ^2-1\right) \epsilon }\mathbfcal{M}_{0,0;0,0,1,2,1}\Bigg)\,.
     \end{split}
 \end{align}
 Using (\ref{MasterM}) we get, 
 \begin{align}
     \begin{split}  \label{eqt20} 
     \chi\Big|_{\mathcal{S}^{0}}=-\frac{m_1^2m_2^2}{m_p^6}\Delta_{\mathfrak{27}}^{\textrm{poly}},\,\,\,\Delta_{\mathfrak{27}}^{\textrm{poly}}=\Bigg(\frac{s_{1}^2 s_{2}^2 \left(x^2+1\right) \left(x^4+1\right)}{2048 \pi ^3 |\boldsymbol{b}|^2 \left(x^2-1\right)^3}\Bigg).&
      \end{split}
 \end{align}
The contribution to the spinning part is given by,
 \begin{align}
     \begin{split}
         i\chi\Big|_{\mathcal{S}^{1}}= -\frac{\gamma s_1^2 s_2^2 m_1^2 m_2^2}{64 m_p^6}\int e^{-i q\cdot b}\hat\delta(q\cdot v_1)\hat\delta(q\cdot v_2)\,&\Bigg(\boldsymbol{\tilde h}_{118}\mathbfcal{M}_{0,0;0,0,1,1,1}+q^2 \boldsymbol{\tilde h}_{119}\mathbfcal{M}_{0,0;0,0,2,1,1} +q^2 \boldsymbol{\tilde h}_{120}\mathbfcal{M}_{0,0;0,0,1,2,1}\Bigg)\\&\Bigg(q\cdot\mathcal{S}_1\cdot v_2- \frac{1-2\epsilon}{4\,\gamma\,(1-\epsilon)}q\cdot\mathcal{S}\cdot v_1\Bigg)\,.
     \end{split}
 \end{align}
 Then using \eqref{MasterM} we get,

 \hfsetfillcolor{gray!10}
\hfsetbordercolor{black!150}
\begin{align}
     \begin{split}  \label{eqt21} 
 \chi\Big|_{\mathcal{S}^{1}}= \frac{m_1^2 m_2^2}{m_p^6}\Bigg(\Delta_{\mathfrak{28}}^{\textrm{poly}}(\hat{b}\cdot\mathcal{S}_1\cdot v_2)+\Delta_{\mathfrak{29}}^{\textrm{poly}} (\hat{b}\cdot\mathcal{S}_2\cdot v_1)\Bigg),\,
     \end{split}
 \end{align}
 where,
 \begin{align}
     \Delta_{\mathfrak{28}}^{\textrm{poly}}=-\Bigg(\frac{s_1^2s_2^2\, x(x^2+1)^2}{1024 \pi ^3 |\boldsymbol{b}|^3 \left(x^2-1\right)^3}\Bigg), \quad \Delta_{\mathfrak{29}}^{\textrm{poly}}= -\frac{x}{2(x^2+1)}\Delta_{\mathfrak{28}}^{\textrm{poly}}.
 \end{align}
 Proceeding as before, for the second diagram, we have only  $\mathcal{O}(\mathcal{S}^0)\,.$  contribution.
 \begin{align}
     \begin{split}  \label{eqt22} 
    \chi \Big|_{\mathcal{S}^0}=\frac{m_1m_2^3}{m_p^6}\Delta_{\mathfrak{30}}^{\textrm{poly}},\quad \Delta_{\mathfrak{30}}^{\textrm{poly}}=-\Bigg(\frac{g_{1} s_{1} s_{2}^3 x^2}{256 \pi ^3 |\boldsymbol{b}|^2 \left(x^2-1\right)^2}\Bigg)\,.
    \end{split}
 \end{align}
 Again, one also needs to add the contributions from 3PM diagrams obtained by interchanging the worldlines one and two in all the diagrams mentioned above. We will not show them explicitly as the contributions from those diagrams will be similar to those we have already computed and can be obtained easily by exchanging labels one and two. 
\par
\textbf{3PM eikonal phase from Chern-Simons interaction vertices:}
Now let us focus on the dCS contribution to the eikonal vertex. The corresponding vertex term is given by,
 \begin{align}
    \begin{split}
        \mathcal{V}_{\textrm{dCS}}(k_1,k_2,q)
        =&\,\epsilon^{\chi\varepsilon\mu\nu}\Bigg(
        k_{1\mu}k_{1\beta}k_{2\chi}k_{2\sigma}h^{\sigma}_{\nu}(-k_1)h_{\delta}^{\beta}(-k_2)\\
        &\qquad\qquad
        -k_{1\mu}(k_1\cdot k_2)\,k_{2\chi}h^{\sigma}_{\nu}(-k_1)h_{\delta\sigma}(-k_2)
        \Bigg)\varphi(-q)\,\hat\delta^{(4)}(k_1+k_2+q)\,.
    \end{split}
\end{align}
With this vertex, we will now proceed with the computation of the contribution of this dCS term to the eikonal phase. Below, we list all possible diagrams for the same. 
\par
\textbullet $\,\,$ The following two diagrams will contribute to the eikonal phase at 3 PM involving Chern-Simons interaction  :
\begin{center}
\includegraphics[width=0.38\linewidth]{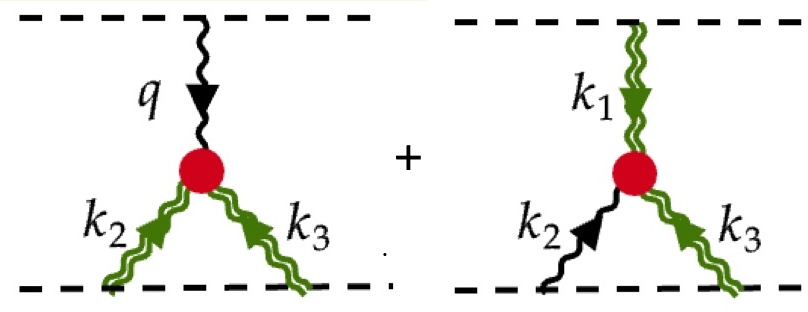}
\end{center}
Contribution to the eikonal phase from the first diagram is given by,
\begin{align}
    \begin{split}
      i  \chi=-l_{dCS}^2\frac{s_1 m_1 m_2^2}{8 m_p^6} \int e^{iq\cdot b}\frac{\hat\delta(q\cdot v_1)\hat\delta(q\cdot v_2)}{q^2}\int_{k_2}\hat\delta(k_2\cdot v_2)\frac{\mathcal{N}}{k_2^2 (k_2+q)^2}
    \end{split}
\end{align}
where the numerator takes the form,
\begin{align}
\begin{split}
    \mathcal{N}&=-\frac{i}{2}\epsilon^{\gamma\delta\tau\sigma}(q+k_2)\cdot k_2(q+k_2)_{\tau}k_{2\gamma}\,\Big(v_{2\sigma}(k_2\cdot \mathcal{S}_2)_{\delta}-[(q+k_2)\cdot \mathcal{S}_2 ]_{\sigma}\,\,v_{2\delta}\Big)\,,\\ &
  \to  \frac{i}{2}\epsilon^{\gamma\delta\tau\sigma}q^2\, q_\tau \,k_{2\gamma} v_{2\sigma}(k_2\cdot \mathcal{S}_2)_{\delta}\,\,(\textrm{ignoring the massless tadpoles})\,.
  \end{split}
\end{align}
Now, doing the $k_2$  and $q$ integral and using the anti-symmetric property of epsilon, it is easy to show that the contribution to the eikonal phase from this diagram is given by,

\begin{align}
\begin{split} \label{eqt23} 
 \chi\Big|_{\mathcal{S}^1}&=l_{dCS}^2\frac{s_1 m_1 m_2^2}{8 m_p^6}\frac{ 3 x\,}{64\pi (1-x^2)\,|\boldsymbol{b}|^4}\,\epsilon^{\gamma\delta\tau\sigma}v_{2\sigma}\mathcal{S}_{2\gamma\delta}\,\hat{b}_\tau\,,\\ &=\Delta_{\mathfrak{1}}^{\textrm{dCS}}\left(\mathcal{S}_2\wedge \hat b\right).
\end{split}
\end{align}
where,
\begin{align}
\Delta_{\mathfrak{1}}^{\textrm{dCS}}=l_{dCS}^2\frac{s_1 }{8 }\frac{ 3 x\,}{64\pi (1-x^2)\,|\boldsymbol{b}|^4}
\end{align}
Correspondingly, the second diagram gives,
\begin{align}
    \begin{split}
      i  \chi=-i\,l_{dCS} ^2\frac{s_2m_1 m_2^2}{m_p^6}\int_{k_1}e^{ik_1\cdot b}\frac{\hat\delta(k_1\cdot v_1)\hat\delta(k_1\cdot v_2)}{k_1^2}\int_{k_2}\hat\delta(k_2\cdot v_2)\frac{\mathcal{N}}{k_2^2(k_1+k_2)^2}
    \end{split}
\end{align}
where the numerator is given by,
\begin{align}
    \begin{split}
        \mathcal{N}=&\epsilon^{\gamma\delta\tau\sigma}\Bigg(v_1^\mu v_1^\nu+i(k_1\cdot \mathcal{S}_1^{(\mu})\,v_1^{\nu)}-\frac{1}{2}(k_1\cdot \mathcal{S}_1)^\mu (k_1\cdot \mathcal{S}_1)^{\nu}\Bigg)\,P_{\mu\nu,\rho\sigma}\Bigg(k_{1\tau}k_{2\gamma}k_1^{\chi}k_2^\rho-\eta^{\rho\chi}(k_1\cdot k_2)\Bigg)\\ &
        \times P_{\chi\varepsilon,\alpha\beta}\Bigg(v_2^\alpha v_2^\beta+i(k_2\cdot \mathcal{S}_2^{(\alpha})\,v_2^{\beta)}-\frac{1}{2}(k_2\cdot \mathcal{S}_2)^\alpha (k_2\cdot \mathcal{S}_2)^{\beta}\Bigg)\,.
    \end{split}
\end{align}
The numerator can be decomposed in different orders of spin and are given by,
\begin{align}
    \begin{split}
&\mathcal{N}\Big|_{\mathcal{S}^0}=\gamma  \left(  k_1 \cdot k_2\right)  \epsilon_{\alpha \beta\rho\sigma} k_1^\alpha k_2^\beta v_1^\rho v_2^\sigma\,, \\ &
\mathcal{N}\Big|_{\mathcal{S}^1}=\frac{1}{4}i\Big(2\gamma(k_1\cdot k_2)(k_1\cdot \mathcal{S}_1)^\mu\epsilon_{\mu \beta\rho\sigma}k_1^\beta k_2^\rho v_2^\sigma+2(k_1\cdot k_2)\epsilon_{\alpha \beta\rho\sigma} k_1^\alpha k_2^\beta v_1^\rho v_2^\sigma (k_1\cdot \mathcal{S}_1\cdot v_2)\,\\&
\hspace{1.2 cm}-2\gamma(k_1\cdot k_2) (k_2\cdot{\mathcal{S}_2})^\alpha\epsilon_{\alpha\beta\rho\sigma}{k_1}^\beta {k_2}^\rho {v_1}^\sigma+2 (k_1\cdot \mathcal{S}_2\cdot k_2)(k_2\cdot v_1)\epsilon_{\alpha\beta\rho\sigma} k_1^\alpha k_2^\beta v_1^\rho v_2^\sigma\\&\hspace{1.0 cm}+2(k_1\cdot k_2)(k_2\cdot \mathcal{S}_2\cdot v_1) \epsilon_{\alpha \beta\rho\sigma} k_1^\alpha k_2^\beta v_1^\rho v_2^\sigma \Big)\,.
\end{split}
\end{align}
Therefore, the contribution to the eikonal phase takes the form, 

\begin{align}
\begin{split}
i\chi\Big|_{\mathcal{S}^0}&
    \to\frac{\gamma}{2}\epsilon_{i j\rho\sigma}\delta_{mn}\,v_1^\rho v_2^\sigma\int_{k_1}e^{ik_1\cdot b}{\hat\delta(k_1\cdot v_1)\hat\delta(k_1\cdot v_2)k_1^i\,k_1^{m}}\underbrace{\int_{\vec k_2}\frac{k_2^j \,k_2^n}{\vec k_2^2(\vec k_2+\vec k_1)^2}}_{\propto ()\delta^{jn}+()k_1^j k_1^n}=0\,.\label{3.79}
    \end{split}
\end{align}
%
Now, let us concentrate on the spin-orbit contribution to the eikonal phase coming from the Chern-Simons interaction vertices. 
\begin{align}
    \begin{split}
        \chi\Big|_{\mathcal{S}^1}=\sum_{i=1}^{3}\chi_{\mathcal{S}^1}^{(i)}
    \end{split}
\end{align}
where,
\begin{align}
    \begin{split}
  \\
      \bullet \,\,\, \chi_{\mathcal{S}^1}^{(1)}&\sim -\frac{i}{4}\epsilon_{\alpha\beta\rho\sigma}(\mathcal{S}_2)^{\eta\tau}\,v_{1\tau}v_1^\rho v_2^\sigma \int_{k_1}e^{ik_1\cdot b}\hat\delta(k_1\cdot v_1)\hat\delta(k_1\cdot v_2)k_1^\alpha \int_{k_2}\frac{\hat\delta(k_2\cdot v_2)k_2^\beta k_{2\eta}}{k_2^2(k_2+k_1)^2}\,,\\ &
      \xrightarrow[]{\epsilon\to 0}\frac{3}{128\pi |\boldsymbol{b}|^4} \Bigg(\frac{2x}{1-x^2}\Bigg)(\overrightarrow{\mathcal{S}_2\cdot v_1})\cdot (\vec v_1\times \hat b)\,.
      \\ &
  \\
 \bullet \,\,\, \chi_{\mathcal{S}^1}^{(2)}&\sim \frac{i}{2}\epsilon_{\alpha\beta \rho\sigma}v_1^\rho v_2^\sigma v_{1\chi}(\mathcal{S}_2)_{\eta\delta}\int e^{i k_1\cdot b}\frac{\hat\delta(k_1\cdot v_1)\hat\delta(k_1\cdot v_2)\,k_1^\eta k_1^\alpha}{k_1^2}\int \frac{\hat\delta(k_2\cdot v_2)k_2^\delta\, k_2^\beta k_2^\chi}{k_2^2\,(k_2+k_1)^2}\,,\\ &
 =-\frac{3}{128\pi |\boldsymbol{b}|^4}\Bigg(\frac{2x}{1-x^2}\Bigg)\Bigg(\frac{1}{2}\vec v_1\cdot \hat b\times (\overrightarrow {v_1\cdot \mathcal{S}_2})\Bigg)=\frac{3}{256\pi |\boldsymbol{b}|^4}(\overrightarrow{\mathcal{S}_2\cdot v_1})\cdot (\vec v_1\times \hat b  )\,.\nonumber
\end{split}
\end{align}
\begin{align}
\begin{split}
\hspace{0.7 cm} \bullet \,\,\, \chi_{\mathcal{S}^1}^{(3)}&\sim -\frac{i \gamma}{2}\epsilon_{\alpha\beta \rho\sigma}v_1^\sigma \mathcal{S}_2^{\eta\alpha}\int e^{ik_1\cdot b}\frac{\hat\delta(k_1\cdot v_1)\hat\delta(k_1\cdot v_2)k_1^\beta}{k_1^2}\int \frac{\hat\delta(k_2\cdot v_2)\,k_1\cdot k_2}{k_2^2(k_2+k_1)^2}k_2^\eta k_{2\rho}\,,\\ &
 \\ &=-\frac{3}{512 \pi|\boldsymbol{b}|^4\sqrt{\gamma^2-1}}\Big(\hat b\cdot (\vec{\mathcal{S}}_2\times \vec v_1)+\gamma \hat{\mathcal{S}}_2\wedge \hat b\Big),\,\,\textrm{with},\,\,\hat{\mathcal{S}}_2\wedge \hat b=\hat b^j\,\epsilon_{jik}\,\mathcal{S}_2^{ik}
      \label{3.81}
    \end{split}
\end{align}
  Therefore, the total contribution coming from the dCS term at $\mathcal{O}(\mathcal{S}^1)$ after inserting all the vertex factors is given by,
\begin{align}
    \begin{split}
 \hspace{0cm}i\chi\Big|_{\mathcal{S}^1}&=-il_{dCS} ^2\frac{s_2m_1 m_2^2}{m_p^6}\Bigg(\frac{2x}{1-x^2}\Bigg)\Bigg[\frac{9}{256 \pi |\boldsymbol{b}|^4} (\overrightarrow{\mathcal{S}_2\cdot v_1})\cdot (\vec v_1\times \hat b)-\frac{3}{512\,\pi |\boldsymbol{b}|^4}\Big(\hat b\cdot (\vec{\mathcal{S}}_2\times \vec v_1)+\frac{1+x^2}{2x} \hat{\mathcal{S}}_2\wedge \hat b\Big)\Bigg]\\ &
 =
 \frac{m_1 m_2^2}{m_p^6}\left[\Delta_{\mathfrak{2}}^{\textrm{dCS}}(\overrightarrow{\mathcal{S}_2\cdot v_1})\cdot (\vec v_1\times \hat b)+\Delta_{\mathfrak{3}}^{\textrm{dCS}}\left(\hat b\cdot \left(\vec{\mathcal{S}}_2\times \vec v_1\right)\right)+\Delta_{\mathfrak{4}}^{\textrm{dCS}}\left(\hat{\mathcal{S}}_2\wedge \hat b\right)\right]
    \end{split} \label{eqt24} 
\end{align}
where, $(\overrightarrow{\mathcal{S}_2\cdot v_1})\equiv (\mathcal{S}_2\cdot v_1)^{i},\,\,\vec {\mathcal{S}}_{1,2}\equiv \mathcal{S}_{1,2}^{0i},\,\textrm{and,} \,\,\mathcal{S}_2\wedge \hat b=\epsilon^{\tau\gamma\delta\sigma}v_{2\sigma}\mathcal{S}_{\gamma\delta}\hat b_{\tau}$ 
and,
\begin{align}
    \begin{split}
\Delta_{\mathfrak{2}}^{\textrm{dCS}}=-l_{\textrm{dCS}}^2 \frac{18 xs_2}{256\pi |\boldsymbol{b}|^4(1-x^2)},\,\Delta_{\mathfrak{3}}^{\textrm{dCS}}=l_{\textrm{dCS}}^2 \frac{6 s_2 x}{512 \pi |\boldsymbol{b}|^4(1-x^2)},\,\textrm{and,}\,\Delta_{\mathfrak{4}}^{\textrm{dCS}}=l_{\textrm{dCS}}^2\frac{3s_2(1+x^2)}{512 \pi|\boldsymbol{b}|^4 (1-x^2)}.
    \end{split}
\end{align}
Let us make some comments on the computation of \eqref{3.81}:
\begin{itemize}
\item Due to the presence of the antisymmetric $\epsilon$ tensor, there is no contribution from the dCS term to the eikonal phase at $\mathcal{O}(\mathcal{S}^0)\,.$ It contributes only in the presence of spin.
    \item We ignore the contribution of the two-loop integral proportional to $k_1^i k_1^j$ because of the antisymmetric properties of the epsilon tensor.
    \item Also it is evident from \eqref{eqt24}, that if we take (anti-)aligned spin, the contribution will again give zero due to the presence of binormal like term. In the case of aligned spins, we define
    \begin{align}
\mathcal{S}_1^{\mu\nu}=\frac{2\,\tilde{s}_1}{|\boldsymbol{b}|\sqrt{\gamma^2-1} }b^{[\mu}(\gamma v_1-v_2)^{\nu]} \,,\hspace{0.4 cm }\mathcal{S}_2^{\mu\nu}=\frac{2\,\tilde{s}_2}{|\boldsymbol{b}|\sqrt{\gamma^2-1} }b^{[\mu}( v_1-\gamma v_2)^{\nu]}, \,\,\, \textrm{with},\,\, {\textrm{tr\,}\,(\mathcal{S}_i\cdot\mathcal{S}_i)=-2 \,\tilde{s}_i^2}.
    \end{align}
      \end{itemize}
      \vspace{-0.89 cm}
   After using this parametrization, one finds that the \textit{3PM contribution from the Chern-Simons-type term (arising at the bulk vertices) vanishes.} Therefore, if the two black holes being scattered have aligned spins in dCS theory, there is no contribution to the eikonal phase and hence no correction to the scattering angle.
Now, adding all spinless contributions at 3PM, we have,

\begin{tcolorbox}[thesisresultbox, title=\textit{Spinless Eikonal at 3PM}]
\restorethesisbodyformat
\begin{align}  \label{dCSne3waaa}
 \chi\Big|_{\mathcal{S}^0}=&\frac{m_1m_2^3}{m_p^6}\left[\Delta_{\mathfrak{01}}^{\textrm{poly}}+\Delta_{\mathfrak{02}}^{\textrm{poly}}+\Delta_{\mathfrak{05}}^{\textrm{poly}}\right]
+\\&\frac{m_1^2 m_2^2}{m_p^6}\Big[\Delta_{\mathfrak{08}}^{\textrm{poly}}+\Delta_{\mathfrak{09}}^{\textrm{poly}}+\Delta_{\mathfrak{12}}^{\textrm{poly}}+\Delta_{\mathfrak{15}}^{\textrm{poly}}+\Delta_{\mathfrak{17}}^{\textrm{poly}}+\Delta_{\mathfrak{17}}^{\textrm{poly}}+\Delta_{\mathfrak{18}}^{\textrm{poly}}+\Delta_{\mathfrak{22}}^{\textrm{poly}}+\Delta_{\mathfrak{27}}^{\textrm{poly}}+\Delta_{\mathfrak{30}}^{\textrm{poly}}\nonumber+\\ &\hspace{1.2 cm}\log(x)\Big(\Delta_{\mathfrak{01}}^{\textrm{log}}+\Delta_{\mathfrak{02}}^{\textrm{log}}+\Delta_{\mathfrak{05}}^{\textrm{log}}+\Delta_{\mathfrak{09}}^{\textrm{log}}+\Delta_{\mathfrak{12}}^{\textrm{log}}+\Delta_{\mathfrak{15}}^{\textrm{log}}\Big)+(\log^2(x)+\textbf{Li}_2(1-x^2))\left(\Delta_{\mathfrak{03}}^{\textrm{Polylog}}\right)\Big]\,.\nonumber
    \end{align}
\end{tcolorbox}
\restorethesisbodyformat
Finally, adding all the spinning contributions to the eikonal phase at 3PM we will get,
\begin{tcolorbox}[thesisresultbox, title=\textit{Spinning Eikonal at 3PM}]
\restorethesisbodyformat
\begin{align}\label{dC7Snew}
 \chi\Big|_{\mathcal{S}^1}=&\frac{m_1 m_2^3}{m_p^6}\Big[(\hat b\cdot\mathcal{S}_1\cdot v_2)\Big\{ \Delta_{\mathfrak{03}}^{\textrm{poly}}+\Delta_{\mathfrak{06}}^{\textrm{poly}} \Big\}+(\hat b\cdot\mathcal{S}_2\cdot v_1)\Big\{ \Delta_{\mathfrak{04}}^{\textrm{poly}}+\Delta_{\mathfrak{07}}^{\textrm{poly}} \Big\}+\Delta_{\mathfrak{25}}^{\textrm{poly}}\Big]+ \\ &
\frac{m_1^2 m_2^2}{m_p^6}\Big[(\hat b\cdot \mathcal{S}_1\cdot v_2)\Big\{\Delta_{\mathfrak{10}}^{\textrm{poly}}+\Delta_{\mathfrak{13}}^{\textrm{poly}}+\Delta_{\mathfrak{16}}^{\textrm{poly}}+\Delta_{\mathfrak{19}}^{\textrm{poly}}+\Delta_{\mathfrak{23}}^{\textrm{poly}}+\Delta_{\mathfrak{26}}^{\textrm{poly}}+\Delta_{\mathfrak{28}}^{\textrm{poly}}\nonumber+\\ &\hspace{0.8cm}\log(x)\Big(\Delta_{\mathfrak{3}}^{\textrm{log}}+\Delta_{\mathfrak{06}}^{\textrm{log}}+\Delta_{\mathfrak{08}}^{\textrm{log}}+\Delta_{\mathfrak{10}}^{\textrm{log}}+\Delta_{\mathfrak{13}}^{\textrm{log}}+\Delta_{\mathfrak{16}}^{\textrm{log}}+\Delta_{\mathfrak{18}}^{\textrm{log}}\Big)\nonumber+\\ &\hspace{0.8cm}
(\log^2 x+\textbf{Li}_2(1-x^2))\Big(\Delta_{\mathfrak{1}}^{\textrm{Polylog}}+\Delta_{\mathfrak{4}}^{\textrm{Polylog}}+\Delta_{\mathfrak{6}}^{\textrm{Polylog}}\Big)+\boldsymbol{\mathfrak{U}_1}^{\textrm{LI}_3}\,\boldsymbol{\mathfrak{(UT)}}_3\Big\}\nonumber+\\ & \hspace{1.2cm}(\hat b\cdot \mathcal{S}_2\cdot v_1)\Big\{\Delta_{\mathfrak{11}}^{\textrm{poly}}+\Delta_{\mathfrak{14}}^{\textrm{poly}}+\Delta_{\mathfrak{20}}^{\textrm{poly}}+\Delta_{\mathfrak{21}}^{\textrm{poly}}+\Delta_{\mathfrak{24}}^{\textrm{poly}}+\Delta_{\mathfrak{29}}^{\textrm{poly}}\nonumber+\\&\hspace{0.8cm}\log(x)\Big(\Delta_{\mathfrak{4}}^{\textrm{log}}+\Delta_{\mathfrak{7}}^{\textrm{log}}+\Delta_{\mathfrak{11}}^{\textrm{log}}+\Delta_{\mathfrak{14}}^{\textrm{log}}+\Delta_{\mathfrak{17}}^{\textrm{log}}\Big)\nonumber+\\ &\hspace{0.8cm}(\log^2 x+\textbf{Li}_2(1-x^2))\Big(\Delta_{\mathfrak{2}}^{\textrm{Polylog}}+\Delta_{\mathfrak{5}}^{\textrm{Polylog}}+\Delta_{\mathfrak{7}}^{\textrm{polylog}}\Big)+\boldsymbol{\mathfrak{U}_2}^{\textrm{LI}_3}\,\boldsymbol{\mathfrak{(UT)}}_3\Big\}\Big]\nonumber+\\ &
 \frac{m_1 m_2^2 }{m_p^6}\left[\Delta_{\mathfrak{1}}^{\textrm{dCS}}(\mathcal{S}_2\wedge \hat b )+\Delta_{\mathfrak{2}}^{\textrm{dCS}}(\overrightarrow{\mathcal{S}_2\cdot v_1})\cdot (\vec v_1\times \hat b)+\Delta_{\mathfrak{3}}^{\textrm{dCS}}\left(\hat b\cdot \left(\vec{\mathcal{S}}_2\times \vec v_1\right)\right)+\Delta_{\mathfrak{4}}^{\textrm{dCS}}\left(\hat{\mathcal{S}}_2\wedge \hat b\right)\right]
\,\nonumber
    \end{align} 
\end{tcolorbox}
\restorethesisbodyformat

\noindent

where, $\boldsymbol{\mathfrak{(UT)}}_3$ a polynomial consisting of a particular combination of functions with transcendental weight three (UT-3) and defined as,
\begin{align}
    \begin{split}
\boldsymbol{\mathfrak{(UT)}}_3&=54\,\textbf{Li}_3\left(x^2\right)-216\, \textbf{Li}_3\left(\frac{x+1}{1-x}\right)
\\ &+432\, \textbf{Li}_3(1-x)+216\, \textbf{Li}_3\left(\frac{x+1}{x-1}\right)+432\, \textbf{Li}_3(x+1)+216 \Big(-\textbf{Li}_2\left(\frac{x+1}{1-x}\right)-\textbf{Li}_2(1-x)\\ &+\textbf{Li}_2\left(\frac{x+1}{x-1}\right)+\textbf{Li}_2(x+1)\Big) \log \left(\frac{2}{x+1}-1\right)-54 \left(\pi ^2-2 \log ^2(x)\right) \log \left(1-x^2\right)-20 \log ^3(x)\\ &+108 i \pi  \log ^2(1-x)+108 i \pi  \log ^2(x+1)-11 \pi ^2 \log (x)-432 \zeta (3)\label{4.133a}\,,
    \end{split}
\end{align}
and, also note that one needs to add the terms ($1\leftrightarrow 2$) with \eqref{dCSne3waaa} and \eqref{dC7Snew} to get the complete answer.
On the first note,  it may seem that the polynomial $\boldsymbol{\mathfrak{(UT)}_3}$ in \eqref{4.133a} is complex, in general. However, it can be easily checked (numerically) that it is actually real in the domain $x\in(0,1)$. \textit{To the best of our knowledge, this is the first example for which, unlike GR, we get UT-3 polynomials in the Eikonal phase at 3PM.} 
Again to remind the readers, $\gamma=\frac{1+x^2}{2\,x}\,.$ As we notice that in the expression of eikonal phase, we encounter $\log(b)$ type terms, and one should put an IR cutoff, $\mu_{IR}$, as $\log(\mu_{IR}b)$ to make the argument dimensionless. However, for computing observables like impulses, one needs to take a derivative with respect to $|b|$, and hence, it is independent of the IR cutoff. Also note that, the $\Delta's$ mentioned above are defined in \cref{eqt1,eqt2,eqt3,eqt4,eqt5,eqt6,eqt7,eqt7b,eqt8,eqt9,eqt10,eqt11,eqt12,eqt13,eqt14,eqt15,eqt16,eqt17,eqt18,eqt19,eqt20,eqt21,eqt22,eqt23}.
  
\section{Conclusion and outlook}\label{con}
In \cite{Jakobsen:2021zvh}, the authors introduced spin degrees of freedom in WQFT by implementing  $\mathcal{N}=2$ supersymmetry as an alternative approach to describe a compact spinning object, considering terms up to quadratic-in-spin. This work motivates us to study the scattering event of two spinning black holes up to $3$PM in a gravity theory modified by a dynamical Chern-Simons (dCS) term using the WQFT approach. 
The introduction of a parity-violating Chern-Simons term in Einstein gravity, which couples to gravity via a scalar field, opens up new avenues in the study of astrophysical events, especially gravitational waves. While the Chern-Simons term does not affect the non-rotating black hole solutions, it induces a correction to the rotating metric. Therefore, it becomes important for the study of spinning black holes. 

\textbullet$\,\,$ We start with the Einstein-Hilbert action modified by the dCS term and formulate the worldline theory for a spinning particle by introducing the anti-commuting worldline vectors $\psi^{a}$. Upon modification, $\mathcal{N}=1$  supersymmetry is broken (approximate), but without loss of generality, we define the classical spin of the body $(\mathcal{S}^{ab}=-2i\psi^{[a}\,\psi^{b]})$. Although, as we have mentioned, the SSC is also approximate due to the broken (approximate) supersymmetry, it holds for small coupling (sensitivity parameters) and leading order in a spin throughout the trajectory, but the asymptotic SUSY is still there. It provides the framework to study the scattering of two spinning black holes using the WQFT formalism. 

\textbullet $\,\,$ Our computation deals with two-loop Feynman integrals. We discuss the mathematical framework \cite{Dlapa:2023hsl,Henn:2014qga} used to solve multi-loop integrals via IBP reduction and differential equations, and then analyze the two-loop integrals required for our computation. \textit{We illustrate these techniques by solving \textit{four} master integrals to higher orders in $\epsilon$}.

\textbullet$\,\,$Next, we explicitly determined the eikonal phase for the spinning black-hole system up to 3PM order. We have shown the correction to the spinning eikonal due to the presence of a scalar field up to 3PM. \textit{To the best of our knowledge, this is the first result for the spinning eikonal in the presence of a massless scalar field at 3PM order. We also find from the eikonal phase that the Chern-Simons interaction does not contribute to non-spinning black holes, nor does it contribute when one considers (anti-)aligned spins.}

\textbf{\textit{$i\varepsilon$ prescription and the IR problem}:}

In the eikonal phase computation, one is expected to use the Feynman $i\varepsilon$ prescription to get sensible results. However,  a naive choice of Feynman $i\varepsilon$ prescription could result in unwanted IR divergences ($\propto \frac{1}{\epsilon}$) in the eikonal phase, which we have also encountered in our computations, specifically for those diagrams with the worldline propagator. More precisely, let's say, after IBP reduction, we face the following master integral (in some diagram): $\mathbfcal{M}_{1,1;0,0,1,1,1}$. Now, using the Feynman prescription amounts to the following substitution:
\begin{align}
    \mathbfcal{M}_{1,1;0,0,1,1,1}\to \frac{1}{2}\left( \mathbfcal{M}_{1,1;0,0,1,1,1}^{(++)}-\mathbfcal{M}_{1,1;0,0,1,1,1}^{(+-)}\right)\,. \label{6.1k}
\end{align}
Then, for some (or most) of the cases, this gives rise to IR divergences. As a possible resolution, the authors of \cite{Kim:2024svw} have proposed the notion of Magnus expansion which fixes the choice of the $i\varepsilon$ prescription for cancelling IR divergences. In general, one needs to consider the following prescription for precise IR cancellation,
\begin{align}
    \mathbfcal{M}_{1,1;0,0,1,1,1}\to \left( w_{(++)}\mathbfcal{M}_{1,1;0,0,1,1,1}^{(++)}+w_{(+-)}\mathbfcal{M}_{1,1;0,0,1,1,1}^{(+-)}\right)\,, \label{6.2k}
\end{align}
where the weight factors ($w$ 's) are determined by the Magnus expansion. However, these weight factors may depend on the underlying input, or it might happen that the prescription also may not cancel the IR divergences (i.e., the theory itself has the IR problem). So, it needs to be properly investigated, especially for the theory considered in this paper. For the moment, we have only considered the finite part of the regulated (dimensionally regularized) integral, giving rise to the eikonal phase and leaving aside a more systematic approach to cancel IR divergences for future studies.

\textbf{\textit{Comments on computation of observables from eikonal:}}

In this paper, we mainly computed the eikonal phase using WQFT. Now, the question is, how does one extract the observables, such as impulse (scattering angle), from the eikonal? A possible answer to this question has been discussed recently in \cite{Kim:2024grz,Kim:2024svw}. The eikonal phase can be understood as a generator of the canonical transformations that can transform the states from in-state to out-state in some scattering event. Any instantaneous change of observable in the classical limit can be written as,
    \begin{align}
        \begin{split}
            \Delta\mathcal{O}=\lim_{\hbar\to0}\left[\langle\psi|S^\dagger\mathcal{O}S|\psi\rangle-\langle\psi|O|\psi\rangle\right]\,.
        \end{split}
    \end{align}
 We know that in the classical limit, the S-matrix enjoys eikonal exponentiation, $S=e^{i\chi/\hbar}$. Therefore, 
 \begin{align}
     \begin{split}
         \mathcal{O}_{out}&=e^{i\chi/\hbar}\mathcal{O}_{in}e^{i\chi/\hbar}\\ &=\mathcal{O}_{in}+\frac{1}{i\hbar}[\chi,\mathcal{O}_{in}]+\frac{1}{2(i\hbar)^2}[\chi,[\chi,\mathcal{O}_{in}]]+\cdots\\ &
         \xrightarrow[]{\hbar\to 0}\mathcal{O}_{in}+\{\chi,\mathcal{O}_{in}\}_{P.B.}+\frac{1}{2}\{\chi,\{\chi,\mathcal{O}_{in}\}\}_{P.B.}+\cdots
     \end{split}
 \end{align}
 Now, in WQFT formalism, the eikonal phase can be identified as the free energy of the theory, $\chi=-i\log\,Z_{WQFT}$. Therefore, the observable, e.g impulse, can be computed as,
 \begin{align}
     \Delta p_i^\mu=\{\chi,p_i^\mu\}_{P.B.}+\frac{1}{2}\{\chi,\{\chi,p_i^\mu\}\}_{P.B.}+\cdots
 \end{align}
 In Post-Minkowskian formalism, one can expand the eikonal perturbatively in $G_N$,
 \begin{align}
     \begin{split}
         \chi=\chi_{1PM}+\chi_{2PM}+\chi_{3PM}+\cdots
     \end{split}
 \end{align}
 Therefore, the impulse can be computed as,
 \begin{align}
     \begin{split}
         &\Delta_{(1PM)}\,p_i^\mu=\{\chi_{(1PM)},p_i^\mu\},\\ &
         \Delta_{(2PM)}p_i^\mu=\{\chi_{(2PM)},p_i^\mu\}+\frac{1}{2}\{\chi_{(1PM)},\{\chi_{(1PM)},p_i^\mu\},\\ &
         \Delta_{(3PM)}p_i^\mu=\{\chi_{(3PM)},p_i^\mu\}+\frac{1}{2}\{\chi_{(2PM)},\{\chi_{(1PM)},p_i^\mu\}+\frac{1}{2}\{\chi_{(1PM)},\{\chi_{(2PM)},p_i^\mu\}+\frac{1}{3!}\{\chi_{1PM},\{\chi_{1PM},\{\chi_{1PM},p_i^\mu\}\}\},\\ &
         \vdots\label{5.5a}
     \end{split}
 \end{align}
 To compute the brackets in \eqref{5.5a}, one needs to use the fundamental Poisson bracket relations mentioned in \cite{Kim:2024grz}. We hope to use this methodology to compute the impulse for our case and report it in the near future.

Our work shows the path of several follow-up opportunities. The immediate one is to find the Eikonal phase in $\mathcal{N}=2$ supersymmetric worldline theory, which deals with the interaction of spin-1 particles or quadratic order in spin, including the scalar dipole term in the action. One can also apply the spinning WQFT formalism to find out observables for the dCS theory for higher PM order. Our work can also be naturally extended to find out the observables, e.g., Impulse $(\Delta P^{\mu}),$ the spin kick $(\Delta S^{\mu\nu})$ from the Eikonal Phase and the waveform up to 3PM. More importantly, one needs to be careful about the $i\varepsilon$ prescription and find a way to handle the IR divergences along the way of \cite{Kim:2024svw}.  We hope to report on this soon.

\appendix 
\section{Useful Feynman integrals} \label{ch3:app:A}
In the main text, we extensively use the following integrals. 
 We take the integrals from \cite{Levi:2011eq}.
\begingroup
\small
\begin{eqnarray}
\label{ch3:tensorfourierindentity}
\hspace{0cm}I&\equiv& \int \frac{d^d \boldsymbol{k}}{(2\pi)^d}\frac{e^{i\bf{k}\cdot\bf{r}}}{({\bf{k}}^2)^\alpha}=\frac{1}{(4\pi)^{d/2}}\frac{\Gamma(d/2-\alpha)}{\Gamma(\alpha)}\Bigg(\frac{\bf{r}^2}{4}\Bigg)^{\alpha-d/2}, \label{int1}\\ 
I^i\equiv\rmint\frac{d^d\bf{k}}{(2\pi)^d}\frac{k^ie^{i\bf{k}\cdot\bf{r}}}{({\bf{k}}^2)^\alpha}&=&\frac{i}{(4\pi)^{d/2}}\frac{\Gamma(d/2-\alpha+1)}{\Gamma(\alpha)}\left(\frac{{\bf{r}}^2}{4}\right)^{\alpha-d/2-1/2}n^i, \label{int}\\
I^{ij}\equiv\rmint\frac{d^d\bf{k}}{(2\pi)^d}\frac{k^ik^je^{i\bf{k}\cdot\bf{r}}}{({\bf{k}}^2)^\alpha}&=&\frac{1}{(4\pi)^{d/2}}\frac{\Gamma(d/2-\alpha+1)}{\Gamma(\alpha)}\left(\frac{{\bf{r}}^2}{4}\right)^{\alpha-d/2-1}\left(\frac{1}{2}\delta^{ij}+(\alpha-1-d/2)n^in^j\right),\\
I^{ijl}\equiv\rmint\frac{d^d\bf{k}}{(2\pi)^d}\frac{k^ik^jk^le^{i\bf{k}\cdot\bf{r}}}{({\bf{k}}^2)^\alpha}&=&\frac{i}{(4\pi)^{d/2}}\frac{\Gamma(d/2-\alpha+2)}{\Gamma(\alpha)}\left(\frac{{\bf{r}}^2}{4}\right)^{\alpha-d/2-3/2}\nonumber\\
&&\quad
\times\left(\frac{1}{2}\left(\delta^{ij}n^l+\delta^{il}n^j+\delta^{jl}n^i\right)+(\alpha-d/2-2)n^in^jn^l\right),\\
I^{ijlm}\textstyle{\equiv\rmint\frac{d^d\bf{k}}{(2\pi)^d}\frac{k^ik^jk^lk^me^{i\bf{k}\cdot\bf{r}}}{({\bf{k}}^2)^\alpha}}&=&\textstyle{\frac{1}{(4\pi)^{d/2}}\frac{\Gamma(d/2-\alpha+2)}{\Gamma(\alpha)}\left(\frac{{\bf{r}}^2}{4}\right)^{\alpha-d/2-2}\left(\frac{1}{4}\left(\delta^{ij}\delta^{lm}+\delta^{il}\delta^{jm}+\delta^{im}\delta^{jl}\right)\right.} \\
&&
\textstyle{+\frac{\alpha-d/2-2}{2}\left(\delta^{ij}n^ln^m+\delta^{il}n^jn^m+\delta^{im}n^jn^l+\delta^{jl}n^in^m+\delta^{jm}n^in^l+\delta^{lm}n^in^j\right) }\nonumber\\
&&\left.\quad
+(\alpha-d/2-2)(\alpha-d/2-3)n^in^jn^ln^m\right).\nonumber
\end{eqnarray}
\endgroup
We use the d-dimensional master formula for one-loop scalar integrals given by ,
\begin{align}
\begin{split}
   & J\equiv\rmint \frac{d^d\bf{k}}{(2\pi)^d}\frac{1}{\left[{\bf{k}}^2\right]^\alpha\left[({\bf{k}-\bf{q}})^2\right]^\beta}=  \frac{1}{(4\pi)^{d/2}}\frac{\Gamma(\alpha+\beta-d/2)}{\Gamma(\alpha)\Gamma(\beta)}\frac{\Gamma(d/2-\alpha)\Gamma(d/2-\beta)}{\Gamma(d-\alpha-\beta)}\left(q^2\right)^{d/2-\alpha-\beta}.\label{ch3:eq:1loop}
    \end{split}
\end{align}
The d-dimensional master formula for one-loop tensor integrals is taken from \cite{Levi:2011eq}. 
Similarly, one can also derive the following d-dimensional formulas for the one-loop tensor integrals: 
\begin{align}
\begin{split}
J^i\equiv\rmint \frac{d^d\bf{k}}{(2\pi)^d}\frac{k^i}{\left[{\bf{k}}^2\right]^\alpha\left[({\bf{k}-\bf{q}})^2\right]^\beta}=\frac{1}{(4\pi)^{d/2}}\frac{\Gamma(\alpha+\beta-d/2)}{\Gamma(\alpha)\Gamma(\beta)}\frac{\Gamma(d/2-\alpha+1)\Gamma(d/2-\beta)}{\Gamma(d-\alpha-\beta+1)}\\ \left(q^2\right)^{d/2-\alpha-\beta}q^i\,,\end{split}
\end{align}
\begin{align}
\begin{split}
J^{ij}\equiv\rmint \frac{d^d\bf{k}}{(2\pi)^d}\frac{k^ik^j}{\left[{\bf{k}}^2\right]^\alpha\left[({\bf{k}-\bf{q}})^2\right]^\beta}=\frac{1}{(4\pi)^{d/2}}\frac{\Gamma(\alpha+\beta-d/2-1)}{\Gamma(\alpha)\Gamma(\beta)}\frac{\Gamma(d/2-\alpha+1)\Gamma(d/2-\beta)}{\Gamma(d-\alpha-\beta+2)}\left(q^2\right)^{d/2-\alpha-\beta}\\
\times\left(\frac{d/2-\beta}{2}q^2\delta^{ij}+(\alpha+\beta-d/2-1)(d/2-\alpha+1)q^iq^j\right),\end{split}
\end{align}
\begin{align}
\begin{split}
J^{ijl}\equiv\rmint \frac{d^d\bf{k}}{(2\pi)^d}\frac{k^ik^jk^l}{\left[{\bf{k}}^2\right]^\alpha\left[({\bf{k}-\bf{q}})^2\right]^\beta}=\frac{1}{(4\pi)^{d/2}}\frac{\Gamma(\alpha+\beta-d/2-1)}{\Gamma(\alpha)\Gamma(\beta)}\frac{\Gamma(d/2-\alpha+2)\Gamma(d/2-\beta)}{\Gamma(d-\alpha-\beta+3)}\left(q^2\right)^{d/2-\alpha-\beta} \\ 
\times\left(\frac{d/2-\beta}{2}q^2\left(\delta^{ij}q^l+\delta^{il}q^j+\delta^{jl}q^i\right)\right.  \left. 
+(\alpha+\beta-d/2-1)(d/2-\alpha+2)q^iq^jq^l\right),
\end{split}
\end{align}

\begin{align}
\begin{split}
&J^{ijlm}\equiv\rmint \frac{d^d\bf{k}}{(2\pi)^d}\frac{k^ik^jk^lk^m}{\left[{\bf{k}}^2\right]^\alpha\left[({\bf{k}-\bf{q}})^2\right]^\beta}=\frac{1}{(4\pi)^{d/2}}\frac{\Gamma(\alpha+\beta-d/2-2)}{\Gamma(\alpha)\Gamma(\beta)}\\& \frac{\Gamma(d/2-\alpha+2)\Gamma(d/2-\beta)}{\Gamma(d-\alpha-\beta+4)}\left(q^2\right)^{d/2-\alpha-\beta}\,\,\,\,
\times\left(\frac{(d/2-\beta)(d/2-\beta+1)}{4}q^4\left(\delta^{ij}\delta^{lm}+\delta^{il}\delta^{jm}+\delta^{jl}\delta^{im}\right)\right.\\
& \textstyle{
+(\alpha+\beta-d/2-2)(d/2-\alpha+2)\frac{d/2-\beta}{2}q^2 
\times\left(\delta^{ij}q^lq^m+\delta^{il}q^jq^m+\delta^{im}q^jq^l+\delta^{jl}q^iq^m+\delta^{jm}q^iq^l+\delta^{lm}q^iq^j\right)}\\
&\left.\,\,\,\,\,\,\,\,\,
+(\alpha+\beta-d/2-2)(\alpha+\beta-d/2-1)(d/2-\alpha+2)(d/2-\alpha+3)\right.
\left.
\times q^iq^jq^lq^m\right).
\end{split}
\end{align}
More details of  Feynman integrals can be found in \cite{smirnov}.
Another important two loop integral that we use to compute in the potential region is given by,
\begin{align}
    \begin{split}
        \mathbfcal{M}^{(++)}=\int_{\ell_{1},\ell_{2}}\frac{1}{(\ell_{1}\cdot n+i\varepsilon)(\ell_{2}\cdot n+i\varepsilon)}\frac{1}{\ell_{1}^2\, \ell_{2}^2\, (\ell_{1}+\ell_{2}-q)^2}\,.\label{intextra}
    \end{split}
\end{align}
Now, introducing an auxiliary delta functio,n we can write the above integral as \cite{Saotome:2012vy, Parra-Martinez:2020dzs},
\begin{align}
    \begin{split}
        \mathbfcal{M}^{(++)}=\frac{1}{3!}\int_{\ell_{1},\ell_{2},l_3}\Bigg(\frac{1}{(\ell_{1}\cdot n+i\varepsilon)(\ell_{2}\cdot n+i\varepsilon)}+\textrm{perms.}\Bigg)\frac{\delta^{(d)}(l_{123}-q)}{\ell_{1}^2\, \ell_{2}^2\, l_3^2}\,.
    \end{split}
\end{align}
Now using the following $\delta$-function identities,
\begin{align}
    \begin{split}
        &\delta(z_1+z_2+z_3)\Bigg(\frac{1}{z_1+i\varepsilon}\frac{1}{z_{12}+i\varepsilon}+\textrm{perms.}\Bigg)=(2\pi i)^2 \delta(z_1)\delta(z_2)\delta(z_3)\,,\\ &
\delta(z_1+z_2+z_3)\Bigg(\frac{1}{z_1+i\varepsilon}\frac{1}{z_{2}+i\varepsilon}+\textrm{perms.}\Bigg)=2(2\pi i)^2 \delta(z_1)\delta(z_2)\delta(z_3),
    \end{split}
\end{align}
one can rewrite the integral as,
\begin{align}
    \begin{split}
        \mathbfcal{M}^{(++)}=\frac{(2\pi i)^2}{6}\int_{\boldsymbol{\ell_1},\boldsymbol{\ell_2}}\frac{1}{\boldsymbol{\ell_1}^2\,\boldsymbol{\ell_2}^2\,(\boldsymbol{\ell_1}+\boldsymbol{\ell_2}-\boldsymbol{q})^2}\sim -\frac{\Gamma^3(-\epsilon)\Gamma(1+2\epsilon)}{\Gamma(-3\epsilon)}\,.
    \end{split}
\end{align}
In principle, the integral \eqref{intextra}, can be computed directly using Schwinger parametrization, but the above-mentioned approach is useful when we go beyond two loops.
\section{Boundary integrals in potential mode relevant for solving the differential equation}\label{ch3:app:B}
\textbf{Computation of boundary integrals in potential region:}\\ Now we compute the boundary integrals relevant to the four master integrals as mentioned previously for our subsequent computations.
We start with the master with two linear propagators:
$\,\,$\begin{align}
    \begin{split}
        \mathbfcal{M}_{1,1;0,0,1,1,1}^{(-+)}\Big|_{v_{\infty}\to 0}^{\textrm{pot.}}&=-v_{\infty}^{-2}\int_{\boldsymbol{\ell_1},\boldsymbol{\ell_2}}\frac{1}{(-\boldsymbol{\ell_1}\cdot \boldsymbol{n}+i\varepsilon)(\boldsymbol{\ell_2}\cdot \boldsymbol{n}+i\varepsilon)(\boldsymbol{\ell_1}+\boldsymbol{\ell_2}-\boldsymbol{q})^2(\boldsymbol{q}-\boldsymbol{\ell_1})^2(\boldsymbol{q}-\boldsymbol{\ell_2})^2}\,.\label{2.30}
    \end{split}
\end{align}
In \eqref{2.30}, the $i\varepsilon$  prescription is very important, and one needs to carefully take care of it. Our goal is to make relation with $\mathcal{M}^{(++)}$ with $\mathcal{M}^{+-}$. Now doing the following change of variable ($\boldsymbol{q}-\boldsymbol{\ell_i}\to \boldsymbol{\ell_i}$) we left with 
\begin{align}
    \begin{split}
        \mathbfcal{M}_{1,1;0,0,1,1,1}^{(-+)}\Big|_{v_{\infty}\to 0}^{\textrm{pot.}}&=-v_{\infty}^{-2}\int_{\boldsymbol{\ell_1},\boldsymbol{\ell_2}}\frac{1}{(\boldsymbol{\ell_1}\cdot \boldsymbol{n}+i\varepsilon)(-\boldsymbol{\ell_2}\cdot \boldsymbol{n}+i\varepsilon)(\boldsymbol{\ell_1}+\boldsymbol{\ell_2}-\boldsymbol{q})^2\,\boldsymbol{\ell_1}^2\,\boldsymbol{\ell_2}^2}\,.
    \end{split}
\end{align}
Now again we do the following relabelling: $\boldsymbol{\ell_1}\to \boldsymbol{\ell_1}+\boldsymbol{\ell_2}-\boldsymbol{q}$, $\boldsymbol{\ell_2}\to -\boldsymbol{\ell_2}$ and, ${\boldsymbol{q}\to -\boldsymbol{q}}\,.$ The we get,
\begin{align}
    \begin{split}
        \mathbfcal{M}_{1,1;0,0,1,1,1}^{(-+)}\Big|_{v_{\infty}\to 0}^{\textrm{pot.}}=-v_\infty^{-2}\int_{\boldsymbol{\ell_1},\boldsymbol{\ell_2}}\frac{1}{(\boldsymbol{\ell_1}\cdot \boldsymbol{n}+\boldsymbol{\ell_2}\cdot \boldsymbol{n}+i\varepsilon)(\boldsymbol{\ell_2}\cdot \boldsymbol{n}+i\varepsilon)(\boldsymbol{\ell_1}+\boldsymbol{\ell_2}-\boldsymbol{q})^2\,\boldsymbol{\ell_1}^2\,\boldsymbol{\ell_2}^2}\,.
    \end{split}
\end{align}
Now interchanging $\boldsymbol{\ell_{1}}\leftrightarrow \boldsymbol{\ell}_2$ and adding we will get,
\begin{align}
    \begin{split}
        & 2\mathbfcal{M}_{1,1;0,0,1,1,1}^{(-+)}\Big|_{v_{\infty}\to 0}^{\textrm{pot.}}=-v_\infty^{-2}\int_{\boldsymbol{\ell_{1},\ell_{2}}}\frac{1}{(\boldsymbol{\ell_1}\cdot \boldsymbol{n}+i\varepsilon)(\boldsymbol{\ell_2}\cdot \boldsymbol{n}+i\varepsilon)(\boldsymbol{\ell_1}+\boldsymbol{\ell_2}-\boldsymbol{q})^2\,\boldsymbol{\ell_1}^2\,\boldsymbol{\ell_2}^2}\equiv\mathbfcal{M}_{1,1;0,0,1,1,1}^{(++)}\Big|_{v_{\infty}\to 0}^{\textrm{pot.}}\,
    \end{split}
\end{align}
and the integral is given by,
\begin{align}
    \mathbfcal{M}^{(++)}_{1,1;0,0,1,1,1}=\frac{1}{v_\infty^2}\frac{1}{(4\pi)^{2-2\epsilon}}\frac{\Gamma^3(-\epsilon)\Gamma(1+2\epsilon)}{3\Gamma(-3\epsilon)}\,.
\end{align}
In the main text, we mainly face the following form of two-loop boundary integral.
\begin{align}
    &\mathbfcal{M}^{(\pm\pm)\textrm{pot.}}_{\alpha_1,\alpha_2;\beta_1,\cdots \beta_5}=\int_{\boldsymbol{\ell_2},\boldsymbol{\ell_2}}\frac{1}{(\pm \boldsymbol{\ell_2}\cdot \boldsymbol{n}+i\varepsilon)^{\alpha_1}(\pm \boldsymbol{\ell_2}\cdot \boldsymbol{n}+i\varepsilon)^{\alpha_2}\,\boldsymbol{D}_1^{\beta_1}\cdots \boldsymbol{D}_5^{\beta_5}}
\end{align}
To compute the integral, one can further use the IBP reduction procedure with the following 7 basis functions
\begin{align}
    \{D\}: \{\boldsymbol{\ell_1}\cdot \boldsymbol{n},\,\boldsymbol{\ell_2}\cdot \boldsymbol{n},\,\boldsymbol{\ell_2}^2,\,\boldsymbol{\ell_2}^2,\,(\boldsymbol{\ell_2}+\boldsymbol{\ell_2}-\boldsymbol{q})^2,\,(\boldsymbol{\ell_2}-\boldsymbol{q})^2,\,(\boldsymbol{\ell_2}-\boldsymbol{q})^2\}\,.
\end{align}
After doing the IBP reduction one will get 15 unique sectors with 22 master integrals.
Particularly, we need the following two loop boundary integrals,
\begin{align}
    \begin{split}
        &\mathbfcal{M}^{(\pm\pm)\textrm{pot.}}_{0,0;1,1,1,1,1}=-\frac{2 \left(1-36 \epsilon ^2\right)}{\left(\boldsymbol q^2\right)^2 (2 \epsilon +1)^2}\mathbfcal{M}^{(\pm\pm)\textrm{pot.}}_{0,0;0,0,1,1,1}-\frac{4 \epsilon }{\boldsymbol q^2 (2 \epsilon +1)}\mathbfcal{M}^{(\pm\pm)\textrm{pot.}}_{0,0;1,1,0,1,1}\,,\\ &
        \mathbfcal{M}^{(\pm\pm)\textrm{pot.}}_{0,0;1,1,2,1,1}=\frac{12 (\epsilon +1) (2 \epsilon -1) (36 \epsilon^2 -1) }{\left(\boldsymbol q^2\right)^3 (2 \epsilon +1) (2 \epsilon +3)^2}\mathbfcal{M}^{(\pm\pm)\textrm{pot.}}_{0,0;0,0,1,1,1}+\frac{8 \epsilon }{\left(\boldsymbol q^2\right)^2 (2 \epsilon +3)}\mathbfcal{M}^{(\pm\pm)\textrm{pot.}}_{0,0;1,1,0,1,1}\,,\\ &
  \mathbfcal{M}^{(\pm\pm)\textrm{pot.}}_{0,0;0,0,2,1,1}=      \frac{2 \epsilon  (6 \epsilon -1)}{\boldsymbol q^2 (2 \epsilon +1)}\mathbfcal{M}^{(\pm\pm)\textrm{pot.}}_{0,0;0,0,1,1,1}=\mathbfcal{M}^{(\pm\pm)\textrm{pot.}}_{0,0;0,0,1,2,1}
    \end{split}
\end{align}
where,
\begin{align}
    \begin{split}
        &\mathbfcal{M}^{(\pm\pm)\textrm{pot.}}_{0,0;0,0,1,1,1}= \frac{1}{(4\pi)^{3-2\epsilon}}\frac{\Gamma(1/2-\epsilon)^3\Gamma(2\epsilon)}{\Gamma(3/2-3\epsilon)}\frac{1}{\boldsymbol{q}^{4\epsilon}}\,,\\ &
        \mathbfcal{M}^{(\pm\pm)\textrm{pot.}}_{0,0;1,1,0,1,1}=\frac{1}{(4\pi)^{3-2\epsilon}}\frac{\Gamma^4(1/2-\epsilon)\Gamma^2(1/2+2\epsilon)}{\Gamma^2(1-2\epsilon)}\,.
    \end{split}
\end{align}
Apart from that we will face the following integrals,
\begin{align}
    \begin{split}
        &
        \mathbfcal{M}^{(\pm\pm)\textrm{pot.}}_{0,0;0,1,1,0,1}=0\,,\\ &
\mathbfcal{M}^{(\pm\pm)\textrm{pot.}}_{0,1;0,0,1,1,1}=-i \frac{\sqrt{\pi}}{(4\pi)^3}\frac{\Gamma(1/2-2\epsilon)\Gamma^2(1/2-\epsilon)\Gamma(-\epsilon)\Gamma(1/2+2\epsilon)}{\Gamma(1/2-3\epsilon)\Gamma(1-2\epsilon)}\,,\\ &    \mathbfcal{M}^{(\pm\pm)\textrm{pot.}}_{0,2;1,0,1,1,0}=-\frac{1}{(4\pi)^3}\frac{4\epsilon\Gamma(2\epsilon)\Gamma^2(-2\epsilon)\Gamma(1/2-\epsilon)\Gamma(1/2+\epsilon)}{\Gamma(-4\epsilon)}\,,\\ &
\mathbfcal{M}^{(\pm\pm)\textrm{pot.}}_{1,1;1,1,0,1,1}=-\frac{\pi}{(4\pi)^3}\frac{\Gamma^4(-\epsilon)\Gamma^2(\epsilon+1)}{\Gamma^2(-2\epsilon)}\,,\\ &
\mathbfcal{M}^{(\pm\pm)\textrm{pot.}}_{0,1;1,0,1,1,0}=-i\frac{\pi}{(4\pi)^3}\frac{\Gamma(\epsilon)\Gamma(1/2-2\epsilon)\Gamma(1/2+2\epsilon)}{\Gamma(1-\epsilon)}\,,\\ &
\mathbfcal{M}^{(++)\textrm{pot.}}_{1,1;0,0,1,1,1}=2\mathbfcal{M}^{(+-)\textrm{pot.}}_{1,1;0,0,1,1,1}=\frac{1}{(4\pi)^{2-2\epsilon}}\frac{\Gamma^3(-\epsilon)\Gamma(1+2\epsilon)}{3\Gamma(-3\epsilon)}\,.
    \end{split}
\end{align}
\section{\texorpdfstring{$A(x,\epsilon)$ in $\epsilon$ factorized form}{A(x, epsilon) in epsilon-factorized form}} \label{ch3:app:C}

As mentioned in the main text, the $A(x,\epsilon)$ matrix constructed from the coupled differential equations required for solving the master integrals reduces to an  $\epsilon$ factorized form ($\epsilon\,\mathbb{S}(x)$) in UT basis. Here, we explicitly  write down the expression for $\mathbb{S}(x)$, which we have used to solve the master integrals. 

\begin{equation}\resizebox{\linewidth}{!}{$
    \mathbb{S}(x)=\left(
\begin{array}
{cccccccccccccccc}
 -\frac{18 x^2+9}{2 x-2 x^3} & -\frac{2 x^2+3}{4 x-4 x^3} & -\frac{6 x}{x^2-1} & 0 & 0 & 0 & 0 & 0 & 0 & 0 & 0 & 0 & 0 & 0 & 0 & 0 \\
 \frac{6 x^2+15}{x-x^3} & \frac{6 x^2+5}{2 x-2 x^3} & \frac{12 x}{x^2-1} & 0 & 0 & 0 & 0 & 0 & 0 & 0 & 0 & 0 & 0 & 0 & 0 & 0 \\
 \frac{6}{x} & -\frac{1}{3 x} & -\frac{2}{x} & 0 & 0 & 0 & 0 & 0 & 0 & 0 & 0 & 0 & 0 & 0 & 0 & 0 \\
 0 & 0 & 0 & \frac{2 (x-4) x+2}{x \left(x^2-1\right)} & 0 & 0 & 0 & 0 & 0 & 0 & 0 & 0 & 0 & 0 & 0 & 0 \\
 0 & 0 & 0 & 0 & \frac{2 \left(x^2+1\right)}{x-x^3} & 0 & 0 & 0 & 0 & 0 & 0 & 0 & 0 & 0 & 0 & 0 \\
 0 & 0 & 0 & 0 & 0 & \frac{2 \left(x^2+1\right)}{x-x^3} & 0 & 0 & 0 & 0 & 0 & 0 & 0 & 0 & 0 & 0 \\
 0 & 0 & 0 & 0 & 0 & 0 & 0 & 0 & 0 & 0 & 0 & 0 & 0 & 0 & 0 & 0 \\
 0 & 0 & 0 & -\frac{1}{x} & 0 & 0 & 0 & 0 & 0 & 0 & 0 & 0 & 0 & 0 & 0 & 0 \\
 0 & 0 & 0 & 0 & 0 & -\frac{1}{x} & 0 & 0 & 0 & 0 & 0 & 0 & 0 & 0 & 0 & 0 \\
 0 & 0 & 0 & 0 & \frac{336}{x^3-x} & 0 & 0 & 0 & 0 & 0 & 0 & 0 & 0 & 0 & 0 & 0 \\
 \frac{211-286 x^2}{15 x-15 x^3} & \frac{117-242 x^2}{30 x-30 x^3} & -\frac{97-122 x^2}{5 x-5 x^3} & 0 & \frac{84}{x-x^3} & 0 & 0 & 0 & 0 & 0 & -\frac{1-5 x^2}{x-x^3} & -\frac{2}{3 x} & 0 & 0 & 0 & 0 \\
 -\frac{3983-3481 x^2}{20 x-20 x^3} & \frac{2687 x^2+231}{40 x-40 x^3} & \frac{1821-4371 x^2}{20 x-20 x^3} & 0 & \frac{126}{x-x^3} & 0 & \frac{24}{x^2-1} & 0 & 0 & 0 & \frac{3-75 x^2}{2 x-2 x^3} & \frac{1-5 x^2}{x-x^3} & 0 & 0 & 0 & 0 \\
 0 & 0 & 0 & -\frac{16 (x (7 x+6)+7)}{5 x \left(x^2-1\right)} & 0 & \frac{64}{x^2-1} & 0 & 0 & 0 & 0 & 0 & 0 & -\frac{4}{x^2-1} & 0 & 0 & 0 \\
 0 & 0 & 0 & 0 & 0 & 0 & 0 & 0 & 0 & 0 & 0 & 0 & 0 & 0 & 0 & 0 \\
 \frac{5}{x} & \frac{25}{6 x} & -\frac{5}{x} & 0 & 0 & 0 & 0 & 0 & 0 & 0 & 0 & 0 & 0 & 0 & 0 & 0 \\
 0 & 0 & 0 & 0 & 0 & 0 & 0 & 0 & 0 & 0 & 0 & 0 & 0 & 0 & 0 & 0 
\end{array}\right)\,.
$}\end{equation}
\section{Iterative UT integrals}\label{ch3:app:D}
\begingroup
\small
\begin{align}
    \begin{split}
     &   \boldsymbol{\mathscr{J}}(\{-1,-1,0\},x)\equiv  -\textbf{Li}_3\left(\frac{1}{x+1}\right)+\frac{1}{6} \log ^3(x+1)-\frac{1}{12} \pi ^2 \log (x+1)+\frac{7 \zeta (3)}{8},\\ &
     \boldsymbol{\mathscr{J}}(\{-1,0,0\},x)\equiv -\textbf{Li}_3(-x)+\textbf{Li}_2(-x) \log (x)+\frac{1}{2} \log (x+1) \log ^2(x)-\frac{3 \zeta (3)}{4},\\ &
    \boldsymbol{\mathscr{J}}(\{-1,1,0\},x)\equiv  \textbf{Li}_3\left(\frac{x+1}{2}\right)+\textbf{Li}_3\left(\frac{1+x}{x}\right)+\textbf{Li}_3(x)-\textbf{Li}_3\left(\frac{x+1}{2 x}\right)+\log (2) \textbf{Li}_2\left(\frac{x+1}{2}\right)\\ &-\log (2) \textbf{Li}_2(x)-\log (2) \textbf{Li}_2\left(\frac{x+1}{2 x}\right)-\textbf{Li}_2\left(\frac{x+1}{2}\right) \log (x+1)+\textbf{Li}_2\left(\frac{1+x}{x}\right) \log \left(\frac{2 x}{x+1}\right)-\textbf{Li}_2(1-x) \log (x+1)\\ &-\textbf{Li}_2(x) \log (x)-\textbf{Li}_2\left(\frac{x+1}{2 x}\right) \log (x)+\textbf{Li}_2\left(\frac{x+1}{2 x}\right) \log (x+1)-\log ^2(2) \log (x)-\frac{1}{2} \log (2) \log ^2(x)\\ &+\log (2) \log (x) \log (x+1)-\log (1-x) \log (x) \log (x+1)-\frac{15 \zeta (3)}{8}+\frac{1}{2} i \pi  \log ^2(2)-\frac{1}{12} \pi ^2 \log (2),\\ &
      \boldsymbol{\mathscr{J}}(\{0,-1,0\},x)\equiv 2 \textbf{Li}_3(-x)-\textbf{Li}_2(-x) \log (x)+\frac{1}{12} \pi ^2 \log (x)+\frac{3 \zeta (3)}{2},\\ &
       \boldsymbol{\mathscr{J}}(\{0,0,0\},x)\equiv \frac{\log ^3(x)}{6},\\ &
       \boldsymbol{\mathscr{J}}(\{0,1,0\},x)\equiv 2 \textbf{Li}_3(x)-\textbf{Li}_2(x) \log (x)-\frac{1}{6} \pi ^2 \log (x)-2 \zeta (3),\\ &
   \boldsymbol{\mathscr{J}}(\{1,0,0\},x)\equiv     -\textbf{Li}_3(x)+\textbf{Li}_2(x) \log (x)+\frac{1}{2} \log (1-x) \log ^2(x)+\zeta (3),\\ &
   \boldsymbol{\mathscr{J}}(\{1,1,0\},x)\equiv-\textbf{Li}_3(1-x),\\ &
   \boldsymbol{\mathscr{J}}(\{1,-1,0\},x)\equiv-i \pi  \textbf{Li}_2\left(\frac{x}{x-1}\right)+i \pi  \textbf{Li}_2\left(\frac{2 x}{x-1}\right)+i \pi  \textbf{Li}_2\left(\frac{1}{x+1}\right)+\textbf{Li}_3\left(\frac{1-x}{2}\right)+\textbf{Li}_3\left(\frac{x}{x-1}\right)-\textbf{Li}_3\left(\frac{2 x}{x-1}\right)\\ &+\textbf{Li}_3(-x)-\log (2) \textbf{Li}_2\left(\frac{x}{x-1}\right)+\log (2) \textbf{Li}_2\left(\frac{2 x}{x-1}\right)-\textbf{Li}_2\left(\frac{2 x}{x-1}\right) \log (1-x)-\textbf{Li}_2\left(\frac{x}{x-1}\right) \log (x)\\ &+\textbf{Li}_2\left(\frac{1-x}{2}\right) \left(\log \left(-\frac{2}{x-1}\right)+i \pi \right)+\textbf{Li}_2\left(\frac{x}{x-1}\right) \log (1-x)+\textbf{Li}_2\left(\frac{2 x}{x-1}\right) \log (x)+\textbf{Li}_2\left(\frac{1}{x+1}\right) \log (1-x)\\ &+\textbf{Li}_2(-x) (-\log (2 x)-2 i \pi )-\log ^2(2) \log (1-x)+\frac{1}{2} \log (2) \log ^2(1-x)+\frac{1}{2} i \pi  \log ^2(x+1)+\frac{1}{2} \log ^2(x+1) \log (1-x)\\ &-i \pi  \log (2) \log (1-x)-i \pi  \log (x) \log (x+1)-\frac{1}{12} \pi ^2 \log (1-x)-\log (x) \log (x+1) \log (1-x)+\frac{3 \zeta (3)}{4}-\frac{i \pi ^3}{4}\\ &+\frac{\log ^3(2)}{3}+\frac{1}{2} i \pi  \log ^2(2)-\frac{1}{4} \pi ^2 \log (2).
    \end{split}
\end{align}
\endgroup
\section{\texorpdfstring{$\boldsymbol{\tilde{h}}$ coefficients}{tilde-h coefficients}} \label{ch3:app:E}
Note that, $\gamma=\frac{x^2+1}{2 x}\,.$
\begin{align}
    \begin{split}
   &  \boldsymbol{\tilde h_1}=4,\,\,
      \boldsymbol{\boldsymbol{\tilde h_2}}   =0,
     \boldsymbol{\tilde h}_3=\frac{72 \epsilon ^2-2}{12 \left(\gamma ^2-1\right)^2},\,\,
        \boldsymbol{\tilde h}_4=\frac{6 \gamma ^2+2 \epsilon -3}{12 \left(\gamma ^2-1\right)},\\& \boldsymbol{\tilde h}_5=\frac{1}{24 \left(\gamma ^2-1\right)^2 (\epsilon -1) \epsilon }\Bigg[6 \gamma ^4+72 \gamma ^2 \epsilon ^4-10 \gamma ^2-36 \left(\gamma ^4+\gamma ^2-1\right) \epsilon ^3\\&+\left(-36 \gamma ^6+150 \gamma ^4-116 \gamma ^2\right) \epsilon ^2+\left(36 \gamma ^6-120 \gamma ^4+110 \gamma ^2-25\right) \epsilon +4\Bigg],
       \boldsymbol{\tilde h}_6=-\frac{4 \gamma ^2 \epsilon ^2+2 \left(\gamma ^2+1\right) \epsilon -4 \gamma ^2+1}{24 \left(\gamma ^2-1\right)^2 (\epsilon -1)}\,,\\ &
       \boldsymbol{\tilde h}_7=\frac{-6 \gamma ^3 (\epsilon  (4 \epsilon -3)+1)-6 \gamma ^2 \epsilon +\gamma  \left(2 \epsilon  \left(\epsilon  \left(-24 \epsilon ^2+22 \epsilon +21\right)-14\right)+7\right)+3 \epsilon  (6 \epsilon +1)}{24 \left(\gamma ^2-1\right)^2 (\epsilon -1) \epsilon }\,,\\ &
   \boldsymbol{\tilde h}_8    =\frac{\gamma  \left(8 \epsilon ^3-13 \epsilon +4\right)}{48 \left(\gamma ^2-1\right) (\epsilon -1) (3 \epsilon -1)}\,,
    \boldsymbol{\tilde h}_9 =\frac{\gamma  \left(-9 \gamma ^2-2 (\epsilon -5) \epsilon +8\right)}{6 \left(\gamma ^2-1\right)^2}\,,
    \\&
    \boldsymbol{\tilde h}_{10}  =-\frac{\gamma}{24 \left(\gamma ^2-1\right)^2 (3 \epsilon -1)}\Bigg[8 \gamma ^4 \left(18 \epsilon ^2-21 \epsilon +5\right) \epsilon +3 \gamma ^2 \left(48 \epsilon ^4-208 \epsilon ^3+216 \epsilon ^2-63 \epsilon +4\right)\\&-288 \epsilon ^5+216 \epsilon ^4+464 \epsilon ^3-558 \epsilon ^2+173 \epsilon -14\Bigg]\,,
    \\ &\boldsymbol{\tilde h}_{11}=\frac{-2 \gamma ^4+72 \gamma ^2 \epsilon ^4+2 \gamma ^2-36 \left(\gamma ^4+\gamma ^2-1\right) \epsilon ^3+\left(12 \gamma ^6+6 \gamma ^4-20 \gamma ^2\right) \epsilon ^2+\left(-12 \gamma ^6+32 \gamma ^4-18 \gamma ^2-1\right) \epsilon }{24 \left(\gamma ^2-1\right)^2 (\epsilon -1) \epsilon}\,,\\ &
   \boldsymbol{\tilde h}_{12}=\frac{-8 \gamma ^4+4 \gamma ^2 \epsilon ^2+8 \gamma ^2+\left(8 \gamma ^4-6 \gamma ^2+2\right) \epsilon -3}{24 \left(\gamma ^2-1\right)^2 (\epsilon -1)}\,,\\&
  \boldsymbol{\tilde h}_{13}  =\frac{\gamma  \left(4 \gamma ^3 (\epsilon -1) (4 \epsilon -1)-12 \gamma ^2 \epsilon +2 \gamma  (\epsilon -1) (4 \epsilon -1) (4 \epsilon  (3 \epsilon +1)-3)+\epsilon  (42 \epsilon +5)\right)}{48 \left(\gamma ^2-1\right)^2 (\epsilon -1) \epsilon }\,,\\ &
 \boldsymbol{\tilde h}_{14}=\frac{\gamma  \left(8 \gamma  (\epsilon -1) \left(\gamma ^2 (8 \epsilon -2)-2 \epsilon  (\epsilon +5)+3\right)-\epsilon \right)}{96 \left(\gamma ^2-1\right) (\epsilon -1) (3 \epsilon -1)}\,,
  \\ &
  \boldsymbol{\tilde h}_{15} =\frac{\gamma  \left(4 \gamma ^4 \epsilon ^2+\gamma ^2 (2 \epsilon  (\epsilon  (12 \epsilon -11)-3)+2)+\epsilon  (16 \epsilon  (\epsilon  (2-3 \epsilon )+1)-1)-1\right)}{12 \left(\gamma ^2-1\right)^2 (\epsilon -1) \epsilon }\,,\\ &
  \boldsymbol{\tilde h}_{16}=\frac{\gamma  \left(\gamma ^4 (4-20 \epsilon )+4 \gamma ^2 (2 \epsilon  (\epsilon +6)-3)-\epsilon  (8 \epsilon +27)+8\right)}{24 \left(\gamma ^2-1\right)^2 (3 \epsilon -1)}\,,\\&\boldsymbol{\tilde h}_{17}=\frac{1}{8 \left(\gamma ^2-1\right)^2 (\epsilon -1)^2 (\epsilon +1) (2 \epsilon +1)}\Bigg[-4 \gamma ^6 \left(108 \epsilon ^5-132 \epsilon ^4-109 \epsilon ^3+142 \epsilon ^2+\epsilon -10\right)\\&+2 \gamma ^4 \left(432 \epsilon ^6-432 \epsilon ^5-792 \epsilon ^4+308 \epsilon ^3+539 \epsilon ^2-16 \epsilon -39\right)\\&+2 \gamma ^2 \left(576 \epsilon ^5+156 \epsilon ^4-596 \epsilon ^3-317 \epsilon ^2+20 \epsilon +21\right)+216 \epsilon ^4+324 \epsilon ^3+138 \epsilon ^2-9 \epsilon -4\Bigg]\,,\nonumber
       \end{split}
\end{align}
\begin{align}
    \begin{split}
&\boldsymbol{\tilde h}_{18}=-\frac{\left(2 \gamma ^2 (\epsilon -1)+1\right) \left(4 \gamma ^4 \left(\epsilon ^2-1\right)-2 \gamma ^2 \left(\epsilon  \left(6 \epsilon ^2+\epsilon -3\right)-2\right)-3 \epsilon  (2 \epsilon +1)+1\right)}{16 \left(\gamma ^2-1\right) (\epsilon -1)^2 (\epsilon +1)}\,,\\&\boldsymbol{\tilde h}_{19}=-\frac{1}{96 (\gamma -1)^2 (\gamma +1)^2 (\epsilon -1)^2 \epsilon  (\epsilon +1) (2 \epsilon +1)}\Bigg[2 \gamma ^5 \epsilon  \big(1152 \epsilon ^7+4312 \epsilon ^6-4644 \epsilon ^5-7706 \epsilon ^4+3361 \epsilon ^3\\&+3347 \epsilon ^2+107 \epsilon +23\big)+2 \gamma ^3 \big(1728 \epsilon ^9-16416 \epsilon ^8+1912 \epsilon ^7+26492 \epsilon ^6+3490 \epsilon ^5-12697 \epsilon ^4-5305 \epsilon ^3\\&+282 \epsilon ^2+35 \epsilon -1\big)+192 \gamma ^2 (\epsilon -1)^2 \epsilon  \left(72 \epsilon ^5+108 \epsilon ^4-70 \epsilon ^3-99 \epsilon ^2+13 \epsilon +6\right)+\gamma  \big(-10368 \epsilon ^9+11520 \epsilon ^8\\&-1888 \epsilon ^7-20272 \epsilon ^6+8000 \epsilon ^5+14004 \epsilon ^4+3092 \epsilon ^3-630 \epsilon ^2-93 \epsilon +1\big)+96 (\epsilon -1)^2 \epsilon  \big(72 \epsilon ^4+172 \epsilon ^3+82 \epsilon ^2\\&-27 \epsilon -9\big)\Bigg]\,,\\&\boldsymbol{\tilde h}_{20}=\frac{1}{96 \left(\gamma ^2-1\right)^2 (\epsilon -1)^2 \epsilon  (\epsilon +1)}\Bigg[\gamma  \big(12 \gamma ^8 \epsilon ^3 \left(\epsilon ^2-1\right)-36 \gamma ^6 \epsilon ^2 \left(2 \epsilon ^4-4 \epsilon ^3-9 \epsilon ^2+4 \epsilon +7\right)\\&+4 \gamma ^4 \epsilon  \left(90 \epsilon ^5-126 \epsilon ^4-276 \epsilon ^3+114 \epsilon ^2+187 \epsilon +11\right)-2 \gamma ^2 \left(252 \epsilon ^6-192 \epsilon ^5-726 \epsilon ^4+204 \epsilon ^3+439 \epsilon ^2+22 \epsilon +1\right)\\&+216 \epsilon ^6-180 \epsilon ^5-576 \epsilon ^4+192 \epsilon ^3+360 \epsilon ^2-11 \epsilon +1\big)\Bigg]\,,\\&\boldsymbol{\tilde h}_{21}=\frac{1}{48 (\gamma -1)^2 (\gamma +1)^2 (\epsilon -1)^2 \epsilon  (\epsilon +1) (2 \epsilon +1)}\Bigg[4 \gamma ^5 \big(1296 \epsilon ^7-360 \epsilon ^6-2348 \epsilon ^5+304 \epsilon ^4+1049 \epsilon ^3+48 \epsilon ^2-6 \epsilon\\& -1\big)+360 \gamma ^4 (\epsilon -1)^2 \epsilon  \left(2 \epsilon ^2+3 \epsilon +1\right)-4 \gamma ^3 \big(2592 \epsilon ^8+864 \epsilon ^7-7644 \epsilon ^6-1168 \epsilon ^5+5281 \epsilon ^4-24 \epsilon ^3-36 \epsilon ^2\\&+100 \epsilon -1\big)-2 \gamma ^2 \left(360 \epsilon ^6-60 \epsilon ^5-694 \epsilon ^4-193 \epsilon ^3+333 \epsilon ^2+253 \epsilon +1\right)-2 \gamma  \epsilon  \big(3456 \epsilon ^6+5136 \epsilon ^5-4484 \epsilon ^4\\&-5268 \epsilon ^3+1339 \epsilon ^2+45 \epsilon -188\big)-408 \epsilon ^5-324 \epsilon ^4+250 \epsilon ^3+323 \epsilon ^2+158 \epsilon +1\Bigg]\,,\\&\boldsymbol{\tilde h}_{22}=\frac{1}{48 \left(\gamma ^2-1\right)^2 (\epsilon -1)^2 \epsilon  (\epsilon +1) (2 \epsilon +1)}\Bigg[8 \gamma ^5 \left(6 \epsilon ^5-3 \epsilon ^4-9 \epsilon ^3+\epsilon ^2+3 \epsilon +2\right)-4 \gamma ^3 \big(36 \epsilon ^6-12 \epsilon ^5-45 \epsilon ^4\\&-21 \epsilon ^3+17 \epsilon ^2+18 \epsilon +7\big)+2 \gamma ^2 (\epsilon -1)^2 \left(2 \epsilon ^2+3 \epsilon +1\right)-6 \gamma  (2 \epsilon +1)^2 \left(3 \epsilon ^3-2 \epsilon ^2-\epsilon -1\right)+2 \epsilon ^3+\epsilon ^2-2 \epsilon -1\Bigg]\,,\\&
        \boldsymbol{\tilde h}_{23}=\frac{4 \gamma ^4 \left(2 \epsilon ^3-3 \epsilon ^2+5 \epsilon -4\right)+2 \gamma ^2 \left(4 \epsilon ^2-8 \epsilon +7\right)+8 \gamma ^6 (\epsilon -1)^2+2 \epsilon -3}{48 \left(\gamma ^2-1\right) (\epsilon -1)^2}\,,\\&\boldsymbol{\tilde h}_{24}-\frac{\left(2 \gamma ^2 (\epsilon -1)+1\right) \left(4 \gamma ^4 \left(\epsilon ^2-1\right)-2 \gamma ^2 \left(\epsilon  \left(6 \epsilon ^2+\epsilon -3\right)-2\right)-3 \epsilon  (2 \epsilon +1)+1\right)}{16 \left(\gamma ^2-1\right) (\epsilon -1)^2 (\epsilon +1)}\,,\\&  \boldsymbol{\tilde h}_{25}= \frac{1}{96 \left(\gamma ^2-1\right)^2 (\epsilon -1)^2 \epsilon  (3 \epsilon -1)}\Bigg[\gamma  \big(-2 \gamma ^2 \left(288 \epsilon ^5-552 \epsilon ^4+534 \epsilon ^3-515 \epsilon ^2+256 \epsilon -27\right) \epsilon\\& +\gamma ^4 \left(372 \epsilon ^4-658 \epsilon ^3+308 \epsilon ^2-28 \epsilon +2\right)-288 \epsilon ^5+288 \epsilon ^4-288 \epsilon ^3+208 \epsilon ^2-33 \epsilon -1\big)\Bigg]\,,\\&  \boldsymbol{\tilde h}_{26}=\frac{\gamma  \left(-4 \gamma ^2 \left(\epsilon  \left(4 \epsilon ^4-3 \epsilon ^3+\epsilon -4\right)+2\right)-2 \gamma ^4 (\epsilon -1)^2 (\epsilon  (6 \epsilon +5)-4)-\epsilon  \left(8 (\epsilon -1) \epsilon ^2+\epsilon +7\right)+4\right)}{96 \left(\gamma ^2-1\right) (\epsilon -1)^2 \epsilon  (3 \epsilon -1)}\,,\\& \boldsymbol{\tilde h}_{27}=\frac{\gamma}{24 \left(\gamma ^2-1\right)^2 (\epsilon -1)^2 \epsilon}\Bigg[-4 \gamma ^4 \left(24 \epsilon ^4-57 \epsilon ^3+41 \epsilon ^2-5 \epsilon -2\right)+\gamma ^2 \left(96 \epsilon ^5-88 \epsilon ^4-224 \epsilon ^3+290 \epsilon ^2-64 \epsilon -8\right)\\&+48 \epsilon ^4+12 \epsilon ^3-90 \epsilon ^2+31 \epsilon +1\Bigg]\,,\nonumber\end{split}\end{align}
        \begin{align}\begin{split}
    & \boldsymbol{\tilde h}_{28}= \frac{\gamma ^5 \left(4 \epsilon ^2-6 \epsilon +2\right)+\gamma ^3 \left(4 \epsilon  \left(4 \epsilon ^2-6 \epsilon +3\right)+1\right)+4 \gamma  \epsilon  (2 \epsilon -3)+\gamma }{48 \left(\gamma ^2-1\right) (\epsilon -1) (3 \epsilon -1)}\,,\\&
   \boldsymbol{\tilde h}_{29}=\frac{4 \epsilon  (2 \epsilon +1)}{\gamma ^2-1}\,, \boldsymbol{\tilde h}_{30}=-\frac{2 (6 \epsilon +5)}{3 \left(\gamma ^2-1\right)}\,, \boldsymbol{\tilde h}_{31}=\frac{2 \left(12 \epsilon ^2+8 \epsilon +1\right)}{3 \left(\gamma ^2-1\right)^2 \epsilon }\,, \boldsymbol{\tilde h}_{32}=-\frac{2 \gamma  (2 \epsilon +3)}{3 \left(\gamma ^2-1\right)}\,,\\& \boldsymbol{\tilde h}_{33}=\frac{18 \epsilon ^3+19 \epsilon ^2+3 \epsilon -1}{4 \left(\gamma ^2-1\right) (\epsilon +1)}\,, \boldsymbol{\tilde h}_{34}=-\frac{54 \epsilon ^3+75 \epsilon ^2+25 \epsilon +1}{24 \left(\gamma ^2-1\right) \epsilon  (\epsilon +1)}\,, \boldsymbol{\tilde h}_{35}=\frac{108 \epsilon ^4+132 \epsilon ^3+37 \epsilon ^2-3 \epsilon -1}{24 \left(\gamma ^2-1\right)^2 \epsilon ^2 (\epsilon +1)}\,,\\& \boldsymbol{\tilde h}_{36}=-\frac{\gamma  \left(18 \epsilon ^2+19 \epsilon +7\right)}{24 \left(\gamma ^2-1\right) (\epsilon +1)}\,,\boldsymbol{\tilde h}_{37}=\frac{\epsilon  (2 \epsilon +1)}{2 \epsilon -1},\,\,\boldsymbol{\tilde h}_{38}=\frac{ (2 \epsilon +1)}{2(1-2 \epsilon) },\,\,\boldsymbol{\tilde h}_{39}=\frac{ (2 \epsilon +1)}{2 \left(\gamma ^2-1\right) \epsilon }\,,\\&\boldsymbol{\tilde h}_{40}=\frac{2 \epsilon +1}{2 (1-3 \epsilon )}\,,\boldsymbol{\tilde h}_{41}=\frac{\epsilon  (2 \epsilon +1) \left(\gamma ^2 (\epsilon -1) (2 \epsilon -1)-\epsilon  (2 \epsilon +5)+1\right)}{\left(\gamma ^2-1\right) (\epsilon -1) (2 \epsilon -1) (3 \epsilon -1)}\,,\boldsymbol{\tilde h}_{42}=\frac{(2 \epsilon +1) (\epsilon  (4 (\epsilon -1) \epsilon +5)-1)}{2 \left(\gamma ^2-1\right) \epsilon  (\epsilon -1) (2 \epsilon -1) (3 \epsilon -1)}\,,\\&\boldsymbol{\tilde h}_{43}=\frac{\gamma ^4 \left(7 \epsilon ^3+4 \epsilon ^2-7 \epsilon -4\right)+\gamma ^2 \left(50 \epsilon ^4-3 \epsilon ^3-49 \epsilon ^2+2\right)+31 \epsilon ^3+31 \epsilon ^2+5 \epsilon +2}{\left(\gamma ^2-1\right) \left(\epsilon ^2-1\right)}\,,\\& \boldsymbol{\tilde h}_{44}=\frac{-21 \gamma ^4 \epsilon  \left(\epsilon ^2-1\right)+\gamma ^2 \left(-150 \epsilon ^4-41 \epsilon ^3+149 \epsilon ^2+44 \epsilon -2\right)-93 \epsilon ^3-135 \epsilon ^2-43 \epsilon +2}{6 \left(\gamma ^2-1\right) \epsilon  \left(\epsilon ^2-1\right)}\,,\\&\boldsymbol{\tilde h}_{45}=\frac{1}{6 \left(\gamma ^2-1\right)^2 \epsilon ^2 \left(\epsilon ^2-1\right)}\Bigg[3 \gamma ^4 \epsilon  \left(14 \epsilon ^3-3 \epsilon ^2-14 \epsilon +3\right)+\gamma ^2 \left(300 \epsilon ^5+32 \epsilon ^4-265 \epsilon ^3-45 \epsilon ^2-20 \epsilon -2\right)\\&+186 \epsilon ^4+225 \epsilon ^3+61 \epsilon ^2+9 \epsilon +2\Bigg]\,,\\& \boldsymbol{\tilde h}_{46}=-\frac{\gamma  \left(\gamma ^2 \left(50 \epsilon ^3+19 \epsilon ^2-44 \epsilon -25\right)+36 \epsilon ^2+43 \epsilon +13\right)}{6 \left(\gamma ^2-1\right) \left(\epsilon ^2-1\right)}\\&\boldsymbol{\tilde h}_{47}=\frac{1}{48 \gamma  \left(\gamma ^2-1\right)^2 (\epsilon -1) \epsilon ^2 (2 \epsilon -1) (3 \epsilon -1)}\Bigg[32 \gamma ^6 \epsilon ^2 \left(5 \epsilon ^2-7 \epsilon +2\right)+2 \gamma ^5 \left(60 \epsilon ^4-116 \epsilon ^3+41 \epsilon ^2+9 \epsilon -4\right)\\&+\gamma ^4 \left(2248 \epsilon ^5-4548 \epsilon ^4+3722 \epsilon ^3-2080 \epsilon ^2+700 \epsilon -96\right)+\gamma ^3 \left(60 \epsilon ^4+42 \epsilon ^3-40 \epsilon ^2-6 \epsilon +4\right)\\&+\gamma ^2 \left(-2248 \epsilon ^5+4388 \epsilon ^4-3434 \epsilon ^3+1884 \epsilon ^2-668 \epsilon +96\right)+\gamma  \left(-180 \epsilon ^4+150 \epsilon ^3+4 \epsilon ^2-31 \epsilon +7\right)\\&-16 \epsilon  \left(7 \epsilon ^2-9 \epsilon +2\right)\Bigg]\,,\\& \boldsymbol{\tilde h}_{48}=\frac{1}{48 (\gamma -1)^2 \gamma  (\gamma +1)^2 (\epsilon -1) \epsilon ^2 (2 \epsilon -1) (3 \epsilon -1)}\Bigg[24 \gamma ^6 \epsilon ^3 \left(30 \epsilon ^3-45 \epsilon ^2+16 \epsilon -1\right)\\&-2 \gamma ^5 \epsilon  \left(180 \epsilon ^4-504 \epsilon ^3+353 \epsilon ^2-81 \epsilon +4\right)+8 \gamma ^4 \epsilon  \left(120 \epsilon ^5-164 \epsilon ^4-9 \epsilon ^3+80 \epsilon ^2-30 \epsilon +3\right)\\&+\gamma ^3 \epsilon  \left(360 \epsilon ^4-648 \epsilon ^3+356 \epsilon ^2-63 \epsilon +1\right)-\gamma ^2 \left(1680 \epsilon ^6-2392 \epsilon ^5+440 \epsilon ^4+520 \epsilon ^3-252 \epsilon ^2+27 \epsilon +1\right)\\&+\gamma  \epsilon  \left(-360 \epsilon ^3+350 \epsilon ^2-99 \epsilon +7\right)+8 \epsilon ^2 \left(7 \epsilon ^2-9 \epsilon +2\right)\Bigg]\,,\\& \boldsymbol{\tilde h}_{49}=\frac{1}{48 (\gamma -1)^2 \gamma  (\gamma +1)^2 (\epsilon -1) \epsilon ^2 (2 \epsilon -1) (3 \epsilon -1)}\Bigg[-60 \gamma ^6 \epsilon ^2 \left(6 \epsilon ^3-11 \epsilon ^2+6 \epsilon -1\right)+30 \gamma ^5 \epsilon ^2 \left(6 \epsilon ^2-5 \epsilon +1\right)\\&+\gamma ^4 \epsilon  \left(240 \epsilon ^4+12 \epsilon ^3-306 \epsilon ^2+77 \epsilon -1\right)+\gamma ^3 \left(-720 \epsilon ^5+1056 \epsilon ^4-448 \epsilon ^3+94 \epsilon ^2-38 \epsilon +8\right)\\&+\gamma ^2 \left(-4800 \epsilon ^6+7864 \epsilon ^5-3904 \epsilon ^4+1058 \epsilon ^3-217 \epsilon ^2-47 \epsilon +24\right)+\gamma  \left(180 \epsilon ^4-82 \epsilon ^3+2 \epsilon ^2-17 \epsilon +7\right)\\&+8 \epsilon  \left(-14 \epsilon ^3+11 \epsilon ^2+5 \epsilon -2\right)\Bigg]\,,\nonumber
 \end{split}
\end{align}
\begin{align}
\begin{split}
&\boldsymbol{\tilde h}_{50}=\frac{-40 \gamma ^3 \left(3 \epsilon ^2-4 \epsilon +1\right)+\gamma ^2 \left(-400 \epsilon ^3+160 \epsilon ^2+372 \epsilon -125\right)+4 \gamma  (25 \epsilon -8)-8 (\epsilon -1) \epsilon }{48 (\gamma -1) (\gamma +1) (\epsilon -1) (3 \epsilon -1)}\,,\\&\boldsymbol{\tilde h}_{51}=\frac{1}{96 \gamma  \left(\gamma ^2-1\right)^2 (\epsilon -1)^2 \epsilon ^2 \left(6 \epsilon ^2-5 \epsilon +1\right)}\Bigg[-16 \gamma ^6 (\epsilon -1)^2 \epsilon ^2 \left(90 \epsilon ^3-75 \epsilon ^2-5 \epsilon +8\right)\\&+\gamma ^4 \left(2880 \epsilon ^7-4960 \epsilon ^6-1448 \epsilon ^5+8824 \epsilon ^4-9099 \epsilon ^3+5167 \epsilon ^2-1568 \epsilon +192\right)\\&-8 \gamma ^3 \epsilon  \left(12 \epsilon ^4-52 \epsilon ^3+67 \epsilon ^2-32 \epsilon +5\right)+\gamma ^2 \big(-1440 \epsilon ^7+880 \epsilon ^6+5080 \epsilon ^5-9562 \epsilon ^4+8769 \epsilon ^3-5245 \epsilon ^2\\&+1720 \epsilon -220\big)-8 \gamma  \left(6 \epsilon ^4-29 \epsilon ^3+39 \epsilon ^2-19 \epsilon +3\right)+2 \epsilon  \left(-103 \epsilon ^3+241 \epsilon ^2-173 \epsilon +32\right)\Bigg]\,,\\&\boldsymbol{\tilde h}_{52}=\frac{1}{96 (\gamma -1)^2 \gamma  (\gamma +1)^2 (\epsilon -1)^2 \epsilon ^2 \left(6 \epsilon ^2-5 \epsilon +1\right)}\Bigg[-4800 \gamma ^2 \left(\gamma ^2-1\right) \epsilon ^7-32 \gamma ^2 \big(15 \gamma ^4-9 \gamma ^3-364 \gamma ^2\\&+9 \gamma +349\big)\epsilon ^6+16 \left(72 \gamma ^6-78 \gamma ^5-384 \gamma ^4+63 \gamma ^3+319 \gamma ^2+15 \gamma -7\right) \epsilon ^5-8 \big(108 \gamma ^6-201 \gamma ^5+500 \gamma ^4\\&+128 \gamma ^3-567 \gamma ^2+73 \gamma -32\big) \epsilon ^4+2 \left(96 \gamma ^6-384 \gamma ^5+2172 \gamma ^4+132 \gamma ^3-2165 \gamma ^2+252 \gamma -91\right) \epsilon ^3\\&+\left(120 \gamma ^5-1120 \gamma ^4+64 \gamma ^3+1109 \gamma ^2-184 \gamma +39\right) \epsilon ^2+3 \left(24 \gamma ^4-8 \gamma ^3-26 \gamma ^2+8 \gamma +1\right) \epsilon -1\Bigg]\,,\\&\boldsymbol{\tilde h}_{53}=\frac{1}{96 (\gamma -1)^2 \gamma  (\gamma +1)^2 (\epsilon -1)^2 \epsilon ^2 \left(6 \epsilon ^2-5 \epsilon +1\right)}\Bigg[-60 \gamma ^8 \epsilon ^3 \left(6 \epsilon ^3-11 \epsilon ^2+6 \epsilon -1\right)\\&+60 \gamma ^6 \epsilon ^2 \left(30 \epsilon ^4-67 \epsilon ^3+52 \epsilon ^2-17 \epsilon +2\right)-3 \gamma ^4 \epsilon ^2 \left(520 \epsilon ^4-836 \epsilon ^3+172 \epsilon ^2+185 \epsilon -48\right)\\&-24 \gamma ^3 \epsilon  \left(24 \epsilon ^5-116 \epsilon ^4+162 \epsilon ^3-87 \epsilon ^2+18 \epsilon -1\right)+\gamma ^2 \big(9600 \epsilon ^7-26696 \epsilon ^6+26436 \epsilon ^5-12388 \epsilon ^4+3812 \epsilon ^3\\&-727 \epsilon ^2-117 \epsilon +60\big)+8 \gamma  \left(12 \epsilon ^5-16 \epsilon ^4+19 \epsilon ^3-29 \epsilon ^2+17 \epsilon -3\right)+\epsilon  \left(224 \epsilon ^4-400 \epsilon ^3+99 \epsilon ^2+107 \epsilon -31\right)\Bigg]\,,\\&\boldsymbol{\tilde h}_{54}=\frac{32 \gamma ^2 (\epsilon -1)^2 \left(25 \epsilon ^2+20 \epsilon -9\right)-40 \gamma  \left(6 \epsilon ^3-17 \epsilon ^2+14 \epsilon -3\right)+16 \epsilon ^3-32 \epsilon ^2+28 \epsilon -5}{96 (\gamma -1) (\gamma +1) (\epsilon -1)^2 (3 \epsilon -1)}\,,\\& \boldsymbol{\tilde h}_{55}=\frac{-2 \gamma ^4 \left(6 \epsilon ^4-7 \epsilon ^3-2 \epsilon ^2+7 \epsilon -4\right)+2 \gamma ^2 \epsilon  \left(104 \epsilon ^4-42 \epsilon ^3-99 \epsilon ^2+33 \epsilon +4\right)+104 \epsilon ^4+40 \epsilon ^3-36 \epsilon ^2+11 \epsilon -8}{2 \left(\gamma ^2-1\right) (\epsilon -1) (\epsilon +1) (2 \epsilon -1)}\,,\\&
       \boldsymbol{\tilde h}_{56}=\frac{1}{12 \left(\gamma ^2-1\right) (\epsilon -1) \epsilon  (\epsilon +1) (2 \epsilon -1)}\Bigg[6 \gamma ^4 \epsilon  \left(6 \epsilon ^3+\epsilon ^2-6 \epsilon -1\right)+\gamma ^2 \big(-624 \epsilon ^5+44 \epsilon ^4+738 \epsilon ^3-34 \epsilon ^2-132 \epsilon\\& +8\big)-312 \epsilon ^4-288 \epsilon ^3+100 \epsilon ^2+77 \epsilon -8\Bigg]\,,\\&\boldsymbol{\tilde h}_{57}=\frac{\gamma ^4 \left(-4 \epsilon ^3+2 \epsilon ^2+4 \epsilon -2\right)+\gamma ^2 \left(-192 \epsilon ^4-28 \epsilon ^3+178 \epsilon ^2+38 \epsilon +4\right)-96 \epsilon ^3-112 \epsilon ^2-22 \epsilon -1}{4 \left(\gamma ^2-1\right)^2 \epsilon  \left(\epsilon ^2-1\right)}\,,\\&\boldsymbol{\tilde h}_{58}=-\frac{\gamma  \left(2 \gamma ^2 \left(8 \epsilon ^3+5 \epsilon ^2-7 \epsilon -6\right)+4 \epsilon ^2+11 \epsilon +8\right)}{4 \left(\gamma ^2-1\right) (\epsilon -1) (\epsilon +1)}\,,\\& \boldsymbol{\tilde h}_{59}=\frac{1}{4 \gamma  \left(\gamma ^2-1\right) (\epsilon -1) (2 \epsilon -1) (3 \epsilon -1)}\Bigg[2 \gamma ^4 \epsilon  \left(4 \epsilon ^3-2 \epsilon ^2-3 \epsilon +1\right)-2 \gamma ^3 \epsilon  \left(4 \epsilon ^3-3 \epsilon ^2-2 \epsilon +1\right)\\&+\gamma ^2 \big(-16 \epsilon ^5+20 \epsilon ^4+48 \epsilon ^3-26 \epsilon ^2+7 \epsilon -1\big)-2 \gamma  \epsilon  \left(5 \epsilon ^2-6 \epsilon +1\right)+\epsilon  \left(-28 \epsilon ^2+31 \epsilon -7\right)\Bigg]\,,\nonumber
    \end{split}
\end{align}
\begin{align}
    \begin{split}
       & 
       \boldsymbol{\tilde h}_{60}=\frac{1}{24 \gamma  \left(\gamma ^2-1\right) (\epsilon -1) \epsilon  (2 \epsilon -1) (3 \epsilon -1)}\Bigg[-6 \gamma ^4 \epsilon  \left(4 \epsilon ^3-2 \epsilon ^2-3 \epsilon +1\right)+6 \gamma ^3 \epsilon  \left(4 \epsilon ^3-3 \epsilon ^2-2 \epsilon +1\right)\\&+\gamma ^2 \left(48 \epsilon ^5-44 \epsilon ^4+48 \epsilon ^3-62 \epsilon ^2+27 \epsilon -5\right)-2 \gamma  \epsilon  \left(5 \epsilon ^2-6 \epsilon +1\right)+\epsilon  \left(-28 \epsilon ^2+31 \epsilon -7\right)\Bigg]\,,
       \\&\boldsymbol{\tilde h}_{61}=-\frac{1}{\left.24 \gamma  \left(\gamma ^2-1\right)^2 (\epsilon -1) \epsilon ^2 (2 \epsilon -1) (3 \epsilon -1)\right)}\Bigg[(2 \epsilon +1) \big(-6 \gamma ^4 \epsilon  \left(4 \epsilon ^3-14 \epsilon ^2+7 \epsilon -1\right)\\&+6 \gamma ^3 (\epsilon -1)^2 \epsilon  (4 \epsilon -1)+\gamma ^2 \left(48 \epsilon ^5-76 \epsilon ^4-60 \epsilon ^3+98 \epsilon ^2-39 \epsilon +5\right)+2 \gamma  \epsilon  \left(5 \epsilon ^2-6 \epsilon +1\right)\\&+\epsilon  \left(28 \epsilon ^2-31 \epsilon +7\right)\big)\Bigg]\,,\\&\boldsymbol{\tilde h}_{62}=\frac{\gamma ^2 \left(4 \epsilon ^3-6 \epsilon +2\right)+2 \gamma  (\epsilon -1) \epsilon +\epsilon  (2 \epsilon -1)}{12 \left(\gamma ^2-1\right) (\epsilon -1) (3 \epsilon -1)}\,,\boldsymbol{\tilde h}_{63}=\frac{\gamma ^2 (4 \epsilon -1)}{3 \left(\gamma ^2-1\right) (3 \epsilon -1)}\,,\\&\boldsymbol{\tilde h}_{64}=\frac{1}{8 \gamma  \left(\gamma ^2-1\right)^2 (\epsilon -1) (\epsilon +1) (2 \epsilon -1) (3 \epsilon -1)}\Bigg[-4 \gamma ^6 \epsilon  \left(4 \epsilon ^4+2 \epsilon ^3-5 \epsilon ^2-2 \epsilon +1\right)\\&+\gamma ^4 \left(256 \epsilon ^6-184 \epsilon ^5-260 \epsilon ^4+27 \epsilon ^3+83 \epsilon ^2-45 \epsilon +7\right)-4 \gamma ^3 \epsilon ^2 \left(2 \epsilon ^2+\epsilon -1\right)\\&-2 \gamma ^2 \left(128 \epsilon ^6-100 \epsilon ^5-138 \epsilon ^4+27 \epsilon ^3+47 \epsilon ^2-29 \epsilon +5\right)+4 \gamma  \epsilon ^2 \left(2 \epsilon ^2+\epsilon -1\right)\\&-8 \epsilon ^4+7 \epsilon ^3+19 \epsilon ^2+3 \epsilon -1\Bigg]\,,\\&\boldsymbol{\tilde h}_{65}=\frac{1}{48 (\gamma -1) \gamma  (\gamma +1) (\epsilon -1) \epsilon  (\epsilon +1) (2 \epsilon -1) (3 \epsilon -1)}\Bigg[12 \gamma ^4 \epsilon  \left(4 \epsilon ^4+2 \epsilon ^3-5 \epsilon ^2-2 \epsilon +1\right)\\&+\gamma ^2 \left(-768 \epsilon ^6+344 \epsilon ^5+676 \epsilon ^4-281 \epsilon ^3-99 \epsilon ^2+39 \epsilon +5\right)+4 \gamma  \epsilon  \left(6 \epsilon ^3+5 \epsilon ^2-2 \epsilon -1\right)\\&+8 \epsilon ^4-7 \epsilon ^3-19 \epsilon ^2-3 \epsilon +1\Bigg]\,,\\& \boldsymbol{\tilde h}_{66}=\frac{1}{48 (\gamma -1)^2 \gamma  (\gamma +1)^2 (\epsilon -1) \epsilon ^2 (\epsilon +1) (2 \epsilon -1) (3 \epsilon -1)}\Bigg[2 \gamma ^4 \big(2400 \epsilon ^6-1180 \epsilon ^5-2616 \epsilon ^4\\&+1385 \epsilon ^3+345 \epsilon ^2-205 \epsilon +15\big) \epsilon +\gamma ^2 \big(6336 \epsilon ^7-4808 \epsilon ^6-4708 \epsilon ^5+3302 \epsilon ^4-401 \epsilon ^3+133 \epsilon ^2-21 \epsilon \\&-13\big)-4 \gamma  \left(12 \epsilon ^4+8 \epsilon ^3-7 \epsilon ^2-2 \epsilon +1\right) \epsilon -44 \epsilon ^5+40 \epsilon ^4+19 \epsilon ^3-77 \epsilon ^2-5 \epsilon +7\Bigg]\,,\\
    &\boldsymbol{\tilde h}_{67}=\frac{\gamma ^2 \left(-8 \epsilon ^3+13 \epsilon -5\right)+2 \gamma  \epsilon -\epsilon +1}{24 (\gamma -1) (\gamma +1) (\epsilon -1) (3 \epsilon -1)}\,,\\&\boldsymbol{\tilde h}_{68}=\frac{-8 \gamma ^2 \left(107 \epsilon ^4+27 \epsilon ^3-114 \epsilon ^2-45 \epsilon +25\right)+100 \epsilon ^3-74 \epsilon ^2-127 \epsilon +47}{24 (\gamma -1) (\gamma +1) (\epsilon -1) (\epsilon +1) (3 \epsilon -1)}\,,\\& \boldsymbol{\tilde h}_{69}=-\frac{\epsilon  \left(\gamma ^2 \left(2 \epsilon ^2-3 \epsilon +1\right)-10 \epsilon ^2-9 \epsilon -5\right)}{2 (\epsilon -1) (2 \epsilon -3) (2 \epsilon -1)}\,,\boldsymbol{\tilde h}_{70}=\frac{ \left(\gamma ^2 (\epsilon -1)-\epsilon -1\right)}{4 (\epsilon -1) (2 \epsilon -3)}\,,\boldsymbol{\tilde h}_{71}=-\frac{ \left(4 \epsilon ^3+5 \epsilon +3\right)}{4 (\epsilon -1) \epsilon  (2 \epsilon -3) (2 \epsilon -1)}\,,\\& \boldsymbol{\tilde h}_{72}=\frac{1}{256 \left(\gamma ^2-1\right)^2 (\epsilon -1)^2 (\gamma  \epsilon  (6 \epsilon -5)+\gamma )}\Bigg[-64 \gamma ^8 (\epsilon -1)^2 \epsilon ^2+\gamma ^6 \big(160 \epsilon ^5+1872 \epsilon ^4-5282 \epsilon ^3+4581 \epsilon ^2\\&-1490 \epsilon +159\big)-2 \gamma ^4 \left(80 \epsilon ^5-3816 \epsilon ^4+8202 \epsilon ^3-5819 \epsilon ^2+1453 \epsilon -120\right)+\gamma ^2 \big(800 \epsilon ^4-2506 \epsilon ^3+1597 \epsilon ^2\\&-184 \epsilon -27\big)-4 \left(22 \epsilon ^2+7 \epsilon -3\right)\Bigg]\,,\nonumber 
 \end{split}
\end{align}
\begin{align}
    \begin{split}
  &\boldsymbol{\tilde h}_{73}=\frac{1}{1536 \gamma  \left(\gamma ^2-1\right) (\epsilon -1)^2 \epsilon  (\epsilon  (6 \epsilon -5)+1)}\Bigg[192 \gamma ^6 (\epsilon -1)^2 \epsilon ^2+\gamma ^4 \big(-480 \epsilon ^5-944 \epsilon ^4+4078 \epsilon ^3-3551 \epsilon ^2\\&+896 \epsilon +1\big)+\gamma ^2 \left(-1440 \epsilon ^4+2762 \epsilon ^3-1733 \epsilon ^2+292 \epsilon -1\right)+4 \left(22 \epsilon ^2+7 \epsilon -3\right)\Bigg]\,,\\&\boldsymbol{\tilde h}_{74}=\frac{1}{1536 \left(\gamma ^2-1\right)^2 (\epsilon -1)^2 \epsilon ^2 (\gamma  \epsilon  (6 \epsilon -5)+\gamma )}\Bigg[(2 \epsilon +1) \big(-192 \gamma ^6 (\epsilon -1)^2 \epsilon ^2+\gamma ^4 \big(480 \epsilon ^5-5360 \epsilon ^4\\&+9646 \epsilon ^3-5663 \epsilon ^2+896 \epsilon +1\big)+\gamma ^2 \left(-480 \epsilon ^4+1850 \epsilon ^3-1493 \epsilon ^2+292 \epsilon -1\right)+4 \left(22 \epsilon ^2+7 \epsilon -3\right)\big)\Bigg]\,,\\&\boldsymbol{\tilde h}_{75}=\frac{4 (\epsilon -3)-\gamma ^2 (\epsilon -1) \left(\gamma ^2 \left(320 \epsilon ^2-418 \epsilon +97\right)-336 \epsilon ^2+466 \epsilon -97\right)}{768 \left(\gamma ^2-1\right) (\epsilon -1)^2 (3 \epsilon -1)}\,,\\&\boldsymbol{\tilde h}_{76}=\frac{\gamma ^2 \left(-200 \epsilon ^6+264 \epsilon ^5+75 \epsilon ^4-184 \epsilon ^3+34 \epsilon ^2+16 \epsilon -5\right)-72 \epsilon ^6-328 \epsilon ^5+117 \epsilon ^4+124 \epsilon ^3-57 \epsilon ^2+3 \epsilon +3}{8 (1-2 \epsilon )^2 (\epsilon -1)^2 (\epsilon +1) (2 \epsilon +1)}\,,\\&\boldsymbol{\tilde h}_{77}=\frac{\gamma ^2 (\epsilon -1)^2 \left(232 \epsilon ^4+90 \epsilon ^3-123 \epsilon ^2-10 \epsilon +19\right)-8 \epsilon ^6+110 \epsilon ^5-49 \epsilon ^4-132 \epsilon ^3+60 \epsilon ^2+16 \epsilon -9}{16 (1-2 \epsilon )^2 (\epsilon -1)^2 \epsilon  (\epsilon +1) (2 \epsilon +1)}\,,\\&\boldsymbol{\tilde h}_{78}=-\frac{64 \epsilon ^7-280 \epsilon ^6+22 \epsilon ^5+115 \epsilon ^4-60 \epsilon ^3+5 \epsilon ^2+7 \epsilon +1}{16 (1-2 \epsilon )^2 (\epsilon -1)^2 \epsilon ^2 (\epsilon +1) (2 \epsilon +1)}\,,\\&\boldsymbol{\tilde h}_{79}=\frac{-2 \gamma ^2 (1-2 \epsilon )^2 \left(10 \epsilon ^3-19 \epsilon ^2+4 \epsilon +8\right)+32 \epsilon ^5-168 \epsilon ^4+228 \epsilon ^3-106 \epsilon ^2+7 \epsilon +7}{32 (1-2 \epsilon )^2 (\epsilon -1)^2 (2 \epsilon +1)}\,,\\&\boldsymbol{\tilde h}_{80}=\frac{\left(4-36 \gamma ^2\right) \epsilon ^4+\left(74 \gamma ^2-26\right) \epsilon ^3+\left(21-48 \gamma ^2\right) \epsilon ^2+2 \left(5 \gamma ^2-2\right) \epsilon -1}{16 (1-2 \epsilon )^2 (\epsilon -1) \epsilon }\,,\boldsymbol{\tilde h}_{81}=-\frac{\left(\gamma ^2-1\right) \left(4 \epsilon ^3+\epsilon ^2-2 \epsilon +1\right)}{16 (1-2 \epsilon )^2 \epsilon  (2 \epsilon +1)}\,,\\&\boldsymbol{\tilde h}_{82}=\frac{1}{128 (\gamma^2 -1) \gamma \left(2 \epsilon ^2-3 \epsilon +1\right)^2 (\epsilon ^2 (6 \epsilon +7)-1)}\Bigg[2 \gamma ^4 \big(3624 \epsilon ^7-5990 \epsilon ^6+1394 \epsilon ^5\\&+2233 \epsilon ^4-1628 \epsilon ^3+362 \epsilon ^2+18 \epsilon -13\big)+16 \gamma ^3 (\epsilon -1)^2 \left(72 \epsilon ^5+36 \epsilon ^4-50 \epsilon ^3-5 \epsilon ^2+8 \epsilon -1\right)\\&+\gamma ^2 \left(-21216 \epsilon ^7+22288 \epsilon ^6+17516 \epsilon ^5-23897 \epsilon ^4+3172 \epsilon ^3+3674 \epsilon ^2-1392 \epsilon +143\right)\\&+3 \left(-3408 \epsilon ^7+4148 \epsilon ^6+1264 \epsilon ^5-2739 \epsilon ^4+652 \epsilon ^3+180 \epsilon ^2-80 \epsilon +7\right)\Bigg]\,,\\&\boldsymbol{\tilde h}_{83}=-\frac{1}{256 (\gamma^2 -1) \gamma \left(2 \epsilon ^2-3 \epsilon +1\right)^2 \epsilon  (\epsilon ^2 (6 \epsilon +7)-1)}\Bigg[2 \gamma ^4 \big(4380 \epsilon ^7-7886 \epsilon ^6+1546 \epsilon ^5\\&+4549 \epsilon ^4-2920 \epsilon ^3+10 \epsilon ^2+402 \epsilon -81\big)+\gamma ^2 \big(-6192 \epsilon ^7+5784 \epsilon ^6+8610 \epsilon ^5-12461 \epsilon ^4+3572 \epsilon ^3\\&+1740 \epsilon ^2-1226 \epsilon +197\big)+888 \epsilon ^7+388 \epsilon ^6-2142 \epsilon ^5+107 \epsilon ^4+1206 \epsilon ^3-514 \epsilon ^2+36 \epsilon +7\Bigg]\,,\\&\boldsymbol{\tilde h}_{84}=-\frac{1}{256 (\gamma^2 -1) \gamma \left(2 \epsilon ^2-3 \epsilon +1\right)^2 \epsilon ^2 (\epsilon ^2 (6 \epsilon +7)-1)}\Bigg[16 \gamma ^3 \left(-2 \epsilon ^2+\epsilon +1\right)^2 \left(6 \epsilon ^3+\epsilon ^2-4 \epsilon +1\right) \epsilon\\& +2 \gamma ^2 \left(-6096 \epsilon ^8+7564 \epsilon ^7+1670 \epsilon ^6-4134 \epsilon ^5+1011 \epsilon ^4+52 \epsilon ^3+36 \epsilon ^2-38 \epsilon +7\right)-9696 \epsilon ^8+10792 \epsilon ^7\\&+4068 \epsilon ^6-6210 \epsilon ^5+801 \epsilon ^4+272 \epsilon ^3+104 \epsilon ^2-66 \epsilon +7\Bigg]\,,\nonumber
    \end{split}
\end{align}
\begin{align}
    \begin{split}
    &\boldsymbol{\tilde h}_{85}=\frac{1}{256 \gamma  \left(\gamma ^2-1\right) (2 \epsilon +1) \left(2 \epsilon ^2-3 \epsilon +1\right)^2}\Bigg[\gamma ^4 (1-2 \epsilon )^2 \left(56 \epsilon ^3-144 \epsilon ^2+32 \epsilon +59\right)\\&+\gamma ^2 \left(-2048 \epsilon ^5+4672 \epsilon ^4-3376 \epsilon ^3+448 \epsilon ^2+444 \epsilon -143\right)-8 \epsilon  (1-2 \epsilon )^2 \left(15 \epsilon ^2-22 \epsilon +7\right)\Bigg]\,,\\&\boldsymbol{\tilde h}_{86}=\frac{1}{256 \gamma  \left(\gamma ^2-1\right) (1-2 \epsilon )^2 (\epsilon -1) \epsilon}\Bigg[4 \gamma ^4 \epsilon  \left(66 \epsilon ^3-133 \epsilon ^2+84 \epsilon -17\right)+\gamma ^2 \big(-720 \epsilon ^4+1104 \epsilon ^3-566 \epsilon ^2\\&+72 \epsilon +14\big)-(1-2 \epsilon )^2 \left(30 \epsilon ^2+\epsilon -7\right)\Bigg]\,,\\&\boldsymbol{\tilde h}_{87}=\frac{(\epsilon +1) \left(2 \gamma ^2 \left(42 \epsilon ^3-60 \epsilon ^2+33 \epsilon -7\right)+(15 \epsilon -7) (1-2 \epsilon )^2\right)}{256 \gamma  (1-2 \epsilon )^2 (\epsilon -1) \epsilon  (2 \epsilon +1)}\,,\\&\boldsymbol{\tilde h}_{88}=-\frac{1}{64 \gamma ^2 \left(\gamma ^2-1\right)^2 (1-2 \epsilon )^4 (\epsilon -1)^3 \left(\epsilon ^2 (6 \epsilon +7)-1\right)^2}\Bigg[\gamma ^8 \big(-156 \epsilon ^6+404 \epsilon ^5-332 \epsilon ^4+43 \epsilon ^3+92 \epsilon ^2\\&-63 \epsilon +12\big)^2+4 \gamma ^7 (\epsilon -1)^2 (2 \epsilon -1)^3 \big(1800 \epsilon ^8+396 \epsilon ^7-3464 \epsilon ^6-1260 \epsilon ^5+1299 \epsilon ^4+345 \epsilon ^3\\&-183 \epsilon ^2-21 \epsilon +8\big)+2 \gamma ^6 \big(112320 \epsilon ^{13}-537984 \epsilon ^{12}+889584 \epsilon ^{11}-486508 \epsilon ^{10}-237128 \epsilon ^9\\&+414152 \epsilon ^8-159781 \epsilon ^7-11177 \epsilon ^6+15554 \epsilon ^5+602 \epsilon ^4+4123 \epsilon ^3-5533 \epsilon ^2+2016 \epsilon -240\big)\\&-4 \gamma ^5 (\epsilon -1)^2 (2 \epsilon -1)^3 \left(7704 \epsilon ^8+8172 \epsilon ^7-5770 \epsilon ^6-5225 \epsilon ^5+2756 \epsilon ^4+1375 \epsilon ^3-511 \epsilon ^2-110 \epsilon +33\right)\\&+\gamma ^4 \big(518400 \epsilon ^{14}-2505600 \epsilon ^{13}+3769056 \epsilon ^{12}-1383360 \epsilon ^{11}-1057208 \epsilon ^{10}+1569336 \epsilon ^9-1608359 \epsilon ^8\\&+817324 \epsilon ^7+308191 \epsilon ^6-483302 \epsilon ^5+121399 \epsilon ^4+31648 \epsilon ^3-15071 \epsilon ^2+162 \epsilon +328\big)\\&+4 \gamma ^3 (\epsilon -1)^2 (2 \epsilon -1)^3 \left(4824 \epsilon ^8+7092 \epsilon ^7+508 \epsilon ^6-1106 \epsilon ^5+1265 \epsilon ^4+316 \epsilon ^3-358 \epsilon ^2-38 \epsilon +25\right)\\&-6 \gamma ^2 \big(172800 \epsilon ^{14}-826560 \epsilon ^{13}+1067520 \epsilon ^{12}+363056 \epsilon ^{11}-1442532 \epsilon ^{10}+407040 \epsilon ^9+477315 \epsilon ^8\\&-177793 \epsilon ^7-64099 \epsilon ^6-4775 \epsilon ^5+24081 \epsilon ^4-135 \epsilon ^3-3513 \epsilon ^2+703 \epsilon -20\big)+12 \gamma  \epsilon  (2 \epsilon -1)^3 (10 \epsilon -17)\\& \left(6 \epsilon ^4+\epsilon ^3-7 \epsilon ^2-\epsilon +1\right)^2+9 \left(-240 \epsilon ^7+620 \epsilon ^6-96 \epsilon ^5-565 \epsilon ^4+212 \epsilon ^3+76 \epsilon ^2-32 \epsilon +1\right)^2\Bigg]\,,\\& \boldsymbol{\tilde h}_{89}=\frac{1}{64 \gamma  \left(\gamma ^2-1\right) (1-2 \epsilon )^2 (\epsilon -1)^2 \epsilon  \left(\epsilon ^2 (6 \epsilon +7)-1\right)}\Bigg[\gamma ^4 \big(1416 \epsilon ^7-2900 \epsilon ^6+596 \epsilon ^5+1956 \epsilon ^4\\&-1131 \epsilon ^3-82 \epsilon ^2+175 \epsilon -30\big)+\gamma ^2 \left(-1344 \epsilon ^7+2432 \epsilon ^6+938 \epsilon ^5-3411 \epsilon ^4+1117 \epsilon ^3+500 \epsilon ^2-291 \epsilon +35\right)\\&+312 \epsilon ^7-716 \epsilon ^6-246 \epsilon ^5+975 \epsilon ^4-116 \epsilon ^3-248 \epsilon ^2+62 \epsilon +1\Bigg]\,,\\&\boldsymbol{\tilde h}_{90}=\frac{1}{64 \gamma  \left(\gamma ^2-1\right) (1-2 \epsilon )^2 (\epsilon -1)^2 \epsilon ^2 \left(\epsilon ^2 (6 \epsilon +7)-1\right)}\Bigg[\gamma ^2 \big(-1536 \epsilon ^8+1528 \epsilon ^7-244 \epsilon ^6-240 \epsilon ^5\\&+696 \epsilon ^4-323 \epsilon ^3-46 \epsilon ^2+19 \epsilon +2\big)-1152 \epsilon ^8+2232 \epsilon ^7+908 \epsilon ^6-2398 \epsilon ^5-303 \epsilon ^4+626 \epsilon ^3+78 \epsilon ^2-64 \epsilon +1\Bigg]\,,\\&\boldsymbol{\tilde h}_{91}=\frac{1}{64 \gamma  \left(\gamma ^2-1\right) (2 \epsilon +1) \left(2 \epsilon ^2-3 \epsilon +1\right)^2}\Bigg[\gamma ^4 \left(-(1-2 \epsilon )^2\right) \left(20 \epsilon ^3-48 \epsilon ^2+7 \epsilon +18\right)+\gamma ^2 \big(304 \epsilon ^5-816 \epsilon ^4\\&+648 \epsilon ^3-72 \epsilon ^2-97 \epsilon +30\big)+8 \epsilon  \left(2 \epsilon ^2-3 \epsilon +1\right)^2\Bigg]\,,\nonumber
    \end{split}
\end{align}
\begin{align}
    \begin{split}
  &\boldsymbol{\tilde h}_{92}=\frac{-4 \gamma ^4 \epsilon  \left(18 \epsilon ^2-19 \epsilon +5\right)+2 \gamma ^2 \left(64 \epsilon ^3-52 \epsilon ^2+11 \epsilon +1\right)+(2 \epsilon +1) (1-2 \epsilon )^2}{64 \gamma  \left(\gamma ^2-1\right) (1-2 \epsilon )^2 \epsilon }\,,\\
   & \boldsymbol{\tilde h}_{93}=-\frac{(\epsilon +1) \left(2 \gamma ^2 \left(6 \epsilon ^2-4 \epsilon +1\right)+(1-2 \epsilon )^2\right)}{64 \gamma  (1-2 \epsilon )^2 \epsilon  (2 \epsilon +1)}\,, \boldsymbol{\tilde h}_{94}=\gamma ^2+\frac{1}{2 (\epsilon -1)}\,,\\&\boldsymbol{\tilde h}_{95}=-\frac{\gamma  (2 \epsilon +1) \left(\gamma ^2 (2 \epsilon -1) (7 \epsilon -2)+10 \epsilon ^2+\epsilon -1\right)}{8 \left(\gamma ^2-1\right) (\epsilon -1) \epsilon  (2 \epsilon -1) (3 \epsilon -1)}\,,\\&\boldsymbol{\tilde h}_{96}=\frac{\gamma  (2 \epsilon +1)^2 (4 \epsilon -1)}{8 \left(\gamma ^2-1\right) (\epsilon -1) \epsilon  (2 \epsilon -1) (3 \epsilon -1)}\,,\boldsymbol{\tilde h}_{97}=\frac{2 \gamma  \epsilon +\gamma }{24 \epsilon ^2-32 \epsilon +8}\,,\\&\boldsymbol{\tilde h}_{98}=-\frac{\gamma^2  (2 \epsilon +1)^2 \left(\gamma ^2 (2 \epsilon -1) (7 \epsilon -2)+10 \epsilon ^2+\epsilon -1\right)}{8 \left(\gamma ^2-1\right) (\epsilon -1)^2 \epsilon  (\epsilon  (6 \epsilon -5)+1)}\,,\boldsymbol{\tilde h}_{99}=\frac{\gamma^2  (2 \epsilon +1)^2}{8 (\epsilon -1)^2 (3 \epsilon -1)}\,,\\&\boldsymbol{\tilde h}_{100}=\frac{\gamma^2  (2 \epsilon +1)^3 (4 \epsilon -1)}{8 \left(\gamma ^2-1\right) (\epsilon -1)^2 \epsilon  (\epsilon  (6 \epsilon -5)+1)}\,,\\&
    \boldsymbol{\tilde h}_{101}=\frac{8 \gamma ^2 (\epsilon -1)^2 \left(6 \epsilon ^2-6 \epsilon +1\right)+36 \epsilon ^4-60 \epsilon ^3+23 \epsilon ^2-\epsilon +1}{4 (\epsilon -1)^3 (2 \epsilon -1)}\,,\boldsymbol{\tilde h}_{102}=\frac{-4 \gamma ^2 (\epsilon -1)^2 (2 \epsilon +1)-12 \epsilon ^3+5 \epsilon +1}{8 (\epsilon -1)^2 \epsilon  (2 \epsilon -1)}\,,\\&\boldsymbol{\tilde h}_{103}=\frac{1}{8 \gamma  \left(\gamma ^2-1\right) (\epsilon -1)^2 (\epsilon  (6 \epsilon -5)+1)}\Bigg[-2 \gamma ^4 \left(228 \epsilon ^4-464 \epsilon ^3+313 \epsilon ^2-85 \epsilon +8\right)\\&+\gamma ^2 \epsilon  \left(-930 \epsilon ^3+1265 \epsilon ^2-454 \epsilon +47\right)-9 \epsilon  \left(18 \epsilon ^3-27 \epsilon ^2+13 \epsilon -2\right)\Bigg]\,,\\&\boldsymbol{\tilde h}_{104}=\frac{2 \gamma ^2 \left(18 \epsilon ^2-23 \epsilon +5\right)+3 \left(9 \epsilon ^2-9 \epsilon +2\right)}{16 \gamma  (\epsilon -1)^2 (3 \epsilon -1)}\,,\\&\boldsymbol{\tilde h}_{105}=\frac{(2 \epsilon +1) \left(4 \gamma ^4 \left(6 \epsilon ^3-11 \epsilon ^2+6 \epsilon -1\right)+2 \gamma ^2 \left(90 \epsilon ^3-119 \epsilon ^2+38 \epsilon -3\right)+54 \epsilon ^3-81 \epsilon ^2+39 \epsilon -6\right)}{16 (\gamma -1) \gamma  (\gamma +1) (\epsilon -1)^2 \epsilon  (2 \epsilon -1) (3 \epsilon -1)}\,,\\&\boldsymbol{\tilde h}_{106}=\frac{1}{32 \left(\gamma ^2-1\right) (\epsilon -1)^3 (\epsilon  (6 \epsilon -5)+1)}\Bigg[2 \gamma ^4 (\epsilon -1)^2 \left(228 \epsilon ^3-314 \epsilon ^2+131 \epsilon -16\right)\\&-4 \gamma ^3 (\epsilon -1)^2 \left(12 \epsilon ^2-8 \epsilon +1\right) \epsilon +\gamma ^2 \left(2640 \epsilon ^5-7108 \epsilon ^4+7052 \epsilon ^3-3179 \epsilon ^2+621 \epsilon -42\right)\\&-4 \gamma  (\epsilon -1)^2 \left(132 \epsilon ^3-76 \epsilon ^2+15 \epsilon -1\right)+2088 \epsilon ^5-4072 \epsilon ^4+2922 \epsilon ^3-1037 \epsilon ^2+162 \epsilon -7\Bigg]\,,\\&\boldsymbol{\tilde h}_{107}=\frac{-2 \gamma ^2 (\epsilon -1)^2 \left(84 \epsilon ^2-98 \epsilon +27\right)+4 \gamma  (\epsilon -1)^2 \left(12 \epsilon ^2-8 \epsilon +1\right)-264 \epsilon ^4+600 \epsilon ^3-482 \epsilon ^2+181 \epsilon -27}{64 (\epsilon -1)^3 \left(6 \epsilon ^2-5 \epsilon +1\right)}\,,\\&\boldsymbol{\tilde h}_{108}=-\frac{1}{\left.64 \left(\gamma ^2-1\right) (\epsilon -1)^3 \epsilon  (\epsilon  (6 \epsilon -5)+1)\right)}\Bigg[(2 \epsilon +1) \big(6 \gamma ^2 \left(64 \epsilon ^4-174 \epsilon ^3+169 \epsilon ^2-68 \epsilon +9\right)\\&-4 \gamma  (\epsilon -1)^2 \left(24 \epsilon ^2-10 \epsilon +1\right)+480 \epsilon ^4-932 \epsilon ^3+648 \epsilon ^2-215 \epsilon +27\big)\Bigg]\,,\\&\boldsymbol{\tilde h}_{109}=\frac{1}{64 \gamma  \left(\gamma ^2-1\right) (\epsilon -1) (2 \epsilon -1) (3 \epsilon -1)}\Bigg[\gamma ^5 \left(36 \epsilon ^3-62 \epsilon ^2+29 \epsilon -4\right)+8 \gamma ^4 \left(14 \epsilon ^2-11 \epsilon +2\right)\\&+\gamma ^3 \left(108 \epsilon ^3-106 \epsilon ^2+25 \epsilon -1\right)-8 \gamma ^2 \left(36 \epsilon ^3-19 \epsilon ^2-5 \epsilon +2\right)+24 \epsilon  (1-3 \epsilon )\Bigg]\,,\\&\boldsymbol{\tilde h}_{110}=\frac{\gamma ^3 \left(-12 \epsilon ^2+18 \epsilon -5\right)+\gamma ^2 (8-16 \epsilon )+24 \epsilon -8}{128 \gamma  (\epsilon -1) (2 \epsilon -1) (3 \epsilon -1)}\,,\\
&\boldsymbol{\tilde h}_{111}=-\frac{(2 \epsilon +1) \left(\gamma ^3 \left(24 \epsilon ^2-24 \epsilon +5\right)-8 \gamma ^2 \left(6 \epsilon ^2-6 \epsilon +1\right)-24 \epsilon +8\right)}{128 \gamma  \left(\gamma ^2-1\right) (\epsilon -1) \epsilon  (2 \epsilon -1) (3 \epsilon -1)}\,,\\
\nonumber
     \end{split}
\end{align}
\begin{align}
\begin{split}
&\boldsymbol{\tilde h}_{112}=\frac{\gamma  \left(-\left(\gamma ^2 (\epsilon -1) (12 \epsilon -5)\right)-5 \epsilon +2\right)}{12 \left(\gamma ^2-1\right) (\epsilon -1)^2}\,,\boldsymbol{\tilde h}_{113}=\frac{\gamma  \left(\gamma ^2 (6 (3-2 \epsilon ) \epsilon -5)-6 \epsilon +2\right)}{96 \left(\gamma ^2-1\right) (\epsilon -1)^2}\,,\\& \boldsymbol{\tilde h}_{114}=\frac{1}{256(\gamma -1)^2\gamma (\gamma +1)^2(\epsilon -1)^2\left(6 \epsilon ^2-5 \epsilon +1\right)}\Bigg[-64 \gamma ^8 (\epsilon -1)^2 \epsilon ^2-32 \gamma ^7 (\epsilon -1)^2 \epsilon  \left(4 \epsilon ^2+4 \epsilon -3\right)\\&+\gamma ^6 \left(160 \epsilon ^5+1872 \epsilon ^4-5282 \epsilon ^3+4581 \epsilon ^2-1490 \epsilon +159\right)+16 \gamma ^5 \left(16 \epsilon ^5-300 \epsilon ^4+556 \epsilon ^3-347 \epsilon ^2+83 \epsilon -8\right)\\&-2 \gamma ^4 \left(80 \epsilon ^5-3816 \epsilon ^4+8202 \epsilon ^3-5819 \epsilon ^2+1453 \epsilon -120\right)-32 \gamma ^3 \left(4 \epsilon ^5-112 \epsilon ^4+213 \epsilon ^3-149 \epsilon ^2+39 \epsilon -4\right)\\&+\gamma ^2 \left(800 \epsilon ^4-2506 \epsilon ^3+1597 \epsilon ^2-184 \epsilon -27\right)+16 \gamma  \epsilon  \left(68 \epsilon ^3-144 \epsilon ^2+69 \epsilon -11\right)-4 \left(22 \epsilon ^2+7 \epsilon -3\right)\Bigg]\,,\\& \boldsymbol{\tilde h}_{115}=\frac{1}{1536 (\gamma -1) \gamma  (\gamma +1) (\epsilon -1)^2 \epsilon  (2 \epsilon -1) (3 \epsilon -1)}\Bigg[192 \gamma ^6 (\epsilon -1)^2 \epsilon ^2+96 \gamma ^5 (\epsilon -1)^2 \epsilon  \left(4 \epsilon ^2+4 \epsilon -3\right)\\&+4 \left(22 \epsilon ^2+7 \epsilon -3\right)-48 \gamma  \epsilon  \left(28 \epsilon ^3-48 \epsilon ^2+23 \epsilon -3\right)+\gamma ^2 \left(-1440 \epsilon ^4+2762 \epsilon ^3-1733 \epsilon ^2+292 \epsilon -1\right)\\&-48 \gamma ^3 \epsilon  \left(8 \epsilon ^4-36 \epsilon ^3+34 \epsilon ^2-3 \epsilon -3\right)+\gamma ^4\left(-944 \epsilon ^4+4078 \epsilon ^3-3551 \epsilon ^2+896 \epsilon -480 \epsilon +1\right)\Bigg]\,,\\&\boldsymbol{\tilde h}_{116}=-\frac{1}{1536 (\gamma -1)^2 \gamma  (\gamma +1)^2 (\epsilon -1)^2 \epsilon ^2 \left(6 \epsilon ^2-5 \epsilon +1\right)}\Bigg[\left(2 \epsilon +1\right)\big(192 \gamma ^6 (\epsilon -1)^2 \epsilon ^2\\&+96 \gamma ^5 \epsilon  \left(4 \epsilon ^4-28 \epsilon ^3+43 \epsilon ^2-22 \epsilon +3\right)-\gamma ^4 \left(480 \epsilon ^5-5360 \epsilon ^4+9646 \epsilon ^3-5663 \epsilon ^2+896 \epsilon +1\right)\\&-48 \gamma ^3 \epsilon  \left(8 \epsilon ^4-36 \epsilon ^3+46 \epsilon ^2-25 \epsilon +3\right)+\gamma ^2 \left(480 \epsilon ^4-1850 \epsilon ^3+1493 \epsilon ^2-292 \epsilon +1\right)\\&+48 \gamma  \epsilon  \left(20 \epsilon ^3-40 \epsilon ^2+19 \epsilon -3\right)-4 \left(22 \epsilon ^2+7 \epsilon -3\right)\big)\Bigg]\,,\\&\boldsymbol{\tilde h}_{117}=\frac{4 (\epsilon -3)-\gamma ^2 (\epsilon -1) \left(\gamma ^2 \left(320 \epsilon ^2-418 \epsilon +97\right)-336 \epsilon ^2+466 \epsilon -97\right)}{768 \left(\gamma ^2-1\right) (\epsilon -1)^2 (3 \epsilon -1)}\,,\\&
\boldsymbol{\tilde h}_{118}=\frac{\epsilon  (2 \epsilon +1) \left(\gamma ^2 (\epsilon -1) (2 \epsilon -1)-\epsilon  (2 \epsilon +5)+1\right)}{2 (\gamma ^2-1) (\epsilon -1) (2 \epsilon -1) (3 \epsilon -1)}\,,\boldsymbol{\tilde h}_{119}=\frac{(2 \epsilon +1)}{4-12 \epsilon }\,,\\&\boldsymbol{\tilde h}_{120}=\frac{(2 \epsilon +1) (\epsilon  (4 (\epsilon -1) \epsilon +5)-1)}{4 (\gamma^2 -1)  (\epsilon -1) \epsilon  (2 \epsilon -1) (3 \epsilon -1)}\,.
\nonumber
   \end{split}
\end{align}
    }%
  }%

  \restorethesisbodyformat
  \restoremainchapterstyle
  \chapter{Heterotic Footprints in Classical Gravity}
  \thesischapterpaperbox{Heterotic Footprints in Classical Gravity: PM dynamics from On-Shell soft amplitudes at one loop}{A.~Bhattacharyya, S.~Ghosh, A.~Mishra, and S.~Pal}{\href{https://doi.org/10.1007/JHEP06(2026)008}{\textit{JHEP} \textbf{06} (2026), 008}, arXiv: \arxivlink{2510.07390}{hep-th}}
  {%
    \restorethesisbodyformat
    \renewcommand{\appendix}{%
      \setcounter{section}{0}%
      \setcounter{subsection}{0}%
      \setcounter{subsubsection}{0}%
      \renewcommand{\thesection}{\thechapter.\Alph{section}}%
      \renewcommand{\thesubsection}{\thesection.\arabic{subsection}}%
      \renewcommand{\theHsection}{chapter.\arabic{chapter}.appendix.\Alph{section}}%
      \renewcommand{\theHsubsection}{\theHsection.\arabic{subsection}}%
    }%
    \ifstrempty{chap4_body.tex}{}{%
      \section{Introduction}
The analysis presented in this chapter is based on Ref.~\cite{Bhattacharyya:2025heterotic}.
{\color{black}
In this chapter we take a complementary approach to the classical two-body scattering problem, namely the scattering-amplitude approach. In the previous chapter, we saw that infrared divergences can arise in the computation of the eikonal phase within a bottom-up effective field theory, formulated in the worldline quantum field theory framework. A common expectation is that such infrared divergences cancel in physical observables once the dissipative contributions are included and the appropriate causal \(i0\) prescription is imposed.

The central question we address here is how robust this infrared finiteness really is. In particular, we would like to understand whether the cancellation of infrared divergences is a consequence of the specific formalism used, or whether it reflects a more general property of the underlying theory. The primary goal of this chapter is therefore to revisit the problem using amplitude methods, rather than the WQFT formalism, and to examine systematically how the infrared subtraction works in this complementary language. 

If, after a systematic subtraction of the known infrared contributions, residual infrared divergences still remain, there are two possible interpretations. One possibility is that the formalism or subtraction procedure is incomplete, and that additional conservative or radiative contributions have not been properly included. The other possibility is more fundamental: the persistence of such divergences may point to a genuine issue in the theory itself, or at least to a limitation in the way the classical observable is being defined. Thus, by comparing the amplitude-based analysis with the WQFT results, we aim to test the robustness of infrared finiteness and clarify whether the cancellation of infrared divergences is formalism-dependent or physically universal.  To sharpen this test, we consider a top-down effective field theory rather than a purely bottom-up model. In particular, we study Einstein--Maxwell--Dilaton theory, which arises as a low-energy effective theory of heterotic string theory. This choice provides a setting in which the interactions are not introduced arbitrarily, but are instead constrained by an ultraviolet completion. It therefore offers a useful arena to investigate whether the cancellation, or persistence, of infrared divergences is a formalism-dependent feature or a more robust property of the underlying theory.
} Therefore, building on the previous chapter, where we studied black-hole scattering in a higher-curvature, parity-violating deformation of GR, we now turn to a different class of beyond-GR effects: additional long-range fields and charges arising from a string-motivated low-energy effective theory. The central example in this chapter is Einstein--Maxwell--Dilaton (EMD) gravity, which provides a  setting in which to track how electromagnetic and dilatonic charges modify conservative scattering observables. As in the previous chapters, the broader motivation comes from precision gravitational-wave physics, where accurate gauge-invariant information about the conservative dynamics is needed to test departures from GR \cite{LIGOScientific:2016aoc,LIGOScientific:2016sjg,LIGOScientific:2016vlm,LIGOScientific:2017bnn,LIGOScientific:2019hgc,Purrer:2019jcp}.

From the effective-theory viewpoint, the origin of EMD is especially attractive because its extra fields are not introduced ad hoc. In the low-energy limit of heterotic string theory, the massless sector necessarily contains the graviton \(g_{\mu\nu}\), dilaton \(\theta\), antisymmetric tensor \(B_{\mu\nu}\), and non-Abelian gauge fields \(A_\mu^{I}\). For backgrounds that are independent of \textcolor{black}{\(d\) of the spacetime directions} and for gauge configurations commuting with \(p\) Abelian generators, the space of classical solutions enjoys a continuous \(O(d)\times O(d+p)\) symmetry that mixes \(g_{\mu\nu}\), \(B_{\mu\nu}\), \(\theta\), and \(A_\mu^{I}\) and acts as a solution-generating transformation \cite{Hassan:1991mq}. Acting with this “twist,” one can, for example, start from a ten-dimensional black 6-brane carrying magnetic charge but no electric or antisymmetric-tensor gauge charge and generate a family of inequivalent solutions with independent electric, magnetic, and antisymmetric-tensor charges \cite{Hassan:1991mq}. In this framework, additional degrees of freedom and higher-derivative corrections arise in tandem rather than as ad hoc additions, yielding symmetry-guided targets for EFT analyses and phenomenology. After restricting to the \(U(1)\) subsector, truncating the antisymmetric tensor, and transforming to Einstein frame, one arrives at the EMD theory used in this chapter; the resulting low-energy effective theory admits exact charged (and spinning) black-hole solutions \cite{GIBBONS1988741,Garfinkle:1990qj,Sen:1992ua}, making it a natural arena for probing possible stringy imprints in two-body dynamics.

For the hyperbolic and high-velocity regime, the post-Minkowskian description is the natural language, and scattering-amplitude methods provide an efficient way to isolate the classical conservative information \cite{Damour:2016gwp,Bini:2017wfr,Bini:2017xzy,Damour:2017zjx,Damour:2019lcq,Bini:2020flp,Bini:2020uiq,Damour:2020tta,Bini:2020rzn,Bini:2021gat,Bini:2022enm,Damour:2022ybd,Bini:2022wrq,Rettegno:2023ghr,Bini:2023fiz,Brandhuber:2022qbk,Brandhuber:2023hhy,Kosower:2018adc,DeAngelis:2023lvf,Cristofoli:2021vyo}. In this chapter we therefore adopt the scattering-amplitude approach to compute the conservative two-body potential, the eikonal phase, and consequently the scattering angle through one loop. Following \cite{Parra-Martinez:2020dzs}, we organize the calculation in terms of soft amplitudes and combine this with the Lippmann--Schwinger analysis of the potential, including the required EFT/Born subtraction of long-range iterations so that the final momentum-space potential is infrared finite. The chapter is organized as follows. In section~\eqref{ch4:sec2}, we motivate Einstein--Maxwell--Dilaton (EMD) theory as the low-energy effective description of the heterotic string, fix conventions and gauge choices, and derive the propagators and minimal set of on-shell Feynman rules required for $2\!\to\!2$ scattering. In section~\eqref{ch4:sec3}, we develop the analytic toolkit for the soft regime: using dimensional regularization, expansion by regions, and integration-by-parts (IBP) reduction, we obtain a compact basis of master integrals and their soft-limit expansions that capture the classical contributions. With these ingredients, we assemble the complete one-loop amplitudes across the relevant topologies (single exchanges, triangles, boxes, penguins) in the classical limit. In section~\eqref{ch4:sec4}, we extract the conservative two-body potential via the Lippmann-Schwinger equation and demonstrate that, after a careful EFT/Born subtraction of long-range iterations, the momentum-space potential is infrared finite. Section~\eqref{ch4:sec5} then determines the eikonal phase by exponentiating the momentum-space amplitude and, consequently, the scattering angle through one loop; we verify the smooth GR limit and delineate the separate roles of electromagnetic and dilatonic charges.

\section{Setup and methodology}\label{ch4:sec2}
\paragraph*{Low energy effective action of Heterotic string and emergence of EMD theory.}
The low-energy effective action of Heterotic string theory in string frame is given by \cite{Metsaev:1987zx,Hassan:1991mq},
\begin{align}
    \begin{split}
        S_{\textrm{low-eff}}=\int d^4x \sqrt{-G}\,e^{-\theta}\left[-R+\frac{1}{12}H_{\mu\nu\rho}H^{\mu\nu \rho}+G^{\mu\nu}\partial_{\mu}\theta \partial_{\nu}\theta-\frac{1}{8}F^2\right]
    \end{split}
\end{align}
    where $G_{\mu\nu}$ is the metric, $F_{\mu\nu}=\partial_\mu A_{\nu}-\partial_{\nu}A_{\mu}$ is the field strength corresponding to the Maxwell field $A_{\mu}$, and $\theta$ is the dilaton field. Moreover, $H_{\mu\nu\rho}=\partial_{\mu} B_{\nu\rho}+\textrm{cyclic per.}-(\Omega_3(A))_{\mu\nu\rho}$, where $B_{\mu\nu}$ is the antisymmetric tensor field and $(\Omega_3(A))_{\mu\nu\rho}=\frac{1}{4}(A_{\mu}F_{\nu\rho}+\textrm{cyclic per.})$ is the gauge Chern-Simons term. The theory has several noteworthy features. First, it arises from heterotic string theory compactified from ten to four dimensions. In the compactification procedure, the massless fields appearing in the effective action arise naturally. Also, only the $U(1)$ subsector of the full $E_{8}\times E_{8}$ or $\texttt{Spin}(32)/\mathbb{Z}_2$ theory has been retained. The metric $G_{\mu\nu}$, which naturally appears in the $\sigma$-model, is conformally related to the Einstein metric $g_{\mu\nu}$. Lastly, the action has been truncated at second order in derivatives of the fields. Now, ignoring the antisymmetric field $H_{\mu\nu\rho}$ and using an appropriate conformal field redefinition, the effective action takes the following form (restoring the factors of $m_p$),
    \begin{align}
        \begin{split}
              S_{\textrm{low-eff}}=\int d^4 x \sqrt{-g}\left[-\frac{m_p^2}{2}R+\frac{1}{2}g^{\mu\nu}\partial_{\mu}\theta\,\partial_\nu \theta-\frac{1}{16}\,e^{-\sqrt{2}\frac{\theta}{m_p}}F^2\right]\,.
        \end{split}
    \end{align}
The dilaton field $\theta$ plays a central role in the theory as its vacuum expectation value is related to the coupling of the theory as $\texttt{g}_{c}=e^{\sqrt{2}\langle\theta\rangle}$. The charged black hole solution (in unit of $G_N=1$) takes the following form \cite{Garfinkle:1990qj,GIBBONS1988741},
\begin{align}
\begin{split}
   & ds^2=\left(1-\frac{2M}{r}\right)dt^2-\left(1-\frac{2M}{r}\right)^{-1}dr^2-r\left(r-\frac{Q^2e^{-\sqrt{2}\theta_0}}{M}\right)d\Omega\,,
  \\ & e^{-\sqrt{2}\theta}=e^{-\sqrt{2}\theta_0}\left(1-\frac{Q^2\, e^{-\sqrt{2}\theta_0}}{Mr}\right)\,.
  \end{split}
\end{align}
Now, expanding the action around $\theta=\langle \theta\rangle:=\theta_0$ we  get,
     \begin{align}
        \begin{split}
              S_{\textrm{low-eff}}&=\int d^4 x \sqrt{-g}\left[-\frac{m_p^2}{2}R+\frac{1}{2}g^{\mu\nu}\partial_{\mu}\delta\theta\,\partial_\nu \delta\theta-\frac{1}{16}e^{-\sqrt{2}\theta_0/m_p}\,e^{-\sqrt{2}\frac{\delta\theta}{m_p}}F^2\right]\,, \\ &
              =\int d^4 x \sqrt{-g}\left[-\frac{m_p^2}{2}R+\frac{1}{2}g^{\mu\nu}\partial_{\mu}\delta\theta\,\partial_\nu \delta\theta-\frac{1}{16 \,\texttt{g}_{c}}\,e^{-\sqrt{2}\frac{\delta\theta}{m_p}}F^2\right]\,. \label{1.3h}
        \end{split}
    \end{align}
The theory described in \eqref{1.3h} has exact charged blackhole solution. The main goal of the paper is to study blackhole scattering and investigate how string theory modifies the fundamental classical observables such as the scattering angle (and the classical potential). As a standard procedure, we model the black holes by a charged scalar field $\Phi$ with dilaton dependent effective mass ($m_i(\theta)$)  , and the matter action is given by,
\begin{align}
    \begin{split}
        S_{\textrm{matter}}=\int d^4x \sqrt{-g}\sum_{i=1}^2\left[g^{\mu\nu}D_{\mu}\Phi^{\dagger}_iD_{\nu}\Phi_i-m_i^2(\theta)\Phi^{\dagger}_i\Phi_i\right]
    \end{split}
\end{align}
where the covariant derivative $D_{\mu}=\partial_{\mu}-ie\alpha_iA_{\mu}$ encodes the information about the gauge field $A_{\mu}$ and the electric charge of the scalar field $\Phi$, \textcolor{black}{ $Q_i=\alpha_i\,e$}.
Note that in the presence of the dilaton, the mass of the scalar field is now a function of the dilaton $\theta$. For computational purposes, the mass can be expanded around the background/asymptotic value of dilaton, i.e. $\theta_0$ as,
\begin{align}
    m_i(\theta)=m_i(\theta_0)+a_i(\theta_0)\,\delta\theta+b_i (\theta_0)\delta\theta^2+\cdots.
\end{align}
For notational simplicity, we use $\delta\theta\to \theta$, for further computation and define the path integral as, $$\int D\delta\theta\, D[A,g,\Phi_i]\cdots\to \int D\theta \,D[A,g,\Phi_i]\cdots\,. $$
\paragraph*{Propagators and Feynman rules.}
The two-point function for the Maxwell field will be modified due to the presence of string coupling $\texttt{g}_c$, but the two-point function of the dilaton and graviton remains unchanged. The gauge fixed action for the Maxwell field is given by,
\begin{align}
    \begin{split}
S_{\texttt{photon}}=\frac{1}{4\texttt{g}_c}\int d^4x \left(-\frac{1}{4}F^2-\frac{1}{2\xi}(\partial\cdot A)^2\right)\,.
    \end{split}
\end{align}
Immediately, we can identify the momentum space propagator in Feynman gauge as,
\begin{align}
\begin{split}
 \hspace{0cm}  D_{\mu\nu}(k)=\thesisinlinefeynman[0.075\linewidth]{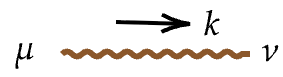}\hspace{0.45cm}=
  4\texttt{g}_c\frac{-i\,\eta_{\mu\nu}}{k^2+i0}.
    \end{split}
\end{align}
Similarly the graviton\footnote{We define $P_{\mu\nu;\rho\sigma}=\left(\frac{1}{2}\eta_{\mu\rho}\eta_{\nu\sigma}+\frac{1}{2}\eta_{\mu\sigma}\eta_{\nu\rho}-\frac{1}{d-2}\eta_{\mu\nu}\eta_{\rho\sigma}\right)$.} and dilaton propagator takes the form,

\begin{align}
\begin{split}
  & 
 D_{\mu\nu;\rho \sigma}(k)= \thesisinlinefeynman[0.075\linewidth]{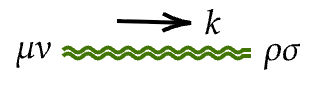}\hspace{0.55cm}=\frac{i}{k^2+i0} \left(\frac{1}{2}\eta_{\mu\rho}\eta_{\nu\sigma}+\frac{1}{2}\eta_{\mu\sigma}\eta_{\nu\rho}-\frac{1}{d-2}\eta_{\mu\nu}\eta_{\rho\sigma}\right),\,\\ &
    D(k)=     \thesisinlinefeynman[0.075\linewidth]{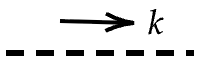}\hspace{0.45cm}=\frac{i}{k^2+i0}.
  \end{split}
\end{align}

Now, expanding the bulk gravitational and matter actions about the background solution and truncating at cubic order yields the three-point interaction vertices needed for our analysis. We enumerate them below.
{\scriptsize
\textbf{\textit{Bulk-Matter interaction vertex:}}
\vspace{0 cm}
\begin{align}
    \begin{split}
      &\hspace{0cm}     \bullet\,\,\thesisinlinefeynman[0.075\linewidth]{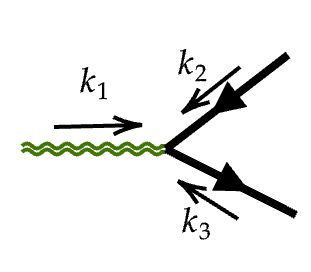}\hspace{0.35cm}
\equiv 
\hspace{0 cm}\,\frac{i}{m_p}\left(k_3^{(\mu}k_2^{\nu)}-\frac{1}{2}\left(k_2\cdot k_3+m_0^2\right)\eta^{\mu\nu}\right),
\bullet\,\, \thesisinlinefeynman[0.075\linewidth]{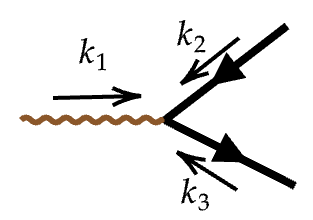}\hspace{0.35cm}
\equiv 
\hspace{0 cm}\, i\,e\, \alpha_i\,(k_3^\mu-k_2^\mu),\\ &
 \hspace{0cm}     \,\,\bullet\,\,\thesisinlinefeynman[0.075\linewidth]{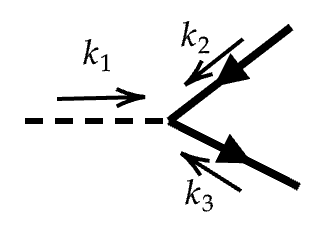}\hspace{0.35cm}
\equiv 
\hspace{0 cm}\, -i \frac{2m_0 a}{m_p},\,\bullet\, \thesisinlinefeynman[0.075\linewidth]{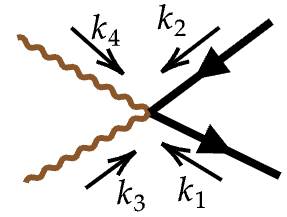}\hspace{0.35cm}
\equiv 
\hspace{0 cm}\,2 {ie}^2 \eta ^{\mu \nu },\,\,\bullet\,\hspace{0cm} \thesisinlinefeynman[0.075\linewidth]{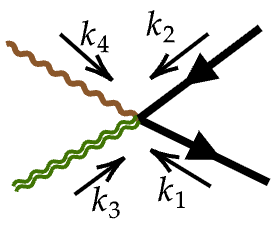}\hspace{0.35cm}
\equiv 
\hspace{0 cm}\,\frac{i e }{m_p}P^{\mu \nu ,\alpha \beta }(k_1-k_2)_{\beta },
    \end{split}
\end{align}
\begin{align}
    \begin{split}
    \\ &
\hspace{0cm}\bullet\, \thesisinlinefeynman[0.075\linewidth]{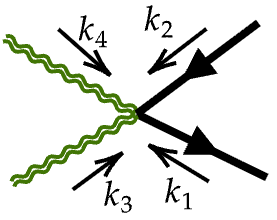}\hspace{0.35cm}
\equiv 
\hspace{0 cm}\,-\frac{i}{2 m_p^2}\Bigg[k_1^{\rho } k_2^{\sigma } \Big(-\eta _{\mu \nu } I_{\alpha \beta ,\rho \sigma }-\eta _{\alpha \beta } I_{\mu \nu ,\rho \sigma }+\eta _{\beta \nu } I_{\alpha \mu ,\rho \sigma }\\ &\hspace{0.8cm}+\eta _{\alpha \mu } I_{\beta \nu ,\rho \sigma }+\eta _{\beta \mu } I_{\alpha \nu ,\rho \sigma }+\eta _{\alpha \nu } I_{\beta \mu ,\rho \sigma }\Big)-P_{\mu \nu ,\alpha \beta }(k_1 \cdot k_2+m^2)\Bigg],\\ &
        \hspace{0cm}\bullet \thesisinlinefeynman[0.075\linewidth]{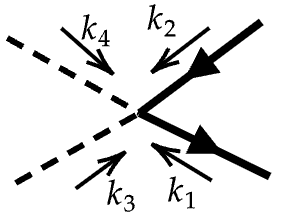}\hspace{0.35cm}
\equiv 
\hspace{0 cm}\,\frac{2 i \left(a^2-2 m_i\,b\right)}{m_p^2},\,\,\bullet \thesisinlinefeynman[0.075\linewidth]{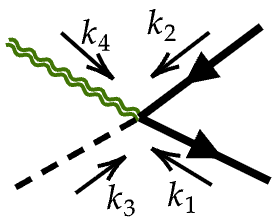}\hspace{0.35cm}
\equiv 
\hspace{0 cm}\,-\frac{i\,a  m_i \eta ^{\mu \nu }}{m_p^2}.
    \end{split}
\end{align}
\textbf{\textit{Bulk  interaction vertex:}}
\begin{align}
    \begin{split}
      &\hspace{0 cm}     \bullet\,\,\thesisinlinefeynman[0.075\linewidth]{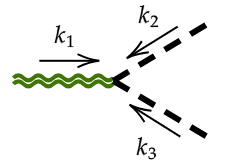}\hspace{0.35cm}
\equiv 
\hspace{0 cm}\,\frac{i}{4m_p}\left( k_1^\mu k_2^\nu + k_1^\nu k_2^\mu - k_1 \cdot k_2 ~\eta^{\mu \nu} \right)  ,
\\ &
\,\bullet\,\,\thesisinlinefeynman[0.075\linewidth]{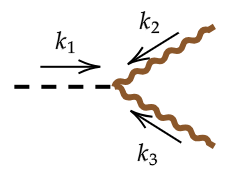}\hspace{0.35cm}
\equiv 
\hspace{0 cm}\, -i\frac{2\sqrt{2} }{16 \texttt{g}_c m_p}\left(\eta^{\mu\nu} k_2 \cdot k_3-k_2^{(\mu}k_3^{\nu)}\right),
\\ &
 \hspace{0 cm}     \bullet\,\,\thesisinlinefeynman[0.075\linewidth]{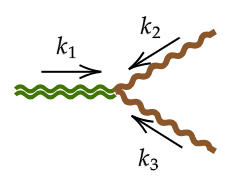}\hspace{0.35cm}
\equiv 
\hspace{0 cm}\,  -\frac{i  }{4\texttt{g}_c m_p}\Bigg[~  (k_2 \cdot k_3 ) P^{\alpha \beta , \mu \nu} +\eta^{\mu \nu} k_2^{ (\alpha} k_3^{\beta)}  
 + 
\frac{1}{2}\Big( \eta^{\alpha \beta}k_2^{\mu}k_3^{\nu} -\eta^{\alpha \mu}k_2^{\beta}k_3^{\nu} - \eta^{\alpha \nu}k_2^{\mu}k_3^{\beta}\\&\hspace{0.8cm} -  \eta^{\beta \mu}k_2^{\alpha}k_3^{\nu} - \eta^{\beta \nu}k_2^{\mu}k_3^{\alpha} \Big)    \Bigg].     
    \end{split}
\end{align}
}
With all the ingredients in hand, we now go on to the computation of the scattering amplitude at the soft limit and discuss how this particular limit of the full scattering amplitude is enough to generate the classically relevant observables. However, we also show that at one loop it can also generate \textit{some part} of the quantum correction to the classical potential. 
\section{Computation of scattering amplitude: Soft-loop expansion and classical limit}\label{ch4:sec3}
We first focus on the computation of the scattering amplitude in the soft limit \cite{Parra-Martinez:2020dzs} and then we go on to the computation of the Post-Minkowskian potential. A general $L$-loop, $n$-point amplitude takes the following form,
\begin{align}
\begin{split}
i \mathbfcal{A}_{n}^{L-\textrm{loop}}\left[\thesisampdiagram[0.15\linewidth]{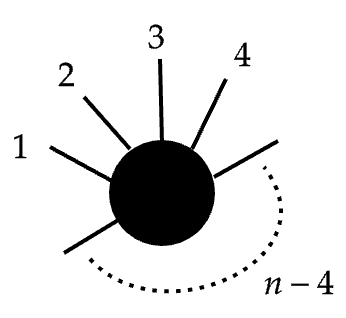}\right]=\sum_{j}\int \left(\prod_{l=1}^L\frac{d^d\ell_{l}}{(2\pi)^D}\right) \frac{1}{S_{j}}\frac{\mathcal{N}_{j}}{\prod_{\gamma_j}\rho_{\gamma_j}}
      \end{split}
\end{align}
where the sum runs over all possible L-loop Feynman diagrams $j$, for each diagram $\ell_{l}$ denotes the L loop momenta, $\gamma_j$ labels the propagators, and $S_j$ is associated with the symmetry factor of the diagram. $n_j's$ are the polynomials constructed from the irreducible scalar products constructed from external and loop momenta. In the subsequent computations, we will focus on the one-loop, four-point scattering amplitude, $\mathbfcal{A}_{4}^{1-\textrm{loop}}$, which is relevant for a binary scattering event. At the classical limit, any one-Loop, four-point amplitude can be written as,
\begin{align}
\begin{split}
   \mathbfcal{A}_{4}^{1-\textrm{loop}}={\mathfrak{c}_{\square}\mathcal{A}_{\square}+\mathfrak{c}_{\Join}\mathcal{A}_{\Join}+\mathfrak{c}_{\triangle}\mathcal{A}_{\triangle}+\mathfrak{c}_{\triangledown}\mathcal{A}_{\triangledown}+(\textrm{purely quantum})}\,.
   \end{split}
\end{align}
After integrating, it takes the form,
\begin{align}
     \mathbfcal{A}_{4}^{1-\textrm{loop}}=A+B \,q^2+\cdots+\frac{\alpha}{\sqrt{-q^2}}+\beta \log(\sqrt{-q^2})+\cdots\,.
\end{align}
Interestingly, it is trivial to show (by taking a suitable Fourier transform) that the non-analytic piece of the amplitude will contribute to the long-range interaction, which is relevant for classical physics. 
\\
\textcolor{black}{
\textbf{Soft expansion and the classical regime:}
Before turning to the detailed computations, we briefly review how the soft expansion emerges and how it encodes the classical limit of the amplitude \cite{Neill:2013wsa,Parra-Martinez:2020dzs}.
For hyperbolic scattering of two compact objects, the classical regime is characterized by an orbital angular momentum that is parametrically large compared to $\hbar$ (Bohr correspondence principle),
\begin{align}
J \gg \hbar \qquad (\text{equivalently, setting }\hbar=1:\; J\gg 1)\, .
\end{align}
Physically, $J\gg 1$ corresponds to a large impact parameter compared to the de Broglie wavelength, so that the scattering process admits a semiclassical description.
In this limit, the momentum transfer $q$ is small compared to the hard external scales set by the Mandelstam invariants and the masses, and classical contributions arise from the non-analytic (long-distance) dependence on $t=q^2$. More precisely, the large-$J$ limit induces a hierarchy of scales,
\begin{align}
s,\; |u|,\; m_i^2 \sim J^2 |t| \gg |t| = |q|^2 \, ,
\end{align}
so that expanding the amplitude at large $J$ is equivalent to expanding in small $|q|$.
This is what we refer to as the \emph{soft expansion}: it is an expansion around small momentum transfer, keeping the hard external kinematics fixed.
Within this expansion, the classical pieces of the amplitude can be systematically isolated (for instance, as the leading terms in $|q|$ at fixed $s$ and masses). It is useful to distinguish three classical limits, depending on how relativistic the scattering is:
\begin{align}
\begin{split}
    &\textrm{Generic classical limit:} \qquad\qquad\qquad\;\; J \gg 1,\\
    &\textrm{Near-static (PN) classical limit:} \qquad\,\,\, J \gg 1,\;\; \sigma \sim 1,\\
    &\textrm{High-energy classical limit:} \qquad\qquad\;\; J \gg 1,\;\; \sigma \gg 1,
\end{split}
\end{align}
where the boost parameter,
\begin{align}
\sigma \equiv \frac{k_1\!\cdot\! k_2}{m_1 m_2}
\end{align}
measures the relative velocity: $\sigma\simeq 1$ corresponds to the near-static/post-Newtonian regime, while $\sigma\gg 1$ captures ultra-relativistic scattering, where we have a stricter scale hierarchy: $s,|u|\gg m_i^2\sim J^2|t|\gg |t|=|q|^2$.
Since our goal is the conservative sector of the dynamics, we further restrict to the \emph{instantaneous} (or potential) long-range kinematics for the momentum transfer,
\begin{align}
q=(q^0,\boldsymbol{q})\,,\qquad |\boldsymbol{q}| \gg q^0\, .
\end{align}
This ensures that the long-range interaction is effectively instantaneous (no on-shell radiation is exchanged), which is precisely the regime relevant for defining a conservative potential from amplitudes. With these scalings in mind, loop integrals are conveniently analyzed using the method of regions.
For an internal graviton (or dilaton) momentum $\ell=(\omega,\boldsymbol{\ell})$, one encounters four standard momentum regions, distinguished by how $(\omega,\boldsymbol{\ell})$ compares to the hard scale $m$ and the soft scale $|\boldsymbol{q}|$:
\begin{align}
\begin{split}
    &\text{hard:}\qquad\qquad\;\; (\omega,\boldsymbol{\ell}) \sim (m,m),\\
    &\text{soft:}\qquad\qquad\;\; (\omega,\boldsymbol{\ell}) \sim (|\boldsymbol{q}|,|\boldsymbol{q}|) 
    \sim J^{-1}(m|\boldsymbol{v}|,\,m|\boldsymbol{v}|),\\
    &\text{potential:}\qquad\;\; (\omega,\boldsymbol{\ell}) \sim (|\boldsymbol{q}|\,|\boldsymbol{v}|,\,|\boldsymbol{q}|)
    \sim J^{-1}(m|\boldsymbol{v}|^2,\,m|\boldsymbol{v}|),\\
    &\text{radiation:}\qquad (\omega,\boldsymbol{\ell}) \sim (|\boldsymbol{q}|\,|\boldsymbol{v}|,\,|\boldsymbol{q}|\,|\boldsymbol{v}|)
    \sim J^{-1}(m|\boldsymbol{v}|^2,\,m|\boldsymbol{v}|^2).
\end{split}
\end{align}
Here we use the reference mass scale $m=m_1+m_2$ and the typical relative velocity scale $|\boldsymbol{v}|$ (which is related to $\sigma$).
The hard region captures short-distance physics and generates local (analytic) contributions, while the soft, potential, and radiation regions encode long-distance physics controlled by the small momentum transfer.
In particular, the \emph{potential} region is the one relevant for conservative dynamics: it corresponds to off-shell exchange with $\omega \ll |\boldsymbol{\ell}|$, matching the instantaneous regime $|\boldsymbol{q}|\gg q^0$.
The \emph{radiation} region, by contrast, is associated with near on-shell modes and is responsible for dissipative effects; we will therefore exclude it when focusing on the conservative potential. Except for the hard region, the remaining regions are governed by two small parameters: the soft scale $|\boldsymbol{q}|\sim J^{-1}$ controlling the classical expansion, and the velocity (equivalently $\sigma$) controlling the non-relativistic/PN/near-static expansion.
This double expansion will be the organizing principle for the computations that follow.}
\\\\
\textbf{Box topologies:}
To compute the amplitude from the box topologies, we encounter the following family of master integrals:
\begin{align}
    G^{\square}_{\alpha_1\alpha_2\alpha_3\alpha_4}=\int \frac{d^D\ell}{(2\pi)^D}\frac{1}{\rho_1^{\alpha_1}\rho_2^{\alpha_2}\rho_3^{\alpha_3}\rho_4^{\alpha_4}}
\end{align}
with, $\rho_1=\ell^2,\,\rho_2=(\ell-q)^2,\,\rho_3=(\ell+k_1)^2-m_1^2, \rho_4=(\ell-k_2)^2-m_2^2$. Now, for convenience, we choose the following kinematics,
\begin{align}
    \begin{split}
        k_1=\bar m_1 u_1-\frac{q}{2}, \,k_2=\bar m_2 u_2+\frac{q}{2}\implies k_1'=\bar m_1 u_1+\frac{q}{2},\,k_2'=\bar m_2 u_2-\frac{q}{2},\,
    \end{split}
\end{align}
Note that, by construction, the momentum transfer $q$ is orthogonal to the velocities $\bar u_is$. We would like to expand the full integral in the  soft region (which is important to take the classical limit), which follows the following hierarchy of scales,
\begin{align}
    |\ell|\sim |q|\ll m_i, \sqrt{s}\to (m_1+m_2)\,.
\end{align}
Now expanding the denominators ($D_2, \,D_4$) in this above mentioned limit we will get,
\begin{align}
\begin{split}
    &\rho_3=\ell^2+2 k_1\cdot \ell+i\epsilon= \ell^2+2 (\bar m_1u_1-\frac{q}{2})\cdot \ell+i\epsilon \sim 2 \bar m_1 u_1\cdot \ell+i\epsilon\,,\\ &
    \rho_4=\ell^2-2 k_2\cdot \ell+i\epsilon= \ell^2-2 (\bar m_2u_2+\frac{q}{2})\cdot \ell+i\epsilon \sim -2 \bar m_2 u_2\cdot \ell+i\epsilon\,.
    \end{split}
\end{align}
We parametrize the velocities as follows,
\begin{align}
    u_1=(1,0,0,0), \,u_2=(\sqrt{1+v^2},0,0,v).
\end{align}
Now in the soft region, the integral family reduces to,
\begin{align}
    \begin{split}
        \mathbfcal{G}^{}
_{\gamma_1\gamma_2\gamma_3\gamma_4}=\int \frac{d^d\ell}{(2\pi)^d}\frac{1}{D_1^{\gamma_1}D_2^{\gamma_2}D_3^{\gamma_3}D_4^{\gamma_4}}
    \end{split}
\end{align}
where the new ISPs (irreducible scalar products) are,
\begin{align}
    D_{i}\equiv \Big(2\bar m_1 u_1\cdot \ell+i\epsilon, -2 \bar m_2u_2\cdot \ell+i\epsilon,\ell^2,(\ell-q)^2\Big)\,.
\end{align}
After solving the IBP identities, one will get the following master integrals:
\begin{align}
    \vec f(x,\epsilon)=\{\mathbfcal{G}_{0,0,1,1},\mathbfcal{G}_{0,1,1,1},\mathbfcal{G}_{1,0,1,1},\mathbfcal{G}_{1,1,1,1}\}
\end{align}
where, $x=\sigma-\sqrt{\sigma^2-1}$ with, \textcolor{black}{$\sigma=\frac{k_1\cdot k_2}{m_1m_2}$}. We solve the master integrals by solving the following differential equation:
\begin{align}
    \partial_{x}\vec f(x,\epsilon)=A(x,\epsilon)\vec f(x,\epsilon)
\end{align}
where, the matrix $A(x,\epsilon)$ (using \cite{Gituliar:2017vzm}) is given by (only scale is $|q|$, so we set $q^2\to -1$, which can be restored from dimensional analysis),
\begin{align}
    \begin{split}
        A(x,\epsilon)=\left(
\begin{array}{ccc}
 0 & 0 & 0 \\
 0 & 0 & 0 \\
 \frac{2 (1-2 \epsilon )}{(x-1) (x+1)} & 0 & \frac{-x^2-1}{(x-1) x (x+1)} \\
\end{array}
\right)\,.
    \end{split}
\end{align}
The main goal is to find the canonical basis $ \vec g(x,\epsilon)=\mathbb{T}^{-1}(x,\epsilon)\vec f(x,\epsilon)$ such that the new differential equation will be $\epsilon$-factorized,   
\begin{align}
    \begin{split}
\partial_{x} \vec g(x,\epsilon)=\mathbb{T}^{-1}(A \mathbb{T}-\partial_x \mathbb{T})\vec g(x,\epsilon)\equiv \epsilon\, \mathbb{S}(x) \vec g(x,\epsilon)\,.
    \end{split}
\end{align}
The transformation matrix $\mathbb{T}$ and the $\epsilon$-factorized matrix $\mathbb{S}$ are given by,
\begin{align}
    \begin{split}
        \mathbb{T}(x,\epsilon)=\left(
\begin{array}{ccc}
 \frac{9 \epsilon }{2 \epsilon -1} & 0 & 0 \\
 2 & 1 & 0 \\
 \frac{2 x}{x^2-1} & -\frac{x}{x^2-1} & -\frac{3 x}{x^2-1} \\
\end{array}
\right),\,\,\,\mathbb{S}(x)=\left(
\begin{array}{ccc}
 0 & 0 & 0 \\
 0 & 0 & 0 \\
 \frac{6  }{x} & 0 & 0 \\
\end{array}
\right)\,.
    \end{split}
\end{align}
Therefore, the differential equation becomes,
\begin{align}
  &  d\,\vec{g}(x,\epsilon)=\epsilon\, \widetilde {\mathbb{S}}\,d\log(x)\,\vec g(x,\epsilon),\, \widetilde {\mathbb{S}}=\left(
\begin{array}{ccc}
 0 & 0 & 0 \\
 0 & 0 & 0 \\
 6 & 0 & 0 \\
\end{array}\right)\,.
\end{align}
  Finally, in  the Laporta basis, the formal solution takes the form,
\begin{center}
\includegraphics[width=0.42\linewidth]{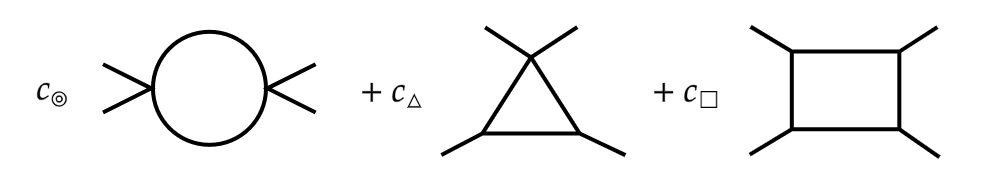}
\end{center}
Structurally, any one-loop soft-box contribution can be represented by the topology shown above.
To solve the master integrals, we need boundary conditions in the near-static ($\boldsymbol{v}\to 0$) limit.
Now we expand the soft-master integrals in the near static limit,
  \begin{align}
      \begin{split}
         \mathbfcal{G}^{\square}_{1,1,1,1}\xrightarrow[]{\boldsymbol{v}\to 0}\int_{-\infty}^{\infty} \frac{d\omega}{2\pi}\int \frac{d^{d-1}\ell}{(2\pi)^{d-1}} \frac{1}{ (2\bar m_1 \omega+i\epsilon)(-2\bar m_2 u_2^0\omega+2\bar m_2\boldsymbol{u_2}\cdot \boldsymbol{\ell}+i\epsilon)(\boldsymbol{-\ell^2}+i\epsilon)(\boldsymbol{-(\ell-q)^2}+i\epsilon)}\,.
      \end{split}
  \end{align}
Now integrating over $\omega$ (closing the contour in upper/lower half of $\omega$ plane) we get,
\begin{align}
    \begin{split}
       \hspace{0cm}\mathbfcal{G}^{\square}_{1,1,1,1}\xrightarrow[]{\boldsymbol{v}\to 0}&-\frac{i}{8\pi\bar m_1\bar m_2}\int \frac{d^d\boldsymbol{\ell}}{(2\pi)^{d-1}}\frac{1}{(\boldsymbol{u}_2\cdot \boldsymbol{\ell}+i\epsilon)(\boldsymbol{\ell}^2-i\epsilon)((\boldsymbol{\ell}-\boldsymbol{q})^2-i\epsilon)}\,,\\ &
        \to -\frac{1}{16\pi\bar m_1 \bar m_2}\int \frac{d^{d-1}\ell}{(2\pi)^{d-1}}\frac{\hat\delta(\boldsymbol{u}_2\cdot \boldsymbol{\ell})}{(\boldsymbol{\ell}^2-i\epsilon)((\boldsymbol{\ell-q})^2-i\epsilon)}\,,\\ &
        =-\frac{4^{\epsilon -2} \pi ^{\epsilon } \left(-q^2\right)^{-\epsilon -1} \Gamma (-\epsilon )^2 \Gamma (\epsilon +1)}{2\pi \bar{m}_1 \bar{m}_2 \Gamma (-2 \epsilon )}\,.
    \end{split}
\end{align}
Similarly, the near static limit of  $\mathbfcal{G}_{0,1,1,1}$ evaluates to,
\begin{align}
    \begin{split}
        \mathbfcal{G}^{\square}_{0,1,1,1}&\xrightarrow[]{v\to 0}-\frac{1}{4\pi\bar m_2}\int \frac{d^{d-1}\ell}{(2\pi)^{d-1}} \frac{1}{(\boldsymbol{\ell}^2-i\epsilon)[(\boldsymbol{\ell-\boldsymbol{q}})^2-i\epsilon]}\int_{-\infty}^{\infty}  \frac{d\omega}{\omega-\boldsymbol{u}_2\cdot \boldsymbol{\ell}-i\epsilon}\,,\\ &
        =-\frac{i\pi}{4\pi\bar m_2}\int 
    \frac{d^{d-1}\boldsymbol{\ell}}{(2\pi)^{d-1}}\frac{1}{(\boldsymbol{\ell}^2-i\epsilon)[(\boldsymbol{\ell-\boldsymbol{q}})^2-i\epsilon]}\,,\\ &
        =-\frac{i 16^{\epsilon -1} \pi ^{\epsilon +1} \left(-q^2\right)^{-\epsilon -\frac{1}{2}} \sec (\pi  \epsilon )}{2\pi\bar{m}_2 \Gamma (1-\epsilon )}
    \end{split}
\end{align}
and,
\begin{align}
    \begin{split}
        \mathbfcal{G}_{0,0,1,1}\xrightarrow[]{v\to 0}0\,.
    \end{split}
\end{align}
Now using (\ref{3.16}), the master integrals in the Laporta basis are given by,

\begin{center}
\begin{tcolorbox}[thesisresultbox, title=\textit{Master Integrals for box topologies}]
\restorethesisbodyformat

\begin{align}
    \begin{split}
     \hspace{0 cm}    &\mathbfcal{G}^{\square}_{0,0,1,1}(x,\epsilon)=0\,,\\&
        \mathbfcal{G}^{\square}_{0,1,1,1}(x,\epsilon)=- \textcolor{black}{\frac{1}{2 \pi}} \frac{i 16^{\epsilon -1} \pi ^{\epsilon +1} \sec (\pi  \epsilon )}{\bar{m}_2 \Gamma (1-\epsilon )}(-q^2)^{-\epsilon-\frac{1}{2}}\,,\\ &
        \mathbfcal{G}^{\square}_{1,0,1,1}(x,\epsilon)=- \textcolor{black}{\frac{1}{2 \pi}} \frac{i 16^{\epsilon -1} \pi ^{\epsilon +1} \sec (\pi  \epsilon )}{\bar{m}_1 \Gamma (1-\epsilon )}(-q^2)^{-\epsilon-\frac{1}{2}}\,,
        \\ &\mathbfcal{G}^{\square}_{1,1,1,1}(x,\epsilon)= \textcolor{black}{\frac{1}{2 \pi}} \frac{x\, 4^{2 \epsilon -1} \pi ^{\epsilon +\frac{3}{2}} \csc (\pi  \epsilon )}{\left(1-x^2\right) \bar{m}_1 \bar{m}_2 \Gamma \left(\frac{1}{2}-\epsilon \right)}(-q^2)^{-\epsilon-1}\,. 
    \end{split}
\end{align}
\end{tcolorbox}
\restorethesisbodyformat
\end{center}
\section*{Crossed-box topologies:}
In the crossed-box topologies, we encounter the following family of master integrals,
\begin{align}
    \begin{split}
        G^{\boxtimes}_{\gamma_1\gamma_2\gamma_3\gamma_4}=\int \frac{d^d\ell}{(2\pi)^d} \frac{1}{{\rho_1}^{\gamma_1}{\rho_2}^{\gamma_2}{\rho_4}^{\gamma_3}\tilde{\rho_4}^{\gamma_4}},
    \end{split}
\end{align}
with,  $\rho_1=\ell^2,\,\rho_2=(\ell-q)^2,\,\rho_3=(\ell+k_1)^2-m_1^2, \tilde\rho_4=(\ell+k_2-q)^2-m_2^2$. As mentioned earlier, we are interested in the soft/potential region of the amplitude. In the soft limit, the family reduces to,
\begin{align}
     \mathbfcal{G}^{\boxtimes}
_{\gamma_1\gamma_2\gamma_3\gamma_4}=\int \frac{d^d\ell}{(2\pi)^d}\frac{1}{\tilde D_1^{\gamma_1}\tilde D_2^{\gamma_2}\tilde D_3^{\gamma_3}\tilde D_4^{\gamma_4}}
\end{align}
where, 
$\tilde D_1=2\ell\cdot u_1+i\epsilon, \,\tilde D_2=2\ell\cdot u_2+i\epsilon,\, \tilde D_3=\ell^2,\,\tilde D_4=(\ell-q)^2. $ As we see in the crossed box topology, the two linearized matter propagators have the same sign, unlike the box integrals. As with the box topology, in this case also we have three master integrals: $\mathcal{G}_{0,0,1,1}^{\boxtimes},\,\mathcal{G}_{0,1,1,1}^{\boxtimes},\,\mathcal{G}_{1,1,1,1}^{\boxtimes}\,.$ Now, one will take a similar procedure as before to solve the integrals. However, one would notice that in the near static limit, the last will be zero.
\begin{align}\label{3.28} \mathbfcal{G}_{1,1,1,1}^{\boxtimes}=\frac{1}{4\bar{m}_1\bar{m}_2}\int \frac{d^{d-1}\boldsymbol{\ell}}{(2\pi)^{d-1}}\frac{1}{(\boldsymbol{\ell}^2-i\epsilon)\left[(\boldsymbol{\ell}-\boldsymbol{q})^2-i\epsilon\right]}\int \frac{d\omega}{2\pi} \frac{1}{(\omega+i\epsilon)(\omega+\boldsymbol{u}_2\cdot\boldsymbol{\ell}+i\epsilon)}\,.
\end{align}
The $\omega$ integral can be shown to be zero by closing the contour in the upper or lower half plane and computing the residues.
\begin{center}
\begin{tcolorbox}[thesisresultbox, title=\textit{Master Integrals for crossed-box topologies}]
\restorethesisbodyformat
\begin{centering}
\begin{align}
    \begin{split}
     \hspace{0 cm}    &\mathbfcal{G}^{\boxtimes}_{0,0,1,1}(x,\epsilon)=0\,,\\&
        \mathbfcal{G}^{\boxtimes}_{0,1,1,1}(x,\epsilon)=- \textcolor{black}{\frac{1}{2 \pi}} \frac{i 16^{\epsilon -1} \pi ^{\epsilon +1} \sec (\pi  \epsilon )}{\bar{m}_2 \Gamma (1-\epsilon )}(-q^2)^{-\epsilon-\frac{1}{2}}\,,\\ &
        \mathbfcal{G}^{\boxtimes}_{1,0,1,1}(x,\epsilon)=- \textcolor{black}{\frac{1}{2 \pi}} \frac{i 16^{\epsilon -1} \pi ^{\epsilon +1} \sec (\pi  \epsilon )}{\bar{m}_1 \Gamma (1-\epsilon )}(-q^2)^{-\epsilon-\frac{1}{2}}\,,\\&\mathbfcal{G}^{\boxtimes}_{1,1,1,1}(x,\epsilon)=0\,. 
    \end{split}
\end{align}
\end{centering}
\end{tcolorbox}
\restorethesisbodyformat
\end{center}
\noindent
Now we are ready to evaluate all the diagrams that contribute to the one-loop scattering amplitude in the soft limit. Below, we give the computational details. Before going to the one-loop computation we first give the results for tree level amplitude as it will play a crucial role to define IR finite potential.\\\\
\textbf{Tree-amplitude:}\\
\begin{center}
\includegraphics[width=0.36\linewidth]{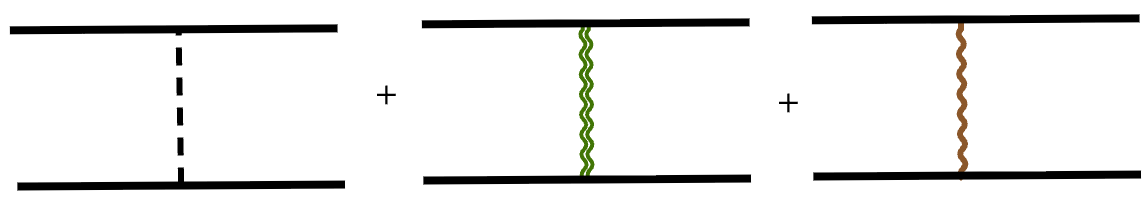}
\end{center}
\begin{align}
\begin{split}
&i\mathbfcal{A}_{4}^{\texttt{tree}}
\\ &=\frac{1}{q^2}\left(-\frac{4  a^2 m_1 m_2}{m_p^2}+16  \alpha _1 \alpha _2 \texttt{g}_ce^2 m_1 m_2 \sigma  +\frac{ m_1^2 m_2^2 \left(-2 \sigma ^2 +1\right)}{2 m_p^2}\right)
    +\left(4  \alpha _1 \alpha _2 \texttt{g}_c e^2 -\frac{ 2 m_2 m_1 \sigma   }{4 m_p^2}\right)\,.    \end{split}
\end{align}
\noindent
We now move on to the computation of one-loop amplitudes in the soft limit and, consequently, use them to compute the classical Post-Minkowskian potential and the conservative scattering angle. \\\\
\textbf{One-loop amplitudes: }\\\\
In this section, we derive the one-loop amplitudes for different topologies and give the necessary details.  Expressions for the numerators of all the Feynman diagrams are given in the ancillary file \texttt{numerators\_emd.nb} submitted to the \texttt{arXiv} along with this paper. 
We start with computing the box (+ cross-box) amplitude.\\\\
\textbf{Soft amplitudes from Box (cross box) topologies:}\\
\textbullet $\,\,$ The amplitude for the box (+cross-box) diagram with double photon exchange is given by,

\begin{flalign}
& i\mathbfcal{A}_{4,\textrm{soft}}^{1-\textrm{loop}}\left[\thesisampdiagram[0.40\linewidth]{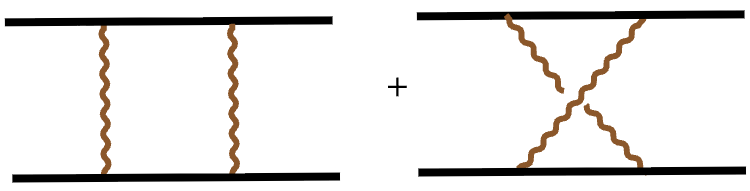}\right]=\int_{\ell}\frac{\mathcal{N}^{(1)}_{\filledsquare{gray}}}{\rho_1 \rho_2 \rho_3 \rho_4}\Bigg|_{\textrm{soft}} +\int_{\ell}\frac{\mathcal{N}^{(1)}_{\Join}}{\rho_1 \rho_2 \rho_3 \tilde\rho_4}\Bigg|_{\textrm{soft}} \nonumber \\
      & = \frac{\alpha _1^2 \alpha _2^2 e^4 m_1 m_2 \sigma ^2 16^{\epsilon +1} \pi ^{\epsilon +\frac{1}{2}} \texttt{g}_c^2 \left(-q^2\right)^{-\epsilon -1} \csc (\pi  \epsilon )}{\sqrt{\sigma ^2-1} \Gamma \left(\frac{1}{2}-\epsilon \right)}\,. &
\end{flalign}
\\ 
\textbullet $\,\,$Amplitude corresponding to the box diagram having photon and graviton exchange is given by,      

      \begin{align}
\begin{split}
&i\mathbfcal{A}_{4,\textrm{soft}}^{1-\textrm{loop}}\left[\thesisampdiagram[0.40\linewidth]{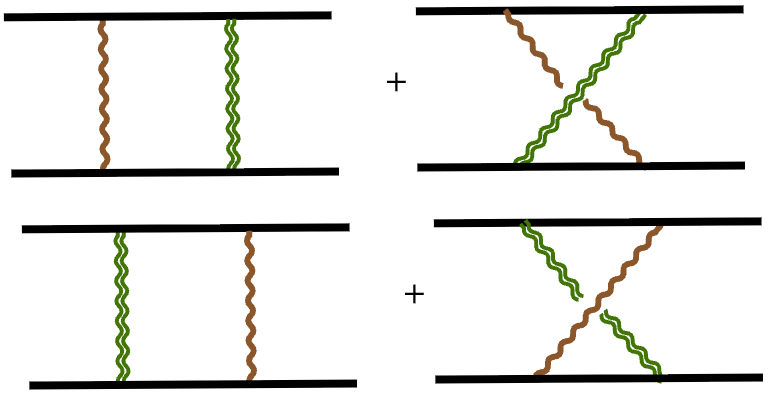}\right]=\sum_{i=a,c}\int_{\ell}\frac{\mathcal{N}^{(2i)}_{\filledsquare{gray}}}{\rho_1 \rho_2 \rho_3 \rho_4}\Bigg|_{\textrm{soft}}+\sum_{j=b,d}\int_{\ell}\frac{\mathcal{N}^{(2j)}_{\Join}}{\rho_1 \rho_2 \rho_3 \tilde\rho_4}\Bigg|_{\textrm{soft}}\\ &
      =    \alpha _1 \alpha _2 e^2 m_1 m_2 2^{4 \epsilon -3} \pi ^{\epsilon } \texttt{g}_c \left(-q^2\right)^{-\epsilon } \left(\frac{-8 \sqrt{\pi } m_1 m_2 \sigma  \left(2 \sigma ^2 (\epsilon -1)+1\right) \csc (\pi  \epsilon )}{-q^2 \sqrt{\sigma ^2-1} (\epsilon -1) m_p^2 \Gamma \left(\frac{1}{2}-\epsilon \right)} \right.\\
      & \left. -\frac{i \left(m_1+m_2\right) \left(6 \sigma ^2 (\epsilon -1)+2 \sigma  \epsilon +1\right) \sec (\pi  \epsilon )}{\sqrt{-q^2} (\epsilon -1) m_p^2 \Gamma (1-\epsilon )}-\frac{i \left(m_1+m_2\right) \left(6 \sigma ^2 (\epsilon -1)-2 \sigma  \epsilon +1\right) \sec (\pi  \epsilon )}{\sqrt{-q^2} m_p^2 \Gamma (2-\epsilon )}\right).
      \end{split}
      \end{align}
Note that, a and b denotes the diagrams of the first row diagram in the first row and c and d denotes the diagrams of the second row.\\\\       
\textbullet $\,\,$The amplitude corresponding to the  box (+ cross-box) diagram with dilaton and photon exchange is given by,

 \begin{flalign}
& i\mathbfcal{A}_{4,\textrm{soft}}^{1-\textrm{loop}}\left[\thesisampdiagram[0.40\linewidth]{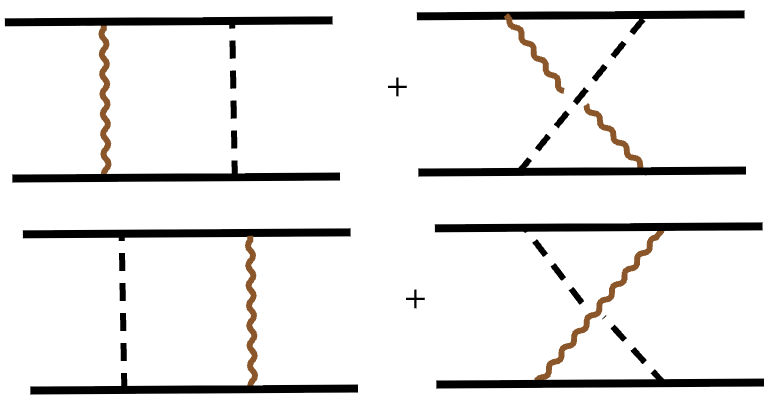}\right] =\sum_{i=a,c}\int_{\ell}\frac{\mathcal{N}^{(3i)}_{\filledsquare{gray}}}{\rho_1 \rho_2 \rho_3 \rho_4}\Bigg|_{\textrm{soft}}+\sum_{j=b,d}\int_{\ell}\frac{\mathcal{N}^{(3j)}_{\Join}}{\rho_1 \rho_2 \rho_3 \tilde\rho_4}\Bigg|_{\textrm{soft}} \nonumber \\
      & = -\frac{a^2 \alpha _1 \alpha _2 e^2 m_1 m_2 \sigma  2^{4 \epsilon +3} \pi ^{\epsilon +\frac{1}{2}} \texttt{g}_c \left(-q^2\right)^{-\epsilon -1} \csc (\pi  \epsilon )}{\sqrt{\sigma ^2-1} m_p^2 \Gamma \left(\frac{1}{2}-\epsilon \right)} . &
      \end{flalign}
\textbullet $\,\,$ The amplitude of the box (+cross-box) diagram corresponding to the double dilaton exchange is given by,
\begin{align}
\begin{split}
&i\mathbfcal{A}_{4,\textrm{soft}}^{1-\textrm{loop}}\left[\thesisampdiagram[0.40\linewidth]{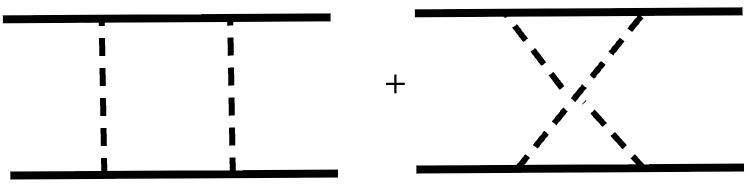}\right] =\int_{\ell}\frac{\mathcal{N}^{(4)}_{\filledsquare{gray}}}{\rho_1 \rho_2 \rho_3 \rho_4}\Bigg|_{\textrm{soft}}+\int_{\ell}\frac{\mathcal{N}^{(4)}_{\Join}}{\rho_1 \rho_2 \rho_3 \tilde\rho_4}\Bigg|_{\textrm{soft}} \\ &= \frac{a^4 2^{4 \epsilon -3} \pi ^{\epsilon +\frac{1}{2}} \left(-q^2\right)^{-\epsilon -1} \csc (\pi  \epsilon ) \left(8 m_1^2 m_2^2 \left(\sigma ^2-1\right)-q^2 \left(2 m_2 m_1 \sigma +m_1^2+m_2^2\right)\right)}{m_1 m_2 \left(\sigma ^2-1\right)^{3/2} m_p^4 \Gamma \left(\frac{1}{2}-\epsilon \right)}\,.
      \end{split}
      \end{align}
      \textbullet $\,\,$ The amplitude of the box (+cross-box) diagram corresponding to the graviton-dilaton  exchange is given by,
\begin{flalign}
& i\mathbfcal{A}_{4,\textrm{soft}}^{1-\textrm{loop}}\left[\thesisampdiagram[0.40\linewidth]{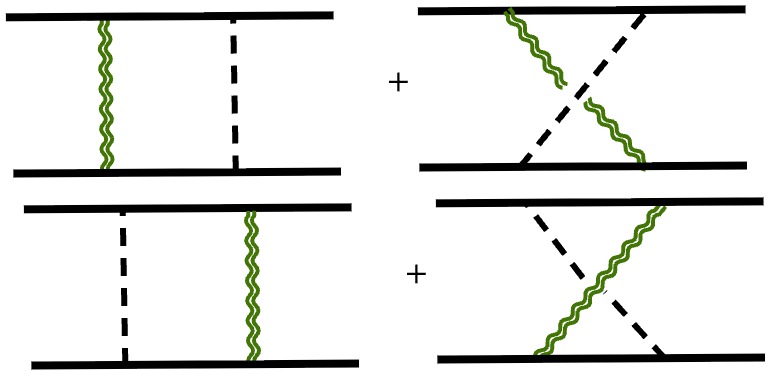}\right]    =\int_{\ell}\frac{\mathcal{N}^{(5)}_{\filledsquare{gray}}}{\rho_1 \rho_2 \rho_3 \rho_4}\Bigg|_{\textrm{soft}}+\int_{\ell}\frac{\mathcal{N}^{(5)}_{\Join}}{\rho_1 \rho_2 \rho_3 \tilde\rho_4}\Bigg|_{\textrm{soft}}\nonumber\\ &
    =\frac{a^2 m_1^2 m_2^2 2^{4 \epsilon -2} \pi ^{\epsilon +\frac{1}{2}} \left(-q^2\right)^{-\epsilon -1} \left(-2 \sigma ^2+2 \sigma ^2 \epsilon +1\right) \csc (\pi  \epsilon )}{\sqrt{\sigma ^2-1} (\epsilon -1) m_p^4 \Gamma \left(\frac{1}{2}-\epsilon \right)}  \nonumber \\
    & \hspace{0 cm} -\frac{i a^2 m_1 m_2 \left(m_1+m_2\right) \sigma  4^{2 \epsilon -1} \pi ^{\epsilon } \left(-q^2\right)^{-\epsilon -\frac{1}{2}} \sec (\pi  \epsilon )}{m_p^4 \Gamma (1-\epsilon )}\,. &
\end{flalign}
\textbullet $\,\,$   The amplitude of the box (+cross-box) diagram corresponding to the graviton-graviton  exchange is given by,  
\begin{flalign}
&\hspace{0cm}i\mathbfcal{A}_{4,\textrm{soft}}^{1-\textrm{loop}}\left[\thesisampdiagram[0.40\linewidth]{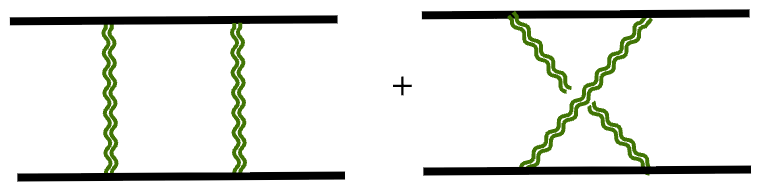}\right]    =\int_{\ell}\frac{\mathcal{N}^{(6)}_{\filledsquare{gray}}}{\rho_1 \rho_2 \rho_3 \rho_4}\Bigg|_{\textrm{soft}}+\int_{\ell}\frac{\mathcal{N}^{(6)}_{\Join}}{\rho_1 \rho_2 \rho_3 \tilde\rho_4}\Bigg|_{\textrm{soft}} \nonumber \\ & =\frac{m_1 m_2 2^{4 \epsilon -9} \pi ^{\epsilon +\frac{1}{2}} \left(-q^2\right)^{-\epsilon }  \csc (\pi  \epsilon ) \left(8 m_1^2 m_2^2 \left(\sigma ^2-1\right)-q^2 \left(2 m_2 m_1 \sigma +m_1^2+m_2^2\right)\right)}{-q^2 \left(2 \sigma ^2 (\epsilon -1)+1\right)^{-2} \left(\sigma ^2-1\right)^{3/2} (\epsilon -1)^2 m_p^4 \Gamma \left(\frac{1}{2}-\epsilon \right)} \nonumber \\ 
    & \hspace{1cm}+\frac{i m_1^2 m_2^2 \left(m_1+m_2\right) 4^{2 \epsilon -3} \pi ^{\epsilon } \epsilon  \left(-q^2\right)^{-\epsilon -\frac{1}{2}} \left(2 \sigma ^2 (\epsilon -1)+1\right) \sec (\pi  \epsilon )}{(\epsilon -1)^2 m_p^4 \Gamma (1-\epsilon )}\,.  &  
\end{flalign}
As one can immediately see by expanding the box(+cross-box) amplitudes, the $\frac{1}{\sqrt{-q^2}}$ terms cancelled, so they do not contribute to the classical potential.
\subsection*{Soft amplitudes from triangle topologies:}   
\textbullet  $\,\,$The soft amplitude corresponding to the triangle topology with photon lines is given by,


\begin{align}
\thesisamplhs{i\mathbfcal{A}_{4,\textrm{soft}}^{1-\textrm{loop}}\left[\thesisampdiagram[0.32\linewidth]{triangle_1.png}\right]}
  &= \int_{\ell}\frac{\mathcal{N}^{(1a)}_{\triangle}\rho_4+\mathcal{N}^{(1b)}_{\triangle}\rho_3}{\rho_1 \rho_2 \rho_3 \rho_4}\Bigg|_{\textrm{soft}} \notag\\
\thesisamplhs{}
  &= \frac{i \alpha _1^2 \alpha _2^2 e^4 \left(m_1+m_2\right) 4^{2 \epsilon +1} \pi ^{\epsilon } g_s^2 \left(-q^2\right)^{-\epsilon -\frac{1}{2}} \sec (\pi  \epsilon )}{\Gamma (1-\epsilon )}\,.
\end{align}

\textbullet $\,\,$ The soft amplitude corresponding to the triangle topology with one graviton and one photon line is given by,   
\begin{align}
\thesisamplhs{i\mathbfcal{A}_{4,\textrm{soft}}^{1-\textrm{loop}}\left[\thesisampdiagram[0.32\linewidth]{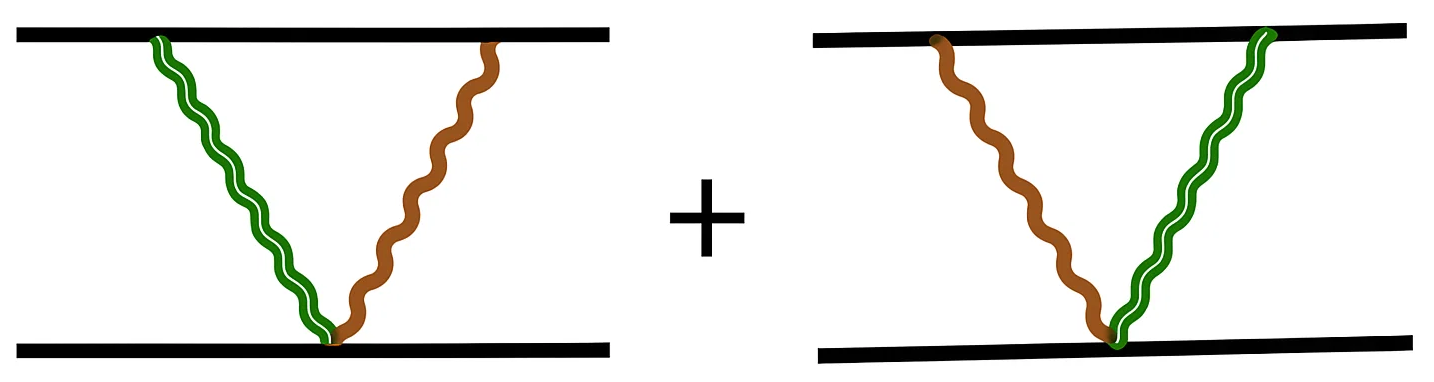}\right]}
  &= \int_{\ell}\frac{\mathcal{N}^{(2a)}_{\triangle}}{\rho_1 \rho_2 \rho_3 } \Bigg|_{\textrm{soft}} \notag\\
\thesisamplhs{}
  &= -\frac{i \alpha _1 \alpha _2 e^2 2^{4 \epsilon -3} \pi ^{\epsilon } g_s \left(-q^2\right)^{-\epsilon -\frac{1}{2}} \sec (\pi  \epsilon )}{(\epsilon -1)^2 m_p^2 \Gamma (1-\epsilon )} \notag\\
\thesisamplhs{}
  &\phantom{=}\times \Big[m_1 q^2 \left(\epsilon ^2-\epsilon +1\right)+m_2 q^2 \sigma  \epsilon \notag\\
\thesisamplhs{}
  &\phantom{=}\qquad\;\; +8 m_1^2 m_2 \sigma  \left(\epsilon ^2-\epsilon +1\right)\Big].
\end{align}
and,
\begin{align}
\thesisamplhs{i\mathbfcal{A}_{4,\textrm{soft}}^{1-\textrm{loop}}\left[\thesisampdiagram[0.32\linewidth]{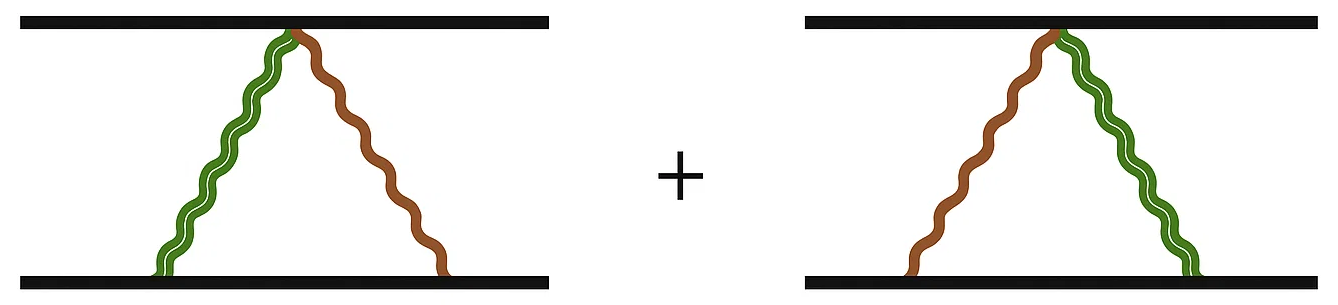}\right]}
  &= \int_{\ell}\frac{\mathcal{N}^{(2b)}_{\triangle}}{\rho_1 \rho_2 \rho_4 } \Bigg|_{\textrm{soft}} \notag\\
\thesisamplhs{}
  &= -\frac{i \alpha _1 \alpha _2 e^2 2^{4 \epsilon -3} \pi ^{\epsilon } g_s \left(-q^2\right)^{-\epsilon -\frac{1}{2}} \sec (\pi  \epsilon )}{(\epsilon -1)^2 m_p^2 \Gamma (1-\epsilon )} \notag\\
\thesisamplhs{}
  &\phantom{=}\times \Big[m_1 \sigma  \left(8 m_2^2 \left(\epsilon ^2-\epsilon +1\right)+q^2 \epsilon \right) \notag\\
\thesisamplhs{}
  &\phantom{=}\qquad\;\; +m_2 q^2 \left(\epsilon ^2-\epsilon +1\right)\Big]\,.
\end{align}
\textbullet $\,\,$ The soft amplitude corresponding to the triangle topology with graviton lines are given by,   
\begin{align}
\thesisamplhs{i\mathbfcal{A}_{4,\textrm{soft}}^{1-\textrm{loop}}\left[\thesisampdiagram[0.32\linewidth]{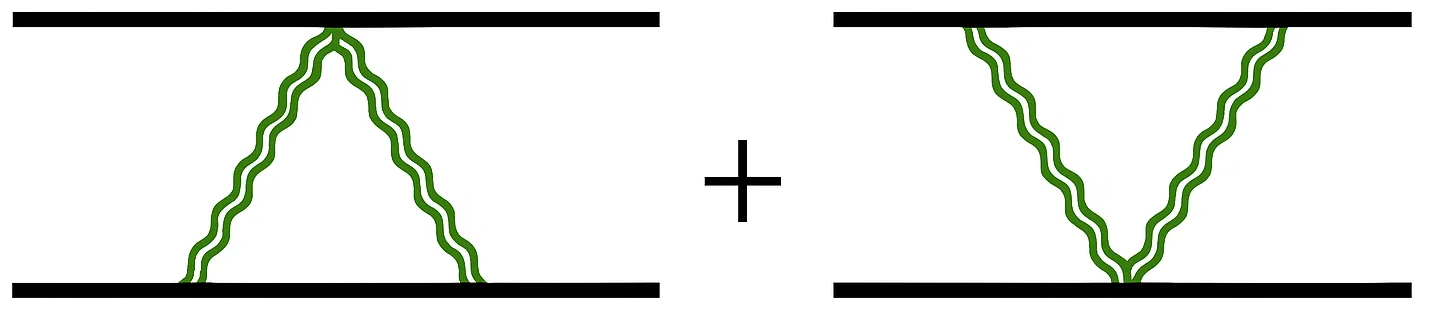}\right]}
  &= \int_{\ell}\frac{\mathcal{N}^{(3a)}_{\triangle}\rho_4+\mathcal{N}^{(3b)}_{\triangle}\rho_3}{\rho_1 \rho_2 \rho_3 } \Bigg|_{\textrm{soft}} \notag\\
\thesisamplhs{}
  &= \frac{i m_1^2 m_2^2 \left(m_1+m_2\right) 2^{4 \epsilon -6} \pi ^{\epsilon } \left(-q^2\right)^{-\epsilon -\frac{1}{2}} \sec (\pi  \epsilon )}{(\epsilon -1)^2 m_p^4 \Gamma (1-\epsilon )} \notag\\
\thesisamplhs{}
  &\phantom{=}\times \left(2 \sigma ^2 (\epsilon -1)^2-\epsilon \right)\,.
\end{align}
\textbullet $\,\,$ The soft amplitude corresponding to the triangle topology with dilaton lines is given by,   
\begin{align}
\thesisamplhs{i\mathbfcal{A}_{4,\textrm{soft}}^{1-\textrm{loop}}\left[\thesisampdiagram[0.32\linewidth]{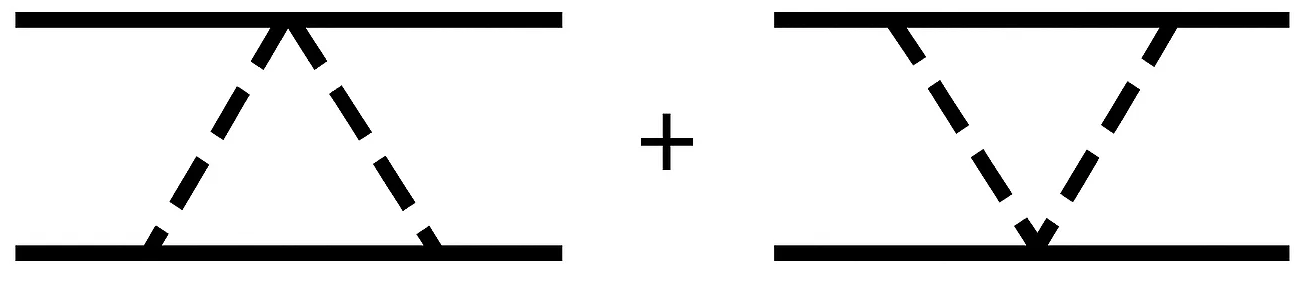}\right]}
  &= \int_{\ell}\frac{\mathcal{N}^{(4a)}_{\triangle}\rho_4+\mathcal{N}^{(4b)}_{\triangle}\rho_3}{\rho_1 \rho_2 \rho_3 \rho_4 } \Bigg|_{\textrm{soft}} \notag\\
\thesisamplhs{}
  &= \frac{i a^2 2^{4 \epsilon -5} \pi ^{\epsilon } \left(-q^2\right)^{-\epsilon -\frac{1}{2}} \sec (\pi  \epsilon )}{m_1 m_2 m_p^4 \Gamma (1-\epsilon )} \notag\\
\thesisamplhs{}
  &\phantom{=}\times \left(8 a^2 m_1 m_2 \left(m_1+m_2\right)-32 b m_1^2 m_2^2\right)\,.
\end{align}
\textbullet $\,\,$ The soft amplitude corresponding to the triangle topology with one dilaton line and one graviton line is given by,   

\begin{align}
\thesisamplhs{i\mathbfcal{A}_{4,\textrm{soft}}^{1-\textrm{loop}}\left[\thesisampdiagram[0.32\linewidth]{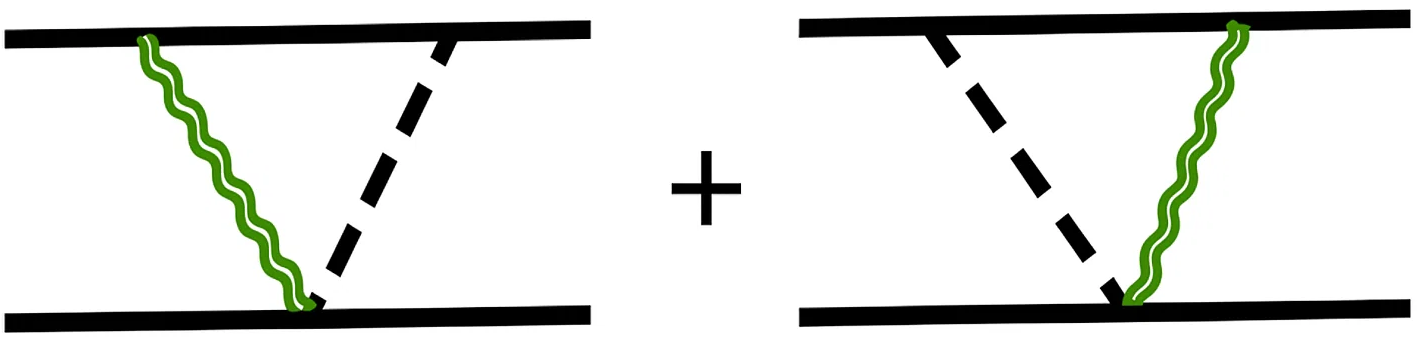}\right]}
  &= \int_{\ell}\frac{\mathcal{N}^{(5a)}_{\triangle}}{\rho_1 \rho_2 \rho_3 } \Bigg|_{\textrm{soft}} \notag\\
\thesisamplhs{}
  &= -\frac{i a^2 m_2 4^{2 \epsilon -3} \pi ^{\epsilon } (\epsilon +1) \left(8 m_1^2+q^2\right) \left(-q^2\right)^{-\epsilon -\frac{1}{2}} \sec (\pi  \epsilon )}{(\epsilon -1) m_p^4 \Gamma (1-\epsilon )}\,.
\end{align}

and,

\begin{align}
\thesisamplhs{i\mathbfcal{A}_{4,\textrm{soft}}^{1-\textrm{loop}}\left[\thesisampdiagram[0.32\linewidth]{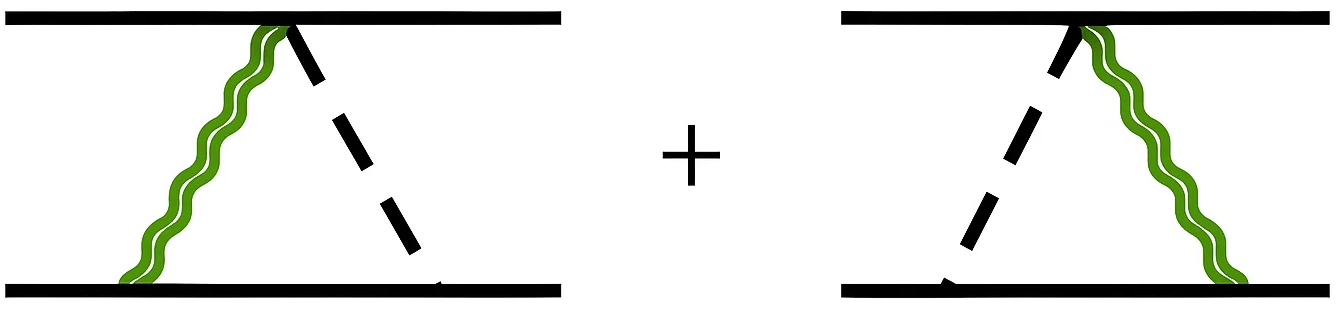}\right]}
  &= \int_{\ell}\frac{\mathcal{N}^{(5b)}_{\triangle}}{\rho_1 \rho_2 \rho_3 } \Bigg|_{\textrm{soft}} \notag\\
\thesisamplhs{}
  &= -\frac{i a^2 m_1 4^{2 \epsilon -3} \pi ^{\epsilon } (\epsilon +1) \left(8 m_2^2+q^2\right) \left(-q^2\right)^{-\epsilon -\frac{1}{2}} \sec (\pi  \epsilon )}{(\epsilon -1) m_p^4 \Gamma (1-\epsilon )}\,.
\end{align}

\begingroup
\setlength{\thesisAmplitudeLhsWidth}{0.59\linewidth}
\setlength{\jot}{1.2pt}
\setlength{\abovedisplayskip}{2.5pt plus 1pt minus 1.5pt}
\setlength{\belowdisplayskip}{2.5pt plus 1pt minus 1.5pt}
\setlength{\abovedisplayshortskip}{1pt plus 1pt minus 0.5pt}
\setlength{\belowdisplayshortskip}{1pt plus 1pt minus 0.5pt}
\subsection*{Soft amplitudes from penguin topologies  }     
\textbullet $\,\,$ The soft amplitudes for penguin diagrams with two dilaton lines and one graviton line are given by,
\begin{align}
\thesisamplhs{i\mathbfcal{A}_{4,\textrm{soft}}^{1-\textrm{loop}}\left[\thesisampdiagram[0.42\linewidth]{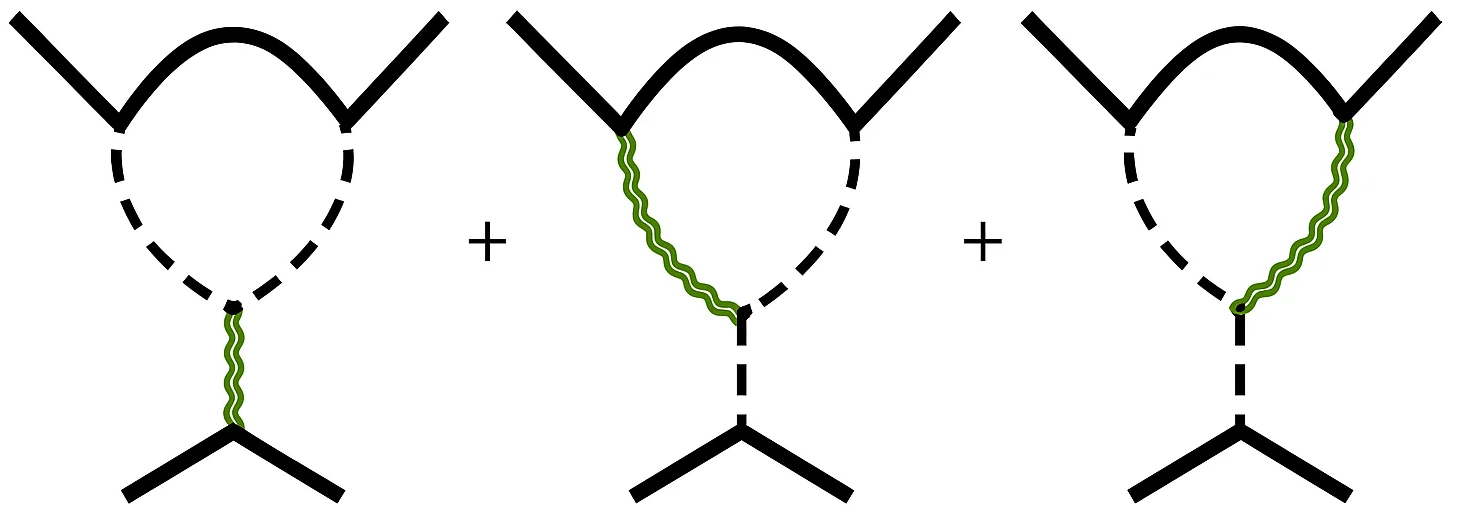}\right]}
  &= \int_{\ell}\frac{\mathcal{N}_{\penguin}^{(1)}[\ell,q,\bar m_i,\sigma]}{q^2 \rho_1\rho_2\rho_3}\Bigg|_{\textrm{soft}} \notag\\
\thesisamplhs{}
  &= -\frac{i a^2 4^{2 \epsilon -5} \pi ^{\epsilon } \left(-q^2\right)^{-\epsilon -\frac{1}{2}} \sec (\pi  \epsilon )}{m_1 m_p^4 \Gamma (2-\epsilon )} \notag\\
\thesisamplhs{}
  &\phantom{=}\times \Big[m_2^2 q^2 \left(3 \sigma ^2-1\right) \notag\\
\thesisamplhs{}
  &\phantom{=}\qquad\;\; +2 m_1^2 \left(4 m_2^2 \left(\sigma ^2-1\right)+q^2\right) \notag\\
\thesisamplhs{}
  &\phantom{=}\qquad\;\; +4 m_1 m_2 q^2 \sigma\Big].
\end{align}
\vspace{-0.35\baselineskip}
\noindent and,
\begin{align}
\thesisamplhs{i\mathbfcal{A}_{4,\textrm{soft}}^{1-\textrm{loop}}\left[\thesisampdiagram[0.42\linewidth]{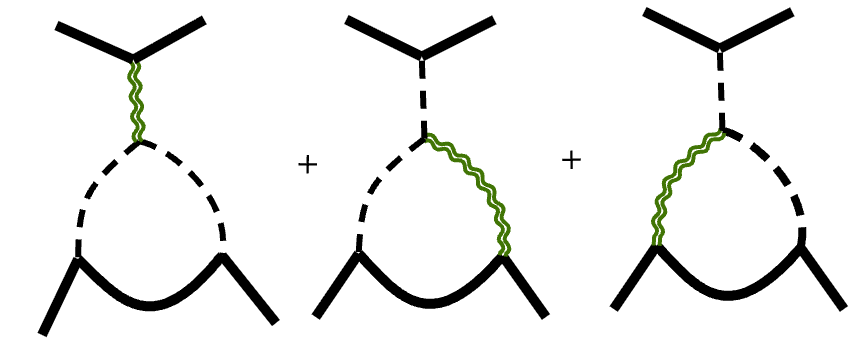}\right]}
  &= \int_{\ell}\frac{\mathcal{N}_{\penguin}^{(2)}[\ell,q,\bar m_i,\sigma]}{q^2 \rho_1\rho_2\rho_4}\Bigg|_{\textrm{soft}} \notag\\
\thesisamplhs{}
  &= -\frac{i a^2 4^{2 \epsilon -5} \pi ^{\epsilon } \left(-q^2\right)^{-\epsilon -\frac{3}{2}} \sec (\pi  \epsilon )}{m_2 m_p^4 \Gamma (2-\epsilon )} \notag\\
\thesisamplhs{}
  &\phantom{=}\times \Big[8 m_1 m_2^2 q^2 \left(m_1 \left(-\sigma ^2+2 \epsilon +1\right)-4 m_2 \epsilon \right) \notag\\
\thesisamplhs{}
  &\phantom{=}\qquad\;\; +q^4 \epsilon  \left(8 m_2^2+q^2\right)\Big]\,.
\end{align}
\vspace{-0.25\baselineskip}
\noindent\textbullet $\,\,$ The soft penguin amplitudes for with two photons, one graviton line are given by,
\begin{align}
\thesisamplhs{i\mathbfcal{A}_{4,\textrm{soft}}^{1-\textrm{loop}}\left[\thesisampdiagram[0.42\linewidth]{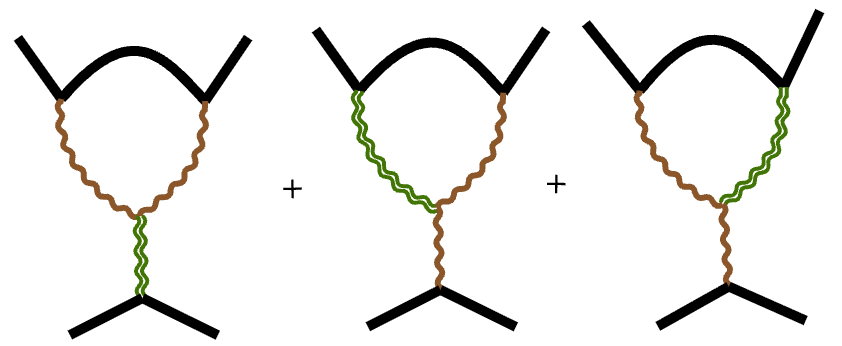}\right]}
  &= \int_{\ell}\frac{\mathcal{N}_{\penguin}^{(3)}[\ell,q,\bar m_i,\sigma]}{q^2 \rho_1\rho_2\rho_3}\Bigg|_{\textrm{soft}} \notag\\
\thesisamplhs{}
  &= -\frac{i \alpha _1 e^2 m_1 m_2 16^{\epsilon -1} \pi ^{\epsilon } \texttt{g}_c \left(-q^2\right)^{-\epsilon -\frac{1}{2}} \sec (\pi  \epsilon ) }{(\epsilon -1) m_p^2 \Gamma (2-\epsilon )} \notag\\
\thesisamplhs{}
  &\phantom{=}\times \Big[ \alpha _1 m_2 \left(-3 \sigma ^2-4 \sigma ^2 \epsilon ^2\right) \notag\\
\thesisamplhs{}
  &\phantom{=}\qquad\;\; +\alpha _1 m_2 \left(\left(7 \sigma ^2-5\right) \epsilon +1\right) \notag\\
\thesisamplhs{}
  &\phantom{=}\qquad\;\; +4 \alpha _2 m_1 \sigma  \left(2 \epsilon ^2-\epsilon +1\right)\Big].
\end{align}
\vspace{-0.35\baselineskip}
\noindent and,
\begin{align}
\thesisamplhs{i\mathbfcal{A}_{4,\textrm{soft}}^{1-\textrm{loop}}\left[\thesisampdiagram[0.42\linewidth]{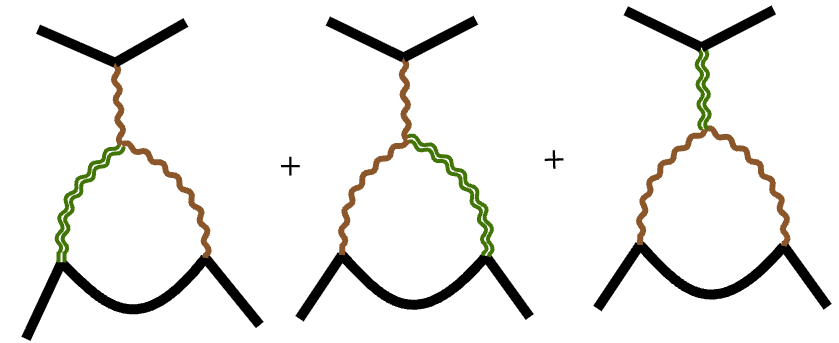}\right]}
  &= \int_{\ell}\frac{\mathcal{N}_{\penguin}^{(4)}[\ell,q,\bar m_i,\sigma]}{q^2 \rho_1\rho_2\rho_4}\Bigg|_{\textrm{soft}} \notag\\
\thesisamplhs{}
  &= -\frac{i \alpha _2 e^2 m_1 m_2 16^{\epsilon -1} \pi ^{\epsilon } \texttt{g}_c \left(-q^2\right)^{-\epsilon -\frac{1}{2}} \sec (\pi  \epsilon ) }{(\epsilon -1)^2 m_p^2 \Gamma (1-\epsilon )} \notag\\
\thesisamplhs{}
  &\phantom{=}\times \Big[ \alpha _2 m_1 \left(3 \sigma ^2+4 \sigma ^2 \epsilon ^2\right) \notag\\
\thesisamplhs{}
  &\phantom{=}\qquad\;\; +\alpha _2 m_1 \left(\left(5-7 \sigma ^2\right) \epsilon -1\right) \notag\\
\thesisamplhs{}
  &\phantom{=}\qquad\;\; -4 \alpha _1 m_2 \sigma  \left(2 \epsilon ^2-\epsilon +1\right)\Big]\,.
\end{align}
\vspace{-0.25\baselineskip}
\noindent\textbullet $\,\,$ The soft amplitudes for the penguin diagram with two photon lines and one dilaton line are given by,
\begin{align}
\thesisamplhs{i\mathbfcal{A}_{4,\textrm{soft}}^{1-\textrm{loop}}\left[\thesisampdiagram[0.42\linewidth]{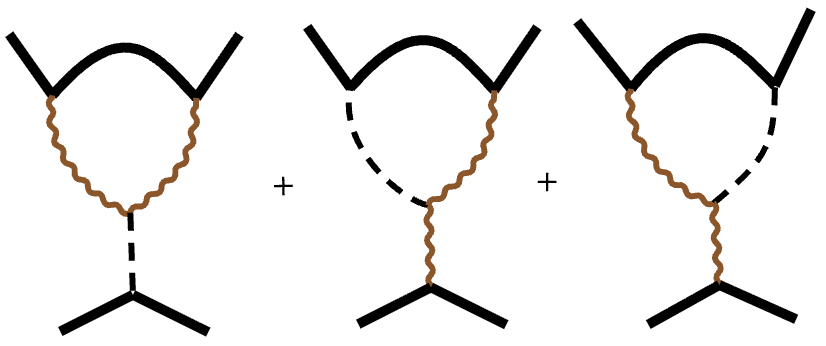}\right]}
  &= \int_{\ell}\frac{\mathcal{N}_{\penguin}^{(5)}[\ell,q,\bar m_i,\sigma]}{q^2 \rho_1\rho_2\rho_3}\Bigg|_{\textrm{soft}} \notag\\
\thesisamplhs{}
  &= \frac{i a \alpha _1 e^2 2^{4 \epsilon -\frac{17}{2}} \pi ^{\epsilon } g_s \left(-q^2\right)^{-\epsilon -\frac{3}{2}} \sec (\pi  \epsilon )}{m_1^3 m_p^2 \Gamma (1-\epsilon )} \notag\\
\thesisamplhs{}
  &\phantom{=}\times \Big[128 m_1^4 m_2 q^2 \left(\alpha _1-2 \alpha _2 \sigma \right) \notag\\
\thesisamplhs{}
  &\phantom{=}\qquad\;\; -q^4 \left(\alpha _1 m_2+2 \alpha _2 m_1\right) \notag\\
\thesisamplhs{}
  &\phantom{=}\qquad\;\; \times \left(8 m_1^2+q^2\right)\Big].
\end{align}
\vspace{-0.35\baselineskip}
\noindent and,
\begin{align}
\thesisamplhs{i\mathbfcal{A}_{4,\textrm{soft}}^{1-\textrm{loop}}\left[\thesisampdiagram[0.42\linewidth]{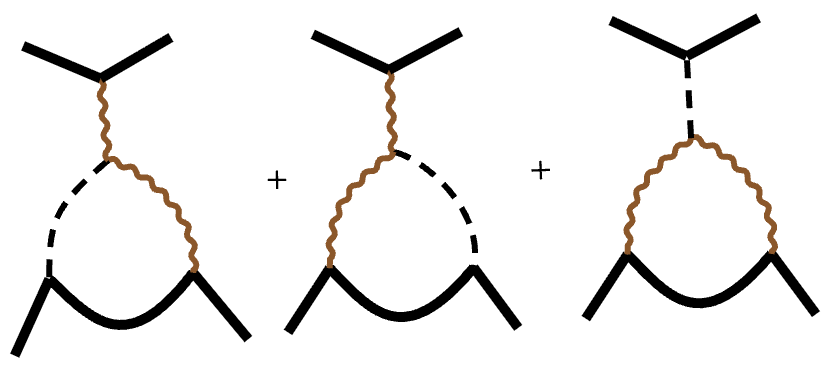}\right]}
  &= \int_{\ell}\frac{\mathcal{N}_{\penguin}^{(6)}[\ell,q,\bar m_i,\sigma]}{q^2 \rho_1\rho_2\rho_4}\Bigg|_{\textrm{soft}} \notag\\
\thesisamplhs{}
  &= -\frac{i a \alpha _2 e^2 2^{4 \epsilon -\frac{17}{2}} \pi ^{\epsilon } g_s \left(-q^2\right)^{-\epsilon -\frac{3}{2}} \sec (\pi  \epsilon )}{m_2^3 m_p^2 \Gamma (1-\epsilon )} \notag\\
\thesisamplhs{}
  &\phantom{=}\times \Big[128 m_1 m_2^4 q^2 \left(2 \alpha _1 \sigma -\alpha _2\right) \notag\\
\thesisamplhs{}
  &\phantom{=}\qquad\;\; +q^4 \left(2 \alpha _1 m_2+\alpha _2 m_1\right) \notag\\
\thesisamplhs{}
  &\phantom{=}\qquad\;\; \times \left(8 m_2^2+q^2\right)\Big]\,.
\end{align}
\vspace{-0.25\baselineskip}
\noindent\textbullet $\,\,$ The soft amplitudes for penguin topology with three graviton lines are given by,
\begin{align}
\thesisamplhs{i\mathbfcal{A}_{4,\textrm{soft}}^{1-\textrm{loop}}\left[\thesisampdiagram[0.42\linewidth]{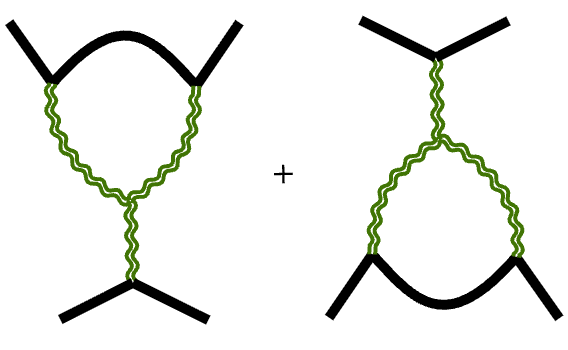}\right]}
  &= \int_{\ell}\frac{\mathcal{N}_{\penguin}^{(7a)}[\ell,q,\bar m_i,\sigma]}{q^2 \rho_1\rho_2\rho_3}\Bigg|_{\textrm{soft}} +\int_{\ell}\frac{\mathcal{N}_{\penguin}^{(7b)}[\ell,q,\bar m_i,\sigma]}{q^2 \rho_1\rho_2 \textcolor{black}{\rho_4}}\Bigg|_{\textrm{soft}} \notag\\
\thesisamplhs{}
  &= -\frac{i m_1^2 m_2^2 \left(m_1+m_2\right) 2^{4 \epsilon -9} \pi ^{\epsilon } \left(-q^2\right)^{-\epsilon -\frac{1}{2}}  \sec (\pi  \epsilon )}{(\epsilon -1)^4 m_p^4 \Gamma (1-\epsilon )} \notag\\
\thesisamplhs{}
  &\phantom{=}\times \Big[ -8 \epsilon ^4+11 \epsilon ^3-15 \epsilon ^2 \notag\\
\thesisamplhs{}
  &\phantom{=}\qquad\;\; +\sigma ^2 (\epsilon -1)^3 (12 \epsilon -1) \notag\\
\thesisamplhs{}
  &\phantom{=}\qquad\;\; +\epsilon +3\Big]\,.
\end{align}
\endgroup
\begin{tcolorbox}[thesisresultbox, title=\textit{Full one loop amplitude in small $|q|$ expansion}, boxsep=0.5mm, left=0.5mm, right=0.5mm, top=0.5mm, bottom=0.5mm]
\restorethesisbodyformat
\begingroup
\footnotesize
\setlength{\jot}{2pt}
\setlength{\abovedisplayskip}{4pt}
\setlength{\belowdisplayskip}{4pt}
\setlength{\abovedisplayshortskip}{2pt}
\setlength{\belowdisplayshortskip}{2pt}
\allowdisplaybreaks[4]
\begin{align*}
&\hspace{0cm}i\,\mathbfcal{A}_{\text{1-loop}}(\sigma,|q|,\ep)
= {} (-q^2)^{-(\tfrac32+\ep)}\;\\ & 
\hspace{0cm}\Biggl[
\;i(-q^2)\;
 \,\frac{2^{-9+4\ep}\,\pi^{\ep}\,\sec(\pi\ep)\,(m_1+m_2)}
             {(-1+\ep)^{3}\,\Gamma(1-\ep)\,\mpow{4}}
\Bigl(
      q^2\bigl(3-9\ep+2\ep^{2}-4\ep^{3}+4(-1+\ep)^{3}\sig-(-1+\ep)^{2}(-1+2\ep)\sig^{2}\bigr)
\\
&\hspace{0cm}
   m_1 m_2 +\,2 q^2\bigl(2 \epsilon ^3-\epsilon ^2+4 \epsilon -1\bigr)\,m_2^{2}
    +\,2 m_1^{2}\bigl(q^2(-1+\ep(4+\ep(-1+2\ep))) 
\\&\hspace{0cm} \,+ 4(-1+\ep)(-\ep+2(-1+\ep)^{2}\sig^{2})m_2^{2}\bigr)
\Bigr)+\; 
 i\,\frac{16^{-3+\ep}\,\pi^{\ep}\,\sec(\pi\ep)\,(m_1+m_2)}
             {(-1+\ep)^{4}\,\Gamma(1-\ep)\,m_1 m_2\,\mpow{4}}
\Bigl(
   2 q^{6}(-1+\ep)^{3}\ep\,m_2^{2} +\, q^{4} m_1 m_2
\\
&\hspace{0cm}
  \Bigl( 2 q^{2} + q^{2}\bigl(4 \sigma +(6 \sigma +5) \epsilon ^4-22 (\sigma +1) \epsilon ^3+3 (10 \sigma +7) \epsilon ^2-2 (9 \sigma +5) \epsilon\bigr)
   - 8\bigl(\epsilon ^4+4 \epsilon ^3-3 \epsilon ^2+4 \epsilon -2\bigr)m_2^{2}\Bigr)
\\
&\hspace{0cm}
 +\,2 q^{4} m_1^{2}\Bigl(q^{2}(-1+\ep)^{3}\ep
   + 4\bigl(-2+2\ep+3\ep^{2}-2\ep^{3}+3\ep^{4}+2(-1+\ep)^{3}(-2+3\ep)\sig\bigr)m_2^{2}\Bigr)
\\
&\hspace{0cm}
 +\,8 m_1^{3} m_2\Bigl(
     -q^{4}\bigl(\epsilon ^4+4 \epsilon ^3-3 \epsilon ^2+4 \epsilon -2\bigr)
     + q^{2}\bigl(3+\ep-15\ep^{2}+11\ep^{3}-8\ep^{4}+(-1+\ep)^{3}(-1+12\ep)\sig^{2}\bigr)m_2^{2}
   \Bigr)
\Bigr)
\\& \hspace{0cm}+\;
(-q^2)\;
\Bigl\{- i\,\frac{16^{-2+\ep}\,\pi^{\ep}\,\sec(\pi\ep)}
             {(-1+\ep)^{2}\,\Gamma(1-\ep)\,\mpow{4}}
\bigl(
  4 q^{2}(1-\ep)\sig(\ep+2\sig-2\ep\,\sig)\,m_1 m_2 (m_1+m_2)
\\
&\hspace{0cm}
 -\,(\ep+2\sig-2\ep\,\sig)(m_1+m_2)\bigl(q^{4}(-1+\ep) + 2(1+2(-1+\ep)\sig^{2})\,m_1^{2}m_2^{2}\bigr)
\\
&\hspace{0cm}
 +\,\frac{q^{2}\bigl(q^{4}(-1+\ep) + 2(1+2(-1+\ep)\sig^{2})\,m_1^{2}m_2^{2}\bigr)}
           {8\,m_1^{2}m_2^{2}}
   \bigl( 2(-1+\ep)\sig\,m_1^{3} + (-8+7\ep)\,m_1^{2} m_2
\\
&\hspace{0cm}
        +\,(-8+7\ep)\,m_1 m_2^{2} + 2(-1+\ep)\sig\,m_2^{3}\bigr)
\Bigr)\Big\}
\\
&\hspace{0cm} +\;
(-q^2)^{\tfrac32}\;
\frac{4^{-5+2\ep}\,\pi^{\ep}\,\bigl(1+2(-1+\ep)\sig^{2}\bigr)}
             {(-1+\ep)^{2}\,\mpow{4}}
\Biggl[
  \frac{i\,\sec(\pi\ep)\,(q^{2}+8 m_1^{2})\,m_2^{2}\,(\ep\,m_1+2(-1+\ep)\sig\,m_2)}{\sqrt{-q^2}\,\Gamma(1-\ep)}
\\
&\hspace{0cm}
 +\,\frac{i\,\sec(\pi\ep)\,m_1^{2}\,(2(-1+\ep)\sig\,m_1+\ep\,m_2)\,(q^{2}+8 m_2^{2})}{\sqrt{-q^2}\,\Gamma(1-\ep)}
\\
&\hspace{0cm}
 +\,\frac{2\sqrt{\pi}\,\bigl(1+2(-1+\ep)\sig^{2}\bigr)\,(-\sig+\SQ)\,\csc(\pi\ep)\,m_1 m_2\,
         \Bigl(8(\sig^{2}-1)m_1^{2}m_2^{2}-q^{2}(m_1^{2}+2\sig\,m_1 m_2 + m_2^{2})\Bigr)}
        {q^{2}\,(\sig^{2}-1)\,\bigl(1+\sig(-\sig+\SQ)\bigr)\,\Gamma(\tfrac12-\ep)}
\Biggr]
\\ &\hspace{0cm}\;+\; a^{4}\;
(-q^2)^{\tfrac12}\,
 \frac{2^{-3+4\ep}\,\pi^{\tfrac12+\ep}\,\csc(\pi\ep)\,
          \Bigl(8(\sig^2-1)\,m_1^{2}m_2^{2} - q^2\,(m_1^{2} + 2\sig\,m_1 m_2 + m_2^{2})\Bigr)}
          {(\sig^2-1)^{3/2}\,\Gamma(\tfrac12-\ep)\,m_1 m_2\,\mpow{4}}
\\ &\hspace{0cm}\;+\; a^{2}\Biggl[
 (-q^2)^{\tfrac32}\;
 \Bigl(-\,\frac{2^{-2+4\ep}\,\pi^{\tfrac12+\ep}\,\sqrt{-q^2}\,\bigl(1-2\sig^2+2\ep\,\sig^2\bigr)\,(-\sig+\SQ)\,\csc(\pi\ep)\,m_1^{2}m_2^{2}}
          {(-1+\ep)\,\bigl(1-\sig^2+\sig\,\SQ\bigr)\,\Gamma(\tfrac12-\ep)\,\mpow{4}}\Bigr)
\\
&\hspace{0cm}
 +\,(-q^2)\;
 \Bigl(-\, i\,\frac{2^{-4+4\ep}\,\pi^{\ep}\,q^{2}\,(\ep-2\sig+2\ep\,\sig)\,\sec(\pi\ep)\,m_1 m_2\,(m_1+m_2)}
             {(-1+\ep)\,\Gamma(1-\ep)\,\mpow{4}}\Bigr)
\\
&\hspace{0cm}
 +\,(-q^2)\;\Bigl[
 \begin{aligned}[t]
 &-\, i\,\frac{4^{-3+2\ep}\,\pi^{\ep}\,(1+\ep)\,\sec(\pi\ep)\,(q^2+8 m_1^{2})\,m_2}
              {(-1+\ep)\,\Gamma(1-\ep)\,\mpow{4}}
 \\
 &-\, i\,\frac{4^{-3+2\ep}\,\pi^{\ep}\,(1+\ep)\,\sec(\pi\ep)\,m_1\,(q^2+8 m_2^{2})}
              {(-1+\ep)\,\Gamma(1-\ep)\,\mpow{4}}
 \end{aligned}
 \Bigr]
\end{align*}
\refstepcounter{equation}\label{3.54j}
\begin{align*}
&\hspace{0cm}
 +\, i\,\frac{4^{-5+2\ep}\,\pi^{\ep}\,\sec(\pi\ep)}
             {\Gamma(2-\ep)\,m_1\,\mpow{4}}
 \Bigl(
   -q^4\,\ep\,(q^2+8 m_1^{2})
   + 8 q^2 m_1^{2} m_2\,(4\ep\,m_1 + (-1-2\ep+\sig^2)\,m_2)
 \Bigr)
\\
&\hspace{0cm}
 +\, i\,\frac{4^{-5+2\ep}\,\pi^{\ep}\,\sec(\pi\ep)}
             {\Gamma(2-\ep)\,m_2\,\mpow{4}}
 \Bigl(
   8 q^2 m_1 m_2^{2}\bigl((1+2\ep-\sig^2)\,m_1 - 4\ep\,m_2\bigr)
   + q^4 \ep\,(q^2+8 m_2^{2})
 \Bigr)
\\
&\hspace{0cm}
 +\,(-q^2)\;
 \Bigl[-\, i\,\frac{2^{-7+4\ep}\,\pi^{\ep}\,\sec(\pi\ep)\,(m_1+m_2)}
             {(-1+\ep)\,\Gamma(1-\ep)\,m_1 m_2\,\mpow{4}}
   \Bigl(q^2(-8+7\ep+2\sig-2\ep\,\sig)\,m_1 m_2
\\
&\hspace{0cm}
   +\,2 q^2(-1+\ep)\,\sig\,m_2^{2}
   +\,2 m_1^{2}\bigl(q^2(-1+\ep)\,\sig + (-4\ep-8\sig+8\ep\,\sig)\,m_2^{2}\bigr)\Bigr)\Bigr]
\\
&\hspace{0cm}
 +\,(-q^2)\;
 \frac{i\,2^{-5+4\ep}\,\pi^{\ep}\,\sec(\pi\ep)}
             {\Gamma(1-\ep)\,m_1 m_2\,\mpow{4}}
\Bigl[
\begin{aligned}
& q^{2} m_2 (a^{2}-2 b m_2)
 + a^{2} m_1 (q^{2}+8 m_2^{2}) \\
& + m_1^{2}\,\bigl(-2 b q^{2} + 8 m_2(a^{2}-4 b m_2)\bigr)
\end{aligned}
\Bigr]
\Biggr]
\\& \hspace{0cm}\;+\; e^{2}\texttt{g}_c\,\alpha_1\alpha_2\Biggl[
 (-q^2)\;
 \Bigl(-\, i\,\frac{2^{-3+4\ep}\,\pi^{\ep}\,\sec(\pi\ep)}
             {(-1+\ep)^{2}\,\Gamma(1-\ep)\,\mpow{2}}\,
   \bigl( q^{2}(1-\ep+\ep^{2})\,m_1 + q^{2}\ep\,\sig\,m_2 + 8(1-\ep+\ep^{2})\,\sig\,m_1^{2} m_2 \bigr)\Bigr)
\\
&\hspace{0cm}
 +\,
 \frac{2^{-3+4\ep}\,\pi^{\ep}\,m_1 m_2}
             {(-1+\ep)\,\mpow{2}}
\Biggl[
  -\,\frac{8\sqrt{\pi}\,\sqrt{-q^2}\,\sig\bigl(1+2(-1+\ep)\sig^{2}\bigr)\,\csc(\pi\ep)\,m_1 m_2}{\SQ\,\Gamma(\tfrac12-\ep)}
  \\ &
  \hspace{0cm}+\,\frac{i\,q^{2}\bigl(1+2\sig(\ep+3(-1+\ep)\sig)\bigr)\,\sec(\pi\ep)\,(m_1+m_2)}{\Gamma(1-\ep)}
\Biggr]
\\
&\hspace{0cm}
 +\,(-q^2)\;
 \frac{i\,2^{-7+4\ep}\,\pi^{\ep}\,\sec(\pi\ep)\,(m_1+m_2)}
             {\Gamma(2-\ep)\,m_1^{3} m_2^{3}\,\mpow{2}}
\Bigl(
 16(1-2\ep\,\sig+6(-1+\ep)\sig^{2})\,m_1^{4} m_2^{4}
\\
&\hspace{0cm}
 +\,2 q^{2} m_1^{2} m_2^{2}\Bigl((1+6(-1+\ep)\sig^{2})\,m_1^{2}
    - \bigl(1-6(-8+\sig)\sig + \ep(8-46\sig+6\sig^{2})\bigr)\,m_1 m_2
    + (1+6(-1+\ep)\sig^{2})
\\
&\hspace{0cm}
\,m_2^{2}\Bigr) +\, q^{4}(-1+\ep)\,\Bigl(8 m_1^{2} m_2^{2} + q^{2}(m_1^{2}-m_1 m_2 + m_2^{2})\Bigr)
\Bigr)
\Biggr]
\\&\hspace{0cm}\;+\; a^{2} e^{2} \texttt{g}_c\,\alpha_1\alpha_2\Biggl[
 \frac{2^{3+4\ep}\,\pi^{\tfrac12+\ep}\,\sqrt{-q^2}\,\sig(\sig-\SQ)\,\csc(\pi\ep)\,m_1 m_2}
      {\bigl(-1+\sig^{2}-\sig\,\SQ\bigr)\,\Gamma(\tfrac12-\ep)\,\mpow{2}}
\\
&\hspace{0cm}
 +\,\frac{i\,2^{4\ep}\,\pi^{\ep}\,q^{2}\,\sec(\pi\ep)\,(m_1+m_2)}{\Gamma(1-\ep)\,\mpow{2}}
 \;+\;
 (-q^2)\;
 \frac{i\,2^{-3+4\ep}\,\pi^{\ep}\,\sec(\pi\ep)\,(m_1+m_2)}
             {\Gamma(1-\ep)\,m_1^{2} m_2^{2}\,\mpow{2}}
\Bigl(-q^{2} m_1 m_2 + q^{2} m_2^{2} 
\\&\hspace{0cm}\;+ m_1^{2}(q^{2}+8 m_2^{2})\Bigr)
\Biggr]+\; e^{4} \texttt{g}_c^{2}\alpha_1^{2}\alpha_2^{2}\Biggl[
 (-q^2)\; i\,\frac{4^{1+2\ep}\,\pi^{\ep}\,\sec(\pi\ep)\,(m_1+m_2)}{\Gamma(1-\ep)}
\\
&\hspace{0cm}
 +\,4^{1+2\ep}\,\pi^{\ep}\,\sig
 \Biggl(
    \frac{4\sqrt{\pi}\,\sqrt{-q^2}\,\sig\,(\sig-\SQ)\,\csc(\pi\ep)\,m_1 m_2}{\bigl(1-\sig^{2}+\sig\,\SQ\bigr)\,\Gamma(\tfrac12-\ep)}
   -\frac{i\,q^{2}\,\sec(\pi\ep)\,(m_1+m_2)}{\Gamma(1-\ep)}
 \Biggr)
\\
&\hspace{0cm}
 +\,(-q^2)\;
 \Bigl[-\, i\,\frac{2^{-1+4\ep}\,\pi^{\ep}\,\sec(\pi\ep)\,(m_1+m_2)}{\Gamma(1-\ep)\,m_1^{2} m_2^{2}}\,
   \bigl(-q^{2}(-4+\sig)\,m_1 m_2 + q^{2}\sig\,m_2^{2} + \sig\,m_1^{2}(q^{2}+8 m_2^{2})\bigr)\Bigr]
\Biggr]
\\&\hspace{0cm}\;+\; e^{2} \texttt{g}_c \Biggl\{
\alpha_1\,
\Bigl[
 \; i\,\frac{4^{-5+2\ep}\,\pi^{\ep}\,\sec(\pi\ep)}
             {\Gamma(2-\ep)\,m_1^{3} m_2^{3}\,\mpow{2}}
 \bigl(
  -q^{2}(-1+3\ep)\,(q^{6}+2 q^{4} m_2^{2})
 \\
 \displaybreak[3]
 &\hspace{0cm}
  + 64 m_1^{4}\Bigl(
      q^{4}\ep(1+\ep)
      + q^{2}(-1+5\ep + (-1+\ep)(-3+4\ep)\sig^{2})\,m_2^{2}
    \Bigr)
\\
 &\hspace{0cm}
  + 8 m_1^{2}\Bigl(
      q^{6}-3 q^{6}\ep
      + q^{4}(q^{2}\ep(1+\ep)+(2-6\ep)m_2^{2})
    \Bigr)
 \bigr)
\Bigr]
\\
& \hspace{0cm} +\;
\alpha_2\,
\Bigl[
 \frac{i\,4^{-5+2\ep}\,\pi^{\ep}\,\sec(\pi\ep)}
      {(-1+\ep)^{2}(\sig-\SQ)^{2}\,\Gamma(1-\ep)\,m_2^{3}\,\mpow{2}}\,
 \bigl(1+2\sig(-\sig+\SQ)\bigr)
\\
&\hspace{0cm}
 \times\Bigl(
   (q^{2}+8 m_2^{2})\bigl(q^{6}(1-3\ep)+8 q^{4}\ep(1+\ep)\,m_2^{2}\bigr)\,\alpha_2
 +\,2 m_1^{2}\bigl(32 q^{2}\,m_2^{4} - q^{4}(1-3\ep)(q^{2}+8 m_2^{2})\bigr)\,\alpha_2
\\
&\hspace{0cm}
 +\,4 \sig\,m_1 m_2\bigl(-64 q^{2}(1+\ep(-1+2\ep))\,m_2^{4}\,\alpha_1
        + q^{4}(-1+\ep)^{2}(q^{2}+8 m_2^{2})(\alpha_1+2\alpha_2)\bigr)
 \Bigr)
\Bigr]
\Biggr\}
\\ &\hspace{0cm}\;+\; a\,e^{2} \texttt{g}_c \Biggl[
 -\, \frac{i\,2^{-\tfrac{17}{2}+4\ep}\,\pi^{\ep}\,\sec(\pi\ep)}
             {\Gamma(1-\ep)\,m_2^{3}\,\mpow{2}}\,
 \alpha_2 \Bigl( 128 q^{2} m_1\,m_2^{4}\,(2\sig\,\alpha_1 - \alpha_2)
       + q^{4}(q^{2}+8 m_2^{2})\,(2 m_2 \alpha_1 + m_1 \alpha_2) \Bigr)
\\
&\hspace{0cm}
 +\, \frac{i\,2^{-\tfrac{17}{2}+4\ep}\,\pi^{\ep}\,\sec(\pi\ep)}
             {\Gamma(1-\ep)\,m_1^{3}\,\mpow{2}}\,
 \alpha_1 \Bigl(
   128 q^{2} m_1^{4} m_2\,(\alpha_1-2\sig\,\alpha_2)
\\
&\hspace{0cm}
       - q^{4}(q^{2}+8 m_1^{2})\,(m_2 \alpha_1 + 2 m_1 \alpha_2)
 \Bigr)
\Biggr]
    +\mathcal{O}(q^2 \log(-q^2))\Biggr]\,.\tag{\theequation}
\end{align*}
\endgroup
\end{tcolorbox}
\restorethesisbodyformat
\section{Post-Minkowskian potential: The Lippmann-Schwinger equation and the Infrared subtraction} \label{ch4:sec4}   
The Post-Minkowskian potential can be derived from the Lippmann-Schwinger equation. The computation will be easier in the center-of-mass coordinate. The kinematics we choose is the following,
\begin{align}
\begin{split}
&   \texttt{incoming momenta:}\,\, k_1^\mu=(E_1,\boldsymbol{k}),\,k_2^\mu=(E_2,-\boldsymbol{k}),\\ &
\texttt{outgoing momenta:} k_1^{'\mu}=(E_1,\boldsymbol{k+q}), \,k_2^{'\mu}=(E_2,\boldsymbol{-k-q})\,.
\end{split}
\end{align}
One can define the potential in terms of scattering amplitude using the Lippmann-Schwinger equation, which states,
\begin{align}
    \mathbfcal{A}(\boldsymbol{k},\boldsymbol{k'})=V(\boldsymbol{k},\boldsymbol{k'})+\int d^3\boldsymbol{\ell} \,\mathbfcal{A}(\boldsymbol{k},\boldsymbol{\ell}) \,G(\boldsymbol{k},\boldsymbol{\ell})\,V(\boldsymbol{\ell},\boldsymbol{k'}),\label{4.2f}
\end{align}
where $G(\boldsymbol{k},\boldsymbol{\ell})$ is the Green function, and the choice is somewhat arbitrary. However, consistency conditions on scattering amplitude, such as unitarity and demanding that the potential should be a real one, can show (using the optical theorem) that,
\begin{align}
\texttt{Im}\,G(\boldsymbol{k},\boldsymbol{\ell})=\frac{\delta(|\boldsymbol{\ell}|^2-|\boldsymbol{k}|^2)}{16\pi^2 \sqrt{s}}.
\end{align}
The simplest choice of the Green function that satisfies the above is the following 

\begin{align}
\begin{split}
G(\boldsymbol{k},\boldsymbol{\ell})=\frac{1}{(2\pi)^3}\frac{1}{2\sqrt{s}}\frac{1}{|\boldsymbol{\ell}|^2-|\boldsymbol{k}|^2-i0}+\cdots\,.\label{4.4g}
   \end{split}
\end{align}
\textcolor{black}{In principle, one may add to \eqref{4.4g} any term that is regular, provided the constraint is satisfied. Different choices correspond to alternative effective potentials $V$ that differ only by off-shell pieces \cite{Correia:2024jgr}. 
Importantly, the constant piece will lead to a modification of the effective potential in the classical regime. In our computation, we fix the constant piece by comparing it with the traditional Lippmann-Schwinger Green's function, and we will comment on that in detail in the subsequent discussions.}
\begin{figure}
 \hspace{0cm}   \centering    \includegraphics[width=0.9\linewidth]{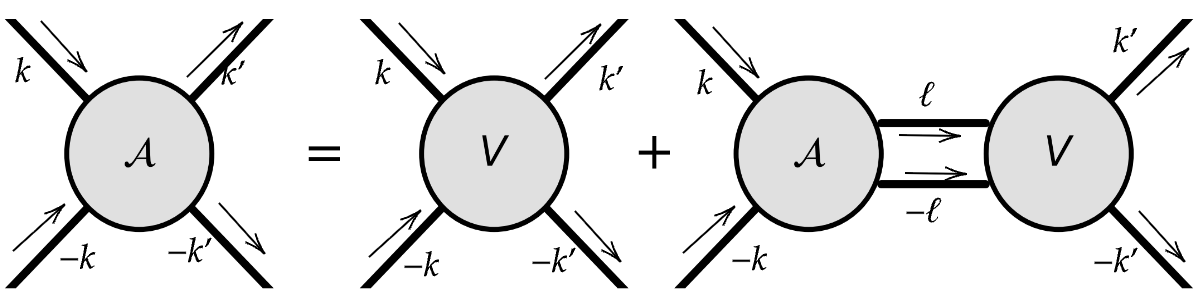}
    \caption{Diagrammatic representation of Lippmann-Schwinger equation}
    \label{fig1}
\end{figure}
Now, by inverting \eqref{4.2f} and truncating at one-loop order, we can write the potential in momentum space iteratively in terms of the scattering amplitude (see Fig.~\ref{fig1} for the diagrammatic representation),
\begin{align}
    V(\boldsymbol k,\boldsymbol k')\Big|_{\texttt{1-loop}}=\mathbfcal{A}_{\texttt{1-loop}}(\boldsymbol k,\boldsymbol k')-\int d^3\boldsymbol{\ell}\,\mathbfcal{A}_{\texttt{tree}}(\boldsymbol{k},\boldsymbol{\ell}) G(\boldsymbol{k},\boldsymbol{\ell})\mathbfcal{A}_{\texttt{tree}}(\boldsymbol{\ell},\boldsymbol{k'})\,. \label{4.5h}
\end{align}
Before proceeding to the explicit computation of the potential, let's pause to discuss some intricacies in the choice of the Green's function.
\\
\paragraph{A digression on Green's functions.}
\textcolor{black}{In many amplitude-based derivations (see e.g.~\cite{Cristofoli:2019neg}) one introduces the two-body Green
function
\begin{equation}
  \mathcal{G}(\boldsymbol{k},\boldsymbol{\ell})
  =\frac{1}{E_{\boldsymbol{k}}-E_{\boldsymbol{\ell}}+i0},
  \qquad
  E_{\boldsymbol{\ell}}=\omega_1(\boldsymbol{\ell})+\omega_2(\boldsymbol{\ell})\, .
\end{equation}
In the main text, we instead employ the representation~\eqref{4.4g}, whose leading singular term comes
with the opposite $i0$-prescription. This apparent mismatch is harmless for the conservative observables
we compute: the two choices become equivalent inside the loop integrals once one is allowed to add
terms that are \emph{regular} at $\boldsymbol{\ell}^{2}=\boldsymbol{k}^{2}$ (the ``regular terms'' in~\eqref{4.4g}).
To see this explicitly, Taylor-expand $\mathcal{G}(\boldsymbol{k},\boldsymbol{\ell})$ around
$\boldsymbol{\ell}\simeq\boldsymbol{k}$. Writing the energy difference as an expansion in
$\Delta\equiv(\boldsymbol{\ell}^{2}-\boldsymbol{k}^{2})$, one finds}
\textcolor{black}{\begin{align}
  E_{\boldsymbol{k}}-E_{\boldsymbol{\ell}}+i0
  &=
  -\left.\frac{\partial E_{\boldsymbol{\ell}}}{\partial \boldsymbol{\ell}^{2}}\right|_{\boldsymbol{\ell}^{2}=\boldsymbol{k}^{2}} \Delta
  -\frac12\left.\frac{\partial}{\partial \boldsymbol{\ell}^{2}}
  \Big(\frac{\partial E_{\boldsymbol{\ell}}}{\partial \boldsymbol{\ell}^{2}}\Big)\right|_{\boldsymbol{\ell}^{2}=\boldsymbol{k}^{2}}\Delta^{2}
  +\cdots + i0\,, \nonumber\\[2pt]
  \left.\frac{\partial E_{\boldsymbol{\ell}}}{\partial \boldsymbol{\ell}^{2}}\right|_{\boldsymbol{\ell}^{2}=\boldsymbol{k}^{2}}
  &=\frac{1}{2\,\xi(\boldsymbol{k})\,E_{\boldsymbol{k}}}\,,\qquad
  \left.\frac{\partial}{\partial \boldsymbol{\ell}^{2}}
  \Big(\frac{\partial E_{\boldsymbol{\ell}}}{\partial \boldsymbol{\ell}^{2}}\Big)\right|_{\boldsymbol{\ell}^{2}=\boldsymbol{k}^{2}}
  =-\frac{1-3\xi(\boldsymbol{k})}{4\,\xi(\boldsymbol{k})^{3}\,E_{\boldsymbol{k}}^{3}}\, .
\end{align}}
\textcolor{black}{
In the large center-of-mass energy regime (large $E_{\boldsymbol{k}}$) this yields
\begin{equation}
  \mathcal{G}(\boldsymbol{k},\boldsymbol{\ell})
  =-\frac{2E_{\boldsymbol{k}}\,\xi(\boldsymbol{k})}{\boldsymbol{\ell}^{2}-\boldsymbol{k}^{2}+i0}
  +\frac{3\xi(\boldsymbol{k})-1}{2\,\xi(\boldsymbol{k})\,E_{\boldsymbol{k}}}
  +\mathcal{O}\!\left(\frac{\boldsymbol{\ell}^{2}-\boldsymbol{k}^{2}}{E_{\boldsymbol{k}}^{3}}\right),
  \qquad
  \xi(\boldsymbol{k})=\frac{\omega_1(\boldsymbol{k})\omega_2(\boldsymbol{k})}{(\omega_1(\boldsymbol{k})+\omega_2(\boldsymbol{k}))^{2}}\, .
  \label{eq:Gexp-large-s}
\end{equation}
On the other hand, our choice~\eqref{4.4g} has the leading behavior}
\textcolor{black}{
\begin{equation}
  G(\boldsymbol{k},\boldsymbol{\ell})
  =\frac{1}{(2\pi)^{3}}\frac{1}{2\sqrt{s}}\,
  \frac{1}{\boldsymbol{\ell}^{2}-\boldsymbol{k}^{2}-i0}+\cdots
  \;\;\longrightarrow\;\;
  \frac{1}{(2\pi)^{3}}\frac{1}{2E_{\boldsymbol{k}}}\,
  \frac{1}{\boldsymbol{\ell}^{2}-\boldsymbol{k}^{2}-i0}+\cdots\, ,
  \label{eq:G-ours-leading}
\end{equation}
which differs from~\eqref{eq:Gexp-large-s} (up to normalization) by an overall sign and by the sign of the
$i0$-prescription. Nevertheless, both prescriptions lead to the same integrated result for the class of
amplitudes we need, namely
\begin{equation}
  \mathcal{A}=\int\!\frac{d^{d}\boldsymbol{\ell}}{(2\pi)^{d}}\,
  \frac{\mathrm{GF}(\boldsymbol{k},\boldsymbol{\ell})}{|\boldsymbol{\ell}-\boldsymbol{k}|^{2}\,|\boldsymbol{\ell}-\boldsymbol{k}'|^{2}}\, .
\end{equation}
Indeed, after the standard manipulations, one encounters an integral of the schematic form
\begin{equation}
  \mathcal{A}[G]=
  \int d^{d}\boldsymbol{\ell}\,
  \frac{1}{\big(\boldsymbol{\ell}\!\cdot\!\tilde{\boldsymbol{q}}_{\perp}-i0\big)\,\boldsymbol{\ell}^{2}\,(\boldsymbol{\ell}-\boldsymbol{q})^{2}}\, .
\end{equation}
Now perform the change of variables $\boldsymbol{\ell}\to-\boldsymbol{\ell}$ together with
$\boldsymbol{q}\to-\boldsymbol{q}$ (which leaves the integration domain invariant). This gives
\begin{equation}
  \mathcal{A}[G]=
  -\int d^{d}\boldsymbol{\ell}\,
  \frac{1}{\big(\boldsymbol{\ell}\!\cdot\!\tilde{\boldsymbol{q}}_{\perp}+i0\big)\,\boldsymbol{\ell}^{2}\,(\boldsymbol{\ell}-\boldsymbol{q})^{2}}
  \equiv \mathcal{A}[\mathcal{G}]\, ,
\end{equation}
showing that the two seemingly different Green functions yield the same result once integrated.
Finally, the ellipsis in~\eqref{eq:G-ours-leading} can be fixed by matching to~\eqref{eq:Gexp-large-s} up to
terms analytic at $\boldsymbol{\ell}^{2}=\boldsymbol{k}^{2}$; a convenient choice is
\begin{equation}
  G(\boldsymbol{k},\boldsymbol{\ell})=\frac{1}{(2\pi)^{3}}
  \left[
    \frac{1}{2E_{\boldsymbol{k}}(\boldsymbol{\ell}^{2}-\boldsymbol{k}^{2}-i0)}
    +\frac{3\xi(\boldsymbol{k})-1}{8\,\xi^2(\boldsymbol{k})\,E_{\boldsymbol{k}}^3}
    +\mathcal{O}\!\left(\frac{\boldsymbol{\ell}^{2}-\boldsymbol{k}^{2}}{E_{\boldsymbol{k}}^{5}}\right)
  \right],
\end{equation}
where the subleading pieces are precisely the ``regular terms'' alluded to below~\eqref{4.4g}.}
\\\\
Therefore, the potential in position space can be computed by taking the Fourier transform of \eqref{4.5h}. We have the one-loop and tree-level amplitude. Then our first task to compute the iterative amplitude  \eqref{4.5h}, which yields (we now work in $d$ dimension to isolate the IR divergence),
\begin{align}
  \mathbfcal{A}_{\texttt{Iterative}}=  \int d^d\boldsymbol{\ell}\,\mathbfcal{A}_{\texttt{tree}}(\boldsymbol{k},\boldsymbol{\ell}) G(\boldsymbol{k},\boldsymbol{\ell})\mathbfcal{A}_{\texttt{tree}}(\boldsymbol{\ell},\boldsymbol{k'})
\end{align}
where the tree amplitude is given by,
\begin{align}
\begin{split}
    \mathbfcal{A}_{\texttt{tree}}(\boldsymbol{k},\boldsymbol{\ell})=&-\frac{1}{|\boldsymbol{\ell}-\boldsymbol{k}|^2}\left(-\frac{4  a^2 m_1 m_2}{m_p^2}+16  \alpha _1 \alpha _2 \texttt{g}_ce^2 m_1 m_2 \sigma  +\frac{ m_1^2 m_2^2 \left(-2 \sigma ^2 +1\right)}{2 m_p^2}\right)\,,\\ &
    +\left(4  \alpha _1 \alpha _2 \texttt{g}_c e^2 -\frac{ 2 m_2 m_1 \sigma   }{4  m_p^2}\right)\,,\\ &
    =-c_{1}(\sigma)\frac{1}{|\boldsymbol{\ell}-\boldsymbol{k}|^2}+c_2(\sigma)\,.
    \end{split}
    \end{align}
Therefore, the iterative Born-subtracted amplitude is given by,
\begin{align}
    \begin{split}
        \mathbfcal{A}_{\texttt{Iterative}}=\mathbfcal{A}^{(1)}_{\texttt{Iterative}}+\mathbfcal{A}^{(2)}_{\texttt{Iterative}}+\mathbfcal{A}^{(3)}_{\texttt{Iterative}}
    \end{split}
\end{align}
where,
\begin{align}
    \begin{split}\mathbfcal{A}^{(1)}_{\texttt{Iterative}}(\boldsymbol{k},\boldsymbol{k'})&= \frac{c_1^2(\sigma)}{2E_{\boldsymbol{k}}} \int \frac{d^d\boldsymbol{\ell}}{(2\pi)^3}\,\frac{1}{|\boldsymbol{\ell}-\boldsymbol{k}|^2(\boldsymbol{\ell}^2-\boldsymbol{k}^2-i0)|\boldsymbol{\ell}-\boldsymbol{k'}|^2}\,,\\ &
    =\frac{c_1^2(\sigma)}{2E_{\boldsymbol{k}}}  \int \frac{d^d\boldsymbol{\ell}}{(2\pi)^3}\frac{1}{(\boldsymbol{\ell}^2+2\boldsymbol{\ell}\cdot \boldsymbol{k}-i0)\boldsymbol{\ell}^2(\boldsymbol{\ell}-\boldsymbol{q})^2}\,.\label{4.20jj}
    \end{split}
\end{align}
As before, the integral can be done using the soft loop expansion.
\begin{align}
    \begin{split}\int_{\boldsymbol{\ell\sim\boldsymbol{q}}} \frac{d^d\boldsymbol{\ell}}{(2\pi)^3}\frac{1}{(\boldsymbol{\ell}^2+2\boldsymbol{\ell}\cdot \boldsymbol{k}-i0)\boldsymbol{\ell}^2(\boldsymbol{\ell}-\boldsymbol{q})^2}&=\int_{\boldsymbol{\ell\sim\boldsymbol{q}}} \frac{d^d\boldsymbol{\ell}}{(2\pi)^3}\frac{1}{(\boldsymbol{\ell}^2+2\boldsymbol{\ell}\cdot (\frac{\tilde{\boldsymbol{q}}_{\perp}-\boldsymbol{q}}{2})-i0)\boldsymbol{\ell}^2(\boldsymbol{\ell}-\boldsymbol{q})^2}\,,\\ &    \hspace{0cm}=\int_{\boldsymbol{\ell\sim\boldsymbol{q}}} \frac{d^d\boldsymbol{\ell}}{(2\pi)^3}\left(1-\frac{\boldsymbol{\ell}^2-\boldsymbol{\ell}\cdot \boldsymbol{q}}{\boldsymbol{\ell}\cdot\tilde{\boldsymbol{q}}_{\perp}}+\cdots\right) \frac{1}{(\boldsymbol{\ell}\cdot\boldsymbol{q_{\perp}}-i0)\boldsymbol{\ell}^2(\boldsymbol{\ell-q})^2}\label{4.10o}
    \end{split}
\end{align}
where, $\tilde{\boldsymbol{q}}_{\perp}=\boldsymbol{q}+2\boldsymbol{k}$ and also note that $\tilde{\boldsymbol{q}}_{\perp}\cdot \boldsymbol{q}=0$. Keeping this in mind the first term in \eqref{4.10o} takes the form,
\begin{align}
    \textstyle{\mathbfcal{A}^{(1)}_{\texttt{Iterative}}(\boldsymbol{k},\boldsymbol{k'})=\frac{c^2_1(\sigma)}{2E_{\boldsymbol{k}}}\int_{\boldsymbol{\ell\sim\boldsymbol{q}}} \frac{d^d\boldsymbol{\ell}}{(2\pi)^3} \frac{1}{(\boldsymbol{\ell}\cdot\tilde{\boldsymbol{q}}_{\perp}-i0)\boldsymbol{\ell}^2(\boldsymbol{\ell-q})^2}}&\xrightarrow[\boldsymbol{q}\to -\boldsymbol{q}]{\boldsymbol{\ell}\to -\boldsymbol{\ell}}-\frac{c^2_{1}(\sigma)}{2E_k}   \int d^{d}\boldsymbol{\ell}\frac{1}{(\boldsymbol{\ell}\cdot \tilde{\boldsymbol{q}}_{\perp}+i0)\boldsymbol{\ell}^2(\boldsymbol{\ell}-\boldsymbol{q})^2},
\end{align}
and adding the two, we get,
\begin{align}
   \begin{split}\mathbfcal{A}^{(1)}_{\texttt{Iterative}}(\boldsymbol{k},\boldsymbol{k'})= \frac{\pi i c_1^2(\sigma)}{2E_{\boldsymbol{k}}}\int \frac{d^d\boldsymbol{\ell}}{(2\pi)^3} \frac{\delta(\boldsymbol{\ell}\cdot\tilde{\boldsymbol{q}}_{\perp})}{\boldsymbol{\ell}^2(\boldsymbol{\ell}-\boldsymbol{q})^2}&=\frac{c_1^2(\sigma)}{2E_{\boldsymbol{k}}}\frac{i\pi}{2\pi |\tilde{\boldsymbol{q}}_{\perp}|}\int \frac{d^{d-1}}{(2\pi)^2}\boldsymbol{\ell} \frac{1}{\boldsymbol{\ell}^2(\boldsymbol{\ell}-\boldsymbol{q})^2}\,,\\ &
 =\frac{c_1^2(\sigma)}{2E_{\boldsymbol{k}}}\frac{i}{2|\boldsymbol{q}_\perp|}\left(\frac{ \log (|\boldsymbol{q}|^2)+\gamma_{E} -\log (4 \pi )}{2 \pi  |\boldsymbol{q}|^2}-\frac{1}{2 \pi |\boldsymbol{ q}|^2 \epsilon }\right)\,.
    \end{split}
\end{align}
Again note that, in the soft limit $|\boldsymbol{q}_\perp|=\sqrt{4\boldsymbol{k}^2-\boldsymbol{q}^2}\sim {2}|\boldsymbol{k}|$.
As we immediately see that the iterative amplitude has an IR divergent piece which is given by,
\begin{align}
    \mathbfcal{A}_{\texttt{Iterative}}^{(1)}(\boldsymbol{k},\boldsymbol{k'})\Bigg|_{\texttt{IR div.}}= -\frac{i c_1^2(\sigma)}{16 \pi E_{\boldsymbol{k}}|\boldsymbol{k}||\boldsymbol{q}|^2\,\epsilon}=-\frac{ic_1^2(\sigma)}{16\pi m_1 m_2 \sqrt{\sigma^2-1}|\boldsymbol{q}|^2\epsilon}\,.
\end{align}
On the other hand,  the one-loop amplitude computed in \eqref{3.54j} has an IR divergent piece,
\begin{align}
    \begin{split}\mathbfcal{A}_{\texttt{1-loop}}\Bigg|_{\texttt{IR div.}}&=-\frac{i m_1 m_2 \left(8 \left(a^2-4 \alpha _1 \alpha _2 e^2 \sigma  \texttt{g}_c m_p^2\right)+m_1 m_2 \left(2 \sigma ^2-1\right)\right){}^2}{64 \pi  \sqrt{\sigma ^2-1} \epsilon \, m_p^4 |\boldsymbol{q}|^2}\,,\\ & =\frac{-ic_1^2(\sigma)}{16\pi m_1 m_2\sqrt{\sigma^2-1}|\boldsymbol{q}|^2\epsilon}\to \mathbfcal{A}_{\texttt{Iterative}}^{(1)}(\boldsymbol{k},\boldsymbol{k'})\Bigg|_{\texttt{IR div.}} \,.
    \end{split}
\end{align}
From the above analysis, \textit{We observe a complete cancellation of infrared singularities: the IR divergence present in the one-loop amplitude is exactly removed by the iterative (Born) contribution. Hence, the Born (IR)-subtracted Post-Minkowskian potential is IR finite. \textcolor{black}{The same subtraction scheme has been first applied for general relativity in~\cite{Cristofoli:2019neg}, where the Born-subtracted potential was also demonstrated to be IR finite.}}   As we will see, the other integrals appearing in the iterative amplitude are free of IR divergence.
Similarly, we have,
\begin{align}
    \begin{split}
        \mathbfcal{A}_{\texttt{Iterative}}^{(2)}(\boldsymbol{k},\boldsymbol{k'})&\sim-c_1(\sigma)c_2(\sigma)\int d^d\boldsymbol{\ell}\frac{1}{|\boldsymbol{\ell}-\boldsymbol{k}|^2(\boldsymbol{\ell}^2-\boldsymbol{k}^2)}\,,\\ &
        =-c_1(\sigma)c_2(\sigma)\int d^d\boldsymbol{\ell}\frac{1}{\boldsymbol{\ell}^2(\boldsymbol{\ell}^2+2\boldsymbol{\ell}\cdot\boldsymbol{k}-i0)}=0\,\, (\texttt{scaleless in soft limit})\,,\\ &
        \to \mathbfcal{A}_{\texttt{Iterative}}^{(3)}(\boldsymbol{k},\boldsymbol{k'}) \,\,(\texttt{in soft limit})\,.
    \end{split}
\end{align}
However, we are not completely done yet. We see that the iterative Born subtraction only has a superclassical contribution. However, we still have the freedom to add a regular term coming from the Taylor expansion mentioned in the footnote. The Green function takes the form,
\begin{align}
    G(\boldsymbol{k},\boldsymbol{\ell})=\frac{1}{(2\pi)^3}\frac{1}{2\sqrt{s}}\frac{1}{|\boldsymbol{\ell}|^2-|\boldsymbol{k}|^2-i0}+\frac{3\xi-1}{64 \pi^3 \xi^2 {s^{\frac{3}{2}}}}+\cdots\,.
\end{align}
Therefore, the only surviving correction to the iterative amplitude in the soft limit is the following,
\begin{align}
    \begin{split}
        \mathbfcal{A}_{\texttt{Iterative}}'(\boldsymbol{k},\boldsymbol{k'})=c_1^2(\sigma)\frac{3\xi-1}{64\pi^3\xi^2s^{\frac{3}{2}}}\int \frac{d^d\boldsymbol{\ell}}{\boldsymbol{\ell}^2(\boldsymbol{\ell}-\boldsymbol{q})^2}=c_1^2(\sigma)\frac{3\xi-1}{64\xi^2s^{\frac{3}{2}}|\boldsymbol{q}|}\,.
    \end{split}
\end{align}
Therefore, the one-loop classical potential in momentum space is given by,
\begin{align}
V(\boldsymbol{k},|\boldsymbol{q}|)=\mathbfcal{A}_{\texttt{1-loop}}(\boldsymbol{k},|\boldsymbol{q}|)\Big|_{\frac{1}{|\boldsymbol{q}|}}-\frac{c_1^2(\sigma)(3\xi-1)}{64\xi^2 E_{\boldsymbol{k}}^3 |\boldsymbol{q}| }\,.
\end{align}
The two-body classical  potential can be computed by taking the Fourier transform of the momentum space amplitude,
\begin{align}   
\begin{split}
V_{\texttt{1-loop}}(\boldsymbol{k},|\boldsymbol{r}|)=-\frac{1}{
4m_1m_2}\int \frac{d^3\boldsymbol{q}}{(2\pi)^3}e^{i\boldsymbol{q}\cdot \boldsymbol{r}}\,V(\boldsymbol{k},|\boldsymbol{q}|) =-\frac{\mathcal{C}_2(\sigma)}{8m_1m_2\pi^2 r^2}\,.
\end{split}
\end{align}
where,
\begin{align}
    \begin{split}
         \mathcal{C}_{2}(\sigma) &= \frac{a^4 \left(m_1+m_2\right)}{4 m_p^4}-\frac{a^2 m_1 m_2 \left(128 b+\left(m_1+m_2\right) \left(\sigma ^2-17\right)\right)}{128 m_p^4}\\
         & -\frac{\sqrt{2} a e^2 m_1 m_2 \left(-4 \alpha _2 \alpha _1 \sigma +\alpha _1^2+\alpha _2^2\right) \texttt{g}_c}{4 m_p^2}  +\frac{3 m_1^2 m_2^2 \left(m_1+m_2\right) \left(5 \sigma ^2-1\right)}{512 m_p^4} \\
         &+ 4 \alpha _1^2 \alpha _2^2 e^4 \left(m_1+m_2\right) \texttt{g}_c^2 -\frac{e^2 m_1 m_2 \texttt{g}_c \left(\left(3 \sigma ^2-1\right) \left(\alpha _1^2 m_2+\alpha _2^2 m_1\right)+12 \alpha _1 \alpha _2 \left(m_1+m_2\right) \sigma \right)}{16 m_p^2}\\
         & -\frac{c_1^2(\sigma)(3\xi-1)}{64\xi^2 E_{\boldsymbol{k}}^3  } \,,\label{4.21g}
    \end{split}
\end{align}
with{\footnote{where $a,b$ has mass dimension 1.}} $\sigma=\frac{k_1\cdot k_2}{m_1m_2}\,$. If one turns off the dilaton field and takes the static limit, then the potential written in \eqref{4.21g} agrees with \cite{Bjerrum-Bohr:2002aqa,Faller:2007sy}. Before closing this section, we briefly comment on a mild ambiguity that arises from the freedom to add \emph{regular} (analytic) terms to the Green's function, as mentioned earlier.\\\\
\textcolor{black}{
\textbf{Comment on adding regular pieces in the Green's function.}  The freedom to add \emph{regular} (analytic) terms to the Green’s function only affects the off-shell extension of the integrand and can at most generate
shorter-range contributions to the conservative potential. In particular, such terms do \emph{not} modify the IR-singular piece that is removed by the Born/iteration subtraction.
To see this explicitly, note that adding analytic regular terms produces (schematically) loop integrals of the form
\begin{align}
\mathcal{I}_n(\boldsymbol{q},\boldsymbol{q}_\perp)
\;\equiv\;
\int d^d\boldsymbol{\ell}\;
\frac{\bigl(2\,\boldsymbol{\ell}\!\cdot\!\boldsymbol{q}_\perp\bigr)^n}{\boldsymbol{\ell}^{\,2}\,(\boldsymbol{\ell}-\boldsymbol{q})^{2}}\,,
\qquad n\ge 1,
\label{eq:In_def_refreply}
\end{align}
where $\boldsymbol{q}_\perp$ is transverse, $\boldsymbol{q}\!\cdot\!\boldsymbol{q}_\perp=0$ (mass on-shell condition: $|\boldsymbol{k}|^2=|\boldsymbol{k}+\boldsymbol{q}|^2$). By symmetry, $\mathcal{I}_n$ vanishes identically for odd $n$.
For even $n$, a standard IBP reduction implies that the numerator can only generate powers of
$\boldsymbol{q}^2$ and $\boldsymbol{q}_\perp^{\,2}$, so that one may write
\begin{align}
\mathcal{I}_n(\boldsymbol{q},\boldsymbol{q}_\perp)
=
C_n(d)\,(\boldsymbol{q}^{2})^{n/2}(\boldsymbol{q}_\perp^{\,2})^{n/2}
\int d^d\boldsymbol{\ell}\,\frac{1}{\boldsymbol{\ell}^{\,2}(\boldsymbol{\ell}-\boldsymbol{q})^{2}}
\;\propto\;
(\boldsymbol{q}^2)^{\frac{n-1}{2}}
\qquad (\text{even } n),
\label{eq:In_scaling_refreply}
\end{align}
where in the last step we used the well-known scaling of the massless bubble integral in $d=3-2\epsilon$ spatial dimensions,
$\int d^d\boldsymbol{\ell}\, [\boldsymbol{\ell}^2(\boldsymbol{\ell}-\boldsymbol{q})^2]^{-1}\propto (\boldsymbol{q}^2)^{d/2-2}\sim (\boldsymbol{q}^2)^{-1/2-\epsilon}$.
Crucially, for $n\ge 1$ the behavior \eqref{eq:In_scaling_refreply} is \emph{less singular} in the soft limit than the $n=0$ term, and therefore it does not contribute to the IR divergence that is cancelled by the iteration (Born) subtraction. Finally, the corresponding contribution to the conservative potential is obtained by Fourier transform.
Using $\int d^3\boldsymbol{q}\,e^{i\boldsymbol{q}\cdot\boldsymbol{r}}\,|\boldsymbol{q}|^{\,m}\sim |r|^{-m-3}$ (up to constants), one finds
\begin{align}
V_n(r)\;\sim\;\int d^3\boldsymbol{q}\;e^{i\boldsymbol{q}\cdot\boldsymbol{r}}\,
(\boldsymbol{q}^2)^{\frac{n-1}{2}}
\;\propto\;\frac{1}{|\boldsymbol{r}|^{\,n+2}}\,, \qquad n=2,4,6,\cdots,
\label{eq:Vn_falloff_refreply}
\end{align}
i.e.\ a contribution that is parametrically shorter-ranged (and more suppressed at large $r$) than the leading long-range terms.
Therefore, purely analytic regular terms cannot alter the IR-subtracted \emph{long-range} potential: their effect is confined to short-distance/contact contributions and, more generally, to off-shell terms that can be absorbed by local field redefinitions (or equivalently by canonical transformations).
It is useful to recall how the relevant non-analytic structures in the momentum-transfer expansion map into the large-distance behavior of the potential at $\mathcal{O}(G^2)$.
In three spatial dimensions, the Fourier transform implies the schematic correspondences
\begin{align}
\frac{1}{|\boldsymbol{q}|^{2}}\;\longleftrightarrow\;\frac{G^2}{r\hbar}\,,\qquad
\frac{1}{|\boldsymbol{q}|}\;\longleftrightarrow\;\frac{G^2 \hbar^0}{r^{2}}\,,\qquad
\log(\boldsymbol{q}^{2})\;\longleftrightarrow\;\frac{G^2\hbar}{r^{3}}\,,
\end{align}
which are naturally interpreted as the super-classical, classical, and quantum contributions, respectively.
In our analysis, the apparent super-classical ($\sim 1/\hbar$) pole is spurious and is removed by the standard leading Born subtraction, leaving an IR-finite long-range result.
Regular terms, on the other hand, only introduce an ambiguity in the definition of the post-Minkowskian potential at the level of off-shell completion.
As emphasized in \cite{Correia:2024jgr,Correia:2024yfx}, different choices of regular terms lead to potentials that differ by off-shell pieces, and hence they do not affect on-shell, gauge-invariant observables (such as the scattering angle or other conservative observables derived from the S-matrix).  }
Next, we discuss eikonal exponentiation in momentum space and deduce the scattering angle.

\section{Eikonal exponentiation and the scattering angle}\label{ch4:sec5}
In this section, we will study the eikonal exponentiation of the conservative scattering amplitude directly in the momentum space following \cite{Parra-Martinez:2020dzs}. Traditionally, in eikonal exponentiation, it is well known that the eikonal phase can be computed by taking the Fourier transform of the scattering amplitude in impact-parameter space.
The primary reason to do the computation in momentum space is to avoid additional IR divergence coming from the tree-level amplitudes arising from the Coulomb interaction. However, one must pay the computational cost in momentum space: simple products in impact parameter space give rise to convolutions in momentum space, which can be computed using the iterative bubble integrals. The statement of eikonal exponentiation reduces to the following: one can write the scattering amplitude as a \textit{convolutional exponential} of the eikonal phase,
\begin{align}
\begin{split}
    i\mathbfcal{A}(\sigma,-q^2)&=\textrm{cexp}\left(i\delta(\sigma,-q^2)\right)-1\,,\\ &
    =i\delta(\sigma,-q^2)-\frac{1}{2!}\delta(\sigma,-q^2)\otimes \delta(\sigma,-q^2)-i\frac{1}{3!}\delta(\sigma,-q^2)\otimes \delta(\sigma,-q^2)\otimes\delta(\sigma,-q^2)+\cdots
    \end{split}
\end{align}
where the convolution is defined as an integral over the $d-2$ dimensional transverse space.
\begin{align}    \varpi_1(\boldsymbol{q}_{\perp})\otimes \varpi_2(\boldsymbol{q}_{\perp}):=\frac{1}{N}\int d\mu_{\boldsymbol{\ell}_{\perp}}^{d-2}\,\varpi_1(\boldsymbol{q}_{\perp})\varpi_2(\boldsymbol{q}_{\perp}-\boldsymbol{\ell}_{\perp})\,,
\end{align}
where $N=4 m_1 m_2 \sqrt{\sigma^2-1}$ is the normalization factor. Similarly, one can write the inverse relation as,
\begin{align}
    \begin{split}
        \delta(\sigma,-q^2)&=-i\textrm{clog}(1+i \mathbfcal{A})\,, \\
        &=\mathbfcal{A}(\sigma ,-q^2)-\frac{i}{2}\mathbfcal{A}(\sigma ,-q^2) \otimes\mathbfcal{A}(\sigma ,-q^2) -\frac{1}{3}\mathbfcal{A}(\sigma ,-q^2) \otimes \mathbfcal{A}(\sigma ,-q^2) \otimes\mathbfcal{A}(\sigma ,-q^2)+\cdots\,.
    \end{split}
\end{align}
This can be written as 
\begin{align}
    \begin{split}
        \delta= \delta^{(0)}+\delta^{(1)}+\delta^{(2)}+\cdots
    \end{split}
\end{align}
where $ \delta^{(L)}$ has $\mathcal{O}(e^{2L+2})$,  $\mathcal{O}(G^{L+1})$ and $\mathcal{O}(e^{2L} G^{L})$ terms and so on hence we can write phases as
\begin{align}
    \begin{split}
    &\delta^{(0)}=\mathbfcal{A}^{\textrm{tree}}\,, \\
    & \delta^{(1)}=\mathbfcal{A}^{\textrm{1-loop}} -\frac{i}{2} \mathbfcal{A}^{\textrm{tree}} \otimes \mathbfcal{A}^{\textrm{tree}}\,.
    \end{split}
\end{align}
So we have
\begin{align}
    \delta^{(0)}(\sigma, -q^2)= \frac{1}{q^2} \left( -\frac{4 a^2 m_1 m_2}{m_p^2} + 16 \alpha _1 \alpha _2 e^2   m_1 m_2 \sigma   \texttt{g}_c+\frac{m_1^2 m_2^2 \left(-2 \sigma ^2+2 \sigma ^2 \epsilon +1\right)}{(2-2 \epsilon ) m_p^2} \right) \label{del0p}
\end{align}
where the $\mathcal{O}(-q^0)$ term of $\mathbfcal{A^{\textrm{tree}}}$ won't contribute as they lead to $\delta(\boldsymbol{b_e})$ (which is simply zero as $\boldsymbol{b_e} \neq 0$) when fourier transformed to impact parameter space.
Now using 
\begin{align}
    \frac{1}{-q^2}\otimes\frac{1}{-q^2}=\frac{1}{\boldsymbol{q^2_\perp}}\otimes\frac{1}{\boldsymbol{q^2_\perp}}=\frac{1}{N} (4 \pi )^{\epsilon -1}  \frac{  \Gamma (-\epsilon )^2 \Gamma (\epsilon +1)}{\Gamma (-2 \epsilon )} (\boldsymbol{q}^2_\perp)^{-\epsilon -1}
\end{align}
we get the following, 
\begin{align}
    \begin{split}
        &\delta^{(1)}(\sigma, -q^2) = \frac{i \left(2 m_2 m_1 \sigma +m_1^2+m_2^2\right) \left(64 a^4+m_1^2 m_2^2 (1-2 \sigma ^2\right)^2)}{128 \pi  m_1 m_2 \left(\sigma ^2-1\right)^{3/2} m_p^4} \log{(\boldsymbol{q}^2_\perp )}  \\
        & + \frac{\left(m_1+m_2\right) \left(128 a^4-4 a^2 m_1 m_2 \left(\sigma ^2-17\right)+3 m_1^2 m_2^2 \left(5 \sigma ^2-1\right)\right)-512 a^2 b m_1 m_2}{512 m_p^4 |\boldsymbol{q}_\perp|} \\
        & -\frac{e^2 m_1 m_2 \texttt{g}_c \left(4 \sqrt{2} a \left(\alpha _1^2+\alpha _2^2\right)+4 \alpha _1 \alpha _2 \sigma  \left(3 \left(m_1+m_2\right)-4 \sqrt{2} a\right)+\left(3 \sigma ^2-1\right) \left(\alpha _1^2 m_2+\alpha _2^2 m_1\right)\right)}{16 m_p^2 |\boldsymbol{q}_\perp|} \\
        &  +\frac{4 \alpha _1^2 \alpha _2^2 e^4 \left(m_1+m_2\right) \texttt{g}_c^2}{|\boldsymbol{q_\perp}|}\,.
        \label{del1p}
    \end{split}
\end{align}
Note that the $\mathcal{O}(|\boldsymbol{q}_\perp|^{-2}) $ term of the 1-loop amplitude is completely canceled at $\epsilon^0$ order by the tree convolution. We also note that the $\delta^{(1)}$ has an imaginary part. However, the imaginary part of the eikonal phase does not contribute to the conservative scattering angle. Physically, the imaginary part of the eikonal phase measures the number of massless particles emitted from the two-body system. We now proceed to the computation of the eikonal phase in impact-parameter space by taking the Fourier transform. 
\begin{align}
    \delta(\sigma, \boldsymbol{b}_e )=\frac{1}{N} \int\frac{d^{D-2}\boldsymbol{q}_\perp}{(2 \pi)^{D-2}} e^{i \boldsymbol{b}_e \cdot \boldsymbol{q}_\perp} \delta(\sigma,\boldsymbol{q}_\perp)\,.
\end{align}
So using \eqref{del0p} and \eqref{del1p} respectively we have
\begin{flalign}
   \hspace{0cm}  \delta^{(0)}(\sigma, \boldsymbol{b}_e )&= \left( -\frac{a^2}{4 \pi  \sqrt{\sigma ^2-1} m_p^2}+\frac{\alpha _1 \alpha _2 e^2 \sigma  \texttt{g}_c}{\pi  \sqrt{\sigma ^2-1}}-\frac{m_1 m_2 \left(2 \sigma ^2-1\right)}{32 \pi  \sqrt{\sigma ^2-1} m_p^2} \right) \log(\boldsymbol{b}^2_e)&
\end{flalign}
and,
\begin{align}
    \begin{split}
        \delta^{(1)}(\sigma, \boldsymbol{b}_e ) &= \frac{128 a^4 \left(m_1+m_2\right)-4 a^2 m_1 m_2 \left(128 b+\left(m_1+m_2\right) \left(\sigma ^2-17\right)\right)+3 m_1^2 m_2^2 \left(m_1+m_2\right) \left(5 \sigma ^2-1\right)}{4096 \pi  m_1 m_2 \sqrt{\sigma ^2-1}  m_p^4 |\boldsymbol{b}_e| } \\ & -\frac{e^2 \texttt{g}_c \left(4 \sqrt{2} a \left(\alpha _1^2+\alpha _2^2\right)+4 \alpha _1 \alpha _2 \sigma  \left(-4 \sqrt{2} a+3 m_1+3 m_2\right)+\left(3 \sigma ^2-1\right) \left(\alpha _1^2 m_2+\alpha _2^2 m_1\right)\right)}{128 \pi  \sqrt{\sigma ^2-1} m_p^2 |\boldsymbol{b}_e|} \\
        & + \frac{\alpha _1^2 \alpha _2^2 e^4 \left(m_1+m_2\right) \texttt{g}_c^2}{2 \pi  m_1 m_2 \sqrt{\sigma ^2-1} |\boldsymbol{b}_e|}\,.
    \end{split}
\end{align}
\textbf{\textit{Scattering angle:}}\\
The phase in impact parameter space and the amplitude in momentum space are related as
\begin{align}
    \mathbfcal{A}(\sigma,-q^2)=\int d^{D-2} \boldsymbol{b}_e \left( e^{i \delta(\sigma,\boldsymbol{b}_e)} -1 \right) e^{-i \boldsymbol{b}_e \cdot \boldsymbol{q}} 
\end{align}
and stationary phase approximation yields the relation
\begin{align}
    \boldsymbol{q}=-\frac{\partial}{\partial\boldsymbol{b}_e} \delta(\sigma,\boldsymbol{b}_e)\,. \label{5.14}
\end{align}
In the centre-of-mass frame, $\boldsymbol{q}$, scattering angle $\chi$ and the COM momentum $\boldsymbol{p}$ are related as 
\begin{align}
    | \boldsymbol{q}|=2 |\boldsymbol{k}| \sin{\frac{\chi}{2}} \label{5.15}
\end{align}
where in terms of centre-of-mass energy
\begin{align}
    |\boldsymbol{k}|=\frac{m_1 m_2 \sqrt{\sigma^2 - 1}}{E} = \frac{m_1 m_2 \sqrt{\sigma^2 - 1}}{\sqrt{m_1^2 + m_2^2 +2 m_1 m_2 \sigma}}\,.
\end{align}
So from \eqref{5.14} and \eqref{5.15} we have
\begin{align}
    \sin{\frac{\chi}{2}}=-\frac{1}{2 |\boldsymbol{k}|} \frac{\partial}{\partial |\boldsymbol{b}_e|} \Re \delta(\sigma,\boldsymbol{b}_e)\,. \label{5.17}
\end{align}
Now we can write RHS of \eqref{5.17} by order
\begin{flalign}
    \frac{\chi^{(0)}}{2}&=\frac{a^2}{4 \pi  |\boldsymbol{k}||\boldsymbol{b}_e| \sqrt{\sigma ^2-1} m_p^2} - \frac{\alpha _1 \alpha _2 e^2 \sigma  \texttt{g}_c}{\pi  |\boldsymbol{k}||\boldsymbol{b}_e| \sqrt{\sigma ^2-1}}+\frac{m_1 m_2 \left(2 \sigma ^2-1\right)}{32 \pi  |\boldsymbol{k}||\boldsymbol{b}_e| \sqrt{\sigma ^2-1} m_p^2}\,, &
\end{flalign}

\begin{align}
    \begin{split}
        \frac{\chi^{(1)}}{2} &= \frac{a^4 \left(m_1+m_2\right)}{64 \pi  m_1 m_2 |\boldsymbol{k}||\boldsymbol{b}^2_e| \sqrt{\sigma ^2-1} m_p^4}-\frac{a^2 \left(128 b+\left(m_1+m_2\right) \left(\sigma ^2-17\right)\right)}{2048 \pi  |\boldsymbol{k}||\boldsymbol{b}^2_e| \sqrt{\sigma ^2-1} m_p^4}  +\frac{3 m_1 m_2 \left(m_1+m_2\right) \left(5 \sigma ^2-1\right)}{8192 \pi  |\boldsymbol{k}||\boldsymbol{b}^2_e| \sqrt{\sigma ^2-1} m_p^4} \\
        & -\frac{e^2 \texttt{g}_c \left(4 \sqrt{2} a \left(\alpha _1^2+\alpha _2^2\right)+4 \alpha _1 \alpha _2 \sigma  \left(-4 \sqrt{2} a+3 m_1+3 m_2\right)+\left(3 \sigma ^2-1\right) \left(\alpha _1^2 m_2+\alpha _2^2 m_1\right)\right)}{256 \pi  |\boldsymbol{k}||\boldsymbol{b}^2_e| \sqrt{\sigma ^2-1} m_p^2} \\
        & + \frac{\alpha _1^2 \alpha _2^2 e^4 \left(m_1+m_2\right) \texttt{g}_c^2}{4 \pi  m_1 m_2 |\boldsymbol{k}||\boldsymbol{b}^2_e| \sqrt{\sigma ^2-1}}\,.
    \end{split}
\end{align}
The above results can be written in terms of angular momentum 
\begin{align}
    J=|\boldsymbol{b} \times \boldsymbol{k}|,
\end{align}
where $\boldsymbol{b}$ is the impact parameter perpendicular to the incoming centre-of-mass momentum $\boldsymbol{p}$. Note, however, this impact parameter is different from the $\boldsymbol{b}_e$ occurring in the eikonal phase, which points in the direction of momentum transfer $\boldsymbol{q}$. The magnitude of $\boldsymbol{b}_e$ and $\boldsymbol{b}$ are related by
\begin{align}
    |\boldsymbol{b}|=|\boldsymbol{b}_e| \cos{\frac{\chi}{2}}.
\end{align}
This difference is unimportant for small-angle scattering, and it will only matter at order $J^{-3}$. So in terms of $J$ and $G$ ($ m_p=(32 \pi G)^{-1/2}$)  we have,
\begin{flalign}
    \frac{\chi^{(0)}}{2}&=\frac{8 a^2 G}{J \sqrt{\sigma ^2-1}}-\frac{\alpha _1 \alpha _2 e^2 \sigma  \texttt{g}_c}{\pi  J \sqrt{\sigma ^2-1}}+\frac{G m_1 m_2 \left(2 \sigma ^2-1\right)}{J \sqrt{\sigma ^2-1}}\,, &
\end{flalign}

\begin{align}
    \begin{split}
        \frac{\chi^{(1)}}{2}&= \frac{16 \pi  a^4 G^2 \left(m_1+m_2\right)}{J^2 \sqrt{2 m_2 m_1 \sigma +m_1^2+m_2^2}} -\frac{\pi  a^2 G^2 m_1 m_2 \left(128 b+\left(m_1+m_2\right) \left(\sigma ^2-17\right)\right)}{2 J^2 \sqrt{2 m_2 m_1 \sigma +m_1^2+m_2^2}} \\
        & + \frac{3 \pi  G^2 m_1^2 m_2^2 \left(m_1+m_2\right) \left(5 \sigma ^2-1\right)}{8 J^2 \sqrt{2 m_2 m_1 \sigma +m_1^2+m_2^2}} +\frac{\alpha _1^2 \alpha _2^2 e^4 \left(m_1+m_2\right) \texttt{g}_c^2}{4 \pi  J^2 \sqrt{2 m_2 m_1 \sigma +m_1^2+m_2^2}} \\
        & -\frac{e^2 G m_1 m_2 \texttt{g}_c \left(4 \sqrt{2} a \left(\alpha _1^2+\alpha _2^2\right)+4 \alpha _1 \alpha _2 \sigma  \left(-4 \sqrt{2} a+3 m_1+3 m_2\right)+\left(3 \sigma ^2-1\right) \left(\alpha _1^2 m_2+\alpha _2^2 m_1\right)\right)}{8 J^2 \sqrt{2 m_2 m_1 \sigma +m_1^2+m_2^2}}\,.
    \end{split}
\end{align}
If we `turn off' the coupling, i.e $G \rightarrow 0$ \cite{Bern:2021xze} or $e \rightarrow0$ \cite{Mogull:2020sak}, and set the screening constants $a_i \, ,  b_i =0$, we retrieve the known gravitational and (scalar) electromagnetic scattering angles. 
\section{Conclusions and Discussion}
\label{sec:conclusions}
We have analyzed classical scattering of charged, non-spinning compact objects in EMD theory using modern scattering–amplitude methods, with cross–checks based on potential theory. Our goal was to extract the conservative two–body potential, establish the infrared (IR) structure and its cancellation in momentum space, and compute the eikonal phase and associated scattering angle through one loop (classical $2{\rm PM}$ order).

\begin{itemize}
  \item  We have motivated EMD as a low–energy effective description inspired by heterotic string theory, in which additional long–range fields (dilaton and gauge bosons) arise from first principles. This provides a UV-motivated setting in which higher–curvature corrections and extra degrees of freedom naturally coexist, while our concrete computations focused on the leading two-derivative EMD sector.
  \item After fixing conventions and gauges, we derived the propagators and Feynman rules needed for $2\!\to\!2$ scattering. This delivers a set of building blocks separating purely gravitational, electromagnetic, and dilatonic exchanges and their interference.
  \item  Using dimensional regularization, expansion by regions, and IBP reduction, we obtained a basis of master integrals and their soft (classical) expansions that capture the long–range, nonanalytic in momentum transfer relevant for conservative dynamics. These ingredients were assembled into the one–loop amplitudes across the relevant topologies (single–exchange, triangle, box, and penguin).
  \item We computed the momentum–space two–body potential via the Lippmann-Schwinger equation. \textit{A careful EFT/Born subtraction removes iterated long–range exchanges, and we explicitly showed that the resulting momentum–space potential is IR finite, mirroring the situation in GR}. This establishes that the soft amplitude, once dressed by the appropriate iterative counter term, is free of IR divergences at the level relevant for conservative observables.
  \item  The conservative scattering angle was obtained by exponentiating the momentum–space amplitude (eikonal exponentiation) and differentiating the eikonal phase with respect to impact parameter, and the result is given as an explicit closed-form function of gravitational, electromagnetic, and dilatonic couplings. We also verify that, in the appropriate limits, our results for the classical potential and the scattering angle agree precisely with those available in the literature.

\end{itemize}
\textbf{Future outlook}:
There are several interesting directions to be done. 
\begin{itemize}
  \item \emph{Higher orders.} \textcolor{black}{Extending to two loops ($3{\rm PM}$) would sharpen the conservative sector and test the robustness of IR cancellations beyond one loop in EMD. However, at two-loop and onward radiative corrections appear to emerge. However, we think that, as long as we consider the conservative scattering angle at two loops, the Born subtraction scheme should work. However if we add the radiative corrections (where one needs to compute the boundary integrals both from potential and radiation region) the subtraction scheme may fail. However, in those cases, the EFT prescription has been checked to work perfectly in the context of GR in \cite{Cheung:2018wkq}. We hope to report this soon \cite{arpan2}.}
  \item \emph{Spin and finite size.} Incorporating spins, spin–orbit/spin–spin couplings, and tidal responses (including dilatonic/electromagnetic polarizabilities) is essential for realistic compact objects and will enrich the phenomenology.
  \item \emph{Radiation and reaction.} Computing real emission (gravitational, electromagnetic, and dilatonic) and the accompanying radiation–reaction forces (in line with \cite{Caron-Huot:2023vxl}) will complete the bridge to waveform modeling for generic scattering and bound–state transitions, which is also work in progress \cite{arpan1}.
  \item \emph{Higher–curvature corrections.} Including $\alpha'$ \cite{Metsaev:1987zx}(higher–derivative) operators from the stringy effective action will test how UV-motivated terms imprint themselves on classical observables and whether symmetry protection persists at higher PM orders.
\end{itemize}

%
    }%
  }%

  \restorethesisbodyformat
  \restoremainchapterstyle
  \chapter{Celestial Amplitude, Shadow, and OPE in Quadratic Gravity}
  \thesischapterpaperbox{The sky remembers everything: Celestial amplitude, shadow and OPE in quadratic EFT of gravity}{A.~Bhattacharyya, S.~Ghosh, and S.~Pal}{\href{https://doi.org/10.21468/SciPostPhys.19.2.041}{\textit{SciPost Phys.} \textbf{19}, no.~2 (2025), 041}; arXiv: \arxivlink{2505.02899}{hep-th}}
  {%
    \restorethesisbodyformat
    \renewcommand{\appendix}{%
      \setcounter{section}{0}%
      \setcounter{subsection}{0}%
      \setcounter{subsubsection}{0}%
      \renewcommand{\thesection}{\thechapter.\Alph{section}}%
      \renewcommand{\thesubsection}{\thesection.\arabic{subsection}}%
      \renewcommand{\theHsection}{chapter.\arabic{chapter}.appendix.\Alph{section}}%
      \renewcommand{\theHsubsection}{\theHsection.\arabic{subsection}}%
    }%
    \ifstrempty{chap5_body.tex}{}{%
      
\section{Introduction}
The analysis presented in this chapter is based on Ref.~\cite{Bhattacharyya:2025sky}.
In conventional quantum field theory, scattering amplitudes are typically computed by considering the external states as momentum eigenstates, ensuring momentum conservation at each interaction vertex. However, an alternative exists, namely the conformal primary basis. In this basis, the external states are labeled by their conformal dimensions and their positions on the celestial two-sphere, parametrized by complex coordinates \( z, \bar{z} \). For massless external states, the conformal primary basis is obtained via a Mellin transformation over the energy variable \( \omega \). Scattering amplitudes expressed in this basis are referred to as \textit{celestial amplitudes}, owing to their kinematic structure and interpretation as correlation functions on the celestial sphere \cite{Pasterski:2016qvg}.\par
A comprehensive understanding of celestial conformal field theory (CCFT) necessitates knowledge of its full operator spectrum. In standard two-dimensional CFTs, the conformal block decomposition~\cite{Osborn:2012vt,Poland:2018epd} serves as a fundamental tool for probing the spectrum. Analogously, the block decomposition of celestial correlators and OPE has been extensively explored in~\cite{Lam:2017ofc, Nandan:2019jas, Law:2020xcf, Fan:2021isc,Fan:2022vbz,Fan:2022kpp,Atanasov:2021cje,De:2022gjn,Garcia-Sepulveda:2022lga, Banerjee:2023jne}. However, unlike standard CFTs, the computation of the block expansion in CCFT differs significantly due to the kinematic constraints imposed by four-dimensional scattering amplitudes, which result in a delta function \(\delta(|z - \bar{z}|)\) in cross-ratio space. This constraint enforces the cross-ratio \(z\) to be real-valued.
To overcome this limitation, one can perform a shadow transformation~\cite{FERRARA1972281,Ferrara:1972kab,Simmons-Duffin:2012juh} on one of the external operators, which relaxes the kinematic constraints and restores complex \(z\) dependence in the celestial correlator~\cite{Fan:2021isc}. This procedure enables the application of standard 2D CFT techniques to analyze the conformal block decomposition, particularly in extracting operator product expansion (OPE) coefficients.
To extract the OPE data for the shadow-transformed amplitude in quadratic EFT, we employ the OPE inversion formula~\cite{Caron-Huot:2017vep,Simmons-Duffin:2017nub,Mazac:2018qmi}. The Euclidean inversion formula, applicable to CCFT, is relatively straightforward compared to the Lorentzian version, which relies on computing the double discontinuity across branch cuts of hypergeometric functions in the complexified cross-ratio space where \(z\) and \(\bar{z}\) are independent variables.
The key idea is to identify OPE coefficients as residues of poles in the analytically continued partial wave expansion coefficients. 
This analytic continuation enables the extraction of OPE data in terms of integrals over partial waves, offering a powerful method for decoding the celestial operator spectrum.

\par 
In spite of these promising ideas and newer developments in CCFT in recent years, Celestial holography has some drawbacks that need to be addressed. Most of the examples of celestial amplitude presented in the literature, especially for the UV incomplete theories, give non-analytic (or purely divergent) results \cite{Arkani-Hamed:2020gyp}, which is because the Mellin transformation of flat space amplitude (at some order of perturbation) integrates over full energy scale (IR to UV) of the external asymptotic states. For example, in Einstein's GR, the Celestial amplitude is purely divergent at any order of perturbation. This issue is, to some extent, resolved in the context of string theory \cite{Stieberger:2018edy, Chang:2021wvv, Donnay:2023kvm} because of the UV softness of the theory. However, in the context of QFT, there are several approaches to handle the problem: the first one is to treat the celestial amplitude as a distribution \cite{Pasterski:2017ylz}. Secondly, by regularizing it using the background field method \cite{Fan:2022vbz,Casali:2022fro, Fan:2022kpp, deGioia:2022fcn, Banerjee:2023rni, Ball:2023ukj, Crawley:2023brz}. Thirdly, by introducing a modified Mellin transformation \cite{Banerjee:2019prz}. More recently, another approach has been proposed in \cite{Adamo:2024mqn} based on the eikonal exponentiation of scattering amplitude (in the context of celestial amplitude see \cite{deGioia:2022fcn}) in both flat and curved space \cite{Adamo:2021rfq}. In the eikonal limit (small angle, high energies), the perturbative series of the scattering amplitude can be exponentiated \cite{Amati:1987uf,Kabat:1992tb} and the full scattering matrix can be written in terms of the Born amplitude dressed by a phase. It has also been shown that because of the oscillating nature of the eikonal amplitudes, they are meromorphic in conformal dimensions with an infinite number of poles in the negative real axis. This provides an example of meromorphic and non-perturbative gravitational celestial amplitude. From the point of view of UV completeness, we expect improved analytic behaviour of celestial amplitude in higher derivative EFTs of gravity, which is one of the main themes of the paper. As previously noted, celestial amplitudes in general relativity (GR) exhibit non-analytic---more precisely, \textit{distributional}--- behaviour. In this work, we aim to explore whether this singular structure persists in higher-derivative EFTs, specifically within the context of \textit{quadratic gravity}. Our investigation centres around the following key questions:

\begin{itemize}
    \item Does the celestial amplitude in quadratic gravity remain distributional, or does it exhibit improved analytic properties?
    \item Is the introduction of \textit{eikonal resummation} necessary in such theories to achieve better analytic behaviour in the celestial amplitude?
    \item Can we get some analytic control to extract the OPE data (including spinning exchanges) from the celestial correlator corresponding to the Born amplitude.
    \item Furthermore, is it possible to extract OPE data from the \textit{non-perturbative (in couplings of the theory) eikonal amplitude}?
\end{itemize}
Through these questions, we aim to understand the interplay between analytic structure, resummation techniques, and CFT data for the celestial amplitudes of massless scalar in higher-derivative theories.
\par
This chapter is organised as follows. In Section~\eqref{ch5:sec3}, we briefly review the key theme of the paper, i.e. eikonal amplitude in the celestial sphere, discuss the boost eigenstate and its connection to the Mellin transform, and then generalise the amplitude analysis to quadratic EFT. In this section, we also describe analyticity and dispersion relation(s) relevant to our context, generalising from pure GR scenario. The nature of the $\gamma$-poles in the celestial amplitude for quadratic EFT is discussed. Further in Section~\eqref{ch5:sec4}, we discuss Celestial \textit{Operator Product Expansion (OPE)} and show the detailed computation of the conformal block decomposition in Celestial conformal field theory (CCFT). To compute the Euclidean OPE coefficients, we used the \textit{OPE-inversion formula} pioneered by Caron-Huot. In Section~\eqref{ch5:sec5}, we calculate the carrollian amplitude from the celestial counterpart and calculate the corrections we obtain from the non-vanishing quadratic EFT coefficients. We also demonstrate how the IR pole in carrollian amplitude shifts its value from that of GR. In Appendix~\ref {ch5:app:A}, we show the derivation and the simplification of the Mellin integration we encountered in terms of the \textit{simplex variables}.
\section{ Celestial eikonal amplitude for quadratic gravity}\label{ch5:sec3}
\textcolor{black}{We consider scattering amplitudes with external massless scalars, with the exchange particle being a graviton which has one massive mode due to non-vanshing coupling constants of quadratic EFT.} The main ingredient for computing the celestial eikonal amplitude is the Born amplitude in quadratic gravity. The gravity theory we are going to consider is the following,
\begin{align}
    \begin{split}
        S_{g}=\rmint d^4 x \sqrt{-g}\left(\kappa \,\mathcal{R}+\alpha \mathcal{R}_{\mu\nu}\mathcal{R}^{\mu\nu}-\frac{1}{3}(\alpha+\beta)\mathcal{R}^2\right)
    \end{split}\label{3.1i}
\end{align}
where, $\kappa\sim m_p^2\sim \frac{1}{G}$ and $\alpha,
\beta$ are dimensionless Wilson coefficients (or coupling constants of the theory) of the EFT. At the tree level, we have the following Feynman diagram(s) as depicted in Fig~\eqref{ch5:fig3}.
To compute the diagram, we need to know the graviton propagator, which is given by \cite{Hamber:2009zz},
\begin{align}
    \begin{split}\label{3.2h}
      \kappa  \langle h_{\mu\nu}(q)\,h_{\eta\delta}(-q)\rangle=2 \mathcal{P}^{(2)}_{\mu\nu;\eta\delta}\left(\frac{1}{q^2}-\frac{1}{q^2+\frac{\kappa}{\alpha}}\right)+\mathcal{P}^{(0)}_{\mu\nu;\eta\delta}\left(-\frac{1}{q^2}+\frac{1}{q^2+\frac{\kappa}{2\beta}}\right)
    \end{split}
\end{align}
where the projectors are defined as,
\begin{align}
    & \mathcal{P}^{(0)}_{\mu\nu;\eta\delta}=-\frac{1}{3q^2}\left(q_\mu q_{\nu}\eta_{\eta\delta}+q_{\eta}q_{\delta}\eta_{\mu\nu}\right)+\frac{1}{3}\eta_{\mu\nu}\eta_{\eta\delta}+\frac{1}{3q^4}q_{\mu}q_{\nu}q_{\eta}q_{\delta}
\end{align}

and
\begin{align}
\begin{split}
    & \mathcal{P}^{(2)}_{\mu\nu;\eta\delta}=\frac{1}{3q^2}\left(q_\mu q_{\nu}\eta_{\eta\delta}+q_{\eta}q_{\delta}\eta_{\mu\nu}\right)-\frac{1}{2 q^2}\Bigg(q_{\mu}q_{\eta}\eta_{\nu\delta}+q_{\mu}q_{\delta}\eta_{\nu\eta}+q_{\nu}q_{\eta}\eta_{\mu\delta}+q_{\nu}q_{\delta}\eta_{\mu\eta}\Bigg)\\&\hspace{0.8cm}+\frac{2}{3q^4}\Bigg(q_\mu q_\nu q_\eta q_\delta\Bigg)+\frac{1}{2}\Bigg(\eta_{\mu\eta}\eta_{\nu\delta}+\eta_{\mu\delta}\eta_{\nu\eta}\Bigg)-\frac{1}{3}\eta_{\mu\nu}\eta_{\eta\delta}\,.
    \end{split}
\end{align}
\noindent
\begin{figure}[t!]
\centering
\scalebox{0.18}{\includegraphics{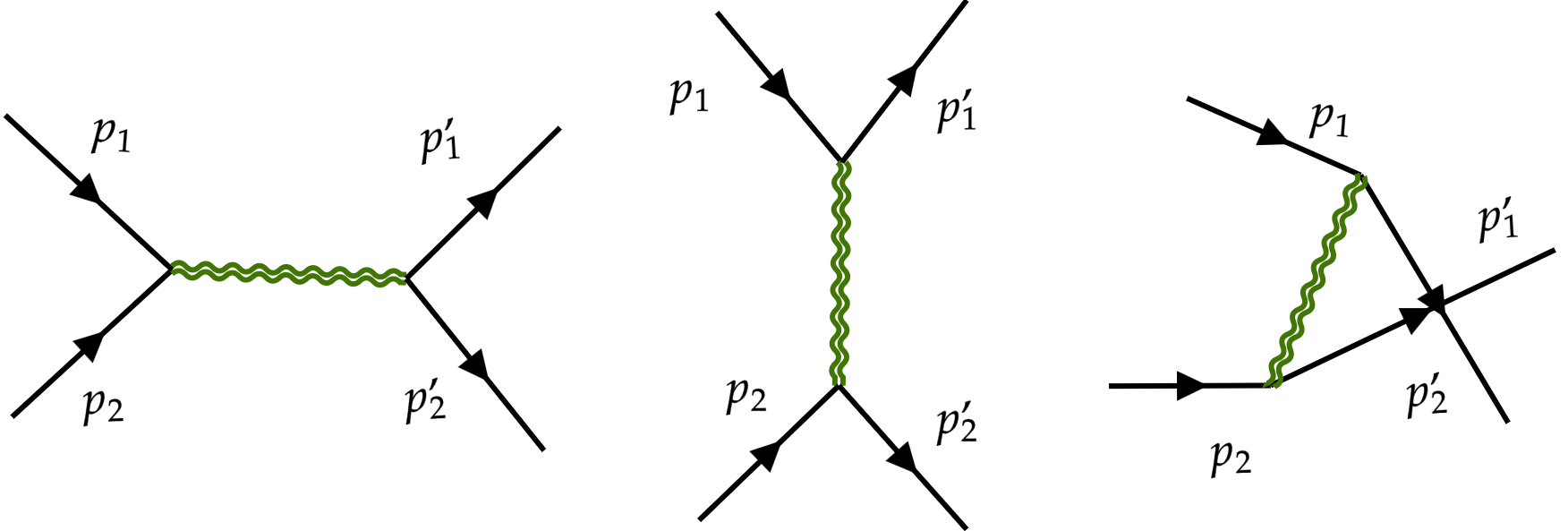}}
\caption{Figure describing the $s,t,u$-channel diagram respectively, relevant for our computation }\label{ch5:fig3}
\end{figure}
\textcolor{black}{Before proceeding to the main discussion, we begin by outlining several foundational aspects of quadratic effective field theories (EFTs) of gravity. It is well understood that quantum corrections to Einstein gravity generically induce higher-derivative terms, including those quadratic in the curvature, such as \( R^2 \). This naturally motivates the study of such terms at the level of the microscopic (bare) action, providing a framework amenable to consistent perturbative renormalization. The inclusion of curvature-squared contributions in classical gravity was first proposed in \cite{weyl}, and their role in rendering gravity power-counting renormalizable was subsequently demonstrated in \cite{1962JMP.....3..608U}, with full renormalizability to all orders in perturbation theory established in \cite{PhysRevD.16.953}. }\par
\textcolor{black}{
Despite these notable achievements, higher-derivative gravitational theories are often regarded as problematic due to issues related to unitarity. This concern becomes manifest at the level of the tree-level propagator \eqref{3.2h}, where the ultraviolet behavior improves from a \( 1/q^2 \) to a \( 1/q^4 \) fall-off, yet this improvement introduces a massive spin-2 ghost with mass \( m = \sqrt{\kappa/\alpha} \), signaling a breakdown of perturbative unitarity. More specifically, while it is possible to quantize the theory such that all excitations have positive-definite energy, this necessitates the inclusion of negative-norm (ghost) states in the Hilbert space. As shown in \cite{PhysRev.79.145}, such states inevitably violate unitarity and cannot be consistently projected out without sacrificing the unitarity of the \( S \)-matrix. Therefore, despite the theory’s favorable UV behavior, its physical consistency remains fuzzy in the absence of a \textit{precise} resolution to the unitarity problem (see also the Appendix of \cite{PhysRevD.16.953} for a detailed discussion).}\par
\textcolor{black}{
However, as emphasized in \cite{FRADKIN1982469}, unitarity is fundamentally a dynamical property, and cannot be definitively assessed within the confines of a purely perturbative treatment or at the level of the tree-level propagator alone. A comprehensive analysis requires accounting for loop corrections and potentially non-perturbative effects at fixed energy. As we will demonstrate in the following sections, the eikonal scattering amplitude constructed within a bottom-up approach—non-perturbative in the coupling—is manifestly unitary by construction.
 } \par
 \textcolor{black}{
Nevertheless, more recent analyses~\cite{Abe:2017abx,Abe:2018rwb,Abe:2020ikj,Abe:2022spe} have shown that Einstein gravity, when treated as a quantum theory, violates tree-level unitarity in the high-energy limit, failing to satisfy the unitarity bound \( |\mathbb{A}(s,t)| < C \), where \( C \) is a finite constant. In particular, in the Regge limit, the amplitude in general relativity exhibits unbounded growth, scaling as \( |\mathbb{A}(s,t)|_{\textrm{GR}} \sim {O}(s^1) \to \infty \). In contrast, the same studies demonstrate that scalar matter scattering mediated by graviton exchange in quadratic gravity satisfies unitarity bound at high energies, despite the presence of negative-norm (ghost) states. This result is in accord with the expectations of the Llewellyn Smith conjecture~\cite{LLEWELLYNSMITH1973233}, which heuristically suggests that renormalizable quantum field theories should also exhibit unitary behavior.}\par
\textcolor{black}{In addition, our subsequent computations of the tree-level scattering amplitude in quadratic gravity yields the following structure:
\begin{align}
\mathbb{A}(s,t) \sim \frac{s^2}{t} + O(s^0) , 
\end{align}
where the leading term arises from the Einstein-Hilbert (GR) sector, while the subleading constant contribution originates from the quadratic curvature corrections in the effective action. When the theory is treated as an effective field theory, the eikonal limit \( t/s \to 0 \) remains well-defined by taking the momentum transfer \( t \) to be sufficiently small and keeping the center-of-mass energy \( s \) bounded by the EFT cutoff \( \Lambda \). Moreover, it is straightforward to observe that the quadratic curvature contributions exhibit improved high-energy behavior compared to the GR term. Consequently, there is no apparent violation of the unitarity bound---at least within the regime where the theory is interpreted as a low-energy effective description of some UV-complete quantum gravity theory.}

\textcolor{black}{
Under this interpretation, Einstein gravity treated as an effective field theory (EFT) remains consistent with unitarity, as the amplitude stays within unitarity bound. Importantly, the constant contribution from the higher-derivative terms does not introduce any violation of the unitarity bound; that is, one finds
\begin{equation}
|\mathbb{A}(s,t)| \leq f(\Lambda) + {O}(s^0),
\end{equation}
for some finite function \( f(\Lambda) \).}
\textcolor{black}{
In summary, when quadratic gravity (or Einstein gravity) is treated as a low-energy EFT, unitarity of the scattering amplitude is not much problematic (rather, in another context, i.e, in the S-matrix bootstrap program, the goal is to put a bound on the EFT coupling, assuming the existence of a high-energy UV completion which is causal and unitary). This contrasts with the situation where one attempts to promote the theory to a full UV-complete quantum field theory, where issues such as ghost modes and associated violations of unitarity may become significant.} 

\textcolor{black}{Keeping these takeaways in mind, we now set up the scattering process for our case. Specifically, we consider the gravitationally mediated scattering of massless scalar matter fields. The dynamics of the scalar field are governed by the following matter action:
\begin{align}
    S_{m} = \rmint d^4x\, \sqrt{-g} \left( -\frac{1}{2} g^{\mu\nu} \nabla_{\mu} \phi \nabla_{\nu} \phi \right),
\end{align}
where \(\phi\) denotes a massless scalar field minimally coupled to gravity and bulk gravitational action is defined in \eqref{3.1i}\,.
}
Now, to compute the amplitude in terms of the Mandelstam variables $s,t,u$ we use the constraint $s+t+u=0$ for external massless on-shell states. Then the amplitude $\mathbb{M}(s,t)$ can be cast as the sum of three distinct channels $(s,t,u)$\footnote{The Mandelstam variables are given by,
\begin{align}
s:=-(p_1+p_2)^2,\,\,\,\,\,\, t:=-(p_1-p_3)^2,\,\,\,\,\,\,\,\,\,\,\,u:=-(p_1-p_4)^2\,.
\end{align}
}, after which we take the \textit{eikonal limit, $\frac{t}{s}\rightarrow 0$}, to extract the dominant contribution.
It is noteworthy that, similar to GR without higher curvature corrections, the dominant eikonal contribution comes from the $t$ channel. The key difference is the contribution from $s$ and $u$ channels, which cancel each other in GR but not if we have higher curvature terms. We can promptly write down the  contributions to the amplitude from the three channels as,
\begin{align}
\mathbb{M}_{\text{Born}}^{\textrm{EFT}}(s,t)=\mathbb{A}_s(s,t)+\mathbb{A}_t(s,t)+\mathbb{A}_u(s,t)\nonumber
\end{align}
where
\begin{align}
\begin{split}
    &\mathbb{A}_s(s,t) =\frac{2}{24} \left(s^2+6 s t+6 t^2\right) \left(\frac{1}{-s}-\frac{1}{\frac{\kappa}{\alpha }-s}\right)-\frac{1}{48} s^2 \left(\frac{1}{-s}-\frac{1}{\frac{\kappa}{2 \beta }-s}\right)\,,\\&
\mathbb{A}_t(s,t)=\frac{2}{24} \left(6 s^2+6 s t+t^2\right) \left(\frac{1}{-t}-\frac{1}{\frac{\kappa}{\alpha }-t}\right)-\frac{1}{48} t^2 \left(\frac{1}{-t}-\frac{1}{\frac{\kappa}{2 \beta }-t}\right)\,,\\&
\mathbb{A}_u(s,t) =\frac{2}{24} \left(s^2-4 s t+t^2\right) \left(-\frac{1}{\frac{\kappa}{\alpha }+(s+t)}-\frac{1}{-s-t}\right)-\frac{1}{48} (s+t)^2 \left(-\frac{1}{\frac{\kappa}{2 \beta }+(s+t)}-\frac{1}{-s-t}\right)\,.
\end{split}
\end{align}

After adding all of them, in the eikonal limit \footnote{ \textcolor{black}{We use the notation \(\mathbb{M}^{\textrm{EFT}}_{\textrm{Born}}\) to denote the Born amplitude in the eikonal limit, as this approximation is employed throughout the manuscript in place of the full Born amplitude. Subsequently, we take the Mellin transformation of the eikonal limit of the Born amplitude to find the celestial amplitude.}}
,  

\begin{align}
\mathbb{M}_{\text{Born}}^{\textrm{EFT}}(s,t)\to-\frac{1}{48} s \left(\frac{ (3 \kappa+\alpha  s-8 \beta  s)}{(\kappa-\alpha  s) (\kappa-2 \beta  s)}-\frac{4 }{\kappa+\alpha  s}+\frac{1}{\kappa+2 \beta  s}+\frac{24 \,(-s) }{-t \kappa }\right)\,.\label{3.9p}
\end{align}
In the limit $\alpha\rightarrow 0,\beta\rightarrow 0\,,$ it gives back the Einstein-GR result, $\mathbb{M}^{\textrm{GR}}\rightarrow \frac{s^2}{-2t}$. For non-zero values of $\alpha$ and $\beta $, we get the tree-level amplitude of the quadratic EFT. This can be written in the following way,
\begin{align}
    \begin{split}
        \mathbb{M}_{\text{Born}}^{\textrm{EFT}} (s=\omega^2,t=-\textcolor{black}{\frac{z-1}{z} }\omega^2)=-\frac{1}{48} \omega ^2 \Bigg(  \frac{3 \kappa+\alpha\omega^2-8 \beta  \omega^2}{(\kappa-\alpha  \omega^2) (\kappa-2 \beta  \omega^2)}-\frac{4 }{\kappa+\alpha  \omega^2}+\frac{1}{\kappa+2 \beta  \omega^2}-\textcolor{black}{\frac{24 \,z}{(z-1)\kappa}}\Bigg)\,.\label{3.8p}
    \end{split}
\end{align}
The variable $z$ should be chosen according to the relevant kinematic region, namely the $12\to 34$, $13\to 24$, and $14\to 23$ channels. \textcolor{black}{One should note that even though we get a non-zero finite piece due to the presence of non vanishing $\alpha,\beta$, in the $s$ and $u$-channel, the $t-$channel contribution in the eikonal limit remains the same as of GR}.  \par


As discussed previously there are three distinct kinematic regions: $\mathbf{12 \to 34}$, $\mathbf{13 \to 24}$, and $\mathbf{14 \to 23}\,.$ These regions differ significantly from one another and are each valid only within specific ranges of the cross-ratios. To go from one kinematical region to another, one typically employs a systematic procedure of analytic continuation.
Now, the celestial amplitude (for external scalar operators) in three different kinematical regions is given by,
\begin{align}
\begin{split}\label{3.10e}
&{\mathbfcal{A}}^{12\to 34}(\mathbf \Delta,z)=2^{3-\gamma}z^2\rmint_0^\infty d\omega\,\omega^{\gamma-1}\,\mathbb{M}_{\textrm{Born}}^{\textrm{EFT}}\left(s=\omega^2,t=-\frac{z-1}{z}\omega^2\right),\,\,\,\,\,(z>1)\\ &\hspace{2.3 cm}
   =\frac{i \pi  2^{-\frac{3 \gamma }{2}-2} z^2 \left(\alpha ^{\frac{\gamma }{2}+1}-2^{\frac{\gamma }{2}+3} \beta ^{\frac{\gamma }{2}+1}\right) \left(\frac{\alpha  \beta }{\kappa }\right)^{-\frac{\gamma }{2}-1}}{3 \left(-1+e^{\frac{i \pi  \gamma }{2}}\right) \kappa }+\underbrace{\frac{2^{3-\gamma}\, \pi\,  z^3 \delta (i (\gamma +2))}{(z-1)\kappa}}_{\text{Einstein gravity}}\,,\\ &
\hspace{2.3 cm}=\delta_1(\alpha,\beta|\gamma)z^2+\delta_2(\gamma)\frac{z^3}{(z-1)\kappa}\,,\\ &
{\mathbfcal{A}}^{13\to 24}(\mathbf \Delta,z)=2^{3-\gamma}z^{2+\frac{\gamma}{2}}\rmint_0^\infty d\omega\,\omega^{\gamma-1}\,\mathbb{M}_{\textrm{Born}}^{\textrm{EFT}}\left(s=-z\omega^2,t=-(1-z)\omega^2\right),\,\,\,\,\,(0<z<1)\\ & \hspace{2.3 cm}\to {\mathbfcal{A}}^{12\to 34}(\mathbf \Delta,z)\,,
   \\ &
{\mathbfcal{A}}^{14\to 23}(\mathbf \Delta,z)=2^{3-\gamma}(-z)^{2+\frac{\gamma}{2}}(1-z)^{-\frac{\gamma}{2}}\rmint_0^\infty d\omega\,\omega^{\gamma-1}\,\mathbb{M}_{\textrm{Born}}^{\textrm{EFT}}\left(s=\frac{z}{1-z}\omega^2,t=\omega^2\right),\,\,\,\,\,(z<0)
   \\ & \hspace{2.3 cm}\to{\mathbfcal{A}}^{12\to 34}(\mathbf \Delta,z)\,\,,
   \end{split}
   \end{align}
where,
\textcolor{black}{
\begin{align}
    \begin{split}
        &\delta_1(\alpha,\beta|\gamma)=\pi \,e^{-\frac{i\pi\gamma}{4}}\frac{  2^{-\frac{3 \gamma }{2}-2} z^2 \kappa^{\gamma/2} \left(\beta ^{-\frac{\gamma }{2}-1}-2^{\frac{\gamma }{2}+3} \alpha ^{-\frac{\gamma }{2}-1}\right) }{6  \sin\left(\frac{\pi \gamma}{4}\right) }, \,\delta_2(\gamma)={2^{3-\gamma}\pi \delta(i(\gamma+2))\,.}
    \end{split}
\end{align}}
\noindent
Collecting the result of the integrals, we can write the full amplitude in the conformal basis as, 

\begin{align}\begin{split}
\centering\label{4.18m}
  \mathcal{A}_4(z,\bar z)\sim\,&  (z-1)^{\frac{\Delta_1-\Delta_2-\Delta_3+\Delta_4}{2}}|z|^{-\Delta_1-\Delta_2}\frac{\delta(|z-\bar z|)}{|z_{13}|^2|z_{24}|^2}\mathbfcal{A}^{ij\to kl}(\mathbf{\Delta},J_i,z)\,.
 \end{split} 
\end{align}

We find that the Mellin-transformed Born amplitude has a better analytic structure than GR, which is purely divergent (or purely distributional), which answers the first question we rise in the introduction. Moreover, we find that the part coming from the quadratic correction in the amplitude is analytic in the whole complex $\gamma$-plane except $\gamma=n,$ where $n\in \mathbb{Z}$. In contrast, for GR the celestial amplitude in the eikonal limit is non-analytic, which replicates the fact that quadratic EFT modifies the analytic behaviour in the whole energy plane, and correspondingly, the amplitude becomes analytic in the $\gamma$ plane with the location of isolated singularities \footnote{ A general proof in \cite{Arkani-Hamed:2020gyp} shows that for QFT's with better UV behaviour(s) leads to analytic amplitudes (rather than purely divergent distributional nature) with poles in the right-half (left-half) complex $\gamma$-plane which we call `UV poles' (IR poles). }.
Next, we proceed to discuss the eikonal amplitude in GR and then generalize it for quadratic EFT.

\noindent
\subsection{Eikonal amplitude in GR} 

\textcolor{black}{The eikonal approximation provides an effective framework for analyzing high-energy scattering processes in the regime where momentum transfer is small compared to the centre-of-mass energy. In gravity, this leads to an eikonal amplitude dominated by ladder-type graviton exchanges, capturing the resummation of leading contributions at each order in perturbation theory \cite{Amati:1987uf,Kabat:1992tb}. Remarkably, the eikonal amplitude encodes both ultraviolet (UV) and infrared (IR) aspects of the theory. The eikonal regime thus serves as a stage for exploring how quantum gravity reconciles high-energy scattering with universal IR behaviour governed by asymptotic symmetries.
}\par

Before going into the eikonal phase calculations of quadratic higher curvature gravity, we briefly describe the computation in the context of GR. The eikonal expression for the amplitude is given by summing the eikonal phase. We take bottom-up approach to compute the eikonal amplitude. In bottom-up approach we define the eikonal phase as \cite{Kabat:1992tb,PhysRev.186.1656, DiVecchia:2023frv},
\begin{align}
    \chi=\frac{2\pi G}{E\,p}\rmint \frac{d^{D-2}\boldsymbol{q}_{\perp}}{(2\pi)^{D-2}}  e^{i \boldsymbol{q}_{\perp}\cdot \boldsymbol{x}}\,\mathbb{M}^{\textrm{EFT}}_{\textrm{Born}}(s,\textcolor{black}{t=- \boldsymbol{q}_{\perp}^2})\,.
\end{align}
Consequently the eikonal amplitude is given by,
\begin{align}
    \begin{split}
    \mathbfcal{M}_{\textrm{eik}}(s,t)=8E\,p\rmint d^{D-2}\boldsymbol{x}_{\perp} e^{-i\boldsymbol{q}_{\perp}\cdot \boldsymbol{x}_{\perp} }(e^{i\chi}-1)\,.
    \end{split}
\end{align}
\noindent
In GR the eikonal amplitude is given by \cite{Amati:1987uf, Kabat:1992tb},
\begin{align}
\begin{split}
    i\mathbfcal{M}^{\textrm{GR}}_{\textrm{eik}}&=\frac{8\pi E p}{\mu^2} \frac{\Gamma\left(1-\frac{i G\frac{s^2}{2}}{2Ep}\right)}{\Gamma\left(\frac{i G\frac{s^2}{2}}{2Ep}\right)}\Bigg(\frac{4\mu^2}{q_{\perp}^2}\Bigg)^{1-\frac{i G\frac{s^2}{2}}{2Ep}}
    =-\underbrace{\frac{16 i \pi  G s^2}{2t}}_{\text{Born Amplitude}}\times\underbrace{\frac{\Gamma\left(-\frac{i G{s^2}}{4Ep}\right)}{\Gamma\left(\frac{i G\frac{s^2}{2}}{2Ep}\right)}\Bigg(\frac{4\mu^2}{q_{\perp}^2}\Bigg)^{-\frac{i G{s^2}}{4Ep}}}_{\text{Phase}}\,.
    \end{split}
\end{align}
This is an important feature that the eikonal (leading) amplitude in GR to all order of $G$ is the product of the born amplitude and a phase factor. 
\subsection{Eikonal amplitude in quadratic gravity}
Now for quadratic EFT of gravity, as the propagator changes, the expression of eikonal phase $\chi$ also changes. \textcolor{black}{ A natural question that arises in the context of high-energy gravitational scattering is whether eikonal exponentiation remains valid in higher-derivative theories such as quadratic gravity. The answer, we argue, is affirmative in the eikonal regime, and we provide the reasoning below. The validity of eikonal exponentiation in quadratic gravity can be addressed by considering the structure of gravitational scattering amplitudes in the high-energy, small-angle (eikonal) regime. In this limit, the dominant contributions arise from the exchange of soft, long-wavelength gravitons. Crucially, this infrared (IR) behavior is governed by the long-distance propagation of the gravitational field, which remains dictated by the Einstein-Hilbert term—even in the presence of higher-curvature corrections such as \( R^2 \), \( R_{\mu\nu}^2 \), or \( R^3 \). These UV modifications primarily affect the short-distance (non-eikonal) part of the interaction and do not interfere with the leading IR dynamics responsible for exponentiation.
Within the effective field theory framework, and under the assumption that the higher-derivative couplings (e.g., \( \alpha, \beta \)) are large compared to the inverse energy scale of interest (which is $m_p$), the eikonal exponentiation is expected to persist. This expectation is supported by the IR universality of the gravitational interaction, and it follows the logic that the eikonal resummation is dominated by ladder diagrams with soft graviton exchanges.
We acknowledge, however, that a full demonstration of exponentiation in quadratic gravity would require explicit computation of loop-level diagrams, particularly the sum of multi-graviton ladder diagrams.  We intend to explore this in future work. This was first shown for Einstein gravity in the seminal work of 't Hooft \cite{tHooft:1987vrq} and then by others (see, e.g., \cite{Kabat:1992tb}). Apart from gravitational field theories, for QFTs involving \textit{massive} (relevant for our case) meson exchange the ladder structure can been nicely be identified and the S-matrix exponentiates \cite{PhysRev.186.1656}. There, the resummation of eikonal ladder diagrams leads to the well-known exponential form of the amplitude:
\begin{align}
&\mathcal{A}_{\text{eik}}(s, q_\perp) \sim 2s \rmint d^2b\, e^{iq_\perp \cdot b} \left( e^{i\chi(s,b)} - 1 \right),
\end{align}
where \( \chi(s, b) \) is the eikonal phase and is given by,
\begin{align}
 &   \chi(s,b)\sim \rmint d^D q\, \delta(2 p_1\cdot b) \delta(2 p_2\cdot b)\,\mathbfcal{A}_{\textrm{tree}}(s,-q^2), \,\textrm{with},\,q^\mu=p_1^\mu-{p_1^\mu}', b^\mu\textrm{ is the impact parameter.}
\end{align}
A recent review \cite{DiVecchia:2023frv} discussed that this structure holds not only in General Relativity but also in UV-complete frameworks such as string theory. Given this, it is natural to expect that intermediate theories like quadratic gravity—lying between Einstein gravity and UV completions—should also admit eikonal exponentiation, at least in the IR regime. We refer specifically to Section 3.1.5 of \cite{DiVecchia:2023frv}, which supports this viewpoint.
While a dedicated study of loop-level exponentiation in quadratic gravity is indeed warranted, we believe the arguments above (as well as supported by our bottom-up computation presented in the manuscript) justify the assumption of exponentiation within the eikonal limit of the effective theory.
}The eikonal phase for our case can be written in the following way,

\begin{align}
\begin{split}
\hspace{0cm}\chi_{\textrm{EFT}}&=\frac{2\pi G}{Ep}\rmint\frac{ d^2k_{\perp}}{(2\pi)^2}e^{i{\bf k_\perp }\cdot {\bf x_\perp}}\Bigg[\widetilde{\mathbb{M}}^{\textrm{EFT}}_{\textrm{Born}}(s)+\frac{s^2}{k_\perp^2+\mu^2-i\epsilon}\Bigg]\,,\\&
    =\underbrace{\frac{2\pi G\,\widetilde{\mathbb{M}}^{\textrm{EFT}}_{\textrm{Born}}(s)}{Ep}\rmint\frac{ d^2k_{\perp}}{(2\pi)^2}e^{i{\bf k_\perp }\cdot {\bf x_\perp}}}_{\text{Contact term contribution in quadratic EFT}}+\,\frac{2\pi G s^2}{Ep}\rmint\frac{ d^2k_{\perp}}{(2\pi)^2}e^{i{\bf k_\perp }\cdot {\bf x_\perp}}\frac{1}{k_\perp^2+\mu^2-i\epsilon}\,,\\&
    =\frac{2\pi G\, \widetilde{\mathbb{M}}^{\textrm{EFT}}_{\textrm{Born}}(s)}{E p}\,\delta^{(2)}(\boldsymbol{x}_\perp)-\frac{G s^2}{2 Ep}\log(\mu \boldsymbol{x}_{\perp})\,.\label{3.19p}
      \end{split}
\end{align}

\begin{tcolorbox}[thesisstatementbox, title=A digression on the functions of Dirac-$\delta$ function (measure)]
\restorethesisbodyformat
We now slightly deviate from our original discussion to briefly outline how one can meaningfully define functions of the \(\delta\)-function. As is well known, the \(\delta\)-function is a tempered distribution rather than a continuous probability distribution, as it has support only at a single point. Consequently, functions of such distributions are generally ill-defined. Nevertheless, as discussed in \cite{maroun2013generalized}, it is possible---albeit with care---to give meaningful interpretations to functions of the \(\delta\)-function (especially the exponential map). We will briefly review this proposition, which, despite its subtleties, also provides a physically reasonable framework for our purposes. \par

The first approach proceeds as follows: one attempts to construct a resolvent-like function of the \(\delta\)-distribution to bypass difficulties associated with the exponential map. Consider the following object, interpreted as a distribution: $\frac{1}{1+\delta(x)}$.

\textbf{Proposition \cite{maroun2013generalized,johnson2000feynman}.} \textit{As a distribution,}
\begin{align}
    \frac{1}{1+\delta(x)} := \mathbb{I},
\end{align}
\textit{where \(\mathbb{I}\) denotes the Lebesgue measure. i.e, for each \(\varphi(x) \in C^{\infty}_c(\mathbb{R})\), one has}
\begin{align}
    \left\langle \frac{1}{1+\delta(x)}, \varphi(x) \right\rangle := \rmint_{\mathbb{R}} \varphi(x).
\end{align}

While instructive, this proposition is of limited use in physical applications: it simply recovers the Lebesgue measure and entirely washes out the singular behaviour of the \(\delta\)-distribution. To address this, one must proceed more carefully. A more refined approach involves invoking the spectral theorem and exploiting the idempotence property of the \(\delta\)-function when treated as a characteristic set function \cite{maroun2013generalized}.

\textbf{Definition (Analytic Linearization) \cite{maroun2013generalized}.} 
\textit{Let \(f(\delta_a)\) be a function of the Dirac measure with singular support at \(x = a\). Its analytic realization is given by the formal series expansion}
\begin{align}
    f(\delta_a) \rightsquigarrow \mathbb{I} + \sum_{n=1}^\infty \frac{f^{(n)}(a)}{n!}\, \delta_a.\label{3.18j}
\end{align}

This construction provides an analytic framework for defining functions of distributions. Importantly, it does not strictly contradict the earlier proposition. To see this, consider:
\begin{align}
    \frac{1}{1+\delta_0} \rightsquigarrow \mathbb{I} - \delta_0 (1-1+1-1+\cdots) \overset{\text{reg}}{=} \mathbb{I} - \frac{1}{2}\delta_0,
    \label{3.23n}
\end{align}
where the alternating sum \(1-1+1-1+\cdots\)—which, according to Riemann’s reordering theorem, can be rearranged to converge to any real value—can be uniquely regularized via Dirichlet Eta summation yielding: $1-1+1-1+\cdots\overset{\text{reg}}{=}\eta(0)=1/2$. Thus, analytic linearization complements (or refines) the earlier proposition by preserving both the regular (Lebesgue) component and the singular structure encoded in Dirac-\(\delta\), offering a physically meaningful extension well-suited to our purpose(s).
\end{tcolorbox}
\restorethesisbodyformat
Now get back to our original discussion: the eikonal amplitude is given by,
\vspace{-0.3 cm}
\begin{align}
    \begin{split}
        \mathbfcal{M}_{\textrm{eik}}&
        =8Ep\rmint d^2 \boldsymbol{x}_{\perp}\,e^{-i \boldsymbol{q}\cdot \boldsymbol{x}_{\perp}}\left[\exp\left(-\frac{iGs^2}{2Ep}\log(\mu \boldsymbol{x}_{\perp})\right)\exp\left(\frac{2\pi i G\,\widetilde{\mathbb{M}}^{\textrm{EFT}}_{\textrm{Born}}(s)}{Ep}\delta^{(2)}(\boldsymbol{x}_{\perp})\right)-1\right]\,,    \label{3.19k}
        \end{split}
\end{align}
where, we have \begin {align}\widetilde{\mathbb{M}}^{\textrm{EFT}}_{\textrm{Born}}(s)=-\frac{1}{48} s \left(\frac{(3 \kappa+\alpha  s-8 \beta  s)}{(\kappa-\alpha  s) (\kappa-2 \beta  s)}-\frac{4}{\kappa+\alpha  s}+\frac{1}{\kappa+2 \beta s}\right)\,.\,
    \label{3.18p}\end{align}
As seen from \eqref{3.19k}, the eikonal amplitude involves an integral over \( e^{\delta(x)} \). Are they really bad or can we get some physically meaningful result out of it? Although functions of distributions are formally ill-defined, with careful treatment, \( e^{\delta(x)} \) can be given a physically intuitive meaning.
We use the `Analytic Linearization' \eqref{3.18j} to address this question. Now using this concept, we can rewrite \eqref{3.19k} as \footnote{\textcolor{black}{We use the following analytic linearization of exponential map to compute the celestial eikonal amplitude: $  e^{\delta_0}\rightsquigarrow\mathbb{I} + \sum_{n=1}^\infty \frac{f^{(n)}(\delta_0=0)}{n!}\, \delta_0=\mathbb{I}+(e-1)\delta_0.$}},
\begin{align}
    \begin{split}
        \mathbfcal{M}_{\textrm{eik}}= 8Ep\rmint d^2 \boldsymbol{x}_{\perp}\,e^{-i \boldsymbol{q}\cdot \boldsymbol{x}_{\perp}}\left[(\mu\,|{x}_{\perp}|)^{\frac{-i G s^2}{2Ep}}\left\{1+\left(e-1\right){\frac{2\pi i G \widetilde{\mathbb{M}}^{\textrm{EFT}}_{\textrm{Born}}(s)}{Ep}}\delta^{(2)}(\boldsymbol{x}_{\perp})\right\}-1\right]\,.\label{3.20k}
    \end{split}
\end{align}
We make use of the \textit{Lebesgue measure and Schwartz bracket} \cite{Borji:2024pvg} to extract the sensible piece.
\noindent
Furthermore, to evaluate the integral of \eqref{3.23j} onwards we use the distributional nature of dirac delta function in terms of sharply picked gaussian distribution. Dividing the integral \eqref{3.20k} into two parts we get,
\begin{align}
    \begin{split}
         \mathbfcal{M}^{\textrm{total}}_{\textrm{eik}} & = 8Ep\rmint d^2 \boldsymbol{x}_{\perp}\,e^{-i \boldsymbol{q}\cdot \boldsymbol{x}_{\perp}}\exp\left(-\frac{iGs^2}{2Ep}\log(\mu \boldsymbol{x}_{\perp})\right)\\ & +8Ep \rmint d^2 \boldsymbol{x}_{\perp}\,e^{-i \boldsymbol{q}\cdot \boldsymbol{x}_{\perp}}\left(e-1\right){\frac{2\pi i G \widetilde{\mathbb{M}}^{\textrm{EFT}}_{\textrm{Born}}(s)}{Ep}}\exp\left(-\frac{iGs^2}{2Ep}\log(\mu |{x}_{\perp}|)\right)\delta^{(2)}(\boldsymbol{x}_{\perp})\,,\\ &
         :=\mathbfcal{M}^{\textrm{GR}}_{\textrm{eik}}+\mathbfcal{M}^{\textrm{EFT}}_{\textrm{eik}}
         \label{3.23j}
    \end{split}
\end{align}
where $\mathbfcal{M}^{\textrm{GR}}_{\textrm{eik}}$ and $\mathbfcal{M}^{\textrm{EFT}}_{\textrm{eik}}$ denote the GR part and the EFT correction respectively.
\textcolor{black}{In \eqref{3.23j}}, the delta function simply evaluates \( f(\boldsymbol{x}_{\perp}) \) at \( \boldsymbol{x}_{\perp} = 0 \), multiplying it by a constant, and has no effect at finite \( \boldsymbol{x}_{\perp} \). However, in the Regge limit, the dominant contribution to the amplitude arises from large transverse separations, \( \boldsymbol{x}_{\perp} \). Since the delta-function term is sharply localized at \( \boldsymbol{x}_{\perp} = 0 \), it contributes only a constant to the eikonal amplitude. To obtain a physically meaningful result, we must smear the delta function around \( \boldsymbol{x}_{\perp} = 0 \), as the logarithmic part of the amplitude is valid only in the regime of large \( \boldsymbol{x}_{\perp} \).

\begin{align}
    \begin{split}
\mathbfcal{M}_{\textrm{eik}}^{\textrm{EFT}}&= {{16\pi i G\, \widetilde{\mathbb{M}}^{\textrm{EFT}}_{\textrm{Born}}(s)}} \left(e-1\right)\,\lim_{\varepsilon\to 0}\rmint  d^2 \boldsymbol{x}_{\perp}\,e^{-i \boldsymbol{q}\cdot \boldsymbol{x}_{\perp}}(\mu\,\boldsymbol{x}_{\perp})^{\frac{-i G s^2}{2Ep}}\delta_{\varepsilon}^{(2)}(\boldsymbol{x}_{\perp})\,,\\ &
        \rightarrow{16\pi i G\, \widetilde{\mathbb{M}}^{\textrm{EFT}}_{\textrm{Born}}(s)} \left(e-1\right)\lim_{\varepsilon\to 0}\rmint dx_{\perp}d\theta\,x_{\perp} e^{-i q  x_{\perp}\cos{\theta}}\,(\mu\,{x}_{\perp})^{\frac{-i G s^2}{2Ep}}\frac{1}{x_{\perp}}\delta(\theta)\delta_{\varepsilon}(x_{\perp})\,,\\ &
       =8Ep \left(e^{\frac{2\pi i G \widetilde{\mathbb{M}}^{\textrm{EFT}}_{\textrm{Born}}(s)}{Ep}}-1\right)\lim_{\varepsilon\to 0}\frac{1}{\sqrt{\pi}\varepsilon}\rmint dx_{\perp} e^{-i q\,x_{\perp}} (\mu\,{x}_{\perp})^{-2iGs} e^{-x^2/\varepsilon^2}\\ &
        =4Ep \left(e^{\frac{2\pi i G \widetilde{\mathbb{M}}^{\textrm{EFT}}_{\textrm{Born}}(s)}{Ep}}-1\right)\lim_{\varepsilon\to 0}\frac{1}{\sqrt{\pi}} \Gamma \left(\frac{1}{2}-i G s\right) (\mu  \epsilon )^{-2 i G s}\\ &
      =\frac{16\pi i G\, \widetilde{\mathbb{M}}^{\textrm{EFT}}_{\textrm{Born}}(s)} {\sqrt{\pi}}(e-1)\Gamma\left(\frac{1}{2}-i G s \right)\left(\frac{1}{\mu\varepsilon}\right)^{2iGs}\,.
    \end{split}
\end{align}
 {From the above equation, it is quite evident that after considering the distributional aspect of the contact delta function, we are able to modify the phase dressing in quadratic EFT along with an IR regulator $\mu_{IR}=\mu\epsilon$.}
Now we proceed to show the analyticity properties, i.e. how the UV (or IR) behaviour of the eikonal amplitude has modified in the presence of the quadratic EFT corrections.
\begin{tcolorbox}[thesisresultbox, title=Eikonal amplitude: GR vs EFT correction]
\restorethesisbodyformat
\begin{align}
    \begin{split}
&\mathbfcal{M}_{\textrm{eik}}^{\textrm{GR}}\sim -\underbrace{\frac{16  \pi  G \xi (s)}{t}}_{\text{Born Amplitude}}\times\underbrace{\frac{\Gamma\left(-iGs\right)}{\Gamma\left(iGs\right)}\Bigg(\frac{4\mu^2}{-t}\Bigg)^{-iGs}}_{\text{Phase}}\,\,\,\,\,\,\\ 
&
\hspace{0.8cm}\oplus
\\&
\mathbfcal{M}_{\textrm{eik}}^{\textrm{EFT}}\sim\underbrace{\widetilde{\mathbb{M}}^{\textrm{EFT}}_{\textrm{Born}}(s|\alpha,\beta)}_{\textrm{Born amplitude}}\times \underbrace{\frac{16\pi  G(e-1)\, } {\sqrt{\pi}}\frac{\Gamma(-2iGs)}{\Gamma(-iGs)}\left(\frac{2}{\mu\varepsilon}\right)^{2iGs}}_{\textrm{phase}}\,.\label{3.27y}
    \end{split}
\end{align}
As demonstrated in ~\eqref{3.27y}, EFT correction to the eikonal amplitude retains the same structural form as in GR. In particular, the eikonal amplitude continues to be expressed as the product of the Born amplitude and a phase factor. The principal distinction lies in the fact that, unlike in GR, the EFT-corrected eikonal amplitude is independent of the Mandelstam variable `\( t \)', thereby eliminating the associated `\( z \)'-dependence in the celestial eikonal amplitude which will be discussed in the next section.

\end{tcolorbox}
\restorethesisbodyformat
\subsection{Analyticity and dispersion relation for celestial eikonal amplitude}
In this section, we explain the  UV behaviour of the eikonal amplitude due to the presence of the higher curvature terms. We also inspect the IR behaviour for the quadratic EFT celestial amplitude. Below, we focus only on the contribution coming from the EFT (curvature squared) part. \textcolor{black}{Although the celestial Born amplitude arising from EFT corrections is meromorphic (as shown in \eqref{3.10e})—rendering the construction of an eikonal amplitude technically unnecessary—employing eikonal exponentiation nonetheless offers a practical framework for capturing non-perturbative (in coupling) structures in the celestial amplitude, thereby addressing the second question posed in the introduction. However, the eikonal amplitude encodes the full non-perturbative scattering amplitude within the given eikonal regime.  }

\noindent
{\textit{\textbf{\,{UV behaviour}}}}:
Thus, the $2\rightarrow2$ scattering amplitude of massless scalars as external states can be casted with the contribution of phase in the eikonal limit as follows \footnote{In the argument of $\Gamma$ function, we neglect terms like $G\alpha\sim \ell_p^2<<1$.}, 
\begin{align}
    \begin{split}\label{3.32a}
    \mathbfcal{A}^{\textrm{EFT}}_{eik}(\gamma,z)&=\frac{16\pi i G \,}{\sqrt{\pi}}(e-1)2^{3-\gamma}z^2\rmint_0^\infty d\omega \,\omega^{\gamma-1}\,{\Gamma\left(\frac{1}{2}-iG \omega^2\right)}\Bigg(\frac{1}{\mu \varepsilon}\Bigg)^{2iG\omega^2}\widetilde{\mathbb{M}}^{\textrm{EFT}}_{\textrm{Born}}(\omega^2)+\textrm{GR part}\,.
    \end{split}
\end{align}
In large $\omega$ limit, the integral reduces to,
\begin{align}
\begin{split}\label{3.27k}
\mathbfcal{A}^{\textrm{EFT}}_{eik}(\gamma,z)&= \frac{i\pi  G g(\alpha,\beta)}{\sqrt{\pi}}(e-1)\,2^{7-\gamma}z^2\rmint_0^\infty d\omega \,\omega^{\gamma-1}\,{\Gamma\left(\frac{1}{2}-iG \omega^2\right)}\Bigg(\frac{1}{\mu \varepsilon}\Bigg)^{2iG\omega^2}+\textrm{GR part}\,,\\ &
        =i\pi G g(\alpha,\beta)(e-1)2^{8-\gamma}z^2 \rmint_{0}^{\infty} d\omega\,\omega^{\gamma-1}\left(\frac{2}{\mu\varepsilon}\right)^{2iG\omega^2}\frac{\Gamma\left(-2iG\omega^2\right)}{\Gamma(-iG\omega^2)}+\cdots\,\,,
    \end{split}
\end{align}where $  g(\alpha,\beta)=\frac{8\beta-\alpha}{2\alpha\beta}+\frac{4}{\alpha}-\frac{1}{2\beta}.$ In obtaining the second line, we used the duplication formula,
$\frac{\Gamma(a)\Gamma(a+\frac{1}{2})}{\Gamma(2a)}= \frac{\sqrt{\pi}}{2^{2a-1}}.$ Now we have,
\begin{align}
\begin{split}
\mathbfcal{A}^{\textrm{EFT}}_{eik}(\gamma,z) &
= i\pi G g(\alpha,\beta)(e-1)2^{7-\gamma}z^2 \rmint_0^\infty dx\, x^{\frac{\gamma - 2}{2}} \left( \frac{2}{\mu \varepsilon} \right)^{2iGx} \frac{\Gamma(-2iGx)}{\Gamma(-iGx)}\,,
\\ &\xrightarrow[]{x\to \infty} i\pi G g(\alpha,\beta)(e-1)2^{7-\gamma}z^2\sqrt{2} \rmint_0^\infty dx\, x^{\frac{\gamma - 2}{2}} \left( \frac{e}{\mu^2 \varepsilon^2} \right)^{iGx}  (-i Gx)^{-iGx}\,.
\end{split}
\end{align}
The integral\footnote{We found a mismatch between signs of $a$ and $b$ in the expression $e^{ax}x^{ibx}$ in \cite{Adamo:2024mqn} where it seems to be of opposite sign according to the asymptotic behaviour of the ratio of gamma functions.} can be done by analytically continuing it in the fourth quadrant of the complex plane.
\begin{align}
    \begin{split}
    \mathbfcal{A}^{\textrm{EFT}}_{eik}(\gamma,z)=i\pi G g(\alpha,\beta)(e-1)2^{\frac{15}{2}-\gamma}z^2  \left(-\frac{i}{G}\right)^{\gamma/2-1}\rmint_0^\infty d\zeta\,\zeta^{\frac{\gamma}{2}-1}\left(\frac{e}{\mu^2\varepsilon^2}\right)^{\zeta}\,(-\zeta)^{-\zeta}\,.\label{3.35b1}
    \end{split}
\end{align}
The integral in \eqref{3.35b1} is analytic for carefully chosen IR cut-off, more precisely the integral is meromorphic in the complex $\gamma$ plane. For large $\gamma$, the integral in \eqref{3.35b1} can be done using the saddle point approximation\footnote{\textcolor{black}{Though it is tough to compare the convergence rate of the integral to that of its GR counterpart, as it is done under the saddle-point approximation, we have checked both integrals numerically and found that the UV behaviour of the part coming from the quadratic gravity is comparatively better. }}
\begin{align}
    \begin{split}
         \mathbfcal{A}^{\textrm{EFT}}_{eik}(\gamma,z)\xrightarrow[]{\gamma\gg1}i\pi G g(\alpha,\beta)(e-1)2^{\frac{15}{2}-\gamma}z^2  \left(-\frac{i}{G}\right)^{\gamma/2}e^{f(\zeta_{*})}\sqrt{\frac{\pi\zeta_{*}^2}{2\zeta_{*}+\gamma}}
    \end{split}
\end{align}
where, $\zeta_{*}=\frac{\gamma }{2 {W}\left(-\frac{1}{2} e \gamma  \mu ^2 \epsilon ^2\right)}$ and $f(\zeta)=\frac{1}{2} \gamma  \log (\zeta )-\zeta  \log (-\zeta )-\zeta  \log \left(\mu ^2 \epsilon ^2\right)+1$, with ${W}$ being the {Lambert}-${W} $ function.

\textcolor{black}{
\textbf{A sidebar:} A quick question that may arise is how one can justify expanding the integrand in Eq.~\eqref{3.32a} in the large-\(\omega\) limit, given that the integration is performed over the entire range \((0, \infty)\).
 The explanation goes as follows: The expansion of the integrand in the large-\(\omega\) limit is employed to extract the ultraviolet (UV) behavior of the corresponding celestial amplitude. Strictly speaking, the appropriate procedure is to first expand the integrand in the large-\(\omega\) regime, perform the indefinite integral, and then analyze the \(\omega \to \infty\) behavior of the result.
}

\textcolor{black}{
Nonetheless, guided by intuition from the saddle point approximation, one expects that in the presence of a large parameter, the dominant contribution to the integral originates near the peak of the integrand, where it is sharply localized. Although the integration domain extends over the full range \((0, \infty)\), the leading contribution arises from the large-\(\omega\) region, which justifies the use of this approximation in capturing the UV behavior. This is why we did not evaluate the integral in Eq.~(3.32) explicitly; instead, we applied a saddle point analysis by choosing the parameter \(\gamma\) to be large, leading to a dominant contribution near the saddle point \(\zeta_*\).}

\textcolor{black}{
For a more rigorous justification, one could invoke the ``strategy of regions'' a well-established technique widely used in the computation of multi-loop Feynman integrals~\cite{Beneke:1997zp}. This method systematically \textit{expands the integrand} in distinct regions characterized by separated scales, allowing for controlled approximations. In our context, this methodology has been applied heuristically, in line with earlier treatments in ~\cite{Adamo:2024mqn}. }

{\textit{\textbf{\,{IR behaviour}}}}: 
Now, to investigate the analytic property of the eikonal amplitude in IR limit, we first consider a general integral of the form,
\begin{align}
    \begin{split}
        \mathcal{I}(a):=\rmint_0^\infty dx\, x^{a-1}\,\phi(x)\, c^{\ell x},\,\quad \textrm{Re}(\ell)=0 \,,
    \end{split}
\end{align}
 where $a\in \mathbb{C}$ and c is a constant. Now $\phi(x)$ is analytic around $x=0$. Also, it doesn't affect the convergence at infinity. Now, to examine the IR behaviour, 
we expand $c^{\ell x}$ and $\phi$ in Taylor series around $x=0$ and see what happens to the integral \footnote{Here, $L>0$ but a small number.}. The integral becomes, 
\begin{align}
    \begin{split}
\mathcal{I}_L(a)&=\sum_{m,n=0}^\infty\phi_m\frac{\ell^n \log(c)^n}{n!}\rmint_0^L dx \,x^{a+m+n-1}\,,\\&
=\sum_{m,n=0}^\infty\phi_m\frac{\ell^n \log(c)^n}{n!}\frac{L^{a+m+n}}{(a+m+n)}
\,\sim\sum_{m,n=0}^\infty\frac{\ell^n \log(c)^n\phi_m}{n!(a+n+m)}L^{a+m+n}+\text{regular terms}\,.
    \end{split}
\end{align}
The integral has poles and admits the following expansion (with $k=m+n$) ,
\begin{align}
\begin{split}
   \mathcal{I}_k(a)\sim\frac{\ell^k \log(c)^kL^{a+k}{\phi_0}}{k!(a+k)}+\frac{\ell^{k-1} \log(c)^{k-1}L^{a+k}{\phi_1}}{(k-1)!(a+k)}+\cdots\text{regular}.\label{3.34p}
\end{split}
\end{align}

\noindent 
Now, in the actual eikonal amplitude given in \eqref{3.27k}, has Gamma functions which can be expanded in a Taylor series (near $\omega=0$) as follows,
\begin{align}
 \begin{split}
\log{\Gamma\left(\frac{1}{2}-iG\omega\right)}=\gamma_E\left(\frac{1}{2}+i G\omega\right)+\sum_{p\geq 2}\frac{\zeta(p)}{p}\left(\frac{1}{2}+iG\omega\right)^{p}
 \end{split}   \end{align}
where $\zeta(p)$ is the Riemann zeta function and $\gamma_E$ is the Euler's constant.
Therefore, the eikonal amplitude for our case can be written as,
 \begin{align}
     \begin{split}
  \mathbfcal{A}^{\textrm{EFT}}_{eik}(\gamma,z)\sim C_E\rmint_0^\infty d\omega'\omega'^{\gamma/2-1}e^{i b \omega'}\exp\Bigg[\sum_{p\geq 2}\frac{\zeta(p)}{p}\left(\frac{1}{2}+iG\omega'\right)^{p}\Bigg]\,{\widetilde{\mathbb{M}}^{\textrm{EFT}}_{\textrm{Born}}(\omega',z)},
     \end{split}
 \end{align}
where \begin{align}
    b=G\left[\log\Bigg(\frac{ 4}{\mu^2 \epsilon^2}\Bigg)+\gamma_E\right],\,\,\, C_E=e^{\gamma_E /2}\,.
\end{align}
 Now by comparing term by term with  \eqref{3.34p}, we can readily identify,

\begin{align}\begin{split}
\label{3.36m}
 &\phi_0=0,\,\,\,\,\,\,\,\phi_1=\frac{1}{12} \sqrt{\pi } G^2 \omega  (\beta -2 \alpha )\,\\&\phi_2=\frac{1}{12} i \sqrt{\pi } G^3 (\gamma_E +\log (4)) (2 \alpha -\beta )\,,\\&\phi_3=\frac{1}{48} \sqrt{\pi } G^4 \left(-8 \alpha ^3+(\gamma_E +\log (4))^2 (4 \alpha -2 \beta )+2 \pi ^2 \alpha +16 \beta^3-\pi ^2 \beta \right)\,,\\&
 \phi_4=\frac{1}{144} i \sqrt{\pi } G^5 \left(24 \psi ^{(0)}\left(\frac{1}{2}\right) \left(\alpha ^3-2 \beta ^3\right)-\left(3 \pi ^2 \psi ^{(0)}\left(\frac{1}{2}\right)+2 \left(\psi ^{(0)}\left(\frac{1}{2}\right)^3+\psi ^{(2)}\left(\frac{1}{2}\right)\right)\right) (2 \alpha -\beta )\right)\,,\\&\,\, \vdots
 \end{split} 
\end{align}
\noindent
where $\psi^{(0)}(x)$ is the digamma function and $\psi^{(2)}(x)$ is the second-order derivative of the digamma function.
Clubbing together all the pieces we get,
\begin{align}
    \begin{split}
\mathbfcal{A}^{\textrm{EFT}}_{eik}(\gamma,z)\sim\Bigg(\frac{L^{\frac{\gamma}{2}+n+1}\left(2i G \log(\frac{2}{\mu\epsilon})\right)^n\,\phi_1}{n!(\frac{\gamma}{2}+n+1)}+\cdots+\text{regular contribution}\Bigg)\,.
    \end{split} \label{eqnnn}
\end{align}

It is evident that, in the IR limit, the leading singularity coming from the EFT contribution differs from that of pure GR. As evident from \eqref{eqnnn}, the location of the leading pole is at $\gamma=-2(n+1)$, and it is now a simple pole, unlike GR, where we get contributions from higher order poles. The introduction of higher curvature terms changes the infrared behaviour by changing only the subleading pole (i.e, simple pole) structure. Now, we proceed to discuss the dispersion relation for the quadratic EFT.

\noindent
\subsection{Dispersion Relations}
Although, like GR \cite{Adamo:2024mqn} we are unable to compute the integral in eikonal amplitude explicitly, we still find out the dispersion relation. The dispersion relation can be figured out from the analytic continuation of the integral \eqref{3.27k} in the full complex plane. To do so, we define the eikonal amplitude with an appropriate  $i\varepsilon$-prescription\footnote{A natural question arises regarding the choice of the \(-i\varepsilon\)-prescription. This choice is essential to impose the correct limiting conditions to recover the results for GR. In contrast, adopting a \(+i\varepsilon\) prescription causes the integration contour to encircle certain unphysical poles whose residues become divergent in the limit \(\alpha, \beta \to 0\). To ensure consistency and recover GR result in the limit  \(\alpha, \beta \to 0\), it is therefore necessary to employ the \(-i\varepsilon\) prescription.
},
\begin{align}
    \begin{split}
\mathbfcal{A}^{\textrm{EFT}}_{eik}(\gamma,z)=\frac{16\pi i G \,}{\sqrt{\pi}}(e-1)2^{3-\gamma}z^2\rmint_{0-i\varepsilon}^{\infty-i\varepsilon} d\omega \,\omega^{\gamma-1}\Gamma\left(\frac{1}{2}-iG\omega^2\right)\Bigg(\frac{1}{\mu \epsilon}\Bigg)^{2iG\omega^2}\,\widetilde{\mathbb{M}}_{\textrm{Born}}^{\textrm{EFT}}(\omega)\,. \label{3.26uu}
    \end{split}
    \end{align}
    where, $\widetilde{\mathbb{M}}_{\textrm{Born}}^{\textrm{EFT}} $ is defined in \eqref{3.18p}. 
   
\begin{figure}[htb!]
    \centering
\includegraphics[width=0.40\linewidth]{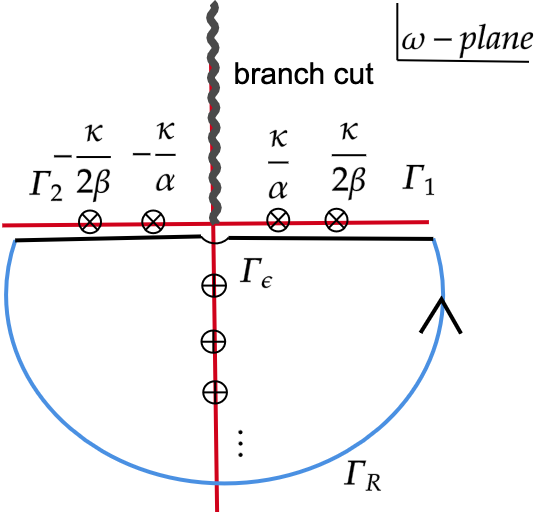}
    \caption{Figure depicting the chosen contour for dispersion relation in the complex-$\omega$ plane. The cross signs depict the location of the poles.}
    \label{fig:o}
\end{figure}

The integral can also be written after a change of variable change in the following way,
\begin{align}
    \begin{split}
\sim z^2\rmint_{0+i\varepsilon}^{ \infty+i\varepsilon} d\omega'\,\omega'^{\gamma/2}\Gamma\left(\frac{1}{2}-iG\omega'\right)\Bigg(\frac{1}{\mu \epsilon}\Bigg)^{2iG\omega'}\frac{\widetilde{\mathbb{M}}_{\textrm{Born}}^{\textrm{EFT}}(\omega')}{\omega'}\,.
    \end{split}
\end{align}
This has poles at,
\begin{align}
    G\omega'=e^{-i\pi/2}\left(n+\frac{1}{2}\right),\,\,\, \mbox{and}\,\, \omega'=\frac{\kappa }{\alpha},-\frac{\kappa }{\alpha},\frac{\kappa }{2\beta},-\frac{\kappa }{2\beta}\,,\,\,\,\,\, n \in \mathbb{Z}_{\geq0}\,.
\end{align}
\noindent
Note that we also have extra poles at $\omega'$ entirely due to the higher curvature contributions. We also have a branch cut, which is running from $0\rightarrow i\infty$. However, as we have poles on positive and negative real axis, we need to change the argument about the branch cut. The integral in $\omega'$  can now be analytically continued to the complex plane, and we use a contour in the clockwise orientation in the complex plane as depicted in Fig.~\ref {fig:o}. The integral can now be written as a sum of four separate pieces as,
\begin{equation}
\begin{split}
    I_{C} 
    &\sim z^2\oint_C d\omega\,\omega^{\gamma/2}\Gamma\left(\frac{1}{2}-iG\omega\right)\Bigg(\frac{1}{\mu \epsilon}\Bigg)^{2iG\omega}\frac{\widetilde{\mathbb{M}}_{\textrm{Born}}^{\textrm{EFT}}(\omega)}{\omega}\,,\\
    &= I_1+I_2 +I_{\epsilon}+I_R 
\end{split}
\end{equation}

\noindent
where the decomposition of the contour $\Gamma=\Gamma_1\cup \Gamma_R\cup \Gamma_2\cup \Gamma_\epsilon$ corresponds to each of the four integrals in the second line, with $\Gamma_R$ the semi-circle part of the contour of radius $R$ and $\Gamma_\epsilon$ the smaller semi-circle part of radius $\epsilon$ as shown in Fig.~\eqref{fig:o}. It can be shown that in the limits where $\epsilon\to0$ and $R\to\infty$, the contributions from $\Gamma_\epsilon$ and $\Gamma_R$ vanishes. Therefore, we have,
\begin{align}
    &I_1+I_2=2\pi i \sum_{n\geq1} \textbf{Res}\Bigg[z^2\,\omega^{\gamma/2}\Gamma\left(\frac{1}{2}-iG\omega\right)\Bigg(\frac{1}{\mu \epsilon}\Bigg)^{2iG\omega}\frac{\widetilde{\mathbb{M}}_{\textrm{Born}}^{\textrm{EFT}}(\omega)}{\omega}\Bigg]\,\nonumber\\&\hspace{0.8cm}\text{at} \,\,\omega=G^{-1}e^{-i\pi/2}(n+1/2)
\end{align}
where the integrals ($I_1,\,I_2$) are given by,
\begin{align}
      I_1=-\mathbfcal{A}^{\textrm{EFT}}_{eik}(\gamma,z),\,\,\,\,\,\,\,\,\,\,\,
      I_2=e^{-i\pi \gamma/2}\,\overline{\mathbfcal{A}^{\textrm{EFT}}_{eik}}(\bar\gamma,-z) \,.
\end{align}
$\overline{\mathbfcal{A}^{\textrm{EFT}}_{eik}}(\bar \gamma) $ denotes the complex conjugate of the eikonal amplitude $\mathbfcal{A}^{\textrm{EFT}}_{eik}(\gamma)$ . Therefore we can write the following \textit{dispersion relation} including the GR contribution as,

\begin{align}
\begin{split}\label{3.53e}
 \hspace{0cm}-&(\mathbfcal{A}_{eik}^{\textrm{GR}}(\gamma,z)+\mathbfcal{A}_{eik}^{\textrm{EFT}}(\gamma,z))+e^{-i\pi \gamma/2}(\overline{\mathbfcal{A}_{eik}^{\textrm{GR}}}\left(\bar \gamma,-z\right)+\overline{\mathbfcal{A}_{eik}^{\textrm{EFT}}}(\bar\gamma,-z)) \\&=z^2\sum_{n\geq 1}\frac{(-1)^n \frac{16\pi i G \,}{\sqrt{\pi}}(e-1)2^{3-\gamma} 2^{-4} \left(-\frac{i \left(n+\frac{1}{2}\right)}{G}\right)^{\gamma /2}}{3 n! (\alpha -2 i  +2 \alpha  n) (\alpha +2 i  +2 \alpha  n) (\beta -i +2 \beta  n) (\beta +i  +2 \beta  n)}\\ &\hspace{2 cm}\times\Big[(2 n+1) \mu _{\text{IR}}^{-2 n-1} \left(4  (\beta -2 \alpha )+\alpha  \beta  (2 n+1)^2 (\alpha -8 \beta )\right)\Big]\\&
\hspace{0.8cm}-\frac{\pi z^3 2^{3-\gamma}}{z-1}\sum_{k\geq 1}\frac{i^k}{k!(k-1)!}\left(-\frac{i k}{G}\right)^{\gamma/2}\Bigg(\frac{k(z-1)}{4zG\mu^2}\Bigg)^k\,.
 \end{split} 
\end{align}
The sum in \eqref{3.53e} does not admit a closed-form expression and is therefore left as a formal sum.
where, $\mu_{IR}=\mu\epsilon$ is a dimensionless IR regulator. As a consistency check, one can verify that setting $\alpha=\beta=0$ recovers the result corresponding to GR \cite{Adamo:2024mqn}. \textcolor{black}{
Up to this point, we have investigated the analyticity and pole structure of the celestial amplitude, highlighting the modifications due to the incorporation of higher curvature terms. In the following section, we focus on the properties of the conformal four-point function that follows from it.}
\par
\textcolor{black}{
Subsequently, we compute the shadow transform of the celestial four-point correlator of primary operators arising from the quadratic effective field theory (EFT). We then carry out the conformal block decomposition and determine the associated operator product expansion (OPE) coefficients. } 
\section{Shadowed correlator and operator product expansion (OPE)}\label{ch5:sec4}
As introduced in \cite{Pasterski:2016qvg, Strominger:2017zoo} and briefly discussed in the introduction, there is significant progress in understanding the 2D holographic description of 4D scattering amplitude in flat space. Specifically, the action of Lorentz group $SL(2,\mathbb{C})$ on the kinematic data can be rescasted as the Mobius transformation on the celestial sphere and the scattering amplitudes (denoted as $\mathcal{A}_{n}(z_i,\bar z_i)$), defined on the boost eigenstate, can be re-interpreted as correlation function in a 2D CFT.\par
In light of this, in this section, our aim is to explore the structure of the celestial conformal field theory (CCFT) associated with the quadratic effective field theory (EFT) amplitude (in the eikonal limit) in flat space. A crucial step in this direction is to determine the full operator spectrum of the conformal theory on the celestial sphere. This, in turn, requires knowledge of the OPE coefficients in various channels. Primarily, we focus on the shadowed amplitude. We first employ the Burchnall-Chaundy (BC) expansion to extract the OPE coefficients by comparing it with the conformal block expansion of Dolan and Osborn \cite{Dolan:2003hv}. However, we find that the BC expansion appears insufficient to capture the complete spectrum of the theory. To go beyond this limitation, we subsequently incorporate the effects of spinning exchanges (which completes the spectrum) in the OPE coefficients using the Euclidean OPE inversion formula \cite{Caron-Huot:2017vep, Simmons-Duffin:2017nub, Mazac:2018qmi, Mukhametzhanov:2018zja} and consequently comment on the subtlety associated with the inversion.

\noindent
\par
The \textit{shadow} of a conformal primary with conformal dimension $h,\bar h$ (and scaling dimension $\Delta=h+\bar h$) can be written as \cite{Simmons-Duffin:2012juh,Fan:2023lky},
\begin{align}
\widetilde{\mathcal{O}}_{\tilde\Delta}(z,\bar z):=\widetilde{{\mathcal{O}}_{\Delta}(y,\overline{y})}=\mathscr{K}_{h,\bar h}\rmint d^2y (z-y)^{2h-2}(\bar z-\bar y)^{2\bar h-2}\mathcal{O}_{\Delta}(y,\bar y),
\end{align}
where the constant is given  by,
\begin{align}
    \mathscr{K}_{h,\bar h}=\frac{(-1)^{2(h-\bar h)}\Gamma(2-2h)}{\pi\Gamma(2\bar h -1)}.
\end{align}
The shadow operator is a primary with conformal dimensions $\{1-h,1-\bar h\}$ which corresponds to a scaling dimension $\tilde \Delta=2-\Delta $.
\subsection{Shadowed amplitude corresponding to eikonal amplitude }
The 4-point eikonal amplitude is given by,
\begin{align}
    \begin{split}
        \mathcal{M}_{\textrm{eik}}
    \end{split}(z_i,\bar z_i)=\lim_{z_4\to \infty}\frac{1}{|z|^{\Delta_1+\Delta_2}|z_4|^{2\Delta_4}}(z-1)^{\frac{\Delta_1-\Delta_2-\Delta_3+\Delta_4}{2}}\delta(iz-i\bar z) \mathbfcal{M}_{\textrm{eik}}(\mathbf \Delta,z)\,.
\end{align}
The normalized amplitude is given by,
\begin{align}
\widehat{\mathcal{M}}_{\textrm{eik}}=\lim_{z_4,\bar z_4\to \infty}z_4^{\Delta_4} \bar z_4^{\Delta_4} \mathcal{A}_{\textrm{eik}}(z_i,\bar z_i)\,.
\end{align}
Therefore the shadowed amplitude is given by,
\begin{align}
    \begin{split} \widetilde{\mathcal{M}}_{\textrm{eik}}^{\textrm{EFT}}(w,\bar w)=&\mathscr{K}_{h_2,\bar h_2}\lim_{z_4\to \infty}|z_4|^{2\Delta_4}\rmint \frac{d^2 z}{(z-w)^{\Delta_2}(\bar z-\bar w)^{2-\Delta_2}}\frac{1}{|z|^{\Delta_1+\Delta_2}}(z-1)^{\frac{\Delta_1-\Delta_2-\Delta_3+\Delta_4}{2}}\delta(iz-i\bar z) \\ &
   \times i\pi G \widetilde{\mathbb{M}}^{\textrm{EFT}}_{\textrm{Born}}(e-1)2^{8-\gamma}z^2 \rmint_{0}^{\infty} d\omega\,\omega^{\gamma-1}\left(\frac{2}{\mu\varepsilon}\right)^{2iG\omega^2}\frac{\Gamma\left(-2iG\omega^2\right)}{\Gamma(-iG\omega^2)}.\label{4.6j}
    \end{split}
\end{align}
Here $\widetilde{M}_{\textrm{Born}}^{\textrm{EFT}}(\omega)$ is defined in \eqref{3.18p}. It contains only the terms arising due the presence of the curvature squared terms. Note that in equation \eqref{4.6j}, the $\omega$-integral is independent of the cross-ratio $z$. This allows us to perform the integrals over $z$ and $\bar z$ explicitly, resulting in a function that depends only on $w$, $\bar w$, and the remaining integral over $\omega$. Consequently, the overall structure of the shadow amplitude (coming from the curvature squared terms) remains unchanged if we replace the full non-perturbative (in coupling) eikonal amplitude with just the Born amplitude. Hence, from the following section onward, we focus on the Born amplitude to study the conformal correlator and its operator product expansion.\par

Before we concluding we note that,  for the GR part, the computation of the shadow amplitude is not so straightforward. The eikonal shadow amplitude for GR has the following form:
\begin{align}
    \begin{split}
        \widetilde{\mathcal{M}}_{\textrm{eik}}^{\textrm{GR}}(w,\bar w)=&\mathscr{K}_{h_2,\bar h_2}\lim_{z_4\to \infty}|z_4|^{2\Delta_4}\rmint \frac{d^2 z}{(z-w)^{\Delta_2}(\bar z-\bar w)^{2-\Delta_2}}\frac{1}{|z|^{\Delta_1+\Delta_2}}(z-1)^{\frac{\Delta_1-\Delta_2-\Delta_3+\Delta_4}{2}}\delta(iz-i\bar z)\\ &
        \times\left(\frac{G z}{1-z}\right)\rmint_{0}^{\infty} d\omega\,\omega^{\gamma+1}\left(\frac{4 z\mu^2}{\omega^2 (z-1)}\right)^{-iG\omega^2}\frac{\Gamma\left(-iG\omega^2\right)}{\Gamma(iG\omega^2)}\,.\label{4.6l}
    \end{split}
\end{align}
As can be seen easily from \eqref{4.6l}, the integral over \( z \) and \( \bar{z} \) can be performed explicitly, leaving us with an integral over \( \omega \). However, this remaining integral cannot be evaluated analytically, making it difficult to obtain the shadowed amplitude non-perturbatively in \( G \) for GR. Our primary objective is to extract the OPE coefficient analytically from the conformal block expansion, which requires a closed-form expression (at least in $w,\,\bar w$) for the amplitude. Fortunately, the contribution due to the  EFT correction in the Born amplitude possesses a well-behaved analytic structure, allowing us to bypass the need for eikonal resummation in both qualitative and quantitative analyses. Therefore, in the following subsections, we will consider only the Born amplitude when computing the OPE coefficients.

\subsection{Shadowed amplitude corresponding to celestial Born amplitude}
 We start by reminding that  the 4-point amplitude is given by,
\begin{align}
    \begin{split}
        \mathcal{A}_{4}(z_i,\bar z_i)=\lim_{z_4\to \infty}\frac{1}{|z|^{\Delta_1+\Delta_2}|z_4|^{2\Delta_4}}(z-1)^{\frac{\Delta_1-\Delta_2-\Delta_3+\Delta_4}{2}}\delta(iz-i\bar z) \mathcal{A}(\mathbf \Delta,z)\,.
    \end{split}
\end{align}
Consequently, one can define the normalized 4-point amplitude as,
\begin{align}
    \begin{split}
         \widehat{\mathcal{A}_{4}}(z_i,\bar z_i)=\lim_{z_4,\bar z_4\to \infty}z_4^{\Delta_4} \bar z_4^{\Delta_4} \mathcal{A}_{4}(z_i,\bar z_i)\,.
    \end{split}
\end{align}
Our goal is to compute the celestial amplitude of the shadowed correlator, find the conformal block expansion and correspondingly calculate the partial wave coefficients using \textit{Euclidean OPE inversion formula}.
 Now, we want to compute the following correlator,
\begin{align}
    \begin{split}
    \widetilde{\mathcal{A}_{4}}(w,\bar w)&:=\lim_{z_4\to \infty}|z_4|^{2\Delta_4}\Big\langle\mathcal{O}_{\Delta_1}(0,0)\widetilde{\mathcal{O}}_{2-\Delta_2}(w,\bar w)\mathcal{O}_{\Delta_3}(1,1)\mathcal{O}_{\Delta_4}(z_4,\bar z_4)\Big\rangle\,,\\ &
          =\mathscr{K}_{h_2,\bar h_2}\lim_{z_4\to \infty}|z_4|^{2\Delta_4}\rmint \frac{d^2 z}{(z-w)^{\Delta_2}(\bar z-\bar w)^{2-\Delta_2}}\Big\langle\mathcal{O}_{\Delta_1}(0,0){\mathcal{O}_{\Delta_2}}(z,\bar z)\mathcal{O}_{\Delta_3}(1,1)\mathcal{O}_{\Delta_4}(\bar z_4,\bar z_4)\Big\rangle\,.
    \end{split}
\end{align}
We can define three celestial amplitudes in the respective kinematic regions by choosing the corresponding integration range of the cross-ratio $z$.
\begin{align}
    \begin{split}\label{4.28i}
 \mathscr{K}_{h_2,\bar h_2}^{-1}\widetilde{\mathcal{A}_{4}}(w,\bar w)&=\rmint \frac{dz}{(z-w)^{2-\Delta_2}(z-\bar w)^{2-\Delta_2}} \frac{(z-1)^{\frac{\Delta_1-\Delta_2-\Delta_3+\Delta_4}{2}}}{|z|^{\Delta_1+\Delta_2}} \mathbfcal{A}(\mathbf{\Delta},z)\,.\end{split}\end{align}
\noindent
The integrals can be expressed in terms of \textit{Appell hypergeometric} functions and the amplitude for the different kinematic channels take the following form:\footnote{ 
\begin{align}
\mathbf{F}_1(a,b_1,b_2,c,x,y)=\sum_{m,n=0}^\infty \frac{(a)_{m+n}(b_1)_m(b_2)_n}{(c)_{m+n}m!n!}x^m y^n \,\,\,\,,\max(|x|,|y|)<1,\nonumber
\end{align}
\noindent
where $(a)_k$ is the Pochhammer symbol is given by,
$
    (a)_k:=\frac{\Gamma(a+k)}{\Gamma(a)}\,.$
Also the integral representation of the same function is given by,
\begin{align}
  \mathbf{F}_1(a,b_1,b_2,c,x,y)=\frac{1}{B(a,c-a)}\rmint_0^1 dm \frac{m^{a-1}(1-m)^{c-a-1}  }{(1-mx)^{b_1}(1-my)^{b_2}}
,\,\,\mathscr{R}(c)>0 \mbox{ and } \mathscr{R}(c-a)>0\nonumber
\end{align}
with,
$B(a,b):=\frac{\Gamma(a)\Gamma(b)}{\Gamma(a+b)}\,, B(a,b)=\rmint_0^1 dx \,x^{a-1}(1-x)^{b-1} $
is the usual Euler Beta function. Given these, we now proceed to compute the shadowed four-point correlator in three different kinematic regimes.}

\hspace{0cm}{\,\,{\textbf{\textit{12$\to$34 kinematics:}}}}

By defining $\textcolor{black}{w=\frac{z_{12'}z_{34}}{z_{13}z_{2'4}}}$ we can cast the integral in \eqref{4.28i} for the $12\to 34$ kinematics in the following way \footnote{For four general points $z_1,z_2',z_3,z_4 $. $z_2'$ is the location of the insertion of the shadow primary.},
\begin{align}
    \begin{split}
    \mathscr{K}_{h_2,\bar h_2}^{-1}\widetilde{\mathcal{A}_{4}}^{12\to 34}(w,\bar w)&=\rmint_{1}^{\infty}\frac{dz}{(z-w)^{2-\Delta_2}(z-\bar w)^{2-\Delta_2}}\frac{(z-1)^{\frac{\Delta_1-\Delta_2-\Delta_3+\Delta_4}{2}}}{|z|^{\Delta_1+\Delta_2}}\mathcal{A}^{12\to 34}(\mathbf{\Delta},z)\,,\\ & =\rmint_{1}^{\infty}\frac{dz}{(z-w)^{2-\Delta_2}(z-\bar w)^{2-\Delta_2}}\frac{(z-1)^{\frac{\Delta_1-\Delta_2-\Delta_3+\Delta_4}{2}}}{|z|^{\Delta_1+\Delta_2}}\textstyle{\left(\delta_{1}(\alpha,\beta|\gamma)z^2+\delta_2(\gamma)\frac{z^3}{z-1}\right)}\,, \\ & =\delta_1(\alpha,\beta|\gamma)B\left(\frac{1}{2}(2+\Delta_1-\Delta_2+\Delta_3-\Delta_4),\frac{1}{2}(2+\Delta_1-\Delta_2-\Delta_3+\Delta_4)\right)\\ &\hspace{0.5 cm}\times F_{1}\left(\frac{1}{2}(2+\Delta_1-\Delta_2+\Delta_3-\Delta_4),2-\Delta_2,2-\Delta_2,2+\Delta_1-\Delta_2,w,\bar w\right)\\&\hspace{0.8cm}+(\textrm{contribution from GR})\,.
    \end{split}
\end{align}
\textcolor{black}{
The contribution from GR can be straightforwardly computed as,
\begingroup\tiny
\begin{align}
    \begin{split}
        (\textrm{contribution from GR})&=\delta_2(\gamma)\rmint_{1}^{\infty}\frac{dz}{(z-w)^{2-\Delta_2}(z-\bar w)^{2-\Delta_2}}\frac{(z-1)^{\frac{\Delta_1-\Delta_2-\Delta_3+\Delta_4}{2}-1}}{|z|^{\Delta_1+\Delta_2-3}}\,, \\ &
        \xrightarrow[]{z \to 1/z} \delta_2(\gamma)\rmint_0^1 \frac{dz}{(1-z\,w)^{2-\Delta_2}(1-z\,\bar{w})^{2-\Delta_2}} (1-z)^{\frac{\Delta_1-\Delta_2-\Delta_3+\Delta_4}{2}-1}z^{\frac{1}{2} \left(\Delta _1-\Delta _2+\Delta _3-\Delta _4\right)}\,,\\ &
        =\delta_2(\gamma)\,B\left(\frac{1}{2} \left(\Delta _1-\Delta _2+\Delta _3-\Delta _4+2\right),\frac{1}{2} \left(\Delta _1-\Delta _2-\Delta _3+\Delta _4\right)\right)\\&
        \hspace{0.6 cm}\times\textrm{F}_1\left(\frac{1}{2} \left(\Delta _1-\Delta _2+\Delta _3-\Delta _4+2\right),2-\Delta_2,2-\Delta_2,\Delta _1-\Delta _2+1,w,\bar w\right)\,.
    \end{split}
\end{align}
\endgroup
In the later computations we will not explicitly show the contributions from GR as our primary goal to examine the contribution coming from the quadratic corrections.}
Hence, finally the 4-point correlator for the single shadowed operator(s) in the $\mathfrak{s}$-channel can be written as,
\begin{align}
    \begin{split}
      \lim_{z_4\to \infty}  |z_4|^{2\Delta_4}\Big\langle\mathcal{O}_{\Delta_1}(0,0)\widetilde{\mathcal{O}_{\Delta_2}}(w,\bar w)\mathcal{O}_{\Delta_3}(1,1)\mathcal{O}_{\Delta_4}(z_4,\bar z_4)\Big\rangle=\frac{1}{|w|^{2+\Delta_1-\Delta_2}}\mathcal{G}^{\mathfrak{s}}
      (w,\bar w)\,.
    \end{split}
\end{align}
Therefore we can identify the conformal block corresponding to quadratic EFT to be,
\begin{align}
    \begin{split} \label{s-channel}
    &\mathcal{G}^{12\to34}
      (w,\bar w)=\delta_1(\alpha,\beta|\gamma)|w|^{2+\Delta_1-\Delta_2}B\left(\frac{1}{2}(2+\Delta_1-\Delta_2+\Delta_3-\Delta_4),\frac{1}{2}(2+\Delta_1-\Delta_2-\Delta_3+\Delta_4)\right)\\ &\times F_{1}\left(\frac{1}{2}(2+\Delta_1-\Delta_2+\Delta_3-\Delta_4),2-\Delta_2,2-\Delta_2,2+\Delta_1-\Delta_2,w,\bar w\right)+(\textrm{contribution from GR})\,.
    \end{split}
\end{align}
{\,\,{\textbf{\textit{13$\to$24 kinematics:}}}}

The $\mathfrak{t}$-channel contribution to the four-point function in quadratic EFT is given by \footnote{From now on we omit writing the GR contribution for each case. One should assume that they are always there inherently. The pure GR contributions impose constraints on the external conformal dimensions which is not there for quadratic gravity.},
\begin{align}
    \begin{split}
    \mathcal{G}^{13\to 24}(w,\bar w)&=\rmint_{0}^{1}\frac{dz}{(z-w)^{2-\Delta_2}(z-\bar w)^{2-\Delta_2}}\frac{(z-1)^{\frac{\Delta_1-\Delta_2-\Delta_3+\Delta_4}{2}}}{|z|^{\Delta_1+\Delta_2}}\mathcal{A}^{13\to 24}(\mathbf{\Delta},z))\\&
 =\delta_1(\alpha,\beta|\gamma) \textcolor{black}{(w\bar w)^{2-
       \Delta_2}B(-\Delta _1-\Delta _2+3,\frac{1}{2} \left(\Delta _1-\Delta _2-\Delta _3+\Delta _4+2\right)})\\&
         \times \mathbf{F}
         _1(3-\Delta_1-\Delta_2,2-\Delta_2,2-\Delta_2,\frac{1}{2} \left(-\Delta _1-3 \Delta _2-\Delta _3+\Delta _4+8\right),\frac{1}{w},\frac{1}{\bar w})\,.
    \end{split}
\end{align}
\noindent
{{\,\,{\textbf{\textit{14$\to$23 kinematics:}}}}}

Similarly for the $14\to 23$ kinematics ($\mathfrak{u}$-channel), four-point function has the following form,
\begin{align}
    \begin{split}
         \mathcal{G}^{14\to 23}(w,\bar w)&=\rmint_{-\infty}^{0}\frac{dz}{(z-w)^{2-\Delta_2}(z-\bar w)^{2-\Delta_2}}\frac{(z-1)^{\frac{\Delta_1-\Delta_2-\Delta_3+\Delta_4}{2}}}{|z|^{\Delta_1+\Delta_2}}\mathcal{A}^{14\to 23}(\mathbf{\Delta},z)\,.
    \end{split}
\end{align}
Now changing the integral variable as $z\to \frac{\omega}{\omega-1}$, we are left with the following result,
\begin{align}
    \begin{split}
         \mathcal{G}^{14\to 23}(w,\bar w)&=\delta_1(\alpha,\beta|\gamma)(w\bar w)^{\Delta_2-2}B(-\Delta _1-\Delta _2+3,\frac{1}{2} \left(\Delta _1-\Delta _2+\Delta _3-\Delta _4+2\right))\\&  \times \mathbf{F}_1(3-\Delta_1-\Delta_2,2-\Delta_2,2-\Delta_2,\frac{1}{2}(8+\Delta_1-\Delta_2+\Delta_3-\Delta_4),\frac{w-1}{w},\frac{\bar w-1}{\bar w})\,.
    \end{split}
\end{align}
Finally, using the identity of $\mathbf{F}_1$, we can write it in the following way, 
\begin{align}
\begin{split}
   &\mathcal{G}^{14\to 23}(w,\bar w) =\delta_1(\alpha,\beta|\gamma)B(-\Delta _1-\Delta _2+3,\frac{1}{2} \left(\Delta _1-\Delta _2+\Delta _3-\Delta _4+2\right))\\&\times\mathbf{F}_1(\frac{1}{2} \left(\Delta _1-\Delta _2+\Delta _3-\Delta _4+2\right),2-\Delta_2,2-\Delta_2,\frac{1}{2}(8-\Delta_1-3\Delta_2+\Delta_3-\Delta_4),1-w,1-\bar w)\,.
    \end{split}
\end{align}
Finally, adding the results from the three kinematic regimes mentioned above, we obtain the following expression,

\begin{align}
    \begin{split}
    \label{4.16o}
       & \textcolor{black}{\mathcal{G}^{12\to 34}(w,\bar w)+\mathcal{G}^{13\to 24}(w,\bar w)+\mathcal{G}^{14\to 23}(w,\bar w)}\\ &
=\delta_1(\alpha,\beta|\gamma)\Bigg[B\Bigg(\frac{1}{2}(2+\Delta_1-\Delta_2+\Delta_3-\Delta_4),\frac{1}{2}(2+\Delta_1-\Delta_2-\Delta_3+\Delta_4)\Bigg)\\&\times \mathbf{F}_{1}\left(\frac{1}{2}(2+\Delta_1-\Delta_2+\Delta_3-\Delta_4),2-\Delta_2,2-\Delta_2,2+\Delta_1-\Delta_2,w,\bar w\right)\\&+(w\bar w)^{2-
       \Delta_2}B(-\Delta _1-\Delta _2+3,\frac{1}{2} \left(\Delta _1-\Delta _2-\Delta _3+\Delta _4+2\right))\\&
         \hspace{2 cm}\times \mathbf{F}
         _1(3-\Delta_1-\Delta_2,2-\Delta_2,2-\Delta_2,\frac{1}{2} \left(-\Delta _1-3 \Delta _2-\Delta _3+\Delta _4+8\right),\frac{1}{w},\frac{1}{\bar w})\\&+B(-\Delta _1-\Delta _2+3,\frac{1}{2} \left(\Delta _1-\Delta _2+\Delta _3-\Delta _4+2\right))\\&\hspace{0.2cm}\times\mathbf{F}_1(\frac{1}{2} \left(\Delta _1-\Delta _2+\Delta _3-\Delta _4+2\right),2-\Delta_2,2-\Delta_2,\frac{1}{2}(8-\Delta_1-3\Delta_2+\Delta_3-\Delta_4),1-w,1-\bar w)\Bigg]\,.
    \end{split}
\end{align}

\noindent
Now our goal is to find the OPE coefficients using the inversion formula derived in \cite{Caron-Huot:2017vep}. For the inversion we need the argument of the blocks to be $w, \bar w$ and here we immediately identify the problem with our four-point function where we have different arguments. Hence, to resolve the problem we need the analytic continuation of the \textbf{Appell function}.

\noindent
\textbf{Analytic continuation:}
\noindent
Here, we describe the analytic continuation of our conformal block that we found in \eqref{4.16o}. In $\mathfrak{t}$ and $\mathfrak{u}$-channel, we use analytic continuation as well as check the properties under the monodromy projection. Some portions of the analytically continued block does not contribute due to their incorrect behaviour under \textit{monodromy projection} \cite{Simmons-Duffin:2012juh} and we discard them. For $\mathfrak{t}$  channel we get, 
\hfsetfillcolor{gray!8}
\hfsetbordercolor{black}
\begin{align}
\begin{split}
\label{4.40o}
    \\& \mathbf{F}_1(\frac{1}{2} \left(\Delta _1-\Delta _2+\Delta _3-\Delta _4+2\right),2-\Delta_2,2-\Delta_2,\frac{1}{2}(8-\Delta_1-3\Delta_2+\Delta_3-\Delta_4),1-w,1-\bar w)\to \\&
   \frac{\Gamma \left(\Delta _1+\Delta _2-2\right) \Gamma \left(\frac{1}{2} \left(\Delta _1-\Delta _2-\Delta _3+\Delta _4+2\right)\right)}{\Gamma \left(\Delta _1-\Delta _2+2\right) \Gamma \left(\frac{1}{2} \left(\Delta _1+3 \Delta _2-\Delta _3+\Delta _4-6\right)\right)}\\&\hspace{0.8cm} \times\mathbf{F}_1(\frac{1}{2} \left(\Delta _1-\Delta _2+\Delta _3-\Delta _4+2\right),2-\Delta_2,2-\Delta_2,2-\Delta_2+\Delta_1,w,\bar w)+\cdots
 \end{split} 
\end{align}
In \eqref{4.40o}, the ellipsis denotes the terms obtained after analytic continuation but do not exhibit correct behaviour under monodromy projection.
The analytic continuation for the $\mathfrak{u}$-channel kinematics can be done similarly.
Now we proceed to compute the OPE coefficients eventually in the $\texttt{s}$-channel kinematics. This choice is generic due to choosing external primary conformal dimension(s) to be equal. Otherwise, one should calculate the OPE coefficients for different channels separately, in euclidean setup. \textcolor{black}{ Though in euclidean scenario there is no sense of time, and therefore operator ordering does not matter. }

\subsection{Conformal block expansion and partial wave coefficients}
For $12\to 34$ kinematics, the four point function of the conformal primaries (with one shadowed operator) can be expanded in $\texttt{s}$-channel OPE in the following way \cite{Dolan:2003hv,Caron-Huot:2017vep},
\begin{align}
    \mathcal{G}^{12\to 34}(w,\bar w)=\sum_{J,\Delta}f_{12\mathbfcal{O}}f_{34\mathbfcal{O}}G_{J,\Delta}(w,\bar w),\label{4.29}
\end{align}
where, $J,\Delta$ are the spin and conformal dimension of the exchanging primary operator $(\mathbfcal{O})$. \textcolor{black}{$f_{ijk}$ are the  OPE coefficients} and $G_{J,\Delta}(z,\bar z)$ is defined as,
\begin{align}
    G_{J,\Delta}=\frac{k_{\Delta-J}(w)k_{\Delta+J}(\bar w)+k_{\Delta+J}(w)k_{\Delta-J}(\bar w)}{1+\delta_{J,0}}, \textrm{\,\,with }\,\,k_{\beta}(w)=w^{\beta/2}\,{}_{2}\mathbf{F}_1\left(\beta/2+a,\beta/2+b,\beta,w\right)
\end{align}
where the constants $a,b$ are identified as $a=\frac{1}{2}(2-\Delta_2-\Delta_1),b=\frac{1}{2}(\Delta_3-\Delta_4)$.
Therefore the block should take the form \cite{Osborn:2012vt,Caron-Huot:2017vep},
\begin{align}    \begin{split}\mathcal{G}^{12\to 34}(w,\bar w)=&\delta_1(\alpha,\beta|\gamma)B\left(\frac{1}{2}(2+\Delta_1-\Delta_2+\Delta_3-\Delta_4),\frac{1}{2}(2+\Delta_1-\Delta_2-\Delta_3+\Delta_4)\right)\\&\sum_{J,\Delta}f_{12\mathbfcal{O}}f_{34\mathbfcal{O}}\left[w^{\frac{\Delta-J}{2}}\bar w^{\frac{\Delta+J}{2}}\right]\times{_2}F_1\left(\genfrac{}{}{0pt}{}{\frac{1}{2}(\Delta-J+2-\Delta_2-\Delta_1),\frac{1}{2}(\Delta-J+\Delta_3-\Delta_4)}{\Delta-J};w\right)\\&\hspace{1.5 cm} \times {_2}F_1\left(\genfrac{}{}{0pt}{}{\frac{1}{2}(\Delta+J+2-\Delta_2-\Delta_1),\frac{1}{2}(\Delta+J+\Delta_3-\Delta_4)}{\Delta+J};\bar w\right)+(J\to -J)\,.\label{4.36k}
  \end{split}
\end{align}
Next, our goal is to compute the OPE coefficients using two different approaches: Burchnall-Chaundy expansion and OPE inversion formula. (Un-)fortunately the first one is only applicable for the four point functions involving the Appell functions.

\noindent
\textbf{Burchnall-Chaundy expansion:}

 We use the Burchnall-Chaundy expansion\footnote {The \textit{Burchnall-Chaundy expansion} of the Appell hypergeometric function enables one to write it in terms of a product of two Gauss hypergeometric functions in the following way,
	\begin{align}
		\label{eq:F1exp}
		\mathbf{F}_1\left(a , b_1, b_2 , c  , x, y\right) =\sum_{n=0}^\infty & {\frac{(a)_n(b_1)_n(b_2)_n(c-a)_n}  {n!(c+n-1)_n (c)_{2n}}}\,x^n y^n\times {}_{2}\mathbf{F}_1\left(\genfrac{}{}{0pt}{}{a+n,b_1+n}{c+2n};x\right) {}_{2}\mathbf{F}_1\left(\genfrac{}{}{0pt}{}{a+n,b_2+n}{c+2n};y\right)\,.
	\end{align}} \cite{Burchnall1940EXPANSIONSOA} (see \cite{Fan:2023lky} for more details) of the {Appell} function in \eqref{s-channel} to get the following,
	\begin{align}
\begin{split}
		\label{5.33u}
		&\mathcal{G}^{12\to 34}(w,\bar w)=\delta_1(\alpha,\beta|\gamma) B\left(\frac{1}{2}(2+\Delta_1-\Delta_2+\Delta_3-\Delta_4),\frac{1}{2}(2+\Delta_1-\Delta_2-\Delta_3+\Delta_4)\right) \\&\hspace{0cm} \times w^{\frac{2+\Delta_1-\Delta_2}{2}}\bar w^{\frac{2+\Delta_1-\Delta_2}{2}}\sum_{n=0}^\infty\frac{(\frac{1}{2}(2+\Delta_1-\Delta_2+\Delta_3-\Delta_4))_n(2-\Delta_2)_n(2-\Delta_2)_n(\frac{1}{2}(2+\Delta_1-\Delta_2-\Delta_3+\Delta_4))_n}{n!(1+n-\Delta_2+\Delta_1 )_n (2+\Delta_1-\Delta_2 )_{2n}}\\
		&\times w^{n}\bar{w}^{n}\,  {}_{2}{F}_1\left(\genfrac{}{}{0pt}{}{\frac{1}{2}(2+\Delta_1-\Delta_2+\Delta_3-\Delta_4)+n,2-\Delta_2+n}{\Delta_1-\Delta_2+2+2n};w\right)\\&
		\hspace{5
        cm}\times{}_{2}{F}_1\left(\genfrac{}{}{0pt}{}{\frac{1}{2}(2+\Delta_1-\Delta_2+\Delta_3-\Delta_4)+n,2-\Delta_2+n}{\Delta_1-\Delta_2+2+2n};\bar w\right).
        \end{split}
	\end{align}     
Now comparing \eqref{4.36k} with \eqref{5.33u} we get,
\begin{align}
   \Delta\pm 
 J=2+2n+\Delta_1-\Delta_2\,.
\end{align}
This is only possible when $J=0$ and immediately implies that the exchange operators have to be scalars.
\textcolor{black}{Therefore the $\textrm{OPE}_{12\mathcal{O}}\otimes \textrm{OPE}_{34\mathcal{O}}$ coefficient (for $J=0$) is given by},
\begin{align}
    \begin{split}
&f_{12\mathbfcal{O}}f_{34\mathbfcal{O}}\sim \delta_1(\alpha,\beta|\gamma) \,B\Big(\frac{1}{2}(2+\Delta_1-\Delta_2+\Delta_3-\Delta_4),\frac{1}{2}(2+\Delta_1-\Delta_2-\Delta_3+\Delta_4)\Big) \\ &\hspace{0 cm}\times\frac{(\frac{1}{2}(2+\Delta_1-\Delta_2+\Delta_3-\Delta_4))_n(2-\Delta_2)_n(2-\Delta_2)_n(\frac{1}{2}(2+\Delta_1-\Delta_2-\Delta_3+\Delta_4))_n}{  n!(1+n-\Delta_2+\Delta_1 )_n (2+\Delta_1-\Delta_2 )_{2n}},\,\,\,\\&\hspace{0.8cm}\textrm{with,\,\,}2n=\Delta-2-\Delta_1+\Delta_2.
    \end{split}
\end{align}
For simplicity we set the conformal dimensions of  the external primaries to be same $\Delta_{\mathcal{O}}$. Therefore the OPE coefficient reduces to (for non-spinning exchange),

\begin{tcolorbox}[thesisresultbox, title=OPE coefficient for EFT correction (scalar exchange)]
\restorethesisbodyformat
\begin{align}
    \begin{split}
    \label{4.46o}
f^2_{\mathcal{O}\mathcal{O}\mathbfcal{O}}=& \delta_1\Big(\alpha,\beta\Big|\textcolor{black}{\gamma\to4(\Delta_{\mathcal{O}}-1)}\Big) B\left(\Delta_{\mathcal{O}},\Delta_{\mathcal{O}}\right)\times\frac{\Gamma \left(\frac{\Delta }{2}\right)^4 \Gamma \left(2 \Delta _\mathcal{O}\right) \Gamma \left(\frac{\Delta }{2}+\Delta _\mathcal{O}-1\right)}{\Gamma (\Delta -1) \Gamma (\Delta ) \Gamma \left(\frac{\Delta }{2}-\Delta _\mathcal{O}+1\right) \Gamma \left(\Delta _\mathcal{O}\right){}^4}\,.
    \end{split}
\end{align}
\end{tcolorbox}
\restorethesisbodyformat

\vspace{0.5 cm}
\noindent
This naturally leads us to the question: what happens if we consider the exchange of spinning primaries, i.e., operators with non-zero spin ($J \ne 0$)?\footnote{It is not entirely clear to us why the Burchnall-Chaundy expansion of the four-point function fails to capture the complete spectrum of the theory.} To address this question, we make use of Caron-Huot's \textit{OPE inversion formula} \cite{Caron-Huot:2017vep}, which applies for  operators with finite, non-zero spin subject to the appropriate unitarity bounds. \textit{If one can demonstrate that the four-point function admits a consistent inversion yielding non-vanishing OPE coefficients for spinning operators, this would imply that the spectrum necessarily includes spinning exchanges. Thus, the problem reduces to examining whether such a consistent inversion is possible. We show that, in this case, the OPE inversion can indeed be performed consistently. Furthermore, we conduct a comparative study of the results obtained for $J=0$ using both approaches. }
\par
\subsection*{OPE Inversion:}
To start with, one needs the following integral representation. The discreteness in $\Delta$ should be converted into an integral form, sometimes known as partial-wave expansion \cite{Dolan:2003hv,Costa:2012cb} and takes the following form \cite{Caron-Huot:2017vep},
\begin{align}
\mathcal{G}^{12\to 34}(w,\bar w)=\mathbf{1}_{12}\mathbf{1}_{34}+\sum_{J=0}^\infty\rmint_{d/2-i\infty}^{d/2+i\infty}\frac{d\Delta}{2\pi i} \,\,c(J,\Delta)F_{J,\Delta}(w,\bar w)\label{5.35}
\end{align}
where, $c(J,\Delta)$ is the partial-wave coefficient and $F_{J,\Delta}$ is given by,
\begin{align}F_{J,\Delta}(w,\bar w)=\frac{1}{2}\Bigg(\textcolor{black}{G_{J,\Delta}(w,\bar w)}\hspace{0.2cm}+\underbrace{\textcolor{black}{\frac{K_{J,d-\Delta}}{K_{J,\Delta}}G_{J,d-\Delta}(w,\bar w)}}_{\text{Shadow contribution}}\Bigg)\end{align}
where the constants can be casted as,
\begin{align}
   & K_{J,\Delta}=\frac{\Gamma(\Delta-1)}{\Gamma(\Delta-d/2)}\kappa_{J+\Delta},\,\,\,\,\,\,\,\,\,\,\,\,\,\,\kappa_{\beta}=\frac{\Gamma(\beta/2-a)\Gamma(\beta/2+a)\Gamma(\beta/2-b)\Gamma(\beta/2+b)}{2\pi^2\Gamma(\beta-1)\Gamma(\beta)}\,,\\& \hspace{0.3cm}a=\frac{1}{2}(2-\Delta_2-\Delta_1)\,,\quad \quad\quad\quad
b=\frac{1}{2}(\Delta_3-\Delta_4)\nonumber
\end{align}
and $w,\bar w$ are the cross-ratio(s). $\Delta_i, \text{ for } \, i=1,\cdots,4$ are conformal dimension of primaries\footnote{Note that for our case we have three primaries and one shadowed primary.}. Under the assumption that  harmonic functions $F(J,\Delta)$ are orthogonal to each other, and Euclidean OPE data(s) can be obtained by inverting \eqref{5.35} in the following way \cite{Caron-Huot:2017vep} ,
\begin{align}
    c_{\texttt{s}}(J,\Delta)=N(J,\Delta)\rmint_{-\infty}^{\infty} d^2w\,\mu(w,\bar w)\,F_{J,\Delta}(w,\bar w)\,\mathcal{G}^{12\to 34}(w,\bar w)\label{4.46p}
\end{align}

where,
\begin{align}
    \begin{split}
  &      \mu(w,\bar w)=\left|\frac{w-\bar w}{w\bar w}\right| ^{d-2}\frac{(1-w)^{a+b}(1-\bar w)^{a+b}}{(w\bar w)^2},\,\textrm{with}\,\,a=\frac{2-\Delta_2-\Delta_1}{2},\,b=\frac{\Delta_3-\Delta_4}{2},\\ &
  G_{J,\Delta}=\frac{k_{\Delta-J}(w)k_{\Delta+J}(\bar w)+k_{\Delta+J}(w)k_{\Delta-J}(\bar w)}{1+\delta_{J,0}}, \textrm{\,\,with }\,\,k_{\beta}(w)=w^{\beta/2}\,{}_{2}\mathbf{F}_1\left(\beta/2+a,\beta/2+b,\beta,w\right)\,.\label{4.47y}
    \end{split}
\end{align}
Now, we can decompose the integral \eqref{4.46p} into three different channels, and as we are interested in the OPE limit, we make the following variable change $$w=\frac{4\rho_w}{(1+\rho_w)^2},\quad {\bar w}=\frac{4\rho_{\bar w}}{(1+\rho_{\bar w})^2}\,\quad |\rho_w|<1 $$ and focus on the $(0,1)$ region as we are interested in the $\texttt{s}$-channel OPE. According to the chosen notion of cross ratio, the $\texttt{s}$-channel OPE in Euclidean case dominates in this specific regime. So finally we get\footnote{We have used the relation $k_{\beta}(w,\bar w)\Bigg|_{a=0,b=0}=(4\rho)^{\beta/2}\,_2F_1(\frac{1}{2},\frac{\beta}{2},\frac{\beta+1}{2},\rho^2)\,.$},
\begin{align}
    \begin{split}
       c_{\texttt{s}}(J,\Delta)&=N(J,\Delta)\rmint_{0}^1\rmint_{0}^1 d\rho_w d\rho_{\bar w}\,\,\mu (\rho_w,\rho_{\bar w})F_{J,\Delta}(\rho_w,\rho_{\bar w}){\mathcal{G}}^{\mathfrak{s}}(\rho_w,\rho_{\bar w})\\&
       = \delta_1(\alpha,\beta|\gamma)N(J,\Delta)\rmint_{|\rho_w|\ll 1} d^2\rho_w \,\mu(\rho_w,\rho_{\bar w })\,\Bigg(\left(\frac{4\rho_w}{(1+\rho_w)^2}\right)^{\frac{\Delta-J}{2}}\left(\frac{4\rho_{\bar w }}{(1+\rho_{\bar w })^2}\right)^{\frac{\Delta+J}{2}}+\rho_w\leftrightarrow\rho_{\bar w}\Bigg)\\&\hspace{0.8cm}\times \mathbf{F}_{1}\left(\Delta_{\mathcal{O}},\Delta_{\mathcal{O}},2\Delta_{\mathcal{O}},\frac{4\rho_w}{(1+\rho_w)^2},\frac{4\rho_{\bar{w}}}{(1+\rho_{\bar{w}})^2}\right)\,,\\& 
        \approx \delta_1(\alpha,\beta|\gamma)\,N(J,\Delta)\textstyle{\rmint_{0}^1\rmint_{0}^1 \frac{d\rho_w d\rho_{\bar w}}{{16 \,\rho_w^2 \,\text{$\rho_{\bar w }$}^2}}{\Bigg(\rho_w ^{\frac{\Delta-J}{2}} \text{$\rho_{\bar w }$}^{\frac{\Delta+J}{2}}+\rho_w ^{\frac{\Delta-J}{2}} \text{$\rho_{\bar w }$}^{\frac{\Delta+J}{2}} \Bigg)\left(\frac{(\text{$\rho_{\bar w} $}+1)^2}{(\text{$\rho_{\bar w }$}-1)^2}\right)^{\Delta_{\mathcal{O}} } \left(1-\rho_w ^2\right) \left(1-\text{$\rho_{\bar w }$}^2\right)}},
        \\& =\delta_1(\alpha,\beta|\gamma) \frac{e^{-2\pi i\Delta_{\mathcal{O}}}N(J,\Delta)}{8 (\Delta +J-2)}\Gamma (1-2 \Delta_{\mathcal{O}} ) \Bigg[\Gamma \left(\frac{1}{2} (-J+\Delta -2)\right)\, \\&
 \hspace{0.8cm}\times_2\tilde{F}_1\left(\frac{1}{2} (-J+\Delta -2),-2 \Delta_{\mathcal{O}} ;\frac{1}{2} (-J+\Delta -4 \Delta_{\mathcal{O}} );-1\right)+J\leftrightarrow -J\Bigg],\\& \hspace{0.8cm} \textrm{with},\, \mathscr{R}[\Delta-J]>2, \textrm{and},\,\mathscr{R}[\Delta+J]>2\,.\label{4.55u}
    \end{split}
\end{align}
In evaluating \eqref{4.55u}, the integral convergence condition of  shadow part of the conformal block violets unitarity bound, hence can be dropped. In simplifying the integral at the Euclidean OPE limit (\textcolor{black}{$\rho_w,\rho_{\bar w}\ll1$}), we have approximated the hypergeometric function to be 1. Similarly we have approximated the whole $\text{Appell}\,\,\mathbf{F}_1$ function as \footnote{ The normalization factor $N(J,\Delta)$ in general dimension is given by,
\begin{align}
   N (J,\Delta)=\frac{4^\Delta \,
   \Gamma(J+\frac{d-2}{2})\Gamma(J+\frac{d}{2})K_{J,\Delta}}{ 2\pi\,\Gamma(J+1)\Gamma(J+d-2)K_{J,d-\Delta}}B\left(\frac{1}{2}(2+\Delta_1-\Delta_2+\Delta_3-\Delta_4),\frac{1}{2}(2+\Delta_1-\Delta_2-\Delta_3+\Delta_4)\right)\,.\nonumber
\end{align}},

\begin{align} \label{approx1}
\mathbf{F}_{1}\left(\Delta_{\mathcal{O}},\Delta_{\mathcal{O}},\Delta_{\mathcal{O}},2\Delta_{\mathcal{O}},\frac{4\rho_w}{(1+\rho_w)^2},\frac{4\rho_{\bar{w}}}{(1+\rho_{\bar{w}})^2}\right)\approx \Bigg(\frac{\rho_{\bar w}+1}{\rho_{\bar w}-1}\Bigg)^{2\Delta_{\mathcal{O}}}\,.
\end{align}

\par 
Now, one can extract the OPE coefficient from the partial-wave coefficient using the following observation that $c(J,\Delta)$ has poles at the real $\Delta$ axis at the location of the physical operators and consequently by computing the residues \cite{Liu:2020tpf,Caron-Huot:2017vep},
\begin{align}
    \begin{split}
        c(J,\Delta')\sim -\sum_{\Delta}\frac{f^2_{\mathcal{O}\mathcal{O}\mathbfcal{O}_{\Delta}}}{\Delta'-\Delta}\,.
    \end{split}
\end{align}
\noindent
In the complex \(\Delta'\)-plane, the integrand has several poles originating from various \(\Gamma\)-functions and potentially from the hypergeometric function as well. By analyzing \eqref{4.55u}, it becomes evident that, to ensure proper convergence of the integral, the contour must be closed on the right side of the \(\Delta'\)-plane i.e. $\textrm{Re}(\Delta')>1$. If we consider the shadow contribution of the block, the contour must be closed on the left. However, due to shadow symmetry, the OPE coefficients remain unchanged, as discussed in \cite{Caron-Huot:2017vep}. Accordingly, we evaluate the residues at the poles that lie in the right half of the complex \(\Delta'\)-plane. We now turn to the function of interest \eqref{4.55u}:
\begin{align}
    \begin{split}
      c_{\texttt{s}}(J,\Delta)=&\delta_1(\alpha,\beta|\gamma) B\left(\Delta_{\mathcal{O}},\Delta_{\mathcal{O}}\right)\Gamma(1-2\Delta_{\mathcal{O}})\\&\times\frac{\Gamma^4 \left(\frac{\Delta'+J}{2}\right)\Gamma(2-\Delta'+J-1)\Gamma(2-\Delta'+J)}{2\pi(\Delta'+J-2)\Gamma(\Delta'+J-1)\Gamma(\Delta'+J)\Gamma^4\left(\frac{2-\Delta'+J}{2}\right)}\\
&\times \Bigg[\Gamma\left(\frac{1}{2}(\Delta'-J-2)\right){_2}\tilde{F}_1\left(\frac{1}{2} (-J+\Delta' -2),-2 \Delta_{\mathcal{O}} ;\frac{1}{2} (-J+\Delta' -4 \Delta_{\mathcal{O}} );-1\right)+J\leftrightarrow-J \Bigg]\,.\label{4.36l}
    \end{split}
\end{align}
We analyze the singularities of \(c_{\mathfrak{s}}(J,\Delta)\) piece by piece. The function has three types of simple poles for \(\mathrm{Re}(\Delta') > 1\) \footnote{One can verify that \({_2}\tilde{F}_1\) does not introduce any poles for \(\mathrm{Re}\,\Delta' > 1\) by observing that, for \(\Delta_{\mathcal{O}} = 1\), it behaves as,
${_2}\tilde{F}_1 \sim -\frac{2 (-\Delta + J + 2)}{\Gamma \left( \frac{1}{2} (-J + \Delta - 2) \right)}$,
which is manifestly analytic in this region. Upon analytically continuing \(\Delta_{\mathcal{O}}\) into the entire complex plane while keeping its real part fixed, no additional pole structure is expected to emerge.
}:
\begin{itemize}
    \item  An infinite tower of simple poles at \(\Delta' = n+J+1\),
\item Another infinite tower of simple poles at \(\Delta' =n+J+2\),
\item There is a single simple pole at \(\Delta' = 2 - J\). However, based on the convergence condition of the integral in \eqref{4.55u}, this pole lies outside the region of convergence and can therefore be excluded from the residue analysis.
\end{itemize}
The corresponding residues are:\\
\textbullet $\,\,$ at $\Delta' = n+J+2:$
\begingroup\tiny
\begin{align}
&\mathop{\mathbf{Res}}_{\Delta' = n+J+2}c_{\texttt{s}}(J,\Delta')
=\frac{e^{-2 i \pi  \Delta_{\mathcal{O}}}\delta_1(\alpha,\beta|\gamma) B\left(\Delta_{\mathcal{O}},\Delta_{\mathcal{O}}\right)\Gamma (1-2 \Delta_{\mathcal{O}}) 2^{2 J+2 n} \Gamma \left(J+\frac{n}{2}+1\right)^4}{ n^3 (2 J+n)^3 \Gamma \left(-\frac{n}{2}\right)^4 \Gamma (n) \Gamma (n+2) \Gamma (2 J+n) \Gamma (2 J+n+2) \Gamma \left(\frac{n}{2}-2 \Delta_{\mathcal{O}}+1\right) \Gamma \left(J+\frac{n}{2}-2 \Delta_{\mathcal{O}}+1\right)}\notag\\
&\quad
 \times\Bigg[-2 n (2 J+n) \Gamma \left(J+\frac{n}{2}+1\right) \Gamma \left(\frac{n}{2}-2 \Delta_{\mathcal{O}}+1\right) \frac{ab}{c^2}{_2}\Theta_1^{(1)} \left( 
\begin{array}{c} 
1, 1 : c, a + 1; b + 1 \\
c + 1 : 2; c + 1 
\end{array} 
; -1, -1 \right)\notag\\
&\quad +2 n (2 J+n) \Bigg(-\Gamma \left(\frac{n}{2}+1\right) \left(\frac{ab}{c^2}{_2}\Theta_1^{(1)} \left( 
\begin{array}{c} 
1, 1 : c, a + 1; b + 1 \\
c + 1 : 2; c + 1 
\end{array} 
; -1, -1 \right)-\frac{a}{c}\Theta_1^{(1)} \left( 
\begin{array}{c} 
1, 1 : a, a + 1; b + 1 \\
a + 1 : 2; c + 1 
\end{array} 
; -1, -1 \right)\right)\notag\\
&\quad \times\Gamma \left(J+\frac{n}{2}-2 \Delta_{\mathcal{O}}+1\right)-\frac{a}{c}\,\Gamma \left(J+\frac{n}{2}+1\right) \Gamma \left(\frac{n}{2}+2 \Delta_{\mathcal{O}}+1\right) \Theta_1^{(1)} \left( 
\begin{array}{c} 
1, 1 : a, a + 1; b + 1 \\
a + 1 : 2; c + 1 
\end{array} 
; -1, -1 \right)\Bigg)\notag\\
&\quad+n \Big(2 \Gamma \left(J+\frac{n}{2}+1\right) \Gamma \left(\frac{n}{2}-2 \Delta_{\mathcal{O}}+1\right) \, _2F_1\left(J+\frac{n}{2},-2 \Delta_{\mathcal{O}};J+\frac{n}{2}-2 \Delta_{\mathcal{O}}+1;-1\right)\notag\\
&\qquad \times\left((2 J+n) \left(H_{J+\frac{n}{2}-2 \Delta_{\mathcal{O}}}-5 H_{J+\frac{n}{2}-1}+4 H_{2 J+n}-4 H_{-\frac{n}{2}-1}+4 H_n-4 \log (2)\right)+\frac{4 J-2}{n+1}-\frac{2}{2 J+n+1}-4\right)\notag\\
&\qquad +n \Gamma \left(\frac{n}{2}\right) \, _2F_1\left(\frac{n}{2},-2 \Delta_{\mathcal{O}};\frac{1}{2} (n-4 \Delta_{\mathcal{O}}+2);-1\right) \Gamma \left(J+\frac{n}{2}-2 \Delta_{\mathcal{O}}+1\right)\notag\\
&\qquad \times\Big(-(2 J+n) \left(4 H_{J+\frac{n}{2}-1}-4 H_{2 J+n}-H_{\frac{n}{2}-2 \Delta_{\mathcal{O}}}+4 H_{-\frac{n}{2}-1}-4 H_n+\psi ^{(0)}\left(\frac{n}{2}\right)+\gamma_E \right)+\frac{4 J-2}{n+1}-\frac{2}{2 J+n+1}\notag\\
&\qquad\qquad -4 \log (2) (2 J+n)-2\Big)\Big)+4 (2 J+n) \Gamma \left(J+\frac{n}{2}+1\right) \Gamma \left(\frac{n}{2}-2 \Delta_{\mathcal{O}}+1\right) \, _2F_1\left(J+\frac{n}{2},-2 \Delta_{\mathcal{O}};J+\frac{n}{2}-2 \Delta_{\mathcal{O}}+1;-1\right)\Bigg]\label{4.62i}
\end{align}
\endgroup
\textbullet $\,\,$ at $\Delta' = n+J+1:$
\begingroup\small
\begin{align}
&\mathop{\mathbf{Res}}_{\Delta' = n+J+1}c_{\texttt{s}}(J,\Delta')
=\delta_1(\alpha,\beta|\gamma) B\left(\Delta_{\mathcal{O}},\Delta_{\mathcal{O}}\right)\notag\\
&\quad\times\frac{1}{(n-1) n! (2 J+n-1) \Gamma \left(\frac{1-n}{2}\right)^4 \Gamma (2 J+n) \Gamma (2 J+n+1)}\notag\\
&\quad\times \Big[e^{-2 i \pi  \Delta_{\mathcal{O}}} (-1)^n \Gamma (1-2 \Delta_{\mathcal{O}}) 4^{J+n} \Gamma (1-n) \Gamma \left(J+\frac{n}{2}+\frac{1}{2}\right)^4 \Big(\Gamma \left(J+\frac{n}{2}+\frac{1}{2}\right)\notag\\
&\qquad\times\, _2\tilde{F}_1\left(\frac{1}{2} (2 J+n-1),-2 \Delta_{\mathcal{O}};\frac{1}{2} (2 J+n-4 \Delta_{\mathcal{O}}+1);-1\right)+\Gamma \left(\frac{n+1}{2}\right) \, _2\tilde{F}_1\left(\frac{n-1}{2},-2 \Delta_{\mathcal{O}};\frac{1}{2} (n-4 \Delta_{\mathcal{O}}+1);-1\right)\Big)\Big]\notag\\
&\quad \,.\label{4.63e1}
\end{align}
\endgroup
Here,  \(H_n = \rmint_{0}^1 dx\,\frac{1 - x^n}{1 - x}\) denotes the harmonic number, and \(\psi^{(0)}\) is the digamma function, which is the logarithmic derivative of the gamma function. Moreover, we write the  derivatives of the hypergeometric functions (appeared in the calculation) can be casted in terms of \textit{Kampé de Fériet-like functions} \cite{Ancarani:2009zz}\footnote{Specifically, $
   {_2}\Theta_1^{(1)} \left( 
\begin{matrix}
a_1, a_2 \; : \; b_1, b_2; b_3 \\
c_1 \; : \; d_1; d_2
\end{matrix}
; x_1, x_2 \right)
= 
\sum_{m_1=0}^{\infty} \sum_{m_2=0}^{\infty} 
\frac{
(a_1)_{m_1} (a_2)_{m_2} (b_1)_{m_1} (b_2)_{m_1 + m_2} (b_3)_{m_1 + m_2}
}{
(c_1)_{m_1} (d_1)_{m_1 + m_2} (d_2)_{m_1 + m_2}
}
\frac{x_1^{m_1} x_2^{m_2}}{m_1! m_2!}\,.\nonumber
$}.
\begin{align}
    \begin{split}    
 &{_2} F_1^{(1,0,0,0)}(a,b,c,-1)=-\frac{a}{c}\Theta_1^{(1)} \left( 
\begin{array}{c} 
1, 1 : a, a + 1; b + 1 \\
a + 1 : 2; c + 1 
\end{array} 
; -1, -1 \right), \\&   \nonumber 
\end{split} \\[0.5em]
\begin{split}
   &  {_2} F_1^{(0,0,1,0)}(a,b,c,-1)=\frac{ab}{c^2}{_2}\Theta_1^{(1)} \left( 
\begin{array}{c} 
1, 1 : c, a + 1; b + 1 \\
c + 1 : 2; c + 1 
\end{array} 
; -1, -1 \right), \\& \hspace{2 cm}\textrm{with}\,\,a=\frac{n}{2}, \,b=-2\Delta_{\mathcal{O}},\,\textcolor{black}{c=\frac{n}{2}-2\Delta_{\mathcal{O}}+1}\,. \label{4.63e}
\end{split}
\end{align}
Collecting all the results from \eqref{4.62i} and \eqref{4.63e1} we find the OPE coefficient to be,

\begin{tcolorbox}[thesisresultbox, title=OPE coefficient for EFT correction (general exchange)]
\restorethesisbodyformat
\begin{align*}
&f^2_{\Delta_\mathcal{O}\Delta_\mathcal{O}\Delta} 
=-\mathop{\mathbf{Res}}_{\Delta' = n+J+2}c_{\texttt{s}}(J,\Delta')\Big|_{n=\Delta-J-2}-\mathop{\mathbf{Res}}_{\Delta' = n+J+1}c_{\texttt{s}}(J,\Delta')\Big|_{n=\Delta-J-1}.\label{4.39}
\end{align*}
\end{tcolorbox}
\restorethesisbodyformat
\vspace{0.5cm}
\textbf{\textit{A comparison of OPE computed from BC expansion and OPE inversion formula}:}
\begin{figure} [htb!]
    \centering
\includegraphics[width=0.60\linewidth]{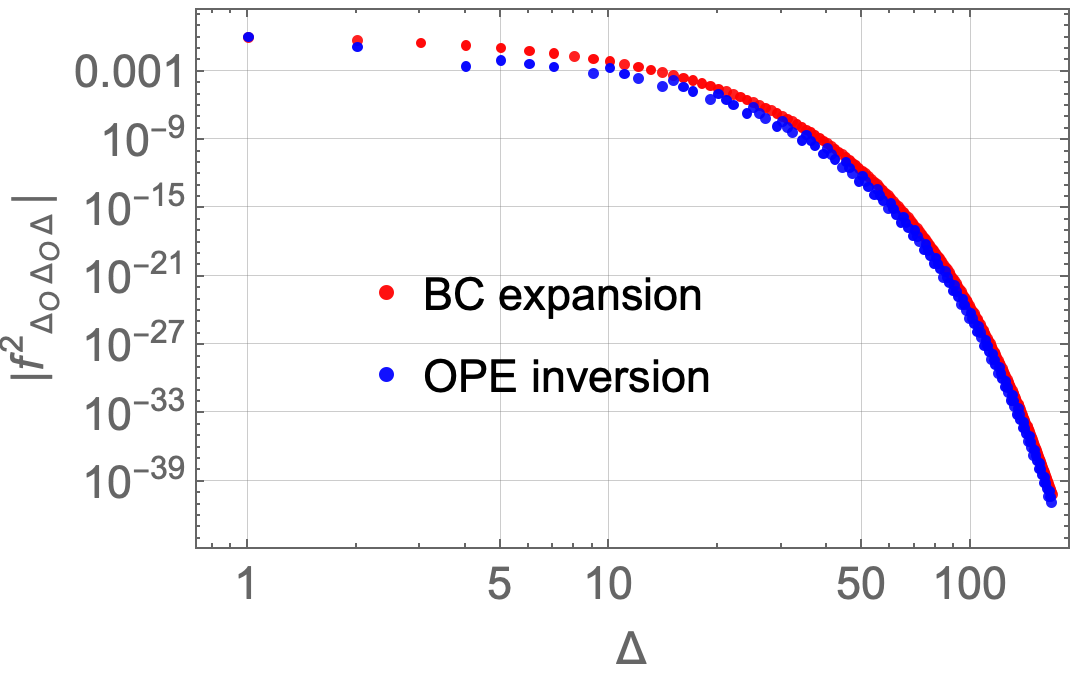}
    \caption{Plot depicting the matching for extraction of OPE coefficient using BC expansion (\textbf{red}) and OPE inversion for spin-0 (\textbf{blue}). We have set the conformal dimension for the external primaries to be: $\Delta_{\mathcal{O}}=1+i$.}
    \label{fig5f}
\end{figure}
Now we present a comparative analysis of the s-channel OPE, which offers valuable insights into the interactions among conformal primary operators on the celestial sphere. In  Fig.~\eqref{fig5f}, we present a comparative study of extracting the OPE coefficients (taking the external conformal dimension to be $\Delta_{\mathcal{O}}=1+i$) from the two approaches: Burchnall-Chaundy (BC) expansion and OPE inversion.
While the OPE coefficients derived via the BC expansion show agreement with those obtained through the OPE inversion formula for $J = 0$ and in the vicinity of: $\textrm{Im}(\Delta_{\mathcal{O}}) \gtrsim 1$, discrepancies begin to emerge at higher values of $\textrm{Im}(\Delta_{\mathcal{O}})$, leaving no major structural differences. These deviations can be attributed to the limit $\rho_w \ll 1$, which approximates the Appell and hypergeometric functions encoding the $\Delta_{\mathcal{O}}$ dependence as shown in \eqref{approx1}. Due to the technical complexity of performing an exact OPE inversion without such approximations, we currently work within a simplified framework. However, it is anticipated that a more complete evaluation of the OPE inversion, incorporating the full integration domain, may yield results that are more consistent with those from the BC expansion. The exact evaluation of OPE coefficients using the inversion formula requires a separate study, which we leave for further investigation.

This calculation demonstrates a way for extracting OPE data from the four-point function within a suitable approximation scheme. Therefore, the answer to the third question raised in the introduction is also affirmative. Once analytic control over the OPE data is established for the Born amplitude, we immediately gain control over the full non-perturbative eikonal celestial amplitude, at least for the EFT correction part, since the $\omega$-integral in \eqref{4.6j} is independent of the cross-ratio $z$. Consequently, the same analysis performed for the Born amplitude case can be applied, with the $\omega$-integral treated as a formal object. However, for the GR part in \eqref{4.6l}, the functional dependence of the shadow amplitude on $z$ is highly sensitive to the $\omega$-integral, for which no closed-form solution is available. As a result, we lose entirely the analytic control over the OPE data for the GR part of the eikonal shadow correlator.

\section{Connection to Carrollian amplitude}
\label{ch5:sec5}
A significant bridge between celestial and Carrollian frameworks is established through the so-called $\mathcal{B}$-transform \cite{Donnay:2022aba,Donnay:2022wvx}. This transformation maps celestial amplitudes defined via Mellin transforms of scattering amplitudes in momentum space into Carrollian amplitudes, which are naturally formulated in a spacetime with Carrollian symmetry arising in the ultra-relativistic limit. The $\mathcal{B}$-transform acts as a change of basis, translating between representations adapted to conformal structures at null infinity and those suited to Carrollian dynamics. Notably, the transformation preserves the essential symmetry content of the theory, shedding light on the interplay between conformal covariance on the celestial sphere and Carrollian symmetries on null hypersurfaces \cite{Donnay:2022aba,Bagchi:2022emh,Donnay:2022wvx}. Now we can compute the unmodified Carrollian amplitude from the celestial amplitude as,
\begin{align}
\begin{split}
   \mathbfcal{C}(u_i,z_i,\bar z_i)&\sim\rmint_0^\infty \prod_id\omega_i e^{-i\sum\epsilon_j\omega_j u_j}\mathbb{M}_{\textrm{eik}}(\omega_i,z_i)\,,\\&
    \sim\prod_{i=1}^4 \rmint_0^1 d\sigma_i \delta{\Bigg(\sum_i\sigma_i-1\Bigg)}[\cdots]\rmint_0^\infty dv \,\textcolor{black}{v^{-1}}e^{-iv^{-1}(\sigma_1u_1+\sigma_2u_2-\sigma_3u_3-\sigma_4u_4)}\mathbb{M}_{eik}(v,z)\,,\\&
    \sim \prod_{i=1}^4 \rmint_0^1 d\sigma_i \delta{\Bigg(\sum_i\sigma_i-1\Bigg)}[\cdots]\Bigg(\rmint_0^\infty dv \,\textcolor{black}{v^{-1}}e^{-iv^{-1}(\sigma_1u_1+\sigma_2u_2-\sigma_3u_3-\sigma_4u_4)}\\&\times\Bigg[ \frac{1}{48\kappa} v^2 \left(\frac{\kappa \left(3 \kappa-v^2 (-\alpha +8 \beta )\right)}{\left(\kappa-\alpha  v^2\right) \left(\kappa-2 \beta  v^2\right)}-\frac{4 \kappa}{\kappa+\alpha  v^2}+\frac{\kappa}{\kappa+2 \beta  v^2}-\frac{24 z}{(z-1)}\right)\Bigg]\Bigg)\,.
       \end{split}
\end{align}
\textcolor{black}{In obtaining the second line from the first one we used the simplex-variable decomposition of the Mellin measure discussed in Appendix~\ref{ch5:app:A}.} Now, changing the variable $v^{-1}=p$ and $(\sigma_1u_1+\sigma_2u_2-\sigma_3u_3-\sigma_4u_4)=h$, we can write the integrand in the parenthesis as,
\begin{align}
\begin{split}&\label{5.2w}
    \tilde{\mathbfcal{C}}(\sigma_i,z_i,\bar z_i)=\textcolor{black}{\frac{h^8z}{{22579200 \sqrt{2 \pi }\,\kappa (z-1)}} \left(-761+280 \Bigg(-\frac{1}{2} (i \pi ) \text{ sign}(h)+\log (h)+\gamma_E \right)\Bigg)}\\&\hspace{1 cm}-(-\beta+2\alpha)\textcolor{black}{\frac{h^{10}}{{109734912000 \sqrt{2 \pi } \,\,\kappa^2 }} \Bigg(-7381+2520 \left(-\frac{1}{2} (i \pi ) \text{ sign}(h)+\log (h)+\gamma_E\right)\Bigg)}
    \end{split}
\end{align}
where $\gamma_E$ is the Euler's constant. The first and second terms in \eqref{5.2w} come from the GR and the first-order correction to GR in quadratic EFT, respectively.  Now, performing the $\sigma$ integral, one can easily find out the Carrollian amplitude corresponding to the celestial amplitude as,
\begin{align}
   \mathbfcal{C}(u_i,z_i,\bar z_i)\sim \prod_{i=1}^4 \rmint_0^1\,d\sigma_i \delta(\sigma_i  - \sigma_{*i})\,\tilde{\mathbfcal{C}}(\sigma_i,z_i,\bar z_i)\,,
\end{align}
where the localization point $\sigma_{i\star}$ can be found in \eqref{5.17u}. \par
One important point to note here is that the Carrollian amplitude has an IR pole in GR, which now shifts due to the non-zero value of $\alpha,\beta$. However, we have found this result by linearizing in $\alpha,\beta$. The IR-pole behaviour is given by,
\begin{align}
  \lim_{\delta\to 0^{+}} \frac{1}{\delta} \frac{h^8(z-1) \left(h^2\frac{ (z-1)}{z} (-\beta +2 \alpha )+540 \kappa\right)}{43545600 \sqrt{2 \pi }   \kappa z}.
\end{align}
Here, the infrared pole is important for boost invariance \cite{Donnay:2022wvx}.
\section{Conclusion and Discussion}\label{ch5:sec7}
Motivated by the case study of celestial eikonal amplitudes and their improved analytic behaviour, we have generalized the study of it for Einstein gravity to quadratic EFT. Below, we list the main findings of our paper,
\begin{enumerate}
\item Motivated by the analysis of \cite{Adamo:2024mqn}, we construct the celestial eikonal amplitude for the quadratic EFT of gravity, despite the fact that the Born amplitude in this case is meromorphic, unlike in GR. We find that the corrections to the eikonal phase arising from the EFT are short-ranged, involving \(\delta\)-function contributions. Nevertheless, by adopting a suitable prescription for handling functions of the \(\delta\)-function, we demonstrate that it is possible to extract physically meaningful results even in the presence of such contact interactions. We find that like GR the eikonal amplitude is multiplication of two parts: Born amplitude and a phase.
\item Furthermore, we analyze the analytic structure of the eikonal amplitude by examining its behavior in both the ultraviolet (UV) and infrared (IR) regimes. In the UV limit, we find that the amplitude exhibits no structural differences compared to GR. However, upon numerical checking we find the EFT correction part converges faster than the GR part. But in the IR regime, the leading singularity is identified as a simple pole, in contrast to GR, where the leading singularity corresponds to an \(n\)-th order pole. Thus, the infrared behavior appears to improve upon the inclusion of EFT corrections. Although we are unable to compute the \(\omega\) integral exactly, we derive the corresponding dispersion relation. This dispersion relation is modified due to the presence of non-vanishing coupling constants \((\alpha, \beta)\). Contributions from poles on the real axis are absent, as they are excluded by the choice of integration contour, which is essential to ensure that the limit \((\alpha, \beta) \to 0\) correctly.

\item  We also compute the celestial operator product expansion (OPE) from a four-point function involving three primary operators and one shadow operator, by expressing the correlator in the basis of \textit{conformal primary wavefunctions}. Upon evaluating the shadowed four-point function, we obtain an Appell function, which we then decompose into a sum of products of hypergeometric functions using the Burchnall-Chaundy expansion. In this process, we identify the celestial conformal blocks and express them in terms of hypergeometric functions. Remarkably, the underlying symmetries fix the conformal dimensions of the exchanged operators in a manner consistent with the Osborn block expansion, allowing the remaining factors to be identified as OPE coefficients.
While such coefficients can alternatively be extracted from the collinear limit of scattering amplitudes, yielding only the leading OPE behaviour; our computation proceeds without invoking this limit. Surprisingly, OPE coefficient obtained using Burchnall-Chaundy 
 does not involve contribution from the spinning exchange. To get the contribution from spin,  we employ the (Euclidean) OPE inversion formula to extract the full set of OPE coefficients. Furthermore, we provide a comparison of OPE coefficients extracted from the two ways mentioned above.
\end{enumerate}
\noindent
Now, we end this section by discussing some possible future outlooks. The first objective is to compute the OPE inversion exactly, without resorting to approximations, for both the GR component and the EFT-corrected component. This task presents a significant challenge, particularly for the GR contribution, as it requires the exact computation of the eikonal amplitude. While the EFT corrections generally allow for a more straightforward  (as the kinematic part factors out form the eikonal amplitude) inversion, the GR component remains nontrivial due to the inherent complexity of exact eikonal amplitude calculations in this context. Moreover, due to the infrared triangle between soft theorems, ward identities, and memory effects, as proposed by Strominger et al. \cite{Strominger:2017zoo,Pate:2017fgt}, it is an open arena for investigating ward identities in quadratic EFT. One can try to find the memory effects in this setup. One can also attempt to compute soft theorems and central charges in this case. \textcolor{black}{In effective field theories (EFTs), gravitons typically possess at least one massive mode. Consequently, when attempting to construct the stress-energy tensor within the celestial framework, one must invoke the shadow transform of the massless graviton mode. However, the presence of residual massive  degrees of freedom complicates this process, rendering the stress tensor's construction nontrivial and subtle}. Last but not the least, One can also consider light transforms in celestial CFT for the quadratic gravity and focus on marginal operator construction \cite{Banerjee:2022hgc,Narayanan:2024qgb,Banerjee:2024hvb}.

\appendix
\section{Few definitions and conventions}\label{ch5:app:A}
\textbullet \,\,Hypergeometric ${}_2\mathbf{F}_1$ satisfies the following identities,

\begin{align}
\begin{split}
   &{}_2\mathbf{F}_1\left(\genfrac{}{}{0pt}{}{a,b}{c};x\right) =(1-x)^{c-a-b} {}_2\mathbf{F}_1\left(\genfrac{}{}{0pt}{}{c-a,c-b}{c};x\right)\,,\\&
    {}_2\mathbf{F}_1\left(\genfrac{}{}{0pt}{}{a,b}{c-1};x\right) =\sum_{m=0}^\infty \frac{(a)_m (b)_m}{(c-1)_{2m}}x^m {}_2\mathbf{F}_1\left(\genfrac{}{}{0pt}{}{a+m,b+m}{c+2m};x\right)\,.
\end{split}
\end{align}

\noindent
\textbullet \,The Appell hypergeometric function $\mathbf{F}_1$ has the  analytic continuation as follows,
	\begin{align}
		\label{eq:F1toOne}
		\begin{aligned}
			& \mathbf{F}_1\left(a, b_1, b_2, c, x, y\right)=\frac{\Gamma(c) \Gamma\left(c-a-b_1-b_2\right)}{\Gamma(c-a) \Gamma\left(c-b_1-b_2\right)}\mathbf{F}_1\left(a, b_1, b_2, 1+a+b_1+b_2-c, 1-x, 1-y\right) \\
			& \quad+\frac{\Gamma(c) \Gamma\left(a+b_2-c\right)}{\Gamma(a) \Gamma\left(b_2\right)}(1-x)^{-b_1}(1-y)^{c-a-b_2} \mathbf{F}_1\left(c-a, b_1, c-b_1-b_2, c-a-b_2+1, \frac{1-y}{1-x}, 1-y\right) \\
			& \quad+\frac{\Gamma(c) \Gamma\left(c-a-b_2\right) \Gamma\left(a+b_1+b_2-c\right)}{\Gamma(a) \Gamma\left(b_1\right) \Gamma(c-a)}(1-x)^{c-a-b_1-b_2} \\
			& \quad \times G_2\left(c-b_1-b_2, b_2, a+b_1+b_2-c, c-a-b_2, x-1, \frac{1-y}{x-1}\right),
		\end{aligned}
	\end{align}
	and 
	\begin{align}
		\label{eq:F1toInf}
		\begin{aligned}
			& \mathbf{F}_1\left(a, b_1, b_2, c, x, y\right)=\frac{\Gamma(c) \Gamma\left(a-b_1-b_2\right)}{\Gamma(a) \Gamma\left(c-b_1-b_2\right)}(-x)^{-b_1}(-y)^{-b_2} \mathbf{F}_1\left(1+b_1+b_2-c, b_1, b_2, 1+b_1+b_2-a, \frac{1}{x}, \frac{1}{y}\right)\\
			& \quad+ \frac{\Gamma(c) \Gamma\left(b_2-a\right)}{\Gamma\left(b_2\right) \Gamma(c-a)}(-y)^{-a} \mathbf{F}_1\left(a, b_1, 1+a-c, 1+a-b_2, \frac{x}{y}, \frac{1}{y}\right) \\
			& \quad+\frac{\Gamma(c) \Gamma\left(a-b_2\right) \Gamma\left(b_1+b_2-a\right)}{\Gamma(a) \Gamma\left(b_1\right) \Gamma(c-a)}(-x)^{b_2-a}(-y)^{-b_2} G_2\left(1+a-c, b_2, b_1+b_2-a, a-b_2,-\frac{1}{x}, -\frac{x}{y}\right).
		\end{aligned}
	\end{align}

\vspace{-0.3 cm}

\noindent
\textbullet \textbf{\, Evaluating the delta function:}
\noindent
The momentum conserving delta functions are always of much importance here because they put strong constraints on the celestial correlators. 
For non-vanishing values of $\alpha$ and $\beta $ we get the tree amplitude as,

\begin{align}
\mathbb{M}_4^{\textrm{tree}}(s,t)\to\frac{1}{48} s \left(\frac{ (3 \kappa-\alpha  s+8 \beta  s)}{(\kappa+\alpha  s) (\kappa+2 \beta  s)}-\frac{4 }{\kappa-\alpha  s}+\frac{1}{\kappa-2 \beta  s}+\frac{24 \,s }{t \kappa}\right)+\mathcal{O}\left(\frac{t}{s}\right)+\cdots
\end{align}

\noindent
The momentum conserving delta function can also be written in terms of the simplex variables \cite{Pasterski:2017ylz} as $\sigma_i=v^{-1}\omega_i  \,\text{with}\,\, \sum_{i=1}^n\sigma_i=1
$,
\begin{align}
  \prod_{i=1}^n\rmint_0^\infty d\omega_i\, \omega_i^{i\lambda_i}[\cdots]=\rmint_0^\infty dv \,v^{\sum i\lambda_i-1}\prod_{i=1}^n\rmint_0^1\,d\sigma_i \sigma_i^{i\lambda_i}\delta^{(4)}\left(\sum_{i=1}^4\epsilon_i\sigma_iq_i\right)\,\delta\left(\sum_{i=1}^4\sigma_{i}-1\right)[\cdots]\,.
\end{align}
We can cast the delta function as \cite{Pasterski:2017ylz},
\begin{align}
\begin{split}\label{5.17u}
&\delta^{(4)}\left(\sum_{i=1}^4\epsilon_i\sigma_iq_i\right)\,\delta\left(\sum_{i=1}^4\sigma_{i}-1\right)
= \frac{1}{4}\delta(|z_{12}z_{34}\bar z_{13}\bar z_{24}-z_{13}z_{24}\bar z_{12}\bar z_{34}|)
\\
&\times\delta\left(\sigma_1+
\frac{\epsilon_1\epsilon_4}{ \mathcal{D}_4 }
	\frac{z_{24}\bar z_{34}}{z_{12}\bar z_{13}}\right)
\delta\left(\sigma_2-
\frac{\epsilon_2\epsilon_4}{ \mathcal{D}_4}
	\frac{z_{34}\bar z_{14}}{z_{23}\bar z_{12}}
\right)\delta\left(\sigma_3+\frac{\epsilon_3\epsilon_4}{ \mathcal{D}_4}
	\frac{z_{24}\bar z_{14}}{z_{23}\bar z_{13}}\right)
	\delta\left(\sigma_4-\frac{1}{\mathcal{D}_4}\right)\,, \\
&\equiv \frac{1}{4}\delta(|z_{12}z_{34}\bar z_{13}\bar z_{24}-z_{13}z_{24}\bar z_{12}\bar z_{34}|)
\prod_{i=1}^4 \delta\left(\sigma_i  - \sigma_{\star i}\right)
\end{split}
\end{align}
where the denominator $\mathcal{D}_4$ is defined as,
\begin{align}
\mathcal{D}_4 = \left( 1-{\epsilon_1\epsilon_4}\right)
	\frac{z_{24} \bar z_{34}}{z_{12}\bar z_{13}}
+\left( {\epsilon_2\epsilon_4}-1\right)
	\frac{z_{34} \bar z_{14}}{z_{23} \bar z_{12}}
+\left(1-{\epsilon_3\epsilon_4}\right)
	\frac{z_{24}\bar z_{14}}{z_{23}\bar z_{13}} \,.
\end{align}
In the above equation, $\sigma_{\star i}$ are the supports of the $\sigma$ integral, using localized Dirac delta functions. 
\noindent
On the support of the delta functions, the Mandelstam variables simplify to $s=v^2, t=-zv^2.$ This parametrization is valid for massless particles as external legs.
Hence the $\sigma_i$ integral becomes,
\begin{align}
\begin{split}
    &\rmint_0^1\,d\sigma_i \sigma_i^{i\lambda_i}\delta{\Bigg(\sum_i\sigma_i-1\Bigg)}\delta^{(4)}(\sum_{i=1}^4\epsilon_i\sigma_iq_i)\,,\\&
    =\frac{1}{4}\delta(|z_{12}z_{34}\bar z_{13}\bar z_{24}-z_{13}z_{24}\bar z_{12}\bar z_{34}|)
\prod_{i=1}^4 \rmint_0^1\,d\sigma_i \sigma_i^{i\lambda_i}\delta\Bigg(\sigma_i  - \sigma_{*i}\Bigg)\,,\\&
\sim (z-1)^{\frac{\Delta_1-\Delta_2-\Delta_3+\Delta_4}{2}}|z|^{-\Delta_1-\Delta_2}\frac{\delta(|z-\bar z|)}{|z_{13}|^2|z_{24}|^2}\prod_{i=1}^4\underbrace{\mathbf{1}_{[0,1]}(\sigma_{*i})}_{\textrm{indicator function}}\,,
\end{split}
\end{align}
where, the indicator function ensures all the $\sigma_{i*}$ are between 0 to 1.

\begin{equation}    \mathbf{1}_{[0,1]}(\sigma_{*i})=\left\{ 
  \begin{array}{ c l }
    1 & \quad \textrm{if } \sigma_{*i} \in[0,1] \\
    0                 & \quad \textrm{otherwise}.
  \end{array}\right.\end{equation}
    }%
  }%

  \restorethesisbodyformat
  \restoremainchapterstyle
  \chapter{Conclusions}
  \thesischapterpaperbox{}{}{}
  {%
    \restorethesisbodyformat
    \renewcommand{\appendix}{%
      \setcounter{section}{0}%
      \setcounter{subsection}{0}%
      \setcounter{subsubsection}{0}%
      \renewcommand{\thesection}{\thechapter.\Alph{section}}%
      \renewcommand{\thesubsection}{\thesection.\arabic{subsection}}%
      \renewcommand{\theHsection}{chapter.\arabic{chapter}.appendix.\Alph{section}}%
      \renewcommand{\theHsubsection}{\theHsection.\arabic{subsection}}%
    }%
    \ifstrempty{}{}{%
      \input{}%
    }%
  }%

This thesis has investigated how modern quantum field theoretic methods can be brought to bear on the classical two-body problem in effective theories of gravity beyond General Relativity. While gravitational-wave physics, black-hole dynamics, and scattering amplitudes are often treated as distinct subjects, one of the central themes of this work is that they are deeply interconnected at a structural level. The conservative and radiative dynamics of compact binaries, the eikonal description of high-energy scattering, worldline formalisms, and amplitude-based methods do not merely coexist; rather, they illuminate and reinforce one another. From this viewpoint, classical gravitational observables should not be regarded as isolated quantities, but as different manifestations of a broader mathematical and physical framework, one that can be efficiently accessed using techniques originally developed in quantum field theory. In this regard, the seminal work of Amati, Ciafaloni, and Veneziano~\cite{Amati:1990xe}, together with later developments by Kabat and Ortiz~\cite{Kabat:1992tb} on eikonal scattering and amplitude re-summation, represents a major conceptual advance. These ideas have proved foundational not only for our understanding of gravitational scattering and its relation to gravitational-wave physics, but also for seemingly distant subjects such as celestial amplitudes.

The first part of the thesis focused on inspiraling black-hole binaries in the presence of additional light degrees of freedom motivated by dark-sector physics. Using worldline effective field theory, we studied the conservative and radiative dynamics of non-spinning binaries coupled to axion-like particles, dark photons, and their interactions with the electromagnetic sector. In this setup, we computed the conservative dynamics in the relevant post-Newtonian regimes and analyzed the corresponding radiation channels. A notable feature of this analysis is that the extra couplings do not merely shift known General Relativistic results; rather, they generate genuinely new structures in both the bound dynamics and the emitted radiation. In particular, the parity-violating axion-photon interaction leads to contributions that vanish for planar orbits but can become non-trivial for genuinely three-dimensional motion. This observation is conceptually important, since it shows that beyond-GR effects can be strongly tied to the geometry of the orbit itself. More broadly, this chapter demonstrates that effective field theory provides a systematic language in which one can isolate possible dark-sector imprints in gravitational-wave observables and organize them in a way suitable for future phenomenological constraints.

We then turned to the scattering regime and studied classical black-hole scattering in scalar-tensor gravity using worldline quantum field theory. Here the goal was to compute directly the classically relevant observables associated with non-spinning compact objects in the presence of an extra scalar degree of freedom. Within thi and e obtained the impulse and waveform up to second post-Minkowskian order, carefully analyzing both the massless and massive scalar cases. An important outcome of this study is that the massive scalar contributions exhibit smooth massless limits, providing a strong consistency check on the formalism and on the treatment of the relevant integrals. We also showed that the waveform problem becomes substantially more intricate when the scalar is massive, and we proposed a stationary phase treatment as a practical analytic tool in that regime. This chapter therefore extends the reach of WQFT beyond pure General Relativity and establishes it as a useful framework for extracting scattering observables in modified theories of gravity, especially when new propagating modes are present.

The analysis was then extended to spinning compact objects in dynamical Chern-Simons gravity. This part of the thesis addressed a much richer problem, since spin introduces new tensorial structures, new worldline degrees of freedom, and new subtleties in the construction of observables. Using the spinning WQFT formalism, we computed the eikonal phase up to third post-Minkowskian order at linear order in spin and investigated the corrections induced by the parity-violating Chern-Simons interaction. One important result is that the dCS interaction does not contribute in the non-spinning limit, and it also does not contribute for aligned or anti-aligned spin configurations, thereby making clear that the observable effects of the theory are tightly connected to genuinely spin-dependent geometry. On the technical side, this chapter required the evaluation of non-trivial two-loop master integrals using IBP reduction and differential equation methods, illustrating that the study of classical spinning dynamics in modified gravity is now inseparable from modern multi-loop technology. At the same time, the appearance of infrared subtleties and prescription dependence in the eikonal phase makes clear that the extraction of physical observables at higher PM order remains a subtle problem, deserving further conceptual attention.

A complementary amplitude-based approach was developed in the study of Einstein-Maxwell-Dilaton theory. In that case, the focus was on classical scattering of charged, non-spinning compact objects and on the extraction of the conservative potential and scattering angle from soft one-loop amplitudes. By combining expansion by regions, IBP reduction, the Lippmann-Schwinger equation, and eikonal exponentiation, we showed explicitly how the infrared structure of the amplitude reorganizes itself into an IR-finite conservative potential after the appropriate Born subtraction. This mirrors a familiar feature of General Relativity, but in a setting enriched by electromagnetic and dilatonic interactions. The resulting scattering angle, written as a closed-form function of the relevant couplings, provides a concrete benchmark for beyond-GR compact-object dynamics. More importantly, this chapter shows that the language of on-shell amplitudes is not merely aesthetically appealing: it gives direct access to physically measurable classical observables, while also clarifying the role of long-range interactions and their cancellations.

The final technical part of the thesis moved to a different but deeply related arena, namely celestial amplitudes in quadratic effective theories of gravity. Here the emphasis was not on binary dynamics directly, but on the asymptotic representation of scattering data and the analytic structure of the corresponding conformal correlators. Motivated by the improved behaviour of eikonal resummation in celestial gravity, we constructed the celestial eikonal amplitude in quadratic gravity and studied its ultraviolet and infrared properties. A particularly interesting outcome is that the infrared behaviour improves relative to Einstein gravity: whereas the GR result exhibits a higher-order pole structure, the EFT-corrected amplitude has a softer leading singularity. We also derived the corresponding dispersion relation and analyzed the shadow-transformed correlator, which allowed us to extract operator product expansion data using both conformal block methods and the Euclidean OPE inversion formula. This chapter therefore extends the scope of the thesis beyond classical observables in spacetime to the celestial encoding of scattering processes, and suggests that higher-derivative corrections may improve not only bulk dynamics but also the analytic behaviour of their asymptotic holographic avatars.

Taken together, the results of this thesis support a common conclusion: quantum field theoretic methods now provide a powerful and flexible framework for studying classical gravity in situations that go far beyond pure General Relativity. Whether one works in the inspiral regime using worldline EFT, in the relativistic scattering regime using WQFT or on-shell amplitudes, or in the asymptotic celestial representation, the same structural themes repeatedly appear. These include the efficient organization of perturbation theory, the importance of infrared consistency, the role of eikonal exponentiation, and the utility of modern integral technology. At a deeper level, the thesis shows that extra degrees of freedom and higher-curvature corrections leave identifiable signatures in classical observables, but these signatures are often subtle: they can depend on orbital geometry, spin orientation, impact-parameter kinematics, or the analytic continuation to celestial space. This means that beyond-GR physics is not captured by a single correction term, but by an interconnected set of effects whose interpretation requires both conceptual clarity and computational control.

There are several natural directions in which this work can be extended. On the gravitational-wave side, an immediate next step is the computation of waveform phases and radiation-reaction effects in the dark-sector models studied here, followed by the inclusion of spin and finite-size effects. In the scattering problem, extending the present analyses to higher PM order, incorporating real radiation, and clarifying the infrared prescriptions by comouting the observables in spinning WQFT will be essential for building a more complete analysis. In Einstein-Maxwell-Dilaton theory, the study of spin, tidal effects, and higher-curvature corrections would help connect the present conservative results to realistic compact-object phenomenology. On the celestial side, it would be important to compute the exact inversion formula without approximation, understand the role of soft theorems and memory in quadratic gravity, and clarify the construction of stress tensors and other distinguished operators in the presence of extra massive modes. Also, extend our previous analysis to the five point radiative amplitude using soft expansion would be an interesting avenue to study. These problems are technically demanding, but they are also precisely the places where the interaction between amplitudes, effective theory, and gravity is likely to be most fruitful.

In conclusion, this thesis has sought to build a coherent bridge between classical black-hole dynamics, effective theories of gravity, and modern quantum field theoretic methods in the study of theories beyond General Relativity. A central lesson that emerges is that amplitudes, worldline formalisms, and effective field theory should not be viewed merely as alternative computational tools; rather, taken together, they provide a unified framework for classical gravity, one that becomes especially powerful in theories extending beyond Einstein gravity. At the same time, venturing beyond General Relativity also sharpens our understanding of the scope, utility, and limitations of these different techniques, while bringing to light new conceptual and technical challenges. As gravitational-wave observations continue to improve in precision, and theoretical interest in modified gravity, ultraviolet completions, and celestial holography keeps expanding, the importance of such a unified perspective will only increase. It is hoped that the results presented in this thesis, as well as the questions they leave open, will contribute in a meaningful way to this broader research program.
\addcontentsline{toc}{chapter}{References}
\providecommand{\href}[2]{#2}\begingroup\raggedright\endgroup

\end{document}